\documentclass[10pt, twoside]{book}

\usepackage[a4paper,inner=3cm,outer=3cm,top=4cm,bottom=3.5cm]{geometry}

\usepackage{amsmath,amssymb,mathtools}
\usepackage{accents}

\usepackage{rotating}

\usepackage[utf8x]{inputenc}

\usepackage[english]{babel}

\usepackage{dsfont}

\usepackage{chapterbib}

\usepackage{appendix}

\usepackage{booktabs}
\usepackage{tabularx}

\usepackage[small,bf]{caption}
\usepackage{subcaption}

\usepackage{graphicx,epsfig}
\usepackage{pstricks-add,pst-pdf}
\usepackage[abs]{overpic}

\usepackage{float}

\usepackage{url}
\usepackage{pdfpages}
\usepackage[colorlinks=true,
	linkcolor=black,
	citecolor=black,
	urlcolor=black,
	filecolor=black,
	hypertexnames=false
]{hyperref}

\usepackage{fancyhdr}

\usepackage{slashed}

\usepackage{setspace}

\usepackage{cancel}

\usepackage{pifont}

\usepackage{listings}

\usepackage{lipsum}

\usepackage{dcolumn}

\usepackage[numbers,compress]{natbib}

\usepackage{emptypage}

\usepackage{multirow}

\usepackage{etex}
\usepackage{bchart}

\newlength{\dhatheight}

\newcommand{\minidiag}[2]{\begin{minipage}{2cm} \includegraphics[height=2cm]{#1/images/#2} \end{minipage}}
\newcommand{\minidiagSize}[3]{\begin{minipage}{#3} \includegraphics[height=#3]{#1/images/#2} \end{minipage}}

\newcommand{\ChPT}{$\chi$PT}

\renewcommand{\O}{\mathcal{O}}
\newcommand{\p}{\partial}
\newcommand{\pvint}{\hspace{0.75mm}\mathcal{P}\hspace{-3.5mm}\int}

\renewcommand{\Re}{\mathrm{Re}}
\renewcommand{\Im}{\mathrm{Im}}

\newcommand{\<}{\langle}
\renewcommand{\>}{\rangle}

\newcommand{\atan}{\mathrm{atan}}
\newcommand{\dprime}{{\prime\prime}}
\newcommand{\tprime}{{\prime\prime\prime}}

\newcommand{\tr}{\mathrm{Tr}}

\newcommand{\mpio}{M_{\pi^0}}
\newcommand{\mpip}{M_{\pi^+}}
\newcommand{\mko}{M_{K^0}}
\newcommand{\mkp}{M_{K^+}}
\newcommand{\meta}{M_\eta}
\newcommand{\ml}{m_\ell}
\newcommand{\mg}{m_\gamma}
\newcommand{\zl}{z_\ell}
\newcommand{\zg}{z_\gamma}
\newcommand{\sg}{s_\gamma}
\newcommand{\tl}{t_\ell}

\newcommand{\dilog}{\mathrm{Li}_2}

\newcommand{\m}{\hphantom{-}}

\newcommand{\cmark}{\ding{51}}
\newcommand{\xmark}{\ding{55}}

\newcommand{\hateq}{\mathrel{\widehat{=}}}

\newcolumntype{.}{D{.}{.}{2} } 
\newcolumntype{d}{D{.}{.}{2.2} }

\makeatletter
\renewcommand{\maketag@@@}[1]{\hbox{\m@th\normalsize\normalfont#1}}%
\makeatother

\makeatletter
\renewcommand\paragraph{\@startsection{paragraph}{4}{\z@}%
  {-3.25ex \@plus -1ex \@minus -0.2ex}%
  {0.01pt}%
  {\bfseries}%
}
\makeatother

\begin{document}

\raggedbottom

\pagestyle{empty}


\newcommand{\disstitle}{Dispersive Treatments of $K_{\ell4}$ Decays and \newline Hadronic Light-by-Light Scattering}
\newcommand{\thesistitle}{Dispersive Treatments of \\ $K_{\ell4}$ Decays \\ and \\ Hadronic Light-by-Light Scattering}
\title{\disstitle}

\begin{titlepage}

	\begin{center}
		{\scalebox{1.39}{
		\begin{minipage}[t][15cm]{0.72\linewidth}
			\centering
			\onehalfspacing
			
			{ \Large \bf
			\thesistitle \\[2cm]
			}
			
			Inauguraldissertation \\
			der Philosophisch-naturwissenschaftlichen Fakultät \\
			der Universität Bern \\[2cm]

			vorgelegt von \\
			{\bf Peter Stoffer} \\
			von Mägenwil AG \\[1.5cm]
			
			2014 \\[1.5cm]

			Leiter der Arbeit: \\
			Prof.~Dr.~Gilberto Colangelo \\
			Albert Einstein Center for Fundamental Physics \\
			Institut für theoretische Physik, Universität Bern
			
			\vfill{}
			
		\end{minipage}}}
	\end{center}

\cleardoublepage

	\begin{center}
		{\scalebox{1.39}{
		\begin{minipage}[t][15cm]{0.72\linewidth}
			\centering
			\onehalfspacing
			
			{ \Large \bf
			\thesistitle \\[1.25cm]
			}
			
			Inauguraldissertation \\
			der Philosophisch-naturwissenschaftlichen Fakultät \\
			der Universität Bern \\[1.25cm]

			vorgelegt von \\
			{\bf Peter Stoffer} \\
			von Mägenwil AG \\[1.25cm]

			Leiter der Arbeit: \\
			Prof.~Dr.~Gilberto Colangelo \\
			Albert Einstein Center for Fundamental Physics \\
			Institut für theoretische Physik, Universität Bern \\[1.25cm]

			{\setlength{\tabcolsep}{-1pt}
			\begin{tabularx}{\linewidth}{p{7cm} X}
				\multicolumn{2}{l}{Von der Philosophisch-naturwissenschaftlichen Fakultät angenommen.} \\
				\\
				Bern, 21.10.2014 & Der Dekan: \\
				\\
				\\
				& Prof.~Dr.~G.~Colangelo
			\end{tabularx}}

			\vfill{}

		\end{minipage}}}
	\end{center}
\end{titlepage}

\cleardoublepage

\mbox{}

\vspace{4cm}

\begin{center}
	{\it in loving memory of my father}
\end{center}

\cleardoublepage

\mbox{}

\vspace{4cm}

	\hspace{7.5cm} \begin{minipage}{9cm}
		{\it On ne voit bien qu'avec le cœur. \\
		L'essentiel est invisible pour les yeux. } \\
		
		{\footnotesize Antoine de Saint-Exupéry, Le Petit Prince }
	\end{minipage}

\cleardoublepage


\mbox{}

\vspace{2cm}

\begin{center}
\Large \bf Abstract
\end{center}

Dispersion relations are a mathematical tool to describe processes in particle physics in terms of the analytic structure of the amplitudes, in particular in terms of the residues at poles and discontinuities along cuts. They are based on the fundamental principles of unitarity and analyticity. Due to the non-perturbative character of dispersion relations, an application of particular interest is the description of processes due to the strong interaction at low and intermediate energies, where perturbative quantum chromodynamics cannot be used.

In this thesis, I present dispersive treatments of two hadronic processes: the semileptonic kaon decay $K_{\ell4}$ and hadronic light-by-light scattering.

The $K_{\ell4}$ decay is one of the best sources of information on some of the parameters of chiral perturbation theory, the low-energy effective theory of quantum chromodynamics. The dispersion relation for $K_{\ell4}$ provides a resummation of $\pi\pi$- and $K\pi$-rescattering effects. In contrast to a pure chiral treatment, it reproduces the observed curvature of one of the form factors. The matching of the dispersion relation to the chiral representation of the form factors allows the extraction of the values of three low-energy constants.

Hadronic light-by-light scattering appears as a virtual process in the calculation of the anomalous magnetic moment of the muon $(g-2)_\mu$. For more than a decade, a discrepancy of about $3\sigma$ has persisted between the experimental determination and the standard-model prediction of the $(g-2)_\mu$. Contributing at $\O(\alpha_\mathrm{QED}^3)$, light-by-light scattering is a sub-leading hadronic effect: hadronic vacuum polarisation already contributes at second order in the electromagnetic coupling. However, it is expected that within a few years hadronic light-by-light scattering will dominate the uncertainty of the theory prediction of the $(g-2)_\mu$.

So far, only model calculations of the hadronic light-by-light contribution are available. However, in view of forthcoming $(g-2)_\mu$ experiments at Fermilab and J-PARC it is crucial that the hadronic light-by-light calculation can be improved systematically. The dispersive description presented here provides a formalism for a data-driven determination of hadronic light-by-light scattering and hence opens up an avenue towards a model-independent evaluation of the $(g-2)_\mu$.

\cleardoublepage


\mbox{}

\vspace{1cm}

\begin{center}
\Large \bf Acknowledgements
\end{center}

First of all, I thank my thesis advisor, Gilberto Colangelo, for his continuous support and advice during the past years. It was an enriching experience to benefit from his knowledge and intuition. Even when we were all stabbing in the dark, his constant optimism never failed.

I thank Bachir Moussallam for being the co-referee of this thesis and the external expert of the defence committee.

It has been a pleasure to collaborate in diverse projects with outstanding physicists, including Vincenzo Cirigliano, Martin Hoferichter, Bastian Kubis, Andreas Nyffeler, Lorenzo Mercolli, Emilie Passemar, Massimo Passera, Massimiliano Procura and Christopher Smith.

I am indebted to Emilie Passemar for all her support during my stay at the Los Alamos National Laboratory. Special thanks go to Vincenzo Cirigliano for his advice and support at LANL and for the chance to start working with him on new interesting projects.

I enjoyed many informative and helpful discussions with Jürg Gasser, Bastian Kubis and Heiri Leutwyler, whom I also cordially thank for their support in diverse application procedures. I further thank Brigitte Bloch-Devaux, Stefan Pislak, Andries van der Schaaf and Peter Truöl for providing me with unpublished data, partially obtained from new analyses of raw data, and for many explanations on the experiments. I thank Hans Bijnens for providing his program code and support concerning his two-loop calculations.

I am very grateful to my office mates and all my friends at the institute for having made the years of my studies a pleasant time around and beyond physics. In particular, I thank \mbox{Jason Aebischer}, Daniel Arnold, David Baumgartner, Michael Bögli, Ahmet Kokulu, Stefan Lanz, \mbox{Vidushi Maillart}, Stefanie Marti, Lorenzo Mercolli, Christof Schüpbach and Kyle Steinhauer. I thank Esther Fiechter, our secretary, for the administrative work in the background.

I thank Daniel Blaschke, Adolfo del Campo, Denise Neudecker, Brian and Kay Newnam, the whole family Koh and Joel Lynn for having made my stay in Los Alamos an enriching and unforgettable experience. For making the daily life back in Switzerland equally enriching, I thank Timon Amstutz, Christof Ebneter, Simone Hirsiger, Martin Straub, Daniela Wyss and Moritz Zaugg.

I thank my mother and my sister, Ursula and Sabine Stoffer, for their love, for all the time of happiness and for their support during difficult periods.

With all my heart, I thank Serena Grädel for her love, her patience and support and for her closeness even in times of geographical distance.

This thesis is dedicated to my father, Walter Stoffer, in memory of his love, wisdom and calmness. His greatest gifts survive oblivion and parting.

\cleardoublepage

\setcounter{page}{1}
	
\pagestyle{fancy}

\setcounter{secnumdepth}{4}
\setcounter{tocdepth}{2}
\tableofcontents

\cleardoublepage


\chapter*{Introduction}
\addcontentsline{toc}{part}{Introduction}
\addtocontents{toc}{\vskip-6pt\par\noindent\protect\hrulefill\par}
\markboth{INTRODUCTION}{INTRODUCTION}

\section*{The Standard Model of Particle Physics}

Research in particle physics is a quest for the unknown. The desire to understand the laws of nature in its innermost part motivates us not only to describe current observations but also to search for phenomena that lie beyond the predictions of established theories. This has pushed physics to smaller and smaller dimensions, corresponding to ever increasing energies that require larger and larger experiments.

The standard model of particle physics was established during the twentieth century as a relativistic quantum field theory, a framework based on two pillars of modern physics: quantum mechanics and special relativity. It incorporates the electromagnetic, the weak and the strong force, describes the building blocks of matter and has had a tremendous success in describing particle physics processes over an energy range of many orders of magnitude. So far, only very few observables have shown some tension with the predictions of the standard model.

The standard model is a gauge theory of the symmetry group
\begin{align}
	\begin{split}
		G_\mathrm{SM} = SU(3)_c \times SU(2)_L \times U(1)_Y .
	\end{split}
\end{align}
The matter content of the standard model consists of fermions, classified into quarks (fermions that interact strongly, i.e.~are charged under $SU(3)_c$) and leptons (fermions that are singlets under $SU(3)_c$). Three generations or families of these fermions exist (see table~\ref{tab:StandardModelFermions}), the second and third being an exact replica of the first concerning quantum numbers, but with increasing masses. The family mass hierarchy is indeed overwhelming: while $u$- and $d$-quarks have a mass of a few MeV, the top-mass is about 173 GeV \cite{PDG2012}.

\begin{table}[H]
	\centering
	\begin{tabular}{l l l l}
		\toprule
		 & 1st family & 2nd family & 3rd family \\
		\midrule
		\multirow{2}{*}{quarks} & $u$ (up) & $c$ (charm) & $t$ (top) \\
		 & $d$ (down) & $s$ (strange) & $b$ (bottom) \\
		\midrule
		\multirow{2}{*}{leptons} & $e$ (electron) & $\mu$ (muon) & $\tau$ (tau) \\
		& $\nu_e$ ($e$-neutrino) & $\nu_\mu$ ($\mu$-neutrino) & $\nu_\tau$ ($\tau$-neutrino) \\
		\bottomrule
	\end{tabular}
	\caption{Elementary fermions (matter content) of the standard model.}
	\label{tab:StandardModelFermions}
\end{table}

The interactions of the fermions are described by the gauge forces (strong, weak and electromagnetic force) and the Yukawa interaction with the Higgs field. The mediator particles of the gauge forces are bosons of spin 1 (see table~\ref{tab:StandardModelBosons}). In contrast, the Higgs particle is the only elementary scalar particle, i.e.~a boson of spin 0. The bosons interact not only with the fermions but also amongst themselves.

\begin{table}[ht]
	\centering
	\begin{tabular}{l l}
		\toprule
		\multirow{3}{*}{\parbox{3cm}{vector bosons \\ (gauge fields)}} & $g$ (gluon) \\
			& $W^\pm$, $Z$ bosons \\
			& $\gamma$ (photon) \\
		\midrule
		scalar boson & $H$ (Higgs particle) \\
		\bottomrule
	\end{tabular}
	\caption{Elementary bosons of the standard model.}
	\label{tab:StandardModelBosons}
\end{table}

In the standard model, explicit mass terms of the fermions and gauge bosons are forbidden by gauge symmetry. The masses of quarks, charged leptons, $W^\pm$, $Z$ and the physical Higgs boson are all generated through spontaneous breaking of the electroweak symmetry
\begin{align}
	\begin{split}
		SU(2)_L \times U(1)_Y \rightarrow U(1)_\mathrm{em}
	\end{split}
\end{align}
by a non-zero vacuum expectation value of the Higgs field, $v \approx 246$~GeV. With the discovery of the Higgs boson by the ATLAS and CMS experiments at the Large Hadron Collider at CERN in 2012, the last missing ingredient of the standard model was experimentally confirmed~\cite{Aad2012,Chatrchyan2012}.

\section*{The Strong Interaction}

While the gauge couplings of the electroweak sector are small enough to allow a perturbative expansion, the strong coupling constant is large at low energies. Hence, perturbative calculations in quantum chromodynamics (QCD), the part of the standard model describing the strong interaction, are only possible for high-energy processes. Quarks and gluons, the elementary particles with an $SU(3)_c$ charge (called `colour'), are not asymptotic states of the theory: QCD exhibits a feature known as confinement. At low energies, quarks and gluons are always bound into hadrons: mesons, which have the quantum numbers of a quark-antiquark pair, and baryons, corresponding to states of three quarks. Hadrons are always `colourless', i.e.~they have no $SU(3)_c$ charge.

Due to confinement, quarks and gluons are never observed as free particles, hence the spectrum of QCD starts with the lightest mesons called pions. Pions can be understood as the Goldstone bosons of the spontaneously broken chiral symmetry of QCD,  $SU(2)_L \times SU(2)_R \rightarrow SU(2)_V$. This symmetry is only approximate: it is explicitly broken by the mass terms of the $u$- and $d$-quarks, which leads to a small but non-zero mass of the pions. The next-heavier hadrons are the kaons, mesons carrying strangeness, and the $\eta$ meson. Together with the pions, they form an octet of Goldstone bosons of the chiral $SU(3)$ symmetry of $u$-, $d$- and $s$-quarks, see table~\ref{tab:Mesons}.

\begin{table}[ht]
	\centering
	\begin{tabular}{l l l l l c c c}
		\toprule
			& & & name &  & quark content & isospin & mass \cite{PDG2012} \\
		\cmidrule{4-8}
			& \multirow{3}{*}{ \hspace{-0.5cm} $\left\{ \begin{matrix} \vphantom{|} \\  \vphantom{|} \\  \vphantom{|} \end{matrix}\right. $ } & \multirow{9}{*}{ \hspace{-0.75cm} $\left\{ \begin{matrix}  \vphantom{|} \\ \vphantom{|} \\ \vphantom{|} \\ \vphantom{|} \\ \vphantom{|} \\ \vphantom{|} \\ \vphantom{|} \\  \vphantom{|} \\  \vphantom{|} \end{matrix}\right. $ } & \multirow{3}{*}{pions} & $\pi^+$ & $u \bar d$ & $+1$ & 139.57 MeV \\
		$SU(2)$ triplet	& & & & $\pi^0$ & $(u\bar u - d \bar d)/\sqrt{2}$ & $0$ & 134.98 MeV \\
			& & & & $\pi^-$ & $\bar u d$ & $-1$ & 139.57 MeV \\
		\cmidrule{4-8}
		\multirow{2}{*}{$SU(3)$ octet} & & & \multirow{4}{*}{kaons} & $K^+$ & $u \bar s$ & $+1/2$ & 493.67 MeV \\
			& & & & $K^-$ & $\bar u s$ & $-1/2$ & 493.67 MeV   \\
			& & & & $K^0$ & $d \bar s$ & $+1/2$ & 497.65 MeV   \\
			& & & & $\bar K^0$ & $\bar d s$ & $-1/2$ & 497.65 MeV  \\
		\cmidrule{4-8}
			& & & eta & $\eta$ & $\approx (u \bar u + d \bar d - 2 s \bar s)/\sqrt{6}$ & $0$ & 547.85 MeV \\
		\cmidrule{4-8}
		$SU(3)$ singlet & & & eta prime	& $\eta^\prime$ & $\approx (u\bar u + d \bar d + s \bar s)/\sqrt{3}$ & $0$ & 957.66 MeV \\
		\bottomrule
	\end{tabular}
	\caption{The lightest hadrons form the pseudoscalar $SU(3)$ octet. The $SU(3)$ singlet $\eta^\prime$ is much heavier due to the chiral anomaly. The physical $\eta$ and $\eta^\prime$ states are actually mixtures of the pure octet and singlet states.}
	\label{tab:Mesons}
\end{table}

The physics of the pions or the octet mesons is described by chiral perturbation theory (\ChPT{}) \cite{Weinberg1968, GasserLeutwyler1984, GasserLeutwyler1985}, the low-energy effective field theory of QCD built on top of the $SU(2)$ or $SU(3)$ chiral symmetry. Since the $s$-quark is much heavier than $u$- and $d$-quarks, $SU(3)$ is not as good a symmetry as $SU(2)$. The kaons and the $\eta$ are heavier than the pions, and $SU(3)$ \ChPT{} does not converge as fast as the $SU(2)$ theory for pions alone.

The $\eta^\prime$ would be the Goldstone boson of the spontaneous breaking of the $U(1)_A$ symmetry, which, however, is already broken by a quantum anomaly. Hence, the $\eta^\prime$ is not a light particle.

The validity of the effective theory breaks down at energy scales around the mass of the next-heavier particle not included in the theory, which is the $\rho$ meson around 770~MeV \cite{PDG2012}. It is only well above this scale that the strong coupling constant becomes small enough to allow perturbative QCD calculations. This leaves us in a peculiar situation: although we know with QCD the underlying theory of the strong interaction, there is a range of intermediate energies where first-principle theory predictions are extremely difficult. The connection between the fundamental particles and the asymptotic states (the observed particles) is not yet completely understood. Some of these problems can be attacked by employing numerical lattice simulations.

Even at low energies, where \ChPT{} provides an effective description of QCD, we face the complication that the effective field theory is parametrised by low-energy constants (LECs), which are in principle calculable from QCD. Due to the non-perturbative nature of QCD at low energies, these constants have to be determined with the help of experimental data or lattice simulations \cite{Aoki2013}. As the number of LECs increases drastically with the order of the effective expansion, at next-to-next-to-leading chiral order only very few LECs are reliably determined \cite{Bijnens2014}.

Before the success of perturbative QCD (pQCD), it was not clear if quantum field theory is the appropriate tool to describe the strong interaction. The growing particle zoo of hadrons suggested that these particles are not elementary. During the sixties, $S$-matrix theory was proposed as the fundamental theory of hadrons. The attempt was to construct the scattering matrix with dispersion relations, using just the basic principles of crossing symmetry, analyticity and unitarity \cite{Eden1966,Chew1966,Barut1967}. While these principles are a natural ingredient of quantum field theories, $S$-matrix theory did not rely on the notion of fields as functions of space-time. It rather tried to reach self-consistency through a bootstrapping principle.

The formal developments of $S$-matrix theory resulted in the first string theories. In the seventies, it was realised that string theories could be understood as a description of gravity rather than the strong interaction. In hadron physics, the $S$-matrix approach was mostly abandoned in favour of pQCD, which successfully described deep inelastic scattering processes.

It seems that the principles and techniques of $S$-matrix theory have been almost forgotten for some time. Their usefulness to describe hadron physics at low and intermediate energies, where pQCD is not applicable, became apparent with the marriage of dispersion relations with \ChPT{} \cite{Donoghue1990}. Nowadays, dispersive methods no longer aim at constructing a fundamental theory of the strong interaction, but rather have become a powerful tool for phenomenological applications by establishing model-independent connections between different processes or energy regions. There are many successful applications of dispersion relations, including $\pi\pi$ scattering \cite{Ananthanarayan2001a,Descotes-Genon2002}, pion and $K\pi$ form factors \cite{Pich2001,Jamin2002,Bernard2006}, $\eta$ decays \cite{Anisovich1996,Colangelo2009a,Lanz2013}, $K\pi$ scattering \cite{Buettiker2004} and photon-hadron processes \cite{Garcia-Martin2010,Hoferichter2011,Hoferichter2012a,Moussallam2013}.

\section*{Search for New Physics}

Despite its tremendous success, we know that the standard model cannot be the fundamental theory of nature. First of all, it does not include gravity. A consistent theory of quantum gravity is not yet known, but it can be expected that quantum gravity effects will set in at energy scales of the Planck mass, i.e.~around $10^{19}$~GeV. However, it is likely that the standard model must be replaced by some new physics already at much smaller energies. There are many indications of physics beyond the standard model (BSM).
\begin{itemize}
	\item The observed oscillation between the neutrino flavours requires non-zero neutrino masses, while in the standard model $\nu_e$, $\nu_\mu$ and $\nu_\tau$ are exactly massless. Neutrino masses are only possible with new degrees of freedom.
	\item The mass of the scalar Higgs boson at the order of the electroweak scale implies a serious problem of fine-tuning, which is regarded as unnatural: one would expect the Higgs mass to lie at some heavy scale of new physics unless there is an extreme cancellation between the quadratic divergences in the Higgs self-energy and the bare mass parameter.
	\item The asymmetry between matter and antimatter in the universe cannot be explained by the standard model.
	\item Astronomical and cosmological observations indicate consistently that only about 15\% of the matter in the universe consists of ordinary matter. The rest is called dark matter and supposed to be made of BSM particles. No particle in the standard model can play the role of dark matter.
\end{itemize}
Several hints to new physics suggest that phenomena beyond the standard model could appear at energy scales of a few TeV. This is the energy regime accessible by the large hadron collider (LHC) at CERN. Indeed, besides investigating the mechanism of electroweak symmetry breaking, one of the main motivations to build the LHC was to discover new physics at the TeV scale. By now, the LHC has discovered the Higgs particle but not yet seen any sign of new physics. Hopes are pinned on the next run of the LHC in 2015, which will almost double the collision energy.

Many models of new physics have been proposed in order to deal with the (possible) shortcomings of the standard model. The majority is based on one of the following two concepts:
\begin{itemize}
	\item \textit{supersymmetry}: the Poincaré symmetry of space-time is extended by a symmetry relating bosonic and fermionic particles. The Higgs hierarchy problem is resolved by a cancellation of the quadratic divergences in the loop corrections between virtual particles and super-partners.
	\item \textit{compositeness}: the Higgs particle is supposed to be a composite instead of an elementary particle. In analogy to the pions in QCD, the Higgs could be a Goldstone boson of a spontaneously broken symmetry of a new strongly coupled sector.
\end{itemize}
Both concepts favour some new phenomena at the (multi-)TeV scale but are not completely ruled out if the LHC does not find new physics.

There is not only the high-energy frontier to search for new physics. Through non-standard virtual effects, new physics will have an impact on observables below the threshold of real production of the BSM particles. Therefore, low-energy precision observables can also constrain or help to identify new physics. This is sometimes called the high-intensity frontier. The investigations on the two frontiers should not be understood as two alternatives but rather as two complementing approaches. They can attack different regions of the parameter space. Very sensitive low-energy probes are e.g.~the $CP$-violating electric dipole moments, which partly already now impose more stringent bounds on new physics than the LHC.

If new physics will be found at the LHC, the low-energy probes will help to constrain or identify its nature. If the LHC does not find any direct signs of BSM physics, the low-energy probes will become even more important as a complement to the high-energy program, which will focus on Higgs phenomenology. An even larger (linear or circular) collider experiment will take a long time to be realised \cite{Behnke2013,Aicheler2012,Bicer2014}.

Instead of blindly building models of new physics and ruling them out one after the other, we can use a more powerful model-independent approach to constrain BSM physics. Let us assume that physics beyond the standard model appears only at a mass scale $\Lambda$ that is much heavier than the electroweak scale $M_\mathrm{EW}$. This assumption is supported by the success of the standard model at energies investigated in present experiments.

The mass gap between $\Lambda$ and $M_\mathrm{EW}$ allows us then to describe new physics in terms of an effective field theory below $\Lambda$, where the Lagrangian is constructed out of standard model fields. The effective theory could be constructed as usual by integrating out the heavy degrees of freedom in the full theory. But even without considering explicit models of new physics, one can study the effective theory in a generic, model-independent way. The non-renormalisable operators in the effective Lagrangian should respect the gauge symmetries of the SM. The order principle of the effective theory is the mass dimension of the new operators. Therefore, the Lagrangian of the effective theory has the form
\begin{align}
	\begin{split}
		\mathcal{L}_\mathrm{eff} &= \mathcal{L}_\mathrm{SM} + \frac{1}{\Lambda} \sum_i C^{(5)}_i \mathcal{Q}^{(5)}_i + \frac{1}{\Lambda^2} \sum_i C^{(6)}_i \mathcal{Q}^{(6)}_i + \O(\Lambda^{-3}) ,
	\end{split}
\end{align}
where the $C^{(n)}_i$ denote the Wilson coefficients of the effective operators $\mathcal{Q}^{(n)}_i$ at the high scale. At mass-dimension five, only the Weinberg operator is present \cite{Weinberg1979}, which can give rise to neutrino masses. The full list of gauge-invariant dimension-six operators built out of standard model fields has been constructed in \cite{Buchmueller1986}. Redundancies in this set of operators have been removed in \cite{Grzadkowski2010}. With renormalisation-group techniques, the running down to low energies and the mixing of operators can be calculated just with standard model physics. The remaining task before the influence of new physics on low-energy observables can be investigated in a model-independent way is then usually the calculation of some hadronic matrix elements, often in a non-perturbative way.

This highlights the importance of a detailed understanding of hadron physics at low energies. While investigations in this field are important to understand the very nature of QCD with its rich spectrum of phenomena on its own, they are crucial as well for the quest for new physics, because the power of low-energy observables is often overshadowed by hadronic uncertainties.

One of the most prominent precision observables is the anomalous magnetic moment of the muon. There, a tension between the standard model and experiment has persisted for more than a decade. The theoretical uncertainty is nowadays completely dominated by hadronic effects.

\section*{Outline}

The present thesis consists of three independent and self-contained parts. The first two parts concern the low-energy structure of strong interaction and aim on the one hand at a better understanding of a decay process in this non-perturbative regime of QCD, on the other hand at a more reliable determination of some low-energy constants of the effective theory. The third part concerns a process where it is important to gain control over hadron physics in order to decide if observed discrepancies with the standard model are a hint to new physics.
\begin{itemize}
	\item In the first part, I present a new dispersive treatment of $K_{\ell4}$ decays. $K_{\ell4}$ denotes the decay of a kaon into two pions and a pair of leptons. This decay is sensitive to some next-to-leading order low-energy constants of \ChPT{} and the best experimental source for their determination. This part is based on a project in collaboration with Gilberto Colangelo and Emilie Passemar and represents the continuation of my master's thesis \cite{Stoffer2010}, including diverse substantial improvements and extensions with respect to this former treatment.
	\item The second part should be understood as a side-project to the first: within \ChPT{} with photons and leptons, I have computed isospin-breaking effects in the $K_{\ell4}$ decay due to the electromagnetism and the quark mass difference $m_u - m_d$. The result of this calculation is used in the first part to correct the fitted experimental data by isospin effects that were not taken into account in the experimental analysis. This part has been published as a paper \cite{Stoffer2014}.
	\item In the third part, I present a dispersive treatment of hadronic light-by-light scattering. This process contributes to the anomalous magnetic moment of the muon and is expected to become the main source of theoretical uncertainty within the next few years. This treatment shows a new path towards a data-driven evaluation of the hadronic light-by-light contribution and therefore a method to reduce the current model dependence. This third part of my thesis is based on a project in collaboration with Gilberto Colangelo, Martin Hoferichter and Massimiliano Procura and signifies a major improvement of our previous treatment \cite{Colangelo2014a}.
\end{itemize}

\clearpage

{
\section*{References}
\renewcommand{\bibfont}{\raggedright}
\renewcommand{\bibsection}{}
\renewcommand\bibname{References}
\bibliographystyle{my-physrev}
\bibliography{Isospin/Literature}
}

\cleardoublepage

%
%


\part[New Dispersive Treatment of $K_{\ell4}$~Decays]{New Dispersive Treatment of $K_{\ell4}$~Decays
                             \\ 
                  \begin{center}
                     \begin{minipage}[l]{11cm}
                     \vspace{2cm}
                     \normalsize
                     \begin{center} 
                     	\textnormal{based on a project in collaboration with \\[0.15cm] Gilberto Colangelo and Emilie Passemar}
                     \end{center}
                     \end{minipage}
                  \end{center}
                 }
\addtocontents{toc}{\vskip-6pt\par\noindent\protect\hrulefill\par}



\chapter{Introduction}

$K_{\ell4}$ denotes the semileptonic decay of a kaon into two pions and a pair of leptons. For several reasons, it is of particular interest for low-energy hadron physics, which is described by chiral perturbation theory (\ChPT{}) \cite{Weinberg1968, GasserLeutwyler1984, GasserLeutwyler1985}. The physical region starts at lower energies than for example $K\pi$ scattering, which gives access to the same low-energy constants as $K_{\ell4}$. Therefore, the chiral expansion, which is an expansion in masses and momenta, should give a better description of $K_{\ell4}$ than $K\pi$ scattering.

Due to its two-pion final state, $K_{\ell4}$ is also one of the cleanest sources of information on $\pi\pi$ interaction \cite{Shabalin1963,Cabibbo1965,Batley2010}.

The latest high-statistics $K_{\ell4}$ experiments E865 at BNL \cite{Pislak2001, Pislak2003} and NA48/2 at CERN \cite{Batley2010, Batley2012} have achieved an impressive accuracy. The statistical errors of the $S$-wave of one form factor reach in both experiments the sub-percent level. This requires a theoretical treatment beyond one-loop order in the chiral expansion. A first treatment beyond one loop, based on dispersion relations, was already done in 1994 \cite{Bijnens1994}. The full two-loop calculation became available in 2000 \cite{Amoros2000}. However, it turns out that even at two loops, \ChPT{} is not able to predict the curvature of one of the form factors.

Here, we present a new dispersive treatment of $K_{\ell4}$ decays, which is valid up to and including $\O(p^6)$ in the chiral expansion. It resummates two-particle final-state rescattering effects, which we expect to be the most important contribution beyond two-loop. Indeed, we observe that the dispersive description is able to reproduce the curvature of the form factor.

The dispersion relation is parametrised by subtraction constants, which are not constrained by unitarity. Therefore, they have to be determined by a fit to data. It turns out that the available data does not constrain all the subtraction constants to a sufficient precision. Therefore, we use the soft-pion theorem, a low-energy theorem for $K_{\ell4}$ that receives only $SU(2)$ chiral corrections, as well as some chiral input to constrain the parameters that are not well determined from data alone.

The presented treatment of $K_{\ell4}$ decays signifies an extension and major improvement of our previous dispersive framework \cite{Stoffer2010, Colangelo2012, Stoffer2013}. The modifications and improvements concern the following aspects.
\begin{itemize}
	\item Instead of a single linear combination of form factors, now we describe the two form factors $F$ and $G$ simultaneously. This allows us to include more experimental data in the fits.
	\item The new framework is valid also for non-vanishing invariant energies of the lepton pair. In the previous treatment, we neglected the dependence on this kinematic variable. This approximation is no longer used and the observed dependence on the lepton invariant energy can be taken into account.
	\item We apply corrections for isospin-breaking effects in the fitted data that have not been taken into account in the experimental analysis.
	\item We perform the matching to \ChPT{} directly on the level of the subtraction constants, which avoids the mixing with the treatment of final-state effects.
	\item Besides a matching to one-loop \ChPT{}, we also study the matching at two-loop level.
\end{itemize}
The first two points required a substantial modification and extension of the dispersive framework from the very start, but rendered it much more powerful. The old treatment can be understood as a limiting case of the new framework.

The outline is as follows: in chapter~\ref{sec:Kl4DispersionRelation}, we derive the dispersion relation for the $K_{\ell4}$ form factors, which has the form of a set of coupled integral equations. In chapter~\ref{sec:Kl4NumericalSolutionDR}, we describe the numerical procedure that is used to solve this system. Chapter~\ref{sec:Kl4SubtractionConstantsDetermination} is devoted to the determination of the free parameters of the dispersion relation and the derivation of matching equations to \ChPT{}. In chapter~\ref{sec:Kl4Results}, we present the results of the fit to data and matching to \ChPT{}. The appendices contain several details on the kinematics, the derivation of the dispersion relation and explicit expressions for the matching equations.


\chapter{Dispersion Relation for $K_{\ell4}$}

\label{sec:Kl4DispersionRelation}

\section{Decay Amplitude and Form Factors}

$K_{\ell4}$ are semileptonic decays of a kaon into two pions and a lepton--neutrino pair:
\begin{align}
	K^+(k) \rightarrow \pi^+(p_1) \pi^-(p_2) \ell^+(p_\ell) \nu_\ell(p_\nu) ,
\end{align}
where $\ell\in\{e,\mu\}$ is either an electron or a muon. There exist other decay modes involving neutral mesons. Their amplitudes are related to the above decay by isospin symmetry -- in our dispersive treatment of $K_{\ell4}$, we will work in the isospin limit. We only consider the above charged mode, or its charge conjugate mode, because in this case the experimental situation is the best.

In the standard model, semileptonic decays are mediated by $W$ bosons. After integrating out the $W$ boson from the standard model Lagrangian, we end up with a Fermi type effective current-current interaction. The matrix element of $K_{\ell4}$ then splits up into a leptonic times a hadronic part. The leptonic matrix element can be treated in a standard way. The hadronic matrix element exhibits the usual $V-A$ structure of weak interaction:
\begin{align}
	{}_\mathrm{out}\<\pi^+(p_1)\pi^-(p_2) \ell^+(p_\ell)\nu_\ell(p_\nu) \big| K^+(k)\>_\mathrm{in} = i (2\pi)^4 \delta^{(4)}(k-p_1-p_2-p_\ell-p_\nu) \mathcal{T} , \\
	\mathcal{T} = \frac{G_F}{\sqrt{2}} V_{us}^* \bar u(p_\nu) \gamma^\mu(1-\gamma_5)v(p_\ell) \< \pi^+(p_1) \pi^-(p_2) \big| V_\mu(0)-A_\mu(0) \big| K^+(k) \> ,
\end{align}
where $V_\mu = \bar s \gamma_\mu u$ and $A_\mu = \bar s \gamma_\mu \gamma_5 u$. Note that although we drop the corresponding labels, the meson states are still in- and out-states with respect to the strong interaction.

The Lorentz structure of the currents allows us to write the two hadronic matrix elements as
\begin{align}
	\mathcal{V}_\mu^{+-} := \big\< \pi^+(p_1) \pi^-(p_2) \big| V_\mu(0) \big| K^+(k)\big\> &= -\frac{H}{M_K^3} \epsilon_{\mu\nu\rho\sigma} L^\nu P^\rho Q^\sigma , \\
	\mathcal{A}_\mu^{+-} := \big\< \pi^+(p_1) \pi^-(p_2) \big| A_\mu(0) \big| K^+(k) \big\> &= -i \frac{1}{M_K} \left( P_\mu F + Q_\mu G + L_\mu R \right) ,
\end{align}
where $P = p_1 + p_2$, $Q = p_1 - p_2$, $L = k - p_1 - p_2$. The form factors $F$, $G$, $R$ and $H$ are dimensionless scalar functions of the Mandelstam variables:
\begin{align}
	\begin{split}
		s &= (p_1 + p_2)^2 = (k - L)^2 , \\
		t &= (k - p_1)^2 = (p_2 + L)^2 , \\
		u &= (k - p_2)^2 = (p_1 + L)^2 .
	\end{split}
\end{align}

We further define the invariant squared energy of the lepton pair $s_\ell = L^2$. For the hadronic matrix element, we regard $s_\ell$ as a fixed external quantity.

\section{Analytic Structure}

\label{sec:AnalyticStructure}

\begin{figure}[H]
	\centering
	\psset{unit=0.248cm}
	\begin{pspicture*}(-30,-30)(30,30)
		\pspolygon[linestyle=none, fillstyle=solid, fillcolor=lightgray](-17.5397,4)(17.5397,4)(0,-26.3797)
		\pspolygon[linewidth=2pt, fillstyle=solid, fillcolor=gray](-1.79221,5.07298)(-1.87635,5.41871)(-1.89766,5.65562)(-1.89763,5.85557)(-1.88566,6.03484)(-1.86574,6.20033)(-1.84,6.35575)(-1.80973,6.50332)(-1.77578,6.64452)(-1.73874,6.78037)(-1.69903,6.91159)(-1.65696,7.03872)(-1.61277,7.16218)(-1.56663,7.28226)(-1.51868,7.39921)(-1.46901,7.51319)(-1.4177,7.62432)(-1.3648,7.73268)(-1.31032,7.83832)(-1.25428,7.94125)(-1.19665,8.04144)(-1.13741,8.13884)(-1.07651,8.23335)(-1.01386,8.32485)(-0.949383,8.41316)(-0.882938,8.49808)(-0.816672,8.5767)(-0.556377,8.82755)(-0.370319,8.94981)(-0.21788,9.01384)(-0.0858075,9.04259)(0.0322829,9.04713)(0.140056,9.0338)(0.239845,9.00664)(0.333238,8.9684)(0.42137,8.92105)(0.505081,8.86604)(0.585009,8.80448)(0.661648,8.73723)(0.735389,8.66495)(0.806542,8.58819)(0.875355,8.50738)(0.942029,8.42286)(1.00672,8.33492)(1.06957,8.24377)(1.13067,8.14959)(1.19009,8.05251)(1.24789,7.95263)(1.30411,7.85001)(1.35876,7.74467)(1.41185,7.63661)(1.46334,7.52579)(1.51319,7.41214)(1.56134,7.29553)(1.57232,7.26788)(1.73916,6.7789)(1.82563,6.42913)(1.87303,6.14704)(1.89519,5.90865)(1.89901,5.70204)(1.88858,5.52011)(1.86661,5.35815)(1.83504,5.21284)(1.79528,5.0817)(1.74843,4.96284)(1.69537,4.85476)(1.63678,4.75624)(1.57324,4.66629)(1.50522,4.5841)(1.43312,4.50899)(1.35726,4.44038)(1.27793,4.37778)(1.19536,4.3208)(1.10974,4.26909)(1.02125,4.22236)(0.930015,4.18038)(0.836147,4.14297)(0.739728,4.10997)(0.640816,4.08129)(0.539446,4.05687)(0.435625,4.03669)(0.329333,4.0208)(0.220519,4.00927)(0.109095,4.00226)(-0.00507317,4.)(-0.122177,4.00284)(-0.242487,4.01122)(-0.36638,4.02581)(-0.494394,4.04753)(-0.627314,4.07776)(-0.766342,4.11856)(-0.913462,4.17338)(-1.07234,4.24856)(-1.25134,4.3586)(-1.4829,4.55967)(-1.50522,4.5841)(-1.57324,4.66629)(-1.63678,4.75624)(-1.69537,4.85476)(-1.74843,4.96284)(-1.79528,5.0817)
		\psline[linewidth=2pt, fillstyle=solid, fillcolor=gray](-10.3668,32.3344)(-10.2517,32.1337)(-10.1366,31.933)(-10.0216,31.7323)(-9.90648,31.5316)(-9.7914,31.3309)(-9.67632,31.1303)(-9.56124,30.9296)(-9.44615,30.7289)(-9.33105,30.5283)(-9.21595,30.3276)(-9.10084,30.127)(-8.98573,29.9264)(-8.87061,29.7258)(-8.75548,29.5252)(-8.64034,29.3246)(-8.52519,29.1241)(-8.41003,28.9235)(-8.29486,28.723)(-8.17967,28.5225)(-8.06447,28.3221)(-7.94926,28.1216)(-7.83404,27.9212)(-7.71879,27.7208)(-7.60353,27.5204)(-7.48825,27.3201)(-7.37295,27.1198)(-7.25763,26.9196)(-7.14228,26.7193)(-7.02691,26.5192)(-6.91151,26.3191)(-6.79607,26.119)(-6.68061,25.919)(-6.56511,25.719)(-6.44958,25.5191)(-6.334,25.3193)(-6.21838,25.1196)(-6.10272,24.9199)(-5.987,24.7203)(-5.87123,24.5209)(-5.75541,24.3215)(-5.63951,24.1222)(-5.52355,23.9231)(-5.40752,23.724)(-5.29141,23.5252)(-5.17521,23.3264)(-5.05891,23.1279)(-4.94251,22.9295)(-4.82601,22.7313)(-4.70938,22.5333)(-4.59262,22.3355)(-4.47572,22.138)(-4.35866,21.9407)(-4.24143,21.7438)(-4.12402,21.5471)(-4.0064,21.3509)(-3.88856,21.155)(-3.77047,20.9595)(-3.6521,20.7645)(-3.53344,20.57)(-3.41445,20.3761)(-3.29509,20.1829)(-3.17532,19.9903)(-3.0551,19.7986)(-2.93436,19.6077)(-2.81305,19.4178)(-2.6911,19.229)(-2.56843,19.0415)(-2.44493,18.8554)(-2.32049,18.6709)(-2.19498,18.4883)(-2.06824,18.3078)(-1.94008,18.1298)(-1.81025,17.9547)(-1.67848,17.7829)(-1.5444,17.6152)(-1.40759,17.4521)(-1.26749,17.2948)(-1.12336,17.1444)(-0.974271,17.0027)(-0.818964,16.8717)(-0.65571,16.7544)(-0.482072,16.6552)(-0.294486,16.5801)(-0.0874935,16.5386)(0.147764,16.5461)(0.426863,16.6295)(0.780557,16.8421)(1.28196,17.3106)(2.18255,18.4704)(2.21315,18.5145)(2.3385,18.6974)(2.46279,18.8821)(2.58617,19.0684)(2.70873,19.2562)(2.83058,19.4451)(2.9518,19.6351)(3.07246,19.8262)(3.19262,20.018)(3.31233,20.2107)(3.43163,20.4041)(3.55057,20.598)(3.66919,20.7926)(3.78751,20.9877)(3.90556,21.1832)(4.02337,21.3791)(4.14096,21.5755)(4.25835,21.7721)(4.37555,21.9691)(4.49258,22.1664)(4.60946,22.364)(4.7262,22.5618)(4.84281,22.7598)(4.9593,22.9581)(5.07569,23.1565)(5.19197,23.3551)(5.30815,23.5538)(5.42426,23.7527)(5.54028,23.9518)(5.65623,24.1509)(5.77211,24.3502)(5.88793,24.5496)(6.00369,24.7491)(6.1194,24.9487)(6.23506,25.1484)(6.35067,25.3481)(6.46624,25.548)(6.58177,25.7479)(6.69726,25.9478)(6.81272,26.1478)(6.92815,26.3479)(7.04354,26.548)(7.15891,26.7482)(7.27426,26.9484)(7.38958,27.1487)(7.50488,27.349)(7.62015,27.5493)(7.73541,27.7497)(7.85065,27.9501)(7.96588,28.1505)(8.08109,28.351)(8.19628,28.5514)(8.31146,28.7519)(8.42663,28.9525)(8.54179,29.153)(8.65694,29.3536)(8.77208,29.5541)(8.88721,29.7547)(9.00233,29.9553)(9.11744,30.1559)(9.23255,30.3566)(9.34765,30.5572)(9.46274,30.7579)(9.57783,30.9585)(9.69292,31.1592)(9.808,31.3599)(9.92307,31.5606)(10.0381,31.7612)(10.1532,31.9619)(10.2683,32.1626)
		\psline[linewidth=2pt, fillstyle=solid, fillcolor=gray](-26.8011,-31.0616)(-26.6878,-30.8616)(-26.5746,-30.6616)(-26.4614,-30.4616)(-26.3482,-30.2616)(-26.235,-30.0616)(-26.1219,-29.8616)(-26.0088,-29.6616)(-25.8958,-29.4616)(-25.7827,-29.2616)(-25.6698,-29.0616)(-25.5568,-28.8616)(-25.4439,-28.6616)(-25.331,-28.4616)(-25.2182,-28.2616)(-25.1054,-28.0616)(-24.9926,-27.8616)(-24.8799,-27.6616)(-24.7672,-27.4616)(-24.6545,-27.2616)(-24.5419,-27.0616)(-24.4294,-26.8616)(-24.3169,-26.6616)(-24.2044,-26.4616)(-24.092,-26.2616)(-23.9796,-26.0616)(-23.8673,-25.8616)(-23.755,-25.6616)(-23.6428,-25.4616)(-23.5306,-25.2616)(-23.4185,-25.0616)(-23.3064,-24.8616)(-23.1944,-24.6616)(-23.0824,-24.4616)(-22.9705,-24.2616)(-22.8587,-24.0616)(-22.7469,-23.8616)(-22.6352,-23.6616)(-22.5236,-23.4616)(-22.412,-23.2616)(-22.3005,-23.0616)(-22.189,-22.8616)(-22.0777,-22.6616)(-21.9664,-22.4616)(-21.8551,-22.2616)(-21.744,-22.0616)(-21.6329,-21.8616)(-21.5219,-21.6616)(-21.411,-21.4616)(-21.3002,-21.2616)(-21.1895,-21.0616)(-21.0789,-20.8616)(-20.9683,-20.6616)(-20.8579,-20.4616)(-20.7476,-20.2616)(-20.6373,-20.0616)(-20.5272,-19.8616)(-20.4172,-19.6616)(-20.3073,-19.4616)(-20.1975,-19.2616)(-20.0878,-19.0616)(-19.9782,-18.8616)(-19.8688,-18.6616)(-19.7595,-18.4616)(-19.6503,-18.2616)(-19.5413,-18.0616)(-19.4324,-17.8616)(-19.3237,-17.6616)(-19.2151,-17.4616)(-19.1067,-17.2616)(-18.9985,-17.0616)(-18.8904,-16.8616)(-18.7825,-16.6616)(-18.6748,-16.4616)(-18.5672,-16.2616)(-18.4599,-16.0616)(-18.3528,-15.8616)(-18.2458,-15.6616)(-18.1391,-15.4616)(-18.0327,-15.2616)(-17.9264,-15.0616)(-17.8204,-14.8616)(-17.7147,-14.6616)(-17.6092,-14.4616)(-17.5041,-14.2616)(-17.3992,-14.0616)(-17.2945,-13.8616)(-17.1903,-13.6616)(-17.0863,-13.4616)(-16.9827,-13.2616)(-16.8794,-13.0616)(-16.7765,-12.8616)(-16.674,-12.6616)(-16.5719,-12.4616)(-16.4703,-12.2616)(-16.369,-12.0616)(-16.2683,-11.8616)(-16.168,-11.6616)(-16.0683,-11.4616)(-15.9691,-11.2616)(-15.8705,-11.0616)(-15.7725,-10.8616)(-15.6751,-10.6616)(-15.5784,-10.4616)(-15.4824,-10.2616)(-15.3871,-10.0616)(-15.2926,-9.86156)(-15.199,-9.66156)(-15.1062,-9.46156)(-15.0144,-9.26156)(-14.9236,-9.06156)(-14.8339,-8.86156)(-14.7453,-8.66156)(-14.6579,-8.46156)(-14.5718,-8.26156)(-14.4871,-8.06156)(-14.4039,-7.86156)(-14.3223,-7.66156)(-14.2424,-7.46156)(-14.1643,-7.26156)(-14.0883,-7.06156)(-14.0144,-6.86156)(-13.9428,-6.66156)(-13.8738,-6.46156)(-13.8075,-6.26156)(-13.7442,-6.06156)(-13.6843,-5.86156)(-13.628,-5.66156)(-13.5757,-5.46156)(-13.5277,-5.26156)(-13.4847,-5.06156)(-13.4471,-4.86156)(-13.4156,-4.66156)(-13.3909,-4.46156)(-13.3739,-4.26156)(-13.3656,-4.06156)(-13.3672,-3.86156)(-13.3802,-3.66156)(-13.4063,-3.46156)(-13.4478,-3.26156)(-13.5072,-3.06156)(-13.5879,-2.86156)(-13.6941,-2.66156)(-13.8313,-2.46156)(-14.0066,-2.26156)(-14.2295,-2.06156)(-14.5135,-1.86156)(-14.8772,-1.66156)(-15.3486,-1.46156)(-15.9707,-1.26156)(-16.8144,-1.06156)(-18.0068,-0.861558)(-19.8031,-0.661558)(-22.8203,-0.461558)(-29.1411,-0.261558)(-57.6626,-0.0615576)(-60,-30)
		\psline[linewidth=2pt, fillstyle=solid, fillcolor=gray](26.8011,-31.0616)(26.6878,-30.8616)(26.5746,-30.6616)(26.4614,-30.4616)(26.3482,-30.2616)(26.235,-30.0616)(26.1219,-29.8616)(26.0088,-29.6616)(25.8958,-29.4616)(25.7827,-29.2616)(25.6698,-29.0616)(25.5568,-28.8616)(25.4439,-28.6616)(25.331,-28.4616)(25.2182,-28.2616)(25.1054,-28.0616)(24.9926,-27.8616)(24.8799,-27.6616)(24.7672,-27.4616)(24.6545,-27.2616)(24.5419,-27.0616)(24.4294,-26.8616)(24.3169,-26.6616)(24.2044,-26.4616)(24.092,-26.2616)(23.9796,-26.0616)(23.8673,-25.8616)(23.755,-25.6616)(23.6428,-25.4616)(23.5306,-25.2616)(23.4185,-25.0616)(23.3064,-24.8616)(23.1944,-24.6616)(23.0824,-24.4616)(22.9705,-24.2616)(22.8587,-24.0616)(22.7469,-23.8616)(22.6352,-23.6616)(22.5236,-23.4616)(22.412,-23.2616)(22.3005,-23.0616)(22.189,-22.8616)(22.0777,-22.6616)(21.9664,-22.4616)(21.8551,-22.2616)(21.744,-22.0616)(21.6329,-21.8616)(21.5219,-21.6616)(21.411,-21.4616)(21.3002,-21.2616)(21.1895,-21.0616)(21.0789,-20.8616)(20.9683,-20.6616)(20.8579,-20.4616)(20.7476,-20.2616)(20.6373,-20.0616)(20.5272,-19.8616)(20.4172,-19.6616)(20.3073,-19.4616)(20.1975,-19.2616)(20.0878,-19.0616)(19.9782,-18.8616)(19.8688,-18.6616)(19.7595,-18.4616)(19.6503,-18.2616)(19.5413,-18.0616)(19.4324,-17.8616)(19.3237,-17.6616)(19.2151,-17.4616)(19.1067,-17.2616)(18.9985,-17.0616)(18.8904,-16.8616)(18.7825,-16.6616)(18.6748,-16.4616)(18.5672,-16.2616)(18.4599,-16.0616)(18.3528,-15.8616)(18.2458,-15.6616)(18.1391,-15.4616)(18.0327,-15.2616)(17.9264,-15.0616)(17.8204,-14.8616)(17.7147,-14.6616)(17.6092,-14.4616)(17.5041,-14.2616)(17.3992,-14.0616)(17.2945,-13.8616)(17.1903,-13.6616)(17.0863,-13.4616)(16.9827,-13.2616)(16.8794,-13.0616)(16.7765,-12.8616)(16.674,-12.6616)(16.5719,-12.4616)(16.4703,-12.2616)(16.369,-12.0616)(16.2683,-11.8616)(16.168,-11.6616)(16.0683,-11.4616)(15.9691,-11.2616)(15.8705,-11.0616)(15.7725,-10.8616)(15.6751,-10.6616)(15.5784,-10.4616)(15.4824,-10.2616)(15.3871,-10.0616)(15.2926,-9.86156)(15.199,-9.66156)(15.1062,-9.46156)(15.0144,-9.26156)(14.9236,-9.06156)(14.8339,-8.86156)(14.7453,-8.66156)(14.6579,-8.46156)(14.5718,-8.26156)(14.4871,-8.06156)(14.4039,-7.86156)(14.3223,-7.66156)(14.2424,-7.46156)(14.1643,-7.26156)(14.0883,-7.06156)(14.0144,-6.86156)(13.9428,-6.66156)(13.8738,-6.46156)(13.8075,-6.26156)(13.7442,-6.06156)(13.6843,-5.86156)(13.628,-5.66156)(13.5757,-5.46156)(13.5277,-5.26156)(13.4847,-5.06156)(13.4471,-4.86156)(13.4156,-4.66156)(13.3909,-4.46156)(13.3739,-4.26156)(13.3656,-4.06156)(13.3672,-3.86156)(13.3802,-3.66156)(13.4063,-3.46156)(13.4478,-3.26156)(13.5072,-3.06156)(13.5879,-2.86156)(13.6941,-2.66156)(13.8313,-2.46156)(14.0066,-2.26156)(14.2295,-2.06156)(14.5135,-1.86156)(14.8772,-1.66156)(15.3486,-1.46156)(15.9707,-1.26156)(16.8144,-1.06156)(18.0068,-0.861558)(19.8031,-0.661558)(22.8203,-0.461558)(29.1411,-0.261558)(57.6626,-0.0615576)(60,-30)
		\psline(66.2747, -100)(-49.1953, 100)
		\psline(57.735, 114.791)(-57.735, -85.2088)
		\psline(-124.01, 0)(106.93, 0)
		\psline[linestyle=dotted](118.786, 9.04788)(-112.154, 9.04788)
		\psline[linestyle=dotted](121.7, 4)(-109.24, 4)
		\psline[linestyle=dotted](114.464, 16.5346)(-116.477, 16.5346)
		\psline[linestyle=dotted](-69.62, 94.2058)(45.85, -105.794)
		\psline[linestyle=dotted](-42.5047, -100)(72.9654, 100)
		\psline[linestyle=dotted](-61.4514, 108.354)(54.0186, -91.6458)
		\psline[linestyle=dotted](-59.085, 112.453)(56.385, -87.5471)
		\psline[linestyle=dotted](-58.842, -100)(56.6281, 100)
		\psline[linestyle=dotted](-63.5747, -100)(51.8954, 100)
		\put(-29.5,0.5){$s=0$}
		\put(-29.5,4.75){$s=4M_\pi^2$}
		\put(-29.5,10){$s=(M_K-\sqrt{s_\ell})^2$}
		\put(-29.5,17.5){$s=(M_K+\sqrt{s_\ell})^2$}
		\rput{-60}(16,-15){$t=0$}
		\rput{-60}(11,-20){$t=(M_K-M_\pi)^2$}
		\rput{-60}(-2.5,-25){$t=(M_K+M_\pi)^2$}
		\rput{60}(3.5,23){$u=0$}
		\rput{60}(9.5,21){$u=(M_K-M_\pi)^2$}
		\rput{60}(21,12.5){$u=(M_K+M_\pi)^2$}
		\put(-4,6){decay region}
		\put(-3.5,27){$s$-channel}
		\put(-27,-10){$t$-channel}
		\put(20,-10){$u$-channel}
		\put(-5,-10){real amplitude}
	\end{pspicture*}
	\caption{Mandelstam diagram for $K_{\ell4}$ for the case $s_\ell > 0$}
	\label{img:MandelstamDiagram2}
\end{figure}
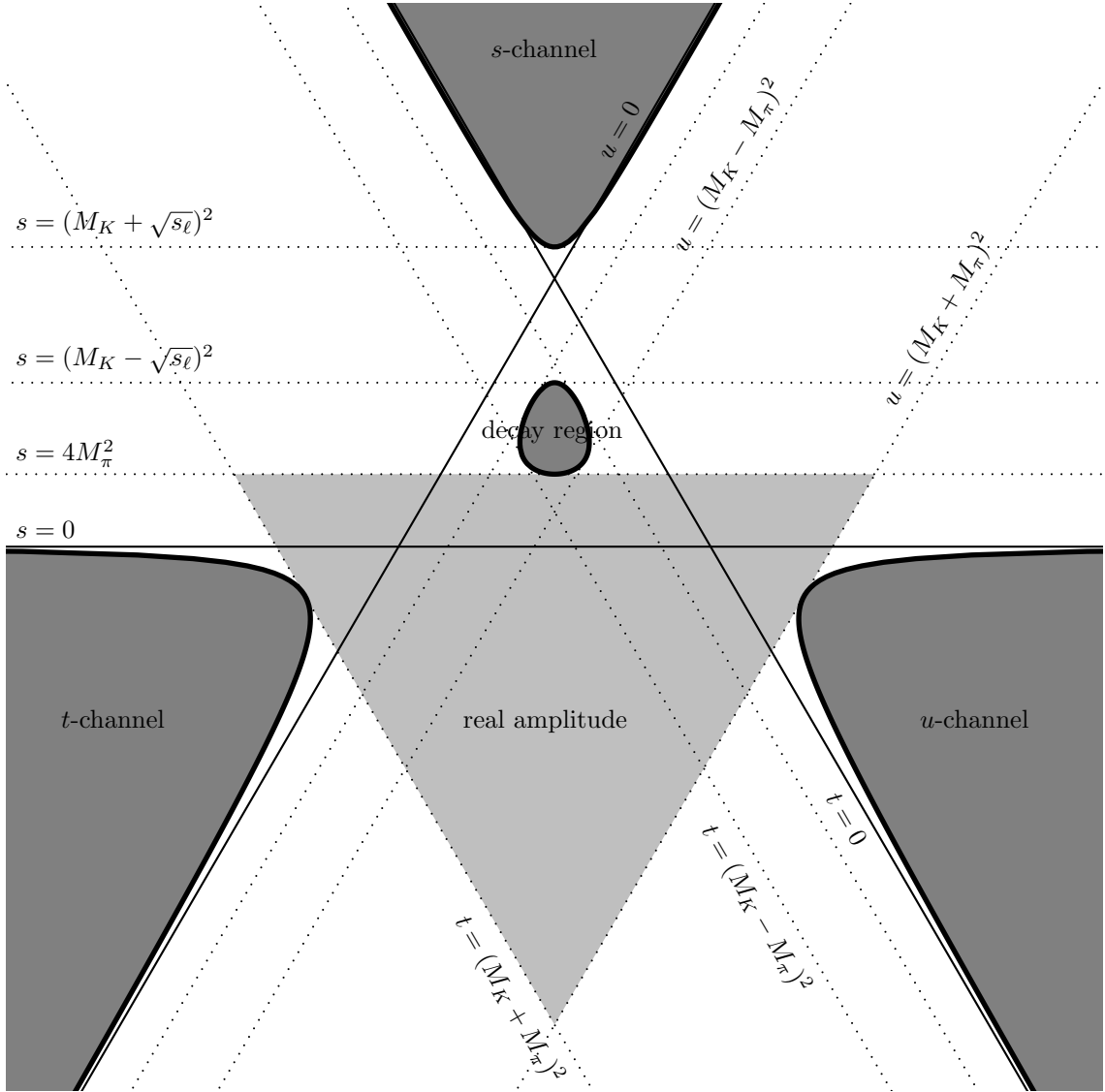

Let us first study the general properties of the hadronic axial vector current matrix element. It is instructive to draw a Mandelstam diagram for the process (see figures~\ref{img:MandelstamDiagram1} and \ref{img:MandelstamDiagram2}): since $s+t+u = M_K^2 + 2 M_\pi^2 + s_\ell =: \Sigma_0$ is constant (for a fixed value of $s_\ell$), the Mandelstam variables can be represented in one plane, using the fact that the sum of distances of a point to the sides of an equilateral triangle is constant.

The same amplitude describes four processes:
\begin{itemize}
	\item the decay $K^+(k) \to \pi^+(p_1) \pi^-(p_2) A_\mu^\dagger(L)$
	\item the $s$-channel scattering $K^+(k) A_\mu(-L) \to \pi^+(p_1) \pi^-(p_2)$
	\item the $t$-channel scattering $K^+(k) \pi^-(-p_1) \to \pi^-(p_2) A_\mu^\dagger(L)$
	\item the $u$-channel scattering $K^+(k) \pi^+(-p_2) \to \pi^+(p_1) A_\mu^\dagger(L)$
\end{itemize}

The physical region of the decay starts at $s = 4 M_\pi^2$ and ends at $s = (M_K - \sqrt{s_\ell})^2$. The $s$-channel scattering starts at $s = (M_K + \sqrt{s_\ell})^2$. If $s_\ell = 0$ is assumed, the two regions touch at $s = M_K^2$ (figure~\ref{img:MandelstamDiagram1}).

The sub-threshold region $s < s_0 := 4 M_\pi^2$, $t < t_0 := (M_K + M_\pi)^2$, $u < u_0 := (M_K + M_\pi)^2$ forms a triangle in the Mandelstam plane where the amplitude is real. Branch cuts of the amplitude start at each threshold $s_0$, $t_0$ and $u_0$. There, physical intermediate states are possible ($\pi\pi$ intermediate states in the $s$-channel, $K\pi$ states in the $t$- and $u$-channel).

\clearpage

\begin{figure}[H]
	\centering
	\psset{unit=0.248cm}
	\begin{pspicture*}(-30,-30)(30,30)
		\pspolygon[linestyle=none, fillstyle=solid, fillcolor=lightgray](-17.7014,4)(17.7014,4)(0,-26.6597)
		\pspolygon[linewidth=2pt, fillstyle=solid, fillcolor=gray](0.,12.5112)(-0.483109,11.4744)(-0.8654,10.6123)(-1.1701,9.88454)(-1.41382,9.26242)(-1.6087,8.72486)(-1.76389,8.25607)(-1.88635,7.84396)(-1.98151,7.47914)(-2.05365,7.15419)(-2.10619,6.86318)(-2.14191,6.60133)(-2.16306,6.3647)(-2.17151,6.15006)(-2.16882,5.95472)(-2.15629,5.77641)(-2.13503,5.61324)(-2.10597,5.46357)(-2.06991,5.32602)(-2.02755,5.19941)(-1.97946,5.08269)(-1.92618,4.97498)(-1.86814,4.87551)(-1.80573,4.78359)(-1.73931,4.69865)(-1.66916,4.62016)(-1.59554,4.54766)(-1.51869,4.48077)(-1.43881,4.41912)(-1.35607,4.36243)(-1.27062,4.31043)(-1.1826,4.26289)(-1.09212,4.21961)(-0.999265,4.18044)(-0.904118,4.14524)(-0.806736,4.11391)(-0.707165,4.08637)(-0.605428,4.06259)(-0.501534,4.04253)(-0.395473,4.02624)(-0.28721,4.01376)(-0.176688,4.00518)(-0.063822,4.00067)(0.0510095,4.00043)(0.164146,4.00447)(0.274925,4.0126)(0.383438,4.02465)(0.489745,4.04052)(0.593881,4.06015)(0.695859,4.08352)(0.795675,4.11063)(0.893304,4.14153)(0.988704,4.1763)(1.08182,4.21502)(1.17257,4.25782)(1.26088,4.30488)(1.34662,4.35637)(1.42967,4.41252)(1.50988,4.47359)(1.58708,4.53988)(1.66107,4.61173)(1.73162,4.68953)(1.79848,4.77373)(1.86135,4.86483)(1.9199,4.96343)(1.97374,5.07017)(2.02243,5.18584)(2.06546,5.3113)(2.10226,5.44756)(2.13214,5.5958)(2.15432,5.75739)(2.16788,5.9339)(2.17173,6.12723)(2.1646,6.33958)(2.14497,6.57359)(2.11099,6.83243)(2.06047,7.11995)(1.99068,7.44082)(1.8983,7.80083)(1.77915,8.20721)(1.62795,8.6691)(1.43792,9.19823)(1.20022,9.80993)(0.90309,10.5246)(0.530513,11.3699)(0.0601512,12.3846)
		\psline[linewidth=2pt, fillstyle=solid, fillcolor=gray](-10.2708,31.5486)(-10.1568,31.346)(-10.0428,31.1435)(-9.92882,30.9409)(-9.81485,30.7383)(-9.70091,30.5357)(-9.58699,30.333)(-9.4731,30.1303)(-9.35922,29.9275)(-9.24537,29.7247)(-9.13154,29.5218)(-9.01774,29.319)(-8.90396,29.116)(-8.7902,28.9131)(-8.67647,28.7101)(-8.56277,28.507)(-8.44909,28.3039)(-8.33544,28.1007)(-8.22182,27.8975)(-8.10822,27.6943)(-7.99466,27.491)(-7.88112,27.2876)(-7.76762,27.0842)(-7.65415,26.8808)(-7.54071,26.6772)(-7.4273,26.4737)(-7.31393,26.27)(-7.20059,26.0663)(-7.08729,25.8626)(-6.97402,25.6588)(-6.8608,25.4549)(-6.74761,25.2509)(-6.63446,25.0469)(-6.52135,24.8428)(-6.40828,24.6387)(-6.29525,24.4344)(-6.18227,24.2301)(-6.06934,24.0257)(-5.95645,23.8213)(-5.84361,23.6167)(-5.73081,23.4121)(-5.61807,23.2074)(-5.50538,23.0025)(-5.39275,22.7976)(-5.28017,22.5926)(-5.16764,22.3875)(-5.05518,22.1823)(-4.94277,21.977)(-4.83043,21.7716)(-4.71815,21.5661)(-4.60594,21.3604)(-4.4938,21.1547)(-4.38173,20.9488)(-4.26973,20.7428)(-4.15781,20.5366)(-4.04596,20.3303)(-3.9342,20.1239)(-3.82252,19.9173)(-3.71093,19.7106)(-3.59943,19.5038)(-3.48802,19.2967)(-3.37671,19.0895)(-3.26549,18.8821)(-3.15439,18.6746)(-3.04339,18.4668)(-2.9325,18.2589)(-2.82173,18.0508)(-2.71108,17.8424)(-2.60056,17.6338)(-2.49017,17.425)(-2.37992,17.216)(-2.26981,17.0067)(-2.15986,16.7972)(-2.05005,16.5873)(-1.94042,16.3772)(-1.83095,16.1668)(-1.72167,15.9561)(-1.61257,15.7451)(-1.50367,15.5337)(-1.39497,15.322)(-1.2865,15.1099)(-1.17825,14.8973)(-1.07025,14.6844)(-0.962501,14.471)(-0.855021,14.2572)(-0.747825,14.0429)(-0.640931,13.828)(-0.534356,13.6126)(-0.42812,13.3966)(-0.322245,13.18)(-0.216753,12.9627)(-0.111671,12.7447)(-0.00702684,12.526)(0.,12.5112)(0.105138,12.7311)(0.21072,12.9502)(0.316717,13.1686)(0.423103,13.3864)(0.529855,13.6035)(0.636949,13.82)(0.744367,14.0359)(0.85209,14.2514)(0.9601,14.4663)(1.06838,14.6807)(1.17692,14.8947)(1.28571,15.1083)(1.39472,15.3215)(1.50396,15.5343)(1.61341,15.7467)(1.72305,15.9588)(1.83289,16.1706)(1.9429,16.382)(2.05309,16.5932)(2.16345,16.804)(2.27396,17.0146)(2.38462,17.2249)(2.49543,17.435)(2.60638,17.6448)(2.71746,17.8544)(2.82867,18.0638)(2.94,18.273)(3.05145,18.4819)(3.16301,18.6907)(3.27469,18.8993)(3.38646,19.1077)(3.49834,19.3159)(3.61032,19.524)(3.72238,19.7319)(3.83454,19.9396)(3.94679,20.1472)(4.05912,20.3546)(4.17154,20.5619)(4.28403,20.7691)(4.39659,20.9761)(4.50924,21.183)(4.62195,21.3898)(4.73473,21.5964)(4.84758,21.803)(4.96049,22.0094)(5.07347,22.2157)(5.18651,22.4219)(5.2996,22.628)(5.41276,22.834)(5.52597,23.04)(5.63923,23.2458)(5.75254,23.4515)(5.86591,23.6572)(5.97933,23.8627)(6.09279,24.0682)(6.2063,24.2736)(6.31985,24.4789)(6.43345,24.6841)(6.5471,24.8893)(6.66078,25.0944)(6.77451,25.2994)(6.88827,25.5044)(7.00208,25.7093)(7.11592,25.9141)(7.2298,26.1188)(7.34371,26.3235)(7.45766,26.5282)(7.57164,26.7328)(7.68566,26.9373)(7.7997,27.1417)(7.91378,27.3461)(8.02789,27.5505)(8.14204,27.7548)(8.25621,27.9591)(8.3704,28.1633)(8.48463,28.3674)(8.59889,28.5715)(8.71317,28.7756)(8.82747,28.9796)(8.94181,29.1836)(9.05616,29.3875)(9.17055,29.5914)(9.28495,29.7952)(9.39938,29.999)(9.51383,30.2028)(9.62831,30.4065)(9.74281,30.6102)(9.85732,30.8138)(9.97186,31.0174)(10.0864,31.221)(10.201,31.4246)
		\psline[linewidth=2pt, fillstyle=solid, fillcolor=gray](-26.3958,-30.4736)(-26.2831,-30.2736)(-26.1704,-30.0736)(-26.0577,-29.8736)(-25.9451,-29.6736)(-25.8325,-29.4736)(-25.7199,-29.2736)(-25.6074,-29.0736)(-25.4949,-28.8736)(-25.3824,-28.6736)(-25.27,-28.4736)(-25.1577,-28.2736)(-25.0454,-28.0736)(-24.9331,-27.8736)(-24.8209,-27.6736)(-24.7087,-27.4736)(-24.5965,-27.2736)(-24.4845,-27.0736)(-24.3724,-26.8736)(-24.2604,-26.6736)(-24.1485,-26.4736)(-24.0366,-26.2736)(-23.9248,-26.0736)(-23.813,-25.8736)(-23.7013,-25.6736)(-23.5896,-25.4736)(-23.478,-25.2736)(-23.3664,-25.0736)(-23.2549,-24.8736)(-23.1435,-24.6736)(-23.0322,-24.4736)(-22.9209,-24.2736)(-22.8096,-24.0736)(-22.6985,-23.8736)(-22.5874,-23.6736)(-22.4764,-23.4736)(-22.3654,-23.2736)(-22.2546,-23.0736)(-22.1438,-22.8736)(-22.0331,-22.6736)(-21.9224,-22.4736)(-21.8119,-22.2736)(-21.7014,-22.0736)(-21.5911,-21.8736)(-21.4808,-21.6736)(-21.3706,-21.4736)(-21.2605,-21.2736)(-21.1505,-21.0736)(-21.0406,-20.8736)(-20.9308,-20.6736)(-20.8212,-20.4736)(-20.7116,-20.2736)(-20.6021,-20.0736)(-20.4928,-19.8736)(-20.3836,-19.6736)(-20.2745,-19.4736)(-20.1655,-19.2736)(-20.0567,-19.0736)(-19.948,-18.8736)(-19.8394,-18.6736)(-19.731,-18.4736)(-19.6227,-18.2736)(-19.5146,-18.0736)(-19.4066,-17.8736)(-19.2988,-17.6736)(-19.1912,-17.4736)(-19.0837,-17.2736)(-18.9764,-17.0736)(-18.8693,-16.8736)(-18.7624,-16.6736)(-18.6557,-16.4736)(-18.5492,-16.2736)(-18.4429,-16.0736)(-18.3369,-15.8736)(-18.231,-15.6736)(-18.1254,-15.4736)(-18.0201,-15.2736)(-17.915,-15.0736)(-17.8102,-14.8736)(-17.7056,-14.6736)(-17.6013,-14.4736)(-17.4974,-14.2736)(-17.3937,-14.0736)(-17.2904,-13.8736)(-17.1874,-13.6736)(-17.0847,-13.4736)(-16.9824,-13.2736)(-16.8805,-13.0736)(-16.779,-12.8736)(-16.6779,-12.6736)(-16.5773,-12.4736)(-16.4771,-12.2736)(-16.3774,-12.0736)(-16.2781,-11.8736)(-16.1794,-11.6736)(-16.0813,-11.4736)(-15.9837,-11.2736)(-15.8868,-11.0736)(-15.7904,-10.8736)(-15.6948,-10.6736)(-15.5998,-10.4736)(-15.5056,-10.2736)(-15.4122,-10.0736)(-15.3197,-9.87356)(-15.228,-9.67356)(-15.1372,-9.47356)(-15.0475,-9.27356)(-14.9587,-9.07356)(-14.8711,-8.87356)(-14.7847,-8.67356)(-14.6996,-8.47356)(-14.6158,-8.27356)(-14.5335,-8.07356)(-14.4527,-7.87356)(-14.3736,-7.67356)(-14.2963,-7.47356)(-14.2209,-7.27356)(-14.1476,-7.07356)(-14.0765,-6.87356)(-14.0079,-6.67356)(-13.9419,-6.47356)(-13.8787,-6.27356)(-13.8187,-6.07356)(-13.7621,-5.87356)(-13.7092,-5.67356)(-13.6605,-5.47356)(-13.6163,-5.27356)(-13.5772,-5.07356)(-13.5436,-4.87356)(-13.5163,-4.67356)(-13.496,-4.47356)(-13.4836,-4.27356)(-13.4801,-4.07356)(-13.4868,-3.87356)(-13.5052,-3.67356)(-13.537,-3.47356)(-13.5844,-3.27356)(-13.6502,-3.07356)(-13.7376,-2.87356)(-13.851,-2.67356)(-13.9959,-2.47356)(-14.1794,-2.27356)(-14.4113,-2.07356)(-14.7048,-1.87356)(-15.0789,-1.67356)(-15.5613,-1.47356)(-16.195,-1.27356)(-17.0504,-1.07356)(-18.2527,-0.873558)(-20.0516,-0.673558)(-23.0417,-0.473558)(-29.1745,-0.273558)(-54.0702,-0.0735576)(-60,-30)
		\psline[linewidth=2pt, fillstyle=solid, fillcolor=gray](26.3958,-30.4736)(26.2831,-30.2736)(26.1704,-30.0736)(26.0577,-29.8736)(25.9451,-29.6736)(25.8325,-29.4736)(25.7199,-29.2736)(25.6074,-29.0736)(25.4949,-28.8736)(25.3824,-28.6736)(25.27,-28.4736)(25.1577,-28.2736)(25.0454,-28.0736)(24.9331,-27.8736)(24.8209,-27.6736)(24.7087,-27.4736)(24.5965,-27.2736)(24.4845,-27.0736)(24.3724,-26.8736)(24.2604,-26.6736)(24.1485,-26.4736)(24.0366,-26.2736)(23.9248,-26.0736)(23.813,-25.8736)(23.7013,-25.6736)(23.5896,-25.4736)(23.478,-25.2736)(23.3664,-25.0736)(23.2549,-24.8736)(23.1435,-24.6736)(23.0322,-24.4736)(22.9209,-24.2736)(22.8096,-24.0736)(22.6985,-23.8736)(22.5874,-23.6736)(22.4764,-23.4736)(22.3654,-23.2736)(22.2546,-23.0736)(22.1438,-22.8736)(22.0331,-22.6736)(21.9224,-22.4736)(21.8119,-22.2736)(21.7014,-22.0736)(21.5911,-21.8736)(21.4808,-21.6736)(21.3706,-21.4736)(21.2605,-21.2736)(21.1505,-21.0736)(21.0406,-20.8736)(20.9308,-20.6736)(20.8212,-20.4736)(20.7116,-20.2736)(20.6021,-20.0736)(20.4928,-19.8736)(20.3836,-19.6736)(20.2745,-19.4736)(20.1655,-19.2736)(20.0567,-19.0736)(19.948,-18.8736)(19.8394,-18.6736)(19.731,-18.4736)(19.6227,-18.2736)(19.5146,-18.0736)(19.4066,-17.8736)(19.2988,-17.6736)(19.1912,-17.4736)(19.0837,-17.2736)(18.9764,-17.0736)(18.8693,-16.8736)(18.7624,-16.6736)(18.6557,-16.4736)(18.5492,-16.2736)(18.4429,-16.0736)(18.3369,-15.8736)(18.231,-15.6736)(18.1254,-15.4736)(18.0201,-15.2736)(17.915,-15.0736)(17.8102,-14.8736)(17.7056,-14.6736)(17.6013,-14.4736)(17.4974,-14.2736)(17.3937,-14.0736)(17.2904,-13.8736)(17.1874,-13.6736)(17.0847,-13.4736)(16.9824,-13.2736)(16.8805,-13.0736)(16.779,-12.8736)(16.6779,-12.6736)(16.5773,-12.4736)(16.4771,-12.2736)(16.3774,-12.0736)(16.2781,-11.8736)(16.1794,-11.6736)(16.0813,-11.4736)(15.9837,-11.2736)(15.8868,-11.0736)(15.7904,-10.8736)(15.6948,-10.6736)(15.5998,-10.4736)(15.5056,-10.2736)(15.4122,-10.0736)(15.3197,-9.87356)(15.228,-9.67356)(15.1372,-9.47356)(15.0475,-9.27356)(14.9587,-9.07356)(14.8711,-8.87356)(14.7847,-8.67356)(14.6996,-8.47356)(14.6158,-8.27356)(14.5335,-8.07356)(14.4527,-7.87356)(14.3736,-7.67356)(14.2963,-7.47356)(14.2209,-7.27356)(14.1476,-7.07356)(14.0765,-6.87356)(14.0079,-6.67356)(13.9419,-6.47356)(13.8787,-6.27356)(13.8187,-6.07356)(13.7621,-5.87356)(13.7092,-5.67356)(13.6605,-5.47356)(13.6163,-5.27356)(13.5772,-5.07356)(13.5436,-4.87356)(13.5163,-4.67356)(13.496,-4.47356)(13.4836,-4.27356)(13.4801,-4.07356)(13.4868,-3.87356)(13.5052,-3.67356)(13.537,-3.47356)(13.5844,-3.27356)(13.6502,-3.07356)(13.7376,-2.87356)(13.851,-2.67356)(13.9959,-2.47356)(14.1794,-2.27356)(14.4113,-2.07356)(14.7048,-1.87356)(15.0789,-1.67356)(15.5613,-1.47356)(16.195,-1.27356)(17.0504,-1.07356)(18.2527,-0.873558)(20.0516,-0.673558)(23.0417,-0.473558)(29.1745,-0.273558)(54.0702,-0.0735576)(60,-30)
		\psline(66.1131, -100)(-49.357, 100)
		\psline(57.735, 114.511)(-57.735, -85.4888)
		\psline(-123.848, 0)(107.092, 0)
		\psline[linestyle=dotted](116.625, 12.5112)(-114.315, 12.5112)
		\psline[linestyle=dotted](121.539, 4)(-109.401, 4)
		\psline[linestyle=dotted](116.625, 12.5112)(-114.315, 12.5112)
		\psline[linestyle=dotted](-69.62, 93.9258)(45.85, -106.074)
		\psline[linestyle=dotted](-42.343, -100)(73.127, 100)
		\psline[linestyle=dotted](-61.4514, 108.074)(54.0186, -91.9258)
		\psline[linestyle=dotted](-58.3124, 113.511)(57.1577, -86.4888)
		\psline[linestyle=dotted](-58.6803, -100)(56.7898, 100)
		\psline[linestyle=dotted](-64.9584, -100)(50.5117, 100)
		\put(-29.5,0.5){$s=0$}
		\put(-29.5,4.75){$s=4M_\pi^2$}
		\put(-29.5,13.25){$s=M_K^2$}
		\rput{-60}(16,-15){$t=0$}
		\rput{-60}(11,-20){$t=(M_K-M_\pi)^2$}
		\rput{-60}(-2.5,-25){$t=(M_K+M_\pi)^2$}
		\rput{60}(4,23){$u=0$}
		\rput{60}(9.5,21){$u=(M_K-M_\pi)^2$}
		\rput{60}(21,12){$u=(M_K+M_\pi)^2$}
		\put(-4,6){decay region}
		\put(-3.5,27){$s$-channel}
		\put(-27,-10){$t$-channel}
		\put(20,-10){$u$-channel}
		\put(-5,-10){real amplitude}
	\end{pspicture*}
	\caption{Mandelstam diagram for $K_{\ell4}$ for the case $s_\ell = 0$}
	\label{img:MandelstamDiagram1}
\end{figure}
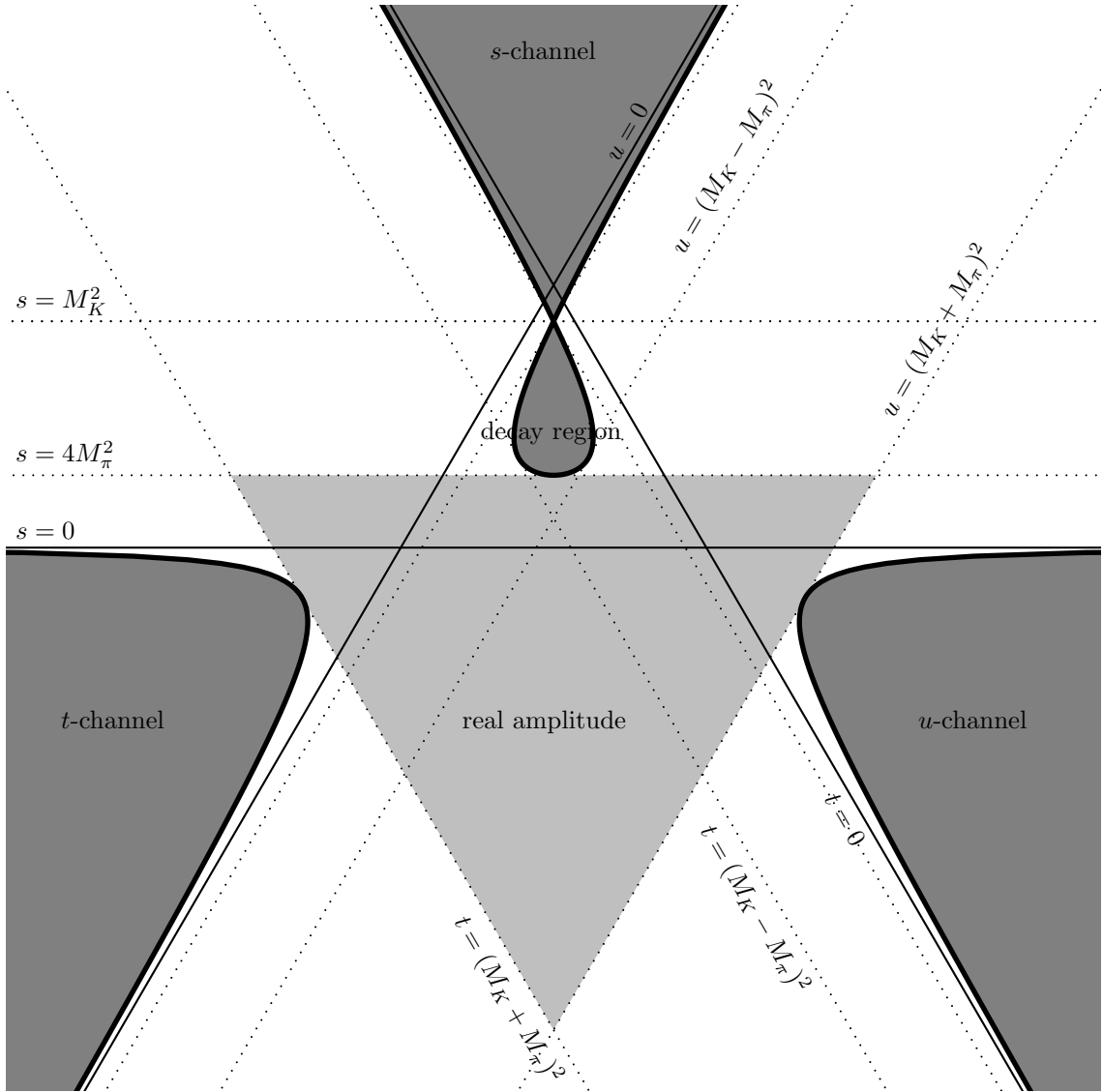

\clearpage

\section{Isospin Decomposition}

Let us study the isospin properties of the hadronic axial vector current matrix element of $K_{\ell4}$ in the different channels: we decompose the physical amplitude into amplitudes with definite isospin.

\subsection{$s$-Channel}

We consider the matrix element
\begin{align}
	\mathcal{A}_\mu^{+-} = \< \pi^+(p_1) \pi^-(p_2) \big| A_\mu(0) \big| K^+(k) \> .
\end{align}
As the weak current satisfies $\Delta I = \frac{1}{2}$, the initial and final states can be decomposed as
\begin{align}
	\begin{split}
		A_\mu(0) \big| K^+(k) \> &= \frac{1}{\sqrt{2}} \big| 1, 0 \> + \frac{1}{\sqrt{2}} \big| 0, 0 \> , \\
		\< \pi^+(p_1) \pi^-(p_2) \big| &= \frac{1}{\sqrt{6}} \< 2, 0 \big| + \frac{1}{\sqrt{2}} \< 1, 0 \big| + \frac{1}{\sqrt{3}} \< 0, 0 \big| , \\
		\< \pi^-(p_1) \pi^+(p_2) \big| &= \frac{1}{\sqrt{6}} \< 2, 0 \big| - \frac{1}{\sqrt{2}} \< 1, 0 \big| + \frac{1}{\sqrt{3}} \< 0, 0 \big| .
	\end{split}
\end{align}
Hence, we can write the following decomposition of the matrix element into pure isospin amplitudes:
\begin{align}
	\begin{split}
		\mathcal{A}_\mu^{+-} = \frac{1}{2} \mathcal{A}_\mu^{(1)} + \frac{1}{\sqrt{6}} \mathcal{A}_\mu^{(0)} , \\
		\mathcal{A}_\mu^{-+} = -\frac{1}{2} \mathcal{A}_\mu^{(1)} + \frac{1}{\sqrt{6}} \mathcal{A}_\mu^{(0)} .
	\end{split}
\end{align}
Using $\mathcal{A}_\mu^{+-}(k, -L \to p_1, p_2) = \mathcal{A}_\mu^{-+}(k, -L \to p_2, p_1)$, we find the following relations:
\begin{align}
	\begin{split}
		\mathcal{A}_\mu^{0}(k, -L \to p_1, p_2) &= \sqrt{\frac{3}{2}} \left( \mathcal{A}_\mu^{+-}(k, -L \to p_1, p_2) + \mathcal{A}_\mu^{+-}(k, -L \to p_2, p_1) \right) , \\
		\mathcal{A}_\mu^{1}(k, -L \to p_1, p_2) &= \left( \mathcal{A}_\mu^{+-}(k, -L \to p_1, p_2) - \mathcal{A}_\mu^{+-}(k, -L \to p_2, p_1) \right) . \\
	\end{split}
\end{align}
The pure isospin form factors are related to the physical ones by
\begin{align}
	\begin{split}
		\label{eq:sChannelIsospinFormFactors}
		F^{(0)}(s,t,u) &= \sqrt{\frac{3}{2}}\left( F(s,t,u) + F(s,u,t) \right) , \\
		G^{(0)}(s,t,u) &= \sqrt{\frac{3}{2}}\left( G(s,t,u) - G(s,u,t) \right) , \\
		R^{(0)}(s,t,u) &= \sqrt{\frac{3}{2}}\left( R(s,t,u) + R(s,u,t) \right) , \\
		F^{(1)}(s,t,u) &= F(s,t,u) - F(s,u,t) , \\
		G^{(1)}(s,t,u) &= G(s,t,u) + G(s,u,t) , \\
		R^{(1)}(s,t,u) &= R(s,t,u) - R(s,u,t) .
	\end{split}
\end{align}
We further note that
\begin{align}
	\begin{split}
		\mathcal{A}_\mu^{(0)}(k, -L \to p_1, p_2) &= \mathcal{A}_\mu^{(0)}(k, -L \to p_2, p_1), \\
		\mathcal{A}_\mu^{(1)}(k, -L \to p_1, p_2) &= - \mathcal{A}_\mu^{(1)}(k, -L \to p_2, p_1) ,
	\end{split}
\end{align}
and that the form factors of the pure isospin amplitudes satisfy
\begin{align}
	\begin{split}
		F^{(0)}(s,t,u) &= F^{(0)}(s,u,t) , \\
		G^{(0)}(s,t,u) &= -G^{(0)}(s,u,t) , \\
		R^{(0)}(s,t,u) &= R^{(0)}(s,u,t) ,
	\end{split}
	\begin{split}
		F^{(1)}(s,t,u) &= -F^{(1)}(s,u,t), \\
		G^{(1)}(s,t,u) &= G^{(1)}(s,u,t), \\
		R^{(1)}(s,t,u) &= -R^{(1)}(s,u,t) .
	\end{split}
\end{align}

\subsection{$t$- and $u$-Channel}

In the crossed $t$-channel, we are concerned with the matrix element
\begin{align}
	\mathcal{A}_\mu^{+-} = \< \pi^-(p_2) \big| A_\mu(0) \big| K^+(k) \pi^-(-p_1)\> .
\end{align}
In the $u$-channel, we analogously look at
\begin{align}
	\mathcal{A}_\mu^{+-} = \< \pi^+(p_1) \big| A_\mu(0) \big| K^+(k) \pi^+(-p_2)\> .
\end{align}
Note that due to crossing, these matrix elements are described by the same function -- or its analytic continuation -- as the corresponding $s$-channel matrix element.

The $t$-channel initial and final states have the isospin decompositions
\begin{align}
	\begin{split}
		\big| K^+(k) \pi^-(-p_1) \> &= \sqrt{\frac{2}{3}} \left| \frac{1}{2}, -\frac{1}{2} \right\> + \sqrt{\frac{1}{3}} \left| \frac{3}{2}, -\frac{1}{2} \right\> , \\
		\< \pi^-(p_2) \big| A_\mu(0) &= \sqrt{\frac{2}{3}} \left\< \frac{1}{2}, -\frac{1}{2} \right| + \sqrt{\frac{1}{3}} \left\< \frac{3}{2}, -\frac{1}{2} \right| ,
	\end{split}
\end{align}
whereas in the $u$-channel, we are concerned with a pure isospin 3/2 scattering:
\begin{align}
	\begin{split}
		\big| K^+(k) \pi^+(-p_2) \> &= \left| \frac{3}{2}, \frac{3}{2} \right\> , \\
		\< \pi^+(p_1) \big| A_\mu(0) &= \left\< \frac{3}{2}, \frac{3}{2} \right| .
	\end{split}
\end{align}
We find the following isospin relation:
\begin{align}
	\begin{split}
		\mathcal{A}_\mu^{(3/2)}(k, -p_2 \to L, p_1) &= \mathcal{A}_\mu^{+-}(k, -L \to p_1, p_2) \\
			&= \frac{2}{3} \mathcal{A}_\mu^{(1/2)}(k, -p_1 \to L, p_2) + \frac{1}{3} \mathcal{A}_\mu^{(3/2)}(k, -p_1 \to L, p_2) .
	\end{split}	
\end{align}
Note that the third component of the isospin does not alter the amplitude: just insert an isospin rotation matrix together with its inverse between in- and out-state to rotate the third component.

The amplitude that describes pure isospin 1/2 scattering in the $t$-channel is then
\begin{align}
	\mathcal{A}_\mu^{(1/2)}(k, -p_1 \to L, p_2) = \frac{3}{2} \mathcal{A}_\mu^{(3/2)}(k, -p_2 \to L, p_1) - \frac{1}{2} \mathcal{A}_\mu^{(3/2)}(k, -p_1 \to L, p_2) .
\end{align}
Defining analogous form factors for the isospin 1/2 amplitude, we find
\begin{align}
	\begin{split}
		\label{eq:Isospin12FormFactors}
		F^{(1/2)}(s,t,u) &= \frac{3}{2} F(s,t,u) - \frac{1}{2} F(s,u,t) , \\
		G^{(1/2)}(s,t,u) &= \frac{3}{2} G(s,t,u) + \frac{1}{2} G(s,u,t) , \\
		R^{(1/2)}(s,t,u) &= \frac{3}{2} R(s,t,u) - \frac{1}{2} R(s,u,t) .
	\end{split}
\end{align}

In the case $s_\ell = 0$, it may be convenient to look at a certain linear combination of the form factors $F$ and $G$, as we did in \cite{Stoffer2010, Colangelo2012, Stoffer2013}:
\begin{align}
	F_1 := X F + (u-t) \frac{PL}{2X} G ,
\end{align}
where $X := \frac{1}{2}\lambda^{1/2}(M_K^2, s, s_\ell)$, $PL := \frac{1}{2}(M_K^2 - s - s_\ell)$ and $\lambda(a,b,c) := a^2 + b^2 + c^2 - 2(ab + bc + ca)$ is the Källén triangle function.

Here, too, we can define the corresponding isospin 1/2 form factor:
\begin{align}
	\begin{split}
		F_1^{(1/2)}(s,t,u) &:= X F^{(1/2)}(s,t,u) + (u-t) \frac{PL}{2X} G^{(1/2)}(s,t,u) \\
			&= \frac{3}{2} \left( X F(s,t,u) + (u-t) \frac{PL}{2X} G(s,t,u) \right) \\
			&\quad - \frac{1}{2} \left( X F(s,u,t) + (t-u) \frac{PL}{2X} G(s,u,t) \right) \\
			&= \frac{3}{2} F_1(s,t,u) - \frac{1}{2} F_1(s,u,t) .
	\end{split}
\end{align}

\clearpage

\section{Unitarity and Partial-Wave Expansion}

In this section, we will investigate the unitarity relations in the different channels and work out expansions of the form factors into partial waves with `nice' properties with respect to unitarity and analyticity: the partial waves shall satisfy Watson's final-state theorem. As we will need analytic continuations of the partial waves, we must also be careful not to introduce kinematic singularities.

The derivation of the partial-wave expansion has been done for the $s$-channel in \cite{Riggenbach1992}. We now apply the same method to all channels.

\subsection{Helicity Amplitudes}

The quantities that have a simple expansion into partial waves are not the form factors but the helicity amplitudes of the $2\to2$ scattering process \cite{Jacob1959}. However, helicity partial waves contain kinematic singularities. In order to determine them, we use the prescriptions of \cite{Martin1970}. 

We obtain the helicity amplitudes by contracting the axial vector current matrix element with the polarisation vectors of the off-shell $W$ boson. In the $W$ rest frame, the polarisation vectors are given by:
\begin{align}
	\begin{split}
		\varepsilon_t^\mu &= \left( 1, 0, 0, 0 \right) , \\
		\varepsilon_{\pm}^\mu &= \frac{1}{\sqrt{2}} \left( 0, 0, \pm 1, i \right) , \\
		\varepsilon_0^\mu &= \left( 0, 1, 0, 0 \right) .
	\end{split}
\end{align}
They are eigenvectors of the spin matrices $S^2$ and $S_1$, defined by
\begin{align}
	\begin{split}
		S_1 = \left(
				\begin{array}{cccc}
				 0 & 0 & 0 & 0 \\
				 0 & 0 & 0 & 0 \\
				 0 & 0 & 0 & -i \\
				 0 & 0 & i & 0 \\
				\end{array}
				\right) , \quad 
		S_2 &= \left(
				\begin{array}{cccc}
				 0 & 0 & 0 & 0 \\
				 0 & 0 & 0 & i \\
				 0 & 0 & 0 & 0 \\
				 0 & -i & 0 & 0 \\
				\end{array}
				\right) , \quad
		S_3 = \left(
				\begin{array}{cccc}
				 0 & 0 & 0 & 0 \\
				 0 & 0 & -i & 0 \\
				 0 & i & 0 & 0 \\
				 0 & 0 & 0 & 0 \\
				\end{array}
				\right) , \\
		S^2 = S_1^2 &+ S_2^2 + S_3^2 = \left(
				\begin{array}{cccc}
				 0 & 0 & 0 & 0 \\
				 0 & 2 & 0 & 0 \\
				 0 & 0 & 2 & 0 \\
				 0 & 0 & 0 & 2 \\
				\end{array}
				\right) .
	\end{split}
\end{align}
The eigenvalues $s(s+1)$ and $s_1$ of $S^2$ and $S_1$ are listed below:
\begin{center}
	\begin{tabular}{l r r r }
		\toprule
		 & $\varepsilon_t^\mu$ & $\varepsilon_{\pm}^\mu$ & $\varepsilon_0^\mu$ \\	
		 \hline
		$s$ & 0 & 1 & 1 \\
		$s_1$ & 0 & $\pm1$ & 0 \\
		\bottomrule
	\end{tabular}
\end{center}
If we boost the polarisation vectors into the frame where the $W$ momentum is given by $L = ( L^0, L^1, 0, 0)$, $L^2 = s_\ell$, we obtain:
\begin{align}
	\begin{split}
		\varepsilon_t^\mu &= \frac{1}{\sqrt{s_\ell}} \left( L^0, L^1, 0, 0 \right) , \\
		\varepsilon_{\pm}^\mu &= \frac{1}{\sqrt{2}} \left( 0, 0, \pm 1, i \right) , \\
		\varepsilon_0^\mu &= \frac{1}{\sqrt{s_\ell}} \left( L^1, L^0, 0, 0 \right) .
	\end{split}
\end{align}
The contractions of these basis vectors with $\mathcal{A}_\mu$ give the different helicity amplitudes:
\begin{align}
	\mathcal{A}_i := \mathcal{A}_\mu \varepsilon_i^\mu .
\end{align}

We extract the kinematic singularities by applying the recipe of \cite{Martin1970}, chapter~7.3.5, to these helicity amplitudes.

\subsection{Partial-Wave Unitarity in the $s$-Channel}

\subsubsection{Helicity Partial Waves}

The unitarity relation for the axial vector current matrix element reads
\begin{align}
	\begin{split}
		\Im\left( i \mathcal{A}_i^{(I)}(k,-L \to p_1, p_2) \right) &= \frac{1}{4} \int \widetilde{dq_1} \widetilde{dq_2} (2\pi)^4 \delta^{(4)}(p_1+p_2-q_1-q_2) \\
		&\qquad {\mathcal{T}^{(I)}}^*(q_1,q_2 \to p_1,p_2) \, i \mathcal{A}_i^{(I)}(k,-L\to q_1,q_2) ,
	\end{split}
\end{align}
where $\widetilde{dq} := \frac{d^3q}{(2\pi)^3 2 q^0}$ is the Lorentz-invariant measure and where a symmetry factor 1/2 for the pions is included. $\mathcal{T}^{(I)}$ denotes the elastic isospin $I$ $\pi\pi$-scattering amplitude. Note that this relation is valid in the physical region and that kinematic singularities have to be removed before an analytic continuation.

We perform the integrals:
\begin{align}
	\begin{split}
		\label{eq:sChannelUnitarity}
		\Im &\Big( i \mathcal{A}_i^{(I)}(k,-L \to p_1, p_2) \Big) = \frac{1}{4} (2\pi)^4 \int \widetilde{dq_1} \frac{1}{(2\pi)^3 2q_1^0} \delta(\sqrt{s} - 2 q_1^0) \\
		&\qquad {\mathcal{T}^{(I)}}^*(p_1,p_2 \to q_1,q_2) \, i \mathcal{A}_i^{(I)}(k,-L\to q_1,q_2) \\
		&= \frac{1}{16} \frac{1}{(2\pi)^2} \int_0^\infty dq \frac{q^2}{M_\pi^2+q^2} \delta( \sqrt{s} - 2\sqrt{M_\pi^2 + q^2} ) \\
		&\qquad \int d\Omega^\dprime {\mathcal{T}^{(I)}}^*(p_1,p_2 \to q_1,q_2) \, i \mathcal{A}_i^{(I)}(k,-L\to q_1,q_2) \\
		&= \frac{1}{16} \frac{1}{(2\pi)^2} \frac{1}{2} \sigma_\pi(s) \int d\Omega^\dprime \; {\mathcal{T}^{(I)}}^*(s,\cos\theta^\prime) \, i \mathcal{A}_i^{(I)}(s,\cos\theta^\dprime,\phi^\dprime) ,
	\end{split}
\end{align}
where $\sigma_\pi(s) = \sqrt{1 - 4M_\pi^2/s}$ and of course $\cos\theta^\prime$ has to be understood as a function of $\cos\theta^\dprime$ and $\phi^\dprime$ through the relation
\begin{align}
	\cos\theta^\prime = \sin\theta \sin\theta^\dprime \cos\phi^\dprime + \cos\theta \cos\theta^\dprime .
\end{align}
If we expand $\mathcal{T}$ and $\mathcal{A}_i$ into appropriate partial waves, we can perform the remaining angular integrals with the help of the relations derived in appendix~\ref{sec:LegendrePolynomials} and find the unitarity relations for the $K_{\ell4}$ partial waves.

We expand the $\pi\pi$-scattering matrix element in the usual way:
\begin{align}
	\mathcal{T}^{(I)}(s, \cos\theta^\prime) = \sum_{l=0}^\infty P_l(\cos\theta^\prime) \, t_l^I(s) 
\end{align}
with
\begin{align}
	t_l^{I}(s) = \left| t_l^{I}(s) \right| e^{i \delta_l^{I}(s)} .
\end{align}

For the expansion of the $K_{\ell4}$ matrix element into partial waves, we first consider the scalar component ($s=0$, $s_1=0$) $\mathcal{A}_t^{(I)}$. According to \cite{Jacob1959, Martin1970}, we can use normal Legendre polynomials for the partial-wave expansion ($d_{00}^{(l)}(\theta)=P_l(\cos\theta)$).

We define the partial wave as
\begin{align}
	i \mathcal{A}_t^{(I)}(s,\cos\theta) = \sum_{l=0}^\infty P_l(\cos\theta) \left( \frac{\lambda^{1/2}_{K\ell}(s) \sigma_\pi(s)}{M_K^2} \right)^l \; a^{(I)}_{t, l}(s) .
\end{align}
The square root of the Källén function cancels exactly the square root branch cut in the Legendre polynomial between $(M_K - \sqrt{s_\ell})^2$ and $(M_K + \sqrt{s_\ell})^2$. Note that $P_l(z)$ contains only even powers of $z$ for even $l$ and vice versa. The factor $M_K^2$ in the denominator appears only for dimensional reasons.

In the region $s>4M_\pi^2$, no kinematic cuts are introduced into the partial waves $a^{(I)}_{t,l}(s)$:
\begin{align}
	\Im \left( i \mathcal{A}_t^{(I)}(s,\cos\theta) \right) = \sum_{l=0}^\infty P_l(\cos\theta) \left( \frac{\lambda^{1/2}_{K\ell}(s) \sigma_\pi(s)}{M_K^2} \right)^l \; \Im \left( a^{(I)}_{t,l}(s) \right) .
\end{align}
We can use this partial-wave expansion for an analytic continuation from the decay region through the unphysical to the scattering region.

We insert the partial-wave expansion into the unitarity equation~(\ref{eq:sChannelUnitarity}) and make use of the relations derived in appendix~\ref{sec:LegendrePolynomials}:
\begin{align}
	\begin{split}
		\Im\Big( i \mathcal{A}_t^{(I)}&(k,-L \to p_1, p_2) \Big) = \sum_{l=0}^\infty P_l(\cos\theta) \left( \frac{\lambda^{1/2}_{K\ell}(s) \sigma_\pi(s)}{M_K^2} \right)^l \; \Im \left( a^{(I)}_{t,l}(s) \right) \\
		&= \frac{1}{16} \frac{1}{(2\pi)^2} \frac{1}{2} \sigma_\pi(s) \int d\Omega^\dprime \; \sum_{l^\prime=0}^\infty P_{l^\prime}(\cos\theta^\prime) \, {t_{l^\prime}^I}^*(s) \\
		&\quad \sum_{l=0}^\infty P_l(\cos\theta^\dprime) \left( \frac{\lambda^{1/2}_{K\ell}(s) \sigma_\pi(s)}{M_K^2} \right)^l \; a^{(I)}_{t,l}(s) \\
		&= \frac{1}{16} \frac{1}{(2\pi)^2} \frac{1}{2} \sigma_\pi(s) \sum_{l^\prime=0}^\infty \, {t_{l^\prime}^I}^*(s) \sum_{l=0}^\infty \left( \frac{\lambda^{1/2}_{K\ell}(s) \sigma_\pi(s)}{M_K^2} \right)^l \; a^{(I)}_{t,l}(s) \\
		&\quad \int d\Omega^\dprime P_{l^\prime}(\cos\theta^\prime) P_l(\cos\theta^\dprime) \\
		&= \sum_{l=0}^\infty P_l(\cos\theta) \left( \frac{\lambda^{1/2}_{K\ell}(s) \sigma_\pi(s)}{M_K^2} \right)^l  \frac{1}{2l+1} \frac{1}{32\pi} \sigma_\pi(s) \, {t_l^I}^*(s)\; a^{(I)}_{t,l}(s) .
	\end{split}
\end{align}
The unitarity relations for the partial waves emerges:
\begin{align}
	\Im \left( a^{(I)}_{t,l}(s) \right) =  \frac{1}{2l+1} \frac{1}{32\pi} \sigma_\pi(s) \, {t_l^I}^*(s)\; a^{(I)}_{t,l}(s) .
\end{align}
In particular, we find that the phases of the $K_{\ell4}$ $s$-channel partial waves are given by the elastic $\pi\pi$-scattering phases (this is Watson's theorem) for all $s$ between $4M_\pi^2$ and some inelastic threshold:
\begin{align}
	a^{(I)}_{t,l}(s) = \left| a^{(I)}_{t,l}(s) \right| e^{i \delta_l^I(s)} .
\end{align}

Let us consider next the longitudinal component $\mathcal{A}_0^{(I)}$. Here, we can still use ordinary Legendre polynomials for the partial-wave expansion. However, the helicity amplitude itself will possess a kinematic singularity: there is a square root branch cut between $(M_K-\sqrt{s_\ell})^2$ and $(M_K+\sqrt{s_\ell})^2$. Hence, we define the partial waves as
\begin{align}
	i \mathcal{A}_0^{(I)}(s,\cos\theta) = i \mathcal{\tilde A}_0^{(I)} \frac{\lambda^{1/2}_{K\ell}(s)}{M_K^2}  = \frac{\lambda^{1/2}_{K\ell}(s)}{M_K^2} \sum_{l=0}^\infty P_l(\cos\theta) \left( \frac{\lambda^{1/2}_{K\ell}(s) \sigma_\pi(s)}{M_K^2} \right)^l \; a^{(I)}_{0, l}(s) .
\end{align}
The partial waves $a^{(I)}_{0,l}(s)$ are again free of kinematic singularities (at least in the region $s>4M_\pi^2$) and satisfy Watson's theorem:
\begin{align}
	\begin{split}
		\Im \left( a^{(I)}_{0,l}(s) \right) &=  \frac{1}{2l+1} \frac{1}{32\pi} \sigma_\pi(s) \, {t_l^I}^*(s)\; a^{(I)}_{0,l}(s) , \\
		a^{(I)}_{0,l}(s) &= \left| a^{(I)}_{0,l}(s) \right| e^{i \delta_l^I(s)} .
	\end{split}
\end{align}

As for the transverse components $\mathcal{A}_{\pm}^{(I)}$, we consider the combination $\mathcal{A}_{2}^{(I)} := \mathcal{A}_{+}^{(I)} - \mathcal{A}_-^{(I)}$. This amplitude should not be expanded into a series of Legendre polynomials but of Wigner $d$-functions $d_{10}^{(l)}(\theta) = -[l(l+1)]^{-1/2} \sin\theta \, P_l^\prime(\cos\theta)$.

In this spin $0,1\to0,0$ scattering amplitude $\mathcal{A}_2^{(I)}$, there is a singularity at the border of the physical region (where $\sin\theta = 0$), which can be removed by defining
\begin{align}
	i \mathcal{A}_2^{(I)} =: i \mathcal{\tilde A}_2^{(I)} \sin\theta .
\end{align}
If we define the partial-wave expansion as
\begin{align}
	\begin{split}
		i \mathcal{A}_2^{(I)}(s,\cos\theta,\phi) &= i \mathcal{\tilde A}_2^{(I)} \sin\theta \\
			&= \sin\theta \sum_{l=1}^\infty P_l^\prime(\cos\theta) \left( \frac{\lambda^{1/2}_{K\ell}(s) \sigma_\pi(s)}{M_K^2} \right)^{l-1} \cos\phi \; a_{2,l}^{(I)}(s) ,
	\end{split}
\end{align}
then the partial waves $a^{(I)}_{2,l}$ will be free of kinematic singularities.

We insert this partial-wave expansion into the unitarity relation:
\begin{align}
	\begin{split}
		\Im&\Big( i \mathcal{A}_2^{(I)}(k,-L \to p_1, p_2) \Big)  =  \sin\theta \sum_{l=1}^\infty P_l^\prime(\cos\theta) \left( \frac{\lambda^{1/2}_{K\ell}(s) \sigma_\pi(s)}{M_K^2} \right)^{l-1} \Im \left( a_{2,l}^{(I)}(s) \right) \\
		&= \frac{1}{16} \frac{1}{(2\pi)^2} \frac{1}{2} \sigma_\pi(s) \int d\Omega^\dprime \; \sum_{l^\prime=0}^\infty P_{l^\prime}(\cos\theta^\prime) \, {t_{l^\prime}^I}^*(s) \\
		&\quad \sin\theta^\dprime \sum_{l=1}^\infty P_l^\prime(\cos\theta^\dprime) \left( \frac{\lambda^{1/2}_{K\ell}(s) \sigma_\pi(s)}{M_K^2} \right)^{l-1} \cos\phi^\dprime \; a_{2,l}^{(I)}(s) \\
		&= \frac{1}{16} \frac{1}{(2\pi)^2} \frac{1}{2} \sigma_\pi(s) \sum_{l^\prime=0}^\infty \, {t_{l^\prime}^I}^*(s) \sum_{l=1}^\infty \left( \frac{\lambda^{1/2}_{K\ell}(s) \sigma_\pi(s)}{M_K^2} \right)^{l-1} \; a_{2,l}^{(I)}(s) \\
		&\quad \int d\Omega^\dprime P_{l^\prime}(\cos\theta^\prime) P_l^\prime(\cos\theta^\dprime) \sin\theta^\dprime \cos\phi^\dprime \\
		&= \sin\theta \sum_{l=1}^\infty P_l^\prime(\cos\theta) \left( \frac{\lambda^{1/2}_{K\ell}(s) \sigma_\pi(s)}{M_K^2} \right)^{l-1} \frac{1}{2l+1} \frac{1}{32\pi} \sigma_\pi(s) \, {t_l^I}^*(s) \; a_{2,l}^{(I)}(s) ,
	\end{split}
\end{align}
hence
\begin{align}
	\Im \left( a_{2,l}^{(I)}(s) \right) =  \frac{1}{2l+1} \frac{1}{32\pi} \sigma_\pi(s) \, {t_l^I}^*(s)\; a_{2,l}^{(I)}(s) .
\end{align}
This is the equation that can be analytically continued to the region $s>4M_\pi^2$ since the kinematic singularities have been factored out.

Again, Watson's theorem holds for $s$ between $4M_\pi^2$ and the inelastic threshold:
\begin{align}
	a_{2,l}^{(I)}(s) = \left| a_{2,l}^{(I)}(s) \right| e^{i \delta_l^I(s)} .
\end{align}

\subsubsection{Partial-Wave Expansion of the Form Factors in the $s$-Channel}

In order to find the partial-wave expansions of the form factors, we write explicitly the components of the axial vector current matrix element in the $\pi\pi$ centre-of-mass frame:
\begin{align}
	\begin{split}
		i \mathcal{A}^\mu_{(I)}(k,-L \to p_1, p_2) = \frac{1}{M_K} \bigg( & \sqrt{s} \; F^{(I)} + \frac{M_K^2 - s - s_\ell}{2\sqrt{s}} \; R^{(I)} , \\
			& \sqrt{s} \sigma_\pi(s) \cos\theta \; G^{(I)} - \frac{1}{2\sqrt{s}} \lambda^{1/2}_{K\ell}(s) \; R^{(I)} , \\
			& \sqrt{s} \sigma_\pi(s) \sin\theta \; G^{(I)} , \\
			& 0 \bigg) ,
	\end{split} \\
		\begin{split}
		i \mathcal{A}^\mu_{(I)}(k,-L \to q_1, q_2) = \frac{1}{M_K} \bigg( & \sqrt{s} \; F^{(I)} + \frac{M_K^2 - s - s_\ell}{2\sqrt{s}} \; R^{(I)} , \\
			& \sqrt{s} \sigma_\pi(s) \cos\theta^\dprime \; G^{(I)} - \frac{1}{2\sqrt{s}} \lambda^{1/2}_{K\ell}(s) \; R^{(I)} , \\
			& \sqrt{s} \sigma_\pi(s) \sin\theta^\dprime \cos\phi^\dprime \; G^{(I)} , \\
			& \sqrt{s} \sigma_\pi(s) \sin\theta^\dprime \sin\phi^\dprime \; G^{(I)} \bigg) .
	\end{split}
\end{align}

By contracting these expressions with the polarisation vectors, we find the helicity amplitudes (generalised to a generic $\phi$):
\begin{align}
	\begin{split}
		i \mathcal{A}_t^{(I)} &= i \mathcal{A}_\mu^{(I)} \varepsilon_t^\mu = i \mathcal{A}_\mu^{(I)} \frac{1}{\sqrt{s_\ell}} L^\mu \\
			&= \frac{1}{M_K \sqrt{s_\ell}} \bigg( \frac{1}{2} ( M_K^2 - s - s_\ell ) \; F^{(I)} + \frac{1}{2} \sigma_\pi(s) \lambda^{1/2}_{K\ell}(s) \cos\theta \; G^{(I)} + s_\ell \; R^{(I)} \bigg) , \\
		i \mathcal{A}_0^{(I)} &= i \mathcal{A}_\mu^{(I)} \varepsilon_0^\mu \\
			&= \frac{-1}{M_K \sqrt{s_\ell}} \bigg( \frac{1}{2} \lambda^{1/2}_{K\ell}(s) \; F^{(I)} + \frac{1}{2} (M_K^2 - s - s_\ell) \sigma_\pi(s) \cos\theta \; G^{(I)} \bigg) , \\
		i \mathcal{A}_2^{(I)} &= i \mathcal{A}_\mu^{(I)} \varepsilon_+^\mu - i \mathcal{A}_\mu^{(I)} \varepsilon_-^\mu \\
			&= \frac{-\sqrt{2}}{M_K} \bigg( \sqrt{s} \sigma_\pi(s) \sin\theta \cos\phi \; G^{(I)} \bigg) .
	\end{split}
\end{align}

Since the contribution of the form factor $R$ to the decay rate is suppressed by $m_\ell^2$, it is invisible in the electron mode and we do not have any data on it. We therefore look only for linear combinations of the form factors $F$ and $G$ that possess a simple partial-wave expansion. We find:
\begin{align}
	\begin{split}
		F^{(I)} &+ \frac{\sigma_\pi(s) PL(s)}{X(s)} \cos\theta \; G^{(I)}  = F^{(I)} + \frac{(M_K^2-s-s_\ell) (u-t)}{\lambda_{K\ell}(s)} \; G^{(I)} \\
			&= - \frac{2 \sqrt{s_\ell}}{M_K} \sum_{l=0}^\infty P_l(\cos\theta) \left( \frac{\lambda^{1/2}_{K\ell}(s) \sigma_\pi(s)}{M_K^2} \right)^l a_{0, l}^{(I)}(s) , \\
		G^{(I)} &= - \frac{M_K}{\sqrt{2s} \sigma_\pi(s)} \sum_{l=1}^\infty P_l^\prime(\cos\theta) \left( \frac{\lambda^{1/2}_{K\ell}(s) \sigma_\pi(s)}{M_K^2} \right)^{l-1} a_{2,l}^{(I)}(s) .
	\end{split}
\end{align}

We write the partial-wave expansions of $F$ and $G$ in the form:
\begin{align}
	\begin{split}
		\label{eq:sChannelFormFactorPartialWaveExpansion}
		F^{(I)} &= \sum_{l=0}^\infty P_l(\cos\theta) \left( \frac{\lambda^{1/2}_{K\ell}(s) \sigma_\pi(s)}{M_K^2}\right)^l f_l^{(I)}(s) - \frac{\sigma_\pi PL}{X} \cos\theta \; G^{(I)}, \\
		G^{(I)} &= \sum_{l=1}^\infty P_l^\prime(\cos\theta) \left( \frac{\lambda^{1/2}_{K\ell}(s) \sigma_\pi(s)}{M_K^2}\right)^{l-1} g_l^{(I)}(s) ,
	\end{split}
\end{align}
where the partial waves $f_l^{(I)}$ and $g_l^{(I)}$ satisfy Watson's theorem in the region $s>4M_\pi^2$:
\begin{align}
	f_l^{(I)}(s) = \left| f_l^{(I)}(s) \right| e^{i \delta_l^I(s)} \quad , \quad g_l^{(I)}(s) = \left| g_l^{(I)}(s) \right| e^{i \delta_l^I(s)} .
\end{align}

\subsection{Partial-Wave Unitarity in the $t$-Channel}

\subsubsection{Helicity Partial Waves}

The discussion in the crossed channels is a bit simpler because we are interested in partial-wave expansions only in the region $t > (M_K+M_\pi)^2$ or $u > (M_K + M_\pi)^2$, i.e.~above all initial and final state thresholds and pseudo-thresholds. Therefore, we do not have to worry about kinematic singularities, since we will not perform analytic continuations into the critical regions.

In the crossed channels, we consider $K\pi$ intermediate states in the unitarity relation:
\begin{align}
	\begin{split}
		\Im\left( i \mathcal{A}_i^{(1/2)}(k,-p_1 \to L, p_2) \right) &= \frac{1}{2} \int \widetilde{dq_K} \widetilde{dq_\pi} (2\pi)^4 \delta^{(4)}(k-p_1-q_K-q_\pi) \\
		& {\mathcal{T}^{(1/2)}}^*(q_K,q_\pi \to k,-p_1) \, i \mathcal{A}_i^{(1/2)}(q_K,q_\pi \to L,p_2) ,
	\end{split}
\end{align}
where $\mathcal{T}^{(1/2)}$ is the isospin $1/2$ elastic $K\pi$-scattering amplitude. By performing the integrals we obtain:
\begin{align}
	\begin{split}
		\label{eq:tChannelUnitarity}
		\Im &\Big( i \mathcal{A}_i^{(1/2)}(k,-p_1 \to L, p_2) \Big) = \frac{1}{2} (2\pi)^4 \int \widetilde{dq_K} \frac{1}{(2\pi)^3 2q_\pi^0} \delta(\sqrt{t} - q_K^0 - q_\pi^0) \\
		&\qquad {\mathcal{T}^{(1/2)}}^*(q_K,q_\pi \to k, -p_1) \, i \mathcal{A}_i^{(1/2)}(q_K,q_\pi \to L,p_2) \\
		&= \frac{1}{8} \frac{1}{(2\pi)^2} \int_0^\infty dq \frac{q^2}{\sqrt{M_\pi^2+q^2}\sqrt{M_K^2+q^2}} \delta( \sqrt{t} - \sqrt{M_\pi^2 + q^2} - \sqrt{M_K^2 + q^2} ) \\
		&\qquad \int d\Omega_t^\dprime {\mathcal{T}^{(1/2)}}^*(q_K,q_\pi \to k, -p_1) \, i \mathcal{A}_i^{(1/2)}(q_K,q_\pi \to L,p_2) \\
		&= \frac{1}{8} \frac{1}{(2\pi)^2} \frac{\lambda^{1/2}_{K\pi}(t)}{2t} \int d\Omega_t^\dprime \; {\mathcal{T}^{(1/2)}}^*(t,\cos\theta_t^\prime) \, i \mathcal{A}_i^{(1/2)}(t,\cos\theta_t^\dprime,\phi_t^\dprime) .
	\end{split}
\end{align}

The $K\pi$ scattering matrix element is expanded in the usual way:
\begin{align}
	\mathcal{T}^{(1/2)}(t,\cos\theta_t) = \sum_{l=0}^\infty P_l(\cos\theta_t) t_l^{1/2}(t)
\end{align}
with
\begin{align}
	t_l^{1/2}(t) = \left| t_l^{1/2}(t) \right| e^{i \delta_l^{1/2}(t)} .
\end{align}

We expand the $K_{\ell4}$ helicity amplitudes as follows:
\begin{align}
	\begin{split}
		i \mathcal{A}_t^{(1/2)}(t,\cos\theta_t) &= \sum_{l=0}^\infty P_l(\cos\theta_t) \left( \frac{\lambda^{1/2}_{K\pi}(t) \lambda^{1/2}_{\ell\pi}(t)}{M_K^4} \right)^l a_{t,l}^{(1/2)}(t) , \\
		i \mathcal{A}_0^{(1/2)}(t,\cos\theta_t) &= \sum_{l=0}^\infty P_l(\cos\theta_t) \left( \frac{\lambda^{1/2}_{K\pi}(t) \lambda^{1/2}_{\ell\pi}(t)}{M_K^4} \right)^l a_{0,l}^{(1/2)}(t) , \\
		i \mathcal{A}_2^{(1/2)}(t,\cos\theta_t,\phi_t) &= i \mathcal{A}_+^{(1/2)}(t,\cos\theta_t,\phi_t) - i \mathcal{A}_-^{(1/2)}(t,\cos\theta_t,\phi_t) \\
			&= \sin\theta_t \cos\phi_t \sum_{l=1}^\infty P_l^\prime(\cos\theta_t) \left( \frac{\lambda^{1/2}_{K\pi}(t) \lambda^{1/2}_{\ell\pi}(t)}{M_K^4} \right)^{l-1} a_{2,l}^{(1/2)}(t) .
	\end{split}
\end{align}
By inserting these expansions into the unitarity relation (\ref{eq:tChannelUnitarity}) we find:
\begin{align}
	\begin{split}
		\Im\Big( i &\mathcal{A}_i^{(1/2)}(k,-p_1\to L,p_2) \Big) = \sum_{l=0}^\infty P_l(\cos\theta_t) \left( \frac{\lambda^{1/2}_{K\pi}(t) \lambda^{1/2}_{\ell\pi}(t)}{M_K^4} \right)^l \Im\left( a_{i,l}^{(1/2)}(t) \right) \\
			&= \frac{1}{8} \frac{1}{(2\pi)^2} \frac{\lambda^{1/2}_{K\pi}(t)}{2t} \int d\Omega_t^\dprime \; \sum_{{l^\prime}=0}^\infty P_{l^\prime}(\cos\theta_t^\prime) {t_{l^\prime}^{1/2}}^*(t) \\
				&\quad \sum_{l=0}^\infty P_l(\cos\theta_t^\dprime) \left( \frac{\lambda^{1/2}_{K\pi}(t) \lambda^{1/2}_{\ell\pi}(t)}{M_K^4} \right)^l a_{i,l}^{(1/2)}(t) \\
			&= \frac{1}{8} \frac{1}{(2\pi)^2} \frac{\lambda^{1/2}_{K\pi}(t)}{2t} \sum_{{l^\prime}=0}^\infty \sum_{l=0}^\infty {t_{l^\prime}^{1/2}}^*(t) \left( \frac{\lambda^{1/2}_{K\pi}(t) \lambda^{1/2}_{\ell\pi}(t)}{M_K^4} \right)^l a_{i,l}^{(1/2)}(t) \\
				&\quad \int d\Omega_t^\dprime \; P_{l^\prime}(\cos\theta_t^\prime) \, P_l(\cos\theta_t^\dprime) \\
			&= \sum_{l=0}^\infty P_l(\cos\theta_t) \left( \frac{\lambda^{1/2}_{K\pi}(t) \lambda^{1/2}_{\ell\pi}(t)}{M_K^4} \right)^l \frac{1}{2l+1}\frac{1}{16\pi} \frac{\lambda^{1/2}_{K\pi}(t)}{t} \; {t_l^{1/2}}^*(t) a_{i,l}^{(1/2)}(t) ,
	\end{split}
\end{align}
where $i=t,0$. For $i=2$, we obtain:
\begin{align}
	\begin{split}
		\Im\Big( i &\mathcal{A}_2^{(1/2)}(k,-p_1\to L,p_2) \Big) = \sin\theta_t \sum_{l=1}^\infty P_l^\prime(\cos\theta_t) \left( \frac{\lambda^{1/2}_{K\pi}(t) \lambda^{1/2}_{\ell\pi}(t)}{M_K^4} \right)^{l-1} \Im\left( a_{2,l}^{(1/2)}(t) \right) \\
			&= \frac{1}{8} \frac{1}{(2\pi)^2} \frac{\lambda^{1/2}_{K\pi}(t)}{2t} \int d\Omega_t^\dprime \; \sum_{{l^\prime}=0}^\infty P_{l^\prime}(\cos\theta_t^\prime) {t_{l^\prime}^{1/2}}^*(t) \\
			& \qquad \sin\theta_t^\dprime \cos\phi_t^\dprime \sum_{l=1}^\infty P_l^\prime(\cos\theta_t^\dprime) \left( \frac{\lambda^{1/2}_{K\pi}(t) \lambda^{1/2}_{\ell\pi}(t)}{M_K^4} \right)^{l-1} a_{2,l}^{(1/2)}(t) \\
			&= \frac{1}{8} \frac{1}{(2\pi)^2} \frac{\lambda^{1/2}_{K\pi}(t)}{2t} \sum_{{l^\prime}=0}^\infty \sum_{l=1}^\infty {t_{l^\prime}^{1/2}}^*(t) \left( \frac{\lambda^{1/2}_{K\pi}(t) \lambda^{1/2}_{\ell\pi}(t)}{M_K^4} \right)^{l-1} a_{2,l}^{(1/2)}(t) \\
			& \qquad \int d\Omega_t^\dprime \; \sin\theta_t^\dprime \cos\phi_t^\dprime P_{l^\prime}(\cos\theta_t^\prime) \, P_l^\prime(\cos\theta_t^\dprime) \\
			&= \sin\theta_t \sum_{l=1}^\infty P_l^\prime(\cos\theta_t) \left( \frac{\lambda^{1/2}_{K\pi}(t) \lambda^{1/2}_{\ell\pi}(t)}{M_K^4} \right)^{l-1}  \frac{1}{2l+1}\frac{1}{16\pi} \frac{\lambda^{1/2}_{K\pi}(t)}{t} \; {t_l^{1/2}}^*(t) a_{2,l}^{(1/2)}(t) .
	\end{split}
\end{align}
Hence, all the partial waves satisfy Watson's theorem ($i=t,0,2$):
\begin{align}
	\begin{split}
		\Im\left( a_{i,l}^{(1/2)}(t) \right) &= \frac{1}{2l+1}\frac{1}{16\pi} \frac{\lambda^{1/2}_{K\pi}(t)}{t} \; {t_l^{1/2}}^*(t) a_{i,l}^{(1/2)}(t) , \\
		a_{i,l}^{(1/2)}(t) &= \left| a_{i,l}^{(1/2)}(t) \right| e^{i\delta_l^{1/2}(t)} .
	\end{split}
\end{align}

\subsubsection{Partial-Wave Expansion of the Form Factors in the $t$-Channel}

The components of the axial vector current matrix element in the $t$-channel $K\pi$ centre-of-mass frame are given by:
\begin{align}
	\begin{split}
		i \mathcal{A}_{(1/2)}^\mu&(k,-p_1 \to L,p_2) =  \\
		\frac{1}{M_K} &\Bigg( \frac{M_K^2 - s_\ell}{2\sqrt{t}} \; F^{(1/2)}  + \frac{M_K^2-2M_\pi^2 - 2t + s_\ell}{2\sqrt{t}} \; G^{(1/2)} + \frac{t+s_\ell - M_\pi^2}{2\sqrt{t}} \; R^{(1/2)} , \\
			&  \frac{\lambda^{1/2}_{\ell\pi}(t)}{2\sqrt{t}} \; \left( F^{(1/2)} - G^{(1/2)} - R^{(1/2)} \right) - \frac{\lambda^{1/2}_{K\pi}(t)}{2\sqrt{t}} \cos\theta_t \; \left( F^{(1/2)} + G^{(1/2)} \right) , \\
			& - \frac{\lambda^{1/2}_{K\pi}(t)}{2\sqrt{t}} \sin\theta_t \; \left( F^{(1/2)} + G^{(1/2)} \right) , \\
			& 0 \Bigg) ,
	\end{split} \\
	\begin{split}
		i \mathcal{A}_{(1/2)}^\mu&(q_K,q_\pi \to L,p_2) =  \\
		\frac{1}{M_K} &\Bigg( \frac{M_K^2 - s_\ell}{2\sqrt{t}} \; F^{(1/2)} + \frac{M_K^2-2M_\pi^2-2t + s_\ell}{2\sqrt{t}} \; G^{(1/2)} + \frac{t+s_\ell - M_\pi^2}{2\sqrt{t}} \; R^{(1/2)} , \\
			& \frac{\lambda^{1/2}_{\ell\pi}(t)}{2\sqrt{t}} \; \left( F^{(1/2)} - G^{(1/2)} - R^{(1/2)} \right)  - \frac{\lambda^{1/2}_{K\pi}(t)}{2\sqrt{t}} \cos\theta_t^\dprime \; \left( F^{(1/2)} + G^{(1/2)} \right) , \\
			& - \frac{\lambda^{1/2}_{K\pi}(t)}{2\sqrt{t}} \sin\theta_t^\dprime \cos\phi_t^\dprime \; \left( F^{(1/2)} + G^{(1/2)} \right) , \\
			& - \frac{\lambda^{1/2}_{K\pi}(t)}{2\sqrt{t}} \sin\theta_t^\dprime \sin\phi_t^\dprime \; \left( F^{(1/2)} + G^{(1/2)} \right) \Bigg) .
	\end{split}
\end{align}
By contraction with the polarisation vectors, we find the helicity amplitudes (again for a generic $\phi_t$). As we are not interested in $R$, we do not need the $\mathcal{A}_t^{(1/2)}$ component:
\begin{align}
	\begin{split}
		i \mathcal{A}_0^{(1/2)} &= i \mathcal{A}^{(1/2)}_\mu \varepsilon_0^\mu \\
			&= \frac{-1}{M_K \sqrt{s_\ell}} \Bigg( \frac{1}{4t} \left( \lambda^{1/2}_{K\pi}(t) (M_\pi^2-s_\ell-t) \cos\theta_t + \lambda^{1/2}_{\ell\pi}(t) (M_K^2 - M_\pi^2 + t) \right) F^{(1/2)} \\
			& \quad + \frac{1}{4t} \left( \lambda^{1/2}_{K\pi}(t) (M_\pi^2-s_\ell-t) \cos\theta_t + \lambda^{1/2}_{\ell\pi}(t) (M_K^2 - M_\pi^2 - 3t) \right) G^{(1/2)} \Bigg) , \\
		i \mathcal{A}_2^{(1/2)} &= i \mathcal{A}^{(1/2)}_\mu \varepsilon_+^\mu - i \mathcal{A}^{(1/2)}_\mu \varepsilon_-^\mu \\
			&= \frac{1}{\sqrt{2}M_K} \left( \frac{\lambda^{1/2}_{K\pi}(t)}{\sqrt{t}} \sin\theta_t \cos\phi_t \; \left( F^{(1/2)} + G^{(1/2)} \right) \right) .
	\end{split}
\end{align}
This results in the following partial-wave expansions of the form factors:
\begin{align}
	\begin{split}
		\label{eq:tChannelFormFactorPartialWaveExpansion}
		F^{(1/2)} - G^{(1/2)} &+ \frac{1}{2t} \left( M_K^2 - M_\pi^2 - t + (M_\pi^2 - s_\ell - t) \frac{\lambda^{1/2}_{K\pi}(t)}{\lambda^{1/2}_{\ell\pi}(t)} \cos\theta_t \right) \left( F^{(1/2)} + G^{(1/2)} \right) \\
		&= \frac{-2M_K \sqrt{s_\ell}}{\lambda^{1/2}_{\ell\pi}(t)} \sum_{l=0}^\infty P_l(\cos\theta_t) \left( \frac{\lambda^{1/2}_{K\pi}(t) \lambda^{1/2}_{\ell\pi}(t)}{M_K^4} \right)^l a_{0,l}^{(1/2)}(t) , \\
		F^{(1/2)} + G^{(1/2)} &= \frac{M_K\sqrt{2t}}{\lambda^{1/2}_{K\pi}(t)} \sum_{l=1}^\infty P_l^\prime(\cos\theta_t) \left( \frac{\lambda^{1/2}_{K\pi}(t) \lambda^{1/2}_{\ell\pi}(t)}{M_K^4} \right)^{l-1} a_{2,l}^{(1/2)}(t) ,
	\end{split}
\end{align}
which we write conveniently in the form
\begin{align}
	\begin{split}
		F^{(1/2)} - G^{(1/2)} &= 2 \sum_{l=0}^\infty P_l(\cos\theta_t) \left( \frac{\lambda^{1/2}_{K\pi}(t) \lambda^{1/2}_{\ell\pi}(t)}{M_K^4} \right)^l f_l^{(1/2)}(t) \\
			&- \frac{1}{2t} \left( M_K^2 - M_\pi^2 - t + (M_\pi^2 - s_\ell - t) \frac{\lambda^{1/2}_{K\pi}(t)}{\lambda^{1/2}_{\ell\pi}(t)} \cos\theta_t \right) \left( F^{(1/2)} + G^{(1/2)} \right) , \\
		F^{(1/2)} + G^{(1/2)} &= 2 \sum_{l=1}^\infty P_l^\prime(\cos\theta_t) \left( \frac{\lambda^{1/2}_{K\pi}(t) \lambda^{1/2}_{\ell\pi}(t)}{M_K^4} \right)^{l-1} g_l^{(1/2)}(t) ,
	\end{split}
\end{align}
or
\begin{align}
	\begin{split}
		\label{eq:tChannelFormFactorPartialWaveExpansion2}
		F^{(1/2)} &= \sum_{l=0}^\infty P_l(\cos\theta_t) \left( \frac{\lambda^{1/2}_{K\pi}(t) \lambda^{1/2}_{\ell\pi}(t)}{M_K^4} \right)^l f_l^{(1/2)}(t) \\
			&\quad - \frac{1}{2t} \left( M_K^2 - M_\pi^2 - 3 t + (M_\pi^2 - s_\ell - t) \frac{\lambda^{1/2}_{K\pi}(t)}{\lambda^{1/2}_{\ell\pi}(t)} \cos\theta_t \right) \\
			&\quad \sum_{l=1}^\infty P_l^\prime(\cos\theta_t) \left( \frac{\lambda^{1/2}_{K\pi}(t) \lambda^{1/2}_{\ell\pi}(t)}{M_K^4} \right)^{l-1} g_l^{(1/2)}(t) , \\
		G^{(1/2)} &= -\sum_{l=0}^\infty P_l(\cos\theta_t) \left( \frac{\lambda^{1/2}_{K\pi}(t) \lambda^{1/2}_{\ell\pi}(t)}{M_K^4} \right)^l f_l^{(1/2)}(t) \\
			&\quad+ \frac{1}{2t} \left( M_K^2 - M_\pi^2 + t + (M_\pi^2 - s_\ell - t) \frac{\lambda^{1/2}_{K\pi}(t)}{\lambda^{1/2}_{\ell\pi}(t)} \cos\theta_t \right) \\
			&\quad \sum_{l=1}^\infty P_l^\prime(\cos\theta_t) \left( \frac{\lambda^{1/2}_{K\pi}(t) \lambda^{1/2}_{\ell\pi}(t)}{M_K^4} \right)^{l-1} g_l^{(1/2)}(t) ,
	\end{split}
\end{align}
where also these new partial waves $f_l^{(1/2)}$ and $g_l^{(1/2)}$ satisfy Watson's theorem in the region $t>(M_K+M_\pi)^2$:
\begin{align}
	f_l^{(1/2)}(t) = \left| f_l^{(1/2)}(t) \right| e^{i \delta_l^{1/2}(t)} \quad , \quad g_l^{(1/2)}(t) = \left| g_l^{(1/2)}(t) \right| e^{i \delta_l^{1/2}(t)} .
\end{align}

\subsection{Partial-Wave Unitarity in the $u$-Channel}

\subsubsection{Helicity Partial Waves}

The $u$-channel (i.e.~the isospin $3/2$ case) can be treated in complete analogy to the $t$-channel. Starting with the unitarity relation
\begin{align}
	\begin{split}
		\Im\left( i \mathcal{A}_i^{(3/2)}(k,-p_2 \to L, p_1) \right) &= \frac{1}{2} \int \widetilde{dq_K} \widetilde{dq_\pi} (2\pi)^4 \delta^{(4)}(k-p_2-q_K-q_\pi) \\
		& {\mathcal{T}^{(3/2)}}^*(q_K,q_\pi \to k,-p_2) \, i \mathcal{A}_i^{(3/2)}(q_K,q_\pi \to L,p_1) ,
	\end{split}
\end{align}
we find again
\begin{align}
	\begin{split}
		\label{eq:uChannelUnitarity}
		\Im &\Big( i \mathcal{A}_i^{(3/2)}(k,-p_2 \to L, p_1) \Big) \\
		&= \frac{1}{8} \frac{1}{(2\pi)^2} \frac{\lambda^{1/2}_{K\pi}(u)}{2u} \int d\Omega_u^\dprime \; {\mathcal{T}^{(3/2)}}^*(u,\cos\theta_u^\prime) \, i \mathcal{A}_i^{(3/2)}(u,\cos\theta_u^\dprime,\phi_u^\dprime) .
	\end{split}
\end{align}
The $K\pi$-scattering matrix element is expanded as
\begin{align}
	\begin{split}
		\mathcal{T}^{(3/2)}(u,\cos\theta_u) &= \sum_{l=0}^\infty P_l(\cos\theta_u) t_l^{3/2}(u) , \\
		t_l^{3/2}(u) &= \left| t_l^{3/2}(u) \right| e^{i \delta_l^{3/2}(u)}
	\end{split}
\end{align}
and the $K_{\ell4}$ helicity amplitudes according to
\begin{align}
	\begin{split}
		i \mathcal{A}_t^{(3/2)}(u,\cos\theta_u) &= \sum_{l=0}^\infty P_l(\cos\theta_u)  \left( \frac{\lambda^{1/2}_{K\pi}(u) \lambda^{1/2}_{\ell\pi}(u)}{M_K^4} \right)^l  a_{t,l}^{(3/2)}(u) , \\
		i \mathcal{A}_0^{(3/2)}(u,\cos\theta_u) &= \sum_{l=0}^\infty P_l(\cos\theta_u) \left( \frac{\lambda^{1/2}_{K\pi}(u) \lambda^{1/2}_{\ell\pi}(u)}{M_K^4} \right)^l a_{0,l}^{(3/2)}(u) , \\
		i \mathcal{A}_2^{(3/2)}(u,\cos\theta_u,\phi_u) &= i \mathcal{A}_+^{(3/2)}(u,\cos\theta_u,\phi_u) - i \mathcal{A}_-^{(3/2)}(u,\cos\theta_u,\phi_u) \\
			&= \sin\theta_u \cos\phi_u \sum_{l=1}^\infty P_l^\prime(\cos\theta_u) \left( \frac{\lambda^{1/2}_{K\pi}(u) \lambda^{1/2}_{\ell\pi}(u)}{M_K^4} \right)^{l-1} a_{2,l}^{(3/2)}(u) .
	\end{split}
\end{align}
Performing the same calculation steps as for the $t$-channel, we find that these partial waves satisfy Watson's theorem:
\begin{align}
	\begin{split}
		\Im\left( a_{i,l}^{(3/2)}(u) \right) &= \frac{1}{2l+1}\frac{1}{16\pi} \frac{\lambda^{1/2}_{K\pi}(u)}{u} \; {t_l^{3/2}}^*(u) a_{i,l}^{(3/2)}(u) , \\
		a_{i,l}^{(3/2)}(u) &= \left| a_{i,l}^{(3/2)}(u) \right| e^{i\delta_l^{3/2}(u)} ,
	\end{split}
\end{align}
where $i=t,0,2$.

\subsubsection{Partial-Wave Expansion of the Form Factors in the $u$-Channel}

The components of the axial vector current matrix element in the $u$-channel $K\pi$ centre-of-mass frame are given by:
\begin{align}
	\begin{split}
		i \mathcal{A}_{(3/2)}^\mu&(k,-p_2 \to L,p_1) =  \\
		\frac{1}{M_K} &\Bigg( \frac{M_K^2 - s_\ell}{2\sqrt{u}} \; F - \frac{M_K^2-2M_\pi^2-2u +s_\ell}{2\sqrt{u}} \; G + \frac{u+s_\ell - M_\pi^2}{2\sqrt{u}} \; R , \\
			& \frac{\lambda^{1/2}_{\ell\pi}(u)}{2\sqrt{u}} \; ( F + G - R ) - \frac{\lambda^{1/2}_{K\pi}(u)}{2\sqrt{u}} \cos\theta_u \; ( F - G ) , \\
			& - \frac{\lambda^{1/2}_{K\pi}(u)}{2\sqrt{u}} \sin\theta_u \; \left( F - G \right) , \\
			& 0 \Bigg) ,
	\end{split} \\
	\begin{split}
		i \mathcal{A}_{(1/2)}^\mu&(q_K,q_\pi \to L,p_1) =  \\
		\frac{1}{M_K} &\Bigg( \frac{M_K^2 - s_\ell}{2\sqrt{u}} \; F - \frac{M_K^2-2M_\pi^2-2u + s_\ell}{2\sqrt{u}} \; G + \frac{u+s_\ell - M_\pi^2}{2\sqrt{u}} \; R , \\
			& \frac{\lambda^{1/2}_{\ell\pi}(u)}{2\sqrt{u}} \; ( F + G - R ) - \frac{\lambda^{1/2}_{K\pi}(u)}{2\sqrt{u}} \cos\theta_u^\dprime \; ( F - G ) , \\
			& - \frac{\lambda^{1/2}_{K\pi}(u)}{2\sqrt{u}} \sin\theta_u^\dprime \cos\phi_u^\dprime \; \left( F - G \right) , \\
			& - \frac{\lambda^{1/2}_{K\pi}(u)}{2\sqrt{u}} \sin\theta_u^\dprime \sin\phi_u^\dprime \; \left( F - G \right)  \Bigg) .
	\end{split}
\end{align}
The contraction with the polarisation vectors yields (for a generic $\phi_u$):
\begin{align}
	\begin{split}
		i \mathcal{A}_0^{(3/2)} &= i \mathcal{A}_\mu^{(3/2)} \varepsilon_0^\mu \\
			&= \frac{-1}{M_K \sqrt{s_\ell}} \Bigg( \frac{1}{4u} \left( \lambda^{1/2}_{K\pi}(u) (M_\pi^2-s_\ell-u) \cos\theta_u + \lambda^{1/2}_{\ell\pi}(u) (M_K^2 - M_\pi^2 + u) \right) F \\
			& \quad - \frac{1}{4u} \left( \lambda^{1/2}_{K\pi}(u) (M_\pi^2-s_\ell-u) \cos\theta_u + \lambda^{1/2}_{\ell\pi}(u) (M_K^2 - M_\pi^2 - 3u) \right) G \Bigg) , \\
		i \mathcal{A}_2^{(3/2)} &= i \mathcal{A}_\mu^{(3/2)} \varepsilon_+^\mu - i \mathcal{A}_\mu^{(3/2)} \varepsilon_-^\mu \\
			&= \frac{1}{\sqrt{2}M_K} \Bigg( \frac{\lambda^{1/2}_{K\pi}(u)}{\sqrt{u}} \sin\theta_u \cos\phi_u \left( F - G \right) \Bigg) .
	\end{split}
\end{align}
Hence, the partial-wave expansion of the form factors is given by
\begin{align}
	\begin{split}
		\label{eq:uChannelFormFactorPartialWaveExpansion}
		F + G &+ \frac{1}{2u} \left( M_K^2 - M_\pi^2 - u + (M_\pi^2 - s_\ell - u ) \frac{\lambda^{1/2}_{K\pi}(u)}{\lambda^{1/2}_{\ell\pi}(u)} \cos\theta_u \right) (F - G) \\
			&= \frac{-2M_K\sqrt{s_\ell}}{\lambda^{1/2}_{\ell\pi}(u)} \sum_{l=0}^\infty P_l(\cos\theta_u) \left( \frac{\lambda^{1/2}_{K\pi}(u) \lambda^{1/2}_{\ell\pi}(u)}{M_K^4} \right)^l a_{0,l}^{(3/2)}(u) , \\
		F - G &= \frac{M_K \sqrt{2u}}{\lambda^{1/2}_{K\pi}(u)} \sum_{l=1}^\infty P_l^\prime(\cos\theta_u) \left( \frac{\lambda^{1/2}_{K\pi}(u) \lambda^{1/2}_{\ell\pi}(u)}{M_K^4} \right)^{l-1} a_{2,l}^{(3/2)}(u) .
	\end{split}
\end{align}
We write this in the form
\begin{align}
	\begin{split}
		F + G &= 2 \sum_{l=0}^\infty P_l(\cos\theta_u) \left( \frac{\lambda^{1/2}_{K\pi}(u) \lambda^{1/2}_{\ell\pi}(u)}{M_K^4} \right)^l f_l^{(3/2)}(u) \\
			&- \frac{1}{2u} \left( M_K^2 - M_\pi^2 - u + (M_\pi^2 - s_\ell - u ) \frac{\lambda^{1/2}_{K\pi}(u)}{\lambda^{1/2}_{\ell\pi}(u)} \cos\theta_u \right) (F - G) , \\
		F - G &= 2 \sum_{l=1}^\infty P_l^\prime(\cos\theta_u) \left( \frac{\lambda^{1/2}_{K\pi}(u) \lambda^{1/2}_{\ell\pi}(u)}{M_K^4} \right)^{l-1} g_l^{(3/2)}(u) ,
	\end{split}
\end{align}
or
\begin{align}
	\begin{split}
		\label{eq:uChannelFormFactorPartialWaveExpansion2}
		F &= \sum_{l=0}^\infty P_l(\cos\theta_u) \left( \frac{\lambda^{1/2}_{K\pi}(u) \lambda^{1/2}_{\ell\pi}(u)}{M_K^4} \right)^l f_l^{(3/2)}(u) \\
			&\quad - \frac{1}{2u} \left( M_K^2 - M_\pi^2 - 3u + (M_\pi^2 - s_\ell - u ) \frac{\lambda^{1/2}_{K\pi}(u)}{\lambda^{1/2}_{\ell\pi}(u)} \cos\theta_u \right) \\
			&\quad \sum_{l=1}^\infty P_l^\prime(\cos\theta_u)  \left( \frac{\lambda^{1/2}_{K\pi}(u) \lambda^{1/2}_{\ell\pi}(u)}{M_K^4} \right)^{l-1} g_l^{(3/2)}(u) , \\
		G &= \sum_{l=0}^\infty P_l(\cos\theta_u) \left( \frac{\lambda^{1/2}_{K\pi}(u) \lambda^{1/2}_{\ell\pi}(u)}{M_K^4} \right)^l f_l^{(3/2)}(u) \\
			&\quad - \frac{1}{2u} \left( M_K^2 - M_\pi^2 + u + (M_\pi^2 - s_\ell - u ) \frac{\lambda^{1/2}_{K\pi}(u)}{\lambda^{1/2}_{\ell\pi}(u)} \cos\theta_u \right) \\
			&\quad \sum_{l=1}^\infty P_l^\prime(\cos\theta_u)  \left( \frac{\lambda^{1/2}_{K\pi}(u) \lambda^{1/2}_{\ell\pi}(u)}{M_K^4} \right)^{l-1} g_l^{(3/2)}(u) ,
	\end{split}
\end{align}
where the partial waves $f_l^{(3/2)}$ and $g_l^{(3/2)}$ satisfy Watson's theorem in the region $u>(M_K+M_\pi)^2$:
\begin{align}
	f_l^{(3/2)}(u) = \left| f_l^{(3/2)}(u) \right| e^{i \delta_l^{3/2}(u)} \quad , \quad g_l^{(3/2)}(u) = \left| g_l^{(3/2)}(u) \right| e^{i \delta_l^{3/2}(u)} .
\end{align}

\subsection{Projection and Analytic Structure of the Partial Waves}

\label{sec:ProjectionAnalyticStructurePartialWaves}

The several partial waves $f_l^{(I)}$ and $g_l^{(I)}$ can be calculated by angular projections:
\begin{align}
	\begin{split}
		\label{eqn:SChannelPartialWaveProjection}
		f_l^{(I)}(s) &= \left( \frac{M_K^2}{\lambda^{1/2}_{K\ell}(s) \sigma_\pi(s)} \right)^{l} \frac{2l+1}{2} \int_{-1}^1 dz \, P_l(z) \left( F^{(I)}(s,z) + \frac{\sigma_\pi(s) PL(s)}{X(s)} z G^{(I)}(s,z) \right) , \\
		g_l^{(I)}(s) &= \left( \frac{M_K^2}{\lambda^{1/2}_{K\ell}(s) \sigma_\pi(s)} \right)^{l-1} \int_{-1}^1 dz \, \frac{P_{l-1}(z) - P_{l+1}(z)}{2} \, G^{(I)}(s,z) ,
	\end{split}
\end{align}
where $X^{(I)}(s,z) := X^{(I)}(s,t(s,z),u(s,z))$, $X\in\{F,G\}$, $I\in\{0,1\}$ and
\begin{align}
	\begin{split}
		t(s,z) &= \frac{1}{2} \left( \Sigma_0 - s - 2 X \sigma_\pi z \right) , \\
		u(s,z) &= \frac{1}{2} \left( \Sigma_0 - s + 2 X \sigma_\pi z \right) .
	\end{split}
\end{align}
Since $t(s,-z)=u(s,z)$, the definition of the pure isospin form factors (\ref{eq:sChannelIsospinFormFactors}) implies
\begin{align}
	\begin{split}
		f_l^{(0)}(s) &= g_l^{(0)}(s) = 0 \quad \forall \; l \text{ odd} , \\
		f_l^{(1)}(s) &= g_l^{(1)}(s) = 0 \quad \forall \; l \text{ even} .
	\end{split}
\end{align}
Hence, we can as well directly use the partial waves of the physical form factors:
\begin{align}
	\begin{split}
		\label{eq:PartialWaveProjectionSChannelPhysicalFF}
		f_l(s) &= \left( \frac{M_K^2}{\lambda^{1/2}_{K\ell}(s) \sigma_\pi(s)} \right)^{l} \frac{2l+1}{2} \int_{-1}^1 dz \, P_l(z) \left( F(s,z) + \frac{\sigma_\pi(s) PL(s)}{X(s)} z G(s,z) \right) , \\
		g_l(s) &= \left( \frac{M_K^2}{\lambda^{1/2}_{K\ell}(s) \sigma_\pi(s)} \right)^{l-1} \int_{-1}^1 dz \, \frac{P_{l-1}(z) - P_{l+1}(z)}{2} \, G(s,z) ,
	\end{split}
\end{align}
which will still fulfil Watson's theorem
\begin{align}
	f_l(s) = \left| f_l(s) \right| e^{i \delta_l^I(s)} \quad , \quad g_l(s) = \left| g_l(s) \right| e^{i \delta_l^I(s)} ,
\end{align}
where $I = (l\mod2)$.

In the crossed channels, the partial wave projections are given by
\begin{align}
	\begin{split}
		\label{eq:PartialWaveProjectionTUChannel}
		f_l^{(1/2)}(t) &= \left( \frac{M_K^4}{\lambda^{1/2}_{K\pi}(t) \lambda^{1/2}_{\ell\pi}(t)} \right)^l \frac{2l+1}{2} \int_{-1}^1 dz_t \, P_l(z_t) \Bigg( \frac{F^{(1/2)}(t,z_t) - G^{(1/2)}(t,z_t)}{2} \\
			+ \frac{1}{2t} \bigg( &M_K^2 - M_\pi^2 - t + (M_\pi^2 - s_\ell - t) \frac{\lambda^{1/2}_{K\pi}(t)}{\lambda^{1/2}_{\ell\pi}(t)} z_t \bigg) \frac{F^{(1/2)}(t,z_t) + G^{(1/2)}(t,z_t)}{2} \Bigg) , \\
		g_l^{(1/2)}(t) &= \left( \frac{M_K^4}{\lambda^{1/2}_{K\pi}(t) \lambda^{1/2}_{\ell\pi}(t)} \right)^{l-1} \\
			&\quad \int_{-1}^1 dz_t \, \frac{P_{l-1}(z_t) - P_{l+1}(z_t)}{2} \, \frac{F^{(1/2)}(t,z_t) + G^{(1/2)}(t,z_t)}{2} , \\
		f_l^{(3/2)}(u) &= \left( \frac{M_K^4}{\lambda^{1/2}_{K\pi}(u) \lambda^{1/2}_{\ell\pi}(u)} \right)^l \frac{2l+1}{2} \int_{-1}^1 dz_u \, P_l(z_u) \Bigg( \frac{F(u,z_u) + G(u,z_u)}{2} \\
			 + \frac{1}{2u} &\left( M_K^2 - M_\pi^2 - u + (M_\pi^2 - s_\ell - u) \frac{\lambda^{1/2}_{K\pi}(u)}{\lambda^{1/2}_{\ell\pi}(u)} z_u \right) \frac{F(u,z_u) - G(u,z_u)}{2} \Bigg) , \\
		g_l^{(3/2)}(u) &= \left( \frac{M_K^4}{\lambda^{1/2}_{K\pi}(u) \lambda^{1/2}_{\ell\pi}(u)} \right)^{l-1} \\
			&\quad \int_{-1}^1 dz_u \, \frac{P_{l-1}(z_u) - P_{l+1}(z_u)}{2} \, \frac{F(u,z_u) - G(u,z_u)}{2} ,
	\end{split}
\end{align}
where $X^{(I)}(t,z_t) := X^{(I)}(s(t,z_t),t,u(t,z_t))$, $X^{(I)}(u,z_u) := X^{(I)}(s(u,z_u),t(u,z_u),u)$, $X\in\{F,G\}$ and
\begin{align}
	\begin{split}
		s(t,z_t) &= \frac{1}{2} \left( \Sigma_0 - t + \frac{1}{t} \left( z_t \, \lambda^{1/2}_{K\pi}(t) \lambda^{1/2}_{\ell\pi}(t) - \Delta_{K\pi}\Delta_{\ell\pi} \right) \right) , \\
		u(t,z_t) &= \frac{1}{2} \left( \Sigma_0 - t - \frac{1}{t} \left( z_t \, \lambda^{1/2}_{K\pi}(t) \lambda^{1/2}_{\ell\pi}(t) - \Delta_{K\pi}\Delta_{\ell\pi} \right) \right)  , \\
		s(u,z_u) &= \frac{1}{2} \left( \Sigma_0 - u + \frac{1}{u} \left( z_u \, \lambda^{1/2}_{K\pi}(u) \lambda^{1/2}_{\ell\pi}(u) - \Delta_{K\pi}\Delta_{\ell\pi} \right) \right) , \\
		t(u,z_u) &= \frac{1}{2} \left( \Sigma_0 - u - \frac{1}{u} \left( z_u \, \lambda^{1/2}_{K\pi}(u) \lambda^{1/2}_{\ell\pi}(u) - \Delta_{K\pi}\Delta_{\ell\pi} \right) \right) .
	\end{split}
\end{align}

The construction of the partial waves has been done in a way that excludes kinematic singularities for $s > 4 M_\pi^2$ and $t,u>(M_K+M_\pi)^2$. There may still be kinematic singularities present below these regions, but they shall not bother us. But also the analytic structure of the partial waves with respect to dynamic singularities is not trivial.

For the $s$-channel partial waves, there is of course the right-hand cut at $s>4M_\pi^2$. Further cuts can appear through the angular integration, i.e.~at points where the integration contour in the $t$- or $u$-plane touches the crossed channel cuts. If $s$ lies in the physical decay region, the integration path is just a horizontal line from one end of the decay region to the other (see the Mandelstam diagram in figure~\ref{img:MandelstamDiagram2}). When we continue analytically into the region $(M_K-\sqrt{s_\ell})^2 < s < (M_K + \sqrt{s_\ell})^2$, the integration path moves into the complex $t$- and $u$-plane and crosses the real Mandelstam plane at $t=u$: the square root of the Källén function $X = \frac{1}{2}\lambda^{1/2}_{K\ell}(s)$ is purely imaginary in this region. One has to know which branch of the square root should be taken. The correct sign is found by taking $s$ real and shifting $M_K \to M_K + i \epsilon$ (see \cite{Kacser1963}). With this prescription, the Källén function turns counterclockwise around $\lambda_{K\ell}=0$ when $s$ runs from $s<(M_K-\sqrt{s_\ell})^2$ to $s>(M_K+\sqrt{s_\ell})^2$. The square root of the Källén function therefore takes the following values:
\begin{align}
	\lambda^{1/2}_{K\ell}(s) = \left\{
	\begin{array}{r c}
		+ | \lambda^{1/2}_{K\ell}(s) | & s < (M_K-\sqrt{s_\ell})^2 , \\
		+ i | \lambda^{1/2}_{K\ell}(s) | & (M_K-\sqrt{s_\ell})^2 < s < (M_K+\sqrt{s_\ell})^2 , \\
		- | \lambda^{1/2}_{K\ell}(s) | & (M_K+\sqrt{s_\ell})^2 < s .
	\end{array} \right. 
\end{align}
In the region $s>(M_K+\sqrt{s_\ell})^2$, the integration path again lies in the real Mandelstam plane from one to the other end of the scattering region.

As we are away from the $t$- and $u$-channel unitarity cuts at $t,u>(M_K+M_\pi)^2$, this extension of the integration path into the complex $t$- and $u$-plane is the only subtlety that has to be taken into account.

In the region $s<4M_\pi^2$, there is a left-hand cut at $s\in(-\infty,0)$: the integration path extends again in the complex $t$- and $u$-plane in the region $0<s<4M_\pi^2$ (due to the second square root). It diverges at $s=0$ and returns to the real axis at $s<0$, but this time it touches the $t$- and $u$-channel unitarity cuts at $t,u>(M_K+M_\pi)^2$ which produces the left-hand cut of the $s$-channel partial waves.

This left-hand cut can be most easily found by looking at the end-points of the integration paths: solving the equation
\begin{align}
	\begin{split}
		t_\pm &= u_\mp = \frac{1}{2} \left( \Sigma_0 - s \mp 2 X(s) \sigma_\pi(s) \right)
	\end{split}
\end{align}
for $t_\pm > (M_K+M_\pi)^2$ gives the left-hand cut $s\in(-\infty,0)$.

Let us consider the crossed $t$-channel (the situation in the $u$-channel is analogous). We have defined the partial-wave expansion in the scattering region $t>(M_K+M_\pi)^2$. Therefore, we also define the square root branches of the Källén functions $\lambda^{1/2}_{K\pi}$ and $\lambda^{1/2}_{\ell\pi}$ in this region. The sign of the square root branch can be absorbed into the definition of the partial waves.

The right-hand $t$-channel unitarity cut at $t>(M_K+M_\pi)^2$ also shows up in the partial waves. A second possibility for singularities in the $t$-channel partial waves arises when the integration path touches the $s$- or $u$-channel unitarity cuts. For $t>(M_K+M_\pi)^2$, the integration path lies  on the negative real axis of the $s$- and $u$-planes (this can be seen in the Mandelstam diagram in figure~\ref{img:MandelstamDiagram2}). In the region $(M_K-M_\pi)^2 < t < (M_K+M_\pi)^2$, the integration path extends into the complex $s$- and $u$-plane. For the value of $t$ fulfilling $\frac{1}{2} \left( \Sigma_0 - t - \frac{1}{t} \Delta_{K\pi}\Delta_{\ell\pi} \right) = 4 M_\pi^2$, the integration path in the $s$-plane touches the $s$-channel branch cut. From this point on towards smaller values of $t$, the integration path has to be deformed in the $s$-plane. Since the $u$-channel cut appears only at $u>(M_K+M_\pi)^2$, such a deformation is not needed in the $u$-plane. At $t=(M_K-M_\pi)^2$, the integration path in the $s$-plane has then the shape of a horseshoe wrapped around the $s$-channel cut. For even smaller values of $t$, the path unwraps itself in a continuous way, such that for $t < \frac{1}{2}(M_K^2 - 2M_\pi^2 + s_\ell)$, the integration path lies completely on the upper side of the $s$-channel cut.

The cut structure in the $t$-channel partial wave is rather complicated, at least for $s_\ell > 0$: The left-hand cuts can be found by solving the equations
\begin{align}
	\begin{split}
		s_\pm &= \frac{1}{2} \left( \Sigma_0 - t + \frac{1}{t} \left( \pm \, \lambda^{1/2}_{K\pi}(t) \lambda^{1/2}_{\ell\pi}(t) - \Delta_{K\pi}\Delta_{\ell\pi} \right) \right) , \\
		u_\pm &= \frac{1}{2} \left( \Sigma_0 - t - \frac{1}{t} \left( \pm \, \lambda^{1/2}_{K\pi}(t) \lambda^{1/2}_{\ell\pi}(t) - \Delta_{K\pi}\Delta_{\ell\pi} \right) \right) ,
	\end{split}
\end{align}
for $s_\pm > 4 M_\pi^2$ and $u_\pm > (M_K+M_\pi)^2$. While the second equation results in a cut along the real axis, the first equation produces an egg-shaped cut structure in the complex $t$-plane with $\Re(t) < (M_K-M_\pi)^2$, shown in figure~\ref{img:tChannelPartialWaveCuts}. The exact shape depends on the value of $s_\ell$.

\input{Kl4Dispersive/sections/Singularities}

\subsection{Simplifications for $s_\ell \to 0$}

\label{sec:PartialWaveSimplificationsZeroSL}

In the experiment, a dependence on $s_\ell$ has been observed only in the first partial wave of the form factor $F$ \cite{Batley2010,Batley2012}. If we neglect this dependence on $s_\ell$ and assume that $s_\ell = 0$, the treatment can be significantly simplified.

\begin{itemize}
	\item The square root of the Källén function simplifies to \[\lim\limits_{s_\ell\to0}\lambda^{1/2}_{K\ell}(s) = M_K^2 - s ,\] the square root branch cut disappears. Hence, the integration path for the angular integrals in the $s$-channel always lies on the real axis.
	\item The left-hand cut structure of $t$- and $u$-channel partial waves simplifies to a straight line along the real axis. The egg-shaped branch cuts disappear in the limit $s_\ell\to0$.
	\item From (\ref{eq:sChannelFormFactorPartialWaveExpansion}), we see that the quantity
	\begin{align}
		\begin{split}
			\lim_{s_\ell\to0} F_1^{(I)} &= \lim_{s_\ell\to0} \left(\frac{1}{2} \lambda^{1/2}_{K\ell}(s) F^{(I)} + \frac{1}{2} \frac{(M_K^2-s-s_\ell) (u-t)}{\lambda^{1/2}_{K\ell}(s)} \; G^{(I)} \right) \\
				&= \frac{M_K^2 - s}{2} F^{(I)} + \frac{u-t}{2} G^{(I)}
		\end{split}
	\end{align}
	has a simple $s$-channel partial-wave expansion into Legendre polynomials. If we consider (\ref{eq:tChannelFormFactorPartialWaveExpansion}) in the limit $s_\ell \to 0$, we find that exactly the same linear combination of the form factors $F^{(1/2)}$ and $G^{(1/2)}$ has a simple $t$-channel partial-wave expansion into Legendre polynomials. The same follows from (\ref{eq:uChannelFormFactorPartialWaveExpansion}) for the $u$-channel. In this limit, the form factor $F_1$ can therefore be treated independently from the other form factors. This is the procedure that has been followed in \cite{Stoffer2010, Colangelo2012, Stoffer2013}.
\end{itemize}

There are several reasons why we abstain here from taking the limit $s_\ell\to0$, which would result in a substantial simplification of the whole treatment. The experiments provide data on both form factors $F$ and $G$. In order to include all the available information, we deal with both form factors at the same time. There is also some data available on the dependence on $s_\ell$, which we include in this treatment. And finally, the matching to \ChPT{} becomes much cleaner if it is performed with $F$ and $G$ directly, since these are the form factors with the simplest chiral representation.

\clearpage

\section{Reconstruction Theorem}

Since the form factors $F$ and $G$ describe a hadronic four-`particle' process, they depend on the three Mandelstam variables $s$, $t$ and $u$ and therefore possess a rather complicated analytic structure. However, it is possible to decompose the form factors into a sum of functions that depend only on a single Mandelstam variable, a procedure known under the name of `reconstruction theorem' \cite{Stern1993, Ananthanarayan2001}. Such a decomposition provides a major simplification of the problem and leads to a powerful dispersive description. 

\subsection{Decomposition of the Form Factors}

In appendix~\ref{sec:AppendixReconstructionTheorem}, we derive explicitly the decomposition of the form factors $F$ and $G$ into functions of a single Mandelstam variable, using fixed-$s$, fixed-$t$ and fixed-$u$ dispersion relations. We have to assume a certain asymptotic behaviour of the form factors, e.g.~for fixed $u$, we assume
\begin{align}
	\label{eq:FroissartInspiredAsymptotics}
	\lim_{|s|\to\infty} \frac{X_s^u(s)}{s^n} = \lim_{|t|\to\infty} \frac{X_t^u(t)}{t^n} = 0 ,
\end{align}
where the Froissart bound \cite{Froissart1961} suggests $n=2$. However, we are also interested in the case $n=3$ in order to meet the asymptotic behaviour of the NNLO \ChPT{} form factors. We therefore write down either a twice- or thrice-subtracted dispersion relation for the form factors. Then, we use the partial-wave expansions derived in the previous section. We neglect the imaginary parts of $D$- and higher waves, an approximation that is violated only at $\O(p^8)$ in the chiral power counting. Therefore, the decomposition is valid up to and including $\O(p^6)$. It implements the case $s_\ell \neq 0$.

The result of the decomposition is the following:
\begin{align}
	\label{eq:FormFactorDecomposition}
	\begin{split}
		F(s,t,u) &= M_0(s) + \frac{u-t}{M_K^2} M_1(s) \\
			&\quad + \frac{2}{3} N_0(t) + \frac{2}{3} \frac{t(s-u) + \Delta_{K\pi}\Delta_{\ell\pi}}{M_K^4} N_1(t) - \frac{2}{3} \frac{\Delta_{K\pi}-3t}{2M_K^2} \tilde N_1(t) \\
			&\quad + \frac{1}{3} R_0(t) + \frac{1}{3} \frac{t(s-u) + \Delta_{K\pi}\Delta_{\ell\pi}}{M_K^4} R_1(t) - \frac{1}{3} \frac{\Delta_{K\pi}-3t}{2M_K^2} \tilde R_1(t) \\
			&\quad + R_0(u) + \frac{u(s-t) + \Delta_{K\pi}\Delta_{\ell\pi}}{M_K^4} R_1(u) - \frac{\Delta_{K\pi}-3u}{2M_K^2} \tilde R_1(u) \\
			&\quad + \O(p^8) , \\
		G(s,t,u) &= \tilde M_1(s) \\
			&\quad - \frac{2}{3} N_0(t) - \frac{2}{3} \frac{t(s-u) + \Delta_{K\pi}\Delta_{\ell\pi}}{M_K^4} N_1(t) + \frac{2}{3} \frac{\Delta_{K\pi}+t}{2M_K^2} \tilde N_1(t) \\
			&\quad - \frac{1}{3} R_0(t) - \frac{1}{3} \frac{t(s-u) + \Delta_{K\pi}\Delta_{\ell\pi}}{M_K^4} R_1(t) + \frac{1}{3} \frac{\Delta_{K\pi}+t}{2M_K^2} \tilde R_1(t) \\
			&\quad + R_0(u) + \frac{u(s-t) + \Delta_{K\pi}\Delta_{\ell\pi}}{M_K^4} R_1(u) - \frac{\Delta_{K\pi} + u}{2M_K^2} \tilde R_1(u) \\
			&\quad + \O(p^8) .
	\end{split}
\end{align}
In the case $n=2$, the various functions of one variable are given by
\begin{align}
	\begin{split}
		\label{eq:FunctionsOfOneVariable}
		M_0(s) &= m_0^0 + m_0^1 \frac{s}{M_K^2} + \frac{s^2}{\pi} \int_{s_0}^\infty \frac{\Im f_0(s^\prime)}{(s^\prime - s - i\epsilon){s^\prime}^2} ds^\prime , \\
		M_1(s) &= m_1^0 + \frac{s}{\pi} \int_{s_0}^\infty \frac{1}{(s^\prime - s - i\epsilon) s^\prime} \Im \left( f_1(s^\prime) - \frac{2PL(s^\prime) M_K^2}{\lambda_{K\ell}(s^\prime)} g_1(s^\prime) \right) ds^\prime , \\
		\tilde M_1(s) &= \tilde m_1^0 + \tilde m_1^1 \frac{s}{M_K^2} + \frac{s^2}{\pi} \int_{s_0}^\infty \frac{\Im g_1(s^\prime)}{(s^\prime - s - i\epsilon){s^\prime}^2} ds^\prime , \\
		N_0(t) &= n_0^1 \frac{t}{M_K^2} + \frac{t^2}{\pi} \int_{t_0}^\infty \frac{\Im  f_0^{(1/2)}(t^\prime)}{(t^\prime-t-i\epsilon){t^\prime}^2} dt^\prime , \\
		N_1(t) &= \frac{1}{\pi} \int_{t_0}^\infty \frac{1}{t^\prime - t - i\epsilon} \Im \left( f_1^{(1/2)}(t^\prime) + \frac{(\Delta_{\ell\pi}+t^\prime)M_K^4}{2t^\prime \lambda_{\ell\pi}(t^\prime)} g_1^{(1/2)}(t^\prime) \right) dt^\prime , \\
		\tilde N_1(t) &= \frac{t}{\pi} \int_{t_0}^\infty \frac{M_K^2}{t^\prime} \frac{\Im g_1^{(1/2)}(t^\prime)}{(t^\prime-t-i\epsilon)t^\prime} dt^\prime , \\
		R_0(t) &= \frac{t^2}{\pi} \int_{t_0}^\infty \frac{\Im  f_0^{(3/2)}(t^\prime)}{(t^\prime-t-i\epsilon){t^\prime}^2} dt^\prime , \\
		R_1(t) &= \frac{1}{\pi} \int_{t_0}^\infty \frac{1}{t^\prime - t - i\epsilon} \Im \left( f_1^{(3/2)}(t^\prime) + \frac{(\Delta_{\ell\pi}+t^\prime)M_K^4}{2t^\prime \lambda_{\ell\pi}(t^\prime)} g_1^{(3/2)}(t^\prime) \right) dt^\prime , \\
		\tilde R_1(t) &= \frac{t}{\pi} \int_{t_0}^\infty \frac{M_K^2}{t^\prime} \frac{\Im g_1^{(3/2)}(t^\prime)}{(t^\prime-t-i\epsilon)t^\prime} dt^\prime ,
	\end{split}
\end{align}
while for $n=3$, the functions of one variable are
\begin{align}
	\begin{split}
		\label{eq:FunctionsOfOneVariable3Subtr}
		M_0(s) &= m_0^0 + m_0^1 \frac{s}{M_K^2} + m_0^2 \frac{s^2}{M_K^4} + \frac{s^3}{\pi} \int_{s_0}^\infty \frac{\Im f_0(s^\prime)}{(s^\prime - s - i\epsilon){s^\prime}^3} ds^\prime , \\
		M_1(s) &= m_1^0 + m_1^1 \frac{s}{M_K^2} + \frac{s^2}{\pi} \int_{s_0}^\infty \frac{1}{(s^\prime - s - i\epsilon) {s^\prime}^2} \Im \left( f_1(s^\prime) - \frac{2PL(s^\prime) M_K^2}{\lambda_{K\ell}(s^\prime)} g_1(s^\prime) \right) ds^\prime , \\
		\tilde M_1(s) &= \tilde m_1^0 + \tilde m_1^1 \frac{s}{M_K^2} + \tilde m_1^2 \frac{s^2}{M_K^4} + \frac{s^3}{\pi} \int_{s_0}^\infty \frac{\Im g_1(s^\prime)}{(s^\prime - s - i\epsilon){s^\prime}^3} ds^\prime , \\
		N_0(t) &= n_0^1 \frac{t}{M_K^2} + n_0^2 \frac{t^2}{M_K^4} + \frac{t^3}{\pi} \int_{t_0}^\infty \frac{\Im  f_0^{(1/2)}(t^\prime)}{(t^\prime-t-i\epsilon){t^\prime}^3} dt^\prime , \\
		N_1(t) &= n_1^0 + \frac{t}{\pi} \int_{t_0}^\infty \frac{1}{(t^\prime - t - i\epsilon)t^\prime} \Im \left( f_1^{(1/2)}(t^\prime) + \frac{(\Delta_{\ell\pi}+t^\prime)M_K^4}{2t^\prime \lambda_{\ell\pi}(t^\prime)} g_1^{(1/2)}(t^\prime) \right) dt^\prime , \\
		\tilde N_1(t) &= \tilde n_1^1 \frac{t}{M_K^2} + \frac{t^2}{\pi} \int_{t_0}^\infty \frac{M_K^2}{t^\prime} \frac{\Im g_1^{(1/2)}(t^\prime)}{(t^\prime-t-i\epsilon){t^\prime}^2} dt^\prime , \\
		R_0(t) &= \frac{t^3}{\pi} \int_{t_0}^\infty \frac{\Im  f_0^{(3/2)}(t^\prime)}{(t^\prime-t-i\epsilon){t^\prime}^3} dt^\prime , \\
		R_1(t) &= \frac{t}{\pi} \int_{t_0}^\infty \frac{1}{(t^\prime - t - i\epsilon){t^\prime}} \Im \left( f_1^{(3/2)}(t^\prime) + \frac{(\Delta_{\ell\pi}+t^\prime)M_K^4}{2t^\prime \lambda_{\ell\pi}(t^\prime)} g_1^{(3/2)}(t^\prime) \right) dt^\prime , \\
		\tilde R_1(t) &= \frac{t^2}{\pi} \int_{t_0}^\infty \frac{M_K^2}{t^\prime} \frac{\Im g_1^{(3/2)}(t^\prime)}{(t^\prime-t-i\epsilon){t^\prime}^2} dt^\prime .
	\end{split}
\end{align}

Actually, since the $P$-wave of isospin $I=3/2$ $K\pi$ scattering is real at $\O(p^6)$, so are the partial waves $f_1^{(3/2)}$ and $g_1^{(3/2)}$. Hence, the functions $R_1(t)$ and $\tilde R_1(t)$ could be dropped altogether in the decomposition. The phase $\delta_1^{3/2}$ is also phenomenologically tiny.

\subsection{Ambiguity of the Decomposition}

We have decomposed the form factors $F$ and $G$ into functions of one variable $M_0(s)$, $\ldots$. However, while the form factors are observable quantities, these functions are not. It is possible to redefine the functions $M_0(s)$, $\ldots$ without changing the form factors and hence without changing the physics.

Therefore, let us study this ambiguity of the decomposition of the form factors. We require the form factors to be invariant under a change of the functions of one variable:
\begin{align}
	\begin{split}
		M_0(s) &\mapsto M_0(s) + \delta M_0(s) , \\
		M_1(s) &\mapsto M_1(s) + \delta M_1(s) , \\
			&\ldots ,
	\end{split}
\end{align}
which we call `gauge transformation'. The shifts have to satisfy
\begin{align}
	\begin{split}
		0 &= \delta M_0(s) + \frac{u-t}{M_K^2} \delta M_1(s) \\
			&\quad + \frac{2}{3} \delta N_0(t) + \frac{2}{3} \frac{t(s-u) + \Delta_{K\pi}\Delta_{\ell\pi}}{M_K^4} \delta N_1(t) - \frac{2}{3} \frac{\Delta_{K\pi}-3t}{2M_K^2} \delta \tilde N_1(t) \\
			&\quad + \frac{1}{3} \delta R_0(t) + \frac{1}{3} \frac{t(s-u) + \Delta_{K\pi}\Delta_{\ell\pi}}{M_K^4} \delta R_1(t) - \frac{1}{3} \frac{\Delta_{K\pi}-3t}{2M_K^2} \delta \tilde R_1(t) \\
			&\quad + \delta R_0(u) + \frac{u(s-t) + \Delta_{K\pi}\Delta_{\ell\pi}}{M_K^4} \delta R_1(u) - \frac{\Delta_{K\pi}-3u}{2M_K^2} \delta \tilde R_1(u) ,
	\end{split} \\
	\begin{split}
		0 &= \delta \tilde M_1(s) \\
			&\quad - \frac{2}{3} \delta N_0(t) - \frac{2}{3} \frac{t(s-u) + \Delta_{K\pi}\Delta_{\ell\pi}}{M_K^4} \delta N_1(t) + \frac{2}{3} \frac{\Delta_{K\pi}+t}{2M_K^2} \delta \tilde N_1(t) \\
			&\quad - \frac{1}{3} \delta R_0(t) - \frac{1}{3} \frac{t(s-u) + \Delta_{K\pi}\Delta_{\ell\pi}}{M_K^4} \delta R_1(t) + \frac{1}{3} \frac{\Delta_{K\pi}+t}{2M_K^2} \delta \tilde R_1(t) \\
			&\quad + \delta R_0(u) + \frac{u(s-t) + \Delta_{K\pi}\Delta_{\ell\pi}}{M_K^4} \delta R_1(u) - \frac{\Delta_{K\pi}+u}{2M_K^2} \delta \tilde R_1(u) .
	\end{split}
\end{align}
The solution to these equations is found in the following way: we substitute one of the three kinematic variables by means of $s+t+u = \Sigma_0$. Then, we take the derivative with respect to one of the two remaining variables and substitute back $\Sigma_0 = s + t + u$. After four or five such differentiations, one gets differential equations for single functions $\delta M_0$, $\ldots$ with the following solution:
\begin{align}
	\begin{split}
		\delta M_0(s) &= c_0^{M_0} + c_1^{M_0} s + c_2^{M_0} s^2 , \\
		\delta M_1(s) &= c_0^{M_1} + c_1^{M_1} s + c_2^{M_1} s^2 , \\
		\delta \tilde M_1(s) &= c_0^{\tilde M_1} + c_1^{\tilde M_1} s + c_2^{\tilde M_1} s^2 + c_3^{\tilde M_1} s^3 , \\
		\delta N_0(t) &= c_{-1}^{N_0} t^{-1} + c_0^{N_0} + c_1^{N_0} t + c_2^{N_0} t^2 , \\
		\delta N_1(t) &= c_{-1}^{N_1} t^{-1} + c_0^{N_1} + c_1^{N_1} t , \\
		\delta \tilde N_1(t) &= c_{-1}^{\tilde N_1} t^{-1} + c_0^{\tilde N_1} + c_1^{\tilde N_1} t + c_2^{\tilde N_1} t^2 , \\
		\delta R_0(t) &= c_{-1}^{R_0} t^{-1} + c_0^{R_0} + c_1^{R_0} t + c_2^{R_0} t^2 , \\
		\delta R_1(t) &= c_{-1}^{R_1} t^{-1} + c_0^{R_1} + c_1^{R_1} t , \\
		\delta \tilde R_1(t) &= c_{-1}^{\tilde R_1} t^{-1} + c_0^{\tilde R_1} + c_1^{\tilde R_1} t + c_2^{\tilde R_1} t^2 .
	\end{split}
\end{align}
Inserting these solutions into the various differential equations results in algebraic equations for the diverse coefficients. In the end, there remain 13 independent parameters. In complete generality, we therefore have a gauge freedom of 13 parameters in the decomposition (\ref{eq:FormFactorDecomposition}). The gauge can be fixed by imposing constraints on the Taylor expansion or the asymptotic behaviour of the functions $M_0(s)$, $\ldots$.

First, let us restrict the gauge freedom by imposing the same vanishing Taylor coefficients as in (\ref{eq:FunctionsOfOneVariable}), i.e.~we exclude all the pole terms, the constants in $N_0$, $\tilde N_1$, $R_0$, $\tilde R_1$ and even a linear term in $R_0$. Then, we further demand that asymptotically the functions behave at most as in (\ref{eq:FunctionsOfOneVariable3Subtr}), i.e.~like $M_1(s) = \O(s)$, $\tilde M_1(s) = \O(s^2)$, $N_1(t) = \O(1)$, $\tilde N_1(t) = \O(t)$, $R_1(t) = \O(1)$ and $\tilde R_1(t) = \O(t)$. After imposing these constraints, we are left with a restricted gauge freedom of 3 parameters, which we call $C^{R_0}$, $A^{R_1}$ and $B^{\tilde R_1}$:
\begin{align}
	\begin{split}
		\label{eq:GaugeTransformation}
		\delta M_0(s) &= \left(2 A^{R_1} - B^{\tilde R_1} + 2 C^{R_0}\right) \frac{\left( \Sigma_0 - s \right)^2 - \Delta_{K\pi}\Delta_{\ell\pi}}{2 M_K^4} , \\
		\delta M_1(s) &= -\left(A^{R_1} + B^{\tilde R_1} + 2 C^{R_0}\right)\frac{\Sigma_0}{M_K^2} + B^{\tilde R_1} \frac{\Delta_{K\pi}}{2 M_K^2} + \left(B^{\tilde R_1} + 2 C^{R_0}\right) \frac{s}{M_K^2} , \\
		\delta \tilde M_1(s) &= \left( \left(B^{\tilde R_1} - 2 C^{R_0}\right) \Sigma_0^2 - \left(2 A^{R_1} + B^{\tilde R_1} - 2 C^{R_0} \right) \Delta_{K\pi} \Delta_{\ell\pi} + B^{\tilde R_1} \Sigma_0 \Delta_{K\pi} \right) \frac{1}{2 M_K^4} \\
			& - \left(B^{\tilde R_1} \Delta_{K\pi} + \left(A^{R_1} + B^{\tilde R_1} - 2 C^{R_0}\right)  2 \Sigma_0 \right) \frac{s}{2 M_K^4} + \left(2 A^{R_1} + B^{\tilde R_1} - 2 C^{R_0}\right) \frac{s^2}{2 M_K^4} , \\
		\delta N_0(t) &= - \left(2 A^{R_1} - B^{\tilde R_1} + 2 C^{R_0} \right) \frac{3 t (\Delta_{K\pi} + 2 \Sigma_0)}{8 M_K^4} + \left(6 A^{R_1} - 3 B^{\tilde R_1} -10 C^{R_0} \right) \frac{t^2}{8 M_K^4} , \\
		\delta N_1(t) &= - \frac{1}{4} \left(2 A^{R_1} + 3 B^{\tilde R_1} - 6 C^{R_0} \right) , \\
		\delta \tilde N_1(t) &= -\left(6 A^{R_1} + 5 B^{\tilde R_1} + 6 C^{R_0} \right)\frac{t}{4 M_K^2} , \\
		\delta R_0(t) &= C^{R_0} \frac{t^2}{M_K^4} , \\
		\delta R_1(t) &= A^{R_1} , \\
		\delta \tilde R_1(t) &= B^{\tilde R_1} \frac{t}{M_K^2}.
	\end{split}
\end{align}
In order to fix the gauge completely, we have to impose further conditions. We will use two different gauges. The first one corresponds to the case of an asymptotic behaviour that needs $n=2$ subtractions. It is most suitable for our numerical dispersive representation of the form factors and for the NLO chiral result. In this case, the asymptotic behaviour excludes quadratic terms in $\delta M_0(s)$ and $\delta \tilde M_1(s)$ or a linear term in $\delta M_1(s)$. Hence, in the representation (\ref{eq:FunctionsOfOneVariable}), the gauge is completely fixed.

The chiral representation, being an expansion in the masses and momenta, does not necessarily reproduce the correct asymptotic behaviour. The $\O(p^6)$ chiral expressions show an asymptotic behaviour that needs $n=3$ subtractions. In this case, one has to fix the gauge rather with the Taylor coefficients, e.g.~by excluding a quadratic term in $R_0$, a constant term in $R_1$ and a linear term in $\tilde R_1$. Therefore, also in the representation (\ref{eq:FunctionsOfOneVariable3Subtr}), the gauge is completely fixed.

Note that the second representation (\ref{eq:FunctionsOfOneVariable3Subtr}) makes less restrictive assumptions about the asymptotic behaviour. Therefore, the first representation (\ref{eq:FunctionsOfOneVariable}) is a special case of the second (\ref{eq:FunctionsOfOneVariable3Subtr}). One can easily switch from the first to the second representation with the help of the gauge transformation (\ref{eq:GaugeTransformation}). In this case, the additional subtraction constants will be given by sum rules.

\pagebreak

\subsection{Simplifications for $s_\ell\to0$}

As a test of the decomposition, let us study the linear combination
\begin{align}
	\begin{split}
		F_1(s,t,u) = \frac{1}{2}(M_K^2 - s) F(s,t,u) + \frac{1}{2}(u-t) G(s,t,u)
	\end{split}
\end{align}
in the limit $s_\ell\to0$. We neglect the contribution of the isospin $3/2$ $P$-wave:
\begin{align}
	\begin{split}
		\lim_{s_\ell\to0} &F_1(s,t,u) = \lim_{s_\ell\to0} \left( \frac{M_K^2 - s}{2} F(s,t,u) + \frac{u-t}{2} G(s,t,u) \right) \\
			&= \frac{M_K^2 - s}{2} M_0(s) \\
			&\quad + \frac{u-t}{2} \left[ \frac{M_K^2 - s}{M_K^2} M_1(s) +  \tilde M_1(s) \right] \\
			&\quad + \frac{2}{3} \left[ (t-M_\pi^2) N_0(t) \right] + \frac{1}{3} \left[ (t-M_\pi^2) R_0(t) \right] + (u-M_\pi^2) R_0(u) \\
			&\quad - \frac{2}{3} \left(t(u-s)+(M_K^2-M_\pi^2)M_\pi^2\right) \left[ \frac{t-M_\pi^2}{M_K^4} N_1(t) - \frac{1}{2 M_K^2} \tilde N_1(t) \right] .
	\end{split}
\end{align}

By identifying
\begin{align}
	\begin{split}
		M_0^{F_1}(s) &= \frac{M_K^2-s}{2} M_0(s) , \\
		M_1^{F_1}(s) &= \frac{1}{2} \left( \frac{M_K^2-s}{M_K^2} M_1(s) + \tilde M_1(s) \right) , \\
		N_0^{F_1}(t) &= (t-M_\pi^2) N_0(t) , \\
		R_0^{F_1}(t) &= (t-M_\pi^2) R_0(t) , \\
		N_1^{F_1}(t) &= \frac{t-M_\pi^2}{M_K^4} N_1(t) - \frac{1}{2M_K^2} \tilde N_1(t) ,
	\end{split}
\end{align}
we recover the decomposition of the form factor $F_1$ used in \cite{Stoffer2010, Colangelo2012, Stoffer2013}. We further note that the imaginary parts of the functions of one variable in this decomposition are given by
\begin{align}
	\begin{split}
		\Im M_0^{F_1}(s) &= \frac{M_K^2-s}{2} \Im f_0(s) , \\
		\Im M_1^{F_1}(s) &= \frac{M_K^2-s}{2M_K^2} \Im f_1(s) , \\
		\Im N_0^{F_1}(t) &= (t-M_\pi^2) \Im f_0^{(1/2)}(t) , \\
		\Im R_0^{F_1}(t) &= (t-M_\pi^2) \Im f_0^{(3/2)}(t) , \\
		\Im N_1^{F_1}(t) &= \frac{t-M_\pi^2}{M_K^4} \Im f_1^{(1/2)}(t) ,
	\end{split}
\end{align}
and repeat the observation of section~\ref{sec:PartialWaveSimplificationsZeroSL} that in the limit $s_\ell\to0$, these partial waves are given by projections of $F_1$ in all three channels. Hence, the form factor $F_1$ decouples in this limit and can be treated independently in the above decomposition.

\clearpage

\section{Integral Equations}

\subsection{Omnès Representation}

The decomposition of the form factors (\ref{eq:FormFactorDecomposition}) signifies a major simplification, since we only have to deal with functions of a single Mandelstam variable. These functions (\ref{eq:FunctionsOfOneVariable}, \ref{eq:FunctionsOfOneVariable3Subtr}) are constructed in such a way that they only contain the right-hand cut of the corresponding partial wave. Their imaginary part on the upper rim of their cut is given by
\begin{align}
	\begin{split}
		\Im M_0(s) &= \Im f_0(s) , \\
		\Im M_1(s) &= \Im \left( f_1(s) - \frac{2PL(s) M_K^2}{\lambda_{K\ell}(s)} g_1(s) \right) , \\
		\Im \tilde M_1(s) &= \Im g_1(s) , \\
		\Im N_0(t) &= \Im  f_0^{(1/2)}(t) , \\
		\Im N_1(t) &= \Im \left( f_1^{(1/2)}(t) + \frac{(\Delta_{\ell\pi}+t)M_K^4}{2t \lambda_{\ell\pi}(t)} g_1^{(1/2)}(t) \right) , \\
		\Im \tilde N_1(t) &= \Im \left(\frac{M_K^2}{t} g_1^{(1/2)}(t) \right) , \\
		\Im R_0(t) &= \Im  f_0^{(3/2)}(t) , \\
		\Im R_1(t) &= \Im \left( f_1^{(3/2)}(t) + \frac{(\Delta_{\ell\pi}+t)M_K^4}{2t \lambda_{\ell\pi}(t)} g_1^{(3/2)}(t) \right) , \\
		\Im \tilde R_1(t) &= \Im \left(\frac{M_K^2}{t} g_1^{(3/2)}(t) \right) .
	\end{split}
\end{align}
Therefore, we can write
\begin{align}
	\begin{split}
		\label{eq:HatFunctionPartialWavesRelation}
		M_0(s) + \hat M_0(s) &= f_0(s) , \\
		M_1(s) + \hat M_1(s) &= f_1(s) - \frac{2PL(s) M_K^2}{\lambda_{K\ell}(s)} g_1(s) , \\
		 \tilde M_1(s) + \hat{\tilde M}_1(s) &= g_1(s) , \\
		N_0(t) + \hat N_0(t) &=  f_0^{(1/2)}(t) , \\
		N_1(t) + \hat N_1(t) &= f_1^{(1/2)}(t) + \frac{(\Delta_{\ell\pi}+t)M_K^4}{2t \lambda_{\ell\pi}(t)} g_1^{(1/2)}(t) , \\
		\tilde N_1(t) + \hat{\tilde N}_1(t) &= \frac{M_K^2}{t} g_1^{(1/2)}(t) , \\
		R_0(t) + \hat R_0(t) &=  f_0^{(3/2)}(t) , \\
		R_1(t) + \hat R_1(t) &= f_1^{(3/2)}(t) + \frac{(\Delta_{\ell\pi}+t)M_K^4}{2t \lambda_{\ell\pi}(t)} g_1^{(3/2)}(t) , \\
		\tilde R_1(t) + \hat{\tilde R}_1(t) &= \frac{M_K^2}{t} g_1^{(3/2)}(t) ,
	\end{split}
\end{align}
where the `hat functions' $\hat M_0(s)$, $\ldots$ are real on the cut: indeed, they do not possess a right-hand cut, but contain the (possibly complicated) left-hand cut structure of the partial waves (see section~\ref{sec:ProjectionAnalyticStructurePartialWaves}). Writing $\Im f_0(s) = f_0(s) e^{-i\delta_0^0(s)} \sin\delta_0^0(s)$, $\ldots$ leads directly to the following equations:
\begin{align}
	\begin{split}
		\Im M_0(s) &= ( M_0(s) + \hat M_0(s) ) e^{-i\delta_0^0(s)} \sin\delta_0^0(s) , \\
		\Im M_1(s) &= ( M_1(s) + \hat M_1(s) ) e^{-i\delta_1^1(s)} \sin\delta_1^1(s) , \\
		\Im \tilde M_1(s) &= ( \tilde M_1(s) + \hat{\tilde M}_1(s) ) e^{-i\delta_1^1(s)} \sin\delta_1^1(s) , \\
		\Im N_0(t) &= ( N_0(t) + \hat N_0(t) ) e^{-i\delta_0^{1/2}(t)} \sin\delta_0^{1/2}(t) , \\
		\Im N_1(t) &= ( N_1(t) + \hat N_1(t) ) e^{-i\delta_1^{1/2}(t)} \sin\delta_1^{1/2}(t) , \\
		\Im \tilde N_1(t) &= ( \tilde N_1(t) + \hat{\tilde N}_1(t) ) e^{-i\delta_1^{1/2}(t)} \sin\delta_1^{1/2}(t) , \\
		\Im R_0(t) &= ( R_0(t) + \hat R_0(t) ) e^{-i\delta_0^{3/2}(t)} \sin\delta_0^{3/2}(t) , \\
		\Im R_1(t) &= ( R_1(t) + \hat R_1(t) ) e^{-i\delta_1^{3/2}(t)} \sin\delta_1^{3/2}(t) , \\
		\Im \tilde R_1(t) &= ( \tilde R_1(t) + \hat{\tilde R}_1(t) ) e^{-i\delta_1^{3/2}(t)} \sin\delta_1^{3/2}(t) ,
	\end{split}
\end{align}
where, below some inelastic threshold, the phases $\delta_l^I$ agree with the elastic $\pi\pi$- or $K\pi$-scattering phase shifts. Therefore, the functions $M_0$, $\ldots$ are given by the solution to the inhomogeneous Omnès problem. The minimal number of subtractions appearing in the Omnès representation is determined by the asymptotic behaviour of the functions $M_0$, $\ldots$ and the phases $\delta_l^I$.

Let us extend these equations even to the region above inelastic thresholds by replacing $\delta \mapsto \omega$,
\begin{align}
	\begin{split}
		\Im M_0(s) &= ( M_0(s) + \hat M_0(s) ) e^{-i\omega_0^0(s)} \sin\omega_0^0(s) , \\
		&\ldots ,
	\end{split}
\end{align}
where $\omega_l^I(s) = \delta_l^I(s) + \eta_l^I(s)$ and $\eta_l^I(s) = 0$ below the inelastic threshold $s = \Lambda^2$.

We define the usual once-subtracted Omnès function
\begin{align}
	\label{eq:OmnesFunction}
	\Omega(s) := \exp\left( \frac{s}{\pi} \int_{s_0}^\infty \frac{\delta(s^\prime)}{(s^\prime - s - i\epsilon) s^\prime} ds^\prime \right) .
\end{align}
If the asymptotic behaviour of the phase is $\lim\limits_{s\to\infty} \delta(s) = m \pi$, the Omnès function behaves asymptotically as $\O(s^{-m})$. Provided that the function $M(s)$ behaves asymptotically as $\O(s^k)$, we can write a dispersion relation for $M/\Omega$ that leads to a modified Omnès solution
\begin{align}
	\begin{split}
		\label{eq:ModifiedOmnesSolution}
		M(s) = \Omega(s) &\left\{ P_{n-1}(s) + \frac{s^n}{\pi} \int_{s_0}^{\Lambda^2} \frac{\hat M(s^\prime) \sin\delta(s^\prime)}{|\Omega(s^\prime)| (s^\prime - s - i\epsilon) {s^\prime}^n} ds^\prime  \right. \\
			&\quad + \frac{s^n}{\pi} \int_{\Lambda^2}^{\infty} \frac{\hat M(s^\prime) \sin\delta(s^\prime)}{|\Omega(s^\prime)| (s^\prime - s - i\epsilon) {s^\prime}^n} ds^\prime \\
			&\quad + \left. \frac{s^n}{\pi} \int_{\Lambda^2}^\infty \frac{(\hat M(s^\prime) + \Re M(s^\prime)) \sin\eta(s^\prime)}{|\Omega(s^\prime)| \cos(\delta(s^\prime) + \eta(s^\prime)) (s^\prime - s - i\epsilon) {s^\prime}^n} ds^\prime \right\} ,
	\end{split}
\end{align}
where the order of the subtraction polynomial is $n-1=k+m$.

Actually, we do not know the phase $\delta$ at high energies. Inelasticities due to multi-Goldstone boson intermediate states, i.e.~more than two Goldstone bosons, appear only at $\O(p^8)$ \cite{Stern1993}, hence the most important inelastic contribution would certainly be a $K\bar K$ intermediate state in the $s$-channel. This could be included by using experimental input on $\eta$ up to $s\approx(1.4 \, \mathrm{GeV})^2$.

We could make a Taylor expansion of the inelasticity integral and neglect terms that only contribute at $\O(p^8)$ to the form factors by applying the power counting $\frac{s}{\Lambda^2} \sim p^2$. This would introduce quite a lot of unknown Taylor coefficients. Here, we follow another strategy: we set $\eta = 0$ and assign a large error to the phases $\delta$ at high energies. We assume further that $\delta$ reaches a multiple of $\pi$ above a certain $s = \Lambda^2$. The two high-energy integrals in (\ref{eq:ModifiedOmnesSolution}) drop in this case.

Assuming that the phases behave asymptotically like $\delta_0^0 \to \pi$ , $\delta_1^1 \to \pi$ and all other $\delta_l^I \to 0$, we find the following solution for the case of $n=2$ subtractions:
\begin{align}
	\begin{alignedat}{2}
		\label{eq:FunctionsOfOneVariableOmnes}
		M_0(s) &= \Omega_0^0(s) & &\bigg\{ a^{M_0} + b^{M_0} \frac{s}{M_K^2} + c^{M_0} \frac{s^2}{M_K^4} + \frac{s^3}{\pi} \int_{s_0}^{\Lambda^2} \frac{\hat M_0(s^\prime) \sin\delta_0^0(s^\prime)}{|\Omega_0^0(s^\prime)| (s^\prime - s - i\epsilon) {s^\prime}^3} ds^\prime \bigg\} , \\
		M_1(s) &= \Omega_1^1(s) & &\bigg\{ a^{M_1} + b^{M_1}  \frac{s}{M_K^2} + \frac{s^2}{\pi} \int_{s_0}^{\Lambda^2} \frac{\hat M_1(s^\prime) \sin\delta_1^1(s^\prime)}{|\Omega_1^1(s^\prime)| (s^\prime - s - i\epsilon) {s^\prime}^2} ds^\prime  \bigg\} , \\
		 \tilde M_1(s) &= \Omega_1^1(s) & &\bigg\{ a^{\tilde M_1} + b^{\tilde M_1}  \frac{s}{M_K^2} + c^{\tilde M_1}  \frac{s^2}{M_K^4} + \frac{s^3}{\pi} \int_{s_0}^{\Lambda^2} \frac{\hat{\tilde M}_1(s^\prime) \sin\delta_1^1(s^\prime)}{|\Omega_1^1(s^\prime)| (s^\prime - s - i\epsilon) {s^\prime}^3} ds^\prime \bigg\} , \\
		N_0(t) &=  \Omega_0^{1/2}(t) & &\bigg\{ b^{N_0} \frac{t}{M_K^2} + \frac{t^2}{\pi} \int_{t_0}^{\Lambda^2} \frac{\hat N_0(t^\prime) \sin\delta_0^{1/2}(t^\prime)}{|\Omega_0^{1/2}(t^\prime)| (t^\prime - t - i\epsilon) {t^\prime}^2} dt^\prime  \bigg\} , \\
		N_1(t) &= \Omega_1^{1/2}(t) & &\bigg\{ \frac{1}{\pi} \int_{t_0}^{\Lambda^2} \frac{\hat N_1(t^\prime) \sin\delta_1^{1/2}(t^\prime)}{|\Omega_1^{1/2}(t^\prime)| (t^\prime - t - i\epsilon)} dt^\prime  \bigg\} , \\
		\tilde N_1(t) &= \Omega_1^{1/2}(t) & &\bigg\{ \frac{t}{\pi} \int_{t_0}^{\Lambda^2} \frac{\hat{\tilde N}_1(t^\prime) \sin\delta_1^{1/2}(t^\prime)}{|\Omega_1^{1/2}(t^\prime)| (t^\prime - t - i\epsilon) t^\prime} dt^\prime  \bigg\} , \\
		R_0(t) &=  \Omega_0^{3/2}(t) & &\bigg\{ \frac{t^2}{\pi} \int_{t_0}^{\Lambda^2} \frac{\hat R_0(t^\prime) \sin\delta_0^{3/2}(t^\prime)}{|\Omega_0^{3/2}(t^\prime)| (t^\prime - t - i\epsilon) {t^\prime}^2} dt^\prime  \bigg\} , \\
		R_1(t) &=  \Omega_1^{3/2}(t) & &\bigg\{ \frac{1}{\pi} \int_{t_0}^{\Lambda^2} \frac{\hat R_1(t^\prime) \sin\delta_1^{3/2}(t^\prime)}{|\Omega_1^{3/2}(t^\prime)| (t^\prime - t - i\epsilon)} dt^\prime  \bigg\} , \\
		\tilde R_1(t) &=   \Omega_1^{3/2}(t) & &\bigg\{ \frac{t}{\pi} \int_{t_0}^{\Lambda^2} \frac{\hat{\tilde R}_1(t^\prime) \sin\delta_1^{3/2}(t^\prime)}{|\Omega_1^{3/2}(t^\prime)| (t^\prime - t - i\epsilon) {t^\prime}} dt^\prime  \bigg\} ,
	\end{alignedat}
\end{align}
where we have fixed some of the subtraction constants in $N_0$, $\tilde N_1$, $R_0$ and $\tilde R_1$ to zero by imposing the same Taylor expansion as in the defining equation (\ref{eq:FunctionsOfOneVariable}).

Note that driving the $K\pi$ phases to 0 is somehow artificial. They are rather supposed to reach $\pi$ at high energies. However, this would introduce three more subtraction constants in our framework. Since the high-energy behaviour of the phases does not have an important influence on our results, we abstain from introducing more subtractions and take these effects into account in the systematic uncertainty.

In the case of $n=3$ subtractions, six additional subtraction constants appear in the Omnès representation. The conversion of a solution for $n=2$ into a solution for $n=3$ requires a gauge transformation in the Omnès representation, as explained in appendix~\ref{sec:AppendixOmnes3Subtractions}.

\subsection{Hat Functions}

The hat functions appearing in the Omnès solution to the functions of one variable (\ref{eq:FunctionsOfOneVariableOmnes}) can be computed through partial-wave projections of the form factors: (\ref{eq:HatFunctionPartialWavesRelation}) should be understood as the defining equation of the hat functions. One has to compute the partial-wave projections of the decomposed form factors $F$ and $G$ (\ref{eq:FormFactorDecomposition}) and subtract the function of one variable ($M_0$, $\ldots$). Finally, one obtains an expression for the hat functions in terms of angular averages of the functions of one variable. The explicit expressions for the hat functions are given in appendix~\ref{sec:AppendixHatFunctions}.


\chapter{Numerical Solution of the Dispersion Relation}

\label{sec:Kl4NumericalSolutionDR}

\section{Iterative Solution of the Dispersion Relation}

We have decomposed the form factors into functions of one variable according to (\ref{eq:FormFactorDecomposition}). The nine functions of one variable are given unambiguously by the Omnès solutions (\ref{eq:FunctionsOfOneVariableOmnes}). The hat functions appearing in the dispersive integrals are given by angular integrals of the nine functions of one variable and link these functions together. Therefore, we face a set of coupled integral equations, parametrised by the nine subtraction constants $a^{M_0}$, $b^{M_0}$, \ldots. For the numerical solution, we assume an asymptotic behaviour of the form factors that requires only $n=2$ subtractions.

It is important to note that this set of equations is linear in the subtraction constants. Any solution can be written as a linear combination of nine basic solutions.

So far, the invariant squared energy of the dilepton system, $s_\ell$, has been treated as an external fixed parameter. On the one hand, it appears in the definition of the hat functions. On the other hand, the subtraction constants have an implicit dependence on $s_\ell$. Let us now turn this dependence into an explicit one by writing the form factors as:
\begin{align}
	\begin{split}
		X(s,t,u) &= \sum_{i=1}^9 a_i(s_\ell) X_i(s,t,u) ,
	\end{split}
\end{align}
where $X\in\{F,G\}$, $\{a_i\}_i = \{ a^{M_0}, b^{M_0}, \ldots, b^{N_0} \}$ and where $X_i$ denotes the basic solution with $a_k = \delta_{ik}$, $k\in\{1,\ldots,9\}$. If $s_\ell$ is allowed to vary, the `functions of one variable' become actually functions of two variables, $M_0(s,s_\ell)$, $\ldots$

The strategy is now as follows. We determine the basic solutions by a numerical iteration of the coupled integral equations. Each basic solution is a function of $s$, $t$ and $u$, where $s+t+u = \Sigma_0 = M_K^2 + 2M_\pi^2 + s_\ell$, or equivalently a function of $s$, $s_\ell$ and $\cos\theta$. Since $s_\ell$ is a fixed external parameter in the integral equations, the iterative solution has to be performed for each value of $s_\ell$ separately. Once the basic solutions are computed, the subtraction constants (or rather functions) have to be determined by suitable means, such as a fit to data, the soft-pion theorem and \ChPT{} input. As the dependence on $s_\ell$ has been found to be rather weak in experiments, the subtraction functions can be assumed to be a low-order polynomial in $s_\ell$.

Note that the iterative procedure has to be performed a couple of times: in order to obtain a basic solution, we fix one of the subtraction constants to 1 and all the other to 0. Further, we choose a fixed value of $s_\ell$. With this setting, we solve the dispersion relation. The procedure has to be repeated for all basic solutions and all values of $s_\ell$ that we are interested in. With a couple of values for $s_\ell$, we then obtain the values of the functions $M_0$, $\ldots$ on a two-dimensional grid in the space $(s,s_\ell)$, $(t,s_\ell)$ or $(u,s_\ell)$. This enables us to reconstruct the basic solutions as functions of three variables $s$, $t$, $u$ or equivalently $s$, $\nu$, $s_\ell$.

We use the following strategy for the iterative procedure:
\begin{enumerate}
	\item set the initial value of the functions $M_0$, $\ldots$ to Omnès function $\times$ subtraction polynomial (the polynomial is in fact either 0 or a simple monomial with coefficient 1 for a particular basic solution);
	\item calculate the hat functions $\hat M_0$, $\ldots$ by means of angular integrals of the functions $M_0$, $\ldots$ ;
	\item calculate the new values of the functions $M_0$, $\ldots$ as Omnès function $\times$ (polynomial + dispersive part), where in the dispersion integral the hat function calculated in step 2 appears;
	\item go to step 2 and iterate this procedure until convergence.
\end{enumerate}
It turns out that this iteration converges quickly. After five or six iterations, the relative changes are of the order $10^{-6}$.

\section{Phase Input}

\subsection{$\pi\pi$ Phase Shifts}

\label{sec:Kl4pipiPhaseShifts}

For the pion scattering phase shifts, we use the parametrisation of \cite{Ananthanarayan2001a, Caprini2012}. The solution depends on 28 para\-meters that have to be varied. The solution~1 in figure~\ref{plot:PiPiPhaseShifts} shows the central solution for the phase shifts as well as the uncertainties due to the parameters (summed in quadrature).

\begin{figure}[H]
	\centering
	\scalebox{0.59}{
		\input{Kl4Dispersive/plots/d00.tex}
		\input{Kl4Dispersive/plots/d11.tex}
		}
	\caption{$\pi\pi$ phase shift inputs}
	\label{plot:PiPiPhaseShifts}
\end{figure}

There are now two aspects that deserve special attention. First, the phase for solution~1 is just taken constant above $\sqrt{s} \approx 1.5$~GeV. Our derivation of the dispersion relation, however, relies on the assumption that $\delta_0^0(s) \to \pi$, $\delta_1^1(s) \to \pi$ for $s\to\infty$. We should therefore change the high-energy behaviour of the phases such that they reach $\pi$ at $s=\Lambda^2$. The exact way how this is achieved must not have an influence on the result at low energies, especially in the physical region. We choose to interpolate smoothly between the value of solution~1 and $\pi$:
\begin{align}
	\begin{split}
		\delta_0^0(s)_\mathrm{sol.2} &:= \left(1 - f_\mathrm{int}(s_1,s_2,s) \right) \delta_0^0(s)_\mathrm{sol.1} + f_\mathrm{int}(s_1,s_2,s) \pi , \\
		\delta_1^1(s)_\mathrm{sol.2} &:= \left(1 - f_\mathrm{int}(s_1,s_2,s) \right) \delta_1^1(s)_\mathrm{sol.1} + f_\mathrm{int}(s_1,s_2,s) \pi , \\
		f_\mathrm{int}(s_1,s_2,s) &:= \left\{ \begin{matrix} 0 & \text{if } s < s_1 , \\ \frac{(s-s_1)^2 (3 s_2 - 2 s - s_1)}{(s_2 - s_1)^3}  & \text{if } s_1 \le s < s_2 , \\ 1 & \text{if } s_2 \le s . \end{matrix} \right. 
	\end{split}
\end{align}
Figure~\ref{plot:PiPiPhaseShifts} shows solution~2 with $s_1 = 68 M_\pi^2$ and $s_2 = 148 M_\pi^2$. These values can be varied and should not have an important influence.

The second subtlety is the problem of the behaviour around the $K\bar K$ threshold \cite{Oller2007}: are the $K_{\ell4}$ partial waves expected to have a peak or a dip in the vicinity of the $K \bar K$ threshold, i.e.~do they rather behave like the strange or the non-strange scalar form factor of the pion? In the latter case, we have to modify the phase such that it follows $\delta_0^0(s)-\pi$ above the $K \bar K$ threshold. The third solution shown in figure~\ref{plot:PiPiPhaseShifts} is given by
\begin{align}
	\begin{split}
		\delta_0^0(s)_\mathrm{sol.3} &:= \left(1 - f_\mathrm{int}(s_1,s_2,s) \right) \left( \delta_0^0(s)_\mathrm{sol.1} - f_\mathrm{int}(\tilde s_1, \tilde s_2, s) \pi \right) + f_\mathrm{int}(s_1,s_2,s) \pi , 
	\end{split}
\end{align}
with $\tilde s_1 = 4 M_K^2$ and $\tilde s_2 = \tilde s_1 + 8 M_\pi^2$.

The solution~4 in figure~\ref{plot:PiPiPhaseShifts} corresponds to solution~2 but with $s_1 = 4 M_K^2$ and $s_2 = s_1 + M_\pi^2$.

As the question of the correct behaviour around the $K\bar K$ threshold is not easy to answer, we declare the solution 3 as the `central' one and use all the other solutions to determine the systematic uncertainty.

\begin{figure}[H]
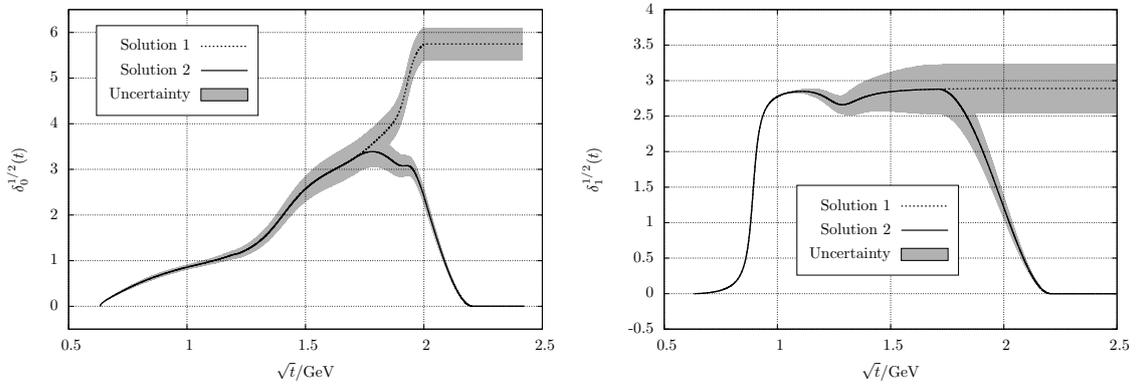

	\centering
	\scalebox{0.59}{
		\input{Kl4Dispersive/plots/d012.tex}
		\input{Kl4Dispersive/plots/d112.tex}
		}
	\caption{$K\pi$ phase shift inputs, isospin $I=1/2$}
	\label{plot:KPiPhaseShifts12}
\end{figure}

\begin{figure}[H]
	\centering
	\scalebox{0.59}{
		\input{Kl4Dispersive/plots/d032.tex}
		\input{Kl4Dispersive/plots/d132.tex}
		}
	\caption{$K\pi$ phase shift inputs, isospin $I=3/2$}
	\label{plot:KPiPhaseShifts32}
\end{figure}

\clearpage

\subsection{$K\pi$ Phase Shifts}

\label{sec:Kl4KpiPhaseShifts}

For the crossed channels, we need the $K\pi$ phase shifts as an input. We use the phase shifts and uncertainties of \cite{Buettiker2004, Boito2010}, but add by hand a more conservative uncertainty that reaches $20^\circ$ at $t=150M_\pi^2$. For the very small phase $\delta_1^{3/2}$, we just assume a 100\% uncertainty. These phase solutions are shown in figures~\ref{plot:KPiPhaseShifts12} and \ref{plot:KPiPhaseShifts32} as `solution~1'.

In the derivation of the dispersion relation, we assume that the $K\pi$ phases go to zero at high energies. We implement this by interpolating smoothly between the solution~1 and zero with $f_\mathrm{int}(t_1,t_2,t)$. These modified phase shifts with $t_1 = 150 M_\pi^2$ and $t_2 = 250 M_\pi^2$ are displayed as `solution~2' in figures~\ref{plot:KPiPhaseShifts12} and \ref{plot:KPiPhaseShifts32}. The difference between `solution~1' and `solution~2' is taken as a measure of the systematic uncertainty due to the high-energy behaviour of the $K\pi$ phases.

\section{Omnès Functions}

\begin{figure}[H]
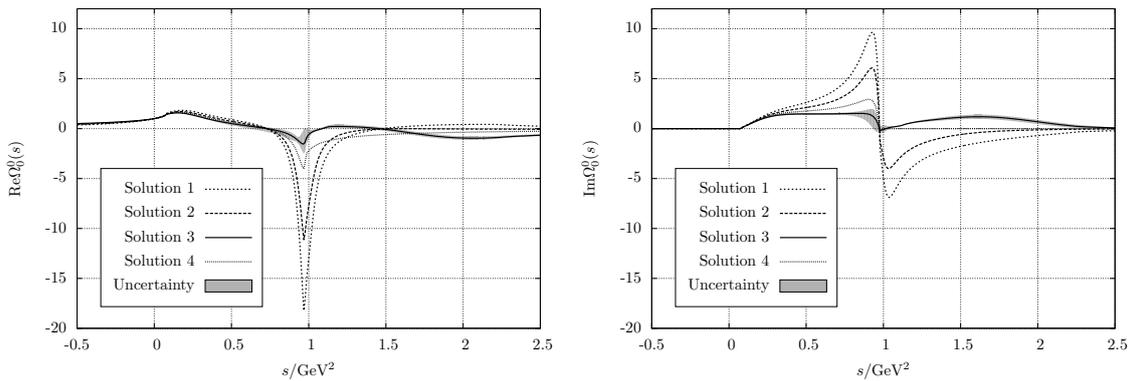

	\centering
	\scalebox{0.59}{
		\input{Kl4Dispersive/plots/ReOm00.tex}
		\input{Kl4Dispersive/plots/ImOm00.tex}
		}
	\caption{$\pi\pi$ $S$-wave Omnès function, isospin $I=0$}
	\label{plot:OmnesFunction00}
\end{figure}

\begin{figure}[H]
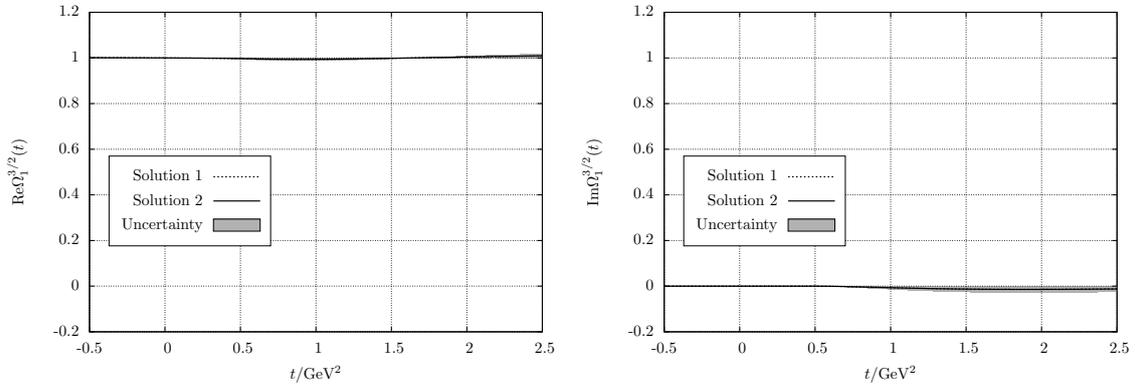

	\centering
	\scalebox{0.59}{
		\input{Kl4Dispersive/plots/ReOm132.tex}
		\input{Kl4Dispersive/plots/ImOm132.tex}
		}
	\caption{$K\pi$ $P$-wave Omnès function, isospin $I=3/2$}
	\label{plot:OmnesFunction132}
\end{figure}

In a first step, the six Omnès functions are computed, defined by
\begin{align}
	\begin{split}
		\Omega_l^I(s) := \exp\left( \frac{s}{\pi} \int_{s_0}^\infty \frac{\delta_l^I(s^\prime)}{(s^\prime - s - i\epsilon) s^\prime} ds^\prime \right) ,
	\end{split}
\end{align}
where $s_0$ denotes the respective threshold. The results are shown in figures~\ref{plot:OmnesFunction00} to \ref{plot:OmnesFunction132}. The error bands are determined as follows.
\begin{itemize}
	\item For the $\pi\pi$ phases, the Omnès function is computed for the phase solution~3 with variation of all the 28 parameters. The differences, appropriately weighted, are summed up in quadrature to give the error band. For the phase solutions~1, 2 and 4, only the central curve is shown.
	\item For the $K\pi$ phases, three curves are computed to give the error band of the Omnès functions: the ones corresponding to the central phase solutions, the upper and the lower border of the phase error bands. The Omnès functions corresponding to both phase solutions are shown.
\end{itemize}

Note that in $\Omega_0^0$, the differences between the different high-energy phase solutions are much larger than the error band due to the phase parameters. However, most of these differences can be absorbed at low energies by the subtraction constants of the dispersion relation.

\begin{figure}[H]
	\centering
	\scalebox{0.59}{
		\input{Kl4Dispersive/plots/ReOm11.tex}
		\input{Kl4Dispersive/plots/ImOm11.tex}
		}
	\caption{$\pi\pi$ $P$-wave Omnès function, isospin $I=1$}
	\label{plot:OmnesFunction11}
\end{figure}

\begin{figure}[H]
	\centering
	\scalebox{0.59}{
		\input{Kl4Dispersive/plots/ReOm012.tex}
		\input{Kl4Dispersive/plots/ImOm012.tex}
		}
	\caption{$K\pi$ $S$-wave Omnès function, isospin $I=1/2$}
	\label{plot:OmnesFunction012}
\end{figure}

\begin{figure}[H]
	\centering
	\scalebox{0.59}{
		\input{Kl4Dispersive/plots/ReOm112.tex}
		\input{Kl4Dispersive/plots/ImOm112.tex}
		}
	\caption{$K\pi$ $P$-wave Omnès function, isospin $I=1/2$}
	\label{plot:OmnesFunction112}
\end{figure}

\begin{figure}[H]
	\centering
	\scalebox{0.59}{
		\input{Kl4Dispersive/plots/ReOm032.tex}
		\input{Kl4Dispersive/plots/ImOm032.tex}
		}
	\caption{$K\pi$ $S$-wave Omnès function, isospin $I=3/2$}
	\label{plot:OmnesFunction032}
\end{figure}

\clearpage

\section{Hat Functions and Angular Projection}

During the iterative solution of the dispersion relation, the hat functions have to be computed by means of angular averages. Since the hat functions appear in the integrand of the dispersive integrals, they have to be known just on the real axis above the threshold of the respective channel.

In the case of $s_\ell = 0$, the calculation of the angular integrals is straightforward. The functions $M_0$, $\ldots$ need to be computed on the real axis, also on the negative part of it. As described in section~\ref{sec:ProjectionAnalyticStructurePartialWaves}, a subtlety arises in the case $s_\ell \neq 0$: in the calculation of the $s$-channel hat functions, we have to know the angular integrals of the $t$- and $u$-channel functions $N_0$, $\ldots$. In the region $(M_K - \sqrt{s_\ell})^2 < s < (M_K + \sqrt{s_\ell})^2$, the angular integration path extends into the complex $t$- or $u$-plane. Therefore, the $t$- and $u$-channel functions $N_0$, $\ldots$ have to be computed not only on the real axis but also in the complex plane. Since the region where this happens is much below the $t$- or $u$-channel cut, we have two options how to perform this:
\begin{itemize}
	\item integrate on a straight line in the complex $t$- or $u$-plane. The functions $N_0(t)$, $\ldots$ have to be known in an egg-shaped region of $s_\ell$-dependent size. The egg lies within $M_\pi^2 - M_K \sqrt{s_\ell} < \Re(t) < M_\pi^2 + M_K \sqrt{s_\ell}$. The functions $N_0(t)$, $\ldots$ can be computed on a two-dimensional grid covering this egg and then e.g.~interpolated with a 2D spline.
	\item since the functions $N_0(t)$, $\ldots$ are analytic in the region of the egg, the angular integration path can be deformed to lie always on the border of the egg. Therefore, the functions $N_0(t)$,~$\ldots$ only have to be computed on points lying on this border (in addition to the real axis) and 1D interpolation methods can be applied.
\end{itemize}
The first method is more straightforward, the second needs less computing time. Let us derive the change of variable needed for the second approach.

\begin{figure}[ht]
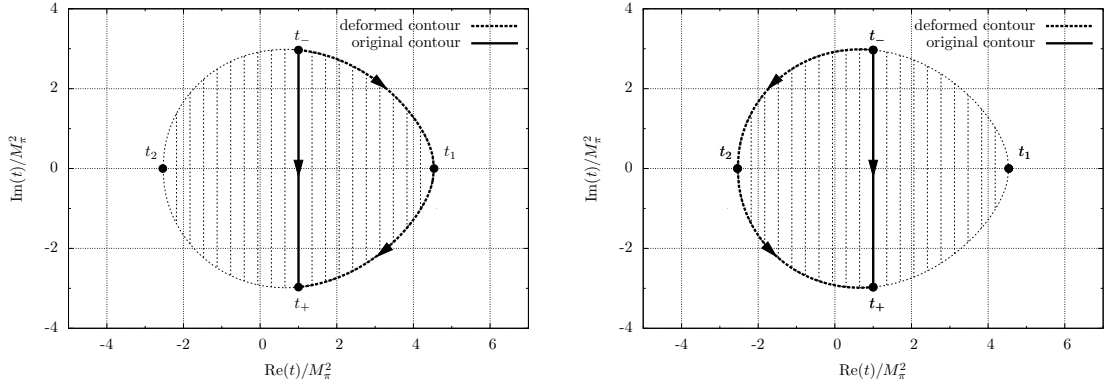

	\centering
	\scalebox{0.59}{
		\input{Kl4Dispersive/plots/egg.tex}
		\input{Kl4Dispersive/plots/egg2.tex}
		}
	\caption{Angular integration contours for $s_\ell = M_\pi^2$}
	\label{fig:Egg}
\end{figure}

We want to compute the angular integral
\begin{align}
	\begin{split}
		\< z^n X \>_{t_s}(s) = \frac{1}{2} \int_{-1}^1 dz \, z^n X(t(s,z)) ,
	\end{split}
\end{align}
where e.g.~$X = N_0$ and
\begin{align}
	\begin{split}
		t(s,z) = \frac{1}{2} \left( \Sigma_0 - s - \lambda^{1/2}_{K\ell}(s) \sigma_\pi(s) z \right) .
	\end{split}
\end{align}
The square root of the Källén function is defined by
\begin{align}
	\lambda^{1/2}_{K\ell}(s) = \left\{
	\begin{array}{r c}
		+ | \lambda^{1/2}_{K\ell}(s) | & s < (M_K-\sqrt{s_\ell})^2 , \\
		+ i | \lambda^{1/2}_{K\ell}(s) | & (M_K-\sqrt{s_\ell})^2 < s < (M_K+\sqrt{s_\ell})^2 , \\
		- | \lambda^{1/2}_{K\ell}(s) | & (M_K+\sqrt{s_\ell})^2 < s 
	\end{array} \right. 
\end{align}
and the critical region is $s_- < s < s_+$, where we define
\begin{align}
	\begin{split}
		s_\pm := (M_K \pm \sqrt{s_\ell})^2 .
	\end{split}
\end{align}
In this region, the angular integration path in the complex $t$-plane runs from $t_- := t(s,-1)$ to $t_+ := t(s,1)$. Due to the analyticity of the function $X(t)$, the straight contour can be deformed along the border of the egg, either to pass $t_1 := t(s_-,z)$ or $t_2 := t(s_+,z)$, see the two plots in figure~\ref{fig:Egg}.
Defining
\begin{align}
	\begin{split}
		z_s(t) = \frac{1}{\lambda^{1/2}_{K\ell}(s) \sigma_\pi(s)} \left( \Sigma_0 - s - 2 t \right) ,
	\end{split}
\end{align}
we rewrite the angular integral as a complex integral:
\begin{align}
	\begin{split}
		\< z^n X \>_{t_s} &= \frac{1}{2} \int_{t_-}^{t_+} \frac{dz_s}{dt} dt \, z_s^n(t) X(t) \\
			&= - \frac{1}{\lambda_{K\ell}^{1/2}(s) \sigma_\pi(s)} \int_{t_-}^{t_+} dt \, z_s^n(t) X(t) \\
			&=  - \frac{1}{\lambda_{K\ell}^{1/2}(s) \sigma_\pi(s)} \left(  \int_{t_-}^{t_1} dt \, z_s^n(t) X(t) -  \int_{t_+}^{t_1} dt \, z_s^n(t) X(t) \right) ,
	\end{split}
\end{align}
or equivalently
\begin{align}
	\begin{split}
		\< z^n X \>_{t_s} &=  - \frac{1}{\lambda_{K\ell}^{1/2}(s) \sigma_\pi(s)} \left(  \int_{t_-}^{t_2} dt \, z_s^n(t) X(t) -  \int_{t_+}^{t_2} dt \, z_s^n(t) X(t) \right) .
	\end{split}
\end{align}
We parametrise the border of the egg by the following curves:
\begin{align}
	\begin{split}
		t_s^\pm(\xi) &:= t(\xi,\pm1) = \frac{1}{2} \left( \Sigma_0 - \xi \mp \lambda_{K\ell}^{1/2}(\xi)\sigma_\pi(\xi) \right) , \quad \xi \in [ s_-, s_+] ,
	\end{split}
\end{align}
hence
\begin{align}
	\begin{split}
		\< z^n X \>_{t_s} &= - \frac{1}{\lambda_{K\ell}^{1/2}(s) \sigma_\pi(s)} \begin{aligned}[t]
			& \Bigg( \int_s^{s_-} d\xi \frac{d t_s^-(\xi)}{d\xi} z_s^n(t_s^-(\xi)) X(t_s^-(\xi)) \\
			& - \int_s^{s_-} d\xi \frac{dt_s^+(\xi)}{d\xi} z_s^n(t_s^+(\xi)) X(t_s^+(\xi)) \Bigg) , \end{aligned}
	\end{split}
\end{align}
or
\begin{align}
	\begin{split}
		\< z^n X \>_{t_s} &= - \frac{1}{\lambda_{K\ell}^{1/2}(s) \sigma_\pi(s)} \begin{aligned}[t]
			& \Bigg( \int_s^{s_+} d\xi \frac{d t_s^-(\xi)}{d\xi} z_s^n(t_s^-(\xi)) X(t_s^-(\xi)) \\
			&  - \int_s^{s_+} d\xi \frac{dt_s^+(\xi)}{d\xi} z_s^n(t_s^+(\xi)) X(t_s^+(\xi)) \Bigg) , \end{aligned}
	\end{split}
\end{align}
where
\begin{align}
	\begin{split}
		\frac{d t_s^\pm(\xi)}{d\xi} &= \frac{1}{2} \left( -1 \mp \frac{d(\lambda_{K\ell}^{1/2}(\xi) \sigma_\pi(\xi))}{d\xi} \right) \\
			&= \frac{1}{2} \left(-1 \mp \frac{2 M_K^4 M_\pi^2 - M_K^2 \left(4 M_\pi^2 s_\ell + \xi^2\right)+(s_\ell-\xi) \left(2 M_\pi^2 (\xi+s_\ell)-\xi^2\right)}{\xi^2 \lambda_{K\ell}^{1/2}(\xi) \sigma_\pi(\xi)} \right) , \\
		z_s(t_s^\pm(\xi)) &= \frac{1}{\lambda_{K\ell}^{1/2}(s) \sigma_\pi(s)} \left( \xi - s \pm \lambda_{K\ell}^{1/2}(\xi) \sigma_\pi(\xi) \right) .
	\end{split}
\end{align}
Note that
\begin{align}
	\begin{split}
		z_s(t_s^+(\xi)) &= - z_s(t_s^-(\xi))^* , \\
		t_s^+(\xi) &= t_s^-(\xi)^* , \\
		\frac{d t_s^+(\xi)}{d\xi} &= \left( \frac{dt_s^-(\xi)}{d\xi} \right)^*
	\end{split}
\end{align}
and hence, due to the Schwarz reflection principle
\begin{align}
	\begin{split}
		X(t_s^+(\xi)) = X(t_s^-(\xi))^* .
	\end{split}
\end{align}
Therefore, the function $X$ has to be computed only on the `upper half-egg':
\begin{align}
	\begin{split}
		\< z^n X \>_{t_s} &= \frac{1}{\lambda_{K\ell}^{1/2}(s) \sigma_\pi(s)} \int_{s_-}^s d\xi \begin{aligned}[t]
			& \Bigg( \frac{dt_s^-(\xi)}{d\xi} z_s^n(t_s^-(\xi)) X(t_s^-(\xi)) \\
			& - (-1)^n \left( \frac{dt_s^-(\xi)}{d\xi} z_s^n(t_s^-(\xi)) X(t_s^-(\xi)) \right)^* \Bigg) \end{aligned}
	\end{split}
\end{align}
or
\begin{align}
	\begin{split}
		\< z^n X \>_{t_s} &= -\frac{1}{\lambda_{K\ell}^{1/2}(s) \sigma_\pi(s)} \int_s^{s_+} d\xi \begin{aligned}[t]
			& \Bigg( \frac{dt_s^-(\xi)}{d\xi} z_s^n(t_s^-(\xi)) X(t_s^-(\xi)) \\
			& - (-1)^n \left( \frac{dt_s^-(\xi)}{d\xi} z_s^n(t_s^-(\xi)) X(t_s^-(\xi)) \right)^* \Bigg) . \end{aligned}
	\end{split}
\end{align}
Although both descriptions are valid in the range $s_- < s < s_+$, one may choose to use the first in the region $s_- < s < s_m$ and the second in the region $s_m < s < s_+$, where $s_m$ lies somewhere in the middle of $s_-$ and $s_+$, e.g.~$s_m = (s_-+s_+)/2$. The motivation to do so is to avoid numerical instabilities: the integral from $s_-$ to $s$ with $s \to s_+$ must tend to zero to give a finite value for the hat function. The integral over the whole half-egg, however, accumulates a numerical uncertainty.


\chapter{Determination of the Subtraction Constants}

\label{sec:Kl4SubtractionConstantsDetermination}

In the previous chapter, we have described how to solve numerically the Omnès dispersion relation for the form factors $F$ and $G$. The solution is parametrised in terms of the subtraction constants $a^{M_0}$,~$\ldots$. The next task is now to determine these subtraction constants in order to fix the parametric freedom. We use three different sources of information for the determination of the subtraction constants:
\begin{itemize}
	\item experimental data on the $K_{\ell4}$ form factors,
	\item the soft-pion theorem, providing relations between $F$, $G$ and the $K_{\ell3}$ vector form factor
	\item and finally input from \ChPT{}.
\end{itemize}

The soft-pion theorem (SPT) is valid up to corrections of $\O(M_\pi^2)$ and hence can be considered as a strong constraint. From the two high-statistics experiments NA48/2 and E865 we have data on the $S$- and $P$-waves of the form factors. Although these experiments have achieved impressive results, the data alone does not determine all the subtraction constants with satisfactory precision. Therefore, we use chiral input to fix some of the subtraction constants that are not well determined by the fit to data.

In the following, we describe in more detail what data we use for the fits and how these fits are performed. For both experiments NA48/2 and E865 we were provided with additional unpublished data, which will be shown below. Therefore, our fits include the maximal amount of experimental information on the $K_{\ell4}$ form factors $F$ and $G$ that is currently available.

\section{Experimental Data}

\label{sec:ExperimentalKl4Data}

The NA48/2 experiment defines the partial wave expansion of the form factors as
\begin{align}
	\begin{split}
		F &= F_s e^{i\delta_s} + F_p e^{i\delta_p} \cos\theta + \ldots , \\
		G &= G_p e^{i\delta_p} + \ldots
	\end{split}
\end{align}
and further defines the linear combination
\begin{align}
	\begin{split}
		\label{eq:Gptilde}
		\tilde G_p := G_p + \frac{X}{\sigma_\pi PL} F_p .
	\end{split}
\end{align}
For us, it is convenient to define the partial wave
\begin{align}
	\begin{split}
		\label{eq:FptildeDefinition}
		\tilde F_p := \frac{M_K^2}{2X\sigma_\pi} F_p .
	\end{split}
\end{align}
In our former treatment of the form factor $F_1$ \cite{Stoffer2010, Colangelo2012, Stoffer2013}, it was most convenient to use the data on $F_s$ and $\tilde G_p$ (which corresponds to the $P$-wave of $F_1$). Now that we describe both form factors $F$ and $G$, we prefer to fit the partial waves $F_s$, $\tilde F_p$ and $G_p$.

The comparison with our definition of the $s$-channel partial-wave expansions
\begin{align}
	\begin{split}
		F &= \sum_{l=0}^\infty P_l(\cos\theta) \left( \frac{2 X \sigma_\pi}{M_K^2}\right)^l f_l - \frac{\sigma_\pi PL}{X} \cos\theta \; G , \\
		G &= \sum_{l=1}^\infty P_l^\prime(\cos\theta) \left( \frac{2 X \sigma_\pi}{M_K^2}\right)^{l-1} g_l ,
	\end{split}
\end{align}
allows us to identify
\begin{align}
	\begin{split}
		\label{eq:Kl4PartialWavesIdentification}
		F_s e^{i \delta_s} &= f_0 , \\
		F_p e^{i \delta_p} &= \frac{2X\sigma_\pi}{M_K^2} f_1 - \frac{\sigma_\pi PL}{X} g_1 , \\
		G_p e^{i \delta_p} &= g_1 .
	\end{split}
\end{align}
The phase shifts are just given by the $\pi\pi$ phases that we use as input. With (\ref{eq:HatFunctionPartialWavesRelation}), we find the fitting equations:
\begin{align}
	\begin{split}
		\label{eqn:PartialWavesFuncOneVariableRelations}
		F_s(s,s_\ell) &= \left( M_0(s,s_\ell) + \hat M_0(s,s_\ell) \right) e^{-i \delta_0^0(s)} , \\
		\tilde F_p(s,s_\ell) &= \left( M_1(s,s_\ell) + \hat M_1(s,s_\ell) \right) e^{-i \delta_1^1(s)} , \\
		G_p(s,s_\ell) &= \left( \tilde M_1(s,s_\ell) + \hat{\tilde M}_1(s,s_\ell) \right) e^{-i \delta_1^1(s)} .
	\end{split}
\end{align}

The NA48/2 collaboration has performed phenomenological fits of the form \cite{Batley2010,Batley2012}
\begin{align}
	\begin{split}
		\label{eq:NA48PhenomenologicalFit}
		\frac{F_s(s,s_\ell)}{f_s} &= 1 + \frac{f_s^\prime}{f_s} q^2 + \frac{f_s^\dprime}{f_s} q^4 + \frac{f_e^\prime}{f_s} \frac{s_\ell}{4M_\pi^2} , \\
		\frac{F_p(s,s_\ell)}{f_s} &= \frac{f_p}{f_s} , \\
		\frac{G_p(s,s_\ell)}{f_s} &= \frac{g_p}{f_s} + \frac{g_p^\prime}{f_s} q^2 ,
	\end{split}
\end{align}
where $q^2 = \frac{s}{4M_\pi^2}-1$. In a first step, only the relative values were measured \cite{Batley2010}. In a second step, the normalisation $f_s$ was determined from the branching ratio measurement and a phase-space integration, using the parametrisation (\ref{eq:NA48PhenomenologicalFit}) and the fitted relative values \cite{Batley2012}.

However, one should note that from (\ref{eq:Kl4PartialWavesIdentification}) it follows that $F_p$ has to vanish at the $\pi\pi$ threshold like $\sim \sqrt{q^2}$. The phenomenological fit (\ref{eq:NA48PhenomenologicalFit}) of \cite{Batley2010,Batley2012}, which assumes $F_p$ to be constant in $q^2$, gives a wrong threshold behaviour. We have not tried to estimate its influence on the determination of the normalisation $f_s$.

For our purpose, we find it convenient to work with $\tilde F_p$, which does not contain kinematic prefactors.

Because all the basic solutions use the same $\pi\pi$ phase as input, the real quantities $F_s$, $\tilde F_p$ and $G_p$ are still linear combinations of the corresponding quantities computed with the basic solutions. Note that the partial waves can be negative, i.e.~one really has to rotate the $\pi\pi$ phase away and not just take the absolute value.

In tables~\ref{tab:NA48Data} and \ref{tab:E865Data}, we quote the experimental values of NA48/2 \cite{Batley2010, Batley2012} and E865 \cite{Pislak2001, Pislak2003} on the partial waves. Some remarks on these numbers are appropriate.
\begin{itemize}
	\item In addition to the published NA48/2 data in 10 bins of $s$, very recently we were provided with unpublished two-dimensional data on $F_s(s,s_\ell)$: in this set, not a single bin but up to 10 bins are used in $s_\ell$-direction.\footnote{We thank Brigitte Bloch-Devaux for providing this two-dimensional data set.} We will include a fit to the full two-dimensional data set in the final publication.
	\item The barycentre values of $s_\ell$ for the bins of NA48/2 are not given in the literature \cite{Batley2010, Batley2012}. They are extracted from the two-dimensional data set. A value of $s_\ell$ could also be extracted from the relation (\ref{eq:Gptilde}) between $F_p$, $G_p$ and $\tilde G_p$ \cite{Stoffer2014}. However, this value does not agree with the barycentre. The reason for this discrepancy is not clear to us.
	\item We compute the value of $\tilde F_p$ with (\ref{eq:FptildeDefinition}) using the values of $F_p$ and the barycentre values of $s$ and $s_\ell$.
	\item There is a discrepancy between \cite{Batley2010} and \cite{Batley2012}. The statistical and systematic uncertainties for $F_s$ in the NA48/2 data have to be calculated from the relative values in \cite{Batley2010}. The numbers in \cite{Batley2012} are not correct.\footnote{We thank Brigitte Bloch-Devaux for confirming this.}
	\item The published values of $F_s$ in the 10 bins of NA48/2 have been normalised in such a way that a fit of the form (\ref{eq:NA48PhenomenologicalFit}) with $f_e^\prime = 0$ results in $F_s(0,0)/f_s = 1$, although a non-zero value of $f_e^\prime$ has been obtained from a fit to the two-dimensional data set. In order to take the $s_\ell$-dependence consistently into account, we have to increase the values of $F_s$ by 0.77\%.
	\item The E865 experiment has assumed in the analysis that the form factors do not depend on $s_\ell$. The values of $s_\ell$ for each bin were not published.\footnote{We thank Peter Truöl and Andries van der Schaaf, who performed a new analysis of the raw data in order to extract the values of $s_\ell$.}
	\item The E865 experiment only provides data on the first partial waves $F_s$ and $G_p$.
	\item The E865 papers \cite{Pislak2001, Pislak2003} include the fully correlated error of the normalisation in their systematic errors (added in quadrature). In table~\ref{tab:E865Data}, we have removed the normalisation error of $1.2\%$\footnote{We thank Stefan Pislak and Peter Truöl for this additional unpublished information.}, because it needs a special treatment for unbiased fitting.
\end{itemize}

\begin{table}[H]
	\centering
	\begin{tabular}{c c c c c c}
		\toprule
		$\sqrt{s}/$MeV & $\sqrt{s_\ell}/$MeV & $F_s$ & $F_p$ & $G_p$ & $\tilde G_p$ \\[0.1cm]
		\hline \\[-0.3cm]
		286.06 &	91.90 &	5.7195(85)(\hphantom{0}88) &				$-0.181(67)(15)$	&	5.053(258)(66) 			&	4.334(74)(19) 	\\
		295.95 &	91.30 &	5.8123(90)(\hphantom{0}45) &				$-0.324(62)(34)$	&	5.186(142)(84) 			&	4.422(53)(31) 	\\
		304.88 &	91.29 &	5.8647(89)(\hphantom{0}50) &				$-0.209(60)(33)$	&	4.941(108)(59) 			&	4.550(46)(25) 	\\
		313.48 &	90.08 &	5.9134(91)(\hphantom{0}50) &				$-0.156(58)(32)$	&	4.896(\hphantom{0}91)(51)	&	4.645(41)(23) 	\\
		322.02 &	88.03 &	5.9496(90)(\hphantom{0}30) &	$-0.366(55)(41)$	&	5.245(\hphantom{0}80)(58) 	&	4.711(38)(28) 	\\
		330.80 &	84.79 &	5.9769(93)(\hphantom{0}44) &				$-0.383(54)(38)$	&	5.283(\hphantom{0}73)(56) 	&	4.767(35)(27) 	\\
		340.17 &	80.83 &	6.0119(92)(\hphantom{0}35) &	$-0.218(55)(46)$	&	5.054(\hphantom{0}68)(59) 	&	4.780(34)(30) 	\\
		350.94 &	76.11 &	6.0354(92)(\hphantom{0}27) &	$-0.302(54)(33)$	&	5.264(\hphantom{0}62)(37) 	&	4.907(34)(20) 	\\
		364.57 &	69.87 &	6.0532(91)(\hphantom{0}32) &	$-0.309(54)(31)$	&	5.357(\hphantom{0}57)(30) 	&	5.019(35)(19) 	\\
		389.95 &	58.75 &	6.1314(93)(159) &				$-0.264(59)(33)$	&	5.418(\hphantom{0}55)(33) 	&	5.163(36)(21) 	\\
		\bottomrule
	\end{tabular}
	\caption[NA48/2 measurements of the partial waves of the form factors.]{NA48/2 measurements \cite{Batley2010,Batley2012} of the partial waves of the form factors. The first error is statistical, the second systematic. In addition, the fully correlated error of the normalisation of $0.62\%$ has to be taken into account.}
	\label{tab:NA48Data}
\end{table}

\begin{table}[H]
	\centering
	\begin{tabular}{c c c c}
		\toprule
		$\sqrt{s}/$MeV & $\sqrt{s_\ell}/$MeV & $F_s$ & $G_p$ \\[0.1cm]
		\hline \\[-0.3cm]
		287.6 &	106.8			&	5.832(13)(39) &	4.703(89)(40) 	\\
		299.5 &	105.7			&	5.875(14)(45) &	4.694(62)(37) 	\\
		311.2 &	103.8			&	5.963(14)(54) &	4.772(54)(41) 	\\
		324.0 &	101.1			&	6.022(16)(60) &	5.000(51)(56)	\\
		339.6 &	\hphantom{0}96.3	&	6.145(17)(61) &	5.003(49)(57) 	\\
		370.0 &	\hphantom{0}84.6	&	6.196(20)(38) &	5.105(50)(42) 	\\
		\bottomrule
	\end{tabular}
	\caption[E865 measurements of the partial waves of the form factors.]{E865 measurements \cite{Pislak2001, Pislak2003} of the partial waves of the form factors. The first error is statistical, the second systematic. In addition, the fully correlated error of the normalisation of $1.2\%$ has to be taken into account.}
	\label{tab:E865Data}
\end{table}

\begin{table}[H]
	\centering
	\begin{tabular}{c c c c c c}
		\toprule
		$\sqrt{s}/$MeV & $\sqrt{s_\ell}/$MeV & $F_s$ & $F_p$ & $G_p$ & $\tilde G_p$ \\[0.1cm]
		\hline \\[-0.3cm]
		286.06 &	91.90 &	5.7448(85)(\hphantom{0}88) 	&	$-0.180(67)(15)$	&	5.036(257)(66) 			&	4.320(74)(19) 	\\
		295.95 &	91.30 &	5.8380(90)(\hphantom{0}45) 	&	$-0.323(62)(34)$	&	5.169(142)(84) 			&	4.407(53)(31) 	\\
		304.88 &	91.29 &	5.8906(89)(\hphantom{0}50) 	&	$-0.208(60)(33)$	&	4.925(108)(59) 			&	4.535(46)(25) 	\\
		313.48 &	90.08 &	5.9395(91)(\hphantom{0}50) 	&	$-0.155(58)(32)$	&	4.880(\hphantom{0}91)(51)	&	4.630(41)(23) 	\\
		322.02 &	88.03 &	5.9759(90)(\hphantom{0}30) 	&	$-0.365(55)(41)$	&	5.228(\hphantom{0}80)(58) 	&	4.695(38)(28) 	\\
		330.80 &	84.79 &	6.0033(93)(\hphantom{0}44) 	&	$-0.382(54)(38)$	&	5.266(\hphantom{0}73)(56) 	&	4.751(35)(27) 	\\
		340.17 &	80.83 &	6.0384(92)(\hphantom{0}35) 	&	$-0.217(55)(46)$	&	5.037(\hphantom{0}68)(59) 	&	4.764(34)(30) 	\\
		350.94 &	76.11 &	6.0621(92)(\hphantom{0}27) 	&	$-0.301(54)(33)$	&	5.247(\hphantom{0}62)(37) 	&	4.891(34)(20) 	\\
		364.57 &	69.87 &	6.0799(91)(\hphantom{0}32) 	&	$-0.308(54)(31)$	&	5.339(\hphantom{0}57)(30) 	&	5.002(35)(19) 	\\
		389.95 &	58.75 &	6.1585(93)(160)			&	$-0.263(59)(33)$	&	5.400(\hphantom{0}55)(33) 	&	5.146(36)(21) 	\\
		\bottomrule
	\end{tabular}
	\caption[NA48/2 data with additional radiative corrections.]{NA48/2 data \cite{Batley2010,Batley2012}, corrected by additional radiative effects \cite{Stoffer2014}. The fully correlated error of the normalisation increases to $0.70\%$. The normalisation of $F_s$ is increased by $0.77\%$ to take the dependence on $s_\ell$ into account (see text).}
	\label{tab:NA48DataRadCorr}
\end{table}

In the data analysis of both experiments, radiative corrections have been applied to some extent. More reliable radiative corrections based on a fixed-order calculation \cite{Stoffer2014} can be applied a posteriori at least to the NA48/2 data, resulting in the numbers of table~\ref{tab:NA48DataRadCorr}. These values also include the mentioned correction of the normalisation of $F_s$ by $0.77\%$ due to the $s_\ell$-dependence.

Furthermore, neither the E865 nor the NA84/2 experiment has corrected the isospin-breaking effects due to the quark and meson mass differences. The calculation of \cite{Stoffer2014} also allows for their correction. The resulting numbers are given in tables~\ref{tab:NA48DataIsoCorr} and \ref{tab:E865DataIsoCorr}. We add the uncertainties of the isospin corrections (without the higher order estimate) in quadrature to the systematic errors.

In addition to the statistical and systematic errors given in the literature, B.~Bloch-Devaux provided us with the correlations between $G_p$ and $\tilde G_p$ of the NA48/2 data, shown in table~\ref{tab:NA48DataGpGptCorrelations}. Only the bin-diagonal correlations are available, hence we assume the bin-to-bin correlation to vanish. We also neglect the correlation with the $S$-wave, which is not (yet) available. With the given correlations, we compute the statistical covariances and correlations of $F_p$ and $G_p$, shown in table~\ref{tab:NA48DataFpGpCorrCov}. Note that $G_p$ is much stronger correlated with $F_p$ than with $\tilde G_p$.

\begin{table}[H]
	\centering
	\begin{tabular}{c c c c c c}
		\toprule
		$\sqrt{s}/$MeV & $\sqrt{s_\ell}/$MeV & $F_s$ & $F_p$ & $G_p$ & $\tilde G_p$ \\[0.1cm]
		\hline \\[-0.3cm]
		286.06 &	91.90 &	5.6941(85)(185) &		$-0.181(67)(15)$	&	5.035(257)(66) 			&	4.317(74)(20) 	\\
		295.95 &	91.30 &	5.7878(90)(170) &		$-0.324(62)(34)$	&	5.168(142)(84) 			&	4.404(53)(32) 	\\
		304.88 &	91.29 &	5.8410(89)(171) &		$-0.209(60)(33)$	&	4.924(108)(59) 			&	4.532(46)(26) 	\\
		313.48 &	90.08 &	5.8905(91)(171) &		$-0.156(58)(32)$	&	4.879(\hphantom{0}91)(51)	&	4.627(41)(24) 	\\
		322.02 &	88.03 &	5.9275(90)(166) &		$-0.366(55)(41)$	&	5.227(\hphantom{0}80)(58) 	&	4.692(38)(29) 	\\
		330.80 &	84.79 &	5.9557(93)(168) &		$-0.383(54)(40)$	&	5.265(\hphantom{0}73)(56) 	&	4.748(35)(28) 	\\
		340.17 &	80.83 &	5.9915(92)(166) &		$-0.218(55)(46)$	&	5.036(\hphantom{0}68)(59) 	&	4.762(34)(31) 	\\
		350.94 &	76.11 &	6.0161(92)(163) &		$-0.302(54)(35)$	&	5.246(\hphantom{0}62)(37) 	&	4.889(34)(21) 	\\
		364.57 &	69.87 &	6.0351(91)(162) &		$-0.309(54)(33)$	&	5.338(\hphantom{0}57)(31) 	&	5.000(35)(20) 	\\
		389.95 &	58.75 &	6.1155(93)(224) &		$-0.264(59)(35)$	&	5.400(\hphantom{0}55)(34) 	&	5.144(36)(22) 	\\
		\bottomrule
	\end{tabular}
	\caption[NA48/2 data with additional radiative corrections and isospin breaking mass effects.]{NA48/2 data \cite{Batley2010,Batley2012}, corrected by additional radiative and isospin-breaking mass effects \cite{Stoffer2014}. The uncertainties of the isospin corrections (without the higher order estimate) are added in quadrature to the systematic error. The fully correlated error of the normalisation is $0.70\%$.}
	\label{tab:NA48DataIsoCorr}
\end{table}

\begin{table}[H]
	\centering
	\begin{tabular}{c c c c}
		\toprule
		$\sqrt{s}/$MeV & $\sqrt{s_\ell}/$MeV & $F_s$ & $G_p$ \\[0.1cm]
		\hline \\[-0.3cm]
		287.6 &		106.8			&	5.781(13)(42) &	4.702(89)(40) 	\\
		299.5 &		105.7			&	5.825(14)(48) &	4.693(62)(37) 	\\
		311.2 &		103.8			&	5.914(14)(56) &	4.771(54)(41) 	\\
		324.0 &		101.1			&	5.974(16)(62) &	4.999(51)(56)	\\
		339.6 &		\hphantom{0}96.3	&	6.097(17)(63) &	5.002(49)(57) 	\\
		370.0 &		\hphantom{0}84.6	&	6.151(20)(41) &	5.104(50)(42) 	\\
		\bottomrule
	\end{tabular}
	\caption[E865 data with isospin breaking mass effects.]{E865 data \cite{Pislak2001, Pislak2003}, corrected by isospin-breaking mass effects \cite{Stoffer2014}. The uncertainties of the isospin corrections (without the higher order estimate) are added in quadrature to the systematic error. The fully correlated error of the normalisation is $1.2\%$.}
	\label{tab:E865DataIsoCorr}
\end{table}

\begin{table}[H]
	\footnotesize
	\centering
	\begin{tabular}{c c c c c c c c c c c}
		\toprule
		$\sqrt{s}/$MeV & 286.06 & 295.95 & 304.88 & 313.48 & 322.02 & 330.80 & 340.17 & 350.94 & 364.57 & 389.95 \\
		Corr($G_p,\tilde G_p$) & 0.04 & 0.09 & 0.13 & 0.16 & 0.20 & 0.21 & 0.22 & 0.23 & 0.22 & 0.24 \\
		\bottomrule
	\end{tabular}
	\caption[Bin-diagonal correlations of $G_p$ with $\tilde G_p$ for the NA48/2 data set.]{Bin-diagonal correlations of $G_p$ with $\tilde G_p$ for the NA48/2 data set.\footnotemark{}}
	\label{tab:NA48DataGpGptCorrelations}
\end{table}
\footnotetext{We are grateful to Brigitte Bloch-Devaux for providing this additional information.}

\begin{table}[H]
	\footnotesize
	\centering
	\tabcolsep=0.107cm
	\begin{tabular}{c c c c c c c c c c c}
		\toprule
		$\sqrt{s}/$MeV & 286.06 & 295.95 & 304.88 & 313.48 & 322.02 & 330.80 & 340.17 & 350.94 & 364.57 & 389.95 \\
		Cov($F_p, G_p$) & $-0.0152$ & $-0.0069$ & $-0.0048$ & $-0.0038$ & $-0.0032$ & $-0.0028$ & $-0.0026$ & $-0.0023$ & $-0.0020$ & $-0.0021$  \\
		Corr($F_p, G_p$) & $-0.89$ & $-0.79$ & $-0.74$ & $-0.73$ & $-0.73$ & $-0.73$ & $-0.71$ & $-0.69$ & $-0.67$ & $-0.64$ \\
		\bottomrule
	\end{tabular}
	\caption{Bin-diagonal statistical covariance and correlation of $F_p$ with $G_p$ for the NA48/2 data set.}
	\label{tab:NA48DataFpGpCorrCov}
\end{table}

\clearpage

\section{Soft-Pion Theorem}

In addition to the experimental input on the partial waves, we use the soft-pion theorem (SPT) \cite{Treiman1972, DeAlfaro1973} as a second source of information to determine the subtraction constants.

There are two different soft-pion theorems for $K_{\ell4}$, depending on which pion is taken to be soft. If the momentum $p_1$ of the positively charged pion is sent to zero, the Mandelstam variables become $s = M_\pi^2$, $t = M_K^2$, $u = s_\ell$. Since the SPT is valid only at $\mathcal{O}(M_\pi^2)$, we set $u = s_\ell + M_\pi^2$, such that the relation $s+t+u = M_K^2 + 2 M_\pi^2 + s_\ell$ remains valid and one does not need to worry about defining an off-shell form factor.

The first SPT states \cite{Stoffer2010}:
\begin{align}
	\label{eqn:SPT1}
	F(M_\pi^2, M_K^2, M_\pi^2+s_\ell) - G(M_\pi^2, M_K^2, M_\pi^2+s_\ell) = \mathcal{O}(M_\pi^2) .
\end{align}

If the momentum $p_2$ of the negatively charged pion is sent to zero, the Mandelstam variables become $s = M_\pi^2$, $t = s_\ell$, $u = M_K^2$. We set $t = s_\ell + M_\pi^2$.

The second SPT gives a relation to the $K_{\ell3}$ vector form factor:
\begin{align}
	\label{eqn:SPT2}
	F(M_\pi^2, M_\pi^2+s_\ell, M_K^2) + G(M_\pi^2, M_\pi^2+s_\ell, M_K^2) - \frac{\sqrt{2} M_K}{F_\pi} f_+(M_\pi^2+s_\ell) = \mathcal{O}(M_\pi^2) .
\end{align}

At leading order in \ChPT{}, the SPTs are exact, i.e.~the right-hand sides of the equations (\ref{eqn:SPT1}) and (\ref{eqn:SPT2}) vanish, at NLO and NNLO, there appear $\O(M_\pi^2)$ corrections.

Numerically, it turns out that the first SPT is fulfilled to a higher precision than the second SPT. At NLO, the correction to the first SPT is about $0.4\%$ for $s_\ell=0$, the second SPT gets a correction of $2.0\%$ if $f_+(M_\pi^2)$ is used. If we make the arbitrary replacement $f_+(M_\pi^2) \mapsto f_+(0)$, again an $\O(M_\pi^2)$ effect, the deviation in the second SPT increases to $4.9\%$.

At NNLO, the corrections become a bit larger.\footnote{We thank Johan Bijnens and Ilaria Jemos for providing the C++ implementation of the NNLO expressions.} If the $\O(p^6)$ LECs $C_i^r$ are all put to zero and $s_\ell = 0$ as well, the first SPT is fulfilled at $1.0\%$, the second at $4.4\%$ with $f_+(M_\pi^2)$ or $7.6\%$ with $f_+(0)$. If the $C_i^r$ parts are replaced by the estimates of \cite{Amoros2000,Bijnens2003} (resonances estimates in the case of $K_{\ell4}$), the accuracy of the first SPT is $1.5\%$, the one of the second SPT $5.4\%$ using $f_+(M_\pi^2)$ or again $7.6\%$ using $f_+(0)$.

We use the size of the NNLO corrections to the SPT as an order-of-magnitude estimate of the SPT tolerance that we allow in the fits.

\section{Results for the Basic Solutions}

\begin{figure}[H]
	\centering
	\scalebox{0.6}{
		\input{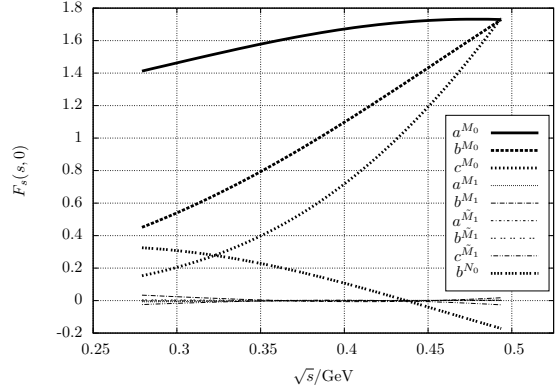}
		}
	\caption{$S$-wave of the form factor $F$ for the different basic solutions for $s_\ell = 0$}
	\label{fig:BasicSolutionsSwave}
\end{figure}

\begin{figure}[H]
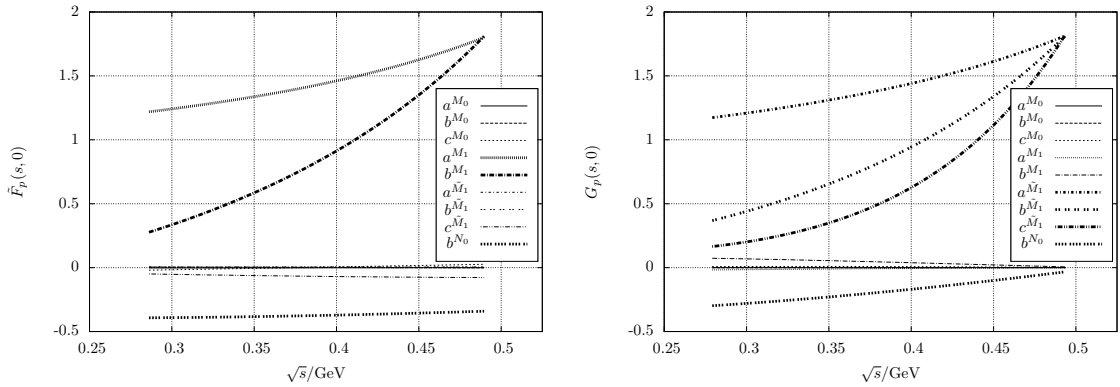

	\centering
	\scalebox{0.59}{
		\input{Kl4Dispersive/plots/Fp.tex}
		\input{Kl4Dispersive/plots/Gp.tex}
		}
	\caption{$P$-waves of the form factors $F$ and $G$ for the different basic solutions for $s_\ell = 0$}
	\label{fig:BasicSolutionsPwaves}
\end{figure}

The figures~\ref{fig:BasicSolutionsSwave} and \ref{fig:BasicSolutionsPwaves} show the partial waves of the basic solutions in the case $s_\ell = 0$. They are computed with the phase solutions that reach the asymptotic values of $\pi$ in the case of the $\pi\pi$ phases and 0 in the case of the $K\pi$ phases. For $\delta_0^0$, the solution with the drop around the $K\bar K$ threshold is used.

The final result will be a linear combination of these partial waves. The figures illustrate what can be learnt also from the definitions (\ref{eq:FunctionsOfOneVariableOmnes}) and (\ref{eqn:PartialWavesFuncOneVariableRelations}): the data on the partial wave $F_s$ will constrain mainly the subtraction constants appearing in $M_0$, the data on $F_p$ mainly the constants in $M_1$ and the data on $G_p$ mainly the constants in $\tilde M_1$. An exception is the constant $b^{N_0}$: through the hat functions, it is constrained by the data on all partial waves.

Table~\ref{tab:BasicSolutionsSPT} shows the values of $(F-G)(M_\pi^2, M_K^2, M_\pi^2)$ and $(F+G)(M_\pi^2, M_\pi^2, M_K^2)$ for the basic solutions. Obviously, the first soft-pion theorem implies mainly a constraint on a linear combination of $a^{M_0}$, $a^{M_1}$, $a^{\tilde M_1}$ and $b^{N_0}$.

\begin{table}[H]
	\centering
	\begin{tabular}{c c c}
		\toprule
		basic solution & $(F-G)_\mathrm{SPP1}$ & $(F+G)_\mathrm{SPP2}$ \\[0.1cm]
		\hline \\[-0.3cm]
		$a^{M_0}$ &		$\m1.06$	& $\m1.05$ \\
		$b^{M_0}$ &		$\m0.08$	& $\m0.09$ \\
		$c^{M_0}$ &		$\m0.03$	& $-0.01$ \\
		$a^{M_1}$ &		$-1.03$	& $\m0.93$ \\
		$b^{M_1}$ &		$\m0.05$	& $\m0.11$ \\
		$a^{\tilde M_1}$ &	$-1.07$	& $\m1.02$ \\
		$b^{\tilde M_1}$ &	$-0.05$	& $\m0.09$ \\
		$c^{\tilde M_1}$ &	$-0.10$	& $-0.01$ \\
		$b^{N_0}$ &		$\m1.62$	& $-0.01$ \\
		\bottomrule
	\end{tabular}
	\caption{Values of the two relevant combinations of the form factors $F$ and $G$ at the soft-pion points, computed for the basic solutions.}
	\label{tab:BasicSolutionsSPT}
\end{table}

\section{Fitting Method}

In the following, we describe how we perform the fit. Basically, we have to deal with a simple linear fit. The only subtlety is the fact that the data contains a fully correlated uncertainty of the normalisation, which is a multiplicative quantity. The fact that we use two experiments with different normalisation errors asks for a special fitting method to avoid a bias \cite{DAgostini1994, Ball2010}. We apply the `$t_0$-method' of \cite{Ball2010}.

First, we construct a covariance matrix for the observations as follows.
\begin{itemize}
	\item For all the partial-wave data that we want to fit we construct the covariance matrix with the squared statistical errors on the diagonal and the statistical covariance between the $P$-waves as off-diagonal elements.
	\item We add the uncorrelated systematic errors, which do not contain the error of the normalisation, in quadrature to the diagonal entries.
	\item We may or may not include the two soft-pion theorems as additional observations. If we do so, we take e.g.~$F-G$ at the first soft-pion point (SPP) and $F+G$ at the second SPP as observations. As uncertainties, we take a value typical for the deviation in \ChPT{} at NNLO, e.g.~$1\%$ or $2\%$ of the LO value of $F$ for the first SPT and a few percent of $\sqrt{2} M_K / F_\pi f_+(0)$ for the second SPT.
	\item We add the errors of the normalisation to the covariance matrix, which are in block-diagonal form for the data of the two experiments:
	\begin{align}
		\begin{split}
			(\mathrm{cov})_{ij} = (\mathrm{rel.cov.})_{ij} + (\mathrm{norm.cov.})_{ij}, \quad (\mathrm{norm.cov.})_{ij} = \Delta_I^2 \, f(s^i,s_\ell^i) f(s^j,s_\ell^j) \delta_{I_i, I_j} ,
		\end{split}
	\end{align}
	where $\Delta_I$ denotes the error of the normalisation for experiment $I$. $I_i$ is the index of the experiment (1 or 2) corresponding to the data point $i$ and $f(s^i,s_\ell^i)$ is the value of the \textit{fitted} partial wave. In a first step, this value has to be computed under the assumption of some starting values for the fit parameters.
\end{itemize}

The fit requires then an iteration. One has to minimise the error function defined by
\begin{align}
	\begin{split}
		\chi^2 = v^T P v ,
	\end{split}
\end{align}
where $v$ is the vector of the residues, i.e.~the differences between the observations and computed values. $P$ is the inverse of the covariance matrix constructed above: $P = (\mathrm{cov})^{-1}$. The minimum of the $\chi^2$ function can be either found with some minimisation routine or, since the fit is linear, directly with the explicit solution
\begin{align}
	\begin{split}
		\mathrm{par} = ( A^T P A )^{-1} A^T P O ,
	\end{split}
\end{align}
where $O$ is the vector of observations and
\begin{align}
	\begin{split}
		A_{ij} = \frac{\p f(s^i,s_\ell^i)}{\p \mathrm{param}_j}
	\end{split}
\end{align}
is the design matrix to be determined with the values of the basic solutions.

With these new values for the fit parameters, one again computes the new covariance matrix (the contribution for the normalisation changes) and iterates this procedure. It turns out that only very few iterations are needed to reach convergence.

If we do not want to determine a parameter through the fit but fix it beforehand to a non-zero value, we have to subtract the fixed contribution from the observations $O$, such that $O$ is purely linear in the parameters and contains no constant contributions.

In the above discussion, we have not specified what we use as fit parameters. One option is to fit the subtraction constants. Since we want to include an $s_\ell$-dependence in the subtraction constants, we write e.g.
\begin{align}
	\begin{split}
		\label{eqn:SlDependenceOmnesSubtractionConstants}
		a^{M_0}(s_\ell) = a^{M_0}_0 + a^{M_0}_1 \frac{s_\ell}{M_K^2} + \ldots ,
	\end{split}
\end{align}
where $a^{M_0}_0$, $a^{M_0}_1$, $\ldots$ are now the parameters collected in the above vector `par'. Another option is to use the matching equations to \ChPT{}, which provide a linear relation between the subtraction constants and the LECs we are interested in, and perform the fit directly with the LECs.


\section{Matching to \ChPT{}}

The final goal of this treatment is the determination of low-energy constants of \ChPT{}. Instead of fitting directly the $K_{\ell4}$ data with the chiral expressions, we use the dispersive representation as an intermediate step. The dispersion relation provides a model-independent resummation of final-state rescattering effects. Therefore, we expect that even the most important effects beyond $\O(p^6)$ are included in the dispersion relation. Of course, in order to extract values for the LECs, one has to perform a matching of the dispersive and the chiral representations. This can be done e.g.~on the level of the form factors \cite{Stoffer2010, Colangelo2012, Stoffer2013}. Since the dispersion relation describes the energy dependence, the matching point can be outside the physical region, i.e.~even at lower energies, where \ChPT{} is expected to converge better.

Here, we follow an improved strategy for the matching: we match the dispersive and the chiral representations not on the level of form factors but directly on the level of subtraction constants. Since the decomposition (\ref{eq:FormFactorDecomposition}) is valid up to terms of $\O(p^8)$, the one-loop and even the two-loop result can be written in this form, which allows us to extract a chiral representation of the subtraction constants. This procedure has the advantage that the matching is performed for each function of one variable $M_0(s)$,~$\ldots$ at its subtraction point, i.e.~at $s=0$, $t=0$ and $u=0$.

\subsection{Matching Equations at $\O(p^4)$}

\subsubsection{Reconstruction of the \ChPT{} Form Factors}

Let us start by reconstructing the NLO form factors in the standard dispersive form (\ref{eq:FunctionsOfOneVariable}).

The LO \ChPT{} form factors are given by
\begin{align}
	\begin{split}
		F_\mathrm{LO} = G_\mathrm{LO} = \frac{M_K}{\sqrt{2} F_\pi} .
	\end{split}
\end{align}
With the partial wave projections (\ref{eq:PartialWaveProjectionSChannelPhysicalFF}), we find
\begin{align}
	\begin{split}
		f_0^\mathrm{LO}(s) &= \frac{M_K}{\sqrt{2} F_\pi} , \\
		f_1^\mathrm{LO}(s) &= \frac{M_K}{\sqrt{2} F_\pi} \frac{M_K^2 PL}{2 X^2} , \\
		g_1^\mathrm{LO}(s) &= \frac{M_K}{\sqrt{2} F_\pi} .
	\end{split}
\end{align}
The isospin 1/2 form factors (\ref{eq:Isospin12FormFactors}) are given by
\begin{align}
	\begin{split}
		F^{(1/2)}_\mathrm{LO} = \frac{M_K}{\sqrt{2} F_\pi} , \quad G^{(1/2)}_\mathrm{LO} = \frac{\sqrt{2} M_K}{F_\pi} .
	\end{split}
\end{align}
Hence, the partial waves in the crossed channels (\ref{eq:PartialWaveProjectionTUChannel}) are
\begin{align}
	\begin{split}
		f_{0,\mathrm{LO}}^{(1/2)}(t) &= \frac{M_K}{\sqrt{2} F_\pi} \frac{3 \Delta_{K\pi} - 5 t}{4t} , \\
		f_{1,\mathrm{LO}}^{(1/2)}(t) &= \frac{M_K}{\sqrt{2} F_\pi} \frac{3 M_K^4( M_\pi^2 - s_\ell - t)}{4t \lambda_{\ell\pi}(t)} , \\
		g_{1,\mathrm{LO}}^{(1/2)}(t) &= \frac{3 M_K}{2\sqrt{2} F_\pi} , \\
		f_{0,\mathrm{LO}}^{(3/2)}(u) &= \frac{M_K}{\sqrt{2} F_\pi} , \\
		f_{1,\mathrm{LO}}^{(3/2)}(u) &= 0 , \\
		g_{1,\mathrm{LO}}^{(3/2)}(u) &= 0 .
	\end{split}
\end{align}

The $\pi\pi$-scattering amplitude can be written as \cite{GasserLeutwyler1984}
\begin{align}
	\begin{split}
		T^{(0)}(s,t,u) &= 3 A(s,t,u) + A(t,u,s) + A(u,s,t) , \\
		T^{(1)}(s,t,u) &= A(t,u,s) - A(u,s,t) ,
	\end{split}
\end{align}
where at LO
\begin{align}
	\begin{split}
		A^\mathrm{LO}(s,t,u) = \frac{s - M_\pi^2}{F_\pi^2} .
	\end{split}
\end{align}
The Mandelstam variables for $\pi\pi$ scattering satisfy
\begin{align}
	\begin{split}
		s+t+u &= 4 M_\pi^2 , \\
		t &= -2 q^2( 1 - z ) ,
	\end{split}
\end{align}
where $q^2 = \frac{s}{4} - M_\pi^2$, $z = \cos\theta$. Hence, the $\pi\pi$ partial waves are
\begin{align}
	\begin{split}
		t_{0,\mathrm{LO}}^0(s) &= \frac{1}{2} \int_{-1}^1 dz \, T^{(0)}_\mathrm{LO}(s,z) = \frac{2s - M_\pi^2}{F_\pi^2} , \\
		t_{1,\mathrm{LO}}^1(s) &= \frac{3}{2} \int_{-1}^1 dz \, z T^{(1)}_\mathrm{LO}(s,z) = \frac{s - 4 M_\pi^2}{F_\pi^2} .
	\end{split}
\end{align}

The $K\pi$-scattering amplitude is given by \cite{Bernard1991}
\begin{align}
	\begin{split}
		T^{(1/2)}(s,t,u) &= \frac{3}{2} T^{(3/2)}(u,t,s) - \frac{1}{2} T^{(3/2)}(s,t,u) ,
	\end{split}
\end{align}
and at LO
\begin{align}
	\begin{split}
		T^{(3/2)}(s,t,u) &= \frac{1}{2 F_\pi^2} ( M_K^2 + M_\pi^2 - s ) .
	\end{split}
\end{align}
Of course, the Mandelstam variables satisfy here $s+t+u = 2 M_K^2 + 2 M_\pi^2$. The partial waves are given by
\begin{align}
	\begin{split}
		t_{0,\mathrm{LO}}^{1/2}(s) &= \frac{1}{8 s F_\pi^2} \left(5 s^2 - 2 s(M_K^2+M_\pi^2) - 3\Delta_{K\pi}^2 \right) , \\
		t_{1,\mathrm{LO}}^{1/2}(s) &= \frac{1}{8 s F_\pi^2} \left(3 s^2 - 6 s(M_K^2+M_\pi^2) + 3\Delta_{K\pi}^2 \right) , \\
		t_{0,\mathrm{LO}}^{3/2}(s) &= \frac{1}{2 F_\pi^2} (M_K^2 + M_\pi^2 - s ) , \\
		t_{1,\mathrm{LO}}^{3/2}(s) &= 0 .
	\end{split}
\end{align}

Using the unitarity relation for the $K_{\ell4}$ partial waves, we can now easily construct their imaginary parts at NLO:
\begin{align}
	\begin{split}
		\Im f_l^\mathrm{NLO}(s) &= \frac{1}{2l+1} \frac{1}{32\pi} \sigma_\pi(s) t_{l,\mathrm{LO}}^{I*}(s) f_l^\mathrm{LO}(s) , \\
		\Im g_l^\mathrm{NLO}(s) &= \frac{1}{2l+1} \frac{1}{32\pi} \sigma_\pi(s) t_{l,\mathrm{LO}}^{I*}(s) g_l^\mathrm{LO}(s) , \\
		\Im f_{l,\mathrm{NLO}}^{(I)}(t) &= \frac{1}{2l+1} \frac{1}{16\pi} \frac{\lambda_{K\pi}^{1/2}(t)}{t} t_{l,\mathrm{LO}}^{I*}(t) f_{l,\mathrm{LO}}^{(I)}(t) , \\
		\Im g_{l,\mathrm{NLO}}^{(I)}(t) &= \frac{1}{2l+1} \frac{1}{16\pi} \frac{\lambda_{K\pi}^{1/2}(t)}{t} t_{l,\mathrm{LO}}^{I*}(t) g_{l,\mathrm{LO}}^{(I)}(t) ,
	\end{split}
\end{align}
hence
\begin{align}
	\begin{split}
		\Im f_0^\mathrm{NLO}(s) &= \frac{1}{32\pi} \sigma_\pi(s) \frac{M_K( 2s - M_\pi^2)}{\sqrt{2} F_\pi^3} , \\
		\Im f_1^\mathrm{NLO}(s) &= \frac{1}{3} \frac{1}{32\pi} \sigma_\pi(s) \frac{M_K( s - 4 M_\pi^2)}{\sqrt{2} F_\pi^3}  \frac{M_K^2 PL}{2 X^2} , \\
		\Im g_1^\mathrm{NLO}(s) &= \frac{1}{3} \frac{1}{32\pi} \sigma_\pi(s) \frac{M_K(s - 4 M_\pi^2)}{\sqrt{2} F_\pi^3} , \\
		\Im f_{0,\mathrm{NLO}}^{(1/2)}(t) &= \frac{1}{16\pi} \frac{\lambda_{K\pi}^{1/2}(t)}{t} \frac{M_K}{32\sqrt{2} t^2 F_\pi^3} \left(5 t^2 - 2 t(M_K^2+M_\pi^2) - 3\Delta_{K\pi}^2 \right) (3 \Delta_{K\pi} - 5 t) , \\
		\Im f_{1,\mathrm{NLO}}^{(1/2)}(t) &= \frac{1}{16\pi} \frac{\lambda_{K\pi}^{1/2}(t)}{t} \frac{M_K}{8 \sqrt{2} t F_\pi^3} \left(3 t^2 - 6 t(M_K^2+M_\pi^2) + 3\Delta_{K\pi}^2 \right) \frac{M_K^4( M_\pi^2 - s_\ell - t)}{4t \lambda_{\ell\pi}(t)} , \\
		\Im g_{1,\mathrm{NLO}}^{(1/2)}(t) &= \frac{1}{16\pi} \frac{\lambda_{K\pi}^{1/2}(t)}{t} \frac{M_K}{16 \sqrt{2} t F_\pi^3} \left(3 t^2 - 6 t(M_K^2+M_\pi^2) + 3\Delta_{K\pi}^2 \right) , \\
		\Im f_{0,\mathrm{NLO}}^{(3/2)}(u) &= \frac{1}{16\pi} \frac{\lambda_{K\pi}^{1/2}(u)}{u} \frac{M_K}{2 \sqrt{2} F_\pi^3} (M_K^2 + M_\pi^2 - u ) , \\
		\Im f_{1,\mathrm{NLO}}^{(3/2)}(u) &= 0 , \\
		\Im g_{1,\mathrm{NLO}}^{(3/2)}(u) &= 0 .
	\end{split}
\end{align}
By inserting these imaginary parts into the dispersion integrals in (\ref{eq:FunctionsOfOneVariable}), we can reconstruct the NLO form factors.
For the comparison with the explicit loop calculation, we rewrite the dispersive integrals in terms of loop functions (see appendix~\ref{sec:AppendixScalarLoopFunctions}):
\begin{align}
	\begin{split}
		M_0^\mathrm{NLO}(s) &= m_{0,\mathrm{NLO}}^0 + m_{0,\mathrm{NLO}}^1 \frac{s}{M_K^2} + \frac{M_K}{2 \sqrt{2} F_\pi^3} \Big( (2s - M_\pi^2) \left( \bar B_{\pi\pi}(s) - \bar B_{\pi\pi}(0) \right) + M_\pi^2 s \, \bar B_{\pi\pi}^\prime(0) \Big) , \\
		M_1^\mathrm{NLO}(s) &= m_{1,\mathrm{NLO}}^0 , \\
		\tilde M_1^\mathrm{NLO}(s) &= \tilde m_{1,\mathrm{NLO}}^0 + \tilde m_{1,\mathrm{NLO}}^1 \frac{s}{M_K^2} + \frac{M_K}{6 \sqrt{2} F_\pi^3} \Big( (s-4M_\pi^2) \left( \bar B_{\pi\pi}(s) - \bar B_{\pi\pi}(0) \right) + 4 M_\pi^2  s \, \bar B_{\pi\pi}^\prime(0) \Big) , \\
		N_0^\mathrm{NLO}(t) &= n_{0,\mathrm{NLO}}^1 \frac{t}{M_K^2} + \frac{M_K}{32\sqrt{2} F_\pi^3} \bigg(  \left( - 25 t  + 5 (5 M_K^2 - M_\pi^2) \right) \left( \bar B_{K\pi}(t) - \bar B_{K\pi}(0) \right) \\
			& + \frac{3 \Delta_{K\pi}}{t^2} \left( t (3 M_K^2 - 7 M_\pi^2) - 3 \Delta_{K\pi}^2 \right) \left( \bar B_{K\pi}(t) - \bar B_{K\pi}(0) - t \,  \bar B_{K\pi}^\prime(0) - \frac{t^2}{2} \,  \bar B_{K\pi}^\dprime(0) \right) \\
			& - 5 t (5 M_K^2 - M_\pi^2) \bar B_{K\pi}^\prime(0) + \frac{3}{2} t \Delta_{K\pi}^3 \bar B_{K\pi}^\tprime(0) \bigg) , \\
		N_1^\mathrm{NLO}(t) &= 0 , \\
		\tilde N_1^\mathrm{NLO}(t) &= \frac{3 M_K^3}{16 \sqrt{2} F_\pi^3} \begin{aligned}[t]
			& \bigg( \frac{1}{t^2} \left( t^2 - 2t(M_K^2+M_\pi^2) + \Delta_{K\pi}^2 \right) \left( \bar B_{K\pi}(t) - \bar B_{K\pi}(0) \right) \\
			& + \frac{1}{t} \left( 2t(M_K^2+M_\pi^2) - \Delta_{K\pi}^2 \right) \bar B_{K\pi}^\prime(0) - \frac{\Delta_{K\pi}^2}{2} \, \bar B_{K\pi}^\dprime(0) \bigg) , \end{aligned} \\
		R_0^\mathrm{NLO}(u) &= \frac{M_K}{2 \sqrt{2} F_\pi^3} \Big(  \left(M_K^2 + M_\pi^2 - u \right) \left( \bar B_{K\pi}(u) - \bar B_{K\pi}(0) \right) - \left(M_K^2 + M_\pi^2 \right) u \, \bar B_{K\pi}^\prime(0) \Big) , \\
		R_1^\mathrm{NLO}(u) &= 0 , \\
		\tilde R_1^\mathrm{NLO}(u) &= 0 .
	\end{split}
\end{align}
Now, we compare this expression with the one-loop calculation \cite{Bijnens1990,Riggenbach1991,Bijnens1994}. As in our dispersive treatment, we only consider $\pi\pi$ intermediate states in the $s$-channel and $K\pi$ intermediate states in the crossed channels, the $K\bar K$ and $\eta\eta$ loops in the $s$-channel and the $K\eta$ loops in the $t$-channel have to be expanded in a Taylor series and absorbed by the subtraction polynomial. The comparison of the dispersive representation with the loop calculation then allows to extract the $\O(p^4)$ values for the subtraction constants.

Note that the only contributions that we neglect when writing the $\O(p^4)$ loop calculation in the dispersive form are the second and higher order Taylor coefficients of the expanded loop functions of higher intermediate states ($K\bar K$, $\eta\eta$ and $K\eta$).

The result for the $\O(p^4)$ subtraction constants can be found in appendix~\ref{sec:AppendixNLOSubtractionConstantsStandardRep}.

\subsubsection{\ChPT{} Form Factors in the Omnès Representation}

The reason why we do not use the standard dispersive form (\ref{eq:FunctionsOfOneVariable}) for the numerical solution of the dispersion relation but rather the Omnès representation (\ref{eq:FunctionsOfOneVariableOmnes}) is mainly the separation of final-state rescattering effects: the Omnès function resummates the most important rescattering effects. The remaining dispersive integrals take the interplay of the different channels into account.

It is therefore desirable to perform the matching to \ChPT{} not on the level of the standard dispersive form but directly with the Omnès representation. This should avoid mixing the final-state resummation with the determination of the LECs.

However, it is not possible to write directly the \ChPT{} representation in the Omnès form, because the chiral expansion of the phase shifts has not the correct asymptotic behaviour. At LO, the phases grow linearly, hence the Omnès dispersion integral (\ref{eq:OmnesFunction}) is logarithmically divergent. Therefore, we subtract the dispersion integral once more:
\begin{align}
	\begin{split}
		\Omega(s) &= \exp\left( \frac{s}{\pi} \int_{s_0}^\infty \frac{\delta(s^\prime)}{(s^\prime - s - i \epsilon) s^\prime} ds^\prime \right) \\
			&= \exp\left(  \frac{s}{\pi} \int_{s_0}^\infty \frac{\delta(s^\prime)}{{s^\prime}^2} ds^\prime +  \frac{s^2}{\pi} \int_{s_0}^\infty \frac{\delta(s^\prime)}{(s^\prime - s - i \epsilon) {s^\prime}^2} ds^\prime \right) \\
			&=: \exp\left( \omega \frac{s}{M_K^2} +  \frac{s^2}{\pi} \int_{s_0}^\infty \frac{\delta(s^\prime)}{(s^\prime - s - i \epsilon) {s^\prime}^2} ds^\prime \right) .
	\end{split}
\end{align}
$\omega$ is divergent if evaluated in \ChPT{}. Let us postpone the determination of this constant for a moment.

Let us now use the Omnès representation to reconstruct the NLO result for the form factors. At LO, the functions of one variable are simply given by
\begin{align}
	\begin{split}
		M_0^\mathrm{LO}(s) &= \tilde M_1^\mathrm{LO}(s) = \frac{M_K}{\sqrt{2} F_\pi} , \\
		M_1^\mathrm{LO}(s) &= N_0^\mathrm{LO}(t) = N_1^\mathrm{LO}(t) = \tilde N_1^\mathrm{LO}(t) = R_0^\mathrm{LO}(u) = R_1^\mathrm{LO}(u) = \tilde R_1^\mathrm{LO}(u) = 0 .
	\end{split}
\end{align}
We start by calculating the hat functions at LO:
\begin{align}
	\begin{split}
		\hat M_0^\mathrm{LO}(s) &= \hat M_1^\mathrm{LO}(s) = \hat {\tilde M}_1^\mathrm{LO}(s) = \hat N_1^\mathrm{LO}(t) = \hat R_1^\mathrm{LO}(u) = \hat {\tilde R}_1^\mathrm{LO}(u) = 0 , \\
		\hat N_0^\mathrm{LO}(t) &= \frac{M_K}{\sqrt{2} F_\pi} \frac{3\Delta_{K\pi} - 5t}{4t} , \\
		\hat {\tilde N}_1^\mathrm{LO}(t) &= \frac{M_K}{\sqrt{2} F_\pi} \frac{3 M_K^2}{2 t} , \\
		\hat R_0^\mathrm{LO}(u) &= \frac{M_K}{\sqrt{2} F_\pi} .
	\end{split}
\end{align}
Further, we need the phase shifts at LO:
\begin{align}
	\begin{split}
		\delta_{0,\mathrm{LO}}^0(s) &= \frac{1}{32 \pi F_\pi^2} ( 2s - M_\pi^2 ) \sigma_\pi(s) , \\
		\delta_{1,\mathrm{LO}}^1(s) &= \frac{1}{96 \pi F_\pi^2} ( s - 4 M_\pi^2 ) \sigma_\pi(s) , \\
		\delta_{0,\mathrm{LO}}^{1/2}(t) &= \frac{1}{128 \pi  F_\pi^2} \left(5 t^2 - 2 t(M_K^2+M_\pi^2) - 3\Delta_{K\pi}^2 \right) \frac{\lambda_{K\pi}^{1/2}(t)}{t^2} , \\
		\delta_{1,\mathrm{LO}}^{1/2}(t) &= \frac{1}{384 \pi F_\pi^2} \left(3 t^2 - 6 t(M_K^2+M_\pi^2) + 3\Delta_{K\pi}^2 \right) \frac{\lambda_{K\pi}^{1/2}(t)}{t^2} , \\
		\delta_{0,\mathrm{LO}}^{3/2}(u) &= \frac{1}{32 \pi F_\pi^2} (M_K^2 + M_\pi^2 - u ) \frac{\lambda_{K\pi}^{1/2}(u)}{u} , \\
		\delta_{1,\mathrm{LO}}^{3/2}(u) &= 0 .
	\end{split}
\end{align}
We expand the Omnès representation (\ref{eq:FunctionsOfOneVariableOmnes}) at NLO:
\begin{align}
	\small
	\begin{split}
		M_0^\mathrm{NLO}(s) &= \left( 1 + \omega_0^0 \frac{s}{M_K^2} +  \frac{s^2}{\pi} \int_{s_0}^\infty \frac{\delta_{0,\mathrm{LO}}^0(s^\prime)}{(s^\prime - s - i \epsilon) {s^\prime}^2} ds^\prime \right) \left( a^{M_0} + b^{M_0} \frac{s}{M_K^2} + c^{M_0} \frac{s^2}{M_K^4} \right), \\
		M_1^\mathrm{NLO}(s) &= \left( 1 + \omega_1^1 \frac{s}{M_K^2} +  \frac{s^2}{\pi} \int_{s_0}^\infty \frac{\delta_{1,\mathrm{LO}}^1(s^\prime)}{(s^\prime - s - i \epsilon) {s^\prime}^2} ds^\prime \right) \left( a^{M_1} + b^{M_1}  \frac{s}{M_K^2}  \right) , \\
		 \tilde M_1^\mathrm{NLO}(s) &= \left( 1 + \omega_1^1 \frac{s}{M_K^2} +  \frac{s^2}{\pi} \int_{s_0}^\infty \frac{\delta_{1,\mathrm{LO}}^1(s^\prime)}{(s^\prime - s - i \epsilon) {s^\prime}^2} ds^\prime \right) \left( a^{\tilde M_1} + b^{\tilde M_1}  \frac{s}{M_K^2} + c^{\tilde M_1}  \frac{s^2}{M_K^4} \right) , \\
		N_0^\mathrm{NLO}(t) &= \left( 1 + \omega_0^{1/2} \frac{t}{M_K^2} +  \frac{t^2}{\pi} \int_{t_0}^\infty \frac{\delta_{0,\mathrm{LO}}^{1/2}(t^\prime)}{(t^\prime - t - i \epsilon) {t^\prime}^2} dt^\prime \right) \left( b^{N_0} \frac{t}{M_K^2} + \frac{t^2}{\pi} \int_{t_0}^\infty \frac{\hat N_0^\mathrm{LO}(t^\prime) \delta_{0,\mathrm{LO}}^{1/2}(t^\prime)}{(t^\prime - t - i\epsilon) {t^\prime}^2} dt^\prime  \right) , \\
		N_1^\mathrm{NLO}(t) &= 0 , \\
		\tilde N_1^\mathrm{NLO}(t) &= \left( 1 + \omega_1^{1/2} \frac{t}{M_K^2} +  \frac{t^2}{\pi} \int_{t_0}^\infty \frac{\delta_{1,\mathrm{LO}}^{1/2}(t^\prime)}{(t^\prime - t - i \epsilon) {t^\prime}^2} dt^\prime \right) \Bigg( \frac{t}{\pi} \int_{t_0}^\infty \frac{\hat{\tilde N}_1^\mathrm{LO}(t^\prime) \delta_{1,\mathrm{LO}}^{1/2}(t^\prime)}{(t^\prime - t - i\epsilon) t^\prime} dt^\prime  \Bigg) , \\
		R_0^\mathrm{NLO}(u) &= \left( 1 + \omega_0^{3/2} \frac{u}{M_K^2} +  \frac{u^2}{\pi} \int_{u_0}^\infty \frac{\delta_{0,\mathrm{LO}}^{3/2}(u^\prime)}{(u^\prime - u - i \epsilon) {u^\prime}^2} du^\prime \right) \left( \frac{u^2}{\pi} \int_{u_0}^\infty \frac{\hat R_0^\mathrm{LO}(u^\prime) \delta_{0,\mathrm{LO}}^{3/2}(u^\prime)}{(u^\prime - u - i\epsilon) {u^\prime}^2} du^\prime  \right) , \\
		R_1^\mathrm{NLO}(u) &= 0 , \\
		\tilde R_1^\mathrm{NLO}(u) &= 0 .
	\end{split}
\end{align}
If we further expand these expressions chirally and neglect higher orders, we obtain (note that only $a^{M_0}$ and $a^{\tilde M_1}$ do not vanish at LO):
\begin{align}
	\small
	\begin{split}
		M_0^\mathrm{NLO}(s) &= a^{M_0}_\mathrm{LO} \left( 1 + \omega_0^0 \frac{s}{M_K^2} +  \frac{s^2}{\pi} \int_{s_0}^\infty \frac{\delta_{0,\mathrm{LO}}^0(s^\prime)}{(s^\prime - s - i \epsilon) {s^\prime}^2} ds^\prime \right) + \Delta a^{M_0}_\mathrm{NLO} + b^{M_0}_\mathrm{NLO} \frac{s}{M_K^2} + c^{M_0}_\mathrm{NLO} \frac{s^2}{M_K^4} , \\
		M_1^\mathrm{NLO}(s) &= a^{M_1}_\mathrm{NLO} + b^{M_1}_\mathrm{NLO}  \frac{s}{M_K^2} , \\
		 \tilde M_1^\mathrm{NLO}(s) &= a^{\tilde M_1}_\mathrm{LO} \left( 1 + \omega_1^1 \frac{s}{M_K^2} +  \frac{s^2}{\pi} \int_{s_0}^\infty \frac{\delta_{1,\mathrm{LO}}^1(s^\prime)}{(s^\prime - s - i \epsilon) {s^\prime}^2} ds^\prime \right) + \Delta a^{\tilde M_1}_\mathrm{NLO} + b^{\tilde M_1}_\mathrm{NLO}  \frac{s}{M_K^2} + c^{\tilde M_1}_\mathrm{NLO}  \frac{s^2}{M_K^4} , \\
		N_0^\mathrm{NLO}(t) &= b^{N_0}_\mathrm{NLO} \frac{t}{M_K^2} + \frac{t^2}{\pi} \int_{t_0}^\infty \frac{\hat N_0^\mathrm{LO}(t^\prime) \delta_{0,\mathrm{LO}}^{1/2}(t^\prime)}{(t^\prime - t - i\epsilon) {t^\prime}^2} dt^\prime , \\
		N_1^\mathrm{NLO}(t) &= 0 , \\
		\tilde N_1^\mathrm{NLO}(t) &= \frac{t}{\pi} \int_{t_0}^\infty \frac{\hat{\tilde N}_1^\mathrm{LO}(t^\prime) \delta_{1,\mathrm{LO}}^{1/2}(t^\prime)}{(t^\prime - t - i\epsilon) t^\prime} dt^\prime , \\
		R_0^\mathrm{NLO}(u) &= \frac{u^2}{\pi} \int_{u_0}^\infty \frac{\hat R_0^\mathrm{LO}(u^\prime) \delta_{0,\mathrm{LO}}^{3/2}(u^\prime)}{(u^\prime - u - i\epsilon) {u^\prime}^2} du^\prime , \\
		R_1^\mathrm{NLO}(u) &= 0 , \\
		\tilde R_1^\mathrm{NLO}(u) &= 0 ,
	\end{split}
\end{align}
where
\begin{align}
	\begin{split}
		a^{M_0}_\mathrm{NLO} &= a^{M_0}_\mathrm{LO} + \Delta a^{M_0}_\mathrm{NLO}, \quad a^{M_0}_\mathrm{LO} = \frac{M_K}{\sqrt{2} F_\pi} , \\
		a^{\tilde M_1}_\mathrm{NLO} &= a^{\tilde M_1}_\mathrm{LO} + \Delta a^{\tilde M_1}_\mathrm{NLO}, \quad a^{\tilde M_1}_\mathrm{LO} = \frac{M_K}{\sqrt{2} F_\pi} .
	\end{split}
\end{align}
Next, we insert the LO phases and hat functions:
\begin{align}
	\begin{split}
		M_0^\mathrm{NLO}(s) &= a^{M_0}_\mathrm{NLO} + \left( b^{M_0}_\mathrm{NLO} + \frac{M_K}{\sqrt{2} F_\pi} \omega_0^0 \right) \frac{s}{M_K^2}  + c^{M_0}_\mathrm{NLO} \frac{s^2}{M_K^4} \\
			&\quad + \frac{s^2}{\pi} \int_{s_0}^\infty \frac{1}{(s^\prime - s - i \epsilon) {s^\prime}^2} \frac{\sigma_\pi(s^\prime)}{32 \pi} \frac{M_K (2s^\prime - M_\pi^2)}{\sqrt{2} F_\pi^3}  ds^\prime , \\
		M_1^\mathrm{NLO}(s) &= a^{M_1}_\mathrm{NLO} + b^{M_1}_\mathrm{NLO}  \frac{s}{M_K^2} , \\
		 \tilde M_1^\mathrm{NLO}(s) &= a^{\tilde M_1}_\mathrm{NLO} + \left( b^{\tilde M_1}_\mathrm{NLO} + \frac{M_K}{\sqrt{2} F_\pi} \omega_1^1 \right) \frac{s}{M_K^2} + c^{\tilde M_1}_\mathrm{NLO}  \frac{s^2}{M_K^4} \\
		 	&\quad + \frac{s^2}{\pi} \int_{s_0}^\infty \frac{1}{(s^\prime - s - i \epsilon) {s^\prime}^2} \frac{\sigma_\pi(s^\prime)}{32 \pi} \frac{M_K(s^\prime - 4 M_\pi^2)}{3\sqrt{2} F_\pi^3} ds^\prime , \\
		N_0^\mathrm{NLO}(t) &= b^{N_0}_\mathrm{NLO} \frac{t}{M_K^2} + \frac{t^2}{\pi} \int_{t_0}^\infty \begin{aligned}[t]
			& \frac{1}{(t^\prime - t - i\epsilon) {t^\prime}^2} \frac{\lambda_{K\pi}^{1/2}(t^\prime)}{16 \pi t^\prime} \frac{M_K(3\Delta_{K\pi} - 5t^\prime)}{32 \sqrt{2}{t^\prime}^2 F_\pi^3} \\
			& \cdot \left(5 {t^\prime}^2 - 2 t^\prime(M_K^2+M_\pi^2) - 3\Delta_{K\pi}^2 \right) dt^\prime , \end{aligned} \\
		N_1^\mathrm{NLO}(t) &= 0 , \\
		\tilde N_1^\mathrm{NLO}(t) &= \frac{t}{\pi} \int_{t_0}^\infty \frac{1}{(t^\prime - t - i\epsilon) t^\prime} \frac{\lambda_{K\pi}^{1/2}(t^\prime)}{16 \pi t^\prime} \frac{M_K^3}{16\sqrt{2} {t^\prime}^2 F_\pi^3} \left(3 {t^\prime}^2 - 6 t^\prime(M_K^2+M_\pi^2) + 3\Delta_{K\pi}^2 \right) dt^\prime , \\
		R_0^\mathrm{NLO}(u) &= \frac{u^2}{\pi} \int_{u_0}^\infty \frac{1}{(u^\prime - u - i\epsilon) {u^\prime}^2} \frac{\lambda_{K\pi}^{1/2}(u^\prime)}{16 \pi u^\prime} \frac{M_K}{2\sqrt{2} F_\pi^3} (M_K^2 + M_\pi^2 - u^\prime ) du^\prime , \\
		R_1^\mathrm{NLO}(u) &= 0 , \\
		\tilde R_1^\mathrm{NLO}(u) &= 0 .
	\end{split}
\end{align}
We see that the form of the Omnès representation is completely equivalent to the standard representation, apart from the presence of the additional subtraction constants $c^{M_0}$, $b^{M_1}$ and $c^{\tilde M_1}$. Therefore, the transformation of the dispersive integrals to loop functions and the matching to the explicit one-loop calculation corresponds to the procedure for the standard representation, apart from the determination of the three additional subtraction constants. We expand the $t$-channel $K\eta$ integrals up to linear terms in $t$ and find:
\begin{align}
	\begin{split}
		\label{eqn:NLORelationOmnesStandardSubtrConst}
		a^{M_0}_\mathrm{NLO} &= m_{0,\mathrm{NLO}}^0 , \\
		b^{M_0}_\mathrm{NLO} &= m_{0,\mathrm{NLO}}^1 - \frac{M_K}{\sqrt{2} F_\pi} \omega_0^0 , \\
		c^{M_0}_\mathrm{NLO} &= \frac{M_K^3}{\sqrt{2}F_\pi^3} \frac{15 M_\eta^4 + M_K^2 M_\pi^2}{1920 \pi^2 M_\eta^4} , \\
		a^{M_1}_\mathrm{NLO} &= m_{1,\mathrm{NLO}}^0 , \\
		b^{M_1}_\mathrm{NLO} &= 0 , \\
		a^{\tilde M_1}_\mathrm{NLO} &= \tilde m_{1,\mathrm{NLO}}^0 , \\
		b^{\tilde M_1}_\mathrm{NLO} &= \tilde m_{1,\mathrm{NLO}}^1 - \frac{M_K}{\sqrt{2} F_\pi} \omega_1^1 , \\
		c^{\tilde M_1}_\mathrm{NLO} &= \frac{M_K^3}{\sqrt{2} F_\pi^3} \frac{1}{1920 \pi^2} , \\
		b^{N_0}_\mathrm{NLO} &= n_{0,\mathrm{NLO}}^1 .
	\end{split}
\end{align}
The constants $\omega_0^0$ and $\omega_1^1$ cannot be evaluated with the chiral phases. If we evaluate them with the physical phases, this leads to exactly the same matching equations for the determination of the $L_i^r$ as if we would match the Taylor expansion of the Omnès representation with the Taylor expansion of the chiral result. Note, however, that the expressions obtained for $c^{M_0}$, $b^{M_1}$ and $c^{\tilde M_1}$ are different. E.g.~for $b^{M_1}$, the chiral expansion leads to $b^{M_1}_\mathrm{NLO} = 0$ while a Taylor expansion of the dispersion relation would require $b^{M_1} = - m_{1,\mathrm{NLO}}^0 {\Omega_1^1}^\prime(0) M_K^2$, where ${\Omega_1^1}^\prime$ is the derivative of the Omnès function calculated with the physical phases. Of course, the difference is a higher order effect in the chiral counting. As higher order effects can be important if due to final state rescattering, we would not like to intermingle them with the matching of the subtraction constants. The matching on the basis of Taylor coefficients would require the linear term of $M_1(s)$ to vanish exactly, while the matching based on the chiral expansion of the dispersion relation gives a non-zero linear term in $M_1(s)$ due to the Omnès function.

\subsection{Matching Equations at $\O(p^6)$}

\subsubsection{Decomposition of the NNLO Form Factors}

In the following, we describe the decomposition of the two-loop result such that the matching can be performed at NNLO. Since the NNLO chiral result has a different asymptotic behaviour than the NLO result and our numerical dispersive representation, we have to use the representation (\ref{eq:FunctionsOfOneVariable3Subtr}), which uses a different gauge and more subtractions than (\ref{eq:FunctionsOfOneVariable}).

The imaginary parts of the $K_{\ell4}$ partial waves at NNLO could again be reconstructed using the unitarity relations, e.g.
\begin{align}
	\begin{split}
		\Im f_l^\mathrm{NNLO}(s) &= \frac{1}{2l+1} \frac{1}{32\pi} \sigma_\pi(s) \left( t_{l,\mathrm{LO}}^{I*}(s) f_l^\mathrm{LO}(s) + \Delta t_{l,\mathrm{NLO}}^{I*}(s) f_l^\mathrm{LO}(s) + t_{l,\mathrm{LO}}^{I*}(s) \Delta f_l^\mathrm{NLO}(s) \right) .
	\end{split}
\end{align}
However, instead of proceeding as for NLO, it is more straightforward to decompose the two-loop result directly into functions of one variable, then to impose the gauge condition and extract the Taylor coefficients of the functions of one variable.

The two-loop result for the form factors $F$ and $G$ was computed in \cite{Amoros2000}. We have the full expressions in form of a C++ program at hand.\footnote{We thank Johan Bijnens and Ilaria Jemos for providing the C++ implementation of the NNLO expressions.} It has the following structure:
\begin{align}
	\begin{split}
		X^\mathrm{NNLO}(s,t,u) = X^\mathrm{LO} & + X^\mathrm{NLO}_L(s,t,u) + X^\mathrm{NLO}_R(s,t,u) \\
			& + X^\mathrm{NNLO}_{C}(s,t,u) + X^\mathrm{NNLO}_{L}(s,t,u) + X^\mathrm{NNLO}_{P}(s,t,u) \\
			& + X^\mathrm{NNLO}_{VS}(s) + X^\mathrm{NNLO}_{VT}(t) + X^\mathrm{NNLO}_{VU}(u) ,
	\end{split}
\end{align}
where $X\in\{F,G\}$ and the different parts denote the following:
\begin{itemize}
	\item $X^\mathrm{NLO}_L$: NLO polynomial containing the LECs $L_i^r$,
	\item $X^\mathrm{NLO}_R$: NLO loops,
	\item $X^\mathrm{NNLO}_C$: NNLO polynomial containing the LECs $C_i^r$,
	\item $X^\mathrm{NNLO}_L$: NNLO part containing $L_i^r \times L_i^r$ and $L_i^r \times \mathrm{loop}$,
	\item $X^\mathrm{NNLO}_P$: NNLO two-loop part without vertex integrals,
	\item $X^\mathrm{NNLO}_{VS}$: NNLO vertex integrals depending on $s$,
	\item $X^\mathrm{NNLO}_{VT}$: NNLO vertex integrals depending on $t$,
	\item $X^\mathrm{NNLO}_{VU}$: NNLO vertex integrals depending on $u$.
\end{itemize}

In appendix~\ref{sec:AppendixTwoLoopDecomposition}, we perform the explicit decomposition of the two-loop result into functions of one Mandelstam variable according to (\ref{eq:FormFactorDecomposition}) and (\ref{eq:FunctionsOfOneVariable3Subtr}) and evaluate numerically the subtraction constants.

\subsubsection{NNLO Form Factors in the Omnès Representation}

As we already pointed out for the NLO matching, it is desirable to use the Omnès representation rather than the standard dispersion relation for the matching and the determination of the LECs. Let us therefore derive the matching equations at NNLO in the Omnès scheme.

We have to use the second gauge for the decomposition of the NNLO representation (\ref{eq:FunctionsOfOneVariable3Subtr}). As a starting point, let us find the NLO Omnès subtraction constants in the second gauge. In the first gauge, we found $R_1^\mathrm{NLO} = \tilde R_1^\mathrm{NLO} = 0$, hence
\begin{align}
	\begin{split}
		c_\mathrm{NLO}^{R_0} &= \frac{M_K}{\sqrt{2} F_\pi^3} \frac{M_K^4}{4} \left((M_K^2 + M_\pi^2) \bar B_{K\pi}^\dprime(0) - 2 \bar B_{K\pi}^\prime(0) \right) , \\
		a_\mathrm{NLO}^{R_1} &= b_\mathrm{NLO}^{\tilde R_1} = 0.
	\end{split}
\end{align}
The gauge-transformation (\ref{eq:GaugeTransformation}) is then defined by
\begin{align}
	\begin{split}
		C_\mathrm{NLO}^{R_0} &= \frac{M_K}{\sqrt{2} F_\pi^3} \frac{1}{32\pi^2} \frac{M_K^4}{\Delta_{K\pi}^4} \Bigg( \frac{(M_K^2 + M_\pi^2)(M_K^4 - 8 M_K^2 M_\pi^2 + M_\pi^4)}{3} + \frac{4 M_K^4 M_\pi^4 \ln\left(\frac{M_K^2}{M_\pi^2}\right)}{\Delta_{K\pi}} \Bigg) , \\
		A_\mathrm{NLO}^{R_1} &= B_\mathrm{NLO}^{\tilde R_1} = 0.
	\end{split}
\end{align}
At NLO, the shifts in the subtraction constants (\ref{eq:OmnesGaugeTransformationParameters}) are therefore given by
\begin{align}
	\begin{alignedat}{4}
		\label{eq:NLOGaugeTransformationSubtractionConstants}
		\delta a_\mathrm{NLO}^{M_0} &= \frac{\Sigma_0^2 - \Delta_{K\pi}\Delta_{\ell\pi}}{M_K^4} C_\mathrm{NLO}^{R_0} , \quad & \delta b_\mathrm{NLO}^{M_0} &= - \frac{2 \Sigma_0}{M_K^2} C_\mathrm{NLO}^{R_0} , \quad & \delta c_\mathrm{NLO}^{M_0} &= C_\mathrm{NLO}^{R_0} , \quad & \delta d_\mathrm{NLO}^{M_0} &= 0 , \\
		\delta a_\mathrm{NLO}^{M_1} &= - \frac{2 \Sigma_0}{M_K^2} C_\mathrm{NLO}^{R_0} , \quad & \delta b_\mathrm{NLO}^{M_1} &= 2 C_\mathrm{NLO}^{R_0} , \quad & \delta c_\mathrm{NLO}^{M_1} &= 0 , \\
		\delta a_\mathrm{NLO}^{\tilde M_1} &= - \frac{\Sigma_0^2 - \Delta_{K\pi} \Delta_{\ell\pi}}{M_K^4} C_\mathrm{NLO}^{R_0} , \quad & \delta b_\mathrm{NLO}^{\tilde M_1} &= \frac{2\Sigma_0}{M_K^2} C_\mathrm{NLO}^{R_0} , \quad & \delta c_\mathrm{NLO}^{\tilde M_1} &= - C_\mathrm{NLO}^{R_0} , \quad & \delta d_\mathrm{NLO}^{\tilde M_1} &= 0 , \\
		\delta b_\mathrm{NLO}^{N_0} &= - \frac{3 (\Delta_{K\pi} + 2 \Sigma_0)}{4 M_K^2} C_\mathrm{NLO}^{R_0} , \quad & \delta c_\mathrm{NLO}^{N_0} &= - \frac{5}{4} C_\mathrm{NLO}^{R_0} , \\
		\delta a_\mathrm{NLO}^{N_1} &= \frac{3}{2} C_\mathrm{NLO}^{R_0} , \quad & \delta b_\mathrm{NLO}^{\tilde N_1} &= -\frac{3}{2} C_\mathrm{NLO}^{R_0} , \\
		\delta c_\mathrm{NLO}^{R_0} &= C_\mathrm{NLO}^{R_0} , \quad & \delta a_\mathrm{NLO}^{R_1} &= 0 , \quad & \delta b_\mathrm{NLO}^{\tilde R_1} &= 0 .
	\end{alignedat}
\end{align}

When studying now the Omnès representation at NNLO, we notice that the asymptotic behaviour of the phases at NNLO is even worse than at NLO, hence, we have to subtract the Omnès function three times:
\begin{align}
	\begin{split}
		\label{eqn:ThriceSubtractedOmnesFunction}
		\Omega(s) &= \exp\left( \frac{s}{\pi} \int_{s_0}^\infty \frac{\delta(s^\prime)}{(s^\prime - s - i \epsilon) s^\prime} ds^\prime \right) \\
			&= \exp\left(  \frac{s}{\pi} \int_{s_0}^\infty \frac{\delta(s^\prime)}{{s^\prime}^2} ds^\prime +  \frac{s^2}{\pi} \int_{s_0}^\infty \frac{\delta(s^\prime)}{{s^\prime}^3} ds^\prime +  \frac{s^3}{\pi} \int_{s_0}^\infty \frac{\delta(s^\prime)}{(s^\prime - s - i \epsilon) {s^\prime}^3} ds^\prime \right) \\
			&=: \exp\left( \omega \frac{s}{M_K^2} +  \bar\omega \frac{s^2}{M_K^4} +  \frac{s^3}{\pi} \int_{s_0}^\infty \frac{\delta(s^\prime)}{(s^\prime - s - i \epsilon) {s^\prime}^3} ds^\prime \right) .
	\end{split}
\end{align}
$\omega$ and $\bar\omega$ are both divergent if evaluated in \ChPT{} at NNLO, hence we will use the physical phases to determine them.

In the case of the NLO matching, we have derived the relation between the standard and the Omnès subtraction constants (\ref{eqn:NLORelationOmnesStandardSubtrConst}) by comparing the Taylor coefficients of the chirally expanded Omnès representation with the Taylor coefficients of the standard dispersive representation. Although it is instructive to understand the chiral expansion of the Omnès representation, a shortcut can be taken. Note that the chiral expansion and the Taylor expansion are interchangeable. Therefore, we easily obtain the relations between the standard subtraction constants $m_0^0$,~$\ldots$ and the Omnès subtraction constants $a^{M_0}$,~$\ldots$ by chirally expanding the Taylor coefficients of the Omnès representation (\ref{eq:FunctionsOfOneVariableOmnes3Subtr}) and comparing it with the Taylor coefficients of (\ref{eq:FunctionsOfOneVariable3Subtr}).

This leads to the following relations between the relevant subtraction constants:
\begin{align}
	\begin{split}
		\label{eqn:NNLORelationOmnesStandardSubtrConst}
		m_{0}^{0,\mathrm{NNLO}} &= a_\mathrm{NNLO}^{M_0} , \\
		m_{0}^{1,\mathrm{NNLO}} &= b_\mathrm{NNLO}^{M_0} + \omega_0^0 a_\mathrm{NLO}^{M_0} , \\
		m_{0}^{2,\mathrm{NNLO}} &= c_\mathrm{NNLO}^{M_0} + \omega_0^0 b_\mathrm{NLO}^{M_0} + \frac{1}{2} {\omega_0^0}^2 a_\mathrm{LO}^{M_0} + a_\mathrm{NLO}^{M_0} \bar\omega_0^0 + h.o. , \\
		m_{1}^{0,\mathrm{NNLO}} &= a_\mathrm{NNLO}^{M_1} , \\
		m_{1}^{1,\mathrm{NNLO}} &= b_\mathrm{NNLO}^{M_1} + \omega_1^1 a_\mathrm{NLO}^{M_1} , \\
		\tilde m_{1}^{0,\mathrm{NNLO}} &= a_\mathrm{NNLO}^{\tilde M_1} , \\
		\tilde m_{1}^{1,\mathrm{NNLO}} &= b_\mathrm{NNLO}^{\tilde M_1} + \omega_1^1 a_\mathrm{NLO}^{\tilde M_1} , \\
		\tilde m_{1}^{2,\mathrm{NNLO}} &= c_\mathrm{NNLO}^{\tilde M_1} + \omega_1^1 b_\mathrm{NLO}^{\tilde M_1} + \frac{1}{2} {\omega_1^1}^2 a_\mathrm{LO}^{\tilde M_1} + a_\mathrm{NLO}^{\tilde M_1} \bar \omega_1^1 + h.o. , \\
		n_{0}^{1,\mathrm{NNLO}} &= b_\mathrm{NNLO}^{N_0} , \\
		n_{0}^{2,\mathrm{NNLO}} &= c_\mathrm{NNLO}^{N_0} + \omega_0^{1/2} b_\mathrm{NLO}^{N_0} , \\
		n_{1}^{0,\mathrm{NNLO}} &= a_\mathrm{NNLO}^{N_1} , \\
		\tilde n_{1}^{1,\mathrm{NNLO}} &= b_\mathrm{NNLO}^{\tilde N_1} .
	\end{split}
\end{align}
The NNLO chiral expansion of the full Omnès representation can be found in appendix~\ref{sec:AppendixNNLOChiralExpansionOmnes} and leads to the same result. It can be used to identify all the imaginary parts and to connect the different dispersive integrals with the discontinuities of the loop diagrams.


\chapter{Results}

\label{sec:Kl4Results}

In this chapter, we discuss the results for the low-energy constants that we determine by fitting the dispersion relation to data and matching it to \ChPT{}. In order to understand the differences between the results at NLO and NNLO and the source of complications that appear at NNLO, it is useful to study in a first step the results of direct \ChPT{} fits. We perform direct fits at NLO and NNLO and compare our results with the literature before using the whole machinery of the dispersive framework at NLO and finally at NNLO.

\section{Comparison of Direct \ChPT{} Fits}

The most recent fits to $K_{\ell4}$ data performed in the literature are \cite{Bijnens2014}. There, a global fit is performed, taking into account the threshold expansion parameters of the $K_{\ell4}$ form factor measurement of NA48/2 \cite{Batley2012}:
\begin{align}
	\begin{alignedat}{3}
		\label{eqn:NA48ThresholdParameters}
		F &= f_s + f_s^\prime q^2 + \ldots, \quad & f_s &= 5.705 \pm 0.035, \quad & f_s^\prime &= 0.867 \pm 0.050 \\
		G &= g_p + g_p^\prime q^2 + \ldots, \quad & g_p &= 4.952 \pm 0.086, \quad & g_p^\prime &= 0.508 \pm 0.122,
	\end{alignedat}
\end{align}
where $q^2 = \frac{s}{4M_\pi^2} - 1$. We have the impression that in \cite{Bijnens2014}, the above quantities are fitted with the form factors at $\cos\theta = 0$ instead of the first partial wave. In addition to the $K_{\ell4}$ form factor data, the global fit of \cite{Bijnens2014} uses many other inputs, like data on the different decay constants and masses, $\pi\pi$- and $K\pi$-scattering parameters, quark mass ratios etc.

We compare now different strategies for direct fits with the results of \cite{Bijnens2014}. We use only $K_{\ell4}$ data for our fits and therefore are only sensitive to the LECs $L_1^r$, $L_2^r$ and $L_3^r$ \cite{Stoffer2010}. The other LECs are taken as a fixed input.

\subsection{Direct Fits at $\O(p^4)$}

\subsubsection{Fits of Threshold Parameters}

In order to make the connection to \cite{Bijnens2014}, we first perform a direct NLO fit to the NA48/2 threshold parameters in (\ref{eqn:NA48ThresholdParameters}). Using $\cos\theta=0$, i.e.~the first Taylor coefficient of an expansion in $z=\cos\theta$, and the LEC inputs $L_4^r=0$ and the fitted value for $L_5^r$ of \cite{Bijnens2014}, we reproduce almost exactly the result of \cite{Bijnens2014} for $L_1^r$, $L_2^r$ and $L_3^r$, see the second and third column in table~\ref{tab:NLOThresholdFitsNA48}. If we use instead the partial-wave projection (\ref{eqn:SChannelPartialWaveProjection}), the fit results for $L_1^r$ and $L_2^r$ change a bit, as shown in the fourth column of table~\ref{tab:NLOThresholdFitsNA48}. The last column uses lattice results \cite{MILC2009,Aoki2013} for the input LECs.

\begin{table}[H]
	\centering
	\begin{tabular}{c c c c c}
		\toprule
						&	Ref.~\cite{Bijnens2014}	& 	Taylor		 &	PWE		&	PWE		 \\[0.1cm]
		\hline \\[-0.3cm]	
		$10^3 \cdot L_1^r$	&	$\m0.98(09)$		&	$\m0.99(09)$	&	$\m1.15(09)$	&	$\m1.17(09)$	\\
		$10^3 \cdot L_2^r$	&	$\m1.56(09)$		&	$\m1.57(09)$	&	$\m1.48(08)$	&	$\m1.50(08)$	\\
		$10^3 \cdot L_3^r$	&	$-3.82(30)$		&	$-3.83(30)$	&	$-3.82(30)$	&	$-3.87(30)$	\\
		$10^3 \cdot L_4^r$	&	$\equiv 0$		&	$\equiv 0$	&	$\equiv 0$	&	$\equiv0.04$	\\
		$10^3 \cdot L_5^r$	&	$\m1.23(06)$		&	$\equiv1.23$	&	$\equiv1.23$	&	$\equiv0.84$	\\
		\hline \\[-0.3cm]
		$\chi^2$			&	16				&	0.3			&	0.3			&	0.3			\\
		dof				& 	5				&	1	 		&	1			&	1			\\
		$\chi^2/$dof		& 	3.2				&	0.3	 		&	0.3			&	0.3			\\
		\bottomrule
	\end{tabular}
	\caption{Comparison of direct NLO fits to the NA48/2 threshold parameters \cite{Batley2012}. The renormalisation scale is $\mu = 770$~MeV. The last column uses the lattice determination of \cite{MILC2009,Aoki2013} for the input LECs. The uncertainties are purely statistical.}
	\label{tab:NLOThresholdFitsNA48}
\end{table}

\subsubsection{Fits of the Complete Form Factor Data}

In a next step, we no longer fit the threshold expansion parameters (\ref{eqn:NA48ThresholdParameters}) of the form factors, but the complete form factor data of NA48/2 \cite{Batley2010,Batley2012} and E865 \cite{Pislak2001, Pislak2003}, as discussed in section~\ref{sec:ExperimentalKl4Data}. The second column of table~\ref{tab:NLODirectFormFactorFits} shows the result of the NLO fit to the NA48/2 data without isospin corrections (table~\ref{tab:NA48Data}, but with the corrected normalisation of $F_s$ to account for the $s_\ell$-dependence). In the third column, isospin corrections are applied to the fitted data (table~\ref{tab:NA48DataIsoCorr}). The fourth and fifth column show the results of a combined fit to NA48/2 and E865 data (tables~\ref{tab:E865Data} and \ref{tab:E865DataIsoCorr}). The smaller $\chi^2$ value in the fits to the data with isospin-breaking corrections is due to the fact that the isospin corrections introduce an additional uncertainty in the data.

\begin{table}[H]
	\centering
	\begin{tabular}{c c c c c c}
		\toprule
						&	NA48/2		 &	NA48/2, \cancel{iso}	&	NA48/2 \& E865	&	NA48/2 \& E865, \cancel{iso}	 \\[0.1cm]
		\hline \\[-0.3cm]	
		$10^3 \cdot L_1^r$	&	$\m0.69(03)$	&	$\m0.71(04)$		&	$\m0.62(03)$		&	$\m0.64(04)$	\\
		$10^3 \cdot L_2^r$	&	$\m1.88(07)$	&	$\m1.80(08)$		&	$\m1.79(06)$		&	$\m1.70(06)$	\\
		$10^3 \cdot L_3^r$	&	$-3.89(13)$	&	$-3.93(14)$		&	$-3.62(11)$		&	$-3.60(12)$	\\
		$10^3 \cdot L_4^r$	&	$\equiv 0.04$	&	$\equiv 0.04$		&	$\equiv 0.04$		&	$\equiv 0.04$	\\
		$10^3 \cdot L_5^r$	&	$\equiv 0.84$	&	$\equiv 0.84$		&	$\equiv 0.84$		&	$\equiv 0.84$	\\
		$10^3 \cdot L_9^r$	&	$\equiv 5.93$	&	$\equiv 5.93$		&	$\equiv 5.93$		&	$\equiv 5.93$	\\
		\hline																					 \\[-0.3cm]
		$\chi^2$			&	159.4		&	67.5				&	199.9			&	117.1		\\
		dof				& 	27	 		&	27				&	39				&	39			\\
		$\chi^2/$dof		& 	5.9	 		&	2.5				&	5.1				&	3.0			\\
		\bottomrule
	\end{tabular}
	\caption{Comparison of direct NLO fits to the NA48/2 and E865 form factor measurements. The renormalisation scale is $\mu = 770$~MeV. For $L_4^r$ and $L_5^r$, we use lattice input \cite{MILC2009,Aoki2013}, for $L_9^r$ the determination of \cite{BijnensTalavera2002}. The uncertainties are purely statistical.}
	\label{tab:NLODirectFormFactorFits}
\end{table}

\begin{figure}[H]
	\centering
	\scalebox{0.75}{
		\small
		\input{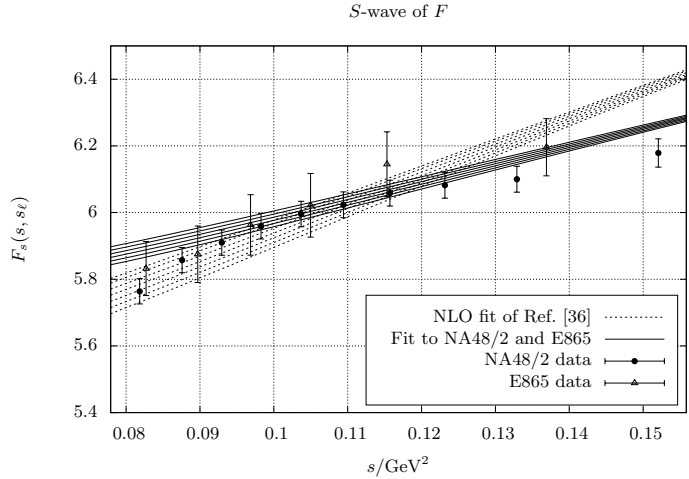}
		}
	\caption{Comparison of different fits for the $S$-wave of the form factor $F$: NA48/2 threshold parameter fit of \cite{Bijnens2014} and a fit to the full data set. The $(s,s_\ell)$ phase space is projected on the $s$-axis. No isospin corrections are applied.}
	\label{fig:FsDirectFits}
\end{figure}

Figure~\ref{fig:FsDirectFits} shows a comparison of the NA48/2 threshold parameter fit of \cite{Bijnens2014} with the result of the fit to the whole form factor data set (forth column of table~\ref{tab:NLODirectFormFactorFits}). It helps to understand the difference between the fitted LECs in the two procedures: in the fit to the threshold parameters, the curvature of the form factor is neglected. Since the NLO chiral representation cannot reproduce the curvature, the data points at higher energies reduce the slope in a fit to the whole data set.

\subsection{Direct Fits at $\O(p^6)$}

\begin{table}[H]
	\centering
	\tabcolsep=0.15cm
	\begin{tabular}{c c c c c c c}
		\toprule
						&	Ref.~\cite{Bijnens2014}	&	Ref.~\cite{Bijnens2014}	&	NA48/2	&	NA48/2 \& E865	&	NA48/2	&	NA48/2 \& E865	 \\[0.1cm]
		\hline \\[-0.3cm]	
		$C_i^r$			&	$\equiv0$		&	BE14		&	$\equiv0$			&	$\equiv0$			&	BE14		&	BE14		\\
		\hline \\[-0.3cm]	
		$10^3 \cdot L_1^r$	&	$\m0.67(06)$	&	$\m0.53(06)$	&	$\m0.34(03)$		&	$\m0.28(02)$		&	$\m0.33(03)$	&	$\m0.27(02)$	\\
		$10^3 \cdot L_2^r$	&	$\m0.17(04)$	&	$\m0.81(04)$	&	$\m0.42(06)$		&	$\m0.35(05)$		&	$\m0.95(06)$	&	$\m0.89(05)$	\\
		$10^3 \cdot L_3^r$	&	$-1.76(21)$	&	$-3.07(20)$	&	$-1.54(14)$		&	$-1.25(11)$		&	$-3.06(14)$	&	$-2.80(11)$	\\
		$10^3 \cdot L_4^r$	&	$\m0.73(10)$	&	$\equiv0.3$	&	$\equiv 0.04$		&	$\equiv 0.04$		&	$\equiv 0.04$	&	$\equiv 0.04$	\\
		$10^3 \cdot L_5^r$	&	$\m0.65(05)$	&	$\m1.01(06)$	&	$\equiv 0.84$		&	$\equiv 0.84$		&	$\equiv 0.84$	&	$\equiv 0.84$	\\
		$10^3 \cdot L_6^r$	&	$\m0.25(09)$	&	$\m0.14(05)$	&	$\equiv 0.07$		&	$\equiv 0.07$		&	$\equiv 0.07$	&	$\equiv 0.07$	\\
		$10^3 \cdot L_7^r$	&	$-0.17(06)$	&	$-0.34(09)$	&	$\equiv -0.34$		&	$\equiv -0.34$		&	$\equiv -0.34$	&	$\equiv -0.34$	\\
		$10^3 \cdot L_8^r$	&	$\m0.22(08)$	&	$\m0.47(10)$	&	$\equiv 0.36$		&	$\equiv 0.36$		&	$\equiv 0.36$	&	$\equiv 0.36$	\\
		$10^3 \cdot L_9^r$	&				&				&	$\equiv 5.93$		&	$\equiv 5.93$		&	$\equiv 5.93$	&	$\equiv 5.93$	\\
		\hline																													 \\[-0.3cm]
		$\chi^2$			&	26			&				&	81.3				&	128.7			&	52.5			&	91.2			\\
		dof				& 	9	 		&		 		&	27				&	39				&	27			&	39			\\
		$\chi^2/$dof		& 	2.9	 		&	1.0	 		&	3.0				&	3.3				&	1.9			&	2.3			\\
		\bottomrule
	\end{tabular}
	\caption{Direct NNLO fits for different choices of the $C_i^r$. The fits of \cite{Bijnens2014} are cited for comparison. The renormalisation scale is $\mu = 770$~MeV. Our results are fits to the entire form factor data including isospin corrections. The uncertainties are purely statistical. The NLO input LECs $L_4^r$, $L_5^r$, $L_6^r$ and $L_8^r$ are lattice determinations \cite{MILC2009,Aoki2013}, $L_7^r$ is the BE14 value \cite{Bijnens2014} and $L_9^r$ is taken from \cite{BijnensTalavera2002}.}
	\label{tab:NNLOThresholdFitsBE14}
\end{table}

\ChPT{} at NNLO suffers from the problem that many new low-energy constants $C_i^r$ appear in the $\O(p^6)$ Lagrangian. In $K_{\ell4}$, in total 24 linearly independent combinations of the $C_i^r$ enter in the NNLO chiral representation of the form factors $F$ and $G$. A fit of so many parameters seems out of question. We would rather like to use some input values for the $C_i^r$. Unfortunately, only very few of the NNLO LECs are known reliably. We could either use determinations of the $C_i^r$ with models like the chiral quark model \cite{Jiang2010}, a resonance estimate \cite{Amoros2000,Bijnens2012} or the educated guess of \cite{Bijnens2014}. These different estimates, however, do not lead to compatible results \cite{Bijnens2014}.

In table~\ref{tab:NNLOThresholdFitsBE14}, we display the results of our direct \ChPT{} fits at NNLO in comparison with the results of \cite{Bijnens2014}. In contrast to \cite{Bijnens2014}, we do not use the threshold parameters but the whole form factor data sets of NA48/2 and E865 corrected by isospin-breaking effects \cite{Stoffer2014}. It turns out that even at NNLO, \ChPT{} has trouble to reproduce the curvature of the $F_s$ data. We also note that the results for the fitted LECs at NNLO differ quite significantly from the results at NLO.

\section{Matching the Dispersion Relation to \ChPT{}}

With the direct \ChPT{} fits, we have encountered some of the problems of \ChPT{}. At NLO and even at NNLO, the energy dependence of the $F_s$ form factor is not very well described. At $\O(p^6)$, the appearance of quite a large number of additional LECs reduces the predictive power of \ChPT{}. Some input values for the $C_i^r$ have to be assumed, as a fit of $K_{\ell4}$ data alone cannot determine these LECs.

Now, we turn to the results using the dispersion relation as an intermediate step in the determination of the LECs: we fit the $K_{\ell4}$ form factor data with the dispersion relation. The matching to \ChPT{} relates the subtraction constants of the dispersion relation to the LECs. As the dispersion relation provides a resummation of final-state rescattering effects, we hope to obtain a better description of the energy dependence of the form factors. However, it is clear that the matching of the dispersion relation to NNLO \ChPT{} will again suffer from the large number of LECs.

\subsection{Matching at $\O(p^4)$}

Our numerical dispersion relation (\ref{eq:FunctionsOfOneVariableOmnes}) is parametrised by nine subtraction constants, which in fact are functions of $s_\ell$. If we use the matching at NLO to identify the subtraction constants with chiral expressions, we see that $a^{M_0}_\mathrm{NLO}$ and $a^{\tilde M_1}_\mathrm{NLO}$ are linear in $s_\ell$, while the other subtraction constants do not depend on $s_\ell$. We therefore introduce this $s_\ell$-dependence according to (\ref{eqn:SlDependenceOmnesSubtractionConstants}) and have to determine in total 11 parameters.

An unconstrained fit with these 11 parameters leads to a low relative $\chi^2$ of 0.80 (19 degrees of freedom) for the NA48/2 data alone or 0.72 (31 dof) for the combined data set of NA48/2 and E865. However, the soft-pion theorem in such a fit is very badly violated. If only the first soft-pion theorem (\ref{eqn:SPT1}) is imposed on the fit at a level of 2\% (the tolerance is inspired by the typical NNLO deviation), the second soft-pion theorem (\ref{eqn:SPT2}) is reproduced to 6.7\% accuracy in the fit to NA48/2 and to 2.4\% in the combined fit to both experiments. If both soft-pion theorems are imposed on the fit, the relative $\chi^2$ is still 0.78 (21 dof) for the NA48/2 fit and 0.71 (33 dof) for the combined fit. This shows that in a fit with all 11 parameters, the soft-pion theorems are not fulfilled automatically but are not a strong additional constraint. In an unconstrained fit, the result for the subtraction constants turns out to be rather unstable: the statistical uncertainties are large and some of the subtraction constants change drastically if the E865 data is included. Therefore, we consider these fits as unphysical and fix to an a priori value those subtraction constants that have the largest statistical uncertainty: these are the subtraction constants of highest order in each function and the ones parametrising the $s_\ell$-dependence, i.e. $c^{M_0}$, $b^{M_1}$, $c^{\tilde M_1}$, $a^{M_0}_1$ and $a^{\tilde M_1}_1$. We fix these subtraction constants to the NLO chiral prediction in the matching (\ref{eqn:NLORelationOmnesStandardSubtrConst}): while $c^{M_0}$, $b^{M_1}$ and $c^{\tilde M_1}$ are purely numerical, $a^{M_0}_1$ depends on $L_9^r$ and $a^{\tilde M_1}_1$ on $L_9^r$ as well as on $L_2^r$. We take those two LECs as input and iterate the fit for $L_2^r$ to reach self-consistency. 

Only six subtraction constants $a^{M_0}_0$, $b^{M_0}$, $a^{M_1}$, $a^{\tilde M_1}_0$, $b^{\tilde M_1}$ and $b^{N_0}$ remain to be fitted to data. In the matching equations (i.e.~(\ref{eqn:NLORelationOmnesStandardSubtrConst}) together with appendix~\ref{sec:AppendixNLOSubtractionConstantsStandardRep}), the three LECs $L_1^r$, $L_2^r$ and $L_3^r$ are overdetermined. Hence, we have to use a second $\chi^2$ minimisation to fix these LECs. As an alternative to this two-step procedure (first fit to data, then matching to \ChPT{}), we can identify the subtraction constants immediately with the NLO chiral expressions and perform the fit of the dispersion relation to data with the LECs as fitting parameters. It turns out that these two strategies lead to almost identical numerical results for the LECs: the differences are much smaller than the statistical errors.

In table~\ref{tab:ResultsDispersionNLOMatching}, we show the results of the fits of the dispersion relation matched to NLO \ChPT{}. For the input LECs, we use lattice results \cite{MILC2009,Aoki2013} and \cite{BijnensTalavera2002}:
\begin{align}
	\begin{split}
		\label{eq:Kl4InputLECsNLOMatching}
		10^3 \cdot L_4^r &= 0.04(14) , \\
		10^3 \cdot L_5^r &= 0.84(38) , \\
		10^3 \cdot L_9^r &= 5.93(43) .
	\end{split}
\end{align}

The $\chi^2$ and degrees of freedom correspond to the strategy of using the LECs as fitting parameters. If we use the two-step fitting/matching strategy instead, the $\chi^2/$dof of the fit of the subtraction constants to data is quite good: between 0.8 and 1.0 for the fit to NA48/2 and between 1.3 and 1.6 for the fit to both experiments. At the same time, the relative $\chi^2/$dof of the matching is bad (between 3.6 and 6.2). This is not surprising because the sum of the total $\chi^2$ of the two steps is approximately equal to the total $\chi^2$ in the one-step procedure, while the dof in the second step is drastically reduced.

The first bracket indicates the statistical uncertainty due to the fitted data. The second bracket gives the systematic uncertainty. In section~\ref{sec:ErrorAnalysis}, we will discuss in more detail the different sources of uncertainty.
\begin{table}[H]
	\footnotesize
	\centering
	\tabcolsep=0.08cm
	\begin{tabular}{c c c c c c c}
		\toprule
						&	NA48/2			&	NA48/2 \& E865	&	NA48/2			&	NA48/2 \& E865	&	NA48/2			&	NA48/2 \& E865	\\[0.1cm]
		\hline																																	\\[-0.3cm]	
		Isospin corr.		&	\xmark			&	\xmark			&	\cmark			&	\cmark			&	\cmark			&	\cmark			\\
		\hline																																	\\[-0.3cm]	
		$\sigma_\mathrm{SPT1}$	&	---			&	---				&	---				&	---				&	2\%				&	2\%				\\
		$\sigma_\mathrm{SPT2}$	&	---			&	---				&	---				&	---				&	5\%				&	5\%				\\
		\hline																																	\\[-0.3cm]	
		$10^3 \cdot L_1^r$	&	$\m0.51(02)(05)$	&	$\m0.47(02)(05)$	&	$\m0.55(03)(05)$	&	$\m0.51(02)(05)$	&	$\m0.56(03)(05)$	&	$\m0.51(02)(05)$	\\
		$10^3 \cdot L_2^r$	&	$\m1.01(05)(08)$	&	$\m0.94(04)(08)$	&	$\m0.92(06)(08)$	&	$\m0.86(05)(08)$	&	$\m0.92(06)(08)$	&	$\m0.86(05)(08)$	\\
		$10^3 \cdot L_3^r$	&	$-3.02(11)(07)$	&	$-2.82(09)(07)$	&	$-2.98(12)(07)$	&	$-2.77(10)(07)$	&	$-2.98(12)(07)$	&	$-2.77(10)(07)$	\\
		\hline																																	\\[-0.3cm]	
		$\chi^2$			&	60.2				&	89.4				&	32.7				&	65.4				&	44.5				&	77.2				\\
		dof				& 	27				&	39				&	27				&	39				&	29				&	41				\\
		$\chi^2/$dof		& 	2.2				&	2.3				&	1.2				&	1.7				&	1.5				&	1.9				\\
		\bottomrule
	\end{tabular}
	\caption{Fit results for the dispersion relation matched to \ChPT{} at NLO. The renormalisation scale is $\mu = 770$~MeV.}
	\label{tab:ResultsDispersionNLOMatching}
\end{table}

While the final results for the LECs do not differ significantly in the one-step and two-step strategies, a difference can be observed concerning the soft-pion theorems. If we use the two-step matching strategy, the soft-pion theorems are not automatically satisfied, but if they are imposed as a fitting constraint, they can be perfectly satisfied with only a slight increase of the $\chi^2$. In contrast, in the one-step strategy, where the subtraction constants have to fulfil the chiral constraints, the accuracy of the soft-pion theorems lies at $\sim4\%$ and $\sim10\%$ respectively. This does not change with the soft-pion constraints added to the fit, which only increases the $\chi^2$ a bit.

The influence of the isospin-breaking corrections of \cite{Stoffer2014} is a bit larger than the statistical uncertainty in the case of $L_1^r$ and $L_2^r$, while $L_3^r$ is less sensitive to the isospin effects.

A plot of the data points indicates that the two experiments NA48/2 and E865 are in agreement, which is confirmed by the fit results. We find it worthwhile to stress that this is only the case if the normalisation of the $F_s$ data points of NA48/2 is increased by $0.77\%$. If the published values are used, which are normalised neglecting the $s_\ell$-dependence, a quite strong tension between the two experiments is observed, resulting in higher $\chi^2$ values for combined fits.

\begin{figure}[H]
	\centering
	\scalebox{0.75}{
		\small
		\input{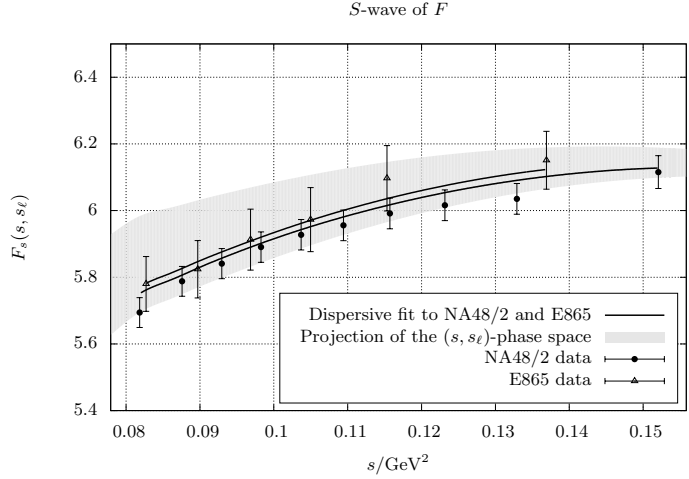}
		}
	\caption{Fit result for the $S$-wave of the form factor $F$. The dispersive description reproduces beautifully the curvature of the form factor. The $(s,s_\ell)$-phase space is projected on the $s$-axis, the plotted lines correspond to splines through the $(s,s_\ell)$-values of the two data sets.}
	\label{fig:FsDispersionRelation}
\end{figure}

\begin{figure}[H]
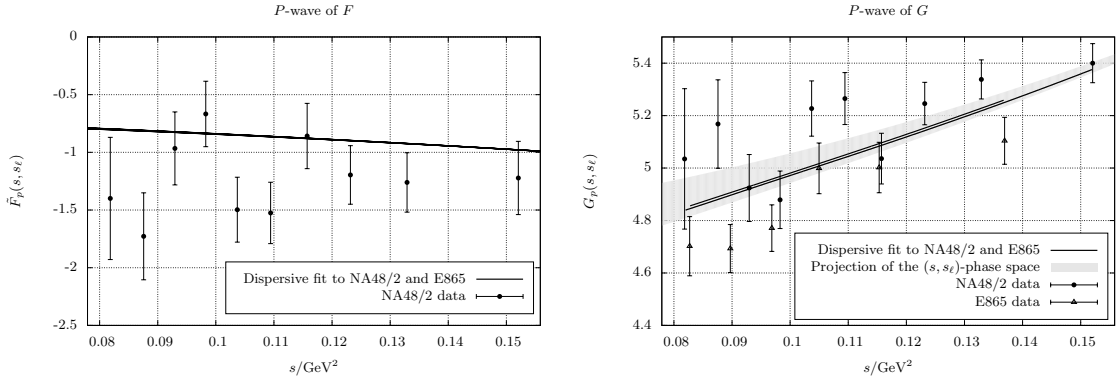

	\centering
	\scalebox{0.59}{
		\small
		\input{Kl4Dispersive/plots/FptDRMatching}
		\input{Kl4Dispersive/plots/GpDRMatching}
		}
	\caption{Fit results for the $P$-waves of the form factors $F$ and $G$. The $(s,s_\ell)$-phase space is again projected on the $s$-axis.}
	\label{fig:FptGpDispersionRelation}
\end{figure}

We note that the $\chi^2$ in the dispersive treatment is clearly improved compared to the direct fit with \ChPT{} at NLO (1.2 instead of 2.5 with 27 dof). This is illustrated in figure~\ref{fig:FsDispersionRelation}: in contrast to a pure chiral treatment, the dispersion relation allows to describe the curvature of the $S$-wave of the form factor $F$. We interpret this as the result of the resummation of final-state rescattering effects. Figure~\ref{fig:FptGpDispersionRelation} shows the fitted $P$-waves of $F$ and $G$.

\subsection{Matching at $\O(p^6)$}

We have seen that in connection with one-loop \ChPT{}, the dispersive treatment clearly exhibits its powers, and the advantage over a pure chiral treatment is evident: the dispersion relation is able to describe the energy-dependence of the form factors, hence the $\chi^2$ of the fit to the whole form factor data is much better. Due to the resummation of final-state rescattering effects, we believe that the dispersion relation captures the most important higher-order contributions and renders the determination of the LECs more robust.

In combination with two-loop \ChPT{}, the treatment becomes more cumbersome. The matching equations at NNLO relate the subtraction constants to chiral expressions that contain the $\O(p^6)$ LECs $C_i^r$. The largest obstacle in a chiral treatment at NNLO is the large number of poorly known $C_i^r$. Unfortunately, the dispersion relation does not help in this regard. It turns out that the determination of the NLO LECs is still strongly affected by the choice of the $C_i^r$, a situation known from direct \ChPT{} fits \cite{Bijnens2012, Bijnens2014}. In table~\ref{tab:ResultsDispersionNNLOMatching}, we present two examples of the matching results at NNLO, using either the $C_i^r$ resonance estimate of \cite{Bijnens2012} or the `preferred values' of \cite{Bijnens2014}. In these fits, we add a tolerance of $10\%$ to the contribution of the $C_i^r$ to the subtraction constants. This tolerance improves the accuracy of the soft-pion theorem but has only a small impact on the values of the $L_i^r$.
\begin{table}[H]
	\centering
	\begin{tabular}{c c c c c}
		\toprule
						&	NA48/2			&	NA48/2 \& E865	&	NA48/2			&	NA48/2 \& E865	\\[0.1cm]
		\hline																							\\[-0.3cm]	
		$C_i^r$ input ($\pm10\%$)	&	resonance \cite{Bijnens2012}	&	resonance \cite{Bijnens2012}	&	BE14 \cite{Bijnens2014}			&	BE14 \cite{Bijnens2014} \\
		\hline																							\\[-0.3cm]	
		$10^3 \cdot L_1^r$	&	$\m0.79(05)$		&	$\m0.73(05)$		&	$\m0.37(10)$		&	$\m0.30(10)$		\\
		$10^3 \cdot L_2^r$	&	$\m0.88(08)$		&	$\m0.78(07)$		&	$\m1.18(10)$		&	$\m1.04(10)$		\\
		$10^3 \cdot L_3^r$	&	$-3.98(27)$		&	$-3.58(25)$		&	$-4.56(36)$		&	$-3.97(33)$		\\
		\hline																							\\[-0.3cm]	
		$\chi^2$			&	40.5				&	72.9				&	37.7				&	74.1				\\
		dof				& 	29				&	41				&	29				&	41				\\
		$\chi^2/$dof		& 	1.4				&	1.8				&	1.3				&	1.8				\\
		\bottomrule
	\end{tabular}
	\caption{Fit results for the dispersion relation matched to \ChPT{} at NNLO. The renormalisation scale is $\mu = 770$~MeV. The uncertainties are purely statistical. As in table \ref{tab:NNLOThresholdFitsBE14}, we use lattice input for $L_4^r$, $L_5^r$, $L_6^r$ and $L_8^r$ \cite{MILC2009,Aoki2013}, $L_7^r$ is the BE14 value \cite{Bijnens2014} and $L_9^r$ is taken from \cite{BijnensTalavera2002}.}
	\label{tab:ResultsDispersionNNLOMatching}
\end{table}
The $\chi^2$ is almost the same as in the NLO fits. Compared to the direct \ChPT{} fits, the corrections from NLO matching to NNLO matching are smaller for $L_1^r$ and especially for $L_2^r$, while the corrections for $L_3^r$ are rather large. Yet the quite strong dependence on the input values for the $C_i^r$ makes it difficult to draw a coherent conclusion concerning the NNLO values of the LECs $L_1^r$, $L_2^r$ and $L_3^r$. Further investigations are needed to stabilise the fit with respect to the $C_i^r$ input. A possible strategy would be to study the chiral convergence of the subtraction constants (or linear combinations of subtraction constants that are independent of the gauge) and to impose a good convergence behaviour, similar to what has been done in \cite{Bijnens2014}.

\section{Error Analysis}

\label{sec:ErrorAnalysis}

We come back to our main result of the NLO matching, shown in table~\ref{tab:ResultsDispersionNLOMatching}. Let us give once more the values for the determined LECs, obtained from the combined fit to the NA48/2 and E865 data:
\begin{align}
	\begin{split}
		10^3 \cdot L_1^r(\mu)  \hphantom{.} &= \hphantom{-}0.51(02)(05) , \\
		10^3 \cdot L_2^r(\mu)  \hphantom{.} &= \hphantom{-}0.86(05)(08) , \\
		10^3 \cdot L_3^r(\mu)  \hphantom{.} &= -2.77(10)(07) , \\
	\end{split}
\end{align}
where $\mu=770$~MeV. The first error indicates the statistical one, i.e.~the error due to the uncertainty of the fitted data. The second error is due to the systematics of our approach, explained in more detail below. The corresponding statistical and systematic correlations are shown in table~\ref{tab:LirCorrelations}.

\begin{table}[H]
	\centering
	\tabcolsep=0.15cm
	\begin{tabular}{r | c c}
		stat.~corr. & $L_2^r$ & $\hphantom{-}L_3^r$ \\
		\hline
		$L_1^r$ & $0.29$ & $-0.59$ \\
		$L_2^r$ & & $-0.93$ \\	
	\end{tabular}
	\hspace{0.5cm}
	\begin{tabular}{r | c c}
		syst.~corr. & $\hphantom{-}L_2^r$ & $\hphantom{-}L_3^r$ \\
		\hline
		$L_1^r$ & $-0.72$ & $\hphantom{-}0.02$ \\
		$L_2^r$ & & $-0.57$ \\	
	\end{tabular}
	\caption{Statistical and systematic correlations of the fitted LECs.}
	\label{tab:LirCorrelations}
\end{table}

\begin{figure}[H]
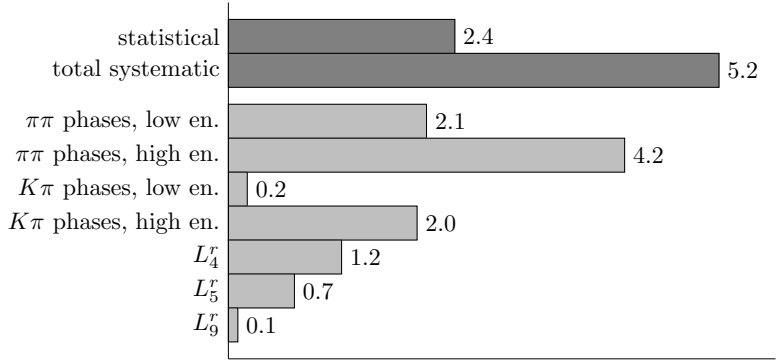

	\centering
	\renewcommand{\bcfontstyle}{\rmfamily}
	\scalebox{0.9}{
	\begin{bchart}[max=5.8,plain]
		\bcbar[label=statistical,color=gray]{2.4}
		\bcbar[label=total systematic,color=gray]{5.2}
		\smallskip
		\bcbar[label={$\pi\pi$ phases, low en.},color=lightgray]{2.1}
		\bcbar[label={$\pi\pi$ phases, high en.},color=lightgray]{4.2}
		\bcbar[label={$K\pi$ phases, low en.},color=lightgray]{0.2}
		\bcbar[label={$K\pi$ phases, high en.},color=lightgray]{2.0}
		\bcbar[label=$L_4^r$,color=lightgray]{1.2}
		\bcbar[label=$L_5^r$,color=lightgray]{0.7}
		\bcbar[label=$L_9^r$,color=lightgray]{0.1}
	\end{bchart} }%
	\caption{Contributions to the uncertainty of $L_1^r$ in the $\O(p^4)$ matching in units of $10^{-5}$.}
	\label{fig:ErrorsL1r}
\end{figure}

\begin{figure}[H]
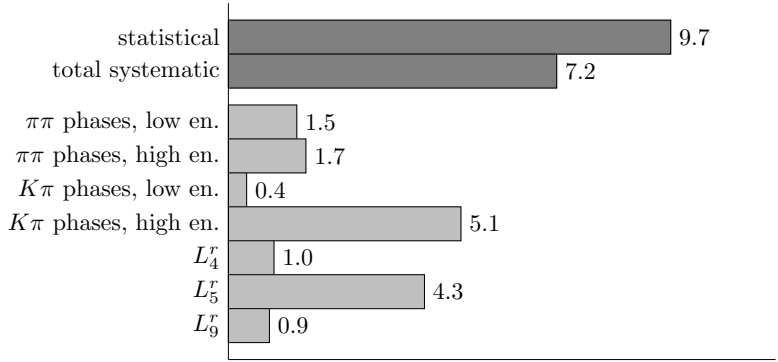

	\centering
	\renewcommand{\bcfontstyle}{\rmfamily}
	\scalebox{0.9}{
	\begin{bchart}[max=12.0,plain]
		\bcbar[label=statistical,color=gray]{9.7}
		\bcbar[label=total systematic,color=gray]{7.2}
		\smallskip
		\bcbar[label={$\pi\pi$ phases, low en.},color=lightgray]{1.5}
		\bcbar[label={$\pi\pi$ phases, high en.},color=lightgray]{1.7}
		\bcbar[label={$K\pi$ phases, low en.},color=lightgray]{0.4}
		\bcbar[label={$K\pi$ phases, high en.},color=lightgray]{5.1}
		\bcbar[label=$L_4^r$,color=lightgray]{1.0}
		\bcbar[label=$L_5^r$,color=lightgray]{4.3}
		\bcbar[label=$L_9^r$,color=lightgray]{0.9}
	\end{bchart} }%
	\caption{Contributions to the uncertainty of $L_3^r$ in the $\O(p^4)$ matching in units of $10^{-5}$.}
	\label{fig:ErrorsL3r}
\end{figure}

Figures~\ref{fig:ErrorsL1r}, \ref{fig:ErrorsL2r} and \ref{fig:ErrorsL3r} show bar charts of the uncertainties of the LECs. In the upper part, statistical and systematic uncertainty are compared. The lower part shows the fractional systematic uncertainties, which we sum in squares.

\begin{figure}[H]
	\centering
	\renewcommand{\bcfontstyle}{\rmfamily}
	\scalebox{0.9}{
	\begin{bchart}[max=9.0,plain]
		\bcbar[label=statistical,color=gray]{4.8}
		\bcbar[label=total systematic,color=gray]{7.7}
		\smallskip
		\bcbar[label={$\pi\pi$ phases, low en.},color=lightgray]{2.0}
		\bcbar[label={$\pi\pi$ phases, high en.},color=lightgray]{5.5}
		\bcbar[label={$K\pi$ phases, low en.},color=lightgray]{0.4}
		\bcbar[label={$K\pi$ phases, high en.},color=lightgray]{2.0}
		\bcbar[label=$L_4^r$,color=lightgray]{3.9}
		\bcbar[label=$L_5^r$,color=lightgray]{2.2}
		\bcbar[label=$L_9^r$,color=lightgray]{0.7}
	\end{bchart} }%
	\caption{Contributions to the uncertainty of $L_2^r$ in the $\O(p^4)$ matching in units of $10^{-5}$.}
	\label{fig:ErrorsL2r}
\end{figure}

The different fractional uncertainties are determined as follows.
\begin{itemize}
	\item For the $\pi\pi$ phases \cite{Ananthanarayan2001a, Caprini2012}, we vary all the 28 parameters and sum the variations of the LECs in squares. In the bar charts, this is the uncertainty labelled by `$\pi\pi$ phases, low energy'.
	\item The second fractional uncertainty is due to the high-energy behaviour of the $\pi\pi$ phases. We sum in squares the differences between the high-energy solutions explained in section~\ref{sec:Kl4pipiPhaseShifts}.
	\item The $K\pi$ phases are simply varied between the centre and upper/lower limit of the error bands. This influence is labelled as `$K\pi$ phases, low energy'.
	\item The uncertainty due to the high-energy behaviour of the $K\pi$ phases is estimated with the two solutions for each of the $K\pi$ phases as explained in section~\ref{sec:Kl4KpiPhaseShifts}.
	\item The three input LECs are varied by their uncertainties given in (\ref{eq:Kl4InputLECsNLOMatching}).
	\item We have checked that the numerical uncertainties due to the discretisation, interpolation and numerical integration of the functions as well as the iteration procedure are completely negligible.
\end{itemize}

We note that the largest contribution to the systematic errors comes from the high-energy behaviour of the phase shifts, either from the $\pi\pi$ phases in the case of $L_1^r$ and $L_2^r$ or the $K\pi$ in the case of $L_3^r$. The uncertainties due to the low-energy parametrisation of the phases are small. The uncertainty due to the input LEC $L_9^r$ is very small as well.

\section{Conclusion and Outlook}

We have presented a new dispersive treatment of $K_{\ell4}$ decays, which provides a very accurate description of the hadronic form factors $F$ and $G$. The dispersion relation is valid up to and including $\O(p^6)$ in the chiral counting. Furthermore, it provides a resummation of final-state $\pi\pi$- and $K\pi$-rescattering effects, which we believe to be the most important contribution beyond $\O(p^6)$.

Our dispersion relation for $K_{\ell4}$ is written in the form of an Omnès representation. It consists of a set of coupled integral equations. We have solved this system numerically in an iteration procedure. The problem is parametrised by subtraction constants, which we have determined in a fit to data and by using the soft-pion theorem as well as chiral input. In contrast to a pure chiral description, the dispersion relation describes perfectly the experimentally observed curvature of the $S$-wave of the form factor $F$, which we interpret as a result of important $\pi\pi$-rescattering effects.

By using the matching equations to \ChPT{} we have extracted the values of the low-energy constants $L_1^r$, $L_2^r$ and $L_3^r$. The matching at two-loop level is quite sensitive to the input for the NNLO LECs and requires some further investigation.

Furthermore, we plan to use the two-dimensional NA48/2 data set for the $S$-wave of $F$, which we were provided with very recently. This will allow us to extract a value for $L_9^r$, which has been an input quantity so far.



\begin{appendices}
	\numberwithin{equation}{chapter}


\chapter{Scalar Loop Functions}

\label{sec:AppendixScalarLoopFunctions}

We use the following conventions for the scalar one-loop functions:
\begin{align}
	\begin{split}
		A_0(m^2) &:= \frac{1}{i} \int \frac{d^nq}{(2\pi)^n} \frac{1}{[ q^2 - m^2 ]} , \\
		B_0(p^2, m_1^2, m_2^2) &:= \frac{1}{i} \int \frac{d^nq}{(2\pi)^n} \frac{1}{[ q^2 - m_1^2 ] [ (q+p)^2 - m_2^2 ]} .
	\end{split}
\end{align}
These loop functions are UV-divergent. We define the renormalised loop functions in the $\overline{MS}$ scheme:
\begin{align}
	\begin{split}
		A_0(m^2) &= -2 m^2 \lambda + \bar A_0(m^2) + \O(4-n) , \\
		B_0(p^2, m_1^2, m_2^2) &= -2\lambda + \bar B_0(p^2, m_1^2, m_2^2) + \O(4-n) ,
	\end{split}
\end{align}
where
\begin{align}
	\begin{split}
		\lambda = \frac{\mu^{n-4}}{16\pi^2} \left( \frac{1}{n-4} - \frac{1}{2} \left( \ln(4\pi) + 1 - \gamma_E \right) \right) .
	\end{split}
\end{align}
$\mu$ denotes the renormalisation scale.

The renormalised loop functions are given by \cite{Amoros2000}
\begin{align}
	\begin{split}
		\bar A_0(m^2) &= -\frac{m^2}{16\pi^2} \ln\left( \frac{m^2}{\mu^2} \right) , \\
		\bar B_0(p^2, m_1^2, m_2^2) &= -\frac{1}{16\pi^2} \frac{m_1^2 \ln\left(\frac{m_1^2}{\mu^2}\right) - m_2^2 \ln\left(\frac{m_2^2}{\mu^2}\right)}{m_1^2 - m_2^2} \\
			&+ \frac{1}{32\pi^2} \left( 2 + \left( -\frac{\Delta}{p^2} + \frac{\Sigma}{\Delta} \right) \ln\left( \frac{m_1^2}{m_2^2} \right) - \frac{\nu}{p^2} \ln\left( \frac{(p^2+\nu)^2 - \Delta^2}{(p^2-\nu)^2 - \Delta^2} \right) \right) ,
	\end{split}
\end{align}
where
\begin{align}
	\begin{split}
		\Delta &:= m_1^2 - m_2^2 , \\
		\Sigma &:= m_1^2 + m_2^2 , \\
		\nu &:= \lambda^{1/2}(s,m_1^2,m_2^2) .
	\end{split}
\end{align}

The renormalised two-point function fulfils a once-subtracted dispersion relation:
\begin{align}
	\begin{split}
		\bar B_0(s, m_1^2, m_2^2) &= \bar B_0(0, m_1^2, m_2^2) + \frac{s}{\pi} \int_{(m_1+m_2)^2}^\infty \frac{\Im \bar B_0(s^\prime,m_1^2,m_2^2)}{(s^\prime - s - i\epsilon)s^\prime} ds^\prime ,
	\end{split}
\end{align}
where the imaginary part is given by
\begin{align}
	\begin{split}
		\Im \bar B_0(s, m_1^2, m_2^2) = \frac{1}{16 \pi} \frac{\lambda^{1/2}(s,m_1^2,m_2^2)}{s}
	\end{split}
\end{align}
and the value at zero is
\begin{align}
	\begin{split}
		B_0(0,m_1^2,m_2^2) =  -\frac{1}{16\pi^2} \frac{m_1^2 \ln\left(\frac{m_1^2}{\mu^2}\right) - m_2^2 \ln\left(\frac{m_2^2}{\mu^2}\right)}{m_1^2 - m_2^2} .
	\end{split}
\end{align}
The first and second derivative at zero are
\begin{align}
	\begin{split}
		B_0^\prime(0,m_1^2,m_2^2) &= \frac{1}{32\pi^2} \frac{\Delta \Sigma - 2 m_1^2 m_2^2 \ln\left(\frac{m_1^2}{m_2^2}\right)}{\Delta^3} , \\
		B_0^\dprime(0,m_1^2,m_2^2) &= \frac{1}{48\pi^2} \frac{\Delta(m_1^4 + 10 m_1^2 m_2^2 + m_2^4) - 6 m_1^2 m_2^2 \Sigma \ln\left(\frac{m_1^2}{m_2^2}\right)}{\Delta^5} .
	\end{split}
\end{align}


\chapter{Kinematics}

For each channel, the partial-wave expansion is performed in the corresponding rest frame, i.e.~in the $\pi\pi$ centre-of-mass frame for the $s$-channel and in one of the $K\pi$ centre-of-mass frames for the $t$- and $u$-channel. Therefore, we work out explicitly the kinematics in the three different frames.

\section{Legendre Polynomials and Spherical Harmonics}

\label{sec:LegendrePolynomials}

For the partial-wave-expansions, we make use of several relations between spherical harmonics and Legendre polynomials.

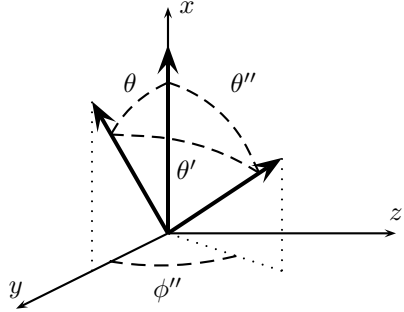
\begin{figure}[ht]
	\centering
	\psset{unit=1cm}
	\begin{pspicture}(-3,-3)(3,3)
		\psline{->}(0,0)(0,3)
		\psline{->}(0,0)(3,0)
		\psline{->}(0,0)(-2,-1)
		\psline[linewidth=1.5pt, arrowsize=8pt]{->}(0,0)(0,2.5)
		\psline[linewidth=1.5pt, arrowsize=8pt]{->}(0,0)(-1,1.75)
		\psline[linewidth=1.5pt, arrowsize=8pt]{->}(0,0)(1.5,1)
		\psline[linestyle=dotted](1.5,1)(1.5,-0.5)
		\psline[linestyle=dotted](0,0)(1.5,-0.5)
		\psline[linestyle=dotted](-1,1.75)(-1,-0.5)
		\psline[linearc=1.75,linestyle=dashed](0,2)(0.75,1.75)(1.2,0.8)
		\psline[linearc=1.5,linestyle=dashed](0,2)(-0.5,1.75)(-0.75,1.31)
		\psline[linearc=3,linestyle=dashed](-0.75,1.31)(0.5,1.25)(1.2,0.8)
		\psline[linearc=4,linestyle=dashed](-0.75,-0.375)(0,-0.5)(0.9,-0.3)
		\rput(0.25,3){$x$}
		\rput(-2,-0.75){$y$}
		\rput(3,0.25){$z$}
		\rput(-0.5,2){$\theta$}
		\rput(0.25,0.85){$\theta^\prime$}
		\rput(1,2){$\theta^\dprime$}
		\rput(0,-0.75){$\phi^\dprime$}
	\end{pspicture}
	\caption{Vectors and angles appearing in the addition theorem for spherical harmonics}
	\label{img:AnglesSphericalHarmonics}
\end{figure}

We use the addition theorem for the spherical harmonics and the relations between Legendre polynomials or derivatives of Legendre polynomials to spherical harmonics:
\begin{align}
	P_l(\cos\theta^\prime) &= \frac{4\pi}{2l+1} \sum_{m=-l}^l Y_l^m(\theta,0) {Y_l^m}^*(\theta^\dprime,\phi^\dprime) , \\
	P_{l^\prime}(\cos\theta^\dprime) &= \sqrt{\frac{4\pi}{2l^\prime+1}} Y_{l^\prime}^0(\theta^\dprime,\phi^\dprime) \qquad \text{(for any $\phi^\dprime$)} , \\
	P_{l^\prime}^\prime(\cos\theta^\dprime) \sin\theta^\dprime &= (-1)\sqrt{\frac{4\pi}{2l^\prime+1}} \sqrt{\frac{(l^\prime+1)!}{(l^\prime-1)!}} \, {Y_{l^\prime}^1}^*(\theta^\dprime,\phi^\dprime) e^{i\phi^\dprime} ,
\end{align}
where $P_l^\prime(z) := \frac{d}{dz}P_l(z)$. The different angles are defined in figure~\ref{img:AnglesSphericalHarmonics}.

We can now easily derive the addition theorem for the Legendre polynomials:
\begin{align}
	\begin{split}
		&\int d\Omega^\dprime P_l(\cos\theta^\prime) P_{l^\prime}(\cos\theta^\dprime) \\
		 &= \int d\Omega^\dprime \frac{4\pi}{2l+1}\sum_{m=-l}^l Y_l^m(\theta,0) {Y_l^m}^*(\theta^\dprime,\phi^\dprime) \sqrt{\frac{4\pi}{2l^\prime+1}} Y_{l^\prime}^0(\theta^\dprime,\phi^\dprime) \\
		 &= \sum_{m=-l}^l Y_l^m(\theta,0) \frac{4\pi}{2l+1} \sqrt{\frac{4\pi}{2l^\prime+1}} \underbrace{\int d\Omega^\dprime {Y_l^m}^*(\theta^\dprime,\phi^\dprime) Y_{l^\prime}^0(\theta^\dprime,\phi^\dprime)}_{\delta_{ll^\prime}\delta_{m0}} \\
		 &= \delta_{ll^\prime} \frac{4\pi}{2l+1} \sqrt{\frac{4\pi}{2l+1}} \, Y_l^0(\theta,0) = \delta_{ll^\prime} \frac{4\pi}{2l+1} P_l(\cos\theta) ,
	\end{split}
\end{align}
as well as the following relation:
\begin{align}
	\begin{split}
		&\int d\Omega^\dprime P_l(\cos\theta^\prime) P_{l^\prime}^\prime(\cos\theta^\dprime) \sin\theta^\dprime e^{-i \phi^\dprime} \\
		 &= \int d\Omega^\dprime \frac{4\pi}{2l+1}\sum_{m=-l}^l {Y_l^m}^*(\theta,0) Y_l^m(\theta^\dprime,\phi^\dprime) (-1) \sqrt{\frac{4\pi}{2l^\prime+1}} \sqrt{\frac{(l^\prime+1)!}{(l^\prime-1)!}} \, {Y_{l^\prime}^1}^*(\theta^\dprime,\phi^\dprime) \\
		 &= \sum_{m=-l}^l {Y_l^m}^*(\theta,0) \frac{4\pi}{2l+1} (-1) \sqrt{\frac{4\pi}{2l^\prime+1}} \sqrt{\frac{(l^\prime+1)!}{(l^\prime-1)!}} \underbrace{\int d\Omega^\dprime Y_l^m(\theta^\dprime,\phi^\dprime) {Y_{l^\prime}^1}^*(\theta^\dprime,\phi^\dprime)}_{\delta_{ll^\prime}\delta_{m1}} \\
		 &= \delta_{ll^\prime} \frac{4\pi}{2l+1} (-1) \sqrt{\frac{4\pi}{2l+1}} \sqrt{\frac{(l+1)!}{(l-1)!}} \, {Y_l^1}^*(\theta,0) = \delta_{ll^\prime} \frac{4\pi}{2l+1} P_l^\prime(\cos\theta) \sin\theta .
	\end{split}
\end{align}
Since the right-hand side is real, we conclude that
\begin{align}
	\begin{split}
		\int d\Omega^\dprime P_l(\cos\theta^\prime) P_{l^\prime}^\prime(\cos\theta^\dprime) \sin\theta^\dprime \cos\phi^\dprime &= \delta_{ll^\prime} \frac{4\pi}{2l+1} P_l^\prime(\cos\theta) \sin\theta , \\
		\int d\Omega^\dprime P_l(\cos\theta^\prime) P_{l^\prime}^\prime(\cos\theta^\dprime) \sin\theta^\dprime \sin\phi^\dprime &= 0 .
	\end{split}
\end{align}

\section{Kinematics in the $s$-Channel}

In the $\pi\pi$ centre-of-mass frame, the four-momenta of the different particles take the following values:
\begin{align}
	\begin{aligned}
		k &= \left( \sqrt{M_K^2 + \vec k^2}, \vec k \right) , \; & q_1 &= \left( \sqrt{ M_\pi^2 + \vec q^2 }, \vec q \right) , \; & p_1 &= \left( \sqrt{ M_\pi^2 + \vec p^2 }, \vec p \right) , \\
		-L &= \left( -\sqrt{ s_\ell + \vec k^2}, -\vec k \right) , \; & q_2 &= \left( \sqrt{ M_\pi^2 + \vec q^2 }, -\vec q \right) , \; & p_2 &= \left( \sqrt{ M_\pi^2 + \vec p^2 }, -\vec p \right) ,
	\end{aligned}
\end{align}
where $q_1$ and $q_2$ will be the momenta of intermediate pions. Note that we choose here the decay region ($L^0$ is positive), but could have equally well chosen the scattering region.

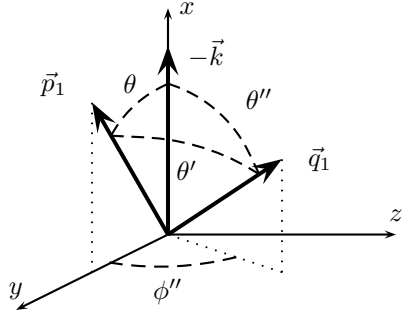
\begin{figure}[ht]
	\centering
	\psset{unit=1cm}
	\begin{pspicture}(-3,-3)(3,3)
		\psline{->}(0,0)(0,3)
		\psline{->}(0,0)(3,0)
		\psline{->}(0,0)(-2,-1)
		\psline[linewidth=1.5pt, arrowsize=8pt]{->}(0,0)(0,2.5)
		\psline[linewidth=1.5pt, arrowsize=8pt]{->}(0,0)(-1,1.75)
		\psline[linewidth=1.5pt, arrowsize=8pt]{->}(0,0)(1.5,1)
		\psline[linestyle=dotted](1.5,1)(1.5,-0.5)
		\psline[linestyle=dotted](0,0)(1.5,-0.5)
		\psline[linestyle=dotted](-1,1.75)(-1,-0.5)
		\psline[linearc=1.75,linestyle=dashed](0,2)(0.75,1.75)(1.2,0.8)
		\psline[linearc=1.5,linestyle=dashed](0,2)(-0.5,1.75)(-0.75,1.31)
		\psline[linearc=3,linestyle=dashed](-0.75,1.31)(0.5,1.25)(1.2,0.8)
		\psline[linearc=4,linestyle=dashed](-0.75,-0.375)(0,-0.5)(0.9,-0.3)
		\rput(0.25,3){$x$}
		\rput(-2,-0.75){$y$}
		\rput(3,0.25){$z$}
		\rput(-0.5,2){$\theta$}
		\rput(0.25,0.85){$\theta^\prime$}
		\rput(1.2,1.8){$\theta^\dprime$}
		\rput(0,-0.75){$\phi^\dprime$}
		\rput(0.5,2.4){$-\vec k$}
		\rput(-1.5,2){$\vec p_1$}
		\rput(2,1){$\vec q_1$}
	\end{pspicture}
	\caption{Vectors and angles in the $s$-channel centre-of-mass frame}
	\label{img:AnglesSChannel}
\end{figure}

Inserting these expressions into $s = (k - L)^2 = (q_1 + q_2)^2 = (p_1 + p_2)^2$ gives the values of $\vec k^2$, $\vec q^2$ and $\vec p^2$. We choose the directions of the three-vectors according to figure~\ref{img:AnglesSChannel}, i.e.~the angles are defined as $\theta := \angle(-\vec k, \vec p_1)$, $\theta^\prime := \angle(\vec p_1, \vec q_1)$, $\theta^\dprime := \angle(-\vec k, \vec q_1)$.

We end up with the following explicit expressions for the four-vectors:
\begin{align}
	\begin{aligned}
		k &= \left( \frac{M_K^2+s-s_\ell}{2\sqrt{s}}, -\frac{\lambda^{1/2}_{K\ell}(s)}{2\sqrt{s}}, 0, 0 \right) , \\
		L &= \left( \frac{M_K^2-s-s_\ell}{2\sqrt{s}}, -\frac{\lambda^{1/2}_{K\ell}(s)}{2\sqrt{s}}, 0, 0 \right) , \\
		q_1 &= \left( \frac{\sqrt{s}}{2}, \sqrt{\frac{s}{4}-M_\pi^2} \cos\theta^\dprime, \sqrt{\frac{s}{4}-M_\pi^2} \sin\theta^\dprime \cos\phi^\dprime, \sqrt{\frac{s}{4}-M_\pi^2} \sin\theta^\dprime \sin\phi^\dprime \right) , \\
		q_2 &= \left( \frac{\sqrt{s}}{2}, -\sqrt{\frac{s}{4}-M_\pi^2} \cos\theta^\dprime, -\sqrt{\frac{s}{4}-M_\pi^2} \sin\theta^\dprime \cos\phi^\dprime, -\sqrt{\frac{s}{4}-M_\pi^2} \sin\theta^\dprime \sin\phi^\dprime \right) , \\
		p_1 &= \left( \frac{\sqrt{s}}{2}, \sqrt{\frac{s}{4}-M_\pi^2} \cos\theta, \sqrt{\frac{s}{4}-M_\pi^2} \sin\theta, 0 \right) , \\
		p_2 &= \left( \frac{\sqrt{s}}{2}, -\sqrt{\frac{s}{4}-M_\pi^2} \cos\theta, -\sqrt{\frac{s}{4}-M_\pi^2} \sin\theta, 0 \right) ,
	\end{aligned}
\end{align}
where $\lambda_{K\ell}(s) := \lambda(M_K^2, s_\ell, s)$. Note that $\lambda_{K\ell}^{1/2}(s)$ has a square root branch cut in the $s$-plane between $(M_K-\sqrt{s_\ell})^2$ and $(M_K+\sqrt{s_\ell})^2$ and changes the sign when we continue it analytically to the scattering region. We will have to pay attention that we do not introduce this kinematic singularity into the partial-wave expansion.

In order to express the $s$-channel scattering angle $\theta$ with the Mandelstam variables, we calculate:
\begin{align}
	\begin{split}
		t - u &= (k-p_1)^2 - (k-p_2)^2 = k^2 + p_1^2 - 2kp_1 - k^2 - p_2^2 + 2kp_2 \\
			&= 2k(p_2-p_1) = 2(k^0 (p_2^0 - p_1^0) - \vec k \cdot (\vec p_2 - \vec p_1) ) \\
			&= \frac{\lambda^{1/2}_{K\ell}(s)}{\sqrt{s}} \left(-2\sqrt{\frac{s}{4}-M_\pi^2}\cos\theta \right) = -\lambda_{K\ell}^{1/2}(s) \sqrt{1-\frac{4M_\pi^2}{s}} \cos\theta \\
			&= - 2 X(s) \sigma_\pi(s) \cos\theta ,
	\end{split}
\end{align}
hence
\begin{align}
	\cos\theta = \frac{u-t}{2X\sigma_\pi} ,
\end{align}
where $\sigma_\pi(s) := \sqrt{1-4M_\pi^2/s}$ and $X(s) = \frac{1}{2}\lambda^{1/2}_{K\ell}(s)$ as before.

\section{Kinematics in the $t$-Channel}

In the $t$-channel, we are in the $K\pi$ centre-of-mass frame and look at the $t$-channel scattering region:
\begin{align}
	\begin{aligned}
		k &= \left( \sqrt{M_K^2 + \vec k^2}, \vec k \right) , \; & q_K &= \left( \sqrt{ M_K^2 + \vec q_K^2 }, \vec q_K \right) , \; & p_2 &= \left( \sqrt{ M_\pi^2 + \vec p_2^2 }, \vec p_2 \right) , \\
		-p_1 &= \left( \sqrt{ M_\pi^2 + \vec k^2}, -\vec k \right) , \; & q_\pi &= \left( \sqrt{ M_\pi^2 + \vec q_K^2 }, -\vec q_K \right) , \; & L &= \left( \sqrt{ s_\ell + \vec p_2^2 }, -\vec p_2 \right) .
	\end{aligned}
\end{align}

Inserting these expressions into $t = (k - p_1)^2 = (q_K + q_\pi)^2 = (p_2 + L)^2$ gives the values of $\vec k^2$, $\vec q_K^2$ and $\vec p_2^2$. We choose the directions of the three-vectors according to figure~\ref{img:AnglesTChannel}, i.e.~the angles are defined as $\theta_t := \angle(-\vec k, \vec p_2)$, $\theta_t^\prime := \angle(\vec k, \vec q_K)$, $\theta_t^\dprime := \angle(-\vec q_K, \vec p_2)$.

\begin{figure}[ht]
	\centering
	\psset{unit=1cm}
	\begin{pspicture}(-3,-3)(3,3)
		\psline{->}(0,0)(0,3)
		\psline{->}(0,0)(3,0)
		\psline{->}(0,0)(-2,-1)
		\psline[linewidth=1.5pt, arrowsize=8pt]{->}(0,0)(0,2.5)
		\psline[linewidth=1.5pt, arrowsize=8pt]{->}(0,0)(-1,1.75)
		\psline[linewidth=1.5pt, arrowsize=8pt]{->}(0,0)(1.5,1)
		\psline[linestyle=dotted](1.5,1)(1.5,-0.5)
		\psline[linestyle=dotted](0,0)(1.5,-0.5)
		\psline[linestyle=dotted](-1,1.75)(-1,-0.5)
		\psline[linearc=1.75,linestyle=dashed](0,2)(0.75,1.75)(1.2,0.8)
		\psline[linearc=1.5,linestyle=dashed](0,2)(-0.5,1.75)(-0.75,1.31)
		\psline[linearc=3,linestyle=dashed](-0.75,1.31)(0.5,1.25)(1.2,0.8)
		\psline[linearc=4,linestyle=dashed](-0.75,-0.375)(0,-0.5)(0.9,-0.3)
		\rput(0.25,3){$x$}
		\rput(-2,-0.75){$y$}
		\rput(3,0.25){$z$}
		\rput(-0.5,2){$\theta_t$}
		\rput(0.25,0.85){$\theta_t^\prime$}
		\rput(1.2,1.8){$\theta_t^\dprime$}
		\rput(0,-0.75){$\phi_t^\dprime$}
		\rput(0.5,2.4){$\vec p_2$}
		\rput(-1.5,2){$-\vec k$}
		\rput(2,1){$-\vec q_K$}
	\end{pspicture}
	\caption{Vectors and angles in the $t$-channel centre-of-mass frame}
	\label{img:AnglesTChannel}
\end{figure}

We find the following results:
\begin{align}
	\begin{aligned}
		k &= \left( \frac{t + M_K^2-M_\pi^2}{2\sqrt{t}}, -\frac{\lambda^{1/2}_{K\pi}(t)}{2\sqrt{t}} \cos\theta_t, -\frac{\lambda^{1/2}_{K\pi}(t)}{2\sqrt{t}} \sin\theta_t, 0 \right) , \\
		p_1 &= \left( \frac{M_K^2-M_\pi^2-t}{2\sqrt{t}}, -\frac{\lambda^{1/2}_{K\pi}(t)}{2\sqrt{t}} \cos\theta_t, -\frac{\lambda^{1/2}_{K\pi}(t)}{2\sqrt{t}} \sin\theta_t, 0 \right) , \\
		q_K &= \left( \frac{t + M_K^2-M_\pi^2}{2\sqrt{t}}, -\frac{\lambda^{1/2}_{K\pi}(t)}{2\sqrt{t}} \cos\theta_t^\dprime, -\frac{\lambda^{1/2}_{K\pi}(t)}{2\sqrt{t}} \sin\theta_t^\dprime \cos\phi_t^\dprime, -\frac{\lambda^{1/2}_{K\pi}(t)}{2\sqrt{t}} \sin\theta_t^\dprime \sin\phi_t^\dprime \right) , \\
		q_\pi &= \left( \frac{t-M_K^2+M_\pi^2}{2\sqrt{t}}, \frac{\lambda^{1/2}_{K\pi}(t)}{2\sqrt{t}} \cos\theta_t^\dprime, \frac{\lambda^{1/2}_{K\pi}(t)}{2\sqrt{t}} \sin\theta_t^\dprime \cos\phi_t^\dprime, \frac{\lambda^{1/2}_{K\pi}(t)}{2\sqrt{t}} \sin\theta_t^\dprime \sin\phi_t^\dprime \right) , \\
		p_2 &= \left( \frac{t-s_\ell+M_\pi^2}{2\sqrt{t}}, \frac{\lambda^{1/2}_{\ell\pi}(t)}{2\sqrt{t}}, 0, 0 \right) , \\
		L &= \left( \frac{t+s_\ell-M_\pi^2}{2\sqrt{t}}, -\frac{\lambda^{1/2}_{\ell\pi}(t)}{2\sqrt{t}}, 0, 0 \right) ,
	\end{aligned}
\end{align}
where $\lambda_{K\pi}(t) := \lambda(M_K^2, M_\pi^2, t)$ and  $\lambda_{\ell\pi}(t) := \lambda(s_\ell, M_\pi^2, t)$. Again, the square root of the first of these Källén functions has in the $t$-plane a branch cut between $(M_K-M_\pi)^2$ and $(M_K+M_\pi)^2$, the second between $(M_\pi-\sqrt{s_\ell})^2$ and $(M_\pi+\sqrt{s_\ell})^2$. Since we need the partial-wave expansion only in the scattering region $t>(M_K+M_\pi)^2$, these branch cuts are not relevant.

We calculate the $t$-channel scattering angle $\theta_t$ as a function of the Mandelstam variables:
\begin{align}
	\begin{split}
		s - u &= (p_1+p_2)^2 - (k-p_2)^2 = p_1^2 + p_2^2 + 2p_1 p_2 - k^2 - p_2^2 + 2kp_2 \\
			&= M_\pi^2 - M_K^2 + 2 p_2 ( k + p_1 ) \\
			&= M_\pi^2 - M_K^2 + 2 \left( \frac{t-s_\ell+M_\pi^2}{2\sqrt{t}} \, \frac{M_K^2 - M_\pi^2}{\sqrt{t}} + \frac{\lambda^{1/2}_{\ell\pi}(t)}{2\sqrt{t}} \frac{\lambda^{1/2}_{K\pi}(t)}{\sqrt{t}} \cos\theta_t \right) ,
	\end{split}
\end{align}
hence
\begin{align}
	\cos\theta_t = \frac{t(s-u) + \Delta_{K\pi} \Delta_{\ell\pi}}{\lambda^{1/2}_{K\pi}(t) \lambda^{1/2}_{\ell\pi}(t)}.
\end{align}

\section{Kinematics in the $u$-Channel}

The $u$-channel is completely analogous to the $t$-channel:
\begin{align}
	\begin{aligned}
		k &= \left( \sqrt{M_K^2 + \vec k^2}, \vec k \right) , \; & q_K &= \left( \sqrt{ M_K^2 + \vec q_K^2 }, \vec q_K \right) , \; & p_1 &= \left( \sqrt{ M_\pi^2 + \vec p_1^2 }, \vec p_1 \right) , \\
		-p_2 &= \left( \sqrt{ M_\pi^2 + \vec k^2}, -\vec k \right) , \; & q_\pi &= \left( \sqrt{ M_\pi^2 + \vec q_K^2 }, -\vec q_K \right) , \; & L &= \left( \sqrt{ s_\ell + \vec p_1^2 }, -\vec p_1 \right) .
	\end{aligned}
\end{align}

Inserting these expressions into $u = (k - p_2)^2 = (q_K + q_\pi)^2 = (p_1 + L)^2$ gives the values of $\vec k^2$, $\vec q_K^2$ and $\vec p_1^2$. We choose the directions of the three-vectors according to figure~\ref{img:AnglesUChannel}, i.e.~the angles are defined as $\theta_u := \angle(-\vec k, \vec p_1)$, $\theta_u^\prime := \angle(\vec k, \vec q_K)$, $\theta_u^\dprime := \angle(-\vec q_K, \vec p_1)$.

\begin{figure}[ht]
	\centering
	\psset{unit=1cm}
	\begin{pspicture}(-3,-3)(3,3)
		\psline{->}(0,0)(0,3)
		\psline{->}(0,0)(3,0)
		\psline{->}(0,0)(-2,-1)
		\psline[linewidth=1.5pt, arrowsize=8pt]{->}(0,0)(0,2.5)
		\psline[linewidth=1.5pt, arrowsize=8pt]{->}(0,0)(-1,1.75)
		\psline[linewidth=1.5pt, arrowsize=8pt]{->}(0,0)(1.5,1)
		\psline[linestyle=dotted](1.5,1)(1.5,-0.5)
		\psline[linestyle=dotted](0,0)(1.5,-0.5)
		\psline[linestyle=dotted](-1,1.75)(-1,-0.5)
		\psline[linearc=1.75,linestyle=dashed](0,2)(0.75,1.75)(1.2,0.8)
		\psline[linearc=1.5,linestyle=dashed](0,2)(-0.5,1.75)(-0.75,1.31)
		\psline[linearc=3,linestyle=dashed](-0.75,1.31)(0.5,1.25)(1.2,0.8)
		\psline[linearc=4,linestyle=dashed](-0.75,-0.375)(0,-0.5)(0.9,-0.3)
		\rput(0.25,3){$x$}
		\rput(-2,-0.75){$y$}
		\rput(3,0.25){$z$}
		\rput(-0.5,2.1){$\theta_u$}
		\rput(0.25,0.85){$\theta_u^\prime$}
		\rput(1.2,1.8){$\theta_u^\dprime$}
		\rput(0,-0.75){$\phi_u^\dprime$}
		\rput(0.5,2.4){$\vec p_1$}
		\rput(-1.5,2){$-\vec k$}
		\rput(2,1){$-\vec q_K$}
	\end{pspicture}
	\caption{Vectors and angles in the $u$-channel centre-of-mass frame}
	\label{img:AnglesUChannel}
\end{figure}

The results for the $u$-channel are then:
\begin{align}
	\begin{aligned}
		k &= \left( \frac{u + M_K^2-M_\pi^2}{2\sqrt{u}}, -\frac{\lambda^{1/2}_{K\pi}(u)}{2\sqrt{u}} \cos\theta_u, -\frac{\lambda^{1/2}_{K\pi}(u)}{2\sqrt{u}} \sin\theta_u, 0 \right) , \\
		p_2 &= \left( \frac{M_K^2-M_\pi^2-u}{2\sqrt{u}}, -\frac{\lambda^{1/2}_{K\pi}(u)}{2\sqrt{u}} \cos\theta_u, -\frac{\lambda^{1/2}_{K\pi}(u)}{2\sqrt{u}} \sin\theta_u, 0 \right) , \\
		q_K &= \left( \frac{u + M_K^2-M_\pi^2}{2\sqrt{u}}, -\frac{\lambda^{1/2}_{K\pi}(u)}{2\sqrt{u}} \cos\theta_u^\dprime, -\frac{\lambda^{1/2}_{K\pi}(u)}{2\sqrt{u}} \sin\theta_u^\dprime \cos\phi_u^\dprime, -\frac{\lambda^{1/2}_{K\pi}(u)}{2\sqrt{u}} \sin\theta_u^\dprime \sin\phi_u^\dprime \right) , \\
		q_\pi &= \left( \frac{u-M_K^2+M_\pi^2}{2\sqrt{u}}, \frac{\lambda^{1/2}_{K\pi}(u)}{2\sqrt{u}} \cos\theta_u^\dprime, \frac{\lambda^{1/2}_{K\pi}(u)}{2\sqrt{u}} \sin\theta_u^\dprime \cos\phi_u^\dprime, \frac{\lambda^{1/2}_{K\pi}(u)}{2\sqrt{u}} \sin\theta_u^\dprime \sin\phi_u^\dprime \right) , \\
		p_1 &= \left( \frac{u-s_\ell+M_\pi^2}{2\sqrt{u}}, \frac{\lambda^{1/2}_{\ell\pi}(u)}{2\sqrt{u}}, 0, 0 \right) , \\
		L &= \left( \frac{u+s_\ell-M_\pi^2}{2\sqrt{u}}, -\frac{\lambda^{1/2}_{\ell\pi}(u)}{2\sqrt{u}}, 0, 0 \right) .
	\end{aligned}
\end{align}

Let us calculate the $u$-channel scattering angle $\theta_u$ as a function of the Mandelstam variables:
\begin{align}
	\begin{split}
		s - t &= (p_1+p_2)^2 - (k-p_1)^2 = p_1^2 + p_2^2 + 2 p_1 p_2 - k^2 - p_1^2 + 2 k p_1 \\
			&= M_\pi^2 - M_K^2 + 2 p_1 ( k + p_2 ) \\
			&= M_\pi^2 - M_K^2 + 2 \left( \frac{u-s_\ell+M_\pi^2}{2\sqrt{u}} \, \frac{M_K^2 - M_\pi^2}{\sqrt{u}} + \frac{\lambda^{1/2}_{\ell\pi}(u)}{2\sqrt{u}} \frac{\lambda^{1/2}_{K\pi}(u)}{\sqrt{u}} \cos\theta_u \right) ,
	\end{split}
\end{align}
hence
\begin{align}
	\cos\theta_u = \frac{u(s-t) + \Delta_{K\pi} \Delta_{\ell\pi}}{\lambda^{1/2}_{K\pi}(u) \lambda^{1/2}_{\ell\pi}(u)}.
\end{align}


\chapter{Derivation of the Reconstruction Theorem}

\label{sec:AppendixReconstructionTheorem}

Consider one of the form factors $X \in \{F, G\}$ on a straight line of fixed $u$ that crosses the triangle where the form factor is real (see figure~\ref{img:MandelstamDiagram2}). For such a fixed $u$, the form factor depends only on one free variable:
\begin{align}
	\begin{split}
		X_s^u(s) &:= X(s,t^u(s),u) \big|_{u \text{ fixed}} , \\
		X_t^u(t) &:= X(s^u(t),t,u) \big|_{u \text{ fixed}} ,
	\end{split}
\end{align}
where $t^u(s) = \Sigma_0 - u - s$ and $s^u(t) = \Sigma_0 - u - t$. Note that
\begin{align}
	\begin{split}
		X_s^u(s^u(t)) &= X_t^u(t) , \\
		X_t^u(t^u(s)) &= X_s^u(s) .
	\end{split}
\end{align}
The function $X_s^u$ exhibits a left-hand cut for $s\in(-\infty,s^u(t_0))$ and a right-hand cut for $s\in(s_0,\infty)$, the function $X_t^u$ analogously a left-hand cut for $t\in(-\infty,t^u(s_0))$ and a right-hand cut for $t\in(t_0,\infty)$.

We assume that
\begin{align}
	\lim_{|s|\to\infty} \frac{X_s^u(s)}{s^n} = \lim_{|t|\to\infty} \frac{X_t^u(t)}{t^n} = 0 ,
\end{align}
where the Froissart bound \cite{Froissart1961} suggests $n=2$. However, we are also interested in the case $n=3$ in order to meet the asymptotic behaviour of the NNLO \ChPT{} form factors.

We write down a twice- or thrice-subtracted dispersion relation for $X_s^u$:
\begin{align}
	X_s^u(s) = P_{n-1}^u(s) + \frac{s^n}{\pi} \int_{s_0}^\infty \frac{\Im X_s^u(s^\prime + i\epsilon)}{(s^\prime-s) {s^\prime}^n} ds^\prime + \frac{s^n}{\pi} \int_{-\infty}^{s^u(t_0)} \frac{\Im X_s^u(s^\prime+i\epsilon)}{(s^\prime-s){s^\prime}^n} ds^\prime ,
\end{align}
where the limit $\epsilon\to0$ is understood implicitly and $s$ is meant somewhere away from the cuts. $P_{n-1}^u(s)$ is a polynomial of order $n-1$ in $s$ with real coefficients (which depend on $u$).

In the second integral, we change the integration variable:
\begin{align}
	s^\prime(t^\prime) := s^u(t^\prime) = \Sigma_0 - u - t^\prime \quad , \quad \frac{ds^\prime}{dt^\prime} = -1 .
\end{align}
Using $s^u(t^\prime) - s = \Sigma_0 - u - s - t^\prime = t^u(s) - t^\prime$, we find:
\begin{align}
	\begin{split}
		\frac{s^n}{\pi} \int_{-\infty}^{s^u(t_0)} &\frac{\Im X_s^u(s^\prime+i\epsilon)}{(s^\prime-s){s^\prime}^n} ds^\prime = - \frac{s^n}{\pi} \int_{\infty}^{t_0} \frac{\Im X_s^u(s^u(t^\prime-i\epsilon))}{(s^u(t^\prime)-s){s^u(t^\prime)}^n} dt^\prime \\
			&= \frac{\left(t^u(s)-(\Sigma_0-u)\right)^n}{\pi} \int_{t_0}^\infty \frac{\Im X_t^u(t^\prime - i\epsilon)}{(t^u(s) - t^\prime)(t^\prime-(\Sigma_0-u))^n} dt^\prime \\
			&= \frac{\left(t^u(s)-(\Sigma_0-u)\right)^n}{\pi} \int_{t_0}^\infty \frac{\Im X_t^u(t^\prime + i\epsilon)}{(t^\prime - t^u(s))(t^\prime-(\Sigma_0-u))^n} dt^\prime \\
			&= P_{n-1}^u(s) + \frac{{t^u(s)}^n}{\pi} \int_{t_0}^\infty \frac{\Im X_t^u(t^\prime + i\epsilon)}{(t^\prime-t^u(s)) {t^\prime}^n} dt^\prime ,
	\end{split}
\end{align}
where $P_{n-1}^u$ stands for some polynomial of order $n-1$ (we use this as a generic name and do not mean the same polynomial whenever $P_{n-1}^u$ appears). Hence, we obtain:
\begin{align}
	X_s^u(s) = P_{n-1}^u(s) + \frac{s^n}{\pi} \int_{s_0}^\infty \frac{\Im X_s^u(s^\prime + i\epsilon)}{(s^\prime-s) {s^\prime}^n} ds^\prime + \frac{{t^u(s)}^n}{\pi} \int_{t_0}^\infty \frac{\Im X_t^u(t^\prime + i\epsilon)}{(t^\prime-t^u(s)) {t^\prime}^n} dt^\prime .
\end{align}
In the end, we will evaluate the form factors in the physical regions on the upper rim of the cuts, i.e. at $s+i\epsilon$, for $s>4M_\pi^2$ or at $t^u(s)+i\epsilon$ for $t^u(s)>(M_K+M_\pi)^2$, where $s$ is real. We therefore conveniently write the dispersion relation in the form
\begin{align}
	\label{eq:FixedUDispersionRelation}
	X_s^u(s) = P_{n-1}^u(s) + \frac{s^n}{\pi} \int_{s_0}^\infty \frac{\Im X_s^u(s^\prime)}{(s^\prime-s-i\epsilon) {s^\prime}^n} ds^\prime + \frac{{t^u(s)}^n}{\pi} \int_{t_0}^\infty \frac{\Im X_t^u(t^\prime)}{(t^\prime-t^u(s)-i\epsilon) {t^\prime}^n} dt^\prime ,
\end{align}
where all the imaginary parts in the dispersion integrals have to be evaluated at the upper rim of the respective cut and the limit $\epsilon\to0$ is implicit.

In a second step, we choose in the Mandelstam plane a line of fixed $t$ that crosses the triangle where the form factors are real. Defining
\begin{align}
	\begin{split}
		X_s^t(s) &:= X(s,t,u^t(s)) \big|_{t \text{ fixed}} , \\
		X_u^t(u) &:= X(s^t(u),t,u) \big|_{t \text{ fixed}} ,
	\end{split}
\end{align}
where $u^t(s) = \Sigma_0 - t - s$ and $s^t(u) = \Sigma_0 - t - u$, such that
\begin{align}
	\begin{split}
		X_s^t(s^t(u)) &= X_u^t(u) , \\
		X_u^t(u^t(s)) &= X_s^t(s) ,
	\end{split}
\end{align}
we find with the same steps as before the dispersion relation
\begin{align}
	\label{eq:FixedTDispersionRelation}
	X_s^t(s) = P_{n-1}^t(s) + \frac{s^n}{\pi} \int_{s_0}^\infty \frac{\Im X_s^t(s^\prime)}{(s^\prime-s - i\epsilon) {s^\prime}^n} ds^\prime + \frac{{u^t(s)}^n}{\pi} \int_{u_0}^\infty \frac{\Im X_u^t(u^\prime)}{(u^\prime-u^t(s) - i\epsilon) {u^\prime}^n} du^\prime .
\end{align}

Finally, we construct a third dispersion relation along a line of fixed $s$ crossing the triangle where the form factors are real. We define
\begin{align}
	\begin{split}
		X_t^s(t) &:= X(s,t,u^s(t)) \big|_{s \text{ fixed}} , \\
		X_u^s(u) &:= X(s,t^s(u),u) \big|_{s \text{ fixed}} ,
	\end{split}
\end{align}
where $u^s(t) = \Sigma_0 - s - t$ and $t^s(u) = \Sigma_0 - s - u$, such that
\begin{align}
	\begin{split}
		X_t^s(t^s(u)) &= X_u^s(u) , \\
		X_u^s(u^s(t)) &= X_t^s(t) ,
	\end{split}
\end{align}
and find the dispersion relation
\begin{align}
	\label{eq:FixedSDispersionRelation}
	X_t^s(t) = P_{n-1}^s(t) + \frac{t^n}{\pi} \int_{t_0}^\infty \frac{\Im X_t^s(t^\prime)}{(t^\prime - t - i\epsilon) {t^\prime}^n} dt^\prime + \frac{{u^s(t)}^n}{\pi} \int_{u_0}^\infty \frac{\Im X_u^s(u^\prime)}{(u^\prime-u^s(t) - i\epsilon) {u^\prime}^n} du^\prime .
\end{align}

In a next step, we apply a partial wave expansion of the form factors. We suppose that $D$- and higher partial waves are real. This assumption is violated only at $\O(p^8)$: the elastic $\pi\pi$- and $K\pi$-scattering amplitudes behave as $\O(p^2)$, hence their imaginary parts are of $\O(p^4)$. The $\O(p^2)$ contribution is therefore real and has no discontinuity. It can be written as a first order polynomial in the Mandelstam variables (higher order polynomials would need coefficients that diverge in the chiral limit), which belongs to $S$- and $P$-waves. $D$- and higher waves appear only at $\O(p^4)$, thus their imaginary part is of $\O(p^8)$. Due to Watson's theorem, the same is true for the partial waves of $K_{\ell4}$. Further, we assume here without proof that inelasticities do not spoil the given argument.

We conclude from the above argument and (\ref{eq:sChannelFormFactorPartialWaveExpansion}) that the imaginary parts of the form factors can be written in the $s$-channel for $s>4M_\pi^2$ as:
\begin{align}
	\begin{split}
		\Im F(s,t,u) &= \Im f_0(s) + \frac{u-t}{M_K^2} \Im f_1(s) - \frac{2PL(s)(u-t)}{\lambda_{K\ell}(s)} \Im g_1(s) + \O(p^8) , \\
		\Im G(s,t,u) &= \Im g_1(s) + \O(p^8) .
	\end{split}
\end{align}
In the $t$-channel, we find with (\ref{eq:tChannelFormFactorPartialWaveExpansion2}) for $t>(M_K+M_\pi)^2$:
\begin{align}
	\begin{split}
		&\Im F^{(1/2)}(s,t,u) = \Im f_0^{(1/2)}(t) + \frac{t(s-u) + \Delta_{K\pi}\Delta_{\ell\pi}}{M_K^4} \Im f_1^{(1/2)}(t) \\
			& - \frac{1}{2t} \left( M_K^2 - M_\pi^2 - 3 t - (\Delta_{\ell\pi} + t) \frac{t(s-u) + \Delta_{K\pi}\Delta_{\ell\pi}}{\lambda_{\ell\pi}(t)} \right) \Im g_1^{(1/2)}(t) + \O(p^8) , \\
		&\Im G^{(1/2)}(s,t,u) = -\Im f_0^{(1/2)}(t) - \frac{t(s-u) + \Delta_{K\pi}\Delta_{\ell\pi}}{M_K^4} \Im f_1^{(1/2)}(t) \\
			& + \frac{1}{2t} \left( M_K^2 - M_\pi^2 + t - (\Delta_{\ell\pi} + t) \frac{t(s-u) + \Delta_{K\pi}\Delta_{\ell\pi}}{\lambda_{\ell\pi}(t)} \right) \Im g_1^{(1/2)}(t) + \O(p^8) .
	\end{split}
\end{align}
Finally, in the $u$-channel for $u>(M_K+M_\pi)^2$, (\ref{eq:uChannelFormFactorPartialWaveExpansion2}) implies:
\begin{align}
	\begin{split}
		&\Im F(s,t,u) = \Im f_0^{(3/2)}(u) + \frac{u(s-t) + \Delta_{K\pi}\Delta_{\ell\pi}}{M_K^4} \Im f_1^{(3/2)}(u) \\
			& - \frac{1}{2u} \left( M_K^2 - M_\pi^2 - 3 u - (\Delta_{\ell\pi} + u) \frac{u(s-t) + \Delta_{K\pi}\Delta_{\ell\pi}}{\lambda_{\ell\pi}(u)} \right) \Im g_1^{(3/2)}(u) + \O(p^8) , \\
		&\Im G(s,t,u) = \Im f_0^{(3/2)}(u) + \frac{u(s-t) + \Delta_{K\pi}\Delta_{\ell\pi}}{M_K^4} \Im f_1^{(3/2)}(u) \\
			& - \frac{1}{2u} \left( M_K^2 - M_\pi^2 + u - (\Delta_{\ell\pi} + u) \frac{u(s-t) + \Delta_{K\pi}\Delta_{\ell\pi}}{\lambda_{\ell\pi}(u)} \right) \Im g_1^{(3/2)}(u) + \O(p^8) .
	\end{split}
\end{align}
The various partial waves satisfy Watson's theorem below the first inelastic threshold, i.e.~their phase is given by the elastic $\pi\pi$- or $K\pi$-scattering phase. Above the inelastic thresholds, the phases are in principle unconstrained.

We insert these expansions of the imaginary parts of the form factors into the fixed-$u$ dispersion relation (\ref{eq:FixedUDispersionRelation}):
\begin{align}
	\footnotesize
	\begin{split}
		F_s^u(s) &= P_{n-1}^u(s) + \frac{s^n}{\pi} \int_{s_0}^\infty \frac{\Im f_0(s^\prime)}{(s^\prime-s - i\epsilon){s^\prime}^n} ds^\prime +  \frac{s^n}{\pi} \int_{s_0}^\infty \frac{u-t^u(s^\prime)}{M_K^2} \frac{\Im f_1(s^\prime)}{(s^\prime-s - i\epsilon){s^\prime}^n} ds^\prime \\
			&\quad  -  \frac{s^n}{\pi} \int_{s_0}^\infty \frac{2PL(s^\prime)(u-t^u(s^\prime))}{\lambda_{K\ell}(s^\prime)} \frac{\Im g_1(s^\prime)}{(s^\prime-s - i\epsilon){s^\prime}^n} ds^\prime \\
			&\quad  + \frac{2}{3} \frac{{t^u(s)}^n}{\pi} \int_{t_0}^\infty \frac{ \Im f_0^{(1/2)}(t^\prime)}{(t^\prime-t^u(s) - i\epsilon) {t^\prime}^n} dt^\prime \\
			&\quad  + \frac{2}{3} \frac{{t^u(s)}^n}{\pi} \int_{t_0}^\infty  \frac{t^\prime(s^u(t^\prime)-u) + \Delta_{K\pi}\Delta_{\ell\pi}}{M_K^4} \frac{ \Im f_1^{(1/2)}(t^\prime) }{(t^\prime-t^u(s) - i\epsilon) {t^\prime}^n} dt^\prime \\
			&\quad  - \frac{2}{3} \frac{{t^u(s)}^n}{\pi} \int_{t_0}^\infty  \frac{1}{2t^\prime} \left( M_K^2 - M_\pi^2 - 3 t^\prime - (\Delta_{\ell\pi} + t^\prime) \frac{t^\prime(s^u(t^\prime)-u) + \Delta_{K\pi}\Delta_{\ell\pi}}{\lambda_{\ell\pi}(t^\prime)} \right) \frac{ \Im g_1^{(1/2)}(t^\prime)}{(t^\prime-t^u(s) - i\epsilon) {t^\prime}^n} dt^\prime \\
			&\quad  + \frac{1}{3} \frac{{t^u(s)}^n}{\pi} \int_{t_0}^\infty \frac{\Im f_0^{(3/2)}(t^\prime)}{(t^\prime-t^u(s) - i\epsilon) {t^\prime}^n} dt^\prime \\
			&\quad  + \frac{1}{3} \frac{{t^u(s)}^n}{\pi} \int_{t_0}^\infty \frac{t^\prime(s^u(t^\prime)-u) + \Delta_{K\pi}\Delta_{\ell\pi}}{M_K^4} \frac{ \Im f_1^{(3/2)}(t^\prime)}{(t^\prime-t^u(s) - i\epsilon) {t^\prime}^n} dt^\prime \\
			&\quad  - \frac{1}{3} \frac{{t^u(s)}^n}{\pi} \int_{t_0}^\infty \frac{1}{2t^\prime} \left( M_K^2 - M_\pi^2 - 3 t^\prime - (\Delta_{\ell\pi} + t^\prime) \frac{t^\prime(s^u(t^\prime)-u) + \Delta_{K\pi}\Delta_{\ell\pi}}{\lambda_{\ell\pi}(t^\prime)} \right) \frac{ \Im g_1^{(3/2)}(t^\prime)}{(t^\prime-t^u(s) - i\epsilon) {t^\prime}^n} dt^\prime \\
			&\quad + \O(p^8) ,
	\end{split}
\end{align}
\begin{align}
	\footnotesize
	\begin{split}
		G_s^u(s) &= P_{n-1}^u(s) + \frac{s^n}{\pi} \int_{s_0}^\infty \frac{\Im g_1(s^\prime)}{(s^\prime-s - i\epsilon){s^\prime}^n} ds^\prime \\
			&\quad  - \frac{2}{3} \frac{{t^u(s)}^n}{\pi} \int_{t_0}^\infty \frac{ \Im f_0^{(1/2)}(t^\prime)}{(t^\prime-t^u(s) - i\epsilon) {t^\prime}^n} dt^\prime \\
			&\quad  - \frac{2}{3} \frac{{t^u(s)}^n}{\pi} \int_{t_0}^\infty  \frac{t^\prime(s^u(t^\prime)-u) + \Delta_{K\pi}\Delta_{\ell\pi}}{M_K^4} \frac{ \Im f_1^{(1/2)}(t^\prime) }{(t^\prime-t^u(s) - i\epsilon) {t^\prime}^n} dt^\prime \\
			&\quad  + \frac{2}{3} \frac{{t^u(s)}^n}{\pi} \int_{t_0}^\infty  \frac{1}{2t^\prime} \left( M_K^2 - M_\pi^2 + t^\prime - (\Delta_{\ell\pi} + t^\prime) \frac{t^\prime(s^u(t^\prime)-u) + \Delta_{K\pi}\Delta_{\ell\pi}}{\lambda_{\ell\pi}(t^\prime)} \right) \frac{ \Im g_1^{(1/2)}(t^\prime)}{(t^\prime-t^u(s) - i\epsilon) {t^\prime}^n} dt^\prime \\
			&\quad  - \frac{1}{3} \frac{{t^u(s)}^n}{\pi} \int_{t_0}^\infty \frac{\Im f_0^{(3/2)}(t^\prime)}{(t^\prime-t^u(s) - i\epsilon) {t^\prime}^n} dt^\prime \\
			&\quad  - \frac{1}{3} \frac{{t^u(s)}^n}{\pi} \int_{t_0}^\infty \frac{t^\prime(s^u(t^\prime)-u) + \Delta_{K\pi}\Delta_{\ell\pi}}{M_K^4} \frac{ \Im f_1^{(3/2)}(t^\prime)}{(t^\prime-t^u(s) - i\epsilon) {t^\prime}^n} dt^\prime \\
			&\quad  + \frac{1}{3} \frac{{t^u(s)}^n}{\pi} \int_{t_0}^\infty \frac{1}{2t^\prime} \left( M_K^2 - M_\pi^2 + t^\prime - (\Delta_{\ell\pi} + t^\prime) \frac{t^\prime(s^u(t^\prime)-u) + \Delta_{K\pi}\Delta_{\ell\pi}}{\lambda_{\ell\pi}(t^\prime)} \right) \frac{ \Im g_1^{(3/2)}(t^\prime)}{(t^\prime-t^u(s) - i\epsilon) {t^\prime}^n} dt^\prime \\
			&\quad + \O(p^8) ,
	\end{split}
\end{align}
where we used the isospin relations
\begin{align}
	\begin{split}
		F(s,t,u) &= \frac{2}{3} F^{(1/2)}(s,t,u) + \frac{1}{3} F(s,u,t) , \\
		G(s,t,u) &= \frac{2}{3} G^{(1/2)}(s,t,u) - \frac{1}{3} G(s,u,t) .
	\end{split}
\end{align}

The partial wave expansions inserted into the fixed-$t$ dispersion relation (\ref{eq:FixedTDispersionRelation}) results in:
\begin{align}
	\footnotesize
	\begin{split}
		F_s^t(s) &= P_{n-1}^t(s) + \frac{s^n}{\pi} \int_{s_0}^\infty \frac{\Im f_0(s^\prime)}{(s^\prime-s - i\epsilon) {s^\prime}^n} ds^\prime + \frac{s^n}{\pi} \int_{s_0}^\infty \frac{u^t(s^\prime)-t}{M_K^2} \frac{ \Im f_1(s^\prime)}{(s^\prime-s - i\epsilon) {s^\prime}^n} ds^\prime \\
			&\quad - \frac{s^n}{\pi} \int_{s_0}^\infty \frac{2PL(s^\prime)(u^t(s^\prime)-t)}{\lambda_{K\ell}(s^\prime)} \frac{ \Im g_1(s^\prime)}{(s^\prime-s - i\epsilon) {s^\prime}^n} ds^\prime \\
			&\quad+ \frac{{u^t(s)}^n}{\pi} \int_{u_0}^\infty \frac{\Im f_0^{(3/2)}(u^\prime)}{(u^\prime-u^t(s) - i\epsilon) {u^\prime}^n} du^\prime \\
			&\quad+ \frac{{u^t(s)}^n}{\pi} \int_{u_0}^\infty \frac{u^\prime(s^t(u^\prime)-t) + \Delta_{K\pi}\Delta_{\ell\pi}}{M_K^4} \frac{\Im f_1^{(3/2)}(u^\prime)}{(u^\prime-u^t(s) - i\epsilon) {u^\prime}^n} du^\prime \\
			&\quad- \frac{{u^t(s)}^n}{\pi} \int_{u_0}^\infty \frac{1}{2u^\prime} \left( M_K^2 - M_\pi^2 - 3 u^\prime - (\Delta_{\ell\pi} + u^\prime) \frac{u^\prime(s^t(u^\prime)-t) + \Delta_{K\pi}\Delta_{\ell\pi}}{\lambda_{\ell\pi}(u^\prime)} \right) \frac{ \Im g_1^{(3/2)}(u^\prime)}{(u^\prime-u^t(s) - i\epsilon) {u^\prime}^n} du^\prime \\
			&\quad+ \O(p^8) ,
	\end{split}
\end{align}
\begin{align}
	\footnotesize
	\begin{split}
		G_s^t(s) &= P_{n-1}^t(s) + \frac{s^n}{\pi} \int_{s_0}^\infty \frac{\Im g_1(s^\prime)}{(s^\prime-s - i\epsilon) {s^\prime}^n} ds^\prime \\
			&\quad+ \frac{{u^t(s)}^n}{\pi} \int_{u_0}^\infty \frac{\Im f_0^{(3/2)}(u^\prime)}{(u^\prime-u^t(s) - i\epsilon) {u^\prime}^n} du^\prime \\
			&\quad+ \frac{{u^t(s)}^n}{\pi} \int_{u_0}^\infty \frac{u^\prime(s^t(u^\prime)-t) + \Delta_{K\pi}\Delta_{\ell\pi}}{M_K^4} \frac{\Im f_1^{(3/2)}(u^\prime)}{(u^\prime-u^t(s) - i\epsilon) {u^\prime}^n} du^\prime \\
			&\quad- \frac{{u^t(s)}^n}{\pi} \int_{u_0}^\infty \frac{1}{2u^\prime} \left( M_K^2 - M_\pi^2 + u^\prime - (\Delta_{\ell\pi} + u^\prime) \frac{u^\prime(s^t(u^\prime)-t) + \Delta_{K\pi}\Delta_{\ell\pi}}{\lambda_{\ell\pi}(u^\prime)} \right) \frac{ \Im g_1^{(3/2)}(u^\prime)}{(u^\prime-u^t(s) - i\epsilon) {u^\prime}^n} du^\prime \\
			&\quad+ \O(p^8) .
	\end{split}
\end{align}

The partial wave expansions inserted into the fixed-$s$ dispersion relation (\ref{eq:FixedSDispersionRelation}) give:
\begin{align}
	\footnotesize
	\begin{split}
		F_t^s(t) &= P_{n-1}^s(t) + \frac{2}{3} \frac{t^n}{\pi} \int_{t_0}^\infty \frac{ \Im f_0^{(1/2)}(t^\prime)}{(t^\prime - t - i\epsilon) {t^\prime}^n} dt^\prime \\
			&\quad + \frac{2}{3} \frac{t^n}{\pi} \int_{t_0}^\infty \frac{t^\prime(s-u^s(t^\prime)) + \Delta_{K\pi}\Delta_{\ell\pi}}{M_K^4} \frac{ \Im f_1^{(1/2)}(t^\prime)}{(t^\prime - t - i\epsilon) {t^\prime}^n} dt^\prime \\
			&\quad - \frac{2}{3} \frac{t^n}{\pi} \int_{t_0}^\infty \frac{1}{2t^\prime} \left( M_K^2 - M_\pi^2 - 3 t^\prime - (\Delta_{\ell\pi} + t^\prime) \frac{t^\prime(s-u^s(t^\prime)) + \Delta_{K\pi}\Delta_{\ell\pi}}{\lambda_{\ell\pi}(t^\prime)} \right) \frac{ \Im g_1^{(1/2)}(t^\prime)}{(t^\prime - t - i\epsilon) {t^\prime}^n} dt^\prime \\
			&\quad + \frac{1}{3} \frac{t^n}{\pi} \int_{t_0}^\infty \frac{ \Im f_0^{(3/2)}(t^\prime)}{(t^\prime - t - i\epsilon) {t^\prime}^n} dt^\prime \\
			&\quad + \frac{1}{3} \frac{t^n}{\pi} \int_{t_0}^\infty \frac{t^\prime(s-u^s(t^\prime)) + \Delta_{K\pi}\Delta_{\ell\pi}}{M_K^4} \frac{ \Im f_1^{(3/2)}(t^\prime)}{(t^\prime - t - i\epsilon) {t^\prime}^n} dt^\prime \\
			&\quad - \frac{1}{3} \frac{t^n}{\pi} \int_{t_0}^\infty \frac{1}{2t^\prime} \left( M_K^2 - M_\pi^2 - 3 t^\prime - (\Delta_{\ell\pi} + t^\prime) \frac{t^\prime(s-u^s(t^\prime)) + \Delta_{K\pi}\Delta_{\ell\pi}}{\lambda_{\ell\pi}(t^\prime)} \right) \frac{ \Im g_1^{(3/2)}(t^\prime)}{(t^\prime - t - i\epsilon) {t^\prime}^n} dt^\prime \\
			&\quad + \frac{{u^s(t)}^n}{\pi} \int_{u_0}^\infty \frac{ \Im f_0^{(3/2)}(u^\prime)}{(u^\prime-u^s(t) - i\epsilon) {u^\prime}^n} du^\prime \\
			&\quad + \frac{{u^s(t)}^n}{\pi} \int_{u_0}^\infty \frac{u^\prime(s-t^s(u^\prime)) + \Delta_{K\pi}\Delta_{\ell\pi}}{M_K^4} \frac{ \Im f_1^{(3/2)}(u^\prime)}{(u^\prime-u^s(t) - i\epsilon) {u^\prime}^n} du^\prime \\
			&\quad - \frac{{u^s(t)}^n}{\pi} \int_{u_0}^\infty \frac{1}{2u^\prime} \left( M_K^2 - M_\pi^2 - 3 u^\prime - (\Delta_{\ell\pi} + u^\prime) \frac{u^\prime(s-t^s(u^\prime)) + \Delta_{K\pi}\Delta_{\ell\pi}}{\lambda_{\ell\pi}(u^\prime)} \right) \frac{ \Im g_1^{(3/2)}(u^\prime)}{(u^\prime-u^s(t) - i\epsilon) {u^\prime}^n} du^\prime \\
			&\quad+ \O(p^8) ,
	\end{split}
\end{align}
\begin{align}
	\footnotesize
	\begin{split}
		G_t^s(t) &= P_{n-1}^s(t) - \frac{2}{3} \frac{t^n}{\pi} \int_{t_0}^\infty \frac{ \Im f_0^{(1/2)}(t^\prime)}{(t^\prime - t - i\epsilon) {t^\prime}^n} dt^\prime \\
			&\quad - \frac{2}{3} \frac{t^n}{\pi} \int_{t_0}^\infty \frac{t^\prime(s-u^s(t^\prime)) + \Delta_{K\pi}\Delta_{\ell\pi}}{M_K^4} \frac{ \Im f_1^{(1/2)}(t^\prime)}{(t^\prime - t - i\epsilon) {t^\prime}^n} dt^\prime \\
			&\quad + \frac{2}{3} \frac{t^n}{\pi} \int_{t_0}^\infty \frac{1}{2t^\prime} \left( M_K^2 - M_\pi^2 + t^\prime - (\Delta_{\ell\pi} + t^\prime) \frac{t^\prime(s-u^s(t^\prime)) + \Delta_{K\pi}\Delta_{\ell\pi}}{\lambda_{\ell\pi}(t^\prime)} \right) \frac{ \Im g_1^{(1/2)}(t^\prime)}{(t^\prime - t - i\epsilon) {t^\prime}^n} dt^\prime \\
			&\quad - \frac{1}{3} \frac{t^n}{\pi} \int_{t_0}^\infty \frac{ \Im f_0^{(3/2)}(t^\prime)}{(t^\prime - t - i\epsilon) {t^\prime}^n} dt^\prime \\
			&\quad - \frac{1}{3} \frac{t^n}{\pi} \int_{t_0}^\infty \frac{t^\prime(s-u^s(t^\prime)) + \Delta_{K\pi}\Delta_{\ell\pi}}{M_K^4} \frac{ \Im f_1^{(3/2)}(t^\prime)}{(t^\prime - t - i\epsilon) {t^\prime}^n} dt^\prime \\
			&\quad + \frac{1}{3} \frac{t^n}{\pi} \int_{t_0}^\infty \frac{1}{2t^\prime} \left( M_K^2 - M_\pi^2 + t^\prime - (\Delta_{\ell\pi} + t^\prime) \frac{t^\prime(s-u^s(t^\prime)) + \Delta_{K\pi}\Delta_{\ell\pi}}{\lambda_{\ell\pi}(t^\prime)} \right) \frac{ \Im g_1^{(3/2)}(t^\prime)}{(t^\prime - t - i\epsilon) {t^\prime}^n} dt^\prime \\
			&\quad + \frac{{u^s(t)}^n}{\pi} \int_{u_0}^\infty \frac{ \Im f_0^{(3/2)}(u^\prime)}{(u^\prime-u^s(t) - i\epsilon) {u^\prime}^n} du^\prime \\
			&\quad + \frac{{u^s(t)}^n}{\pi} \int_{u_0}^\infty \frac{u^\prime(s-t^s(u^\prime)) + \Delta_{K\pi}\Delta_{\ell\pi}}{M_K^4} \frac{ \Im f_1^{(3/2)}(u^\prime)}{(u^\prime-u^s(t) - i\epsilon) {u^\prime}^n} du^\prime \\
			&\quad - \frac{{u^s(t)}^n}{\pi} \int_{u_0}^\infty \frac{1}{2u^\prime} \left( M_K^2 - M_\pi^2 + u^\prime - (\Delta_{\ell\pi} + u^\prime) \frac{u^\prime(s-t^s(u^\prime)) + \Delta_{K\pi}\Delta_{\ell\pi}}{\lambda_{\ell\pi}(u^\prime)} \right) \frac{ \Im g_1^{(3/2)}(u^\prime)}{(u^\prime-u^s(t) - i\epsilon) {u^\prime}^n} du^\prime \\
			&\quad+ \O(p^8) .
	\end{split}
\end{align}

In order to further simplify these various expressions, we use the relation
\begin{align}
	\frac{s}{(s^\prime-s)s^\prime} = \frac{1}{s^\prime-s} - \frac{1}{s^\prime}
\end{align}
for the following calculations:
\begin{align}
	\footnotesize
	\begin{split}
		\frac{s^n}{\pi} &\int_{s_0}^\infty (u-t^u(s^\prime)) \frac{\Im A(s^\prime)}{(s^\prime-s){s^\prime}^n} ds^\prime = \frac{s^n}{\pi} \int_{s_0}^\infty (2u-\Sigma_0 +s^\prime) \frac{\Im A(s^\prime)}{(s^\prime-s){s^\prime}^n} ds^\prime \\
			&= (2u-\Sigma_0) \frac{s^n}{\pi} \int_{s_0}^\infty \frac{\Im A(s^\prime)}{(s^\prime-s){s^\prime}^n} ds^\prime + \frac{s^n}{\pi} \int_{s_0}^\infty \frac{\Im A(s^\prime)}{(s^\prime-s){s^\prime}^{n-1}} ds^\prime \\
			&= (2u-\Sigma_0) \frac{s^{n-1}}{\pi} \int_{s_0}^\infty \frac{\Im A(s^\prime)}{(s^\prime-s){s^\prime}^{n-1}} ds^\prime - (2u-\Sigma_0) \frac{s^{n-1}}{\pi} \int_{s_0}^\infty \frac{\Im A(s^\prime)}{{s^\prime}^n} ds^\prime + \frac{s^n}{\pi} \int_{s_0}^\infty \frac{\Im A(s^\prime)}{(s^\prime-s){s^\prime}^{n-1}} ds^\prime \\
			&= (2u-\Sigma_0+s) \frac{s^{n-1}}{\pi} \int_{s_0}^\infty \frac{\Im A(s^\prime)}{(s^\prime-s){s^\prime}^{n-1}} ds^\prime + P_{n-1}^u(s) \\
			&= (u - t^u(s)) \frac{s^{n-1}}{\pi} \int_{s_0}^\infty  \frac{\Im A(s^\prime)}{(s^\prime-s){s^\prime}^{n-1}} ds^\prime + P_{n-1}^u(s) .
	\end{split}
\end{align}
\begin{align}
	\footnotesize
	\begin{split}
		\frac{s^n}{\pi} &\int_{s_0}^\infty (u^t(s^\prime)-t) \frac{\Im A(s^\prime)}{(s^\prime-s){s^\prime}^n} ds^\prime = \frac{s^n}{\pi} \int_{s_0}^\infty (\Sigma_0 - 2t - s^\prime) \frac{\Im A(s^\prime)}{(s^\prime-s){s^\prime}^n} ds^\prime \\
			&= (\Sigma_0-2t) \frac{s^n}{\pi} \int_{s_0}^\infty \frac{\Im A(s^\prime)}{(s^\prime-s){s^\prime}^n} ds^\prime - \frac{s^n}{\pi} \int_{s_0}^\infty \frac{\Im A(s^\prime)}{(s^\prime-s){s^\prime}^{n-1}} ds^\prime \\
			&= (\Sigma_0-2t) \frac{s^{n-1}}{\pi} \int_{s_0}^\infty \frac{\Im A(s^\prime)}{(s^\prime-s){s^\prime}^{n-1}} ds^\prime - (\Sigma_0-2t) \frac{s^{n-1}}{\pi} \int_{s_0}^\infty \frac{\Im A(s^\prime)}{{s^\prime}^n} ds^\prime - \frac{s^n}{\pi} \int_{s_0}^\infty \frac{\Im A(s^\prime)}{(s^\prime-s){s^\prime}^{n-1}} ds^\prime \\
			&= (\Sigma_0-2t-s) \frac{s^{n-1}}{\pi} \int_{s_0}^\infty \frac{\Im A(s^\prime)}{(s^\prime-s){s^\prime}^{n-1}} ds^\prime + P_{n-1}^t(s) \\
			&= (u^t(s) - t) \frac{s^{n-1}}{\pi} \int_{s_0}^\infty  \frac{\Im A(s^\prime)}{(s^\prime-s){s^\prime}^{n-1}} ds^\prime + P_{n-1}^t(s) .
	\end{split}
\end{align}
We calculate further:
\begin{align}
	\begin{split}
		&\frac{t^u(s)^n}{\pi} \int_{t_0}^\infty (t^\prime(s^u(t^\prime)-u) + \Delta_{K\pi}\Delta_{\ell\pi}) \frac{\Im A(t^\prime)}{(t^\prime-t^u(s)){t^\prime}^n} dt^\prime \\
			&= \Delta_{K\pi}\Delta_{\ell\pi} \frac{t^u(s)^n}{\pi} \int_{t_0}^\infty \frac{\Im A(t^\prime)}{(t^\prime-t^u(s)){t^\prime}^n} dt^\prime \\
			&\quad + \frac{t^u(s)^n}{\pi} \int_{t_0}^\infty (\Sigma_0 - 2u - t^\prime) \frac{\Im A(t^\prime)}{(t^\prime-t^u(s)){t^\prime}^{n-1}} dt^\prime \\
			&= \Delta_{K\pi}\Delta_{\ell\pi} \frac{t^u(s)^{n-1}}{\pi} \int_{t_0}^\infty \frac{\Im A(t^\prime)}{(t^\prime-t^u(s)){t^\prime}^{n-1}} dt^\prime - \Delta_{K\pi}\Delta_{\ell\pi} \frac{t^u(s)^{n-1}}{\pi} \int_{t_0}^\infty \frac{\Im A(t^\prime)}{{t^\prime}^n} dt^\prime \\
			&\quad +  (\Sigma_0 - 2u) \frac{t^u(s)^n}{\pi} \int_{t_0}^\infty \frac{\Im A(t^\prime)}{(t^\prime-t^u(s)){t^\prime}^{n-1}} dt^\prime  - \frac{t^u(s)^n}{\pi} \int_{t_0}^\infty \frac{\Im A(t^\prime)}{(t^\prime-t^u(s)){t^\prime}^{n-2}} dt^\prime \\
			&= \Delta_{K\pi}\Delta_{\ell\pi} \frac{t^u(s)^{n-2}}{\pi} \int_{t_0}^\infty \frac{\Im A(t^\prime)}{(t^\prime-t^u(s)){t^\prime}^{n-2}} dt^\prime  - \Delta_{K\pi}\Delta_{\ell\pi} \frac{t^u(s)^{n-2}}{\pi} \int_{t_0}^\infty \frac{\Im A(t^\prime)}{{t^\prime}^{n-1}} dt^\prime \\
			&\quad - \Delta_{K\pi}\Delta_{\ell\pi} \frac{t^u(s)^{n-1}}{\pi} \int_{t_0}^\infty \frac{\Im A(t^\prime)}{{t^\prime}^n} dt^\prime  +  (\Sigma_0 - 2u) \frac{t^u(s)^{n-1}}{\pi} \int_{t_0}^\infty \frac{\Im A(t^\prime)}{(t^\prime-t^u(s)){t^\prime}^{n-2}} dt^\prime \\
			&\quad -  (\Sigma_0 - 2u) \frac{t^u(s)^{n-1}}{\pi} \int_{t_0}^\infty \frac{\Im A(t^\prime)}{{t^\prime}^{n-1}} dt^\prime - \frac{t^u(s)^n}{\pi} \int_{t_0}^\infty \frac{\Im A(t^\prime)}{(t^\prime-t^u(s)){t^\prime}^{n-2}} dt^\prime \\
			&= ((\Sigma_0 - 2u - t^u(s)) t^u(s) + \Delta_{K\pi}\Delta_{\ell\pi}) \frac{t^u(s)^{n-2}}{\pi} \int_{t_0}^\infty \frac{\Im A(t^\prime)}{(t^\prime-t^u(s)){t^\prime}^{n-2}} dt^\prime + P_{n-1}^u(s) \\
			&= (t^u(s)(s-u) + \Delta_{K\pi}\Delta_{\ell\pi}) \frac{t^u(s)^{n-2}}{\pi} \int_{t_0}^\infty \frac{\Im A(t^\prime)}{(t^\prime-t^u(s)){t^\prime}^{n-2}} dt^\prime + P_{n-1}^u(s) ,
	\end{split}
\end{align}
and analogously
\begin{align}
	\begin{split}
		\frac{u^t(s)^n}{\pi} &\int_{u_0}^\infty (u^\prime(s^t(u^\prime)-t) + \Delta_{K\pi}\Delta_{\ell\pi}) \frac{\Im A(u^\prime)}{(u^\prime-u^t(s)){u^\prime}^n} du^\prime \\
			&= (u^t(s)(s-t) + \Delta_{K\pi}\Delta_{\ell\pi}) \frac{u^t(s)^{n-2}}{\pi} \int_{u_0}^\infty \frac{\Im A(u^\prime)}{(u^\prime-u^t(s)){u^\prime}^{n-2}} du^\prime + P_{n-1}^t(s) .
	\end{split}
\end{align}
The same transformations in the fixed-$s$ representation produces again only a polynomial of order $n-1$:
\begin{align}
	\begin{split}
		\frac{t^n}{\pi} &\int_{t_0}^\infty (t^\prime(s-u^s(t^\prime)) + \Delta_{K\pi}\Delta_{\ell\pi}) \frac{\Im A(t^\prime)}{(t^\prime-t){t^\prime}^n} dt^\prime \\
			&= (t(s-u^s(t)) + \Delta_{K\pi}\Delta_{\ell\pi}) \frac{t^{n-2}}{\pi} \int_{t_0}^\infty \frac{\Im A(t^\prime)}{(t^\prime-t){t^\prime}^{n-2}} dt^\prime + P_{n-1}^s(t) ,
	\end{split} \\
	\begin{split}
		\frac{u^s(t)^n}{\pi} &\int_{u_0}^\infty (u^\prime(s-t^s(u^\prime)) + \Delta_{K\pi}\Delta_{\ell\pi}) \frac{\Im A(u^\prime)}{(u^\prime-u^s(t)){u^\prime}^n} du^\prime \\
			&= (u^s(t)(s-t) + \Delta_{K\pi}\Delta_{\ell\pi}) \frac{u^s(t)^{n-2}}{\pi} \int_{u_0}^\infty \frac{\Im A(u^\prime)}{(u^\prime-u^s(t)){u^\prime}^{n-2}} du^\prime + P_{n-1}^s(t) .
	\end{split}
\end{align}
We further use:
\begin{align}
	\begin{split}
			&\frac{{t^u(s)}^n}{\pi} \int_{t_0}^\infty  \frac{\Delta_{K\pi} + \alpha t^\prime}{2t^\prime} \frac{ \Im A(t^\prime)}{(t^\prime-t^u(s) - i\epsilon) {t^\prime}^n} dt^\prime \\
			&= \frac{\Delta_{K\pi}}{2M_K^2} \frac{t^u(s)^{n-1}}{\pi} \int_{t_0}^\infty  \frac{M_K^2}{t^\prime} \frac{ \Im A(t^\prime)}{(t^\prime-t^u(s) - i\epsilon) {t^\prime}^{n-1}} dt^\prime  - \frac{\Delta_{K\pi}}{2M_K^2} \frac{t^u(s)^{n-1}}{\pi} \int_{t_0}^\infty  \frac{M_K^2}{t^\prime} \frac{ \Im A(t^\prime)}{{t^\prime}^n} dt^\prime \\
			&\quad + \frac{\alpha t^u(s)}{2M_K^2} \frac{t^u(s)^{n-1}}{\pi} \int_{t_0}^\infty  \frac{M_K^2}{t^\prime} \frac{ \Im A(t^\prime)}{(t^\prime-t^u(s) - i\epsilon) {t^\prime}^{n-1}} dt^\prime \\
			&= \frac{\Delta_{K\pi} + \alpha t^u(s)}{2M_K^2} \frac{t^u(s)^{n-1}}{\pi} \int_{t_0}^\infty  \frac{M_K^2}{t^\prime} \frac{ \Im A(t^\prime)}{(t^\prime-t^u(s) - i\epsilon) {t^\prime}^{n-1}} dt^\prime + P_{n-1}^u(s) ,
	\end{split}
\intertext{and similarly}
	\begin{split}
			&\frac{{u^t(s)}^n}{\pi} \int_{u_0}^\infty  \frac{\Delta_{K\pi} + \alpha u^\prime}{2u^\prime} \frac{ \Im A(u^\prime)}{(u^\prime-u^t(s) - i\epsilon) {u^\prime}^n} du^\prime \\
			&= \frac{\Delta_{K\pi} + \alpha u^t(s)}{2M_K^2} \frac{u^t(s)^{n-1}}{\pi} \int_{u_0}^\infty  \frac{M_K^2}{u^\prime} \frac{ \Im A(u^\prime)}{(u^\prime-u^t(s) - i\epsilon) {u^\prime}^{n-1}} du^\prime + P_1^t(s) .
	\end{split}
\end{align}

All these transformations only generate polynomials or order $n-1$, when we apply them in the dispersion relations for the form factors. These polynomials can be reabsorbed into the subtraction polynomials:
\begin{align}
	\begin{split}
		F_s^u(s) &= P_{n-1}^u(s) + \frac{s^n}{\pi} \int_{s_0}^\infty \frac{\Im f_0(s^\prime)}{(s^\prime-s - i\epsilon){s^\prime}^n} ds^\prime + \frac{u-t^u(s)}{M_K^2} \frac{s^{n-1}}{\pi} \int_{s_0}^\infty \frac{\Im f_1(s^\prime)}{(s^\prime-s - i\epsilon){s^\prime}^{n-1}} ds^\prime \\
			&\quad  - \frac{u-t^u(s)}{M_K^2} \frac{s^{n-1}}{\pi} \int_{s_0}^\infty \frac{2PL(s^\prime)M_K^2}{\lambda_{K\ell}(s^\prime)} \frac{\Im g_1(s^\prime)}{(s^\prime-s - i\epsilon){s^\prime}^{n-1}} ds^\prime \\
			&\quad  + \frac{2}{3} \frac{{t^u(s)}^n}{\pi} \int_{t_0}^\infty \frac{ \Im f_0^{(1/2)}(t^\prime)}{(t^\prime-t^u(s) - i\epsilon) {t^\prime}^n} dt^\prime \\
			&\quad  + \frac{2}{3}  \frac{t^u(s)(s-u) + \Delta_{K\pi}\Delta_{\ell\pi}}{M_K^4} \frac{t^u(s)^{n-2}}{\pi} \int_{t_0}^\infty \frac{ \Im f_1^{(1/2)}(t^\prime) }{(t^\prime-t^u(s) - i\epsilon){t^\prime}^{n-2}} dt^\prime \\
			&\quad  + \frac{2}{3}  \frac{t^u(s)(s-u) + \Delta_{K\pi}\Delta_{\ell\pi}}{M_K^4} \frac{t^u(s)^{n-2}}{\pi} \int_{t_0}^\infty  \frac{(\Delta_{\ell\pi} + t^\prime)M_K^4}{2t^\prime \lambda_{\ell\pi}(t^\prime)} \frac{ \Im g_1^{(1/2)}(t^\prime)}{(t^\prime-t^u(s) - i\epsilon){t^\prime}^{n-2}} dt^\prime \\
			&\quad  - \frac{2}{3} \frac{\Delta_{K\pi} - 3 t^u(s)}{2M_K^2} \frac{t^u(s)^{n-1}}{\pi} \int_{t_0}^\infty  \frac{M_K^2}{t^\prime} \frac{ \Im g_1^{(1/2)}(t^\prime)}{(t^\prime-t^u(s) - i\epsilon) {t^\prime}^{n-1}} dt^\prime \\
			&\quad  + \frac{1}{3} \frac{{t^u(s)}^n}{\pi} \int_{t_0}^\infty \frac{\Im f_0^{(3/2)}(t^\prime)}{(t^\prime-t^u(s) - i\epsilon) {t^\prime}^n} dt^\prime \\
			&\quad  + \frac{1}{3} \frac{t^u(s)(s-u) + \Delta_{K\pi}\Delta_{\ell\pi}}{M_K^4} \frac{t^u(s)^{n-2}}{\pi} \int_{t_0}^\infty \frac{ \Im f_1^{(3/2)}(t^\prime)}{(t^\prime-t^u(s) - i\epsilon){t^\prime}^{n-2}} dt^\prime \\
			&\quad  + \frac{1}{3} \frac{t^u(s)(s-u) + \Delta_{K\pi}\Delta_{\ell\pi}}{M_K^4} \frac{t^u(s)^{n-2}}{\pi} \int_{t_0}^\infty \frac{(\Delta_{\ell\pi} + t^\prime) M_K^4}{2t^\prime \lambda_{\ell\pi}(t^\prime)} \frac{ \Im g_1^{(3/2)}(t^\prime)}{(t^\prime-t^u(s) - i\epsilon){t^\prime}^{n-2}} dt^\prime \\
			&\quad  - \frac{1}{3} \frac{\Delta_{K\pi} - 3 t^u(s)}{2M_K^2} \frac{t^u(s)^{n-1}}{\pi} \int_{t_0}^\infty \frac{M_K^2}{t^\prime} \frac{ \Im g_1^{(3/2)}(t^\prime)}{(t^\prime-t^u(s) - i\epsilon) {t^\prime}^{n-1}} dt^\prime \\
			&\quad + \O(p^8) ,
	\end{split}
\end{align}
\begin{align}
	\begin{split}
		G_s^u(s) &= P_{n-1}^u(s) + \frac{s^n}{\pi} \int_{s_0}^\infty \frac{\Im g_1(s^\prime)}{(s^\prime-s - i\epsilon){s^\prime}^n} ds^\prime \\
			&\quad  - \frac{2}{3} \frac{{t^u(s)}^n}{\pi} \int_{t_0}^\infty \frac{ \Im f_0^{(1/2)}(t^\prime)}{(t^\prime-t^u(s) - i\epsilon) {t^\prime}^n} dt^\prime \\
			&\quad  - \frac{2}{3} \frac{t^u(s)(s-u) + \Delta_{K\pi}\Delta_{\ell\pi}}{M_K^4}  \frac{t^u(s)^{n-2}}{\pi} \int_{t_0}^\infty \frac{ \Im f_1^{(1/2)}(t^\prime) }{(t^\prime-t^u(s) - i\epsilon){t^\prime}^{n-2}} dt^\prime \\
			&\quad  - \frac{2}{3} \frac{t^u(s)(s-u) + \Delta_{K\pi}\Delta_{\ell\pi}}{M_K^4} \frac{t^u(s)^{n-2}}{\pi} \int_{t_0}^\infty \frac{(\Delta_{\ell\pi} + t^\prime)M_K^4}{2t^\prime \lambda_{\ell\pi}(t^\prime)} \frac{ \Im g_1^{(1/2)}(t^\prime)}{(t^\prime-t^u(s) - i\epsilon){t^\prime}^{n-2}} dt^\prime \\
			&\quad  + \frac{2}{3} \frac{\Delta_{K\pi} + t^u(s)}{2M_K^2} \frac{t^u(s)^{n-1}}{\pi} \int_{t_0}^\infty  \frac{M_K^2}{t^\prime} \frac{ \Im g_1^{(1/2)}(t^\prime)}{(t^\prime-t^u(s) - i\epsilon) {t^\prime}^{n-1}} dt^\prime \\
			&\quad  - \frac{1}{3} \frac{{t^u(s)}^n}{\pi} \int_{t_0}^\infty \frac{\Im f_0^{(3/2)}(t^\prime)}{(t^\prime-t^u(s) - i\epsilon) {t^\prime}^n} dt^\prime \\
			&\quad  - \frac{1}{3} \frac{t^u(s)(s-u) + \Delta_{K\pi}\Delta_{\ell\pi}}{M_K^4} \frac{t^u(s)^{n-2}}{\pi} \int_{t_0}^\infty \frac{ \Im f_1^{(3/2)}(t^\prime)}{(t^\prime-t^u(s) - i\epsilon){t^\prime}^{n-2}} dt^\prime \\
			&\quad  - \frac{1}{3} \frac{t^u(s)(s-u) + \Delta_{K\pi}\Delta_{\ell\pi}}{M_K^4} \frac{t^u(s)^{n-2}}{\pi} \int_{t_0}^\infty  \frac{(\Delta_{\ell\pi} + t^\prime) M_K^4}{2t^\prime \lambda_{\ell\pi}(t^\prime)} \frac{ \Im g_1^{(3/2)}(t^\prime)}{(t^\prime-t^u(s) - i\epsilon){t^\prime}^{n-2}} dt^\prime \\
			&\quad  + \frac{1}{3} \frac{\Delta_{K\pi} + t^u(s)}{2M_K^2} \frac{t^u(s)^{n-1}}{\pi} \int_{t_0}^\infty \frac{M_K^2}{t^\prime} \frac{ \Im g_1^{(3/2)}(t^\prime)}{(t^\prime-t^u(s) - i\epsilon) {t^\prime}^{n-1}} dt^\prime \\
			&\quad + \O(p^8) ,
	\end{split}
\end{align}

\begin{align}
	\begin{split}
		F_s^t(s) &= P_{n-1}^t(s) + \frac{s^n}{\pi} \int_{s_0}^\infty \frac{\Im f_0(s^\prime)}{(s^\prime-s - i\epsilon) {s^\prime}^n} ds^\prime + \frac{u^t(s)-t}{M_K^2} \frac{s^{n-1}}{\pi} \int_{s_0}^\infty \frac{ \Im f_1(s^\prime)}{(s^\prime-s - i\epsilon) {s^\prime}^{n-1}} ds^\prime \\
			&\quad - \frac{u^t(s)-t}{M_K^2} \frac{s^{n-1}}{\pi} \int_{s_0}^\infty \frac{2PL(s^\prime)M_K^2}{\lambda_{K\ell}(s^\prime)} \frac{ \Im g_1(s^\prime)}{(s^\prime-s - i\epsilon) {s^\prime}^{n-1}} ds^\prime \\
			&\quad+ \frac{{u^t(s)}^n}{\pi} \int_{u_0}^\infty \frac{\Im f_0^{(3/2)}(u^\prime)}{(u^\prime-u^t(s) - i\epsilon) {u^\prime}^n} du^\prime \\
			&\quad+ \frac{u^t(s)(s-t) + \Delta_{K\pi}\Delta_{\ell\pi}}{M_K^4} \frac{{u^t(s)}^{n-2}}{\pi} \int_{u_0}^\infty \frac{\Im f_1^{(3/2)}(u^\prime)}{(u^\prime-u^t(s) - i\epsilon){u^\prime}^{n-2}} du^\prime \\
			&\quad + \frac{u^t(s)(s-t) + \Delta_{K\pi}\Delta_{\ell\pi}}{M_K^4} \frac{{u^t(s)}^{n-2}}{\pi} \int_{u_0}^\infty \frac{(\Delta_{\ell\pi} + u^\prime) M_K^4}{2u^\prime \lambda_{\ell\pi}(u^\prime)} \frac{ \Im g_1^{(3/2)}(u^\prime)}{(u^\prime-u^t(s) - i\epsilon){u^\prime}^{n-2}} du^\prime \\
			&\quad- \frac{\Delta_{K\pi} - 3 u^t(s)}{2M_K^2} \frac{{u^t(s)}^{n-1}}{\pi} \int_{u_0}^\infty \frac{M_K^2}{u^\prime} \frac{ \Im g_1^{(3/2)}(u^\prime)}{(u^\prime-u^t(s) - i\epsilon) {u^\prime}^{n-1}} du^\prime \\
			&\quad+ \O(p^8) ,
	\end{split}
\end{align}
\begin{align}
	\begin{split}
		G_s^t(s) &= P_{n-1}^t(s) + \frac{s^n}{\pi} \int_{s_0}^\infty \frac{\Im g_1(s^\prime)}{(s^\prime-s - i\epsilon) {s^\prime}^n} ds^\prime \\
			&\quad+ \frac{{u^t(s)}^n}{\pi} \int_{u_0}^\infty \frac{\Im f_0^{(3/2)}(u^\prime)}{(u^\prime-u^t(s) - i\epsilon) {u^\prime}^n} du^\prime \\
			&\quad+ \frac{u^t(s)(s-t) + \Delta_{K\pi}\Delta_{\ell\pi}}{M_K^4} \frac{{u^t(s)}^{n-2}}{\pi} \int_{u_0}^\infty \frac{\Im f_1^{(3/2)}(u^\prime)}{(u^\prime-u^t(s) - i\epsilon){u^\prime}^{n-2}} du^\prime \\
			&\quad + \frac{u^t(s)(s-t) + \Delta_{K\pi}\Delta_{\ell\pi}}{M_K^4} \frac{{u^t(s)}^{n-2}}{\pi} \int_{u_0}^\infty \frac{(\Delta_{\ell\pi} + u^\prime) M_K^4}{2u^\prime \lambda_{\ell\pi}(u^\prime)} \frac{ \Im g_1^{(3/2)}(u^\prime)}{(u^\prime-u^t(s) - i\epsilon){u^\prime}^{n-2}} du^\prime \\
			&\quad- \frac{\Delta_{K\pi} + u^t(s)}{2M_K^2} \frac{{u^t(s)}^{n-1}}{\pi} \int_{u_0}^\infty \frac{M_K^2}{u^\prime} \frac{ \Im g_1^{(3/2)}(u^\prime)}{(u^\prime-u^t(s) - i\epsilon) {u^\prime}^{n-1}} du^\prime \\
			&\quad+ \O(p^8) ,
	\end{split}
\end{align}

\begin{align}
	\begin{split}
		F_t^s(t) &= P_{n-1}^s(t) + \frac{2}{3} \frac{t^n}{\pi} \int_{t_0}^\infty \frac{ \Im f_0^{(1/2)}(t^\prime)}{(t^\prime - t - i\epsilon) {t^\prime}^n} dt^\prime \\
			&\quad + \frac{2}{3} \frac{t(s-u^s(t)) + \Delta_{K\pi}\Delta_{\ell\pi}}{M_K^4} \frac{t^{n-2}}{\pi} \int_{t_0}^\infty \frac{ \Im f_1^{(1/2)}(t^\prime)}{(t^\prime - t - i\epsilon){t^\prime}^{n-2}} dt^\prime \\
			&\quad + \frac{2}{3} \frac{t(s-u^s(t)) + \Delta_{K\pi}\Delta_{\ell\pi}}{M_K^4} \frac{t^{n-2}}{\pi} \int_{t_0}^\infty \frac{(\Delta_{\ell\pi} + t^\prime) M_K^4}{2t^\prime \lambda_{\ell\pi}(t^\prime)} \frac{ \Im g_1^{(1/2)}(t^\prime)}{(t^\prime - t - i\epsilon){t^\prime}^{n-2}} dt^\prime \\
			&\quad - \frac{2}{3} \frac{\Delta_{K\pi} - 3 t}{2M_K^2} \frac{t^{n-1}}{\pi} \int_{t_0}^\infty \frac{M_K^2}{t^\prime} \frac{ \Im g_1^{(1/2)}(t^\prime)}{(t^\prime - t - i\epsilon) {t^\prime}^{n-1}} dt^\prime \\
			&\quad + \frac{1}{3} \frac{t^n}{\pi} \int_{t_0}^\infty \frac{ \Im f_0^{(3/2)}(t^\prime)}{(t^\prime - t - i\epsilon) {t^\prime}^n} dt^\prime \\
			&\quad + \frac{1}{3} \frac{t(s-u^s(t)) + \Delta_{K\pi}\Delta_{\ell\pi}}{M_K^4} \frac{t^{n-2}}{\pi} \int_{t_0}^\infty \frac{ \Im f_1^{(3/2)}(t^\prime)}{(t^\prime - t - i\epsilon){t^\prime}^{n-2}} dt^\prime \\
			&\quad + \frac{1}{3} \frac{t(s-u^s(t)) + \Delta_{K\pi}\Delta_{\ell\pi}}{M_K^4} \frac{t^{n-2}}{\pi} \int_{t_0}^\infty \frac{(\Delta_{\ell\pi} + t^\prime) M_K^4}{2t^\prime \lambda_{\ell\pi}(t^\prime)} \frac{ \Im g_1^{(3/2)}(t^\prime)}{(t^\prime - t - i\epsilon){t^\prime}^{n-2}} dt^\prime \\
			&\quad - \frac{1}{3} \frac{\Delta_{K\pi} - 3 t}{2M_K^2} \frac{t^{n-1}}{\pi} \int_{t_0}^\infty \frac{M_K^2}{t^\prime} \frac{ \Im g_1^{(3/2)}(t^\prime)}{(t^\prime - t - i\epsilon) {t^\prime}^{n-1}} dt^\prime \\
			&\quad + \frac{{u^s(t)}^n}{\pi} \int_{u_0}^\infty \frac{ \Im f_0^{(3/2)}(u^\prime)}{(u^\prime-u^s(t) - i\epsilon) {u^\prime}^n} du^\prime \\
			&\quad + \frac{u^s(t)(s-t) + \Delta_{K\pi}\Delta_{\ell\pi}}{M_K^4} \frac{{u^s(t)}^{n-2}}{\pi} \int_{u_0}^\infty \frac{ \Im f_1^{(3/2)}(u^\prime)}{(u^\prime-u^s(t) - i\epsilon){u^\prime}^{n-2}} du^\prime \\
			&\quad + \frac{u^s(t)(s-t) + \Delta_{K\pi}\Delta_{\ell\pi}}{M_K^4}  \frac{{u^s(t)}^{n-2}}{\pi} \int_{u_0}^\infty \frac{(\Delta_{\ell\pi} + u^\prime) M_K^4}{2u^\prime \lambda_{\ell\pi}(u^\prime)} \frac{ \Im g_1^{(3/2)}(u^\prime)}{(u^\prime-u^s(t) - i\epsilon){u^\prime}^{n-2}} du^\prime \\
			&\quad -  \frac{\Delta_{K\pi} - 3 u^s(t)}{2M_K^2} \frac{{u^s(t)}^{n-1}}{\pi} \int_{u_0}^\infty \frac{M_K^2}{u^\prime} \frac{ \Im g_1^{(3/2)}(u^\prime)}{(u^\prime-u^s(t) - i\epsilon) {u^\prime}^{n-1}} du^\prime \\
			&\quad+ \O(p^8) ,
	\end{split}
\end{align}
\begin{align}
	\begin{split}
		G_t^s(t) &= P_{n-1}^s(t) - \frac{2}{3} \frac{t^n}{\pi} \int_{t_0}^\infty \frac{ \Im f_0^{(1/2)}(t^\prime)}{(t^\prime - t - i\epsilon) {t^\prime}^n} dt^\prime \\
			&\quad - \frac{2}{3} \frac{t(s-u^s(t)) + \Delta_{K\pi}\Delta_{\ell\pi}}{M_K^4} \frac{t^{n-2}}{\pi} \int_{t_0}^\infty \frac{ \Im f_1^{(1/2)}(t^\prime)}{(t^\prime - t - i\epsilon){t^\prime}^{n-2}} dt^\prime \\
			&\quad - \frac{2}{3} \frac{t(s-u^s(t)) + \Delta_{K\pi}\Delta_{\ell\pi}}{M_K^4} \frac{t^{n-2}}{\pi} \int_{t_0}^\infty \frac{(\Delta_{\ell\pi} + t^\prime) M_K^4}{2t^\prime \lambda_{\ell\pi}(t^\prime)} \frac{ \Im g_1^{(1/2)}(t^\prime)}{(t^\prime - t - i\epsilon){t^\prime}^{n-2}} dt^\prime \\
			&\quad + \frac{2}{3} \frac{\Delta_{K\pi} + t}{2M_K^2} \frac{t^{n-1}}{\pi} \int_{t_0}^\infty \frac{M_K^2}{t^\prime} \frac{ \Im g_1^{(1/2)}(t^\prime)}{(t^\prime - t - i\epsilon) {t^\prime}^{n-1}} dt^\prime \\
			&\quad - \frac{1}{3} \frac{t^n}{\pi} \int_{t_0}^\infty \frac{ \Im f_0^{(3/2)}(t^\prime)}{(t^\prime - t - i\epsilon) {t^\prime}^n} dt^\prime \\
			&\quad - \frac{1}{3} \frac{t(s-u^s(t)) + \Delta_{K\pi}\Delta_{\ell\pi}}{M_K^4} \frac{t^{n-2}}{\pi} \int_{t_0}^\infty \frac{ \Im f_1^{(3/2)}(t^\prime)}{(t^\prime - t - i\epsilon){t^\prime}^{n-2}} dt^\prime \\
			&\quad - \frac{1}{3} \frac{t(s-u^s(t)) + \Delta_{K\pi}\Delta_{\ell\pi}}{M_K^4} \frac{t^{n-2}}{\pi} \int_{t_0}^\infty \frac{(\Delta_{\ell\pi} + t^\prime) M_K^4}{2t^\prime \lambda_{\ell\pi}(t^\prime)} \frac{ \Im g_1^{(3/2)}(t^\prime)}{(t^\prime - t - i\epsilon){t^\prime}^{n-2}} dt^\prime \\
			&\quad + \frac{1}{3} \frac{\Delta_{K\pi} + t}{2M_K^2} \frac{t^{n-1}}{\pi} \int_{t_0}^\infty \frac{M_K^2}{t^\prime} \frac{ \Im g_1^{(3/2)}(t^\prime)}{(t^\prime - t - i\epsilon) {t^\prime}^{n-1}} dt^\prime \\
			&\quad + \frac{{u^s(t)}^n}{\pi} \int_{u_0}^\infty \frac{ \Im f_0^{(3/2)}(u^\prime)}{(u^\prime-u^s(t) - i\epsilon) {u^\prime}^n} du^\prime \\
			&\quad + \frac{u^s(t)(s-t) + \Delta_{K\pi}\Delta_{\ell\pi}}{M_K^4} \frac{{u^s(t)}^{n-2}}{\pi} \int_{u_0}^\infty \frac{ \Im f_1^{(3/2)}(u^\prime)}{(u^\prime-u^s(t) - i\epsilon){u^\prime}^{n-2}} du^\prime \\
			&\quad + \frac{u^s(t)(s-t) + \Delta_{K\pi}\Delta_{\ell\pi}}{M_K^4} \frac{{u^s(t)}^{n-2}}{\pi} \int_{u_0}^\infty \frac{(\Delta_{\ell\pi} + u^\prime) M_K^4}{2u^\prime \lambda_{\ell\pi}(u^\prime)} \frac{ \Im g_1^{(3/2)}(u^\prime)}{(u^\prime-u^s(t) - i\epsilon){u^\prime}^{n-2}} du^\prime \\
			&\quad - \frac{ \Delta_{K\pi} + u^s(t)}{2M_K^2} \frac{{u^s(t)}^{n-1}}{\pi} \int_{u_0}^\infty \frac{M_K^2}{u^\prime} \frac{ \Im g_1^{(3/2)}(u^\prime)}{(u^\prime-u^s(t) - i\epsilon) {u^\prime}^{n-1}} du^\prime \\
			&\quad+ \O(p^8) .
	\end{split}
\end{align}

We have now found three representations of the form factors, where in each one of the Mandelstam variables $s$, $t$ and $u$ is fixed. We would like to have a representation of the form factors that is valid in the whole Mandelstam plane, i.e.~we have to match and continue analytically the three representations. For this purpose, we note that all the integrals appear in two of the three representations and can be understood as a part of the polynomial in the third representation. We can split these integrals off the polynomial such that the three representations agree except for the remaining polynomial parts.

Explicitly, in the fixed-$u$ representation, we separate from the polynomial the integrals that appear in the fixed-$t$ and fixed-$s$ representations, and which are constant or linear terms in $s$ with $u$-dependent coefficients (hence, this procedure is valid for $n\ge 2$):
\begin{align}
	\begin{split}
		F_s^u(s) &= P_{n-1}^u(s) + \frac{s^n}{\pi} \int_{s_0}^\infty \frac{\Im f_0(s^\prime)}{(s^\prime-s - i\epsilon){s^\prime}^n} ds^\prime + \frac{u-t^u(s)}{M_K^2} \frac{s^{n-1}}{\pi} \int_{s_0}^\infty \frac{\Im f_1(s^\prime)}{(s^\prime-s - i\epsilon){s^\prime}^{n-1}} ds^\prime \\
			&\quad  - \frac{u-t^u(s)}{M_K^2} \frac{s^{n-1}}{\pi} \int_{s_0}^\infty \frac{2PL(s^\prime)M_K^2}{\lambda_{K\ell}(s^\prime)} \frac{\Im g_1(s^\prime)}{(s^\prime-s - i\epsilon){s^\prime}^{n-1}} ds^\prime \\
			&\quad  + \frac{2}{3} \frac{{t^u(s)}^n}{\pi} \int_{t_0}^\infty \frac{ \Im f_0^{(1/2)}(t^\prime)}{(t^\prime-t^u(s) - i\epsilon) {t^\prime}^n} dt^\prime \\
			&\quad  + \frac{2}{3}  \frac{t^u(s)(s-u) + \Delta_{K\pi}\Delta_{\ell\pi}}{M_K^4} \frac{{t^u(s)}^{n-2}}{\pi} \int_{t_0}^\infty \frac{ \Im f_1^{(1/2)}(t^\prime) }{(t^\prime-t^u(s) - i\epsilon){t^\prime}^{n-2}} dt^\prime \\
			&\quad  + \frac{2}{3}  \frac{t^u(s)(s-u) + \Delta_{K\pi}\Delta_{\ell\pi}}{M_K^4} \frac{{t^u(s)}^{n-2}}{\pi} \int_{t_0}^\infty  \frac{(\Delta_{\ell\pi} + t^\prime)M_K^4}{2t^\prime \lambda_{\ell\pi}(t^\prime)} \frac{ \Im g_1^{(1/2)}(t^\prime)}{(t^\prime-t^u(s) - i\epsilon){t^\prime}^{n-2}} dt^\prime \\
			&\quad  - \frac{2}{3}  \frac{\Delta_{K\pi} - 3 t^u(s)}{2M_K^2} \frac{{t^u(s)}^{n-1}}{\pi} \int_{t_0}^\infty  \frac{M_K^2}{t^\prime} \frac{ \Im g_1^{(1/2)}(t^\prime)}{(t^\prime-t^u(s) - i\epsilon) {t^\prime}^{n-1}} dt^\prime \\
			&\quad  + \frac{1}{3} \frac{{t^u(s)}^n}{\pi} \int_{t_0}^\infty \frac{\Im f_0^{(3/2)}(t^\prime)}{(t^\prime-t^u(s) - i\epsilon) {t^\prime}^n} dt^\prime \\
			&\quad  + \frac{1}{3} \frac{t^u(s)(s-u) + \Delta_{K\pi}\Delta_{\ell\pi}}{M_K^4} \frac{{t^u(s)}^{n-2}}{\pi} \int_{t_0}^\infty \frac{ \Im f_1^{(3/2)}(t^\prime)}{(t^\prime-t^u(s) - i\epsilon){t^\prime}^{n-2}} dt^\prime \\
			&\quad  + \frac{1}{3} \frac{t^u(s)(s-u) + \Delta_{K\pi}\Delta_{\ell\pi}}{M_K^4} \frac{{t^u(s)}^{n-2}}{\pi} \int_{t_0}^\infty \frac{(\Delta_{\ell\pi} + t^\prime) M_K^4}{2t^\prime \lambda_{\ell\pi}(t^\prime)} \frac{ \Im g_1^{(3/2)}(t^\prime)}{(t^\prime-t^u(s) - i\epsilon){t^\prime}^{n-2}} dt^\prime \\
			&\quad  - \frac{1}{3} \frac{\Delta_{K\pi} - 3 t^u(s)}{2M_K^2} \frac{{t^u(s)}^{n-1}}{\pi} \int_{t_0}^\infty \frac{M_K^2}{t^\prime} \frac{ \Im g_1^{(3/2)}(t^\prime)}{(t^\prime-t^u(s) - i\epsilon) {t^\prime}^{n-1}} dt^\prime \\
			&\quad + \frac{u^n}{\pi} \int_{u_0}^\infty \frac{\Im f_0^{(3/2)}(u^\prime)}{(u^\prime-u - i\epsilon) {u^\prime}^n} du^\prime \\
			&\quad + \frac{u(s-t^u(s)) + \Delta_{K\pi}\Delta_{\ell\pi}}{M_K^4} \frac{u^{n-2}}{\pi} \int_{u_0}^\infty \frac{\Im f_1^{(3/2)}(u^\prime)}{(u^\prime-u - i\epsilon){u^\prime}^{n-2}} du^\prime \\
			&\quad + \frac{u(s-t^u(s)) + \Delta_{K\pi}\Delta_{\ell\pi}}{M_K^4} \frac{u^{n-2}}{\pi} \int_{u_0}^\infty \frac{(\Delta_{\ell\pi} + u^\prime) M_K^4}{2u^\prime \lambda_{\ell\pi}(u^\prime)} \frac{ \Im g_1^{(3/2)}(u^\prime)}{(u^\prime-u - i\epsilon){u^\prime}^{n-2}} du^\prime \\
			&\quad -  \frac{\Delta_{K\pi} - 3 u}{2M_K^2} \frac{u^{n-1}}{\pi} \int_{u_0}^\infty \frac{M_K^2}{u^\prime} \frac{ \Im g_1^{(3/2)}(u^\prime)}{(u^\prime-u - i\epsilon) {u^\prime}^{n-1}} du^\prime \\
			&\quad+ \O(p^8) ,
	\end{split}
\end{align}
\begin{align}
	\begin{split}
		G_s^u(s) &= P_{n-1}^u(s) + \frac{s^n}{\pi} \int_{s_0}^\infty \frac{\Im g_1(s^\prime)}{(s^\prime-s - i\epsilon){s^\prime}^n} ds^\prime \\
			&\quad  - \frac{2}{3} \frac{{t^u(s)}^n}{\pi} \int_{t_0}^\infty \frac{ \Im f_0^{(1/2)}(t^\prime)}{(t^\prime-t^u(s) - i\epsilon) {t^\prime}^n} dt^\prime \\
			&\quad  - \frac{2}{3} \frac{t^u(s)(s-u) + \Delta_{K\pi}\Delta_{\ell\pi}}{M_K^4}  \frac{{t^u(s)}^{n-2}}{\pi} \int_{t_0}^\infty \frac{ \Im f_1^{(1/2)}(t^\prime) }{(t^\prime-t^u(s) - i\epsilon){t^\prime}^{n-2}} dt^\prime \\
			&\quad  - \frac{2}{3} \frac{t^u(s)(s-u) + \Delta_{K\pi}\Delta_{\ell\pi}}{M_K^4} \frac{{t^u(s)}^{n-2}}{\pi} \int_{t_0}^\infty \frac{(\Delta_{\ell\pi} + t^\prime)M_K^4}{2t^\prime \lambda_{\ell\pi}(t^\prime)} \frac{ \Im g_1^{(1/2)}(t^\prime)}{(t^\prime-t^u(s) - i\epsilon){t^\prime}^{n-2}} dt^\prime \\
			&\quad  + \frac{2}{3} \frac{\Delta_{K\pi} + t^u(s)}{2M_K^2} \frac{{t^u(s)}^{n-1}}{\pi} \int_{t_0}^\infty \frac{M_K^2}{t^\prime} \frac{ \Im g_1^{(1/2)}(t^\prime)}{(t^\prime-t^u(s) - i\epsilon) {t^\prime}^{n-1}} dt^\prime \\
			&\quad  - \frac{1}{3} \frac{{t^u(s)}^n}{\pi} \int_{t_0}^\infty \frac{\Im f_0^{(3/2)}(t^\prime)}{(t^\prime-t^u(s) - i\epsilon) {t^\prime}^n} dt^\prime \\
			&\quad  - \frac{1}{3} \frac{t^u(s)(s-u) + \Delta_{K\pi}\Delta_{\ell\pi}}{M_K^4} \frac{{t^u(s)}^{n-2}}{\pi} \int_{t_0}^\infty \frac{ \Im f_1^{(3/2)}(t^\prime)}{(t^\prime-t^u(s) - i\epsilon){t^\prime}^{n-2}} dt^\prime \\
			&\quad  - \frac{1}{3} \frac{t^u(s)(s-u) + \Delta_{K\pi}\Delta_{\ell\pi}}{M_K^4} \frac{{t^u(s)}^{n-2}}{\pi} \int_{t_0}^\infty  \frac{(\Delta_{\ell\pi} + t^\prime) M_K^4}{2t^\prime \lambda_{\ell\pi}(t^\prime)} \frac{ \Im g_1^{(3/2)}(t^\prime)}{(t^\prime-t^u(s) - i\epsilon){t^\prime}^{n-2}} dt^\prime \\
			&\quad  + \frac{1}{3} \frac{\Delta_{K\pi} + t^u(s)}{2M_K^2} \frac{{t^u(s)}^{n-1}}{\pi} \int_{t_0}^\infty \frac{M_K^2}{t^\prime} \frac{ \Im g_1^{(3/2)}(t^\prime)}{(t^\prime-t^u(s) - i\epsilon) {t^\prime}^{n-1}} dt^\prime \\
			&\quad + \frac{u^n}{\pi} \int_{u_0}^\infty \frac{\Im f_0^{(3/2)}(u^\prime)}{(u^\prime-u - i\epsilon) {u^\prime}^n} du^\prime \\
			&\quad + \frac{u(s-t^u(s)) + \Delta_{K\pi}\Delta_{\ell\pi}}{M_K^4} \frac{u^{n-2}}{\pi} \int_{u_0}^\infty \frac{\Im f_1^{(3/2)}(u^\prime)}{(u^\prime-u - i\epsilon){u^\prime}^{n-2}} du^\prime \\
			&\quad + \frac{u(s-t^u(s)) + \Delta_{K\pi}\Delta_{\ell\pi}}{M_K^4} \frac{u^{n-2}}{\pi} \int_{u_0}^\infty \frac{(\Delta_{\ell\pi} + u^\prime) M_K^4}{2u^\prime \lambda_{\ell\pi}(u^\prime)} \frac{ \Im g_1^{(3/2)}(u^\prime)}{(u^\prime-u - i\epsilon){u^\prime}^{n-2}} du^\prime \\
			&\quad-  \frac{\Delta_{K\pi} + u}{2M_K^2} \frac{u^{n-1}}{\pi} \int_{u_0}^\infty \frac{M_K^2}{u^\prime} \frac{ \Im g_1^{(3/2)}(u^\prime)}{(u^\prime-u - i\epsilon) {u^\prime}^{n-1}} du^\prime \\
			&\quad+ \O(p^8) .
	\end{split}
\end{align}

Similarly, in the fixed-$t$ representation, we separate from the polynomial the integrals appearing in the fixed-$s$ and fixed-$u$ representations, which again are only constant and linear in $s$ but depend on $t$:
\begin{align}
	\begin{split}
		F_s^t(s) &= P_{n-1}^t(s) + \frac{s^n}{\pi} \int_{s_0}^\infty \frac{\Im f_0(s^\prime)}{(s^\prime-s - i\epsilon) {s^\prime}^n} ds^\prime + \frac{u^t(s)-t}{M_K^2} \frac{s^{n-1}}{\pi} \int_{s_0}^\infty \frac{ \Im f_1(s^\prime)}{(s^\prime-s - i\epsilon) {s^\prime}^{n-1}} ds^\prime \\
			&\quad - \frac{u^t(s)-t}{M_K^2} \frac{s^{n-1}}{\pi} \int_{s_0}^\infty \frac{2PL(s^\prime)M_K^2}{\lambda_{K\ell}(s^\prime)} \frac{ \Im g_1(s^\prime)}{(s^\prime-s - i\epsilon) {s^\prime}^{n-1}} ds^\prime \\
			&\quad  + \frac{2}{3} \frac{t^n}{\pi} \int_{t_0}^\infty \frac{ \Im f_0^{(1/2)}(t^\prime)}{(t^\prime-t - i\epsilon) {t^\prime}^n} dt^\prime \\
			&\quad  + \frac{2}{3}  \frac{t(s-u^t(s)) + \Delta_{K\pi}\Delta_{\ell\pi}}{M_K^4} \frac{t^{n-2}}{\pi} \int_{t_0}^\infty \frac{ \Im f_1^{(1/2)}(t^\prime) }{(t^\prime-t - i\epsilon){t^\prime}^{n-2}} dt^\prime \\
			&\quad  + \frac{2}{3}  \frac{t(s-u^t(s)) + \Delta_{K\pi}\Delta_{\ell\pi}}{M_K^4} \frac{t^{n-2}}{\pi} \int_{t_0}^\infty  \frac{(\Delta_{\ell\pi} + t^\prime)M_K^4}{2t^\prime \lambda_{\ell\pi}(t^\prime)} \frac{ \Im g_1^{(1/2)}(t^\prime)}{(t^\prime-t - i\epsilon){t^\prime}^{n-2}} dt^\prime \\
			&\quad  - \frac{2}{3} \frac{\Delta_{K\pi} - 3 t}{2M_K^2} \frac{t^{n-1}}{\pi} \int_{t_0}^\infty  \frac{M_K^2}{t^\prime} \frac{ \Im g_1^{(1/2)}(t^\prime)}{(t^\prime-t - i\epsilon) {t^\prime}^{n-1}} dt^\prime \\
			&\quad  + \frac{1}{3} \frac{t^n}{\pi} \int_{t_0}^\infty \frac{\Im f_0^{(3/2)}(t^\prime)}{(t^\prime-t - i\epsilon) {t^\prime}^n} dt^\prime \\
			&\quad  + \frac{1}{3} \frac{t(s-u^t(s)) + \Delta_{K\pi}\Delta_{\ell\pi}}{M_K^4} \frac{t^{n-2}}{\pi} \int_{t_0}^\infty \frac{ \Im f_1^{(3/2)}(t^\prime)}{(t^\prime-t - i\epsilon){t^\prime}^{n-2}} dt^\prime \\
			&\quad  + \frac{1}{3} \frac{t(s-u^t(s)) + \Delta_{K\pi}\Delta_{\ell\pi}}{M_K^4} \frac{t^{n-2}}{\pi} \int_{t_0}^\infty \frac{(\Delta_{\ell\pi} + t^\prime) M_K^4}{2t^\prime \lambda_{\ell\pi}(t^\prime)} \frac{ \Im g_1^{(3/2)}(t^\prime)}{(t^\prime-t - i\epsilon){t^\prime}^{n-2}} dt^\prime \\
			&\quad  - \frac{1}{3} \frac{\Delta_{K\pi} - 3 t}{2M_K^2} \frac{t^{n-1}}{\pi} \int_{t_0}^\infty \frac{M_K^2}{t^\prime} \frac{ \Im g_1^{(3/2)}(t^\prime)}{(t^\prime-t - i\epsilon) {t^\prime}^{n-1}} dt^\prime \\
			&\quad + \frac{{u^t(s)}^n}{\pi} \int_{u_0}^\infty \frac{\Im f_0^{(3/2)}(u^\prime)}{(u^\prime-u^t(s) - i\epsilon) {u^\prime}^n} du^\prime \\
			&\quad + \frac{u^t(s)(s-t) + \Delta_{K\pi}\Delta_{\ell\pi}}{M_K^4} \frac{{u^t(s)}^{n-2}}{\pi} \int_{u_0}^\infty \frac{\Im f_1^{(3/2)}(u^\prime)}{(u^\prime-u^t(s) - i\epsilon){u^\prime}^{n-2}} du^\prime \\
			&\quad + \frac{u^t(s)(s-t) + \Delta_{K\pi}\Delta_{\ell\pi}}{M_K^4} \frac{{u^t(s)}^{n-2}}{\pi} \int_{u_0}^\infty \frac{(\Delta_{\ell\pi} + u^\prime) M_K^4}{2u^\prime \lambda_{\ell\pi}(u^\prime)} \frac{ \Im g_1^{(3/2)}(u^\prime)}{(u^\prime-u^t(s) - i\epsilon){u^\prime}^{n-2}} du^\prime \\
			&\quad- \frac{\Delta_{K\pi} - 3 u^t(s)}{2M_K^2} \frac{{u^t(s)}^{n-1}}{\pi} \int_{u_0}^\infty \frac{M_K^2}{u^\prime} \frac{ \Im g_1^{(3/2)}(u^\prime)}{(u^\prime-u^t(s) - i\epsilon) {u^\prime}^{n-1}} du^\prime \\
			&\quad+ \O(p^8) ,
	\end{split}
\end{align}
\begin{align}
	\begin{split}
		G_s^t(s) &= P_{n-1}^t(s) + \frac{s^n}{\pi} \int_{s_0}^\infty \frac{\Im g_1(s^\prime)}{(s^\prime-s - i\epsilon) {s^\prime}^n} ds^\prime \\
			&\quad  - \frac{2}{3} \frac{t^n}{\pi} \int_{t_0}^\infty \frac{ \Im f_0^{(1/2)}(t^\prime)}{(t^\prime-t - i\epsilon) {t^\prime}^n} dt^\prime \\
			&\quad  - \frac{2}{3} \frac{t(s-u^t(s)) + \Delta_{K\pi}\Delta_{\ell\pi}}{M_K^4}  \frac{t^{n-2}}{\pi} \int_{t_0}^\infty \frac{ \Im f_1^{(1/2)}(t^\prime) }{(t^\prime-t - i\epsilon){t^\prime}^{n-2}} dt^\prime \\
			&\quad  - \frac{2}{3} \frac{t(s-u^t(s)) + \Delta_{K\pi}\Delta_{\ell\pi}}{M_K^4} \frac{t^{n-2}}{\pi} \int_{t_0}^\infty \frac{(\Delta_{\ell\pi} + t^\prime)M_K^4}{2t^\prime \lambda_{\ell\pi}(t^\prime)} \frac{ \Im g_1^{(1/2)}(t^\prime)}{(t^\prime-t - i\epsilon){t^\prime}^{n-2}} dt^\prime \\
			&\quad  + \frac{2}{3} \frac{\Delta_{K\pi} + t}{2M_K^2} \frac{t^{n-1}}{\pi} \int_{t_0}^\infty  \frac{M_K^2}{t^\prime} \frac{ \Im g_1^{(1/2)}(t^\prime)}{(t^\prime-t - i\epsilon) {t^\prime}^{n-1}} dt^\prime \\
			&\quad  - \frac{1}{3} \frac{t^n}{\pi} \int_{t_0}^\infty \frac{\Im f_0^{(3/2)}(t^\prime)}{(t^\prime-t - i\epsilon) {t^\prime}^n} dt^\prime \\
			&\quad  - \frac{1}{3} \frac{t(s-u^t(s)) + \Delta_{K\pi}\Delta_{\ell\pi}}{M_K^4} \frac{t^{n-2}}{\pi} \int_{t_0}^\infty \frac{ \Im f_1^{(3/2)}(t^\prime)}{(t^\prime-t - i\epsilon){t^\prime}^{n-2}} dt^\prime \\
			&\quad  - \frac{1}{3} \frac{t(s-u^t(s)) + \Delta_{K\pi}\Delta_{\ell\pi}}{M_K^4} \frac{t^{n-2}}{\pi} \int_{t_0}^\infty  \frac{(\Delta_{\ell\pi} + t^\prime) M_K^4}{2t^\prime \lambda_{\ell\pi}(t^\prime)} \frac{ \Im g_1^{(3/2)}(t^\prime)}{(t^\prime-t - i\epsilon){t^\prime}^{n-2}} dt^\prime \\
			&\quad  + \frac{1}{3} \frac{\Delta_{K\pi} + t}{2M_K^2} \frac{t^{n-1}}{\pi} \int_{t_0}^\infty \frac{M_K^2}{t^\prime} \frac{ \Im g_1^{(3/2)}(t^\prime)}{(t^\prime-t - i\epsilon) {t^\prime}^{n-1}} dt^\prime \\
			&\quad + \frac{{u^t(s)}^n}{\pi} \int_{u_0}^\infty \frac{\Im f_0^{(3/2)}(u^\prime)}{(u^\prime-u^t(s) - i\epsilon) {u^\prime}^n} du^\prime \\
			&\quad+ \frac{u^t(s)(s-t) + \Delta_{K\pi}\Delta_{\ell\pi}}{M_K^4} \frac{{u^t(s)}^{n-2}}{\pi} \int_{u_0}^\infty \frac{\Im f_1^{(3/2)}(u^\prime)}{(u^\prime-u^t(s) - i\epsilon){u^\prime}^{n-2}} du^\prime \\
			&\quad + \frac{u^t(s)(s-t) + \Delta_{K\pi}\Delta_{\ell\pi}}{M_K^4} \frac{{u^t(s)}^{n-2}}{\pi} \int_{u_0}^\infty \frac{(\Delta_{\ell\pi} + u^\prime) M_K^4}{2u^\prime \lambda_{\ell\pi}(u^\prime)} \frac{ \Im g_1^{(3/2)}(u^\prime)}{(u^\prime-u^t(s) - i\epsilon){u^\prime}^{n-2}} du^\prime \\
			&\quad-  \frac{\Delta_{K\pi} + u^t(s)}{2M_K^2} \frac{{u^t(s)}^{n-1}}{\pi} \int_{u_0}^\infty \frac{M_K^2}{u^\prime} \frac{ \Im g_1^{(3/2)}(u^\prime)}{(u^\prime-u^t(s) - i\epsilon) {u^\prime}^{n-1}} du^\prime \\
			&\quad+ \O(p^8) .
	\end{split}
\end{align}

Finally, in the fixed-$s$ representation, we separate from the polynomial the integrals appearing in the fixed-$t$ and fixed-$u$ representations, which are indeed constant and linear in $t$ but with arbitrary $s$-dependence:
\begin{align}
	\begin{split}
		F_t^s(t) &= P_{n-1}^s(t) + \frac{s^n}{\pi} \int_{s_0}^\infty \frac{\Im f_0(s^\prime)}{(s^\prime-s - i\epsilon) {s^\prime}^n} ds^\prime + \frac{u^s(t)-t}{M_K^2} \frac{s^{n-1}}{\pi} \int_{s_0}^\infty \frac{ \Im f_1(s^\prime)}{(s^\prime-s - i\epsilon) {s^\prime}^{n-1}} ds^\prime \\
			&\quad - \frac{u^s(t)-t}{M_K^2} \frac{s^{n-1}}{\pi} \int_{s_0}^\infty \frac{2PL(s^\prime)M_K^2}{\lambda_{K\ell}(s^\prime)} \frac{ \Im g_1(s^\prime)}{(s^\prime-s - i\epsilon) {s^\prime}^{n-1}} ds^\prime \\
			&\quad + \frac{2}{3} \frac{t^n}{\pi} \int_{t_0}^\infty \frac{ \Im f_0^{(1/2)}(t^\prime)}{(t^\prime - t - i\epsilon) {t^\prime}^n} dt^\prime \\
			&\quad + \frac{2}{3} \frac{t(s-u^s(t)) + \Delta_{K\pi}\Delta_{\ell\pi}}{M_K^4} \frac{t^{n-2}}{\pi} \int_{t_0}^\infty \frac{ \Im f_1^{(1/2)}(t^\prime)}{(t^\prime - t - i\epsilon){t^\prime}^{n-2}} dt^\prime \\
			&\quad + \frac{2}{3} \frac{t(s-u^s(t)) + \Delta_{K\pi}\Delta_{\ell\pi}}{M_K^4} \frac{t^{n-2}}{\pi} \int_{t_0}^\infty \frac{(\Delta_{\ell\pi} + t^\prime) M_K^4}{2t^\prime \lambda_{\ell\pi}(t^\prime)} \frac{ \Im g_1^{(1/2)}(t^\prime)}{(t^\prime - t - i\epsilon){t^\prime}^{n-2}} dt^\prime \\
			&\quad - \frac{2}{3} \frac{\Delta_{K\pi} - 3 t}{2M_K^2} \frac{t^{n-1}}{\pi} \int_{t_0}^\infty \frac{M_K^2}{t^\prime} \frac{ \Im g_1^{(1/2)}(t^\prime)}{(t^\prime - t - i\epsilon) {t^\prime}^{n-1}} dt^\prime \\
			&\quad + \frac{1}{3} \frac{t^n}{\pi} \int_{t_0}^\infty \frac{ \Im f_0^{(3/2)}(t^\prime)}{(t^\prime - t - i\epsilon) {t^\prime}^n} dt^\prime \\
			&\quad + \frac{1}{3} \frac{t(s-u^s(t)) + \Delta_{K\pi}\Delta_{\ell\pi}}{M_K^4} \frac{t^{n-2}}{\pi} \int_{t_0}^\infty \frac{ \Im f_1^{(3/2)}(t^\prime)}{(t^\prime - t - i\epsilon){t^\prime}^{n-2}} dt^\prime \\
			&\quad + \frac{1}{3} \frac{t(s-u^s(t)) + \Delta_{K\pi}\Delta_{\ell\pi}}{M_K^4} \frac{t^{n-2}}{\pi} \int_{t_0}^\infty \frac{(\Delta_{\ell\pi} + t^\prime) M_K^4}{2t^\prime \lambda_{\ell\pi}(t^\prime)} \frac{ \Im g_1^{(3/2)}(t^\prime)}{(t^\prime - t - i\epsilon){t^\prime}^{n-2}} dt^\prime \\
			&\quad - \frac{1}{3} \frac{\Delta_{K\pi} - 3 t}{2M_K^2} \frac{t^{n-1}}{\pi} \int_{t_0}^\infty \frac{M_K^2}{t^\prime} \frac{ \Im g_1^{(3/2)}(t^\prime)}{(t^\prime - t - i\epsilon) {t^\prime}^{n-1}} dt^\prime \\
			&\quad + \frac{{u^s(t)}^n}{\pi} \int_{u_0}^\infty \frac{ \Im f_0^{(3/2)}(u^\prime)}{(u^\prime-u^s(t) - i\epsilon) {u^\prime}^n} du^\prime \\
			&\quad + \frac{u^s(t)(s-t) + \Delta_{K\pi}\Delta_{\ell\pi}}{M_K^4} \frac{{u^s(t)}^{n-2}}{\pi} \int_{u_0}^\infty \frac{ \Im f_1^{(3/2)}(u^\prime)}{(u^\prime-u^s(t) - i\epsilon){u^\prime}^{n-2}} du^\prime \\
			&\quad + \frac{u^s(t)(s-t) + \Delta_{K\pi}\Delta_{\ell\pi}}{M_K^4}  \frac{{u^s(t)}^{n-2}}{\pi} \int_{u_0}^\infty \frac{(\Delta_{\ell\pi} + u^\prime) M_K^4}{2u^\prime \lambda_{\ell\pi}(u^\prime)} \frac{ \Im g_1^{(3/2)}(u^\prime)}{(u^\prime-u^s(t) - i\epsilon){u^\prime}^{n-2}} du^\prime \\
			&\quad - \frac{\Delta_{K\pi} - 3 u^s(t)}{2M_K^2} \frac{{u^s(t)}^{n-1}}{\pi} \int_{u_0}^\infty \frac{M_K^2}{u^\prime} \frac{ \Im g_1^{(3/2)}(u^\prime)}{(u^\prime-u^s(t) - i\epsilon) {u^\prime}^{n-1}} du^\prime \\
			&\quad+ \O(p^8) ,
	\end{split}
\end{align}
\begin{align}
	\begin{split}
		G_t^s(t) &= P_{n-1}^s(t) + \frac{s^n}{\pi} \int_{s_0}^\infty \frac{\Im g_1(s^\prime)}{(s^\prime-s - i\epsilon) {s^\prime}^n} ds^\prime \\
			&\quad - \frac{2}{3} \frac{t^n}{\pi} \int_{t_0}^\infty \frac{ \Im f_0^{(1/2)}(t^\prime)}{(t^\prime - t - i\epsilon) {t^\prime}^n} dt^\prime \\
			&\quad - \frac{2}{3} \frac{t(s-u^s(t)) + \Delta_{K\pi}\Delta_{\ell\pi}}{M_K^4} \frac{t^{n-2}}{\pi} \int_{t_0}^\infty \frac{ \Im f_1^{(1/2)}(t^\prime)}{(t^\prime - t - i\epsilon){t^\prime}^{n-2}} dt^\prime \\
			&\quad - \frac{2}{3} \frac{t(s-u^s(t)) + \Delta_{K\pi}\Delta_{\ell\pi}}{M_K^4} \frac{t^{n-2}}{\pi} \int_{t_0}^\infty \frac{(\Delta_{\ell\pi} + t^\prime) M_K^4}{2t^\prime \lambda_{\ell\pi}(t^\prime)} \frac{ \Im g_1^{(1/2)}(t^\prime)}{(t^\prime - t - i\epsilon){t^\prime}^{n-2}} dt^\prime \\
			&\quad + \frac{2}{3} \frac{\Delta_{K\pi} + t}{2M_K^2} \frac{t^{n-1}}{\pi} \int_{t_0}^\infty \frac{M_K^2}{t^\prime} \frac{ \Im g_1^{(1/2)}(t^\prime)}{(t^\prime - t - i\epsilon) {t^\prime}^{n-1}} dt^\prime \\
			&\quad - \frac{1}{3} \frac{t^n}{\pi} \int_{t_0}^\infty \frac{ \Im f_0^{(3/2)}(t^\prime)}{(t^\prime - t - i\epsilon) {t^\prime}^n} dt^\prime \\
			&\quad - \frac{1}{3} \frac{t(s-u^s(t)) + \Delta_{K\pi}\Delta_{\ell\pi}}{M_K^4} \frac{t^{n-2}}{\pi} \int_{t_0}^\infty \frac{ \Im f_1^{(3/2)}(t^\prime)}{(t^\prime - t - i\epsilon){t^\prime}^{n-2}} dt^\prime \\
			&\quad - \frac{1}{3} \frac{t(s-u^s(t)) + \Delta_{K\pi}\Delta_{\ell\pi}}{M_K^4} \frac{t^{n-2}}{\pi} \int_{t_0}^\infty \frac{(\Delta_{\ell\pi} + t^\prime) M_K^4}{2t^\prime \lambda_{\ell\pi}(t^\prime)} \frac{ \Im g_1^{(3/2)}(t^\prime)}{(t^\prime - t - i\epsilon){t^\prime}^{n-2}} dt^\prime \\
			&\quad + \frac{1}{3} \frac{\Delta_{K\pi} + t}{2M_K^2} \frac{t^{n-1}}{\pi} \int_{t_0}^\infty \frac{M_K^2}{t^\prime} \frac{ \Im g_1^{(3/2)}(t^\prime)}{(t^\prime - t - i\epsilon) {t^\prime}^{n-1}} dt^\prime \\
			&\quad + \frac{{u^s(t)}^n}{\pi} \int_{u_0}^\infty \frac{ \Im f_0^{(3/2)}(u^\prime)}{(u^\prime-u^s(t) - i\epsilon) {u^\prime}^n} du^\prime \\
			&\quad + \frac{u^s(t)(s-t) + \Delta_{K\pi}\Delta_{\ell\pi}}{M_K^4} \frac{{u^s(t)}^{n-2}}{\pi} \int_{u_0}^\infty \frac{ \Im f_1^{(3/2)}(u^\prime)}{(u^\prime-u^s(t) - i\epsilon){u^\prime}^{n-2}} du^\prime \\
			&\quad + \frac{u^s(t)(s-t) + \Delta_{K\pi}\Delta_{\ell\pi}}{M_K^4} \frac{{u^s(t)}^{n-2}}{\pi} \int_{u_0}^\infty \frac{(\Delta_{\ell\pi} + u^\prime) M_K^4}{2u^\prime \lambda_{\ell\pi}(u^\prime)} \frac{ \Im g_1^{(3/2)}(u^\prime)}{(u^\prime-u^s(t) - i\epsilon){u^\prime}^{n-2}} du^\prime \\
			&\quad - \frac{ \Delta_{K\pi} + u^s(t)}{2M_K^2} \frac{{u^s(t)}^{n-1}}{\pi} \int_{u_0}^\infty \frac{M_K^2}{u^\prime} \frac{ \Im g_1^{(3/2)}(u^\prime)}{(u^\prime-u^s(t) - i\epsilon) {u^\prime}^{n-1}} du^\prime \\
			&\quad+ \O(p^8) .
	\end{split}
\end{align}

Except for the polynomials, the dependence on all the Mandelstam variables in the three representations is now explicit. We fix the polynomial by imposing the condition
\begin{align}
	\begin{split}
		P_{n-1}^s(t) &= P_{n-1}^t(s) = P_{n-1}^u(s)\big|_{u=u^t(s)} .
	\end{split}
\end{align}
In the case of $n=2$, this fixes the polynomial contribution to linear terms in $s$, $t$ and $u$. Note that for this step, we need the three dispersion relations: equating only two of the three polynomials allows also some quadratic terms.

In the case of $n=3$, the above condition allows for seven parameters and a term of the form $c(s^2 t + s t^2)$. Only if we exclude cubic terms in all directions (e.g.~for fixed $t-u$), we can exclude a term of this form and reduce the number of parameters to six per polynomial.

We finally find the following representations as the analytic continuation of the form factors to the whole Mandelstam plane:
\begin{align}
	\begin{split}
		F(s,t,u) &= P_{n-1}^F(s,t,u) + \frac{s^n}{\pi} \int_{s_0}^\infty \frac{\Im f_0(s^\prime)}{(s^\prime-s - i\epsilon) {s^\prime}^n} ds^\prime + \frac{u-t}{M_K^2} \frac{s^{n-1}}{\pi} \int_{s_0}^\infty \frac{ \Im f_1(s^\prime)}{(s^\prime-s - i\epsilon) {s^\prime}^{n-1}} ds^\prime \\
			&\quad - \frac{u-t}{M_K^2} \frac{s^{n-1}}{\pi} \int_{s_0}^\infty \frac{2PL(s^\prime)M_K^2}{\lambda_{K\ell}(s^\prime)} \frac{ \Im g_1(s^\prime)}{(s^\prime-s - i\epsilon) {s^\prime}^{n-1}} ds^\prime \\
			&\quad  + \frac{2}{3} \frac{t^n}{\pi} \int_{t_0}^\infty \frac{ \Im f_0^{(1/2)}(t^\prime)}{(t^\prime-t - i\epsilon) {t^\prime}^n} dt^\prime \\
			&\quad  + \frac{2}{3}  \frac{t(s-u) + \Delta_{K\pi}\Delta_{\ell\pi}}{M_K^4} \frac{t^{n-2}}{\pi} \int_{t_0}^\infty \frac{ \Im f_1^{(1/2)}(t^\prime) }{(t^\prime-t - i\epsilon){t^\prime}^{n-2}} dt^\prime \\
			&\quad  + \frac{2}{3}  \frac{t(s-u) + \Delta_{K\pi}\Delta_{\ell\pi}}{M_K^4} \frac{t^{n-2}}{\pi} \int_{t_0}^\infty  \frac{(\Delta_{\ell\pi} + t^\prime)M_K^4}{2t^\prime \lambda_{\ell\pi}(t^\prime)} \frac{ \Im g_1^{(1/2)}(t^\prime)}{(t^\prime-t - i\epsilon){t^\prime}^{n-2}} dt^\prime \\
			&\quad  - \frac{2}{3} \frac{\Delta_{K\pi} - 3 t}{2M_K^2} \frac{t^{n-1}}{\pi} \int_{t_0}^\infty  \frac{M_K^2}{t^\prime} \frac{ \Im g_1^{(1/2)}(t^\prime)}{(t^\prime-t - i\epsilon) {t^\prime}^{n-1}} dt^\prime \\
			&\quad  + \frac{1}{3} \frac{t^n}{\pi} \int_{t_0}^\infty \frac{\Im f_0^{(3/2)}(t^\prime)}{(t^\prime-t - i\epsilon) {t^\prime}^n} dt^\prime \\
			&\quad  + \frac{1}{3} \frac{t(s-u) + \Delta_{K\pi}\Delta_{\ell\pi}}{M_K^4} \frac{t^{n-2}}{\pi} \int_{t_0}^\infty \frac{ \Im f_1^{(3/2)}(t^\prime)}{(t^\prime-t - i\epsilon){t^\prime}^{n-2}} dt^\prime \\
			&\quad  + \frac{1}{3} \frac{t(s-u) + \Delta_{K\pi}\Delta_{\ell\pi}}{M_K^4} \frac{t^{n-2}}{\pi} \int_{t_0}^\infty \frac{(\Delta_{\ell\pi} + t^\prime) M_K^4}{2t^\prime \lambda_{\ell\pi}(t^\prime)} \frac{ \Im g_1^{(3/2)}(t^\prime)}{(t^\prime-t - i\epsilon){t^\prime}^{n-2}} dt^\prime \\
			&\quad  - \frac{1}{3} \frac{\Delta_{K\pi} - 3 t}{2M_K^2} \frac{t^{n-1}}{\pi} \int_{t_0}^\infty \frac{M_K^2}{t^\prime} \frac{ \Im g_1^{(3/2)}(t^\prime)}{(t^\prime-t - i\epsilon) {t^\prime}^{n-1}} dt^\prime \\
			&\quad + \frac{{u}^n}{\pi} \int_{u_0}^\infty \frac{\Im f_0^{(3/2)}(u^\prime)}{(u^\prime-u - i\epsilon) {u^\prime}^n} du^\prime \\
			&\quad + \frac{u(s-t) + \Delta_{K\pi}\Delta_{\ell\pi}}{M_K^4} \frac{u^{n-2}}{\pi} \int_{u_0}^\infty \frac{\Im f_1^{(3/2)}(u^\prime)}{(u^\prime-u - i\epsilon){u^\prime}^{n-2}} du^\prime \\
			&\quad + \frac{u(s-t) + \Delta_{K\pi}\Delta_{\ell\pi}}{M_K^4} \frac{u^{n-2}}{\pi} \int_{u_0}^\infty \frac{(\Delta_{\ell\pi} + u^\prime) M_K^4}{2u^\prime \lambda_{\ell\pi}(u^\prime)} \frac{ \Im g_1^{(3/2)}(u^\prime)}{(u^\prime-u - i\epsilon){u^\prime}^{n-2}} du^\prime \\
			&\quad- \frac{\Delta_{K\pi} - 3 u}{2M_K^2} \frac{u^{n-1}}{\pi} \int_{u_0}^\infty \frac{M_K^2}{u^\prime} \frac{ \Im g_1^{(3/2)}(u^\prime)}{(u^\prime-u - i\epsilon) {u^\prime}^{n-1}} du^\prime \\
			&\quad+ \O(p^8) ,
	\end{split}
\end{align}
\begin{align}
	\begin{split}
		G(s,t,u) &= P_{n-1}^G(s,t,u) + \frac{s^n}{\pi} \int_{s_0}^\infty \frac{\Im g_1(s^\prime)}{(s^\prime-s - i\epsilon) {s^\prime}^n} ds^\prime \\
			&\quad  - \frac{2}{3} \frac{t^n}{\pi} \int_{t_0}^\infty \frac{ \Im f_0^{(1/2)}(t^\prime)}{(t^\prime-t - i\epsilon) {t^\prime}^n} dt^\prime \\
			&\quad  - \frac{2}{3} \frac{t(s-u) + \Delta_{K\pi}\Delta_{\ell\pi}}{M_K^4}  \frac{t^{n-2}}{\pi} \int_{t_0}^\infty \frac{ \Im f_1^{(1/2)}(t^\prime) }{(t^\prime-t - i\epsilon){t^\prime}^{n-2}} dt^\prime \\
			&\quad  - \frac{2}{3} \frac{t(s-u) + \Delta_{K\pi}\Delta_{\ell\pi}}{M_K^4} \frac{t^{n-2}}{\pi} \int_{t_0}^\infty \frac{(\Delta_{\ell\pi} + t^\prime)M_K^4}{2t^\prime \lambda_{\ell\pi}(t^\prime)} \frac{ \Im g_1^{(1/2)}(t^\prime)}{(t^\prime-t - i\epsilon){t^\prime}^{n-2}} dt^\prime \\
			&\quad  + \frac{2}{3} \frac{\Delta_{K\pi} + t}{2M_K^2} \frac{t^{n-1}}{\pi} \int_{t_0}^\infty  \frac{M_K^2}{t^\prime} \frac{ \Im g_1^{(1/2)}(t^\prime)}{(t^\prime-t - i\epsilon) {t^\prime}^{n-1}} dt^\prime \\
			&\quad  - \frac{1}{3} \frac{t^n}{\pi} \int_{t_0}^\infty \frac{\Im f_0^{(3/2)}(t^\prime)}{(t^\prime-t - i\epsilon) {t^\prime}^n} dt^\prime \\
			&\quad  - \frac{1}{3} \frac{t(s-u) + \Delta_{K\pi}\Delta_{\ell\pi}}{M_K^4} \frac{t^{n-2}}{\pi} \int_{t_0}^\infty \frac{ \Im f_1^{(3/2)}(t^\prime)}{(t^\prime-t - i\epsilon){t^\prime}^{n-2}} dt^\prime \\
			&\quad  - \frac{1}{3} \frac{t(s-u) + \Delta_{K\pi}\Delta_{\ell\pi}}{M_K^4} \frac{t^{n-2}}{\pi} \int_{t_0}^\infty  \frac{(\Delta_{\ell\pi} + t^\prime) M_K^4}{2t^\prime \lambda_{\ell\pi}(t^\prime)} \frac{ \Im g_1^{(3/2)}(t^\prime)}{(t^\prime-t - i\epsilon){t^\prime}^{n-2}} dt^\prime \\
			&\quad  + \frac{1}{3} \frac{\Delta_{K\pi} + t}{2M_K^2} \frac{t^{n-1}}{\pi} \int_{t_0}^\infty \frac{M_K^2}{t^\prime} \frac{ \Im g_1^{(3/2)}(t^\prime)}{(t^\prime-t - i\epsilon) {t^\prime}^{n-1}} dt^\prime \\
			&\quad + \frac{{u}^n}{\pi} \int_{u_0}^\infty \frac{\Im f_0^{(3/2)}(u^\prime)}{(u^\prime-u - i\epsilon) {u^\prime}^n} du^\prime \\
			&\quad+ \frac{u(s-t) + \Delta_{K\pi}\Delta_{\ell\pi}}{M_K^4} \frac{u^{n-2}}{\pi} \int_{u_0}^\infty \frac{\Im f_1^{(3/2)}(u^\prime)}{(u^\prime-u - i\epsilon){u^\prime}^{n-2}} du^\prime \\
			&\quad + \frac{u(s-t) + \Delta_{K\pi}\Delta_{\ell\pi}}{M_K^4} \frac{u^{n-2}}{\pi} \int_{u_0}^\infty \frac{(\Delta_{\ell\pi} + u^\prime) M_K^4}{2u^\prime \lambda_{\ell\pi}(u^\prime)} \frac{ \Im g_1^{(3/2)}(u^\prime)}{(u^\prime-u - i\epsilon){u^\prime}^{n-2}} du^\prime \\
			&\quad- \frac{\Delta_{K\pi} + u}{2M_K^2} \frac{u^{n-1}}{\pi} \int_{u_0}^\infty \frac{M_K^2}{u^\prime} \frac{ \Im g_1^{(3/2)}(u^\prime)}{(u^\prime-u - i\epsilon) {u^\prime}^{n-1}} du^\prime \\
			&\quad+ \O(p^8) ,
	\end{split}
\end{align}
where, of course, the constraint $s+t+u=\Sigma_0$ is still valid and $P_{n-1}^F$ and $P_{n-1}^G$ are polynomials of the order $n-1$ in all three Mandelstam variables. For $n=2$, they can be written as
\begin{align}
	\begin{split}
		P_1^F(s,t,u) &= p_0^F + p_1^F \frac{s}{M_K^2} + q_0^F \frac{t-u}{M_K^2} , \\
		P_1^G(s,t,u) &= p_0^G + p_1^G \frac{s}{M_K^2} + q_0^G \frac{t-u}{M_K^2} .
	\end{split}
\end{align}
In the case $n=3$, they have the form
\begin{align}
	\begin{split}
		P_2^F(s,t,u) &= p_0^F + p_1^F \frac{s}{M_K^2} + p_2^F \frac{s^2}{M_K^4} + q_0^F \frac{t-u}{M_K^2} + q_1^F \frac{s(t-u)}{M_K^4} + p_3^F \frac{(t-u)^2}{M_K^4} , \\
		P_2^G(s,t,u) &= p_0^G + p_1^G \frac{s}{M_K^2} + p_2^G \frac{s^2}{M_K^4} + q_0^G \frac{t-u}{M_K^2} + q_1^G \frac{s(t-u)}{M_K^4} + p_3^G \frac{(t-u)^2}{M_K^4}  .
	\end{split}
\end{align}

The form factors $F$ and $G$ are now decomposed into functions of only one Mandelstam variable as shown in (\ref{eq:FormFactorDecomposition}), with the functions defined in (\ref{eq:FunctionsOfOneVariable}) for $n=2$ or (\ref{eq:FunctionsOfOneVariable3Subtr}) for $n=3$. For $n=2$, the six free parameters are in one-to-one correspondence with the coefficients of the two subtraction polynomials:
\begin{align}
	\begin{split}
		m_0^0 &= p_0^F + \frac{\Sigma_0}{M_K^2} q_0^G , \\
		m_0^1 &= p_1^F - q_0^G , \\
		m_1^0 &= -q_0^F - q_0^G , \\
		\tilde m_1^0 &= p_0^G - \frac{\Sigma_0}{M_K^2} q_0^G , \\
		\tilde m_1^1 &= p_1^G + q_0^G , \\
		n_0^1 &= -3 q_0^G .
	\end{split}
\end{align}
For $n=3$, the twelve free parameters in (\ref{eq:FunctionsOfOneVariable3Subtr}) correspond again to the coefficients of the subtraction polynomials. The explicit relation is rather large but easy to find.


\chapter{Omnès Solution to the Dispersion Relation}

\section{Solution for $n=3$ Subtractions}

\label{sec:AppendixOmnes3Subtractions}

For $n=3$ subtractions, the Omnès representation reads
\begin{align}
	\begin{alignedat}{2}
		\label{eq:FunctionsOfOneVariableOmnes3Subtr}
		M_0(s) &= \Omega_0^0(s) & &\bigg\{ a^{M_0} + b^{M_0} \frac{s}{M_K^2} + c^{M_0} \frac{s^2}{M_K^4} + d^{M_0} \frac{s^3}{M_K^6} + \frac{s^4}{\pi} \int_{s_0}^{\Lambda^2} \frac{\hat M_0(s^\prime) \sin\delta_0^0(s^\prime)}{|\Omega_0^0(s^\prime)| (s^\prime - s - i\epsilon) {s^\prime}^4} ds^\prime \bigg\} , \\
		M_1(s) &= \Omega_1^1(s) & &\bigg\{ a^{M_1} + b^{M_1}  \frac{s}{M_K^2} + c^{M_1} \frac{s^2}{M_K^4} + \frac{s^3}{\pi} \int_{s_0}^{\Lambda^2} \frac{\hat M_1(s^\prime) \sin\delta_1^1(s^\prime)}{|\Omega_1^1(s^\prime)| (s^\prime - s - i\epsilon) {s^\prime}^3} ds^\prime  \bigg\} , \\
		 \tilde M_1(s) &= \Omega_1^1(s) & &\bigg\{ a^{\tilde M_1} + b^{\tilde M_1}  \frac{s}{M_K^2} + c^{\tilde M_1}  \frac{s^2}{M_K^4} + d^{\tilde M_1} \frac{s^3}{M_K^6} + \frac{s^4}{\pi} \int_{s_0}^{\Lambda^2} \frac{\hat{\tilde M}_1(s^\prime) \sin\delta_1^1(s^\prime)}{|\Omega_1^1(s^\prime)| (s^\prime - s - i\epsilon) {s^\prime}^4} ds^\prime \bigg\} , \\
		N_0(t) &=  \Omega_0^{1/2}(t) & &\bigg\{ b^{N_0} \frac{t}{M_K^2} + c^{N_0} \frac{t^2}{M_K^4} + \frac{t^3}{\pi} \int_{t_0}^{\Lambda^2} \frac{\hat N_0(t^\prime) \sin\delta_0^{1/2}(t^\prime)}{|\Omega_0^{1/2}(t^\prime)| (t^\prime - t - i\epsilon) {t^\prime}^3} dt^\prime  \bigg\} , \\
		N_1(t) &= \Omega_1^{1/2}(t) & &\bigg\{ a^{N_1} + \frac{t}{\pi} \int_{t_0}^{\Lambda^2} \frac{\hat N_1(t^\prime) \sin\delta_1^{1/2}(t^\prime)}{|\Omega_1^{1/2}(t^\prime)| (t^\prime - t - i\epsilon)t^\prime} dt^\prime  \bigg\} , \\
		\tilde N_1(t) &= \Omega_1^{1/2}(t) & &\bigg\{ b^{\tilde N_1} \frac{t}{M_K^2} + \frac{t^2}{\pi} \int_{t_0}^{\Lambda^2} \frac{\hat{\tilde N}_1(t^\prime) \sin\delta_1^{1/2}(t^\prime)}{|\Omega_1^{1/2}(t^\prime)| (t^\prime - t - i\epsilon) {t^\prime}^2} dt^\prime  \bigg\} , \\
		R_0(t) &=  \Omega_0^{3/2}(t) & &\bigg\{ \frac{t^3}{\pi} \int_{t_0}^{\Lambda^2} \frac{\hat R_0(t^\prime) \sin\delta_0^{3/2}(t^\prime)}{|\Omega_0^{3/2}(t^\prime)| (t^\prime - t - i\epsilon) {t^\prime}^3} dt^\prime  \bigg\} , \\
		R_1(t) &=  \Omega_1^{3/2}(t) & &\bigg\{ \frac{t}{\pi} \int_{t_0}^{\Lambda^2} \frac{\hat R_1(t^\prime) \sin\delta_1^{3/2}(t^\prime)}{|\Omega_1^{3/2}(t^\prime)| (t^\prime - t - i\epsilon)t^\prime} dt^\prime  \bigg\} , \\
		\tilde R_1(t) &=   \Omega_1^{3/2}(t) & &\bigg\{ \frac{t^2}{\pi} \int_{t_0}^{\Lambda^2} \frac{\hat{\tilde R}_1(t^\prime) \sin\delta_1^{3/2}(t^\prime)}{|\Omega_1^{3/2}(t^\prime)| (t^\prime - t - i\epsilon) {t^\prime}^2} dt^\prime  \bigg\} .
	\end{alignedat}
\end{align}

Let us work out how to transform the Omnès representation (\ref{eq:FunctionsOfOneVariableOmnes}) into the one with more subtractions (\ref{eq:FunctionsOfOneVariableOmnes3Subtr}).
We start by subtracting all the dispersive integrals once more, using the relation
\begin{align}
	\begin{split}
		\frac{1}{s^\prime-s} = \frac{1}{s^\prime} + \frac{s}{(s^\prime-s)s^\prime} .
	\end{split}
\end{align}
This generates nine additional subtraction constants:
\begin{align}
	\footnotesize
	\begin{alignedat}{2}
		M_0(s) &= \Omega_0^0(s) & &\bigg\{ a^{M_0} + b^{M_0} \frac{s}{M_K^2} + c^{M_0} \frac{s^2}{M_K^4} + d^{M_0} \frac{s^3}{M_K^6} + \frac{s^4}{\pi} \int_{s_0}^{\Lambda^2} \frac{\hat M_0(s^\prime) \sin\delta_0^0(s^\prime)}{|\Omega_0^0(s^\prime)| (s^\prime - s - i\epsilon) {s^\prime}^4} ds^\prime \bigg\} , \\
		M_1(s) &= \Omega_1^1(s) & &\bigg\{ a^{M_1} + b^{M_1}  \frac{s}{M_K^2} + c^{M_1} \frac{s^2}{M_K^4} + \frac{s^3}{\pi} \int_{s_0}^{\Lambda^2} \frac{\hat M_1(s^\prime) \sin\delta_1^1(s^\prime)}{|\Omega_1^1(s^\prime)| (s^\prime - s - i\epsilon) {s^\prime}^3} ds^\prime  \bigg\} , \\
		 \tilde M_1(s) &= \Omega_1^1(s) & &\bigg\{ a^{\tilde M_1} + b^{\tilde M_1}  \frac{s}{M_K^2} + c^{\tilde M_1}  \frac{s^2}{M_K^4} + d^{\tilde M_1} \frac{s^3}{M_K^6} + \frac{s^4}{\pi} \int_{s_0}^{\Lambda^2} \frac{\hat{\tilde M}_1(s^\prime) \sin\delta_1^1(s^\prime)}{|\Omega_1^1(s^\prime)| (s^\prime - s - i\epsilon) {s^\prime}^4} ds^\prime \bigg\} , \\
		N_0(t) &=  \Omega_0^{1/2}(t) & &\bigg\{ b^{N_0} \frac{t}{M_K^2} + c^{N_0} \frac{t^2}{M_K^4} + \frac{t^3}{\pi} \int_{t_0}^{\Lambda^2} \frac{\hat N_0(t^\prime) \sin\delta_0^{1/2}(t^\prime)}{|\Omega_0^{1/2}(t^\prime)| (t^\prime - t - i\epsilon) {t^\prime}^3} dt^\prime  \bigg\} , \\
		N_1(t) &= \Omega_1^{1/2}(t) & &\bigg\{ a^{N_1} + \frac{t}{\pi} \int_{t_0}^{\Lambda^2} \frac{\hat N_1(t^\prime) \sin\delta_1^{1/2}(t^\prime)}{|\Omega_1^{1/2}(t^\prime)| (t^\prime - t - i\epsilon){t^\prime}} dt^\prime  \bigg\} , \\
		\tilde N_1(t) &= \Omega_1^{1/2}(t) & &\bigg\{ b^{\tilde N_1} \frac{t}{M_K^2} + \frac{t^2}{\pi} \int_{t_0}^{\Lambda^2} \frac{\hat{\tilde N}_1(t^\prime) \sin\delta_1^{1/2}(t^\prime)}{|\Omega_1^{1/2}(t^\prime)| (t^\prime - t - i\epsilon) {t^\prime}^2} dt^\prime  \bigg\} , \\
		R_0(t) &=  \Omega_0^{3/2}(t) & &\bigg\{ c^{R_0} \frac{t^2}{M_K^4} + \frac{t^3}{\pi} \int_{t_0}^{\Lambda^2} \frac{\hat R_0(t^\prime) \sin\delta_0^{3/2}(t^\prime)}{|\Omega_0^{3/2}(t^\prime)| (t^\prime - t - i\epsilon) {t^\prime}^3} dt^\prime  \bigg\} , \\
		R_1(t) &=  \Omega_1^{3/2}(t) & &\bigg\{ a^{R_1} +  \frac{t}{\pi} \int_{t_0}^{\Lambda^2} \frac{\hat R_1(t^\prime) \sin\delta_1^{3/2}(t^\prime)}{|\Omega_1^{3/2}(t^\prime)| (t^\prime - t - i\epsilon) t^\prime} dt^\prime  \bigg\} , \\
		\tilde R_1(t) &=   \Omega_1^{3/2}(t) & &\bigg\{ b^{\tilde R_1} \frac{t}{M_K^2} + \frac{t^2}{\pi} \int_{t_0}^{\Lambda^2} \frac{\hat{\tilde R}_1(t^\prime) \sin\delta_1^{3/2}(t^\prime)}{|\Omega_1^{3/2}(t^\prime)| (t^\prime - t - i\epsilon) {t^\prime}^2} dt^\prime  \bigg\} .
	\end{alignedat}
\end{align}
To get rid of the subtraction constants in the $R$-functions, we apply a gauge transformation (\ref{eq:GaugeTransformation}). To this end, let us write the gauge transformation in the Omnès representation:
\begin{align}
	\footnotesize
	\begin{alignedat}{2}
		\label{eq:OmnesGaugeTransformation}
		\delta M_0(s) &= \Omega_0^0(s) & &\bigg\{ \delta a^{M_0} + \delta b^{M_0} \frac{s}{M_K^2} + \delta c^{M_0} \frac{s^2}{M_K^4} + \delta d^{M_0} \frac{s^3}{M_K^6} + \frac{s^4}{\pi} \int_{s_0}^{\Lambda^2} \frac{\delta \hat M_0(s^\prime) \sin\delta_0^0(s^\prime)}{|\Omega_0^0(s^\prime)| (s^\prime - s - i\epsilon) {s^\prime}^4} ds^\prime \bigg\} , \\
		\delta M_1(s) &= \Omega_1^1(s) & &\bigg\{ \delta a^{M_1} + \delta b^{M_1}  \frac{s}{M_K^2} + \delta c^{M_1} \frac{s^2}{M_K^4} + \frac{s^3}{\pi} \int_{s_0}^{\Lambda^2} \frac{\delta \hat M_1(s^\prime) \sin\delta_1^1(s^\prime)}{|\Omega_1^1(s^\prime)| (s^\prime - s - i\epsilon) {s^\prime}^3} ds^\prime  \bigg\} , \\
		\delta  \tilde M_1(s) &= \Omega_1^1(s) & &\bigg\{ \delta a^{\tilde M_1} + \delta b^{\tilde M_1}  \frac{s}{M_K^2} + \delta c^{\tilde M_1}  \frac{s^2}{M_K^4} + \delta d^{\tilde M_1} \frac{s^3}{M_K^6} + \frac{s^4}{\pi} \int_{s_0}^{\Lambda^2} \frac{\delta \hat{\tilde M}_1(s^\prime) \sin\delta_1^1(s^\prime)}{|\Omega_1^1(s^\prime)| (s^\prime - s - i\epsilon) {s^\prime}^4} ds^\prime \bigg\} , \\
		\delta N_0(t) &=  \Omega_0^{1/2}(t) & &\bigg\{ \delta b^{N_0} \frac{t}{M_K^2} + \delta c^{N_0} \frac{t^2}{M_K^4} + \frac{t^3}{\pi} \int_{t_0}^{\Lambda^2} \frac{\delta \hat N_0(t^\prime) \sin\delta_0^{1/2}(t^\prime)}{|\Omega_0^{1/2}(t^\prime)| (t^\prime - t - i\epsilon) {t^\prime}^3} dt^\prime  \bigg\} , \\
		\delta N_1(t) &= \Omega_1^{1/2}(t) & &\bigg\{ \delta a^{N_1} + \frac{t}{\pi} \int_{t_0}^{\Lambda^2} \frac{\delta \hat N_1(t^\prime) \sin\delta_1^{1/2}(t^\prime)}{|\Omega_1^{1/2}(t^\prime)| (t^\prime - t - i\epsilon){t^\prime}} dt^\prime  \bigg\} , \\
		\delta \tilde N_1(t) &= \Omega_1^{1/2}(t) & &\bigg\{ \delta b^{\tilde N_1} \frac{t}{M_K^2} + \frac{t^2}{\pi} \int_{t_0}^{\Lambda^2} \frac{\delta \hat{\tilde N}_1(t^\prime) \sin\delta_1^{1/2}(t^\prime)}{|\Omega_1^{1/2}(t^\prime)| (t^\prime - t - i\epsilon) {t^\prime}^2} dt^\prime  \bigg\} , \\
		\delta R_0(t) &=  \Omega_0^{3/2}(t) & &\bigg\{ \delta c^{R_0} \frac{t^2}{M_K^4} + \frac{t^3}{\pi} \int_{t_0}^{\Lambda^2} \frac{\delta \hat R_0(t^\prime) \sin\delta_0^{3/2}(t^\prime)}{|\Omega_0^{3/2}(t^\prime)| (t^\prime - t - i\epsilon) {t^\prime}^3} dt^\prime  \bigg\} , \\
		\delta R_1(t) &=  \Omega_1^{3/2}(t) & &\bigg\{ \delta a^{R_1} +  \frac{t}{\pi} \int_{t_0}^{\Lambda^2} \frac{\delta \hat R_1(t^\prime) \sin\delta_1^{3/2}(t^\prime)}{|\Omega_1^{3/2}(t^\prime)| (t^\prime - t - i\epsilon) t^\prime} dt^\prime  \bigg\} , \\
		\delta \tilde R_1(t) &=   \Omega_1^{3/2}(t) & &\bigg\{ \delta b^{\tilde R_1} \frac{t}{M_K^2} + \frac{t^2}{\pi} \int_{t_0}^{\Lambda^2} \frac{\delta \hat{\tilde R}_1(t^\prime) \sin\delta_1^{3/2}(t^\prime)}{|\Omega_1^{3/2}(t^\prime)| (t^\prime - t - i\epsilon) {t^\prime}^2} dt^\prime  \bigg\} .
	\end{alignedat}
\end{align}
Since the gauge transformation is a polynomial and has no discontinuity, the changes in the hat functions are given by $\delta \hat M_0 = - \delta M_0$ etc., which assures that the partial waves are unchanged. The shifts in the subtraction constants are most easily found by comparing the Taylor expansion of (\ref{eq:OmnesGaugeTransformation}) with (\ref{eq:GaugeTransformation}):
\begin{align}
\label{eq:OmnesGaugeTransformationParameters}
\scalebox{0.8}{
\begin{minipage}{1.1\textwidth}
$	\begin{split}
		\delta a^{M_0} &= \left(2 A^{R_1} - B^{\tilde R_1} + 2 C^{R_0}\right) \frac{\Sigma_0^2 - \Delta_{K\pi}\Delta_{\ell\pi}}{2M_K^4} , \\
		\delta b^{M_0} &= - \left(2 A^{R_1} - B^{\tilde R_1} + 2 C^{R_0}\right) \left( \frac{\Sigma_0}{M_K^2} + \omega_0^0 \frac{\Sigma_0^2 - \Delta_{K\pi}\Delta_{\ell\pi}}{2 M_K^4} \right) , \\
		\delta c^{M_0} &= \left(2 A^{R_1} - B^{\tilde R_1} + 2 C^{R_0}\right) \left(\frac{1}{2} + \omega_0^0 \frac{\Sigma_0}{M_K^2} + \left( \frac{{\omega_0^0}^2}{2} - \bar \omega_0^0 \right) \frac{\Sigma_0^2 - \Delta_{K\pi}\Delta_{\ell\pi}}{2M_K^4} \right) , \\
		\delta d^{M_0} &= - \left(2 A^{R_1} - B^{\tilde R_1} + 2 C^{R_0}\right) \left( \frac{\omega_0^0}{2} - \left( \bar\omega_0^0 - \frac{{\omega_0^0}^2}{2} \right) \frac{\Sigma_0}{M_K^2} + \left( {\omega_0^0}^3 - 6 \omega_0^0 \bar\omega_0^0 + 6 \bar{\bar\omega}_0^0 \right) \frac{\Sigma_0^2 - \Delta_{K\pi} \Delta_{\ell\pi}}{12 M_K^4} \right) , \\
		\delta a^{M_1} &= -\left(A^{R_1} + B^{\tilde R_1} + 2 C^{R_0}\right)\frac{\Sigma_0}{M_K^2} + B^{\tilde R_1} \frac{\Delta_{K\pi}}{2 M_K^2} , \\
		\delta b^{M_1} &=  \left(B^{\tilde R_1} + 2 C^{R_0}\right) + \omega_1^1 \left( \left(A^{R_1} + B^{\tilde R_1} + 2 C^{R_0}\right)\frac{\Sigma_0}{M_K^2} - B^{\tilde R_1} \frac{\Delta_{K\pi}}{2 M_K^2} \right) , \\
		\delta c^{M_1} &= - \left( \left(A^{R_1} + B^{\tilde R_1} + 2 C^{R_0}\right)\frac{\Sigma_0}{M_K^2} - B^{\tilde R_1} \frac{\Delta_{K\pi}}{2 M_K^2}\right) \left(\frac{{\omega_1^1}^2}{2} - \bar\omega_1^1 \right) - \omega_1^1 \left(B^{\tilde R_1} + 2 C^{R_0} \right) , \\
		\delta a^{\tilde M_1} &= \left(B^{\tilde R_1} - 2 C^{R_0}\right) \frac{\Sigma_0^2}{2 M_K^4} - \left(2 A^{R_1} + B^{\tilde R_1} - 2 C^{R_0} \right) \frac{\Delta_{K\pi} \Delta_{\ell\pi}}{2 M_K^4} + B^{\tilde R_1} \frac{\Sigma_0 \Delta_{K\pi}}{2 M_K^4} , \\
		\delta b^{\tilde M_1} &=  - \left(B^{\tilde R_1} \frac{\Delta_{K\pi}}{2 M_K^2} + \left(A^{R_1} + B^{\tilde R_1} - 2 C^{R_0}\right) \frac{\Sigma_0}{M_K^2} \right) \\
			&\quad - \omega_1^1 \left(\left(B^{\tilde R_1} - 2 C^{R_0}\right) \frac{\Sigma_0^2}{2 M_K^4} - \left(2 A^{R_1} + B^{\tilde R_1} - 2 C^{R_0} \right) \frac{\Delta_{K\pi} \Delta_{\ell\pi}}{2 M_K^4} + B^{\tilde R_1} \frac{\Sigma_0 \Delta_{K\pi}}{2 M_K^4} \right) , \\
		\delta c^{\tilde M_1} &= \frac{1}{2} \left(2 A^{R_1} + B^{\tilde R_1} - 2 C^{R_0}\right) + \omega_1^1 \left(B^{\tilde R_1} \frac{\Delta_{K\pi}}{2 M_K^2} + \left(A^{R_1} + B^{\tilde R_1} - 2 C^{R_0}\right) \frac{\Sigma_0}{M_K^2} \right) \\
			&\quad + \left( \frac{{\omega_1^1}^2}{2} - \bar \omega_1^1 \right) \left( \left(B^{\tilde R_1} - 2 C^{R_0}\right) \frac{\Sigma_0^2}{2 M_K^4} - \left(2 A^{R_1} + B^{\tilde R_1} - 2 C^{R_0} \right) \frac{\Delta_{K\pi} \Delta_{\ell\pi}}{2 M_K^4} + B^{\tilde R_1} \frac{\Sigma_0 \Delta_{K\pi}}{2 M_K^4} \right) , \\
		\delta d^{\tilde M_1} &= - \frac{1}{2} \omega_1^1 \left(2 A^{R_1} + B^{\tilde R_1} - 2 C^{R_0}\right) - \left( \frac{{\omega_1^1}^2}{2} - \bar\omega_1^1 \right) \left(B^{\tilde R_1} \frac{\Delta_{K\pi}}{2 M_K^2} + \left(A^{R_1} + B^{\tilde R_1} - 2 C^{R_0}\right) \frac{\Sigma_0}{M_K^2} \right) \\
			&\quad - \frac{1}{6} \left( {\omega_1^1}^3 - 6 \omega_1^1 \bar\omega_1^1 + 6 \bar{\bar\omega}_1^1 \right) \left( \left(B^{\tilde R_1} - 2 C^{R_0}\right) \frac{\Sigma_0^2}{2 M_K^4} - \left(2 A^{R_1} + B^{\tilde R_1} - 2 C^{R_0} \right) \frac{\Delta_{K\pi} \Delta_{\ell\pi}}{2 M_K^4} + B^{\tilde R_1} \frac{\Sigma_0 \Delta_{K\pi}}{2 M_K^4} \right) , \\
		\delta b^{N_0} &= - \left(2 A^{R_1} - B^{\tilde R_1} + 2 C^{R_0} \right) \frac{3 (\Delta_{K\pi} + 2 \Sigma_0)}{8 M_K^2} , \\
		\delta c^{N_0} &= \frac{1}{8} \left(6 A^{R_1} - 3 B^{\tilde R_1} -10 C^{R_0} \right) + \omega_0^{1/2} \left(2 A^{R_1} - B^{\tilde R_1} + 2 C^{R_0} \right) \frac{3 (\Delta_{K\pi} + 2 \Sigma_0)}{8 M_K^2} , \\
		\delta a^{N_1} &= - \frac{1}{4} \left(2 A^{R_1} + 3 B^{\tilde R_1} - 6 C^{R_0} \right) , \\
		\delta b^{\tilde N_1} &= -\frac{1}{4} \left(6 A^{R_1} + 5 B^{\tilde R_1} + 6 C^{R_0} \right) , \\
		\delta c^{R_0} &= C^{R_0} , \\
		\delta a^{R_1} &= A^{R_1} , \\
		\delta b^{\tilde R_1} &= B^{\tilde R_1} ,
	\end{split}
	$
\end{minipage}}
\end{align}
where $\omega$, $\bar\omega$ and $\bar{\bar\omega}$ are defined by applying subtractions to the Omnès functions:
\begin{align}
	\footnotesize
	\begin{split}
		\Omega(s) &= \exp\left( \frac{s}{\pi} \int_{s_0}^\infty \frac{\delta(s^\prime)}{(s^\prime - s - i \epsilon) s^\prime} ds^\prime \right) \\
			&= \exp\left(  \frac{s}{\pi} \int_{s_0}^\infty \frac{\delta(s^\prime)}{{s^\prime}^2} ds^\prime +  \frac{s^2}{\pi} \int_{s_0}^\infty \frac{\delta(s^\prime)}{{s^\prime}^3} ds^\prime +  \frac{s^3}{\pi} \int_{s_0}^\infty \frac{\delta(s^\prime)}{{s^\prime}^4} ds^\prime +  \frac{s^4}{\pi} \int_{s_0}^\infty \frac{\delta(s^\prime)}{(s^\prime - s - i \epsilon) {s^\prime}^4} ds^\prime \right) \\
			&=: \exp\left( \omega \frac{s}{M_K^2} +  \bar\omega \frac{s^2}{M_K^4} + \bar{\bar\omega} \frac{s^3}{M_K^6} + \frac{s^4}{\pi} \int_{s_0}^\infty \frac{\delta(s^\prime)}{(s^\prime - s - i \epsilon) {s^\prime}^4} ds^\prime \right) .
	\end{split}
\end{align}
In order to obtain the form (\ref{eq:FunctionsOfOneVariableOmnes3Subtr}), the subtraction constants in the $R$-functions can now be removed with the gauge transformation
\begin{align}
	\begin{split}
		C^{R_0} &= -c^{R_0} , \\
		A^{R_1} &= -a^{R_1} , \\
		B^{\tilde R_1} &= -b^{\tilde R_1} .
	\end{split}
\end{align}

\section{Hat Functions}

\label{sec:AppendixHatFunctions}

In the following, we provide the explicit expressions for the hat functions that appear in the Omnès solution to the dispersion relation.
\begin{align}
	\footnotesize
	\begin{split}
		\hat M_0(s) &=  \frac{2}{3} \Big( \<N_0\>_{t_s}+2\<R_0\>_{t_s} \Big) - \Big( \<z N_0\>_{t_s} + 2 \<z R_0\>_{t_s} \Big) \frac{2 \sigma_\pi PL }{3 X} \\
			&\quad - \Big( \<N_1\>_{t_s} + 2 \<R_1\>_{t_s} \Big) \frac{3 s^2-4 s \Sigma_0+\Sigma_0^2-4 \Delta_{K\pi} \Delta_{\ell\pi}}{6 M_K^4} \\
			&\quad + \Big( \<z N_1\>_{t_s} + 2 \<z R_1\>_{t_s} \Big) \frac{\sigma _{\pi } \left(-4 PL \Delta_{K\pi} \Delta_{\ell\pi} + PL \left(3 s^2-4 s \Sigma_0+\Sigma_0^2\right)-4 s X^2\right)}{6 M_K^4 X} \\
			&\quad + \Big( \<z^2 N_1\>_{t_s} + 2 \<z^2 R_1\>_{t_s} \Big) \frac{2 \sigma_\pi^2 \left( PL s + X^2 \right)}{3 M_K^4} - \Big( \<z^3 N_1\>_{t_s} + 2 \<z^3 R_1\>_{t_s} \Big) \frac{2 \sigma_\pi^3 PL \, X}{3 M_K^4} \\
			&\quad - \Big( \<\tilde N_1\>_{t_s} + 2 \<\tilde R_1\>_{t_s} \Big) \frac{2 \Delta_{K\pi} + 3 s - 3 \Sigma_0}{6 M_K^2} \\
			&\quad + \Big( \<z \tilde N_1\>_{t_s} + 2 \<z \tilde R_1\>_{t_s} \Big) \frac{\sigma_\pi \left(PL \left(2 \Delta_{K\pi} - s + \Sigma_0\right)-6 X^2\right)}{6 M_K^2 X}  - \Big( \<z^2 \tilde N_1\>_{t_s} + 2 \<z^2 \tilde R_1\>_{t_s} \Big) \frac{\sigma_\pi^2 PL}{3 M_K^2} , \\
	\end{split}
\end{align}
\begin{align}
	\footnotesize
	\begin{split}
		\hat M_1(s) &= \Big( \<N_0\>_{t_s} - \<R_0\>_{t_s} \Big) \frac{M_K^2 PL}{2 X^2}  + \Big( \< z N_0 \>_{t_s} - \<z R_0\>_{t_s} \Big) \frac{M_K^2}{\sigma_\pi X} - \Big( \< z^2 N_0\>_{t_s} - \< z^2 R_0\>_{t_s} \Big) \frac{3 M_K^2 PL}{2 X^2} \\
			&\quad + \Big( \<N_1\>_{t_s} - \<R_1\>_{t_s} \Big) \frac{\left(4 \Delta_{K\pi} \Delta_{\ell\pi} -3 s^2+4 s \Sigma_0 - \Sigma_0^2\right) PL}{8 M_K^2 X^2} \\
			&\quad - \Big( \<z N_1\>_{t_s} - \<z R_1\>_{t_s} \Big) \frac{ 3 s^2 + 2 \sigma_\pi^2 PL s - 4 s \Sigma_0+\Sigma_0^2 - 4 \Delta_{K\pi} \Delta_{\ell\pi} }{4 M_K^2 \sigma_\pi X} \\
			&\quad + \Big( \<z^2 N_1\>_{t_s} - \<z^2 R_1\>_{t_s} \Big) \Bigg( \frac{ 3 PL \left(3 s^2 - 4 s \Sigma_0 + \Sigma_0^2\right) -12 PL \Delta_{K\pi} \Delta_{\ell\pi} }{8 M_K^2 X^2} +  \frac{ \sigma_\pi^2 PL - 2 s}{2 M_K^2} \Bigg) \\
			&\quad + \Big( \<z^3 N_1\>_{t_s} - \<z^3 R_1\>_{t_s} \Big) \frac{\sigma_\pi \left(3 s PL + 2 X^2\right)}{2 X M_K^2}  - \Big( \<z^4 N_1\>_{t_s} - \<z^4 R_1\>_{t_s} \Big) \frac{3 \sigma_\pi^2 PL}{2 M_K^2} \\
			&\quad - \Big( \<\tilde N_1\>_{t_s} - \<\tilde R_1\>_{t_s} \Big) \frac{PL \left(2 \Delta_{K\pi} - s + \Sigma_0\right)}{8 X^2} + \Big( \<z \tilde N_1\>_{t_s} - \<z \tilde R_1\>_{t_s} \Big) \frac{3 \Sigma_0 - 2 \Delta_{K\pi} + \sigma_\pi^2 PL - 3 s}{4 \sigma_\pi X} \\
			&\quad + \Big( \<z^2 \tilde N_1\>_{t_s} - \<z^2 \tilde R_1\>_{t_s} \Big) \left(\frac{3 PL \left(2 \Delta_{K\pi} - s + \Sigma_0 \right)}{8 X^2} - \frac{3}{2}\right) - \Big( \<z^3 \tilde N_1\>_{t_s} - \<z^3 \tilde R_1\>_{t_s} \Big) \frac{3 PL \sigma_\pi }{4 X} , \\
	\end{split}
\end{align}
\begin{align}
	\footnotesize
	\begin{split}
		\hat{\tilde M}_1(s) &= - \Big( \<N_0\>_{t_s} - \< R_0 \>_{t_s} \Big) + \Big( \<z^2 N_0\>_{t_s} - \< z^2 R_0 \>_{t_s} \Big) \\
			&\quad + \Big( \<N_1\>_{t_s} - \< R_1 \>_{t_s} \Big) \frac{3 s^2 - 4 s \Sigma_0 + \Sigma_0^2-4 \Delta_{K\pi} \Delta_{\ell\pi}}{4 M_K^4} + \Big( \<z N_1\>_{t_s} - \<z R_1 \>_{t_s} \Big) \frac{s \sigma_\pi X}{M_K^4} \\
			&\quad - \Big( \<z^2 N_1\>_{t_s} - \<z^2 R_1 \>_{t_s} \Big) \frac{3 s^2 - 4 s \Sigma_0 + \Sigma_0^2 + 4 \sigma_\pi^2 X^2 -4 \Delta_{K\pi} \Delta_{\ell\pi}}{4 M_K^4} \\
			&\quad - \Big( \<z^3 N_1\>_{t_s} - \<z^3 R_1 \>_{t_s} \Big) \frac{s \sigma_\pi X}{M_K^4} + \Big( \<z^4 N_1\>_{t_s} - \<z^4 R_1 \>_{t_s} \Big) \frac{\sigma_\pi^2 X^2}{M_K^4} \\
			&\quad + \Big( \<\tilde N_1\>_{t_s} - \<\tilde R_1 \>_{t_s} \Big) \frac{2 \Delta_{K\pi} - s + \Sigma_0}{4 M_K^2} - \Big( \<z \tilde N_1\>_{t_s} - \<z \tilde R_1 \>_{t_s} \Big) \frac{\sigma_\pi X}{2 M_K^2} \\
			&\quad - \Big( \<z^2 \tilde N_1\>_{t_s} - \<z^2 \tilde R_1 \>_{t_s} \Big) \frac{2 \Delta_{K\pi} - s + \Sigma_0}{4 M_K^2} + \Big( \<z^3 \tilde N_1\>_{t_s} - \<z^3 \tilde R_1 \>_{t_s} \Big) \frac{\sigma_\pi X}{2 M_K^2} , \\
	\end{split}
\end{align}
\begin{align}
	\footnotesize
	\begin{split}
		\hat N_0(t) &= \< M_0 \>_{s_t} \frac{\Delta_{K\pi}+t}{4 t}  - \< z M_0\>_{s_t} \frac{\lambda_{K\pi}^{1/2}(t) \left(\Delta_{\ell\pi} + t\right)}{4 t \lambda_{\ell\pi}^{1/2}(t)} + \< M_1 \>_{s_t} \frac{\left(\Delta_{K\pi} + t\right) \left(\Delta_{K\pi} \Delta_{\ell\pi} + t \left(\Sigma_0 - 3 t\right)\right)}{4 t^2 M_K^2} \\
			&\quad - \< z M_1 \>_{s_t} \frac{\lambda_{K\pi}^{1/2}(t) \left(\Delta_{K\pi} \left(\lambda_{\ell\pi}(t) + \Delta_{\ell\pi} \left(\Delta_{\ell\pi} + t\right)\right) + t \left(\lambda_{\ell\pi}(t) + \left(\Sigma_0-3 t\right) \left(\Delta_{\ell\pi}+t\right)\right)\right)}{4 t^2 M_K^2 \lambda_{\ell\pi}^{1/2}(t)} \\
			&\quad + \< z^2 M_1 \>_{s_t} \frac{\lambda_{K\pi}(t) \left(\Delta_{\ell\pi}+t\right)}{4 t^2 M_K^2} + \< \tilde M_1 \>_{s_t} \frac{\Delta_{K\pi}-3t}{2t}  - \< z \tilde M_1 \>_{s_t} \frac{\lambda_{K\pi}^{1/2}(t) \left(\Delta_{\ell\pi}+t\right)}{2 t \lambda_{\ell\pi}^{1/2}(t)} \\
			&\quad + \Big( \< N_0 \>_{u_t} - 4 \< R_0 \>_{u_t} \Big) \frac{t-\Delta_{K\pi}}{6 t}  + \Big( \< z N_0 \>_{u_t} - 4 \< z R_0 \>_{u_t} \Big) \frac{ \lambda_{K\pi}^{1/2}(t) \left(\Delta_{\ell\pi}+t\right)}{6 t \lambda_{\ell\pi}^{1/2}(t)} \\
			&\quad + \Big( \< N_1 \>_{u_t} - 4 \< R_1 \>_{u_t} \Big) \frac{\left(\Delta_{K\pi}-t\right) \left(\Delta_{K\pi} \Delta_{\ell\pi}+t \left(t-\Sigma_0\right)\right)\left(\Delta_{K\pi} \Delta_{\ell\pi}+t \left(\Sigma_0-3 t\right)\right)}{24 t^3 M_K^4} \\
			&\quad - \Big( \< z N_1 \>_{u_t} - 4 \< z R_1 \>_{u_t} \Big) \begin{aligned}[t]
				& \Bigg( \frac{\lambda_{K\pi}^{1/2}(t) \left(\Delta_{\ell\pi}+t\right) \left(\Delta_{K\pi} \Delta_{\ell\pi}+t \left(t-\Sigma_0\right)\right) \left(\Delta_{K\pi} \Delta_{\ell\pi}+t \left(\Sigma_0-3 t\right)\right)}{24 t^3 M_K^4 \lambda_{\ell\pi}^{1/2}(t) } \\
				& + \frac{\left(\Delta_{K\pi}-t\right) \lambda_{K\pi}^{1/2}(t) \lambda_{\ell\pi}^{1/2}(t) \left(\Delta_{K\pi} \Delta_{\ell\pi}+t^2\right)}{12 t^3 M_K^4} \Bigg) \end{aligned} \\
			&\quad + \Big( \< z^2 N_1 \>_{u_t} - 4 \< z^2 R_1 \>_{u_t} \Big) \Bigg( \frac{\left(\Delta_{K\pi}-t\right) \lambda_{K\pi}(t) \lambda_{\ell\pi}(t)}{24 t^3 M_K^4}+\frac{\lambda_{K\pi}(t) \left(\Delta_{\ell\pi}+t\right) \left(\Delta_{K\pi} \Delta_{\ell\pi}+t^2\right)}{12 t^3 M_K^4} \Bigg) \\
			&\quad - \Big( \< z^3 N_1 \>_{u_t} - 4 \< z^3 R_1 \>_{u_t} \Big) \frac{\lambda_{K\pi}^{3/2}(t) \lambda_{\ell\pi}^{1/2}(t) \left(\Delta_{\ell\pi}+t\right)}{24 t^3 M_K^4} \\
			&\quad - \Big( \< \tilde N_1 \>_{u_t} - 4 \< \tilde R_1 \>_{u_t} \Big) \frac{t \Delta_{K\pi} \left(3 \Delta_{\ell\pi}+\Sigma_0+t\right)+\Delta_{K\pi}^2 \left(\Delta_{\ell\pi}-2 t\right)+3 t^2 \left(\Sigma_0-t\right)}{24 t^2 M_K^2} \\
			&\quad + \Big( \< z \tilde N_1 \>_{u_t} - 4 \< z \tilde R_1 \>_{u_t} \Big) \begin{aligned}[t]
				& \Bigg( \frac{\lambda_{K\pi}^{1/2}(t) \left(\Delta_{\ell\pi}+t\right) \left(\Delta_{K\pi} \left(\Delta_{\ell\pi}-2 t\right)+t \left(\Sigma_0-t\right)\right)}{24 t^2 M_K^2 \lambda_{\ell\pi}^{1/2}(t)} \\
				& +\frac{\left(\Delta_{K\pi}+3 t\right) \lambda_{K\pi}^{1/2}(t) \lambda_{\ell\pi}^{1/2}(t)}{24 t^2 M_K^2} \Bigg) \end{aligned} \\
			&\quad - \Big( \< z^2 \tilde N_1 \>_{u_t} - 4 \< z^2 \tilde R_1 \>_{u_t} \Big) \frac{\lambda_{K\pi}(t) \left(\Delta_{\ell\pi}+t\right)}{24 t^2 M_K^2} , \\
	\end{split}
\end{align}

\begin{align}
	\footnotesize
	\begin{split}
		\hat N_1(t) &= \< M_0 \>_{s_t} \frac{3 M_K^4 \left(\Delta_{\ell\pi}+t\right)}{8 t \lambda_{\ell\pi}(t)} + \< z M_0 \>_{s_t} \frac{3 M_K^4 \left(\Delta_{K\pi}+t\right)}{4 t \lambda_{K\pi}^{1/2}(t) \lambda_{\ell\pi}^{1/2}(t)}  - \< z^2 M_0 \>_{s_t} \frac{9 M_K^4 \left(\Delta_{\ell\pi}+t\right)}{8 t \lambda_{\ell\pi}(t)} \\
			&\quad + \< M_1 \>_{s_t} \frac{3 M_K^2 \left(\Delta_{\ell\pi}+t\right) \left(\Delta_{K\pi} \Delta_{\ell\pi}+t \left(\Sigma_0-3 t\right)\right)}{8 t^2 \lambda_{\ell\pi}(t)} \\
			&\quad + \< z M_1 \>_{s_t} \Bigg( \frac{3 M_K^2 \left(\Delta_{K\pi}+t\right) \left(\Delta_{K\pi} \Delta_{\ell\pi}+t \left(\Sigma_0-3 t\right)\right)}{4 t^2 \lambda_{K\pi}^{1/2}(t) \lambda_{\ell\pi}^{1/2}(t)}-\frac{3 M_K^2 \lambda_{K\pi}^{1/2}(t) \left(\Delta_{\ell\pi}+t\right)}{8 t^2 \lambda_{\ell\pi}^{1/2}(t)} \Bigg) \\
			&\quad - \< z^2 M_1 \>_{s_t} \Bigg( \frac{9 M_K^2 \left(\Delta_{\ell\pi}+t\right) \left(\Delta_{K\pi} \Delta_{\ell\pi}+t \left(\Sigma_0-3t\right)\right)}{8 t^2 \lambda_{\ell\pi}(t)} + \frac{3 M_K^2 \left(\Delta_{K\pi}+t\right)}{4 t^2} \Bigg) \\
			&\quad + \< z^3 M_1 \>_{s_t} \frac{9 M_K^2 \lambda_{K\pi}^{1/2}(t) \left(\Delta_{\ell\pi}+t\right)}{8 t^2 \lambda_{\ell\pi}^{1/2}(t)} + \< \tilde M_1 \>_{s_t} \frac{3 M_K^4 \left(\Delta_{\ell\pi}+t\right)}{4 t \lambda_{\ell\pi}(t)} + \< z \tilde M_1 \>_{s_t} \frac{3 M_K^4 \left(\Delta_{K\pi}-3 t\right)}{2 t \lambda_{K\pi}^{1/2}(t) \lambda_{\ell\pi}^{1/2}(t)} \\
			&\quad - \< z^2 \tilde M_1 \>_{s_t} \frac{9 M_K^4 \left(\Delta_{\ell\pi}+t\right)}{4 t \lambda_{\ell\pi}(t)} - \Big( \< N_0 \>_{u_t} - 4 \< R_0 \>_{u_t} \Big) \frac{M_K^4 \left(\Delta_{\ell\pi}+t\right)}{4 t \lambda_{\ell\pi}(t)} \\
			&\quad + \Big( \< z N_0 \>_{u_t} - 4 \< z R_0 \>_{u_t} \Big) \frac{M_K^4 \left(t-\Delta_{K\pi}\right)}{2 t \lambda_{K\pi}^{1/2}(t) \lambda_{\ell\pi}^{1/2}(t) } + \Big( \< z^2 N_0 \>_{u_t} - 4 \< z^2 R_0 \>_{u_t} \Big) \frac{3 M_K^4 \left(\Delta_{\ell\pi}+t\right)}{4 t \lambda_{\ell\pi}(t)} \\
			&\quad + \Big( \< N_1 \>_{u_t} - 4 \< R_1 \>_{u_t} \Big) \frac{\left(\Delta_{\ell\pi}+t\right) \left(\Delta_{K\pi} \Delta_{\ell\pi}+t \left(t-\Sigma_0\right)\right)
   \left(\Delta_{K\pi} \Delta_{\ell\pi}+t \left(\Sigma_0-3 t\right)\right)}{16 t^3 \lambda_{\ell\pi}(t)} \\
			&\quad + \Big( \< z N_1 \>_{u_t} - 4 \< z R_1 \>_{u_t} \Big) \begin{aligned}[t]
				& \Bigg( \frac{\left(\Delta_{K\pi}-t\right) \left(\Delta_{K\pi} \Delta_{\ell\pi}+t \left(t-\Sigma_0\right)\right) \left(\Delta_{K\pi} \Delta_{\ell\pi}+t \left(\Sigma_0-3 t\right)\right)}{8 t^3 \lambda_{K\pi}^{1/2}(t) \lambda_{\ell\pi}^{1/2}(t)} \\
				& -\frac{\lambda_{K\pi}^{1/2}(t) \left(\Delta_{\ell\pi}+t\right) \left(\Delta_{K\pi} \Delta_{\ell\pi}+t^2\right)}{8 t^3 \lambda_{\ell\pi}^{1/2}(t) }  \Bigg) \end{aligned} \\
			&\quad - \Big( \< z^2 N_1 \>_{u_t} - 4 \< z^2 R_1 \>_{u_t} \Big) \begin{aligned}[t]
				& \Bigg( \frac{3 \left(\Delta_{\ell\pi}+t\right) \left(\Delta_{K\pi} \Delta_{\ell\pi}+t \left(t-\Sigma_0\right)\right) \left(\Delta_{K\pi} \Delta_{\ell\pi}+t \left(\Sigma_0-3 t\right)\right)}{16 t^3 \lambda_{\ell\pi}(t)} \\
				& - \frac{\lambda_{K\pi}(t) \left(\Delta_{\ell\pi}+t\right)}{16 t^3} + \frac{\left(\Delta_{K\pi}-t\right) \left(\Delta_{K\pi} \Delta_{\ell\pi}+t^2\right)}{4 t^3} \Bigg) \end{aligned} \\
			&\quad + \Big( \< z^3 N_1 \>_{u_t} - 4 \< z^3 R_1 \>_{u_t} \Big) \Bigg( \frac{\left(\Delta_{K\pi}-t\right) \lambda_{K\pi}^{1/2}(t) \lambda_{\ell\pi}^{1/2}(t) }{8 t^3}+\frac{3 \lambda_{K\pi}^{1/2}(t) \left(\Delta_{\ell\pi}+t\right) \left(\Delta_{K\pi} \Delta_{\ell\pi}+t^2\right)}{8 t^3 \lambda_{\ell\pi}^{1/2}(t)} \Bigg) \\
			&\quad - \Big( \< z^4 N_1 \>_{u_t} - 4 \< z^4 R_1 \>_{u_t} \Big) \frac{3 \lambda_{K\pi}(t) \left(\Delta_{\ell\pi}+t\right)}{16 t^3} \\
			&\quad + \Big( \< \tilde N_1 \>_{u_t} - 4 \< \tilde R_1 \>_{u_t} \Big) \frac{M_K^2 \left(\Delta_{\ell\pi}+t\right) \left(t^2+2 t \Delta_{K\pi} -\Sigma_0 t-\Delta_{K\pi} \Delta_{\ell\pi}\right)}{16 t^2 \lambda_{\ell\pi}(t)} \\
			&\quad + \Big( \< z \tilde N_1 \>_{u_t} - 4 \< z \tilde R_1 \>_{u_t} \Big) \begin{aligned}[t]
				& \Bigg( \frac{M_K^2 \lambda_{K\pi}^{1/2}(t) \left(\Delta_{\ell\pi}+t\right)}{16 t^2 \lambda_{\ell\pi}^{1/2}(t) } \\
				& -\frac{M_K^2 \left(t \Delta_{K\pi} \left(3 \Delta_{\ell\pi}+\Sigma_0+t\right)+\Delta_{K\pi}^2 \left(\Delta_{\ell\pi}-2 t\right)+3 t^2 \left(\Sigma_0-t\right)\right)}{8 t^2 \lambda_{K\pi}^{1/2}(t) \lambda_{\ell\pi}^{1/2}(t) } \Bigg) \end{aligned} \\
			&\quad + \Big( \< z^2 \tilde N_1 \>_{u_t} - 4 \< z^2 \tilde R_1 \>_{u_t} \Big) \begin{aligned}[t]
				& \Bigg( \frac{3 M_K^2 \left(\Delta_{\ell\pi}+t\right) \left(\Delta_{K\pi} \left(\Delta_{\ell\pi}-2 t\right)+t \left(\Sigma_0-t\right)\right)}{16 t^2 \lambda_{\ell\pi}(t)} \\
				& +\frac{M_K^2 \left(\Delta_{K\pi}+3 t\right)}{8 t^2} \Bigg) \end{aligned} \\
			&\quad - \Big( \< z^3 \tilde N_1 \>_{u_t} - 4 \< z^3 \tilde R_1 \>_{u_t} \Big) \frac{3 M_K^2 \lambda_{K\pi}^{1/2}(t) \left(\Delta_{\ell\pi}+t\right)}{16 t^2 \lambda_{\ell\pi}^{1/2}(t) } , \\
	\end{split}
\end{align}

\begin{align}
	\footnotesize
	\begin{split}
		\hat{\tilde N}_1(t) &= \< (1-z^2) M_0 \>_{s_t} \frac{3 M_K^2}{4 t} + \< (1-z^2) M_1 \>_{s_t} \frac{3 \left(\Delta_{K\pi} \Delta_{\ell\pi}+t \left(\Sigma_0 - 3 t\right)\right)}{4 t^2} \\
			&\quad - \< (1-z^2) z M_1 \>_{s_t} \frac{3 \lambda_{K\pi}^{1/2}(t) \lambda_{\ell\pi}^{1/2}(t)}{4 t^2} + \< (1-z^2) \tilde M_1 \>_{s_t} \frac{3 M_K^2}{2 t} \\
			&\quad - \Big( \< (1-z^2) N_0 \>_{u_t} - 4 \< (1-z^2) R_0 \>_{u_t} \Big) \frac{M_K^2}{2 t} \\
			&\quad + \Big( \< (1-z^2) N_1 \>_{u_t} - 4 \< (1-z^2) R_1 \>_{u_t} \Big) \frac{\left(\Delta_{K\pi} \Delta_{\ell\pi}+t \left(t-\Sigma_0\right)\right) \left(\Delta_{K\pi} \Delta_{\ell\pi}+t \left(\Sigma_0-3 t\right)\right)}{8 t^3 M_K^2} \\
			&\quad - \Big( \< (1-z^2) z N_1 \>_{u_t} - 4 \< (1-z^2) z R_1 \>_{u_t} \Big) \frac{\lambda_{K\pi}^{1/2}(t) \lambda_{\ell\pi}^{1/2}(t) \left(\Delta_{K\pi} \Delta_{\ell\pi}+t^2\right)}{4 t^3 M_K^2} \\
			&\quad + \Big( \< (1-z^2) z^2 N_1 \>_{u_t} - 4 \< (1-z^2) z^2 R_1 \>_{u_t} \Big) \frac{\lambda_{K\pi}(t) \lambda_{\ell\pi}(t)}{8 t^3 M_K^2} \\
			&\quad + \Big( \< (1-z^2) \tilde N_1 \>_{u_t} - 4 \< (1-z^2) \tilde R_1 \>_{u_t} \Big) \frac{t^2 + 2 t \Delta_{K\pi}-\Sigma_0 t-\Delta_{K\pi} \Delta_{\ell\pi}}{8 t^2} \\
			&\quad + \Big( \< (1-z^2) z \tilde N_1 \>_{u_t} - 4 \< (1-z^2) z \tilde R_1 \>_{u_t} \Big) \frac{\lambda_{K\pi}^{1/2}(t) \lambda_{\ell\pi}^{1/2}(t) }{8 t^2} , \\
	\end{split}
\end{align}

\begin{align}
	\footnotesize
	\begin{split}
		\hat R_0(u) &= \< M_0 \>_{s_u}  \frac{u+\Delta_{K\pi}}{4 u} - \< z M_0 \>_{s_u} \frac{\left(u+\Delta_{\ell\pi}\right) \lambda_{K\pi}^{1/2}(u)}{4 u \lambda_{\ell\pi}^{1/2}(u)} - \< M_1 \>_{s_u} \frac{\left(u+\Delta_{K\pi}\right) \left(u \left(\Sigma_0-3 u\right)+\Delta_{K\pi} \Delta_{\ell\pi}\right)}{8u^2 M_K^2} \\
			&\quad + \< z M_1 \>_{s_u} \Bigg( \frac{\lambda_{K\pi}^{1/2}(u) \lambda_{\ell\pi}^{1/2}(u) \left(u+\Delta_{K\pi}\right)}{8 u^2 M_K^2}+\frac{\left(u+\Delta_{\ell\pi}\right) \left(u \left(\Sigma_0-3 u\right)+\Delta_{K\pi} \Delta_{\ell\pi}\right) \lambda_{K\pi}^{1/2}(u)}{8 u^2 M_K^2 \lambda_{\ell\pi}^{1/2}(u)} \Bigg) \\
			&\quad - \< z^2 M_1 \>_{s_u} \frac{\left(u+\Delta_{\ell\pi}\right) \lambda_{K\pi}(u)}{8 u^2 M_K^2} + \< \tilde M_1 \>_{s_u}  \frac{3u-\Delta_{K\pi}}{4 u} + \< z \tilde M_1 \>_{s_u} \frac{\left(u+\Delta_{\ell\pi}\right) \lambda_{K\pi}^{1/2}(u)}{4 u \lambda_{\ell\pi}^{1/2}(u)} \\
			&\quad + \Big( 2 \< N_0 \>_{t_u} + \< R_0 \>_{t_u} \Big) \frac{\Delta_{K\pi} - u}{6u} - \Big( 2 \< z N_0 \>_{t_u} + \< z R_0 \>_{t_u} \Big) \frac{\left(u+\Delta_{\ell\pi}\right) \lambda_{K\pi}^{1/2}(u)}{6 u \lambda_{\ell\pi}^{1/2}(u)} \\
			&\quad - \Big( 2 \< N_1 \>_{t_u} + \< R_1 \>_{t_u} \Big) \frac{\left(\Delta_{K\pi}-u\right) \left(u \left(u-\Sigma_0\right)+\Delta_{K\pi} \Delta_{\ell\pi}\right) \left(u \left(\Sigma_0-3 u\right)+\Delta_{K\pi} \Delta_{\ell\pi}\right)}{24 u^3 M_K^4} \\
			&\quad + \Big( 2 \< z N_1 \>_{t_u} + \< z R_1 \>_{t_u} \Big) \begin{aligned}[t]
				&\Bigg( \frac{\left(\Delta_{K\pi}-u\right) \lambda_{K\pi}^{1/2}(u) \lambda_{\ell\pi}^{1/2}(u) \left(u^2+\Delta_{K\pi} \Delta_{\ell\pi}\right)}{12 u^3 M_K^4} \\
				& \hspace{-0.5cm} + \frac{\left(u+\Delta_{\ell\pi}\right) \left(u \left(u-\Sigma_0\right)+\Delta_{K\pi} \Delta_{\ell\pi}\right) \left(u \left(\Sigma_0-3 u\right)+\Delta_{K\pi} \Delta_{\ell\pi}\right) \lambda_{K\pi}^{1/2}(u)}{24 u^3 M_K^4 \lambda_{\ell\pi}^{1/2}(u)} \Bigg) \end{aligned} \\
			&\quad + \Big( 2 \< z^2 N_1 \>_{t_u} + \< z^2 R_1 \>_{t_u} \Big) \Bigg( \frac{\left(u-\Delta_{K\pi}\right) \lambda_{K\pi}(u) \lambda_{\ell\pi}(u)}{24 u^3 M_K^4}-\frac{\left(u+\Delta_{\ell\pi}\right) \left(u^2+\Delta_{K\pi} \Delta_{\ell\pi}\right) \lambda_{K\pi}(u)}{12 u^3 M_K^4} \Bigg) \\
			&\quad + \Big( 2 \< z^3 N_1 \>_{t_u} + \< z^3 R_1 \>_{t_u} \Big) \frac{\left(u+\Delta_{\ell\pi}\right) \lambda_{K\pi}^{3/2}(u) \lambda_{\ell\pi}^{1/2}(u)}{24 u^3 M_K^4} \\
			&\quad + \Big( 2 \< \tilde N_1 \>_{t_u} + \< \tilde R_1 \>_{t_u} \Big) \frac{3 \left(\Sigma_0-u\right) u^2+\Delta_{K\pi} \left(u+\Sigma_0+3 \Delta_{\ell\pi}\right) u+\Delta_{K\pi}^2 \left(\Delta_{\ell\pi}-2 u\right)}{24 u^2 M_K^2} \\
			&\quad + \Big( 2 \< z \tilde N_1 \>_{t_u} + \< z \tilde R_1 \>_{t_u} \Big) \begin{aligned}[t]
				& \Bigg( \frac{\left(u+\Delta_{\ell\pi}\right) \left(u^2-\Sigma_0 u+2 \Delta_{K\pi} u-\Delta_{K\pi} \Delta_{\ell\pi}\right) \lambda_{K\pi}^{1/2}(u)}{24 u^2 M_K^2 \lambda_{\ell\pi}^{1/2}(u)} \\
				& -\frac{\left(3 u+\Delta_{K\pi}\right) \lambda_{K\pi}^{1/2}(u) \lambda_{\ell\pi}^{1/2}(u)}{24 u^2 M_K^2} \Bigg) \end{aligned} \\
			&\quad + \Big( 2 \< z^2 \tilde N_1 \>_{t_u} + \< z^2 \tilde R_1 \>_{t_u} \Big) \frac{\left(u+\Delta_{\ell\pi}\right) \lambda_{K\pi}(u)}{24 u^2 M_K^2} , \\
	\end{split}
\end{align}

\begin{align}
	\footnotesize
	\begin{split}
		\hat R_1(u) &= \< M_0 \>_{s_u}  \frac{3 M_K^4 \left(u+\Delta_{\ell\pi}\right)}{8 u \lambda_{\ell\pi}(u)} + \< z M_0 \>_{s_u} \frac{3 M_K^4 \left(u+\Delta_{K\pi}\right)}{4 u \lambda_{K\pi}^{1/2}(u) \lambda_{\ell\pi}^{1/2}(u)} - \< z^2 M_0 \>_{s_u} \frac{9 M_K^4 \left(u+\Delta_{\ell\pi}\right)}{8 u \lambda_{\ell\pi}(u)} \\
			&\quad - \< M_1 \>_{s_u} \frac{3 M_K^2 \left(u+\Delta_{\ell\pi}\right) \left(u \left(\Sigma_0-3 u\right)+\Delta_{K\pi} \Delta_{\ell\pi}\right)}{16 u^2 \lambda_{\ell\pi}(u)} \\
			&\quad + \< z M_1 \>_{s_u} \Bigg( \frac{3 M_K^2 \lambda_{K\pi}^{1/2}(u) \left(\Delta_{\ell\pi}+u\right)}{16 u^2 \lambda_{\ell\pi}^{1/2}(u)}-\frac{3 M_K^2 \left(\Delta_{K\pi}+u\right) \left(\Delta_{K\pi} \Delta_{\ell\pi}+u \left(\Sigma_0-3 u\right)\right)}{8u^2 \lambda_{K\pi}^{1/2}(u) \lambda_{\ell\pi}^{1/2}(u) } \Bigg) \\
			&\quad + \< z^2 M_1 \>_{s_u} \Bigg( \frac{9 M_K^2 \left(\Delta_{\ell\pi}+u\right) \left(\Delta_{K\pi} \Delta_{\ell\pi}+u \left(\Sigma_0-3u\right)\right)}{16 u^2 \lambda_{\ell\pi}(u)}+\frac{3 M_K^2 \left(\Delta_{K\pi}+u\right)}{8 u^2} \Bigg) \\
			&\quad - \< z^3 M_1 \>_{s_u} \frac{9 M_K^2 \left(u+\Delta_{\ell\pi}\right) \lambda_{K\pi}^{1/2}(u)}{16 u^2 \lambda_{\ell\pi}^{1/2}(u)} - \< \tilde M_1 \>_{s_u} \frac{3 M_K^4 \left(u+\Delta_{\ell\pi}\right)}{8 u \lambda_{\ell\pi}(u)} - \< z \tilde M_1 \>_{s_u} \frac{3 M_K^4 \left(\Delta_{K\pi}-3 u\right)}{4 u \lambda_{K\pi}^{1/2}(u) \lambda_{\ell\pi}^{1/2}(u)} \\
			&\quad + \< z^2 \tilde M_1 \>_{s_u} \frac{9 M_K^4 \left(u+\Delta_{\ell\pi}\right)}{8 u \lambda_{\ell\pi}(u)} + \Big( 2 \< N_0 \>_{t_u} + \< R_0 \>_{t_u} \Big) \frac{M_K^4 \left(u+\Delta_{\ell\pi}\right)}{4 u \lambda_{\ell\pi}(u)} \\
			&\quad + \Big( 2 \< z N_0 \>_{t_u} + \< z R_0 \>_{t_u} \Big) \frac{M_K^4 \left(\Delta_{K\pi}-u\right)}{2 u \lambda_{K\pi}^{1/2}(u) \lambda_{\ell\pi}^{1/2}(u)} - \Big( 2 \< z^2 N_0 \>_{t_u} + \< z^2 R_0 \>_{t_u} \Big) \frac{3 M_K^4 \left(u+\Delta_{\ell\pi}\right)}{4 u \lambda_{\ell\pi}(u)} \\
			&\quad - \Big( 2 \< N_1 \>_{t_u} + \< R_1 \>_{t_u} \Big) \frac{\left(u+\Delta_{\ell\pi}\right) \left(u \left(u-\Sigma_0\right)+\Delta_{K\pi} \Delta_{\ell\pi}\right) \left(u \left(\Sigma_0-3 u\right)+\Delta_{K\pi} \Delta_{\ell\pi}\right)}{16 u^3 \lambda_{\ell\pi}(u)} \\
			&\quad + \Big( 2 \< z N_1 \>_{t_u} + \< z R_1 \>_{t_u} \Big) \begin{aligned}[t]
				& \Bigg( \frac{\lambda_{K\pi}^{1/2}(u) \left(\Delta_{\ell\pi}+u\right) \left(\Delta_{K\pi} \Delta_{\ell\pi}+u^2\right)}{8u^3 \lambda_{\ell\pi}^{1/2}(u)} \\
				& -\frac{\left(\Delta_{K\pi}-u\right) \left(\Delta_{K\pi} \Delta_{\ell\pi}+u \left(u-\Sigma_0\right)\right) \left(\Delta_{K\pi} \Delta_{\ell\pi}+u \left(\Sigma_0-3 u\right)\right)}{8 u^3 \lambda_{K\pi}^{1/2}(u) \lambda_{\ell\pi}^{1/2}(u)} \Bigg) \end{aligned} \\
			&\quad + \Big( 2 \< z^2 N_1 \>_{t_u} + \< z^2 R_1 \>_{t_u} \Big) \begin{aligned}[t]
				& \Bigg( \frac{3 \left(\Delta_{\ell\pi}+u\right) \left(\Delta_{K\pi} \Delta_{\ell\pi}+u \left(u-\Sigma_0\right)\right) \left(\Delta_{K\pi} \Delta_{\ell\pi}+u \left(\Sigma_0-3 u\right)\right)}{16 u^3 \lambda_{\ell\pi}(u)} \\
				& -\frac{\lambda_{K\pi}(u) \left(\Delta_{\ell\pi}+u\right)}{16 u^3}+\frac{\left(\Delta_{K\pi}-u\right) \left(\Delta_{K\pi} \Delta_{\ell\pi}+u^2\right)}{4 u^3} \Bigg) \end{aligned} \\
			&\quad + \Big( 2 \< z^3 N_1 \>_{t_u} + \< z^3 R_1 \>_{t_u} \Big) \begin{aligned}[t]
				& \Bigg( \frac{\left(u-\Delta_{K\pi}\right) \lambda_{K\pi}^{1/2}(u) \lambda_{\ell\pi}^{1/2}(u)}{8 u^3} \\
				& -\frac{3 \lambda_{K\pi}^{1/2}(u) \left(\Delta_{\ell\pi}+u\right) \left(\Delta_{K\pi} \Delta_{\ell\pi}+u^2\right)}{8 u^3 \lambda_{\ell\pi}^{1/2}(u)} \Bigg) \end{aligned} \\
			&\quad + \Big( 2 \< z^4 N_1 \>_{t_u} + \< z^4 R_1 \>_{t_u} \Big) \frac{3 \left(u+\Delta_{\ell\pi}\right) \lambda_{K\pi}(u)}{16 u^3} \\
			&\quad + \Big( 2 \< \tilde N_1 \>_{t_u} + \< \tilde R_1 \>_{t_u} \Big) \frac{M_K^2 \left(u+\Delta_{\ell\pi}\right) \left(u \left(\Sigma_0-u\right)+\Delta_{K\pi} \left(\Delta_{\ell\pi}-2u\right)\right)}{16 u^2 \lambda_{\ell\pi}(u)} \\
			&\quad + \Big( 2 \< z \tilde N_1 \>_{t_u} + \< z \tilde R_1 \>_{t_u} \Big) \begin{aligned}[t]
				&\Bigg( \frac{M_K^2 \left(u \Delta_{K\pi} \left(3 \Delta_{\ell\pi}+\Sigma_0+u\right)+\Delta_{K\pi}^2 \left(\Delta_{\ell\pi}-2 u\right)+3 u^2 \left(\Sigma_0-u\right)\right)}{8 u^2 \lambda_{K\pi}^{1/2}(u) \lambda_{\ell\pi}^{1/2}(u)} \\
				& -\frac{M_K^2 \lambda_{K\pi}^{1/2}(u) \left(\Delta_{\ell\pi}+u\right)}{16 u^2 \lambda_{\ell\pi}^{1/2}(u)} \Bigg) \end{aligned} \\
			&\quad + \Big( 2 \< z^2 \tilde N_1 \>_{t_u} + \< z^2 \tilde R_1 \>_{t_u} \Big) \begin{aligned}[t]
				& \Bigg( \frac{3 M_K^2 \left(\Delta_{\ell\pi}+u\right) \left(-\Delta_{K\pi} \Delta_{\ell\pi}+2 u \Delta_{K\pi}+u^2-\Sigma_0 u\right)}{16 u^2 \lambda_{\ell\pi}(u)} \\
				& -\frac{M_K^2 \left(\Delta_{K\pi}+3 u\right)}{8 u^2} \Bigg) \end{aligned} \\
			&\quad + \Big( 2 \< z^3 \tilde N_1 \>_{t_u} + \< z^3 \tilde R_1 \>_{t_u} \Big) \frac{3 M_K^2 \lambda_{K\pi}^{1/2}(u) \left(\Delta_{\ell\pi}+u\right)}{16 u^2 \lambda_{\ell\pi}^{1/2}(u)} , \\
	\end{split}
\end{align}

\begin{align}
	\footnotesize
	\begin{split}
		\hat{\tilde R}_1(u) &= \< (1-z^2) M_0 \>_{s_u} \frac{3 M_K^2}{4 u} - \< (1-z^2) M_1 \>_{s_u} \frac{3 \left(\Delta_{K\pi} \Delta_{\ell\pi}+u \left(\Sigma_0-3 u\right)\right)}{8 u^2} \\
			&\quad + \< (1-z^2) z M_1 \>_{s_u} \frac{3 \lambda_{K\pi}^{1/2}(u) \lambda_{\ell\pi}^{1/2}(u) }{8 u^2} - \< (1-z^2) \tilde M_1 \>_{s_u} \frac{3 M_K^2}{4 u} \\
			&\quad + \Big( 2 \< (1-z^2) N_0 \>_{t_u} + \< (1-z^2) R_0 \>_{t_u} \Big) \frac{M_K^2}{2 u} \\
			&\quad - \Big( 2 \< (1-z^2) N_1 \>_{t_u} + \< (1-z^2) R_1 \>_{t_u} \Big) \frac{\left(\Delta_{K\pi} \Delta_{\ell\pi}+u \left(u-\Sigma_0\right)\right) \left(\Delta_{K\pi} \Delta_{\ell\pi}+u \left(\Sigma_0-3 u\right)\right)}{8 u^3 M_K^2} \\
			&\quad + \Big( 2 \< (1-z^2) z N_1 \>_{t_u} + \< (1-z^2) z R_1 \>_{t_u} \Big) \frac{\lambda_{K\pi}^{1/2}(u) \lambda_{\ell\pi}^{1/2}(u) \left(\Delta_{K\pi} \Delta_{\ell\pi}+u^2\right)}{4 u^3 M_K^2} \\
			&\quad - \Big( 2 \< (1-z^2) z^2 N_1 \>_{t_u} + \< (1-z^2) z^2 R_1 \>_{t_u} \Big) \frac{\lambda_{K\pi}(u) \lambda_{\ell\pi}(u)}{8 u^3 M_K^2} \\
			&\quad + \Big( 2 \< (1-z^2) \tilde N_1 \>_{t_u} + \< (1-z^2) \tilde R_1 \>_{t_u} \Big) \frac{\Delta_{K\pi} \left(\Delta_{\ell\pi}-2 u\right)+u \left(\Sigma_0-u\right)}{8 u^2} \\
			&\quad - \Big( 2 \< (1-z^2) z \tilde N_1 \>_{t_u} + \< (1-z^2) z \tilde R_1 \>_{t_u} \Big) \frac{\lambda_{K\pi}^{1/2}(u) \lambda_{\ell\pi}^{1/2}(u)}{8 u^2} , \\
	\end{split}
\end{align}
where
\begin{align}
	\begin{split}
		\< z^n X \>_{t_s} := \frac{1}{2} \int_{-1}^1 z^n X(t(s,z)) dz , \\
		\< z^n X \>_{s_t} := \frac{1}{2} \int_{-1}^1 z^n X(s(t,z)) dz , \\
		\< z^n X \>_{u_t} := \frac{1}{2} \int_{-1}^1 z^n X(u(t,z)) dz , \\
		\< z^n X \>_{s_u} := \frac{1}{2} \int_{-1}^1 z^n X(s(u,z)) dz , \\
		\< z^n X \>_{t_u} := \frac{1}{2} \int_{-1}^1 z^n X(t(u,z)) dz , \\
	\end{split}
\end{align}
and
\begin{align}
	\begin{split}
		t(s,z) &= \frac{1}{2} \left( \Sigma_0 - s - 2 X \sigma_\pi z \right) , \\
		s(t,z) &= \frac{1}{2} \left( \Sigma_0 - t + \frac{1}{t} \left( z \, \lambda^{1/2}_{K\pi}(t) \lambda^{1/2}_{\ell\pi}(t) - \Delta_{K\pi}\Delta_{\ell\pi} \right) \right) , \\
		u(t,z) &= \frac{1}{2} \left( \Sigma_0 - t - \frac{1}{t} \left( z \, \lambda^{1/2}_{K\pi}(t) \lambda^{1/2}_{\ell\pi}(t) - \Delta_{K\pi}\Delta_{\ell\pi} \right) \right) , \\
		s(u,z) &= \frac{1}{2} \left( \Sigma_0 - u + \frac{1}{u} \left( z \, \lambda^{1/2}_{K\pi}(u) \lambda^{1/2}_{\ell\pi}(u) - \Delta_{K\pi}\Delta_{\ell\pi} \right) \right) , \\
		t(u,z) &= \frac{1}{2} \left( \Sigma_0 - u - \frac{1}{u} \left( z \, \lambda^{1/2}_{K\pi}(u) \lambda^{1/2}_{\ell\pi}(u) - \Delta_{K\pi}\Delta_{\ell\pi} \right) \right) .
	\end{split}
\end{align}
We recall the abbreviations
\begin{align}
	\begin{split}
		\Delta_{K\pi} &= M_K^2 - M_\pi^2, \quad \Delta_{\ell\pi} = s_\ell - M_\pi^2, \quad \Sigma_0 = M_K^2 + 2 M_\pi^2 + s_\ell , \\
		PL &= \frac{1}{2}(M_K^2 - s - s_\ell) , \quad X = \frac{1}{2} \lambda^{1/2}(M_K^2, s, s_\ell), \quad \sigma_\pi = \sqrt{1 - \frac{4M_\pi^2}{s}} .
	\end{split}
\end{align}



\chapter{Matching Equations}

\section{Subtraction Constants at $\O(p^4)$ in \ChPT{}}

\label{sec:AppendixNLOSubtractionConstantsStandardRep}

In the following expressions for the subtraction constants at NLO, we have used the Gell-Mann--Okubo (GMO) formula $M_\eta^2 = (4 M_K^2 - M_\pi^2)/3$ to simplify the analytic expressions considerably. This introduces an error only at NNLO. In practise, we use the physical $\eta$ mass and not the GMO relation. We do not show the analytic expressions for this case because they are much larger.
\begin{align}
	\footnotesize
	\begin{split}
		m_{0,\mathrm{NLO}}^0 &= \frac{M_K}{\sqrt{2} F_\pi} \Bigg( 1 + \frac{1}{F_\pi^2} \begin{aligned}[t]
			&\bigg(-64 L_1^r M_\pi^2 + 16 L_2^r (M_K^2 + M_\pi^2) + 4 L_3^r (M_K^2 - 3 M_\pi^2) + 32 L_4^r M_\pi^2 + 4 L_5^r M_\pi^2 + 2 L_9^r s_\ell \\
   			& - \frac{161 M_K^6 + 42 M_K^4 M_\pi^2 - 27 M_K^2 M_\pi^4 + 4 M_\pi^6}{384 \pi^2 \Delta_{K\pi}^2} - s_\ell \frac{73 M_K^4 - 14 M_K^2 M_\pi^2 + M_\pi^4}{384 \pi^2 \Delta_{K\pi}^2} \\
			& +  \ln\left(\frac{M_\pi^2}{\mu^2}\right) \bigg(\frac{3 M_\pi^2 (3 M_K^6 - 8 M_K^4 M_\pi^2 + 2 M_K^2 M_\pi^4 + M_\pi^6)}{128 \pi^2 \Delta_{K\pi}^3} - s_\ell \frac{M_\pi^4 (3 M_K^2 - M_\pi^2)}{128 \pi^2 \Delta_{K\pi}^3}\bigg)\\
			& - \ln\left(\frac{M_K^2}{\mu^2}\right) \bigg(\frac{M_K^2 (92 M_K^6 - 15 M_K^2 M_\pi^4 + M_\pi^6)}{64 \pi^2 \Delta_{K\pi}^3} + s_\ell \frac{M_K^4 (41 M_K^2 - 15 M_\pi^2)}{64 \pi^2 \Delta_{K\pi}^3}\bigg) \\
			& +  \ln\left(\frac{M_\eta^2}{\mu^2}\right) \begin{aligned}[t]
				& \bigg(\frac{172 M_K^8 + 17 M_K^6 M_\pi^2 - 12 M_K^4 M_\pi^4 - 22 M_K^2 M_\pi^6 + 7 M_\pi^8}{128 \pi^2 \Delta_{K\pi}^3} \\
				& + s_\ell \frac{(4 M_K^2 - M_\pi^2)^2 (5 M_K^2 + M_\pi^2)}{128 \pi^2 \Delta_{K\pi}^3}\bigg) \bigg) \Bigg) , \end{aligned} \end{aligned}
	\end{split}
\end{align}
\begin{align}
	\footnotesize
	\begin{split}
		m_{0,\mathrm{NLO}}^1 &=  \frac{M_K}{\sqrt{2} F_\pi^3} \begin{aligned}[t]
			& \Bigg( 32 L_1^r M_K^2 + 8 L_3^r M_K^2 \\
			& + \frac{M_K^2 (116 M_K^6 + 273 M_K^4 M_\pi^2 - 258 M_K^2 M_\pi^4 + 49 M_\pi^6)}{384 \pi^2 \Delta_{K\pi}^2 (4 M_K^2 - M_\pi^2)} \\
			& - \ln\left(\frac{M_\pi^2}{\mu^2}\right) \frac{M_K^2 (8 M_K^6 - 24 M_K^4 M_\pi^2 + 21 M_K^2 M_\pi^4 - 7 M_\pi^6)}{128 \pi^2 \Delta_{K\pi}^3} \\
			& + \ln\left(\frac{M_K^2}{\mu^2}\right) \frac{M_K^2 (38 M_K^6 - 6 M_K^4 M_\pi^2 - 9 M_K^2 M_\pi^4 + 3 M_\pi^6)}{64 \pi^2 \Delta_{K\pi}^3} \\
			& - \ln\left( \frac{M_\eta^2}{\mu^2}\right) \frac{M_K^2 (4 M_K^2 - M_\pi^2)^2 (5 M_K^2 + M_\pi^2)}{128 \pi^2 \Delta_{K\pi}^3} \Bigg) , \end{aligned} \\
	\end{split}
\end{align}
\begin{align}
	\footnotesize
	\begin{split}
		m_{1,\mathrm{NLO}}^0 &= \frac{M_K}{\sqrt{2} F_\pi^3} \begin{aligned}[t]
			& \Bigg( -8 L_2^r M_K^2 \\
			& + \frac{M_K^2 (79 M_K^4 - 2 M_K^2 M_\pi^2 + 7 M_\pi^4)}{384 \pi^2 \Delta_{K\pi}^2} \\
			& + \ln\left(\frac{M_\pi^2}{\mu^2}\right) \frac{5 M_K^2 M_\pi^4 (3 M_K^2 - M_\pi^2)}{128 \pi^2 \Delta_{K\pi}^3} \\
			& + \ln\left(\frac{M_K^2}{\mu^2}\right) \frac{M_K^6 (43 M_K^2 - 21 M_\pi^2)}{64 \pi^2 \Delta_{K\pi}^3} \\
			& - \ln\left(\frac{M_\eta^2}{\mu^2}\right) \frac{M_K^2 (4 M_K^2 - M_\pi^2)^2 (5 M_K^2 + M_\pi^2)}{128 \pi^2 \Delta_{K\pi}^3} \Bigg) , \end{aligned} \\
	\end{split}
\end{align}
\begin{align}
	\footnotesize
	\begin{split}
		\tilde m_{1,\mathrm{NLO}}^0 &= \frac{M_K}{\sqrt{2} F_\pi} \Bigg( 1 + \frac{1}{F_\pi^2} \begin{aligned}[t]
			& \bigg(  - 8 L_2^r (M_K^2 + 2 M_\pi^2 + s_\ell) - 4 L_3^r (M_K^2 + M_\pi^2) + 4 L_5^r M_\pi^2 + 2 L_9^r s_\ell \\
			& + \frac{16 M_K^6 - 3 M_K^4 M_\pi^2 + 3 M_K^2 M_\pi^4 + 2 M_\pi^6}{96 \pi^2 \Delta_{K\pi}^2} + s_\ell \frac{(M_K^2 + M_\pi^2)^2}{64 \pi^2 \Delta_{K\pi}^2} \\
			& - \ln\left(\frac{M_\pi^2}{\mu^2}\right) \bigg(\frac{M_\pi^2 (3 M_K^2 - M_\pi^2) (M_K^4 - 8 M_K^2 M_\pi^2 - 5 M_\pi^4)}{128 \pi^2 \Delta_{K\pi}^3} -  s_\ell \frac{M_\pi^4 (3 M_K^2 - M_\pi^2)}{32 \pi^2 \Delta_{K\pi}^3}\bigg) \\
			& + \ln\left(\frac{M_K^2}{\mu^2}\right) \bigg(\frac{M_K^2 (37 M_K^6 - 35 M_K^4 M_\pi^2 - 17 M_K^2 M_\pi^4 + 3 M_\pi^6)}{64 \pi^2 \Delta_{K\pi}^3} + s_\ell \frac{M_K^4 (M_K^2 - 3 M_\pi^2)}{32 \pi^2 \Delta_{K\pi}^3}\bigg) \\
			& - \ln\left(\frac{M_\eta^2}{\mu^2}\right) \frac{68 M_K^6 + 7 M_K^4 M_\pi^2 - 2 M_K^2 M_\pi^4 - M_\pi^6}{128 \pi^2 \Delta_{K\pi}^2} \bigg) \Bigg) , \end{aligned}
	\end{split}
\end{align}
\begin{align}
	\footnotesize
	\begin{split}
		\tilde m_{1,\mathrm{NLO}}^1 &= \frac{M_K}{\sqrt{2} F_\pi^3} \begin{aligned}[t]
			& \Bigg(8 L_2^r M_K^2 \\
			& - \frac{M_K^2 (M_K^4 + M_\pi^4)}{32 \pi^2 \Delta_{K\pi}^2} \\
			& - \ln\left(\frac{M_\pi^2}{\mu^2}\right) \frac{M_K^2 (M_K^6 - 3 M_K^4 M_\pi^2 + 12 M_K^2 M_\pi^4 - 4 M_\pi^6)}{96 \pi^2 \Delta_{K\pi}^3} \\
			& - \ln\left(\frac{M_K^2}{\mu^2}\right) \frac{M_K^2 (7 M_K^6 - 21 M_K^4 M_\pi^2 + 3 M_K^2 M_\pi^4 - M_\pi^6)}{192 \pi^2 \Delta_{K\pi}^3} \Bigg) , \end{aligned} \\
	\end{split}
\end{align}
\begin{align}
	\footnotesize
	\begin{split}
		n_{0,\mathrm{NLO}}^1 &= \frac{M_K}{\sqrt{2} F_\pi^3} \begin{aligned}[t]
			& \Bigg(-24 L_2^r M_K^2 - 6 L_3^r M_K^2 \\
			& - \frac{M_K^2 (16613 M_K^6 - 2179 M_K^4 M_\pi^2 + 29 M_K^2 M_\pi^4 + 69 M_\pi^6)}{2048 \pi^2 \Delta_{K\pi}^3} \\
			& + \ln\left(\frac{M_\pi^2}{\mu^2}\right) \frac{3 M_K^2 M_\pi^4 (37 M_K^4 - 80 M_K^2 M_\pi^2 + 20 M_\pi^4)}{512 \pi^2 \Delta_{K\pi}^4} \\
			& + \ln\left(\frac{M_K^2}{\mu^2}\right) \frac{3 M_K^6 (-4840 M_K^4 + 1216 M_K^2 M_\pi^2 + 83 M_\pi^4)}{512 \pi^2 \Delta_{K\pi}^4} \\
			& + \ln\left(\frac{M_\eta^2}{\mu^2}\right) \frac{3 M_K^2 (4 M_K^2 - M_\pi^2) (304 M_K^6 - 6 M_K^4 M_\pi^2 - M_\pi^6)}{128 \pi^2 \Delta_{K\pi}^4} \Bigg) . \end{aligned}
	\end{split}
\end{align}

\section{Matching at NNLO}

\subsection{Decomposition of the Two-Loop Result}

\label{sec:AppendixTwoLoopDecomposition}

\subsubsection{NLO Contribution}

We have already decomposed the NLO contributions. We apply a gauge transformation to convert the expressions to the second gauge and evaluate the result numerically:
\begin{align}
	\begin{split}
		m_{0,L}^{0,\mathrm{NLO}} &= \frac{M_K}{\sqrt{2} F_\pi} \begin{aligned}[t]
			& \Big( -0.1466 \cdot 10^3 L_1^r + 0.4953 \cdot 10^3 L_2^r + 0.0872 \cdot 10^3 L_3^r + 0.0733 \cdot 10^3 L_4^r \\
			& + 0.0092 \cdot 10^3 L_5^r + 0.0573 \cdot 10^3 L_9^r \frac{s_\ell}{M_K^2} \Big) , \end{aligned} \\
		m_{0,L}^{1,\mathrm{NLO}} &= \frac{M_K}{\sqrt{2} F_\pi} \begin{aligned}[t]
			& \Big( 0.9173 \cdot 10^3 L_1^r + 0.2293 \cdot 10^3 L_3^r \Big) , \end{aligned} \\
		m_{0,L}^{2,\mathrm{NLO}} &= 0 , \\
		m_{1,L}^{0,\mathrm{NLO}} &= \frac{M_K}{\sqrt{2} F_\pi} \begin{aligned}[t]
			& \Big( -0.2293 \cdot 10^3 L_2^r \Big) , \end{aligned} \\
		m_{1,L}^{1,\mathrm{NLO}} &= 0 , \\
		\tilde m_{1,L}^{0,\mathrm{NLO}} &= \frac{M_K}{\sqrt{2} F_\pi} \begin{aligned}[t]
			& \Big( -0.2660 \cdot 10^3 L_2^r - 0.1238 \cdot 10^3 L_3^r + 0.0092 \cdot 10^3 L_5^r \\
			&- ( 0.2293 \cdot 10^3 L_2^r - 0.0573 \cdot 10^3 L_9^r) \frac{s_\ell}{M_K^2} \Big) , \end{aligned} \\
		\tilde m_{1,L}^{1,\mathrm{NLO}} &= \frac{M_K}{\sqrt{2} F_\pi} \begin{aligned}[t]
			& \Big( 0.2293 \cdot 10^3 L_2^r \Big) , \end{aligned} \\
		\tilde m_{1,L}^{2,\mathrm{NLO}} &= 0 , \\
		n_{0,L}^{1,\mathrm{NLO}} &= \frac{M_K}{\sqrt{2} F_\pi} \begin{aligned}[t]
			& \Big( -0.6880 \cdot 10^3 L_2^r - 0.1720 \cdot 10^3 L_3^r \Big) , \end{aligned} \\
		n_{0,L}^{2,\mathrm{NLO}} &= n_{1,L}^{0,\mathrm{NLO}} = \tilde n_{1,L}^{1,\mathrm{NLO}} = 0 , 
	\end{split}
\end{align}
\begin{align}
	\begin{split}
		m_{0,R}^{0,\mathrm{NLO}} &= \frac{M_K}{\sqrt{2} F_\pi} \left( 0.1393 + 0.0444 \frac{s_\ell}{M_K^2} + 0.0256 \frac{s_\ell^2}{M_K^4} \right) , \\
		m_{0,R}^{1,\mathrm{NLO}} &= \frac{M_K}{\sqrt{2} F_\pi} \left( 0.3413 - 0.0512 \frac{s_\ell}{M_K^2} \right) , \\
		m_{0,R}^{2,\mathrm{NLO}} &= \frac{M_K}{\sqrt{2} F_\pi} \left( 0.4080 \right) , \\
		m_{1,R}^{0,\mathrm{NLO}} &= \frac{M_K}{\sqrt{2} F_\pi} \left( -0.0916 - 0.0512 \frac{s_\ell}{M_K^2} \right) , \\
		m_{1,R}^{1,\mathrm{NLO}} &= \frac{M_K}{\sqrt{2} F_\pi} \left( 0.0512 \right) , \\
		\tilde m_{1,R}^{0,\mathrm{NLO}} &= \frac{M_K}{\sqrt{2} F_\pi} \left( -0.0902 - 0.0595 \frac{s_\ell}{M_K^2} - 0.0256 \frac{s_\ell^2}{M_K^4} \right) , \\
		\tilde m_{1,R}^{1,\mathrm{NLO}} &= \frac{M_K}{\sqrt{2} F_\pi} \left( 0.1545 + 0.0512 \frac{s_\ell}{M_K^2} \right) , \\
		\tilde m_{1,R}^{2,\mathrm{NLO}} &= \frac{M_K}{\sqrt{2} F_\pi} \left( 0.0137 \right) , \\
		n_{0,R}^{1,\mathrm{NLO}} &= \frac{M_K}{\sqrt{2} F_\pi} \left( -0.1376 - 0.0384 \frac{s_\ell}{M_K^2} \right) , \\
		n_{0,R}^{2,\mathrm{NLO}} &= \frac{M_K}{\sqrt{2} F_\pi} \left( -0.0796 \right) , \\
		n_{1,R}^{0,\mathrm{NLO}} &= \frac{M_K}{\sqrt{2} F_\pi} \left( 0.0384 \right) , \\
		\tilde n_{1,R}^{1,\mathrm{NLO}} &= \frac{M_K}{\sqrt{2} F_\pi} \left( -0.0282 \right) .
	\end{split}
\end{align}

\subsubsection{NNLO LECs}

First, we consider the contribution of the NNLO LECs, the $C_i^r$. We decompose this contribution into the form of the polynomial part in (\ref{eq:FunctionsOfOneVariable3Subtr}):
\begin{align}
	\begin{split}
		m_{0,C}^{0,\mathrm{NNLO}} &= \frac{M_K}{\sqrt{2}F_\pi} \frac{1}{F_\pi^4} \begin{aligned}[t]
			&\bigg(4 M_K^4 \begin{aligned}[t]
				& \big(C_{1}^r-2 C_{3}^r-2C_{4}^r+2C_{5}^r +4 C_{6}^r \\
				& +2 C_{10}^r +8 C_{11}^r-4 C_{12}^r-8 C_{13}^r+2C_{22}^r+4 C_{23}^r-2C_{34}^r\big) \end{aligned} \\
			& - M_\pi^2 M_K^2 \begin{aligned}[t]
				&\big(4 C_{1}^r+64 C_{2}^r+56 C_{3}^r+34 C_{4}^r-8 C_{5}^r+40 C_{6}^r \\
				& +64 C_{7}^r+24 C_{8}^r -16 C_{10}^r-48 C_{11}^r-8 C_{12}^r+112 C_{13}^r \\
				& +16 C_{14}^r-80 C_{15}^r-64 C_{17}^r +8 C_{22}^r-16 C_{23}^r+16 C_{25}^r \\
				& +16 C_{26}^r+32 C_{29}^r+64 C_{30}^r+32 C_{36}^r \\
				& +C_{66}^r+2 C_{67}^r-C_{69}^r-C_{88}^r+C_{90}^r\big) \end{aligned} \\
			& + M_\pi^4 \begin{aligned}[t]
				&\big(-24 C_{1}^r-128 C_{2}^r-32 C_{3}^r-18 C_{4}^r-32 C_{5}^r-24 C_{6}^r -64 C_{7}^r \\
				& +8 C_{8}^r +8 C_{10}^r+16 C_{11}^r-80 C_{12}^r-80 C_{13}^r+32 C_{14}^r+8 C_{15}^r \\
				& +128 C_{16}^r-48 C_{17}^r +16 C_{22}^r+32 C_{23}^r+32 C_{26}^r+128 C_{28}^r \\
				&-5 (C_{66}^r+2 C_{67}^r - C_{69}^r -C_{88}^r +C_{90}^r) \big) \end{aligned} \\
			& + s_\ell \begin{aligned}[t]
				&\Big( M_K^2 \begin{aligned}[t]
					&\big( 4 C_{1}^r-8 C_{3}^r-6 C_{4}^r-8 C_{12}^r-32 C_{13}^r \\
					& - 8 C_{63}^r-8 C_{64}^r+C_{66}^r+2 C_{67}^r+3 C_{69}^r-C_{88}^r-3 C_{90}^r\big) \end{aligned} \\
				& - M_\pi^2 \begin{aligned}[t]
					&\big(12 C_{1}^r+64 C_{2}^r+72 C_{3}^r+10 C_{4}^r-48 C_{13}^r-8 C_{22}^r \\
					& -32 C_{23}^r+8 C_{25}^r + 4 C_{64}^r+4 C_{65}^r+9 C_{66}^r+2 C_{67}^r \\
					& +16 C_{68}^r+3 C_{69}^r + 8 C_{83}^r+16 C_{84}^r+3 C_{88}^r+C_{90}^r \big) \Big) \end{aligned} \end{aligned} \\
			& - 2 s_\ell^2 \big(8 C_{3}^r+2 C_{4}^r+C_{66}^r+2 C_{67}^r-C_{69}^r+C_{88}^r-C_{90}^r \big) \bigg) , \end{aligned} \\
		m_{0,C}^{1,\mathrm{NNLO}} &= \frac{M_K}{\sqrt{2} F_\pi} \frac{1}{F_\pi^4} \begin{aligned}[t]
			& \bigg( 8 M_K^4 \begin{aligned}[t]
				&\big(C_{1}^r+4 C_{2}^r+4 C_{3}^r+2 C_{4}^r+4 C_{6}^r+4 C_{7}^r+2 C_{8}^r \\
				& +2 C_{12}^r+8 C_{13}^r-2 C_{23}^r+C_{25}^r\big) \end{aligned} \\
				&+ 2 M_\pi^2 M_K^2 \begin{aligned}[t]
					&\big(24 C_{1}^r+96 C_{2}^r+32 C_{3}^r-6 C_{4}^r+8 C_{5}^r \\
					& +8 C_{6}^r+16 C_{7}^r -16 C_{13}^r-16 C_{23}^r+8 C_{25}^r \\
					& +C_{66}^r+2 C_{67}^r-C_{69}^r-C_{88}^r+C_{90}^r \big) \end{aligned} \\
			&+2 s_\ell M_K^2 \begin{aligned}[t]
				&\big(4 C_{1}^r+16 C_{2}^r+16 C_{3}^r-2 C_{4}^r \\
				& +3 C_{66}^r+2 C_{67}^r+4 C_{68}^r+C_{69}^r+C_{88}^r-C_{90}^r \big) \bigg) , \end{aligned} \end{aligned} \\
		m_{0,C}^{2,\mathrm{NNLO}} &= \frac{M_K}{\sqrt{2}F_\pi} \frac{1}{F_\pi^4} 16 M_K^4 \big(-C_{1}^r-4 C_{2}^r-C_{3}^r+C_{4}^r\big), \\
		m_{1,C}^{0,\mathrm{NNLO}} &= \frac{M_K}{\sqrt{2} F_\pi} \frac{1}{F_\pi^4} 2 M_K^2 \begin{aligned}[t]
			& \bigg( M_K^2  \begin{aligned}[t]
					& \big(4 C_{3}^r+2 C_{4}^r-4 C_{10}^r-16 C_{11}^r+12 C_{12}^r+32 C_{13}^r \\
					& -2 C_{22}^r-4 C_{23}^r+2 C_{63}^r+C_{66}^r+C_{67}^r-C_{69}^r-2 C_{83}^r+C_{90}^r \big) \end{aligned} \\
				& + M_\pi^2 \begin{aligned}[t]
					& \big(16 C_{3}^r+16 C_{4}^r-4 C_{10}^r-8 C_{11}^r+4 C_{12}^r+16 C_{13}^r \\
					& -10 C_{22}^r-8 C_{23}^r+6 C_{25}^r-2 C_{63}^r+4 C_{67}^r+2 C_{83}^r-C_{88}^r \big) \end{aligned} \\
			& + s_\ell \big(12 C_{3}^r+2 C_{4}^r+C_{66}^r+C_{67}^r-C_{69}^r+C_{88}^r-C_{90}^r \big) \bigg), \end{aligned} \\
		m_{1,C}^{1,\mathrm{NNLO}} &= \frac{M_K}{\sqrt{2} F_\pi} \frac{1}{F_\pi^4} 2 M_K^4\big(-16 C_{3}^r-6 C_{4}^r+C_{66}^r-C_{67}^r-C_{69}^r-C_{88}^r+C_{90}^r\big) , \\
	\end{split}
\end{align}
\begin{align}
	\begin{split}
		\tilde m_{1,C}^{0,\mathrm{NNLO}} &= \frac{M_K}{\sqrt{2} F_\pi} \frac{1}{F_\pi^4}  \begin{aligned}[t]
			& \bigg(M_\pi^4  \begin{aligned}[t]
				& \big(-8 C_{1}^r+10 C_{4}^r-8 C_{6}^r-8 C_{8}^r-8 C_{10}^r-16 C_{11}^r+16 C_{12}^r \\
				& +16 C_{13}^r+8 C_{15}^r+16 C_{17}^r -16 C_{22}^r-32 C_{23}^r \\
				& +C_{66}^r+2 C_{67}^r-C_{69}^r-C_{88}^r+C_{90}^r\big) \end{aligned} \\
			& + M_\pi^2 M_K^2 \begin{aligned}[t]
				& \big(-12 C_{1}^r+24 C_{3}^r+34 C_{4}^r-8 C_{5}^r-24 C_{6}^r-8 C_{8}^r \\
				& -8 C_{10}^r-32 C_{11}^r+16 C_{12}^r+16 C_{13}^r+16 C_{14}^r+16 C_{15}^r \\
				& -36 C_{22}^r-32 C_{23}^r+20 C_{25}^r+16 C_{26}^r-32 C_{29}^r \\
				& +4 C_{63}^r+C_{66}^r-6 C_{67}^r-C_{69}^r-4 C_{83}^r+C_{88}^r+C_{90}^r\big) \end{aligned} \\
			& -2 M_K^4 \begin{aligned}[t]
				& \big(2 C_{1}^r-2 C_{4}^r+4 C_{5}^r+8 C_{6}^r+12 C_{12}^r+16 C_{13}^r+2 C_{22}^r \\
				& +4 C_{23}^r+4 C_{34}^r +2 C_{63}^r+C_{66}^r+C_{67}^r-C_{69}^r-2 C_{83}^r+C_{90}^r\big) \end{aligned} \\
			& +s_\ell  \begin{aligned}[t]
				& \Big( M_\pi^2 \begin{aligned}[t]
					& \big(-4 C_{1}^r+40 C_{3}^r+26 C_{4}^r -8 C_{10}^r-16 C_{11}^r+8 C_{12}^r+16 C_{13}^r \\
					& -12 C_{22}^r-16 C_{23}^r+4 C_{25}^r -4 C_{63}^r-4 C_{64}^r-4 C_{65}^r+C_{66}^r \\
					& -6 C_{67}^r-5 C_{69}^r-4 C_{83}^r+C_{88}^r-7 C_{90}^r\big) \end{aligned} \\
				& - M_K^2 \begin{aligned}[t]
					& \big(4 C_{1}^r+8 C_{3}^r-6 C_{4}^r+8 C_{10}^r+32 C_{11}^r-16 C_{12}^r-32 C_{13}^r \\
					& +4 C_{22}^r+8 C_{23}^r +4 C_{63}^r+8 C_{64}^r+C_{66}^r+2 C_{67}^r+3 C_{69}^r \\
					& +4 C_{83}^r+C_{88}^r+C_{90}^r\big) \Big) \end{aligned} \end{aligned} \\
			& + s_\ell^2 \big(8 C_{3}^r-2 C_{67}^r\big) \bigg) , \end{aligned} \\
		\tilde m_{1,C}^{1,\mathrm{NNLO}} &= \frac{M_K}{\sqrt{2} F_\pi} \frac{1}{F_\pi^4} 2 M_K^2 \begin{aligned}[t]
			& \bigg( -M_K^2 \begin{aligned}[t]
				& \big(4 C_{3}^r+8 C_{4}^r-4 C_{10}^r-16 C_{11}^r +4 C_{12}^r+32 C_{13}^r -6 C_{22}^r \\
				& -4 C_{23}^r+4 C_{25}^r -2 C_{63}^r-2 C_{67}^r+2 C_{83}^r-C_{88}^r \big) \end{aligned} \\
			& -M_\pi^2 \begin{aligned}[t]
				& \big(16 C_{3}^r+10 C_{4}^r-4 C_{10}^r-8 C_{11}^r \\
				& +12 C_{12}^r+16 C_{13}^r-6 C_{22}^r-8 C_{23}^r+2 C_{25}^r \\
				& +2 C_{63}^r+C_{66}^r-2 C_{67}^r-C_{69}^r-2 C_{83}^r+C_{90}^r\big) \end{aligned} \\
			& -s_\ell \big(12 C_{3}^r+2 C_{4}^r+C_{66}^r-2 C_{67}^r-C_{69}^r+C_{88}^r-C_{90}^r\big) \bigg), \end{aligned} \\
		\tilde m_{1,C}^{2,\mathrm{NNLO}} &= \frac{M_K}{\sqrt{2} F_\pi} \frac{1}{F_\pi^4} 2 M_K^4 \big(8 C_{3}^r+2 C_{4}^r+C_{66}^r-C_{67}^r-C_{69}^r-C_{88}^r+C_{90}^r\big) , \\
		n_{0,C}^{1,\mathrm{NNLO}} &= \frac{M_K}{\sqrt{2}F_\pi} \frac{1}{F_\pi^4} 3 M_K^2 \begin{aligned}[t]
			& \bigg( - 2M_K^2 \begin{aligned}[t]
				& \big(3 C_{1}^r -4 C_{4}^r + 2C_{5}^r +4 C_{6}^r \\
				& +2C_{10}^r +8 C_{11}^r -2C_{12}^r -8 C_{13}^r +2C_{22}^r +4 C_{23}^r \big) \end{aligned} \\
			& - \frac{1}{2} M_\pi^2 \begin{aligned}[t]
				& \big(16 C_{1}^r +8 C_{3}^r -18 C_{4}^r +8 C_{6}^r +8 C_{8}^r \\
				& +8 C_{10}^r +16 C_{11}^r -16 C_{12}^r -16 C_{13}^r +16 C_{22}^r +32 C_{23}^r \\
				& -C_{66}^r -2 C_{67}^r +C_{69}^r +C_{88}^r -C_{90}^r \big) \end{aligned} \\
			& - \frac{1}{2} s_\ell \begin{aligned}[t]
				&\big(4 C_{1}^r -8 C_{3}^r -6 C_{4}^r +C_{66}^r +2 C_{67}^r +3 C_{69}^r \\
				& -C_{88}^r +C_{90}^r \big) \bigg) , \end{aligned} \end{aligned} \\
		n_{0,C}^{2,\mathrm{NNLO}} &= \frac{M_K}{\sqrt{2}F_\pi} \frac{1}{F_\pi^4} 12 M_K^4 \big(C_{1}^r +C_{3}^r -C_{4}^r \big), \\
		n_{1,C}^{0,\mathrm{NNLO}} &= \frac{M_K}{\sqrt{2}F_\pi} \frac{1}{F_\pi^4} \frac{-3 M_K^4}{2} \big(16 C_{3}^r +6 C_{4}^r -C_{66}^r -2 C_{67}^r +C_{69}^r +C_{88}^r -C_{90}^r \big) , \\
		\tilde n_{1,C}^{1,\mathrm{NNLO}} &= \frac{M_K}{\sqrt{2}F_\pi} \frac{1}{F_\pi^4} 3 M_K^4 \big(8 C_{3}^r +2 C_{4}^r +C_{66}^r +2 C_{67}^r -C_{69}^r -C_{88}^r +C_{90}^r \big) .
	\end{split}
\end{align}
Unfortunately, a lot of NNLO LECs enter the polynomial. In total, there appear 24 linearly independent combinations of the $C_i^r$.

If we use the resonance estimate of \cite{Bijnens2012}, we obtain the following values for the NNLO counterterm contribution:
\begin{align}
	\begin{split}
		m_{0,\mathrm{reso}}^{0,\mathrm{NNLO}} &= \frac{M_K}{\sqrt{2} F_\pi} \left( -0.1546 - 0.1716 \frac{s_\ell}{M_K^2} + 0.0316 \frac{s_\ell^2}{M_K^4} \right) , \\
		m_{0,\mathrm{reso}}^{1,\mathrm{NNLO}} &= \frac{M_K}{\sqrt{2} F_\pi} \left( 0.1747 - 0.0316 \frac{s_\ell}{M_K^2} \right) , \\
		m_{0,\mathrm{reso}}^{2,\mathrm{NNLO}} &= \frac{M_K}{\sqrt{2} F_\pi} \left( 0.0310 \right) , \\
		m_{1,\mathrm{reso}}^{0,\mathrm{NNLO}} &= \frac{M_K}{\sqrt{2} F_\pi} \left( 0.1657 - 0.0316 \frac{s_\ell}{M_K^2} \right) , \\
		m_{1,\mathrm{reso}}^{1,\mathrm{NNLO}} &= \frac{M_K}{\sqrt{2} F_\pi} \left( -0.0104 \right) , \\
		\tilde m_{1,\mathrm{reso}}^{0,\mathrm{NNLO}} &= \frac{M_K}{\sqrt{2} F_\pi} \left( -0.0900 - 0.0135 \frac{s_\ell}{M_K^2} \right) , \\
		\tilde m_{1,\mathrm{reso}}^{1,\mathrm{NNLO}} &= \frac{M_K}{\sqrt{2} F_\pi} \left( -0.1712 - 0.0316 \frac{s_\ell}{M_K^2} \right) , \\
		\tilde m_{1,\mathrm{reso}}^{2,\mathrm{NNLO}} &= \frac{M_K}{\sqrt{2} F_\pi} \left( 0.1805 \right) , \\
		n_{0,\mathrm{reso}}^{1,\mathrm{NNLO}} &= \frac{M_K}{\sqrt{2} F_\pi} \left( 0.1502 - 0.0237 \frac{s_\ell}{M_K^2} \right) , \\
		n_{0,\mathrm{reso}}^{2,\mathrm{NNLO}} &= \frac{M_K}{\sqrt{2} F_\pi} \left( -0.0233 \right) , \\
		n_{1,\mathrm{reso}}^{0,\mathrm{NNLO}} &= \frac{M_K}{\sqrt{2} F_\pi} \left( -0.0078 \right) , \\
		\tilde n_{1,\mathrm{reso}}^{1,\mathrm{NNLO}} &= \frac{M_K}{\sqrt{2} F_\pi} \left( 0.2707 \right) .
	\end{split}
\end{align}

Alternatively, if we use the `preferred values' of the BE14 fit \cite{Bijnens2014} (complemented with $C_{88}^r-C_{90}^r = -55 \cdot 10^{-6}$ \cite{BijnensTalavera2002} and the remaining LECs that appear in the $s_\ell$-dependence set to zero), we obtain the following values for the NNLO counterterm contribution:
\begin{align}
	\begin{split}
		m_{0,\mathrm{BE14}}^{0,\mathrm{NNLO}} &= \frac{M_K}{\sqrt{2} F_\pi} \left( -0.4108 - 0.1823 \frac{s_\ell}{M_K^2} - 0.0033 \frac{s_\ell^2}{M_K^4} \right) , \\
		m_{0,\mathrm{BE14}}^{1,\mathrm{NNLO}} &= \frac{M_K}{\sqrt{2} F_\pi} \left( 0.7959 + 0.0986 \frac{s_\ell}{M_K^2} \right) , \\
		m_{0,\mathrm{BE14}}^{2,\mathrm{NNLO}} &= \frac{M_K}{\sqrt{2} F_\pi} \left( -0.1709 \right) , \\
		m_{1,\mathrm{BE14}}^{0,\mathrm{NNLO}} &= \frac{M_K}{\sqrt{2} F_\pi} \left( 0.2627 + 0.0296 \frac{s_\ell}{M_K^2} \right) , \\
		m_{1,\mathrm{BE14}}^{1,\mathrm{NNLO}} &= \frac{M_K}{\sqrt{2} F_\pi} \left( -0.1709 \right) , \\
		\tilde m_{1,\mathrm{BE14}}^{0,\mathrm{NNLO}} &= \frac{M_K}{\sqrt{2} F_\pi} \left( 0.0356 + 0.1050 \frac{s_\ell}{M_K^2} + 0.0263 \frac{s_\ell^2}{M_K^4} \right) , \\
		\tilde m_{1,\mathrm{BE14}}^{1,\mathrm{NNLO}} &= \frac{M_K}{\sqrt{2} F_\pi} \left( -0.2942 - 0.0296 \frac{s_\ell}{M_K^2} \right) , \\
		\tilde m_{1,\mathrm{BE14}}^{2,\mathrm{NNLO}} &= \frac{M_K}{\sqrt{2} F_\pi} \left( 0.1841 \right) , \\
		n_{0,\mathrm{BE14}}^{1,\mathrm{NNLO}} &= \frac{M_K}{\sqrt{2} F_\pi} \left( 0.3505 + 0.0296 \frac{s_\ell}{M_K^2} \right) , \\
		n_{0,\mathrm{BE14}}^{2,\mathrm{NNLO}} &= \frac{M_K}{\sqrt{2} F_\pi} \left( 0.0099 \right) , \\
		n_{1,\mathrm{BE14}}^{0,\mathrm{NNLO}} &= \frac{M_K}{\sqrt{2} F_\pi} \left( -0.1282 \right) , \\
		\tilde n_{1,\mathrm{BE14}}^{1,\mathrm{NNLO}} &= \frac{M_K}{\sqrt{2} F_\pi} \left( 0.2761 \right) .
	\end{split}
\end{align}

\subsubsection{Vertex Integrals}

Let us study in more detail the contribution of the vertex integrals: the $u$-channel vertex integrals fulfil $F^\mathrm{NNLO}_{VU} = G^\mathrm{NNLO}_{VU}$. In the following, we treat them numerically. The contribution to $R_0$ is obtained by subtracting the constant, linear and quadratic terms:
\begin{align}
	\begin{split}
		R_0^V(u,s_\ell) &= F_{VU}^\mathrm{NNLO}(u,s_\ell) - P_{VU}^\mathrm{NNLO}(u,s_\ell) , \\
		P_{VU}^\mathrm{NNLO}(u,s_\ell) &= F_{VU}^\mathrm{NNLO}(0,s_\ell) + u {F_{VU}^\mathrm{NNLO}}^\prime(0,s_\ell) + \frac{1}{2} u^2 {F_{VU}^\mathrm{NNLO}}^\dprime(0,s_\ell) ,
	\end{split}
\end{align}
where ${}^\prime$ stands for the derivative with respect to the first argument ($u$). The polynomial $P_{VU}^\mathrm{NNLO}$ has to be lumped into the overall polynomial and finally reshuffled into the subtraction constants. Numerically, we find
\begin{align}
	\begin{split}
		P_{VU}^\mathrm{NNLO}(u,s_\ell) &\approx \frac{M_K}{\sqrt{2} F_\pi} \left( 0.4001 + 0.0045 \frac{s_\ell}{M_K^2} + \left( -0.2518 - 0.0047 \frac{s_\ell}{M_K^2} \right) \frac{u}{M_K^2} + 0.0558 \frac{u^2}{M_K^4} \right).
		\raisetag{-0.1cm}
	\end{split}
\end{align}
As we have checked again numerically, the polynomial-subtracted $u$-channel contribution of the vertex integrals fulfils the dispersion relation
\begin{align}
	\begin{split}
		R_0^V(u,s_\ell) &= \frac{u^3}{\pi} \int_{u_0}^\infty \frac{\Im R_0^V(u^\prime,s_\ell)}{(u^\prime-u-i\epsilon){u^\prime}^3} du^\prime .
	\end{split}
\end{align}

Next, we consider the $s$-channel vertex integrals: apart from a polynomial, they belong to either $M_0$ or $\tilde M_1$. Again, we subtract the constant, linear and quadratic terms:
\begin{align}
	\begin{split}
		M_0^V(s,s_\ell) &= F_{VS}^\mathrm{NNLO}(s,s_\ell) - P_{F,VS}^\mathrm{NNLO}(s,s_\ell) , \\
		\tilde M_1^V(s,s_\ell) &= G_{VS}^\mathrm{NNLO}(s,s_\ell) - P_{G,VS}^\mathrm{NNLO}(s,s_\ell) , \\
		P_{F,VS}^\mathrm{NNLO}(s,s_\ell) &= F_{VS}^\mathrm{NNLO}(0,s_\ell) + s {F_{VS}^\mathrm{NNLO}}^\prime(0,s_\ell) + \frac{1}{2} s^2 {F_{VS}^\mathrm{NNLO}}^\dprime(0,s_\ell) , \\
		P_{G,VS}^\mathrm{NNLO}(s,s_\ell) &= G_{VS}^\mathrm{NNLO}(0,s_\ell) + s {G_{VS}^\mathrm{NNLO}}^\prime(0,s_\ell) + \frac{1}{2} s^2 {G_{VS}^\mathrm{NNLO}}^\dprime(0,s_\ell) .
	\end{split}
\end{align}
We find numerically
\begin{align}
	\begin{split}
		P_{F,VS}^\mathrm{NNLO}(s,s_\ell) &\approx \frac{M_K}{\sqrt{2} F_\pi} \left( 0.2640 + 0.1007 \frac{s_\ell}{M_K^2} + \left( -1.7836 - 0.0723 \frac{s_\ell}{M_K^2} \right) \frac{s}{M_K^2} - 0.5435 \frac{s^2}{M_K^4} \right) , \\
		P_{G,VS}^\mathrm{NNLO}(s,s_\ell) &\approx \frac{M_K}{\sqrt{2} F_\pi} \left( -0.3170 - 0.0703 \frac{s_\ell}{M_K^2} + \left( 0.1461 + 0.0031 \frac{s_\ell}{M_K^2} \right) \frac{s}{M_K^2} - 0.0013 \frac{s^2}{M_K^4} \right) .
		\raisetag{-0.1cm}
	\end{split}
\end{align}
A numerical check shows that the polynomial-subtracted $s$-channel contributions of the vertex integrals fulfil the dispersion relations
\begin{align}
	\begin{split}
		M_0^V(s,s_\ell) &= \frac{s^3}{\pi} \int_{s_0}^\infty \frac{\Im M_0^V(s^\prime,s_\ell)}{(s^\prime - s - i\epsilon) {s^\prime}^3} ds^\prime , \\
		\tilde M_1^V(s,s_\ell) &= \frac{s^3}{\pi} \int_{s_0}^\infty \frac{\Im \tilde M_1^V(s^\prime,s_\ell)}{(s^\prime - s - i\epsilon) {s^\prime}^3} ds^\prime .
	\end{split}
\end{align}
Finally, we consider the $t$-channel, which is a bit more intricate: the reason is that not all linear and quadratic terms of a simple Taylor expansion in $t$ belong to the subtraction polynomial. Since the $t$-channel contributions depend only on $t$, we can write them as
\begin{align}
	\begin{split}
		F_{VT}^\mathrm{NNLO}(t,s_\ell) &= \frac{2}{3} N_0^V(t,s_\ell) - \frac{2}{3} \frac{\Delta_{K\pi} - 3 t}{2M_K^2} \tilde N_1^V(t,s_\ell) + \frac{1}{3} R_0^V(t,s_\ell) + P_{F,VT}^\mathrm{NNLO}(t,s_\ell) , \\
		G_{VT}^\mathrm{NNLO}(t,s_\ell) &= -\frac{2}{3} N_0^V(t,s_\ell) + \frac{2}{3} \frac{\Delta_{K\pi} + t}{2M_K^2} \tilde N_1^V(t,s_\ell) - \frac{1}{3} R_0^V(t,s_\ell) + P_{G,VT}^\mathrm{NNLO}(t,s_\ell) , \\
	\end{split}
\end{align}
where $P_{F,VT}^\mathrm{NNLO}$ and $P_{G,VT}^\mathrm{NNLO}$ are second order polynomials. The Taylor expansion of $N_0^V$ starts with a cubic and the one of $\tilde N_1^V$ with a quadratic term. Therefore, in the sum
\begin{align}
	\begin{split}
		F_{VT}^\mathrm{NNLO}(t,s_\ell) + G_{VT}^\mathrm{NNLO}(t,s_\ell) &= \frac{4 t}{3 M_K^2} \tilde N_1^V(t,s_\ell) + P_{F,VT}^\mathrm{NNLO}(t,s_\ell) + P_{G,VT}^\mathrm{NNLO}(t,s_\ell) ,
	\end{split}
\end{align}
we can easily separate $\tilde N_1^V$ from the sum of the polynomials. After having identified $\tilde N_1^V$ (in particular the quadratic term of its Taylor expansion), we can also separate the difference of the polynomials using
\begin{align}
	\begin{split}
		F_{VT}^\mathrm{NNLO}(t,s_\ell) - G_{VT}^\mathrm{NNLO}(t,s_\ell) &= \frac{4}{3} N_0^V(t,s_\ell) - \frac{2}{3} \frac{\Delta_{K\pi} - t}{M_K^2} \tilde N_1^V(t,s_\ell) + \frac{2}{3} R_0^V(t,s_\ell) \\
			&\quad + P_{F,VT}^\mathrm{NNLO}(t,s_\ell) - P_{G,VT}^\mathrm{NNLO}(t,s_\ell) .
	\end{split}
\end{align}
Numerically, we find
\begin{align}
	\begin{split}
		P_{F,VT}^\mathrm{NNLO}(t,s_\ell) &\approx \frac{M_K}{\sqrt{2} F_\pi} \left( -0.6964 - 0.1186 \frac{s_\ell}{M_K^2} + \left( 0.2873 + 0.0107 \frac{s_\ell}{M_K^2} \right) \frac{t}{M_K^2} + 0.4183 \frac{t^2}{M_K^4} \right) , \\
		P_{G,VT}^\mathrm{NNLO}(t,s_\ell) &\approx \frac{M_K}{\sqrt{2} F_\pi} \left( -0.0069 - 0.0104 \frac{s_\ell}{M_K^2} + \left( 0.0134 - 0.0037 \frac{s_\ell}{M_K^2} \right) \frac{t}{M_K^2} - 0.4460 \frac{t^2}{M_K^4} \right) .
		\raisetag{-0.1cm}
	\end{split}
\end{align}
Again, we test numerically that the following dispersion relations are fulfilled:
\begin{align}
	\begin{split}
		N_0^V(t,s_\ell) &= \frac{t^3}{\pi} \int_{t_0}^\infty \frac{\Im N_0^V(t^\prime,s_\ell)}{(t^\prime - t - i\epsilon) {t^\prime}^3} dt^\prime , \\
		\tilde N_1^V(t,s_\ell) &= \frac{t^2}{\pi} \int_{t_0}^\infty \frac{\Im \tilde N_1^V(t^\prime,s_\ell)}{(t^\prime - t - i\epsilon) {t^\prime}^2} dt^\prime .
	\end{split}
\end{align}
Reshuffling the polynomial contributions into the subtraction constants leads to
\begin{align}
	\begin{split}
		m_{0,V}^{0,\mathrm{NNLO}} &= \frac{M_K}{\sqrt{2} F_\pi} \left( 0.0629 + 0.0629 \frac{s_\ell}{M_K^2} - 0.0523 \frac{s_\ell^2}{M_K^4} \right) , \\
		m_{0,V}^{1,\mathrm{NNLO}} &= \frac{M_K}{\sqrt{2} F_\pi} \left( -1.8045 + 0.0358 \frac{s_\ell}{M_K^2} \right) , \\
		m_{0,V}^{2,\mathrm{NNLO}} &= \frac{M_K}{\sqrt{2} F_\pi} \left( -0.5993 \right) , \\
		m_{1,V}^{0,\mathrm{NNLO}} &= \frac{M_K}{\sqrt{2} F_\pi} \left( -0.2727 + 0.1034 \frac{s_\ell}{M_K^2} \right) , \\
		m_{1,V}^{1,\mathrm{NNLO}} &= \frac{M_K}{\sqrt{2} F_\pi} \left( -0.1116 \right) , \\
		\tilde m_{1,V}^{0,\mathrm{NNLO}} &= \frac{M_K}{\sqrt{2} F_\pi} \left( -0.1367 - 0.2554 \frac{s_\ell}{M_K^2} + 0.0511 \frac{s_\ell^2}{M_K^4} \right) , \\
		\tilde m_{1,V}^{1,\mathrm{NNLO}} &= \frac{M_K}{\sqrt{2} F_\pi} \left( 0.2685 - 0.1038 \frac{s_\ell}{M_K^2} \right) , \\
		\tilde m_{1,V}^{2,\mathrm{NNLO}} &= \frac{M_K}{\sqrt{2} F_\pi} \left( 0.0545 \right) , \\
		n_{0,V}^{1,\mathrm{NNLO}} &= \frac{M_K}{\sqrt{2} F_\pi} \left( -0.2718 + 0.0822 \frac{s_\ell}{M_K^2} \right) , \\
		n_{0,V}^{2,\mathrm{NNLO}} &= \frac{M_K}{\sqrt{2} F_\pi} \left( 0.7005 \right) , \\
		n_{1,V}^{0,\mathrm{NNLO}} &= \frac{M_K}{\sqrt{2} F_\pi} \left( -0.0837 \right) , \\
		\tilde n_{1,V}^{1,\mathrm{NNLO}} &= \frac{M_K}{\sqrt{2} F_\pi} \left( 0.0629 \right) .
	\end{split}
\end{align}

\subsubsection{Remaining Two-Loop Integrals}

Next, we consider the remaining two-loop parts, $X_P^\mathrm{NNLO}$. It is easy to decompose them into functions of one Mandelstam variable:
\begin{align}
	\begin{split}
		F_P^\mathrm{NNLO}(s,t,u) &= F_{PS}^\mathrm{NNLO}(s,s_\ell) + F_{PT,0}^\mathrm{NNLO}(t,s_\ell) + \frac{s-u}{M_K^2} F_{PT,1}^\mathrm{NNLO}(t,s_\ell) \\
			&\quad + F_{PU}^\mathrm{NNLO}(u,s_\ell) + P_{F,P}^\mathrm{NNLO}(s,t,u) , \\
		G_P^\mathrm{NNLO}(s,t,u) &= G_{PS}^\mathrm{NNLO}(s,s_\ell) + G_{PT,0}^\mathrm{NNLO}(t,s_\ell) + \frac{s-u}{M_K^2} G_{PT,1}^\mathrm{NNLO}(t,s_\ell) \\
			&\quad + G_{PU}^\mathrm{NNLO}(u,s_\ell) + P_{G,P}^\mathrm{NNLO}(s,t,u) , \\
	\end{split}
\end{align}
where $P_{F,P}^\mathrm{NNLO}$ and $P_{G,P}^\mathrm{NNLO}$ are second order polynomials. Again, we apply subtractions to the different functions:
\begin{align}
	\begin{split}
		M_0^P(s,s_\ell) &= F_{PS}^\mathrm{NNLO}(s,s_\ell) - P_{F,PS}^\mathrm{NNLO}(s,s_\ell) , \\
		\tilde M_1^P(s,s_\ell) &= G_{PS}^\mathrm{NNLO}(s,s_\ell) - P_{G,PS}^\mathrm{NNLO}(s,s_\ell) , \\
		R_0^P(u,s_\ell) &= F_{PU}^\mathrm{NNLO}(u,s_\ell) - P_{F,PU}^\mathrm{NNLO}(u,s_\ell) \\
			&= G_{PU}^\mathrm{NNLO}(u,s_\ell) - P_{G,PU}^\mathrm{NNLO}(u,s_\ell) , \\
	\end{split}
\end{align}
where
\begin{align}
	\begin{split}
		P_{F,PS}^\mathrm{NNLO}(s,s_\ell) &= F_{PS}^\mathrm{NNLO}(0,s_\ell) + s {F_{PS}^\mathrm{NNLO}}^\prime(0,s_\ell) + \frac{1}{2} s^2 {F_{PS}^\mathrm{NNLO}}^\dprime(0,s_\ell) , \\
		P_{G,PS}^\mathrm{NNLO}(s,s_\ell) &= G_{PS}^\mathrm{NNLO}(0,s_\ell) + s {G_{PS}^\mathrm{NNLO}}^\prime(0,s_\ell) + \frac{1}{2} s^2 {G_{PS}^\mathrm{NNLO}}^\dprime(0,s_\ell) , \\
		P_{F,PU}^\mathrm{NNLO}(u,s_\ell) &= F_{PU}^\mathrm{NNLO}(0,s_\ell) + u {F_{PU}^\mathrm{NNLO}}^\prime(0,s_\ell) + \frac{1}{2} u^2 {F_{PU}^\mathrm{NNLO}}^\dprime(0,s_\ell) , \\
		P_{G,PU}^\mathrm{NNLO}(u,s_\ell) &= G_{PU}^\mathrm{NNLO}(0,s_\ell) + u {G_{PU}^\mathrm{NNLO}}^\prime(0,s_\ell) + \frac{1}{2} u^2 {G_{PU}^\mathrm{NNLO}}^\dprime(0,s_\ell) .
	\end{split}
\end{align}
Numerically, we find
\begin{align}
	\begin{split}
		P_{F,PS}^\mathrm{NNLO}(s,s_\ell) &\approx \frac{M_K}{\sqrt{2} F_\pi} \left( -0.1676 + 0.0002 \frac{s_\ell}{M_K^2} + \left( 1.2215 + 0.0337 \frac{s_\ell}{M_K^2} \right) \frac{s}{M_K^2} + 0.8344 \frac{s^2}{M_K^4} \right) , \\
		P_{G,PS}^\mathrm{NNLO}(s,s_\ell) &\approx \frac{M_K}{\sqrt{2} F_\pi} \left( 0.0623 + 0.0118 \frac{s_\ell}{M_K^2} + \left( -0.0619 - 0.0057 \frac{s_\ell}{M_K^2} \right) \frac{s}{M_K^2} \right) , \\
		P_{F,PU}^\mathrm{NNLO}(u,s_\ell) &\approx \frac{M_K}{\sqrt{2} F_\pi} \left( 0.0588 - 0.0089 \frac{s_\ell}{M_K^2} + \left( 0.0157 + 0.0018 \frac{s_\ell}{M_K^2} \right) \frac{u}{M_K^2} \right) , \\
		P_{G,PU}^\mathrm{NNLO}(u,s_\ell) &\approx \frac{M_K}{\sqrt{2} F_\pi} \left( 0.0588 - 0.0089 \frac{s_\ell}{M_K^2} + \left( 0.0780 + 0.0018 \frac{s_\ell}{M_K^2} \right) \frac{u}{M_K^2} \right) , \\
	\end{split}
\end{align}
where we neglect tiny quadratic terms.

For the $t$-channel contribution, we first have to make some reshuffling:
\begin{align}
	\begin{split}
		\frac{s-u}{M_K^2} F_{PT,1}^\mathrm{NNLO}(t,s_\ell) &= \frac{s-u}{M_K^2} F_{PT,1}^\mathrm{NNLO}(0,s_\ell) - \frac{2}{3} \frac{\Delta_{K\pi} \Delta_{\ell\pi}}{M_K^4} \bar F_{PT,1}^\mathrm{NNLO}(t,s_\ell) \\
				&\quad + \frac{2}{3} \frac{t(s-u) + \Delta_{K\pi} \Delta_{\ell\pi}}{M_K^4} \bar F_{PT,1}^\mathrm{NNLO}(t,s_\ell) , \\
		\frac{s-u}{M_K^2} G_{PT,1}^\mathrm{NNLO}(t,s_\ell) &= \frac{s-u}{M_K^2} G_{PT,1}^\mathrm{NNLO}(0,s_\ell) + \frac{2}{3} \frac{\Delta_{K\pi} \Delta_{\ell\pi}}{M_K^4} \bar G_{PT,1}^\mathrm{NNLO}(t,s_\ell) \\
				&\quad - \frac{2}{3} \frac{t(s-u) + \Delta_{K\pi} \Delta_{\ell\pi}}{M_K^4} \bar G_{PT,1}^\mathrm{NNLO}(t,s_\ell) ,
	\end{split}
\end{align}
where
\begin{align}
	\begin{split}
		\bar F_{PT,1}^\mathrm{NNLO}(t,s_\ell) &= \frac{3 M_K^2}{2t} \left( F_{PT,1}^\mathrm{NNLO}(t,s_\ell) - F_{PT,1}^\mathrm{NNLO}(0,s_\ell) \right) , \\
		\bar G_{PT,1}^\mathrm{NNLO}(t,s_\ell) &= - \frac{3 M_K^2}{2t} \left( G_{PT,1}^\mathrm{NNLO}(t,s_\ell) - G_{PT,1}^\mathrm{NNLO}(0,s_\ell) \right) .
	\end{split}
\end{align}
In fact, it turns out that $F_{PT,1}^\mathrm{NNLO}(t,s_\ell) = - G_{PT,1}^\mathrm{NNLO}(t,s_\ell)$, hence $\bar F_{PT,1}^\mathrm{NNLO}(t,s_\ell) = \bar G_{PT,1}^\mathrm{NNLO}(t,s_\ell)$ .

Next, we identify
\begin{align}
	\begin{split}
		N_1^P(t,s_\ell) &= \bar F_{PT,1}^\mathrm{NNLO}(t,s_\ell) - P_{F,PT1}^\mathrm{NNLO}(t,s_\ell) \\
			&= \bar G_{PT,1}^\mathrm{NNLO}(t,s_\ell) - P_{G,PT1}^\mathrm{NNLO}(t,s_\ell) , \\
		P_{F,PT1}^\mathrm{NNLO}(t,s_\ell) &= \bar F_{PT,1}^\mathrm{NNLO}(0,s_\ell) , \\
		P_{G,PT1}^\mathrm{NNLO}(t,s_\ell) &= \bar G_{PT,1}^\mathrm{NNLO}(0,s_\ell) , \\
	\end{split}
\end{align}
numerically
\begin{align}
	\begin{split}
		P_{F,PT1}^\mathrm{NNLO}(t,s_\ell) = P_{G,PT1}^\mathrm{NNLO}(t,s_\ell) &\approx \frac{M_K}{\sqrt{2} F_\pi} \left( -0.0009 \right) , \\
		F_{PT,1}^\mathrm{NNLO}(0,s_\ell) = - G_{PT,1}^\mathrm{NNLO}(0,s_\ell) &\approx \frac{M_K}{\sqrt{2} F_\pi} \left( -0.0043 \right) .
	\end{split}
\end{align}
The only missing pieces are the contributions to $N_0$ and $\tilde N_1$:
\begin{align}
	\begin{split}
		F_{PT,0}^\mathrm{NNLO}(t,s_\ell) - \frac{2}{3} \frac{\Delta_{K\pi} \Delta_{\ell\pi}}{M_K^4} \bar F_{PT,1}^\mathrm{NNLO}(t,s_\ell) &= \frac{2}{3} N_0^P(t,s_\ell) - \frac{2}{3} \frac{\Delta_{K\pi} - 3 t}{2M_K^2} \tilde N_1^P(t,s_\ell) + \frac{1}{3} R_0^P(t,s_\ell) \\
			&\quad + P_{F,PT0}^\mathrm{NNLO}(t,s_\ell) , \\
		G_{PT,0}^\mathrm{NNLO}(t,s_\ell) + \frac{2}{3} \frac{\Delta_{K\pi} \Delta_{\ell\pi}}{M_K^4} \bar G_{PT,1}^\mathrm{NNLO}(t,s_\ell) &= -\frac{2}{3} N_0^P(t,s_\ell) + \frac{2}{3} \frac{\Delta_{K\pi} + t}{2M_K^2} \tilde N_1^P(t,s_\ell) - \frac{1}{3} R_0^P(t,s_\ell) \\
			&\quad  + P_{G,PT0}^\mathrm{NNLO}(t,s_\ell) .
	\end{split}
\end{align}
The sum and difference of these equations give
\begin{align}
	\begin{split}
		F_{PT,0}^\mathrm{NNLO}(t,s_\ell) + G_{PT,0}^\mathrm{NNLO}(t,s_\ell) &= \frac{4 t}{3 M_K^2} \tilde N_1^P(t,s_\ell) + P_{F,PT0}^\mathrm{NNLO}(t,s_\ell) + P_{G,PT0}^\mathrm{NNLO}(t,s_\ell) , \\
		F_{PT,0}^\mathrm{NNLO}(t,s_\ell) - G_{PT,0}^\mathrm{NNLO}(t,s_\ell) &- \frac{4}{3} \frac{\Delta_{K\pi} \Delta_{\ell\pi}}{M_K^4} \bar F_{PT,1}^\mathrm{NNLO}(t,s_\ell) \\
			&= \frac{4}{3} N_0^P(t,s_\ell) - \frac{2}{3} \frac{\Delta_{K\pi} - t}{M_K^2} \tilde N_1^P(t,s_\ell) + \frac{2}{3} R_0^P(t,s_\ell) \\
			&\quad + P_{F,PT0}^\mathrm{NNLO}(t,s_\ell) - P_{G,PT0}^\mathrm{NNLO}(t,s_\ell) .
	\end{split}
\end{align}
We find numerically
\begin{align}
	\begin{split}
		P_{F,PT0}^\mathrm{NNLO}(t,s_\ell) &\approx \frac{M_K}{\sqrt{2} F_\pi} \left( 0.2089 + 0.0343 \frac{s_\ell}{M_K^2} + \left( -0.2095 + 0.0020 \frac{s_\ell}{M_K^2} \right) \frac{t}{M_K^2} \right) , \\
		P_{G,PT0}^\mathrm{NNLO}(t,s_\ell) &\approx \frac{M_K}{\sqrt{2} F_\pi} \left( -0.0650 + 0.0085 \frac{s_\ell}{M_K^2} + \left( 0.0829 - 0.0054 \frac{s_\ell}{M_K^2} \right) \frac{t}{M_K^2} \right) .
	\end{split}
\end{align}
Finally, the additional polynomials are given by
\begin{align}
	\begin{split}
		P_{F,P}^\mathrm{NNLO}(s,t,u) &\approx \frac{M_K}{\sqrt{2} F_\pi} \begin{aligned}[t]
			&\bigg( 0.2852 - 0.0443 \frac{s_\ell}{M_K^2} + 0.0018 \frac{s}{M_K^2} - 0.0473 \frac{t}{M_K^2} \bigg) , \end{aligned} \\
		P_{G,P}^\mathrm{NNLO}(s,t,u) &\approx \frac{M_K}{\sqrt{2} F_\pi} \begin{aligned}[t]
			&\bigg( 0.0061 - 0.0254 \frac{s_\ell}{M_K^2} + \left(0.0245 - 0.0003 \frac{s_\ell}{M_K^2}\right) \frac{s}{M_K^2} \\
			& + \left(0.0652 - 0.0003 \frac{s_\ell}{M_K^2}\right) \frac{t}{M_K^2} + 0.0002 \frac{(s+t)^2}{M_K^4} \bigg) , \end{aligned}
	\end{split}
\end{align}
where we truncate the Taylor expansion in $s_\ell$ after the linear terms.

The polynomials contribute to the subtraction constants as
\begin{align}
	\begin{split}
		m_{0,P}^{0,\mathrm{NNLO}} &= \frac{M_K}{\sqrt{2} F_\pi} \left( 0.2911 - 0.1019 \frac{s_\ell}{M_K^2} - 0.0019 \frac{s_\ell^2}{M_K^4} \right) , \\
		m_{0,P}^{1,\mathrm{NNLO}} &= \frac{M_K}{\sqrt{2} F_\pi} \left( 1.3004 + 0.0358 \frac{s_\ell}{M_K^2} \right) , \\
		m_{0,P}^{2,\mathrm{NNLO}} &= \frac{M_K}{\sqrt{2} F_\pi} \left( 0.8342 \right) , \\
		m_{1,P}^{0,\mathrm{NNLO}} &= \frac{M_K}{\sqrt{2} F_\pi} \left( 0.1012 + 0.0037 \frac{s_\ell}{M_K^2} \right) , \\
		m_{1,P}^{1,\mathrm{NNLO}} &= \frac{M_K}{\sqrt{2} F_\pi} \left( -0.0002 \right) , \\
		\tilde m_{1,P}^{0,\mathrm{NNLO}} &= \frac{M_K}{\sqrt{2} F_\pi} \left( 0.1477 + 0.0616 \frac{s_\ell}{M_K^2} + 0.0018 \frac{s_\ell^2}{M_K^4} \right) , \\
		\tilde m_{1,P}^{1,\mathrm{NNLO}} &= \frac{M_K}{\sqrt{2} F_\pi} \left( -0.1068 - 0.0078 \frac{s_\ell}{M_K^2} \right) , \\
		\tilde m_{1,P}^{2,\mathrm{NNLO}} &= \frac{M_K}{\sqrt{2} F_\pi} \left( 0.0002 \right) , \\
		n_{0,P}^{1,\mathrm{NNLO}} &= \frac{M_K}{\sqrt{2} F_\pi} \left( -0.1119 + 0.0110 \frac{s_\ell}{M_K^2} \right) , \\
		n_{0,P}^{2,\mathrm{NNLO}} &= \frac{M_K}{\sqrt{2} F_\pi} \left( 0.0001 \right) , \\
		n_{1,P}^{0,\mathrm{NNLO}} &= \frac{M_K}{\sqrt{2} F_\pi} \left( -0.0012 \right) , \\
		\tilde n_{1,P}^{1,\mathrm{NNLO}} &= \frac{M_K}{\sqrt{2} F_\pi} \left( 0.0002 \right) .
	\end{split}
\end{align}

\subsubsection{NNLO One-Loop Integrals}

The last NNLO piece that we have to decompose is the part containing the $L_i^r$. Similar to the two-loop part without vertex integrals, it can be easily decomposed into functions of one variables. Since this contribution contains only one-loop integrals, we can express it in terms of $A_0$ and $B_0$ functions, which can be treated analytically. After decomposing the NNLO one-loop part according to
\begin{align}
	\begin{split}
		F_L^\mathrm{NNLO}(s,t,u) &= F_{LS,0}^\mathrm{NNLO}(s,s_\ell) + \frac{u-t}{M_K^2} F_{LS,1}^\mathrm{NNLO}(s,s_\ell) + F_{LT,0}^\mathrm{NNLO}(t,s_\ell) + \frac{s-u}{M_K^2} F_{LT,1}^\mathrm{NNLO}(t,s_\ell) \\
			&\quad + F_{LU}^\mathrm{NNLO}(u,s_\ell) + P_{F,L}^\mathrm{NNLO}(s,t,u) , \\
		G_L^\mathrm{NNLO}(s,t,u) &= G_{LS}^\mathrm{NNLO}(s,s_\ell) + G_{LT,0}^\mathrm{NNLO}(t,s_\ell) + \frac{s-u}{M_K^2} G_{LT,1}^\mathrm{NNLO}(t,s_\ell) \\
			&\quad + G_{LU}^\mathrm{NNLO}(u,s_\ell) + P_{G,L}^\mathrm{NNLO}(s,t,u) ,
	\end{split}
\end{align}
the polynomial contribution is found in analogy to the two-loop part. Reshuffling the polynomial gives very lengthy expressions for the subtraction constants. We perform a Taylor expansion in $s_\ell$ and evaluate the expressions numerically, using the physical masses and $\mu=770$~MeV:
\begin{align}
	\footnotesize
	\begin{split}
		m_{0,L}^{0,\mathrm{NNLO}} &= \frac{M_K}{\sqrt{2} F_\pi} \begin{aligned}[t]
			&\bigg( (0.0243 + 0.0155 \cdot 10^3 L_5^r) \cdot 10^3 L_1^r + (0.3528 - 0.0523 \cdot 10^3 L_5^r) \cdot 10^3 L_2^r \\
			& + (0.0831 -0.0092 \cdot 10^3 L_5^r) \cdot 10^3 L_3^r + (0.0400 + 0.0350 \cdot 10^3 L_4^r - 0.0020 \cdot 10^3 L_5^r ) \cdot 10^3 L_4^r \\
			& + ( 0.0066 + 0.0048 \cdot 10^3 L_5^r ) \cdot 10^3 L_5^r - (0.0012 + 0.0699 \cdot 10^3 L_4^r + 0.0087 \cdot 10^3 L_5^r) \cdot 10^3 L_6^r \\
			& + 0.0213 \cdot 10^3 L_7^r + ( 0.0100 - 0.0027 \cdot 10^3 L_4^r - 0.0003 \cdot 10^3 L_5^r ) \cdot 10^3 L_8^r \\
			&+ \frac{s_\ell}{M_K^2}  \begin{aligned}[t]
				& \Big( 0.0213 \cdot 10^3 L_1^r - 0.0161 \cdot 10^3 L_2^r + 0.0230 \cdot 10^3 L_3^r + 0.0139 \cdot 10^3 L_4^r + 0.0018 \cdot 10^3 L_5^r \\
				&- 0.0017 \cdot 10^3 L_6^r - 0.0008 \cdot 10^3 L_8^r + (0.0229 - 0.0060 \cdot 10^3 L_5^r ) \cdot 10^3 L_9^r \Big) \end{aligned} \\
			&+ \frac{s_\ell^2}{M_K^4} \begin{aligned}[t]
				& \Big(-0.0053 \cdot 10^3 L_1^r - 0.0029 \cdot 10^3 L_2^r - 0.0029 \cdot 10^3 L_3^r + 0.0065 \cdot 10^3 L_4^r \\
				& + 0.0010 \cdot 10^3 L_5^r - 0.0012 \cdot 10^3 L_6^r - 0.0006 \cdot 10^3 L_8^r + 0.0025 \cdot 10^3 L_9^r \Big)  \bigg) ,\end{aligned} \end{aligned} \\
		m_{0,L}^{1,\mathrm{NNLO}} &= \frac{M_K}{\sqrt{2} F_\pi} \begin{aligned}[t]
			&\bigg(  - ( 0.1644 + 0.0968 \cdot 10^3 L_5^r) \cdot 10^3 L_1^r - 0.2921 \cdot 10^3 L_2^r - ( 0.1665 + 0.0242 \cdot 10^3 L_5^r) \cdot 10^3 L_3^r \\
			& - 0.0353 \cdot 10^3 L_4^r + 0.0049 \cdot 10^3 L_5^r + 0.0185 \cdot 10^3 L_6^r - 0.0033 \cdot 10^3 L_7^r + 0.0076 \cdot 10^3 L_8^r \\
			& + \frac{s_\ell}{M_K^2} \begin{aligned}[t]
				&\Big(0.0138 \cdot 10^3 L_1^r - 0.0575 \cdot 10^3 L_2^r - 0.0087 \cdot 10^3 L_3^r - 0.0130 \cdot 10^3 L_4^r \\
				& - 0.0020 \cdot 10^3 L_5^r + 0.0024 \cdot 10^3 L_6^r + 0.0012 \cdot 10^3 L_8^r + 0.0196 \cdot 10^3 L_9^r \Big) \bigg) , \end{aligned} \end{aligned} \\
		m_{0,L}^{2,\mathrm{NNLO}} &= \frac{M_K}{\sqrt{2} F_\pi} \begin{aligned}[t]
			&\bigg( 0.3345 \cdot 10^3 L_1^r + 0.2734 \cdot 10^3 L_2^r + 0.1618 \cdot 10^3 L_3^r + 0.0863 \cdot 10^3 L_4^r \\
			&+ 0.0096 \cdot 10^3 L_5^r + 0.0067 \cdot 10^3 L_6^r - 0.0003 \cdot 10^3 L_7^r + 0.0032 \cdot 10^3 L_8^r \bigg) ,\end{aligned} \\
		m_{1,L}^{0,\mathrm{NNLO}} &= \frac{M_K}{\sqrt{2} F_\pi} \begin{aligned}[t]
			&\bigg( - 0.1203 \cdot 10^3 L_1^r + ( - 0.2247 + 0.0242 \cdot 10^3 L_5^r) \cdot 10^3 L_2^r - 0.0727 \cdot 10^3 L_3^r \\
			& - 0.0241 \cdot 10^3 L_4^r - 0.0046 \cdot 10^3 L_5^r + 0.0078 \cdot 10^3 L_6^r + 0.0039 \cdot 10^3 L_8^r \\
			& + \frac{s_\ell}{M_K^2}  \begin{aligned}[t] 
				& \Big( 0.0121 \cdot 10^3 L_1^r + 0.0044 \cdot 10^3 L_2^r + 0.0063 \cdot 10^3 L_3^r - 0.0130 \cdot 10^3 L_4^r \\
				& - 0.0020 \cdot 10^3 L_5^r + 0.0024 \cdot 10^3 L_6^r + 0.0012 \cdot 10^3 L_8^r - 0.0053 \cdot 10^3 L_9^r \Big) \bigg) , \end{aligned} \end{aligned} \\
		m_{1,L}^{1,\mathrm{NNLO}} &= \frac{M_K}{\sqrt{2} F_\pi} \begin{aligned}[t]
			&\bigg( -0.0198 \cdot 10^3 L_1^r - 0.0059 \cdot 10^3 L_2^r - 0.0056 \cdot 10^3 L_3^r + 0.0130 \cdot 10^3 L_4^r \\
			& + 0.0020 \cdot 10^3 L_5^r - 0.0024 \cdot 10^3 L_6^r - 0.0012 \cdot 10^3 L_8^r \bigg) ,\end{aligned} \\
		\tilde m_{1,L}^{0,\mathrm{NNLO}} &= \frac{M_K}{\sqrt{2} F_\pi} \begin{aligned}[t]
			&\bigg( 0.0440 \cdot 10^3 L_1^r + ( - 0.1488 + 0.0281 \cdot 10^3 L_5^r) \cdot 10^3 L_2^r + ( - 0.0140 + 0.0131 \cdot 10^3 L_5^r) \cdot 10^3 L_3^r \\
			& + (0.0001 + 0.0044 \cdot 10^3 L_5^r) \cdot 10^3 L_4^r + ( - 0.0033 + 0.0048 \cdot 10^3 L_5^r ) \cdot 10^3 L_5^r \\
			& + ( 0.0186 - 0.0087 \cdot 10^3 L_5^r ) \cdot 10^3 L_6^r - 0.0135 \cdot 10^3 L_7^r + ( 0.0026 - 0.0003 \cdot 10^3 L_5^r ) \cdot 10^3 L_8^r \\
			& + \frac{s_\ell}{M_K^2} \begin{aligned}[t]
				& \Big( -0.0957 \cdot 10^3 L_1^r - ( 0.2423 - 0.0242 \cdot 10^3 L_5^r) \cdot 10^3 L_2^r - 0.0520 \cdot 10^3 L_3^r - 0.0134 \cdot 10^3 L_4^r \\
				& - 0.0033 \cdot 10^3 L_5^r + 0.0067 \cdot 10^3 L_6^r + 0.0033 \cdot 10^3 L_8^r + ( 0.0098 - 0.0060 \cdot 10^3 L_5^r ) \cdot 10^3 L_9^r \Big) \end{aligned} \\
			& + \frac{s_\ell^2}{M_K^4} \begin{aligned}[t]
				& \Big( 0.0057 \cdot 10^3 L_1^r + 0.0010 \cdot 10^3 L_2^r + 0.0029 \cdot 10^3 L_3^r - 0.0065 \cdot 10^3 L_4^r \\
				& - 0.0010 \cdot 10^3 L_5^r + 0.0012 \cdot 10^3 L_6^r + 0.0006 \cdot 10^3 L_8^r - 0.0034 \cdot 10^3 L_9^r \Big) \bigg) , \end{aligned} \end{aligned} \\
	\end{split}
\end{align}
\begin{align}
	\footnotesize
	\begin{split}
		\tilde m_{1,L}^{1,\mathrm{NNLO}} &= \frac{M_K}{\sqrt{2} F_\pi} \begin{aligned}[t]
			&\bigg( 0.0987 \cdot 10^3 L_1^r + (0.2328 - 0.0242 \cdot 10^3 L_5^r) \cdot 10^3 L_2^r + 0.0581 \cdot 10^3 L_3^r + 0.0213 \cdot 10^3 L_4^r \\
			& + 0.0062 \cdot 10^3 L_5^r - 0.0078 \cdot 10^3 L_6^r - 0.0039 \cdot 10^3 L_8^r \\
			& + \frac{s_\ell}{M_K^2} \begin{aligned}[t]
				& \Big(-0.0138 \cdot 10^3 L_1^r - 0.0029 \cdot 10^3 L_2^r - 0.0026 \cdot 10^3 L_3^r + 0.0130 \cdot 10^3 L_4^r \\
				& + 0.0020 \cdot 10^3 L_5^r - 0.0024 \cdot 10^3 L_6^r - 0.0012 \cdot 10^3 L_8^r + 0.0089 \cdot 10^3 L_9^r \Big) \bigg) , \end{aligned} \end{aligned} \\
		\tilde m_{1,L}^{2,\mathrm{NNLO}} &= \frac{M_K}{\sqrt{2} F_\pi} \begin{aligned}[t]
			&\bigg( -0.0070 \cdot 10^3 L_1^r + 0.0114 \cdot 10^3 L_2^r - 0.0097 \cdot 10^3 L_3^r - 0.0044 \cdot 10^3 L_4^r \\
			& + 0.0001 \cdot 10^3 L_5^r + 0.0012 \cdot 10^3 L_6^r + 0.0006 \cdot 10^3 L_8^r \bigg) ,\end{aligned} \\
		n_{0,L}^{1,\mathrm{NNLO}} &= \frac{M_K}{\sqrt{2} F_\pi} \begin{aligned}[t]
			&\bigg( - 0.0796 \cdot 10^3 L_1^r + ( - 0.4712 + 0.0726 \cdot 10^3 L_5^r) \cdot 10^3 L_2^r + ( - 0.1097 + 0.0181 \cdot 10^3 L_5^r) \cdot 10^3 L_3^r \\
			& - 0.0262 \cdot 10^3 L_4^r - 0.0049 \cdot 10^3 L_5^r + 0.0075 \cdot 10^3 L_6^r - 0.0117 \cdot 10^3 L_7^r - 0.0021 \cdot 10^3 L_8^r \\
			& + \frac{s_\ell}{M_K^2} \begin{aligned}[t]
				& \Big( 0.0095 \cdot 10^3 L_1^r + 0.0035 \cdot 10^3 L_2^r + 0.0040 \cdot 10^3 L_3^r - 0.0098 \cdot 10^3 L_4^r \\
				& - 0.0015 \cdot 10^3 L_5^r + 0.0018 \cdot 10^3 L_6^r + 0.0009 \cdot 10^3 L_8^r - 0.0079 \cdot 10^3 L_9^r \Big)  \bigg) , \end{aligned} \end{aligned} \\
		n_{0,L}^{2,\mathrm{NNLO}} &= \frac{M_K}{\sqrt{2} F_\pi} \begin{aligned}[t]
			&\bigg( -0.0003 \cdot 10^3 L_1^r - 0.0010 \cdot 10^3 L_2^r + 0.0101 \cdot 10^3 L_3^r - 0.0007 \cdot 10^3 L_4^r \\
			& - 0.0022 \cdot 10^3 L_5^r + 0.0003 \cdot 10^3 L_6^r - 0.0010 \cdot 10^3 L_7^r - 0.0004 \cdot 10^3 L_8^r \bigg) ,\end{aligned} \\
		n_{1,L}^{0,\mathrm{NNLO}} &= \frac{M_K}{\sqrt{2} F_\pi} \begin{aligned}[t]
			&\bigg( -0.0125 \cdot 10^3 L_1^r - 0.0051 \cdot 10^3 L_2^r - 0.0069 \cdot 10^3 L_3^r + 0.0098 \cdot 10^3 L_4^r \\
			& + 0.0015 \cdot 10^3 L_5^r - 0.0018 \cdot 10^3 L_6^r - 0.0009 \cdot 10^3 L_8^r \bigg) ,\end{aligned} \\
		\tilde n_{1,L}^{1,\mathrm{NNLO}} &= \frac{M_K}{\sqrt{2} F_\pi} \begin{aligned}[t]
			&\bigg( 0.0059 \cdot 10^3 L_1^r + 0.0096 \cdot 10^3 L_2^r + 0.0051 \cdot 10^3 L_3^r - 0.0072 \cdot 10^3 L_4^r \\
			& - 0.0012 \cdot 10^3 L_5^r + 0.0018 \cdot 10^3 L_6^r + 0.0009 \cdot 10^3 L_8^r \bigg) . \end{aligned}
	\end{split}
\end{align}
Note that there are no quadratic terms in $L_1^r$, $L_2^r$ or $L_3^r$.

\subsection{Chiral Expansion of the Omnès Representation}

\label{sec:AppendixNNLOChiralExpansionOmnes}

In order to derive the NNLO chiral expansion of the Omnès representation (\ref{eq:FunctionsOfOneVariableOmnes3Subtr}), we first expand the Omnès function chirally:
\begin{align}
	\begin{split}
		\Omega^\mathrm{NNLO}(s) &= 1 + \omega \frac{s}{M_K^2} +  \bar\omega \frac{s^2}{M_K^4} +  \frac{s^3}{\pi} \int_{s_0}^\infty \frac{\delta_\mathrm{NLO}(s^\prime)}{(s^\prime - s - i \epsilon) {s^\prime}^3} ds^\prime \\
			&\quad + \frac{1}{2} \left( \omega \frac{s}{M_K^2} + \frac{s^2}{\pi} \int_{s_0}^\infty \frac{\delta_\mathrm{LO}(s^\prime)}{(s^\prime - s - i \epsilon) {s^\prime}^2} ds^\prime \right)^2 ,
	\end{split}
\end{align}
where the subtraction terms $\omega$ and $\bar \omega$ are defined in (\ref{eqn:ThriceSubtractedOmnesFunction}).

In the quadratic term of the expansion, only the LO phase enters and therefore only two subtractions are needed. The NLO expansion of the modulus of the inverse Omnès function is given by
\begin{align}
	\begin{split}
		\frac{1}{| \Omega^\mathrm{NLO}(s)|} &= 1 - \omega \frac{s}{M_K^2} - \frac{s^2}{\pi} \pvint_{s_0}^\infty \frac{\delta_\mathrm{NLO}(s^\prime)}{(s^\prime - s - i \epsilon) {s^\prime}^2} ds^\prime .
	\end{split}
\end{align}
Therefore, the NNLO chiral expansion of the argument of the dispersive integrals reads
\begin{align}
	\begin{split}
		\frac{\hat M(s) \sin\delta(s)}{|\Omega(s)|}\Bigg|_\mathrm{NNLO} &= \hat M^\mathrm{LO}(s) \delta_\mathrm{NLO}(s) + \hat M^\mathrm{NLO}(s) \delta_\mathrm{LO}(s) \\
			&\quad - \hat M^\mathrm{LO}(s) \delta_\mathrm{LO}(s) \left( 1 + \omega \frac{s}{M_K^2} + \frac{s^2}{\pi} \pvint_{s_0}^\infty \frac{\delta_\mathrm{LO}(s^\prime)}{(s^\prime - s - i\epsilon) {s^\prime}^2} ds^\prime \right) .
	\end{split}
\end{align}

This leads to
\begin{align}
	\scalebox{0.8}{
	\begin{minipage}{1.1\textwidth}
	$
	\begin{split}
		M_0^\mathrm{NNLO}(s) &= a_\mathrm{LO}^{M_0} \Bigg( \omega_0^0 \frac{s}{M_K^2} +  \bar\omega_0^0 \frac{s^2}{M_K^4} +  \frac{s^3}{\pi} \int_{s_0}^\infty \frac{\delta_{0,\mathrm{NLO}}^0(s^\prime)}{(s^\prime - s - i \epsilon) {s^\prime}^3} ds^\prime + \frac{1}{2} \bigg( \omega_0^0 \frac{s}{M_K^2} + \frac{s^2}{\pi} \int_{s_0}^\infty \frac{\delta_{0,\mathrm{LO}}^0(s^\prime)}{(s^\prime - s - i \epsilon) {s^\prime}^2} ds^\prime \bigg)^2  \Bigg) \\
			&\quad + \bigg( \Delta a_\mathrm{NLO}^{M_0} + b_\mathrm{NLO}^{M_0} \frac{s}{M_K^2} + c_\mathrm{NLO}^{M_0} \frac{s^2}{M_K^4} \bigg) \Bigg( \omega_0^0 \frac{s}{M_K^2} + \frac{s^2}{\pi} \int_{s_0}^\infty \frac{\delta_{0,\mathrm{LO}}^0(s^\prime)}{(s^\prime - s - i \epsilon) {s^\prime}^2} ds^\prime \Bigg) \\
			&\quad + a_\mathrm{NNLO}^{M_0} + b_\mathrm{NNLO}^{M_0} \frac{s}{M_K^2} + c_\mathrm{NNLO}^{M_0} \frac{s^2}{M_K^4} + d_\mathrm{NNLO}^{M_0} \frac{s^3}{M_K^6} + \frac{s^4}{\pi} \int_{s_0}^\infty \frac{\hat M_0^\mathrm{NLO}(s^\prime) \delta_{0,\mathrm{LO}}^0(s^\prime)}{(s^\prime - s - i\epsilon) {s^\prime}^4} ds^\prime , \\
		M_1^\mathrm{NNLO}(s) &= \bigg( a_\mathrm{NLO}^{M_1} + b_\mathrm{NLO}^{M_1} \frac{s}{M_K^2} \bigg) \Bigg( \omega_1^1 \frac{s}{M_K^2} + \frac{s^2}{\pi} \int_{s_0}^\infty \frac{\delta_{1,\mathrm{LO}}^1(s^\prime)}{(s^\prime - s - i \epsilon) {s^\prime}^2} ds^\prime \Bigg) \\
			& \quad + a_\mathrm{NNLO}^{M_1} + b_\mathrm{NNLO}^{M_1}  \frac{s}{M_K^2} + c_\mathrm{NNLO}^{M_1} \frac{s^2}{M_K^4} + \frac{s^3}{\pi} \int_{s_0}^\infty \frac{\hat M_1^\mathrm{NLO}(s^\prime) \delta_{1,\mathrm{LO}}^1(s^\prime)}{(s^\prime - s - i\epsilon) {s^\prime}^3} ds^\prime , \\
		 \tilde M_1^\mathrm{NNLO}(s) &= a_\mathrm{LO}^{\tilde M_1} \Bigg( \omega_1^1 \frac{s}{M_K^2} +  \bar\omega_1^1 \frac{s^2}{M_K^4} +  \frac{s^3}{\pi} \int_{s_0}^\infty \frac{\delta_{1,\mathrm{NLO}}^1(s^\prime)}{(s^\prime - s - i \epsilon) {s^\prime}^3} ds^\prime + \frac{1}{2} \bigg( \omega_1^1 \frac{s}{M_K^2} + \frac{s^2}{\pi} \int_{s_0}^\infty \frac{\delta_{1,\mathrm{LO}}^1(s^\prime)}{(s^\prime - s - i \epsilon) {s^\prime}^2} ds^\prime \bigg)^2 \Bigg) \\
			& \quad + \bigg( \Delta a_\mathrm{NLO}^{\tilde M_1} + b_\mathrm{NLO}^{\tilde M_1}  \frac{s}{M_K^2} + c_\mathrm{NLO}^{\tilde M_1}  \frac{s^2}{M_K^4} \bigg) \Bigg( \omega_1^1 \frac{s}{M_K^2} + \frac{s^2}{\pi} \int_{s_0}^\infty \frac{\delta_{1,\mathrm{LO}}^1(s^\prime)}{(s^\prime - s - i \epsilon) {s^\prime}^2} ds^\prime \Bigg) \\
			& \quad + a_\mathrm{NNLO}^{\tilde M_1} + b_\mathrm{NNLO}^{\tilde M_1}  \frac{s}{M_K^2} + c_\mathrm{NNLO}^{\tilde M_1}  \frac{s^2}{M_K^4} + d_\mathrm{NNLO}^{\tilde M_1} \frac{s^3}{M_K^6} + \frac{s^4}{\pi} \int_{s_0}^\infty \frac{\hat{\tilde M}_1^\mathrm{NLO}(s^\prime) \delta_{1,\mathrm{LO}}^1(s^\prime)}{(s^\prime - s - i\epsilon) {s^\prime}^4} ds^\prime , \\
		N_0^\mathrm{NNLO}(t) &= \Bigg( b_\mathrm{NLO}^{N_0} \frac{t}{M_K^2} + c_\mathrm{NLO}^{N_0} \frac{t^2}{M_K^4} + \frac{t^3}{\pi} \int_{t_0}^\infty \frac{\hat N_0^\mathrm{LO}(t^\prime) \delta_{0,\mathrm{LO}}^{1/2}(t^\prime)}{(t^\prime - t - i\epsilon) {t^\prime}^3} dt^\prime \Bigg) \Bigg( \omega_0^{1/2} \frac{t}{M_K^2} + \frac{t^2}{\pi} \int_{t_0}^\infty \frac{\delta_{0,\mathrm{LO}}^{1/2}(t^\prime)}{(t^\prime - t - i \epsilon) {t^\prime}^2} dt^\prime \Bigg) \\
			&\quad + b_\mathrm{NNLO}^{N_0} \frac{t}{M_K^2} + c_\mathrm{NNLO}^{N_0} \frac{t^2}{M_K^4} + \frac{t^3}{\pi} \int_{t_0}^\infty \frac{\hat N_0^\mathrm{LO}(t^\prime) \delta_{0,\mathrm{NLO}}^{1/2}(t^\prime)}{(t^\prime - t - i\epsilon) {t^\prime}^3} dt^\prime + \frac{t^3}{\pi} \int_{t_0}^\infty \frac{\hat N_0^\mathrm{NLO}(t^\prime) \delta_{0,\mathrm{LO}}^{1/2}(t^\prime)}{(t^\prime - t - i\epsilon) {t^\prime}^3} dt^\prime \\
			&\quad - \frac{t^3}{\pi} \int_{t_0}^\infty \frac{\hat N_0^\mathrm{LO}(t^\prime) \delta_{0,\mathrm{LO}}^{1/2}(t^\prime)}{(t^\prime - t - i\epsilon) {t^\prime}^3} \bigg( 1 + \omega_0^{1/2} \frac{t^\prime}{M_K^2} + \frac{{t^\prime}^2}{\pi} \pvint_{t_0}^\infty \frac{\delta_{0,\mathrm{LO}}^{1/2}(t^\dprime)}{(t^\dprime - t^\prime - i \epsilon){t^\dprime}^2} dt^\dprime \bigg) dt^\prime , \\
		N_1^\mathrm{NNLO}(t) &= a_\mathrm{NLO}^{N_1} \Bigg( \omega_1^{1/2} \frac{t}{M_K^2} + \frac{t^2}{\pi} \int_{t_0}^\infty \frac{\delta_{1,\mathrm{LO}}^{1/2}(t^\prime)}{(t^\prime - t - i \epsilon) {t^\prime}^2} dt^\prime \Bigg) + a_\mathrm{NNLO}^{N_1} + \frac{t}{\pi} \int_{t_0}^\infty \frac{\hat N_1^\mathrm{NLO}(t^\prime) \delta_{1,\mathrm{LO}}^{1/2}(t^\prime)}{(t^\prime - t - i\epsilon){t^\prime}} dt^\prime, \\
		\tilde N_1^\mathrm{NNLO}(t) &= \Bigg( b_\mathrm{NLO}^{\tilde N_1} \frac{t}{M_K^2} + \frac{t^2}{\pi} \int_{t_0}^\infty \frac{\hat{\tilde N}_1^\mathrm{LO}(t^\prime) \delta_{1,\mathrm{LO}}^{1/2}(t^\prime)}{(t^\prime - t - i\epsilon) {t^\prime}^2} dt^\prime \Bigg) \Bigg( \omega_1^{1/2} \frac{t}{M_K^2} + \frac{t^2}{\pi} \int_{t_0}^\infty \frac{\delta_{1,\mathrm{LO}}^{1/2}(t^\prime)}{(t^\prime - t - i \epsilon) {t^\prime}^2} dt^\prime \Bigg) \\
			&\quad +  b_\mathrm{NNLO}^{\tilde N_1} \frac{t}{M_K^2} + \frac{t^2}{\pi} \int_{t_0}^\infty \frac{\hat{\tilde N}_1^\mathrm{LO}(t^\prime) \delta_{1,\mathrm{NLO}}^{1/2}(t^\prime)}{(t^\prime - t - i\epsilon) {t^\prime}^2} dt^\prime + \frac{t^2}{\pi} \int_{t_0}^\infty \frac{\hat{\tilde N}_1^\mathrm{NLO}(t^\prime) \delta_{1,\mathrm{LO}}^{1/2}(t^\prime)}{(t^\prime - t - i\epsilon) {t^\prime}^2} dt^\prime \\
			&\quad - \frac{t^2}{\pi} \int_{t_0}^\infty \frac{\hat{\tilde N}_1^\mathrm{LO}(t^\prime) \delta_{1,\mathrm{LO}}^{1/2}(t^\prime)}{(t^\prime - t - i\epsilon) {t^\prime}^2} \bigg( 1 + \omega_1^{1/2} \frac{t^\prime}{M_K^2} + \frac{{t^\prime}^2}{\pi} \pvint_{t_0}^\infty \frac{\delta_{1,\mathrm{LO}}^{1/2}(t^\dprime)}{(t^\dprime-t^\prime-i\epsilon){t^\dprime}^2} dt^\dprime \bigg) dt^\prime , \\
		R_0^\mathrm{NNLO}(t) &= \Bigg( \frac{t^3}{\pi} \int_{t_0}^\infty \frac{\hat R_0^\mathrm{LO}(t^\prime) \delta_{0,\mathrm{LO}}^{3/2}(t^\prime)}{(t^\prime - t - i\epsilon) {t^\prime}^3} dt^\prime \Bigg) \Bigg( \omega_0^{3/2} \frac{t}{M_K^2} + \frac{t^2}{\pi} \int_{t_0}^\infty \frac{\delta_{0,\mathrm{LO}}^{3/2}(t^\prime)}{(t^\prime - t - i \epsilon) {t^\prime}^2} dt^\prime \Bigg) \\
			&\quad +  \frac{t^3}{\pi} \int_{t_0}^\infty \frac{\hat R_0^\mathrm{LO}(t^\prime) \delta_{0,\mathrm{NLO}}^{3/2}(t^\prime)}{(t^\prime - t - i\epsilon) {t^\prime}^3} dt^\prime +  \frac{t^3}{\pi} \int_{t_0}^\infty \frac{\hat R_0^\mathrm{NLO}(t^\prime) \delta_{0,\mathrm{LO}}^{3/2}(t^\prime)}{(t^\prime - t - i\epsilon) {t^\prime}^3} dt^\prime \\
			& \quad - \frac{t^3}{\pi} \int_{t_0}^\infty \frac{\hat R_0^\mathrm{LO}(t^\prime) \delta_{0,\mathrm{LO}}^{3/2}(t^\prime)}{(t^\prime - t - i\epsilon) {t^\prime}^3} \bigg( 1 + \omega_0^{3/2} \frac{t^\prime}{M_K^2} + \frac{{t^\prime}^2}{\pi} \pvint_{t_0}^\infty \frac{\delta_{0,\mathrm{LO}}^{3/2}(t^\dprime)}{(t^\dprime-t^\prime-i\epsilon){t^\dprime}^2} dt^\dprime \bigg) dt^\prime  , \\
		R_1^\mathrm{NNLO}(t) &= 0 , \\
		\tilde R_1^\mathrm{NNLO}(t) &= 0 ,
	\end{split}
	$
	\end{minipage}
	}
\end{align}
where we use the following notation for the contributions to the subtraction constants:
\begin{align}
	\begin{split}
		a_\mathrm{NLO} &= a_\mathrm{LO} + \Delta a_\mathrm{NLO} , \\
		a_\mathrm{NNLO} &= a_\mathrm{LO} + \Delta a_\mathrm{NLO} + \Delta a_\mathrm{NNLO} .
	\end{split}
\end{align}
Remember that $b_\mathrm{NLO}^{M_1}$ and $a_\mathrm{NLO}^{N_1}$ are non-zero after the gauge transformation.

We further define
\begin{align}
	\begin{split}
		b^{M_0}_\mathrm{NLO} &=: -\omega_0^0 \frac{M_K}{\sqrt{2} F_\pi} + \bar b^{M_0}_\mathrm{NLO} , \\
		b^{M_0}_\mathrm{NNLO} &=: -\omega_0^0 \frac{M_K}{\sqrt{2} F_\pi} + \bar b^{M_0}_\mathrm{NNLO} , \\
		b^{\tilde M_1}_\mathrm{NLO} &=: -\omega_1^1 \frac{M_K}{\sqrt{2} F_\pi} + \bar b^{\tilde M_1}_\mathrm{NLO} , \\
		b^{\tilde M_1}_\mathrm{NNLO} &=: -\omega_1^1 \frac{M_K}{\sqrt{2} F_\pi} + \bar b^{\tilde M_1}_\mathrm{NNLO} ,
	\end{split}
\end{align}
which allows the simplifications
\begin{align}
	\scalebox{0.75}{
	\begin{minipage}{1.2\textwidth}
	$
	\begin{split}
		\label{eq:OmnesRepresentationChirallyExpanded3Subtr}
		M_0^\mathrm{NNLO}(s) &= \frac{M_K}{\sqrt{2} F_\pi} \begin{aligned}[t]
				& \Bigg( \left( \bar\omega_0^0 - \frac{1}{2}{\omega_0^0}^2 \right) \frac{s^2}{M_K^4} +  \frac{s^3}{\pi} \int_{s_0}^\infty \frac{\delta_{0,\mathrm{NLO}}^0(s^\prime)}{(s^\prime - s - i \epsilon) {s^\prime}^3} ds^\prime + \frac{1}{2}\left(\frac{s^2}{\pi} \int_{s_0}^\infty \frac{\delta_{0,\mathrm{LO}}^0(s^\prime)}{(s^\prime - s - i \epsilon) {s^\prime}^2} ds^\prime \right)^2  \Bigg) \end{aligned} \\
			&\quad + \bigg( \Delta a_\mathrm{NLO}^{M_0} + \bar b^{M_0}_\mathrm{NLO} \frac{s}{M_K^2} + c_\mathrm{NLO}^{M_0} \frac{s^2}{M_K^4} \bigg) \Bigg( \omega_0^0 \frac{s}{M_K^2} + \frac{s^2}{\pi} \int_{s_0}^\infty \frac{\delta_{0,\mathrm{LO}}^0(s^\prime)}{(s^\prime - s - i \epsilon) {s^\prime}^2} ds^\prime \Bigg) \\
			&\quad + a_\mathrm{NNLO}^{M_0} + \bar b_\mathrm{NNLO}^{M_0} \frac{s}{M_K^2} + c_\mathrm{NNLO}^{M_0} \frac{s^2}{M_K^4} + d_\mathrm{NNLO}^{M_0} \frac{s^3}{M_K^6} + \frac{s^4}{\pi} \int_{s_0}^\infty \frac{\hat M_0^\mathrm{NLO}(s^\prime) \delta_{0,\mathrm{LO}}^0(s^\prime)}{(s^\prime - s - i\epsilon) {s^\prime}^4} ds^\prime , \\
		M_1^\mathrm{NNLO}(s) &= \bigg( a_\mathrm{NLO}^{M_1} + b_\mathrm{NLO}^{M_1} \frac{s}{M_K^2} \bigg) \Bigg( \omega_1^1 \frac{s}{M_K^2} + \frac{s^2}{\pi} \int_{s_0}^\infty \frac{\delta_{1,\mathrm{LO}}^1(s^\prime)}{(s^\prime - s - i \epsilon) {s^\prime}^2} ds^\prime \Bigg) \\
			& \quad + a_\mathrm{NNLO}^{M_1} + b_\mathrm{NNLO}^{M_1}  \frac{s}{M_K^2} + c_\mathrm{NNLO}^{M_1}  \frac{s^2}{M_K^4} + \frac{s^3}{\pi} \int_{s_0}^\infty \frac{\hat M_1^\mathrm{NLO}(s^\prime) \delta_{1,\mathrm{LO}}^1(s^\prime)}{(s^\prime - s - i\epsilon) {s^\prime}^3} ds^\prime , \\
		\tilde M_1^\mathrm{NNLO}(s) &= \frac{M_K}{\sqrt{2} F_\pi} \begin{aligned}[t]
		 		& \Bigg( \left( \bar\omega_1^1 - \frac{1}{2} {\omega_1^1}^2 \right) \frac{s^2}{M_K^4} +  \frac{s^3}{\pi} \int_{s_0}^\infty \frac{\delta_{1,\mathrm{NLO}}^1(s^\prime)}{(s^\prime - s - i \epsilon) {s^\prime}^3} ds^\prime + \frac{1}{2} \bigg( \frac{s^2}{\pi} \int_{s_0}^\infty \frac{\delta_{1,\mathrm{LO}}^1(s^\prime)}{(s^\prime - s - i \epsilon) {s^\prime}^2} ds^\prime \bigg)^2 \Bigg) \end{aligned} \\
			& \quad + \bigg( \Delta a_\mathrm{NLO}^{\tilde M_1} + \bar b^{\tilde M_1}_\mathrm{NLO} \frac{s}{M_K^2} + c_\mathrm{NLO}^{\tilde M_1}  \frac{s^2}{M_K^4} \bigg) \Bigg( \omega_1^1 \frac{s}{M_K^2} + \frac{s^2}{\pi} \int_{s_0}^\infty \frac{\delta_{1,\mathrm{LO}}^1(s^\prime)}{(s^\prime - s - i \epsilon) {s^\prime}^2} ds^\prime \Bigg) \\
			& \quad + a_\mathrm{NNLO}^{\tilde M_1} + \bar b_\mathrm{NNLO}^{\tilde M_1}  \frac{s}{M_K^2} + c_\mathrm{NNLO}^{\tilde M_1}  \frac{s^2}{M_K^4} + d_\mathrm{NNLO}^{\tilde M_1}  \frac{s^3}{M_K^6} + \frac{s^4}{\pi} \int_{s_0}^\infty \frac{\hat{\tilde M}_1^\mathrm{NLO}(s^\prime) \delta_{1,\mathrm{LO}}^1(s^\prime)}{(s^\prime - s - i\epsilon) {s^\prime}^4} ds^\prime , \\
		N_0^\mathrm{NNLO}(t) &= b_\mathrm{NNLO}^{N_0} \frac{t}{M_K^2} + c_\mathrm{NNLO}^{N_0} \frac{t^2}{M_K^4} + \omega_0^{1/2} \frac{t}{M_K^2} \Bigg( b_\mathrm{NLO}^{N_0} \frac{t}{M_K^2} + \delta c_\mathrm{NLO}^{N_0} \frac{t^2}{M_K^4} \Bigg) \\
			&\quad +  \Bigg(\frac{t^2}{\pi} \int_{t_0}^\infty \frac{\delta_{0,\mathrm{LO}}^{1/2}(t^\prime)}{(t^\prime - t - i \epsilon) {t^\prime}^2} dt^\prime \Bigg) \Bigg( b_\mathrm{NLO}^{N_0} \frac{t}{M_K^2} + c_\mathrm{NLO}^{N_0} \frac{t^2}{M_K^4} + \frac{t^3}{\pi} \int_{t_0}^\infty \frac{\hat N_0^\mathrm{LO}(t^\prime) \delta_{0,\mathrm{LO}}^{1/2}(t^\prime)}{(t^\prime - t - i\epsilon) {t^\prime}^3} dt^\prime \Bigg) \\
			&\quad + \frac{t^3}{\pi} \int_{t_0}^\infty \frac{\hat N_0^\mathrm{LO}(t^\prime) \delta_{0,\mathrm{NLO}}^{1/2}(t^\prime)}{(t^\prime - t - i\epsilon) {t^\prime}^3} dt^\prime + \frac{t^3}{\pi} \int_{t_0}^\infty \frac{\hat N_0^\mathrm{NLO}(t^\prime) \delta_{0,\mathrm{LO}}^{1/2}(t^\prime)}{(t^\prime - t - i\epsilon) {t^\prime}^3} dt^\prime \\
			&\quad - \frac{t^3}{\pi} \int_{t_0}^\infty \frac{\hat N_0^\mathrm{LO}(t^\prime) \delta_{0,\mathrm{LO}}^{1/2}(t^\prime)}{(t^\prime - t - i\epsilon) {t^\prime}^3} \bigg( 1 + \frac{{t^\prime}^2}{\pi} \pvint_{t_0}^\infty \frac{\delta_{0,\mathrm{LO}}^{1/2}(t^\dprime)}{(t^\dprime - t^\prime - i \epsilon){t^\dprime}^2} dt^\dprime \bigg) dt^\prime , \\
		N_1^\mathrm{NNLO}(t) &= a_\mathrm{NNLO}^{N_1} + a_\mathrm{NLO}^{N_1} \omega_1^{1/2} \frac{t}{M_K^2} \\
			&\quad + a_\mathrm{NLO}^{N_1} \frac{t^2}{\pi} \int_{t_0}^\infty \frac{\delta_{1,\mathrm{LO}}^{1/2}(t^\prime)}{(t^\prime - t - i \epsilon) {t^\prime}^2} dt^\prime + \frac{t}{\pi} \int_{t_0}^\infty \frac{\hat N_1^\mathrm{NLO}(t^\prime) \delta_{1,\mathrm{LO}}^{1/2}(t^\prime)}{(t^\prime - t - i\epsilon){t^\prime}} dt^\prime, \\
		\tilde N_1^\mathrm{NNLO}(t) &= b_\mathrm{NNLO}^{\tilde N_1} \frac{t}{M_K^2} + \omega_1^{1/2} \delta b_\mathrm{NLO}^{\tilde N_1} \frac{t^2}{M_K^4} \\
			&\quad + \Bigg( \frac{t^2}{\pi} \int_{t_0}^\infty \frac{\delta_{1,\mathrm{LO}}^{1/2}(t^\prime)}{(t^\prime - t - i \epsilon) {t^\prime}^2} dt^\prime \Bigg) \bigg( b_\mathrm{NLO}^{\tilde N_1} \frac{t}{M_K^2} + \frac{t^2}{\pi} \int_{t_0}^\infty \frac{\hat{\tilde N}_1^\mathrm{LO}(t^\prime) \delta_{1,\mathrm{LO}}^{1/2}(t^\prime)}{(t^\prime - t - i\epsilon) {t^\prime}^2} dt^\prime \bigg) + \frac{t^2}{\pi} \int_{t_0}^\infty \frac{\hat{\tilde N}_1^\mathrm{LO}(t^\prime) \delta_{1,\mathrm{NLO}}^{1/2}(t^\prime)}{(t^\prime - t - i\epsilon) {t^\prime}^2} dt^\prime \\
			&\quad + \frac{t^2}{\pi} \int_{t_0}^\infty \frac{\hat{\tilde N}_1^\mathrm{NLO}(t^\prime) \delta_{1,\mathrm{LO}}^{1/2}(t^\prime)}{(t^\prime - t - i\epsilon) {t^\prime}^2} dt^\prime - \frac{t^2}{\pi} \int_{t_0}^\infty \frac{\hat{\tilde N}_1^\mathrm{LO}(t^\prime) \delta_{1,\mathrm{LO}}^{1/2}(t^\prime)}{(t^\prime - t - i\epsilon) {t^\prime}^2} \bigg( 1 + \frac{{t^\prime}^2}{\pi} \pvint_{t_0}^\infty \frac{\delta_{1,\mathrm{LO}}^{1/2}(t^\dprime)}{(t^\dprime-t^\prime-i\epsilon){t^\dprime}^2} dt^\dprime \bigg) dt^\prime , \\
		R_0^\mathrm{NNLO}(t) &= - \omega_0^{3/2} \frac{t}{M_K^2} \frac{t^2}{\pi} \int_{t_0}^\infty \frac{\hat R_0^\mathrm{LO}(t^\prime) \delta_{0,\mathrm{LO}}^{3/2}(t^\prime)}{{t^\prime}^3} dt^\prime \\
			&\quad + \Bigg( \frac{t^2}{\pi} \int_{t_0}^\infty \frac{\delta_{0,\mathrm{LO}}^{3/2}(t^\prime)}{(t^\prime - t - i \epsilon) {t^\prime}^2} dt^\prime \Bigg) \bigg( \frac{t^3}{\pi} \int_{t_0}^\infty \frac{\hat R_0^\mathrm{LO}(t^\prime) \delta_{0,\mathrm{LO}}^{3/2}(t^\prime)}{(t^\prime - t - i\epsilon) {t^\prime}^3} dt^\prime \bigg) + \frac{t^3}{\pi} \int_{t_0}^\infty \frac{\hat R_0^\mathrm{LO}(t^\prime) \delta_{0,\mathrm{NLO}}^{3/2}(t^\prime)}{(t^\prime - t - i\epsilon) {t^\prime}^3} dt^\prime \\
			&\quad +  \frac{t^3}{\pi} \int_{t_0}^\infty \frac{\hat R_0^\mathrm{NLO}(t^\prime) \delta_{0,\mathrm{LO}}^{3/2}(t^\prime)}{(t^\prime - t - i\epsilon) {t^\prime}^3} dt^\prime - \frac{t^3}{\pi} \int_{t_0}^\infty \frac{\hat R_0^\mathrm{LO}(t^\prime) \delta_{0,\mathrm{LO}}^{3/2}(t^\prime)}{(t^\prime - t - i\epsilon) {t^\prime}^3} \bigg( 1 + \frac{{t^\prime}^2}{\pi} \pvint_{t_0}^\infty \frac{\delta_{0,\mathrm{LO}}^{3/2}(t^\dprime)}{(t^\dprime-t^\prime-i\epsilon){t^\dprime}^2} dt^\dprime \bigg) dt^\prime , \\
		R_1^\mathrm{NNLO}(t) &= 0 , \\
		\tilde R_1^\mathrm{NNLO}(t) &= 0 ,
	\end{split}
	$
	\end{minipage}
	}
\end{align}
where $\delta c_\mathrm{NLO}^{N_0}$ and $\delta b_\mathrm{NLO}^{\tilde N_1}$ are given by (\ref{eq:NLOGaugeTransformationSubtractionConstants}) and the remaining subtraction constants denote the quantities after the gauge transformation. Note that $\omega$ and $\bar\omega$ appear only in polynomial terms. In $M_0$, $M_1$ and $\tilde M_1$, they can be reabsorbed into the NNLO subtraction constants. However, this is not the case for $N_0$, $N_1$, $\tilde N_1$ and $R_0$. Here, we are required to fix the $\omega$-terms by imposing that the chirally expanded Omnès representation agrees with the standard dispersive representation (or finally the two-loop representation). This somewhat awkward situation is just another manifestation of the fact that we identify the chiral representation with the Omnès dispersion relation although the phase shifts of the former have a wrong asymptotic behaviour.

The comparison of the Taylor expansions of (\ref{eq:OmnesRepresentationChirallyExpanded3Subtr}) and (\ref{eq:FunctionsOfOneVariable3Subtr}) leads to the relation (\ref{eqn:NNLORelationOmnesStandardSubtrConst}) for the subtraction constants.

\end{appendices}

\clearpage
\renewcommand\bibname{References}
\renewcommand{\bibfont}{\raggedright}
\bibliographystyle{my-physrev}
\phantomsection
\addcontentsline{toc}{chapter}{References}
\bibliography{Isospin/Literature}


\part[Isospin-Breaking Effects in $K_{\ell4}$~Decays]{Isospin-Breaking Effects in $K_{\ell4}$~Decays 
                             \\ 
                  \begin{center}
                     \begin{minipage}[l]{11cm}
                     \vspace{2cm}
                     \normalsize
                     \begin{center} 
                     	\textnormal{published in \\[0.5cm] Eur.Phys.J. {\bf C74} (2014) 2749} \\[2cm]
                     	Abstract
                     \end{center}
                     \textnormal{In the framework of chiral perturbation theory with photons and leptons, the one-loop isospin-breaking effects in $K_{\ell4}$ decays due to both the photonic contribution and the quark and meson mass differences are computed. \\
A comparison with the isospin-breaking corrections applied by recent high statistics $K_{e4}$ experiments is performed. \\
The calculation can be used to correct the existing form factor measurements by isospin-breaking effects that have not yet been taken into account in the experimental analysis. Based on the present work, possible forthcoming experiments on $K_{e4}$ decays could correct the isospin-breaking effects in a more consistent way.}
                     \end{minipage}
                  \end{center}
                 }
\addtocontents{toc}{\vskip-6pt\par\noindent\protect\hrulefill\par}
                 
\setcounter{chapter}{0}
                 

\chapter{Introduction}

High-precision hadron physics at low energies pursues mainly two aims: a better understanding of the strong interaction in its non-perturbative regime and the indirect search for new physics beyond the standard model. As perturbative QCD is not applicable, one has to use non-perturbative methods like effective field theories, lattice simulations or dispersion relations. The effective theory describing the strong interaction at low energy is chiral perturbation theory (\ChPT{},  \cite{Weinberg1968, GasserLeutwyler1984, GasserLeutwyler1985}). In order to render it predictive, one has to determine the parameters of the theory, the low-energy constants (LECs), either by comparison with experiments or with the help of lattice calculations. Dispersion relations and sum rules have proven to be useful to perform this task.

The $K_{\ell4}$ decay is for several reasons a particularly interesting process. The physical region starts at the $\pi\pi$ threshold, i.e.~at lower energies than $K\pi$ scattering, which gives access to the same low-energy constants. \ChPT{}, being an expansion in the meson masses and momenta, should therefore give a better description of $K_{\ell4}$ than $K\pi$ scattering. Besides providing a very clean access to some of the LECs, $K_{\ell4}$ is, due to its final state, one of the best sources of information on $\pi\pi$ interaction \cite{Shabalin1963,Cabibbo1965,Batley2010}.

The recent high-statistics  $K_{\ell4}$ experiments E865 at BNL \cite{Pislak2001, Pislak2003} and NA48/2 at CERN \cite{Batley2010, Batley2012} have achieved an impressive accuracy. The statistical errors of the form factor measurements of both experiments reach the sub-percent level (at least for the $S$-waves) and ask for a consistent treatment of isospin-breaking effects. Usually, theoretical calculations are performed in an ideal world with intact isospin, the $SU(2)$ symmetry of up- and down-quarks. The quark mass difference and the electromagnetism break isospin symmetry at the percent level.

Isospin-breaking effects in $K_{\ell4}$ have been studied in the previous literature and played a major role concerning the success of standard \ChPT{}. In \cite{Colangelo2009}, the effect of quark and meson mass differences on the phase shifts was studied. The inclusion of this effect brought the NA48/2 measurement of the scattering lengths into perfect agreement with the prediction of the \ChPT{}/Roy equation analysis \cite{Colangelo2000}. For a review of these developments, see \cite{Gasser2009}. The mass effects on the phases at two-loop order have been recently studied in an elaborate dispersive framework \cite{Bernard2013}, which confirms the previous results. In both works, the photonic effects are assumed to be treated consistently in the experimental analysis. The earlier work \cite{Cuplov2003, Cuplov2004} treats both mass and photonic effects. However, the calculation of virtual photon effects is incomplete and real photon corrections are taken into account only in the soft approximation.

The experimental analysis of the latest experiment \cite{Batley2010, Batley2012} treats photonic corrections with the semi-classical Gamow-Sommerfeld factor and PHOTOS Monte Carlo \cite{Barberio1994}, which assumes phase space factorisation.

The need for a theoretical treatment of the full radiative corrections was pointed out in \cite{Colangelo2009} and considered as a long-term project. With the present work, I intend to fill this gap. The obtained results enable a better correction of isospin effects in the data:
\begin{itemize}
	\item as I will explain below, one can improve already now the handling of isospin effects in the data analysis;
	\item in the future, an event generator which incorporates the matrix element calculated here should be written and used to perform the data analysis.
\end{itemize}

The paper is organised as follows. In section~\ref{sec:Kinematics}, I define the kinematics, matrix elements and form factors of $K_{\ell4}$ and the radiative decay $K_{\ell4\gamma}$. In section~\ref{sec:ChPTCalculation}, I calculate the matrix elements within \ChPT{} including leptons and photons \cite{Urech1995, Knecht2000}. In section~\ref{sec:ExtractionOfIsospinCorrections}, I present the strategy of extracting the isospin corrections and perform the phase space integration for the radiative decay. The cancellation of both infrared and mass divergences is demonstrated. In section~\ref{sec:Numerics}, the isospin corrections are evaluated numerically. I compare the full radiative process with the soft photon approximation and with the strategy used in the experimental analysis. The appendices give details on the calculation and explicit results for the matrix elements.

It should be noted that large parts of this work assume a small lepton mass and are therefore not applicable to the muonic mode of the process.


\chapter{Kinematics and Decay Rate}

\label{sec:Kinematics}

\section{The $K_{\ell4}$ Decay}

\subsection{Definition of the Decay}

$K_{\ell4}$ is the semileptonic decay of a kaon into two pions, a lepton and a neutrino. Let us consider here the following charged mode:
\begin{align}
	\label{eqn:Kl4Decay}
	K^+(p) \rightarrow \pi^+(p_1) \pi^-(p_2) \ell^+(p_\ell) \nu_\ell(p_\nu),
\end{align}
where $\ell\in\{e,\mu\}$ is either an electron or a muon.

\begin{figure}[ht]
	\centering
	\setlength{\unitlength}{1cm}
	\begin{pspicture}(-1,-1)(13,5)
		\pspolygon[linestyle=none,fillstyle=solid](-1,-1)(13,-1)(13,5)(-1,5)
		\psline[linestyle=dotted](6,2)(7,4)(5,4)
		\psline[linestyle=dashed](1,2)(11,2)
		\psline[linestyle=dashed](2.5,0)(4.5,4)
		\psline[linestyle=dashed](9.5,0)(7.4,4.2)
		\psTextFrame[linestyle=none,fillstyle=solid](5.98,1.7)(6.9,2.3){$K^+$}
		\psline(5,4)(2,4)(0,0)(5,0)(6,2)
		\pspolygon(7,0)(12,0)(9.9,4.2)(4.9,4.2)
		\psline[arrowsize=6pt]{<->}(4,0.5)(3,3.5)
		\rput(4.3,0.5){$\pi^-$}
		\rput(3.3,3.7){$\pi^+$}
		\psline[arrowsize=6pt]{<->}(7,1.5)(10,2.5)
		\rput(10,2.8){$\ell^+$}
		\rput(7,1.2){$\nu_\ell$}
		\psarc{->}(6,2){1}{63.43}{116.57}
		\rput(6,2.7){$\phi$}
		\psarc{<-}(3.5,2){1}{108.43}{180}
		\rput(3,2.4){$\theta_\pi$}
		\psarc{->}(8.5,2){1}{0}{18.43}
		\psline[linecolor=white,linewidth=2pt](8.95,2)(9.45,2)
		\rput(9.25,2){$\theta_\ell$}
		\rput(1,0.5){$\Sigma_{2\pi}$}
		\rput(11,0.5){$\Sigma_{\ell\nu}$}
		\psline[arrowsize=4pt]{->}(3.5,2)(3.875,2.75)
		\rput(4.1,2.6){$\vec c$}
		\psline[arrowsize=4pt]{->}(8.5,2)(8.125,2.75)
		\rput(8.5,2.6){$\vec d$}
		\psline[arrowsize=4pt]{->}(6,2)(5.25,2)
		\rput(5.3,2.3){$\vec v$}
	\end{pspicture}
	\caption{The reference frames and the kinematic variables for the $K_{\ell4}$ decay.}
	\label{img:Kl4Kinematics}
\end{figure}
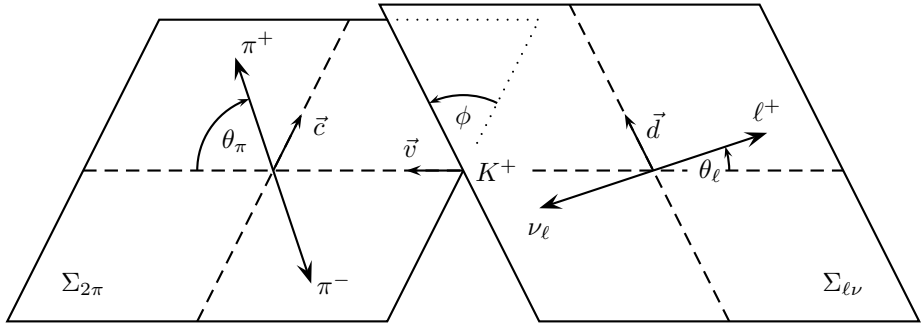

The kinematics of the decay (\ref{eqn:Kl4Decay}) can be described by five variables. The same conventions as in \cite{Bijnens1994} will be used, first introduced by Cabibbo and Maksymowicz \cite{Cabibbo1965}. There appear three different reference frames: the rest frame of the kaon $\Sigma_K$, the $\pi^+\pi^-$ centre-of-mass frame $\Sigma_{2\pi}$ and the $\ell^+\nu$ centre-of-mass frame $\Sigma_{\ell\nu}$. The situation is sketched in figure \ref{img:Kl4Kinematics}.

The five kinematic variables are then:
\begin{itemize}
	\item $s$, the total centre-of-mass squared energy of the two pions,
	\item $s_\ell$, the total centre-of-mass squared energy of the two leptons,
	\item $\theta_\pi$, the angle between the $\pi^+$ in $\Sigma_{2\pi}$ and the dipion line of flight in $\Sigma_K$,
	\item $\theta_\ell$, the angle between the $\ell^+$ in $\Sigma_{\ell\nu}$ and the dilepton line of flight in $\Sigma_K$,
	\item $\phi$, the angle between the dipion plane and the dilepton plane in $\Sigma_K$.
\end{itemize}

The (physical) ranges of these variables are:
\begin{align}
	\begin{alignedat}{2}
		4 \mpip^2 &\le s &&\le (\mkp - m_\ell)^2 , \\
		m_\ell^2 &\le s_\ell &&\le (\mkp - \sqrt{s})^2 , \\
		0 &\le \theta_\pi &&\le \pi , \\
		0 &\le \theta_\ell &&\le \pi , \\
		0 &\le \phi &&\le 2\pi.
	\end{alignedat}
\end{align}

Following \cite{Bijnens1994}, I define the four-momenta:
\begin{align}
	\label{eqn:FourMomenta}
	P := p_1 + p_2 , \quad Q := p_1 - p_2 , \quad L := p_\ell + p_\nu , \quad N := p_\ell - p_\nu .
\end{align}
Total momentum conservation implies $p = P + L$.

I will use the Mandelstam variables
\begin{align}
	s := (p_1 + p_2)^2 , \quad t := (p - p_1)^2 , \quad u := (p - p_2)^2
\end{align}
and the abbreviation
\begin{align}
	\begin{split}
		\label{eqn:DefinitionsXSigmaZ}
		z_\ell &:= m_\ell^2 / s_\ell , \\
		X &:= \frac{1}{2}\lambda^{1/2}_{K\ell}(s) := \frac{1}{2}\lambda^{1/2}(\mkp^2,s,s_\ell) , \quad \lambda(a,b,c) := a^2 + b^2 + c^2 - 2(ab+bc+ca) , \\
		\sigma_\pi &:= \sqrt{1-\frac{4\mpip^2}{s}} , \\
		\nu &:= t - u = -2\sigma_\pi X \cos\theta_\pi , \\
		\Sigma_0 &:= s + t + u = \mkp^2 + 2 \mpip^2 + s_\ell.
	\end{split}
\end{align}
In the appendix \ref{sec:LorentzTransformationsKl4}, the Lorentz transformations between the three reference frames are determined and the Lorentz invariant products of the momenta are computed.

\subsection{Matrix Element, Form Factors and Decay Rate}

\subsubsection{$K_{\ell4}$ in the Isospin Limit}

After integrating out the $W$ boson in the standard model, we end up with a Fermi type current-current interaction. If we switch off the electromagnetic interaction, the matrix element of $K_{\ell4}$
\begin{align}
	\begin{split}
		{\vphantom{\big\<}}_\mathrm{out}\big\< \pi^+(p_1) \pi^-(p_2) \ell^+(p_\ell) \nu_\ell(p_\nu) \big| K^+(p) \big\>_\mathrm{in} &= \big\< \pi^+(p_1) \pi^-(p_2) \ell^+(p_\ell) \nu_\ell(p_\nu) \big| i T \big| K^+(p)\big\> \\
		&= i (2\pi)^4 \delta^{(4)}(p - P - L ) \; \mathcal{T}
	\end{split}
\end{align}
splits up into a leptonic times a hadronic part:
\begin{align}
	\begin{split}
		\mathcal{T} &= \frac{G_F}{\sqrt{2}} V_{us}^* \bar u(p_\nu) \gamma_\mu (1-\gamma^5)v(p_\ell) \; \big\< \pi^+(p_1) \pi^-(p_2) \big| \bar s \gamma^\mu(1-\gamma^5) u \big| K^+(p) \big\> .
	\end{split}
\end{align}
The hadronic matrix element exhibits the usual $V-A$ structure of weak interaction. Its Lorentz structure allows us to write the two contributions as
\begin{align}
	\big\< \pi^+(p_1) \pi^-(p_2) \big| V_\mu(0) \big| K^+(p)\big\> &= -\frac{H}{\mkp^3} \epsilon_{\mu\nu\rho\sigma} L^\nu P^\rho Q^\sigma , \\
	\big\< \pi^+(p_1) \pi^-(p_2) \big| A_\mu(0) \big| K^+(p) \big\> &= -i \frac{1}{\mkp} \left( P_\mu F + Q_\mu G + L_\mu R \right) ,
\end{align}
where $V_\mu = \bar s \gamma_\mu u$ and $A_\mu = \bar s \gamma_\mu \gamma^5 u$. The form factors $F$, $G$, $R$ and $H$ are functions of $s$, $s_\ell$ and $\cos\theta_\pi$ (or $s$, $t$ and $u$).

In order to write the decay rate in a compact form, it is convenient to introduce new form factors as linear combinations of $F$, $G$, $R$ and $H$ (following \cite{Pais1968, Bijnens1994}) that correspond to definite helicity amplitudes:
\begin{align}
	\label{eqn:Kl4NewFormFactors}
	\begin{split}
		F_1 &:= X F + \sigma_\pi (PL) \cos\theta_\pi G , \\
		F_2 &:= \sigma_\pi \sqrt{s s_\ell} G , \\
		F_3 &:= \sigma_\pi X \sqrt{s s_\ell} \frac{H}{\mkp^2} , \\
		F_4 &:= -(PL) F - s_\ell R - \sigma_\pi X \cos\theta_\pi G.
	\end{split}
\end{align}

The partial decay rate for the $K_{\ell4}$ decay is given by
\begin{align}
	d\Gamma = \frac{1}{2\mkp(2\pi)^8} \sum_\mathrm{spins} | \mathcal{T} |^2 \delta^{(4)}(p - P - L) \frac{d^3 p_1}{2p_1^0} \frac{d^3 p_2}{2p_2^0} \frac{d^3 p_\ell}{2p_\ell^0} \frac{d^3 p_\nu}{2p_\nu^0}.
\end{align}
Since the kinematics is described by five phase space variables, seven integrals can be performed. This leads to
\begin{align}
	\label{eqn:FiveDimensionalPhaseSpace}
	d\Gamma = G_F^2 |V_{us}|^2 \frac{(1-z_\ell) \sigma_\pi(s)X}{2^{13}\pi^6\mkp^5} J_5(s,s_\ell,\theta_\pi,\theta_\ell,\phi) \, ds \, ds_\ell \, d\cos\theta_\pi \, d\cos\theta_\ell \, d\phi .
\end{align}

The explicit expression for $J_5$ is derived in the appendix \ref{sec:DecayRateIsospinLimit}. 

$F_4$ corresponds to the helicity amplitude of the spin 0 or temporal polarisation of the virtual $W$ boson. Its contribution to the decay rate is therefore helicity suppressed by a factor $m_\ell^2$ and invisible in the electron mode. In the chiral expansion, $F_3$ appears due to the chiral anomaly, which is at the level of the Lagrangian an $\mathcal{O}(p^4)$ effect. Thus, the important form factors for the experiment are $F_1$ and $F_2$, or equivalently $F$ and $G$.

\subsubsection{$K_{\ell4}$ in the Case of Broken Isospin}

In the presence of electromagnetism, the above factorisation of the $K_{\ell4}$ matrix element into a hadronic and a leptonic part is no longer valid. In addition to the $V-A$ structure, a tensorial form factor has to be taken into account \cite{Cuplov2003,Cuplov2004}:
\begin{align}
	\begin{split}
		\label{eqn:TMatrixBrokenIsospin}
		\mathcal{T} ={}& \frac{G_F}{\sqrt{2}} V_{us}^* \left( \bar u(p_\nu) \gamma_\mu (1-\gamma^5)v(p_\ell) \; ( \mathcal{V}^\mu - \mathcal{A}^\mu ) + \bar u(p_\nu) \sigma_{\mu\nu}(1+\gamma^5) v(p_\ell) \mathcal{T}^{\mu\nu} \right), \\
		\mathcal{V}_\mu :={}& -\frac{H}{\mkp^3} \epsilon_{\mu\nu\rho\sigma} L^\nu P^\rho Q^\sigma , \\
		\mathcal{A}_\mu :={}& -i \frac{1}{\mkp} \left( P_\mu F + Q_\mu G + L_\mu R \right) , \\
		\mathcal{T}^{\mu\nu} :={}& \frac{1}{\mkp^2} p_1^\mu p_2^\nu T ,
	\end{split}
\end{align}
where $\sigma_{\mu\nu} = \frac{i}{2} [ \gamma_\mu, \gamma_\nu ]$. The form factors $F$, $G$, $R$, $H$ and $T$ depend now on all five kinematic variables $s$, $s_\ell$, $\theta_\pi$, $\theta_\ell$ and $\phi$.

I follow \cite{Cuplov2004} and introduce in addition to (\ref{eqn:Kl4NewFormFactors}) the form factor $F_5$ (with a slightly different normalisation):
\begin{align}
	\label{eqn:TensorialFormFactorF5}
	F_5 := \frac{\sigma_\pi(s) s s_\ell}{2 \mkp \ml} \, T .
\end{align}
Still, the phase space is parametrised by the same five kinematic variables and the differential decay rate can be written as in (\ref{eqn:FiveDimensionalPhaseSpace}). In the isospin broken case, the presence of the additional tensorial form factor changes the function $J_5$. We will see that $F_5$ is finite in the limit $\ml\to0$. Its contribution to the decay rate is suppressed by $\ml^2$. Details are given in the appendix \ref{sec:DecayRateIsospinBroken}.

It is convenient to define the following additional Lorentz invariants \cite{Cuplov2004}:
\begin{align}
		t_\ell &:= (p - p_\ell)^2 , \quad u_\ell := (p-p_\nu)^2 , \quad s_{1\ell} := (p_1 + p_\ell)^2, \quad s_{2\ell} := (p_2 + p_\ell)^2 .
\end{align}


\section{The Radiative Decay $K_{\ell4\gamma}$}

\subsection{Definition of the Decay}

If we consider electromagnetic corrections to $K_{\ell4}$, we have to take into account contributions from both virtual photons and real photon emission, because only an appropriate inclusive observable is free of infrared singularities. As long as we restrict ourselves to $\O(e^2)$ corrections, the radiative process with just one additional final state photon is the only one of interest (each additional photon comes along with a factor $e^2$ in the decay rate).

The radiative process $K_{\ell4\gamma}$ is defined as
\begin{align}
	\begin{split}
		\label{eqn:Kl4gDecay}
		K^+(p) \to \pi^+(p_1) \pi^-(p_2) \ell^+(p_\ell) \nu_\ell(p_\nu) \gamma(q) .
	\end{split}
\end{align}
There are several possibilities to parametrise the phase space. I find it most convenient to replace the dilepton sub-phase space of $K_{\ell4}$ by a convenient three particle phase space.

I describe the kinematics still in three reference frames: the rest frame of the kaon $\Sigma_K$, the dipion centre-of-mass frame $\Sigma_{2\pi}$ and the dilepton-photon centre-of-mass frame $\Sigma_{\ell\nu\gamma}$. In total, we need eight phase space variables:
\begin{itemize}
	\item $s$, the total centre-of-mass squared energy of the two pions,
	\item $s_\ell$, the total centre-of-mass squared energy of the dilepton-photon system,
	\item $\theta_\pi$, the angle between the $\pi^+$ in $\Sigma_{2\pi}$ and the dipion line of flight in $\Sigma_K$,
	\item $\theta_\gamma$, the angle between the photon in $\Sigma_{\ell\nu\gamma}$ and the $\ell\nu\gamma$ line of flight in $\Sigma_K$,
	\item $\phi$, the angle between the dipion plane and the $(\ell\nu)\gamma$ plane in $\Sigma_K$.
	\item $q^0$, the photon energy in $\Sigma_{\ell\nu\gamma}$,
	\item $p_\ell^0$, the lepton energy in $\Sigma_{\ell\nu\gamma}$,
	\item $\phi_\ell$, the angle between the $\ell\nu$ plane in $\Sigma_{\ell\nu\gamma}$ and the $(\ell\nu)\gamma$ plane in $\Sigma_K$.
\end{itemize}
The variables $s$, $s_\ell$, $\theta_\pi$ are defined in analogy to the $K_{\ell4}$ decay. The reason for the chosen parametrisation of the $\ell\nu\gamma$ subsystem is that $p_\ell^0$ and $\phi_\ell$ are of purely kinematic nature, i.e.~the dynamics depends only on the six other variables.

Instead of the $q^0$ and $p_\ell^0$, I will mostly use the dimensionless variables
\begin{align}
	\begin{split}
		x := \frac{2 Lq}{s_\ell} , \quad y := \frac{2 L p_\ell}{s_\ell},
	\end{split}
\end{align}
where $L := p_\ell + p_\nu + q$ and $s_\ell = L^2$. They are related to $q^0$ and $p_\ell^0$ by
\begin{align}
	\begin{split}
		x = \frac{2 q^0}{\sqrt{s_\ell}} , \quad y = \frac{2 p_\ell^0}{\sqrt{s_\ell}} .
	\end{split}
\end{align}

I give the photon an artificial small mass $m_\gamma$ in order to regulate the infrared divergences. The ranges of the phase space variables are:
\begin{align}
	\begin{alignedat}{2}
		4 \mpip^2 &\le s &&\le (\mkp - \ml - m_\gamma)^2 , \\
		(\ml + m_\gamma)^2 &\le s_\ell &&\le (\mkp - \sqrt{s})^2 , \\
		0 &\le \theta_\pi &&\le \pi , \\
		0 &\le \theta_\gamma &&\le \pi , \\
		0 &\le \phi &&\le 2\pi , \\
		0 &\le \phi_\ell &&\le 2\pi .
	\end{alignedat}
\end{align}

Let us determine in the following the ranges of the two variables $x$ and $y$. Introducing the variable $s_{\ell\nu} := (p_\ell+p_\nu)^2$, I find the relations
\begin{align}
	\begin{split}
		q^0 = \frac{s_\ell + \mg^2 - s_{\ell\nu}}{2\sqrt{s_\ell}} , \quad x = 1 + \frac{\mg^2}{s_\ell} - \frac{s_{\ell\nu}}{s_\ell} .
	\end{split}
\end{align}
The range of $s_{\ell\nu}$ is obviously
\begin{align}
	\begin{split}
		\ml^2 \le s_{\ell\nu} \le (\sqrt{s_\ell} - \mg)^2 ,
	\end{split}
\end{align}
which leads to
\begin{align}
	\begin{split}
		2 \sqrt{\zg} \le x \le 1 + \zg - \zl ,
	\end{split}
\end{align}
where I have defined
\begin{align}
	\begin{split}
		\zl := \frac{\ml^2}{s_\ell}, \quad \zg = \frac{\mg^2}{s_\ell} .
	\end{split}
\end{align}

The range of $y$ for a given value of $x$ can be found as follows. Determine the boost from $\Sigma_{\ell\nu\gamma}$ to the $\ell\nu$ centre-of-mass frame $\Sigma_{\ell\nu}$ by considering the vector $p_\ell + p_\nu$ in both frames. Define $z = \cos\hat\theta_\ell$ with $\hat\theta_\ell$ being the angle between the lepton momentum in $\Sigma_{\ell\nu}$ and the dilepton line of flight in $\Sigma_{\ell\nu\gamma}$. Then, with the help of the inverse boost, you will find $y$ in terms of $z$ and $x$:
\begin{align}
	\begin{split}
		y = \frac{z\sqrt{x^2-4\zg}(1+\zg-\zl-x) + (2 - x)(1+\zg+\zl-x)}{2(1+\zg-x)} .
	\end{split}
\end{align}
In the limit $\mg\to0$, I obtain the following range:
\begin{align}
	\begin{split}
		\label{eqn:yRange}
		1 - x + \frac{\zl}{1-x} \le y \le 1 + \zl .
	\end{split}
\end{align}

Similar to $K_{\ell4}$, I introduce for the radiative process the momenta
\begin{align}
	\label{eqn:FourMomentaKl4g}
	P := p_1 + p_2 , \quad Q := p_1 - p_2 , \quad L := q + p_\ell + p_\nu , \quad N := q + p_\ell - p_\nu .
\end{align}
It will be useful to define also the momenta
\begin{align}
	\begin{split}
		\label{eqn:AlternativeFourMomentaKl4g}
		\hat L := p_\ell + p_\nu = L - q , \quad \hat N := p_\ell - p_\nu = N - q .
	\end{split}
\end{align}
Total momentum conservation implies $p = P + L$. I will use the Lorentz invariants
\begin{align}
	s := (p_1 + p_2)^2 , \quad t := (p - p_1)^2 , \quad u := (p - p_2)^2 , \quad \sg := (p_\ell + q)^2 = s_\ell( x + y - 1) .
\end{align}
In the appendix \ref{sec:LorentzTransformationsKl4g}, the Lorentz transformations between the three reference frames are determined and all the Lorentz invariant products are computed.

\subsection{Matrix Element, Form Factors and Decay Rate}

\label{sec:Kl4gMatrixElement}

The matrix element of the radiative decay (\ref{eqn:Kl4gDecay}) has the form (in analogy to $K_{\ell3\gamma}$ \cite{Gasser2005})
{\small
\begin{align}
	\begin{split}
		\label{eqn:Kl4gTMatrix}
		\mathcal{T}_\gamma &= - \frac{G_F}{\sqrt{2}} e V_{us}^* \epsilon_\mu(q)^*
				\begin{aligned}[t]
					&\bigg[ \mathcal{H}^{\mu\nu} \; \bar u(p_\nu) \gamma_\nu (1-\gamma^5)v(p_\ell) + \mathcal{H}^\nu \; \frac{1}{2 p_\ell q} \bar u(p_\nu) \gamma_\nu (1-\gamma^5)(m_\ell - \slashed p_\ell - \slashed q) \gamma^\mu v(p_\ell) \bigg]
				\end{aligned} \\
			&=: \epsilon_\mu(q)^* \mathcal{M}^\mu ,
	\end{split}
\end{align} }%
where $\mathcal{H}^\nu = \mathcal{V}^\nu-\mathcal{A}^\nu$ is the hadronic part of the $K_{\ell4}$ matrix element. The second part of the matrix element stems from diagrams where the photon is radiated off the lepton line, the first part contains all the rest. The hadronic tensor $\mathcal{H}^{\mu\nu} = \mathcal{V}^{\mu\nu} - \mathcal{A}^{\mu\nu}$ is defined by
\begin{align}
	\begin{split}
		\mathcal{I}^{\mu\nu} &= i \int d^4x \, e^{iqx} \< \pi^+(p_1) \pi^-(p_2) | T \{ V^\mu_\mathrm{em}(x) I^\nu(0) \} | K^+(p) \> , \\
		\mathcal{I} &= \mathcal{V}, \mathcal{A} , \quad I = V, A,
	\end{split}
\end{align}
and satisfies the Ward identity
\begin{align}
	\begin{split}
		q_\mu \mathcal{H}^{\mu\nu} = \mathcal{H}^\nu ,
	\end{split}
\end{align}
such that the condition $q_\mu \mathcal{M}^\mu = 0$ required by gauge invariance is fulfilled.

If the contributions from the anomalous sector are neglected, the hadronic tensor can be decomposed into dimensionless form factors as (the photon is taken on-shell)
\begin{align}
	\begin{split}
		\mathcal{H}^{\mu\nu} &= \frac{i}{\mkp} g^{\mu\nu} \Pi + \frac{i}{\mkp^2}\left( P^\mu \Pi_0^\nu + Q^\mu \Pi_1^\nu + L^\mu \Pi_2^\nu \right) , \\
		\Pi_i^\nu &= \frac{1}{\mkp} \left( P^\nu \Pi_{i0} + Q^\nu \Pi_{i1} + L^\nu \Pi_{i2} + q^\nu \Pi_{i3}  \right) .
	\end{split}
\end{align}
Gauge invariance requires the following relations:
\begin{align}
	\begin{split}
		\label{eqn:GaugeInvarianceFFRelations}
		\mkp^2 \, F - Pq \, \Pi_{00} - Qq \, \Pi_{10} - Lq \, \Pi_{20} &= 0 , \\
		\mkp^2 \, G - Pq \, \Pi_{01} - Qq \, \Pi_{11} - Lq \, \Pi_{21} &= 0 , \\
		\mkp^2 \, R - Pq \, \Pi_{02} - Qq \, \Pi_{12} - Lq \, \Pi_{22} &= 0 , \\
		\mkp^2 \, \Pi + Pq \, \Pi_{03} + Qq \, \Pi_{13} + Lq \, \Pi_{23} &= 0 , \\
	\end{split}
\end{align}
where $F$, $G$ and $R$ are the $K_{\ell4}$ form factors.

The partial decay rate for $K_{\ell4\gamma}$ is given by
\begin{align}
	d\Gamma_\gamma = \frac{1}{2\mkp(2\pi)^{11}} \sum_{\substack{\mathrm{spins} \\ \mathrm{polar.}}} | \mathcal{T}_\gamma |^2 \delta^{(4)}(p - P - L) \frac{d^3 p_1}{2p_1^0} \frac{d^3 p_2}{2p_2^0} \frac{d^3 p_\ell}{2p_\ell^0} \frac{d^3 p_\nu}{2p_\nu^0} \frac{d^3 q}{2q^0}.
\end{align}
Seven integrals can be performed, leaving the integrals over the eight phase space variables:
\begin{align}
	\label{eqn:EightDimensionalPhaseSpace}
	d\Gamma_\gamma &= G_F^2 |V_{us}|^2 e^2 \frac{s_\ell \, \sigma_\pi(s) X}{2^{20}\pi^9 \mkp^7} J_8(s,s_\ell,\theta_\pi,\theta_\gamma,\phi,x,y,\phi_\ell) \, ds \, ds_\ell \, d\cos\theta_\pi \, d\cos\theta_\gamma \, d\phi \, dx \, dy \, d\phi_\ell .
\end{align}
The procedure how to find the explicit expression for $J_8$ in terms of the form factors is explained in appendix~\ref{sec:RadiativeDecayRate}.


\chapter{\ChPT{} Calculation of the Amplitudes}

\label{sec:ChPTCalculation}

Isospin symmetry is the symmetry under $SU(2)$ transformations of up- and down-quarks. In nature, this symmetry is realised only approximately. The source of isospin symmetry breaking is twofold: on the one hand, $u$- and $d$-quarks do not have the same mass, on the other hand, their electric charge is different. On the fundamental level of the standard model, we can therefore distinguish between quark mass effects and electromagnetic effects.

Usually, calculations of processes can be simplified substantially if isospin symmetry is assumed to be exact. In order to link such calculations to real word measurements, the effects of isospin breaking have to be known. The aim of this work is to compute such isospin-breaking corrections to the $K_{\ell4}$ decay.

As $K_{\ell4}$ is a process that happens at low energies, the hadronic part of the matrix element can not be computed perturbatively in QCD. The low-energy effective theory of QCD, chiral perturbation theory (\ChPT{}) \cite{Weinberg1968, GasserLeutwyler1984, GasserLeutwyler1985}, does not treat quarks and gluons but the Goldstone bosons of the spontaneously broken chiral symmetry of QCD as the degrees of freedom. In this effective theory, the isospin-breaking effects show up as differences in the masses of the charged and neutral mesons and in form of photonic corrections. The meson mass differences are due to both isospin-breaking sources, the quark mass difference as well as electromagnetism. I compute the isospin-breaking effects in $K_{\ell4}$ within \ChPT{} including virtual photons and leptons \cite{Urech1995, Knecht2000}. As this is a well-known framework, I abstain from giving a review but only collect the most important formulae in appendix~\ref{sec:AppendixChPT} in order to settle the conventions.

I take into account only first order corrections in the isospin-breaking parameters and effects up to one loop. The leading-order form factors behave as $\O(p)$, i.e.~I consider effects of $\O(p^3)$, $\O(\epsilon \, p^3)$, $\O(e^2 \, p)$, where $e=+|e|$ is the electric unit charge and
\begin{align}
	\begin{split}
		\epsilon := \frac{\sqrt{3}}{4} \frac{m_u - m_d}{\hat m - m_s} , \quad \hat m := \frac{m_u + m_d}{2} .
	\end{split}
\end{align}

Since the chiral anomaly shows up first at next-to-leading chiral order, I do not compute isospin-breaking corrections to the form factor $H$.


\section{Mass Effects}

In contrast to the photonic effects that appear as $\O(e^2)$ corrections in my calculation, the `non-photonic' electromagnetic effects due to the different meson masses in the loops give corrections of the order $\O(Z e^2)$, where $Z$ is the low-energy constant in $\mathcal{L}_{e^2}$. This allows for a separation of the mass effects from purely photonic corrections (a subtlety concerning the counterterms will be discussed later). Let us thus first discuss the mass effects, i.e.~the isospin corrections in the absence of virtual photons.

These $\O(\epsilon \, p^3)$ and $\O(Z e^2 p)$ effects have been considered in \cite{Cuplov2003, Cuplov2004, Colangelo2009}. The present calculation agrees with the results given in \cite{Cuplov2003, Cuplov2004}. For completeness, I give the explicit expressions in my conventions.

\subsection{Leading Order}

\begin{figure}[ht]
	\centering
	\begin{subfigure}[b]{0.3\textwidth}
		\centering
		\scalebox{0.8}{
			\begin{pspicture}(0,-0.5)(3,3)
				\put(0,1.6){$K^+$}
				\put(2.2,3.1){$\pi^+$}
				\put(3,2.3){$\pi^-$}
				\put(3.1,0.7){$\ell^+$}
				\put(2.3,0){$\nu_\ell$}
				\includegraphics[width=3cm]{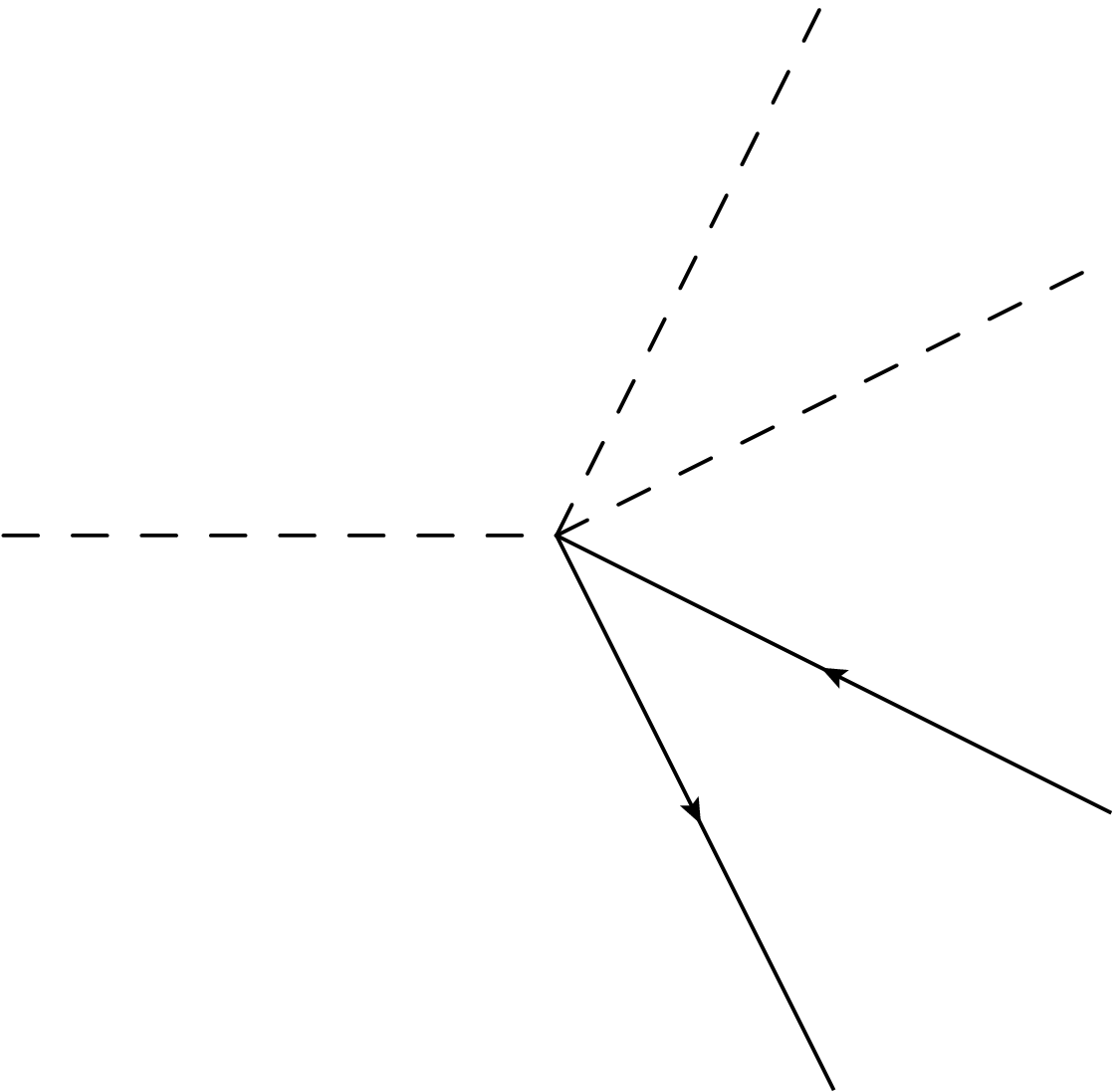}
			\end{pspicture}
			}
		\caption{}
		\label{img:Kl4_LO1}
	\end{subfigure}
	\begin{subfigure}[b]{0.3\textwidth}
		\centering
		\scalebox{0.8}{
			\begin{pspicture}(0,-0.5)(3,3)
				\put(0,1.8){$K^+$}
				\put(1.75,3.0){$\pi^+$}
				\put(2.5,2.3){$\pi^-$}
				\put(3.1,0.5){$\ell^+$}
				\put(2.5,0){$\nu_\ell$}
				\put(1.0,1.0){$K^+$}
				\includegraphics[width=3cm]{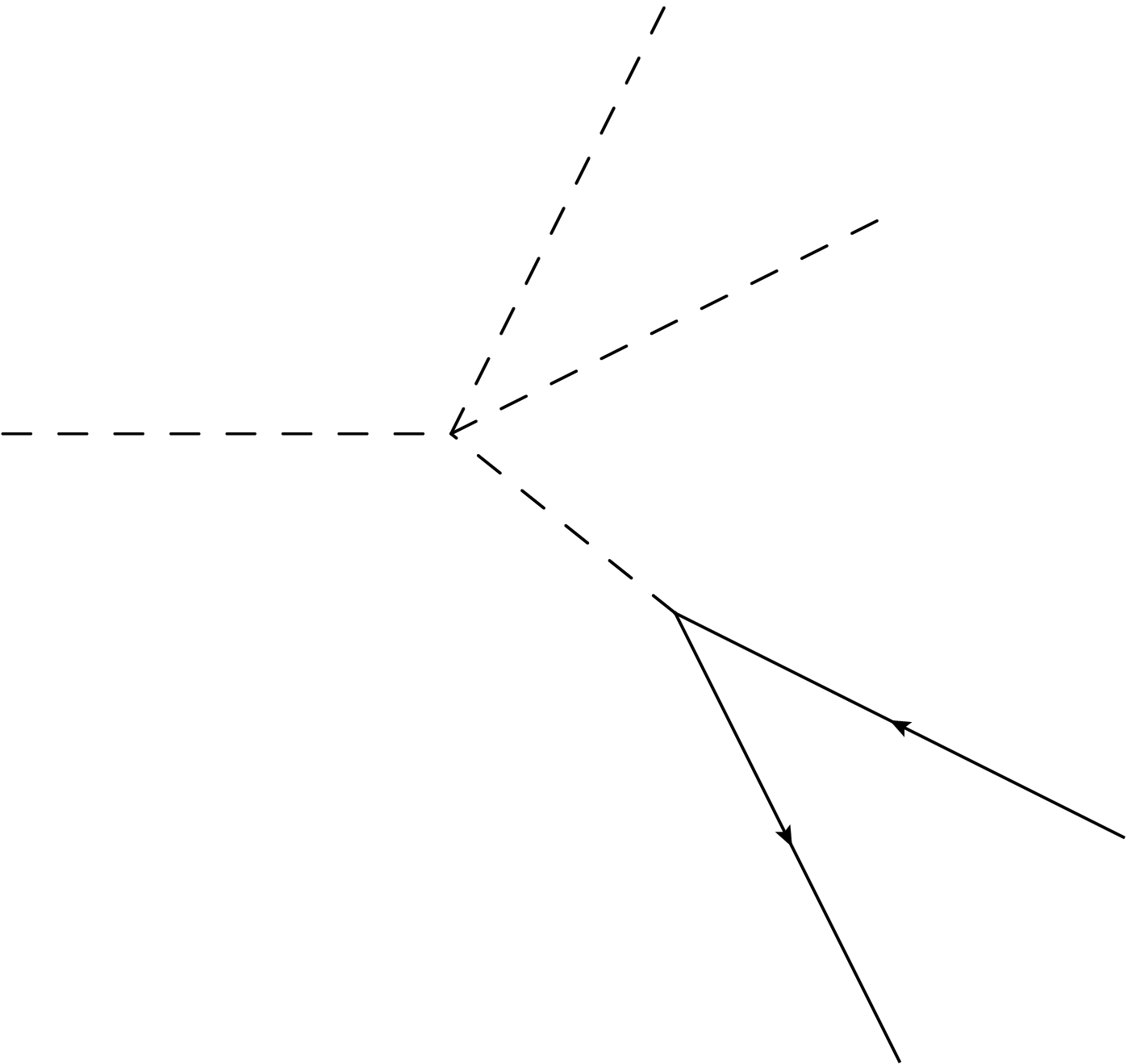}
			\end{pspicture}
			}
		\caption{}
		\label{img:Kl4_LO2}
	\end{subfigure}
	\caption{Tree-level diagrams for the $K_{\ell4}$ decay.}
	\label{img:Kl4LO}
\end{figure}

At leading order, we have to compute two tree-level diagrams, shown in figure~\ref{img:Kl4LO}.

Diagram~\ref{img:Kl4_LO1} contributes to the form factors $F$, $G$ and $R$, whereas diagram~\ref{img:Kl4_LO2} only contributes to the form factor $R$. This is true for all diagrams with an intermediate kaon pole, also at one-loop level.

The leading-order results for the form factors are:
\begin{align}
	\begin{split}
		F^\mathrm{LO}_\mathrm{ME} &= G^\mathrm{LO}_\mathrm{ME} = \frac{\mkp}{\sqrt{2} F_0} , \\
		R^\mathrm{LO}_\mathrm{ME} &= \frac{\mkp}{2 \sqrt{2}F_0} \frac{\mkp^2 - s -s_\ell - \nu - 4 \Delta_\pi}{\mkp^2 - s_\ell} , \\
		T^\mathrm{LO}_\mathrm{ME} &= 0.
	\end{split}
\end{align}
Only the form factor $R$ gets at leading order an isospin correction due to the mass effects.

\subsection{Next-to-Leading Order}

Since the contributions of both $R$ and $T$ to the decay rate are suppressed by $\ml^2$ and experimentally inaccessible in the electron mode, I will calculate only corrections to the form factors $F$ and $G$. Thus, I neglect at next-to-leading order all diagrams that contribute only to the form factor $R$, i.e.~diagrams with a kaon pole in the $s_\ell$-channel. It is convenient to write the NLO expressions for the form factors as
\begin{align}
	\begin{split}
		\label{eqn:NLOFormFactorsMassEffects}
		F^\mathrm{NLO}_\mathrm{ME} &= F^\mathrm{LO}_\mathrm{ME} \left( 1 + \delta F^\mathrm{NLO}_\mathrm{ME} \right) , \\
		G^\mathrm{NLO}_\mathrm{ME} &= G^\mathrm{LO}_\mathrm{ME} \left( 1 + \delta G^\mathrm{NLO}_\mathrm{ME} \right) .
	\end{split}
\end{align}
Since the LO contribution is of $\O(p)$, the order of the NLO corrections considered here is
\begin{align}
	\begin{split}
		\delta F^\mathrm{NLO}_\mathrm{ME}, \delta G^\mathrm{NLO}_\mathrm{ME} = \O(p^2) + \O(\epsilon \, p^2) + \O(Z e^2) .
	\end{split}
\end{align}

Of course, the loop integrals appearing at NLO are UV-divergent. I will regularise them dimensionally and renormalise the theory as usual in the $\overline{MS}$ scheme. The divergent parts of the loop integrals are cancelled by the divergent parts of the LECs.

The explicit NLO results are rather lengthy and can be found in appendix~\ref{sec:AppendixDiagramsMassEffects}.

\subsubsection{Loop Diagrams}

\begin{figure}[ht]
	\centering
	\begin{subfigure}[b]{0.24\textwidth}
		\centering
		\scalebox{0.7}{
			\begin{pspicture}(0,-0.5)(4,5)
				\put(0,2.25){$K^+$}
				\put(2.25,3.75){$\pi^+$}
				\put(3.75,3.25){$\pi^-$}
				\put(3.75,1.25){$\ell^+$}
				\put(2.5,0){$\nu_\ell$}
				\includegraphics[width=4cm]{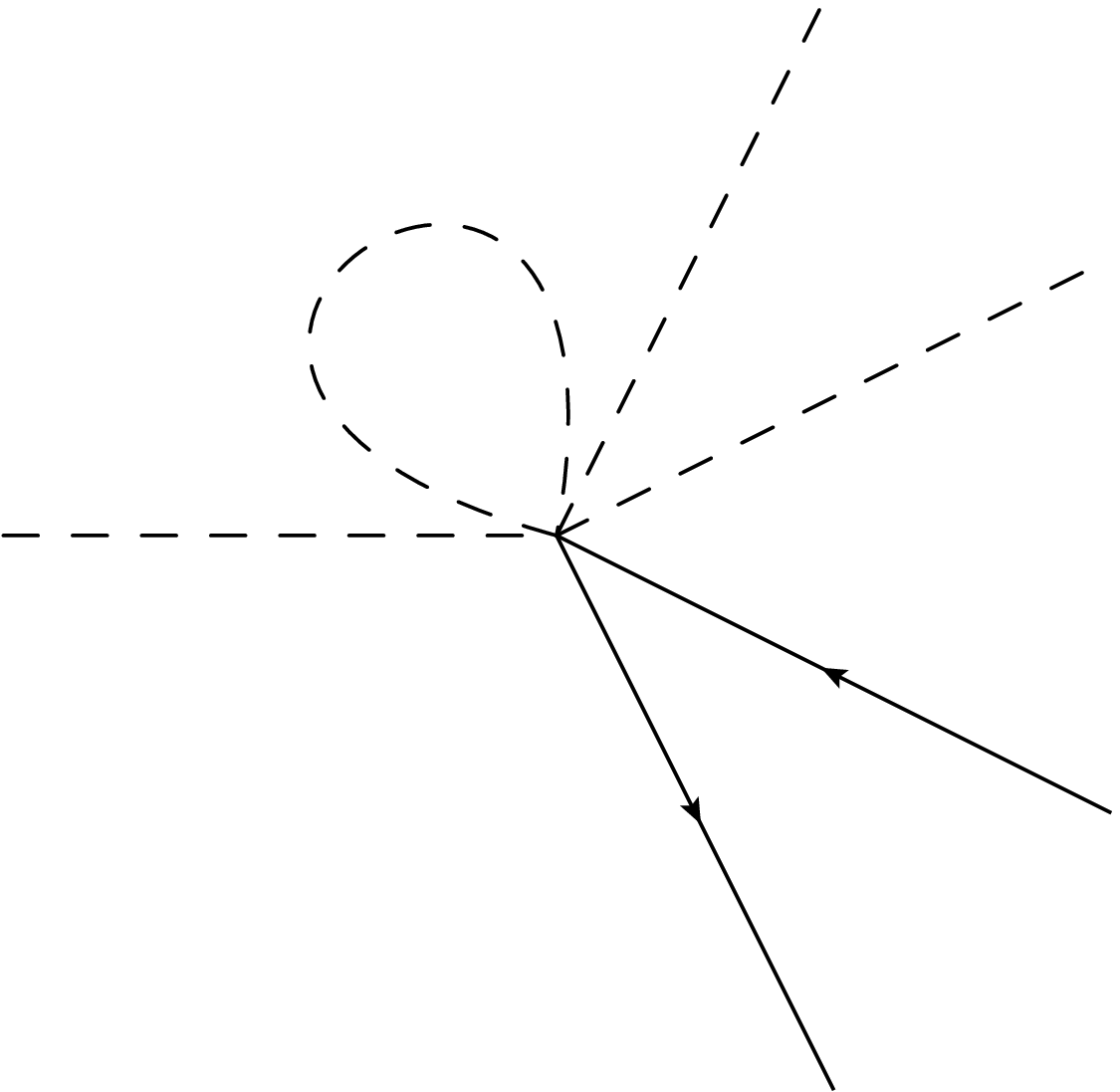}
			\end{pspicture}
			}
		\caption{}
		\label{img:Kl4_NLOTadpole}
	\end{subfigure}
	\begin{subfigure}[b]{0.24\textwidth}
		\centering
		\scalebox{0.7}{
			\begin{pspicture}(0,-0.5)(4,5)
				\put(0,2.0){$K^+$}
				\put(2.5,4.25){$\pi^+$}
				\put(3.75,3.75){$\pi^-$}
				\put(3.25,1.0){$\ell^+$}
				\put(1.75,0){$\nu_\ell$}
				\includegraphics[width=4cm]{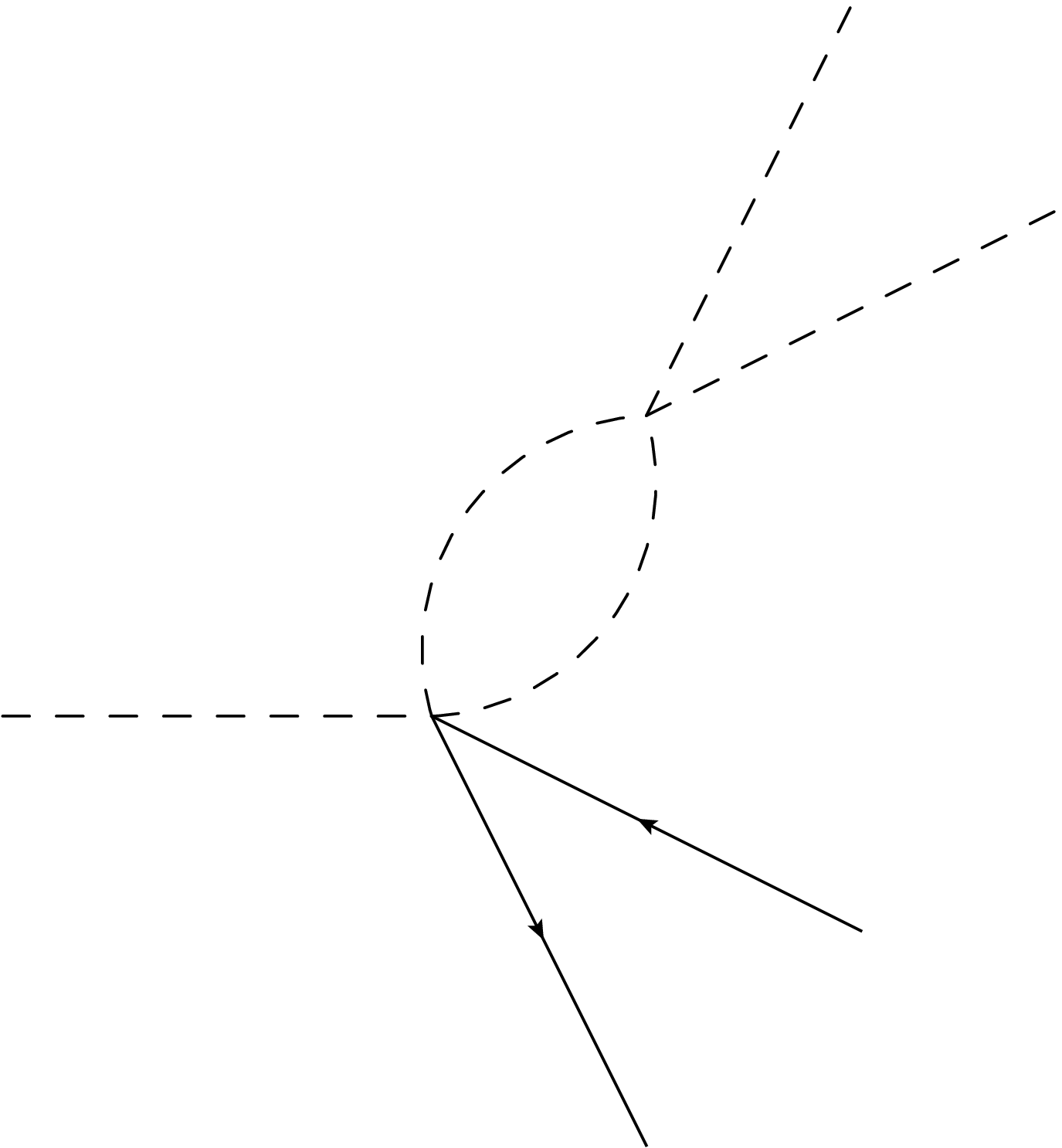}
			\end{pspicture}
			}
		\caption{}
		\label{img:Kl4_NLOSloop}
	\end{subfigure}
	\begin{subfigure}[b]{0.24\textwidth}
		\centering
		\scalebox{0.7}{
			\begin{pspicture}(0,-0.5)(4,5)
				\put(0,2.5){$K^+$}
				\put(2.0,4.0){$\pi^+$}
				\put(3.75,2.5){$\pi^-$}
				\put(3.75,1.0){$\ell^+$}
				\put(2.75,0){$\nu_\ell$}
				\includegraphics[width=4cm]{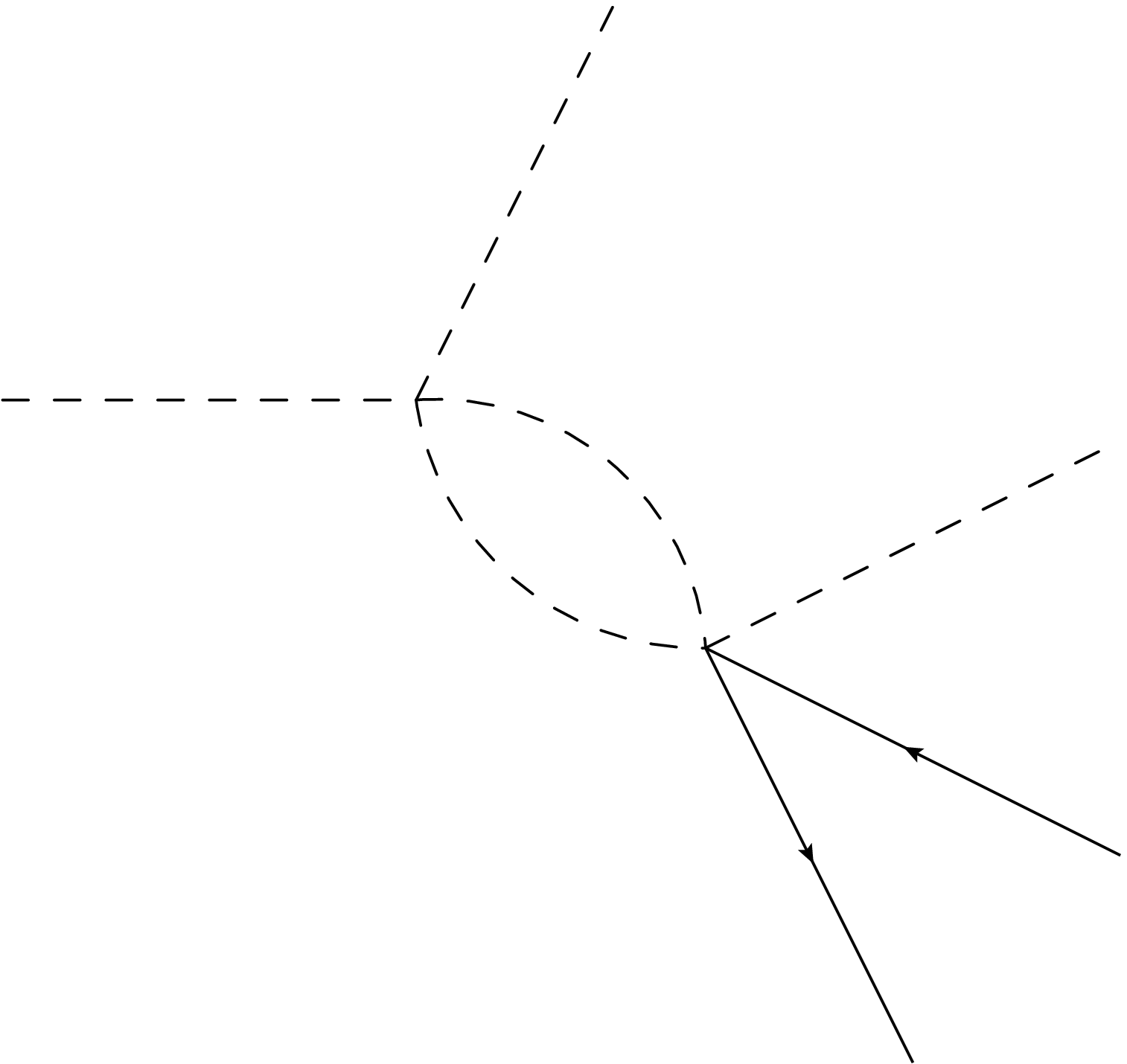}
			\end{pspicture}
			}
		\caption{}
		\label{img:Kl4_NLOTloop}
	\end{subfigure}
	\begin{subfigure}[b]{0.24\textwidth}
		\centering
		\scalebox{0.7}{
			\begin{pspicture}(0,-0.5)(4,5)
				\put(0,2.5){$K^+$}
				\put(3.5,2.75){$\pi^+$}
				\put(2.5,3.25){$\pi^-$}
				\put(3.75,1.0){$\ell^+$}
				\put(2.75,0){$\nu_\ell$}
				\includegraphics[width=4cm]{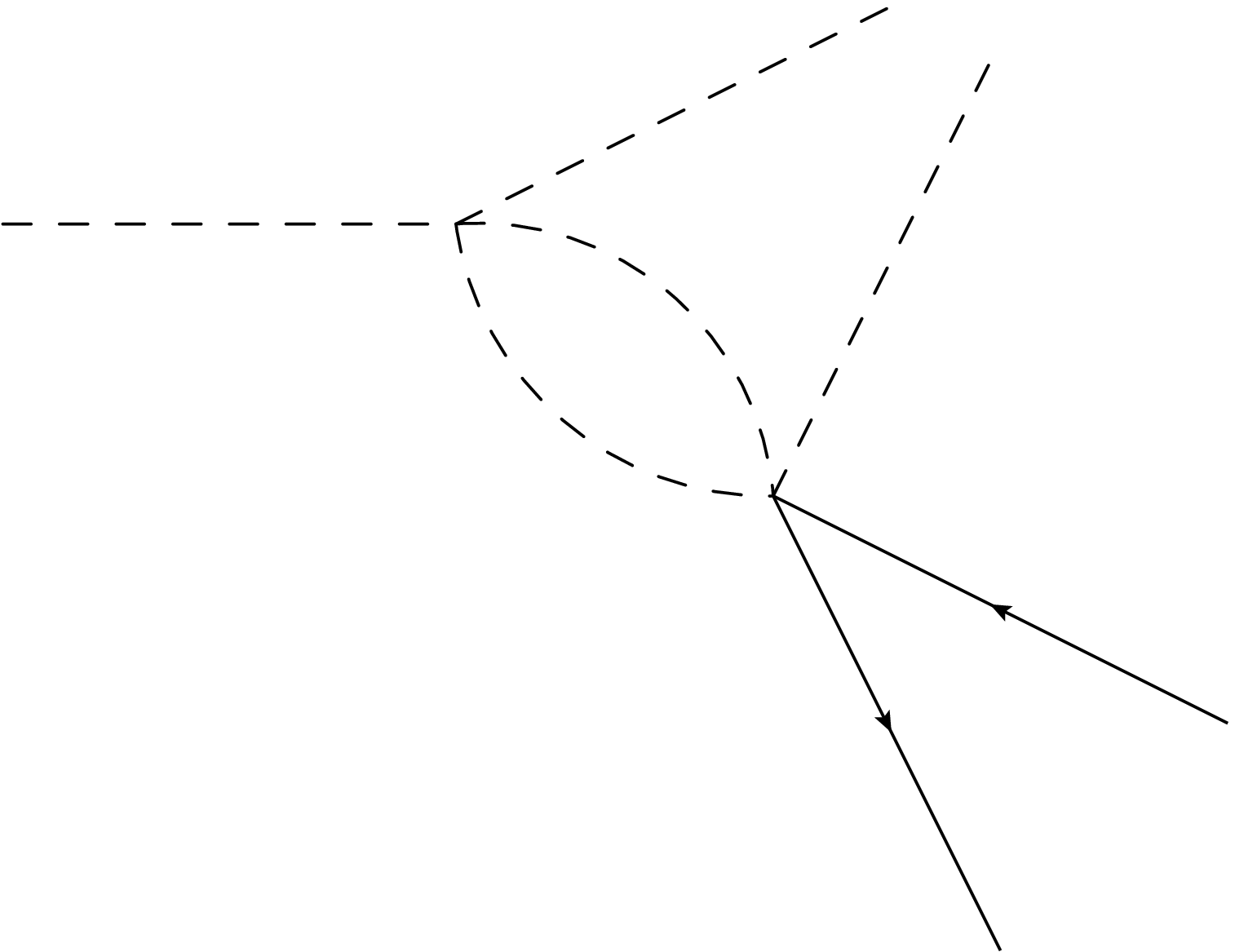}
			\end{pspicture}
			}
		\caption{}
		\label{img:Kl4_NLOUloop}
	\end{subfigure}
	\caption{One-loop diagrams contributing to the $K_{\ell4}$ form factors $F$ and $G$.}
	\label{img:Kl4_Loops}
\end{figure}

At NLO, we have to compute the tadpole diagram~\ref{img:Kl4_NLOTadpole} with all possible mesons ($\pi^0$, $\pi^+$, $K^0$, $K^+$ and $\eta$) in the loop as well as the diagrams~\ref{img:Kl4_NLOSloop}-\ref{img:Kl4_NLOUloop} with two-meson intermediate states in the $s$-, $t$- and $u$-channel.

The contributions of the meson loop diagrams can be expressed in terms of the scalar loop functions $A_0$ and $B_0$ (which should not be confused with the low-energy constant $B_0$).

\subsubsection{Counterterms}

\begin{figure}[ht]
	\centering
		\scalebox{0.7}{
			\begin{pspicture}(0,-0.5)(4,5)
				\put(0,2.25){$K^+$}
				\put(2.25,3.75){$\pi^+$}
				\put(3.75,3.25){$\pi^-$}
				\put(3.75,1.25){$\ell^+$}
				\put(2.5,0){$\nu_\ell$}
				\includegraphics[width=4cm]{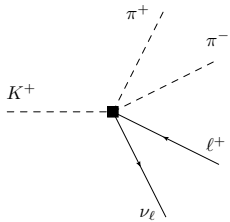}
			\end{pspicture}
			}
	\caption{Counterterm diagram contributing to the $K_{\ell4}$ form factors $F$ and $G$.}
	\label{img:Kl4_CT1}
\end{figure}

I express the one-loop corrections in terms of the scalar loop functions $A_0$ and $B_0$. These loop functions contain UV divergences that have to cancel against the UV divergences in the counterterms and the field strength renormalisation. The only relevant counterterm diagram is shown in figure~\ref{img:Kl4_CT1}. It contains a vertex from the NLO Lagrangian. Now, a subtlety arises here. As we are interested in the mass effects, we have neglected pure $\O(e^2)$ loop corrections, but kept $\O(Z e^2)$ contributions. If we used the same prescription for the counterterms, the UV divergences would not cancel. The reason is that some of the electromagnetic LECs $K_i$ contain a factor $Z$ in their beta function, hence their divergent part is multiplied by $Z$ and contributes at $\O(Z e^2)$. We therefore have to assign also these LECs to the mass effects.

\clearpage

\subsubsection{External Leg Corrections}

\begin{figure}[H]
	\centering
	\begin{subfigure}[b]{0.25\textwidth}
		\centering
		\scalebox{0.8}{
			\begin{pspicture}(0,0)(4,3)
				\includegraphics[width=3cm]{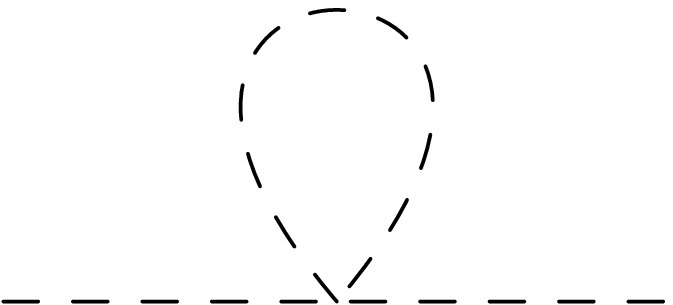}
			\end{pspicture}
			}
		\caption{}
		\label{img:Kl4_SELoop}
	\end{subfigure}
	\begin{subfigure}[b]{0.25\textwidth}
		\centering
		\scalebox{0.8}{
			\begin{pspicture}(0,0)(4,3)
				\includegraphics[width=3cm]{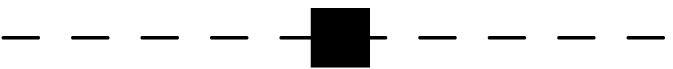}
			\end{pspicture}
			}
		\caption{}
		\label{img:Kl4_SECT}
	\end{subfigure}
	\caption{Meson self-energy diagrams.}
	\label{img:Kl4_ExternalLegs}
\end{figure}

The last contribution at NLO are the external leg corrections. We have to compute only the field strength renormalisation of the mesons (the lepton propagators get no corrections). For the self-energy of the mesons at NLO, the corrections to the propagator shown in figure~\ref{img:Kl4_ExternalLegs} have to be taken into account. All the Goldstone bosons $\pi^+$, $\pi^0$, $K^+$, $K^0$ and $\eta$ have to be inserted in the tadpole diagram.

\subsubsection{Renormalisation}

The complete expressions for the form factors at NLO including the mass effects are
\begin{align}
	\begin{split}
		X^\mathrm{NLO}_\mathrm{ME} &= X^\mathrm{LO}_\mathrm{ME} \left( 1 + \delta X^\mathrm{NLO}_\mathrm{ME} \right) , \\
	\end{split}		\tag{\ref{eqn:NLOFormFactorsMassEffects}}
\end{align}
with
\begin{align}
	\begin{split}
		\delta X^\mathrm{NLO}_\mathrm{ME} &= \delta X^\mathrm{NLO}_\mathrm{tadpole} + \delta X^\mathrm{NLO}_\text{$s$-loop} + \delta X^\mathrm{NLO}_\text{$t$-loop} + \delta X^\mathrm{NLO}_\text{$u$-loop} + \delta X^\mathrm{NLO}_\mathrm{ct} + \delta X_Z^\mathrm{NLO} ,
	\end{split}
\end{align}
where $X \in \{ F, G \}$. The explicit expressions for the individual contributions can be found in the appendix~\ref{sec:AppendixDiagramsMassEffects}. The form factors have to be UV-finite, hence, we should check that in the above sum, all the UV divergences cancel. If I replace the LECs with the help of (\ref{eqn:RenormalisedLECs}) and the loop functions with (\ref{eqn:RenormalisedLoopFunctions}), I find indeed that all the terms proportional to the UV divergence $\lambda$ (\ref{eqn:UVDivergenceLambda}) cancel.


\section{Photonic Effects}

In a next step, I calculate in the effective theory the effects due to the presence of photons. I include virtual photon corrections up to NLO, i.e.~I have to compute again one-loop diagrams, counterterms and external leg corrections. The sum of these contributions will be UV-finite but contain IR and collinear (in the limit $m_\ell\to0$) singularities. As it is well known, the IR divergences will cancel in the sum of the decay rates of $K_{\ell4}$ and the soft real photon emission process $K^+ \to \pi^+ \pi^- \ell^+ \nu_\ell \gamma_\mathrm{soft}$. The collinear divergence is in the physical case regulated by the lepton mass, which plays the role of a natural cut-off. It cancels in the sum of the decay rates of $K_{\ell4}$ and the (soft and hard) collinear real photon emission process. (Note that at $\O(e^2)$, the emission of only one photon has to be taken into account.)
The fully inclusive decay rate $K^+ \to \pi^+ \pi^- \ell^+ \nu_\ell (\gamma)$ is free of IR and mass divergences and does not depend on a cut-off, in accordance with the KLN theorem \cite{KinoshitaSirlin1959,Kinoshita1962,Lee1964}.

As in the case of the mass effects, also the photonic effects have already been computed in \cite{Cuplov2003, Cuplov2004}. However, in these works a whole gauge invariant class of diagrams appearing at NLO has been overlooked\footnote{I thank V.~Cuplov for confirming this.}. The present calculation confirms the results for the diagrams calculated in \cite{Cuplov2004} (in \cite{Cuplov2003}, eq.~(72) gives a wrong result for one of the diagrams) and completes it with the missing class of diagrams.

For the calculation of the photonic effects, I take into account NLO corrections of $\O(e^2)$, but I neglect contributions of $\O(Z e^2)$ as well as $\O(\epsilon \, p^2)$, since they are treated by the mass effects.

\subsection{Leading Order}

\begin{figure}[ht]
	\centering
	\begin{subfigure}[b]{0.25\textwidth}
		\centering
		\scalebox{0.7}{
			\begin{pspicture}(0,-0.5)(4,5)
				\put(0,1.75){$K^+$}
				\put(2.75,3.0){$\pi^+$}
				\put(3.75,2.5){$\pi^-$}
				\put(2.85,0.85){$\ell^+$}
				\put(1.65,0){$\nu_\ell$}
				\includegraphics[width=4cm]{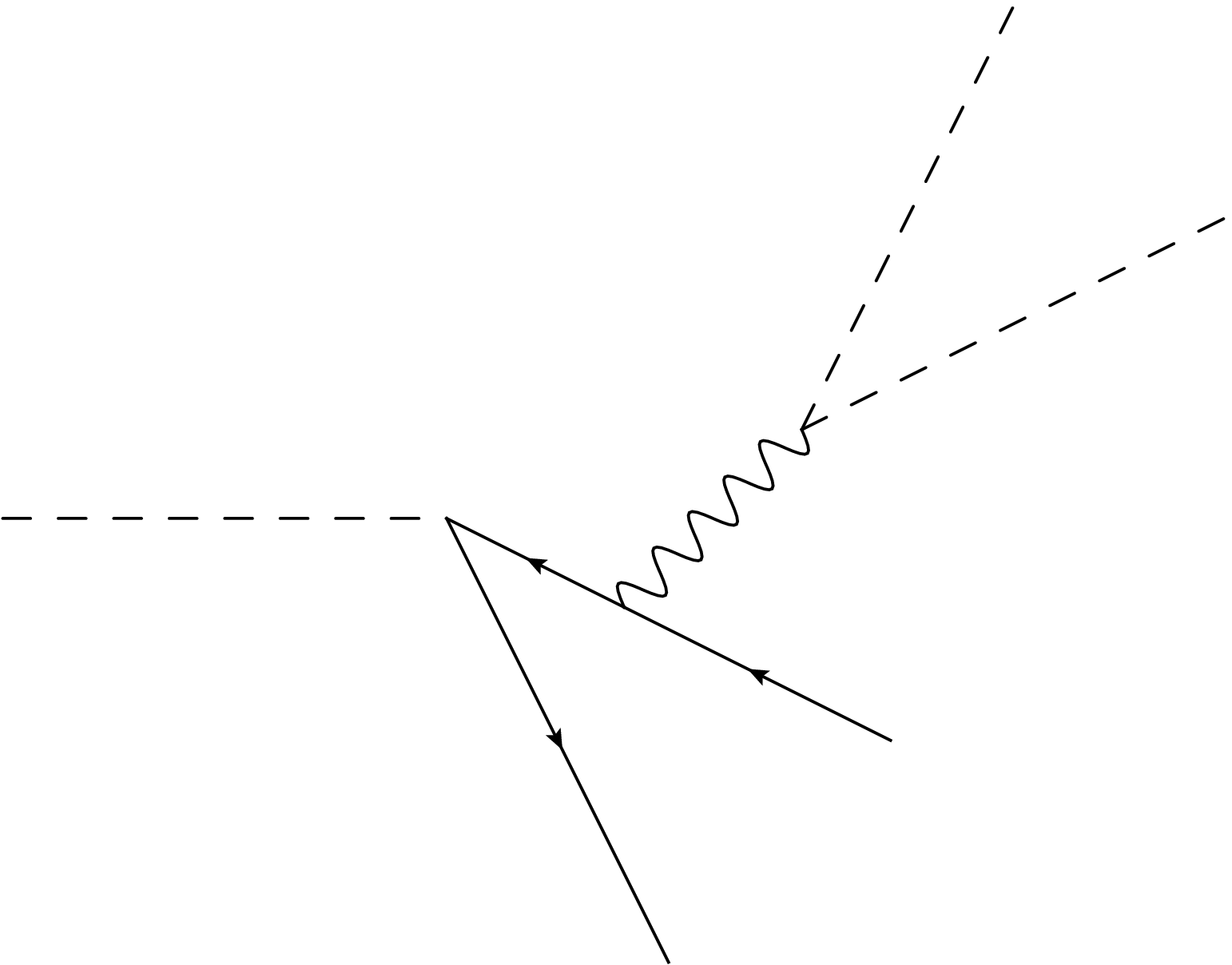}
			\end{pspicture}
			}
		\caption{}
		\label{img:Kl4_LO3}
	\end{subfigure}
	\begin{subfigure}[b]{0.25\textwidth}
		\centering
		\scalebox{0.7}{
			\begin{pspicture}(0,-0.5)(4,5)
				\put(0,2.0){$K^+$}
				\put(2.75,4.0){$\pi^+$}
				\put(3.75,3.25){$\pi^-$}
				\put(3.25,1.0){$\ell^+$}
				\put(2.0,0){$\nu_\ell$}
				\includegraphics[width=4cm]{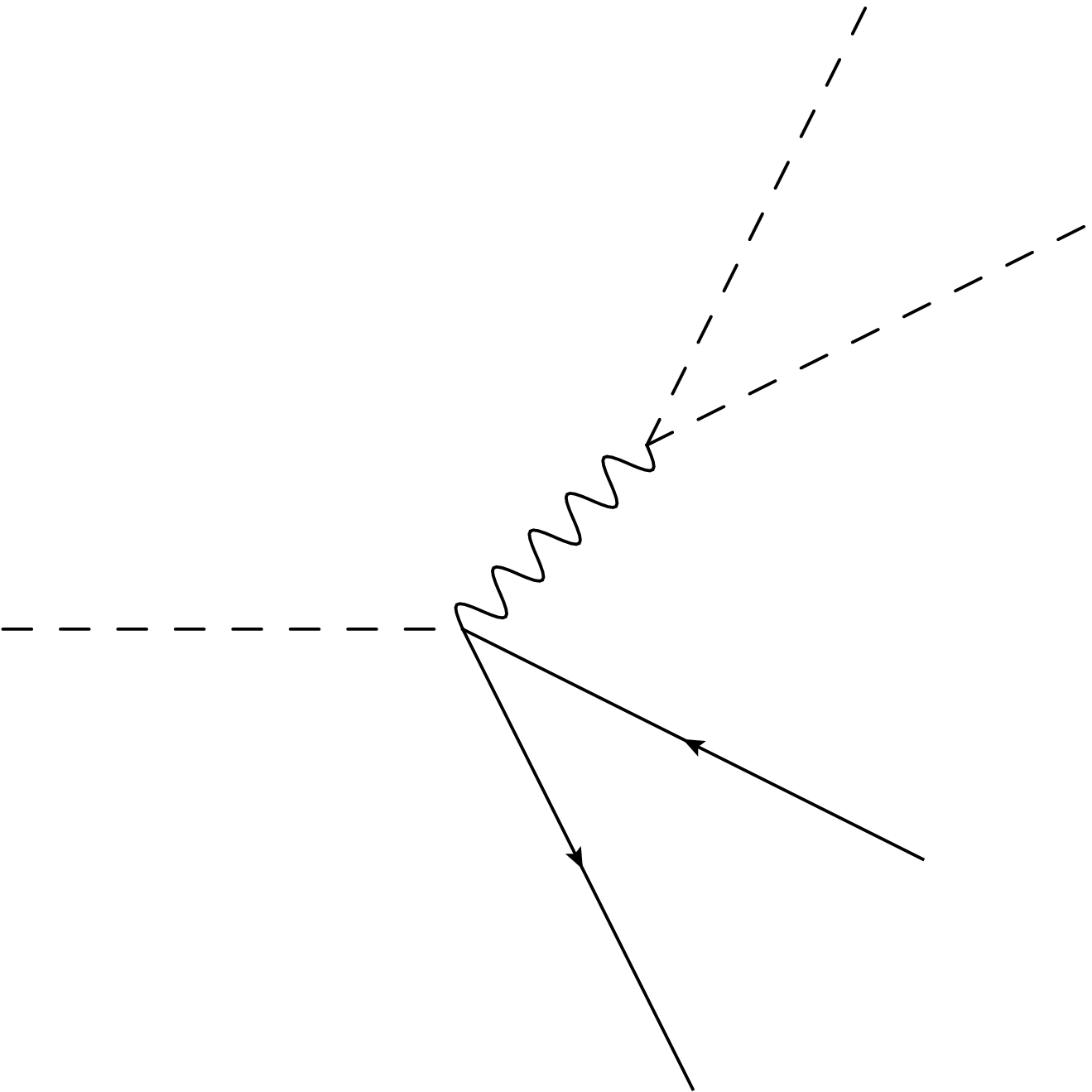}
			\end{pspicture}
			}
		\caption{}
		\label{img:Kl4_LO4}
	\end{subfigure}
	\begin{subfigure}[b]{0.25\textwidth}
		\centering
		\scalebox{0.7}{
			\begin{pspicture}(0,-0.5)(4,5)
				\put(0,2.5){$K^+$}
				\put(2.5,4.25){$\pi^+$}
				\put(3.5,3.75){$\pi^-$}
				\put(4.0,1.0){$\ell^+$}
				\put(2.75,0){$\nu_\ell$}
				\includegraphics[width=4cm]{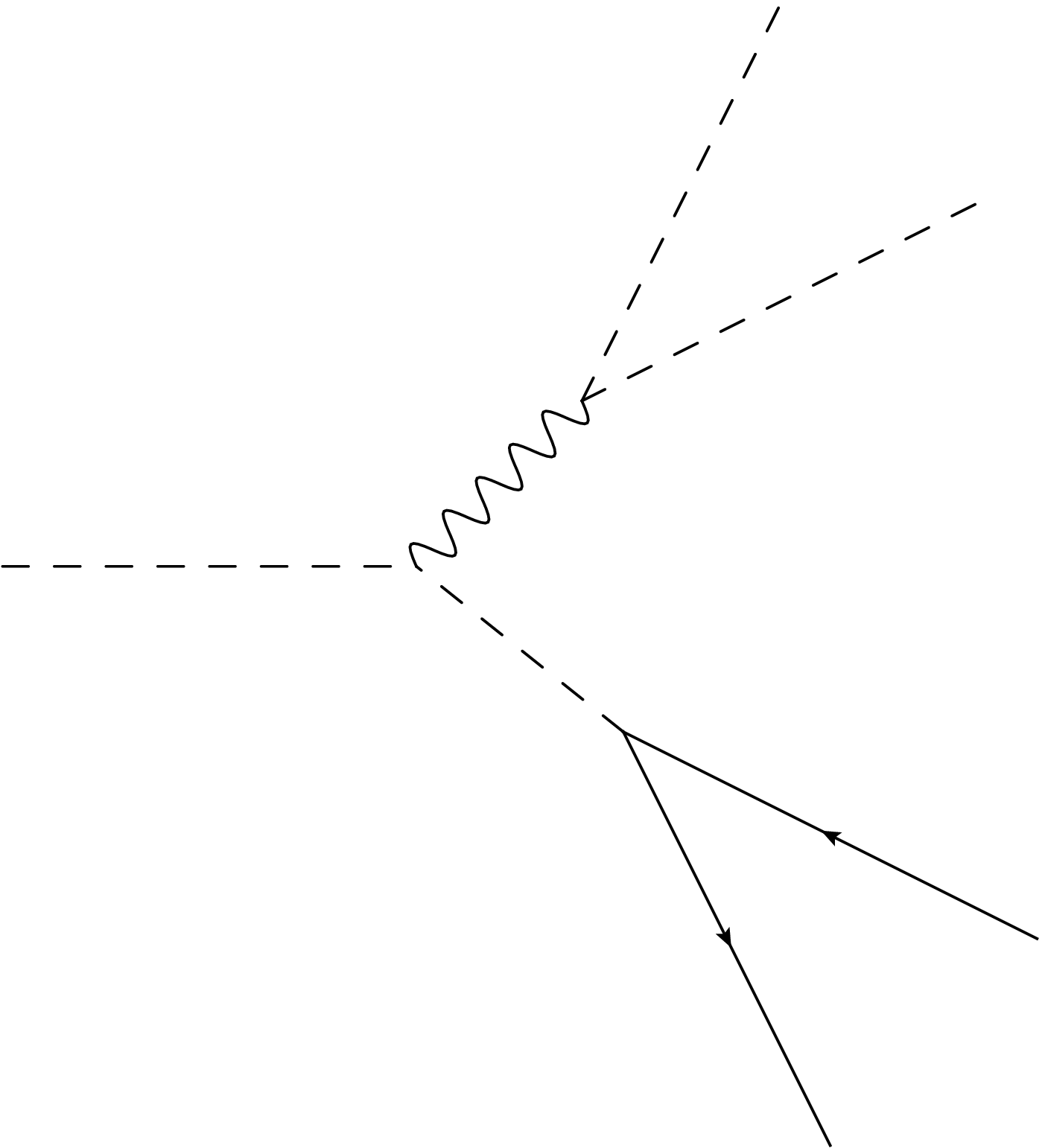}
			\end{pspicture}
			}
		\caption{}
		\label{img:Kl4_LO5}
	\end{subfigure}
	\caption{Tree-level diagrams for the $K_{\ell4}$ decay with a virtual photon.}
	\label{img:Kl4LO_Photons}
\end{figure}

Photonic effects appear already at leading order in the effective theory, i.e.~at $\O(e^2 p^{-1})$, as diagrams with a virtual photon splitting into two pions. In addition to the $\O(e^0 p)$ tree-level diagrams in figure~\ref{img:Kl4LO}, the diagrams shown in figure~\ref{img:Kl4LO_Photons} have to be calculated.

The diagram~\ref{img:Kl4_LO3} contributes to the form factors $G$, $R$ and the tensorial form factor $T$. However, the contribution to $G$ gets exactly cancelled by the diagram~\ref{img:Kl4_LO4}. Diagram~\ref{img:Kl4_LO5} only contributes to $R$.

Therefore, the contribution of the diagrams in figure~\ref{img:Kl4LO_Photons} does not alter the form factors $F$ and $G$:
\begin{align}
	\begin{split}
		F^\mathrm{LO}_{\mathrm{virt.}\gamma} &= \frac{\mkp}{\sqrt{2} F_0} , \quad G^\mathrm{LO}_{\mathrm{virt.}\gamma} =  \frac{\mkp}{\sqrt{2} F_0} .
	\end{split}
\end{align}
The other form factors read (in agreement with \cite{Cuplov2003})
\begin{align}
	\begin{split}
		R^\mathrm{LO}_{\mathrm{virt.}\gamma} &=  \frac{\mkp}{2 \sqrt{2}F_0} \left( \frac{\mkp^2 - s -s_\ell - \nu}{\mkp^2 - s_\ell} + \frac{4 e^2 F_0}{s} \left( \frac{s_{1\ell} - s_{2\ell}}{u_\ell - \ml^2} + \frac{\nu}{\mkp^2 - s_\ell} \right) \right) , \\
		T^\mathrm{LO}_{\mathrm{virt.}\gamma} &= 2 \sqrt{2} e^2 F_0 \frac{\mkp^2 \ml}{s(u_\ell - \ml^2)} .
	\end{split}
\end{align}
We see that the tensorial form factor $F_5$, which was defined above,
\begin{align}
	\begin{split}
		F_5 = \frac{\sigma_\pi(s) s s_\ell}{2 \mkp \ml} \, T ,
	\end{split} \tag{\ref{eqn:TensorialFormFactorF5}}
\end{align}
stays finite in the limit $\ml\to0$. This shows that its contribution to the decay rate (see (\ref{eqn:DecayRateFormFactorsIsoBrokenTensorial}) and (\ref{eqn:DecayRateFormFactorsIsoBrokenInterference}) in the appendix) is suppressed by $\ml^2$. In the following, I will therefore only consider the form factors $F$ and $G$.

\subsection{Next-to-Leading Order}

In order to regularise the IR divergence of loop diagrams with virtual photons, I introduce an artificial photon mass $\mg$ and use the propagator of a massive vector field:
\begin{align}
	\begin{split}
		\tilde G^{\mu\nu}(k) = \frac{-i}{k^2 - \mg^2 + i\epsilon} \left( g^{\mu\nu} - \frac{k^\mu k^\nu}{ \mg^2} \right) .
	\end{split}
\end{align}
The same regulator has to be used in the calculation of the radiative process. In the end, one has to take the limit $\mg\to0$, which restores gauge invariance. Terms that do not contribute in this limit are neglected.

For the NLO calculation of photonic effects, I consider all contributions to the form factors $F$ and $G$ of $\O(e^2 p)$. They consist of loop diagrams with virtual photons, counter\-terms and external leg corrections for $K_{\ell4}$. On the other hand, tree diagrams for the radiative process $K_{\ell4\gamma}$ have to be included at the level of the decay rate.

It is again convenient to write the NLO contribution in the form
\begin{align}
	\begin{split}
		F^\mathrm{NLO}_{\mathrm{virt.}\gamma} &= F^\mathrm{LO}_{\mathrm{virt.}\gamma} \left( 1 + \delta F^\mathrm{NLO}_{\mathrm{virt.}\gamma} \right) , \\
		G^\mathrm{NLO}_{\mathrm{virt.}\gamma} &= G^\mathrm{LO}_{\mathrm{virt.}\gamma} \left( 1 + \delta G^\mathrm{NLO}_{\mathrm{virt.}\gamma} \right) .
	\end{split}
\end{align}
The explicit results are collected in the appendix~\ref{sec:AppendixDiagramsPhotonicEffects}.

\subsubsection{Loop Diagrams}

\begin{figure}[H]
	\centering
	\begin{subfigure}[b]{0.16\textwidth}
		\centering
		\scalebox{0.75}{
			\begin{pspicture}(0,-0.5)(4,5)
				\includegraphics[width=3cm]{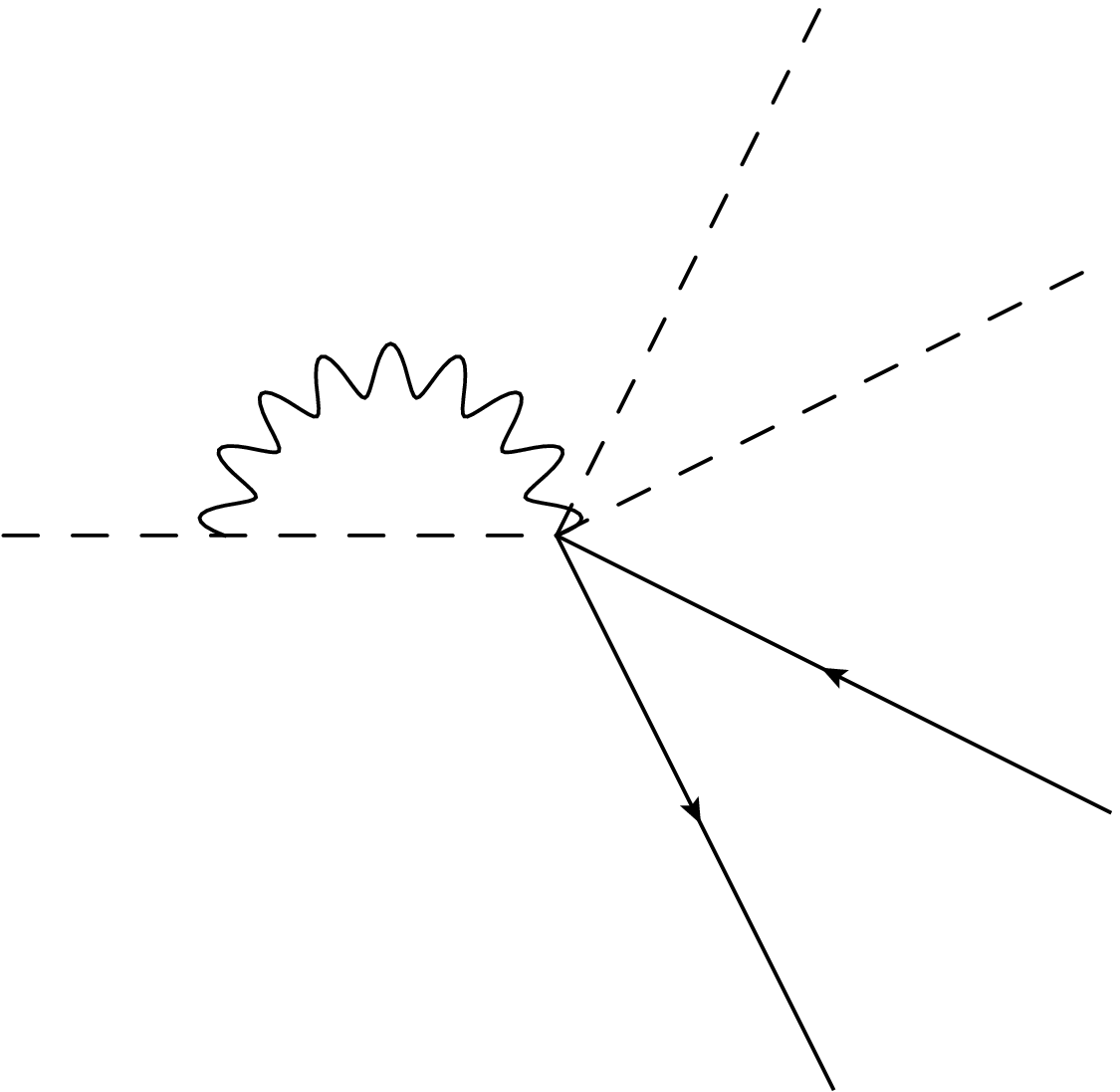}
			\end{pspicture}
			}
		\caption{}
		\label{img:Kl4_NLOgLoop1}
	\end{subfigure}
	\begin{subfigure}[b]{0.16\textwidth}
		\centering
		\scalebox{0.75}{
			\begin{pspicture}(0,-0.5)(4,5)
				\includegraphics[width=3cm]{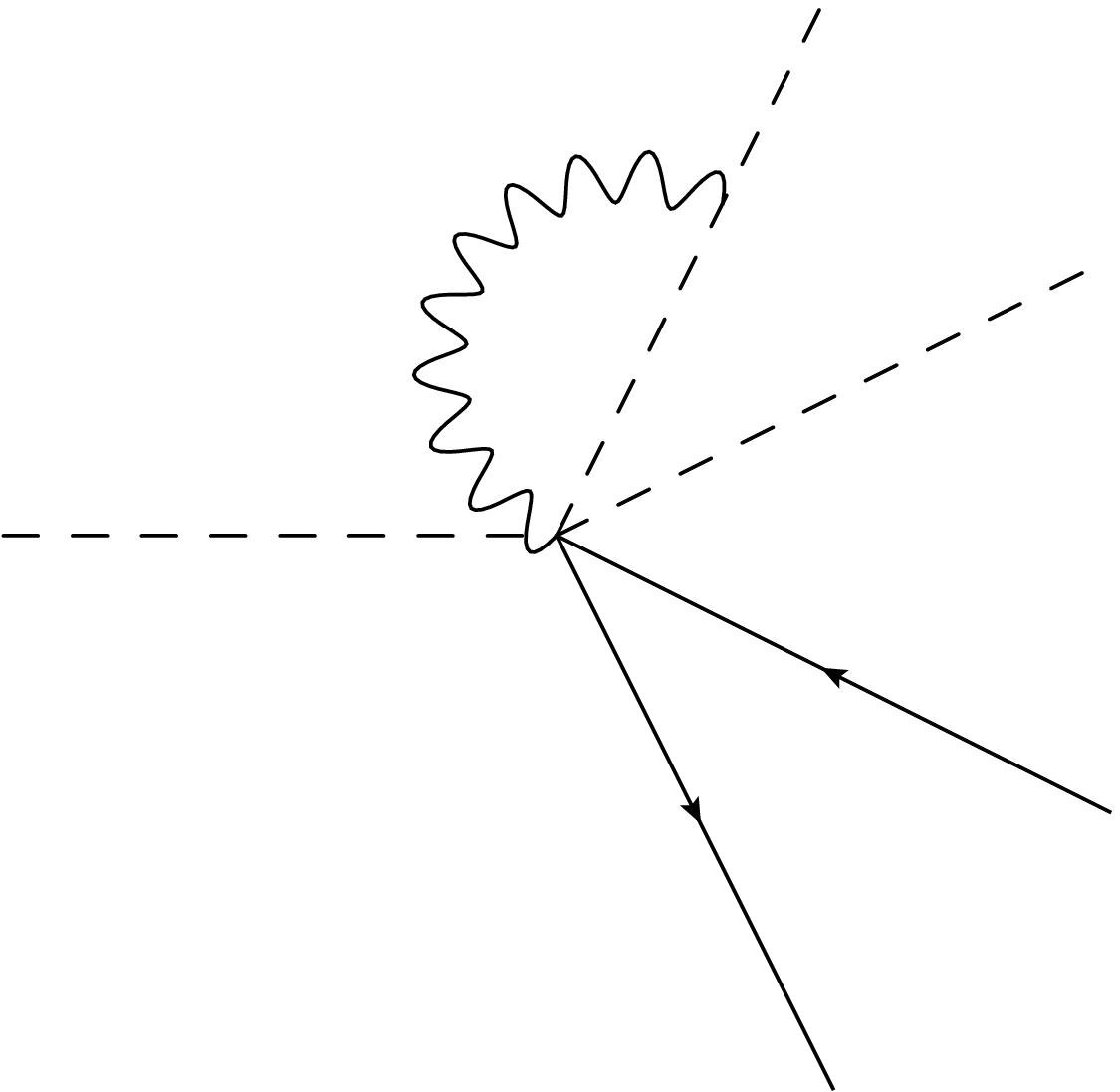}
			\end{pspicture}
			}
		\caption{}
		\label{img:Kl4_NLOgLoop2}
	\end{subfigure}
	\begin{subfigure}[b]{0.16\textwidth}
		\centering
		\scalebox{0.75}{
			\begin{pspicture}(0,-0.5)(4,5)
				\includegraphics[width=3cm]{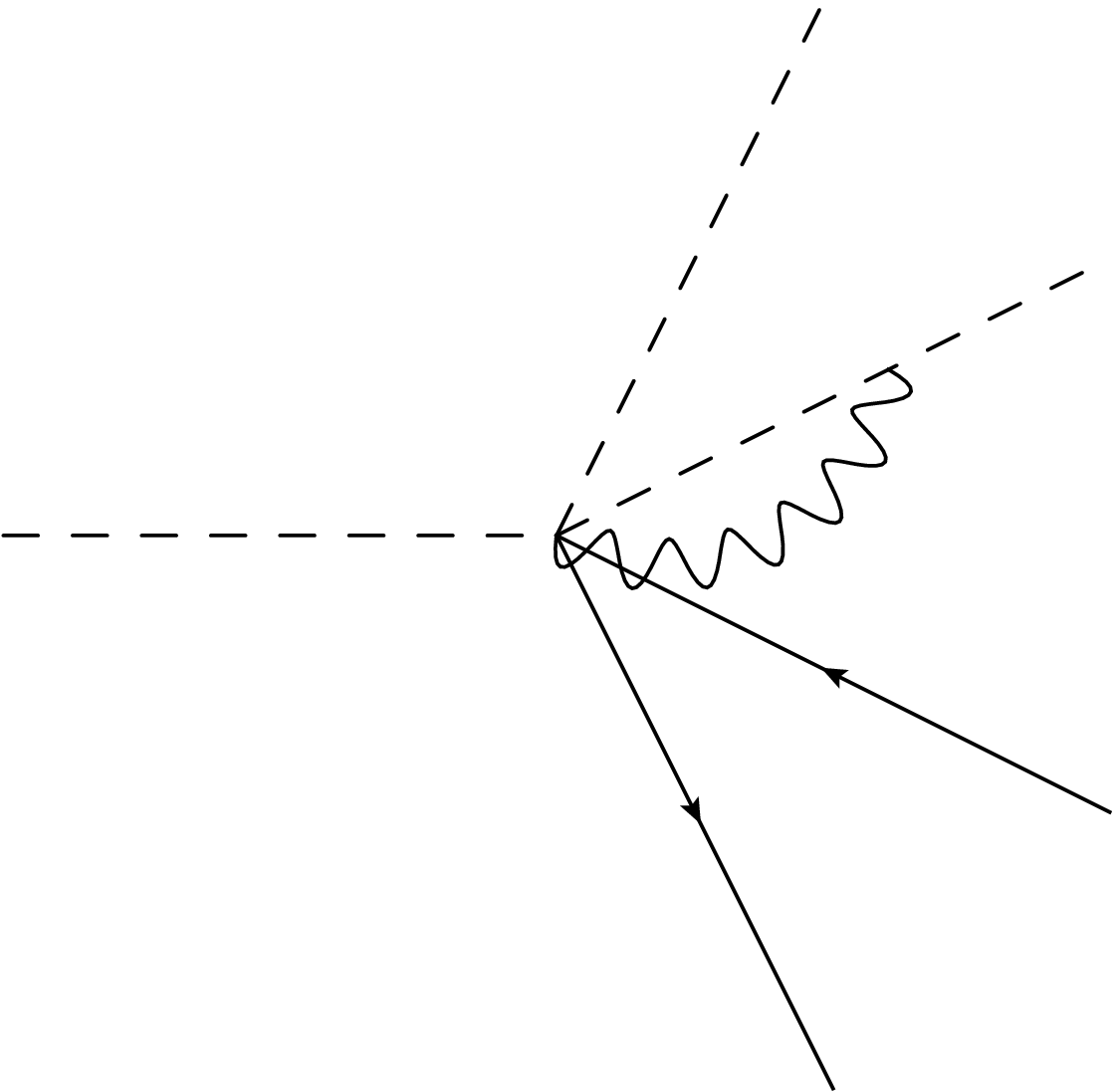}
			\end{pspicture}
			}
		\caption{}
		\label{img:Kl4_NLOgLoop3}
	\end{subfigure}
	\begin{subfigure}[b]{0.16\textwidth}
		\centering
		\scalebox{0.75}{
			\begin{pspicture}(0,-0.5)(4,5)
				\includegraphics[width=3cm]{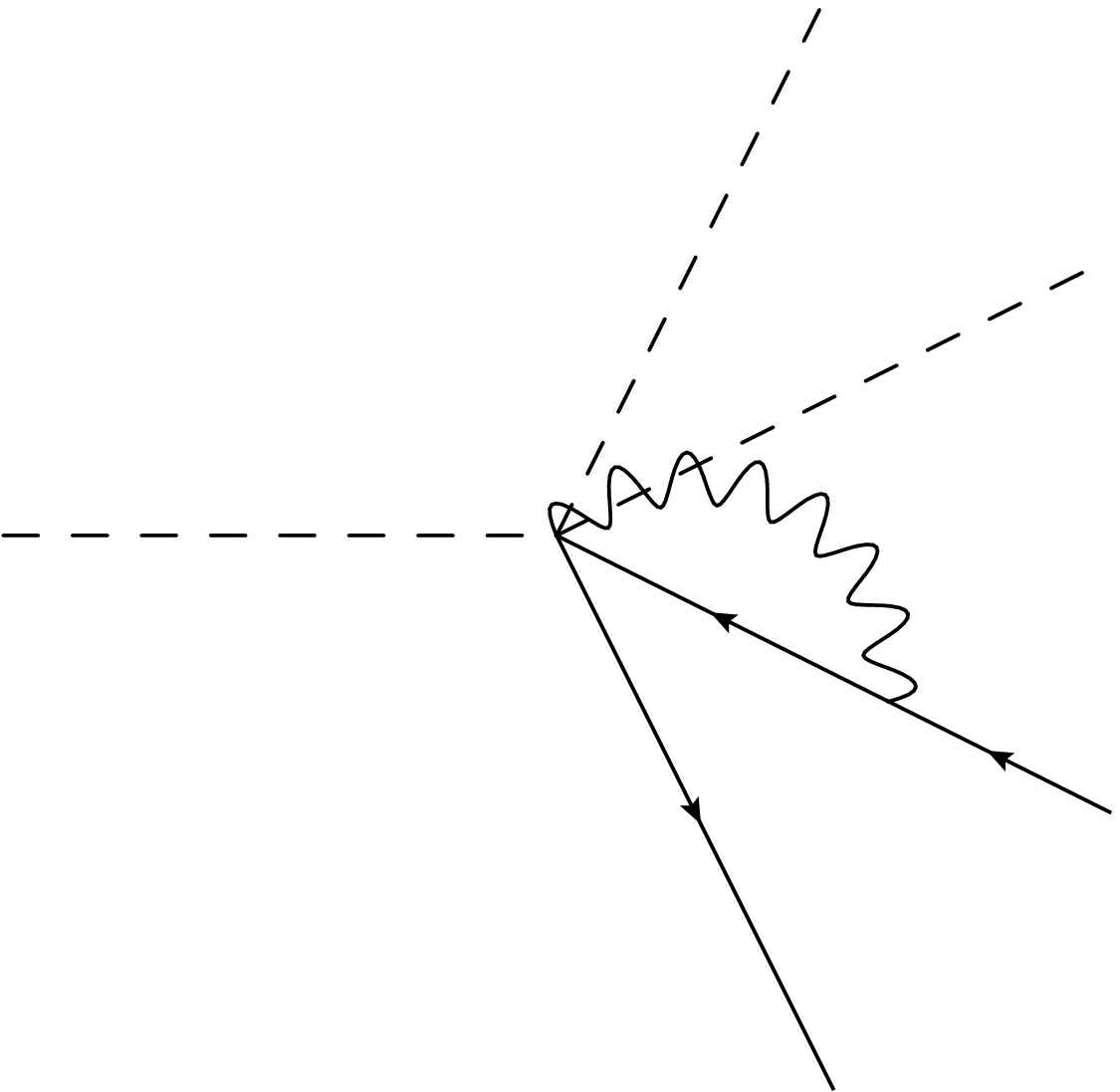}
			\end{pspicture}
			}
		\caption{}
		\label{img:Kl4_NLOgLoop4}
	\end{subfigure}
	
	\vspace{0.5cm}

	\begin{subfigure}[b]{0.16\textwidth}
		\centering
		\scalebox{0.75}{
			\begin{pspicture}(0,-0.5)(4,5)
				\includegraphics[width=3cm]{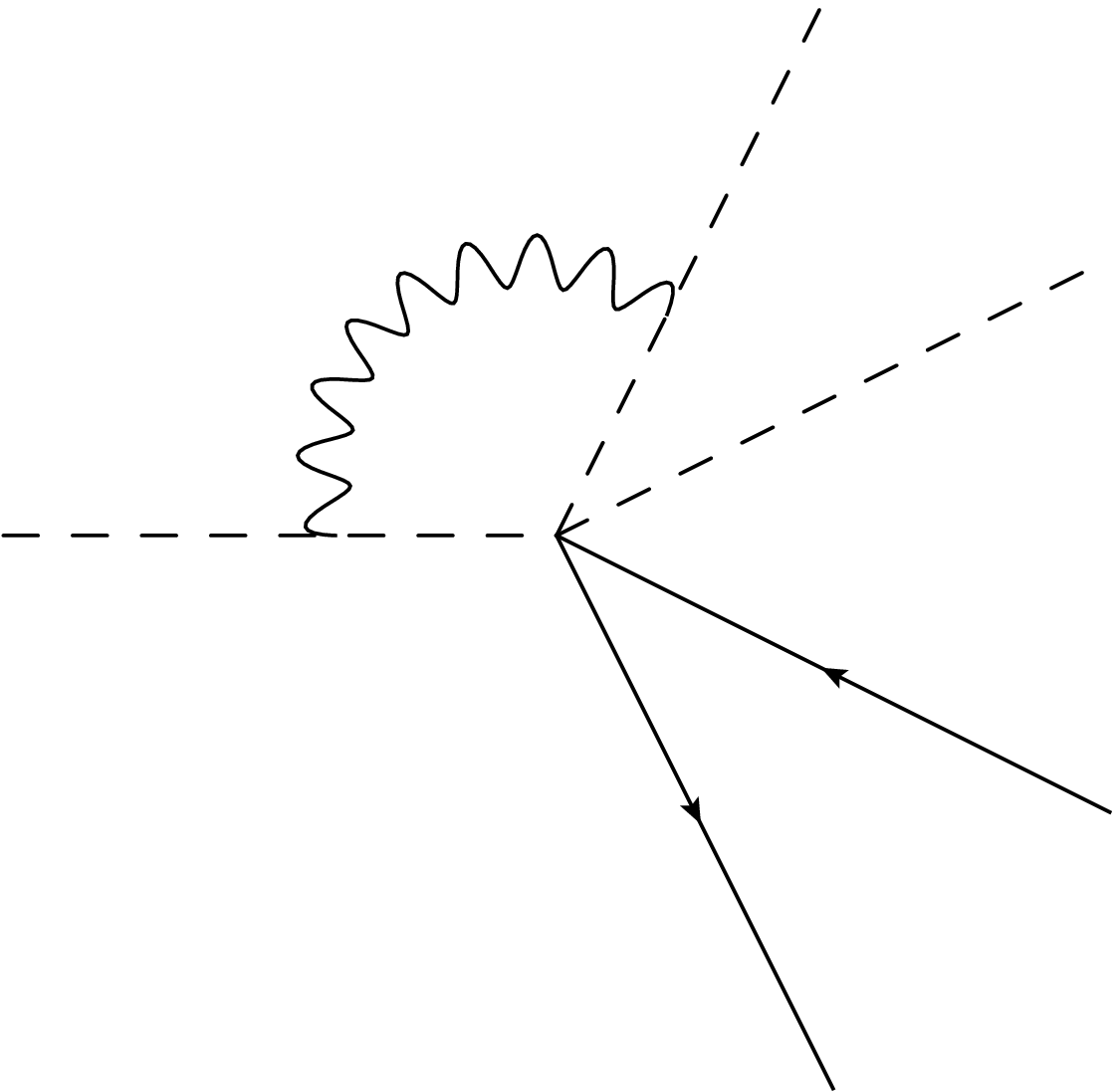}
			\end{pspicture}
			}
		\caption{}
		\label{img:Kl4_NLOgLoop5}
	\end{subfigure}
	\begin{subfigure}[b]{0.16\textwidth}
		\centering
		\scalebox{0.75}{
			\begin{pspicture}(0,-0.5)(4,5)
				\includegraphics[width=3cm]{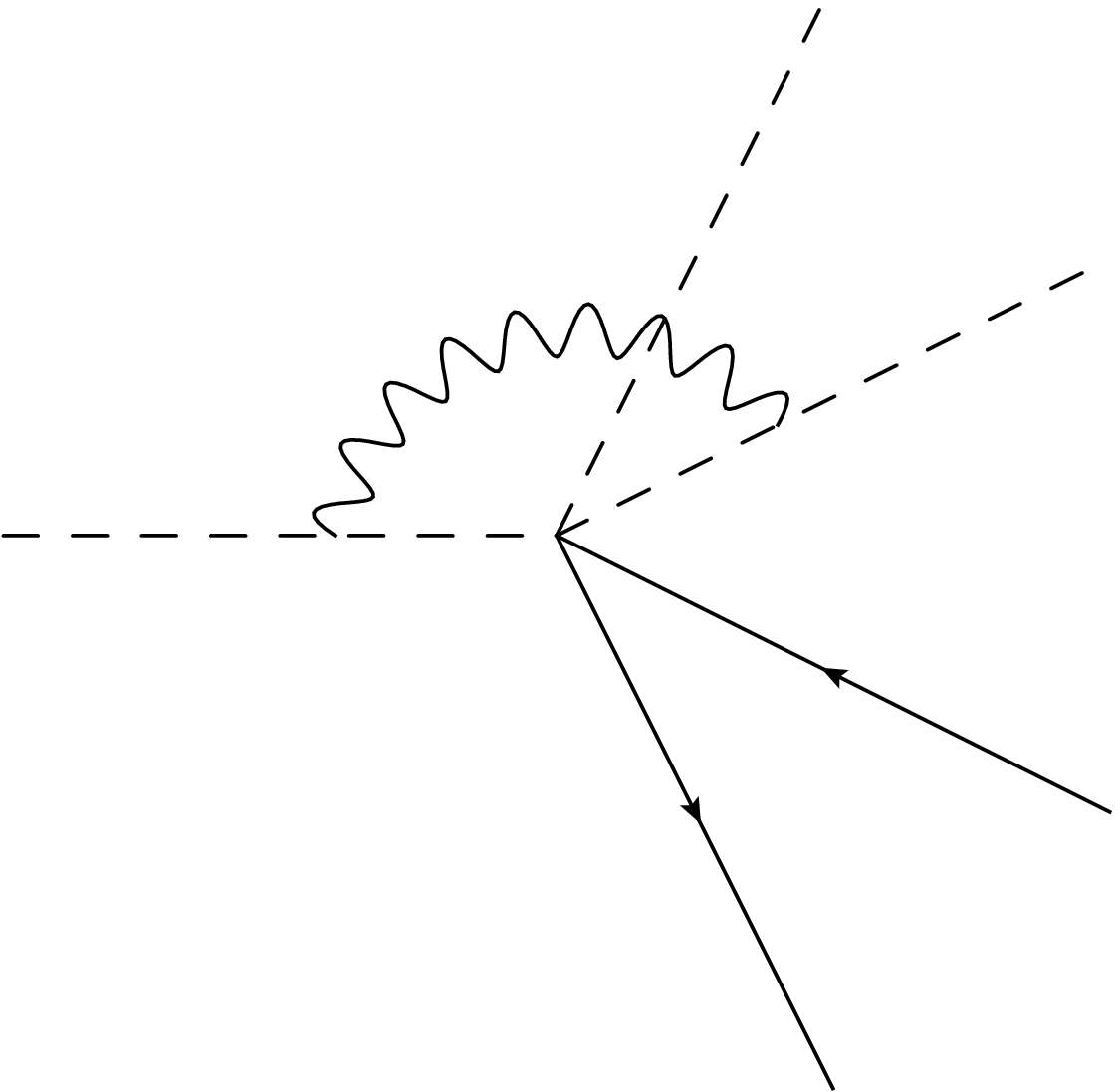}
			\end{pspicture}
			}
		\caption{}
		\label{img:Kl4_NLOgLoop6}
	\end{subfigure}
	\begin{subfigure}[b]{0.16\textwidth}
		\centering
		\scalebox{0.75}{
			\begin{pspicture}(0,-0.5)(4,5)
				\includegraphics[width=3cm]{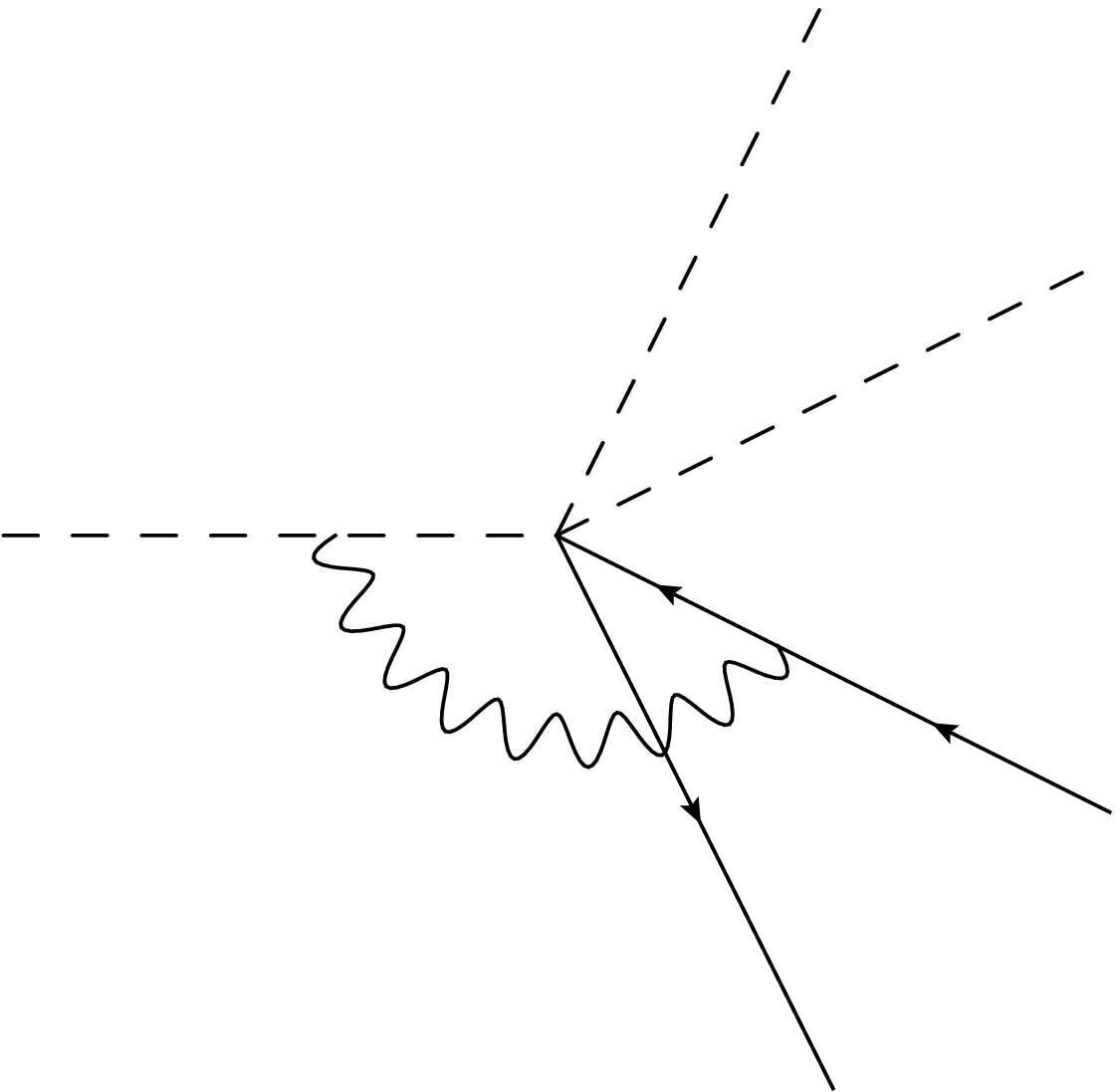}
			\end{pspicture}
			}
		\caption{}
		\label{img:Kl4_NLOgLoop7}
	\end{subfigure}
	\begin{subfigure}[b]{0.16\textwidth}
		\centering
		\scalebox{0.75}{
			\begin{pspicture}(0,-0.5)(4,5)
				\includegraphics[width=3cm]{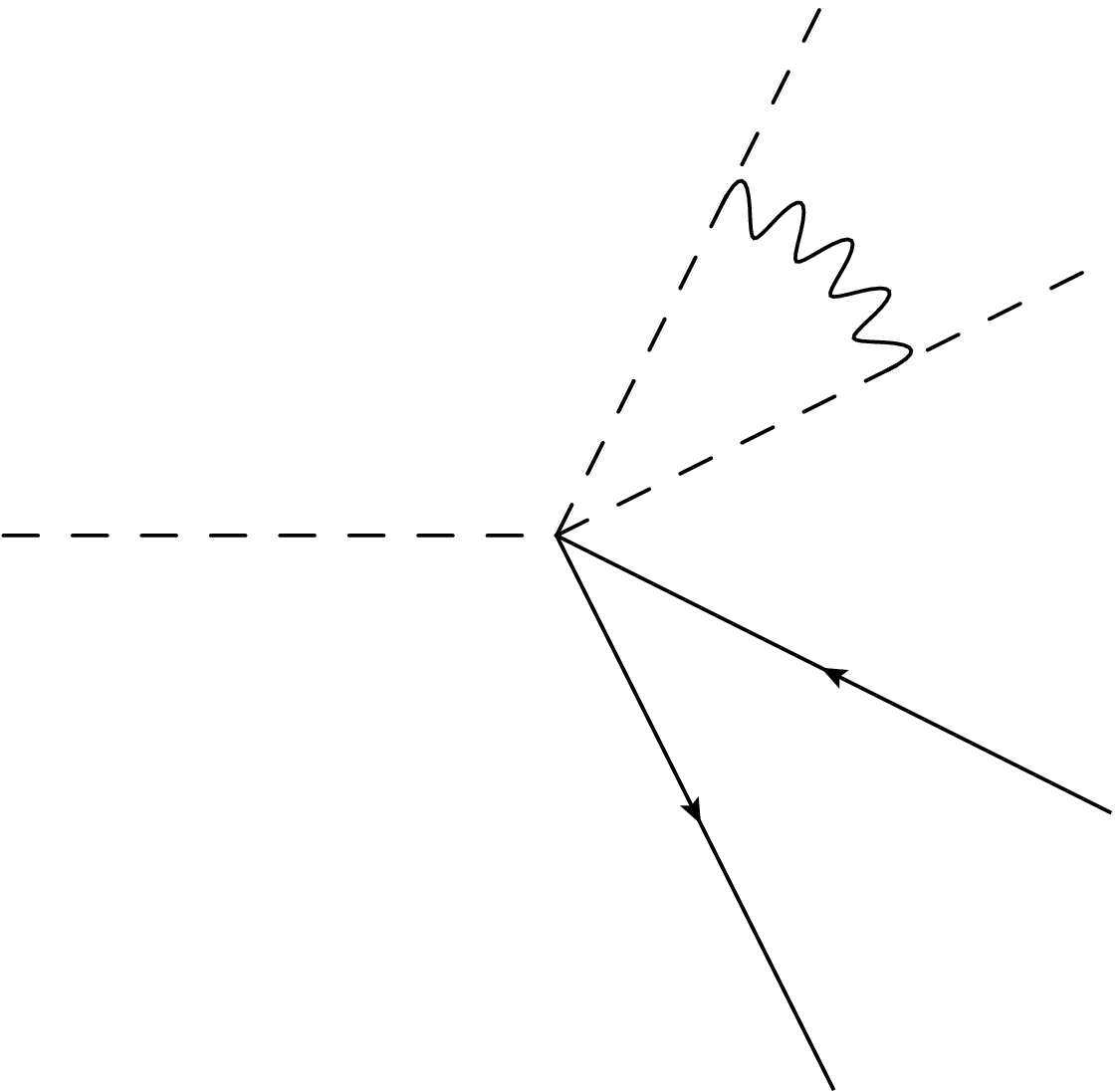}
			\end{pspicture}
			}
		\caption{}
		\label{img:Kl4_NLOgLoop8}
	\end{subfigure}
	\begin{subfigure}[b]{0.16\textwidth}
		\centering
		\scalebox{0.75}{
			\begin{pspicture}(0,-0.5)(4,5)
				\includegraphics[width=3cm]{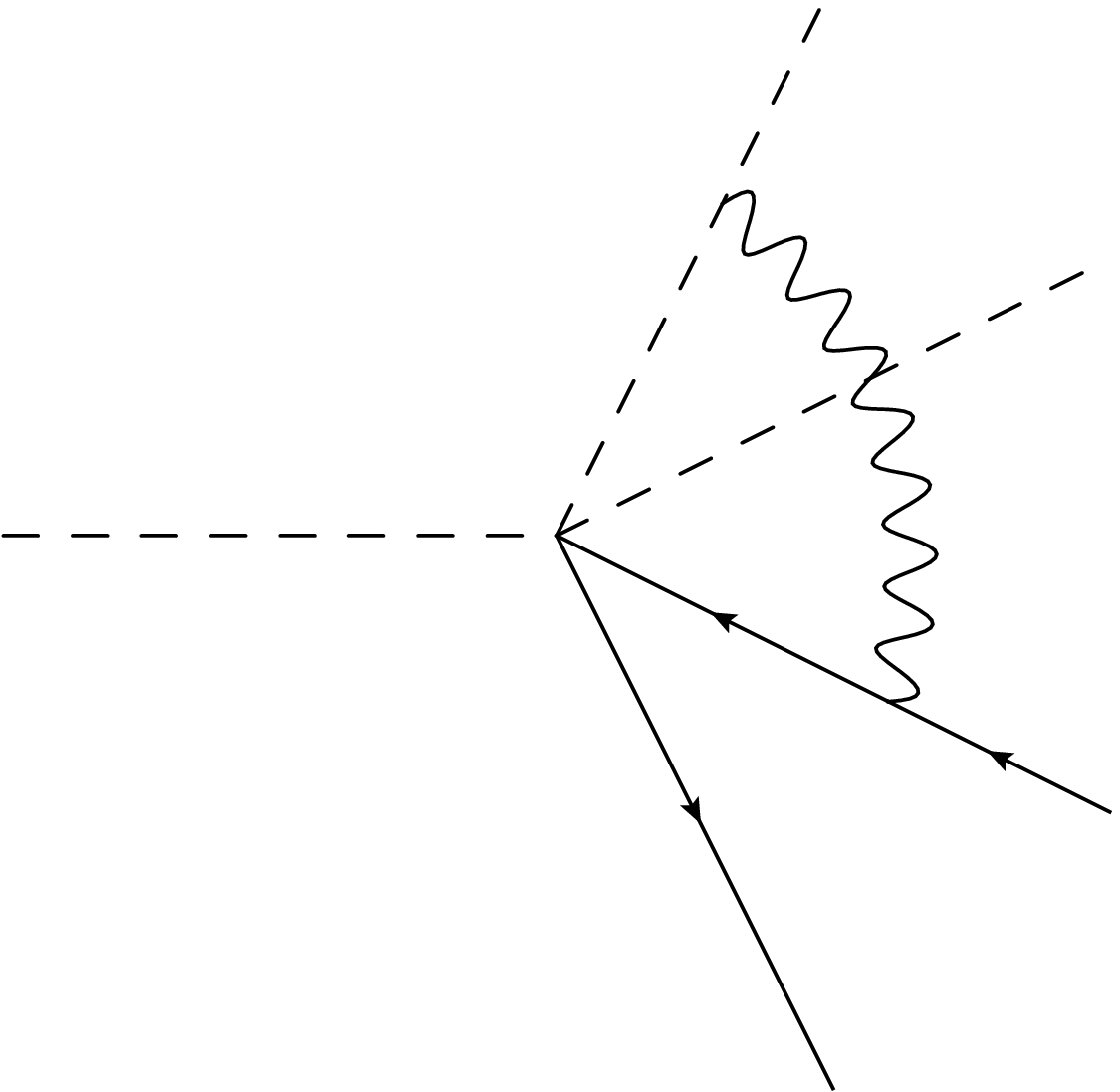}
			\end{pspicture}
			}
		\caption{}
		\label{img:Kl4_NLOgLoop9}
	\end{subfigure}
	\begin{subfigure}[b]{0.16\textwidth}
		\centering
		\scalebox{0.75}{
			\begin{pspicture}(0,-0.5)(4,5)
				\includegraphics[width=3cm]{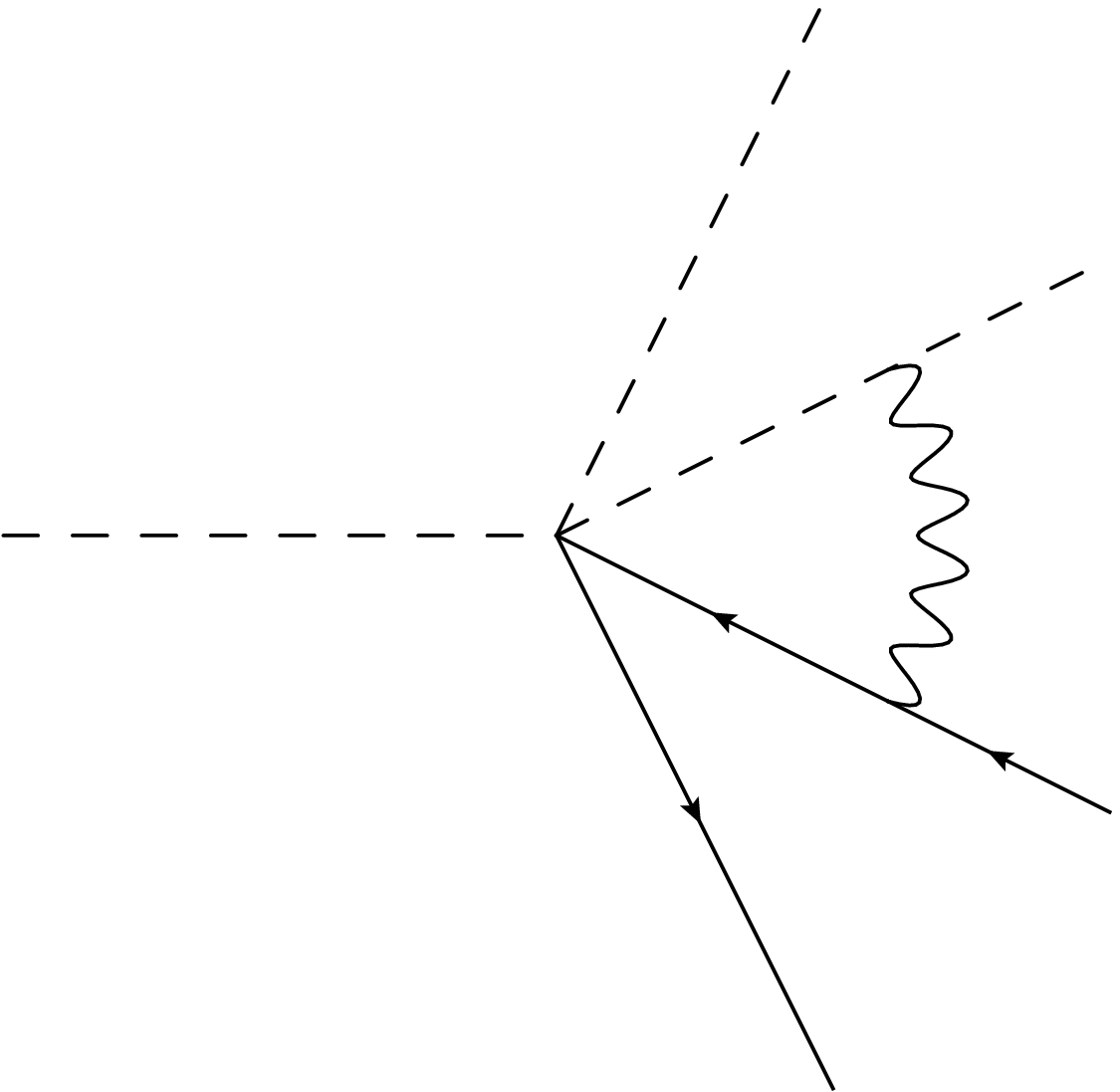}
			\end{pspicture}
			}
		\caption{}
		\label{img:Kl4_NLOgLoop10}
	\end{subfigure}

	\vspace{0.5cm}

	\begin{subfigure}[b]{0.16\textwidth}
		\centering
		\scalebox{0.75}{
			\begin{pspicture}(0,-0.5)(4,5)
				\includegraphics[width=3cm]{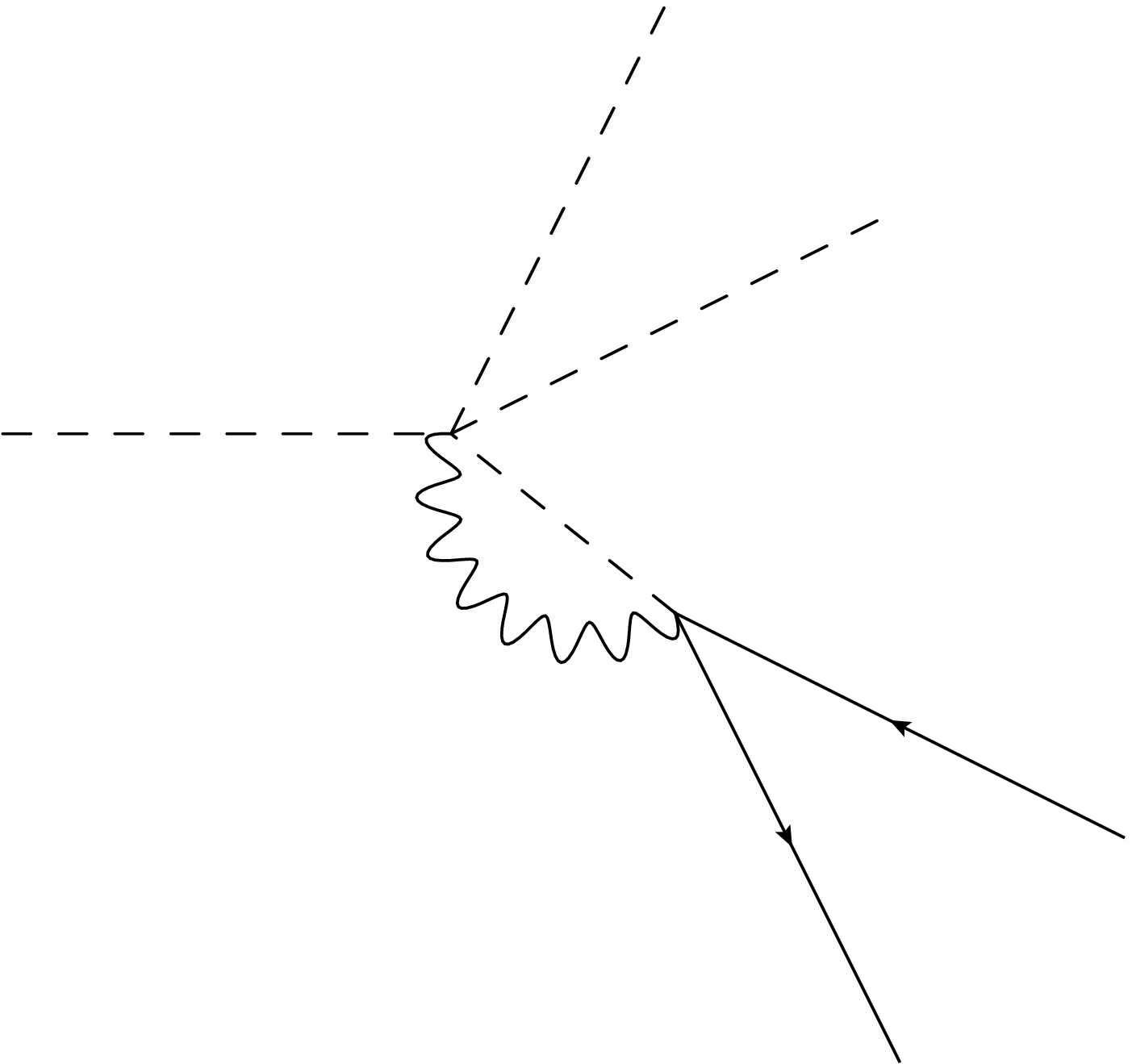}
			\end{pspicture}
			}
		\caption{}
		\label{img:Kl4_NLOgLoop11}
	\end{subfigure}
	\begin{subfigure}[b]{0.16\textwidth}
		\centering
		\scalebox{0.75}{
			\begin{pspicture}(0,-0.5)(4,5)
				\includegraphics[width=3cm]{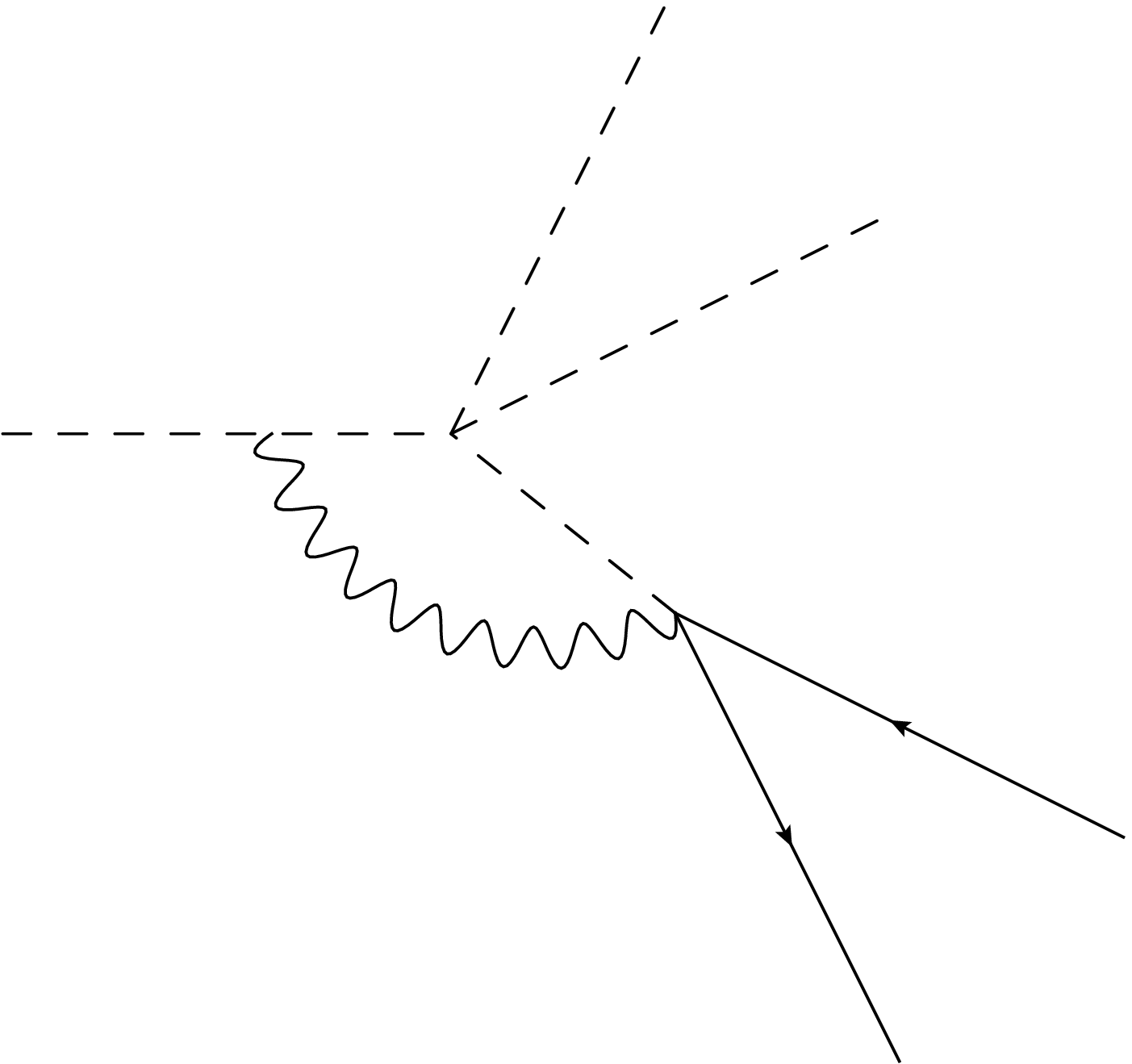}
			\end{pspicture}
			}
		\caption{}
		\label{img:Kl4_NLOgLoop12}
	\end{subfigure}
	\begin{subfigure}[b]{0.16\textwidth}
		\centering
		\scalebox{0.75}{
			\begin{pspicture}(0,-0.5)(4,5)
				\includegraphics[width=3cm]{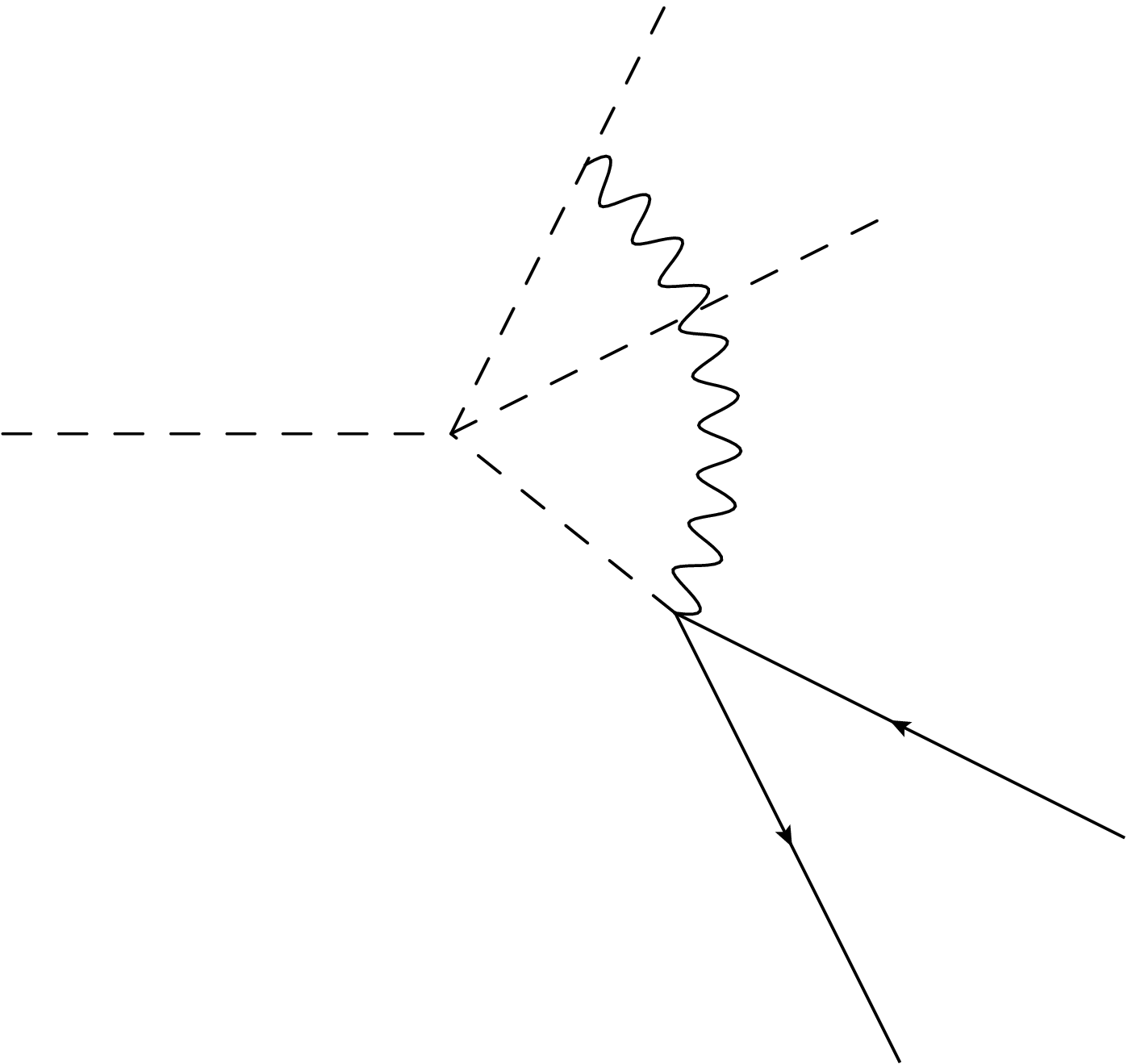}
			\end{pspicture}
			}
		\caption{}
		\label{img:Kl4_NLOgLoop13}
	\end{subfigure}
	\begin{subfigure}[b]{0.16\textwidth}
		\centering
		\scalebox{0.75}{
			\begin{pspicture}(0,-0.5)(4,5)
				\includegraphics[width=3cm]{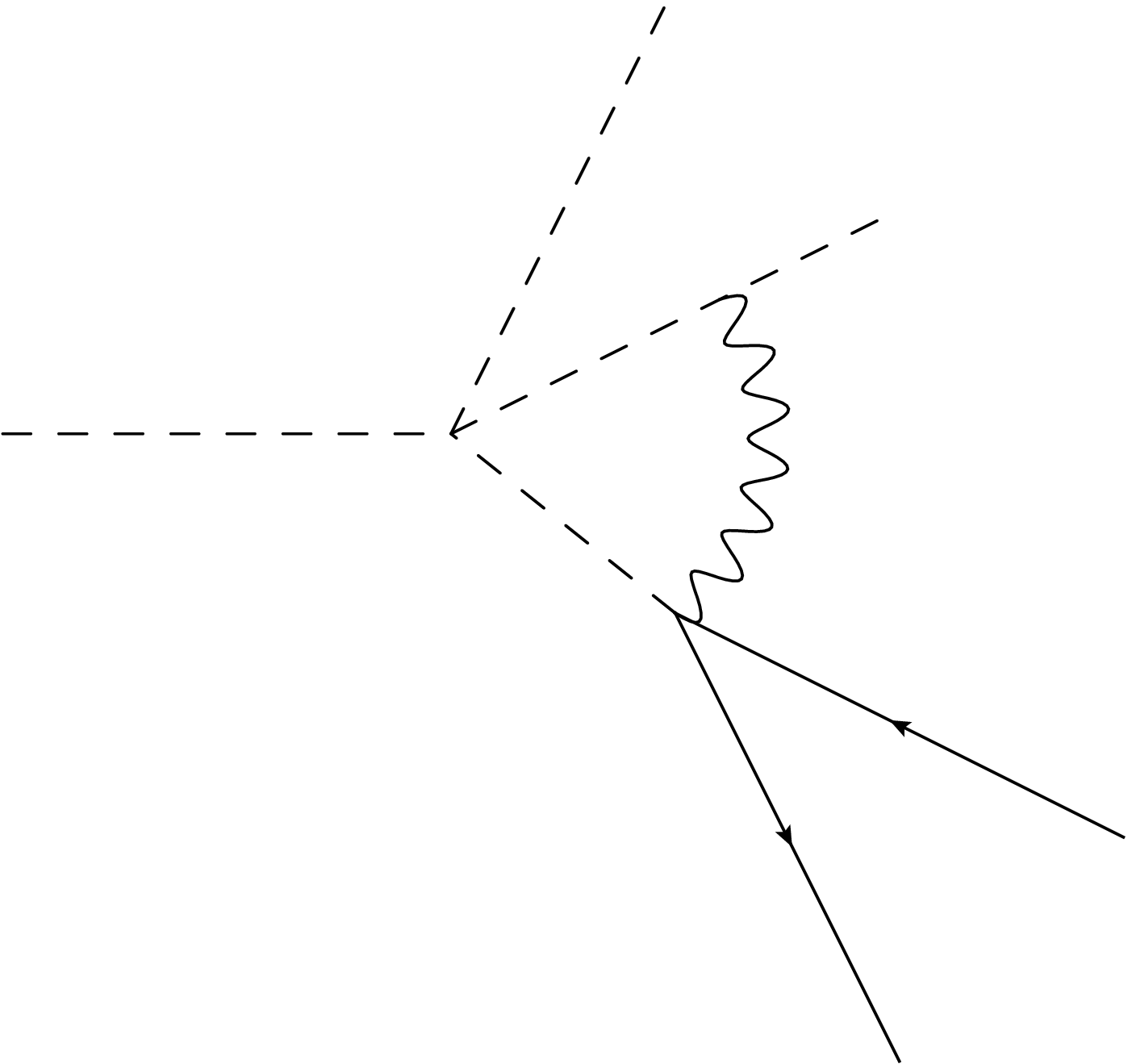}
			\end{pspicture}
			}
		\caption{}
		\label{img:Kl4_NLOgLoop14}
	\end{subfigure}
	\begin{subfigure}[b]{0.16\textwidth}
		\centering
		\scalebox{0.75}{
			\begin{pspicture}(0,-0.5)(4,5)
				\includegraphics[width=3cm]{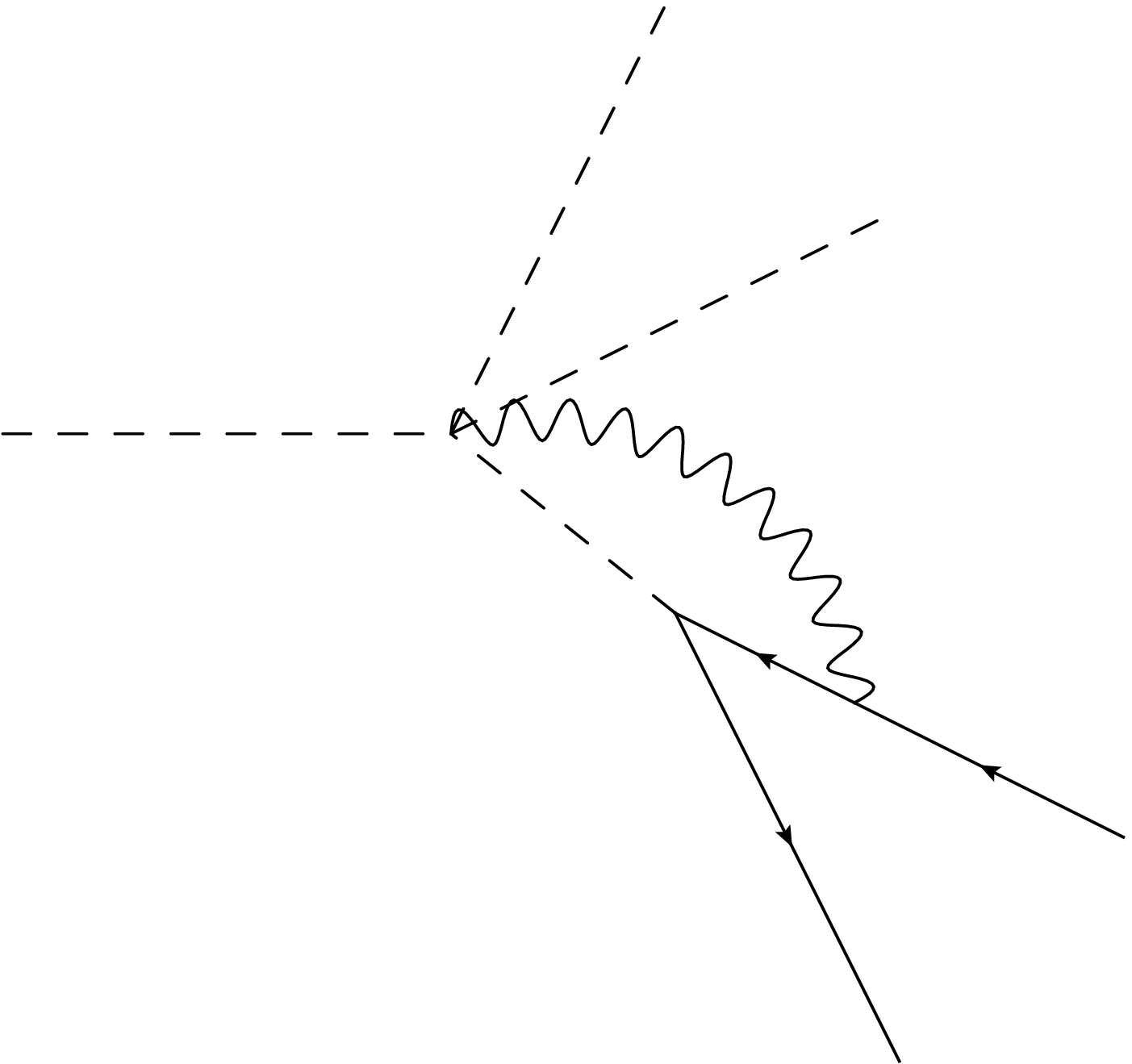}
			\end{pspicture}
			}
		\caption{}
		\label{img:Kl4_NLOgLoop15}
	\end{subfigure}

	\vspace{0.5cm}

	\begin{subfigure}[b]{0.16\textwidth}
		\centering
		\scalebox{0.75}{
			\begin{pspicture}(0,-0.5)(4,5)
				\includegraphics[width=3cm]{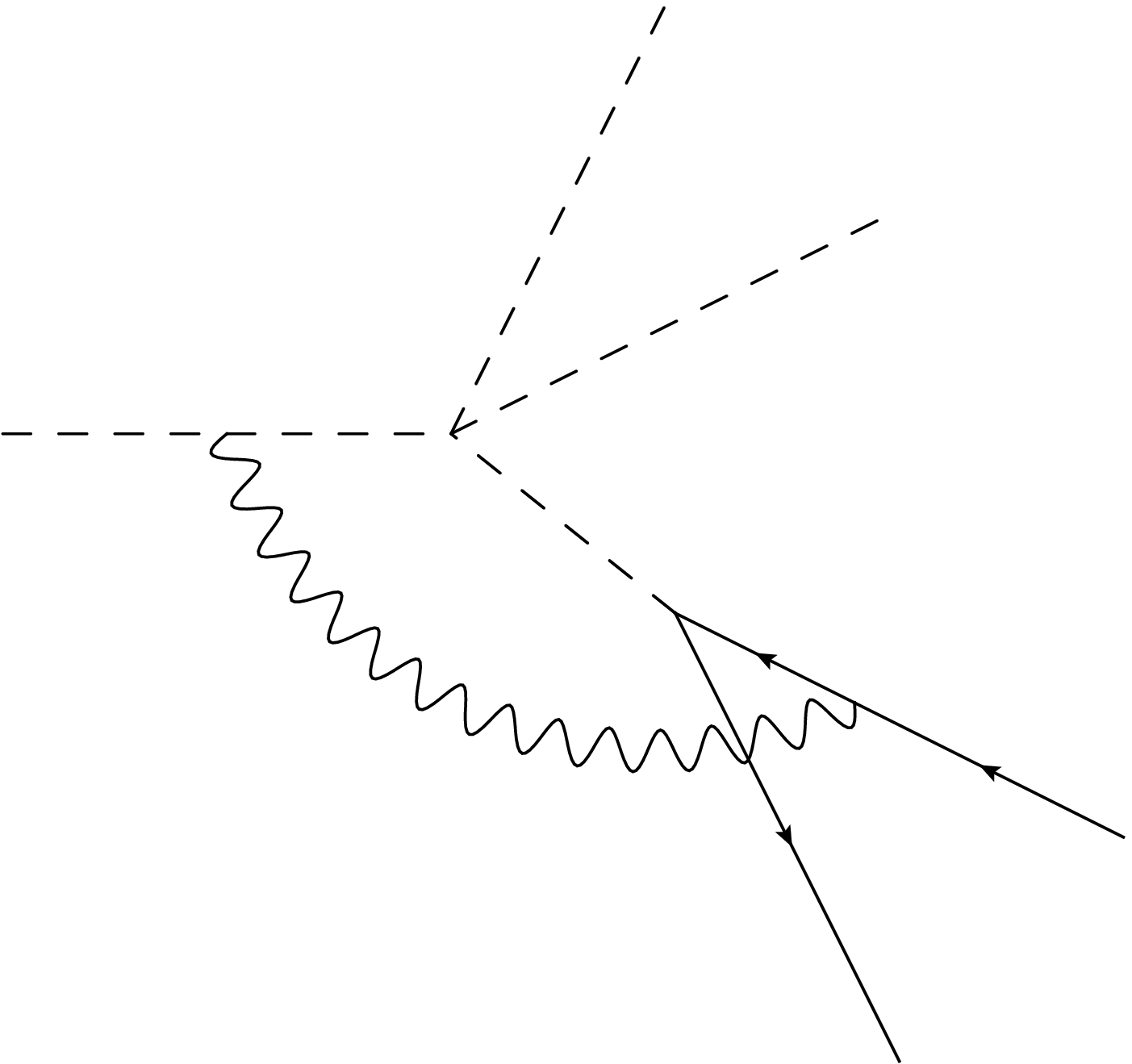}
			\end{pspicture}
			}
		\caption{}
		\label{img:Kl4_NLOgLoop16}
	\end{subfigure}
	\begin{subfigure}[b]{0.16\textwidth}
		\centering
		\scalebox{0.75}{
			\begin{pspicture}(0,-0.5)(4,5)
				\includegraphics[width=3cm]{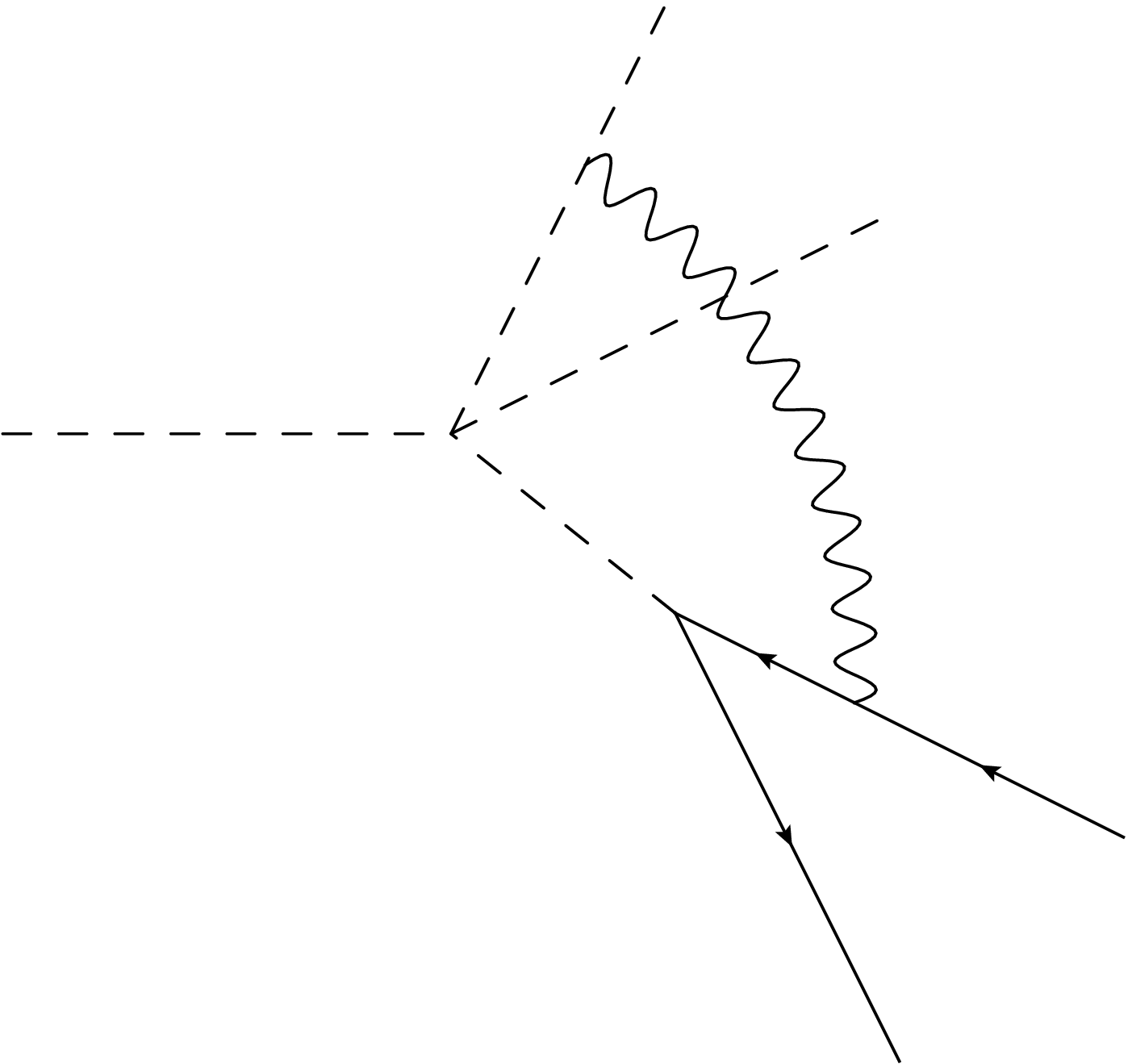}
			\end{pspicture}
			}
		\caption{}
		\label{img:Kl4_NLOgLoop17}
	\end{subfigure}
	\begin{subfigure}[b]{0.16\textwidth}
		\centering
		\scalebox{0.75}{
			\begin{pspicture}(0,-0.5)(4,5)
				\includegraphics[width=3cm]{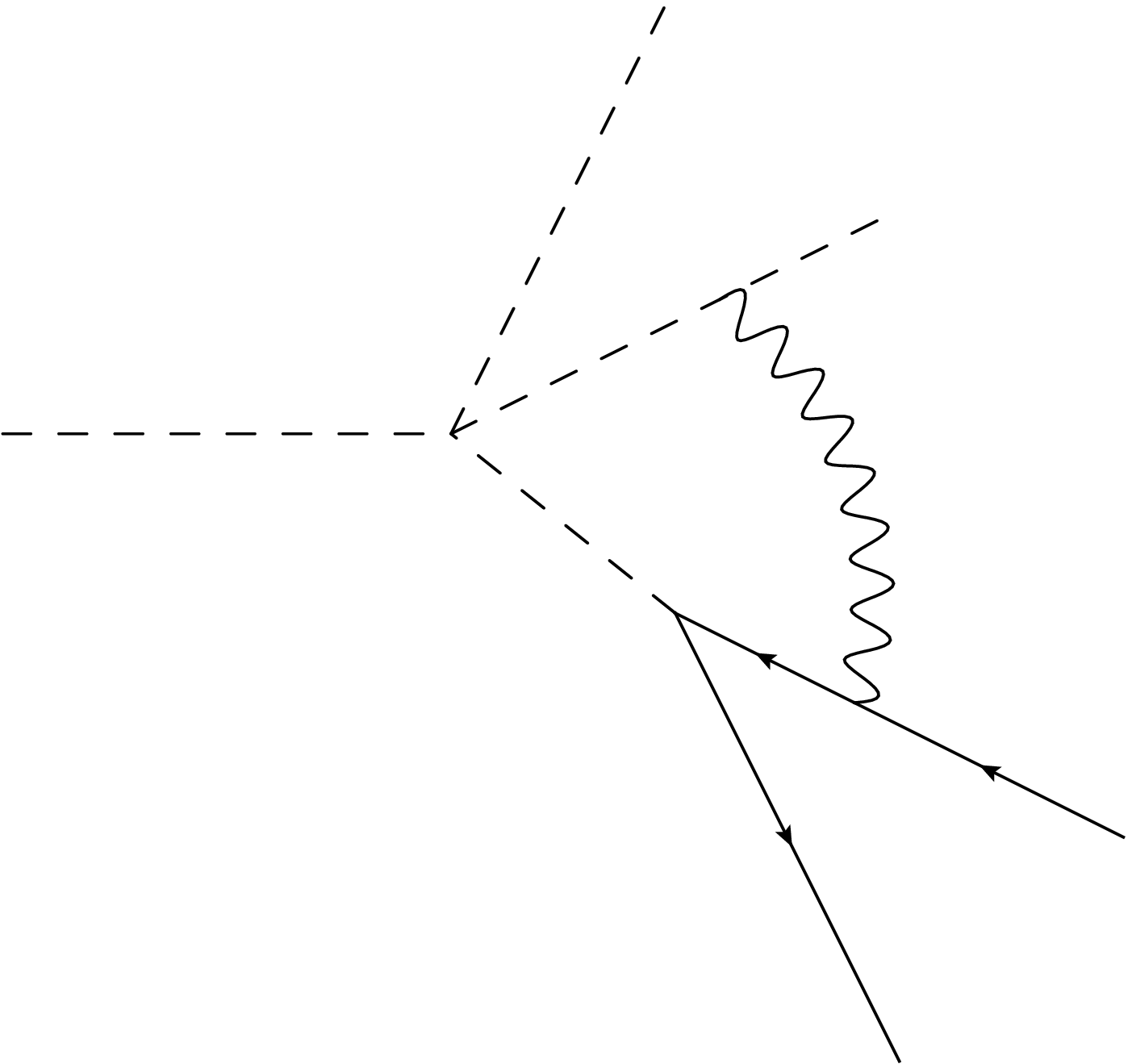}
			\end{pspicture}
			}
		\caption{}
		\label{img:Kl4_NLOgLoop18}
	\end{subfigure}

	\caption{First set of one-loop diagrams with virtual photons: they are obtained by a virtual photon insertion in the tree diagrams in figure~\ref{img:Kl4LO} (I drop the labels for the external particles as they are always the same). Diagrams contributing only to the form factor $R$ are omitted.}
	\label{img:Kl4_gLoops}
\end{figure}

A first class of loop diagrams is obtained by adding a virtual photon to the tree diagrams in figure~\ref{img:Kl4LO}. All diagrams contributing to $F$ and $G$ are shown in figure~\ref{img:Kl4_gLoops}. Again, diagrams with a virtual kaon pole are omitted, as they contribute only to $R$.

I choose to express most of the results in terms of the basic scalar loop functions $A_0$, $B_0$, $C_0$ and $D_0$. 

\begin{figure}[H]
	\centering
	\begin{subfigure}[b]{0.2\textwidth}
		\centering
		\scalebox{0.75}{
			\begin{pspicture}(0,-0.5)(4,5)
				\includegraphics[height=3cm]{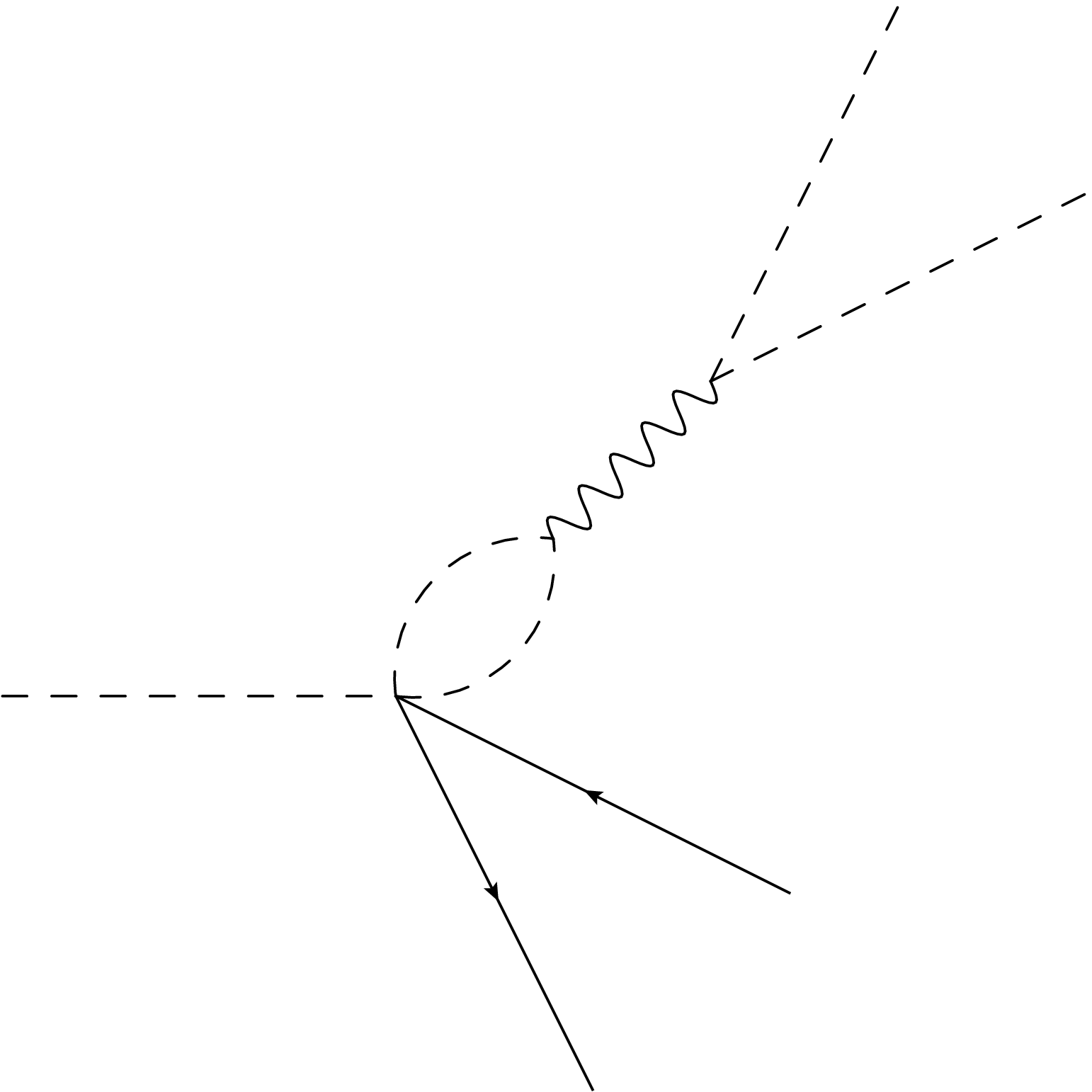}
			\end{pspicture}
			}
		\caption{}
		\label{img:Kl4_NLOmLoop1}
	\end{subfigure}
	\begin{subfigure}[b]{0.2\textwidth}
		\centering
		\scalebox{0.75}{
			\begin{pspicture}(0,-0.5)(4,5)
				\includegraphics[height=3cm]{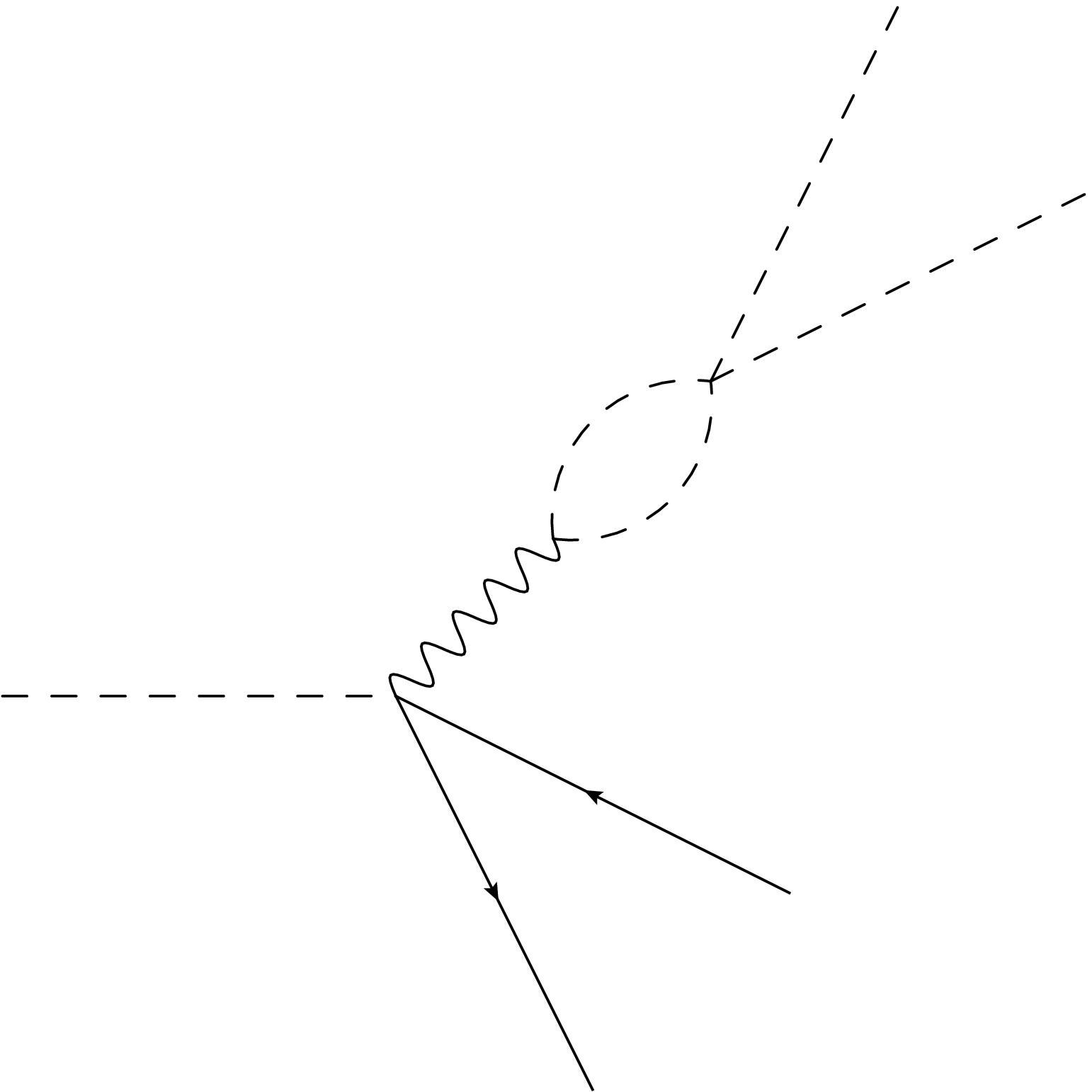}
			\end{pspicture}
			}
		\caption{}
		\label{img:Kl4_NLOmLoop2}
	\end{subfigure}
	\begin{subfigure}[b]{0.2\textwidth}
		\centering
		\scalebox{0.75}{
			\begin{pspicture}(0,-0.5)(4,5)
				\includegraphics[height=3cm]{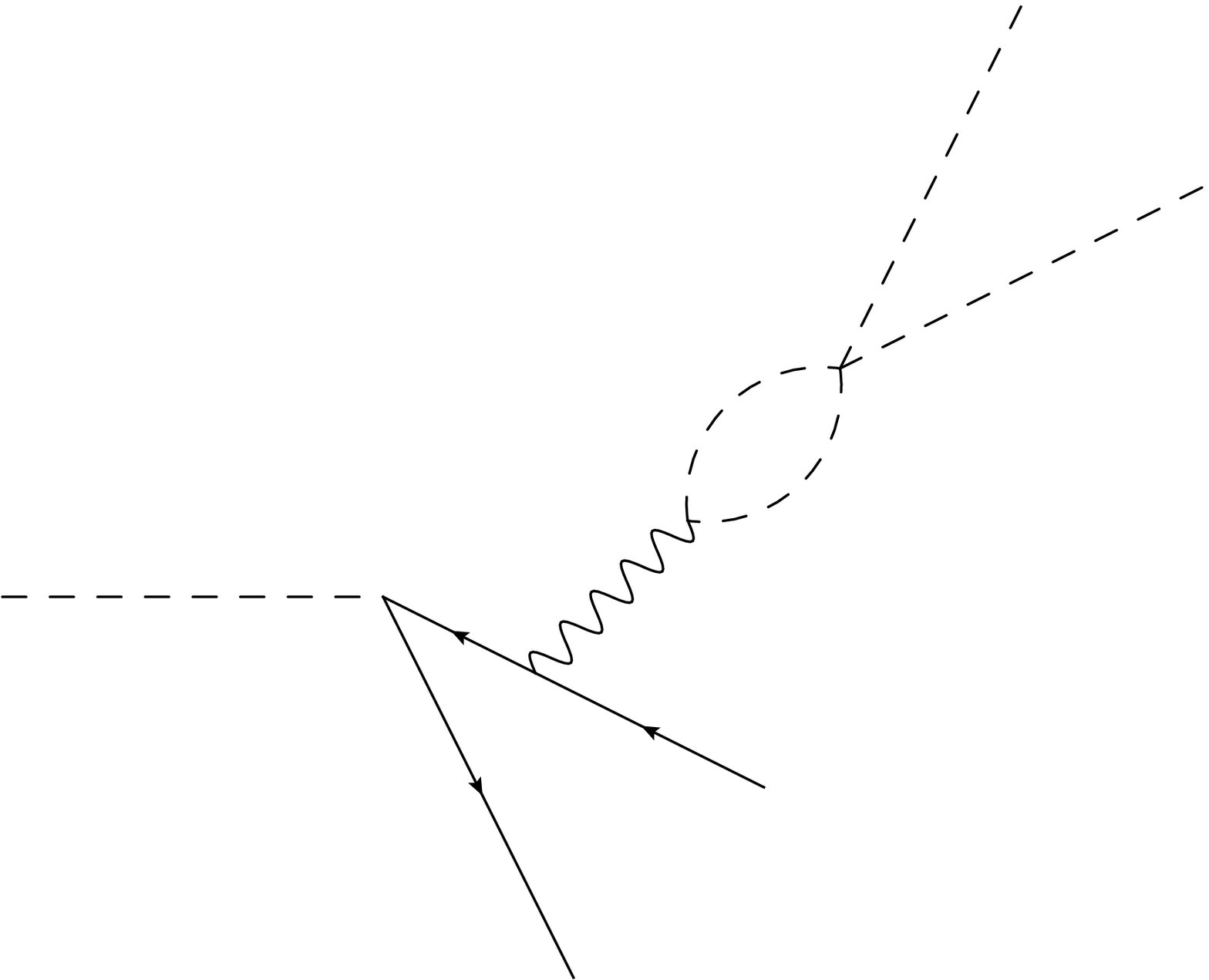}
			\end{pspicture}
			}
		\caption{}
		\label{img:Kl4_NLOmLoop3}
	\end{subfigure}
	\begin{subfigure}[b]{0.2\textwidth}
		\centering
		\scalebox{0.75}{
			\begin{pspicture}(0,-0.5)(4,5)
				\includegraphics[height=3cm]{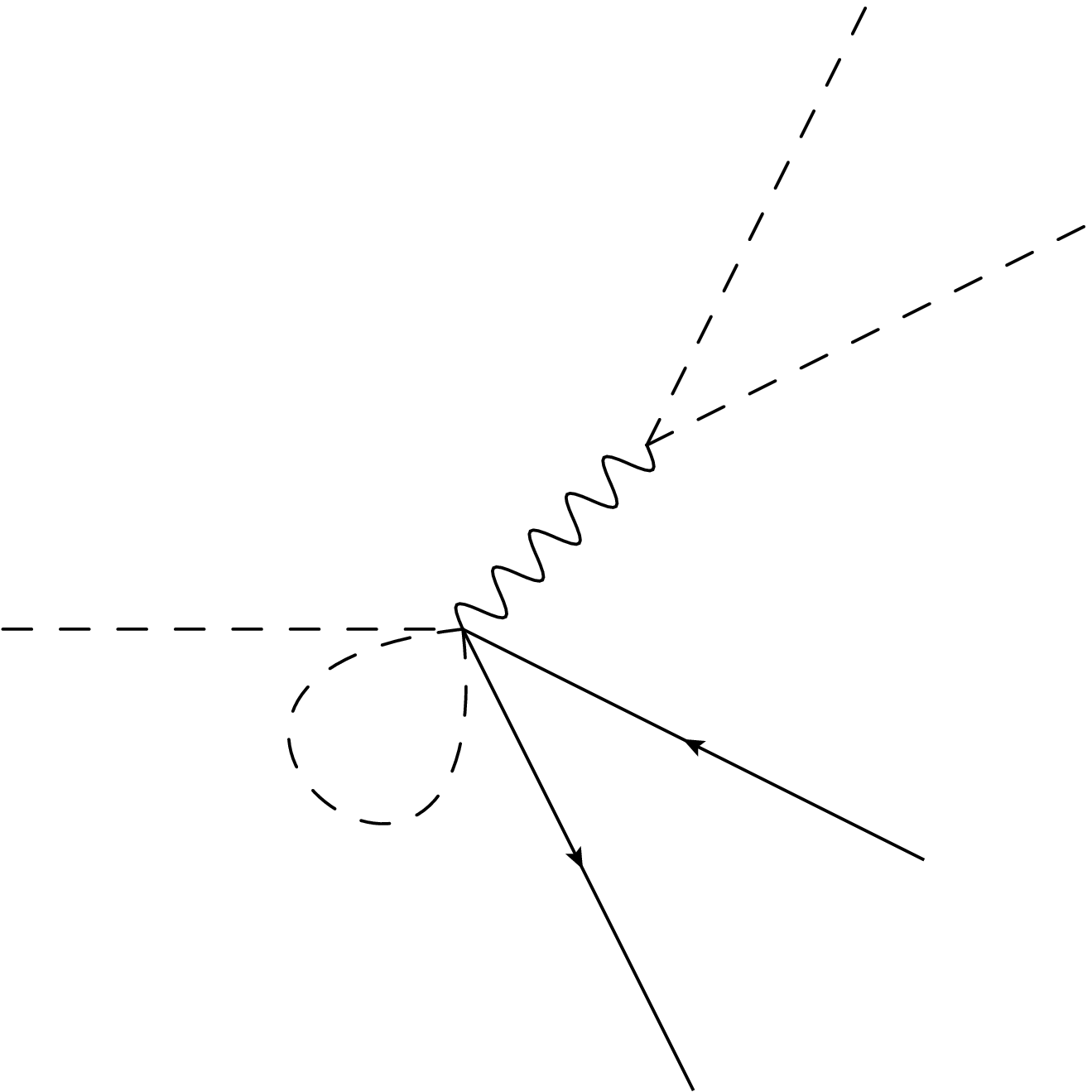}
			\end{pspicture}
			}
		\caption{}
		\label{img:Kl4_NLOmLoop4}
	\end{subfigure}
	
	\vspace{0.5cm}

	\begin{subfigure}[b]{0.2\textwidth}
		\centering
		\scalebox{0.75}{
			\begin{pspicture}(0,-0.5)(4,5)
				\includegraphics[height=3cm]{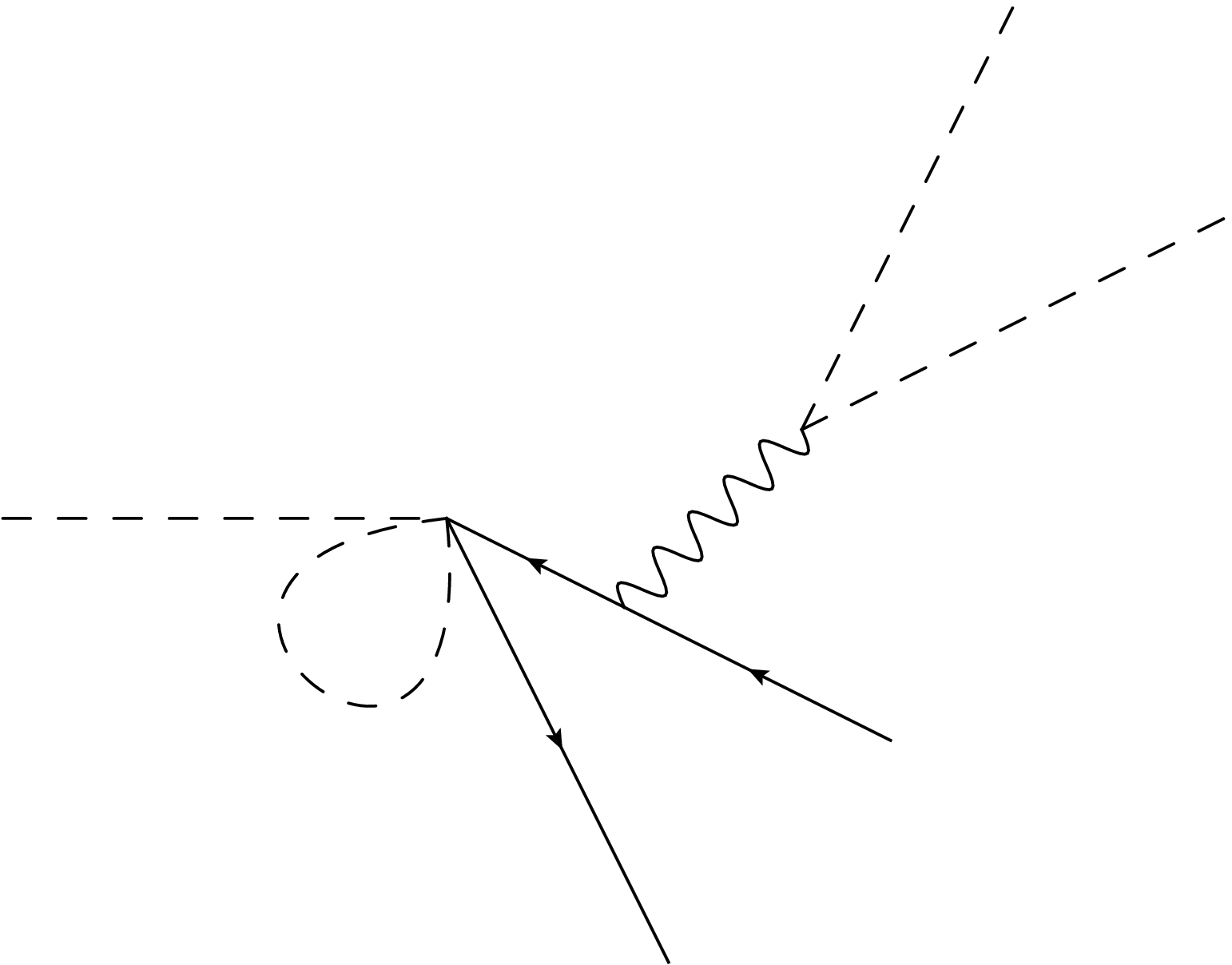}
			\end{pspicture}
			}
		\caption{}
		\label{img:Kl4_NLOmLoop5}
	\end{subfigure}
	\begin{subfigure}[b]{0.2\textwidth}
		\centering
		\scalebox{0.75}{
			\begin{pspicture}(0,-0.5)(4,5)
				\includegraphics[height=3cm]{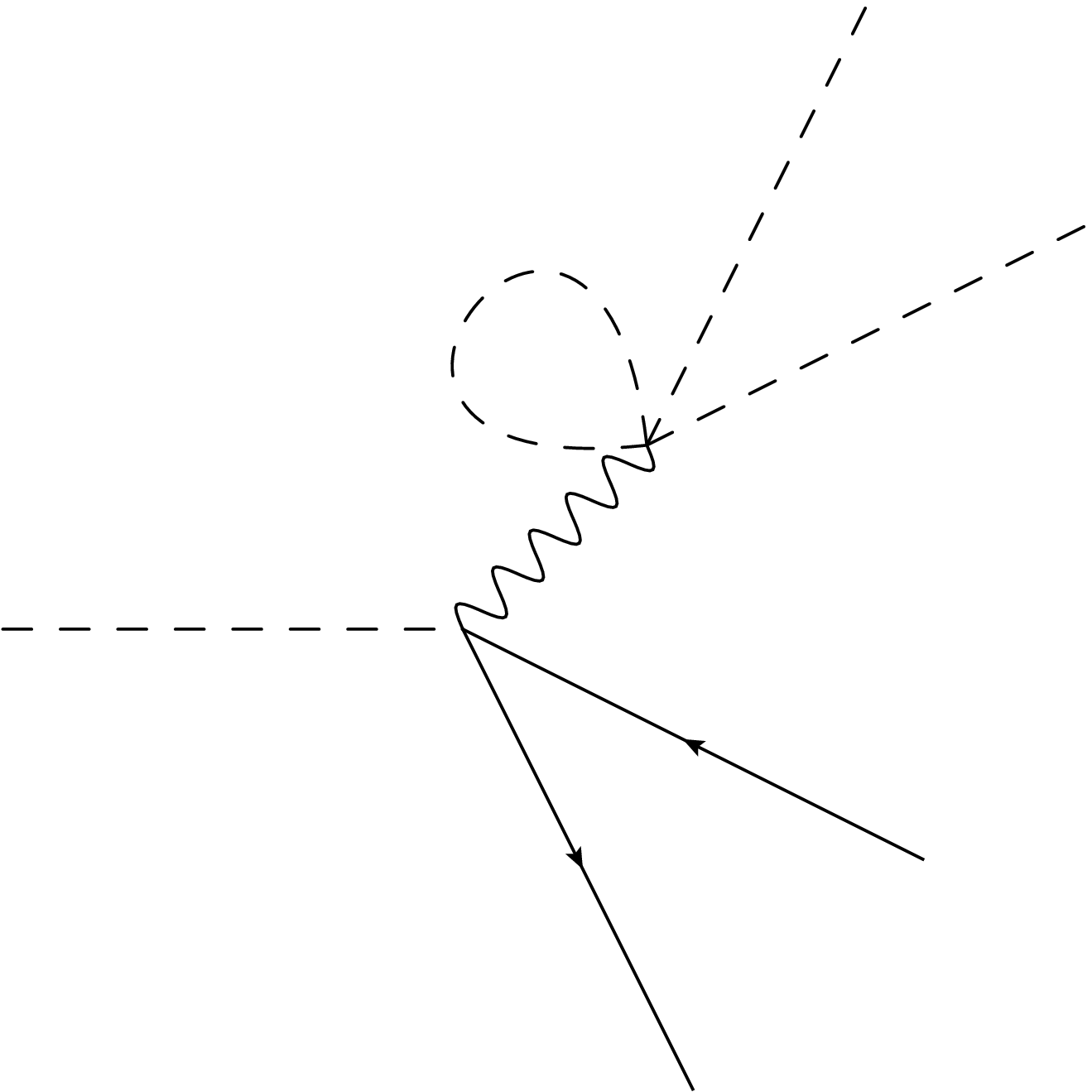}
			\end{pspicture}
			}
		\caption{}
		\label{img:Kl4_NLOmLoop6}
	\end{subfigure}
	\begin{subfigure}[b]{0.2\textwidth}
		\centering
		\scalebox{0.75}{
			\begin{pspicture}(0,-0.5)(4,5)
				\includegraphics[height=3cm]{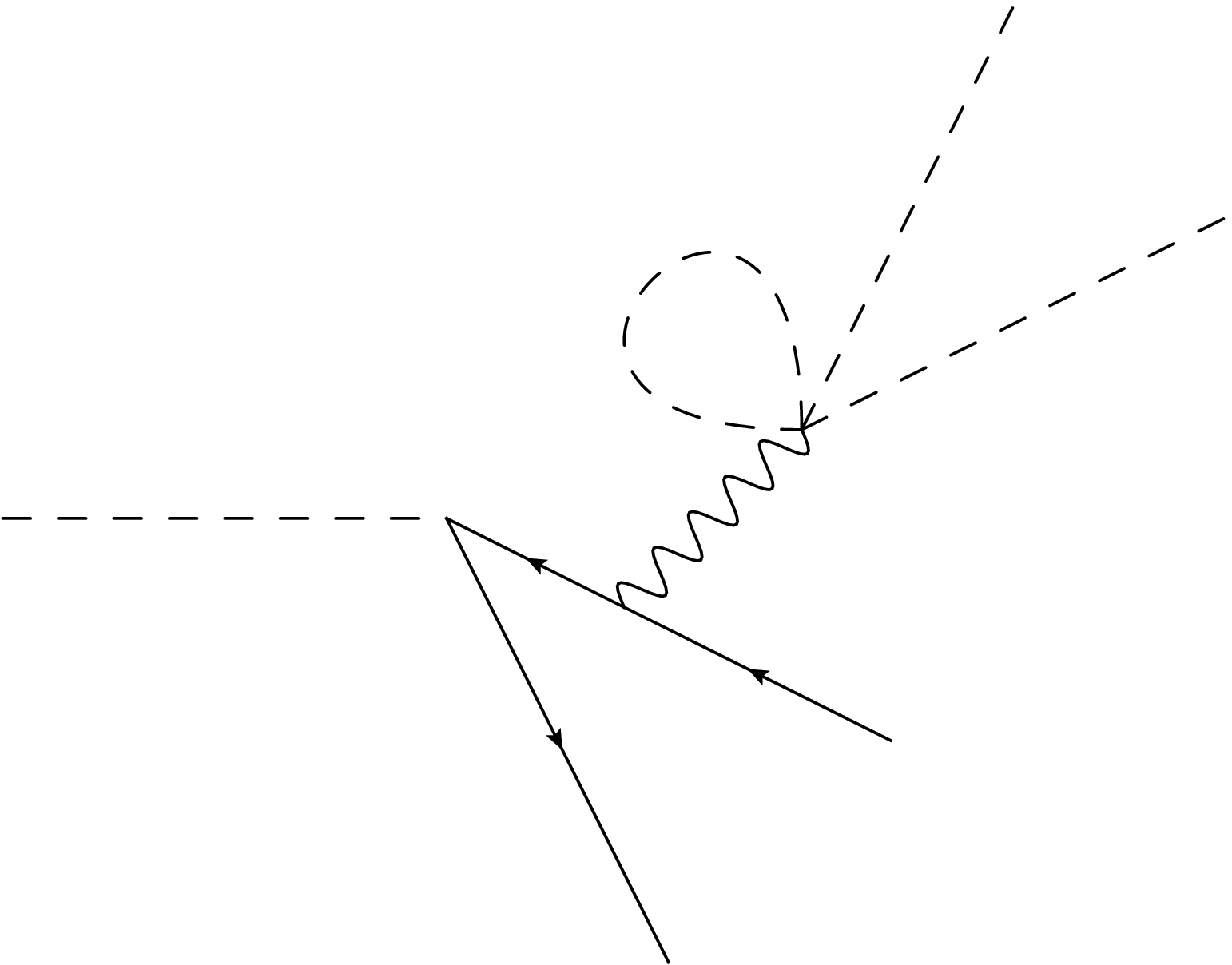}
			\end{pspicture}
			}
		\caption{}
		\label{img:Kl4_NLOmLoop7}
	\end{subfigure}

	\caption{Second set of one-loop diagrams with virtual photons: they are obtained by a meson loop insertion in the tree diagrams in figure~\ref{img:Kl4LO_Photons}. Diagrams contributing only to the form factor $R$ are omitted.}
	\label{img:Kl4_mLoops}
\end{figure}


The contributions of the diagrams~\ref{img:Kl4_NLOgLoop1} - \ref{img:Kl4_NLOgLoop4}, where one end of the photon line is attached to a charged external line and the other end to the vertex, are all IR-finite.

The next six (triangle) diagrams, obtained by attaching a virtual photon to two external lines, generate an IR divergence. My results differ from \cite{Cuplov2004} only by the contribution of the additional term in the propagator for the massive vector boson. This contribution will cancel in the sum with the external leg corrections.

The remaining eight diagrams in this first set consist of one bulb, four triangle and finally three box diagrams that are obtained by an insertion of a virtual photon into diagram~\ref{img:Kl4_LO2}.

A second set of loop diagrams, shown in figure~\ref{img:Kl4_mLoops}, is obtained by inserting virtual mesons into the tree-level diagrams in figure~\ref{img:Kl4LO_Photons}. Although the contributions of the LO diagrams in figure~\ref{img:Kl4LO_Photons} to the form factors $F$ and $G$ vanish, the NLO diagrams give a finite contribution to $G$. To my knowledge, they have not been considered in the previous literature.

In diagrams \ref{img:Kl4_NLOmLoop1} - \ref{img:Kl4_NLOmLoop3}, we have to insert charged mesons in the loop. In the tadpole loops, all octet mesons have to be included.

\subsubsection{Counterterms}

\begin{figure}[ht]
	\centering
	\begin{subfigure}[b]{0.19\textwidth}
		\centering
		\scalebox{0.8}{
			\begin{pspicture}(0,-0.5)(4,5)
				\includegraphics[width=3cm]{images/NLO_CT1}
			\end{pspicture}
			}
		\caption{}
		\label{img:Kl4_NLOCT1}
	\end{subfigure}
	\begin{subfigure}[b]{0.19\textwidth}
		\centering
		\scalebox{0.8}{
			\begin{pspicture}(0,-0.5)(4,5)
				\includegraphics[width=3cm]{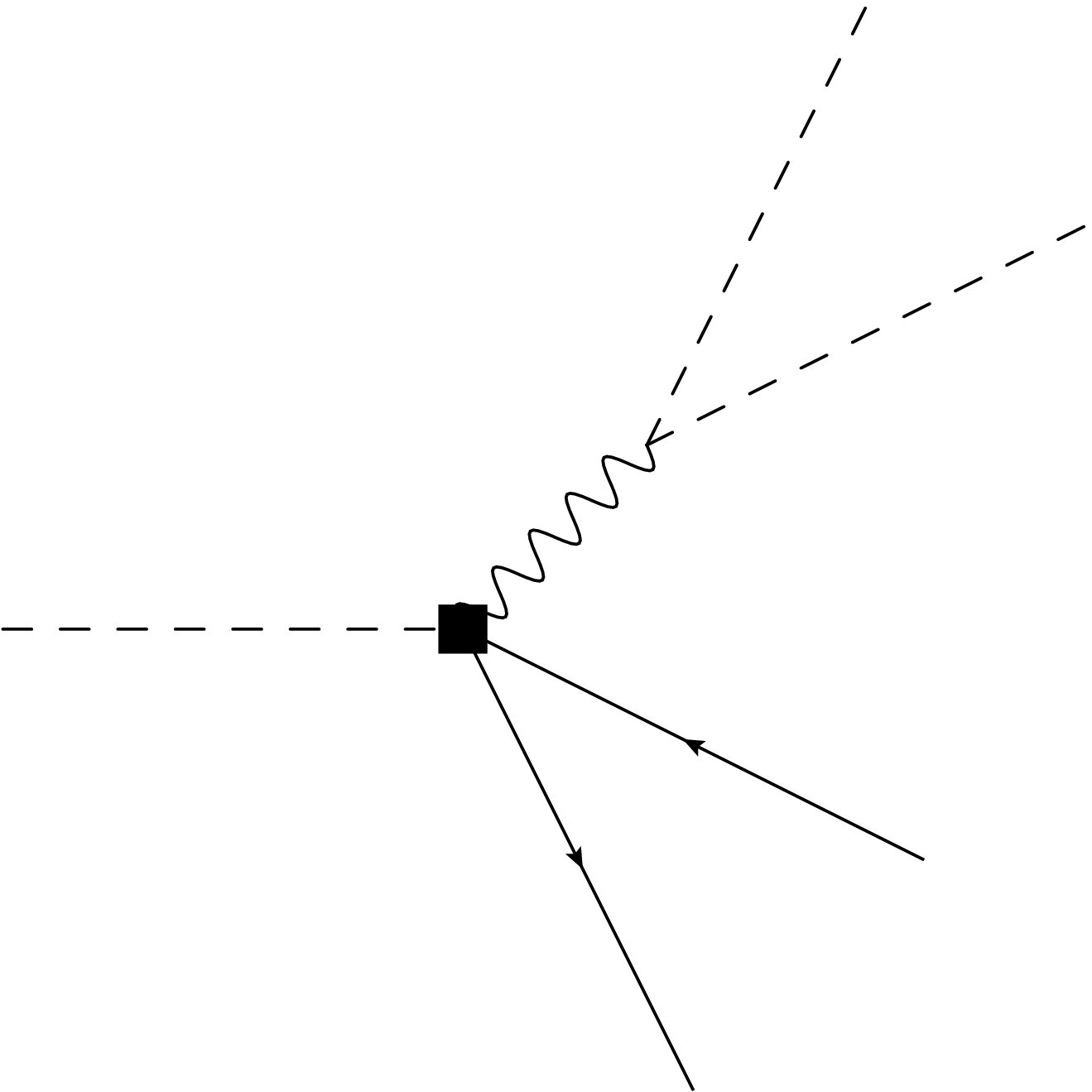}
			\end{pspicture}
			}
		\caption{}
		\label{img:Kl4_NLOCT2}
	\end{subfigure}
	\begin{subfigure}[b]{0.19\textwidth}
		\centering
		\scalebox{0.8}{
			\begin{pspicture}(0,-0.5)(4,5)
				\includegraphics[width=3cm]{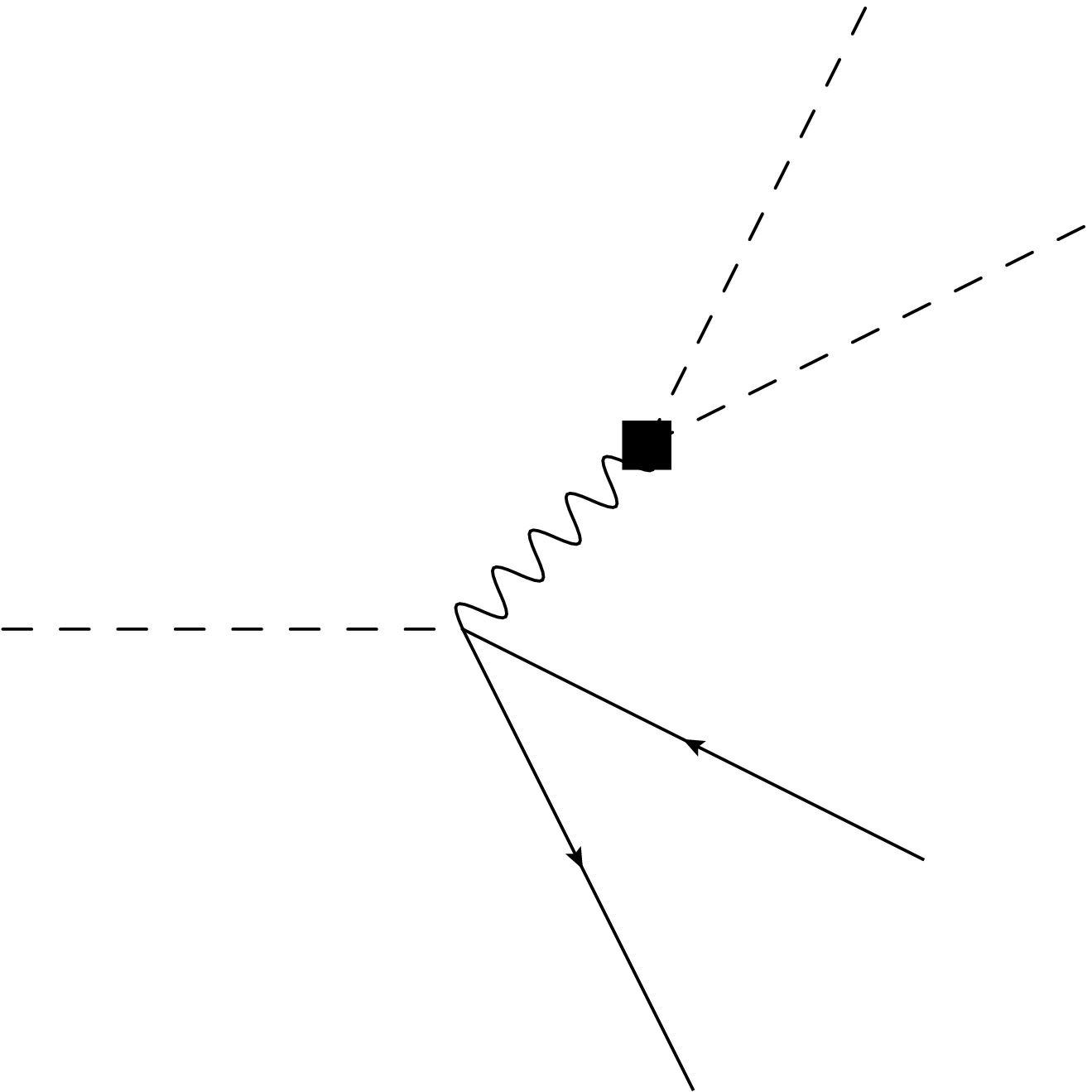}
			\end{pspicture}
			}
		\caption{}
		\label{img:Kl4_NLOCT3}
	\end{subfigure}
	\begin{subfigure}[b]{0.19\textwidth}
		\centering
		\scalebox{0.8}{
			\begin{pspicture}(0,-0.5)(4,5)
				\includegraphics[width=3cm]{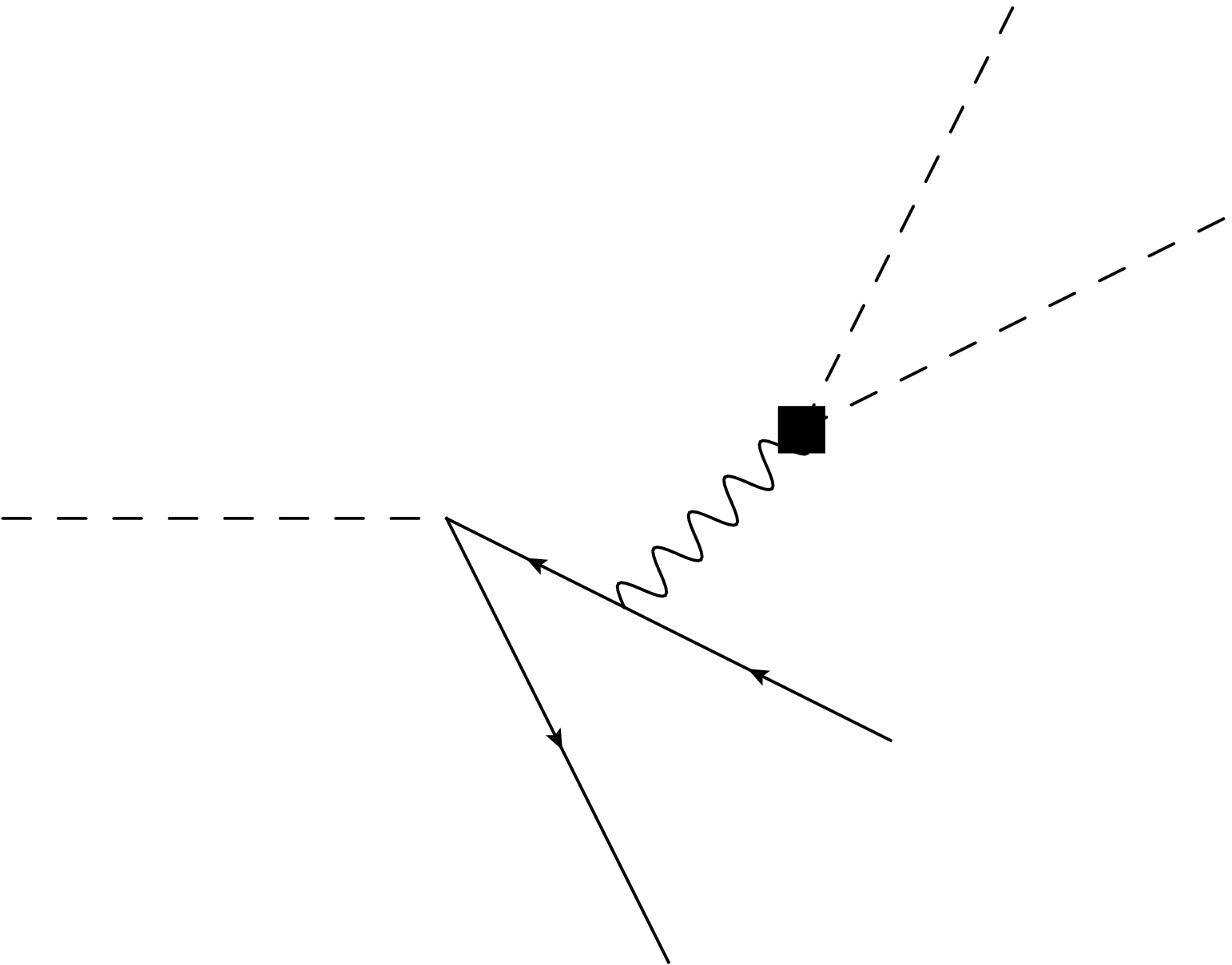}
			\end{pspicture}
			}
		\caption{}
		\label{img:Kl4_NLOCT4}
	\end{subfigure}
	\begin{subfigure}[b]{0.19\textwidth}
		\centering
		\scalebox{0.8}{
			\begin{pspicture}(0,-0.5)(4,5)
				\includegraphics[width=3cm]{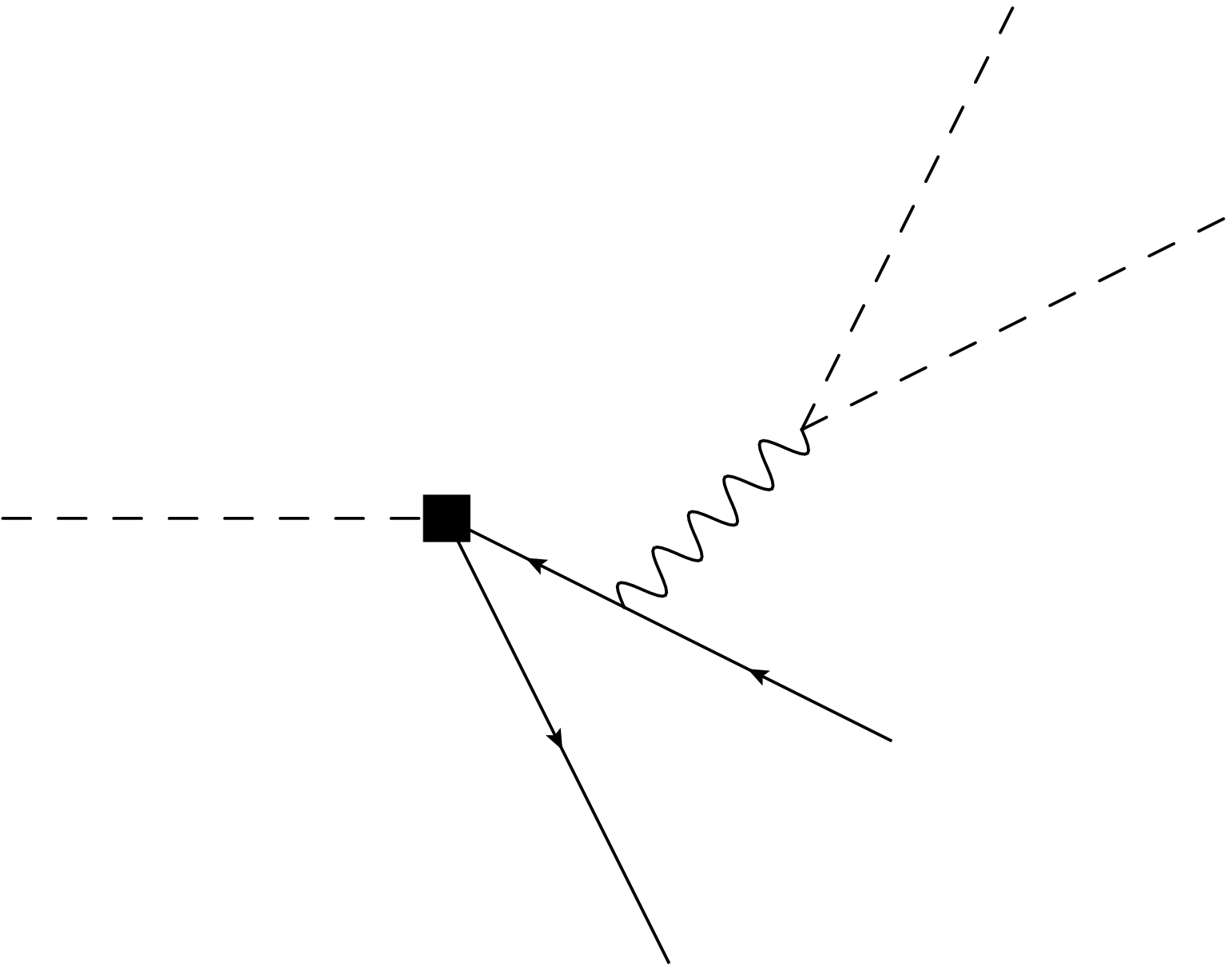}
			\end{pspicture}
			}
		\caption{}
		\label{img:Kl4_NLOCT5}
	\end{subfigure}

	\caption{Counterterms needed to renormalise the loops with virtual photons.}
	\label{img:Kl4_gCT}
\end{figure}

In order to renormalise the UV divergences in the loop functions, we have to compute the counterterm contribution, i.e.~tree-level diagrams with one vertex from $\mathcal{L}_{p^4}$, $\mathcal{L}_{e^2p^2}$ or $\mathcal{L}_\mathrm{lept}$. These diagrams are shown in figure~\ref{img:Kl4_gCT}. The loop diagrams of the first class, figure~\ref{img:Kl4_gLoops}, need only the counterterm~\ref{img:Kl4_NLOCT1}, the remaining four counterterm diagrams renormalise the meson loops of the second class, figure~\ref{img:Kl4_mLoops}.

\subsubsection{External Leg Corrections}

In order to complete the NLO calculation, we need the external leg corrections at $\O(e^2 p)$. At this order, the corrections for both charged mesons and lepton have to be taken into account.

\begin{figure}[H]
	\centering
	\begin{subfigure}[b]{0.2\textwidth}
		\centering
		\scalebox{0.8}{
			\begin{pspicture}(0,0)(4,3)
				\includegraphics[width=3cm]{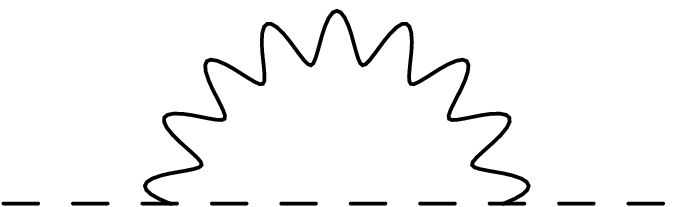}
			\end{pspicture}
			}
		\caption{}
		\label{img:Kl4_mSEgLoop}
	\end{subfigure}
	\begin{subfigure}[b]{0.2\textwidth}
		\centering
		\scalebox{0.8}{
			\begin{pspicture}(0,0)(4,3)
				\includegraphics[width=3cm]{images/SE_CT}
			\end{pspicture}
			}
		\caption{}
		\label{img:Kl4_mSECT}
	\end{subfigure}
	\hspace{1cm}
	\begin{subfigure}[b]{0.2\textwidth}
		\centering
		\scalebox{0.8}{
			\begin{pspicture}(0,0)(4,3)
				\includegraphics[width=3cm]{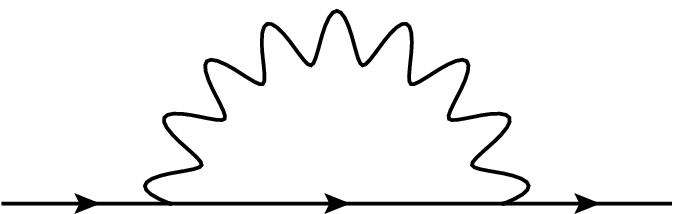}
			\end{pspicture}
			}
		\caption{}
		\label{img:Kl4_lSEgLoop}
	\end{subfigure}
	\begin{subfigure}[b]{0.2\textwidth}
		\centering
		\scalebox{0.8}{
			\begin{pspicture}(0,0)(4,3)
				\includegraphics[width=3cm]{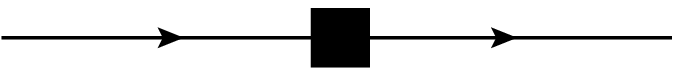}
			\end{pspicture}
			}
		\caption{}
		\label{img:Kl4_lSECT}
	\end{subfigure}
	\caption{Meson and lepton self-energy diagrams.}
	\label{img:Kl4_gExternalLegs}
\end{figure}

The calculation of the field strength renormalisation and its contribution to the form factors can be found in the appendix~\ref{sec:AppendixExternalLegCorrectionsPhotonicEffects}.

\subsubsection{Renormalisation}

The form factors at $\O(e^2 p)$ are given by
\begin{align}
	\begin{split}
		X^\mathrm{NLO}_{\mathrm{virt.}\gamma} &= X^\mathrm{LO}_{\mathrm{virt.}\gamma} \left( 1 + \delta X^\mathrm{NLO}_{\mathrm{virt.}\gamma}  \right) , \quad X\in\{F,G\} ,
	\end{split}
\end{align}
where the NLO corrections consists of
\begin{align}
	\begin{split}
		\delta X^\mathrm{NLO}_{\mathrm{virt.}\gamma} &= \delta X^\mathrm{NLO}_{\gamma-\mathrm{loop}} + \delta X^\mathrm{NLO}_{\gamma-\mathrm{pole}} + \delta X^\mathrm{NLO}_{\gamma-\mathrm{ct}} + \delta X^\mathrm{NLO}_{\gamma-Z} .
	\end{split}
\end{align}
The individual contributions are all given explicitly in the appendix~\ref{sec:AppendixDiagramsPhotonicEffects}. With these expressions, it can be easily verified that the contributions stemming from the additional term $k^\mu k^\nu / \mg^2$ in the propagator for a massive gauge boson (with respect to a massless propagator in Feynman gauge) cancel in the above sum (in the limit $\mg\to0$). In appendix~\ref{sec:RadiativeDecayRate}, I show that the radiative decay rate only gets $\O(\mg^2)$ contributions from the additional term in the propagator. Hence, in the limit $\mg\to0$, the longitudinal modes decouple and gauge invariance is restored~\cite{Ticciati1999}.

Next, let us check that the UV-divergent parts cancel in the sum of all NLO contributions. Working in the $\overline{MS}$ scheme, I replace the LECs according to (\ref{eqn:RenormalisedLECs}) with their renormalised counterparts. Introducing also the renormalised loop functions (\ref{eqn:RenormalisedLoopFunctions}) and tensor coefficient functions (\ref{eqn:RenormalisedTensorCoefficients}), I find that all the terms proportional to the UV divergence $\lambda$ cancel.

\subsection{Real Photon Emission}

\label{sec:MatrixElementRealPhotonEmission}

As explained before, an IR-finite result can only be obtained for a sufficiently inclusive observable. In the present case, we have to add the $\O(e^2)$ contribution of the radiative process at the decay rate level. Let us therefore compute the $\O(e)$ tree-level amplitude for $K_{\ell4\gamma}$.

\begin{figure}[H]
	\centering
	\begin{subfigure}[b]{0.19\textwidth}
		\centering
		\scalebox{0.75}{
			\begin{pspicture}(0,0)(4,3)
				\includegraphics[width=3.5cm]{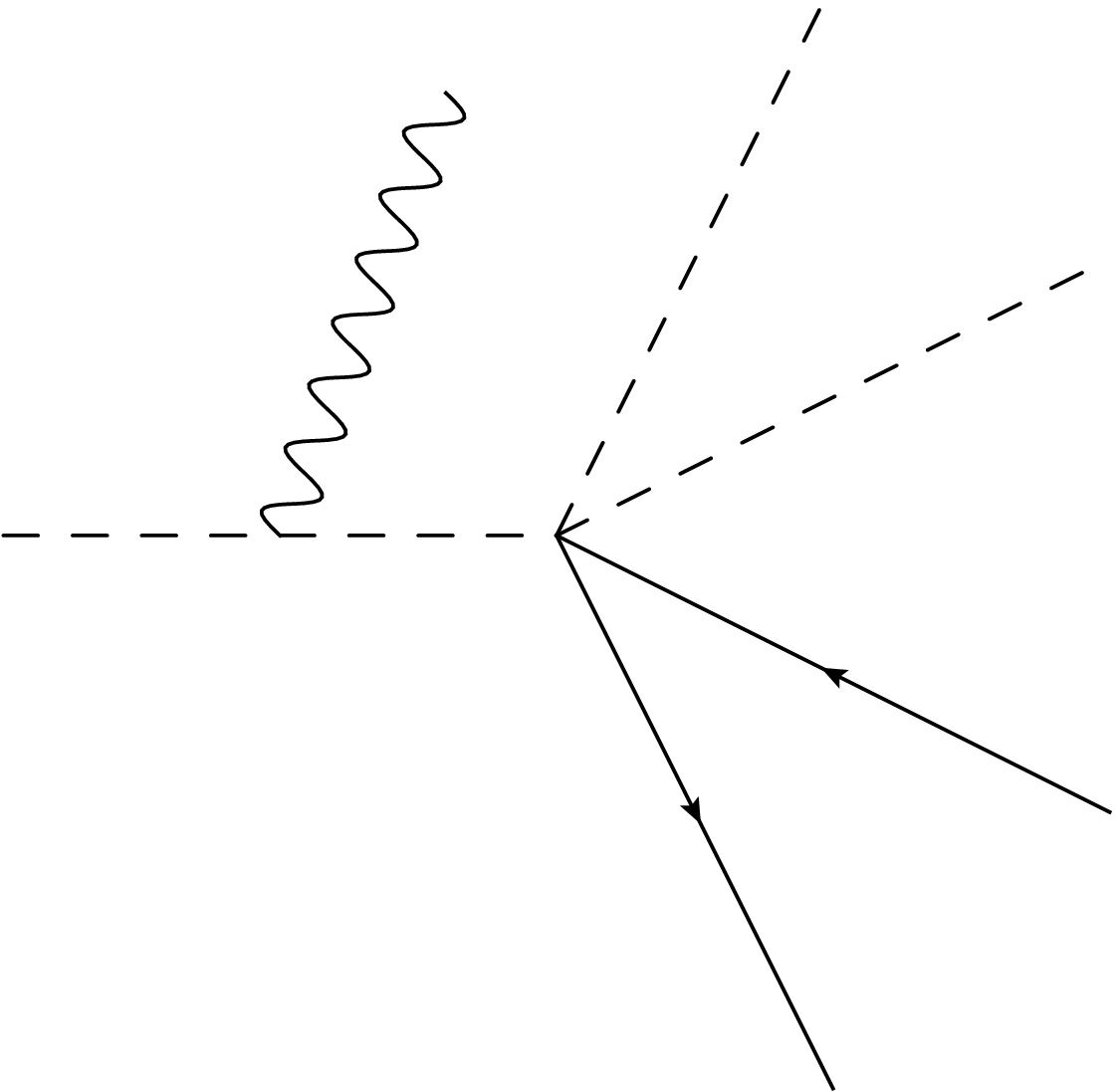}
			\end{pspicture}
			}
		\caption{}
		\label{img:Kl4g1}
	\end{subfigure}
	\begin{subfigure}[b]{0.19\textwidth}
		\centering
		\scalebox{0.75}{
			\begin{pspicture}(0,0)(4,3)
				\includegraphics[width=3.5cm]{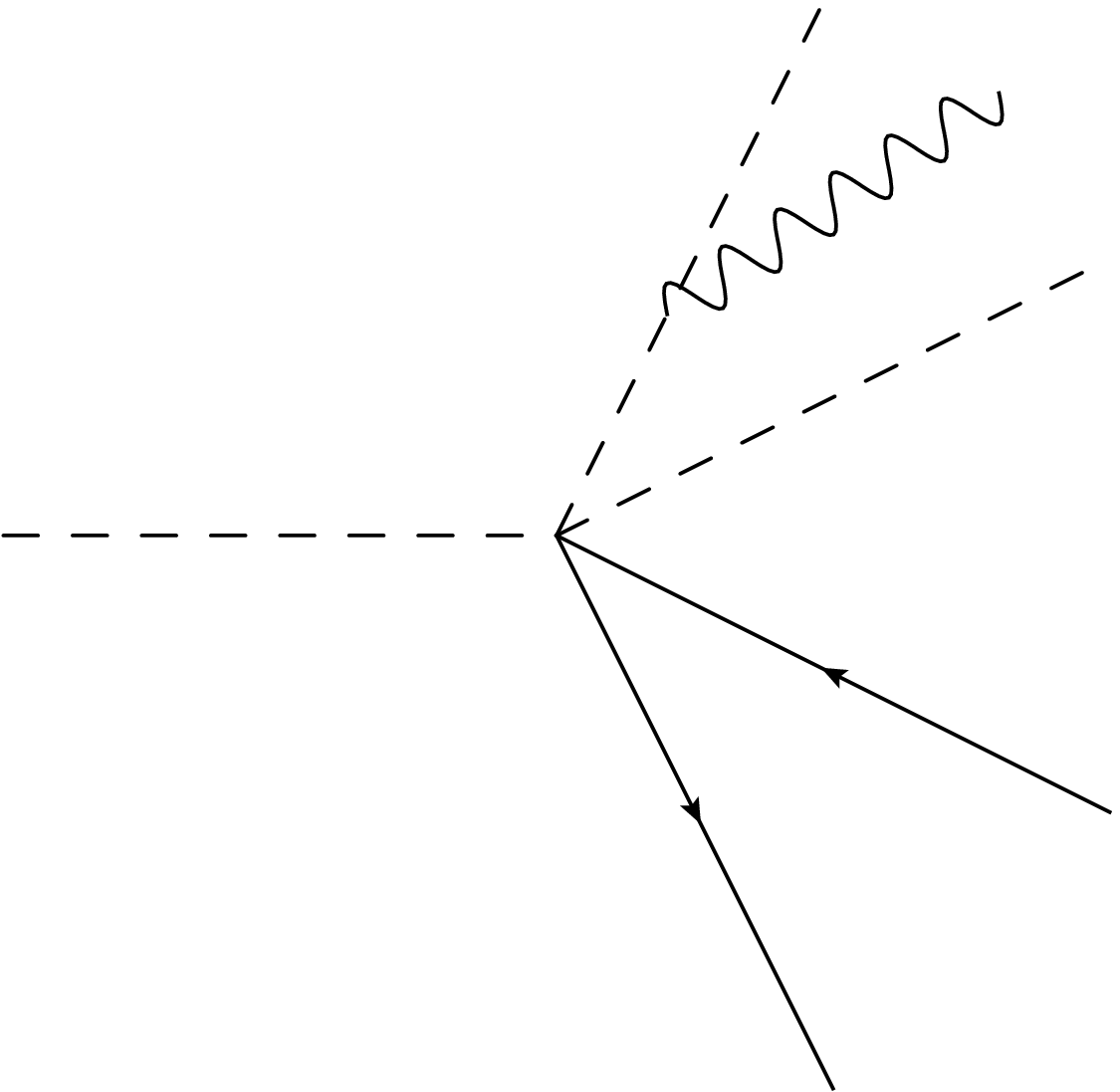}
			\end{pspicture}
			}
		\caption{}
		\label{img:Kl4g2}
	\end{subfigure}
	\begin{subfigure}[b]{0.19\textwidth}
		\centering
		\scalebox{0.75}{
			\begin{pspicture}(0,0)(4,3)
				\includegraphics[width=3.5cm]{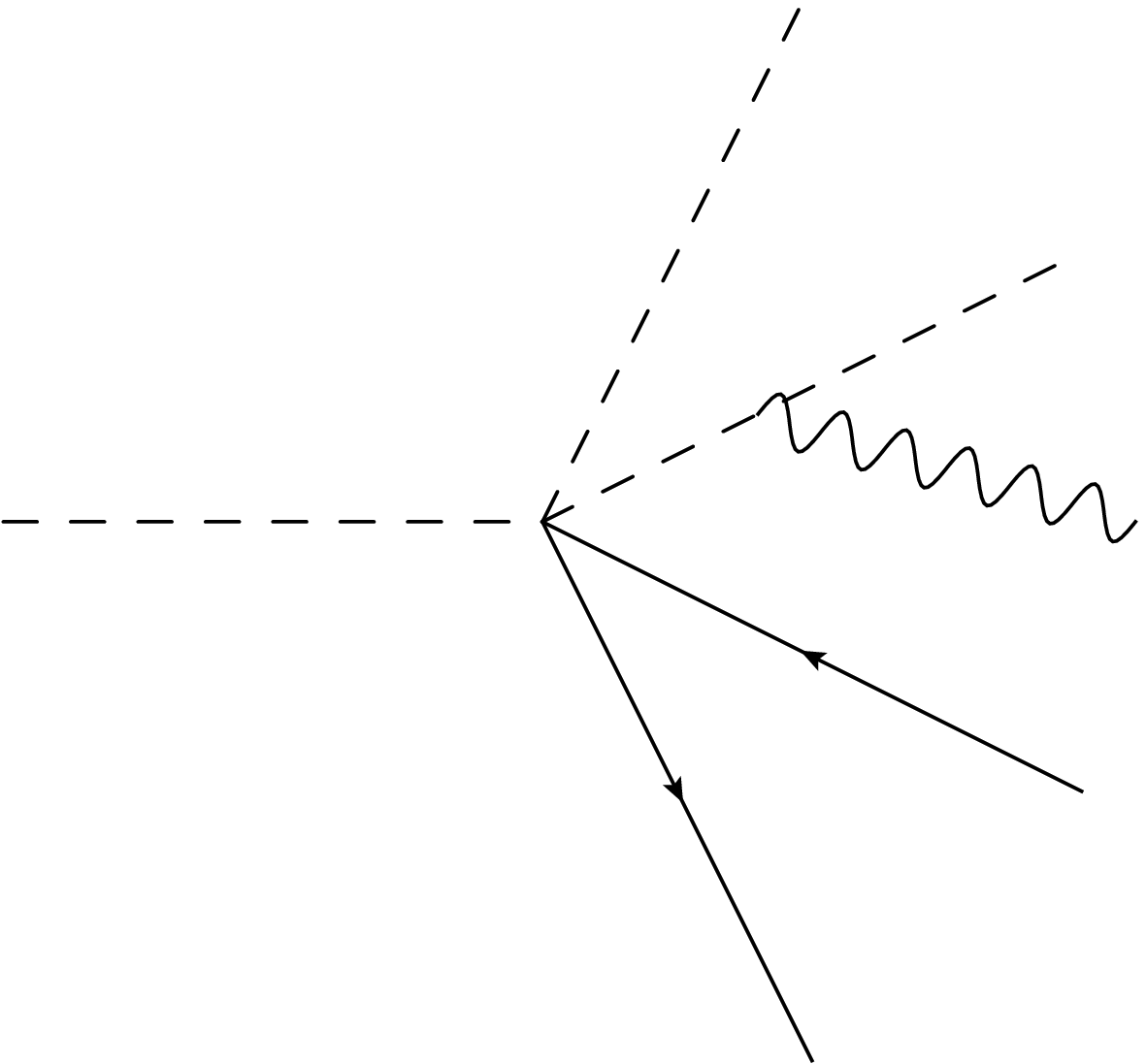}
			\end{pspicture}
			}
		\caption{}
		\label{img:Kl4g3}
	\end{subfigure}
	\begin{subfigure}[b]{0.19\textwidth}
		\centering
		\scalebox{0.75}{
			\begin{pspicture}(0,0)(4,3)
				\includegraphics[width=3.5cm]{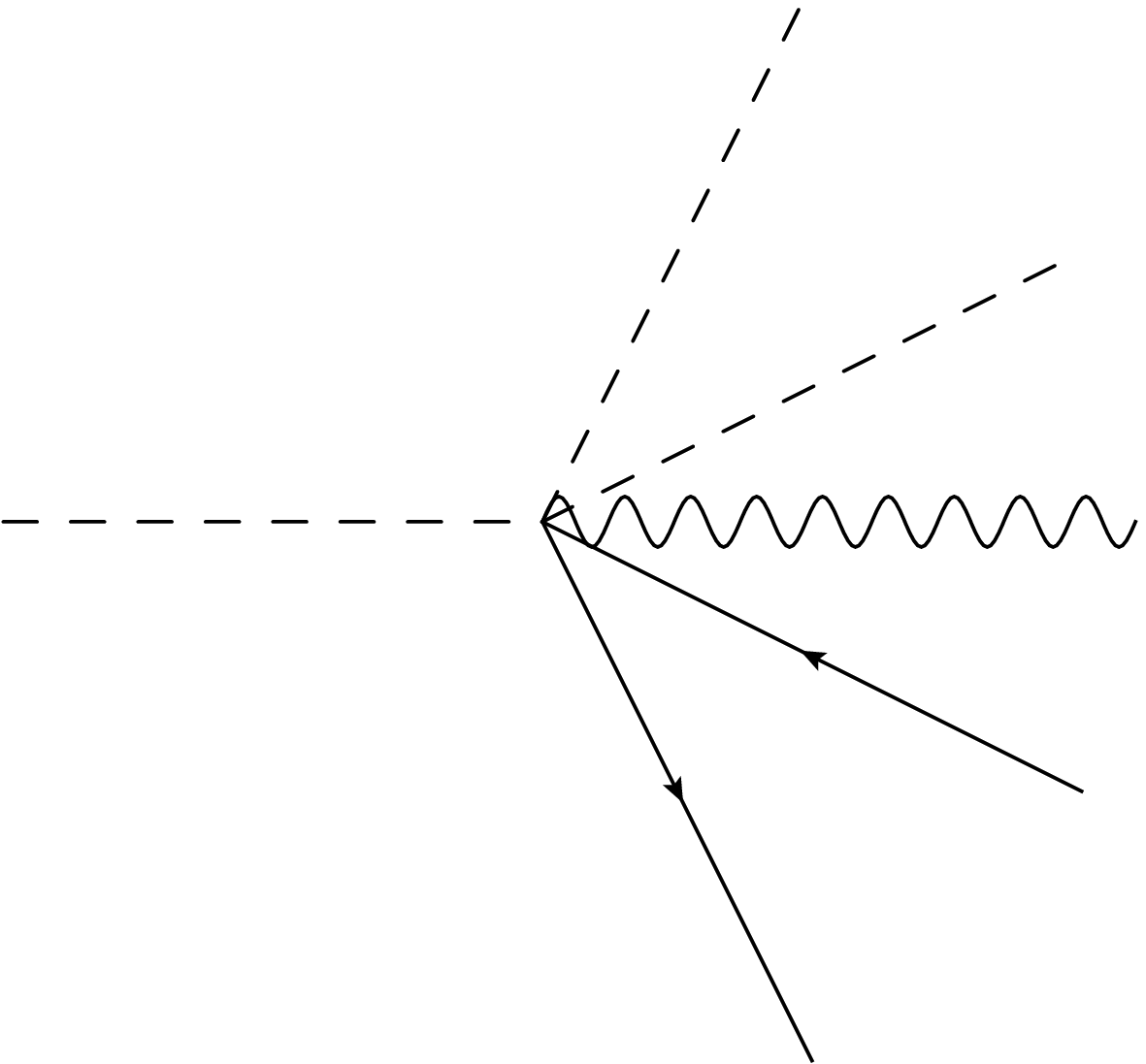}
			\end{pspicture}
			}
		\caption{}
		\label{img:Kl4g4}
	\end{subfigure}
	\begin{subfigure}[b]{0.19\textwidth}
		\centering
		\scalebox{0.75}{
			\begin{pspicture}(0,0)(4,3)
				\includegraphics[width=3.5cm]{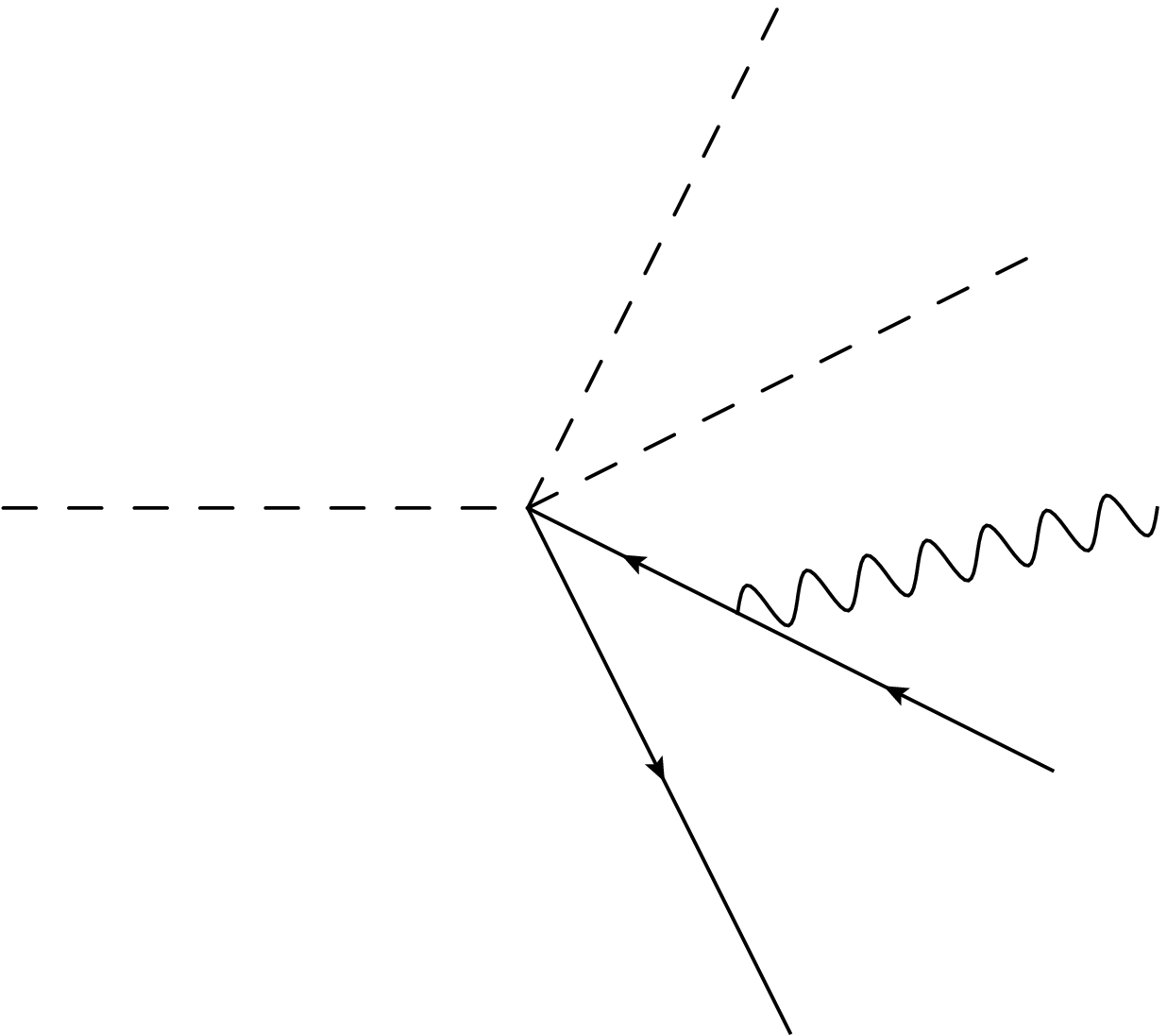}
			\end{pspicture}
			}
		\caption{}
		\label{img:Kl4g5}
	\end{subfigure}
	
	\vspace{0.5cm}
	
	\begin{subfigure}[b]{0.19\textwidth}
		\centering
		\scalebox{0.75}{
			\begin{pspicture}(0,0)(4,3)
				\includegraphics[width=3.5cm]{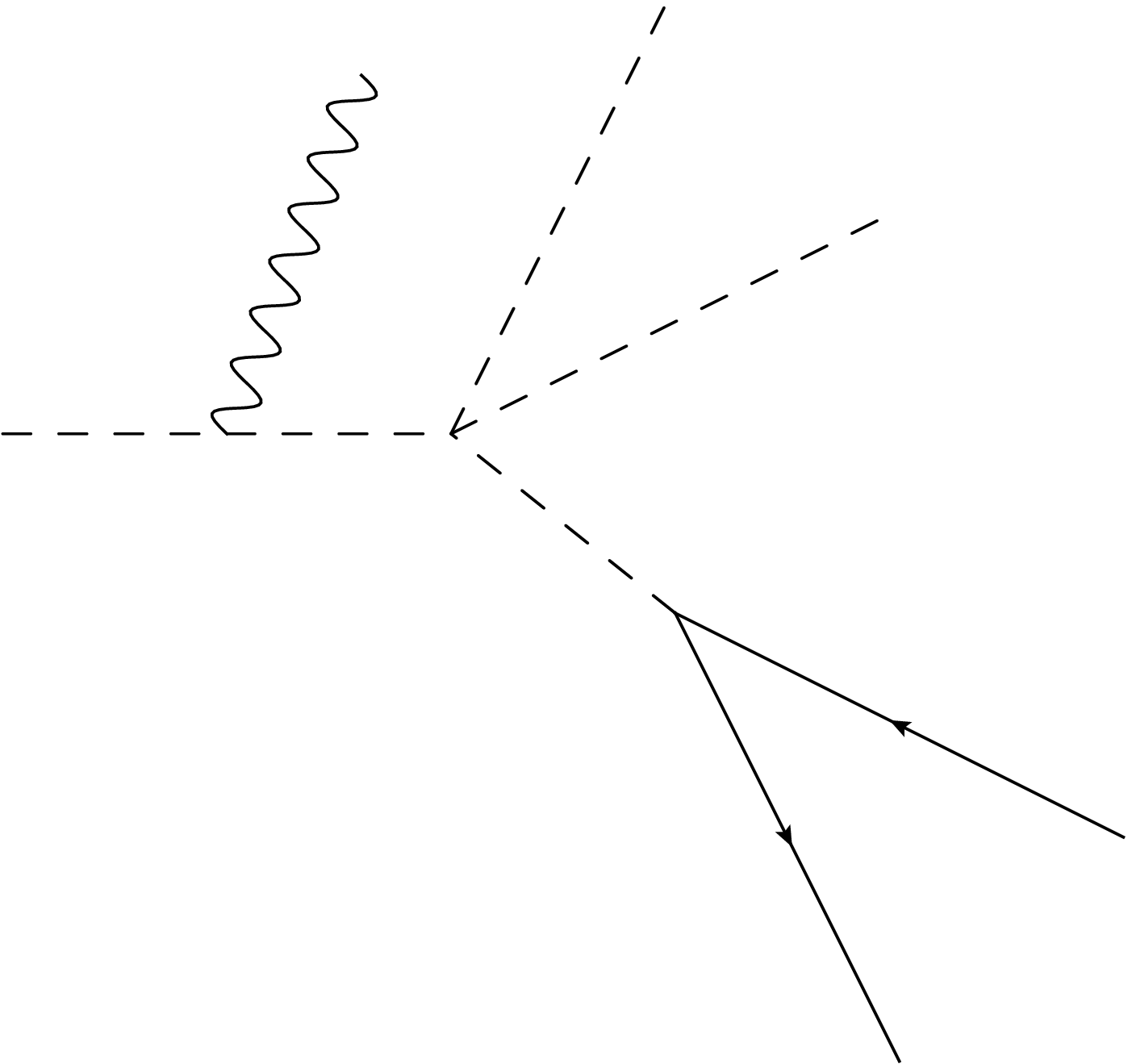}
			\end{pspicture}
			}
		\caption{}
		\label{img:Kl4g6}
	\end{subfigure}
	\begin{subfigure}[b]{0.19\textwidth}
		\centering
		\scalebox{0.75}{
			\begin{pspicture}(0,0)(4,3)
				\includegraphics[width=3.5cm]{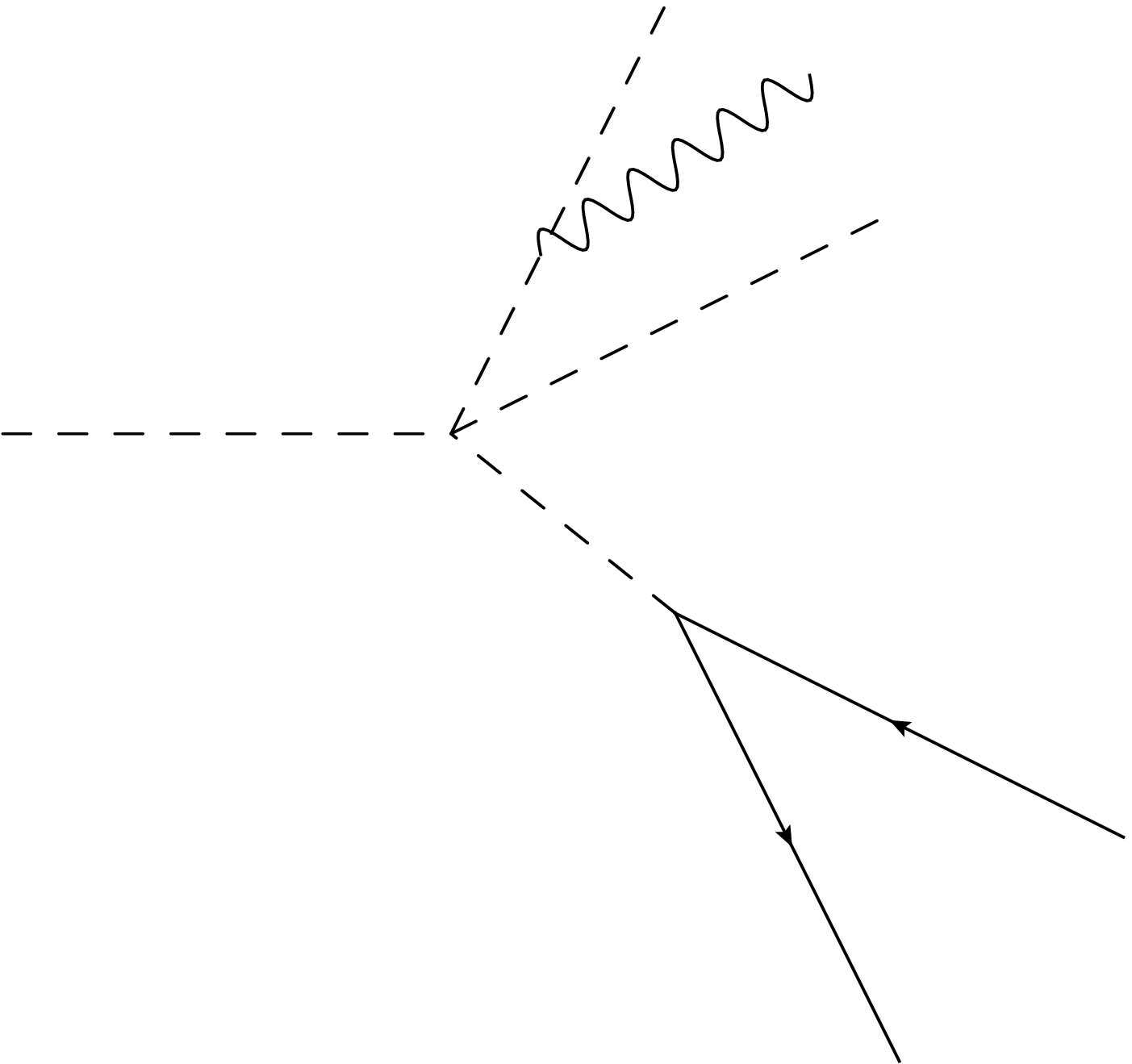}
			\end{pspicture}
			}
		\caption{}
		\label{img:Kl4g7}
	\end{subfigure}
	\begin{subfigure}[b]{0.19\textwidth}
		\centering
		\scalebox{0.75}{
			\begin{pspicture}(0,0)(4,3)
				\includegraphics[width=3.5cm]{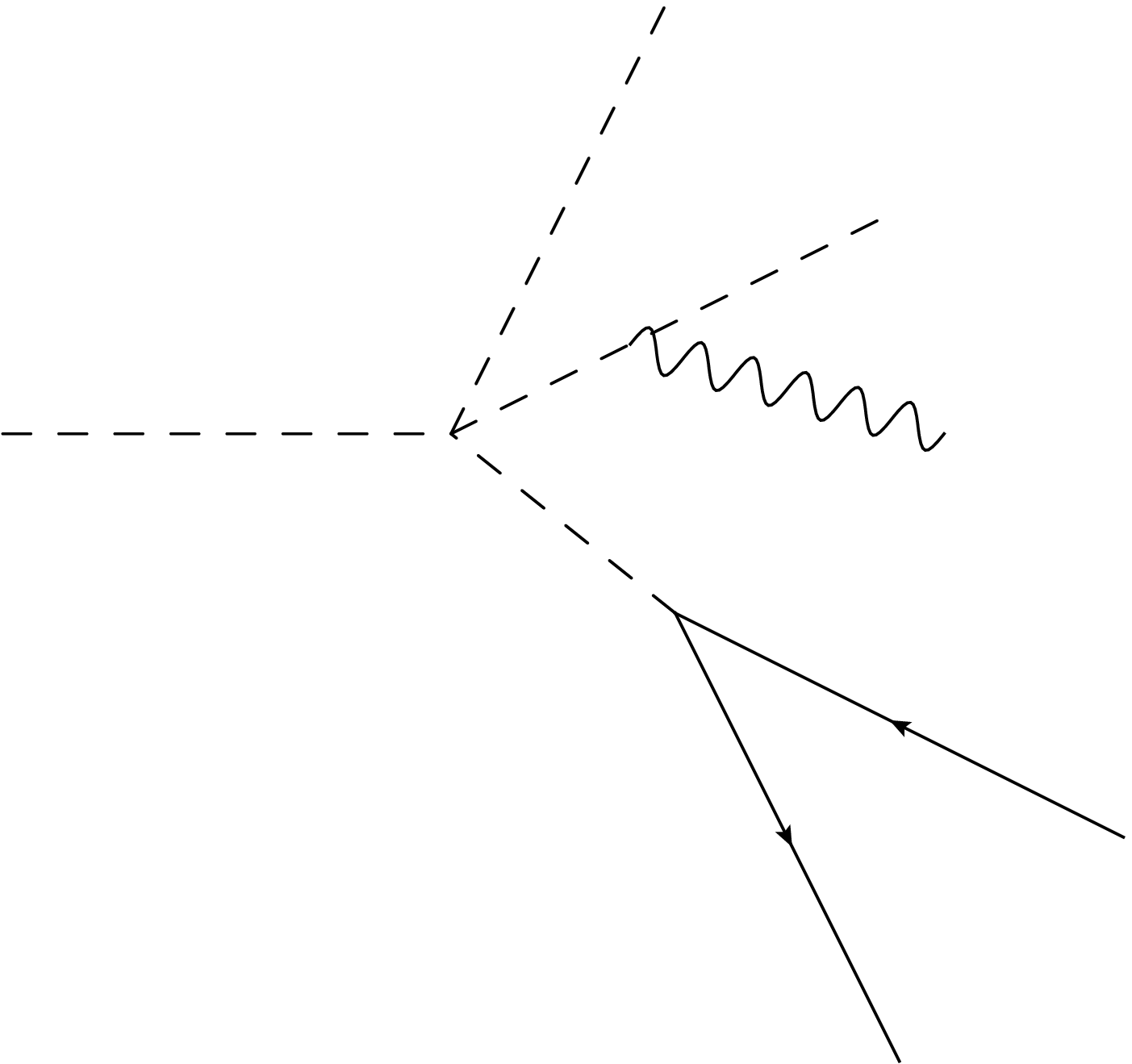}
			\end{pspicture}
			}
		\caption{}
		\label{img:Kl4g8}
	\end{subfigure}
	
	\vspace{0.5cm}
	
	\begin{subfigure}[b]{0.19\textwidth}
		\centering
		\scalebox{0.75}{
			\begin{pspicture}(0,0)(4,3)
				\includegraphics[width=3.5cm]{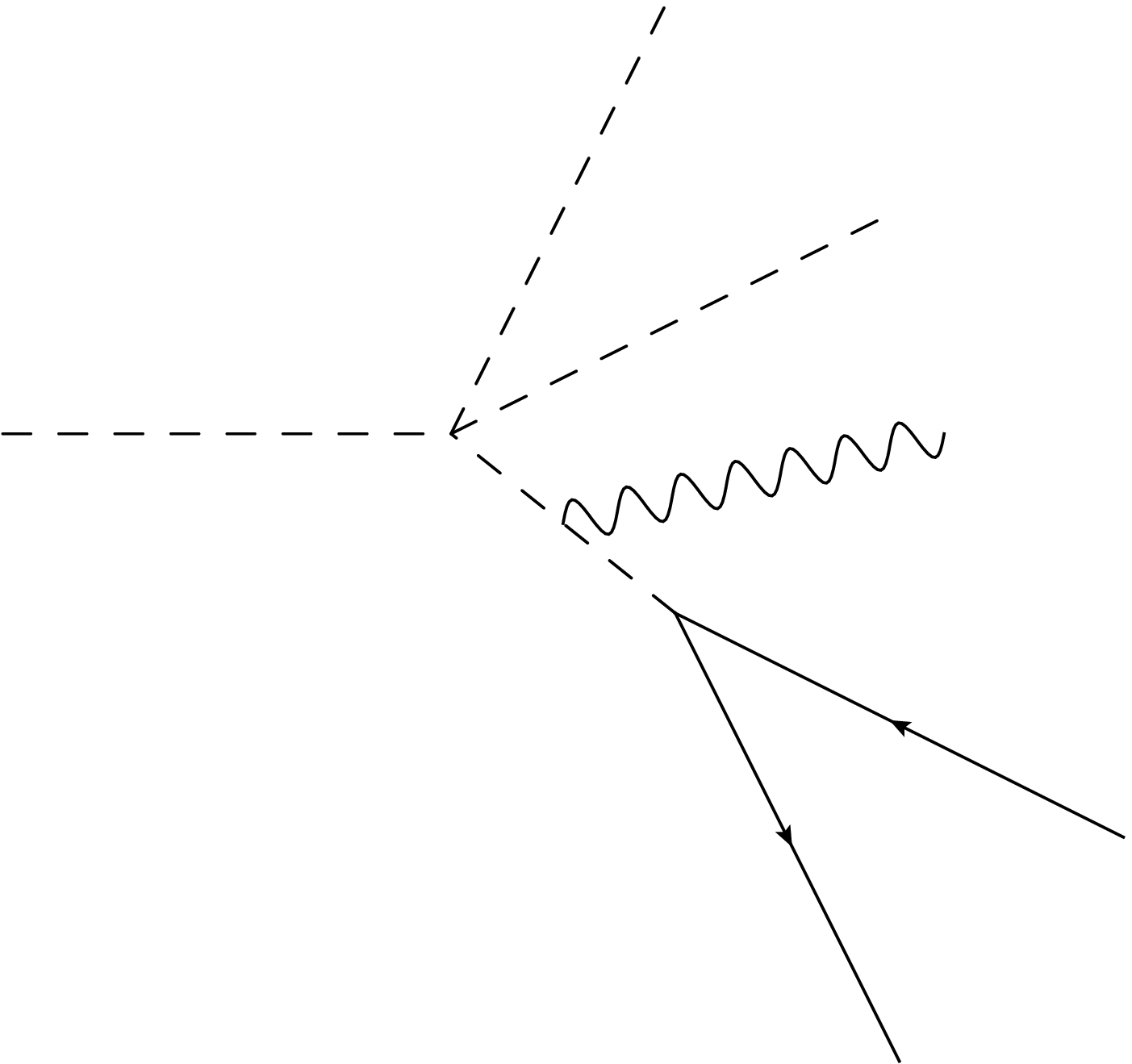}
			\end{pspicture}
			}
		\caption{}
		\label{img:Kl4g9}
	\end{subfigure}
	\begin{subfigure}[b]{0.19\textwidth}
		\centering
		\scalebox{0.75}{
			\begin{pspicture}(0,0)(4,3)
				\includegraphics[width=3.5cm]{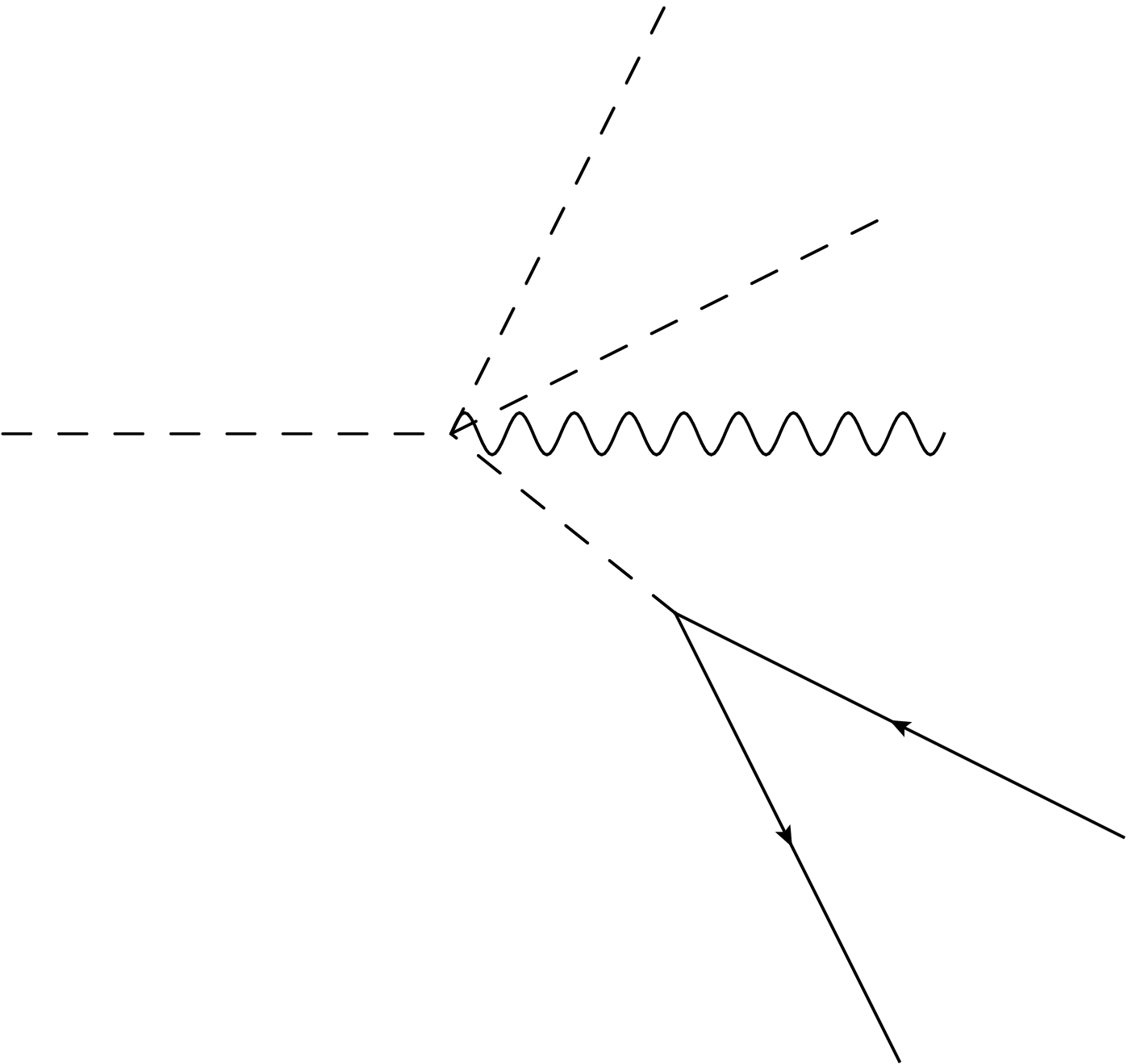}
			\end{pspicture}
			}
		\caption{}
		\label{img:Kl4g10}
	\end{subfigure}
	\begin{subfigure}[b]{0.19\textwidth}
		\centering
		\scalebox{0.75}{
			\begin{pspicture}(0,0)(4,3)
				\includegraphics[width=3.5cm]{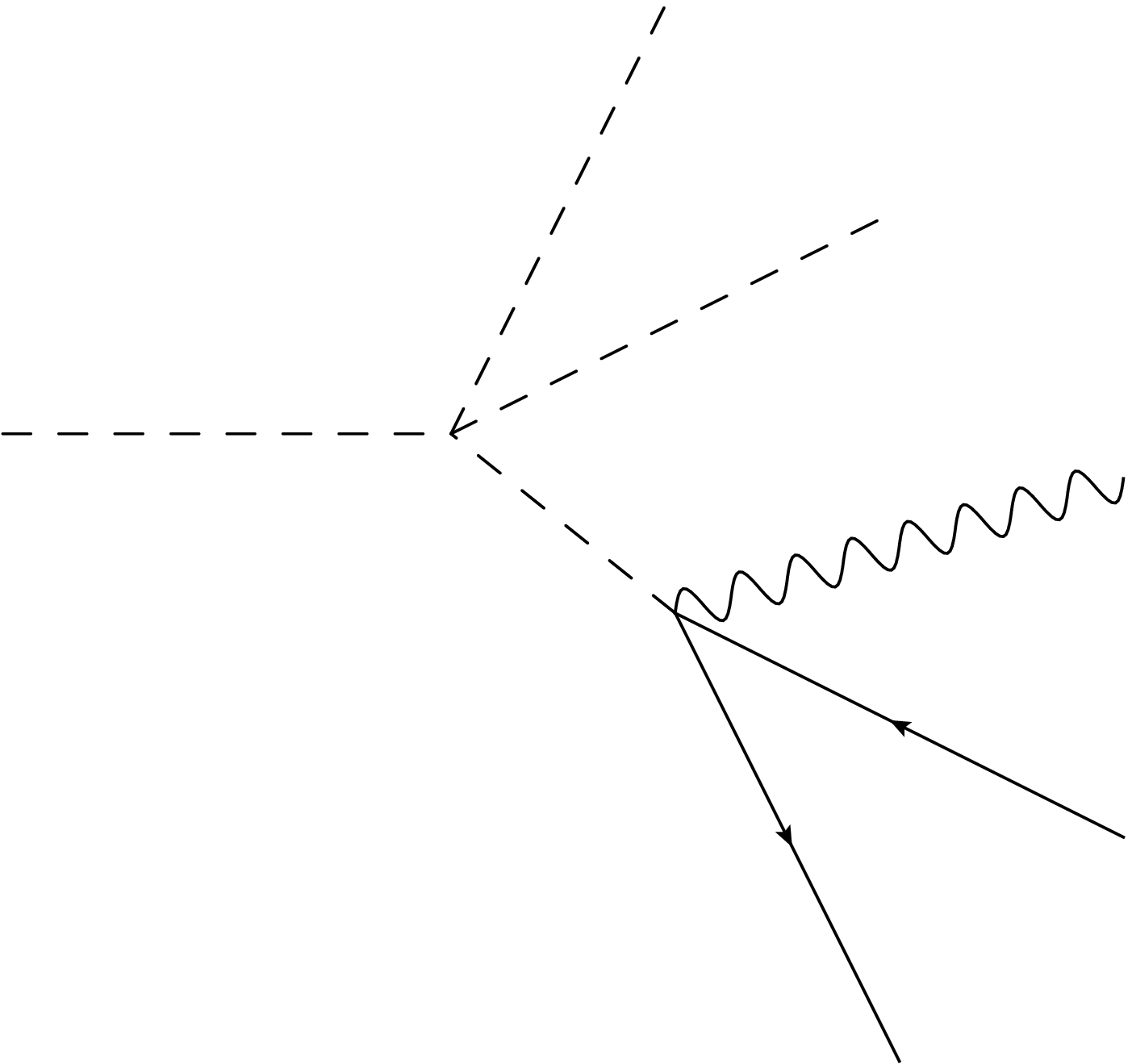}
			\end{pspicture}
			}
		\caption{}
		\label{img:Kl4g11}
	\end{subfigure}
	\begin{subfigure}[b]{0.19\textwidth}
		\centering
		\scalebox{0.75}{
			\begin{pspicture}(0,0)(4,3)
				\includegraphics[width=3.5cm]{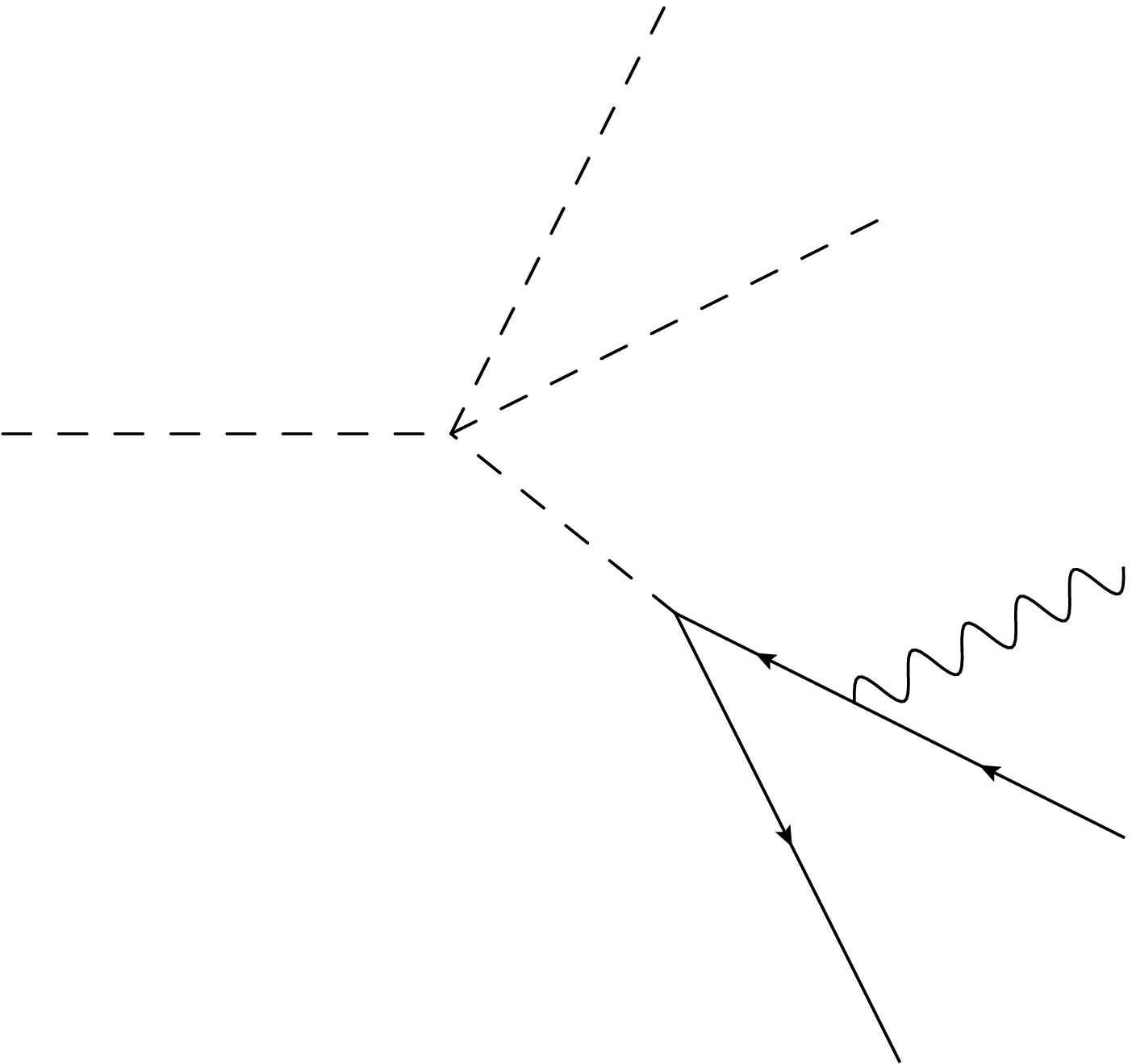}
			\end{pspicture}
			}
		\caption{}
		\label{img:Kl4g12}
	\end{subfigure}
	\caption{Tree-level diagrams for the decay $K_{\ell4\gamma}$.}
	\label{img:Kl4g}
\end{figure}

The relevant diagrams are shown in figure~\ref{img:Kl4g}. If we use the decomposition of the matrix element defined in section~\ref{sec:Kl4gMatrixElement}, the diagrams~\ref{img:Kl4g5} and \ref{img:Kl4g12} just reproduce the second term in (\ref{eqn:Kl4gTMatrix}), where the hadronic part is given by the LO $K_{\ell4}$ form factors in the isospin limit.

The diagrams~\ref{img:Kl4g4} and \ref{img:Kl4g11}, where the photon is emitted off the vertex, correspond to the form factor $\Pi$:
\begin{align}
	\begin{split}
		\Pi =  \frac{\mkp}{2\sqrt{2} F_0} \left( 5 - \frac{ s + \nu }{\mkp^2 - s_\ell} \right) ,
	\end{split}
\end{align}
where $\nu = t-u$.

The form factors $\Pi_{ij}$ correspond to the remaining 8 diagrams, where the photon is emitted off a meson line or a mesonic vertex. The form factors multiplying $\bar u(p_\nu) \slashed P (1 - \gamma^5) v(p_\ell)$ or ${\bar u(p_\nu) \slashed Q (1 - \gamma^5) v(p_\ell)}$ have a simple form:
\begin{align}
	\begin{split}
		\Pi_{00} &= \Pi_{01} = - \frac{\mkp^3}{\sqrt{2} F_0} \left( \frac{2}{\mg^2 - 2 pq} + \frac{1}{\mg^2 + 2 p_1q} - \frac{1}{\mg^2 + 2 p_2q} \right) , \\
		\Pi_{10} &= \Pi_{11} = - \frac{\mkp^3}{\sqrt{2} F_0} \left( \frac{1}{\mg^2 + 2 p_1q} + \frac{1}{\mg^2 + 2 p_2q} \right) , \\
		\Pi_{20} &= \Pi_{21} = - \frac{\mkp^3}{\sqrt{2} F_0} \frac{2}{\mg^2 - 2 pq} .
	\end{split}
\end{align}
The remaining form factors are a bit more complicated. They satisfy the relations
\begin{align}
	\begin{split}
		\Pi_{03} &= - \Pi_{02} - \frac{\mkp^3}{\sqrt{2} F_0} \frac{2}{\mg^2 + 2 p_1q} , \\
		\Pi_{13} &= - \Pi_{12} - \frac{\mkp^3}{\sqrt{2} F_0} \frac{2}{\mg^2 + 2 p_1q} , \\
		\Pi_{23} &= - \Pi_{22} .
	\end{split}
\end{align}
In the limit $\mg\to0$, I find
{\small
\begin{align}
	\begin{split}
		\Pi_{02} &= \frac{\mkp^3}{\sqrt{2} F_0} \frac{1}{\mkp^2 - s_\ell + 2 Lq} \begin{aligned}[t]
				&\Bigg( \frac{\mkp^2 - s - t + u - s_\ell}{4} \left( \frac{2}{pq} - \frac{1}{p_1q} + \frac{1}{p_2q} \right) + \frac{Lq -  Qq}{pq} - \frac{Lq}{p_1q}  - \frac{Qq}{p_2q} \Bigg) ,
				\end{aligned} \\
		\Pi_{12} &= \frac{\mkp^3}{\sqrt{2} F_0} \frac{1}{\mkp^2 - s_\ell + 2 Lq} \begin{aligned}[t]
				&\Bigg( \frac{\mkp^2 - s - t + u - s_\ell}{4} \left( - \frac{1}{p_1q} - \frac{1}{p_2q} \right) - \frac{Lq}{p_1q} + \frac{Qq}{p_2q} + 3 \Bigg) ,
				\end{aligned} \\
		\Pi_{22} &= \frac{\mkp^3}{\sqrt{2} F_0} \frac{1}{\mkp^2 - s_\ell + 2 Lq} \begin{aligned}[t]
				& \Bigg( \frac{\mkp^2 - s - t + u - s_\ell}{2} \left( \frac{1}{pq} + \frac{2}{\mkp^2 - s_\ell} \right) + \frac{Lq - Qq}{pq} + 1 \Bigg) .
				\end{aligned}
	\end{split}
\end{align} }%
These expressions fulfil the relations (\ref{eqn:GaugeInvarianceFFRelations}) as required by gauge invariance.

The contribution of the diagrams~\ref{img:Kl4g6}-\ref{img:Kl4g10} to the decay rate is helicity suppressed by a factor of $\ml^2$. The suppression at leading chiral order also works for the diagrams~\ref{img:Kl4g11} and \ref{img:Kl4g12}. One could therefore omit all diagrams with a kaon pole in the limit $\ml\to0$. However, from a technical point of view, this barely reduces the complexity of the calculation. Hence, I have given here the results for the form factors using the complete set of diagrams. Moreover, at higher chiral order, one has to expect structure dependent contributions not suppressed by $\ml^2$.


\chapter{Extraction of the Isospin Corrections}

\label{sec:ExtractionOfIsospinCorrections}

This section discusses the extraction of the isospin-breaking corrections to the $K_{\ell4}$ form factors and decay rate. While the experiments are performed in our real world, where isospin is broken, it is useful to translate measured quantities into the context of an ideal world with conserved isospin, i.e.~a world with no electromagnetism and identical up- and down-quark masses. The motivation for doing such a transformation is that in an isospin symmetric world, calculations may become much easier. The isospin-breaking corrections for $K_{\ell4}$ will be used in a forthcoming dispersive treatment of this decay \cite{Colangelo2012, Stoffer2013, Colangelo2013}, which is performed in the isospin limit.

Correcting the isospin-breaking effects in existing experimental data on the $K_{\ell4}$ form factors is a delicate matter: the $K_{\ell4}$ form factors are in the real world themselves not observable quantities, because they are not infrared-safe. As explained above, any experiment will measure a semi-inclusive decay rate of $K_{\ell4}$ and $K_{\ell4+n\gamma}$, typically with some cuts on the real photons. These cuts are detector specific and naturally defined in the lab frame. It is almost impossible to handle such cuts in an analytic way. Therefore, one must rely on a Monte Carlo simulation of the semi-inclusive decay that models the detector geometry and all the applied cuts in order to extract isospin corrected quantities. I suggest for future experiments the inclusion of the here presented amplitudes for $K_{\ell4}$ and $K_{\ell4\gamma}$ in a Monte Carlo simulation like PHOTOS \cite{Barberio1994}.

The isospin corrections due to the mass effects can be extracted directly for the form factors. For the photonic effects, I calculate the radiative corrections to the (semi-)inclusive decay rate.

\section{Mass Effects}

I define the isospin-breaking corrections to the form factors as follows.

The measured semi-inclusive differential decay rate $d\Gamma_{(\gamma, \mathrm{cut})}^\mathrm{exp}$ (neglecting experimental uncertainties) equals the result from the presented NLO calculation up to higher order in the chiral expansion or the isospin-breaking parameters:
\begin{align}
	\begin{split}
		d\Gamma_{(\gamma, \mathrm{cut})}^\mathrm{exp} = d\Gamma_{(\gamma, \mathrm{cut})}^\mathrm{NLO} + h.o. &= d\Gamma^\mathrm{NLO}_\mathrm{iso} + \Delta d\Gamma^\mathrm{NLO}_\mathrm{ME} + \Delta d\Gamma^\mathrm{NLO}_{\mathrm{virt.}\gamma} + \int_\mathrm{cut} d\Gamma_{\gamma} \\
			& + \O(p^6, \epsilon \, p^6, Ze^2 p^4, e^2 p^4) + \O(\epsilon^2, \epsilon \, e^2, e^4),
	\end{split}
\end{align}
where the real photon in the radiative decay rate is integrated using the same cuts as in the experiment. I expect the contribution of higher order in the breaking parameters to be negligible, while the $\O(p^6)$ contribution is certainly not. The different terms are of the following order:
\begin{align}
	\begin{split}
		d\Gamma^\mathrm{NLO}_\mathrm{iso} &= \O( p^4 ), \quad \Delta d\Gamma^\mathrm{NLO}_\mathrm{ME} = \O( \epsilon \, p^4, Z e^2 p^2 ) , \\
		\Delta d\Gamma^\mathrm{NLO}_{\mathrm{virt.}\gamma} &= \O( e^2 p^2 ) , \quad \int_\mathrm{cut} d\Gamma_{\gamma} = \O( e^2 p^2 ) .
	\end{split}
\end{align}
The NA48/2 analysis assumes the following isospin-breaking effects:
\begin{align}
	\begin{split}
		d\Gamma^\mathrm{exp}_{(\gamma,\mathrm{cut})} &= d\Gamma^\mathrm{exp} + \Delta d\Gamma_\mathrm{Coulomb} + \Delta d\Gamma_\mathrm{PHOTOS}^\mathrm{cut} .
	\end{split}
\end{align}
If I assume
\begin{align}
	\begin{split}
		\Delta d\Gamma_\mathrm{Coulomb} + \Delta d\Gamma_\mathrm{PHOTOS}^\mathrm{cut} \approx \Delta d\Gamma^\mathrm{NLO}_{\mathrm{virt.}\gamma} + \int_\mathrm{cut} d\Gamma_{\gamma} + \O(e^2 p^4) ,
	\end{split}
\end{align}
(an approximation that I will test later), the form factors given by the experiment contain only the isospin-breaking mass effects (note that $X^\mathrm{LO} = \O(p)$):
\begin{align}
	\begin{split}
		X^\mathrm{exp} &= X^\mathrm{NLO}_\mathrm{ME} + \O(p^5, \epsilon \, p^5, Z e^2 p^3).
	\end{split}
\end{align}
The relative isospin corrections to the form factors due to the mass effects are
\begin{align}
	\begin{split}
		\delta_\mathrm{ME} X &:=  1 - \frac{X_\mathrm{iso}}{X_\mathrm{ME}} =  1 - \frac{X_\mathrm{iso}^\mathrm{NLO}}{X_\mathrm{ME}^\mathrm{NLO}}  + \O(\epsilon \, p^4, Z e^2 p^2) .
	\end{split}
\end{align}
The uncertainty can be naïvely estimated to be $\O(\epsilon \, p^4 , Z e^2 p^2)  \approx 0.2\%$. The mass effects at NNLO in the chiral expansion have been studied in a dispersive treatment \cite{Bernard2013} and found to be small given the present experimental accuracy.

The definition of the isospin limit is a convention. I choose here
\begin{align}
	\begin{split}
		X^\mathrm{NLO}_\mathrm{iso} := \lim_{\substack{\epsilon\to0, \\ e^2\to0}} \; \lim_{\substack{\mpio\to\mpip^\mathrm{exp}, \\ \mko\to\mkp^\mathrm{exp}}} X^\mathrm{NLO}_\mathrm{ME} .
	\end{split}
\end{align}

\section{Photonic Effects}

In this section, I calculate the (semi-)inclusive decay rate for $K_{\ell4(\gamma)}$. This will allow on the one hand for a more precise treatment of photonic corrections in future experiments (compared to previous treatments that do not make use of the matrix elements). On the other hand, I will be able to study the approximation
\begin{align}
	\begin{split}
		\Delta d\Gamma_\mathrm{Coulomb} + \Delta d\Gamma_\mathrm{PHOTOS}^\mathrm{cut} \approx \Delta d\Gamma^\mathrm{NLO}_{\mathrm{virt.}\gamma} + \int_\mathrm{cut} d\Gamma_{\gamma} + \O(e^2 p^4) ,
	\end{split}
\end{align}
although not for the experimental cuts, but for a simplified cut that can be handled analytically.

\subsection{Strategy for the Phase Space Integration}

I have introduced a finite photon mass as a regulator and will eventually send this regulator to zero (in the inclusive decay rate). We are not interested in the full dependence of the decay rate on the photon mass, but only in terms that do not vanish in the limit $m_\gamma\to0$, i.e.~in the IR-singular and finite pieces.

I use this fact to simplify the calculation as follows. I split the phase space of the radiative decay into a soft photon and a hard photon region. In the soft region, I use the soft photon approximation (SPA) to simplify the amplitude. This region will produce the IR singularity, which cancels against the divergence in the virtual corrections. The hard region gives an IR-finite result. Here, the limit $m_\gamma\to0$ can be taken immediately. The dependence on the photon energy cut $\Delta\varepsilon$ that separates the soft from the hard region must cancel in the sum of the two contributions. The hard region either covers the whole hard photon phase space, or alternatively, an additional cut on the maximum photon energy in the dilepton-photon system can be introduced rather easily.

\subsection{Soft Region}

Let us calculate the soft photon amplitude. In the real emission amplitude
\begin{align}
	\begin{split}
		\mathcal{T}_\gamma &= - \frac{G_F}{\sqrt{2}} e V_{us}^* \epsilon_\mu(q)^* \bigg[ \mathcal{H}^{\mu\nu} \; \mathcal{L}_\nu +  \mathcal{H}_\nu \; \mathcal{\tilde L}^{\mu\nu} \bigg]
	\end{split}
\end{align}
where
\begin{align}
	\begin{split}
		\mathcal{L}_\nu &:=  \bar u(p_\nu) \gamma_\nu (1-\gamma^5)v(p_\ell) , \\
		\mathcal{\tilde L}^{\mu\nu} &:= \frac{1}{2 p_\ell q} \bar u(p_\nu) \gamma^\nu (1-\gamma^5)(m_\ell - \slashed p_\ell - \slashed q) \gamma^\mu v(p_\ell) , \\
		\mathcal{H}_\nu &:= \frac{i}{\mkp} \left( P_\nu F + Q_\nu G + L_\nu R \right) , \\
		\mathcal{H}^{\mu\nu} &:= \frac{i}{\mkp} g^{\mu\nu} \Pi + \frac{i}{\mkp^2}\left( P^\mu \Pi_0^\nu + Q^\mu \Pi_1^\nu + L^\mu \Pi_2^\nu \right) , \\
		\Pi_i^\nu &:= \frac{1}{\mkp} \left( P^\nu \Pi_{i0} + Q^\nu \Pi_{i1} + L^\nu \Pi_{i2} + q^\nu \Pi_{i3}  \right) ,
	\end{split}
\end{align}
I neglect according to the SPA terms with a $q$ in the numerator, i.e.~the $\slashed q$ in $\mathcal{\tilde L}^{\mu\nu}$ and the $q^\nu$ in $\Pi_i^\nu$. If I insert the tree-level expressions for the form factors and consistently keep only terms that diverge as $q^{-1}$, I find that the soft photon amplitude factorises as
\begin{align}
	\label{eqn:SoftPhotonFactorisation}
	\begin{split}
		\mathcal{T}_\gamma^\mathrm{soft} &= e \mathcal{T}^\mathrm{LO}_\mathrm{iso} \left( - \frac{p \epsilon^*(q)}{p q} + \frac{p_\ell \epsilon^*(q)}{p_\ell q} + \frac{p_1 \epsilon^*(q)}{p_1 q} - \frac{p_2 \epsilon^*(q)}{p_2 q} \right) ,
	\end{split}
\end{align}
where $\mathcal{T}^\mathrm{LO}_\mathrm{iso}$ is the tree-level $K_{\ell4}$ matrix element in the isospin limit.

In the SPA, also the photon momentum in the delta function of the phase space measure is neglected. This means that we can essentially use $K_{\ell4}$ kinematics to describe the other momenta:
\begin{align}
	\label{eqn:DecayRateSoftRegion}
	\begin{split}
		d\Gamma_\gamma^\mathrm{soft} &= \frac{1}{2\mkp} \widetilde{dp_1} \widetilde{dp_2} \widetilde{dp_\ell} \widetilde{dp_\nu} \widetilde{dq_{\;}} \delta^{(4)}(p - p_1 - p_2 - p_\ell - p_\nu) \sum_{\substack{\mathrm{spins},\\ \mathrm{polar.}}} \left| \mathcal{T}_\gamma^\mathrm{soft} \right|^2 \\
			&= e^2 d\Gamma_\mathrm{iso}^\mathrm{LO} \int\limits_{|\vec q| \le \Delta \varepsilon} \widetilde{dq_{\;}} \sum_\mathrm{polar.} \left| - \frac{p \epsilon^*(q)}{p q} + \frac{p_\ell \epsilon^*(q)}{p_\ell q} + \frac{p_1 \epsilon^*(q)}{p_1 q} - \frac{p_2 \epsilon^*(q)}{p_2 q} \right|^2 \\
			&= - e^2 d\Gamma_\mathrm{iso}^\mathrm{LO} \int\limits_{|\vec q| \le \Delta \varepsilon} \widetilde{dq_{\;}} \begin{aligned}[t]
				&\bigg[ \frac{\mkp^2}{(pq)^2} + \frac{\ml^2}{(p_\ell q)^2} + \frac{\mpip^2}{(p_1 q)^2} + \frac{\mpip^2}{(p_2 q)^2} \\
				 & - \frac{2 p p_\ell}{(pq)(p_\ell q)} - \frac{2 p p_1}{(pq)(p_1 q)} + \frac{2 p p_2}{(pq)(p_2 q)} \\
				 & + \frac{2 p_1 p_\ell}{(p_1 q)(p_\ell q)} - \frac{2 p_2 p_\ell}{(p_2 q)(p_\ell q)} - \frac{2 p_1 p_2}{(p_1 q)(p_2 q)} \bigg] , \end{aligned}
	\end{split}
\end{align}
where I use the abbreviation
\begin{align}
	\begin{split}
		\widetilde{dk} := \frac{d^3 k}{(2\pi)^3 2k^0} .
	\end{split}
\end{align}
These are standard bremsstrahlung integrals, which have been computed in \cite{Hooft1979} (see also \cite{Itzykson1980}). They amount to
\begin{align}
	\begin{split}
		I_1(k) :={}& \int\limits_{|\vec q| \le \Delta\varepsilon} \widetilde{dq_{\;}} \frac{1}{(k q)^2} = \frac{1}{8\pi^2} \frac{1}{k^2} \Bigg[ 2 \ln\left( \frac{2 \Delta\varepsilon}{\mg} \right) - \frac{k^0}{|\vec k|} \ln\left( \frac{k^0 + |\vec k|}{k^0 - |\vec k|} \right) \Bigg] + \O(\mg^2) .
	\end{split}
\end{align}
The integrals with two different momenta are more complicated:
{\small
\begin{align}
	\begin{split}
		I_2(k_1,k_2) :={}& \int\limits_{|\vec q| \le \Delta\varepsilon} \widetilde{dq_{\;}} \frac{1}{(k_1 q)(k_2 q)} = \frac{\alpha}{8\pi^2} \left[ \frac{2}{k_1^2 - k^2} \ln\left( \frac{k_1^2}{k^2} \right) \ln\left( \frac{2\Delta\varepsilon}{\mg} \right) + \tilde I_2(k_1,k_2) \right] + \O(\mg^2) , \\
		\tilde I_2(k_1,k_2) ={}& \frac{1}{k_1^0 - k^0} \frac{1}{v} \begin{aligned}[t]
			& \Bigg[ \frac{1}{4} \ln^2\left( \frac{u^0 - |\vec u|}{u^0 + |\vec u|} \right) + \dilog\left( \frac{v-u^0 + |\vec u|}{v} \right) + \dilog\left( \frac{v-u^0-|\vec u|}{v} \right) \Bigg] \Bigg|_{u=k}^{u=k_1} , \end{aligned}
	\end{split}
\end{align} }%
where $k = \alpha k_2$ and $\alpha$ is the solution of $(k_1 - \alpha k_2)^2=0$ such that $\alpha k_2^0 - k_1^0$ has the same sign as $k_1^0$. Further, $v$ is defined as
\begin{align}
	\begin{split}
		v := \frac{k_1^2 - k^2}{2(k_1^0 - k^0)} .
	\end{split}
\end{align}

I find it most convenient to evaluate the soft photon contribution in the rest frame of the two leptons and the photon, $\Sigma_{\ell\nu\gamma}$. The particle momenta in this frame are given by
\begin{align}
	\begin{split}
		\begin{aligned}[t]
			p^0 &= \frac{\mkp^2 - s + s_\ell}{2\sqrt{s_\ell}} , \quad & |\vec p| &= \frac{\lambda_{K\ell}^{1/2}(s)}{2 \sqrt{s_\ell}} , \\
			p_\ell^0 &= \frac{\sqrt{s_\ell}}{2}(1+\zl) , \quad & |\vec p_\ell| &= \frac{\sqrt{s_\ell}}{2}(1-\zl) , \\
			p_1^0 &= \frac{PL + \sigma_\pi X \cos\theta_\pi}{2\sqrt{s_\ell}} , \quad & |\vec p_1| &= \sqrt{(p_1^0)^2 - \mpip^2} , \\
			p_2^0 &= \frac{PL - \sigma_\pi X \cos\theta_\pi}{2\sqrt{s_\ell}} , \quad & |\vec p_2| &= \sqrt{(p_2^0)^2 - \mpip^2} .
		\end{aligned}
	\end{split}
\end{align}

The bremsstrahlung integrals become
\begin{align}
	\begin{split}
		I_1(p) &= \frac{1}{8\pi^2} \frac{1}{\mkp^2} \Bigg[ 2 \ln\left( \frac{2 \Delta\varepsilon}{\mg} \right) - \frac{\mkp^2-s+s_\ell}{\lambda_{K\ell}^{1/2}(s)} \ln\left( \frac{\mkp^2 - s + s_\ell + \lambda_{K\ell}^{1/2}(s)}{\mkp^2 - s + s_\ell - \lambda_{K\ell}^{1/2}(s)} \right) \Bigg] , \\
		I_1(p_\ell) &= \frac{1}{8\pi^2} \frac{1}{\ml^2} \Bigg[ 2 \ln\left( \frac{2 \Delta\varepsilon}{\mg} \right) + \frac{1+\zl}{1-\zl} \ln(\zl) \Bigg] , \\
		I_1(p_1) &= \frac{1}{8\pi^2} \frac{1}{\mpip^2} \Bigg[ 2 \ln\left( \frac{2 \Delta\varepsilon}{\mg} \right) - \frac{p_1^0}{|\vec p_1|} \ln\left( \frac{p_1^0 + |\vec p_1|}{p_1^0 - |\vec p_1|} \right) \Bigg], \\
		I_1(p_2) &= \frac{1}{8\pi^2} \frac{1}{\mpip^2} \Bigg[ 2 \ln\left( \frac{2 \Delta\varepsilon}{\mg} \right) - \frac{p_2^0}{|\vec p_2|} \ln\left( \frac{p_2^0 + |\vec p_2|}{p_2^0 - |\vec p_2|} \right) \Bigg] .
	\end{split}
\end{align}

The evaluation of the integrals with two momenta is straightforward but a bit tedious. I give here the respective values of $\alpha(k_1,k_2)$:
{\small
\begin{align}
	\begin{alignedat}{2}
		\alpha(p, p_\ell) &= \frac{\lambda^{1/2}(\tl,\mkp^2,\ml^2) + \mkp^2 + \ml^2 - \tl}{2 \ml^2}, \quad & \alpha(p, p_1) &= \frac{\lambda_{K\pi}^{1/2}(t) + \mkp^2 + \mpip^2 - t}{2\mpip^2} , \\
		\alpha(p_1, p_\ell) &= \frac{\lambda^{1/2}(s_{1\ell},\mpip^2, \ml^2) -\ml^2 - \mpip^2 + s_{1\ell}}{2\ml^2} , \quad & \alpha(p, p_2) &= \frac{\lambda_{K\pi}^{1/2}(u) + \mkp^2 + \mpip^2 - u}{2\mpip^2} , \\
		\alpha(p_2, p_\ell) &= \frac{\lambda^{1/2}(s_{2\ell},\mpip^2, \ml^2) -\ml^2 - \mpip^2 + s_{2\ell}}{2\ml^2} , \quad & \alpha(p_1, p_2) &= \frac{s \sigma_\pi + s - 2 \mpip^2}{2\mpip^2} .
	\end{alignedat}
\end{align} }%

\pagebreak

\subsection{Hard Region}

The hard region is defined as the phase space region where $|\vec q| > \Delta \varepsilon$, i.e.
\begin{align}
	\begin{split}
		x > x_\mathrm{min} = \frac{2\Delta\varepsilon}{\sqrt{s_\ell}}  =: \tilde x_\mathrm{min} (1-\zl) ,
	\end{split}
\end{align}
where the variable $\tilde x_\mathrm{min}$ is independent of $s_\ell$.

Here, the full $K_{\ell4\gamma}$ kinematics has to be applied. However, as the hard region does not produce any IR singularity, the limit $\mg\to0$ can be taken at the very beginning.

In the appendix~\ref{sec:RadiativeDecayRate}, I have derived the expression for the decay rate
\begin{align}
	\begin{split}
		d\Gamma_\gamma^\mathrm{hard} &= G_F^2 |V_{us}|^2 e^2 \frac{s_\ell \, \sigma_\pi(s) X}{2^{20}\pi^9 \mkp^7} J_8 \, ds \, ds_\ell \, d\cos\theta_\pi \, d\cos\theta_\gamma \, d\phi \, dx \, dy \, d\phi_\ell ,
	\end{split}
\end{align}
where
\begin{align}
	\begin{split}
		J_8 &= \mkp^4 \sum_\mathrm{polar.} \epsilon_\mu(q)^* \epsilon_\rho(q)
			\begin{aligned}[t]
				&\Bigg[ \mathcal{H}_\nu \mathcal{H}^*_\sigma \sum_\mathrm{spins} \mathcal{\tilde L}^{\mu\nu} \mathcal{\tilde L}^{*\rho\sigma} + \mathcal{H}^{\mu\nu} \mathcal{H}^{*\rho\sigma} \sum_\mathrm{spins} \mathcal{L}_\nu \mathcal{L}^*_\sigma \\
				& + 2 \Re\bigg( \mathcal{H}^{\mu\nu} \mathcal{H}^{*\sigma} \sum_\mathrm{spins} \mathcal{L}_\nu \mathcal{\tilde L}^{*\rho}{}_{\sigma} \bigg) \Bigg] .
			\end{aligned}
	\end{split}
\end{align}
Since the form factors only depend on the first six phase space variables, the integrals over $y$ and $\phi_\ell$ can be performed without knowledge of the dynamics. The $K_{\ell4}$ form factors and the form factor $\Pi$ depend on $s$, $s_\ell$ and $\cos\theta_\pi$ only (at the order we consider). I therefore split the hadronic tensor into two pieces
\begin{align}
	\begin{split}
		\mathcal{H}^{\mu\nu} = \frac{i}{\mkp} g^{\mu\nu} \Pi + \frac{i}{\mkp^2} \mathcal{\tilde H}^{\mu\nu} , \quad \mathcal{\tilde H}^{\mu\nu} = P^\mu \Pi_0^\nu + Q^\mu \Pi_1^\nu + L^\mu \Pi_2^\nu
	\end{split}
\end{align}
and write $J_8$ as follows:
\begin{align}
	\begin{split}
		J_8 &= J_8^{\ell\ell} + J_8^{hh} + J_8^\mathrm{int} , \\
		J_8^{\ell\ell} &= \mkp^4 \sum_\mathrm{polar.} \epsilon_\mu(q)^* \epsilon_\rho(q) 
			\begin{aligned}[t]
				& \Bigg[ \mathcal{H}_\nu \mathcal{H}^*_\sigma \sum_\mathrm{spins} \mathcal{\tilde L}^{\mu\nu} \mathcal{\tilde L}^{*\rho\sigma} + \frac{1}{\mkp^2} g^{\mu\nu} g^{\rho\sigma} | \Pi |^2 \sum_\mathrm{spins} \mathcal{L}_\nu \mathcal{L}^*_\sigma \\
				&+ \frac{i}{\mkp} \bigg( g^{\mu\nu} \Pi \, \mathcal{H}^{*\sigma} \sum_\mathrm{spins} \mathcal{L}_\nu \mathcal{\tilde L}^{*\rho}{}_{\sigma} - g^{\mu\nu} \Pi^* \, \mathcal{H}^{\sigma} \sum_\mathrm{spins} \mathcal{L}_\nu^* \mathcal{\tilde L}^{\rho}{}_{\sigma} \bigg) \Bigg] ,
			\end{aligned} \\
		J_8^{hh} &= \sum_\mathrm{polar.} \epsilon_\mu(q)^* \epsilon_\rho(q) \Bigg[ \mathcal{\tilde H}^{\mu\nu} \mathcal{\tilde H}^{*\rho\sigma} \sum_\mathrm{spins} \mathcal{L}_\nu \mathcal{L}^*_\sigma \Bigg] , \\
		J_8^\mathrm{int} &= \mkp^2 \sum_\mathrm{polar.} \epsilon_\mu(q)^* \epsilon_\rho(q)
			\begin{aligned}[t]
				&\Bigg[ \frac{1}{\mkp} \left( g^{\mu\nu} \mathcal{\tilde H}^{*\rho\sigma} \Pi + \mathcal{\tilde H}^{\mu\nu} g^{\rho\sigma} \Pi^*\right) \sum_\mathrm{spins} \mathcal{L}_\nu \mathcal{L}^*_\sigma \\
				&+ i \bigg( \mathcal{\tilde H}^{\mu\nu} \mathcal{H}^{*\sigma} \sum_\mathrm{spins} \mathcal{L}_\nu \mathcal{\tilde L}^{*\rho}{}_{\sigma} - \mathcal{\tilde H}^{*\mu\nu} \mathcal{H}^{\sigma} \sum_\mathrm{spins} \mathcal{L}_\nu^* \mathcal{\tilde L}^{\rho}{}_{\sigma} \bigg) \Bigg] .
			\end{aligned}
	\end{split}
\end{align}
The first term, $J_8^{\ell\ell}$, denotes the absolute square of the contributions where the photon is attached to the lepton line (either the external line or the vertex). Here, the hadronic part is described by the $K_{\ell4}$ form factors and $\Pi$. I can therefore integrate directly over the five phase space variables $\cos\theta_\gamma$, $\phi$, $x$, $y$ and $\phi_\ell$.

The second term, $J_8^{hh}$, is the absolute square of the contributions with the photon emitted off the hadrons. The form factors $\Pi_{ij}$ describe here the hadronic part. As they depend on six phase space variables, I perform first the integral over $\phi_\ell$ and $y$, then insert the explicit tree-level expressions for the form factors $\Pi_{ij}$, given in section~\ref{sec:MatrixElementRealPhotonEmission}. I further integrate the decay rate and keep it differential only with respect to $s$, $s_\ell$ and $\cos\theta_\pi$. The same strategy applies to the third term, $J_8^\mathrm{int}$, the interference of off-lepton and off-hadron emission.

It is important to note that for a vanishing lepton mass $\ml$, the phase space integrals containing $\mathcal{H}^\mu$ produce a singularity for collinear photons. The lepton mass plays the role of a natural cut-off for this collinear divergence, which emerges as a $\ln\ml^2$ mass singularity. In those integrals, the limit $\ml\to0$ must not be taken before the integration.

Let us now consider the three parts separately.

I perform the five phase space integrals in the $\ell\ell$-part and apply an expansion for small values of $\tilde x_\mathrm{min}$, keeping only the logarithmic term. Only after the integration, it is safe to expand the result for small values of $\ml$:
\begin{align}
	\begin{split}
		\frac{d\Gamma_\gamma^{\mathrm{hard},\ell\ell}}{ds ds_\ell d\cos\theta_\pi} &= e^2 G_F^2 |V_{us}|^2 \frac{\sigma_\pi(s) X}{9 \cdot 2^{15} \, \pi^7 \mkp^5} \begin{aligned}[t]
			& \bigg( 2 \left( |F_1|^2 + \sin^2\theta_\pi |F_2|^2 \right) \left( 12 \ln \tilde x_\mathrm{min} - 3 \ln\zl + 5 \right) \\
			& + 3 \left| F_4 + s_\ell \Pi \right|^2 \bigg) + \O(\zl \ln \zl) . \end{aligned}
	\end{split}
\end{align}
The soft photon contribution corresponding to the square of the off-lepton emission amplitude is given by $I_1(p_\ell)$. In the sum of the soft and the hard photon emission, the dependence on $\Delta\varepsilon$ drops out:
\begin{align}
	\label{eqn:RadiativeDecayRateLL}
	\begin{split}
		\frac{d\Gamma_\gamma^{\ell\ell}}{ds ds_\ell d\cos\theta_\pi} &= e^2 G_F^2 |V_{us}|^2 \frac{\sigma_\pi(s) X}{9 \cdot 2^{15} \, \pi^7 \mkp^5} \begin{aligned}[t]
			& \bigg( 2 \left( |F_1|^2 + \sin^2\theta_\pi |F_2|^2 \right) \left( 5 + 6 \ln \zg  - 9 \ln\zl \right) \\
			& + 3 \left|F_4+s_\ell \Pi \right|^2  \bigg) + \O(\zl \ln \zl) . \end{aligned}
	\end{split}
\end{align}
I can introduce an additional cut on the photon energy in $\Sigma_{\ell\nu\gamma}$ by integrating $x$ only over a part of the hard region:
\begin{align}
	\begin{split}
		\tilde x_\mathrm{min} (1-\zl) < x < \tilde x_\mathrm{max}(1 - \zl) .
	\end{split}
\end{align}
Instead of (\ref{eqn:RadiativeDecayRateLL}), I find then
{\small
\begin{align}
	\label{eqn:RadiativeDecayRateLLCut}
	\begin{split}
		\frac{d\Gamma_{\gamma,\mathrm{cut}}^{\ell\ell}}{ds ds_\ell d\cos\theta_\pi} &= e^2 G_F^2 |V_{us}|^2 \frac{\sigma_\pi(s) X}{9 \cdot 2^{15} \, \pi^7 \mkp^5} \begin{aligned}[t]
			& \bigg( 2  (|F_1|^2 + \sin^2\theta_\pi |F_2|^2) \\
			& \cdot \begin{aligned}[t] & \Big( \tilde x_\mathrm{max} (9 - \tilde x_\mathrm{max}(3 + \tilde x_\mathrm{max}))  + 6 \ln\zg  - 3 (2 + \tilde x_\mathrm{max}^2) \ln\zl \\
				& - 3 (1 - \tilde x_\mathrm{max}^2) \ln(1 - \tilde x_\mathrm{max}) - 12 \ln(\tilde x_\mathrm{max}) \Big) \end{aligned} \\
			& + 3 \tilde x_\mathrm{max}^2 (3 - 2 \tilde x_\mathrm{max}) |F_4 + s_\ell \Pi|^2
  \bigg) + \O(\zl \ln \zl) . \end{aligned}
	\end{split}
\end{align} }%

The integration of the $hh$-part is more involved. I perform the integrals over $\phi_\ell$ and $y$ analytically, insert the explicit form factors $\Pi_{ij}$ and integrate over $x$ analytically, too (either with or without the energy cut $\tilde x_\mathrm{max}$). Although, with some effort, the integrals over $\phi$ and $\cos\theta_\gamma$ could be performed analytically, I choose to integrate these two angles numerically: since they only describe the orientation of the dilepton-photon three-body system with respect to the pions, these two integrals contain nothing delicate. The dependence on the cuts $\tilde x_\mathrm{min}$ and $\tilde x_\mathrm{max}$ is manifest after the integration over $x$ and collinear singularities cannot show up in the remaining integrals. I therefore write the $hh$-part as
{\small
\begin{align}
	\begin{split}
		\frac{d\Gamma_{\gamma,\mathrm{cut}}^{\mathrm{hard},hh}}{ds ds_\ell d\cos\theta_\pi} &= e^2 G_F^2 |V_{us}|^2 \frac{s_\ell \sigma_\pi(s) X}{2^{20} \pi^9 \mkp^7}  \begin{aligned}[t]
				& \Bigg( \ln\left(\frac{\tilde x_\mathrm{min}}{\tilde x_\mathrm{max}}\right) \int_{-1}^1  d\cos\theta_\gamma \int_0^{2\pi} d\phi \, j_1^{hh}(s,s_\ell,\cos\theta_\pi,\cos\theta_\gamma,\phi) \\
				& + \int_{-1}^1  d\cos\theta_\gamma \int_0^{2\pi} d\phi \, j_{2,\mathrm{cut}}^{hh}(s,s_\ell,\cos\theta_\pi,\cos\theta_\gamma,\phi) \Bigg) .
			\end{aligned}
	\end{split}
\end{align} }%
The function $j_1^{hh}$ is given by
\begin{align}
	\begin{split}
		j_1^{hh}(s,s_\ell,\cos\theta_\pi,\cos\theta_\gamma,\phi) &= \frac{32\pi \mkp^4}{3 F_0^2} \left( (PL + X \sigma_\pi \cos\theta_\pi)^2 - 4 s_\ell \mpip^2 \right) \\
			& \quad \cdot \begin{aligned}[t]
				&\Bigg( \frac{s}{A_1^2} + \frac{s}{A_2^2} + \frac{2 PL + s + s_\ell}{(PL + s_\ell + \cos\theta_\gamma X)^2} + \frac{2 (PL + s)}{A_1 (PL + s_\ell + \cos\theta_\gamma X)} \\
				& - \frac{2 (PL + s)}{A_2 (PL + s_\ell + \cos\theta_\gamma X)}  - \frac{2s + 4 \cos\theta_\pi X \sigma_\pi}{A_1 A_2} \\
				& + \frac{ 4 \cos\theta_\pi X s_\ell \sigma_\pi}{A_1 A_2 (PL + s_\ell + \cos\theta_\gamma X)}  - \frac{4 s  \sigma_\pi^2 (PL + \cos\theta_\gamma X)^2}{A_1^2 A_2^2} \Bigg) , \end{aligned}
	\end{split}
\end{align}
where the $\phi$-dependence is hidden in
\begin{align}
	\begin{split}
		A_1 &= PL + \cos\theta_\gamma X - \cos\theta_\gamma \cos\theta_\pi PL \sigma_\pi - \cos\theta_\pi X \sigma_\pi \\
			& \quad + \cos\phi \, \sigma_\pi \sqrt{(1 - \cos\theta_\gamma^2) (1 - \cos\theta_\pi^2) s s_\ell} , \\
		A_2 &= PL + \cos\theta_\gamma X + \cos\theta_\gamma \cos\theta_\pi PL \sigma_\pi + \cos\theta_\pi X \sigma_\pi \\
			& \quad - \cos\phi \, \sigma_\pi \sqrt{(1 - \cos\theta_\gamma^2) (1 - \cos\theta_\pi^2) s s_\ell} .
	\end{split}
\end{align}

The integrand $j_{2,\mathrm{cut}}^{hh}$ of the second numerical integral is a lengthy expression that I do not state here explicitly.

The soft photon contribution to this second part contains the six bremsstrahlung integrals $I_1(p)$, $I_1(p_1)$, $I_1(p_2)$, $I_2(p,p_1)$, $I_2(p,p_2)$ and $I_2(p_1,p_2)$. It is easy to verify numerically that in the sum of the contributions from soft and hard region, the dependence on $\Delta\varepsilon$ (i.e.~on $\tilde x_\mathrm{min}$) again drops out. The analytic result of the integral over $j_1^{hh}$ can therefore be inferred from the soft photon $hh$-part (note that these bremsstrahlung integrals do not depend on $\phi$ or $\cos\theta_\ell$).

The interference term of off-lepton and off-hadron photon emission is the last and most intricate part of the phase space integral calculation. On the one hand, the explicit form factors $\Pi_{ij}$ have to be inserted after the $\phi_\ell$- and $y$-integration. On the other hand, while the part of the interference term containing $\Pi$ is free of collinear singularities and independent of $\tilde x_\mathrm{min}$, the contrary is true for the part involving the $K_{\ell4}$ form factors. I again integrate over $\phi_\ell$, $y$ and $x$ analytically, expand the result for small $\ml^2$ and obtain the structure
\begin{align}
	\label{eqn:HardPhotonInterferenceTerms}
	\begin{split}
		\frac{d\Gamma_{\gamma,\mathrm{cut}}^{\mathrm{hard,int}}}{ds ds_\ell d\cos\theta_\pi} &= e^2 G_F^2 |V_{us}|^2 \frac{s_\ell \sigma_\pi(s) X}{2^{20} \pi^9 \mkp^7} \\
			& \quad \cdot \begin{aligned}[t]
				& \Bigg( \ln \zl \left(\tilde x_\mathrm{max} + \ln\left(\frac{ \tilde x_\mathrm{min}}{ \tilde x_\mathrm{max}}\right) \right) \int_{-1}^1 d\cos\theta_\gamma \int_0^{2\pi} d\phi \, j_1^\mathrm{int}(s,s_\ell,\cos\theta_\pi,\cos\theta_\gamma,\phi) \\
				& + \ln \left( \frac{\tilde x_\mathrm{min}}{\tilde x_\mathrm{max}}\right) \int_{-1}^1 d\cos\theta_\gamma \int_0^{2\pi} d\phi \, j_2^\mathrm{int}(s,s_\ell,\cos\theta_\pi,\cos\theta_\gamma,\phi) \\
				& + \int_{-1}^1  d\cos\theta_\gamma \int_0^{2\pi} d\phi \, j_{3,\mathrm{cut}}^\mathrm{int}(s,s_\ell,\cos\theta_\pi,\cos\theta_\gamma,\phi) \Bigg) .
			\end{aligned}
	\end{split}
\end{align}
I perform the integrals over $\phi$ and $\cos\theta_\gamma$ numerically. The expressions for the integrands $j_i^\mathrm{int}$ are too lengthy to be given explicitly. $j_{3,\mathrm{cut}}^\mathrm{int}$ depends on the cut $\tilde x_\mathrm{max}$.

Again, the sum of the soft and hard photon contribution must not depend on $\Delta\varepsilon$. I expand the soft contribution, given by the remaining bremsstrahlung integrals $I_2(p,p_\ell)$, $I_2(p_1,p_\ell)$ and $I_2(p_2,p_\ell)$, in $\ml$ and neglect terms that vanish for $\ml\to0$:
{\small
\begin{align}
	\begin{split}
		\frac{d\Gamma_\gamma^\mathrm{soft,int}}{ds ds_\ell d\cos\theta_\pi} &= - e^2  \int_{-1}^1  d\cos\theta_\ell \int_0^{2\pi} d\phi \; d\Gamma_\mathrm{iso}^\mathrm{LO} \frac{1}{16\pi^2} \begin{aligned}[t]
				&\Bigg( \ln\left( \frac{2\Delta\varepsilon}{\mg} \right) \begin{aligned}[t]
					& \bigg[ 4 \ln \zl + b_1^\mathrm{int}(s,s_\ell,\cos\theta_\pi,\cos\theta_\ell,\phi) \bigg] \end{aligned} \\
				& + \ln^2 \zl + b_2^\mathrm{int}(s,s_\ell,\cos\theta_\pi,\cos\theta_\ell,\phi) \Bigg) , \end{aligned} 
	\end{split}
\end{align} }%
where the $b_i^\mathrm{int}$ are again rather lengthy expressions.

I perform the integrals over $\cos\theta_\ell$ and $\phi$ numerically and find that the dependence on $\Delta\varepsilon$ drops out indeed in the sum of soft and hard photon contribution.

\subsection{Cancellation of Divergences}

Both the virtual corrections and the real emission contain infrared divergences. These divergences, which are regulated by the artificial photon mass $m_\gamma$, must vanish in the inclusive decay rate. In the radiative process, the IR divergence is generated in the soft region, which I have treated in the soft photon approximation.

Furthermore, collinear (or mass) divergences arise in the virtual corrections and in the soft and hard region of the radiative process. They are regulated by the lepton mass $m_\ell$ that acts as a natural cutoff. According to the KLN theorem \cite{KinoshitaSirlin1959,Kinoshita1962,Lee1964}, there must not be any divergences in the fully inclusive decay rate. Since the limit $\ml\to0$ is usually taken in experimental analyses, I apply the same approximation to the inclusive decay rate. Here, however, it is crucial that the collinear divergences indeed cancel.

Note that I use everywhere the physical lepton mass, which can be identified (up to higher order effects) with the renormalised mass. A necessary condition for the KLN theorem to hold in this representation is that the mass renormalisation does not diverge in the limit $\ml\to0$. This condition is fulfilled by (\ref{eqn:LeptonMassRenormalisation}).

\subsubsection{Infrared Singularities}

In the virtual corrections, the six triangle diagrams~\ref{img:Kl4_NLOgLoop5}-\ref{img:Kl4_NLOgLoop10} and the external leg corrections are IR-divergent. The relevant loop functions are given in appendix~\ref{sec:IRdivergentLoopFunctions}.

\begin{figure}[ht]
	\centering
	\scalebox{0.8}{
		\begin{pspicture}(0,0)(8,6)
			\put(4.5,5.35){(a)}
			\put(2.5,5.35){(b)}
			\put(1.75,4.75){(c)}
			\put(5.75,5.2){(d)}
			\includegraphics[width=8cm]{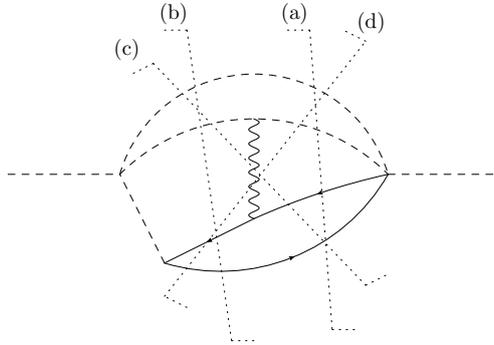}
		\end{pspicture}
		}
	\caption{Four-loop kaon self-energy diagram with four- or five-particle cuts.}
	\label{img:KaonSEFourLoop}
\end{figure}

\begin{figure}[ht]
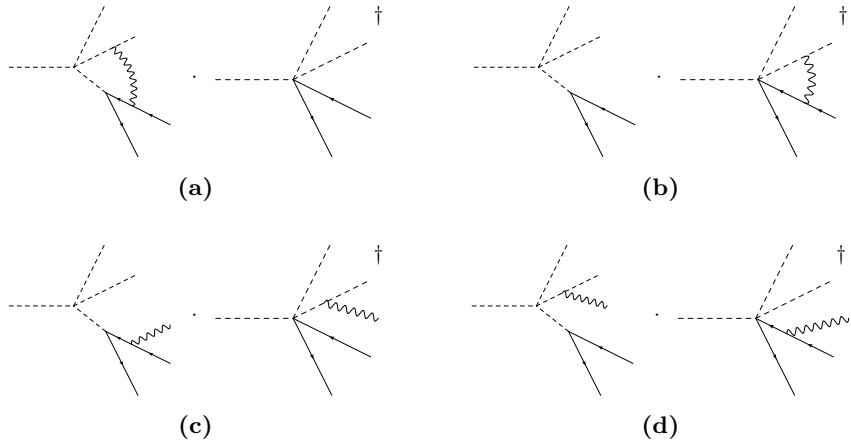

	\centering
	\begin{subfigure}[b]{0.4\textwidth}
		\centering
		\scalebox{0.8}{
			\begin{pspicture}(0,0)(6,3)
				\put(3,1.25){$\cdot$}
				\put(6,2.25){$\dagger$}
				\includegraphics[height=2.5cm]{images/NLO_gLoop18}
				\hspace{0.5cm}
				\includegraphics[height=2.5cm]{images/LO1}
			\end{pspicture}
			}
		\caption{}
		\label{img:PhaseSpaceProductA}
	\end{subfigure}
	\begin{subfigure}[b]{0.4\textwidth}
		\centering
		\scalebox{0.8}{
			\begin{pspicture}(0,0)(6,3)
				\put(3,1.25){$\cdot$}
				\put(6,2.25){$\dagger$}
				\includegraphics[height=2.5cm]{images/LO2}
				\hspace{0.5cm}
				\includegraphics[height=2.5cm]{images/NLO_gLoop10}
			\end{pspicture}
			}
		\caption{}
		\label{img:PhaseSpaceProductB}
	\end{subfigure}
	
	\vspace{0.5cm}

	\begin{subfigure}[b]{0.4\textwidth}
		\centering
		\scalebox{0.8}{
			\begin{pspicture}(0,0)(6,3)
				\put(3,1.25){$\cdot$}
				\put(6,2.25){$\dagger$}
				\includegraphics[height=2.5cm]{images/RE12}
				\hspace{0.5cm}
				\includegraphics[height=2.5cm]{images/RE3}
			\end{pspicture}
			}
		\caption{}
		\label{img:PhaseSpaceProductC}
	\end{subfigure}
	\begin{subfigure}[b]{0.4\textwidth}
		\centering
		\scalebox{0.8}{
			\begin{pspicture}(0,0)(6,3)
				\put(3,1.25){$\cdot$}
				\put(6,2.25){$\dagger$}
				\includegraphics[height=2.5cm]{images/RE8}
				\hspace{0.5cm}
				\includegraphics[height=2.5cm]{images/RE5}
			\end{pspicture}
			}
		\caption{}
		\label{img:PhaseSpaceProductD}
	\end{subfigure}
	\caption{Phase space products corresponding to the four cuts of the kaon self-energy diagram.}
	\label{img:PhaseSpaceProducts}
\end{figure}

A priori, one would expect that the box diagrams~\ref{img:Kl4_NLOgLoop16}-\ref{img:Kl4_NLOgLoop18} also give rise to an IR singularity, because the scalar four point loop function $D_0$ is IR-divergent as well. However, as can be shown with Passarino-Veltman reduction techniques  \cite{Hooft1979, Passarino1979} and the explicit expressions for the IR-divergent scalar box integral \cite{Beenakker1990}, the contribution of the box diagrams to the form factors $F$ and $G$ are IR-finite. This can be understood rather easily: consider the four-loop kaon self-energy diagram in figure~\ref{img:KaonSEFourLoop}. This diagram is an IR-finite quantity and so must be the sum of its four- and five-particle cuts. Each of the four cuts corresponds to a phase space integral of the product of two diagrams, shown in figure~\ref{img:PhaseSpaceProducts}. Now, as the IR divergence in the radiative process is generated in the soft region, where the matrix element factorises into the LO non-radiative process times the soft photon factor (\ref{eqn:SoftPhotonFactorisation}), the IR divergence has to drop out already in the differential inclusive decay rate, where the photon is integrated. The phase space products~\ref{img:PhaseSpaceProductB}-\ref{img:PhaseSpaceProductD} can only contribute to the term $R F^*$, $R G^*$ and $|R|^2$. Therefore, the phase space product~\ref{img:PhaseSpaceProductA} cannot give an IR-divergent contribution to $|F|^2$ or $|G|^2$. Hence, the box diagram on the left-hand side of the product can only give IR-divergent contributions to $R$. An analogous argument works for the two other box diagrams.

\pagebreak

Let us now turn our attention to the IR divergences of the virtual corrections. Summing all the IR-divergent contributions (after UV renormalisation), I find
\begin{align}
	\begin{split}
		\delta F^\mathrm{NLO,IR}_{\mathrm{virt.}\gamma} &= \delta F^\mathrm{NLO,IR}_{\gamma-\mathrm{loop},e-j} + \delta F_{\gamma-Z}^\mathrm{NLO,IR} = \delta G^\mathrm{NLO,IR}_{\mathrm{virt.}\gamma} = \delta G^\mathrm{NLO,IR}_{\gamma-\mathrm{loop},e-j} + \delta G_{\gamma-Z}^\mathrm{NLO,IR} \\
		&= 2 e^2 \begin{aligned}[t]
			& \Bigg( (\mkp^2+\mpip^2-t) C_0^\mathrm{IR}(\mpip^2,t, \mkp^2,\mg^2,\mpip^2,\mkp^2) \\
			& - (\mkp^2+\mpip^2-u) C_0^\mathrm{IR}(\mpip^2,u,\mkp^2,\mg^2,\mpip^2,\mkp^2) \\
			& + (\mkp^2+\ml^2-\tl) C_0^\mathrm{IR}(\ml^2,\tl,\mkp^2,\mg^2,\ml^2,\mkp^2) \\
			& + (2\mpip^2-s) C_0^\mathrm{IR}(\mpip^2,s,\mpip^2,\mg^2,\mpip^2,\mpip^2) \\
			& - (\mpip^2 + \ml^2 - s_{1\ell} ) C_0^\mathrm{IR}(\ml^2, s_{1\ell}, \mpip^2, \mg^2, \ml^2, \mpip^2) \\
			& + (\mpip^2 + \ml^2 - s_{2\ell}) C_0^\mathrm{IR}(\ml^2, s_{2\ell}, \mpip^2, \mg^2, \ml^2, \mpip^2) \\
			& - \frac{1}{8\pi^2} \ln\zg \Bigg) =: \delta X^\mathrm{NLO,IR}_{\mathrm{virt.}\gamma} , \end{aligned} \\
	\end{split}
\end{align}
where
\begin{align}
	\begin{split}
		C_0^\mathrm{IR}(m^2, s, M^2, \mg^2, m^2, M^2) &= - \frac{1}{16\pi^2} \frac{x_s}{m M(1-x_s^2)} \ln x_s \ln \zg , \\
			x_s &=  -\frac{1 - \sqrt{1 - \frac{4 m M}{s - (m-M)^2}}}{1 + \sqrt{1 - \frac{4 m M}{s - (m-M)^2}}} .
	\end{split}
\end{align}

The infrared-divergent part of the NLO decay rate is given by
\begin{align}
	\begin{split}
		d\Gamma^\mathrm{NLO,IR} = d\Gamma^\mathrm{LO}_\mathrm{iso} \; 2 \Re(\delta X^\mathrm{NLO,IR}_{\mathrm{virt.}\gamma}) + \O(\zl \ln\zl) .
	\end{split}
\end{align}
By extracting the IR divergence (terms proportional to $\ln\zg$) out of the soft photon contribution to the radiative decay rate (\ref{eqn:DecayRateSoftRegion}), it is now easy to verify that the sum of virtual corrections and soft bremsstrahlung (where the photon is integrated) and hence the inclusive decay rate is free of infrared divergences:
\begin{align}
	\begin{split}
		d\Gamma^\mathrm{NLO,IR} + d\Gamma_\gamma^\mathrm{soft,IR} = 0 .
	\end{split}
\end{align}

\subsubsection{Collinear Singularities}

Both the soft and the hard region of the radiative process give rise to collinear singularities, terms proportional to $\ln\zl$. Let us now check that these mass divergences cancel in the fully inclusive decay rate (the cut on the photon energy must be removed for this purpose, i.e.~I take the limit $\tilde x_\mathrm{max} \to 1$). Virtual photon corrections can produce a collinear divergence if one end of the photon line is attached to the lepton line. Since the mass divergence in the radiative process is produced in the collinear region of the phase space (soft and hard), where the matrix element could be factorised similarly to the soft region \cite{Bohm2001}, one can argue in an analogous way as for the IR divergences that the contribution of the box diagrams to the form factors $F$ and $G$ has no mass divergence. This is confirmed by the explicit expressions for the diagrams. The only collinear divergent contributions stem from the external leg correction for the lepton and the three diagrams~\ref{img:Kl4_NLOgLoop7}, \ref{img:Kl4_NLOgLoop9} and \ref{img:Kl4_NLOgLoop10}.

The external leg correction for the lepton contains the following collinear divergence:
\begin{align}
	\begin{split}
		\delta F_{\gamma-Z}^{\mathrm{NLO,coll}} &= \delta G_{\gamma-Z}^{\mathrm{NLO,coll}} = \frac{3 e^2}{32\pi^2} \ln\zl ,
	\end{split}
\end{align}
contributing to the decay rate as
\begin{align}
	\begin{split}
		d\Gamma^\mathrm{NLO,coll}_{Z} &= d\Gamma^{\mathrm{LO}}_\mathrm{iso} \; \frac{3e^2}{16\pi^2} \ln\zl .
	\end{split}
\end{align}
This cancels exactly the mass divergence in the $\ell\ell$-part of the real photon corrections (\ref{eqn:RadiativeDecayRateLL}).

Next, I collect the mass divergent terms contained in the three relevant loop diagrams:
\begin{align}
	\begin{split}
		\delta F^\mathrm{NLO,coll}_{\gamma-\mathrm{loop}} &= \delta G^\mathrm{NLO,coll}_{\gamma-\mathrm{loop}} = \frac{e^2}{16\pi^2} \ln\zl \left( \frac{1}{2} \ln \zl - \ln \zg - 2 \right) , 
	\end{split}
\end{align}
resulting in a collinear divergence in the decay rate of
\begin{align}
	\begin{split}
		d\Gamma^\mathrm{NLO,coll}_{\mathrm{loop}} &= d\Gamma^{\mathrm{LO}}_\mathrm{iso} \; \frac{e^2}{16\pi^2} \ln\zl \left( \ln \zl - 2 \ln \zg - 4 \right) .
	\end{split}
\end{align}
This singularity must cancel with the mass divergence in the interference term of the radiative decay rate. The divergent contribution from the soft photon region is given by
\begin{align}
	\begin{split}
		d\Gamma^\mathrm{soft,int}_{\gamma,\mathrm{coll}} &= - d\Gamma^\mathrm{LO}_\mathrm{iso} \frac{e^2}{16\pi^2} \ln\zl \left( \ln\zl + 4 \ln\left( \frac{2\Delta\varepsilon}{\mg} \right) \right) \\
			&=  - d\Gamma^\mathrm{LO}_\mathrm{iso} \frac{e^2}{16\pi^2} \ln\zl \left( \ln\zl - 2 \ln\zg + 4 \ln\left( \frac{2\Delta\varepsilon}{\sqrt{s_\ell}} \right) \right) .
	\end{split}
\end{align}
In the sum of virtual and soft real corrections, the double divergences (double collinear and soft-collinear) cancel:
\begin{align}
	\begin{split}
		d\Gamma^\mathrm{NLO,coll}_{\mathrm{loop}} + d\Gamma^\mathrm{soft,int}_{\gamma,\mathrm{coll}} &= - d\Gamma^\mathrm{LO}_\mathrm{iso} \frac{e^2}{4\pi^2} \ln\zl \left( 1 + \ln \tilde x_\mathrm{min} \right) .
	\end{split}
\end{align}
This single divergence must cancel against the one in the hard real corrections (\ref{eqn:HardPhotonInterferenceTerms}). By evaluating numerically the integral over $j_1^\mathrm{int}$, I have checked that this cancellation takes place.

I have now verified that the fully inclusive decay rate
\begin{align}
	\begin{split}
		\frac{d\Gamma_{(\gamma)}}{ds ds_\ell d\cos\theta_\pi} &= \frac{d\Gamma_{\mathrm{virt.}\gamma}^\mathrm{NLO}}{ds ds_\ell d\cos\theta_\pi} + \frac{d\Gamma_{\gamma}^\mathrm{soft}}{ds ds_\ell d\cos\theta_\pi} + \frac{d\Gamma_{\gamma}^\mathrm{hard}}{ds ds_\ell d\cos\theta_\pi}
	\end{split}
\end{align}
does not depend on the energy cut separating the soft from the hard region and contains neither infrared nor collinear (mass) singularities. The calculation is therefore in accordance with the KLN theorem. Note that this is a necessary but highly non-trivial consistency check, since the two regions of the radiative phase space are parametrised differently.


\chapter{Numerical Evaluation}

\label{sec:Numerics}

The existing high statistics experiments on $K_{\ell4}$, E865 \cite{Pislak2001, Pislak2003} and NA48/2 \cite{Batley2010, Batley2012}, have applied isospin corrections to a certain extent and with different approximations. In the NA48/2 experiment, the data was corrected by the semi-classical Gamow-Sommerfeld (or Coulomb) factor and with the help of PHOTOS \cite{Barberio1994}. The E865 experiment used the same analytic prescription by Diamant-Berger \cite{DiamantBerger1976} as the older Geneva-Saclay experiment \cite{Rosselet1977}. Both treatments did not make use of the full matrix element and relied on factorisation of the tree-level amplitude as it happens in a soft and collinear photon approximation. The isospin breaking due to the mass effects was not taken into account.

Unfortunately, in the case of NA48/2, an analysis without the effect of PHOTOS is not available. Hence, it seems almost impossible to make use of the here calculated photonic effects for a full a posteriori correction of the form factors. Nevertheless, I have a program at hand that calculates the effect of PHOTOS on the (partially) inclusive decay rate\footnote{I am very grateful to B.~Bloch-Devaux for providing me with this program.}. This enables me to perform a comparison of the here presented calculation with the effect of PHOTOS, using the simple photon energy cut in $\Sigma_{\ell\nu\gamma}$ described in the previous section.

I pursue therefore two aims in the following sections. First, the isospin corrections due to the mass effects can be extracted directly for the form factors. Second, for the photonic effects, I calculate the radiative corrections to the (semi-)inclusive decay rate. These isospin-breaking effects are then compared with the correction applied by NA48/2.

\section{Corrections due to the Mass Effects}

As explained in the previous chapter, the isospin-breaking effects due to the quark and meson mass differences can be extracted on the level of the amplitude or form factors. I now evaluate these corrections numerically.

The form factors have the partial wave expansions \cite{Bijnens1994}
\begin{align}
	\begin{split}
		F + \frac{\sigma_\pi PL}{X} \cos\theta_\pi G &= \sum_{l=0}^\infty P_l(\cos\theta_\pi) f_l(s,s_\ell) , \\
		G &= \sum_{l=1}^\infty P_l^\prime(\cos\theta_\pi) g_l(s,s_\ell) ,
	\end{split}
\end{align}
where $P_l$ are the Legendre polynomials. The NA48/2 experiment \cite{Batley2012} uses the expansion
\begin{align}
	\begin{split}
		F &= F_s e^{i\delta_s} + F_p e^{i\delta_p} \cos\theta_\pi + \ldots, \\
		G &= G_p e^{i\delta_p} + \ldots
	\end{split}
\end{align}
and defines
\begin{align}
	\begin{split}
		\tilde G_p &= G_p + \frac{X}{\sigma_\pi PL} F_p .
	\end{split}
\end{align}
Hence, I identify
\begin{align}
	\begin{split}
		F_s &= | f_0 |, \quad \tilde G_p = \frac{X}{\sigma_\pi PL} | f_1 |, \quad G_p = | g_1 |
	\end{split}
\end{align}
and calculate the partial wave projections
\begin{align}
	\begin{split}
		f_l &= \frac{2l+1}{2} \int_{-1}^1 d\cos\theta_\pi P_l(\cos\theta_\pi) \left( F + \frac{\sigma_\pi PL}{X} \cos\theta_\pi G \right) , \\
		g_l &= \int_{-1}^1 d\cos\theta_\pi \frac{P_{l-1}(\cos\theta_\pi) - P_{l+1}(\cos\theta_\pi)}{2} \, G .
	\end{split}
\end{align}
At the order that I consider, the isospin correction due to the mass effects to the norms and phases of the partial waves is then given by
\begin{align}
	\begin{split}
		\delta_\mathrm{ME} F_s :={}& 1 - \frac{1}{|f_0|}\lim\limits_{\mathrm{isospin}}  |f_0| = 1 - \frac{1}{| \Re(f_0) |} \lim_\mathrm{isospin} |\Re(f_0)| + \O(p^4) , \\
		\delta_\mathrm{ME} \tilde G_p :={}& 1 - \frac{1}{|f_1|} \lim_{\mathrm{isospin}}  |f_1| = 1 - \frac{1}{| \Re(f_1) |} \lim_\mathrm{isospin} |\Re(f_1)| + \O(p^4) , \\
		\delta_\mathrm{ME} G_p :={}& 1 - \frac{1}{ |g_1|} \lim_{\mathrm{isospin}}  |g_1| = 1 - \frac{1}{| \Re(g_1) |} \lim_\mathrm{isospin} |\Re(g_1) | + \O(p^4) , \\
		\Delta_\mathrm{ME} \delta_0^0 :={}& \arg(f_0) - \lim_\mathrm{isospin} \arg(f_0) = \frac{\Im(f_0)}{f_0^\mathrm{LO}} - \lim_\mathrm{isospin} \frac{\Im(f_0)}{f_0^\mathrm{LO}} + \O(p^4) , \\
		\Delta_\mathrm{ME} \delta_1^1 :={}& \arg(f_1) - \lim_\mathrm{isospin} \arg(f_1) = \frac{\Im(f_1)}{f_1^\mathrm{LO}} - \lim_\mathrm{isospin} \frac{\Im(f_1)}{f_1^\mathrm{LO}} + \O(p^4) \\
			={}& \arg(g_1) - \lim_\mathrm{isospin} \arg(g_1) = \frac{\Im(g_1)}{g_1^\mathrm{LO}} - \lim_\mathrm{isospin} \frac{\Im(g_1)}{g_1^\mathrm{LO}} + \O(p^4) .
	\end{split}
\end{align}
The isospin correction to the $P$-wave phase shift vanishes at this order. Using the inputs described in \cite{Colangelo2009,Bernard2013}, I reproduce their NLO results for the $S$-wave phase shift.

The correction to the phase depends on the pion decay constant and the breaking parameters. In the correction to the norm of the partial waves, also the low-energy constants $L_4^r$, $K_2^r$, $K_4^r$ and $K_6^r$ appear ($K_4^r$ only appears in the correction to the $P$-wave).

I have presented the analytic results of the loop calculation in terms of the decay constant in the chiral limit $F_0$. Unfortunately, different lattice determinations do not yet agree on its value \cite{Aoki2013}. For the numerics, I convert the results to an expansion in $1/F_\pi$ using the relation between $F_0$ and $F_\pi$ in pure QCD at $\O(p^4, \epsilon p^4)$ \cite{Neufeld1996},
\begin{align}
	\begin{split}
		F_\pi &= F_0 \begin{aligned}[t]
			& \Bigg[ 1 + \frac{4}{F_0^2} \Big( L_4^r(\mu) (M_\pi^2 + 2 M_K^2) + L_5^r(\mu) M_\pi^2 \Big) \\
			& - \frac{1}{2(4\pi)^2 F_0^2} \left( 2 M_\pi^2 \ln\left( \frac{M_\pi^2}{\mu^2}\right) + M_K^2 \ln\left( \frac{M_K^2}{\mu^2} \right) \right)  \Bigg], \end{aligned}
	\end{split}
\end{align}
where $M_{\pi,K}$ denote the masses in the isospin limit, defined as
\begin{align}
	\begin{split}
		M_\pi^2 = \mpio^2 , \quad M_K^2 = \frac{1}{2}\left( \mkp^2 + \mko^2 - \mpip^2 + \mpio^2 \right) .
	\end{split}
\end{align}
For $F_\pi$ and the meson masses, I use the current PDG values \cite{Beringer2012}.

Another strategy would be to work directly with $F_0$ and assign a large error that covers the different determinations, as done in~\cite{Bernard2013}. I use the solution based on the expansion in $1/F_0$ with a central value of $F_0 = 75 \; \mathrm{MeV}$ for a very rough estimate of higher order corrections.

The correction to the norms of the partial waves depends rather strongly on the value of $L_4^r$. The $\O(p^4)$ fits in \cite{BijnensTalavera2002, Bijnens2012} give the large value $L_4^r = 1.5 \cdot 10^{-3}$. I decide however, to rely on the lattice estimate of \cite{MILC2009}, recommended in \cite{Aoki2013}, but to use a more conservative uncertainty of $\pm 0.5\cdot 10^{-3}$ (see table~\ref{tab:InputsMassEffects}).

For the NLO constants of the electromagnetic sector, I use the estimates of \cite{Ananthanarayan2004} and assign a 100\% error. For the isospin-breaking parameter $\epsilon$, I take the latest recommendation in the FLAG report~\cite{Aoki2013},
\begin{align}
	\begin{split}
		\epsilon = \frac{\sqrt{3}}{4 R} , \quad R = 35.8 \pm 2.6 ,
	\end{split}
\end{align}
where I added the lattice and electromagnetic errors in quadrature.

I fix the electromagnetic low-energy constant $Z$ with the LO relation to the pion mass difference (\ref{eqn:LOMassDifferences}).

\begin{table}[H]
	\centering
	\begin{tabular}{c c r}
		\toprule
		$10^3 \cdot L_4^r(\mu)$ & $0.04 \pm 0.50$ & \cite{Aoki2013} \\
		$10^3 \cdot L_5^r(\mu)$ & $0.84 \pm 0.50$ & \cite{Aoki2013} \\[0.05cm]
		\hline \\[-0.35cm]
		$10^3 \cdot K_2^r(\mu)$ & $0.69 \pm 0.69$ & \cite{Ananthanarayan2004} \\
		$10^3 \cdot K_4^r(\mu)$ & $1.38 \pm 1.38$ & \cite{Ananthanarayan2004} \\
		$10^3 \cdot K_6^r(\mu)$ & $2.77 \pm 2.77$ & \cite{Ananthanarayan2004} \\[0.05cm]
		\hline \\[-0.35cm]
		$F_\pi$ & $(92.21 \pm 0.14)$~MeV & \cite{Beringer2012} \\[0.05cm]
		\hline \\[-0.35cm]
		$R$ & $35.8 \pm 2.6\hphantom{0}$ & \cite{Aoki2013} \\
		\bottomrule
	\end{tabular}
	\caption{Input parameters for the evaluation of the mass effects ($\mu = 770$~MeV).}
	\label{tab:InputsMassEffects}
\end{table}

\begin{figure}[ht]
	\centering
	\large
	\scalebox{0.63}{
		\input{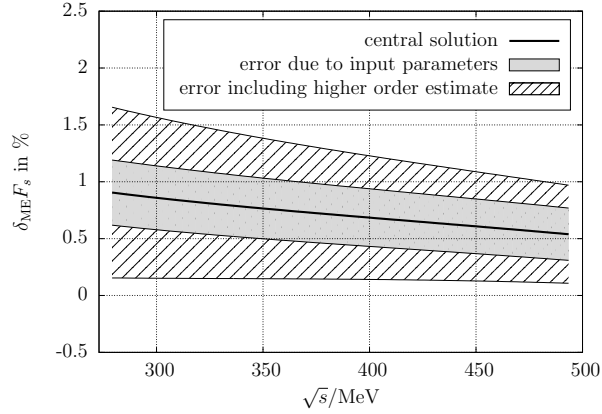}
		}
	\caption{Relative value of the mass effect correction to the $S$-wave $F_s$ for $s_\ell = 0$. The exact meaning of the error bands is explained in the text.}
	\label{plot:SWave}
\end{figure}

\begin{figure}[ht]
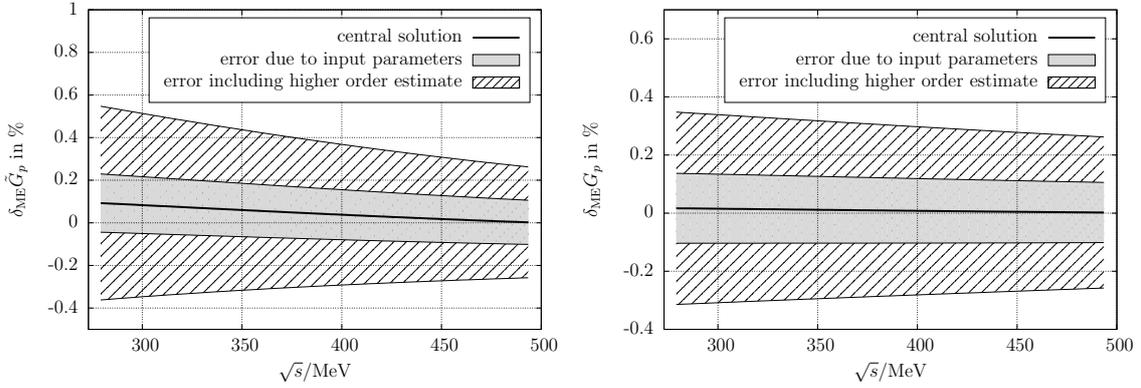

	\centering
	\large
	\scalebox{0.59}{
		\input{plots/gpt.tex}
		\input{plots/gp.tex}
		}
	\caption{Relative value of the mass effect corrections to the $P$-waves $\tilde G_p$ and $G_p$ for $s_\ell = 0$.}
	\label{plot:PWaves}
\end{figure}

The plots in figures~\ref{plot:SWave} and \ref{plot:PWaves} show the relative isospin correction due to the mass effects for the norm of the partial waves. I separately show the error band due to the variation of the input parameters and the error band that also includes the estimate of higher order corrections, given by the difference between the $F_\pi$- and the $F_0$-solution, added in quadrature. The error due to the input parameters is dominated by the uncertainty of the low-energy constant $L_4^r$. The LECs of the electromagnetic sector and the isospin-breaking parameter $R$ play a minor role.

In contrast to the $S$-wave, where the isospin corrections are at the percent level, the effect in the two $P$-waves is within the uncertainty compatible with zero. The dependence on $s_\ell$ is rather weak and covered by the error bands.

To conclude this section, I suggest to apply the additional isospin-breaking corrections to the NA48/2 measurement \cite{Batley2012} shown in table~\ref{tab:IsospinCorrectionsMassEffectsNA48}. In order to obtain the partial waves of the form factors in the isospin limit, one has to subtract the given corrections. The corrections to the $P$-waves are certainly negligible. However, for the $S$-wave, the isospin correction (and also its uncertainty, unfortunately) is much larger than the experimental errors.

\begin{table}[H]
	\footnotesize
	\centering
	\begin{tabular}{c c c c c c c c}
		\toprule
		$\sqrt{s}/$MeV & $\sqrt{s_\ell}/$MeV & $F_s$ \cite{Batley2010,Batley2012} & $\delta_\mathrm{ME} F_s \cdot F_s$ & $\tilde G_p$ \cite{Batley2010,Batley2012} & $\delta_\mathrm{ME}\tilde G_p \cdot \tilde G_p$ & $G_p$ \cite{Batley2010,Batley2012} & $\delta_\mathrm{ME} G_p \cdot G_p$ \\[0.1cm]
		\hline \\[-0.3cm]
		286.06 &	126.44 &				5.7195(122) &				$0.050(16)(38)$ &	4.334(76) &	$0.003(6)(16)$	&	5.053(266) 				& $0.001(6)(15)$ \\
		295.95 &	142.60 &				5.8123(101) &				$0.050(16)(37)$ &	4.422(61) &	$0.002(6)(16)$	&	5.186(165) 				& $0.001(6)(16)$ \\
		304.88 &	141.31 &				5.8647(102) &				$0.049(16)(36)$ &	4.550(52) &	$0.002(6)(16)$	&	4.941(123) 				& $0.001(6)(15)$ \\
		313.48 &	137.47 &				5.9134(104) &				$0.048(16)(36)$ &	4.645(47) &	$0.002(6)(16)$	&	4.896(104) 				& $0.001(6)(14)$ \\
		322.02 &	130.92 &				5.9496(\hphantom{0}95) &	$0.048(16)(35)$ &	4.711(47) &	$0.002(6)(16)$	&	5.245(\hphantom{0}99) 		& $0.001(6)(15)$ \\
		330.80 &	124.14 &				5.9769(103) &				$0.047(16)(34)$ &	4.767(44) &	$0.002(6)(15)$	&	5.283(\hphantom{0}92) 		& $0.001(6)(15)$ \\
		340.17 &	116.91 &				6.0119(\hphantom{0}98) &	$0.046(16)(34)$ &	4.780(45) &	$0.002(6)(15)$	&	5.054(\hphantom{0}90) 		& $0.001(6)(14)$ \\
		350.94 &	108.19 &				6.0354(\hphantom{0}96) &	$0.046(16)(33)$ &	4.907(39) &	$0.002(6)(15)$	&	5.264(\hphantom{0}72) 		& $0.001(6)(15)$ \\
		364.57 &	\hphantom{0}98.53 &	6.0532(\hphantom{0}96) &	$0.044(16)(32)$ &	5.019(40) &	$0.002(6)(15)$	&	5.357(\hphantom{0}64) 		& $0.001(6)(15)$ \\
		389.95 &	\hphantom{0}80.62 &	6.1314(184) &				$0.043(16)(30)$ &	5.163(42) &	$0.001(6)(15)$	&	5.418(\hphantom{0}64) 		& $0.001(6)(15)$ \\
		\bottomrule
	\end{tabular}
	\caption[Isospin-breaking corrections due to the mass effects.]{Isospin-breaking corrections due to the mass effects, calculated for the bins of the NA48/2 measurement \cite{Batley2010,Batley2012}. For comparison, I quote the values of the partial waves with their uncertainties (statistical and systematic errors added in quadrature) without including the dominant error of the normalisation. Note that the uncertainties of $F_s$ are taken from \cite{Batley2010}, as the values displayed in \cite{Batley2012} are not correct\footnotemark{}. The first error to the isospin correction is due to the input parameters, the second is a rough estimate of higher order corrections.}
	\label{tab:IsospinCorrectionsMassEffectsNA48}
\end{table}
\footnotetext{I thank B.~Bloch-Devaux for the confirmation thereof.}

\section{Discussion of the Photonic Effects}

For the numerical evaluation of the photonic effects, I compute the (semi-)inclusive decay rate, differential with respect to $s$, $s_\ell$ and $\cos\theta_\pi$. After some general considerations and tests, I compare the resulting $\O(e^2)$ correction to the one applied in the NA48/2 experiment \cite{Batley2012}, i.e.~the Gamow-Sommerfeld factor combined with PHOTOS~\cite{Barberio1994}.

For the numerical evaluation of the inclusive decay rate $d\Gamma_{(\gamma)}$, I need several input parameters. As I am interested in $\O(e^2)$ effects but work only at leading chiral order, I directly replace $F_0$ by the physical pion decay constant $F_\pi$. When calculating the fully inclusive decay rate, I take advantage of the cancellation of collinear singularities and send the lepton mass $\ml$ to zero, while I use the physical masses of the charged mesons \cite{Beringer2012}. In the calculation of the semi-inclusive decay rate with the photon energy cut $\Delta x$, I neglect terms that vanish in the limit $\ml\to0$ and evaluate the large logarithm $\ln\zl$ with the physical electron mass \cite{Beringer2012}.

In the NLO counterterm corrections, the low-energy constants $L_9^r$ and $L_{10}^r$ of the strong sector enter. The lattice determinations of these LECs have not yet reached `green status' in the FLAG report \cite{Aoki2013}. For $L_9^r$, I use the value of \cite{BijnensTalavera2002}, for $L_{10}^r$, I take the $\O(p^4)$ fit of \cite{Gonzalez-Alonso2008}, which is compatible with the available lattice determinations.

As for the case of the mass effects, I again use the estimates of \cite{Ananthanarayan2004,Moussallam1997} for the electromagnetic LECs with a 100\% error assigned to them.

The `leptonic' LECs $X_1^r$ and $X_6^r$ are unknown. $X_6^r$ contains the universal short-distance contribution \cite{Marciano1993}, which I split off following the treatment in \cite{Cirigliano2002}:
\begin{align}
	\begin{split}
		X_6^r(\mu) = \tilde X_6^r(\mu) + X_6^\mathrm{SD}, \quad e^2 X_6^\mathrm{SD} = 1 - S_\mathrm{EW}(M_\rho, M_Z) = - \frac{e^2}{4\pi^2} \ln\left( \frac{M_Z^2}{M_\rho^2} \right) ,
	\end{split}
\end{align}
such that $\tilde X_6^r$ is of the typical size of a LEC in \ChPT{}. I use the naïve dimensional estimate that those LECs are of the order $1/(4\pi)^2$. For the short-distance contribution, I take the value that includes leading logarithmic and QCD corrections \cite{Marciano1993}.

\begin{table}[H]
	\centering
	\begin{tabular}{c r r}
		\toprule
		$10^3 \cdot L_9^r(\mu)$ & $5.93 \pm 0.43\hphantom{0} \qquad$ & \cite{BijnensTalavera2002} \\
		$10^3 \cdot L_{10}^r(\mu)$ & $-5.22 \pm 0.06\hphantom{0} \qquad$ & \cite{Gonzalez-Alonso2008} \\[0.05cm]
		\hline \\[-0.35cm]
		$10^3 \cdot K_1^r(\mu)$ & $ -2.71 \pm 2.71\hphantom{0} \qquad$ & \cite{Ananthanarayan2004} \\
		$10^3 \cdot K_3^r(\mu)$ & $ 2.71 \pm 2.71\hphantom{0}  \qquad$ & \cite{Ananthanarayan2004} \\
		$10^3 \cdot K_5^r(\mu)$ & $ 11.59 \pm 11.59 \qquad$ & \cite{Ananthanarayan2004} \\
		$10^3 \cdot K_{12}^r(\mu)$ & $ -4.25 \pm 4.25\hphantom{0} \qquad$ & \cite{Moussallam1997}  \\[0.05cm]
		\hline \\[-0.35cm]
		$10^3 \cdot X_1^r(\mu)$ & $ 0 \pm 6.3\hphantom{00}  \qquad$ & \\
		$10^3 \cdot \tilde X_6^r(\mu)$ & $ 0 \pm 6.3\hphantom{00} \qquad$ & \\[0.05cm]
		\hline \\[-0.35cm]
		$S_\mathrm{EW}$ & $1.0232\qquad\quad\,$ & \cite{Marciano1993} \\[0.05cm]
		\hline \\[-0.35cm]
		$F_\pi$ & $(92.21 \pm 0.14)$~MeV & \cite{Beringer2012} \\
		\bottomrule
	\end{tabular}
	\caption{Input parameters for the evaluation of the photonic effects ($\mu = 770$~MeV).}
\end{table}

\subsection{Soft Photon Approximation vs.~Full Matrix Element}

\begin{figure}[H]
	\centering
	\large
	\scalebox{0.63}{
		\input{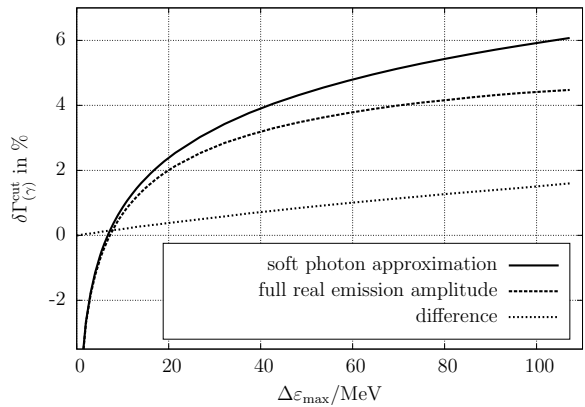}
		}
	\caption{Comparison of the $\O(e^2)$ photonic correction (virtual and real photons) to the semi-inclusive total decay rate as a function of the photon energy cut in $\Sigma_{\ell\nu\gamma}$, using the soft photon approximation vs.~the full radiative matrix element.}
	\label{plot:SPAvsHard}
\end{figure}

In a first step, I want to quantify the importance of considering the full (hard) matrix element for the radiative process instead of relying on the soft photon approximation. To this end, I compare the semi-inclusive total and differential decay rates (using the photon energy cut $\tilde x_\mathrm{max}$) with the decay rate, where the radiative process is just given by the SPA with a finite $\Delta \varepsilon$. The same energy cut in the two descriptions is obtained by setting
\begin{align}
	\begin{split}
		\tilde x_\mathrm{min} = \tilde x_\mathrm{max} \quad \Rightarrow \quad \Delta \varepsilon =  \frac{\sqrt{s_\ell}}{2} \tilde x_\mathrm{max} ( 1 - \zl).
	\end{split}
\end{align}
In this prescription, the photon energy cut is not constant but respects the bounds given by the phase space. The maximum photon energy is
\begin{align}
	\begin{split}
		\Delta\varepsilon_\mathrm{max} = \tilde x_\mathrm{max} \frac{(\mkp-2\mpip)^2 - \ml^2}{2(\mkp-2\mpip)} .
	\end{split}
\end{align}
I compare in the following the corrections to the total decay rate, defined by
\begin{align}
	\begin{split}
		\Gamma_{(\gamma)}^\mathrm{cut} = \Gamma^\mathrm{LO} \left( 1 + \delta \Gamma_{(\gamma)}^\mathrm{cut} \right) .
	\end{split}
\end{align}
In figure~\ref{plot:SPAvsHard}, the correction to the decay rate $\delta \Gamma_{(\gamma)}^\mathrm{cut}$ is shown as a function of the photon energy cut. The virtual corrections are evaluated using the central values of the input parameters. The soft photon approximation depends logarithmically on the energy cut (reflecting the IR divergence at low energies), whereas the correction using the full matrix element is somewhat smaller. Since I use a cut in the dilepton-photon rest frame, the result cannot be applied directly to the experiment, where an energy cut is present in the lab frame. However, I expect that the picture of the difference between full matrix element and soft photon approximation will look similar in the kaon centre-of-mass frame. In the relative form factor measurement of NA48/2, a 3~GeV photon energy cut was applied in the lab frame \cite{Batley2010}. This translates into a minimal detectable photon energy of 11.7~MeV in the centre-of-mass frame. For such a low photon energy, the soft approximation can be expected to still work well (the deviation in $\Sigma_{\ell\nu\gamma}$ is $\approx0.2\%$ of the total rate). However, the experimental cut is not sharp: at the outer edge of the calorimeter, the minimal detectable centre-of-mass photon energy is about 36.8~MeV and of course, only photons flying in the direction of the calorimeter can be detected. At larger photon energies, the error introduced by using a SPA is quite substantial (up to 1.6\% of the total rate for hard photons). This can be understood in terms of the collinear singularity: the SPA alone does not produce the correct dependence on the lepton mass, hence, the large logarithm does not cancel.

As explained before, the gauge invariant class of loop diagrams in figure~\ref{img:Kl4_mLoops} together with the corresponding counterterms has been neglected in the previous literature \cite{Cuplov2003, Cuplov2004}. To judge the influence of these diagrams, I compute the total inclusive decay rate, remove the cut ($\tilde x_\mathrm{max} = 1$) and sum the uncertainties due to the input parameters in quadrature. Using all the diagrams for the virtual corrections, I find
\begin{align}
	\begin{split}
		\delta \Gamma_{(\gamma)} = (4.53 \pm 0.66) \% ,
	\end{split}
\end{align}
whereas neglecting the mentioned class of diagrams results in
\begin{align}
	\begin{split}
		\delta \Gamma_{(\gamma)}^\mathrm{negl.} = (4.70 \pm 0.66) \% .
	\end{split}
\end{align}
The uncertainty is completely dominated by $X_1^r(\mu)$. Note that approximately half of the correction (2.32\%) is due to the short-distance enhancement.

\subsection{Comparison with Coulomb Factor $\times$ PHOTOS}

The Gamow-Sommerfeld (or Coulomb) factor is defined by
\begin{align}
	\begin{split}
		d\Gamma_\mathrm{Coulomb} &= d\Gamma \cdot \prod_{i < j} \frac{\omega_{ij}}{e^{\omega_{ij}}-1} ,
	\end{split}
\end{align}
where $i,j$ run over the three charged final state particles, $\pi^+$, $\pi^-$ and $\ell^+$, and where
\begin{align}
	\begin{split}
		\omega_{ij} := \frac{q_i q_j e^2}{2\beta_{ij}} , \quad \beta_{ij} := \sqrt{ 1 - \frac{4 m_i^2 m_j^2}{(s_{ij}-m_i^2-m_j^2)^2}}, \quad s_{ij} := (p_i + p_j)^2 .
	\end{split}
\end{align}
$q_{i,j}$ denote the charges of the particles in units of $e$.

The Coulomb factor is a semiclassical approximation of the final state interactions. However, it is non-perturbative and includes contributions to all orders in $e^2$. In $K_{e4}$, the factors involving the electron are negligible, the important contribution is the $\pi^+\pi^-$ interaction. An expansion of the Coulomb factor in $e^2$ gives
\begin{align}
	\begin{split}
		\frac{\omega_{\pi^+\pi^-}}{e^{\omega_{\pi^+\pi^-}}-1} = 1 + e^2 \frac{1 + \sigma_\pi^2(s)}{8\sigma_\pi(s)} + \O(e^4) .
	\end{split}
\end{align}
If one expands the triangle diagram~\ref{img:Kl4_NLOgLoop8} for $s$ near the threshold (i.e.~for small values of $\sigma_\pi$), exactly the same contribution to the correction of the decay rate is found, up to terms that are finite for $\sigma_\pi\to0$ (but contain e.g.~the IR divergence). The Coulomb factor is therefore an approximation of a part of the virtual corrections, resummed to all orders. It increases the fully inclusive total decay rate by 3.25\%, the $\O(e^2)$ part being responsible for 3.17\%.

The effect of PHOTOS can be described by a multiplicative factor on the decay rate, too,
\begin{align}
	\begin{split}
		d\Gamma_\mathrm{PHOTOS} = d\Gamma \cdot f_\mathrm{PHOTOS}(s,s_\ell,\cos\theta_\pi,\cos\theta_\ell,\phi) ,
	\end{split}
\end{align}
where I determine $f_\mathrm{PHOTOS}$ numerically through a simulation.

Note that PHOTOS assumes the virtual corrections to take such a value that the divergences cancel but that the fully inclusive total decay rate does not change \cite{Nanava2007}. The NA48/2 experiment however claims that PHOTOS has been used even in the determination of the form factor normalisation, i.e.~to take the effect of real photons on the total decay rate into account \cite{Batley2012}. The inclusion of PHOTOS increased the simulated decay rate by 0.69\%\footnote{B. Bloch-Devaux, private communication.}. I was not able to reproduce this number and suspect it to be only an effect due to finite resolution or statistical fluctuations. The results of my own simulations with a large statistics of $8 \cdot 10^{10}$ events are compatible with the assumption that PHOTOS does not change the fully inclusive total decay rate.

I compare now the results for the fully inclusive as well as for the semi-inclusive differential rate with a photon energy cut of $\Delta \varepsilon_\mathrm{max} = 40$~MeV in $\Sigma_{\ell\nu\gamma}$. I include only the $\O(e^2)$ contribution of the Coulomb factor.

\begin{figure}[H]
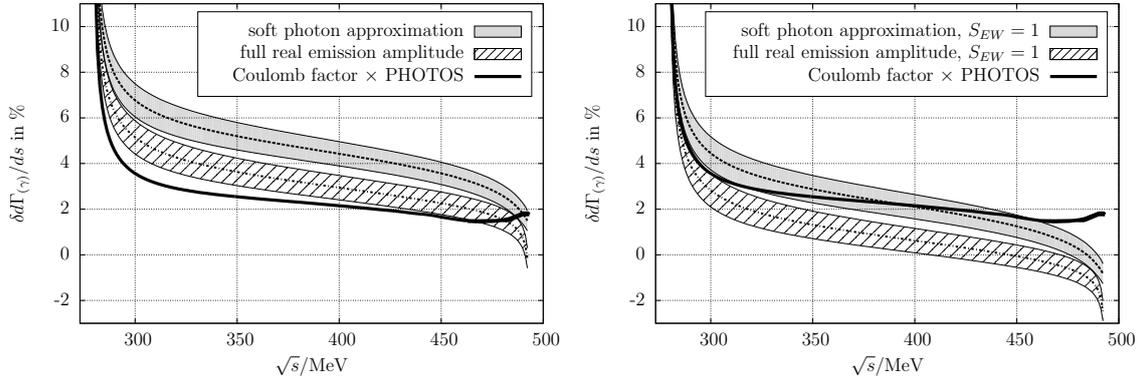

	\centering
	\large
	\scalebox{0.59}{
		\input{plots/SPAvsHardvsCoulPHDiffS.tex}
		\input{plots/SPAvsHardvsCoulPHDiffSNoSD.tex}
		}
	\caption{Comparison of the photonic corrections to the fully inclusive differential decay rate. The right plot excludes the short-distance enhancement factor.}
	\label{plot:SPAvsHardvsCoulPHDiffS}
\end{figure}

\begin{figure}[H]
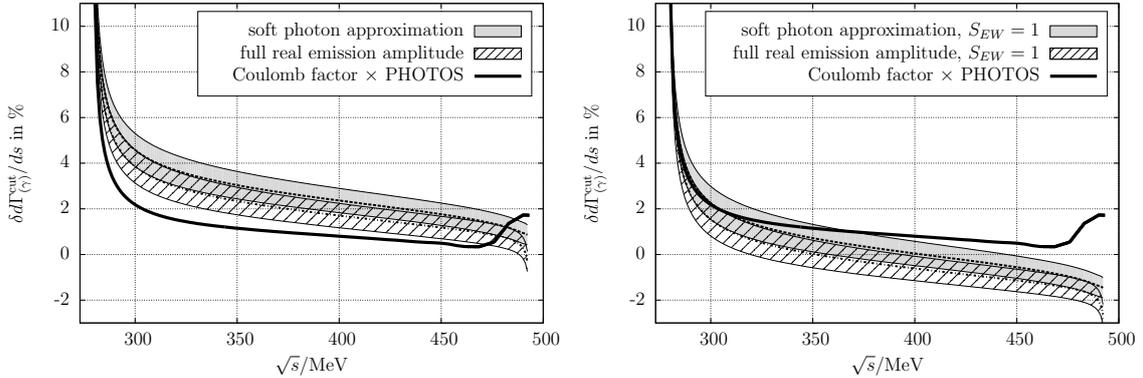

	\centering
	\large
	\scalebox{0.59}{
		\input{plots/SPAvsHardvsCoulPHDiffSCut.tex}
		\input{plots/SPAvsHardvsCoulPHDiffSCutNoSD.tex}
		}
	\caption{Comparison of the photonic corrections to the semi-inclusive total decay rate with a photon energy cut of $\Delta \varepsilon_\mathrm{max} = 40$~MeV in $\Sigma_{\ell\nu\gamma}$. The rise of the PHOTOS factor at large $s$ could be a numerical artefact, as the decay rate approaches zero in this phase space region.}
	\label{plot:SPAvsHardvsCoulPHDiffSCut}
\end{figure}

The plots in figures~\ref{plot:SPAvsHardvsCoulPHDiffS} and \ref{plot:SPAvsHardvsCoulPHDiffSCut} show the corrections to the differential decay rate. The divergence at the $\pi\pi$ threshold is the Coulomb singularity, reproduced in all descriptions. The rise of the PHOTOS factor at large values of $s$, however, could be a numerical artefact, because the differential decay rate drops to zero at the upper border of the phase space.

The comparison without the short-distance enhancement shows that the Coulomb factor $\times$ PHOTOS approach is relatively close to the soft photon approximation, which overestimates the radiative corrections. However, the short-distance factor has not been included in the experimental analysis, such that in total, the radiative corrections are underestimated.

Unfortunately, it is not possible to calculate the radiative corrections for a realistic setup with the experi\-mental cuts. Nevertheless, as NA48/2 determined the branching ratio in a fully inclusive measurement, it is possible to correct the normalisation of the form factors. For the relative values of the form factors, one has to assume that the Coulomb factor $\times$ PHOTOS approach is an acceptable description of the radiative corrections (a free normalisation factor corresponds to a free additive constant in the correction, hence the slopes of the corrections have to be compared).

I suggest to replace in a matching procedure the $\O(e^2)$ part of the Coulomb factor and the 0.69\% PHOTOS effect (or rather artefact) with the result of the here presented fixed order calculation, i.e.~to apply the following correction to the norm of the form factors $X\in\{F,G\}$:
\begin{align}
	\begin{split}
		| X | &= | X^\mathrm{exp} | \left( 1 + \frac{1}{2} \left( \delta\Gamma_\mathrm{Coulomb}^{e^2} + \delta\Gamma_\mathrm{PHOTOS} - \delta \Gamma_{(\gamma)} \right) \right) \\
			&= | X^\mathrm{exp} | \left( 0.9967 \pm 0.0033 \right) , \\
		\Rightarrow \delta |X| &= ( -0.33 \pm 0.33 ) \%.
	\end{split}
\end{align}
Note that replacing the systematic PHOTOS uncertainty with the above error increases the $0.62\%$ uncertainty of the NA48/2 norm measurement \cite{Batley2012} to 0.70\%.

The fact that the a posteriori correction is so small is at least partly accidental: as argued above, I have the strong suspicion that the estimate $\delta \Gamma_\mathrm{PHOTOS}=0.69\%$ is simply the outcome of statistical fluctuations. By chance, this number leads to a result close to the estimate obtained by Diamant-Berger in his analytic treatment of radiative corrections. For this reason, it has been considered so far as a reliable estimate\footnote{B. Bloch-Devaux, private communication.}.


\chapter{Discussion and Conclusion}

In the present work, I have computed the one-loop isospin-breaking corrections to the $K_{\ell4}$ decay within \ChPT{} including leptons and photons. The corrections can be separated into mass effects and photonic effects. The mass effects for the $S$-wave are quite substantial but the result for the norm of the form factors suffers from large uncertainties, on the one hand due to the uncertainty in the LEC $L_4^r$, on the other hand due to higher order corrections. The mass effects for the $P$-waves are negligible.

For the photonic corrections, I have compared the fixed order calculation with the Coulomb $\times$ PHOTOS approach used in the experimental analysis of NA48/2. An a posteriori correction of the data is possible for the normalisation but not for the relative values of the form factors. The present calculation includes for the first time a treatment of the full radiative process and compares it with the soft photon approximation.

For possible forthcoming experiments on $K_{\ell4}$, I suggest that photonic corrections are applied in a Monte Carlo simulation that includes the exact matrix element. This can be done e.g.~with PHOTOS. The mass effects can be easily corrected a posteriori.

This work goes either beyond the isospin-breaking treatments in previous literature or is complementary: I confirm the largest part of the amplitude calculation of \cite{Cuplov2003, Cuplov2004}, but correct their results by a neglected gauge invariant class of diagrams. I have included the full radiative process and shown that the soft photon approximation is not necessarily trustworthy and certainly not applicable for the fully inclusive decay.

I reproduce the NLO mass effect calculation for phases of the form factors done in \cite{Colangelo2009, Bernard2013}, but concentrate here on the absolute values of the form factors. As the NLO mass effect calculation suffers from large uncertainties, an extension of the dispersive framework of \cite{Bernard2013} to the norm of the form factors would be desirable.

To judge the reliability of the photonic corrections, one should ideally calculate them to higher chiral orders, which is however prohibitive (and would bring in many unknown low-energy constants). Here, I have assumed implicitly that the photonic corrections factorise and therefore modify the higher chiral orders with the same multiplicative correction as the lowest order. It is hard to judge if this assumption is justified: for this reason, I have attached a rather conservative estimate of the uncertainties to the photonic corrections presented here.

\clearpage

\section*{Acknowledgements}
\addcontentsline{toc}{chapter}{Acknowledgements}

I cordially thank my thesis advisor, Gilberto Colangelo, for his continuous support during the last years (in many respects and also at unearthly hours) and for having drawn my attention to the isospin-breaking corrections to $K_{\ell4}$, which became a side-project to my thesis. I am very grateful to Brigitte Bloch-Devaux for her time and effort to provide me additional data and information and help me to understand diverse aspects of the experiment. I have enjoyed discussions with many people: Thomas Becher, Vincenzo Cirigliano, Vesna Cuplov, Jürg Gasser, Bastian Kubis, Stefan Lanz, Heiri Leutwyler, Lorenzo Mercolli, Emilie Passemar, Massimo Passera, Stefan Pislak, Christopher Smith, Peter Truöl and Zbigniew~W\c{a}s. I thank Brigitte Bloch-Devaux, Vincenzo Cirigliano, Gilberto Colangelo, Jürg Gasser, Serena Grädel and Emilie Passemar for corrections and many useful comments on the manuscript. I thank the Los Alamos National Laboratory, where part of this work was done, for the hospitality. The Albert Einstein Center for Fundamental Physics is supported by the ‘Innovations- und Kooperationsprojekt C--13’ of the ‘Schweizerische Universitätskonferenz SUK/CRUS’. This work was supported in part by the Swiss National Science Foundation.

\begin{appendices}
	\numberwithin{equation}{chapter}


\chapter{Loop Functions}

\section{Scalar Functions}

I use the following conventions for the scalar loop functions:
\begin{align}
	\begin{split}
		A_0&(m^2) \\
			&:= \frac{1}{i} \int \frac{d^nq}{(2\pi)^n} \frac{1}{[ q^2 - m^2 ]} , \\
		B_0&(p^2, m_1^2, m_2^2) \\
			&:= \frac{1}{i} \int \frac{d^nq}{(2\pi)^n} \frac{1}{[ q^2 - m_1^2 ] [ (q+p)^2 - m_2^2 ]} , \\
		C_0&(p_1^2, (p_1-p_2)^2, p_2^2, m_1^2, m_2^2, m_3^2) \\
			&:= \frac{1}{i} \int \frac{d^nq}{(2\pi)^n} \frac{1}{[ q^2 - m_1^2 ] [ (q+p_1)^2 - m_2^2 ] [ (q+p_2)^2 - m_3^2 ]} , \\
		D_0&(p_1^2, (p_1-p_2)^2,(p_2-p_3)^2,p_3^2,p_2^2,(p_1-p_3)^2, m_1^2, m_2^2, m_3^2, m_4^2) \\
			&:= \frac{1}{i} \int \frac{d^nq}{(2\pi)^n} \frac{1}{[ q^2 - m_1^2 ] [ (q+p_1)^2 - m_2^2 ] [ (q+p_2)^2 - m_3^2 ] [ (q+p_3)^2 - m_4^2 ]} .
	\end{split}
\end{align}

The loop functions $A_0$ and $B_0$ are UV-divergent. The renormalised loop functions are defined in the $\overline{MS}$ scheme by
\begin{align}
	\begin{split}
		\label{eqn:RenormalisedLoopFunctions}
		A_0(m^2) &= -2 m^2 \lambda + \bar A_0(m^2) + \O(4-n) , \\
		B_0(p^2, m_1^2, m_2^2) &= -2\lambda + \bar B_0(p^2, m_1^2, m_2^2) + \O(4-n) ,
	\end{split}
\end{align}
where
\begin{align}
	\begin{split}
		\lambda = \frac{\mu^{n-4}}{16\pi^2} \left( \frac{1}{n-4} - \frac{1}{2} \left( \ln(4\pi) + 1 - \gamma_E \right) \right) .
	\end{split}
\end{align}
$\mu$ denotes the renormalisation scale.

The renormalised loop functions are given by \cite{Amoros2000}
\begin{align}
	\begin{split}
		\bar A_0(m^2) &= -\frac{m^2}{16\pi^2} \ln\left( \frac{m^2}{\mu^2} \right) , \\
		\bar B_0(p^2, m_1^2, m_2^2) &= -\frac{1}{16\pi^2} \frac{m_1^2 \ln\left(\frac{m_1^2}{\mu^2}\right) - m_2^2 \ln\left(\frac{m_2^2}{\mu^2}\right)}{m_1^2 - m_2^2} \\
			&+ \frac{1}{32\pi^2} \left( 2 + \left( -\frac{\Delta}{p^2} + \frac{\Sigma}{\Delta} \right) \ln\left( \frac{m_1^2}{m_2^2} \right) - \frac{\nu}{p^2} \ln\left( \frac{(p^2+\nu)^2 - \Delta^2}{(p^2-\nu)^2 - \Delta^2} \right) \right) ,
	\end{split}
\end{align}
where
\begin{align}
	\begin{split}
		\Delta &:= m_1^2 - m_2^2 , \\
		\Sigma &:= m_1^2 + m_2^2 , \\
		\nu &:= \sqrt{ (s - (m_1 + m_2)^2)(s-(m_1-m_2)^2) } = \lambda^{1/2}(s,m_1^2,m_2^2) .
	\end{split}
\end{align}

\section{Tensor-Coefficient Functions}

\label{sec:AppendixTensorCoefficientFunctions}

Although all the loop integrals can be expressed in terms of the basic scalar loop functions by means of a Passarino-Veltman reduction \cite{Hooft1979, Passarino1979}, this produces sometimes very long polynomial coefficients. I therefore also use the tensor coefficient functions. The tensor integrals that I use are defined by
{ \small
\begin{align}
	\begin{split}
		B^{\mu\nu}(p; m_1^2, m_2^2)
			&:= \frac{1}{i} \int \frac{d^nq}{(2\pi)^n} \frac{q^\mu q^\nu}{[ q^2 - m_1^2 ] [ (q+p)^2 - m_2^2 ]} , \\
		C^\mu(p_1,p_2; m_1^2, m_2^2, m_3^2)
			&:= \frac{1}{i} \int \frac{d^nq}{(2\pi)^n} \frac{q^\mu}{[ q^2 - m_1^2 ] [ (q+p_1)^2 - m_2^2 ] [ (q+p_2)^2 - m_3^2 ]} , \\
		C^{\mu\nu}(p_1,p_2; m_1^2, m_2^2, m_3^2)
			&:= \frac{1}{i} \int \frac{d^nq}{(2\pi)^n} \frac{q^\mu q^\nu}{[ q^2 - m_1^2 ] [ (q+p_1)^2 - m_2^2 ] [ (q+p_2)^2 - m_3^2 ]} , \\
		D^\mu(p_1,p_2,p_3; m_1^2, m_2^2, m_3^2,m_4^2)
			&:= \frac{1}{i} \int \frac{d^nq}{(2\pi)^n} \frac{q^\mu}{[ q^2 - m_1^2 ] [ (q+p_1)^2 - m_2^2 ] [ (q+p_2)^2 - m_3^2 ] [ (q+p_3)^2 - m_4^2 ]} , \\
		D^{\mu\nu}(p_1,p_2,p_3; m_1^2, m_2^2, m_3^2,m_4^2)
			&:= \frac{1}{i} \int \frac{d^nq}{(2\pi)^n} \frac{q^\mu q^\nu}{[ q^2 - m_1^2 ] [ (q+p_1)^2 - m_2^2 ] [ (q+p_2)^2 - m_3^2 ] [ (q+p_3)^2 - m_4^2 ]} .
		\raisetag{-0.25cm}
	\end{split}
\end{align} }%
The tensor coefficients are then given by a Lorentz decomposition:
{ \small
\begin{align}
	\begin{split}
		B^{\mu\nu}(p; m_1^2, m_2^2) &= g^{\mu\nu} B_{00}(p^2, m_1^2, m_2^2) + p^\mu p^\nu B_{11}(p^2, m_1^2, m_2^2) , \\
		C^\mu(p_1, p_2; m_1^2, m_2^2, m_3^2) &= p_1^\mu C_1(p_1^2, (p_1-p_2)^2, p_2^2, m_1^2, m_2^2, m_3^2) \\
			& + p_2^\mu C_2(p_1^2, (p_1-p_2)^2, p_2^2, m_1^2, m_2^2, m_3^2) , \\
		C^{\mu\nu}(p_1, p_2; m_1^2, m_2^2, m_3^2) &= g^{\mu\nu} C_{00}(p_1^2, (p_1-p_2)^2, p_2^2, m_1^2, m_2^2, m_3^2) \\
			& + \sum_{i,j=1}^2 p_i^\mu p_j^\nu C_{ij}(p_1^2, (p_1-p_2)^2, p_2^2, m_1^2, m_2^2, m_3^2) , \\
		D^\mu(p_1, p_2, p_3; m_1^2, m_2^2, m_3^2, m_4^2) &= \sum_{i=1}^3 p_i^\mu D_i(p_1^2, (p_1-p_2)^2,(p_2-p_3)^2,p_3^2,p_2^2,(p_1-p_3)^2, m_1^2, m_2^2, m_3^2, m_4^2) , \\
		D^{\mu\nu}(p_1, p_2, p_3; m_1^2, m_2^2, m_3^2, m_4^2) &= g^{\mu\nu} D_{00}(p_1^2, (p_1-p_2)^2,(p_2-p_3)^2,p_3^2,p_2^2,(p_1-p_3)^2, m_1^2, m_2^2, m_3^2, m_4^2) \\
			& \hspace{-0.5cm} + \sum_{i,j=1}^3 p_i^\mu p_j^\nu D_{ij}(p_1^2, (p_1-p_2)^2,(p_2-p_3)^2,p_3^2,p_2^2,(p_1-p_3)^2, m_1^2, m_2^2, m_3^2, m_4^2) .
	\end{split}
\end{align} }%
Only some of those tensor coefficient functions are UV-divergent:
\begin{align}
	\begin{split}
		\label{eqn:RenormalisedTensorCoefficients}
		B_{00}(p^2, m_1^2, m_2^2) &= -\frac{\lambda}{2} \left( m_1^2 + m_2^2 - \frac{p^2}{3} \right) + \bar B_{00}(p^2, m_1^2, m_2^2) + \O(4-n) , \\
		B_{11}(p^2, m_1^2, m_2^2) &= -\frac{2}{3}\lambda + \bar B_{11}(p^2, m_1^2, m_2^2) + \O(4-n) , \\
		C_{00}(p_1^2, (p_1-p_2)^2, p_2^2, m_1^2, m_2^2, m_3^2) &= -\frac{\lambda}{2} + \bar C_{00}(p_1^2, (p_1-p_2)^2, p_2^2, m_1^2, m_2^2, m_3^2) + \O(4-n) .
	\end{split}
\end{align}

\pagebreak

\section{Infrared Divergences in Loop Functions}

\label{sec:IRdivergentLoopFunctions}

The following explicit formulae are used to extract the IR divergence in the loop functions.

The derivative of the two-point function is IR-divergent:
\begin{align}
	\begin{split}
		\bar B_0^\prime(M^2, M^2, \mg^2) &= - \frac{1}{16\pi^2} \frac{1}{M^2} \left( 1 + \frac{1}{2} \ln\left( \frac{\mg^2}{M^2} \right) \right) + \O(\mg) , \\
		\bar B_0(M^2, M^2, \mg^2) &= \frac{1}{16\pi^2} \left( 1 - \ln\left( \frac{M^2}{\mu^2} \right) \right) + \O(\mg), \\
		\bar B_0(0, M^2, \mg^2) &= - \frac{1}{16\pi^2} \ln\left( \frac{M^2}{\mu^2} \right) + \O(\mg) .
	\end{split}
\end{align}
The IR-divergent three-point function is given by \cite{Beenakker1990}
{\small
\begin{align}
	\begin{split}
		C_0(m^2, s, M^2, \mg^2, m^2, M^2)
			&= \frac{1}{16\pi^2} \frac{x_s}{m M(1-x_s^2)} \Bigg( \ln x_s \left( -\frac{1}{2} \ln x_s + 2 \ln(1-x_s^2) + \ln\left( \frac{m M}{\mg^2} \right) \right) \\
				& - \frac{\pi^2}{6} + \dilog(x_s^2) + \frac{1}{2} \ln^2\left( \frac{m}{M} \right) + \dilog\left( 1 - x_s \frac{m}{M} \right) + \dilog\left( 1 - x_s \frac{M}{m} \right) \Bigg) \\
				& + \O(m_\gamma^2) ,
	\end{split}
\end{align} }%
where
\begin{align}
	\begin{split}
		x_s = -\frac{1 - \sqrt{1 - \frac{4 m M}{s - (m-M)^2}}}{1 + \sqrt{1 - \frac{4 m M}{s - (m-M)^2}}} .
	\end{split}
\end{align}



\chapter{Kinematics}

\section{Lorentz Frames and Transformations in $K_{\ell4}$}

\label{sec:LorentzTransformationsKl4}

Let us first look at the kaon rest frame $\Sigma_K$. From the relations
\begin{align}
	\begin{split}
		P &= p_1 + p_2 = \left(\sqrt{ s + \vec P^2}, \vec P \right) , \\
		L &= p_\ell + p_\nu = \left(\sqrt{ s_\ell + \vec P^2}, - \vec P \right) , \\
		p &= P + L = \left( \mkp, \vec 0 \right) ,
	\end{split}
\end{align}
one finds
\begin{align}
	\vec P^2 = \frac{\lambda_{K\ell}(s)}{4\mkp^2},
\end{align}
where $\lambda_{K\ell}(s) := \lambda(\mkp^2,s,s_\ell)$ and $\lambda(a,b,c):=a^2+b^2+c^2-2(ab+bc+ca)$.

I choose the $x$-axis along the dipion line of flight:
\begin{align}
	\begin{split}
		P &= \left( \frac{\mkp^2-s_\ell+s}{2\mkp}, \frac{\lambda_{K\ell}^{1/2}(s)}{2\mkp}, 0, 0 \right) , \\
		L &= \left( \frac{\mkp^2+s_\ell-s}{2\mkp}, - \frac{\lambda_{K\ell}^{1/2}(s)}{2\mkp}, 0, 0 \right) .
	\end{split}
\end{align}
In the dipion centre-of-mass frame $\Sigma_{2\pi}$, the boosted dipion four-momentum is
\begin{align}
	P^\prime = \Lambda_1^{-1} P = \left( \sqrt{s}, \vec 0 \right) .
\end{align}
$\Lambda_1$ is just a boost in the $x$-direction. Thus, I find
\begin{align}
	\Lambda_1 = \left(\begin{array}{cccc} \frac{\mkp^2+s-s_\ell}{2\mkp\sqrt{s}} & \frac{\lambda_{K\ell}^{1/2}(s)}{2\mkp\sqrt{s}} & 0 & 0 \\ \frac{\lambda_{K\ell}^{1/2}(s)}{2\mkp\sqrt{s}} & \frac{\mkp^2+s-s_\ell}{2\mkp\sqrt{s}} & 0 & 0 \\0 & 0 & 1 & 0 \\0 & 0 & 0 & 1\end{array}\right) .
\end{align}
Analogously, in the dilepton centre-of-mass frame $\Sigma_{\ell\nu}$, the boosted dilepton four-momentum is
\begin{align}
	L^\dprime = \Lambda_2^{-1} L = \left( \sqrt{s_\ell}, \vec 0 \right) .
\end{align}
$\Lambda_2$ is given by a rotation around the $x$-axis and a subsequent boost in the $x$-direction. I find
\begin{align}
	\label{eqn:LorentzTransformation2}
	\Lambda_2 = \left(\begin{array}{cccc} \frac{\mkp^2-s+s_\ell}{2\mkp\sqrt{s_\ell}} & -\frac{\lambda_{K\ell}^{1/2}(s)}{2\mkp\sqrt{s_\ell}} & 0 & 0 \\ -\frac{\lambda_{K\ell}^{1/2}(s)}{2\mkp\sqrt{s_\ell}} & \frac{\mkp^2-s+s_\ell}{2\mkp\sqrt{s_\ell}} & 0 & 0 \\0 & 0 & \cos\phi & \sin\phi \\0 & 0 & -\sin\phi & \cos\phi \end{array}\right) .
\end{align}

Let us determine the momenta of the four final-state particles in the kaon rest frame. In $\Sigma_{2\pi}$, the pion momenta
\begin{align}
	\begin{split}
		p_1^\prime &= \left( \sqrt{\mpip^2 + \vec p^2}, \vec p \right) , \\
		p_2^\prime &= \left( \sqrt{\mpip^2 + \vec p^2}, -\vec p \right)
	\end{split}
\end{align}
satisfy
\begin{align}
	\begin{split}
		P^\prime &= p_1^\prime + p_2^\prime = \left( \sqrt{s}, \vec 0 \right) .
	\end{split}
\end{align}
Therefore, we find
\begin{align}
	\vec p^2 = \frac{s}{4} - \mpip^2 ,
\end{align}
leading to
\begin{align}
	\begin{split}
		p_1^\prime &= \left( \frac{\sqrt{s}}{2}, \sqrt{\frac{s}{4} - \mpip^2} \cos\theta_\pi, \sqrt{\frac{s}{4} - \mpip^2} \sin\theta_\pi, 0 \right) , \\
		p_2^\prime &= \left( \frac{\sqrt{s}}{2}, -\sqrt{\frac{s}{4} - \mpip^2} \cos\theta_\pi, -\sqrt{\frac{s}{4} - \mpip^2} \sin\theta_\pi, 0 \right) .
	\end{split}
\end{align}
The pion momenta in $\Sigma_K$ are then given by
\begin{align}
	\begin{alignedat}{2}
		p_1 &= \Lambda_1 p_1^\prime = &&\Bigg( \frac{\mkp^2+s-s_\ell}{4\mkp} + \frac{\lambda_{K\ell}^{1/2}(s)}{4\mkp}\sigma_\pi(s)\cos\theta_\pi, \\
			& & &\quad \frac{\lambda_{K\ell}^{1/2}(s)}{4\mkp} + \frac{\mkp^2+s-s_\ell}{4\mkp}\sigma_\pi(s)\cos\theta_\pi, \sqrt{\frac{s}{4} - \mpip^2} \sin\theta_\pi, 0 \Bigg) , \\
		p_2 &= \Lambda_1 p_2^\prime = &&\Bigg( \frac{\mkp^2+s-s_\ell}{4\mkp} - \frac{\lambda_{K\ell}^{1/2}(s)}{4\mkp}\sigma_\pi(s)\cos\theta_\pi, \\
			& & &\quad \frac{\lambda_{K\ell}^{1/2}(s)}{4\mkp} - \frac{\mkp^2+s-s_\ell}{4\mkp}\sigma_\pi(s)\cos\theta_\pi, -\sqrt{\frac{s}{4} - \mpip^2} \sin\theta_\pi, 0 \Bigg) ,
	\end{alignedat}
\end{align}
where $\sigma_\pi(s) = \sqrt{1 - \frac{4\mpip^2}{s}}$.

Again, the analogous procedure for the dilepton system leads to the lepton momenta in the kaon system. In $\Sigma_{\ell\nu}$, the lepton momenta are
\begin{align}
	\begin{split}
		p_\ell^\dprime &= \left( \sqrt{\ml^2 + \vec p_\ell^2}, \vec p_\ell \right) , \quad p_\nu^\dprime = \left( |\vec p_\ell |, -\vec p_\ell \right) ,
	\end{split}
\end{align}
satisfying
\begin{align}
	L^\dprime = p_\ell^\dprime + p_\nu^\dprime = \left( \sqrt{s_\ell}, \vec 0 \right) ,
\end{align}
with the solution
\begin{align}
	\vec p_\ell^2 = \frac{\left(s_\ell-\ml^2\right)^2}{4s_\ell} ,
\end{align}
hence
\begin{align}
	\begin{split}
		p_\ell^\dprime &= \left( \frac{s_\ell+\ml^2}{2\sqrt{s_\ell}}, -\frac{s_\ell-\ml^2}{2\sqrt{s_\ell}} \cos\theta_\ell, \frac{s_\ell-\ml^2}{2\sqrt{s_\ell}} \sin\theta_\ell, 0 \right) , \\
		p_\nu^\dprime &= \left( \frac{s_\ell-\ml^2}{2\sqrt{s_\ell}}, \frac{s_\ell-\ml^2}{2\sqrt{s_\ell}} \cos\theta_\ell, -\frac{s_\ell-\ml^2}{2\sqrt{s_\ell}} \sin\theta_\ell, 0 \right) .
	\end{split}
\end{align}
I obtain the lepton momenta in $\Sigma_K$ by applying the Lorentz transformation $\Lambda_2$:
\begin{align}
	\begin{alignedat}{2}
		p_\ell &= \Lambda_2 p_\ell^\dprime = &&\Bigg( (1+z_\ell) \frac{\mkp^2-s+s_\ell}{4\mkp} + (1-z_\ell) \frac{\lambda_{K\ell}^{1/2}(s)}{4\mkp}\cos\theta_\ell, \\
			& & &\quad -(1+z_\ell)\frac{\lambda_{K\ell}^{1/2}(s)}{4\mkp} - (1-z_\ell) \frac{\mkp^2-s+s_\ell}{4\mkp}\cos\theta_\ell , \\
			& & &\quad \frac{s_\ell-\ml^2}{2\sqrt{s_\ell}} \sin\theta_\ell \cos\phi, -\frac{s_\ell-\ml^2}{2\sqrt{s_\ell}} \sin\theta_\ell \sin\phi \Bigg) , \\
		p_\nu &= \Lambda_2 p_\nu^\dprime = &&\Bigg( (1-z_\ell) \left( \frac{\mkp^2-s+s_\ell}{4\mkp} - \frac{\lambda_{K\ell}^{1/2}(s)}{4\mkp}\cos\theta_\ell \right), \\
			& & &\quad -(1-z_\ell) \left( \frac{\lambda_{K\ell}^{1/2}(s)}{4\mkp} - \frac{\mkp^2-s+s_\ell}{4\mkp}\cos\theta_\ell \right), \\
			& & &\quad -\frac{s_\ell-\ml^2}{2\sqrt{s_\ell}} \sin\theta_\ell \cos\phi, \frac{s_\ell-\ml^2}{2\sqrt{s_\ell}} \sin\theta_\ell \sin\phi \Bigg) ,
	\end{alignedat}
\end{align}
where $z_\ell = \ml^2/s_\ell$.

With these explicit expressions for the particle momenta, I calculate in the following all the Lorentz invariant products in terms of the five phase space variables.

The Lorentz invariant squares of the vectors (\ref{eqn:FourMomenta}) are given by
\begin{align}
	\begin{split}
		P^2 &= p_1^2 + 2p_1p_2 + p_2^2 = 2\mpip^2 + 2p_1p_2 = s , \\
		Q^2 &= p_1^2 - 2 p_1 p_2 + p_2^2 = 4\mpip^2 - s , \\
		L^2 &= p_\ell^2 + 2 p_\ell p_\nu + p_\nu^2 = m_\ell^2 + 2p_\ell p_\nu = s_\ell , \\
		N^2 &= p_\ell^2 - 2 p_\ell p_\nu + p_\nu^2 = 2m_\ell^2 - s_\ell.
	\end{split}
\end{align}
The remaining Lorentz invariant products are:
{\small
\begin{align}
	\begin{split}
		PQ ={}& p_1^2 - p_2^2 = 0 , \\
		PL ={}& \frac{1}{2}\left(p^2 - P^2 - L^2\right) = \frac{1}{2}\left(\mkp^2 - s - s_\ell \right) , \\
		PN ={}& \frac{1}{2}\left((p-2p_\nu)^2-P^2-N^2\right) = \frac{1}{2}z_\ell \left(\mkp^2 - s - s_\ell\right) + (1-z_\ell)X\cos\theta_\ell , \\
		QL ={}& Qp = \sigma_\pi X \cos\theta_\pi , \\
		QN ={}& z_\ell \sigma_\pi X \cos\theta_\pi + \sigma_\pi (1-z_\ell) \left\{ \frac{1}{2}\left(\mkp^2-s - s_\ell\right) \cos\theta_\pi \cos\theta_\ell - \sqrt{s s_\ell} \sin\theta_\pi \sin\theta_\ell \cos\phi \vphantom{\frac{1}{2}}\right\} , \\
		LN ={}& (p_\ell+p_\nu)(p_\ell-p_\nu) = m_\ell^2 , \\
		\< LNPQ \> :={}& \epsilon_{\mu\nu\rho\sigma} L^\mu N^\nu P^\rho Q^\sigma = -(1-z_\ell) \sigma_\pi X \sqrt{s_\ell s} \sin\theta_\pi \sin\theta_\ell \sin\phi .
	\end{split}
\end{align} }%

\section{Lorentz Frames and Transformations in $K_{\ell4\gamma}$}

\label{sec:LorentzTransformationsKl4g}

For the radiative process, I copy the results for the dipion subsystem from the $K_{\ell4}$ kinematics and therefore find the following expressions for the momenta in the kaon rest frame $\Sigma_K$:
\begin{align}
	\begin{split}
		P &= \left( \frac{\mkp^2-s_\ell+s}{2\mkp}, \frac{\lambda_{K\ell}^{1/2}(s)}{2\mkp}, 0, 0 \right) , \\
		L &= \left( \frac{\mkp^2+s_\ell-s}{2\mkp}, - \frac{\lambda_{K\ell}^{1/2}(s)}{2\mkp}, 0, 0 \right) .
	\end{split}
\end{align}
\begin{align}
	\begin{split}
		p_1 &= \Bigg( \frac{\mkp^2+s-s_\ell}{4\mkp} + \frac{\lambda_{K\ell}^{1/2}(s)}{4\mkp}\sigma_\pi(s)\cos\theta_\pi, \\
			&\qquad \frac{\lambda_{K\ell}^{1/2}(s)}{4\mkp} + \frac{\mkp^2+s-s_\ell}{4\mkp}\sigma_\pi(s)\cos\theta_\pi, \sqrt{\frac{s}{4} - \mpip^2} \sin\theta_\pi, 0 \Bigg) , \\
		p_2 &= \Bigg( \frac{\mkp^2+s-s_\ell}{4\mkp} - \frac{\lambda_{K\ell}^{1/2}(s)}{4\mkp}\sigma_\pi(s)\cos\theta_\pi, \\
			&\qquad \frac{\lambda_{K\ell}^{1/2}(s)}{4\mkp} - \frac{\mkp^2+s-s_\ell}{4\mkp}\sigma_\pi(s)\cos\theta_\pi, -\sqrt{\frac{s}{4} - \mpip^2} \sin\theta_\pi, 0 \Bigg) .
	\end{split}
\end{align}

We still need to determine the momenta of the photon and the two leptons. The photon and charged lepton momenta in $\Sigma_{\ell\nu\gamma}$ are given by
\begin{align}
	\begin{split}
		q^\dprime &= \Bigg( \frac{\sqrt{s_\ell}}{2} x, - \frac{\sqrt{s_\ell}}{2}\sqrt{x^2 - 4 \zg} \cos\theta_\gamma, \frac{\sqrt{s_\ell}}{2}\sqrt{x^2 - 4 \zg} \sin\theta_\gamma, 0 \Bigg) , \\
		p_\ell^\dprime &= \Bigg( \frac{\sqrt{s_\ell}}{2} y, \begin{aligned}[t]
			& \frac{\sqrt{s_\ell}}{2} \sqrt{y^2 - 4 \zl} \left( \sin\theta_\gamma \sin\theta_{\ell\gamma} \cos\phi_\ell - \cos\theta_\gamma \cos\theta_{\ell\gamma} \right) , \\
			& \frac{\sqrt{s_\ell}}{2} \sqrt{y^2 - 4 \zl} \left( \cos\theta_\gamma \sin\theta_{\ell\gamma} \cos\phi_\ell + \sin\theta_\gamma \cos\theta_{\ell\gamma} \right) , \\
			& \frac{\sqrt{s_\ell}}{2} \sqrt{y^2 - 4 \zl} \sin\theta_{\ell\gamma} \sin\phi_\ell \Bigg) ,
			\end{aligned}
	\end{split}
\end{align}
where $\theta_{\ell\gamma}$ denotes the angle between photon and lepton in $\Sigma_{\ell\nu\gamma}$:
\begin{align}
	\begin{split}
		\cos\theta_{\ell\gamma} = \frac{x(y-2) + 2(1-y + \zl +\zg)}{\sqrt{x^2 - 4 \zg} \sqrt{y^2 - 4 \zl}} .
	\end{split}
\end{align}
The neutrino momentum is then easily found by $p_\nu^\dprime = L^\dprime - q^\dprime - p_\ell^\dprime$.

The momenta in the kaon rest frame $\Sigma_K$ are given by
\begin{align}
	\begin{split}
		q &= \Lambda_2 q^\dprime , \quad p_\ell = \Lambda_2 p_\ell^\dprime , \quad p_\nu = \Lambda_2 p_\nu^\dprime ,
	\end{split}
\end{align}
where $\Lambda_2$ is defined in (\ref{eqn:LorentzTransformation2}). I do not state here the expressions explicitly, as they are rather long. I use them to calculate in the following all the Lorentz invariant products in terms of the eight phase space variables.

The Lorentz invariant squares of the vectors (\ref{eqn:FourMomentaKl4g}) are
\begin{align}
	\begin{split}
		P^2 &= p_1^2 + 2p_1p_2 + p_2^2 = 2\mpip^2 + 2p_1p_2 = s , \\
		Q^2 &= p_1^2 - 2 p_1 p_2 + p_2^2 = 4\mpip^2 - s , \\
		L^2 &= (p_\ell+q)^2 + 2 (p_\ell+q) p_\nu + p_\nu^2 = \sg + 2(p_\ell+q) p_\nu = s_\ell , \\
		N^2 &= (p_\ell+q)^2 - 2 (p_\ell+q) p_\nu + p_\nu^2 = 2 \sg - s_\ell = s_\ell( 2x + 2y - 3) .
	\end{split}
\end{align}

The remaining Lorentz invariant products involving the vectors (\ref{eqn:FourMomentaKl4g}) are given by:
{\small
\begin{align}
	\begin{split}
		PQ ={}& 0 , \quad PL = \frac{1}{2}\left(\mkp^2 - s - s_\ell \right) , \quad QL = \sigma_\pi X \cos\theta_\pi , \quad LN = s_\ell(x+y-1) , \\
		PN ={}& (x+y-1) \frac{1}{2} \left(\mkp^2 - s - s_\ell\right) + X \Big( \sqrt{x^2 - 4 \zg} \cos\theta_\gamma \\
			&+ \sqrt{y^2 - 4 \zl} \left( \cos\theta_{\ell\gamma} \cos\theta_\gamma - \sin\theta_{\ell\gamma} \sin\theta_\gamma \cos\phi_\ell \right) \Big) , \\
		QN ={}& (x+y-1) \sigma_\pi X \cos\theta_\pi \\
			&+ \sigma_\pi \begin{aligned}[t] & \bigg\{ \frac{1}{2}\left(\mkp^2-s - s_\ell\right) \cos\theta_\pi \Big( \sqrt{x^2-4\zg} \cos\theta_\gamma \\
				&\qquad + \sqrt{y^2-4\zl} \left( \cos\theta_{\ell\gamma}\cos\theta_\gamma - \sin\theta_{\ell\gamma}\sin\theta_\gamma\cos\phi_\ell \right) \Big) \\
				& - \sqrt{s s_\ell} \sin\theta_\pi \begin{aligned}[t] &\bigg[ \cos\phi \Big( \sqrt{x^2 - 4 \zg} \sin\theta_\gamma \\
					& \qquad + \sqrt{y^2 - 4\zl} \left( \cos\theta_{\ell\gamma} \sin\theta_\gamma + \sin\theta_{\ell\gamma} \cos\theta_\gamma \cos\phi_\ell \right) \Big) \\
				& + \sin\phi \sqrt{y^2 - 4 \zl} \sin\theta_{\ell\gamma} \sin\phi_\ell \bigg] \bigg\} ,
				\end{aligned} 
			\end{aligned} \\
		\< LNPQ \> :={}& \epsilon_{\mu\nu\rho\sigma} L^\mu N^\nu P^\rho Q^\sigma \\
			={}& - \sigma_\pi X \sqrt{s s_\ell} \sin\theta_\pi 
				\begin{aligned}[t]
					&\Big( \sqrt{x^2 - 4 \zg} \sin\phi \sin\theta_\gamma \\
					&+ \sqrt{y^2 - 4 \zl}
					\begin{aligned}[t]
						& \big( \sin\phi \left( \cos\theta_{\ell\gamma} \sin\theta_\gamma + \sin\theta_{\ell\gamma}\cos\theta_\gamma \cos\phi_\ell \right) \\
						& - \cos\phi \sin\theta_{\ell\gamma} \sin\phi_\ell \big) \Big) .
					\end{aligned}
				\end{aligned}
	\end{split}
\end{align} }%
In addition, we need the Lorentz invariant products involving $q$:
{\small
\begin{align}
	\begin{split}
		Pq &= \frac{x}{4} \left( \mkp^2 - s - s_\ell \right) + \frac{X}{2} \sqrt{x^2 - 4 \zg} \cos\theta_\gamma , \\
		Qq &= \frac{\sigma_\pi}{2}
				\begin{aligned}[t]
					&\bigg[ x X \cos\theta_\pi + \sqrt{x^2 - 4 \zg} 
						\begin{aligned}[t]
							&\bigg( \cos\theta_\pi  \frac{1}{2} (\mkp^2-s-s_\ell) \cos\theta_\gamma  - \sin\theta_\pi \sqrt{s s_\ell} \sin\theta_\gamma \cos\phi \bigg) \bigg] ,
						\end{aligned} \\
				\end{aligned} \\
		Lq &= \frac{s_\ell}{2} x , \\
		Nq &= \frac{s_\ell}{2} \left( x + 2( y - 1 + \zg - \zl ) \right) , \\
		\< LNPq \> &= \frac{1}{2} X s_\ell \sqrt{x^2 - 4\zg} \sqrt{y^2 - 4 \zl} \sin\theta_{\ell\gamma} \sin\theta_\gamma \sin\phi_\ell , \\
		\< LNQq \> &= \frac{1}{2} \sigma_\pi s_\ell \sqrt{x^2-4\zg} \sqrt{y^2-4\zl} \sin\theta_{\ell\gamma} \\
			&\quad \cdot \begin{aligned}[t] 
				& \bigg( \frac{1}{2} ( \mkp^2 - s - s_\ell) \cos\theta_\pi \sin\theta_\gamma \sin\phi_\ell  - \sqrt{s s_\ell} \sin\theta_\pi \left( \sin\phi \cos\phi_\ell - \cos\phi \sin\phi_\ell \cos\theta_\gamma \right) \bigg) ,
				\end{aligned} \\
		\< LPQq \> &= - \frac{1}{2} \sigma_\pi X \sqrt{s s_\ell} \sin\theta_\pi \sqrt{x^2 - 4\zg} \sin\theta_\gamma \sin\phi , \\
		\< NPQq \> &= \frac{1}{2} \sigma_\pi \sqrt{s s_\ell} \\
				& \cdot 
				\begin{aligned}[t]
					&\bigg\{ \sqrt{x^2 - 4\zg} \sqrt{y^2-4\zl} \sin\theta_{\ell\gamma}  \bigg(  - \sqrt{s s_\ell} \cos\theta_\pi \sin\theta_\gamma \sin\phi_\ell \\
					& \quad  + \frac{1}{2}(\mkp^2-s-s_\ell) \sin\theta_\pi \left( \sin\phi \cos\phi_\ell - \cos\phi \sin\phi_\ell \cos\theta_\gamma \right) \bigg) \\
					& + X \sin\theta_\pi
					\begin{aligned}[t]
						&\bigg( x \sqrt{y^2-4\zl}  \big( - \sin\theta_{\ell\gamma} \left( \cos\phi \sin\phi_\ell - \sin\phi \cos\phi_\ell \cos\theta_\gamma \right)  + \cos\theta_{\ell\gamma} \sin\phi \sin\theta_\gamma \big) \\
						& - (y-1) \sqrt{x^2-4\zg} \sin\phi \sin\theta_\gamma \bigg) \bigg\} .
					\end{aligned}
				\end{aligned}
	\end{split}
\end{align} }%


\chapter{Decay Rates}

\label{sec:DecayRates}

\section{Decay Rate for $K_{\ell4}$}

\subsection{Isospin Limit}

\label{sec:DecayRateIsospinLimit}

The partial decay rate for the $K_{\ell4}$ decay is given by
\begin{align}
	d\Gamma = \frac{1}{2\mkp(2\pi)^8} \sum_\mathrm{spins} | \mathcal{T} |^2 \delta^{(4)}(p - P - L) \frac{d^3 p_1}{2p_1^0} \frac{d^3 p_2}{2p_2^0} \frac{d^3 p_\ell}{2p_\ell^0} \frac{d^3 p_\nu}{2p_\nu^0}.
\end{align}
The kinematics of the decay is described by the 5 variables $s$, $s_\ell$, $\theta_\pi$, $\theta_\ell$ and $\phi$. The remaining 7 integrals can be performed explicitly \cite{Cabibbo1965}. Let us review the reduction of the partial decay rate to the five-dimensional phase space integral.

The spin summed square of the matrix element
\begin{align}
	\begin{split}
		\mathcal{T} &= \frac{G_F}{\sqrt{2}} V_{us}^* \bar u(p_\nu) \gamma_\mu (1-\gamma^5)v(p_\ell) \; \big\< \pi^+(p_1) \pi^-(p_2) \big| \bar s \gamma^\mu(1-\gamma^5) u \big| K^+(p) \big\> \\
			&= \frac{G_F}{\sqrt{2}} V_{us}^* \; \mathcal{L}_\mu \mathcal{H}^\mu ,
	\end{split}
\end{align}
where $\mathcal{L}_\mu :=  \bar u(p_\nu) \gamma_\mu (1-\gamma^5)v(p_\ell)$ and $\mathcal{H}^\mu := \big\< \pi^+(p_1) \pi^-(p_2) \big| \bar s \gamma^\mu(1-\gamma^5) u \big| K^+(p) \big\>$, can be written as
\begin{align}
	\begin{split}
		\sum_\mathrm{spins} | \mathcal{T} |^2 &= \frac{G_F^2 |V_{us}|^2}{2} \mathcal{H}^\mu {\mathcal{H}^*}^\nu \sum_\mathrm{spins} \mathcal{L}_\mu \mathcal{L}_\nu^* .
	\end{split}
\end{align}
The spin sum can be performed with standard trace techniques:
\begin{align}
	\begin{split}
		4 \mathcal{L}_{\mu\nu} := \sum_\mathrm{spins} \mathcal{L}_\mu \mathcal{L}_\nu^* &= \sum_\mathrm{spins} \bar u(p_\nu) \gamma_\mu (1-\gamma^5)v(p_\ell) \bar v(p_\ell) \gamma_\nu (1-\gamma^5)u(p_\nu) \\
			&= \tr\left[ \slashed p_\nu \gamma_\mu (1-\gamma^5) (\slashed p_\ell - m_\ell) \gamma_\nu (1-\gamma^5) \right] \\
			&= -2 g_{\mu\nu} ( L^2 - N^2 ) + 4 (L_\mu L_\nu - N_\mu N_\nu) + 4 i \epsilon_{\mu\nu\rho\sigma} L^\rho N^\sigma \\
			&= 4 \left( g_{\mu\nu} ( m_\ell^2 - s_\ell ) + L_\mu L_\nu - N_\mu N_\nu  + i \epsilon_{\mu\nu\rho\sigma} L^\rho N^\sigma \right) .
	\end{split}
\end{align}
After the contraction with the hadronic matrix element, expressed in terms of the form factors,
\begin{align}
	\mathcal{H}^\mu &= -\frac{H}{\mkp^3} \epsilon_{\mu\nu\rho\sigma} L^\nu P^\rho Q^\sigma  + i \frac{1}{\mkp} \left( P_\mu F + Q_\mu G + L_\mu R \right) ,
\end{align}
all the scalar products can be expressed in terms of the five phase space variables $s$, $s_\ell$, $\theta_\pi$, $\theta_\ell$ and $\phi$.

Let us now consider the phase space measure:
\begin{align}
	\begin{split}
		dI :={}& \delta^{(4)}(p-p_1-p_2-p_\ell-p_\nu) \frac{d^3 p_1}{2p_1^0} \frac{d^3 p_2}{2p_2^0} \frac{d^3 p_\ell}{2p_\ell^0} \frac{d^3 p_\nu}{2p_\nu^0} \\
		={}& \delta^{(4)}(p-p_1-p_2-p_\ell-p_\nu) \frac{d^3 p_1}{2p_1^0} \frac{d^3 p_2}{2p_2^0} \frac{d^3 p_\ell}{2p_\ell^0} \frac{d^3 p_\nu}{2p_\nu^0} \\
		&\qquad\cdot \delta^{(4)}(p_1 + p_2 - P) \delta^{(4)}(p_\ell+p_\nu - L) d^4P d^4L \; \theta(P^0) \theta(L^0) \\
		={}& ds \, ds_\ell \, \delta^{(4)}(p-P-L) \, d^4P \delta(s - P^2) \theta(P^0)  d^4L \delta(s_\ell - L^2) \theta(L^0) \\
		&\qquad\cdot \delta^{(4)}(p_1+p_2-P)  \frac{d^3 p_1}{2p_1^0} \frac{d^3 p_2}{2p_2^0} \, \delta^{(4)}(p_\ell+p_\nu - L) \frac{d^3 p_\ell}{2p_\ell^0} \frac{d^3 p_\nu}{2p_\nu^0} .
	\end{split}
\end{align}
The phase space integral can be split into three separately Lorentz invariant pieces:
\begin{align}
	\begin{split}
		dI ={}& dI_1 dI_2 dI_3 , \\
		dI_1 :={}& ds \, ds_\ell \, \delta^{(4)}(p-P-L) \, d^4P \delta(s - P^2) \theta(P^0)  d^4L \delta(s_\ell - L^2) \theta(L^0) , \\
		dI_2 :={}& \delta^{(4)}(p_1+p_2-P)  \frac{d^3 p_1}{2p_1^0} \frac{d^3 p_2}{2p_2^0} , \\
		dI_3 :={}& \delta^{(4)}(p_\ell+p_\nu - L) \frac{d^3 p_\ell}{2p_\ell^0} \frac{d^3 p_\nu}{2p_\nu^0} .
	\end{split}
\end{align}
Each of these three pieces can be evaluated in a convenient frame. For $dI_1$, I choose the kaon rest frame:
\begin{align}
	\begin{split}
		dI_1 &= ds \, ds_\ell \, \delta^{(3)}(\vec p - \vec P - \vec L) \delta\left(p^0 - \sqrt{\vec P^2 + s} - \sqrt{\vec L^2 + s_\ell} \right) \frac{d^3P}{2\sqrt{\vec P^2 + s}} \frac{d^3L}{2\sqrt{\vec L^2 + s_\ell}} \\
		&= ds \, ds_\ell \, \delta^{(3)}(\vec P + \vec L) \delta\left(\mkp - \sqrt{\vec P^2 + s} - \sqrt{\vec L^2 + s_\ell} \right) \frac{d^3P}{2\sqrt{\vec P^2 + s}} \frac{d^3L}{2\sqrt{\vec L^2 + s_\ell}} \\
		&= ds \, ds_\ell \, \delta\left(\mkp - \sqrt{\vec P^2 + s} - \sqrt{\vec P^2 + s_\ell} \right) \frac{d^3P}{2\sqrt{\vec P^2 + s}} \frac{1}{2\sqrt{\vec P^2 + s_\ell}} \\
		&= \pi ds \, ds_\ell \, \delta\left(\mkp - \sqrt{\vec P^2 + s} - \sqrt{\vec P^2 + s_\ell} \right) \frac{\vec P^2}{\sqrt{\vec P^2 + s} \sqrt{\vec P^2 + s_\ell}} d|\vec P| \\
		&= \pi ds \, ds_\ell \, \delta\left( |\vec P| - \frac{\lambda^{1/2}(\mkp^2,s,s_\ell)}{2 \mkp} \right) \frac{|\vec P|}{\sqrt{\vec P^2 + s}+\sqrt{\vec P^2 + s_\ell}}  d|\vec P| \\
		&= \pi ds \, ds_\ell \, \frac{\lambda^{1/2}_{K\ell}(s)}{2\mkp^2} = \pi ds \, ds_\ell \, \frac{X}{\mkp^2} .
	\end{split}
\end{align}
I have used that the integrand depends on $\vec P$ only through $\vec P^2$.

The second piece is evaluated in the dipion frame:
\begin{align}
	\begin{split}
		dI_2 &= \delta^{(3)}(\vec p_1 + \vec p_2 - \vec P) \delta\left(\sqrt{\vec p_1^2 + \mpip^2} + \sqrt{\vec p_2^2 + \mpip^2} - P^0\right) \frac{d^3 p_1}{2\sqrt{\vec p_1^2 + \mpip^2}} \frac{d^3 p_2}{2\sqrt{\vec p_2^2 + \mpip^2}} \\
			&= \delta^{(3)}(\vec p_1 + \vec p_2) \delta\left(\sqrt{\vec p_1^2 + \mpip^2} + \sqrt{\vec p_2^2 + \mpip^2} - \sqrt{s}\right) \frac{d^3 p_1}{2\sqrt{\vec p_1^2 + \mpip^2}} \frac{d^3 p_2}{2\sqrt{\vec p_2^2 + \mpip^2}} \\
			&= \delta\left(2\sqrt{\vec p_1^2 + \mpip^2} - \sqrt{s}\right) \frac{d^3 p_1}{4 (\vec p_1^2 + \mpip^2)} = \delta\left( |\vec p_1| - \sqrt{\frac{s}{4} - \mpip^2} \right) \frac{\pi}{4} \sigma_\pi(s) d\cos\theta_\pi d |\vec p_1| \\
			&= \frac{\pi}{4} \sigma_\pi(s) d\cos\theta_\pi ,
	\end{split}
\end{align}
and the third piece analogously in the dilepton frame:
\begin{align}
	\begin{split}
		dI_3 &= \delta^{(3)}(\vec p_\ell + \vec p_\nu - \vec L) \delta\left(\sqrt{\vec p_\ell^2 + \ml^2} + |\vec p_\nu| - L^0\right) \frac{d^3 p_\ell}{2\sqrt{\vec p_\ell^2 + \ml^2}} \frac{d^3 p_\nu}{2|\vec p_\nu|} \\
			&= \delta^{(3)}(\vec p_\ell + \vec p_\nu) \delta\left(\sqrt{\vec p_\ell^2 + \ml^2} + |\vec p_\nu| - \sqrt{s_\ell}\right) \frac{d^3 p_\ell}{2\sqrt{\vec p_\ell^2 + \ml^2}} \frac{d^3 p_\nu}{2|\vec p_\nu|} \\
			&= \delta\left(\sqrt{\vec p_\ell^2 + \ml^2} + |\vec p_\ell| - \sqrt{s_\ell}\right) \frac{d^3 p_\ell}{4|\vec p_\ell|\sqrt{\vec p_\ell^2 + \ml^2}} = \delta\left( |\vec p_\ell| - \frac{s_\ell-\ml^2}{2\sqrt{s_\ell}} \right) \frac{1}{8}(1 - z_\ell) d\cos\theta_\ell d\phi d |\vec p_\ell| \\
			&= \frac{1}{8} (1-z_\ell) d\cos\theta_\ell d\phi .
	\end{split}
\end{align}
Putting the three pieces together, I find
\begin{align}
	\begin{split}
		dI = \frac{\lambda^{1/2}_{K\ell}(s)}{\mkp^2} \frac{\pi^2}{64} (1-z_\ell) \sigma_\pi(s) \, ds \, ds_\ell \, d\cos\theta_\pi \, d\cos\theta_\ell \, d\phi ,
	\end{split}
\end{align}
and for the differential decay rate
\begin{align}
	\begin{split}
		d\Gamma &= \frac{1}{2^{15}\pi^6} \frac{\lambda^{1/2}_{K\ell}(s)}{\mkp^3} (1-z_\ell) \sigma_\pi(s)  \sum_\mathrm{spins} | \mathcal{T} |^2 \, ds \, ds_\ell \, d\cos\theta_\pi \, d\cos\theta_\ell \, d\phi \\
		&= G_F^2 |V_{us}|^2 \frac{(1-z_\ell) \sigma_\pi(s)X}{2^{13}\pi^6\mkp^3} \mathcal{H}^\mu {\mathcal{H}^*}^\nu \mathcal{L}_{\mu\nu} \, ds \, ds_\ell \, d\cos\theta_\pi \, d\cos\theta_\ell \, d\phi  \\
		&=: G_F^2 |V_{us}|^2 \frac{(1-z_\ell) \sigma_\pi(s)X}{2^{13}\pi^6\mkp^5} J_5(s,s_\ell,\theta_\pi,\theta_\ell,\phi) \, ds \, ds_\ell \, d\cos\theta_\pi \, d\cos\theta_\ell \, d\phi .
	\end{split}
\end{align}
A rather tedious calculation yields (in accordance with \cite{Bijnens1994})
\begin{align}
	\begin{split}
		J_5(s,s_\ell,\theta_\pi,\theta_\ell,\phi) &= \mkp^2 \mathcal{H}^\mu {\mathcal{H}^*}^\nu \mathcal{L}_{\mu\nu} \\
		&= 2(1-z_\ell) \begin{aligned}[t]
			& \bigg[ I_1 + I_2 \cos(2\theta_\ell) + I_3 \sin^2(\theta_\ell) \cos(2\phi) + I_4 \sin(2\theta_\ell) \cos(\phi) \\
			& + I_5 \sin(\theta_\ell) \cos(\phi) + I_6 \cos(\theta_\ell) + I_7 \sin(\theta_\ell) \sin(\phi) + I_8 \sin(2\theta_\ell) \sin(\phi) \\
			& + I_9 \sin^2(\theta_\ell) \sin(2\phi) \bigg] , \end{aligned}
	\end{split}
\end{align}
where
\begin{align}
	\begin{split}
		\label{eqn:DecayRateFormFactorsIsoLimit}
		I_1 :={}& \frac{1}{4} \left((1+z_\ell) |F_1|^2+\frac{1}{2} (3+z_\ell) \sin^2(\theta_\pi) \left(|F_2|^2+|F_3|^2\right)+2 z_\ell |F_4|^2\right) , \\
		I_2 :={}& -\frac{1}{4}(1-z_\ell) \left(|F_1|^2-\frac{1}{2} \sin^2(\theta_\pi)\left(|F_2|^2+|F_3|^2\right)\right) , \\
		I_3 :={}& -\frac{1}{4} (1-z_\ell) \sin^2(\theta_\pi) \left(|F_2|^2-|F_3|^2\right) , \\
		I_4 :={}& \frac{1}{2} (1-z_\ell) \sin(\theta_\pi) \Re\left(F_1^*F_2\right) , \\
		I_5 :={}& -\sin(\theta_\pi) \left(\Re\left(F_1^* F_3\right)+z_\ell \Re\left(F_2^*F_4\right)\right) , \\
		I_6 :={}& z_\ell \Re\left(F_1^*F_4\right)-\sin^2(\theta_\pi) \Re\left(F_2^*F_3\right) , \\
		I_7 :={}& \sin(\theta_\pi) \left(z_\ell \Im\left(F_3^*F_4\right)-\Im\left(F_1^*F_2\right)\right) , \\
		I_8 :={}& \frac{1}{2} (1-z_\ell) \sin(\theta_\pi) \Im\left(F_1^*F_3\right) , \\
		I_9 :={}& -\frac{1}{2} (1-z_\ell) \sin^2(\theta_\pi) \Im\left(F_2^*F_3\right) .
	\end{split}
\end{align}

\subsection{Broken Isospin}

\label{sec:DecayRateIsospinBroken}

In the case of broken isospin, the Lorentz structure of the $K_{\ell4}$ matrix element is modified by the presence of the additional tensorial form factor. The expression for the spin sum has to be adapted. This is, however, the only necessary modification. The phase space is still parametrised by the same five kinematic variables.

The $T$-matrix element is given by (see also (\ref{eqn:TMatrixBrokenIsospin}))
\begin{align}
	\begin{split}
		\mathcal{T} &= \frac{G_F}{\sqrt{2}} V_{us}^* \left( \bar u(p_\nu) \gamma_\mu (1-\gamma^5)v(p_\ell) \mathcal{H}^\mu + \bar u(p_\nu) \sigma_{\mu\nu}(1+\gamma^5) v(p_\ell) \mathcal{T}^{\mu\nu} \right), \\
		\mathcal{H}^\mu &= \mathcal{V}^\mu - \mathcal{A}^\mu , \quad \mathcal{T}^{\mu\nu} = \frac{1}{\mkp^2} p_1^\mu p_2^\nu T .
	\end{split}
\end{align}

Let us calculate the spin sum of the squared $T$-matrix:
{\small
\begin{align}
	\begin{split}
		\sum_\mathrm{spins} | \mathcal{T} |^2 = \frac{G_F^2 |V_{us}|^2}{2} &\Bigg( \mathcal{H}^\mu \mathcal{H^*}^\nu\sum_\mathrm{spins} \mathcal{L}_\mu \mathcal{L}_\nu^* + \mathcal{T}^{\mu\nu} \mathcal{T^*}^{\rho\sigma} \sum_\mathrm{spins} \mathcal{\hat L}_{\mu\nu} \mathcal{\hat L^*}_{\rho\sigma} + 2 \Re \bigg[ \mathcal{H}^\mu \mathcal{T^*}^{\rho\sigma} \sum_\mathrm{spins} \mathcal{L}_\mu \mathcal{\hat L^*}_{\rho\sigma} \bigg]  \Bigg) ,
	\end{split}
\end{align} }%
where again $\mathcal{L}_\mu =  \bar u(p_\nu) \gamma_\mu (1-\gamma^5)v(p_\ell)$ and $\mathcal{\hat L}_{\mu\nu} := \bar u(p_\nu) \sigma_{\mu\nu}(1+\gamma^5) v(p_\ell)$.

The differential decay rate is given by
\begin{align}
	\begin{split}
		d\Gamma &= \frac{1}{2^{15}\pi^6} \frac{\lambda^{1/2}_{K\ell}(s)}{\mkp^3} (1-z_\ell) \sigma_\pi(s)  \sum_\mathrm{spins} | \mathcal{T} |^2 \, ds \, ds_\ell \, d\cos\theta_\pi \, d\cos\theta_\ell \, d\phi \\
		&=: G_F^2 |V_{us}|^2 \frac{(1-z_\ell) \sigma_\pi(s)X}{2^{13}\pi^6\mkp^5} J_5(s,s_\ell,\theta_\pi,\theta_\ell,\phi) \, ds \, ds_\ell \, d\cos\theta_\pi \, d\cos\theta_\ell \, d\phi ,
	\end{split}
\end{align}
where now
\begin{align}
	\begin{split}
		J_5 :={}& J_5^{V-A} + J_5^T + J_5^\mathrm{int} , \\
		J_5^{V-A} :={}&  \frac{\mkp^2}{4} \mathcal{H}^\mu \mathcal{H^*}^\nu \sum_\mathrm{spins} \mathcal{L}_\mu \mathcal{L}_\nu^* , \\
		J_5^{T} :={}&  \frac{\mkp^2}{4} \mathcal{T}^{\mu\nu} \mathcal{T^*}^{\rho\sigma} \sum_\mathrm{spins} \mathcal{\hat L}_{\mu\nu} \mathcal{\hat L^*}_{\rho\sigma} , \\
		J_5^\mathrm{int} :={}&  \frac{\mkp^2}{2} \Re \bigg[ \mathcal{H}^\mu \mathcal{T^*}^{\rho\sigma} \sum_\mathrm{spins} \mathcal{L}_\mu \mathcal{\hat L^*}_{\rho\sigma} \bigg] .
	\end{split}
\end{align}
$J_5^{V-A}$ agrees with $J_5$ in the isospin limit, but with the form factors $F_1$, \ldots, $F_4$ replaced by the isospin corrected ones. $J_5^T$ is due to the tensorial form factor only, $J_5^\mathrm{int}$ is the interference of the tensorial and the $V-A$ part.

$J_5$ can still be written in the form
\begin{align}
	\begin{split}
		J_5(s,s_\ell,\theta_\pi,\theta_\ell,\phi)
		&= 2(1-z_\ell) \begin{aligned}[t]
			&\bigg[ I_1 + I_2 \cos(2\theta_\ell) + I_3 \sin^2(\theta_\ell) \cos(2\phi) + I_4 \sin(2\theta_\ell) \cos(\phi) \\
			& + I_5 \sin(\theta_\ell) \cos(\phi) + I_6 \cos(\theta_\ell) + I_7 \sin(\theta_\ell) \sin(\phi) + I_8 \sin(2\theta_\ell) \sin(\phi) \\
			& + I_9 \sin^2(\theta_\ell) \sin(2\phi) \bigg] , \end{aligned}
	\end{split}
\end{align}
where $I_i = I_i^{V-A} + I_i^T + I_i^\mathrm{int}$. $I_i^{V-A}$ correspond to the functions $I_i$ in the isospin limit (\ref{eqn:DecayRateFormFactorsIsoLimit}). The additional pieces are given by
\begin{align}
	\begin{split}
		\label{eqn:DecayRateFormFactorsIsoBrokenTensorial}
		I_1^T ={}& \frac{1}{4} z_\ell \left((1+z_\ell)+\sin^2(\theta_\pi) \left( (1+3z_\ell) \frac{X^2}{s s_\ell} - \frac{1}{2}(1-z_\ell) \right) \right) |F_5|^2 , \\
		I_2^T ={}& \frac{1}{4} z_\ell (1-z_\ell) \left( 1 - \sin^2(\theta_\pi) \left( \frac{X^2}{s s_\ell} + \frac{3}{2} \right) \right) |F_5|^2 , \\
		I_3^T ={}& \frac{1}{4} z_\ell (1-z_\ell) \sin^2(\theta_\pi) |F_5|^2 , \\
		I_4^T ={}& -\frac{1}{4} z_\ell (1-z_\ell) \sin(2\theta_\pi) \frac{PL}{\sqrt{s s_\ell}} |F_5|^2 , \\
		I_5^T ={}& -\frac{1}{2} z_\ell^2 \sin(2\theta_\pi) \frac{X}{\sqrt{s s_\ell}} |F_5|^2 , \\
		I_6^T ={}& -z_\ell^2 \sin^2(\theta_\pi) \frac{PL \; X}{s s_\ell} |F_5|^2 , \\
		I_7^T ={}& I_8^T = I_9^T = 0
	\end{split}
\end{align}
and
\begin{align}
	\begin{split}
		\label{eqn:DecayRateFormFactorsIsoBrokenInterference}
		I_1^\mathrm{int} ={}& z_\ell \bigg( -\cos(\theta_\pi) \Re(F_1^* F_5) - \frac{PL}{\sqrt{s s_\ell}} \sin^2(\theta_\pi) \Re(F_2^* F_5) - \frac{X}{\sqrt{s s_\ell}} \sin^2(\theta_\pi) \Re(F_3^* F_5) \bigg)  , \\
		I_2^\mathrm{int} ={}& I_3^\mathrm{int} = I_4^\mathrm{int} = 0 , \\
		I_5^\mathrm{int} ={}& z_\ell \bigg( \frac{X}{\sqrt{s s_\ell}} \sin(\theta_\pi) \Re(F_1^* F_5) + \sin(\theta_\pi) \cos(\theta_\pi) \Re(F_3^* F_5) - \frac{PL}{\sqrt{s s_\ell}} \sin(\theta_\pi) \Re(F_4^* F_5) \bigg)  , \\
		I_6^\mathrm{int} ={}& z_\ell \bigg( \frac{X}{\sqrt{s s_\ell}} \sin^2(\theta_\pi) \Re(F_2^* F_5) + \frac{PL}{\sqrt{s s_\ell}} \sin^2(\theta_\pi) \Re(F_3^* F_5) - \cos(\theta_\pi) \Re(F_4^* F_5) \bigg) , \\
		I_7^\mathrm{int} ={}& z_\ell \bigg( \frac{PL}{\sqrt{s s_\ell}} \sin(\theta_\pi) \Im(F_1^* F_5) - \sin(\theta_\pi)\cos(\theta_\pi) \Im(F_2^* F_5) + \frac{X}{\sqrt{s s_\ell}} \sin(\theta_\pi) \Im(F_4^* F_5) \bigg) , \\
		I_8^\mathrm{int} ={}& I_9^\mathrm{int} = 0 .
	\end{split}
\end{align}
These results agree with \cite{Cuplov2004} apart from the different normalisation of $F_5$.

\section{Decay Rate for $K_{\ell4\gamma}$}

\label{sec:RadiativeDecayRate}

The partial decay rate for the $K_{\ell4\gamma}$ decay is given by
\begin{align}
	d\Gamma_\gamma = \frac{1}{2\mkp(2\pi)^{11}} \sum_{\substack{\mathrm{spins} \\ \mathrm{polar.}}} | \mathcal{T}_\gamma |^2 \delta^{(4)}(p - P - L) \frac{d^3 p_1}{2p_1^0} \frac{d^3 p_2}{2p_2^0} \frac{d^3 p_\ell}{2p_\ell^0} \frac{d^3 p_\nu}{2p_\nu^0} \frac{d^3 q}{2q^0}.
\end{align}
The kinematics of the decay is described by the 8 variables $s$, $s_\ell$, $\theta_\pi$, $\theta_\gamma$, $\phi$, $x$, $y$ and $\phi_\ell$. The remaining 7 integrals can be performed explicitly. The reduction of the partial decay rate to the eight-dimensional phase space integral is performed in the following.

The spin summed square of the matrix element
\begin{align}
	\begin{split}
		\mathcal{T}_\gamma &= - \frac{G_F}{\sqrt{2}} e V_{us}^* \epsilon_\mu(q)^* \bigg[ \mathcal{H}^{\mu\nu} \; \mathcal{L}_\nu +  \mathcal{H}_\nu \; \mathcal{\tilde L}^{\mu\nu} \bigg] ,
	\end{split}
\end{align}
where
\begin{align}
	\begin{split}
		\mathcal{L}_\nu &:=  \bar u(p_\nu) \gamma_\nu (1-\gamma^5)v(p_\ell) , \\
		\mathcal{\tilde L}^{\mu\nu} &:= \frac{1}{2 p_\ell q} \bar u(p_\nu) \gamma^\nu (1-\gamma^5)(m_\ell - \slashed p_\ell - \slashed q) \gamma^\mu v(p_\ell) ,
	\end{split}
\end{align}
can be written as
\begin{align}
	\begin{split}
		\sum_{\substack{\mathrm{spins}\\ \mathrm{polar.}}} | \mathcal{T}_\gamma |^2 &= \frac{e^2 G_F^2 |V_{us}|^2}{2} \sum_\mathrm{polar.} \epsilon_\mu(q)^* \epsilon_\rho(q)
			\begin{aligned}[t]
				&\Bigg[ \mathcal{H}_\nu \mathcal{H}^*_\sigma \sum_\mathrm{spins} \mathcal{\tilde L}^{\mu\nu} \mathcal{\tilde L}^{*\rho\sigma} + \mathcal{H}^{\mu\nu} \mathcal{H}^{*\rho\sigma} \sum_\mathrm{spins} \mathcal{L}_\nu \mathcal{L}^*_\sigma \\
				&+ 2 \Re\bigg( \mathcal{H}^{\mu\nu} \mathcal{H}^{*\sigma} \sum_\mathrm{spins} \mathcal{L}_\nu \mathcal{\tilde L}^{*\rho}{}_{\sigma} \bigg) \Bigg] .
			\end{aligned}
	\end{split}
\end{align}
All the spin sums can be performed with standard trace techniques. As I give the photon an artificial small mass $m_\gamma$, I have to use the polarisation sum formula for a massive vector boson:
\begin{align}
	\begin{split}
		\sum_\mathrm{polar.}  \epsilon_\mu(q)^* \epsilon_\rho(q) &= - g_{\mu\rho} + \frac{q_\mu q_\rho}{m_\gamma^2} .
	\end{split}
\end{align}
Using the Ward identity, I find that the second term in the polarisation sum formula does only contribute at~$\O(m_\gamma^2)$:
\begin{align}
	\begin{split}
		& \frac{q_\mu q_\rho}{m_\gamma^2}
			\begin{aligned}[t]
				\Bigg[ \mathcal{H}_\nu \mathcal{H}^*_\sigma & \sum_\mathrm{spins} \mathcal{\tilde L}^{\mu\nu} \mathcal{\tilde L}^{*\rho\sigma} + \mathcal{H}^{\mu\nu} \mathcal{H}^{*\rho\sigma} \sum_\mathrm{spins} \mathcal{L}_\nu \mathcal{L}^*_\sigma + 2 \Re\bigg( \mathcal{H}^{\mu\nu} \mathcal{H}^{*\sigma} \sum_\mathrm{spins} \mathcal{L}_\nu \mathcal{\tilde L}^{*\rho}{}_{\sigma} \bigg) \Bigg]
			\end{aligned} \\
			&= \frac{1}{m_\gamma^2} \Re \Bigg[ \mathcal{H}^\nu \mathcal{H}^{*\sigma} \sum_\mathrm{spins} \left( q^\mu q^\rho \mathcal{\tilde L}_{\mu\nu} \mathcal{\tilde L}_{\rho\sigma}^* + \mathcal{L}_\nu \mathcal{L}_\sigma^* + 2 q^\rho \mathcal{L}_\nu \mathcal{\tilde L}_{\rho\sigma}^* \right) \Bigg] \\
			&= \frac{4 m_\gamma^2}{( \hat L q + \hat N q )^2} \Re \Bigg[ \mathcal{H}^\nu \mathcal{H}^{*\sigma} \left( g_{\nu\sigma} \frac{\hat N^2 - \hat L^2}{2} + \hat L_\nu \hat L_\sigma - \hat N_\nu \hat N_\sigma + i \epsilon_{\nu\sigma\alpha\beta} \hat L^\alpha \hat N^\beta \right) \Bigg].
	\end{split}
\end{align}
I therefore find the following results for the spin and polarisation sums:
{\small
\begin{align}
	\begin{split}
		\sum_{\substack{\mathrm{spins} \\ \mathrm{polar.}}}  \epsilon_\mu(q)^* \epsilon_\rho(q) \mathcal{\tilde L}^{\mu\nu} \mathcal{\tilde L}^{*\rho\sigma} &= \frac{8}{Lq + Nq}
			\begin{aligned}[t]
				&\bigg( g^{\nu\sigma} (Nq - Lq) + q^\nu L^\sigma + q^\sigma L^\nu - q^\nu N^\sigma - q^\sigma N^\nu \\
				&+ i \epsilon^{\nu\sigma\alpha\beta} L_\alpha q_\beta - i \epsilon^{\nu\sigma\alpha\beta} N_\alpha q_\beta \bigg) \\
			\end{aligned} \\
			& - \frac{16 \ml^2}{(Lq + Nq)^2} \cdot \Big(g^{\nu\sigma} \frac{N^2-L^2}{2} + L^\nu L^\sigma - N^\nu N^\sigma + i \epsilon^{\nu\sigma\alpha\beta} L_\alpha N_\beta \Big) + \O(m_\gamma^2) ,
	\end{split}
\end{align}
\begin{align}
	\begin{split}
		\sum_{\substack{\mathrm{spins} \\ \mathrm{polar.}}}  \epsilon_\mu(q)^* \epsilon_\rho(q) \mathcal{L}_\nu \mathcal{L}_\sigma^* &= - 4 g_{\mu\rho} \bigg( g_{\nu\sigma} \frac{\hat N^2 - \hat L^2}{2} + \hat L_\nu \hat L_\sigma - \hat N_\nu \hat N_\sigma + i \epsilon_{\nu\sigma\alpha\beta} \hat L^\alpha \hat N^\beta \bigg) + \O(m_\gamma^2) , \\
	\end{split}
\end{align}
\begin{align}
	\begin{split}
		\sum_{\substack{\mathrm{spins} \\ \mathrm{polar.}}}  \epsilon_\mu(q)^* \epsilon_\rho(q) \mathcal{L}_\nu \mathcal{\tilde L}^{*\rho}{}_\sigma &= \frac{4}{Lq + Nq} \Bigg[ L_\mu L_\nu L_\sigma - N_\mu N_\nu N_\sigma + N_\mu L_\nu L_\sigma - L_\mu N_\nu N_\sigma \\
			& - q_\mu L_\nu L_\sigma + q_\mu N_\nu N_\sigma - q_\nu L_\mu L_\sigma + q_\nu N_\mu N_\sigma + q_\sigma L_\mu N_\nu - q_\sigma L_\nu N_\mu \\
			& + g_{\mu\nu} \left( \frac{N^2-L^2}{2} q_\sigma - Nq \, L_\sigma + Lq \, N_\sigma \right) + g_{\mu\sigma} \left( \frac{L^2-N^2}{2} q_\nu - Lq \, L_\nu + Nq \, N_\nu \right) \\
			& + g_{\nu\sigma} \left( \frac{N^2-L^2}{2} (L_\mu + N_\mu - q_\mu) + Lq \, L_\mu - Nq \, N_\mu \right) - i g_{\nu\sigma} \epsilon_{\mu\alpha\beta\gamma} L^\alpha N^\beta q^\gamma \\
			& + (L_\sigma - N_\sigma) \frac{i}{2} \epsilon_{\mu\nu\alpha\beta} (L^\alpha + N^\alpha) q^\beta + (L_\nu - N_\nu) \frac{i}{2} \epsilon_{\mu\sigma\alpha\beta} (L^\alpha + N^\alpha) q^\beta \\
			& + (L_\mu + N_\mu) \frac{i}{2} \epsilon_{\nu\sigma\alpha\beta} (-L^\alpha + N^\alpha) q^\beta + (L_\mu + N_\mu - q_\mu) i \epsilon_{\nu\sigma\alpha\beta} L^\alpha N^\beta \\
			& + \frac{i}{2}\epsilon_{\mu\nu\sigma\alpha} (L^\alpha - N^\alpha) (Lq + Nq)  \Bigg] + \O(m_\gamma^2) .
	\end{split}
\end{align} }%

I perform the contraction with the hadronic part and express all the scalar products in terms of the eight phase space variables. Neglecting the contribution form the anomalous sector, one can express the hadronic matrix elements in terms of the following form factors:
\begin{align}
	\begin{split}
		\mathcal{H}^\mu &= \frac{i}{\mkp} \left( P^\mu F + Q^\mu G + L^\mu R \right) , \\
		\mathcal{H}^{\mu\nu} &= \frac{i}{\mkp} g^{\mu\nu} \Pi + \frac{i}{\mkp^2}\left( P^\mu \Pi_0^\nu + Q^\mu \Pi_1^\nu + L^\mu \Pi_2^\nu \right) , \\
		\Pi_i^\nu &= \frac{1}{\mkp} \left( P^\nu \Pi_{i0} + Q^\nu \Pi_{i1} + L^\nu \Pi_{i2} + q^\nu \Pi_{i3}  \right) .
	\end{split}
\end{align}
The $K_{\ell4}$ form factors $F$, $G$, $R$ depend on scalar products of $P$, $Q$ and $L$, hence, they can be expressed as functions of $s$, $s_\ell$ and $\theta_\pi$. The $K_{\ell4\gamma}$ form factors $\Pi$ and $\Pi_{ij}$ depend on the scalar products of $P$, $Q$, $L$ and $q$. They are therefore functions of the six phase space variables $s$, $s_\ell$, $\theta_\pi$, $\theta_\gamma$, $\phi$ and $x$.

I consider now the phase space measure:
\begin{align}
	\begin{split}
		dI_\gamma :={}& \delta^{(4)}(p-p_1-p_2-p_\ell-p_\nu-q) \frac{d^3 p_1}{2p_1^0} \frac{d^3 p_2}{2p_2^0} \frac{d^3 p_\ell}{2p_\ell^0} \frac{d^3 p_\nu}{2p_\nu^0} \frac{d^3q}{2q^0} \\
		={}& \delta^{(4)}(p-p_1-p_2-p_\ell-p_\nu-q) \frac{d^3 p_1}{2p_1^0} \frac{d^3 p_2}{2p_2^0} \frac{d^3 p_\ell}{2p_\ell^0} \frac{d^3 p_\nu}{2p_\nu^0}\frac{d^3q}{2q^0} \\
		&\qquad \cdot \delta^{(4)}(p_1 + p_2 - P) \delta^{(4)}(p_\ell+p_\nu+q - L) d^4P d^4L \; \theta(P^0) \theta(L^0) \\
		={}& ds \, ds_\ell \, \delta^{(4)}(p-P-L) \, d^4P \delta(s - P^2) \theta(P^0)  d^4L \delta(s_\ell - L^2) \theta(L^0) \\
		&\qquad \cdot \delta^{(4)}(p_1+p_2-P)  \frac{d^3 p_1}{2p_1^0} \frac{d^3 p_2}{2p_2^0} \, \delta^{(4)}(p_\ell+p_\nu+q - L) \frac{d^3 p_\ell}{2p_\ell^0} \frac{d^3 p_\nu}{2p_\nu^0} \frac{d^3q}{2q^0} .
	\end{split}
\end{align}
The phase space integral can again be split into three separately Lorentz invariant pieces:
\begin{align}
	\begin{split}
		dI_\gamma ={}& dI_1^\gamma dI_2^\gamma dI_3^\gamma , \\
		dI_1^\gamma :={}& ds \, ds_\ell \, \delta^{(4)}(p-P-L) \, d^4P \delta(s - P^2) \theta(P^0)  d^4L \delta(s_\ell - L^2) \theta(L^0) , \\
		dI_2^\gamma :={}& \delta^{(4)}(p_1+p_2-P)  \frac{d^3 p_1}{2p_1^0} \frac{d^3 p_2}{2p_2^0} , \\
		dI_3^\gamma :={}& \delta^{(4)}(p_\ell+p_\nu+q - L) \frac{d^3 p_\ell}{2p_\ell^0} \frac{d^3 p_\nu}{2p_\nu^0} \frac{d^3q}{2q^0} .
	\end{split}
\end{align}
Each of these three pieces can be evaluated in a convenient frame. $dI_1^\gamma$ and $dI_2^\gamma$ can be evaluated in complete analogy to $K_{\ell4}$, i.e.~in the kaon and dipion rest frames:
\begin{align}
	\begin{split}
		dI_1^\gamma &=  \pi ds \, ds_\ell \, \frac{\lambda^{1/2}_{K\ell}(s)}{2\mkp^2} = \pi ds \, ds_\ell \, \frac{X}{\mkp^2} , \quad dI_2^\gamma = \frac{\pi}{4} \sigma_\pi(s) d\cos\theta_\pi .
	\end{split}
\end{align}

The third piece represents now a three body decay. I first perform the neutrino momentum integrals in the three body rest frame:
\begin{align}
	\begin{split}
		dI_3^\gamma &= \delta^{(3)}(\vec p_\ell + \vec p_\nu + \vec q - \vec L) \delta\left(\sqrt{\vec p_\ell^2 + \ml^2} + |\vec p_\nu| + \sqrt{\vec q^2 + \mg^2} - L^0\right) \frac{d^3 p_\ell}{2\sqrt{\vec p_\ell^2 + \ml^2}} \frac{d^3 p_\nu}{2|\vec p_\nu|} \frac{d^3q}{2\sqrt{\vec q^2 + \mg^2}} \\
			&= \delta\left(\sqrt{\vec p_\ell^2 + \ml^2} + |\vec p_\ell + \vec q| + \sqrt{\vec q^2 + \mg^2} - \sqrt{s_\ell} \right) \frac{d^3 p_\ell}{2\sqrt{\vec p_\ell^2 + \ml^2}} \frac{1}{2|\vec p_\ell + \vec q|} \frac{d^3q}{2\sqrt{\vec q^2 + \mg^2}} \\
			&= \delta\left(\sqrt{|\vec p_\ell|^2 + \ml^2} + \sqrt{ |\vec p_\ell|^2 + |\vec q|^2 + 2 |\vec p_\ell| |\vec q| \cos\theta_{\ell\gamma} } + \sqrt{|\vec q|^2 + \mg^2} - \sqrt{s_\ell} \right) \\
				&\quad \cdot \frac{|\vec p_\ell|^2 d |\vec p_\ell| d\cos\theta_{\ell\gamma} d\phi_\ell |\vec q|^2 d|\vec q| d\cos\theta_\gamma d\phi}{8\sqrt{|\vec p_\ell|^2 + \ml^2} \sqrt{ |\vec p_\ell|^2 + |\vec q|^2 + 2 |\vec p_\ell| |\vec q| \cos\theta_{\ell\gamma} } \sqrt{|\vec q|^2 + \mg^2}} \\
			&= \frac{|\vec p_\ell| |\vec q|}{8\sqrt{|\vec p_\ell|^2 + \mg^2}\sqrt{|\vec q|^2 + \mg^2}} d|\vec p_\ell| d|\vec q| d\phi_\ell d\cos\theta_\gamma d\phi \\
			&= \frac{1}{8} d p_\ell^0 dq^0 d\phi_\ell d\cos\theta_\gamma d\phi = \frac{s_\ell}{32} dx dy d\phi_\ell d\cos\theta_\gamma d\phi ,
	\end{split}
\end{align}
where I have used the angle $\theta_{\ell\gamma}$ between the photon and the lepton.

Putting the three pieces together, I find
\begin{align}
	\begin{split}
		dI_\gamma &= \frac{\lambda^{1/2}_{K\ell}(s)}{\mkp^2} \frac{\pi^2}{256} \sigma_\pi(s) s_\ell \, ds \, ds_\ell \, d\cos\theta_\pi \, d\cos\theta_\gamma \, d\phi \, dx \, dy \, d\phi_\ell , 
	\end{split}
\end{align}
and for the differential decay rate
\begin{align}
	\begin{split}
		d\Gamma_\gamma &= \frac{1}{2\mkp(2\pi)^{11}} \sum_{\substack{\mathrm{spins} \\ \mathrm{polar.}}} | \mathcal{T}_\gamma |^2 dI_\gamma \\
			&= \frac{1}{2^{20}\pi^9} \frac{\lambda^{1/2}_{K\ell}(s)}{\mkp^3} \sigma_\pi(s) s_\ell  \sum_{\substack{\mathrm{spins} \\ \mathrm{polar.}}} | \mathcal{T}_\gamma |^2 \, ds \, ds_\ell \, d\cos\theta_\pi \, d\cos\theta_\gamma \, d\phi \, dx \, dy \, d\phi_\ell \\
			&= G_F^2 |V_{us}|^2 e^2 \frac{s_\ell \, \sigma_\pi(s) X}{2^{20}\pi^9 \mkp^7} J_8 \, ds \, ds_\ell \, d\cos\theta_\pi \, d\cos\theta_\gamma \, d\phi \, dx \, dy \, d\phi_\ell ,
	\end{split}
\end{align}
where
\begin{align}
	\begin{split}
		J_8 &= \mkp^4 \sum_\mathrm{polar.} \epsilon_\mu(q)^* \epsilon_\rho(q)
			\begin{aligned}[t]
				&\Bigg[ \mathcal{H}_\nu \mathcal{H}^*_\sigma \sum_\mathrm{spins} \mathcal{\tilde L}^{\mu\nu} \mathcal{\tilde L}^{*\rho\sigma} + \mathcal{H}^{\mu\nu} \mathcal{H}^{*\rho\sigma} \sum_\mathrm{spins} \mathcal{L}_\nu \mathcal{L}^*_\sigma \\
				& + 2 \Re\bigg( \mathcal{H}^{\mu\nu} \mathcal{H}^{*\sigma} \sum_\mathrm{spins} \mathcal{L}_\nu \mathcal{\tilde L}^{*\rho}{}_{\sigma} \bigg) \Bigg] .
			\end{aligned}
	\end{split}
\end{align}


\chapter{\ChPT{} with Photons and Leptons}

\label{sec:AppendixChPT}

In order to settle the conventions, I collect here the most important formulae needed to define \ChPT{} with photons and leptons \cite{Weinberg1968, GasserLeutwyler1984, GasserLeutwyler1985, Urech1995,Knecht2000}.

We consider $SU(3)$ \ChPT{}, where the Goldstone bosons are collected in the $SU(3)$ matrix
\begin{align}
	\begin{split}
		U = \exp\left( \frac{i \sqrt{2}}{F_0} \phi \right) ,
	\end{split}
\end{align}
with
{\small
\begin{align}
	\begin{split}
		\phi &= \sum_{a=1}^8 \lambda_a \phi_a = \left( \begin{matrix}
				\pi^0 \left( \frac{1}{\sqrt{2}} + \frac{\epsilon}{\sqrt{6}} \right) + \eta \left( \frac{1}{\sqrt{6}} - \frac{\epsilon}{\sqrt{2}} \right) & \pi^+ & K^+ \\
				\pi^- & \pi^0 \left(  \frac{\epsilon}{\sqrt{6}} - \frac{1}{\sqrt{2}} \right) + \eta \left( \frac{1}{\sqrt{6}} + \frac{\epsilon}{\sqrt{2}} \right) & K^0 \\
				K^- & \bar K^0 & - \eta \sqrt{\frac{2}{3}} - \pi^0 \sqrt{\frac{2}{3}} \epsilon
			 \end{matrix} \right) .
		\raisetag{-0.1cm}
	\end{split}
\end{align} }%
At leading order, the Lagrangian is given by\footnote{I denote by $\< \cdot \>$ the flavour trace.}
\begin{align}
	\begin{split}
		\mathcal{L}_\mathrm{eff}^\mathrm{LO} &= \mathcal{L}_{p^2} + \mathcal{L}_{e^2} + \mathcal{L}_\mathrm{QED}, \\
		\mathcal{L}_{p^2} &= \frac{F_0^2}{4} \< D_\mu U D^\mu U^\dagger + \chi U^\dagger + U \chi^\dagger \> , \\
		\mathcal{L}_{e^2} &= e^2 F_0^4 Z \< U Q U^\dagger Q \> , \\
		\mathcal{L}_\mathrm{QED} &= - \frac{1}{4} F_{\mu\nu} F^{\mu\nu} + \sum_{\ell} \left[ \bar \ell (i \slashed \p + e \slashed A - m_\ell) \ell + \bar \nu_{\ell L} i \slashed \p \nu_{\ell L} \right] ,
	\end{split}
\end{align}
where
\begin{align}
	\begin{split}
		D_\mu U &= \p_\mu U - i r_\mu U + i U l_\mu , \\
		\chi &= 2 B_0 (s + i p) , \quad r_\mu = v_\mu + a_\mu , \quad l_\mu = v_\mu - a_\mu , \\
		F_{\mu\nu} &= \p_\mu A_\nu - \p_\nu A_\mu , \\
		\nu_{\ell L} &= \frac{1-\gamma_5}{2} \nu_\ell .
	\end{split}
\end{align}
The external fields are fixed by
\begin{align}
	\begin{split}
		s + i p &= \mathcal{M} = \mathrm{diag}( m_u, m_d, m_s ) , \\
		r_\mu &= -e A_\mu Q , \\
		l_\mu &= -e A_\mu Q  +  \sum_{\ell} \left( \bar \ell \gamma_\mu \nu_{\ell L} Q_L^w + \bar \nu_{\ell L} \gamma_\mu \ell Q_L^{w\dagger} \right) , \\
		Q &= \frac{1}{3} \mathrm{diag}( 2, -1, -1 ) , \\
		Q_L^w &= -2 \sqrt{2} G_F T, \quad T = \left( \begin{matrix} 0 & V_{ud} & V_{us} \\ 0 & 0 & 0 \\ 0 & 0 & 0 \end{matrix} \right) .
	\end{split}
\end{align}

By expanding $\mathcal{L}_\mathrm{eff}^\mathrm{LO}$ in the meson fields, we can extract the mass terms. At leading order, I find:
\begin{align}
	\begin{split}
		\mpio^2 &= 2 B_0 \hat m , \\
		\mpip^2 &= 2 B_0 \hat m + 2 e^2 Z F_0^2 , \\
		\mko^2 &= B_0\left( m_s + \hat m + \frac{2\epsilon}{\sqrt{3}}( m_s - \hat m ) \right) , \\
		\mkp^2 &= B_0\left( m_s + \hat m - \frac{2\epsilon}{\sqrt{3}}( m_s - \hat m ) \right) + 2 e^2 Z F_0^2 , \\
		\meta^2 &= \frac{4}{3} B_0 \left( m_s + \frac{\hat m}{2} \right) .
	\end{split}
\end{align}
At this order, the masses obey the Gell-Mann -- Okubo relation:
\begin{align}
	\begin{split}
		2 \mkp^2 + 2 \mko^2 - 2 \mpip^2 + \mpio^2 = 3 \meta^2.
	\end{split}
\end{align}
Let us define
\begin{align}
	\label{eqn:LOMassDifferences}
	\begin{split}
		\Delta_\pi &:= \mpip^2 - \mpio^2 = 2 e^2 Z F_0^2 , \\
		\Delta_K &:= \mkp^2 - \mko^2 = 2 e^2 Z F_0^2 + B_0 (m_u - m_d) .
	\end{split}
\end{align}

The next-to-leading-order Lagrangian is given by
\begin{align}
	\begin{split}
		\mathcal{L}_\mathrm{eff}^\mathrm{NLO} &= \mathcal{L}_\mathrm{eff}^\mathrm{LO} + \mathcal{L}_{p^4} + \mathcal{L}_{e^2 p^2} + \mathcal{L}_\mathrm{lept} + \mathcal{L}_\gamma,
	\end{split}
\end{align}
where
\begin{align}
	\begin{split}
		\mathcal{L}_{p^4} &= L_1 \< D_\mu U D^\mu U^\dagger \> \<D_\nu U D^\nu U^\dagger \> + L_2 \< D_\mu U D_\nu U^\dagger \> \< D^\mu U D^\nu U^\dagger \> \\
			&+ L_3 \< D_\mu U D^\mu U^\dagger D_\nu U D^\nu U^\dagger \> + L_4 \< D_\mu U D^\mu U^\dagger \> \< \chi U^\dagger + U \chi^\dagger \> \\
			&+ L_5 \< D_\mu U D^\mu U^\dagger ( \chi U^\dagger + U \chi^\dagger ) \> + L_6 \< \chi U^\dagger + U \chi^\dagger \>^2 + L_7 \< \chi U^\dagger - U \chi^\dagger \>^2 \\
			&+ L_8 \< U \chi^\dagger U \chi^\dagger + \chi U^\dagger \chi U^\dagger \> - i L_9  \< F_R^{\mu\nu} D_\mu U D_\nu U^\dagger + F_L^{\mu\nu} D_\mu U^\dagger D_\nu U \> + L_{10} \< U F_L^{\mu\nu} U^\dagger F^R_{\mu\nu} \> \\
			&+ H_1 \< F_R^{\mu\nu} F^R_{\mu\nu} + F_L^{\mu\nu} F^L_{\mu\nu} \> + H_2 \< \chi \chi^\dagger \> , \\
	\end{split}
\end{align}
\begin{align}
	\begin{split}
		\mathcal{L}_{e^2 p^2} &= e^2 F_0^2 \begin{aligned}[t]
			&\bigg\{ K_1 \< Q Q \> \< D_\mu U D^\mu U^\dagger \>  + K_2 \< Q U^\dagger Q U \> \< D_\mu U D^\mu U^\dagger \> \\
			& + K_3 \left( \< Q U^\dagger D_\mu U \> \< Q U^\dagger D^\mu U \> + \< Q U D_\mu U^\dagger \> \< Q U D^\mu U^\dagger \> \right) \\
			& + K_4 \< Q U^\dagger D_\mu U \> \< Q U D^\mu U^\dagger \>  + K_5 \< Q Q ( D_\mu U^\dagger D^\mu U + D_\mu U D^\mu U^\dagger) \> \\
			& + K_6 \< U Q U^\dagger Q D_\mu U D^\mu U^\dagger + U^\dagger Q U Q D_\mu U^\dagger D^\mu U \> + K_7 \< Q Q \> \< \chi U^\dagger + U \chi^\dagger \> \\
			& + K_8 \< Q U^\dagger Q U \> \< \chi U^\dagger + U \chi^\dagger \>  + K_9 \< Q Q ( U^\dagger \chi + \chi^\dagger U + \chi U^\dagger + U \chi^\dagger ) \> \\
			& + K_{10} \< Q U^\dagger Q \chi + Q U Q \chi^\dagger + Q U^\dagger Q U \chi^\dagger U + Q U Q U^\dagger \chi U^\dagger \> \\
			& - K_{11} \< Q U^\dagger Q \chi + Q U Q \chi^\dagger - Q U^\dagger Q U \chi^\dagger U - Q U Q U^\dagger \chi U^\dagger \> \\
			& + i K_{12} \< \big[ [ l_\mu , Q ] , Q \big] D^\mu U^\dagger U + \big[ [ r_\mu , Q ] , Q \big] D^\mu U U^\dagger \> \\
			& - K_{13} \< [ l_\mu , Q ] U^\dagger [ r^\mu , Q ] U \> + 2 K_{14} \< l_\mu [ l^\mu , Q ] Q + r_\mu [ r^\mu , Q ] Q \> \bigg\} ,
			\end{aligned} \\
	\end{split}
\end{align}
\begin{align}
	\begin{split}
		\mathcal{L}_\mathrm{lept} &= e^2 \sum_\ell \begin{aligned}[t]
			&\bigg\{ F_0^2 \begin{aligned}[t]
				&\Big[ X_1 \bar \ell \gamma_\mu \nu_{\ell L} i \< D^\mu U Q_L^w U^\dagger Q - D^\mu U^\dagger Q U Q_L^w \> \\
				& - X_2 \bar \ell \gamma_\mu \nu_{\ell L} i \< D^\mu U Q_L^w U^\dagger Q + D^\mu U^\dagger Q U Q_L^w \> \\
				& + X_3 m_\ell \bar \ell \nu_{\ell L} \< Q_L^w U^\dagger Q U \>  + X_4 \bar \ell \gamma_\mu \nu_{\ell L} \< Q_L^w l^\mu Q - Q_L^w Q l^\mu \> \\
				& + X_5 \bar \ell \gamma_\mu \nu_{\ell L} \< Q_L^w U^\dagger r^\mu Q U - Q_L^w U^\dagger Q r^\mu U \> + h.c. \Big]  \end{aligned} \\
			& + X_6 \bar \ell ( i \slashed \p + e \slashed A) \ell + X_7 m_\ell \bar \ell \ell \bigg\} ,
			\end{aligned}
	\end{split}
\end{align}
\begin{align}
	\begin{split}
		\mathcal{L}_\gamma &= e^2 X_8 F_{\mu\nu} F^{\mu\nu} .
	\end{split}
\end{align}

The low-energy constants (LECs) are UV-divergent. Their finite part is defined by
\begin{align}
	\begin{split}
		\label{eqn:RenormalisedLECs}
		L_i &= \Gamma_i \lambda + L_i^r(\mu) , \\
		H_i &= \Delta_i \lambda + H_i^r(\mu) , \\
		K_i &= \Sigma_i \lambda + K_i^r(\mu) , \\
		X_i &= \Xi_i \lambda + X_i^r(\mu) ,
	\end{split}
\end{align}
where
\begin{align}
	\begin{split}
		\label{eqn:UVDivergenceLambda}
		\lambda = \frac{\mu^{n-4}}{16\pi^2} \left( \frac{1}{n-4} - \frac{1}{2} \left( \ln(4\pi) + 1 - \gamma_E \right) \right) .
	\end{split}
\end{align}
The coefficients $\Gamma_i$, $\Delta_i$, $\Sigma_i$ and $\Xi_i$ can be found in \cite{GasserLeutwyler1985, Urech1995, Knecht2000}.


\chapter{Feynman Diagrams}

\label{sec:AppendixDiagrams}

\section{Mass Effects}

\label{sec:AppendixDiagramsMassEffects}

\subsection{Loop Diagrams}
The meson loop diagrams contribute as follows to the form factors $F$ and $G$:
\begin{align}
	\begin{split}
		\delta F^\mathrm{NLO}_\mathrm{tadpole} &= \frac{1}{12 F_0^2} \left[ A_0(\mpio^2) + 4 A_0(\mpip^2) + 8 A_0(\mko^2) + 8 A_0(\mkp^2) + 9 A_0(\meta^2) \right] , \\
		\delta G^\mathrm{NLO}_\mathrm{tadpole} &= \frac{1}{4 F_0^2} \left[ A_0(\mpio^2) + 4 A_0(\mpip^2) + 4 A_0(\mkp^2) + A_0(\meta^2) \right] ,
	\end{split}
\end{align}
{ \small
\begin{align}
	\begin{split}
		\delta F^\mathrm{NLO}_\text{$s$-loop} &=  \frac{1}{F_0^2} \begin{aligned}[t]
				& \bigg[ 3(s-\mpio^2) B_0(s,\mpio^2,\mpio^2)  + 3( s + 4 \Delta_\pi) B_0(s,\mpip^2,\mpip^2) \\
				& + \left( \frac{3}{2}s + \Delta_K - \Delta_\pi \right) B_0(s,\mko^2,\mko^2) + 3(4 \Delta_\pi + s) B_0(s,\mkp^2,\mkp^2) \\
				& + 3 \mpio^2 B_0(s,\meta^2,\meta^2) - 2 A_0(\mpio^2) - 2 A_0(\mpip^2) - A_0(\mko^2) - 2 A_0(\mkp^2) \\
				& + 2 \sqrt{3} \epsilon \begin{aligned}[t]
					&\Big( 3(s-\mpio^2) B_0(s,\mpio^2,\mpio^2) + \frac{2}{3}(\mko^2 - \mpio^2) B_0(s, \mko^2, \mko^2) \\
					& + ( 4 \mpio^2 - 3s) B_0(s,\meta^2,\mpio^2) - \mpio^2 B_0(s,\meta^2,\meta^2) - A_0(\mpio^2) + A_0(\meta^2) \Big) \bigg] ,
					\end{aligned}
			\end{aligned} \\
		\delta G^\mathrm{NLO}_\text{$s$-loop} &= \frac{1}{6 F_0^2} \begin{aligned}[t]
			& \bigg[ (s - 4 \mkp^2) B_0(s,\mkp^2,\mkp^2) - \frac{1}{2}(s - 4 \mko^2) B_0(s,\mko^2,\mko^2) \\
			& + (s - 4 \mpip^2) B_0(s,\mpip^2,\mpip^2) - 2 A_0(\mkp^2) + A_0(\mko^2) - 2 A_0(\mpip^2) \\
			& + \frac{2 \mko^2 - 4 \mkp^2 - 4 \mpip^2 + s}{16\pi^2} \bigg] ,
			\end{aligned}
	\end{split}
\end{align}
\begin{align}
	\begin{split}
		\delta F^\mathrm{NLO}_\text{$t$-loop} &= \frac{1}{6 F_0^2} \Bigg[ \frac{1}{4 t^2} \begin{aligned}[t]
				& \bigg( \mkp^2 \left(2 t-6\meta^2\right)+6 \meta^2 \mpip^2+3 \meta^2 t \\
				& +6 \mko^2 \left(\mkp^2-\mpip^2-t\right)-3\mpio^2 t-2 \mpip^2 t \bigg)  \left(\mko^2-\meta^2\right) B_0\left(0,\meta^2,\mko^2\right) \end{aligned} \\
			& + \frac{1}{4 t^2} \begin{aligned}[t] 
				& \bigg(\mko^2 \left(2\mkp^2 \left(6 \meta^2-t\right)-3 \meta^2 \left(4 \mpip^2+3 t\right)+t \left(3 \mpio^2+2\mpip^2-12 t\right)\right) \\
				& + \left(\meta^2-t\right) \left(\mkp^2 \left(2 t-6 \meta^2\right)+3 \meta^2\left(2 \mpip^2+t\right)-t \left(3 \mpio^2+2 \mpip^2\right)\right) \\
				& + 6 \mko^4\left(-\mkp^2+\mpip^2+t\right)\bigg) B_0\left(t,\meta^2,\mko^2\right) \end{aligned} \\
			& + \frac{1}{2t^2} \begin{aligned}[t]
				& \bigg(\mko^2 \left(\mkp^2-\mpip^2\right)-\mkp^2\left(\mpio^2+2 t\right) \\
				& + \mpio^2 \mpip^2+3 \mpio^2 t+2 \mpip^2 t \bigg) \left(\mko^2-\mpio^2\right) B_0\left(0,\mko^2,\mpio^2\right) \end{aligned} \\
			& + \frac{1}{2t^2} \begin{aligned}[t] 
				& \bigg(\mko^2\left(2 \mkp^2 \left(\mpio^2+t\right)-\mpio^2 \left(2 \mpip^2+3 t\right)+t \left(3 t-2\mpip^2\right)\right) \\
				& + \mko^4 \left(\mpip^2-\mkp^2\right) - \left(\mpio^4+\mpio^2 t-2 t^2\right) \left(\mkp^2-\mpip^2-3 t\right)\bigg) B_0\left(t,\mko^2,\mpio^2\right) \end{aligned} \\
			& + \frac{1}{t^2} \begin{aligned}[t]
				& \bigg(\mkp^4-2 \mkp^2\left(\mpip^2+t\right)  + 3 \mpio^2 t+\mpip^4-\mpip^2 t\bigg) \left(\mkp^2-\mpip^2\right) B_0\left(0,\mkp^2,\mpip^2\right) \end{aligned} \\
			& + \frac{1}{t^2} \begin{aligned}[t]
				& \bigg(-\mkp^6+\mkp^4\left(3 \mpip^2+2 t\right)-\mkp^2 \left(t \left(3 \mpio^2+t\right)+3 \mpip^4\right) \\
				& + \mpip^2 t \left(3\mpio^2-5 t\right)+3 \mpio^2 t^2+\mpip^6-2 \mpip^4 t \bigg) B_0\left(t,\mkp^2,\mpip^2\right) \end{aligned} \\
			& - \frac{3 \left(\mkp^2-\mpip^2+t\right)}{2 t} A_0\left(\meta^2\right) + \frac{\left(-\mkp^2+\mpip^2+3t\right)}{2 t} A_0\left(\mpio^2\right) \\
			& + \frac{\left(\mpip^2-\mkp^2\right)}{t} A_0\left(\mpip^2\right) - A_0\left(\mkp^2\right) \\
			& + \frac{\left(\mkp^2-\mpip^2-3 t\right)\left(3 \meta^2+4 \mko^2+2 \mkp^2+\mpio^2+2 \mpip^2-2 t\right)}{64 \pi ^2 t}  \Bigg] \\
			& + \frac{1}{6 F_0^2} \sqrt{3} \epsilon \Bigg[ \frac{1}{9 t^2} \bigg(\mko^4-\mko^2 \left(2 \mpio^2+t\right)+\mpio^4-2\mpio^2 t\bigg) (\mko^2-\mpio^2) B_0(0,\meta^2,\mko^2)  \\
			& + \frac{1}{9 t^2} \begin{aligned}[t]
				& \bigg(-\mko^6+\mko^4 \left(3\mpio^2+13 t\right)-\mko^2 \left(3 \mpio^4+14 \mpio^2 t+57 t^2\right) \\
				& + \mpio^6+\mpio^4 t+3\mpio^2 t^2+27 t^3 \bigg) B_0(t,\meta^2,\mko^2) \end{aligned} \\
			& - \frac{1}{t^2} \bigg(\mko^4-\mko^2 \left(2 \mpio^2+t\right)+\mpio^4+2\mpio^2 t \bigg) (\mko^2-\mpio^2) B_0(0,\mko^2,\mpio^2) \\
			& + \frac{1}{t^2} \begin{aligned}[t]
				& \bigg(\mko^6-\mko^4 \left(3\mpio^2+t\right)+\mko^2 \left(3 \mpio^4+2 \mpio^2 t+t^2\right) \\
				& - \left(\mpio^2-t\right)^2\left(\mpio^2+3 t\right)\bigg) B_0(t,\mko^2,\mpio^2) \end{aligned} \\
			& - \frac{\left(\mko^2-\mpio^2+t\right)}{t} A_0(\meta^2) + \frac{\left(\mko^2-\mpio^2+t\right)}{t} A_0(\mpio^2) \\
			& + \frac{\left(\mko^2-\mpio^2\right) \left(\mko^2-\mpio^2-3t\right)}{24 \pi ^2 t} \Bigg] , \\
	\end{split}
\end{align}
\begin{align}
	\begin{split}
		\delta G^\mathrm{NLO}_\text{$t$-loop} &= \frac{1}{6 F_0^2} \Bigg[ \frac{1}{4 t^2} \begin{aligned}[t]
			& \bigg( \mkp^2 \left(6 \meta^2-2t\right)-6 \meta^2 \mpip^2+3 \meta^2 t-6 \mko^2 \left(\mkp^2-\mpip^2\right) \\
			& + 3 \mpio^2 t+2\mpip^2 t-6 t^2 \bigg) \left(\mko^2-\meta^2\right) B_0\left(0,\meta^2,\mko^2\right) \end{aligned} \\
		& + \frac{1}{4 t^2} \begin{aligned}[t]
			& \bigg( -\mko^2 \left(2 \mkp^2\left(6 \meta^2-t\right)+3 \meta^2 \left(t-4 \mpip^2\right)+t \left(3 \mpio^2+2\mpip^2\right)\right) \\
			& + \left(\meta^2-t\right) \left(\mkp^2 \left(6 \meta^2-2 t\right)+\meta^2 \left(3t-6 \mpip^2\right)+t \left(3 \mpio^2+2 \mpip^2-6 t\right)\right) \\
			& + 6 \mko^4\left(\mkp^2-\mpip^2\right) \bigg) B_0\left(t,\meta^2,\mko^2\right) \end{aligned} \\
		& - \frac{1}{2t^2} \begin{aligned}[t] 
			& \bigg( \mko^2 \left(\mkp^2-\mpip^2+t\right)-\mkp^2\left(\mpio^2+2 t\right) \\
			& + \mpio^2 \mpip^2+2 \mpio^2 t+2 \mpip^2 t+t^2 \bigg) \left(\mko^2-\mpio^2\right)B_0\left(0,\mko^2,\mpio^2\right) \end{aligned} \\
		& + \frac{1}{2t^2} \begin{aligned}[t]
			& \bigg( \mko^2\left(-2 \mkp^2 \left(\mpio^2+t\right)+\mpio^2 \left(2 \mpip^2+t\right)+t \left(2 \mpip^2-5t\right)\right) \\
			& - \left(\mpio^2-t\right) \left(\mkp^2 \left(-\left(\mpio^2+2 t\right)\right)+\mpio^2\left(\mpip^2+2 t\right)+t \left(2 \mpip^2+7 t\right)\right) \\
			& + \mko^4 \left(\mkp^2-\mpip^2+t\right) \bigg) B_0\left(t,\mko^2,\mpio^2\right) \end{aligned} \\
		& - \frac{1}{t^2} \begin{aligned}[t]
			& \bigg( \mkp^4-\mkp^2\left(2 \mpip^2+t\right)+t \left(3 \mpio^2+t\right) \\
			& + \mpip^4-2 \mpip^2t \bigg) \left(\mkp^2-\mpip^2\right) B_0\left(0,\mkp^2,\mpip^2\right) \end{aligned} \\
		& + \frac{1}{t^2} \begin{aligned}[t]
			& \bigg( \mkp^6-\mkp^4 \left(3 \mpip^2+t\right)+\mkp^2 \left(t \left(3\mpio^2-t\right)+3 \mpip^4-2 \mpip^2 t\right) \\
			& + 3 \mpip^2 t \left(t-\mpio^2\right)+t^2 \left(t-3\mpio^2\right)-\mpip^6+3 \mpip^4 t \bigg) B_0\left(t,\mkp^2,\mpip^2\right) \end{aligned} \\
		& - \frac{3\left(-\mkp^2+\mpip^2+t\right)}{2 t} A_0\left(\meta^2\right) + \frac{\left(\mkp^2-\mpip^2-5 t\right)}{2 t} A_0\left(\mpio^2\right) \\
		& + \frac{\left(\mkp^2-\mpip^2-2t\right)}{t} A_0\left(\mpip^2\right) + A_0\left(\mkp^2\right) \\
		& - \frac{\left(\mkp^2-\mpip^2+t\right) \left(3 \meta^2+4\mko^2+2 \mkp^2+\mpio^2+2 \mpip^2-2 t\right)}{64 \pi ^2 t} \Bigg] \\
		& + \frac{1}{6 F_0^2} \sqrt{3} \epsilon \Bigg[ \frac{1}{9 t^2} \bigg( \mko^4-2 \mko^2 \mpio^2+\mpio^4-3 \mpio^2 t-3 t^2 \bigg) \left(\mpio^2-\mko^2\right) B_0\left(0,\meta^2,\mko^2\right) \\
		& + \frac{1}{9 t^2} \begin{aligned}[t]
			& \bigg( \mko^6-3 \mko^4 \left(\mpio^2+4 t\right)+3 \mko^2 \left(\mpio^4+4 \mpio^2 t+5 t^2\right) \\
			& - \mpio^6+3 \mpio^2 t^2-18 t^3 \bigg) B_0\left(t,\meta^2,\mko^2\right) \end{aligned} \\
		& + \frac{1}{t^2} \bigg( \mko^4-2 \mko^2 \mpio^2+\mpio^4+\mpio^2 t+t^2 \bigg) \left(\mko^2-\mpio^2\right) B_0\left(0,\mko^2,\mpio^2\right) \\
		& + \frac{1}{t^2} \bigg( -\mko^6+3 \mko^4 \mpio^2+\mko^2 \left(t^2-3 \mpio^4\right)+\left(\mpio^2-t\right)^2 \left(\mpio^2+2 t\right)  \bigg) B_0\left(t,\mko^2,\mpio^2\right) \\
		& - \frac{\left(-\mko^2+\mpio^2+t\right)}{t} A_0\left(\meta^2\right) + \frac{\left(-\mko^2+\mpio^2+t\right)}{t} A_0\left(\mpio^2\right) \\
		& - \frac{\left(\mko^2-\mpio^2\right) \left(\mko^2-\mpio^2+t\right)}{24 \pi ^2 t} \Bigg] , 
	\end{split}
\end{align}
}
\begin{align}
	\begin{split}
		\delta F^\mathrm{NLO}_\text{$u$-loop} &= \delta G^\mathrm{NLO}_\text{$u$-loop} \\
			&= \frac{1}{2F_0^2} \begin{aligned}[t]
				& \bigg[ B_0(u,\mkp^2,\mpip^2) ( \mkp^2 + 3 \mpip^2 - 2 \mpio^2 - u) + \frac{1}{3} A_0(\mkp^2) + \frac{1}{3} A_0(\mpip^2) \bigg] .
				\end{aligned}
	\end{split}
\end{align}

\subsection{Counterterms}

The counterterm contribution to the form factors is given by:
\begin{align}
	\begin{split}
		\delta F^\mathrm{NLO}_\mathrm{ct} &= \frac{1}{F_0^2} \begin{aligned}[t]
				& \Bigg[ 32(s-2\mpip^2) L_1 + 8(\mkp^2+s-s_\ell) L_2 + 4(\mkp^2 - 3 \mpip^2 + 2 s -t) L_3 \\
				& + 8\bigg(2 \mko^2 + 5 \mpio^2 - \frac{4 \sqrt{3} \epsilon}{3} (\mko^2-\mpio^2)\bigg) L_4 \\
				& + 4(\mkp^2 + 2 \mpip^2 - 3 \Delta_\pi) L_5 + 2 s_\ell L_9 \Bigg]  + \frac{2}{9} e^2 \left( 84 K_2 + 37 K_6 \right) ,
				\end{aligned} \\
		\delta G^\mathrm{NLO}_\mathrm{ct} &= \frac{1}{F_0^2} \begin{aligned}[t]
				& \Bigg[ 8(t-u) L_2 - 4(\mkp^2 + \mpip^2 - t) L_3  + 8\bigg(2 \mko^2 + \mpio^2 - \frac{4 \sqrt{3} \epsilon}{3} (\mko^2 - \mpio^2) \bigg) L_4 \\
				& + 4(\mkp^2 + 2 \mpip^2 - 3 \Delta_\pi) L_5 + 2 s_\ell L_9 \Bigg] + \frac{2}{9} e^2 \left( 12 K_2 + 18 K_4 + 25 K_6 \right) .
				\end{aligned}
	\end{split}
\end{align}

\subsection{External Leg Corrections}
Let us first consider the pion self-energy: it is given by
\begin{align}
	\begin{split}
		\Sigma_{\pi^+}(p^2) = i ( \mathcal{D}_{\pi^+}^\mathrm{loop} + \mathcal{D}_{\pi^+}^\mathrm{ct} ) ,
	\end{split}
\end{align}
where $p$ denotes the external pion momentum.

The value of the tadpole diagram is
\begin{align}
	\begin{split}
		\mathcal{D}_{\pi^+}^\mathrm{loop} &= \frac{i}{6 F_0^2} \bigg[  p^2 \left( A_0(\mkp^2) + A_0(\mko^2) + 2 A_0(\mpip^2) + 2 A_0(\mpio^2) \right) \\
			& - \mpip^2 \left( A_0(\mkp^2) + A_0(\mko^2) + A_0(\meta^2) + 2 A_0(\mpip^2) - A_0(\mpio^2) \right) \bigg] \\
			& - \frac{i}{3} e^2 Z \left( 6 A_0(\mkp^2) - A_0(\meta^2) + 12 A_0(\mpip^2) + 3 A_0(\mpio^2) \right) ,
	\end{split}
\end{align}
and the counterterm is given by
{\small
\begin{align}
	\begin{split}
		\mathcal{D}_{\pi^+}^\mathrm{ct} &= p^2 \Bigg[ \frac{8 i}{F_0^2} \bigg( (2\mkp^2 - 2 \mpip^2 + 3 \mpio^2) L_4 + \mpio^2 L_5 + \frac{4 \sqrt{3} \epsilon}{3} (\mkp^2-\mpip^2) L_4 \bigg) + \frac{4i}{9} e^2 (6 K_2 + 5 K_6) \Bigg] \\
			& + \frac{16 i}{F_0^2} \begin{aligned}[t]
				& \bigg( (-2 \mpio^2 \mkp^2 + 3 \mpip^4 - 4 \mpio^2 \mpip^2) L_6 + \mpip^2( \mpip^2 - 2 \mpio^2) L_8 \\
				& - \frac{4 \sqrt{3} \epsilon}{3} \mpip^2(\mkp^2-\mpip^2) L_6 \bigg) - \frac{4i}{9} e^2 \left( 3(6\mkp^2+5\mpip^2) K_8 + 23 \mpip^2 K_{10} \right) . \end{aligned}
	\end{split}
\end{align} }%
Since the full propagator is
\begin{align}
	\begin{split}
		\frac{i}{p^2 - \mpip^2 - \Sigma_{\pi^+}(p^2)} = \frac{i Z_{\pi^+}}{p^2 - M_{\pi^+,\mathrm{ph}}^2} + \mathrm{regular} ,
	\end{split}
\end{align}
the field strength renormalisation $Z_{\pi^+}$ can be computed as
\begin{align}
	\begin{split}
		Z_{\pi^+} &= 1 + \Sigma_{\pi^+}^\prime(M_{\pi^+,\mathrm{ph}}^2) + h.o. =  1 + \Sigma_{\pi^+}^\prime(\mpip^2) + h.o. ,
	\end{split}
\end{align}
where $h.o.$ denotes higher order terms.

The physical mass, i.e.~the position of the pole is given by
\begin{align}
	\begin{split}
		M_{\pi^+,\mathrm{ph}}^2 = \mpip^2 + \delta \mpip^2 , \quad \delta \mpip^2 = \Sigma_{\pi^+}(M_{\pi^+,\mathrm{ph}}^2) = \Sigma_{\pi^+}(\mpip^2) + h.o.
	\end{split}
\end{align}

I find the following expression for the field strength renormalisation of the pion:
\begin{align}
	\begin{split}
		Z_{\pi^+} &= 1 - \frac{1}{F_0^2} \begin{aligned}[t] 
			& \bigg( \frac{1}{6} \left( A_0(\mko^2) + A_0(\mkp^2) + 2 A_0(\mpip^2) + 2 A_0(\mpio^2) \right) \\
			& + 8(2 \mkp^2 - 2 \mpip^2 + 3 \mpio^2) L_4 + 8 \mpio^2 L_5 \\
			& + \frac{32\sqrt{3} \epsilon}{3} (\mkp^2-\mpip^2) L_4 \bigg) - \frac{4}{9} e^2 \left( 6 K_2 + 5 K_6 \right) . \end{aligned}
	\end{split}
\end{align}

We still have to compute the external leg correction for the kaon. The values of the two self-energy diagrams for a charged kaon are given by
{ \small
\begin{align}
	\begin{split}
		\mathcal{D}_{K^+}^\mathrm{loop} &= p^2 \begin{aligned}[t] 
				&\Bigg[ \frac{i}{12F_0^2} \left( 2 A_0(\mko^2) + 4 A_0(\mkp^2) + 3 A_0(\meta^2) + 2 A_0(\mpip^2) + A_0(\mpio^2) \right) \\
				& - \frac{i \sqrt{3} \epsilon}{6 F_0^2} \left( A_0(\meta^2) - A_0(\mpio^2) \right) \Bigg] \end{aligned} \\
			& - \frac{i \mkp^2}{12 F_0^2} \begin{aligned}[t] 
				& \bigg( 2 A_0(\mko^2) + 4 A_0(\mkp^2) - A_0(\meta^2) + 2 A_0(\mpip^2) + A_0(\mpio^2) \\
				& + 2 \sqrt{3}\epsilon \left( A_0(\meta^2) - A_0(\mpio^2) \right) \bigg)  - \frac{2i}{3} e^2 Z \left( 6 A_0(\mkp^2) + A_0(\meta^2) + 3 A_0(\mpip^2) \right) , \end{aligned}
	\end{split}
\end{align}
\begin{align}
	\begin{split}
		\mathcal{D}_{K^+}^\mathrm{ct} &= p^2 \Bigg[ \frac{8 i}{F_0^2} \bigg( (2\mkp^2 - 2 \mpip^2 + 3 \mpio^2) L_4 + (\mkp^2 - \mpip^2 + \mpio^2) L_5 \\
			& + \frac{4 \sqrt{3} \epsilon}{3} (\mkp^2-\mpip^2) L_4 \bigg) + \frac{4i}{9} e^2 (6 K_2 + 5 K_6) \Bigg] \\
			& + \frac{16 i}{F_0^2} \bigg( (\mkp^2(4 \mpip^2 - 2 \mkp^2 - 5\mpio^2) + \mpip^2 \Delta_\pi) L_6 \\
			& - \mkp^2( \mkp^2 - 2 \Delta_\pi) L_8 - \frac{4 \sqrt{3} \epsilon}{3} \mkp^2(\mkp^2-\mpip^2) L_6 \bigg) \\
			& - \frac{4i}{9} e^2 \left( 3(8 \mkp^2 + 3 \mpip^2) K_8 + (20\mkp^2 + 3 \mpip^2) K_{10} \right) .
	\end{split}
\end{align}
}
The field strength renormalisation of the kaon is given by
\begin{align}
	\begin{split}
		Z_{K^+} &= 1 \begin{aligned}[t]
			&- \frac{1}{F_0^2} \begin{aligned}[t] 
				& \bigg( \frac{1}{12} \left( 2 A_0(\mko^2) + 4 A_0(\mkp^2) + 3 A_0(\meta^2) + 2 A_0(\mpip^2) + A_0(\mpio^2) \right) \\
				& + 8(2 \mkp^2 - 2 \mpip^2 + 3 \mpio^2) L_4 + 8 ( \mkp^2 - \Delta_\pi) L_5 \\
				& + \frac{32\sqrt{3} \epsilon}{3} (\mkp^2-\mpip^2) L_4 - \frac{\sqrt{3}\epsilon}{6} \left(A_0(\meta^2) - A_0(\mpio^2)\right) \bigg) \end{aligned} \\
			&- \frac{4}{9} e^2 \left( 6 K_2 + 5 K_6 \right) . \end{aligned}
	\end{split}
\end{align}

The contribution of the field strength renormalisation to the amplitude consists of the LO tree diagrams multiplied by a factor of $\sqrt{Z_i}$ for every external particle $i$. Therefore, the NLO external leg corrections to the form factors are given by
\begin{align}
	\begin{split}
		\delta F_Z^\mathrm{NLO} &= \delta G_Z^\mathrm{NLO} =  Z_{\pi^+} \sqrt{Z_{K^+}} - 1 \\
			&= \begin{aligned}[t]
				&- \frac{1}{F_0^2} \begin{aligned}[t]
					& \bigg( \frac{1}{24} \left( 6 A_0(\mko^2) + 8 A_0(\mkp^2) + 3 A_0(\meta^2) + 10 A_0(\mpip^2) + 9 A_0(\mpio^2) \right) \\
					& + 12(2\mkp^2 - 2 \mpip^2 + 3 \mpio^2) L_4 + 4(\mkp^2 - \mpip^2 + 3 \mpio^2) L_5 \\
					& - \frac{\sqrt{3} \epsilon}{12} \left( A_0(\meta^2) - A_0(\mpio^2) \right) + 16\sqrt{3} \epsilon (\mkp^2-\mpip^2) L_4 \bigg)
					\end{aligned} \\
				& - \frac{2}{3} e^2 ( 6 K_2 + 5 K_6 ) .
				\end{aligned}
	\end{split}
\end{align}

\section{Photonic Effects}

\label{sec:AppendixDiagramsPhotonicEffects}

\subsection{Loop Diagrams}

Here, I give the explicit expressions for the contributions of the loop diagrams shown in figures~\ref{img:Kl4_gLoops} and \ref{img:Kl4_mLoops} to the form factors $F$ and $G$.

The first four diagrams only contain bulb topologies. Their contribution, expressed in terms of the scalar loop function $B_0$, is given by
\begin{align}
	\begin{split}
		\delta F^\mathrm{NLO}_{\gamma-\mathrm{loop},a} &= \frac{4}{3} e^2 \left( B_0(0,\mkp^2,\mg^2) - 4 B_0(\mkp^2,\mkp^2,\mg^2) \right), \\
		\delta G^\mathrm{NLO}_{\gamma-\mathrm{loop},a} &= 0 , \\
		\delta F^\mathrm{NLO}_{\gamma-\mathrm{loop},b} &= \delta G^\mathrm{NLO}_{\gamma-\mathrm{loop},b} = -\delta F^\mathrm{NLO}_{\gamma-\mathrm{loop},c} = \delta G^\mathrm{NLO}_{\gamma-\mathrm{loop},c} \\
			&=  \frac{2}{3} e^2 \left( B_0(0,\mpip^2,\mg^2) - 4 B_0(\mpip^2, \mpip^2, \mg^2) \right) , \\
		\delta F^\mathrm{NLO}_{\gamma-\mathrm{loop},d} &= \delta G^\mathrm{NLO}_{\gamma-\mathrm{loop},d} = 0 ,
	\end{split}
\end{align}
hence, in total
\begin{align}
	\begin{split}
		\delta F^\mathrm{NLO}_{\gamma-\mathrm{loop},a-d} &= \frac{4}{3} e^2 \left( B_0(0,\mkp^2,\mg^2) - 4 B_0(\mkp^2,\mkp^2,\mg^2) \right), \\
		\delta G^\mathrm{NLO}_{\gamma-\mathrm{loop},a-d} &= \frac{4}{3} e^2 \left( B_0(0,\mpip^2,\mg^2) - 4 B_0(\mpip^2, \mpip^2, \mg^2) \right) .
	\end{split}
\end{align}

The next six diagrams consist of triangle topologies. My results agree with \cite{Cuplov2004} up to the contribution of the additional term in the massive gauge boson propagator (which cancels in the end), though I choose to employ Passarino-Veltman reduction techniques to write everything in terms of the basic scalar loop functions $A_0$, $B_0$ and $C_0$, even if this results in longer expressions.
{
\small
\begin{align}
	\begin{split}
		\delta F^\mathrm{NLO}_{\gamma-\mathrm{loop},e} &= \frac{1}{2} e^2 \begin{aligned}[t]
			&\Bigg( 4(\mkp^2+\mpip^2-t) C_0(\mpip^2,t, \mkp^2,\mg^2,\mpip^2,\mkp^2) \\
			& + \begin{aligned}[t]
				& \bigg(3 \mkp^4 - 5 \mpip^4 - 6 \mkp^2 \mpip^2 - 2 t(3\mkp^2-\mpip^2) + 3 t^2 \bigg) \frac{2 B_0(\mpip^2,\mpip^2,\mg^2)}{\lambda(t,\mpip^2,\mkp^2)} \end{aligned} \\
			& + \bigg(\mkp^4+2\mkp^2(\mpip^2-3t) + 5(\mpip^2-t)^2 \bigg) \frac{2 B_0(\mkp^2,\mkp^2,\mg^2)}{\lambda(t,\mpip^2,\mkp^2)} \\
			& - \begin{aligned}[t]
				& \bigg( (\mkp^2-\mpip^2)^3 + t(\mkp^4 - 3 \mpip^4 + 2\mkp^2\mpip^2) \\
				& - t^2(13\mkp^2 + 7 \mpip^2) + 11 t^3 \bigg) \frac{B_0(t,\mpip^2,\mkp^2)}{t \lambda(t,\mpip^2,\mkp^2)} - B_0(0,\mpip^2,\mg^2) \end{aligned} \\
			& - 2 B_0(0,\mkp^2,\mg^2) + \frac{\mkp^2-\mpip^2}{t} B_0(0,\mpip^2,\mkp^2) \Bigg)  - e^2 \frac{A_0(\mg^2)}{\mg^2} , \end{aligned}
	\end{split}
\end{align}
\begin{align}
	\begin{split}
		\delta G^\mathrm{NLO}_{\gamma-\mathrm{loop},e} &= \frac{1}{2} e^2 \begin{aligned}[t]
			&\Bigg( 4(\mkp^2+\mpip^2-t) C_0(\mpip^2,t, \mkp^2,\mg^2,\mpip^2,\mkp^2) \\
			& + \bigg( 3 \mkp^4 + 2 \mkp^2(\mpip^2 - 3t) + 3(\mpip^2-t)^2 \bigg) \frac{2 B_0(\mpip^2,\mpip^2,\mg^2)}{\lambda(t,\mpip^2,\mkp^2)} \\
			& -  \bigg( 3\mkp^4+2\mkp^2(3\mpip^2-t) - (\mpip^2-t)^2 \bigg) \frac{2B_0(\mkp^2,\mkp^2,\mg^2)}{\lambda(t,\mpip^2,\mkp^2)} \\
			& + \begin{aligned}[t]
				& \bigg( (\mkp^2-\mpip^2)^3 + t(\mkp^4 - 3 \mpip^4 + 2\mkp^2\mpip^2) \\
				& + 3t^2(\mkp^2 + 3 \mpip^2) - 5 t^3 \bigg) \frac{B_0(t,\mpip^2,\mkp^2)}{t \lambda(t,\mpip^2,\mkp^2)} \end{aligned} \\
			& - B_0(0,\mpip^2,\mg^2) - \frac{\mkp^2-\mpip^2}{t} B_0(0,\mpip^2,\mkp^2) \Bigg) - e^2 \frac{A_0(\mg^2)}{\mg^2} , \end{aligned}
	\end{split}
\end{align}
\begin{align}
	\begin{split}
		\delta F^\mathrm{NLO}_{\gamma-\mathrm{loop},f} &= \delta G^\mathrm{NLO}_{\gamma-\mathrm{loop},f} \\
			&= - e^2 \begin{aligned}[t]
			& \bigg( 2(\mkp^2+\mpip^2-u) C_0(\mpip^2,u,\mkp^2,\mg^2,\mpip^2,\mkp^2) \\
			& + B_0(\mpip^2,\mpip^2,\mg^2) + B_0(\mkp^2,\mkp^2,\mg^2) - B_0(u,\mpip^2,\mkp^2)  \bigg)  + e^2 \frac{A_0(\mg^2)}{\mg^2} , \end{aligned}
	\end{split}
\end{align}
\begin{align}
	\begin{split}
		\delta F^\mathrm{NLO}_{\gamma-\mathrm{loop},g} &= e^2 \begin{aligned}[t]
			& \Bigg( 2(\mkp^2+\ml^2-\tl) C_0(\ml^2,\tl,\mkp^2,\mg^2,\ml^2,\mkp^2) \\
			& + \bigg( 3 \mkp^4 - \mkp^2 (5\ml^2 + 6 \tl) - \ml^2 \tl - 2 \ml^4 + 3 \tl^2 \bigg) \frac{2 B_0(\ml^2,\ml^2,\mg^2)}{3 \lambda(\tl,\ml^2,\mkp^2)} \\
			& + \bigg( \mkp^4 - 8 \mkp^2 \tl  + 7(\ml^2-\tl)^2 \bigg) \frac{B_0(\mkp^2,\mkp^2,\mg^2)}{3 \lambda(\tl,\ml^2,\mkp^2)} \\
			& + \bigg( \mkp^2(3\tl - 2 \ml^2) + \ml^2 \tl + 2 \ml^4 - 3 \tl^2 \bigg) \frac{2 B_0(\tl,\ml^2,\mkp^2)}{3 \lambda(\tl,\ml^2,\mkp^2)} \\
			& - \frac{1}{3} B_0(0,\mkp^2,\mg^2) \Bigg)  - e^2 \frac{A_0(\mg^2)}{\mg^2} , \end{aligned}
	\end{split}
\end{align}
\begin{align}
	\begin{split}
		\delta G^\mathrm{NLO}_{\gamma-\mathrm{loop},g} &= e^2 \begin{aligned}[t]
			& \Bigg( 2(\mkp^2+\ml^2-\tl) C_0(\ml^2,\tl,\mkp^2,\mg^2,\ml^2,\mkp^2) \\
			& + \bigg( \mkp^4 - \mkp^2 (\ml^2 + 2 \tl) + \tl(\tl-\ml^2) \bigg) \frac{2 B_0(\ml^2,\ml^2,\mg^2)}{\lambda(\tl,\ml^2,\mkp^2)} \\
			& - \bigg( \mkp^4  - (\ml^2-\tl)^2 \bigg) \frac{B_0(\mkp^2,\mkp^2,\mg^2)}{\lambda(\tl,\ml^2,\mkp^2)} \\
			& + \bigg( \mkp^2 + \ml^2 - \tl \bigg) \frac{2 \tl B_0(\tl,\ml^2,\mkp^2)}{\lambda(\tl,\ml^2,\mkp^2)} \Bigg)  - e^2 \frac{A_0(\mg^2)}{\mg^2} , \end{aligned}
	\end{split}
\end{align}
\begin{align}
	\begin{split}
		\delta F^\mathrm{NLO}_{\gamma-\mathrm{loop},h} &= e^2 \begin{aligned}[t]
			& \Bigg( 2(2\mpip^2-s) C_0(\mpip^2,s,\mpip^2,\mg^2,\mpip^2,\mpip^2) \\
			& + 2 B_0(\mpip^2,\mpip^2,\mg^2) - B_0(s,\mpip^2,\mpip^2) \Bigg)  - e^2 \frac{A_0(\mg^2)}{\mg^2} , \end{aligned}
	\end{split}
\end{align}
\begin{align}
	\begin{split}
		\delta G^\mathrm{NLO}_{\gamma-\mathrm{loop},h} &= e^2 \begin{aligned}[t]
			& \Bigg( 2(2\mpip^2-s) C_0(\mpip^2,s,\mpip^2,\mg^2,\mpip^2,\mpip^2) \\
			& + \frac{2(8\mpip^2-3s)}{4\mpip^2-s} B_0(\mpip^2,\mpip^2,\mg^2) \\
			& - \frac{4(2\mpip^2-s)}{4\mpip^2-s} B_0(s,\mpip^2,\mpip^2) - B_0(0,\mpip^2,\mg^2) \Bigg)  - e^2 \frac{A_0(\mg^2)}{\mg^2} , \end{aligned}
	\end{split}
\end{align}
\begin{align}
	\begin{split}
		\delta F^\mathrm{NLO}_{\gamma-\mathrm{loop},i} &= \delta G^\mathrm{NLO}_{\gamma-\mathrm{loop},i} \\
			&= e^2 \begin{aligned}[t]
			& \Bigg( -2(\mpip^2 + \ml^2 - s_{1\ell} ) C_0(\ml^2, s_{1\ell}, \mpip^2, \mg^2, \ml^2, \mpip^2) \\
			& + \bigg( 2 \ml^4 + \ml^2(s_{1\ell} + 5 \mpip^2) - 3(\mpip^2-s_{1\ell})^2  \bigg) \frac{2 B_0(\ml^2, \ml^2, \mg^2)}{3\lambda(s_{1\ell}, \ml^2, \mpip^2)} \\
			& + \bigg( \mpip^4 + 7 \ml^4 - 2 s_{1\ell}( 4 \mpip^2 + 7 \ml^2 ) + 7 s_{1\ell}^2 \bigg) \frac{B_0(\mpip^2, \mpip^2, \mg^2)}{3\lambda(s_{1\ell}, \ml^2, \mpip^2)} \\
			& - \bigg( 4 \ml^4 - 2 \ml^2(2 \mpip^2 - s_{1\ell}) - 6 s_{1\ell}(s_{1\ell}-\mpip^2) \bigg) \frac{B_0(s_{1\ell}, \ml^2, \mpip^2)}{3\lambda(s_{1\ell}, \ml^2, \mpip^2)} \\
			& + \frac{1}{3} B_0(0, \mpip^2, \mg^2) \Bigg) +  e^2 \frac{A_0(\mg^2)}{\mg^2} , \end{aligned}
	\end{split}
\end{align}
\begin{align}
	\begin{split}
		\delta F^\mathrm{NLO}_{\gamma-\mathrm{loop},j} &= e^2 \begin{aligned}[t]
			& \Bigg( 2 (\mpip^2 + \ml^2 - s_{2\ell}) C_0(\ml^2, s_{2\ell}, \mpip^2, \mg^2, \ml^2, \mpip^2) \\
			& + \bigg( 4 \ml^4 + \ml^2(\mpip^2 - 7 s_{2\ell}) + 3 (\mpip^2-s_{2\ell})^2 \bigg) \frac{2 B_0(\ml^2, \ml^2, \mg^2)}{3 \lambda(s_{2\ell}, \ml^2, \mpip^2)} \\
			& + \bigg( \ml^4 - 5 \mpip^4 - 12 \ml^2 \mpip^2 - 2s_{2\ell}(\ml^2 - 2 \mpip^2) + s_{2\ell}^2 \bigg) \frac{B_0(\mpip^2, \mpip^2, \mg^2)}{3\lambda(s_{2\ell}, \ml^2, \mpip^2)} \\
			& - \bigg( 8 \ml^4  - 2 \ml^2(7s_{2\ell} + 4\mpip^2) - 6 s_{2\ell}(\mpip^2-s_{2\ell}) \bigg) \frac{B_0(s_{2\ell}, \ml^2, \mpip^2)}{3\lambda(s_{2\ell}, \ml^2, \mpip^2)} \\
			& + \frac{1}{6} B_0(0, \mpip^2, \mg^2) \Bigg) - e^2 \frac{A_0(\mg^2)}{\mg^2} , \end{aligned}
	\end{split}
\end{align}
\begin{align}
	\begin{split}
		\delta G^\mathrm{NLO}_{\gamma-\mathrm{loop},j} &= e^2 \begin{aligned}[t]
			& \Bigg( 2 (\mpip^2 + \ml^2 - s_{2\ell}) C_0(\ml^2, s_{2\ell}, \mpip^2, \mg^2, \ml^2, \mpip^2) \\
			& + \bigg( 2 \ml^4 - \ml^2(\mpip^2 + 5 s_{2\ell}) + 3 (\mpip^2-s_{2\ell})^2 \bigg) \frac{2 B_0(\ml^2, \ml^2, \mg^2)}{3 \lambda(s_{2\ell}, \ml^2, \mpip^2)} \\
			& + \bigg( 5 \ml^4 - 2 \ml^2 ( 5 s_{2\ell} + 6 \mpip^2) + (s_{2\ell} - \mpip^2)(5 s_{2\ell} + \mpip^2) \bigg) \frac{B_0(\mpip^2, \mpip^2, \mg^2)}{3\lambda(s_{2\ell}, \ml^2, \mpip^2)} \\
			& - \bigg( 4 \ml^4  - 2 \ml^2(5s_{2\ell} + 2\mpip^2) - 6 s_{2\ell}(\mpip^2-s_{2\ell}) \bigg) \frac{B_0(s_{2\ell}, \ml^2, \mpip^2)}{3\lambda(s_{2\ell}, \ml^2, \mpip^2)} \\
			& - \frac{1}{6} B_0(0, \mpip^2, \mg^2) \Bigg) - e^2 \frac{A_0(\mg^2)}{\mg^2} . \end{aligned}
	\end{split}
\end{align}
}

The remaining eight diagrams in this first set are loop corrections to the diagram~\ref{img:Kl4_LO2}. Here, the Passarino-Veltman reduction \cite{Hooft1979, Passarino1979} produces too lengthy expressions, hence, I use the tensor-coefficient functions (see appendix~\ref{sec:AppendixTensorCoefficientFunctions}):
\begin{align}
	\small
	\begin{split}
		\delta F^\mathrm{NLO}_{\gamma-\mathrm{loop},k} &= \delta G^\mathrm{NLO}_{\gamma-\mathrm{loop},k} = -\frac{4}{3} e^2 B_0(s_\ell, \mkp^2, \mg^2) + \frac{4}{3} e^2 \frac{B_{00}(s_\ell, \mkp^2, \mg^2)}{\mg^2} ,
	\end{split}
\end{align}
\begin{align}
	\small
	\begin{split}
		\delta F^\mathrm{NLO}_{\gamma-\mathrm{loop},l} &= e^2 \begin{aligned}[t]
			& \Bigg( \frac{1}{3} B_0(s_\ell, \mkp^2, \mg^2) + \frac{1}{3} B_0(\mkp^2, \mkp^2, \mg^2) - \frac{1}{12} B_0(0, \mkp^2, \mg^2) \\
			& - (s + \nu) C_0(s_\ell, s, \mkp^2, \mg^2, \mkp^2, \mkp^2) - \nu C_1(s_\ell, s, \mkp^2, \mg^2, \mkp^2, \mkp^2) \\
			& - \frac{s + 3\nu}{2} C_2(s_\ell, s, \mkp^2, \mg^2, \mkp^2, \mkp^2)  - \frac{\nu}{2} C_{12}(s_\ell, s, \mkp^2, \mg^2, \mkp^2, \mkp^2) \\
			& - \frac{\nu}{2} C_{22}(s_\ell, s, \mkp^2, \mg^2, \mkp^2, \mkp^2)  \Bigg) - \frac{1}{3} e^2 \frac{B_{00}(s_\ell, \mkp^2, \mg^2)}{\mg^2} ,
			\end{aligned}
	\end{split}
\end{align}
\begin{align}
	\small
	\begin{split}
		\delta G^\mathrm{NLO}_{\gamma-\mathrm{loop},l} &= e^2 C_{00}(s_\ell, s, \mkp^2, \mg^2, \mkp^2, \mkp^2) - e^2 \frac{B_{00}(s_\ell, \mkp^2, \mg^2)}{\mg^2} ,
	\end{split}
\end{align}
\begin{align}
	\small
	\begin{split}
		\delta F^\mathrm{NLO}_{\gamma-\mathrm{loop},m} &= \delta G^\mathrm{NLO}_{\gamma-\mathrm{loop},m} \\
			&= -e^2 \begin{aligned}[t]
				& \Bigg( \frac{1}{3} B_0(s_\ell, \mkp^2, \mg^2) + \frac{1}{3} B_0(\mpip^2, \mpip^2, \mg^2) - \frac{1}{12} B_0(0, \mpip^2, \mg^2) \\
				& + (\mkp^2 + \mpip^2 - u) C_0(\mpip^2, u, s_\ell, \mg^2, \mpip^2, \mkp^2) \\
				& + \frac{\mkp^2 + \mpip^2 - u}{2} C_1(\mpip^2, u, s_\ell, \mg^2, \mpip^2, \mkp^2) \Bigg)  + \frac{1}{3} e^2 \frac{B_{00}(s_\ell, \mkp^2, \mg^2)}{\mg^2} , \end{aligned}
	\end{split}
\end{align}
\begin{align}
	\small
	\begin{split}
		\delta F^\mathrm{NLO}_{\gamma-\mathrm{loop},n} &= e^2 \begin{aligned}[t]
			& \Bigg( - \frac{2}{3} B_0(\mpip^2, \mpip^2, \mg^2) + \frac{1}{3} B_0(s_\ell, \mkp^2, \mg^2) + \frac{1}{6} B_0(0, \mpip^2, \mg^2) \\
			& + (s_\ell + \mpip^2 - u) C_0(\mpip^2, t, s_\ell, \mg^2, \mpip^2, \mkp^2) \\
			& - \frac{\mkp^2 + 5 \mpip^2 - 3s - t}{2} C_1(\mpip^2, t, s_\ell, \mg^2, \mpip^2, \mkp^2) \\
			& + (s_\ell + \mpip^2 - u) C_2(\mpip^2, t, s_\ell, \mg^2, \mpip^2, \mkp^2) \\
			& + C_{00}(\mpip^2, t, s_\ell, \mg^2, \mpip^2, \mkp^2) + \frac{s - 2\mpip^2}{2} C_{11}(\mpip^2, t, s_\ell, \mg^2, \mpip^2, \mkp^2) \\
			& + \frac{s_\ell + \mpip^2 - u}{2} C_{12}(\mpip^2, t, s_\ell, \mg^2, \mpip^2, \mkp^2) \Bigg) - \frac{4}{3} e^2 \frac{B_{00}(s_\ell, \mkp^2, \mg^2)}{\mg^2} , \end{aligned}
	\end{split}
\end{align}
\begin{align}
	\small
	\begin{split}
		\delta G^\mathrm{NLO}_{\gamma-\mathrm{loop},n} &= - \delta F^\mathrm{NLO}_{\gamma-\mathrm{loop},n} + 2 e^2 C_{00}(\mpip^2, t, s_\ell, \mg^2, \mpip^2, \mkp^2) - 2 e^2 \frac{B_{00}(s_\ell, \mkp^2, \mg^2)}{\mg^2} ,
	\end{split}
\end{align}
\begin{align}
	\small
	\begin{split}
		\delta F^\mathrm{NLO}_{\gamma-\mathrm{loop},o} &= \delta G^\mathrm{NLO}_{\gamma-\mathrm{loop},o} \\
			&= \frac{4}{3} e^2 \frac{1}{\ml^2-s_\ell} \begin{aligned}[t]
				& \Bigg( \ml^2 B_0(\ml^2, \ml^2, \mg^2) - s_\ell B_0(s_\ell, \mkp^2, \mg^2) \\
				& + \ml^2 (\mkp^2 - s_\ell) C_0(\ml^2, 0, s_\ell, \mg^2, \ml^2, \mkp^2) \Bigg)  - \frac{4}{3} e^2 \frac{B_{00}(s_\ell, \mkp^2, \mg^2)}{\mg^2} , \end{aligned}
	\end{split}
\end{align}
\begin{align}
	\small
	\begin{split}
		\delta F^\mathrm{NLO}_{\gamma-\mathrm{loop},p} &= e^2 \begin{aligned}[t]
			& \Bigg( \frac{1}{12} B_0(0, \mkp^2, \mg^2) - \frac{1}{3} B_0(\mkp^2, \mkp^2, \mg^2) - \frac{1}{3} B_0(s_\ell, \mkp^2, \mg^2) \\
			& + (\nu + s) C_0(\mkp^2, s, s_\ell, \mg^2, \mkp^2, \mkp^2) + \frac{1}{2} (\nu + s) C_1(\mkp^2, s, s_\ell, \mg^2, \mkp^2, \mkp^2) \\
			& + \nu C_{1+2}(\mkp^2, s, s_\ell, \mg^2, \mkp^2, \mkp^2) + \frac{1}{2} \nu C_{11+12}(\mkp^2, s, s_\ell, \mg^2, \mkp^2, \mkp^2) \\
			& + \frac{1}{3} \ml^2 C_{1+2}(\ml^2, 0, s_\ell, \mg^2, \ml^2, \mkp^2) + \frac{1}{3} \ml^2 C_2(\mkp^2, \tl, \ml^2, \mg^2, \mkp^2, \ml^2) \\
			& - \ml^2 (\nu + s) D_{2+3}(\mkp^2, \tl, 0, s_\ell, \ml^2, s, \mg^2, \mkp^2, \ml^2, \mkp^2) \\
			& - \ml^2 \nu D_{12+13}(\mkp^2, \tl, 0, s_\ell, \ml^2, s, \mg^2, \mkp^2, \ml^2, \mkp^2) \\
			& + \ml^2 (s_{1\ell} - s_{2\ell}) D_{22+23}(\mkp^2, \tl, 0, s_\ell, \ml^2, s, \mg^2, \mkp^2, \ml^2, \mkp^2) \\
			& - \ml^2 \nu D_{23+33}(\mkp^2, \tl, 0, s_\ell, \ml^2, s, \mg^2, \mkp^2, \ml^2, \mkp^2) \Bigg) + \frac{1}{3} e^2 \frac{B_{00}(s_\ell, \mkp^2, \mg^2)}{\mg^2} , \end{aligned}
	\end{split}
\end{align}
where I use the abbreviation
\begin{align}
	\begin{split}
		C_{i+j}(X) &:= C_i(X) + C_j(X) , \\
		D_{i+j}(X) &:= D_i(X) + D_j(X) .
	\end{split}
\end{align}

\begin{align}
	\small
	\begin{split}
		\delta G^\mathrm{NLO}_{\gamma-\mathrm{loop},p} &= - e^2 \begin{aligned}[t]
			& \Bigg( C_{00}(\mkp^2, s, s_\ell, \mg^2, \mkp^2, \mkp^2) \\
			& + 2 \ml^2 D_{00}(\ml^2, \tl, s, s_\ell, \mkp^2, 0, \mg^2, \ml^2, \mkp^2, \mkp^2) \Bigg)  + e^2 \frac{B_{00}(s_\ell, \mkp^2, \mg^2)}{\mg^2} , \end{aligned}
	\end{split}
\end{align}
\begin{align}
	\small
	\begin{split}
		\delta F^\mathrm{NLO}_{\gamma-\mathrm{loop},q} &= \delta G^\mathrm{NLO}_{\gamma-\mathrm{loop},q} \\
			&= e^2 \begin{aligned}[t]
			& \Bigg( - \frac{1}{12} B_0(0, \mpip^2, \mg^2) + \frac{1}{3} B_0(\mpip^2, \mpip^2, \mg^2) + \frac{1}{3} B_0(s_\ell, \mkp^2, \mg^2) \\
			& + (\mkp^2 + \mpip^2 - u) C_0(\mpip^2, u, s_\ell, \mg^2, \mpip^2, \mkp^2) \\
			& + \frac{1}{2} (\mkp^2 + \mpip^2 - u) C_1(\mpip^2, u, s_\ell, \mg^2, \mpip^2, \mkp^2) \\
			& - \frac{1}{3} \ml^2 C_{1+2}(\ml^2, 0, s_\ell, \mg^2, \ml^2, \mkp^2) - \frac{1}{3} \ml^2 C_1(\ml^2, s_{1\ell}, \mpip^2, \mg^2, \ml^2, \mpip^2) \\
			& - \ml^2 (\mkp^2 + \mpip^2 - u) D_{2+3}(\mpip^2, s_{1\ell}, 0, s_\ell, \ml^2, u, \mg^2, \mpip^2, \ml^2, \mkp^2) \Bigg) \end{aligned} \\
			&  - \frac{1}{3} e^2 \frac{B_{00}(s_\ell, \mkp^2, \mg^2)}{\mg^2} ,
	\end{split}
\end{align}
\begin{align}
	\small
	\begin{split}
		\delta F^\mathrm{NLO}_{\gamma-\mathrm{loop},r} &= e^2 \begin{aligned}[t]
			& \Bigg(- \frac{1}{6} B_0(0, \mpip^2, \mg^2) + \frac{2}{3} B_0(\mpip^2, \mpip^2, \mg^2) - \frac{1}{3} B_0(s_\ell, \mkp^2, \mg^2) \\
			& - (\mpip^2 + s_\ell - u) C_0(\mpip^2, t, s_\ell, \mg^2, \mpip^2, \mkp^2) \\
			& + \frac{1}{2} (3 \mpip^2 - 2 s - s_\ell + u) C_1(\mpip^2, t, s_\ell, \mg^2, \mpip^2, \mkp^2) \\
			& - (\mpip^2 + s_\ell - u) C_2(\mpip^2, t, s_\ell, \mg^2, \mpip^2, \mkp^2) - C_{00}(\mpip^2, t, s_\ell, \mg^2, \mpip^2, \mkp^2) \\
			& + \frac{1}{2}(2\mpip^2 - s) C_{11}(\mpip^2, t, s_\ell, \mg^2, \mpip^2, \mkp^2) \\
			& - \frac{1}{2} (\mpip^2 + s_\ell - u) C_{12}(\mpip^2, t, s_\ell, \mg^2, \mpip^2, \mkp^2) \\
			& + \frac{1}{3} \ml^2 C_{1+2}(\ml^2, 0, s_\ell, \mg^2, \ml^2, \mkp^2) - \frac{2}{3} \ml^2 C_1(\ml^2, s_{2\ell}, \mpip^2, \mg^2, \ml^2, \mpip^2) \\
			& + \ml^2 (\mpip^2 + s_\ell - u) D_1(s_\ell, t, s_{2\ell}, \ml^2, \mpip^2, 0, \mg^2, \mkp^2, \mpip^2, \ml^2) \\
			& + \ml^2 (\mpip^2 + s_\ell - u) D_3(s_\ell, t, s_{2\ell}, \ml^2, \mpip^2, 0, \mg^2, \mkp^2, \mpip^2, \ml^2) \\
			& + \ml^2 (\mpip^2 + s_\ell - u) D_{11}(s_\ell, t, s_{2\ell}, \ml^2, \mpip^2, 0, \mg^2, \mkp^2, \mpip^2, \ml^2) \\
			& + \ml^2 (s - 2 \mpip^2) D_{12}(s_\ell, t, s_{2\ell}, \ml^2, \mpip^2, 0, \mg^2, \mkp^2, \mpip^2, \ml^2) \\
			& + \ml^2 (\ml^2 + 2 \mpip^2 - s_{1\ell} + s_\ell - u) D_{13}(s_\ell, t, s_{2\ell}, \ml^2, \mpip^2, 0, \mg^2, \mkp^2, \mpip^2, \ml^2) \\
			& + \ml^2 (s - 2 \mpip^2) D_{23}(s_\ell, t, s_{2\ell}, \ml^2, \mpip^2, 0, \mg^2, \mkp^2, \mpip^2, \ml^2) \\
			& + \ml^2 (\ml^2 + \mpip^2 - s_{1\ell}) D_{33}(s_\ell, t, s_{2\ell}, \ml^2, \mpip^2, 0, \mg^2, \mkp^2, \mpip^2, \ml^2) \Bigg) \end{aligned} \\
			& + \frac{4}{3} e^2 \frac{B_{00}(s_\ell, \mkp^2, \mg^2)}{\mg^2} ,
	\end{split}
\end{align}
\begin{align}
	\small
	\begin{split}
		\delta G^\mathrm{NLO}_{\gamma-\mathrm{loop},r} &= - \delta F^\mathrm{NLO}_{\gamma-\mathrm{loop},r} - 2 e^2 C_{00}(\mpip^2, t, s_\ell, \mg^2, \mpip^2, \mkp^2) + 2 e^2 \frac{B_{00}(s_\ell, \mkp^2, \mg^2)}{\mg^2} .
	\end{split}
\end{align}
I use the notation $\nu = t - u$, $\lambda_{K\ell}(s) = \lambda(\mkp^2, s, s_\ell)$, $\lambda_{\pi\ell}(s) = \lambda(\mpip^2, s, s_\ell)$.

Next, I give the explicit expressions for the diagrams of the second set, shown in figure~\ref{img:Kl4_mLoops}. These diagrams contain a photon pole in the $s$-channel and mesonic loops.
\begin{align}
	\begin{split}
		\delta F^\mathrm{NLO}_{\gamma-\mathrm{pole},a} &= 0 , \\
		\delta G^\mathrm{NLO}_{\gamma-\mathrm{pole},a} &= -e^2 \frac{1}{3s} \begin{aligned}[t]
			& \Bigg( 2( s - 4\mkp^2) B_0(s, \mkp^2, \mkp^2) + (s - 4\mpip^2) B_0(s,\mpip^2, \mpip^2) \\
			& - 4 A_0(\mkp^2) - 2 A_0(\mpip^2) - \frac{4\mkp^2 + 2\mpip^2 -s}{8\pi^2}  \Bigg) , \end{aligned}
	\end{split}
\end{align}
\begin{align}
	\begin{split}
		\delta F^\mathrm{NLO}_{\gamma-\mathrm{pole},b} &= -\delta F^\mathrm{NLO}_{\gamma-\mathrm{pole},c} = -e^2 \frac{s - 6 \mpip^2}{144 \pi^2 \mg^2} , \\
		\delta G^\mathrm{NLO}_{\gamma-\mathrm{pole},b} &= -\delta G^\mathrm{NLO}_{\gamma-\mathrm{pole},c} =  -e^2 \frac{1}{6s} \begin{aligned}[t]
			& \Bigg( ( s - 4\mkp^2) B_0(s, \mkp^2, \mkp^2) \\
			& + 2(s - 4\mpip^2) B_0(s,\mpip^2, \mpip^2) \\
			& - 2 A_0(\mkp^2) - 4 A_0(\mpip^2) - \frac{2\mkp^2 + 4\mpip^2 -s}{8\pi^2}  \Bigg) , \end{aligned}
	\end{split}
\end{align}
\begin{align}
	\begin{split}
		\delta F^\mathrm{NLO}_{\gamma-\mathrm{pole},d} &= 0 , \\
		\delta G^\mathrm{NLO}_{\gamma-\mathrm{pole},d} &= -e^2 \frac{1}{3s} \bigg( A_0(\mpio^2) + 8 A_0(\mpip^2) + 2 A_0(\mko^2) + 16 A_0(\mkp^2) + 3 A_0(\meta^2) \bigg) ,
	\end{split}
\end{align}
\begin{align}
	\begin{split}
		\delta F^\mathrm{NLO}_{\gamma-\mathrm{pole},e} &= 0 , \\
		\delta G^\mathrm{NLO}_{\gamma-\mathrm{pole},e} &= e^2 \frac{1}{3s} \bigg( A_0(\mpio^2) + 2 A_0(\mpip^2) + 2 A_0(\mko^2) + 4 A_0(\mkp^2) + 3 A_0(\meta^2) \bigg) ,
	\end{split}
\end{align}
\begin{align}
	\begin{split}
		\delta F^\mathrm{NLO}_{\gamma-\mathrm{pole},f} &= \delta F^\mathrm{NLO}_{\gamma-\mathrm{pole},g} = 0 , \\
		\delta G^\mathrm{NLO}_{\gamma-\mathrm{pole},f} &= -\delta G^\mathrm{NLO}_{\gamma-\mathrm{pole},g} = -e^2 \frac{1}{3s} \bigg( 2 A_0(\mpio^2) + 8 A_0(\mpip^2) + A_0(\mko^2) + 4 A_0(\mkp^2) \bigg) .
	\end{split}
\end{align}
In the sum of these diagrams, the contribution to $F$ vanishes:
\begin{align}
	\begin{split}
		\delta F^\mathrm{NLO}_{\gamma-\mathrm{pole}} &= 0 , \\
		\delta G^\mathrm{NLO}_{\gamma-\mathrm{pole}} &= - e^2 \frac{1}{3s} \begin{aligned}[t]
			& \Bigg( 2 ( s - 4\mkp^2) B_0(s, \mkp^2, \mkp^2) + (s - 4\mpip^2) B_0(s,\mpip^2, \mpip^2) \\
			& + 8 A_0(\mkp^2) + 4 A_0(\mpip^2) - \frac{4\mkp^2 + 2\mpip^2 -s}{8\pi^2}  \Bigg) . \end{aligned}
	\end{split}
\end{align}

\subsection{Counterterms}

The individual contributions of the counterterm diagrams, shown in figure~\ref{img:Kl4_gCT}, are given by
{\small
\begin{align}
	\begin{split}
		\delta F^\mathrm{NLO}_{\gamma-\mathrm{ct},a} &= \frac{2}{9} e^2 \left( 12 K_1 + 19 K_5 + 9 K_{12} - 30 X_1 \right) , \\
		\delta G^\mathrm{NLO}_{\gamma-\mathrm{ct},a} &= \frac{2}{9} e^2 \left( 12 K_1 + 36 K_3 + 7 K_5 + 9 K_{12} + 6 X_1 \right) , \\
		\delta F^\mathrm{NLO}_{\gamma-\mathrm{ct},b} &= - e^2 \frac{4(t-u)}{s} \left( L_9 + L_{10} \right) , \\
		\delta G^\mathrm{NLO}_{\gamma-\mathrm{ct},b} &= - e^2 \frac{4}{s} \begin{aligned}[t]
													& \bigg( (\mkp^2 + s - s_\ell) L_9 + (\mkp^2 - s - s_\ell) L_{10}  + 4( 2 \mkp^2 + \mpip^2) L_4 + 4 \mkp^2 L_5  \bigg) , \end{aligned} \\
		\delta F^\mathrm{NLO}_{\gamma-\mathrm{ct},c} &= \delta F^\mathrm{NLO}_{\gamma-\mathrm{ct},d} = 0 , \\
		\delta G^\mathrm{NLO}_{\gamma-\mathrm{ct},c} &= -\delta G^\mathrm{NLO}_{\gamma-\mathrm{ct},d} = - e^2 \frac{4}{s} \left( 4(2\mkp^2 + \mpip^2) L_4 + 4 \mpip^2 L_5 + s L_9 \right) , \\
		\delta F^\mathrm{NLO}_{\gamma-\mathrm{ct},e} &= 0 , \\
		\delta G^\mathrm{NLO}_{\gamma-\mathrm{ct},e} &= e^2 \frac{16}{s} \left( (2\mkp^2 + \mpip^2) L_4 + \mkp^2 L_5 \right) .
	\end{split}
\end{align} }%
Their sum is
{\small
\begin{align}
	\begin{split}
		\delta F^\mathrm{NLO}_{\gamma-\mathrm{ct}} &= \frac{2}{9} e^2 \left( 12 K_1 + 19 K_5 + 9 K_{12} - 30 X_1 \right) - e^2 \frac{4(t-u)}{s}(L_9 + L_{10}) , \\
		\delta G^\mathrm{NLO}_{\gamma-\mathrm{ct}} &= \frac{2}{9} e^2 \left( 12 K_1 + 36 K_3 + 7 K_5 + 9 K_{12} + 6 X_1 \right) - e^2 \frac{4}{s} \left( (\mkp^2 + s - s_\ell) L_9 + (\mkp^2 - s - s_\ell) L_{10} \right) .
	\end{split}
\end{align} }%

\subsection{External Leg Corrections}

\label{sec:AppendixExternalLegCorrectionsPhotonicEffects}
I first compute the external leg corrections for the mesons (figures~\ref{img:Kl4_mSEgLoop} and \ref{img:Kl4_mSECT}). The field strength renormalisation of a charged meson is related to the self-energy by
\begin{align}
	\begin{split}
		Z_{m^+}^\gamma &= 1 + \Sigma_{m^+}^{\gamma\prime}(M_{m^+,\mathrm{ph}}^2) + h.o. =  1 + \Sigma_{m^+}^{\gamma\prime}(M_{m^+}^2) + h.o. , \\
		\Sigma_{m^+}^\gamma(p^2) &= i ( \mathcal{D}_{m^+}^{\gamma-\mathrm{loop}} + \mathcal{D}_{m^+}^{\gamma-\mathrm{ct}} ) , \quad \Sigma_{m^+}^{\gamma\prime}(p^2) = \frac{d}{d p^2} \Sigma_{m^+}^\gamma(p^2) ,
	\end{split}
\end{align}
where $p$ denotes the meson momentum and $h.o.$ stands for higher order terms.

I find the following field strength renormalisations:
{\small
\begin{align}
	\begin{split}
		Z_{\pi^+}^\gamma &= 1 + e^2 \left( \frac{A_0(\mg^2)}{\mg^2} + 2 B_0(\mpip^2, \mpip^2, \mg^2) + 4 \mpip^2 B_0^\prime(\mpip^2, \mpip^2, \mg^2) \right) - \frac{4}{9} e^2 \left( 6 K_1 + 5 K_5 \right) , \\
		Z_{K^+}^\gamma &= 1 + e^2 \left( \frac{A_0(\mg^2)}{\mg^2} + 2 B_0(\mkp^2, \mkp^2, \mg^2) + 4 \mkp^2 B_0^\prime(\mkp^2, \mkp^2, \mg^2) \right) - \frac{4}{9} e^2 \left( 6 K_1 + 5 K_5 \right) .
	\end{split}
\end{align} }%

Finally, we need the field strength renormalisation of the lepton. The two diagrams~\ref{img:Kl4_lSEgLoop} and \ref{img:Kl4_lSECT} contribute to the self-energy:
\begin{align}
	\begin{split}
		Z_\ell^\gamma &= 1 + \Sigma_\ell^{\gamma\prime}(\ml) + h.o. , \\
		\Sigma_\ell^\gamma(\slashed p) &=  i ( \mathcal{D}_\ell^{\gamma-\mathrm{loop}} + \mathcal{D}_\ell^{\gamma-\mathrm{ct}} ) , \quad \Sigma_\ell^{\gamma\prime}(\slashed p) = \frac{d}{d \slashed p} \Sigma_\ell^\gamma(\slashed p) .
	\end{split}
\end{align}
Up to terms that vanish for $\mg\to0$, the lepton self-energy is given by
\begin{align}
	\begin{split}
		Z_\ell^\gamma &= 1 + e^2 \left( \frac{3 A_0(\mg^2)}{\mg^2} - \frac{3 A_0(\ml^2)}{\ml^2} - X_6 - \frac{3}{16\pi^2}  \right) .
	\end{split}
\end{align}
The contribution of the field strength renormalisation to the form factors is therefore
\begin{align}
	\begin{split}
		\delta F_{\gamma-Z}^\mathrm{NLO} &= \delta G_{\gamma-Z}^\mathrm{NLO} =  Z_{\pi^+}^\gamma \sqrt{Z_{K^+}^\gamma Z_\ell^\gamma} - 1 \\
			&= e^2 \begin{aligned}[t]
				& \bigg( B_0(\mkp^2, \mkp^2, \mg^2) + 2 B_0(\mpip^2, \mpip^2, \mg^2) \\
				& - \frac{A_0(\mkp^2)}{\mkp^2} - \frac{2 A_0(\mpip^2)}{\mpip^2} - \frac{3 A_0(\ml^2)}{2 \ml^2} + \frac{6 A_0(\mg^2)}{\mg^2}  - 4 K_1 - \frac{10}{3} K_5 - \frac{1}{2} X_6 - \frac{15}{32\pi^2} \bigg) .
				\end{aligned}
	\end{split}
\end{align}
The mass renormalisation of the lepton is given by
\begin{align}
	\label{eqn:LeptonMassRenormalisation}
	\begin{split}
		\ml^\mathrm{ph} = \ml^\mathrm{NLO} + h.o., \quad \ml^\mathrm{NLO} &= \ml^\mathrm{bare} + \delta \ml = \ml^\mathrm{bare} + \Sigma_\ell^\gamma(\slashed p = \ml) \\
			&= e^2 \ml \left( \frac{1}{16\pi^2} - \frac{3 A_0(\ml^2)}{\ml^2} - X_6 - X_7 \right) .
	\end{split}
\end{align}

\end{appendices}

\cleardoublepage
\addtocontents{toc}{\protect\enlargethispage*{1cm}}
\renewcommand\bibname{References}
\renewcommand{\bibfont}{\raggedright}
\bibliographystyle{my-physrev}
\phantomsection
\addcontentsline{toc}{chapter}{References}
\bibliography{Isospin/Literature}


\part[Dispersive Approach to HLbL~Scattering]{Dispersive Approach to Hadronic Light-by-Light Scattering \\ and the Muon $g-2$
                             \\ 
                  \begin{center}
                     \begin{minipage}[l]{11cm}
                     \vspace{2cm}
                     \normalsize
                     \begin{center} 
                     	\textnormal{based on a project in collaboration with \\[0.15cm] Gilberto Colangelo, Martin Hoferichter and Massimiliano Procura}
                     \end{center}
                     \end{minipage}
                  \end{center}
                 }

\addtocontents{toc}{\vskip-6pt\par\noindent\protect\hrulefill\par}

\setcounter{chapter}{0}


\chapter{Introduction}

Anomalous magnetic moments have played a crucial role in the history of particle physics. Already in 1928, Dirac predicted that the relation between the intrinsic spin of a lepton and its magnetic moment should be
\begin{align}
	\begin{split}
		\vec \mu_\ell &= g_\ell \frac{e}{2m_\ell} \vec s
	\end{split}
\end{align}
with the proportionality factor $g_\ell = 2$, called Landé $g$-factor or gyromagnetic ratio. This result of the relativistic Dirac theory \cite{Dirac1928,Dirac1928a} was unexpected, as the gyromagnetic ratio of a classical charged body (and the one associated with orbital angular momentum) is $g=1$. In 1934, the $g$-factor of the electron was measured to be compatible with Dirac's prediction \cite{Kinster1934}.

In 1947, almost twenty years after Dirac's prediction, measurements of the hyperfine structure of atomic spectra suggested a deviation from $g_e = 2$ \cite{Nafe1947,Nagle1947,Breit1947}. In 1948 (submitted on 26th~December, 1947), Kusch and Foley published the first precise determination of the $g$-factor of the electron, finding a deviation from $g_e=2$ \cite{Foley1948}. Four days later, Schwinger submitted the result of his one-loop calculation of the magnetic moment in quantum electrodynamics -- one of the first loop calculations in QED \cite{Schwinger1948}. This quantum correction explained the deviation from $g_\ell=2$, called anomalous magnetic moment:
\begin{align}
	\begin{split}
		a_\ell := \frac{g_\ell - 2}{2} = \frac{\alpha}{2\pi} + \O(\alpha^2) .
	\end{split}
\end{align}
The agreement with the observed value signified an enormous success of quantum electrodynamics and helped to establish not only QED but also quantum field theory in general as the framework to describe particle physics.

Experimentally, the anomalous magnetic moment of the muon $(g-2)_\mu$ is a much bigger challenge than the one of the electron, because the muon is unstable and has a lifetime of $\sim 2.2 \cdot 10^{-6}$~s \cite{PDG2012}. The first measurement of the $(g-2)_\mu$ was performed in 1960 at Columbia University \cite{Garwin1960}. The precision experiment at CERN in 1961 established the muon as a heavy sibling of the electron \cite{Charpak1961}. A series of $(g-2)_\mu$ experiments at CERN improved the precision, the second muon storage ring (1969--1976) reached 7~ppm \cite{Bailey1977}. Nowadays, the world average is completely dominated by the measurement of the E821 experiment at Brookhaven National Laboratory, which reached in the years 2000--2004 an enormous precision of 0.54~ppm \cite{Bennett2006}. Therefore, nowadays the $(g-2)_\mu$ probes not only QED but the whole standard model (SM).

Apart from possible logarithmic corrections, the influence of heavy virtual particles of mass $M$ on the anomalous magnetic moment scales like \cite{Jegerlehner2009}
\begin{align}
	\begin{split}
		\frac{\Delta a_\ell}{a_\ell} \propto \frac{m_\ell^2}{M^2} .
	\end{split}
\end{align}
Although the electron $g$-factor is known to a much higher precision than the one of the muon, this scaling implies that the $(g-2)_\mu$ is still about 20 times more sensitive to contributions of new physics than the $(g-2)_e$, because of the much heavier muon mass, $(m_\mu/m_e)^2 \sim 4 \cdot 10^4$. However, for the same reason the muon is also much more sensitive to virtual effects due to electroweak and hadronic states.

It is very interesting that for more than a decade a discrepancy of about $3\sigma$ between the SM prediction and the experimental measurement of the $(g-2)_\mu$ has persisted. The origin of this discrepancy is still unresolved: could it be a systematic error in the experiment that has not been taken into account or is the interpretation of the experiment problematic? Or is the reason to be found on the theory side? If this is the case, the discrepancy could either be a hint to new physics or just as well due to badly understood SM effects. An answer to this thrilling question might be found in a couple of years: new experiments at Fermilab and J-PARC aim at reducing the experimental uncertainty by a factor of 4. In order to take advantage of this increased experimental precision, the theory prediction should improve by a similar factor because at present, theoretical and experimental uncertainty are about the same size.

\begin{table}[H]
	\centering
	\begin{tabular}{l . d r}
		\toprule
		& \multicolumn{1}{c}{$10^{11} \cdot a_\mu$} & \multicolumn{1}{c}{$10^{11} \cdot \Delta a_\mu$}  & $\quad$ Ref. \\
		\midrule
		BNL E821 & 116\,592\,091 & 63 & \cite{PDG2012}\\
		\midrule
		QED $\O(\alpha)$ & 116\,140\,973.32 & 0.08 \\
		QED $\O(\alpha^2)$ & 413\,217.63 & 0.01 \\
		QED $\O(\alpha^3)$ & 30\,141.90 & 0.00 \\
		QED $\O(\alpha^4)$ & 381.01 & 0.02 \\
		QED $\O(\alpha^5)$ & 5.09 & 0.01 \\
		QED total & 116\,584\,718.95 & 0.08 & \cite{Aoyama2012} \\
		\midrule
		EW & 153.6 & 1.0 & \\
		\midrule
		LO HVP & 6\,949 & 43 & \cite{Hagiwara2011} \\
		NLO HVP & -98 & 1 & \cite{Hagiwara2011} \\
		NNLO HVP & 12.4 & 0.1 & \cite{Kurz2014} \\
		LO HLbL & 116 & 40 & \cite{Jegerlehner2009} \\
		NLO HLbL & 3 & 2 & \cite{Colangelo2014} \\
		Hadronic total & 6982 & 59 & \\
		\midrule
		Theory total & 116\,591\,855 & 59 \\
		\bottomrule
	\end{tabular}
	\caption{Comparison of the experimental determination of $a_\mu$ with the theory prediction in the standard model.}
	\label{tab:SMContributionsAmu}
\end{table}

\begin{figure}[ht]
	\centering
	\includegraphics[width=4cm]{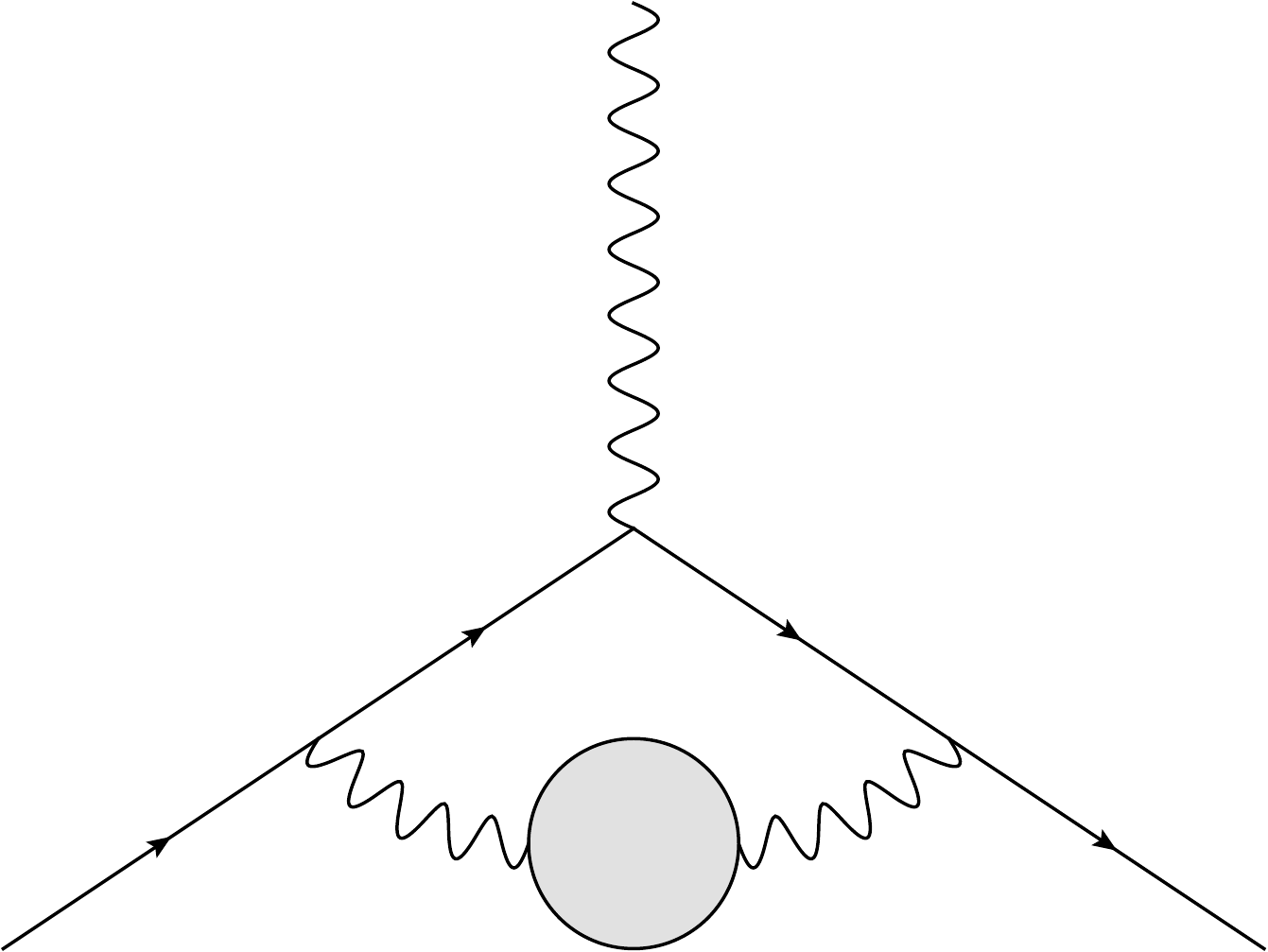}

	\vspace{-0.2cm}\hspace{.1cm}${}_{had.}$

	\caption{Diagram representing the hadronic vacuum polarisation contribution to the $(g-2)_\mu$.}
	\label{img:HVP}
\end{figure}

Table~\ref{tab:SMContributionsAmu} shows a comparison of the experimental value for $a_\mu$ and the different contributions to the theoretical calculation within the standard model, together with the associated uncertainties. An interesting pattern is visible concerning the origin of the theory uncertainties:
\begin{itemize}
	\item QED and electroweak (EW) contributions are known to very high precision, the total theory uncertainty is almost exclusively due to hadronic effects;
	\item the largest uncertainty comes from hadronic vacuum polarisation (HVP);
	\item the second largest uncertainty is due to the hadronic light-by-light contribution (HLbL).
\end{itemize}
Let us therefore focus on the hadronic effects. The leading hadronic contribution is due to hadronic vacuum polarisation, shown in figure~\ref{img:HVP}. The state-of-the-art evaluation of the hadronic `blob' inside the loop diagram is based on unitarity and analyticity: the optical theorem relates the imaginary part of the vacuum polarisation function to the total hadronic cross-section in $e^+e^-$ annihilation (properly `undressed' by radiative effects). A dispersion integral over this imaginary part reconstructs the real part of the renormalised vacuum polarisation function. Therefore, the HVP contribution to the $(g-2)_\mu$ is given as an integral over an experimentally accessible quantity. Given a dedicated $e^+e^-$ program (BaBar, Belle, BESIII, CMD3, KLOE2, SND), it can be expected that the data input to the dispersion integral will be improved significantly and therefore the uncertainty of the HVP contribution to the $(g-2)_\mu$ will be much reduced in the next couple of years, see e.g.~\cite{Blum2013}.

\begin{figure}[ht]
	\centering
	\includegraphics[width=4cm]{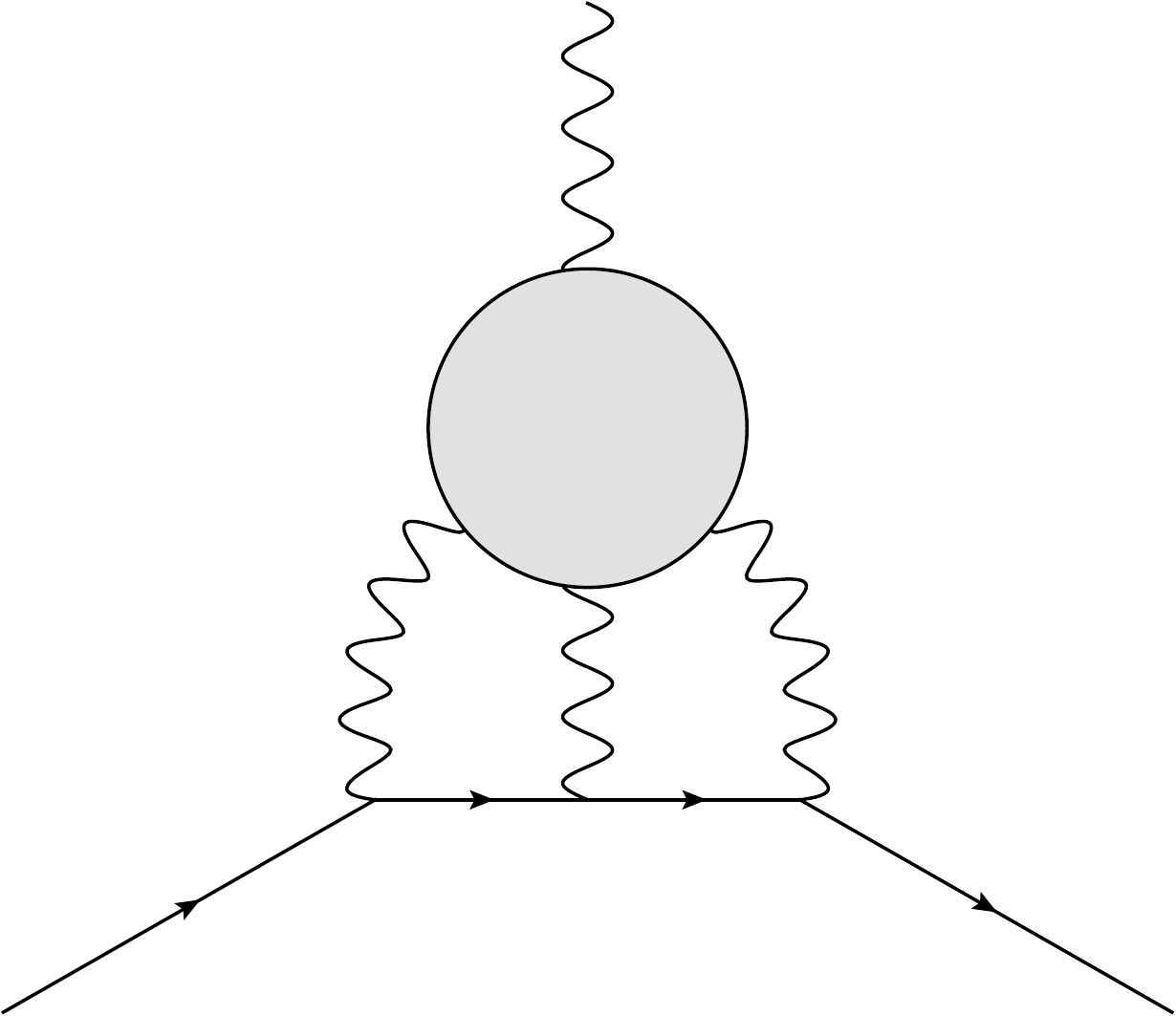}

	\vspace{-2.4cm}\hspace{.15cm}${}_{had.}$
	\vspace{2cm}

	\caption{Diagram representing the hadronic light-by-light contribution to the $(g-2)_\mu$.}
	\label{img:HLbLContribution}
\end{figure}

This means that the hadronic light-by-light contribution, illustrated in figure~\ref{img:HLbLContribution}, will soon dominate the theory uncertainty. So far, only model calculations of the HLbL contribution are available \cite{Rafael1994, Bijnens1995, Bijnens1996, Bijnens2002, Hayakawa1995, Hayakawa1996, Hayakawa1998, Knecht2002a, Knecht2002, Ramsey-Musolf2002, Melnikov2004, Goecke2011}. A systematic improvement of these model calculations and their uncertainty estimate seems difficult. Also, it is unclear if first-principle calculations based on lattice QCD will become competitive \cite{Hayakawa2006, Blum2012}.

While the use of dispersion relations as a model-independent framework is standard for evaluating the HVP contribution, it has been repeatedly claimed that such an approach is not possible in the case of HLbL scattering \cite{Kinoshita1985, Bijnens1995, Blum2012, Rafael2013}. Indeed, the object in question is certainly much more complicated: in the case of HVP, a hadronic matrix element of two electromagnetic currents must be evaluated, which is determined due to gauge invariance by a single scalar function of only one kinematic variable:
\begin{align}
	\begin{split}
			\minidiagSize{HLbL}{HVP2}{0.75cm} \hspace{1.85cm} &= - \int d^4x e^{iqx} \<0 | T j_\mu^\mathrm{em}(x) j_\nu^\mathrm{em}(0) | 0 \> = - i (q^2 g_{\mu\nu} - q_\mu q_\nu) \Pi(q^2) .
	\end{split}
\end{align}
In contrast, HLbL scattering is described by a hadronic four-point function. Here, we present a formalism based on the fundamental principles of unitarity, analyticity, crossing symmetry and gauge invariance that will allow a dispersive description of the HLbL contribution to the $(g-2)_\mu$ by establishing a relation to quantities that are either experimentally accessible or can again be reconstructed dispersively with data input.

We have presented a first step in this direction in \cite{Colangelo2014a}. The present treatment overcomes the main difficulties of our previous formalism by the construction of a generating set for the Lorentz structures of the HLbL tensor that is explicitly free of kinematic singularities and zeros. This technique simplifies some parts of the calculation considerably and allows for the inclusion of higher partial waves than $S$-waves, which was not feasible in the approach of \cite{Colangelo2014a}.

The outline is as follows: in chapter~\ref{sec:SubProcess}, we first discuss some aspects of the process $\gamma^*\gamma^*\to\pi\pi$, which will be a crucial input for the dispersive treatment of HLbL scattering. It also allows us to illustrate in a simpler example some techniques that we apply afterwards to HLbL. In chapter~\ref{sec:LorentzStructureHLbLTensor}, we derive the decomposition of the HLbL tensor into a set of scalar functions that are free of kinematics. In chapter~\ref{sec:HLbLContributionToGminus2}, we use this decomposition to derive a master formula for the HLbL contribution to the $(g-2)_\mu$ that is parametrised by the scalar functions. This formula could be used for any given calculation of the HLbL tensor and is suited for a direct numerical implementation. In chapter~\ref{sec:MandelstamRepresentation}, we use the Mandelstam representation to derive a dispersive description of the scalar functions that parametrise the HLbL tensor. This framework relates the HLbL tensor to the pion transition form factor, the pion electromagnetic form factor and $\gamma^*\gamma^*\to\pi\pi$ partial waves. In chapter~\ref{sec:HLbLDiscussionConclusion}, we conclude with a discussion of the input and an outlook. In the appendices, we collect some of the lengthier analytic expressions.


\chapter{The Sub-Process $\gamma^*\gamma^*\to\pi\pi$}

\label{sec:SubProcess}

As a prelude, we discuss in this chapter the process $\gamma^*\gamma^*\to\pi\pi$, which will become important as a sub-process when we write down a dispersion relation for the HLbL tensor. Because the Lorentz structure of $\gamma^*\gamma^*\to\pi\pi$ is much simpler than the one of light-by-light scattering, it also allows us to illustrate a technique for the construction of kinematic-free amplitudes, which we will apply afterwards to the more complicated case of HLbL.

\section{Kinematics and Matrix Element}

\begin{figure}[ht]
	\centering
	\includegraphics[width=4cm]{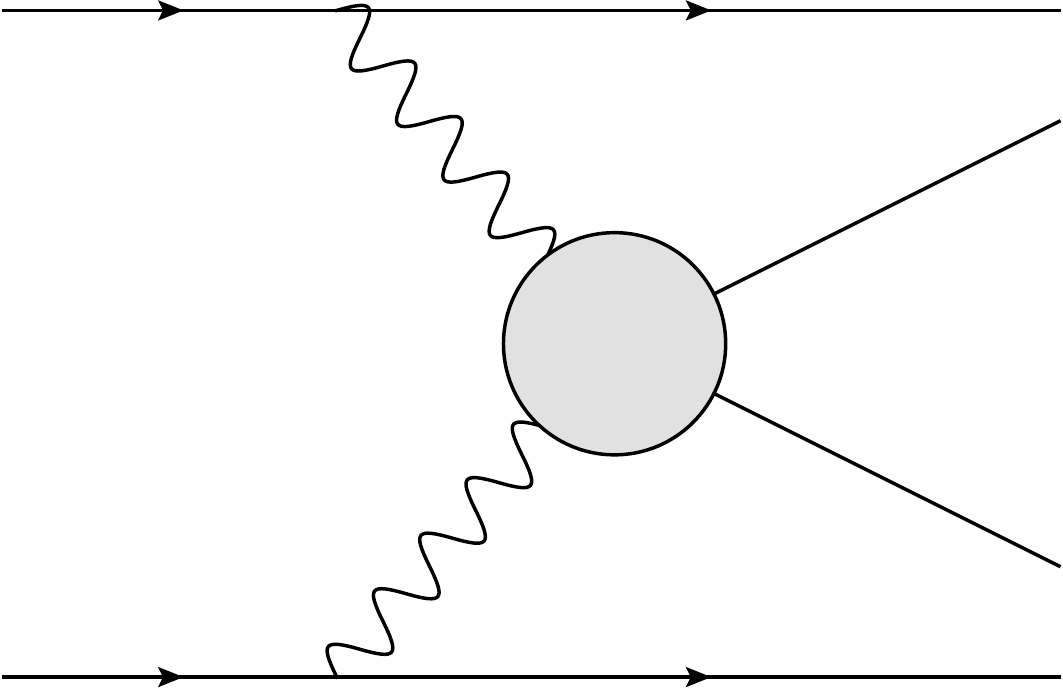}
	\caption{$\gamma^*\gamma^*\to\pi\pi$ as a sub-process of $e^-e^-\to e^-e^-\pi\pi$.}
	\label{img:eetopipi}
\end{figure}

Consider the process
\begin{align}
	\begin{split}
		e^-(k_1) e^-(k_2) \to e^-(k_3) e^-(k_4) \gamma^*(q_1) \gamma^*(q_2) \to e^-(k_3) e^-(k_4) \pi^a(p_1) \pi^b(p_2),
	\end{split}
\end{align}
shown in figure~\ref{img:eetopipi}. At $\O(e^4)$, the amplitude for this process is given by
\begin{align}
	\begin{split}
		i \mathcal{T} &= \bar u(k_3) (-i e \gamma_\alpha) u(k_1) \bar u(k_4) (-i e \gamma_\beta) u(k_2) \\
			& \quad \cdot \frac{-i}{q_1^2} \left( g^{\alpha\mu} - (1-\xi) \frac{q_1^\alpha q_1^\mu}{q_1^2} \right) \frac{-i}{q_2^2} \left( g^{\beta\nu} - (1-\xi) \frac{q_2^\alpha q_2^\mu}{q_2^2} \right) i e^2 W_{\mu\nu}^{ab}(p_1,p_2,q_1) ,
	\end{split}
\end{align}
where $\xi$ is an arbitrary gauge parameter for the photon propagators and the tensor $W_{\mu\nu}^{ab}$ is defined as the pure QCD matrix element
\begin{align}
	\begin{split}
		W^{\mu\nu}_{ab}(p_1,p_2,q_1) = i \int d^4x \, e^{-i q_1 x} \<\pi^a(p_1) \pi^b(p_2) | T \{ j_\mathrm{em}^\mu(x) j_\mathrm{em}^\nu(0) \} | 0 \> .
	\end{split}
\end{align}
The contraction thereof with appropriate polarisation vectors may be understood as an amplitude for the artificial process
\begin{align}
	\begin{split}
		\gamma^*(q_1,\lambda_1) \gamma^*(q_2,\lambda_2) \to \pi^a(p_1) \pi^b(p_2) ,
	\end{split}
\end{align}
where $\lambda_{1,2}$ denote the helicities of the off-shell photons. We define the connected part of such a matrix element by
\begin{align}
	\begin{split}
		& \< \pi^a(p_1) \pi^b(p_2) | \gamma^*(q_1,\lambda_1) \gamma^*(q_2,\lambda_2) \> \\
		&  = - e^2 \epsilon_\mu^{\lambda_1}(q_1) \epsilon_\nu^{\lambda_2}(q_2) \int d^4x \, d^4y \, e^{-i (q_1 x + q_2 y)} \<\pi^a(p_1) \pi^b(p_2)| T \{ j_\mathrm{em}^\mu(x) j_\mathrm{em}^\nu(y) \} | 0 \> \\
		&= - e^2 (2\pi)^4 \delta^{(4)}(p_1+p_2-q_1-q_2) \epsilon_\mu^{\lambda_1}(q_1) \epsilon_\nu^{\lambda_2}(q_2) \\
			& \quad \cdot \int d^4x \, e^{-i q_1 x} \<\pi^a(p_1) \pi^b(p_2)| T \{ j_\mathrm{em}^\mu(x) j_\mathrm{em}^\nu(0) \} | 0 \>  \\
		&= i e^2 (2\pi)^4 \delta^{(4)}(p_1+p_2-q_1-q_2) \epsilon_\mu^{\lambda_1}(q_1) \epsilon_\nu^{\lambda_2}(q_2) W^{\mu\nu}_{ab}(p_1,p_2,q_1) .
	\end{split}
\end{align}
The helicity amplitudes are given by the contraction with polarisation vectors:
\begin{align}
	\begin{split}
		 \epsilon_\mu^{\lambda_1}(q_1) \epsilon_\nu^{\lambda_2}(q_2) W^{\mu\nu}_{ab}(p_1,p_2,q_1) = e^{i(\lambda_1-\lambda_2)\phi} H_{\lambda_1\lambda_2}^{ab} .
	\end{split}
\end{align}
We introduce the following kinematic variables:
\begin{align}
	\begin{split}
		s &:= (q_1+q_2)^2 = (p_1+p_2)^2 , \\
		t &:= (q_1-p_1)^2 = (q_2-p_2)^2 , \\
		u &:= (q_1-p_2)^2 = (q_2-p_1)^2 ,
	\end{split}
\end{align}
which satisfy $s+t+u = q_1^2 + q_2^2 + 2 M_\pi^2$. (In this chapter~\ref{sec:SubProcess} only, $s$, $t$ and $u$ refer to the Mandelstam variables of $\gamma^*\gamma^*\to\pi\pi$ and not to the ones of the HLbL tensor.)

\section{Tensor Decomposition}

\label{sec:ggtopipiTensorDecomposition}

The tensor $W^{\mu\nu}_{ab}$ can be decomposed based on Lorentz covariance as (we drop isospin indices)
\begin{align}
	\begin{split}
		W^{\mu\nu} = g^{\mu\nu} W_1 + q_i^\mu q_j^\nu W_2^{ij},
	\end{split}
\end{align}
where we abbreviate $q_i = \{ q_1, q_2, p_2-p_1 \}$ and where double indices are summed. The ten coefficient functions $\{W_1, W_2^{ij}\}$ cannot contain any kinematic but only dynamic singularities. However, they have to fulfil kinematic constraints that are required e.g.~by gauge invariance, hence they contain kinematic zeros. Conservation of the electromagnetic current implies the Ward identities
\begin{align}
	\begin{split}
		q_1^\mu W_{\mu\nu} = q_2^\nu W_{\mu\nu} = 0 ,
	\end{split}
\end{align}
which impose five linearly independent relations on the scalar functions and reduce the set to five independent functions.

Let us construct now a set of scalar functions which are free of both kinematic singularities and zeros. We follow the recipe given by Bardeen and Tung \cite{Bardeen1968} and Tarrach \cite{Tarrach1975}. As shown in \cite{Tarrach1975}, there exists no minimal basis (consisting of five functions) free of kinematic singularities and zeros. However, a redundant set of six structures can be constructed, which fulfils the requirements.

We define the projector
\begin{align}
	\begin{split}
		I^{\mu\nu} = g^{\mu\nu} - \frac{q_2^\mu q_1^\nu}{q_1 \cdot q_2} ,
	\end{split}
\end{align}
which satisfies
\begin{align}
	\begin{split}
		{I_\mu}^\lambda W_{\lambda\nu} &= W_{\mu\lambda} {I^\lambda}_\nu = W_{\mu\nu} , \\
		q_1^\mu I_{\mu\nu} &= q_2^\nu I_{\mu\nu} = 0 ,
	\end{split}
\end{align}
i.e.~the tensor $W^{\mu\nu}$ is invariant under contraction with the projector, but the contraction of every Lorentz structure produces a gauge-invariant structure.

Let us apply this projector for both photons:
\begin{align}
	\begin{split}
		W_{\mu\nu} &= I_{\mu\mu^\prime} I_{\nu^\prime\nu} W^{\mu^\prime\nu^\prime} = \sum_{i=1}^5 \bar T_{\mu\nu}^i \bar A_i ,
	\end{split}
\end{align}
where
\begin{align}
	\begin{alignedat}{2}
		\bar T_1^{\mu\nu} &= g^{\mu\nu} - \frac{q_2^\mu q_1^\nu}{q_1 \cdot q_2} , \quad & \bar A_1 &= W_1 , \\
		\bar T_2^{\mu\nu} &= q_1^\mu q_2^\nu - \frac{q_1^2 q_2^\mu q_2^\nu}{q_1\cdot q_2} - \frac{q_2^2 q_1^\mu q_1^\nu}{q_1 \cdot q_2} + \frac{q_1^2 q_2^2 q_2^\mu q_1^\nu}{(q_1 \cdot q_2)^2} , \quad & \bar A_2 &= W_2^{12} , \\
		\bar T_3^{\mu\nu} &= q_1^\mu q_3^\nu - \frac{q_1^2 q_2^\mu q_3^\nu}{q_1 \cdot q_2} - \frac{q_2 \cdot q_3 q_1^\mu q_1^\nu}{q_1 \cdot q_2} + \frac{q_1^2 q_2 \cdot q_3 q_2^\mu q_1^\nu}{(q_1 \cdot q_2)^2} , \quad & \bar A_3 &= W_2^{13} , \\
		\bar T_4^{\mu\nu} &= q_3^\mu q_2^\nu - \frac{q_2^2 q_3^\mu q_1^\nu}{q_1 \cdot q_2} - \frac{q_1 \cdot q_3 q_2^\mu q_2^\nu}{q_1 \cdot q_2} + \frac{q_2^2 q_1 \cdot q_3 q_2^\mu q_1^\nu}{(q_1 \cdot q_2)^2} , \quad & \bar A_4 &= W_2^{32} , \\
		\bar T_5^{\mu\nu} &= q_3^\mu q_3^\nu - \frac{q_1 \cdot q_3 q_2^\mu q_3^\nu}{q_1 \cdot q_2} - \frac{q_2 \cdot q_3 q_3^\mu q_1^\nu}{q_1 \cdot q_2} + \frac{q_1 \cdot q_3 q_2 \cdot q_3 q_2^\mu q_1^\nu}{(q_1 \cdot q_2)^2} , \quad & \bar A_5 &= W_2^{33} .
	\end{alignedat}
\end{align}
As the functions $\bar A_i$ are a subset of the original scalar functions, they are still free of kinematic singularities, but contain zeros, because the Lorentz structures contain singularities. We have to remove now the single and double poles in $q_1 \cdot q_2$ from the Lorentz structures $\bar T_i^{\mu\nu}$. This is achieved as follows \cite{Bardeen1968}:
\begin{itemize}
	\item remove as many double poles as possible by adding to the structures linear combinations of other structures with non-singular coefficients.
	\item if no more double poles can be removed in this way, multiply the structures that still contain double poles by $q_1 \cdot q_2$.
	\item proceed in the same way with single poles.
\end{itemize}

It turns out that no double poles in $q_1 \cdot q_2$ can be removed by adding to the structures multiples of the other structures, hence $\bar T_2^{\mu\nu}$, \ldots, $\bar T_5^{\mu\nu}$ have to be multiplied by $q_1 \cdot q_2$. The resulting simple poles can be removed by adding multiples of $\bar T_1^{\mu\nu}$. In the end, we have to multiply $\bar T_1^{\mu\nu}$ by $q_1\cdot q_2$ in order to remove the last pole. We then arrive at the following representation:
\begin{align}
	\begin{split}
		W_{\mu\nu} &= \sum_{i=1}^5 T_{\mu\nu}^i \tilde A_i ,
	\end{split}
\end{align}
where
\begin{align}
	\begin{split}
		T_1^{\mu\nu} &= q_1 \cdot q_2 g^{\mu\nu} - q_2^\mu q_1^\nu , \\
		T_2^{\mu\nu} &= q_1^2 q_2^2 g^{\mu\nu} + q_1 \cdot q_2 q_1^\mu q_2^\nu - q_1^2 q_2^\mu q_2^\nu - q_2^2 q_1^\mu q_1^\nu , \\
		T_3^{\mu\nu} &= q_1^2 q_2 \cdot q_3 g^{\mu\nu} + q_1 \cdot q_2 q_1^\mu q_3^\nu - q_1^2 q_2^\mu q_3^\nu - q_2 \cdot q_3 q_1^\mu q_1^\nu , \\
		T_4^{\mu\nu} &= q_2^2 q_1 \cdot q_3 g^{\mu\nu} + q_1 \cdot q_2 q_3^\mu q_2^\nu - q_2^2 q_3^\mu q_1^\nu - q_1 \cdot q_3 q_2^\mu q_2^\nu , \\
		T_5^{\mu\nu} &= q_1 \cdot q_3 q_2 \cdot q_3  g^{\mu\nu} + q_1 \cdot q_2 q_3^\mu q_3^\nu - q_1 \cdot q_3 q_2^\mu q_3^\nu - q_2 \cdot q_3 q_3^\mu q_1^\nu ,
	\end{split}
\end{align}
and
\begin{align}
	\begin{split}
		\tilde A_1 &= \frac{1}{q_1 \cdot q_2} W_1 - \frac{q_1^2 q_2^2}{(q_1 \cdot q_2)^2} W_2^{12} - \frac{q_1^2 q_2 \cdot q_3}{(q_1\cdot q_2)^2} W_2^{13} - \frac{q_2^2 q_1\cdot q_3}{(q_1\cdot q_2)^2} W_2^{32} - \frac{q_1 \cdot q_3 q_2 \cdot q_3}{(q_1 \cdot q_2)^2} W_2^{33} , \\
		\tilde A_2 &= \frac{1}{q_1 \cdot q_2} W_2^{12} , \\
		\tilde A_3 &= \frac{1}{q_1 \cdot q_2} W_2^{13} , \\
		\tilde A_4 &= \frac{1}{q_1 \cdot q_2} W_2^{32} , \\
		\tilde A_5 &= \frac{1}{q_1 \cdot q_2} W_2^{33} .
	\end{split}
\end{align}
As shown by Tarrach, it is not possible to construct a minimal basis free of kinematic singularities and zeros \cite{Tarrach1975}. The structures $T_i^{\mu\nu}$ form a basis for $q_1 \cdot q_2 \neq 0$, but are degenerate for $q_1 \cdot q_2 = 0$. This degeneracy implies that there is a linear combination of the structures $T_i^{\mu\nu}$ that is proportional to $q_1 \cdot q_2$. Hence, a sixth structure
\begin{align}
	\begin{split}
		T_6^{\mu\nu} = \left(q_1^2 q_3^\mu - q_1 \cdot q_3 q_1^\mu \right) \left(q_2^2 q_3^\nu - q_2 \cdot q_3 q_2^\nu \right) 
	\end{split}
\end{align}
has to be added by hand. Projected on the basis, it gives coefficients with poles in $q_1 \cdot q_2$. Although there is no basis free of kinematic singularities and zeros, we have found the exact form of the singularities:
\begin{align}
	\begin{split}
		W_{\mu\nu} &= \sum_{i=1}^5 T_{\mu\nu}^i \tilde A_i =  \sum_{i=1}^6 T_{\mu\nu}^i A_i ,
	\end{split}
\end{align}
where
\begin{align}
	\begin{split}
		\tilde A_1 &= A_1 , \\
		\tilde A_2 &= A_2 + \frac{q_1 \cdot q_3 q_2 \cdot q_3}{q_1 \cdot q_2} A_6 , \\
		\tilde A_3 &= A_3 - \frac{q_2^2 q_1 \cdot q_3}{q_1 \cdot q_2} A_6 , \\
		\tilde A_4 &= A_4 - \frac{q_1^2 q_2 \cdot q_3}{q_1 \cdot q_2} A_6 , \\
		\tilde A_5 &= A_5 + \frac{q_1^2 q_2^2}{q_1 \cdot q_2} A_6 .
	\end{split}
\end{align}
The functions $A_i$ are now free of both kinematic singularities and zeros. The tensor structures can be written in terms of the Mandelstam variables $t$ and $u$ as
\begin{align}
	\begin{split}
		T_1^{\mu\nu} &= q_1 \cdot q_2 g^{\mu\nu} - q_2^\mu q_1^\nu , \\
		T_2^{\mu\nu} &= q_1^2 q_2^2 g^{\mu\nu} + q_1 \cdot q_2 q_1^\mu q_2^\nu - q_1^2 q_2^\mu q_2^\nu - q_2^2 q_1^\mu q_1^\nu , \\
		T_3^{\mu\nu} &= q_1 \cdot q_2 q_1^\mu q_3^\nu - q_1^2 q_2^\mu q_3^\nu - \frac{1}{2} (t-u) q_1^2 g^{\mu\nu} + \frac{1}{2} (t-u) q_1^\mu q_1^\nu , \\
		T_4^{\mu\nu} &= q_1 \cdot q_2 q_3^\mu q_2^\nu - q_2^2 q_3^\mu q_1^\nu + \frac{1}{2} (t-u) q_2^2 g^{\mu\nu} - \frac{1}{2} (t-u) q_2^\mu q_2^\nu , \\
		T_5^{\mu\nu} &= q_1 \cdot q_2 q_3^\mu q_3^\nu - \frac{1}{4} (t-u)^2  g^{\mu\nu} + \frac{1}{2} (t-u) \left( q_3^\mu q_1^\nu - q_2^\mu q_3^\nu \right) .
	\end{split}
\end{align}
Since the two-photon state is even under charge conjugation, so must be the two-pion state. Therefore, the isospin $I=1$ amplitude vanishes. Bose symmetry implies that the amplitude and hence the tensor $W^{\mu\nu}$ is invariant under $p_1 \leftrightarrow p_2$ or equivalently $q_3 \leftrightarrow -q_3$. Under this transformation, the tensor structures $T_i^{\mu\nu}$ are even for $i=1,2,5$ and odd for $i=3,4$. It is natural to choose $A_6$ to be even under $t\leftrightarrow u$, hence, $A_{3,4}$ must be odd. We can therefore write
\begin{align}
	\begin{split}
		A_{3,4} = (t-u) \hat A_{3,4} ,
	\end{split}
\end{align}
where $\hat A_{3,4}$ are free of kinematic singularities.

Crossing symmetry of the photons requires the invariance of $W^{\mu\nu}$ under $q_1 \leftrightarrow q_2$, $\mu \leftrightarrow \nu$. While $T_{1,2,5}^{\mu\nu}$ are invariant under this transformation, we observe the crossing relation $T_3^{\mu\nu} \leftrightarrow T_4^{\mu\nu}$. Hence, for fixed Mandelstam variables, $A_{1,2,5}$ and $\hat A_3 - \hat A_4$ are even under $q_1^2 \leftrightarrow q_2^2$, while $\hat A_3 + \hat A_4$ is odd.

\section{Helicity Amplitudes and Soft-Photon Zeros}

Let us construct the helicity amplitudes with the momenta and polarisation vectors in the centre-of-mass frame. We define the particle momenta as
\begin{align}
	\begin{split}
		q_1 &= ( E_{q_1}, 0, 0, |\vec q| ) , \quad q_2 = ( E_{q_2}, 0, 0, -|\vec q| ) , \\
		p_1 &= ( E_p, |\vec p| \sin \theta \cos\phi, |\vec p| \sin \theta \sin\phi, |\vec p| \cos\theta ) , \\
		p_2 &= ( E_p, -|\vec p| \sin \theta \cos\phi, -|\vec p| \sin \theta \sin\phi, -|\vec p| \cos\theta ) , \\
	\end{split}
\end{align}
where
\begin{align}
	\begin{split}
		E_{q_1} &= \sqrt{ q_1^2 + \vec q^2} = \frac{s + q_1^2 - q_2^2}{2\sqrt{s}}, \quad E_{q_2} = \sqrt{ q_2^2 + \vec q^2} = \frac{s - q_1^2 + q_2^2}{2\sqrt{s}}, \quad |\vec q| = \frac{\lambda^{1/2}(s,q_1^2,q_2^2)}{2\sqrt{s}} , \\
		E_p &= \sqrt{ M_\pi^2 + \vec p^2} = \frac{\sqrt{s}}{2} , \quad |\vec p| = \sqrt{\frac{s}{4} - M_\pi^2} = \frac{\sqrt{s}}{2} \sigma_\pi(s) .
	\end{split}
\end{align}
The scattering angle is then given by
\begin{align}
	\begin{split}
		z := \cos\theta = \frac{t-u}{4 |\vec q| |\vec p|} = \frac{t-u}{\sigma_\pi(s) \lambda^{1/2}(s,q_1^2,q_2^2)} .
	\end{split}
\end{align}
We define the polarisation vectors by
\begin{align}
	\begin{split}
		\epsilon_\pm(q_1) &= \mp \frac{1}{\sqrt{2}}( 0, 1, \pm i, 0 ) , \\
		\epsilon_0(q_1) &= \frac{1}{\sqrt{q_1^2}}( |\vec q|, 0, 0, E_{q_1} ) , \\
		\epsilon_\pm(q_2) &= \mp \frac{1}{\sqrt{2}}( 0, 1, \mp i, 0 ) , \\
		\epsilon_0(q_2) &= \frac{1}{\sqrt{q_2^2}}( -|\vec q|, 0, 0, E_{q_2} ) .
	\end{split}
\end{align}
The helicity amplitudes are then given by
\begin{align}
	\begin{split}
		\label{eq:ggpipiHelicityAmplitudes}
		H_{++} = H_{--} &= -\frac{1}{2}(s-q_1^2-q_2^2) A_1 - q_1^2 q_2^2 A_2 + \frac{1}{2}(t-u)( q_1^2 A_3 - q_2^2 A_4) \\
			& \quad + \frac{1}{4} (s-4M_\pi^2)\left( (s - q_1^2 - q_2^2) + \left( \frac{(q_1^2-q_2^2)^2}{s} - (q_1^2 + q_2^2)\right) z^2 \right) A_5 \\
			& \quad + \frac{1}{2} q_1^2 q_2^2 (s-4M_\pi^2) (1-z^2) A_6 , \\
		H_{+-} = H_{-+} &= - \frac{1}{4} (s-4M_\pi^2)(s-q_1^2-q_2^2)(1-z^2) A_5 - \frac{1}{2} q_1^2 q_2^2 (s-4M_\pi^2) (1-z^2) A_6 , \\
		H_{+0} = - H_{-0} &= -\frac{\sqrt{q_2^2}}{4} \begin{aligned}[t]
			&\bigg( \sqrt{2s} \sigma_\pi(s) \lambda^{1/2}(s,q_1^2,q_2^2)  \sqrt{1-z^2} A_4 \\
			& + \sqrt{\frac{2}{s}}(s-4M_\pi^2)(s+q_1^2-q_2^2)z\sqrt{1-z^2} A_5 \\
			& - q_1^2 \sqrt{\frac{2}{s}} (s-4M_\pi^2)(s-q_1^2 + q_2^2) z \sqrt{1-z^2} A_6 \bigg) , \end{aligned} \\
		H_{0+} = - H_{0-} &= \frac{\sqrt{q_1^2}}{4} \begin{aligned}[t]
			&\bigg(  \sqrt{2s} \sigma_\pi(s) \lambda^{1/2}(s,q_1^2,q_2^2) \sqrt{1-z^2} A_3 \\
			& - \sqrt{\frac{2}{s}}(s-4M_\pi^2)(s-q_1^2+q_2^2)z\sqrt{1-z^2} A_5  \\
			& - q_2^2 \sqrt{\frac{2}{s}} (s-4M_\pi^2)(s + q_1^2 - q_2^2) z \sqrt{1-z^2} A_6 \bigg) , \end{aligned} \\
		H_{00} &= \sqrt{q_1^2 q_2^2} \begin{aligned}[t]
			&\bigg( - A_1 - \frac{1}{2}(s - q_1^2 - q_2^2) A_2 - \frac{1}{2}(t-u)(A_3 - A_4) \\
			& + (s-4M_\pi^2) z^2 A_5 + \frac{1}{4s} (s - 4M_\pi^2) \left(s^2 - (q_1^2 - q_2^2)^2\right) z^2 A_6 \bigg) . \end{aligned}
	\end{split}
\end{align}
Since the functions $A_i$ are free of kinematic singularities and zeros, we can read off from these equations the soft-photon zeros \cite{Low1958, Moussallam2013}. Let us consider the limit $q_1\to0$. The Mandelstam variables become
\begin{align}
	\begin{split}
		s = q_2^2, \quad t = u = M_\pi^2 .
	\end{split}
\end{align}
We conclude that the helicity amplitudes must vanish at this point apart from terms containing a dynamic singularity, which will be discussed in the next section. The second soft-photon limit, $q_2\to0$, leads to
\begin{align}
	\begin{split}
		s = q_1^2, \quad t = u = M_\pi^2 ,
	\end{split}
\end{align}
and the same arguments apply for the helicity amplitudes.

Crossing symmetry of the photons implies that under the transformation $q_1^2\leftrightarrow q_2^2$ (and fixed Mandelstam variables), $H_{++}$, $H_{+-}$ and $H_{00}$ remain invariant, the other two helicity amplitudes transform as $H_{+0}\leftrightarrow H_{0+}$.

\clearpage

\section{Pion-Pole Contribution}

\label{sec:PionPoleggpipi}

Let us define the contribution to $\gamma^*\gamma^*\to\pi^+\pi^-$ due to the exchange of a single pion. As we employ a dispersive picture for this definition, the pion-pole contribution coincides with what we understand as the Born contribution.

We assume that the asymptotic behaviour of $\gamma^*\gamma^*\to\pi^+\pi^-$ in the crossed Mandelstam variables $t$ and $u$ is such that an unsubtracted fixed-$s$ dispersion relation can be written for the scalar functions:
\begin{align}
	\begin{split}
		\label{eq:FixedSDispRelggpipi}
		A_i^s(s,t,u) &= \frac{\hat\rho_{i;t}^s(s)}{t-M_\pi^2} + \frac{\hat\rho_{i;u}^s(s)}{u-M_\pi^2} + \frac{1}{\pi} \int_{4M_\pi^2}^\infty dt^\prime \frac{\hat D_{i;t}^s(t^\prime;s)}{t^\prime - t} + \frac{1}{\pi} \int_{4M_\pi^2}^\infty du^\prime \frac{\hat D_{i;u}^s(u^\prime;s)}{u^\prime - u} ,
	\end{split}
\end{align}
where $\hat\rho_{i;t,u}^s$ denote the pole residues and $\hat D_{i;t,u}^s$ the discontinuities along the $t$- and $u$-channel cuts. Both are determined by unitarity. Consider the $t$-channel unitarity relation:
\begin{align}
	\begin{split}
		\Im^t &\left( e^2 (2\pi)^4 \delta^{(4)}( p_1 + p_2 - q_1 - q_2) \epsilon_\mu^{\lambda_1}(q_1) {\epsilon_\nu^{\lambda_2}}^*(-q_2) W^{\mu\nu}(p_1,p_2,q_1) \right) \\
			&= \sum_n \frac{1}{2S_n} \left( \prod_{i=1}^n \int \widetilde{dk_i} \right) \< n;\{k_i\} | \pi^-(p_2) \gamma^*(-q_2,\lambda_2) \>^* \< n;\{k_i\} | \pi^-(-p_1) \gamma^*(q_1,\lambda_1) \> ,
	\end{split}
\end{align}
where $S_n$ is the symmetry factor for the intermediate state $|n\>$. If we single out the one-pion state from the sum over intermediate states, we find
\begin{align}
	\begin{split}
		\Im_\pi^t W^{\mu\nu} &= \frac{1}{2} \int \widetilde{dk} \, (2\pi)^4 \delta^{(4)}(q_1 - p_1 - k) (k-p_1)^\mu (k+p_2)^\nu F_\pi^V(q_1^2) F_\pi^V(q_2^2) ,
	\end{split}
\end{align}
where $F_\pi^V$ is the electromagnetic form factor of the pion, defined by
\begin{align}
	\begin{split}
		\< \pi^+(k) | j^\mu_\mathrm{em}(0) | \pi^+(p) \> = (k + p)^\mu \; F_\pi^V\big((k-p)^2\big) .
	\end{split}
\end{align}
After performing the trivial integral, we obtain
\begin{align}
	\begin{split}
		\label{eq:ggpipiPionPoleUnitarityTChannel}
		\Im_\pi^t W^{\mu\nu} &= F_\pi^V(q_1^2) F_\pi^V(q_2^2) \, \pi \delta( t - M_\pi^2 ) \left( q_3^\mu q_1^\nu - q_2^\mu q_3^\nu - q_2^\mu q_1^\nu + q_3^\mu q_3^\nu \right) .
	\end{split}
\end{align}
$t$- and $u$-channel are related by $p_1 \leftrightarrow p_2$:
\begin{align}
	\begin{split}
		\Im_\pi^u W^{\mu\nu} &= F_\pi^V(q_1^2) F_\pi^V(q_2^2) \, \pi \delta( u - M_\pi^2 ) \left( q_2^\mu q_3^\nu - q_3^\mu q_1^\nu - q_2^\mu q_1^\nu + q_3^\mu q_3^\nu \right) .
	\end{split}
\end{align}
Note that these expressions are only gauge-invariant due to the presence of the delta function: the contraction of the bracket in (\ref{eq:ggpipiPionPoleUnitarityTChannel}) with $q_1^\mu$ or $q_2^\nu$ is proportional to $t-M_\pi^2$.

If we project the imaginary parts of $W^{\mu\nu}$ onto the scalar functions, making use of the delta function we obtain
\begin{align}
	\begin{split}
		\Im_\pi^t A_1 &= F_\pi^V(q_1^2) F_\pi^V(q_2^2) \, \pi \delta( t - M_\pi^2 ) , \\
		\Im_\pi^t A_5 &= \frac{2}{s-q_1^2-q_2^2}F_\pi^V(q_1^2) F_\pi^V(q_2^2) \, \pi \delta( t - M_\pi^2 ) , \\
		\Im_\pi^u A_1 &= F_\pi^V(q_1^2) F_\pi^V(q_2^2) \, \pi \delta( u - M_\pi^2 ) , \\
		\Im_\pi^u A_5 &= \frac{2}{s-q_1^2-q_2^2}F_\pi^V(q_1^2) F_\pi^V(q_2^2) \, \pi \delta( u - M_\pi^2 ) ,
	\end{split}
\end{align}
while the one-pion contribution to the imaginary parts of the remaining scalar functions vanishes.

\begin{figure}[ht]
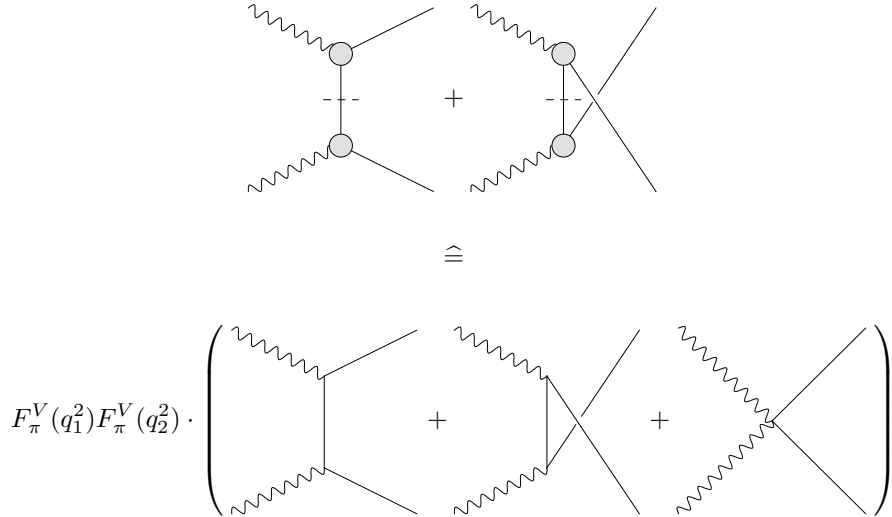

	\centering
	\begin{gather*}
			\minidiagSize{HLbL}{SubTPole}{2.5cm} + \minidiagSize{HLbL}{SubUPole}{2.5cm} \\
			 \\
			\hateq \\
			 \\
			F_\pi^V(q_1^2) F_\pi^V(q_2^2) \cdot \left( \; \minidiagSize{HLbL}{SubTPoleFeyn}{2.5cm} +  \minidiagSize{HLbL}{SubUPoleFeyn}{2.5cm} + \minidiagSize{HLbL}{SubSeagullFeyn}{2.5cm} \; \right)
	\end{gather*}
	\caption{Correspondence of the pion-pole contribution to $\gamma^*\gamma^*\to\pi^+\pi^-$ in terms of unitarity diagrams and the Born contribution in terms of sQED Feynman diagrams. The dashed lines in the unitarity diagrams indicate a cut line, hence the internal pion is on-shell.}
	\label{img:Poleggpipi}
\end{figure}

The pion-pole contribution to the scalar functions is therefore
\begin{align}
	\begin{split}
		A_1^\pi &= - F_\pi^V(q_1^2) F_\pi^V(q_2^2) \left( \frac{1}{t - M_\pi^2} + \frac{1}{u - M_\pi^2} \right) , \\
		A_5^\pi &= - F_\pi^V(q_1^2) F_\pi^V(q_2^2) \frac{2}{s - q_1^2 - q_2^2} \left( \frac{1}{t - M_\pi^2} + \frac{1}{u - M_\pi^2} \right) , \\
		A_2^\pi &= A_3^\pi = A_4^\pi = A_6^\pi = 0 .
	\end{split}
\end{align}
This is exactly the Born contribution in scalar QED (sQED) multiplied by electromagnetic pion form factors for the two off-shell photons, see appendix~\ref{sec:AppendixScalarQEDBornggpipi}. Note that, if we think in terms of unitarity diagrams, we have now considered the pure pole contribution to the scalar functions. However, in terms of Feynman diagrams in sQED this corresponds to a sum of two pole diagrams and the seagull diagram. It is important to be aware of the different meaning of a topology in the sense of unitarity and a Feynman diagram, see figure~\ref{img:Poleggpipi}.

\section{Partial-Wave Expansion}

The helicity amplitudes are the appropriate quantities for a partial-wave expansion. Using the formalism of \cite{Jacob1959}, we write the partial-wave expansions as
\begin{align}
	\begin{split}
		H_{\lambda_1\lambda_2}(s,t,u) = \sum_l (2l + 1) d_{m0}^l(z) h^l_{\lambda_1\lambda_2}(s) ,
	\end{split}
\end{align}
where $d_{m0}^l$ is the Wigner $d$-function and $m=\lambda_1-\lambda_2$. Of course, the helicity partial waves $h^l_{\lambda_1\lambda_2}$ depend implicitly on the photon virtualities $q_1^2$ and $q_2^2$. Since the isospin of the two-pion system is $I=0,2$, only even partial waves are allowed. For $m=0$, i.e.~$H_{++}$ and $H_{00}$, the partial-wave expansion starts at $l=0$, otherwise at $l=2$. Note also that $d_{00}^l(z) = P_l(z)$ are the Legendre polynomials.

The partial waves can be obtained by projection:
\begin{align}
	\begin{split}
		h^l_{\lambda_1\lambda_2}(s) = \frac{1}{2} \int_1^1 dz \, d_{m0}^l(z) H_{\lambda_1\lambda_2}(s,z) .
	\end{split}
\end{align}

Since the functions $A_i$ are free of kinematic singularities and zeros, they are appropriate for a dispersive description. In a forthcoming publication, we will generalise the Roy-Steiner treatment of \cite{Garcia-Martin2010,Hoferichter2011,Moussallam2013} to the doubly-virtual case \cite{Hoferichter2013}. We start by writing down hyperbolic dispersion relations
\begin{align}
	\begin{split}
		A_i(s,t,u) &= A_i^{\text{Born}}(s,t,u) + \frac{1}{\pi} \int_{4M_\pi^2}^\infty ds^\prime \frac{\Im A_i(s^\prime, z^\prime)}{s^\prime - s} \\
			&\quad + \frac{1}{\pi} \int_{t_0}^\infty dt^\prime \Im A_i(t^\prime,u^\prime) \left( \frac{1}{t^\prime-t} + \frac{1}{t^\prime-u} - \frac{1}{t^\prime-a} \right) .
	\end{split}
\end{align}
If we invert (\ref{eq:ggpipiHelicityAmplitudes}) to express the $A_i$ in terms of the helicity amplitudes and insert the partial-wave expansion of the helicity amplitudes both on the left- and right-hand side of the dispersion relation, we obtain a set of Roy-Steiner equations
\begin{align}
	\begin{split}
		h^l_\chi(s) = \sum_{l^\prime} \sum_{\chi^\prime} \frac{1}{\pi} \int_{4M_\pi^2}^\infty ds^\prime K_{l l^\prime}^{\chi\chi^\prime}(s,s^\prime) \Im h^{l^\prime}_{\chi^\prime}(s^\prime) + \ldots ,
	\end{split}
\end{align}
where $\chi,\chi^\prime \in \{++,+-,+0,0+,00\} =: \{1,2,3,4,5\}$ and $K_{ll^\prime}^{\chi\chi^\prime}$ are integral kernels. The dots stand for the contribution of the partial waves of the crossed channels.

As we will see later, the process $\gamma^*\gamma^*\to\pi\pi$ will serve as an input in the dispersion relation for HLbL scattering. In this context, we will truncate the partial-wave expansion either at $S$- or $D$-waves.

If only $S$-waves are taken into account, the relation between the scalar functions and the helicity partial waves in the $s$-channel is given by
\begin{align}
	\begin{split}
		\label{eq:SWavesSubProcessScalarFunctions}
		A_1 &= \frac{2}{\lambda(s,q_1^2,q_2^2)} \left( 2 \sqrt{q_1^2 q_2^2} h_{00}^0(s) - (s - q_1^2 - q_2^2) h_{++}^0(s) \right) , \\
		A_2 &= \frac{2}{\lambda(s,q_1^2,q_2^2)} \left( - \frac{s - q_1^2 - q_2^2}{\sqrt{q_1^2 q_2^2}} h_{00}^0(s) + 2 h_{++}^0(s) \right) .
	\end{split}
\end{align}
In this case, the scalar functions depend only on $s$.

If we take into account $D$-waves as well, the scalar functions become
\begin{align}
	\begin{split}
		\label{eq:DWavesSubProcessScalarFunctions}
		A_1 &= \frac{4 \sqrt{ q_1^2 q_2^2}}{\lambda_{12}(s)} \left( h_{00}^0(s) + \frac{5}{2}(3z^2-1) h_{00}^2(s) \right) - \frac{2 (s - q_1^2 - q_2^2)}{\lambda_{12}(s)} \left( h_{++}^0(s) + \frac{5}{2} (3z^2-1) h_{++}^2(s) \right) \\
			& \quad - \frac{10 \sqrt{3} \sqrt{q_1^2} \left( s - q_1^2 + q_2^2 \right) z^2}{\sqrt{s} \lambda_{12}(s)} h_{0+}^2(s) +\frac{10 \sqrt{3} \sqrt{q_2^2} ( s + q_1^2 - q_2^2 ) z^2}{\sqrt{s} \lambda_{12}(s)} h_{+0}^2(s) \\
			& \quad + 5 \sqrt{\frac{3}{2}} \left( \frac{z^2}{s} - (1+z^2) \frac{s - q_1^2 - q_2^2}{\lambda_{12}(s)} \right) h_{+-}^2(s) , \\
		A_2 &= - \frac{2 (s - q_1^2 - q_2^2)}{\sqrt{ q_1^2 q_2^2} \lambda_{12}(s)} \left( h_{00}^0(s) + \frac{5}{2}(3z^2-1) h_{00}^2(s) \right) + \frac{4}{\lambda_{12}(s)} \left( h_{++}^0(s) + \frac{5}{2} \left(3 z^2-1\right) h_{++}^2(s) \right) \\
			& \quad + \frac{10 \sqrt{3} \left( s + q_1^2 - q_2^2 \right) z^2}{\sqrt{s q_1^2} \lambda_{12}(s)} h_{0+}^2(s) - \frac{10 \sqrt{3} \left( s - q_1^2 + q_2^2\right) z^2}{ \sqrt{s q_2^2} \lambda_{12}(s)} h_{+0}^2(s) \\
   			& \quad + 5 \sqrt{6} \left(\frac{1 + z^2}{\lambda_{12}(s)} - \frac{z^2}{s (s - q_1^2 - q_2^2)}\right) h_{+-}^2(s) , \\
		A_3 &= - \frac{10 \sqrt{3}z}{\sqrt{q_1^2 (s - 4M_\pi^2) \lambda_{12}(s)}} h_{0+}^2(s) - \frac{5 \sqrt{6} z ( s - q_1^2 + q_2^2)}{s (s - q_1^2 - q_2^2) \sigma_\pi(s) \sqrt{\lambda_{12}(s)}} h_{+-}^2(s) , \\
		A_4 &= -\frac{10 \sqrt{3} z}{\sqrt{q_2^2 (s-4M_\pi^2) \lambda_{12}(s)}} h_{+0}^2(s) + \frac{5 \sqrt{6}  z (s + q_1^2 - q_2^2)}{s (s - q_1^2 - q_2^2) \sigma_\pi(s) \sqrt{\lambda_{12}(s)}} h_{+-}^2(s) , \\
		A_5 &= - \frac{5 \sqrt{6}}{(s-4M_\pi^2)(s - q_1^2 - q_2^2)} h_{+-}^2(s) , \\
	\end{split}
\end{align}
where $\lambda_{12}(s) := \lambda(s,q_1^2,q_2^2)$. Now, the scalar functions are polynomials in $t$ and $u$ of at most second order.

We note that under $q_1^2 \leftrightarrow q_2^2$, the partial waves transform as $h_{+0}^2 \leftrightarrow - h_{0+}^2$, while all the other $S$- and $D$-waves are invariant (the sign comes from the Wigner $d$-function).

\begin{figure}[H]
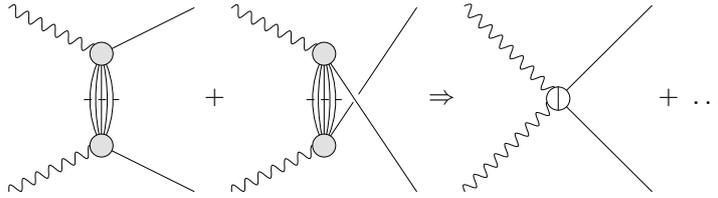

	\centering
	$\minidiagSize{HLbL}{SubTDisc}{2.5cm} + \minidiagSize{HLbL}{SubUDisc}{2.5cm} \Rightarrow \minidiagSize{HLbL}{SubPoly}{2.5cm} + \; \ldots$
	\caption{Approximation of higher intermediate states in $t$- and $u$-channel by a polynomial through a partial-wave expansion.}
	\label{img:ggpipiPWApprox}
\end{figure}

Of course, if we replace the scalar functions with a truncated partial-wave expansion in the $s$-channel, we neglect the structure in the crossed channels. In a first approximation, we will take into account explicitly the $t$- and $u$-channel pole terms and apply a partial-wave expansion to the remainder, i.e.~we approximate the cut due to higher intermediate states by a polynomial as illustrated in figure~\ref{img:ggpipiPWApprox}.


\chapter{Lorentz Structure of the HLbL Tensor}

\label{sec:LorentzStructureHLbLTensor}

\section{Definitions}

In order to study the contribution of hadronic light-by-light scattering to the anomalous magnetic moment of the muon, we need first of all a description of the hadronic light-by-light (HLbL) tensor. The object in question is the hadronic Green's function of four electromagnetic currents, evaluated in pure QCD (i.e.~with $\alpha_\mathrm{QED} = 0$):
\begin{align}
	\begin{split}
		\label{eq:HLbLTensorDefinition}
		\Pi^{\mu\nu\lambda\sigma}(q_1,q_2,q_3) = -i \int d^4x \, d^4y \, d^4z \, e^{-i(q_1 x + q_2 y + q_3 z)} \< 0 | T \{ j_\mathrm{em}^\mu(x) j_\mathrm{em}^\nu(y) j_\mathrm{em}^\lambda(z) j_\mathrm{em}^\sigma(0) \} | 0 \> .
	\end{split}
\end{align}
The electromagnetic current includes only the three lightest quarks:
\begin{align}
	\begin{split}
		j_\mathrm{em}^\mu := \bar q Q \gamma^\mu q ,
	\end{split}
\end{align}
where $q = ( u , d, s )^T$ and $Q = \mathrm{diag}(\frac{2}{3}, -\frac{1}{3}, -\frac{1}{3})$.

\begin{figure}[H]
	\centering
	\begin{pspicture}(5,4)
		\psframe(-1,1.999)(5,2.001) 
		\psframe[linecolor=white,linewidth=5pt,fillstyle=solid,fillcolor=white](-1.5,0.0)(5.5,4.0) 
		\put(0,3){$q_1$}
		\put(0,1){$q_2$}
		\put(3.75,3){$-q_3$}
		\put(3.75,1){$k=q_4$}
		\psline[arrows=->](0,0.5)(0.5,1)
		\psline[arrows=->](0,3.5)(0.5,3)
		\psline[arrows=->](3.5,1)(4,0.5)
		\psline[arrows=->](3.5,3)(4,3.5)
		\includegraphics[width=4cm]{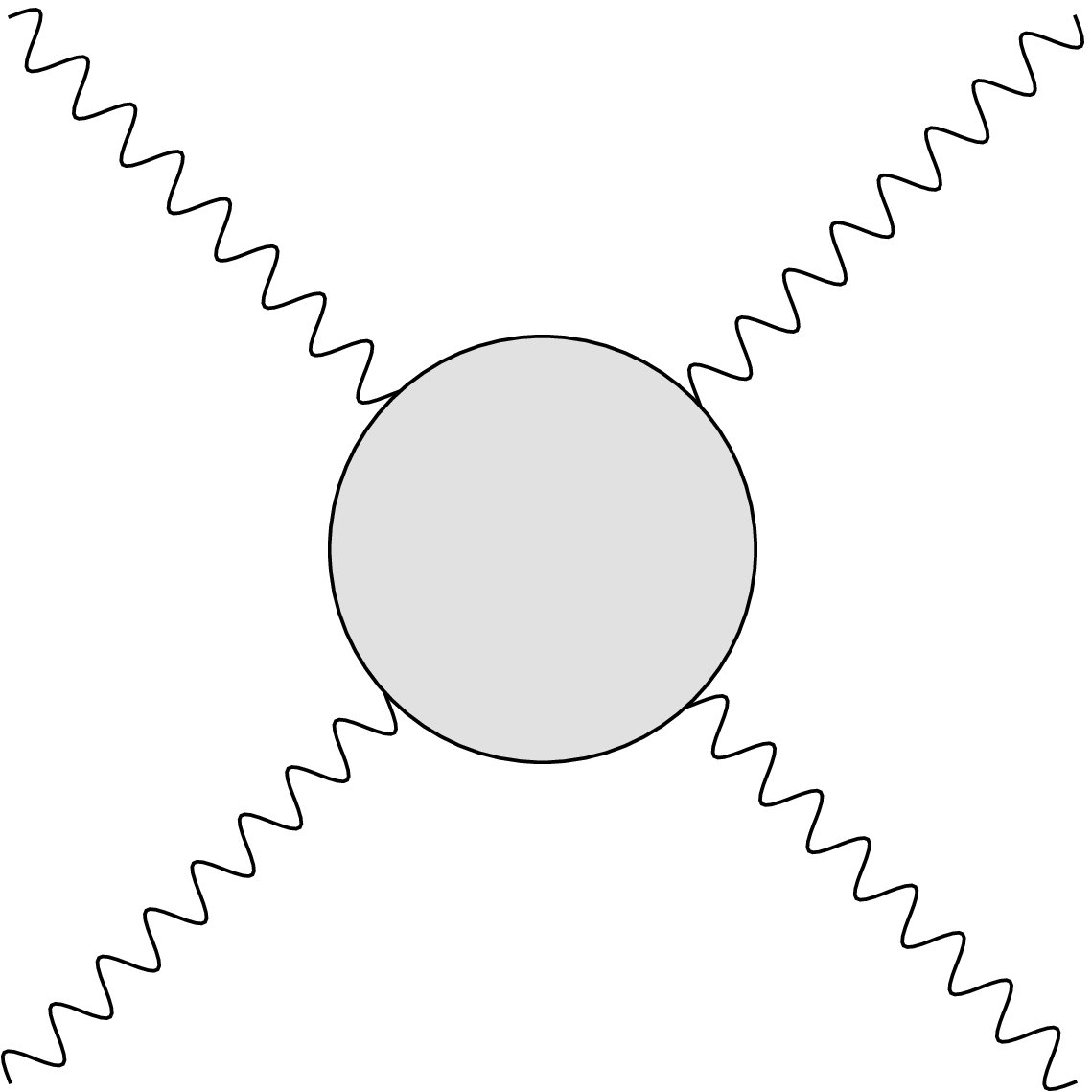}
	\end{pspicture}
	\caption{Kinematics of the light-by-light scattering amplitude.}
	\label{img:FullHLbL}
\end{figure}

The contraction of the HLbL tensor with polarisation vectors gives the hadronic contribution to the helicity amplitude for (off-shell) photon-photon scattering:
\begin{align}
	\begin{split}
		H_{\lambda_1\lambda_2,\lambda_3\lambda_4} = \epsilon_\mu^{\lambda_1}(q_1) \epsilon_\nu^{\lambda_2}(q_2) {\epsilon_\lambda^{\lambda_3}}^*(-q_3) {\epsilon_\sigma^{\lambda_4}}^*(k) \Pi^{\mu\nu\lambda\sigma}(q_1,q_2,q_3) .
	\end{split}
\end{align}
For notational convenience, we define
\begin{align}
	\begin{split}
		q_4 := k = q_1 + q_2 + q_3 .
	\end{split}
\end{align}
The kinematics is illustrated in figure~\ref{img:FullHLbL}.

We use the following Lorentz scalars as kinematic variables -- these are the usual Mandelstam variables:
\begin{align}
	\begin{split}
		s := (q_1+q_2)^2, \quad t := (q_1+q_3)^2, \quad u := (q_1 - q_4)^2,
	\end{split}
\end{align}
which fulfil (we will take $k^2 = 0$ at some later point)
\begin{align}
	\begin{split}
		s + t + u = \sum_{i=1}^4 q_i^2 =: \Sigma .
	\end{split}
\end{align}

Gauge invariance requires the HLbL tensor to satisfy the Ward-Takahashi identities
\begin{align}
	\begin{split}
		\label{eq:WardIdentitiesHLbLTensor}
		\{q_1^\mu, q_2^\nu, q_3^\lambda, q_4^\sigma\} \Pi_{\mu\nu\lambda\sigma}(q_1,q_2,q_3) = 0 .
	\end{split}
\end{align}

\section{Tensor Decomposition}

\label{sec:HLbLTensorBTTDecomposition}

In general, the HLbL tensor can be decomposed into 138 Lorentz structures \cite{Karplus1950, Leo1975, Bijnens1996}:
\begin{align}
	\begin{split}
		\label{eq:HLbLTensor138StructuresLSM}
		\Pi^{\mu\nu\lambda\sigma} &= g^{\mu\nu} g^{\lambda\sigma} \, \Pi^1 + g^{\mu\lambda} g^{\nu\sigma} \, \Pi^2 + g^{\mu\sigma} g^{\nu\lambda} \, \Pi^3 \\
			& + \sum_{\substack{i=2,3,4 \\ j=1,3,4}} \sum_{\substack{k=1,2,4 \\ l=1,2,3}} q_i^\mu q_j^\nu q_k^\lambda q_l^\sigma \, \Pi^4_{ijkl} \\
			& + \sum_{\substack{i=2,3,4 \\ j=1,3,4}} g^{\lambda\sigma} q_i^\mu q_j^\nu \, \Pi^5_{ij} + \sum_{\substack{i=2,3,4 \\ k=1,2,4}} g^{\nu\sigma} q_i^\mu q_k^\lambda \, \Pi^6_{ik} + \sum_{\substack{i=2,3,4 \\ l=1,2,3}} g^{\nu\lambda} q_i^\mu q_l^\sigma \, \Pi^7_{il} \\
			& + \sum_{\substack{j=1,3,4 \\ k=1,2,4}} g^{\mu\sigma} q_j^\nu q_k^\lambda \, \Pi^8_{jk} + \sum_{\substack{j=1,3,4 \\ l=1,2,3}} g^{\mu\lambda} q_j^\nu q_l^\sigma \, \Pi^9_{jl} + \sum_{\substack{k=1,2,4 \\ l=1,2,3}} g^{\mu\nu} q_k^\lambda q_l^\sigma \, \Pi^{10}_{kl} \\
			&=: \sum_{i=1}^{138} L_i^{\mu\nu\lambda\sigma} \, \Xi_i.
	\end{split}
\end{align}
The 138 scalar functions
\begin{align}
	\begin{split}
		\{\Xi_i\} := \{ \Pi^1, \Pi^2, \Pi^3, \Pi^4_{ijkl}, \Pi^5_{ij}, \Pi^6_{ik}, \Pi^7_{il}, \Pi^8_{jk}, \Pi^9_{jl}, \Pi^{10}_{kl}\}
	\end{split}
\end{align}
depend on six independent kinematic variables, e.g.~on two Mandelstam variables $s$ and $t$ and the virtualities $q_1^2$, $q_2^2$, $q_3^2$ and $q_4^2$. They are free of kinematic singularities but contain kinematic zeros, because they have to fulfil kinematic constraints required by gauge invariance. The Ward identities (\ref{eq:WardIdentitiesHLbLTensor}) impose 95 linearly independent relations on the scalar functions, reducing the set to 43 functions.

As we did in section~\ref{sec:ggtopipiTensorDecomposition} for the case of $\gamma^*\gamma^*\to\pi\pi$, we will now construct a set of Lorentz structures and scalar functions, such that the scalar functions contain neither kinematic singularities nor zeros. Compared to $\gamma^*\gamma^*\to\pi\pi$, the application of the recipe given by Bardeen, Tung \cite{Bardeen1968} and Tarrach \cite{Tarrach1975} is much more involved. Again, there exists no kinematic-free minimal basis (which would consist here of 43 scalar functions). We will construct a redundant set of 54 structures, which is free of kinematic singularities and zeros.

In a first step, we define the two projectors
\begin{align}
	\begin{split}
		I_{12}^{\mu\nu} := g^{\mu\nu} - \frac{q_2^\mu q_1^\nu}{q_1 \cdot q_2} , \quad I_{34}^{\lambda\sigma} := g^{\lambda\sigma} - \frac{q_4^\lambda q_3^\sigma}{q_3 \cdot q_4} ,
	\end{split}
\end{align}
which have the following properties:
\begin{align}
	\begin{alignedat}{2}
		q_1^\mu I^{12}_{\mu\nu} &= 0 , \quad &  q_2^\nu I^{12}_{\mu\nu} &= 0 , \\
		q_3^\lambda I^{34}_{\lambda\sigma} &= 0 , \quad &  q_4^\sigma I^{34}_{\lambda\sigma} &= 0 , \\
		I_{12}^{\mu\mu^\prime} \Pi_{\mu^\prime\nu\lambda\sigma} &= \Pi^\mu{}_{\nu\lambda\sigma} , \quad & I_{12}^{\nu^\prime\nu} \Pi_{\mu\nu^\prime\lambda\sigma} &= \Pi_\mu{}^\nu{}_{\lambda\sigma} , \\
		I_{34}^{\lambda\lambda^\prime} \Pi_{\mu\nu\lambda^\prime\sigma} &= \Pi_{\mu\nu}{}^\lambda{}_\sigma , \quad & I_{34}^{\sigma^\prime\sigma} \Pi_{\mu\nu\lambda\sigma^\prime} &= \Pi_{\mu\nu\lambda}{}^\sigma ,
	\end{alignedat}
\end{align}
i.e.~the HLbL tensor is invariant under contraction with the projectors, but the contraction of every Lorentz structure produces a gauge-invariant structure. Hence, we project the tensor 
\begin{align}
	\begin{split}
		\Pi^{\mu\nu\lambda\sigma} &= I_{12}^{\mu\mu^\prime} I_{12}^{\nu^\prime\nu} I_{34}^{\lambda\lambda^\prime} I_{34}^{\sigma^\prime\sigma} \Pi_{\mu^\prime\nu^\prime\lambda^\prime\sigma^\prime} \\
			&= \sum_{i=1}^{138} I_{12}^{\mu\mu^\prime} I_{12}^{\nu^\prime\nu} I_{34}^{\lambda\lambda^\prime} I_{34}^{\sigma^\prime\sigma} L^i_{\mu^\prime\nu^\prime\lambda^\prime\sigma^\prime} \, \Xi_i \\
			&=: \sum_{i=1}^{138} \bar L_i^{\mu\nu\lambda\sigma} \, \Xi_i =  \sum_{j=1}^{43} \bar L_{i_j}^{\mu\nu\lambda\sigma} \, \Xi_{i_j} .
	\end{split}
\end{align}
Only 43 of the 138 projected structures $\bar L_i^{\mu\nu\lambda\sigma}$ are non-zero, i.e.~all constraints imposed by gauge invariance are already manifestly implemented. Since the projected structures are still multiplied by the original scalar functions $\Xi_i$, no kinematic singularities have been introduced into the scalar functions. We now have to remove the kinematic zeros from the scalar functions by removing the single and double poles in $q_1 \cdot q_2$ and $q_3 \cdot q_4$, which are present in the structures $\bar L_i^{\mu\nu\lambda\sigma}$. We adapt the recipe of \cite{Bardeen1968}:
\begin{itemize}
	\item remove as many ($q_1 \cdot q_2$, $q_3 \cdot q_4$) double-double poles as possible by adding to the structures linear combinations of other structures with coefficients containing no poles.
	\item if no more double-double poles can be removed in this way, multiply the structures that still contain double-double poles by either $q_1 \cdot q_2$ or $q_3 \cdot q_4$ (the choice is irrelevant in the end).
	\item proceed in the same way with double-single, single-double poles, etc.~until no poles at all are left in the structures.
\end{itemize}
As already mentioned, is is again impossible to avoid introducing kinematic singularities into the scalar functions by applying this procedure \cite{Tarrach1975}. However, the only step where kinematic singularities can be introduced is the multiplication of the structures by $q_1\cdot q_2$ or $q_3 \cdot q_4$ (i.e.~the division of the scalar functions by these terms). This means that the only possible singularities are (double or single) poles in $q_1\cdot q_2$ or $q_3 \cdot q_4$. The precise form of these poles can be easily determined: they correspond to degeneracies of the obtained basis of Lorentz structures in the limit $q_1\cdot q_2 \to 0$ and/or $q_3 \cdot q_4 \to 0$. Therefore, the 43-dimensional basis has to be extended by additional structures, which are easily found by studying the null-space of the present structures in the mentioned limits. 11 such structures can be found. The extended generating set of 54 structures exhibits all possible crossing symmetries in a manifest way (but is of course no longer a basis).

Explicitly, the resulting representation of the HLbL tensor reads
\begin{align}
	\begin{split}
		\label{eqn:HLbLTensorKinematicFreeStructures}
		\Pi^{\mu\nu\lambda\sigma} &= \sum_{i=1}^{54} T_i^{\mu\nu\lambda\sigma} \Pi_i , 
	\end{split}
\end{align}
where
\begin{align}
	\begin{split}
		\label{eq:HLbLBTTStructures}
		T_1^{\mu\nu\lambda\sigma} &= \epsilon^{\mu\nu\alpha\beta} \epsilon^{\lambda\sigma\gamma\delta} {q_1}_\alpha {q_2}_\beta {q_3}_\gamma {q_4}_\delta , \\
		T_4^{\mu\nu\lambda\sigma} &= \Big(q_2^\mu q_1^\nu - q_1 \cdot q_2 g^{\mu \nu} \Big) \Big( q_4^\lambda q_3^\sigma - q_3 \cdot q_4 g^{\lambda \sigma} \Big) , \\
		T_7^{\mu\nu\lambda\sigma} &= \Big(q_2^\mu q_1^\nu - q_1 \cdot q_2 g^{\mu \nu } \Big) \Big( q_1 \cdot q_4 \left(q_1^\lambda q_3^\sigma -q_1 \cdot q_3 g^{\lambda \sigma} \right) + q_4^\lambda q_1^\sigma q_1 \cdot q_3 - q_1^\lambda q_1^\sigma q_3 \cdot q_4 \Big) , \\
		 T_{19}^{\mu\nu\lambda\sigma} &= \Big( q_2^\mu q_1^\nu - q_1 \cdot q_2 g^{\mu \nu } \Big) \Big(q_2 \cdot q_4 \left(q_1^\lambda q_3^\sigma - q_1 \cdot q_3 g^{\lambda\sigma} \right)+q_4^\lambda q_2^\sigma q_1 \cdot q_3 - q_1^\lambda q_2^\sigma q_3 \cdot q_4 \Big) , \\
		T_{31}^{\mu\nu\lambda\sigma} &= \Big(q_2^\mu q_1^\nu - q_1\cdot q_2 g^{\mu\nu}\Big) \Big(q_2^\lambda q_1\cdot q_3 - q_1^\lambda q_2\cdot q_3\Big) \Big(q_2^\sigma q_1\cdot q_4 - q_1^\sigma q_2\cdot q_4\Big) , \\
		T_{37}^{\mu\nu\lambda\sigma} &= \Big( q_3^\mu q_1\cdot q_4 - q_4^\mu q_1\cdot q_3\Big) \begin{aligned}[t]
			& \Big( q_3^\nu q_4^\lambda q_2^\sigma - q_4^\nu q_2^\lambda q_3^\sigma + g^{\lambda\sigma} \left(q_4^\nu q_2\cdot q_3 - q_3^\nu q_2\cdot q_4\right) \\
			& + g^{\nu\sigma} \left( q_2^\lambda q_3\cdot q_4 - q_4^\lambda q_2\cdot q_3 \right) + g^{\lambda\nu} \left( q_3^\sigma q_2\cdot q_4 - q_2^\sigma q_3\cdot q_4 \right) \Big) , \end{aligned} \\
		T_{49}^{\mu\nu\lambda\sigma} &= q_3^\sigma  \begin{aligned}[t]
				& \Big( q_1\cdot q_3 q_2\cdot q_4 q_4^\mu g^{\lambda\nu} - q_2\cdot q_3 q_1\cdot q_4 q_4^\nu g^{\lambda\mu} + q_4^\mu q_4^\nu \left( q_1^\lambda q_2\cdot q_3 - q_2^\lambda q_1\cdot q_3 \right) \\
				& + q_1\cdot q_4 q_3^\mu q_4^\nu q_2^\lambda - q_2\cdot q_4 q_4^\mu q_3^\nu q_1^\lambda + q_1\cdot q_4 q_2\cdot q_4 \left(q_3^\nu g^{\lambda\mu} - q_3^\mu g^{\lambda\nu}\right) \Big) \end{aligned} \\
			& - q_4^\lambda \begin{aligned}[t]
				& \Big( q_1\cdot q_4 q_2\cdot q_3 q_3^\mu g^{\nu\sigma} - q_2\cdot q_4 q_1\cdot q_3 q_3^\nu g^{\mu\sigma} + q_3^\mu q_3^\nu \left(q_1^\sigma q_2\cdot q_4 - q_2^\sigma q_1\cdot q_4\right) \\
				& + q_1\cdot q_3 q_4^\mu q_3^\nu q_2^\sigma - q_2\cdot q_3 q_3^\mu q_4^\nu q_1^\sigma + q_1\cdot q_3 q_2\cdot q_3 \left( q_4^\nu g^{\mu\sigma} - q_4^\mu g^{\nu\sigma} \right) \Big) \end{aligned} \\
			& + q_3\cdot q_4 \begin{aligned}[t]
				& \Big(\left(q_1^\lambda q_4^\mu - q_1\cdot q_4 g^{\lambda\mu}\right) \left(q_3^\nu q_2^\sigma - q_2\cdot q_3 g^{\nu\sigma}\right) \\
				& - \left(q_2^\lambda q_4^\nu - q_2\cdot q_4 g^{\lambda\nu}\right) \left(q_3^\mu q_1^\sigma - q_1\cdot q_3 g^{\mu\sigma}\right)\Big) . \end{aligned}
	\end{split}
\end{align}
These structures satisfy the following crossing symmetries:
\begin{align}
	\begin{split}
		T_1^{\mu\nu\lambda\sigma} &= \mathcal{C}_{12}[ T_1^{\mu\nu\lambda\sigma} ] = \mathcal{C}_{34}[ T_1^{\mu\nu\lambda\sigma} ] = \mathcal{C}_{34}[ \mathcal{C}_{12}[ T_1^{\mu\nu\lambda\sigma} ] ] = \mathcal{C}_{24}[ \mathcal{C}_{13}[ T_1^{\mu\nu\lambda\sigma} ]] \\
			&= \mathcal{C}_{23}[ \mathcal{C}_{14}[ T_1^{\mu\nu\lambda\sigma} ] ] = \mathcal{C}_{24}[ \mathcal{C}_{13}[ \mathcal{C}_{34}[ T_1^{\mu\nu\lambda\sigma} ] ] ] = \mathcal{C}_{23}[ \mathcal{C}_{14}[ \mathcal{C}_{34}[ T_1^{\mu\nu\lambda\sigma} ] ] ] , \\
		T_4^{\mu\nu\lambda\sigma} &= \mathcal{C}_{12}[ T_4^{\mu\nu\lambda\sigma} ] = \mathcal{C}_{34}[ T_4^{\mu\nu\lambda\sigma} ] = \mathcal{C}_{34}[ \mathcal{C}_{12}[ T_4^{\mu\nu\lambda\sigma} ] ] = \mathcal{C}_{24}[ \mathcal{C}_{13}[ T_4^{\mu\nu\lambda\sigma} ]] \\
			&= \mathcal{C}_{23}[ \mathcal{C}_{14}[ T_4^{\mu\nu\lambda\sigma} ] ] = \mathcal{C}_{24}[ \mathcal{C}_{13}[ \mathcal{C}_{34}[ T_4^{\mu\nu\lambda\sigma} ] ] ] = \mathcal{C}_{23}[ \mathcal{C}_{14}[ \mathcal{C}_{34}[ T_4^{\mu\nu\lambda\sigma} ] ] ] , \\
		T_7^{\mu\nu\lambda\sigma} &= \mathcal{C}_{34}[ T_7^{\mu\nu\lambda\sigma} ] , \\
		T_{19}^{\mu\nu\lambda\sigma} &= \mathcal{C}_{34}[ \mathcal{C}_{12}[ T_{19}^{\mu\nu\lambda\sigma} ] ] , \\
		T_{31}^{\mu\nu\lambda\sigma} &= \mathcal{C}_{12}[ T_{31}^{\mu\nu\lambda\sigma} ] = \mathcal{C}_{34}[ T_{31}^{\mu\nu\lambda\sigma} ] = \mathcal{C}_{34}[ \mathcal{C}_{12}[ T_{31}^{\mu\nu\lambda\sigma} ] ]  , \\
		T_{37}^{\mu\nu\lambda\sigma} &= \mathcal{C}_{34}[ T_{37}^{\mu\nu\lambda\sigma} ] , \\
		T_{49}^{\mu\nu\lambda\sigma} &= - \mathcal{C}_{12}[ T_{49}^{\mu\nu\lambda\sigma} ] = - \mathcal{C}_{34}[ T_{49}^{\mu\nu\lambda\sigma} ] = \mathcal{C}_{34}[ \mathcal{C}_{12}[ T_{49}^{\mu\nu\lambda\sigma} ] ]  ,
	\end{split}
\end{align}
where the crossing operators $\mathcal{C}_{ij}$ exchange momenta and Lorentz indices of the photons $i$ and $j$, e.g.
\begin{align}
	\begin{split}
		\mathcal{C}_{12}[f] := f( \mu \leftrightarrow \nu, q_1 \leftrightarrow q_2 ) .
	\end{split}
\end{align}

The remaining structures are given by crossed versions of the above seven structures:
\begin{align}
	\begin{alignedat}{3}
		T_2^{\mu\nu\lambda\sigma} &= \mathcal{C}_{14}[ T_1^{\mu\nu\lambda\sigma} ] , \quad & T_3^{\mu\nu\lambda\sigma} &= \mathcal{C}_{13}[ T_1^{\mu\nu\lambda\sigma} ] , \\
		T_5^{\mu\nu\lambda\sigma} &= \mathcal{C}_{14}[ T_4^{\mu\nu\lambda\sigma} ] , \quad & T_6^{\mu\nu\lambda\sigma} &= \mathcal{C}_{13}[ T_4^{\mu\nu\lambda\sigma} ] , \\
		T_8^{\mu\nu\lambda\sigma} &= \mathcal{C}_{12}[ T_7^{\mu\nu\lambda\sigma} ] , \quad & T_9^{\mu\nu\lambda\sigma} &= \mathcal{C}_{13}[ \mathcal{C}_{23}[ T_7^{\mu\nu\lambda\sigma} ] ] , \quad & T_{10}^{\mu\nu\lambda\sigma} &= \mathcal{C}_{23}[ T_7^{\mu\nu\lambda\sigma} ] , \\
		T_{11}^{\mu\nu\lambda\sigma} &= \mathcal{C}_{24}[ T_7^{\mu\nu\lambda\sigma} ] , \quad & T_{12}^{\mu\nu\lambda\sigma} &= \mathcal{C}_{14}[ \mathcal{C}_{24}[ T_7^{\mu\nu\lambda\sigma} ] ]  , \quad & T_{13}^{\mu\nu\lambda\sigma} &= \mathcal{C}_{13}[ T_7^{\mu\nu\lambda\sigma} ] , \\
		T_{14}^{\mu\nu\lambda\sigma} &= \mathcal{C}_{23}[  \mathcal{C}_{13}[ T_7^{\mu\nu\lambda\sigma} ] ] , \quad & T_{15}^{\mu\nu\lambda\sigma} &= \mathcal{C}_{14}[ T_7^{\mu\nu\lambda\sigma} ] , \quad & T_{16}^{\mu\nu\lambda\sigma} &= \mathcal{C}_{24}[ \mathcal{C}_{14}[ T_7^{\mu\nu\lambda\sigma} ] ] , \\
		T_{17}^{\mu\nu\lambda\sigma} &= \mathcal{C}_{24}[  \mathcal{C}_{13}[ T_7^{\mu\nu\lambda\sigma} ] ] , \quad & T_{18}^{\mu\nu\lambda\sigma} &= \mathcal{C}_{23}[  \mathcal{C}_{14}[ T_7^{\mu\nu\lambda\sigma} ] ] , \\
		T_{20}^{\mu\nu\lambda\sigma} &= \mathcal{C}_{34}[ T_{19}^{\mu\nu\lambda\sigma} ] , \quad & T_{21}^{\mu\nu\lambda\sigma} &= \mathcal{C}_{23}[ T_{19}^{\mu\nu\lambda\sigma} ] , \quad & T_{22}^{\mu\nu\lambda\sigma} &= \mathcal{C}_{24}[ \mathcal{C}_{23}[ T_{19}^{\mu\nu\lambda\sigma} ] ] , \\
		T_{23}^{\mu\nu\lambda\sigma} &= \mathcal{C}_{23}[ \mathcal{C}_{24}[ T_{19}^{\mu\nu\lambda\sigma} ] ] , \quad & T_{24}^{\mu\nu\lambda\sigma} &= \mathcal{C}_{24}[ T_{19}^{\mu\nu\lambda\sigma} ] , \quad & T_{25}^{\mu\nu\lambda\sigma} &= \mathcal{C}_{23}[ \mathcal{C}_{13}[ T_{19}^{\mu\nu\lambda\sigma} ] ] , \\
		T_{26}^{\mu\nu\lambda\sigma} &= \mathcal{C}_{13}[ T_{19}^{\mu\nu\lambda\sigma} ] , \quad & T_{27}^{\mu\nu\lambda\sigma} &= \mathcal{C}_{14}[ T_{19}^{\mu\nu\lambda\sigma} ] , \quad & T_{28}^{\mu\nu\lambda\sigma} &= \mathcal{C}_{24}[ \mathcal{C}_{14}[ T_{19}^{\mu\nu\lambda\sigma} ] ] , \\
		T_{29}^{\mu\nu\lambda\sigma} &= \mathcal{C}_{24}[ \mathcal{C}_{13}[ T_{19}^{\mu\nu\lambda\sigma} ] ] , \quad & T_{30}^{\mu\nu\lambda\sigma} &= \mathcal{C}_{34}[ \mathcal{C}_{24}[ \mathcal{C}_{13}[ T_{19}^{\mu\nu\lambda\sigma} ]]] , \\
		T_{32}^{\mu\nu\lambda\sigma} &= \mathcal{C}_{24}[ \mathcal{C}_{13}[ T_{31}^{\mu\nu\lambda\sigma} ] ] , \quad & T_{33}^{\mu\nu\lambda\sigma} &= \mathcal{C}_{23}[ T_{31}^{\mu\nu\lambda\sigma} ] , \quad & T_{34}^{\mu\nu\lambda\sigma} &= \mathcal{C}_{13}[ T_{31}^{\mu\nu\lambda\sigma} ] , \\
		T_{35}^{\mu\nu\lambda\sigma} &= \mathcal{C}_{24}[ T_{31}^{\mu\nu\lambda\sigma} ] , \quad & T_{36}^{\mu\nu\lambda\sigma} &= \mathcal{C}_{14}[ T_{31}^{\mu\nu\lambda\sigma} ] , \\
		T_{38}^{\mu\nu\lambda\sigma} &= \mathcal{C}_{34}[ \mathcal{C}_{14}[ T_{37}^{\mu\nu\lambda\sigma} ] ] , \quad & T_{39}^{\mu\nu\lambda\sigma} &= \mathcal{C}_{14}[ T_{37}^{\mu\nu\lambda\sigma} ] , \quad & T_{40}^{\mu\nu\lambda\sigma} &= \mathcal{C}_{12}[ \mathcal{C}_{14}[ T_{37}^{\mu\nu\lambda\sigma} ] ] , \\
		T_{41}^{\mu\nu\lambda\sigma} &= \mathcal{C}_{23}[ \mathcal{C}_{12}[ T_{37}^{\mu\nu\lambda\sigma} ] ] , \quad & T_{42}^{\mu\nu\lambda\sigma} &= \mathcal{C}_{12}[ \mathcal{C}_{24}[ T_{37}^{\mu\nu\lambda\sigma} ] ] , \quad & T_{43}^{\mu\nu\lambda\sigma} &= \mathcal{C}_{24}[ T_{37}^{\mu\nu\lambda\sigma} ] , \\
		T_{44}^{\mu\nu\lambda\sigma} &= \mathcal{C}_{12}[ \mathcal{C}_{23}[ T_{37}^{\mu\nu\lambda\sigma} ] ] , \quad & T_{45}^{\mu\nu\lambda\sigma} &= \mathcal{C}_{23}[ T_{37}^{\mu\nu\lambda\sigma} ] , \quad & T_{46}^{\mu\nu\lambda\sigma} &= \mathcal{C}_{14}[ \mathcal{C}_{23}[ T_{37}^{\mu\nu\lambda\sigma} ] ] , \\
		T_{47}^{\mu\nu\lambda\sigma} &= \mathcal{C}_{24}[ \mathcal{C}_{13}[ T_{37}^{\mu\nu\lambda\sigma} ] ] , \quad & T_{48}^{\mu\nu\lambda\sigma} &= \mathcal{C}_{12}[ T_{37}^{\mu\nu\lambda\sigma} ] , \\
		T_{50}^{\mu\nu\lambda\sigma} &= \mathcal{C}_{12}[ \mathcal{C}_{24}[ T_{49}^{\mu\nu\lambda\sigma} ] ] , \quad & T_{51}^{\mu\nu\lambda\sigma} &= \mathcal{C}_{24}[ T_{49}^{\mu\nu\lambda\sigma} ] , \quad & T_{52}^{\mu\nu\lambda\sigma} &= \mathcal{C}_{13}[ T_{49}^{\mu\nu\lambda\sigma} ] , \\
		T_{53}^{\mu\nu\lambda\sigma} &= \mathcal{C}_{12}[ \mathcal{C}_{13}[ T_{49}^{\mu\nu\lambda\sigma} ] ] , \quad & T_{54}^{\mu\nu\lambda\sigma} &= \mathcal{C}_{23}[ \mathcal{C}_{14}[ T_{49}^{\mu\nu\lambda\sigma} ] ] .
	\end{alignedat}
\end{align}

Since the HLbL tensor $\Pi^{\mu\nu\lambda\sigma}$ is totally crossing symmetric, the scalar functions $\Pi_i$ have to fulfil exactly the same crossing properties as the Lorentz structures. Therefore, only seven different scalar functions $\Pi_i$ appear, together with their crossed versions. These scalar functions are free of kinematic singularities and zeros and hence fulfil a Mandelstam representation (note however the antisymmetric crossing relations in $\Pi_{49}$). They are suitable quantities for a dispersive description.

The subset consisting of the following 43 Lorentz structures forms a basis:
\begin{align}
	\begin{split}
		\label{eq:HLbLBasisStructures}
		\{\mathcal{B}^{\mu\nu\lambda\sigma}_i\} := \Big\{ T^{\mu\nu\lambda\sigma}_i \big| i \in\{ 1,\ldots,21,23,25,27,29,30,33,\ldots,36,38,\ldots,45,49,\ldots,53 \} \Big\} ,
	\end{split}
\end{align}
The corresponding scalar coefficient functions $\tilde \Pi_i$, defined by
\begin{align}
	\begin{split}
		\label{eq:HLbLTensor43Basis}
		\Pi^{\mu\nu\lambda\sigma} &= \sum_{i=1}^{43} \mathcal{B}_i^{\mu\nu\lambda\sigma} \tilde\Pi_i ,
	\end{split}
\end{align}
exhibit kinematic singularities in $q_1 \cdot q_2$ and $q_3 \cdot q_4$. The exact form of these kinematic singularities can be determined by projecting (\ref{eqn:HLbLTensorKinematicFreeStructures}) on this basis:
\begin{align}
	\begin{alignedat}{3}
		\label{eq:HLbLBTTProjectedOnBasis}
		\tilde\Pi_1 &= \Pi_1 , & \quad
		\tilde\Pi_2 &= \Pi_2 , & \quad
		\tilde\Pi_3 &= \Pi_3 , \\
		\tilde\Pi_4 &= \Pi_4 , & \quad
		\tilde\Pi_5 &= \Pi_5 , & \quad
		\tilde\Pi_6 &= \Pi_6 , \\
		\tilde\Pi_7 &= \Pi_7 - \frac{ q_2\cdot q_3 q_2\cdot q_4}{q_3\cdot q_4} \Pi_{31} , & \quad
		\tilde\Pi_8 &= \Pi_8-\frac{q_1\cdot q_3 q_1\cdot q_4}{q_3\cdot q_4} \Pi_{31} , \\
		\tilde\Pi_9 &= \Pi_9+\frac{q_1\cdot q_4}{q_3\cdot q_4} \Pi_{22} , & \quad
		\tilde\Pi_{10} &= \Pi_{10}+\frac{q_2\cdot q_3}{q_1\cdot q_2} \Pi_{22} , & \\
		\tilde\Pi_{11} &= \Pi_{11}-\frac{q_2\cdot q_4}{q_1\cdot q_2} \Pi_{24} , & \quad
		\tilde\Pi_{12} &= \Pi_{12}-\frac{q_1\cdot q_3}{q_3\cdot q_4} \Pi_{24} , \\
		\tilde\Pi_{13} &= \Pi_{13}+\frac{q_2\cdot q_4}{q_3\cdot q_4} \Pi_{26} , & \quad
		\tilde\Pi_{14} &= \Pi_{14}+\frac{q_1\cdot q_3}{q_1\cdot q_2} \Pi_{26} , \\
		\tilde\Pi_{15} &= \Pi_{15}-\frac{q_2\cdot q_3}{q_3\cdot q_4} \Pi_{28} , & \quad
		\tilde\Pi_{16} &= \Pi_{16}-\frac{q_1\cdot q_4}{q_1\cdot q_2} \Pi_{28} , \\
		\tilde\Pi_{17} &= \Pi_{17}-\frac{q_1\cdot q_4 q_2\cdot q_4}{q_1\cdot q_2} \Pi_{32} , & \quad
		\tilde\Pi_{18} &= \Pi_{18}-\frac{q_1\cdot q_3 q_2\cdot q_3}{q_1\cdot q_2} \Pi_{32} , \\
		\tilde\Pi_{19} &= \Pi_{19}+\frac{q_1\cdot q_4 q_2\cdot q_3}{q_3\cdot q_4} \Pi_{31} , & \quad
		\tilde\Pi_{20} &= \Pi_{20}+\frac{q_1\cdot q_3 q_2\cdot q_4}{q_3\cdot q_4} \Pi_{31} , \\
		\tilde\Pi_{21} &= \Pi_{21}-\frac{q_1\cdot q_4 q_2\cdot q_3}{q_1\cdot q_2 q_3\cdot q_4} \Pi_{22} , & \quad
		\tilde\Pi_{22} &= \Pi_{23}-\frac{q_1\cdot q_3 q_2\cdot q_4}{q_1\cdot q_2 q_3\cdot q_4} \Pi_{24} , \\
		\tilde\Pi_{23} &= \Pi_{25}-\frac{q_1\cdot q_3 q_2\cdot q_4}{q_1\cdot q_2 q_3\cdot q_4} \Pi_{26} , & \quad
		\tilde\Pi_{24} &= \Pi_{27}-\frac{q_1\cdot q_4 q_2\cdot q_3}{q_1\cdot q_2 q_3\cdot q_4} \Pi_{28} , \\
		\tilde\Pi_{25} &= \Pi_{29}-\frac{q_1\cdot q_4 q_2\cdot q_3}{q_1\cdot q_2} \Pi_{32} , & \quad
		\tilde\Pi_{26} &= \Pi_{30}-\frac{q_1\cdot q_3 q_2\cdot q_4}{q_1\cdot q_2} \Pi_{32} , \\
		\tilde\Pi_{27} &= \Pi_{33} + \frac{q_2\cdot q_4}{q_1\cdot q_2 q_3\cdot q_4} \Pi_{22} , & \quad
		\tilde\Pi_{28} &= \Pi_{34} + \frac{q_1\cdot q_4}{q_1\cdot q_2 q_3\cdot q_4} \Pi_{26} , \\
		\tilde\Pi_{29} &= \Pi_{35}-\frac{q_2\cdot q_3}{q_1\cdot q_2 q_3\cdot q_4} \Pi_{24} , & \quad
		\tilde\Pi_{30} &= \Pi_{36}-\frac{q_1\cdot q_3}{q_1\cdot q_2 q_3\cdot q_4} \Pi_{28} , \\
		\tilde\Pi_{31} &= \Pi_{38}+\frac{q_2\cdot q_3}{q_3\cdot q_4} \Pi_{47} , & \quad
		\tilde\Pi_{32} &= \Pi_{39}-\frac{q_2\cdot q_4}{q_3\cdot q_4} \Pi_{46} , & \quad
		\tilde\Pi_{33} &= \Pi_{40}-\frac{q_1\cdot q_4}{q_3\cdot q_4} \Pi_{46} , \\
		\tilde\Pi_{34} &= \Pi_{41}+\frac{q_1\cdot q_3}{q_3\cdot q_4} \Pi_{47} , & \quad
		\tilde\Pi_{35} &= \Pi_{42}+\frac{q_2\cdot q_4}{q_1\cdot q_2} \Pi_{48} , & \quad
		\tilde\Pi_{36} &= \Pi_{43} + \frac{q_1\cdot q_4}{q_1\cdot q_2} \Pi_{37} , \\
		\tilde\Pi_{37} &= \Pi_{44}-\frac{q_2\cdot q_3}{q_1\cdot q_2} \Pi_{48} , & \quad
		\tilde\Pi_{38} &= \Pi_{45}-\frac{q_1\cdot q_3}{q_1\cdot q_2} \Pi_{37} , \\
		\tilde\Pi_{39} &= \Pi_{49}+\frac{q_1\cdot q_2}{q_3\cdot q_4} \Pi_{54} , & \quad
		\tilde\Pi_{40} &= \Pi_{50}-\frac{q_2\cdot q_4}{q_3\cdot q_4} \Pi_{54} , & \quad
		\tilde\Pi_{41} &= \Pi_{51}+\frac{q_1\cdot q_4}{q_3\cdot q_4} \Pi_{54} , \\
		\tilde\Pi_{42} &= \Pi_{52}-\frac{q_2\cdot q_3}{q_3\cdot q_4} \Pi_{54} , & \quad
		\tilde\Pi_{43} &= \Pi_{53}+\frac{q_1\cdot q_3}{q_3\cdot q_4} \Pi_{54} .
	\end{alignedat}
\end{align}


\chapter{HLbL Contribution to $(g-2)_\mu$}

\label{sec:HLbLContributionToGminus2}

In section~\ref{sec:ProjectorTechniquesg-2}, we review the definition and calculation of the anomalous magnetic moment of the muon $(g-2)_\mu$ and the HLbL contribution to $(g-2)_\mu$ (see e.g.~\cite{Jegerlehner2008}). The well-known general formula requires still a rather long calculation before a number can be finally obtained. For the pion-pole contribution, these steps have been worked out long ago \cite{Knecht2002}. With our complete set of 54 kinematic-free structures (\ref{eqn:HLbLTensorKinematicFreeStructures}), this procedure can be repeated for the whole HLbL contribution in full generality, as we explain in sections~\ref{sec:LoopIntegration} and \ref{sec:MasterFormula}.

\section{Projector Techniques}

\label{sec:ProjectorTechniquesg-2}

Consider the interaction of a muon with the electromagnetic field:
\begin{align}
	\begin{split}
		\mathcal{M}^\mu(x;p_1,p_2) := -i e \< \mu^-(p_2,s_2) | j_\mathrm{em}^\mu(x) | \mu^-(p_1, s_1) \> ,
	\end{split}
\end{align}
where $p_1$ and $s_1$ ($p_2$ and $s_2$) are momentum and spin of the incoming (outgoing) muon and $j_\mathrm{em}^\mu$ is the electromagnetic current. We pull out explicitly the electric unit charge $e = |e|$.

Using translation invariance
\begin{align}
	\begin{split}
		j_\mathrm{em}^\mu(x) = e^{i P x} j_\mathrm{em}^\mu(0) e^{-iPx},
	\end{split}
\end{align}
we find in momentum space
\begin{align}
	\begin{split}
		\mathcal{\tilde M}^\mu(k;p_1,p_2) :={}& \int d^4x e^{ikx} \mathcal{M}^\mu(x;p_1,p_2) \\
			={}& -\int d^4x e^{i(k - p_1 + p_2)x} i e \< \mu^-(p_2, s_2) | j_\mathrm{em}^\mu(0) | \mu^-(p_1,s_1)\> \\
			={}& -(2\pi)^4 \delta^{(4)}(k - p_1 + p_2) i e \< \mu^-(p_2,s_2) | j_\mathrm{em}^\mu(0) | \mu^-(p_1,s_1) \> .
	\end{split}
\end{align}

The $T$-matrix element is defined by
\begin{align}
	\begin{split}
		\mathcal{T}^\mu(p_1,p_2) := - e \<\mu^-(p_2,s_2) | j_\mathrm{em}^\mu(0) | \mu^-(p_1,s_1) \>,
	\end{split}
\end{align}
diagrammatically
\begin{align}
	\begin{split}
		\minidiag{HLbL}{MuMuGamma} = i \mathcal{T}^\mu(p_1,p_2) = (-ie) \bar u(p_2) \Gamma^\mu(p_1, p_2) u(p_1) .
	\end{split}
\end{align}
The vertex function $\Gamma^\mu$ can be decomposed into form factors as
\begin{align}
	\begin{split}
		\Gamma^\mu(p_1,p_2) &= \gamma^\mu F_E(k^2) - i \frac{\sigma^{\mu\nu} k_\nu}{2 m_\mu} F_M(k^2) \\
			&+ \left( \gamma^\mu + \frac{2m_\mu k^\mu}{k^2} \right) \gamma_5 F_A(k^2) - \frac{\sigma^{\mu\nu} k_\nu}{2m_\mu} \gamma_5 F_D(k^2) ,
	\end{split}
\end{align}
where $F_E$ is the electric charge or Dirac form factor, $F_M$ the magnetic or Pauli form factor, $F_A$ the anapole form factor and $F_D$ the electric dipole form factor. We use
\begin{align}
	\begin{split}
		k = p_1 - p_2
	\end{split}
\end{align}
and $\sigma^{\mu\nu} := \frac{i}{2} [ \gamma^\mu, \gamma^\nu]$. In principle, the form factors depend on $p_1$ and $p_2$, hence on the scalar products $p_1^2$, $p_2^2$ and $p_1 \cdot p_2$. Given the fact that the muon is on-shell:
\begin{align}
	\begin{split}
		p_1^2 &= p_2^2 = m_\mu^2, \\
		p_1 \cdot p_2 &= \frac{1}{2}( 2 m_\mu^2 - k^2),
	\end{split}
\end{align}
the form factors depend on a single variable $k^2$ only.

The anomalous magnetic moment of the muon is given by
\begin{align}
	\begin{split}
		a_\mu = \frac{1}{2}(g - 2)_\mu = F_M(0) .
	\end{split}
\end{align}
It can be projected out by
\begin{align}
	\begin{split}
		\label{eq:AmuProjection}
		F_M(k^2) = \mathrm{Tr}\left( (\slashed p_1 + m_\mu) \Lambda_2^\mu(p_1,p_2)(\slashed p_2 + m_\mu) \Gamma_\mu(p_1,p_2) \right) ,
	\end{split}
\end{align}
where the projector is defined by
\begin{align}
	\begin{split}
		\Lambda_2^\mu(p_1,p_2) := - \frac{m_\mu^2}{k^2(k^2-4m_\mu^2)} \left( \gamma^\mu + \frac{k^2 + 2 m_\mu^2}{k^2 - 4 m_\mu^2} \frac{p_1^\mu + p_2^\mu}{m_\mu} \right) .
	\end{split}
\end{align}
Next, we use the identity \cite{Knecht2002}
\begin{align}
	\begin{split}
		(\slashed p_1 + m_\mu) \gamma^\mu (\slashed p_2 + m_\mu) = (\slashed p_1 + m_\mu) \left( \frac{p^\mu}{m_\mu} + \frac{i}{2m_\mu} \sigma^{\mu\nu} k_\nu \right) (\slashed p_2 + m_\mu)
	\end{split}
\end{align}
to simplify and expand the projector
\begin{align}
	\begin{split}
		& (\slashed p_1 + m_\mu) \Lambda_2^\mu(p_1,p_2) (\slashed p_2 + m_\mu) \\
		&= - (\slashed p_1 + m_\mu) \left(\frac{m_\mu}{k^2(k^2-4m_\mu^2)} \frac{i}{2} \sigma^{\mu\nu} k_\nu + \frac{3m_\mu}{(k^2-4m_\mu^2)^2} p^\mu \right) (\slashed p_2 + m_\mu) \\
		&= - (\slashed p + m_\mu) \left( -\frac{1}{4m_\mu k^2} \frac{i}{2} \sigma^{\mu\nu} k_\nu + \frac{3}{16 m_\mu^3} p^\mu \right) (\slashed p + m_\mu) + \frac{1}{8} \left(\gamma^\mu - \frac{k^\mu \slashed k}{k^2} + \frac{p^\mu}{m_\mu}\right) + \O(k) .
	\end{split}
\end{align}

We expand the vertex function $\Gamma^\mu$ to first order in powers of $k^\mu$:
\begin{align}
	\begin{split}
		\Gamma^\mu(p_1,p_2) &= \Gamma^\mu(p,p) + k_\nu \frac{\p}{\p k_\nu} \Gamma^\mu(p_1,p_2) \bigg|_{k=0} + \ldots  \\
			&=: V^\mu(p) + k_\nu \Gamma^{\mu\nu}(p) + \ldots ,
	\end{split}
\end{align}
where $p := \frac{1}{2}(p_1 + p_2)$.

This leads to
\begin{align}
	\begin{split}
		F_M(k^2) &= \mathrm{Tr}\left( (\slashed p_1 + m_\mu) \Lambda_2^\mu(p_1,p_2)(\slashed p_2 + m_\mu) \Gamma_\mu(p_1,p_2) \right) \\
			&= \mathrm{Tr}\left( (\slashed p + m_\mu) \left( \frac{1}{4m_\mu k^2} \frac{i}{2} \sigma^{\mu\nu} k_\nu - \frac{3}{16 m_\mu^3} p^\mu \right) (\slashed p + m_\mu) V_\mu(p) \right) \\
			&+ \mathrm{Tr}\left( \frac{1}{8} \left(\gamma^\mu - \frac{k^\mu \slashed k}{k^2} + \frac{p^\mu}{m_\mu}\right) V_\mu(p) \right) \\
			&+ \mathrm{Tr}\left( (\slashed p + m_\mu) \left( \frac{1}{4m_\mu k^2} \frac{i}{2} \sigma^{\mu\nu} k_\nu k^\rho \right) (\slashed p + m_\mu) \Gamma_{\mu\rho}(p) \right) + \O(k) .
	\end{split}
\end{align}
Since $F_M$ depends only on $k^2$, we can average both sides of the equation over all spatial directions of $k$ with respect to $p$:
\begin{align}
	\begin{split}
		F_M(k^2) = \int \frac{d\Omega(p,k)}{4\pi} F_M(k^2) .
	\end{split}
\end{align}
We therefore need to calculate the angular averages of $k^\mu$ and $k^\mu k^\nu$:
\begin{align}
	\begin{split}
		\int \frac{d\Omega(p,k)}{4\pi} k^\mu = 0 ,
	\end{split}
\end{align}
because the integrand is odd, and
\begin{align}
	\begin{split}
		\int \frac{d\Omega(p,k)}{4\pi} k^\mu k^\nu = \alpha g^{\mu\nu} + \beta \frac{p^\mu p^\nu}{p^2} .
	\end{split}
\end{align}
Since $k \cdot p = 0$, the contraction of the above equation with $p^\mu$ gives $\alpha = - \beta$, while the contraction with $g^{\mu\nu}$ gives $\alpha = \frac{k^2}{3}$, hence
\begin{align}
	\begin{split}
		\int \frac{d\Omega(p,k)}{4\pi} \frac{k^\mu k^\nu}{k^2} = \frac{1}{3} \left(g^{\mu\nu} - \frac{p^\mu p^\nu}{p^2} \right) .
	\end{split}
\end{align}
We find for the anomalous magnetic moment:
\begin{align}
	\begin{split}
		a_\mu &= \mathrm{Tr}\left( \left(\frac{1}{12} \gamma^\mu - \frac{1}{3} \left( \frac{p^\mu \slashed p}{m_\mu^2} \right) - \frac{1}{4} \frac{p^\mu}{m_\mu} \right) V_\mu(p) \right) \\
			 &- \frac{1}{48 m_\mu} \mathrm{Tr}\left( (\slashed p + m_\mu) [\gamma^\mu,\gamma^\rho] (\slashed p + m_\mu) \Gamma_{\mu\rho}(p) \right) ,
	\end{split}
\end{align}
where now $p^2 = m_\mu^2$. (Note that we have defined $k$ as outgoing, resulting in the different sign of the second term with respect to \cite{Jegerlehner2008}.) If there appear UV divergences in the calculation of the contribution to $a_\mu$ that are regulated dimensionally, the above formula should be adapted to $n$ dimensions \cite{Jegerlehner2008}.

We are interested in the contribution of the HLbL tensor to $a_\mu$, diagrammatically
\begin{align}
	\begin{split}
		\minidiag{HLbL}{MuMuGammaHLbL} = (-ie) \bar u(p_2) \Gamma^\mu_\mathrm{HLbL}(p_1,p_2) u(p_1) ,
	\end{split}
\end{align}
where
\begin{align}
	\begin{split}
		\Gamma^\sigma_\mathrm{HLbL}(p_1,p_2) &= \int \frac{d^4q_1}{(2\pi)^4} \frac{d^4q_2}{(2\pi)^4} (-ie\gamma_\mu) \frac{i(\slashed p_2 + \slashed q_1 + m_\mu)}{(p_2+q_1)^2 - m_\mu^2}(-ie\gamma_\lambda) \\
			& \quad \cdot \frac{i(\slashed p_1 - \slashed q_2 + m_\mu)}{(p_1-q_2)^2 - m_\mu^2}(-ie\gamma_\nu) \frac{(-i)^3}{q_1^2 q_2^2 (p_1-p_2-q_1-q_2)^2} (-i e)^3 \\
			& \quad \cdot i \Pi^{\mu\nu\lambda\sigma}(q_1,q_2,p_1-p_2-q_1-q_2) \\
			&= - e^6 \int \frac{d^4q_1}{(2\pi)^4} \frac{d^4q_2}{(2\pi)^4} \gamma_\mu \frac{(\slashed p_2 + \slashed q_1 + m_\mu)}{(p_2+q_1)^2 - m_\mu^2} \gamma_\lambda  \frac{(\slashed p_1 - \slashed q_2 + m_\mu)}{(p_1-q_2)^2 - m_\mu^2} \gamma_\nu \\
			& \quad \cdot \frac{1}{q_1^2 q_2^2 (p_1-p_2-q_1-q_2)^2} \Pi^{\mu\nu\lambda\sigma}(q_1,q_2,p_1-p_2-q_1-q_2) .
	\end{split}
\end{align}
The HLbL tensor is defined in (\ref{eq:HLbLTensorDefinition}). Let us recall again the Ward-Takahashi identities
\begin{align}
	\begin{split}
		\{q_1^\mu, q_2^\nu, q_3^\lambda, k^\sigma\} \Pi_{\mu\nu\lambda\sigma}(q_1,q_2,q_3) = 0 ,
	\end{split}
	\tag{\ref{eq:WardIdentitiesHLbLTensor}}
\end{align}
where $k=q_1+q_2+q_3$. Differentiating the fourth Ward identity with respect to $k_\rho$ yields
\begin{align}
	\begin{split}
		\label{eq:DiffWardIdentity}
		\Pi_{\mu\nu\lambda\rho}(q_1,q_2,k-q_1-q_2) = - k^\sigma \frac{\p}{\p k^\rho} \Pi_{\mu\nu\lambda\sigma}(q_1,q_2,k-q_1-q_2) .
	\end{split}
\end{align}
It can be argued \cite{Aldins1970} that $\Pi_{\mu\nu\lambda\sigma}$ vanishes linearly with $k$ (i.e.~the derivative contains no singularity), and so must $\Gamma_\sigma^\mathrm{HLbL}$. The HLbL contribution to the anomalous magnetic moment is therefore given by
\begin{align}
	\begin{split}
		\label{eq:amuTraceFormula}
		a_\mu^\mathrm{HLbL} = - \frac{1}{48 m_\mu} \mathrm{Tr}\left( (\slashed p + m_\mu) [\gamma^\rho,\gamma^\sigma] (\slashed p + m_\mu) \Gamma_{\rho\sigma}^\mathrm{HLbL}(p) \right) ,
	\end{split}
\end{align}
where
\begin{align}
	\begin{split}
		\Gamma^\mathrm{HLbL}_{\rho\sigma}(p) = \frac{\p}{\p k^\sigma} \Gamma^\mathrm{HLbL}_\rho(p_1,p_2) \bigg|_{k=0} .
	\end{split}
\end{align}
We use the Ward identity (\ref{eq:DiffWardIdentity}) to write
\begin{align}
	\begin{split}
		\Gamma_\rho^\mathrm{HLbL}(p_1,p_2) &= e^6 \int \frac{d^4q_1}{(2\pi)^4} \frac{d^4q_2}{(2\pi)^4} \gamma^\mu \frac{(\slashed p_2 + \slashed q_1 + m_\mu)}{(p_2+q_1)^2 - m_\mu^2} \gamma^\lambda  \frac{(\slashed p_1 - \slashed q_2 + m_\mu)}{(p_1-q_2)^2 - m_\mu^2} \gamma^\nu \\
			& \quad \cdot \frac{1}{q_1^2 q_2^2 (p_1-p_2-q_1-q_2)^2} k^\sigma \frac{\p}{\p k^\rho} \Pi_{\mu\nu\lambda\sigma}(q_1,q_2,k-q_1-q_2) .
	\end{split}
\end{align}
Taking the derivative and limit, we obtain
\begin{align}
	\begin{split}
		\label{eq:GammaHLbLTwoLoop}
		\Gamma_{\rho\sigma}^\mathrm{HLbL}(p) &= e^6 \int \frac{d^4q_1}{(2\pi)^4} \frac{d^4q_2}{(2\pi)^4} \gamma^\mu \frac{(\slashed p + \slashed q_1 + m_\mu)}{(p+q_1)^2 - m_\mu^2} \gamma^\lambda  \frac{(\slashed p - \slashed q_2 + m_\mu)}{(p-q_2)^2 - m_\mu^2} \gamma^\nu \\
			& \quad \cdot \frac{1}{q_1^2 q_2^2 (q_1+q_2)^2} \frac{\p}{\p k^\rho} \Pi_{\mu\nu\lambda\sigma}(q_1,q_2,k-q_1-q_2) \bigg|_{k=0}.
	\end{split}
\end{align}

\clearpage

\section{Loop Integration}

\label{sec:LoopIntegration}

In order to compute the contribution to $(g-2)_\mu$, one has to take the trace in (\ref{eq:amuTraceFormula}) and perform the two-loop integral of equation (\ref{eq:GammaHLbLTwoLoop}). Five of the eight integrals can be carried out analytically with the help of Gegenbauer polynomial techniques \cite{Rosner1967}. To this end, let us employ the representation of the HLbL tensor in terms of the 54 Lorentz structures $T_i^{\mu\nu\lambda\sigma}$:
\begin{align}
	\begin{split}
		a_\mu^\mathrm{HLbL} &= - \frac{e^6}{48 m_\mu}  \int \frac{d^4q_1}{(2\pi)^4} \frac{d^4q_2}{(2\pi)^4} \frac{1}{q_1^2 q_2^2 (q_1+q_2)^2} \frac{1}{(p+q_1)^2 - m_\mu^2} \frac{1}{(p-q_2)^2 - m_\mu^2} \\
			& \quad \cdot \mathrm{Tr}\left( (\slashed p + m_\mu) [\gamma^\rho,\gamma^\sigma] (\slashed p + m_\mu) \gamma^\mu (\slashed p + \slashed q_1 + m_\mu) \gamma^\lambda (\slashed p - \slashed q_2 + m_\mu) \gamma^\nu \right)  \\
			& \quad \cdot \frac{\p}{\p k^\rho} \Pi_{\mu\nu\lambda\sigma}(q_1,q_2,k-q_1-q_2) \bigg|_{k=0} \\
			&= - \frac{e^6}{48 m_\mu}  \int \frac{d^4q_1}{(2\pi)^4} \frac{d^4q_2}{(2\pi)^4} \frac{1}{q_1^2 q_2^2 (q_1+q_2)^2} \frac{1}{(p+q_1)^2 - m_\mu^2} \frac{1}{(p-q_2)^2 - m_\mu^2} \\
			& \quad \cdot \mathrm{Tr}\left( (\slashed p + m_\mu) [\gamma^\rho,\gamma^\sigma] (\slashed p + m_\mu) \gamma^\mu (\slashed p + \slashed q_1 + m_\mu) \gamma^\lambda (\slashed p - \slashed q_2 + m_\mu) \gamma^\nu \right)  \\
			& \quad \cdot  \sum_{i=1}^{54} \left( \frac{\p}{\p k^\rho} T^i_{\mu\nu\lambda\sigma}(q_1,q_2,k-q_1-q_2) \right) \bigg|_{k=0} \Pi_i(q_1,q_2,-q_1-q_2) .
	\end{split}
\end{align}
It turns out that there are only 19 independent linear combinations of the structures $T_i^{\mu\nu\lambda\sigma}$, which contribute to the $(g-2)_\mu$. It is possible to make a basis change in the 54 structures
\begin{align}
	\begin{split}
		\Pi^{\mu\nu\lambda\sigma} = \sum_{i=1}^{54} T_i^{\mu\nu\lambda\sigma} \Pi_i =  \sum_{i=1}^{54} \hat T_i^{\mu\nu\lambda\sigma} \hat \Pi_i ,
	\end{split}
\end{align}
such that in the limit $k\to0$, the derivative of 35 structures $\hat T_i^{\mu\nu\lambda\sigma}$ vanishes. Although this change of basis does not introduce kinematic singularities into the scalar functions $\hat \Pi_i$, it somewhat obscures crossing symmetry. The 19 structures $\hat T_i^{\mu\nu\lambda\sigma}$ that contribute to the $(g-2)_\mu$ can be chosen as follows:
\begin{align}
	\begin{split}
		\left\{ \hat T_i^{\mu\nu\lambda\sigma} \Big| i = 1, \ldots, 19 \right\} &= \left\{ T_i^{\mu\nu\lambda\sigma} \Big| i = 1,\ldots,11,13,14,16,17 \right\} \\
			& \qquad \cup \left\{ T_{39}^{\mu\nu\lambda\sigma} + T_{40}^{\mu\nu\lambda\sigma}, T_{42}^{\mu\nu\lambda\sigma}, T_{43}^{\mu\nu\lambda\sigma}, T_{50}^{\mu\nu\lambda\sigma} - T_{51}^{\mu\nu\lambda\sigma} \right\}.
	\end{split}
\end{align}
The corresponding 19 scalar functions $\hat \Pi_i$ are linear combinations of 33 scalar functions $\Pi_i$:
\begin{align}
	\begin{split}
		\hat \Pi_1 &= \Pi_1 + q_1 \cdot q_2 \Pi_{47} , \\
		\hat \Pi_2 &= \Pi_2 - \frac{1}{2} \left(q_1^2 + q_1 \cdot q_2 \right) \left( 2 \Pi_{47} - \Pi_{50} - \Pi_{51} - \Pi_{54} \right) , \\
		\hat \Pi_3 &= \Pi_3 - \frac{1}{2} \left(q_1 \cdot q_2 + q_2^2\right) \left( 2 \Pi_{47} - \Pi_{50} - \Pi_{51} + \Pi_{54} \right) , \\
		\hat \Pi_4 &= \Pi_4 + \left(q_1^2 + q_1 \cdot q_2 \right) \Pi_{19} + \left(q_1 \cdot q_2 + q_2^2 \right) \Pi_{20} \\
			& \quad + \left(q_1^2 + q_1 \cdot q_2 \right) \left(q_1 \cdot q_2+q_2^2\right) \Pi_{31} - \frac{s}{2} \left( 2 \Pi_{47} - \Pi_{50} - \Pi_{51} \right) + \frac{1}{2} \left( q_1^2 - q_2^2\right) \Pi_{54} , \\
		\hat \Pi_5 &= \Pi_5 - q_1 \cdot q_2 \Pi_{21} + \frac{1}{2} \left(q_1 \cdot q_2+q_2^2\right) \left( 2 \Pi_{22} - 2 q_1 \cdot q_2 \Pi_{33}  + \Pi_{50}  + \Pi_{51} - \Pi_{54}\right) - q_2^2 \Pi_{47} , \\
		\hat \Pi_6 &= \Pi_6 - q_1 \cdot q_2 \Pi_{25} + \frac{1}{2} \left(q_1^2+q_1 \cdot q_2\right) \left( 2 \Pi_{26} - 2 q_1 \cdot q_2 \Pi_{34} + \Pi_{50} + \Pi_{51} + \Pi_{54} \right) - q_1^2 \Pi_{47} , \\
		\hat \Pi_7 &= \Pi_7 - \Pi_{19} - \left( q_1 \cdot q_2 + q_2^2\right) \Pi_{31} , \\
		\hat \Pi_8 &= \Pi_8 - \Pi_{20} - \left(q_1^2 + q_1 \cdot q_2\right) \Pi_{31} , \\
		\hat \Pi_9 &= \Pi_9 - \Pi_{22} + q_1 \cdot q_2 \Pi_{33} , \\
		\hat \Pi_{10} &= \Pi_{10} - \Pi_{21} - \left( q_1 \cdot q_2 + q_2^2 \right) \Pi_{33} , \\
		\hat \Pi_{11} &= \Pi_{11} + \Pi_{47} - \Pi_{54} , \\
		\hat \Pi_{12} &= \Pi_{13} - \Pi_{26} + q_1 \cdot q_2 \Pi_{34} , \\
		\hat \Pi_{13} &= \Pi_{14} - \Pi_{25} - \left( q_1^2 + q_1 \cdot q_2\right) \Pi_{34} , \\
		\hat \Pi_{14} &= \Pi_{16} + \Pi_{47} + \Pi_{54} , \\
		\hat \Pi_{15} &= \Pi_{17} + \Pi_{47} - \Pi_{50} - \Pi_{51} , \\
		\hat \Pi_{16} &= \frac{1}{2} \left( \Pi_{39} + \Pi_{40} + \Pi_{46} \right) , \\
		\hat \Pi_{17} &= \Pi_{42} - \Pi_{47} + \frac{1}{2} \left( \Pi_{50} + \Pi_{51} + \Pi_{54} \right) , \\
		\hat \Pi_{18} &= \Pi_{43} - \Pi_{47} + \frac{1}{2} \left( \Pi_{50} + \Pi_{51} - \Pi_{54} \right) , \\
		\hat \Pi_{19} &= \frac{1}{2} \left( \Pi_{50} - \Pi_{51} + \Pi_{54} \right) .
   	\end{split}
\end{align}
This means that the 21 scalar functions
\begin{align}
	\begin{split}
		\left\{ \Pi_i \Big| i = 12, 15, 18, 23, 24, 27, 28, 29, 30, 32, 35, 36, 37, 38, 41, 44, 45, 48, 49, 52, 53 \right\}
	\end{split}
\end{align}
are irrelevant for the calculation of the $(g-2)_\mu$.

The HLbL contribution to the $(g-2)_\mu$ can now be written as
\begin{align}
	\begin{split}
		a_\mu^\mathrm{HLbL} &= - e^6  \int \frac{d^4q_1}{(2\pi)^4} \frac{d^4q_2}{(2\pi)^4} \frac{1}{q_1^2 q_2^2 (q_1+q_2)^2} \frac{1}{(p+q_1)^2 - m_\mu^2} \frac{1}{(p-q_2)^2 - m_\mu^2} \\
			& \quad \cdot  \sum_{i=1}^{19} \hat T_i(q_1,q_2;p) \hat\Pi_i(q_1,q_2,-q_1-q_2) ,
	\end{split}
\end{align}
where
\begin{align}
	\begin{split}
		\label{eq:DefinitionIntermediateKernels}
		\hat T_i(q_1,q_2;p) := \frac{1}{48 m_\mu} \mathrm{Tr}&\left( (\slashed p + m_\mu) [\gamma^\rho,\gamma^\sigma] (\slashed p + m_\mu) \gamma^\mu (\slashed p + \slashed q_1 + m_\mu) \gamma^\lambda (\slashed p - \slashed q_2 + m_\mu) \gamma^\nu \right)  \\
			& \cdot \left( \frac{\p}{\p k^\rho} \hat T^i_{\mu\nu\lambda\sigma}(q_1,q_2,k-q_1-q_2) \right) \bigg|_{k=0}
	\end{split}
\end{align}
and the $\hat\Pi_i$ are needed for the reduced kinematics
\begin{align}
	\begin{split}
		\label{eqn:ReducedKinematics}
		s = (q_1 + q_2)^2 , \quad t = q_2^2 , \quad u = q_1^2 , \quad q_1^2, \quad q_2^2, \quad q_3^2 = (q_1 + q_2)^2 , \quad k^2 = q_4^2 = 0.
	\end{split}
\end{align}
The explicit result of the trace calculation and the contraction of the Lorentz indices is given in appendix~\ref{sec:AppendixIntermediateKernels}.

We can reduce the number of terms contributing to $(g-2)_\mu$ further by using the symmetry under the exchange of the momenta $q_1 \leftrightarrow -q_2$: The loop integration measure and the product of propagators are invariant under this transformation, while the kernels $\hat T_i$ transform under $q_1 \leftrightarrow -q_2$ as
\begin{align}
	\begin{alignedat}{4}
		\hat T_1 &\longleftrightarrow \hat T_1 , \quad & \hat T_2 &\longleftrightarrow \hat T_3 , \quad & \hat T_4 &\longleftrightarrow \hat T_4 , \quad & \hat T_5 &\longleftrightarrow \hat T_6 , \\
		\hat T_7 &\longleftrightarrow \hat T_8 , \quad & \hat T_9 &\longleftrightarrow \hat T_{12} , \quad & \hat T_{10} &\longleftrightarrow \hat T_{13} , \quad & \hat T_{11} &\longleftrightarrow \hat T_{14} , \\
		\hat T_{15} &\longleftrightarrow \hat T_{15} , \quad & \hat T_{16} &\longleftrightarrow \hat T_{16} , \quad & \hat T_{17} &\longleftrightarrow \hat T_{18} , \quad & \hat T_{19} &\longleftrightarrow -\hat T_{19} .
	\end{alignedat}
\end{align}
For the reduced kinematics (\ref{eqn:ReducedKinematics}) the exchange $q_1 \leftrightarrow -q_2$ is equivalent to the crossing transformation $t\leftrightarrow u$, $q_1^2 \leftrightarrow q_2^2$. With the help of the crossing relations of the scalar functions $\Pi_i$, it is easy to check that the $\vphantom{\Big|}\hat \Pi_i$ transform analogously to the kernels $\hat T_i$, i.e.
\begin{align}
	\begin{alignedat}{4}
		\hat \Pi_1 &\longleftrightarrow \hat \Pi_1 , \quad & \hat \Pi_2 &\longleftrightarrow \hat \Pi_3 , \quad & \hat \Pi_4 &\longleftrightarrow \hat \Pi_4 , \quad & \hat \Pi_5 &\longleftrightarrow \hat \Pi_6 , \\
		\hat \Pi_7 &\longleftrightarrow \hat \Pi_8 , \quad & \hat \Pi_9 &\longleftrightarrow \hat \Pi_{12} , \quad & \hat \Pi_{10} &\longleftrightarrow \hat \Pi_{13} , \quad & \hat \Pi_{11} &\longleftrightarrow \hat \Pi_{14} , \\
		\hat \Pi_{15} &\longleftrightarrow \hat \Pi_{15} , \quad & \hat \Pi_{16} &\longleftrightarrow \hat \Pi_{16} , \quad & \hat \Pi_{17} &\longleftrightarrow \hat \Pi_{18} , \quad & \hat \Pi_{19} &\longleftrightarrow -\hat \Pi_{19} .
	\end{alignedat}
\end{align}
Therefore, it is convenient to write the HLbL contribution to the $(g-2)_\mu$ as a sum of 12 terms:
\begin{align}
	\begin{split}
		\label{eq:HLbLMasterFormula8dim}
		a_\mu^\mathrm{HLbL} &= - e^6  \int \frac{d^4q_1}{(2\pi)^4} \frac{d^4q_2}{(2\pi)^4} \frac{1}{q_1^2 q_2^2 (q_1+q_2)^2} \frac{1}{(p+q_1)^2 - m_\mu^2} \frac{1}{(p-q_2)^2 - m_\mu^2} \\
			& \quad \cdot  \sum_{j=1}^{12} \xi_j \hat T_{i_j}(q_1,q_2;p) \hat\Pi_{i_j}(q_1,q_2,-q_1-q_2) ,
	\end{split}
\end{align}
where
\begin{align}
	\begin{split}
		\{ i_j | j=1,\ldots,12\} &= \{ 1, 2, 4, 5, 7, 9, 10, 14, 15, 16, 17, 19 \} , \\
		\{ \xi_j | j=1, \ldots, 12\} &= \{ 1, 2, 1, 2, 2, 2, 2, 2, 1, 1, 2, 1 \} .
	\end{split}
\end{align}
Note that the first two terms in this sum correspond to the well-known result for the pion-pole contribution \cite{Knecht2002} (up to some conventions: exchange of $\hat T_1$ and $\hat T_2$, the explicit factor $\xi_2=2$ and symmetrisation of $\hat T_1$).

In (\ref{eq:HLbLMasterFormula8dim}), the integrand depends on the five scalar products $q_1^2$, $q_2^2$, $q_1 \cdot q_2$, $p \cdot q_1$ and $p \cdot q_2$, where the dependence on the last two is explicitly given (the scalar functions only depend on $q_1^2$, $q_2^2$ and $q_1 \cdot q_2$). Therefore, five of the eight integrals can be performed without knowledge of the scalar functions. The same integrals as in the case of the pion-pole contribution occur \cite{Jegerlehner2009}, which have been solved using the technique of Gegenbauer polynomials \cite{Rosner1967}.

Let us assume that we can safely perform a Wick rotation for the momenta $q_1$, $q_2$ and $p$. We denote the Wick-rotated Euclidean momenta by capital letters $Q_1$, $Q_2$ and $P$. Note that $Q_1^2 = -q_1^2$, $Q_2^2 = -q_2^2$, $P^2 = -m_\mu^2$. Since $a_\mu^\mathrm{HLbL}$ is a pure number, it does not depend on the direction of the momentum $P$ of the muon, hence we can take the angular average by integrating over the four-dimensional hypersphere:
\begin{align}
	\begin{split}
		a_\mu^\mathrm{HLbL} = \int  \frac{d\Omega_4(P)}{2\pi^2} a_\mu^\mathrm{HLbL} .
	\end{split}
\end{align}
The kernels $\hat T_i$ are at most quadratic in $p$, therefore we need the following angular integrals \cite{Jegerlehner2009}:
\begin{align}
	\label{eq:GegenbauerIntegrals}
	\begin{split}
		 \int  \frac{d\Omega_4(P)}{2\pi^2} \frac{1}{(P+Q_1)^2 + m_\mu^2}\frac{1}{(P-Q_2)^2 + m_\mu^2} &= \frac{1}{m_\mu^2 R_{12}} \atan\left(\frac{z x}{1 - z \tau}\right) , \\
		 \int  \frac{d\Omega_4(P)}{2\pi^2} \frac{1}{(P+Q_1)^2 + m_\mu^2} &= - \frac{1 - \sigma^E_1}{2m_\mu^2} , \\
		 \int  \frac{d\Omega_4(P)}{2\pi^2} \frac{1}{(P-Q_2)^2 + m_\mu^2} &= - \frac{1 - \sigma^E_2}{2m_\mu^2} , \\
		 \int  \frac{d\Omega_4(P)}{2\pi^2} \frac{P \cdot Q_2}{(P+Q_1)^2 + m_\mu^2} &= Q_1 \cdot Q_2 \frac{(1 - \sigma^E_1)^2}{8 m_\mu^2} , \\
		 \int  \frac{d\Omega_4(P)}{2\pi^2} \frac{P \cdot Q_1}{(P-Q_2)^2 + m_\mu^2} &= - Q_1 \cdot Q_2 \frac{(1 - \sigma^E_2)^2}{8 m_\mu^2} , \\
	\end{split}
\end{align}
where $\tau = \cos\theta_4$, defined by $Q_1 \cdot Q_2 = |Q_1| |Q_2| \tau$, is the cosine of the angle between the Euclidean four-momenta $Q_1$ and $Q_2$, and further
\begin{align}
	\begin{split}
		\sigma^E_i &:= \sqrt{ 1 + \frac{4 m_\mu^2}{Q_i^2} } , \quad R_{12} := |Q_1| |Q_2| x , \quad x := \sqrt{1 - \tau^2} , \\
		z &:= \frac{|Q_1||Q_2|}{4m_\mu^2} (1-\sigma^E_1)(1-\sigma^E_2) .
	\end{split}
\end{align}

\clearpage

\section{Master Formula}

\label{sec:MasterFormula}

After using the angular integrals (\ref{eq:GegenbauerIntegrals}), we can perform immediately five of the eight loop integrals by changing to spherical coordinates in four dimensions. This leads us to a master formula for the HLbL contribution to the anomalous magnetic moment of the muon:
\begin{align}
	\begin{split}
		\label{eq:MasterFormula3Dim}
		a_\mu^\mathrm{HLbL} &= \frac{2 \alpha^3}{3 \pi^2} \int_0^\infty dQ_1 \int_0^\infty dQ_2 \int_{-1}^1 d\tau \sqrt{1-\tau^2} Q_1^3 Q_2^3 \sum_{i=1}^{12} T_i(Q_1,Q_2,\tau) \bar \Pi_i(Q_1,Q_2,\tau) ,
	\end{split}
\end{align}
where $\alpha = e^2/(4\pi)$, $Q_1 := |Q_1|$, $Q_2 := |Q_2|$. The hadronic scalar functions $\bar \Pi_i$ are just a subset of the $\hat \Pi_i$ and defined by
\begin{align}
	\begin{split}
		\bar \Pi_1 &= \Pi_1 + q_1 \cdot q_2 \Pi_{47} , \\
		\bar \Pi_2 &= \Pi_2 - \frac{1}{2} \left(q_1^2 + q_1 \cdot q_2 \right) \left( 2 \Pi_{47} - \Pi_{50} - \Pi_{51} - \Pi_{54} \right) , \\
		\bar \Pi_3 &= \Pi_4 + \left(q_1^2 + q_1 \cdot q_2 \right) \Pi_{19} + \left(q_1 \cdot q_2 + q_2^2 \right) \Pi_{20} \\
			& \quad + \left(q_1^2 + q_1 \cdot q_2 \right) \left(q_1 \cdot q_2+q_2^2\right) \Pi_{31} - \frac{s}{2} \left( 2 \Pi_{47} - \Pi_{50} - \Pi_{51} \right) + \frac{1}{2} \left( q_1^2 - q_2^2\right) \Pi_{54} , \\
		\bar \Pi_4 &= \Pi_5 - q_1 \cdot q_2 \Pi_{21} + \frac{1}{2} \left(q_1 \cdot q_2+q_2^2\right) \left( 2 \Pi_{22} - 2 q_1 \cdot q_2 \Pi_{33}  + \Pi_{50}  + \Pi_{51} - \Pi_{54}\right) - q_2^2 \Pi_{47} , \\
		\bar \Pi_5 &= \Pi_7 - \Pi_{19} - \left( q_1 \cdot q_2 + q_2^2\right) \Pi_{31} , \\
		\bar \Pi_6 &= \Pi_9 - \Pi_{22} + q_1 \cdot q_2 \Pi_{33} , \\
		\bar \Pi_7 &= \Pi_{10} - \Pi_{21} - \left( q_1 \cdot q_2 + q_2^2 \right) \Pi_{33} , \\
		\bar \Pi_8 &= \Pi_{16} + \Pi_{47} + \Pi_{54} , \\
		\bar \Pi_9 &= \Pi_{17} + \Pi_{47} - \Pi_{50} - \Pi_{51} , \\
		\bar \Pi_{10} &= \frac{1}{2} \left( \Pi_{39} + \Pi_{40} + \Pi_{46} \right) , \\
		\bar \Pi_{11} &= \Pi_{42} - \Pi_{47} + \frac{1}{2} \left( \Pi_{50} + \Pi_{51} + \Pi_{54} \right) , \\
		\bar \Pi_{12} &= \frac{1}{2} \left( \Pi_{50} - \Pi_{51} + \Pi_{54} \right) .
   	\end{split}
\end{align}
They have to be evaluated for the reduced kinematics
\begin{align}
	\begin{split}
		s &= - Q_3^2 = -Q_1^2 - 2 Q_1 Q_2 \tau - Q_2^2 , \quad t = -Q_2^2 , \quad u = -Q_1^2 , \\
		q_1^2 &= -Q_1^2, \quad q_2^2 = -Q_2^2, \quad q_3^2 = - Q_3^2 = - Q_1^2 - 2 Q_1 Q_2 \tau - Q_2^2 , \quad k^2 = q_4^2 = 0.
	\end{split}
\end{align}
The integral kernels $T_i$ are listed in appendix~\ref{sec:AppendixMasterFormulaKernels}. The scalar functions $\Pi_i$ parametrise the hadronic content of the master formula.

Note that (\ref{eq:MasterFormula3Dim}) is the generalisation of the three-dimensional integral formula for the pion-pole contribution \cite{Jegerlehner2009}. It is valid for the whole HLbL contribution and completely generic, i.e.~it can be used to compute the HLbL contribution to the $(g-2)_\mu$ given any representation of the HLbL tensor (even a model calculation). If the HLbL tensor is known, the scalar functions $\Pi_i$ can be easily obtained by projection and identification of the kinematic singularities, see appendix~\ref{sec:AppendixProjection}.

As (\ref{eq:MasterFormula3Dim}) only contains a three-dimensional integral, this formula is also suited for a direct numerical implementation.

Our main task is the calculation of the scalar functions $\Pi_i$ in a model-independent way by making use of dispersion relations.


\chapter{Mandelstam Representation}

\label{sec:MandelstamRepresentation}

In the previous chapter, we have obtained a master formula (\ref{eq:MasterFormula3Dim}) for the HLbL contribution to the anomalous magnetic moment of the muon, where the hadronic dynamics is parametrised in terms of the scalar functions $\Pi_i$. Since these functions are free of kinematic singularities and zeros, they are the quantities that should satisfy a Mandelstam representation \cite{Mandelstam1958}. We need to determine seven scalar functions that are not related to each other by crossing symmetry.

Due to the complexity of the problem, we cannot obtain an exact solution for the scalar functions but have to rely on some approximations to be able to write down a dispersion relation. These are the following:
\begin{enumerate}
	\item we limit ourselves to the lightest intermediate states in the direct channel, specifically one- or two-pion intermediate states;
	\item we take into account the double-spectral contributions of only two-pion intermediate states in the crossed channel and we approximate higher intermediate states in the crossed channel by a partial-wave expansion.
\end{enumerate}
We will discuss later how to improve our treatment in view of these two approximations.

\section{Derivation of the Double-Spectral Representation}

\begin{figure}[H]
	\centering
	\includegraphics[width=2.5cm]{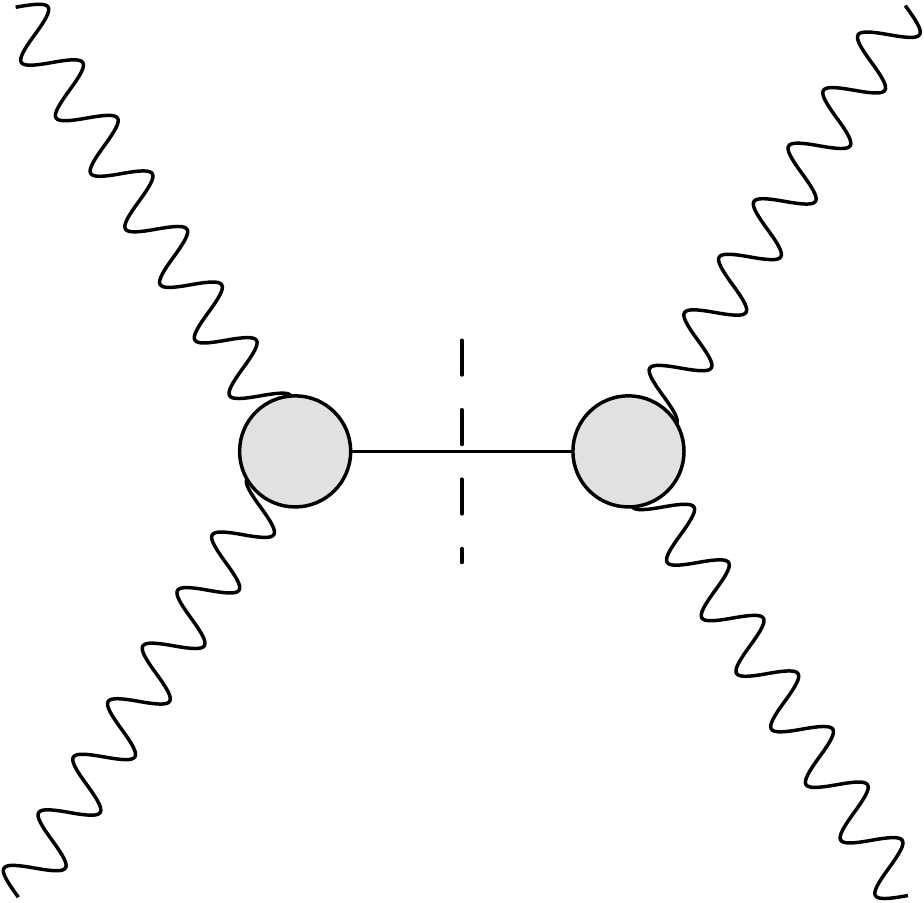}
	\hspace{1cm}
	\includegraphics[width=2.5cm]{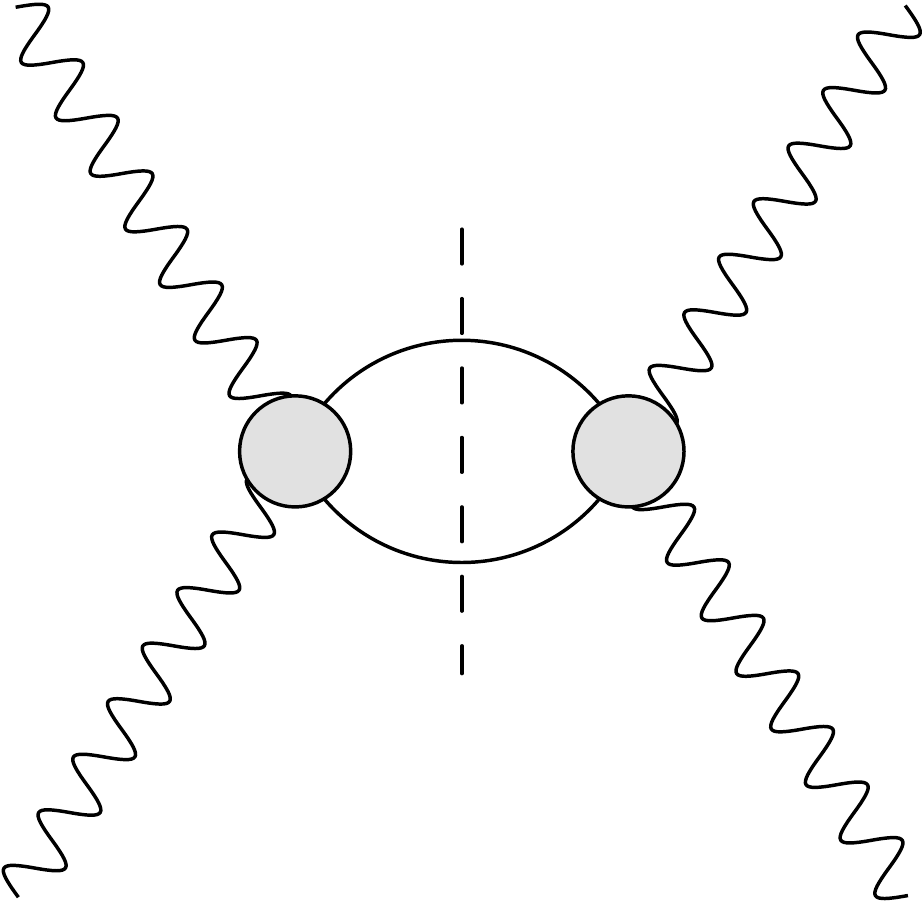}
	\caption{Intermediate states in the direct channel: pion pole and two-pion cut.}
	\label{img:HLbLIntermediateStates}
\end{figure}

For the derivation of a Mandelstam representation of the scalar functions, we follow the discussion in \cite{Martin1970}. We assume that the photon virtualities $q_i^2$ are fixed and small enough such that no anomalous thresholds are present. As we need a parameter-free description of the HLbL tensor, we assume that the asymptotic behaviour of the scalar functions allows us to write down a fixed-$t$ dispersion relation without any subtractions. This assumption is supported by the behaviour of the imaginary parts, which is determined by the asymptotics of the sub-processes. Hence, for a generic scalar function $\Pi_i$, we write
\begin{align}
	\begin{split}
		\Pi_i^t(s,t,u) &= c_i^t + \frac{\rho_{i;s}^t}{s - M_\pi^2} + \frac{\rho_{i;u}^t}{u - M_\pi^2} \\
			&\quad + \frac{1}{\pi} \int_{4M_\pi^2}^\infty ds^\prime \frac{ \Im_s \Pi_i^t(s^\prime,t,u^\prime)}{s^\prime - s} + \frac{1}{\pi} \int_{4M_\pi^2}^\infty du^\prime \frac{\Im_u \Pi_i^t(s^\prime,t,u^\prime)}{u^\prime-u} ,
	\end{split}
\end{align}
where $c_i^t$ is supposed to behave as $\lim\limits_{t\to0} c_i^t = 0$ and takes into account the $t$-channel pole. The imaginary parts are understood to be evaluated just above the corresponding cut. The primed variables fulfil
\begin{align}
	\begin{split}
		s^\prime + t + u^\prime = \Sigma := \sum_{i=1}^4 q_i^2 .
	\end{split}
\end{align}
If we continue the fixed-$t$ dispersion relation analytically in $t$, we have to replace the imaginary parts by the discontinuities, defined by
\begin{align}
	\begin{split}
		D_{i;s}^t(s^\prime) &:= \frac{1}{2i} \left( \Pi_i^t(s^\prime+i\epsilon,t,u^\prime) - \Pi_i^t(s^\prime-i\epsilon,t,u^\prime) \right) , \\
		D_{i;u}^t(u^\prime) &:= \frac{1}{2i} \left( \Pi_i^t(s^\prime,t,u^\prime+i\epsilon) - \Pi_i^t(s^\prime,t,u^\prime-i\epsilon) \right) ,
	\end{split}
\end{align}
hence
\begin{align}
	\begin{split}
		\label{eq:HLbLUnsubtractedDRforScalarFunctions}
		\Pi_i^t(s,t,u) &= c_i^t + \frac{\rho_{i;s}^t}{s - M_\pi^2} + \frac{\rho_{i;u}^t}{u - M_\pi^2} + \frac{1}{\pi} \int_{4M_\pi^2}^\infty ds^\prime \frac{ D_{i;s}^t(s^\prime)}{s^\prime - s} + \frac{1}{\pi} \int_{4M_\pi^2}^\infty du^\prime \frac{D_{i;u}^t(u^\prime)}{u^\prime-u} .
	\end{split}
\end{align}
Both, the discontinuities as well as the pole residues are determined by $s$- or $u$-channel unitarity, which also defines their analytic continuation in $t$. While $\rho^t_{i;s,u}$ are due to a one-pion intermediate state, $D_{i;s,u}^t$ are due to multi-particle intermediate states, see figure~\ref{img:HLbLIntermediateStates}. We limit ourselves to two-pion intermediate states and neglect the contribution of heavier intermediate states to the discontinuities.

Let us first study the pion-pole contribution by analysing the unitarity relation:
\begin{align}
	\begin{split}
		\Im_s &\left( e^4 (2\pi)^4 \delta^{(4)}(q_1 + q_2 + q_3 - q_4) H_{\lambda_1\lambda_2,\lambda_3\lambda_4} \right) \\
			&= \sum_n \frac{1}{2S_n} \left( \prod_{i=1}^n \int \widetilde{dp_i} \right) \< \gamma^*(-q_3,\lambda_3) \gamma^*(q_4,\lambda_3) | n; \{p_i\} \> \< \gamma^*(q_1,\lambda_1) \gamma^*(q_2,\lambda_2) | n; \{p_i\} \>^* ,
	\end{split}
\end{align}
where $S_n$ is the symmetry factor of the intermediate state $|n\>$. We consider now only the $\pi^0$ intermediate state in the sum:
\begin{align}
	\begin{split}
		\Im_s^\pi &\left( e^4 (2\pi)^4 \delta^{(4)}(q_1 + q_2 + q_3 - q_4) H_{\lambda_1\lambda_2,\lambda_3\lambda_4} \right) \\
			&= \frac{1}{2} \int \widetilde{dp} \; \< \gamma^*(-q_3,\lambda_3) \gamma^*(q_4,\lambda_3) | \pi^0(p) \> \< \gamma^*(q_1,\lambda_1) \gamma^*(q_2,\lambda_2) | \pi^0(p) \>^* .
	\end{split}
\end{align}
After reducing the matrix elements and using the definition of the pion transition form factor
\begin{align}
	\begin{split}
		i \int d^4x \; e^{i q x} \< 0 | T \{ j^\mu_\mathrm{em}(x) j^\nu_\mathrm{em}(0) \} | \pi^0(p) \> = \epsilon^{\mu\nu\alpha\beta} q_\alpha p_\beta \mathcal{F}_{\pi^0\gamma^*\gamma^*}\big(q^2, (q-p)^2\big) ,
	\end{split}
\end{align}
we find
\begin{align}
	\begin{split}
		\Im_s^\pi \Pi^{\mu\nu\lambda\sigma} &= - \frac{1}{2} \begin{aligned}[t]
				& \int \widetilde{dp} \; (2\pi)^4 \delta^{(4)}(q_1+q_2-p) \epsilon^{\mu\nu\alpha\beta} \epsilon^{\lambda\sigma\gamma\delta} {q_1}_\alpha {q_2}_\beta {q_3}_\gamma {q_4}_\delta \\
				&\quad \cdot \mathcal{F}_{\pi^0\gamma^*\gamma^*}\big(q_1^2,q_2^2\big) \mathcal{F}_{\pi^0\gamma^*\gamma^*}\big(q_3^2,q_4^2\big) \end{aligned} \\
			&= - \pi \delta( s - M_\pi^2 ) \epsilon^{\mu\nu\alpha\beta} \epsilon^{\lambda\sigma\gamma\delta} {q_1}_\alpha {q_2}_\beta {q_3}_\gamma {q_4}_\delta \mathcal{F}_{\pi^0\gamma^*\gamma^*}\big(q_1^2,q_2^2\big) \mathcal{F}_{\pi^0\gamma^*\gamma^*}\big(q_3^2,q_4^2\big) .
	\end{split}
\end{align}
By projecting onto the scalar functions $\Pi_i$, we find
\begin{align}
	\begin{split}
		\rho_{i;s}^t &= \left\{ \begin{matrix} \mathcal{F}_{\pi^0\gamma^*\gamma^*}\big(q_1^2,q_2^2\big) \mathcal{F}_{\pi^0\gamma^*\gamma^*}\big(q_3^2,q_4^2\big) & i = 1 , \\
									0 & i \neq 1 . \end{matrix} \right.
	\end{split}
\end{align}
Analogously
\begin{align}
	\begin{split}
		\rho_{i;u}^t &= \left\{ \begin{matrix} \mathcal{F}_{\pi^0\gamma^*\gamma^*}\big(q_1^2,q_4^2\big) \mathcal{F}_{\pi^0\gamma^*\gamma^*}\big(q_2^2,q_3^2\big) & i = 3 , \\
									0 & i \neq 3 . \end{matrix} \right.
	\end{split}
\end{align}

In order to identify the discontinuities, we project the unitarity relation selecting two-pion intermediate states:
\begin{align}
	\begin{split}
		& \Im_s^{\pi\pi} \left( e^4 (2\pi)^4 \delta^{(4)}(q_1 + q_2 + q_3 - q_4) H_{\lambda_1\lambda_2,\lambda_3\lambda_4} \right) \\
			&= \frac{1}{2} \int \widetilde{dp}_1 \widetilde{dp}_2 \< \pi^+(p_1) \pi^-(p_2) | \gamma^*(-q_3,\lambda_3) \gamma^*(q_4,\lambda_3) \>^* \< \pi^+(p_1) \pi^-(p_2) | \gamma^*(q_1,\lambda_1) \gamma^*(q_2,\lambda_2) \> \\
			&\quad + \frac{1}{4} \int \widetilde{dp}_1 \widetilde{dp}_2 \< \pi^0(p_1) \pi^0(p_2) | \gamma^*(-q_3,\lambda_3) \gamma^*(q_4,\lambda_3) \>^* \< \pi^0(p_1) \pi^0(p_2) | \gamma^*(q_1,\lambda_1) \gamma^*(q_2,\lambda_2) \> ,
	\end{split}
\end{align}
hence
\begin{align}
	\begin{split}
		\Im_s^{\pi\pi} \Pi^{\mu\nu\lambda\sigma} &= \frac{1}{32\pi^2} \frac{\sigma_\pi(s)}{2} \int d\Omega_s^\dprime \begin{aligned}[t]
			& \bigg( W_{+-}^{\mu\nu}(p_1,p_2,q_1) {W_{+-}^{\lambda\sigma}}^*(p_1,p_2,-q_3) \\
			& + \frac{1}{2} W_{00}^{\mu\nu}(p_1,p_2,q_1) {W_{00}^{\lambda\sigma}}^*(p_1,p_2,-q_3) \bigg) . \end{aligned}
	\end{split}
\end{align}
The analytic continuation of the unitarity relation can be obtained if the $\gamma^*\gamma^*\to\pi\pi$ matrix element $W^{\mu\nu}$ is expressed in terms of the fixed-$s$ dispersion relation (\ref{eq:FixedSDispRelggpipi}) for its scalar functions:
\begin{align}
	\begin{split}
		W_{+-}^{\mu\nu} &= \sum_{i=1}^6 T^{\mu\nu}_i \begin{aligned}[t]
			& \Bigg( \frac{\hat\rho_{i;t}^{s;+-}(s)}{t-M_\pi^2} + \frac{\hat\rho_{i;u}^{s;+-}(s)}{u-M_\pi^2} \\
			& + \frac{1}{\pi} \int_{4M_\pi^2}^\infty dt_1 \frac{\hat D_{i;t}^{s;+-}(t_1;s)}{t_1 - t} + \frac{1}{\pi} \int_{4M_\pi^2}^\infty du_1 \frac{\hat D_{i;u}^{s;+-}(u_1;s)}{u_1 - u} \Bigg) , \end{aligned} \\
		W_{00}^{\mu\nu} &= \sum_{i=1}^6 T^{\mu\nu}_i \begin{aligned}[t]
			& \Bigg( \frac{1}{\pi} \int_{4M_\pi^2}^\infty dt_1 \frac{\hat D_{i;t}^{s;00}(t_1;s)}{t_1 - t} + \frac{1}{\pi} \int_{4M_\pi^2}^\infty du_1 \frac{\hat D_{i;u}^{s;00}(u_1;s)}{u_1 - u} \Bigg) . \end{aligned}
	\end{split}
\end{align}
Note that $W_{00}^{\mu\nu}$ does not contain any pole terms because the photon does not couple to two neutral pions due to parity conservation and Bose symmetry.

If we pick the contribution of the pole terms on both sides of the cut, we single out box topologies:
\begin{align}
	\begin{split}
		\Im_s^{\pi\pi} \Pi^{\mu\nu\lambda\sigma} \Big|_\mathrm{box} &= \frac{1}{32\pi^2} \frac{\sigma_\pi(s)}{2} \int d\Omega_s^\dprime  \sum_{i,j=1,5} T^{\mu\nu}_i T^{\lambda\sigma}_j \begin{aligned}[t]
			& \Bigg( \frac{\hat\rho_{i;t}^{s;+-}(s)}{t^\prime-M_\pi^2} + \frac{\hat\rho_{i;u}^{s;+-}(s)}{u^\prime-M_\pi^2} \Bigg) \\
			& \cdot \Bigg( \frac{\hat\rho_{j;t}^{s;+-}(s)}{t^\dprime-M_\pi^2} + \frac{\hat\rho_{j;u}^{s;+-}(s)}{u^\dprime-M_\pi^2} \Bigg)^* , \end{aligned}
	\end{split}
\end{align}
where the primed variables belong to the sub-process on the left-hand side and the double-primed variables to the sub-process on the right-hand side of the cut.

We could now apply a tensor reduction and perform the phase-space integrals for the reduced scalar quantities. The projection on the scalar functions $\Pi_i$ would allow us then to identify the discontinuities $D_{i;s}^t$ due to box structures. However, already now it can be noted that the non-zero pole residues $\hat\rho_{i;t,u}^{s;+-}$ contain two electromagnetic pion form factors for the off-shell photons. These form factors can be factored out and multiply then the discontinuity that would be obtained by applying Cutkosky's rules \cite{Cutkosky1960} to the sQED pion loop calculation. This becomes clear from the relation between the pole terms and the sQED Born terms as discussed in section~\ref{sec:PionPoleggpipi} and appendix~\ref{sec:AppendixScalarQEDBornggpipi}. Therefore, the box contribution is nothing else but the sQED contribution multiplied by a form factor $F_\pi^V(q_i^2)$ for each of the off-shell photons. As in the case of the sub-process, the difference between unitarity diagrams and Feynman diagrams should be noted: the sQED contribution consists of boxes, triangles and bulb Feynman diagrams but corresponds to the pure box topology in terms of unitarity.

Finally, there are the contributions with discontinuities either in one or in both of the sub-processes:
\begin{align}
	\small
	\begin{split}
		\Im_s^{\pi\pi} \Pi^{\mu\nu\lambda\sigma} \Big|_\mathrm{1disc} &= \frac{1}{32\pi^2} \frac{\sigma_\pi(s)}{2} \int d\Omega_s^\dprime  \sum_{i,j=1}^6 T^{\mu\nu}_i T^{\lambda\sigma}_j \\
			& \cdot \begin{aligned}[t]
			& \Bigg[ \bigg( \frac{\hat\rho_{i;t}^{s;+-}(s)}{t^\prime-M_\pi^2} + \frac{\hat\rho_{i;u}^{s;+-}(s)}{u^\prime-M_\pi^2} \bigg) \bigg( \frac{1}{\pi} \int_{4M_\pi^2}^\infty dt_2 \frac{\hat D_{j;t}^{s;+-}(t_2;s)}{t_2 - t^\dprime} + \frac{1}{\pi} \int_{4M_\pi^2}^\infty du_2 \frac{\hat D_{j;u}^{s;+-}(u_2;s)}{u_2 - u^\dprime} \bigg)^* \\
			& + \bigg(\frac{1}{\pi} \int_{4M_\pi^2}^\infty dt_1 \frac{\hat D_{i;t}^{s;+-}(t_1;s)}{t_1 - t^\prime} + \frac{1}{\pi} \int_{4M_\pi^2}^\infty du_1 \frac{\hat D_{i;u}^{s;+-}(u_1;s)}{u_1 - u^\prime} \bigg) \bigg( \frac{\hat\rho_{j;t}^{s;+-}(s)}{t^\dprime-M_\pi^2} + \frac{\hat\rho_{j;u}^{s;+-}(s)}{u^\dprime-M_\pi^2} \bigg)^* \Bigg] , \end{aligned} \\
		\Im_s^{\pi\pi} \Pi^{\mu\nu\lambda\sigma} \Big|_\mathrm{2disc} &= \frac{1}{32\pi^2} \frac{\sigma_\pi(s)}{2} \int d\Omega_s^\dprime \sum_{i,j=1}^6 T^{\mu\nu}_i T^{\lambda\sigma}_j \\
			&\cdot \Bigg[ \begin{aligned}[t]
				& \bigg( \frac{1}{\pi} \int_{4M_\pi^2}^\infty dt_1 \frac{\hat D_{i;t}^{s;+-}(t_1;s)}{t_1 - t^\prime} + \frac{1}{\pi} \int_{4M_\pi^2}^\infty du_1 \frac{\hat D_{i;u}^{s;+-}(u_1;s)}{u_1 - u^\prime} \bigg) \\
				& \cdot \bigg( \frac{1}{\pi} \int_{4M_\pi^2}^\infty dt_2 \frac{\hat D_{j;t}^{s;+-}(t_2;s)}{t_2 - t^\dprime} + \frac{1}{\pi} \int_{4M_\pi^2}^\infty du_2 \frac{\hat D_{j;u}^{s;+-}(u_2;s)}{u_2 - u^\dprime} \bigg)^* \\
				& + \frac{1}{2} \bigg( \frac{1}{\pi} \int_{4M_\pi^2}^\infty dt_1 \frac{\hat D_{i;t}^{s;00}(t_1;s)}{t_1 - t^\prime} + \frac{1}{\pi} \int_{4M_\pi^2}^\infty du_1 \frac{\hat D_{i;u}^{s;00}(u_1;s)}{u_1 - u^\prime} \bigg) \\
				& \cdot \bigg( \frac{1}{\pi} \int_{4M_\pi^2}^\infty dt_2 \frac{\hat D_{j;t}^{s;00}(t_2;s)}{t_2 - t^\dprime} + \frac{1}{\pi} \int_{4M_\pi^2}^\infty du_2 \frac{\hat D_{j;u}^{s;00}(u_2;s)}{u_2 - u^\dprime} \bigg)^* \Bigg] . \end{aligned}
	\end{split}
\end{align}
If the order of phase-space and dispersive integrals are exchanged, the phase-space integrals can be performed by applying a tensor reduction to the quantities
\begin{align}
	\begin{split}
		\int d\Omega_s^\dprime \sum_{i,j=1}^6 T_i^{\mu\nu} T_j^{\lambda\sigma} \frac{1}{t_1-t^\prime}\frac{1}{t_2-t^\dprime} .
	\end{split}
\end{align}
The reduced scalar integrals can then be transformed into another dispersive integral. Together with the dispersion integral $ds^\prime$ of the primary cut, this produces the double-spectral representation. The case of the simplest scalar phase-space integral is explained in appendix~\ref{sec:AppendixPhaseSpaceIntegration}.

\section{Classification into Topologies}

In the previous section, we have explained how the double-spectral representation can be derived from a fixed-$t$ dispersion relation by taking the analytic continuation in $t$, which is defined by the unitarity relation. In the $s$-channel unitarity relation, a fixed-$s$ dispersion relation of the sub-process is inserted (in the unitarity relation for the $u$-channel contribution, the variable $u$ is kept fixed, which, however, plays again the role of $s$ in the sub-process). Of course, one could have started with a fixed-$u$ or fixed-$s$ dispersion relation in the first place. The requirement that this leads to the same result allows us to identify a symmetric representation, which treats the Mandelstam variables on an equal footing and therefore implements crossing. In this symmetric representation, we classify the different contributions in terms of topologies. Note that in the case of HLbL, we get the two other possibilities (i.e.~taking fixed-$u$ and fixed-$s$ dispersion relations as the starting point) for free, because we consider a totally crossing symmetric process.

If we compare the three different representations of unsubtracted dispersion relations (\ref{eq:HLbLUnsubtractedDRforScalarFunctions}), we immediately see that the contributions in one representation are either contained explicitly in the other representations or can be understood as part of the respective constant $c_i^t$ or as a contribution of neglected higher intermediate states.

\subsection{Pion-Pole Contribution}

\begin{figure}[H]
	\centering
	\includegraphics[width=3cm]{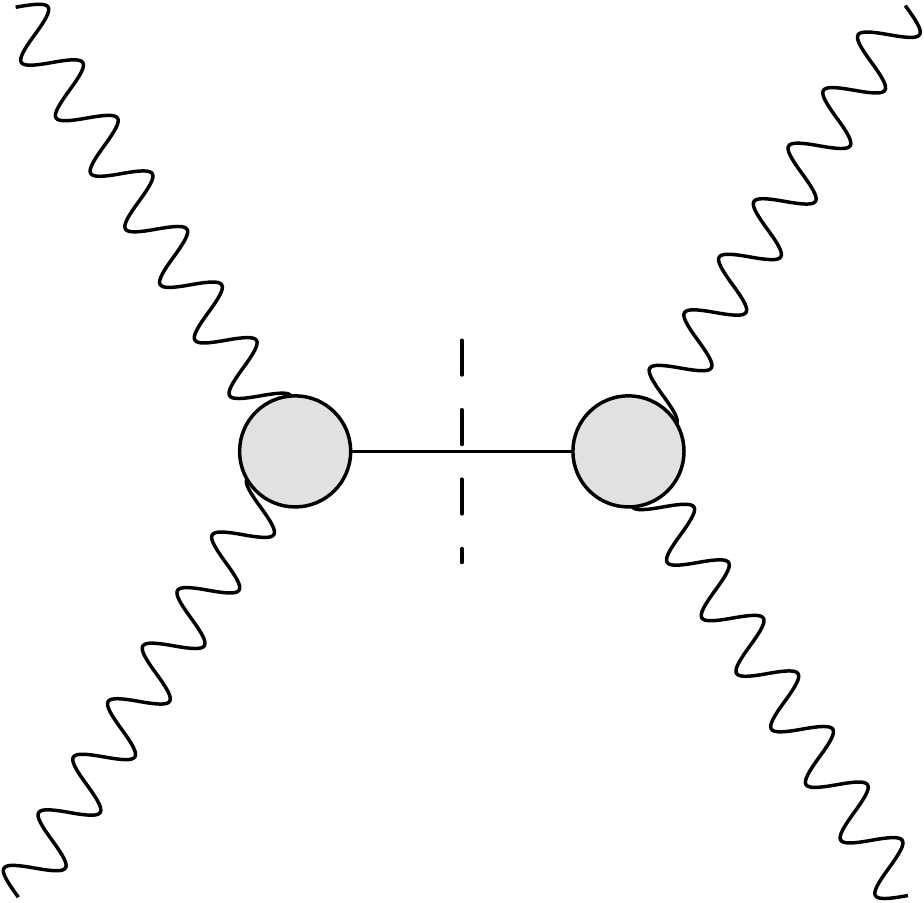}
	\caption{Unitarity diagram representing the pion-pole contribution in one channel.}
	\label{img:HLbLPionPole}
\end{figure}

The fixed-$t$ dispersion relation contains explicitly the poles in the $s$- and $u$-channel. Analogously, the fixed-$s$ (fixed-$u$) dispersion relation contains the poles in the $t$- and $u$-channel ($s$- and $t$-channel). The $t$-channel pole contribution, which is not explicit in the fixed-$t$ representation, can be identified with $c_i^t$ as it vanishes in the limit $t\to\infty$. Hence, the total pion-pole contribution is just given by
\begin{align}
	\begin{split}
		\Pi_i^{\pi^0\text{-pole}}(s,t,u) &= \frac{\rho_{i;s}}{s-M_\pi^2} + \frac{\rho_{i;t}}{t-M_\pi^2} + \frac{\rho_{i;u}}{u-M_\pi^2} ,
	\end{split}
\end{align}
where the pole residues are products of pion transition form factors:
\begin{align}
	\begin{split}
		\rho_{i,s} &= \delta_{i1} \; \mathcal{F}_{\pi^0\gamma^*\gamma^*}\big(q_1^2,q_2^2\big) \mathcal{F}_{\pi^0\gamma^*\gamma^*}\big(q_3^2,q_4^2\big) , \\
		\rho_{i,t} &= \delta_{i2} \; \mathcal{F}_{\pi^0\gamma^*\gamma^*}\big(q_1^2,q_3^2\big) \mathcal{F}_{\pi^0\gamma^*\gamma^*}\big(q_2^2,q_4^2\big) , \\
		\rho_{i,u} &= \delta_{i3} \; \mathcal{F}_{\pi^0\gamma^*\gamma^*}\big(q_1^2,q_4^2\big) \mathcal{F}_{\pi^0\gamma^*\gamma^*}\big(q_2^2,q_3^2\big) , \\
	\end{split}
\end{align}
where $\delta_{ij}$ is the Kronecker delta.

Using the master formula (\ref{eq:MasterFormula3Dim}), we find the well-known result for the pion-pole contribution to $a_\mu$ \cite{Knecht2002}:
\begin{align}
	\begin{split}
		a_\mu^{\pi^0\text{-pole}} &= \frac{2 \alpha^3}{3 \pi^2} \int_0^\infty dQ_1 \int_0^\infty dQ_2 \int_{-1}^1 d\tau \sqrt{1-\tau^2} Q_1^3 Q_2^3 \\
			&\quad \cdot \left( T_1(Q_1,Q_2,\tau) \bar \Pi_1^{\pi^0\text{-pole}}(Q_1,Q_2,\tau) + T_2(Q_1,Q_2,\tau) \bar \Pi_2^{\pi^0\text{-pole}}(Q_1,Q_2,\tau) \right) ,
	\end{split}
\end{align}
with
\begin{align}
	\begin{split}
		\bar \Pi_1^{\pi^0\text{-pole}} &= - \frac{\mathcal{F}_{\pi^0\gamma^*\gamma^*}\big({-Q_1^2},-Q_2^2\big) \mathcal{F}_{\pi^0\gamma^*\gamma^*}\big({-Q_3^2},0\big)}{Q_3^2+M_\pi^2} , \\
		\bar \Pi_2^{\pi^0\text{-pole}} &= - \frac{\mathcal{F}_{\pi^0\gamma^*\gamma^*}\big( {-Q_1^2},-Q_3^2\big) \mathcal{F}_{\pi^0\gamma^*\gamma^*}\big({-Q_2^2},0\big)}{Q_2^2+M_\pi^2} , \\
	\end{split}
\end{align}
where $Q_3^2 = Q_1^2 + 2 Q_1 Q_2 \tau + Q_2^2$ and the integral kernels $T_i$ are given in appendix~\ref{sec:AppendixMasterFormulaKernels}.

\subsection{Box Contribution}

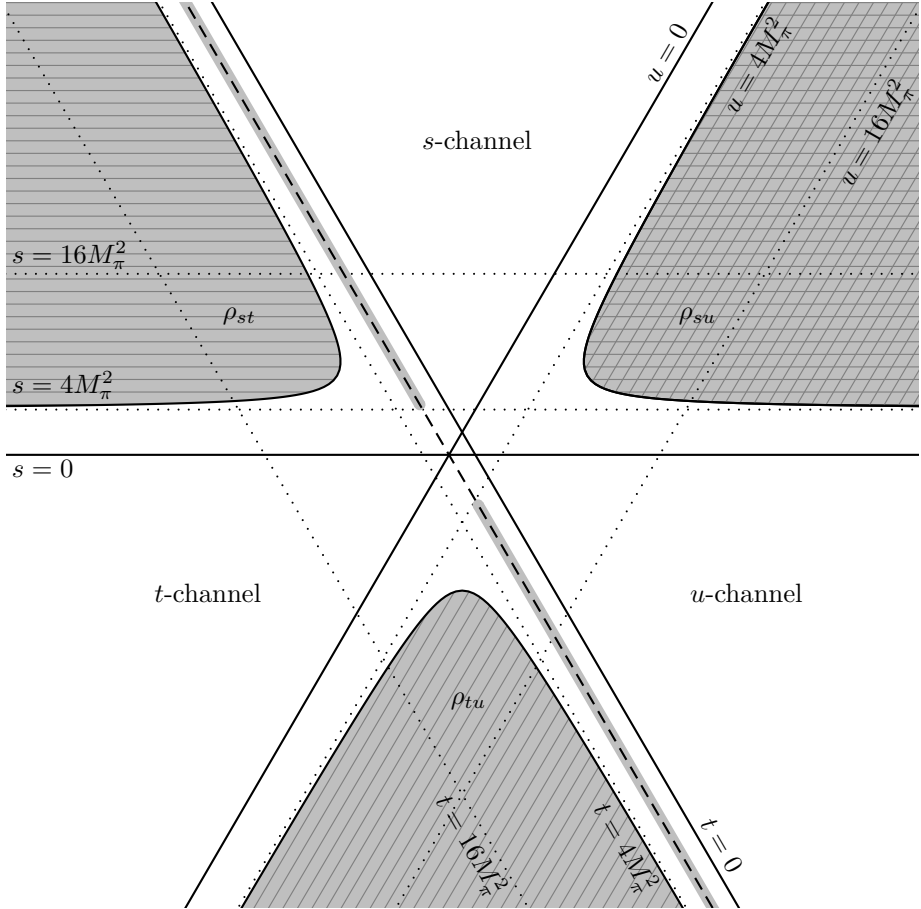
\begin{figure}[H]
	\centering
	\psset{unit=0.15cm}
	\begin{pspicture*}(-40,-40)(40,40)
		\pnode(0,0){orig}
		\pnode(0,2){twomp2}

		\pnode([angle=-30,nodesep=4]twomp2){u4mp2}
		\pnode([angle=-30,nodesep=16]twomp2){u16mp2}

		\pnode(-3.4641,4.){tconst}
		\pnode([angle=210,nodesep=4]twomp2){t4mp2}
		\pnode([angle=210,nodesep=16]twomp2){t16mp2}

		\pnode([angle=120,nodesep=4.618802153517007]t4mp2){stcorner}
		\pnode([angle=60,nodesep=4.618802153517007]u4mp2){sucorner}
		\pnode(0,-6){tucorner}

		\psline[fillstyle=hlines*, hatchangle=0, hatchwidth=0.1, hatchsep=1, hatchcolor=gray, fillcolor=lightgray](-70.5295,4.16071)(-69.953,4.16216)(-69.3765,4.16364)(-68.8,4.16514)(-68.2236,4.16667)(-67.6471,4.16822)(-67.0707,4.16981)(-66.4943,4.17143)(-65.9179,4.17308)(-65.3415,4.17476)(-64.7651,4.17647)(-64.1888,4.17822)(-63.6125,4.18)(-63.0362,4.18182)(-62.4599,4.18367)(-61.8836,4.18557)(-61.3074,4.1875)(-60.7312,4.18947)(-60.155,4.19149)(-59.5788,4.19355)(-59.0027,4.19565)(-58.4266,4.1978)(-57.8505,4.2)(-57.2744,4.20225)(-56.6984,4.20455)(-56.1224,4.2069)(-55.5465,4.2093)(-54.9705,4.21176)(-54.3946,4.21429)(-53.8188,4.21687)(-53.243,4.21951)(-52.6672,4.22222)(-52.0914,4.225)(-51.5157,4.22785)(-50.9401,4.23077)(-50.3644,4.23377)(-49.7889,4.23684)(-49.2133,4.24)(-48.6379,4.24324)(-48.0624,4.24658)(-47.4871,4.25)(-46.9117,4.25352)(-46.3365,4.25714)(-45.7613,4.26087)(-45.1861,4.26471)(-44.6111,4.26866)(-44.0361,4.27273)(-43.4612,4.27692)(-42.8863,4.28125)(-42.3115,4.28571)(-41.7368,4.29032)(-41.1622,4.29508)(-40.5877,4.3)(-40.0133,4.30508)(-39.439,4.31034)(-38.8648,4.31579)(-38.2907,4.32143)(-37.7167,4.32727)(-37.1429,4.33333)(-36.5691,4.33962)(-35.9956,4.34615)(-35.4221,4.35294)(-34.8489,4.36)(-34.2758,4.36735)(-33.7028,4.375)(-33.1301,4.38298)(-32.5575,4.3913)(-31.9852,4.4)(-31.4131,4.40909)(-30.8412,4.4186)(-30.2696,4.42857)(-29.6983,4.43902)(-29.1273,4.45)(-28.5566,4.46154)(-27.9863,4.47368)(-27.4163,4.48649)(-26.8468,4.5)(-26.2777,4.51429)(-25.7091,4.52941)(-25.141,4.54545)(-24.5735,4.5625)(-24.0066,4.58065)(-23.4404,4.6)(-22.875,4.62069)(-22.3105,4.64286)(-21.7469,4.66667)(-21.1843,4.69231)(-20.623,4.72)(-20.0629,4.75)(-19.5044,4.78261)(-18.9476,4.81818)(-18.3927,4.85714)(-17.8401,4.9)(-17.2901,4.94737)(-16.7432,5.)(-16.1998,5.05882)(-15.6606,5.125)(-15.1266,5.2)(-14.5987,5.28571)(-14.0785,5.38462)(-13.5677,5.5)(-13.0691,5.63636)(-12.5862,5.8)(-12.1244,6.)(-11.6913,6.25)(-11.2996,6.57143)(-10.9697,7.)(-10.7635,7.5)(-10.681,8.)(-10.681,8.5)(-10.7387,9.)(-10.8384,9.5)(-10.9697,10.)(-11.1251,10.5)(-11.2996,11.)(-11.4893,11.5)(-11.6913,12.)(-11.9036,12.5)(-12.1244,13.)(-12.3523,13.5)(-12.5862,14.)(-12.8254,14.5)(-13.0691,15.)(-13.3167,15.5)(-13.5677,16.)(-13.8218,16.5)(-14.0785,17.)(-14.3375,17.5)(-14.5987,18.)(-14.8618,18.5)(-15.1266,19.)(-15.3929,19.5)(-15.6606,20.)(-15.9296,20.5)(-16.1998,21.)(-16.471,21.5)(-16.7432,22.)(-17.0162,22.5)(-17.2901,23.)(-17.5648,23.5)(-17.8401,24.)(-18.1161,24.5)(-18.3927,25.)(-18.6699,25.5)(-18.9476,26.)(-19.2258,26.5)(-19.5044,27.)(-19.7835,27.5)(-20.0629,28.)(-20.3428,28.5)(-20.623,29.)(-20.9035,29.5)(-21.1843,30.)(-21.4654,30.5)(-21.7469,31.)(-22.0285,31.5)(-22.3105,32.)(-22.5926,32.5)(-22.875,33.)(-23.1576,33.5)(-23.4404,34.)(-23.7234,34.5)(-24.0066,35.)(-24.29,35.5)(-24.5735,36.)(-24.8571,36.5)(-25.141,37.)(-25.425,37.5)(-25.7091,38.)(-25.9933,38.5)(-26.2777,39.)(-26.5622,39.5)(-26.8468,40.)(-27.1315,40.5)(-27.4163,41.)(-27.7013,41.5)(-27.9863,42.)(-28.2714,42.5)(-28.5566,43.)(-28.8419,43.5)(-29.1273,44.)(-29.4128,44.5)(-29.6983,45.)(-29.984,45.5)(-30.2696,46.)(-30.5554,46.5)(-30.8412,47.)(-31.1271,47.5)(-31.4131,48.)(-31.6991,48.5)(-31.9852,49.)(-32.2713,49.5)(-32.5575,50.)(-32.8438,50.5)(-33.1301,51.)(-33.4164,51.5)(-33.7028,52.)(-33.9893,52.5)(-34.2758,53.)(-34.5623,53.5)(-34.8489,54.)(-35.1355,54.5)(-35.4221,55.)(-35.7088,55.5)(-35.9956,56.)(-36.2823,56.5)(-36.5691,57.)(-36.856,57.5)(-37.1429,58.)(-37.4298,58.5)(-37.7167,59.)(-38.0037,59.5)(-38.2907,60.)
		
		\psline[fillstyle=hlines*, hatchangle=0, hatchwidth=0.1, hatchsep=1, hatchcolor=gray, fillcolor=lightgray](70.5295,4.16071)(69.953,4.16216)(69.3765,4.16364)(68.8,4.16514)(68.2236,4.16667)(67.6471,4.16822)(67.0707,4.16981)(66.4943,4.17143)(65.9179,4.17308)(65.3415,4.17476)(64.7651,4.17647)(64.1888,4.17822)(63.6125,4.18)(63.0362,4.18182)(62.4599,4.18367)(61.8836,4.18557)(61.3074,4.1875)(60.7312,4.18947)(60.155,4.19149)(59.5788,4.19355)(59.0027,4.19565)(58.4266,4.1978)(57.8505,4.2)(57.2744,4.20225)(56.6984,4.20455)(56.1224,4.2069)(55.5465,4.2093)(54.9705,4.21176)(54.3946,4.21429)(53.8188,4.21687)(53.243,4.21951)(52.6672,4.22222)(52.0914,4.225)(51.5157,4.22785)(50.9401,4.23077)(50.3644,4.23377)(49.7889,4.23684)(49.2133,4.24)(48.6379,4.24324)(48.0624,4.24658)(47.4871,4.25)(46.9117,4.25352)(46.3365,4.25714)(45.7613,4.26087)(45.1861,4.26471)(44.6111,4.26866)(44.0361,4.27273)(43.4612,4.27692)(42.8863,4.28125)(42.3115,4.28571)(41.7368,4.29032)(41.1622,4.29508)(40.5877,4.3)(40.0133,4.30508)(39.439,4.31034)(38.8648,4.31579)(38.2907,4.32143)(37.7167,4.32727)(37.1429,4.33333)(36.5691,4.33962)(35.9956,4.34615)(35.4221,4.35294)(34.8489,4.36)(34.2758,4.36735)(33.7028,4.375)(33.1301,4.38298)(32.5575,4.3913)(31.9852,4.4)(31.4131,4.40909)(30.8412,4.4186)(30.2696,4.42857)(29.6983,4.43902)(29.1273,4.45)(28.5566,4.46154)(27.9863,4.47368)(27.4163,4.48649)(26.8468,4.5)(26.2777,4.51429)(25.7091,4.52941)(25.141,4.54545)(24.5735,4.5625)(24.0066,4.58065)(23.4404,4.6)(22.875,4.62069)(22.3105,4.64286)(21.7469,4.66667)(21.1843,4.69231)(20.623,4.72)(20.0629,4.75)(19.5044,4.78261)(18.9476,4.81818)(18.3927,4.85714)(17.8401,4.9)(17.2901,4.94737)(16.7432,5.)(16.1998,5.05882)(15.6606,5.125)(15.1266,5.2)(14.5987,5.28571)(14.0785,5.38462)(13.5677,5.5)(13.0691,5.63636)(12.5862,5.8)(12.1244,6.)(11.6913,6.25)(11.2996,6.57143)(10.9697,7.)(10.7635,7.5)(10.681,8.)(10.681,8.5)(10.7387,9.)(10.8384,9.5)(10.9697,10.)(11.1251,10.5)(11.2996,11.)(11.4893,11.5)(11.6913,12.)(11.9036,12.5)(12.1244,13.)(12.3523,13.5)(12.5862,14.)(12.8254,14.5)(13.0691,15.)(13.3167,15.5)(13.5677,16.)(13.8218,16.5)(14.0785,17.)(14.3375,17.5)(14.5987,18.)(14.8618,18.5)(15.1266,19.)(15.3929,19.5)(15.6606,20.)(15.9296,20.5)(16.1998,21.)(16.471,21.5)(16.7432,22.)(17.0162,22.5)(17.2901,23.)(17.5648,23.5)(17.8401,24.)(18.1161,24.5)(18.3927,25.)(18.6699,25.5)(18.9476,26.)(19.2258,26.5)(19.5044,27.)(19.7835,27.5)(20.0629,28.)(20.3428,28.5)(20.623,29.)(20.9035,29.5)(21.1843,30.)(21.4654,30.5)(21.7469,31.)(22.0285,31.5)(22.3105,32.)(22.5926,32.5)(22.875,33.)(23.1576,33.5)(23.4404,34.)(23.7234,34.5)(24.0066,35.)(24.29,35.5)(24.5735,36.)(24.8571,36.5)(25.141,37.)(25.425,37.5)(25.7091,38.)(25.9933,38.5)(26.2777,39.)(26.5622,39.5)(26.8468,40.)(27.1315,40.5)(27.4163,41.)(27.7013,41.5)(27.9863,42.)(28.2714,42.5)(28.5566,43.)(28.8419,43.5)(29.1273,44.)(29.4128,44.5)(29.6983,45.)(29.984,45.5)(30.2696,46.)(30.5554,46.5)(30.8412,47.)(31.1271,47.5)(31.4131,48.)(31.6991,48.5)(31.9852,49.)(32.2713,49.5)(32.5575,50.)(32.8438,50.5)(33.1301,51.)(33.4164,51.5)(33.7028,52.)(33.9893,52.5)(34.2758,53.)(34.5623,53.5)(34.8489,54.)(35.1355,54.5)(35.4221,55.)(35.7088,55.5)(35.9956,56.)(36.2823,56.5)(36.5691,57.)(36.856,57.5)(37.1429,58.)(37.4298,58.5)(37.7167,59.)(38.0037,59.5)(38.2907,60.)

		\psline[fillstyle=hlines, hatchangle=60, hatchwidth=0.1, hatchsep=1, hatchcolor=gray](70.5295,4.16071)(69.953,4.16216)(69.3765,4.16364)(68.8,4.16514)(68.2236,4.16667)(67.6471,4.16822)(67.0707,4.16981)(66.4943,4.17143)(65.9179,4.17308)(65.3415,4.17476)(64.7651,4.17647)(64.1888,4.17822)(63.6125,4.18)(63.0362,4.18182)(62.4599,4.18367)(61.8836,4.18557)(61.3074,4.1875)(60.7312,4.18947)(60.155,4.19149)(59.5788,4.19355)(59.0027,4.19565)(58.4266,4.1978)(57.8505,4.2)(57.2744,4.20225)(56.6984,4.20455)(56.1224,4.2069)(55.5465,4.2093)(54.9705,4.21176)(54.3946,4.21429)(53.8188,4.21687)(53.243,4.21951)(52.6672,4.22222)(52.0914,4.225)(51.5157,4.22785)(50.9401,4.23077)(50.3644,4.23377)(49.7889,4.23684)(49.2133,4.24)(48.6379,4.24324)(48.0624,4.24658)(47.4871,4.25)(46.9117,4.25352)(46.3365,4.25714)(45.7613,4.26087)(45.1861,4.26471)(44.6111,4.26866)(44.0361,4.27273)(43.4612,4.27692)(42.8863,4.28125)(42.3115,4.28571)(41.7368,4.29032)(41.1622,4.29508)(40.5877,4.3)(40.0133,4.30508)(39.439,4.31034)(38.8648,4.31579)(38.2907,4.32143)(37.7167,4.32727)(37.1429,4.33333)(36.5691,4.33962)(35.9956,4.34615)(35.4221,4.35294)(34.8489,4.36)(34.2758,4.36735)(33.7028,4.375)(33.1301,4.38298)(32.5575,4.3913)(31.9852,4.4)(31.4131,4.40909)(30.8412,4.4186)(30.2696,4.42857)(29.6983,4.43902)(29.1273,4.45)(28.5566,4.46154)(27.9863,4.47368)(27.4163,4.48649)(26.8468,4.5)(26.2777,4.51429)(25.7091,4.52941)(25.141,4.54545)(24.5735,4.5625)(24.0066,4.58065)(23.4404,4.6)(22.875,4.62069)(22.3105,4.64286)(21.7469,4.66667)(21.1843,4.69231)(20.623,4.72)(20.0629,4.75)(19.5044,4.78261)(18.9476,4.81818)(18.3927,4.85714)(17.8401,4.9)(17.2901,4.94737)(16.7432,5.)(16.1998,5.05882)(15.6606,5.125)(15.1266,5.2)(14.5987,5.28571)(14.0785,5.38462)(13.5677,5.5)(13.0691,5.63636)(12.5862,5.8)(12.1244,6.)(11.6913,6.25)(11.2996,6.57143)(10.9697,7.)(10.7635,7.5)(10.681,8.)(10.681,8.5)(10.7387,9.)(10.8384,9.5)(10.9697,10.)(11.1251,10.5)(11.2996,11.)(11.4893,11.5)(11.6913,12.)(11.9036,12.5)(12.1244,13.)(12.3523,13.5)(12.5862,14.)(12.8254,14.5)(13.0691,15.)(13.3167,15.5)(13.5677,16.)(13.8218,16.5)(14.0785,17.)(14.3375,17.5)(14.5987,18.)(14.8618,18.5)(15.1266,19.)(15.3929,19.5)(15.6606,20.)(15.9296,20.5)(16.1998,21.)(16.471,21.5)(16.7432,22.)(17.0162,22.5)(17.2901,23.)(17.5648,23.5)(17.8401,24.)(18.1161,24.5)(18.3927,25.)(18.6699,25.5)(18.9476,26.)(19.2258,26.5)(19.5044,27.)(19.7835,27.5)(20.0629,28.)(20.3428,28.5)(20.623,29.)(20.9035,29.5)(21.1843,30.)(21.4654,30.5)(21.7469,31.)(22.0285,31.5)(22.3105,32.)(22.5926,32.5)(22.875,33.)(23.1576,33.5)(23.4404,34.)(23.7234,34.5)(24.0066,35.)(24.29,35.5)(24.5735,36.)(24.8571,36.5)(25.141,37.)(25.425,37.5)(25.7091,38.)(25.9933,38.5)(26.2777,39.)(26.5622,39.5)(26.8468,40.)(27.1315,40.5)(27.4163,41.)(27.7013,41.5)(27.9863,42.)(28.2714,42.5)(28.5566,43.)(28.8419,43.5)(29.1273,44.)(29.4128,44.5)(29.6983,45.)(29.984,45.5)(30.2696,46.)(30.5554,46.5)(30.8412,47.)(31.1271,47.5)(31.4131,48.)(31.6991,48.5)(31.9852,49.)(32.2713,49.5)(32.5575,50.)(32.8438,50.5)(33.1301,51.)(33.4164,51.5)(33.7028,52.)(33.9893,52.5)(34.2758,53.)(34.5623,53.5)(34.8489,54.)(35.1355,54.5)(35.4221,55.)(35.7088,55.5)(35.9956,56.)(36.2823,56.5)(36.5691,57.)(36.856,57.5)(37.1429,58.)(37.4298,58.5)(37.7167,59.)(38.0037,59.5)(38.2907,60.)
		
		\psline[fillstyle=hlines*, hatchangle=60, hatchwidth=0.1, hatchsep=1, hatchcolor=gray, fillcolor=lightgray](-32.2388,-62.1607)(-31.9493,-61.6622)(-31.6598,-61.1636)(-31.3702,-60.6651)(-31.0807,-60.1667)(-30.7911,-59.6682)(-30.5015,-59.1698)(-30.2119,-58.6714)(-29.9223,-58.1731)(-29.6326,-57.6748)(-29.343,-57.1765)(-29.0533,-56.6782)(-28.7636,-56.18)(-28.4739,-55.6818)(-28.1841,-55.1837)(-27.8944,-54.6856)(-27.6046,-54.1875)(-27.3147,-53.6895)(-27.0249,-53.1915)(-26.735,-52.6935)(-26.4452,-52.1957)(-26.1552,-51.6978)(-25.8653,-51.2)(-25.5753,-50.7022)(-25.2853,-50.2045)(-24.9953,-49.7069)(-24.7052,-49.2093)(-24.4151,-48.7118)(-24.125,-48.2143)(-23.8348,-47.7169)(-23.5446,-47.2195)(-23.2544,-46.7222)(-22.9641,-46.225)(-22.6738,-45.7278)(-22.3834,-45.2308)(-22.093,-44.7338)(-21.8026,-44.2368)(-21.5121,-43.74)(-21.2215,-43.2432)(-20.9309,-42.7466)(-20.6403,-42.25)(-20.3496,-41.7535)(-20.0588,-41.2571)(-19.768,-40.7609)(-19.4771,-40.2647)(-19.1861,-39.7687)(-18.8951,-39.2727)(-18.604,-38.7769)(-18.3128,-38.2813)(-18.0216,-37.7857)(-17.7302,-37.2903)(-17.4388,-36.7951)(-17.1473,-36.3)(-16.8557,-35.8051)(-16.564,-35.3103)(-16.2722,-34.8158)(-15.9802,-34.3214)(-15.6882,-33.8273)(-15.396,-33.3333)(-15.1037,-32.8396)(-14.8113,-32.3462)(-14.5187,-31.8529)(-14.2259,-31.36)(-13.933,-30.8673)(-13.6399,-30.375)(-13.3466,-29.883)(-13.0531,-29.3913)(-12.7594,-28.9)(-12.4655,-28.4091)(-12.1713,-27.9186)(-11.8769,-27.4286)(-11.5822,-26.939)(-11.2872,-26.45)(-10.9919,-25.9615)(-10.6962,-25.4737)(-10.4001,-24.9865)(-10.1036,-24.5)(-9.80671,-24.0143)(-9.5093,-23.5294)(-9.21136,-23.0455)(-8.91284,-22.5625)(-8.61369,-22.0806)(-8.31384,-21.6)(-8.01322,-21.1207)(-7.71175,-20.6429)(-7.40933,-20.1667)(-7.10585,-19.6923)(-6.80119,-19.22)(-6.49519,-18.75)(-6.18769,-18.2826)(-5.87848,-17.8182)(-5.56731,-17.3571)(-5.25389,-16.9)(-4.93786,-16.4474)(-4.6188,-16.)(-4.29617,-15.5588)(-3.96928,-15.125)(-3.63731,-14.7)(-3.29914,-14.2857)(-2.95337,-13.8846)(-2.59808,-13.5)(-2.23067,-13.1364)(-1.84752,-12.8)(-1.44338,-12.5)(-1.01036,-12.25)(-0.536111,-12.0714)(0.,-12.)(0.536111,-12.0714)(1.01036,-12.25)(1.44338,-12.5)(1.84752,-12.8)(2.23067,-13.1364)(2.59808,-13.5)(2.95337,-13.8846)(3.29914,-14.2857)(3.63731,-14.7)(3.96928,-15.125)(4.29617,-15.5588)(4.6188,-16.)(4.93786,-16.4474)(5.25389,-16.9)(5.56731,-17.3571)(5.87848,-17.8182)(6.18769,-18.2826)(6.49519,-18.75)(6.80119,-19.22)(7.10585,-19.6923)(7.40933,-20.1667)(7.71175,-20.6429)(8.01322,-21.1207)(8.31384,-21.6)(8.61369,-22.0806)(8.91284,-22.5625)(9.21136,-23.0455)(9.5093,-23.5294)(9.80671,-24.0143)(10.1036,-24.5)(10.4001,-24.9865)(10.6962,-25.4737)(10.9919,-25.9615)(11.2872,-26.45)(11.5822,-26.939)(11.8769,-27.4286)(12.1713,-27.9186)(12.4655,-28.4091)(12.7594,-28.9)(13.0531,-29.3913)(13.3466,-29.883)(13.6399,-30.375)(13.933,-30.8673)(14.2259,-31.36)(14.5187,-31.8529)(14.8113,-32.3462)(15.1037,-32.8396)(15.396,-33.3333)(15.6882,-33.8273)(15.9802,-34.3214)(16.2722,-34.8158)(16.564,-35.3103)(16.8557,-35.8051)(17.1473,-36.3)(17.4388,-36.7951)(17.7302,-37.2903)(18.0216,-37.7857)(18.3128,-38.2813)(18.604,-38.7769)(18.8951,-39.2727)(19.1861,-39.7687)(19.4771,-40.2647)(19.768,-40.7609)(20.0588,-41.2571)(20.3496,-41.7535)(20.6403,-42.25)(20.9309,-42.7466)(21.2215,-43.2432)(21.5121,-43.74)(21.8026,-44.2368)(22.093,-44.7338)(22.3834,-45.2308)(22.6738,-45.7278)(22.9641,-46.225)(23.2544,-46.7222)(23.5446,-47.2195)(23.8348,-47.7169)(24.125,-48.2143)(24.4151,-48.7118)(24.7052,-49.2093)(24.9953,-49.7069)(25.2853,-50.2045)(25.5753,-50.7022)(25.8653,-51.2)(26.1552,-51.6978)(26.4452,-52.1957)(26.735,-52.6935)(27.0249,-53.1915)(27.3147,-53.6895)(27.6046,-54.1875)(27.8944,-54.6856)(28.1841,-55.1837)(28.4739,-55.6818)(28.7636,-56.18)(29.0533,-56.6782)(29.343,-57.1765)(29.6326,-57.6748)(29.9223,-58.1731)(30.2119,-58.6714)(30.5015,-59.1698)(30.7911,-59.6682)(31.0807,-60.1667)(31.3702,-60.6651)(31.6598,-61.1636)(31.9493,-61.6622)(32.2388,-62.1607)
		
		\psline(-100, 0)(100, 0)
		\psline([angle=60,nodesep=-60]twomp2)([angle=60,nodesep=60]twomp2)
		\psline([angle=120,nodesep=-60]twomp2)([angle=120,nodesep=60]twomp2)

		\psline[linestyle=dotted](-50, 4)(50, 4)
		\psline[linestyle=dotted](-50, 16)(50, 16)

		\psline[linestyle=dotted]([angle=60,nodesep=60]u4mp2)([angle=60,nodesep=-60]u4mp2)
		\psline[linestyle=dotted]([angle=60,nodesep=60]u16mp2)([angle=60,nodesep=-60]u16mp2)

		\psline[linestyle=dotted]([angle=120,nodesep=60]t4mp2)([angle=120,nodesep=-60]t4mp2)
		\psline[linestyle=dotted]([angle=120,nodesep=60]t16mp2)([angle=120,nodesep=-60]t16mp2)

		\psline[linestyle=solid,linewidth=1.,linecolor=lightgray,arrows=cc-cc]([angle=120,nodesep=0]tconst)([angle=120,nodesep=60]tconst)
		\psline[linestyle=solid,linewidth=1.,linecolor=lightgray,arrows=cc-cc]([angle=120,nodesep=-9.2376]tconst)([angle=120,nodesep=-60]tconst)
		\psline[linestyle=dashed]([angle=120,nodesep=60]tconst)([angle=120,nodesep=-60]tconst)

		\put(-39.5,-2){$s=0$}
		\put(-39.5,5.5){$s=4M_\pi^2$}
		\put(-39.5,17){$s=16M_\pi^2$}
		
		\rput{-60}([angle=-58,nodesep=43]twomp2){$t=0$}
		\rput{-60}([angle=-63,nodesep=39]t4mp2){$t=4M_\pi^2$}
		\rput{-60}([angle=-63,nodesep=32]t16mp2){$t=16M_\pi^2$}
		
		\rput{60}([angle=62,nodesep=38]twomp2){$u=0$}
		\rput{60}([angle=57,nodesep=41]u4mp2){$u=4M_\pi^2$}
		\rput{60}([angle=57,nodesep=41]u16mp2){$u=16M_\pi^2$}

		\put(-3.5,27){$s$-channel}
		\put(-27,-13){$t$-channel}
		\put(20,-13){$u$-channel}

		\put(-21,12){$\rho_{st}$}
		\put(19,12){$\rho_{su}$}
		\put(-1,-22){$\rho_{tu}$}

	\end{pspicture*}
	\caption{Mandelstam diagram for HLbL scattering for the case $q_i^2=0.5M_\pi^2$ with double-spectral regions for box topologies. The dashed line marks a line of fixed $t$ with its $s$- and $u$-channel cuts highlighted in grey.}
	\label{img:HLbLMandelstamDiagramBox}
\end{figure}

We consider now the contribution of box topologies. Here, Mandelstam diagrams are very useful for the discussion of double-spectral regions. In figure~\ref{img:HLbLMandelstamDiagramBox}, such a diagram is shown for the case $q_i^2 = 0.5 M_\pi^2$. A dashed line indicates a line of fixed $t$, used for writing the fixed-$t$ dispersion relation. The two cuts are highlighted in grey. The discontinuities along these cuts can be written again as a dispersive integral over double-spectral functions. The three regions of non-vanishing double-spectral functions are labelled in figure~\ref{img:HLbLMandelstamDiagramBox} by $\rho_{st}$, $\rho_{su}$ and $\rho_{tu}$.

\begin{figure}[H]
	\centering
	\begin{subfigure}[b]{0.3\textwidth}
		\centering
		\includegraphics[width=3cm]{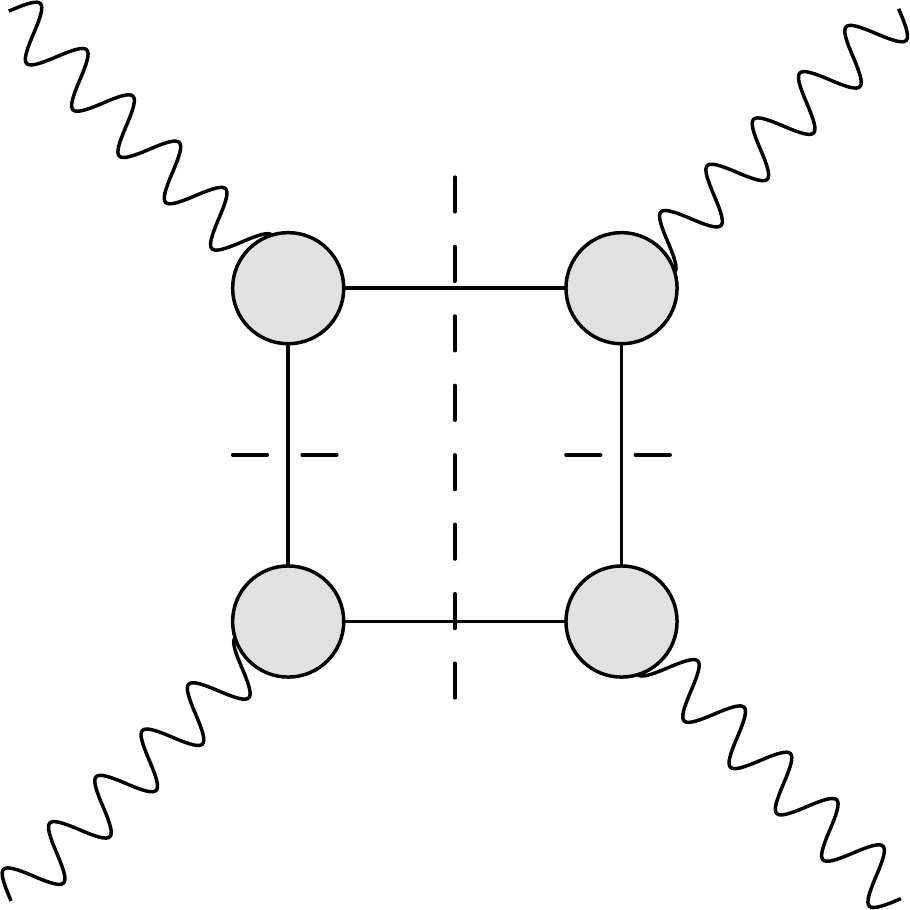}
		\caption{}
		\label{img:HLbLBoxA}
	\end{subfigure}
	\begin{subfigure}[b]{0.3\textwidth}
		\centering
		\includegraphics[width=3cm]{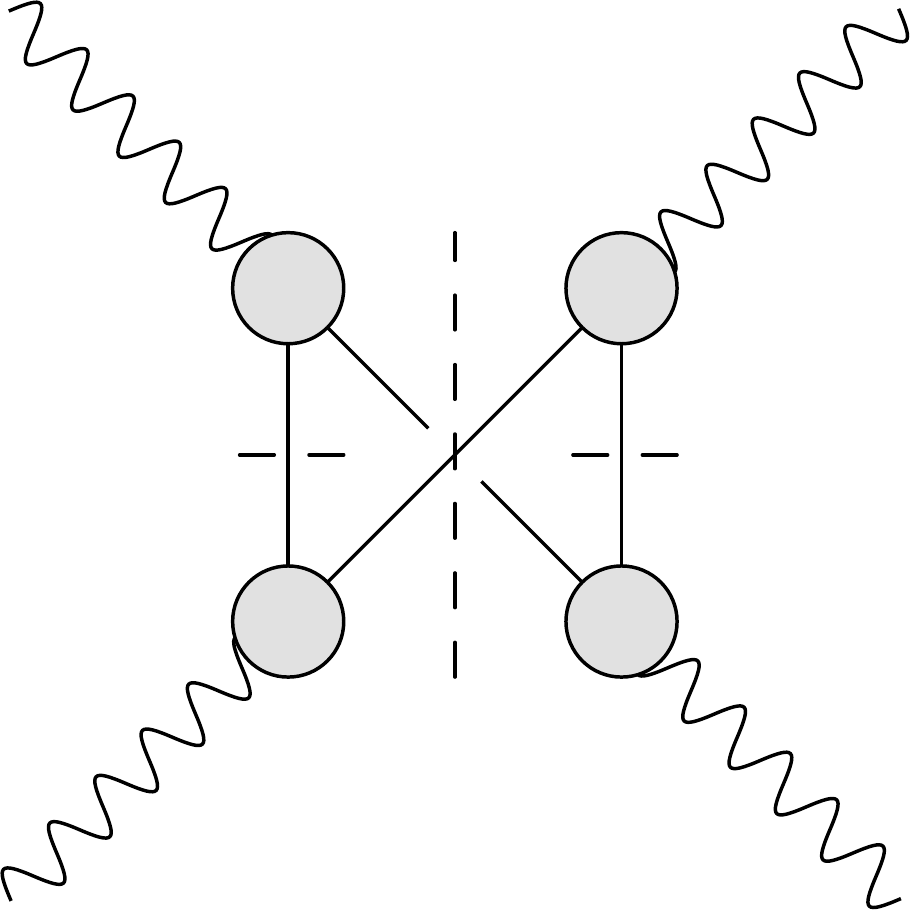}
		\caption{}
		\label{img:HLbLBoxB}
	\end{subfigure}
	\begin{subfigure}[b]{0.3\textwidth}
		\centering
		\includegraphics[width=3cm]{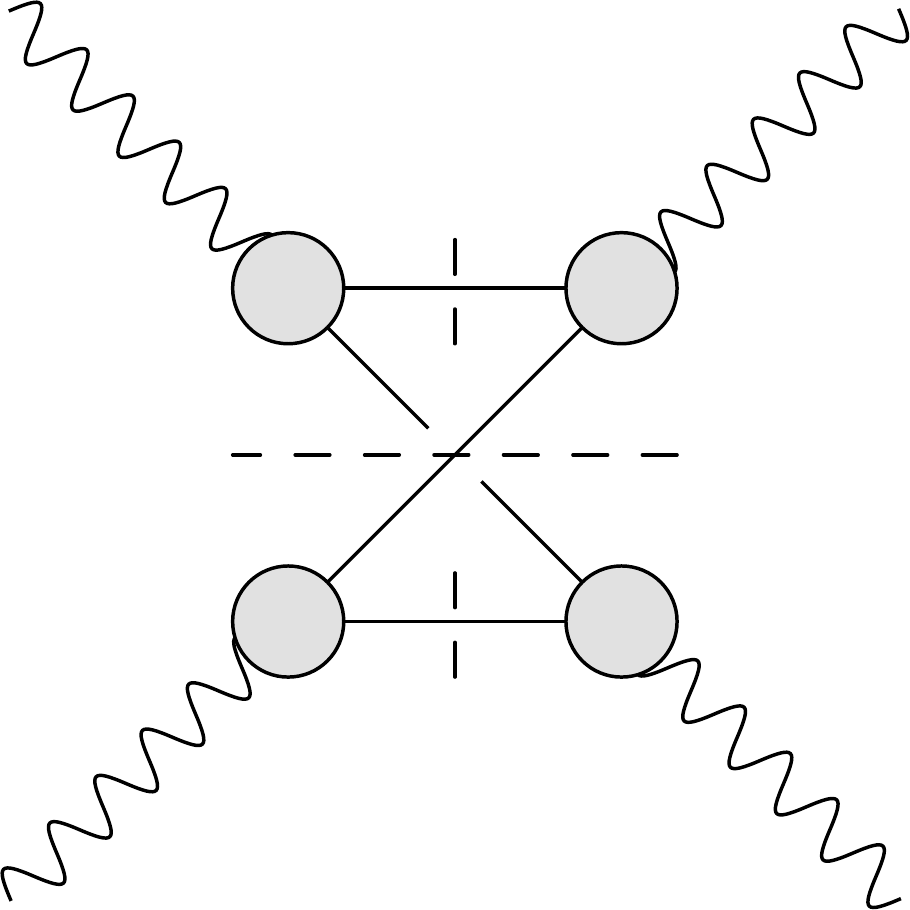}
		\caption{}
		\label{img:HLbLBoxC}
	\end{subfigure}
	\caption{Unitarity diagrams representing the box contributions.}
	\label{img:HLbLBox}
\end{figure}

The $s$-channel cut gets contributions from the double-spectral regions $\rho_{st}$ and $\rho_{su}$, according to the unitarity diagrams \ref{img:HLbLBoxA} and \ref{img:HLbLBoxB}, where first the vertical cut, then the horizontal cut is applied. (In fact, each of the shown diagrams corresponds to two topologies because the pion is charged and its line has a direction.) The $u$-channel cut gets contributions again from $\rho_{su}$ and from $\rho_{tu}$, according to the unitarity diagrams \ref{img:HLbLBoxB} and \ref{img:HLbLBoxC}. In diagram \ref{img:HLbLBoxB}, the horizontal cut is now applied first.

Hence, the fixed-$t$ dispersion relation leads to a priori four double-spectral integrals: one for each of the regions $\rho_{st}$ and $\rho_{tu}$ and two for the region $\rho_{su}$. However, it turns out that the sum of the two double-spectral integrals for the region $\rho_{su}$ equals the crossed version of one of the other double-spectral integrals. This is illustrated for the example of a simple scalar box diagram in appendix~\ref{sec:AppendixScalar4PointFunction}. Therefore, the box contributions constructed from a fixed-$t$ dispersion relation are already crossing symmetric and identical to the box contributions that would be obtained from a fixed-$s$ or fixed-$u$ dispersion relation.

As explained in the previous section, the box contribution is equal to the scalar QED loop calculation, multiplied by electromagnetic pion form factors for each off-shell photon. For this reason, we called this contribution FsQED (form factor scalar QED) in \cite{Colangelo2014a}. While the unitarity diagrams consist only of boxes, the scalar QED calculation contains Feynman diagrams of box, triangle and bulb type.

\subsection{Higher Intermediate States and $\pi\pi$-Rescattering Contribution}

\subsubsection{Derivation of a Partial-Wave Dispersion Relation}

The remaining contributions are box topologies where the pole in one or both of the sub-processes is replaced by a discontinuity. The symmetrisation procedure is identical in both cases. We discuss the Mandelstam diagram for the case of a discontinuity in both sub-processes.

\begin{figure}[H]
	\centering
	\begin{subfigure}[b]{0.24\textwidth}
		\centering
		\includegraphics[width=3cm]{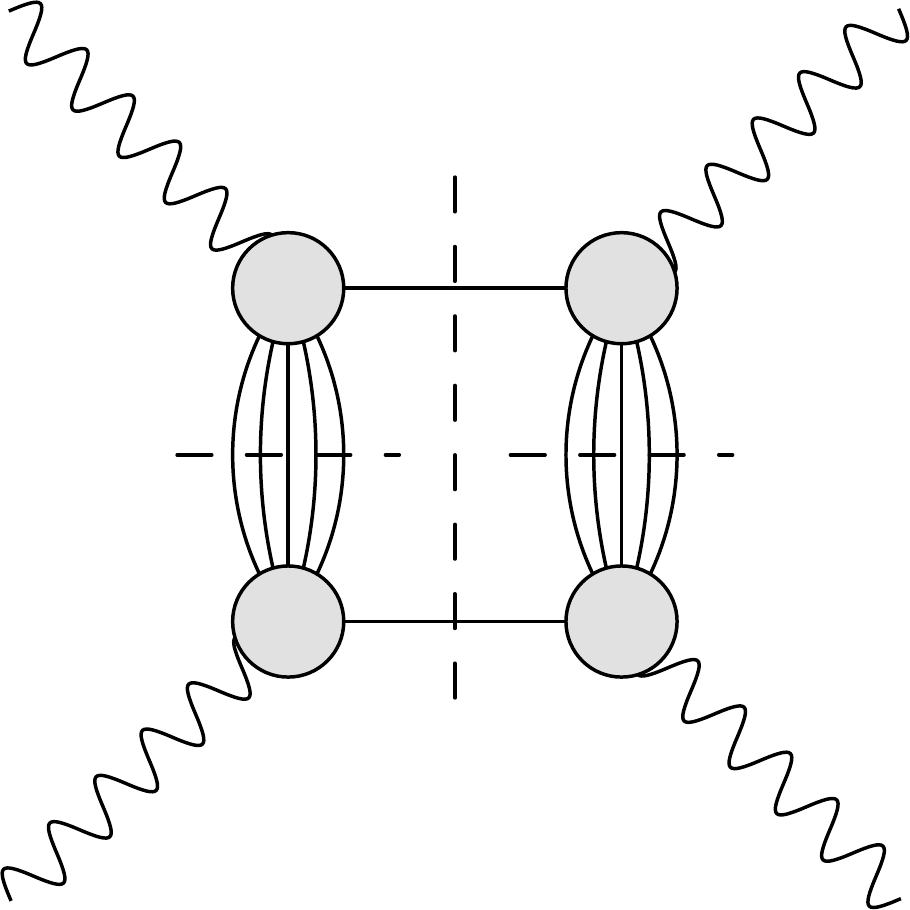}
		\caption{$\rho_{st}$}
		\label{img:HLbLBox2DiscA}
	\end{subfigure}
	\begin{subfigure}[b]{0.24\textwidth}
		\centering
		\includegraphics[width=3cm]{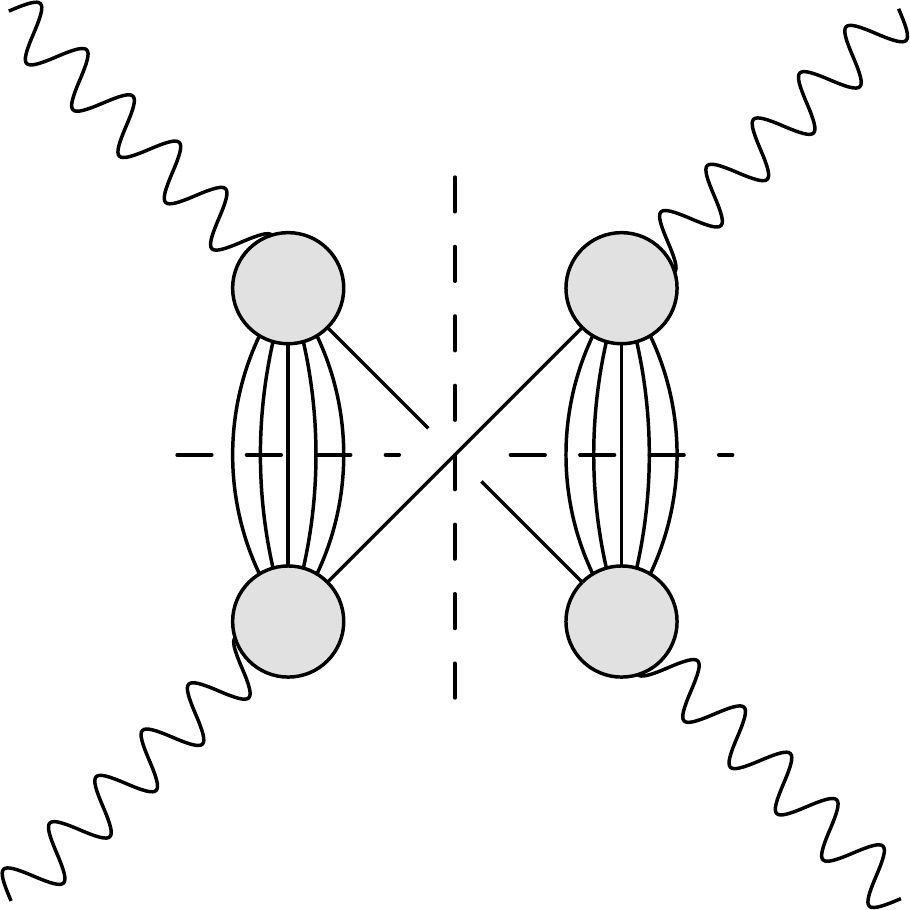}
		\caption{$\rho_{su}$}
		\label{img:HLbLBox2DiscB}
	\end{subfigure}
	\begin{subfigure}[b]{0.24\textwidth}
		\centering
		\includegraphics[width=3cm]{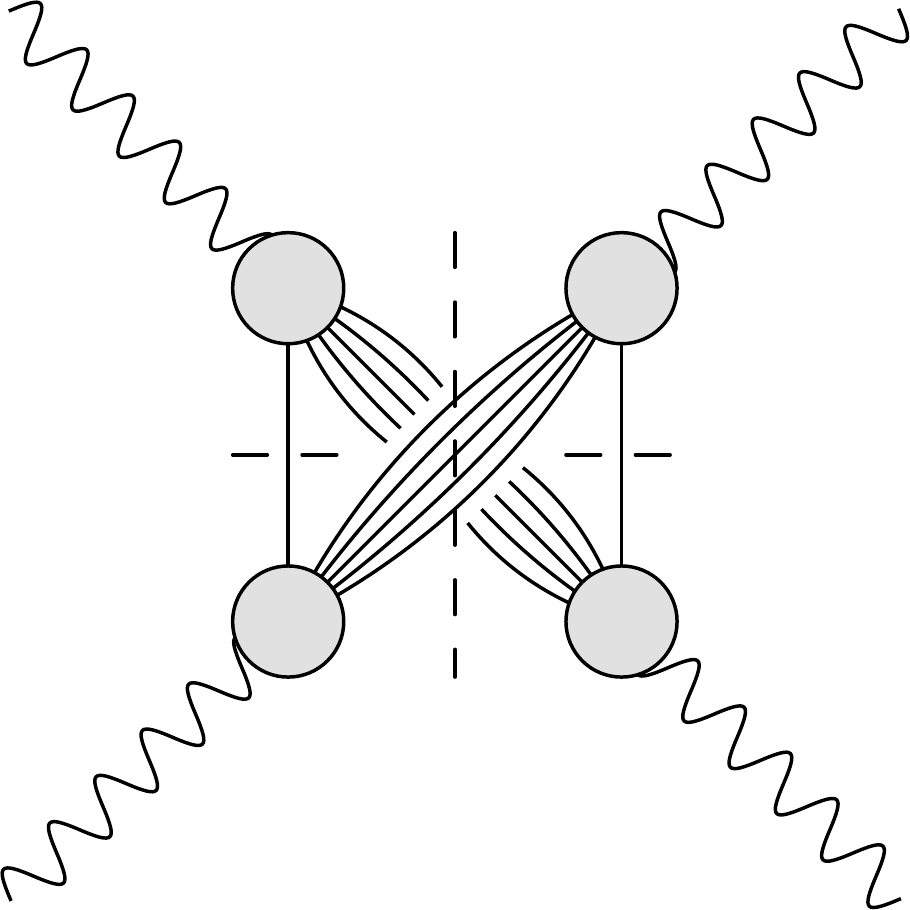}
		\caption{$\rho_{us}$}
		\label{img:HLbLBox2DiscC}
	\end{subfigure}
	\begin{subfigure}[b]{0.24\textwidth}
		\centering
		\includegraphics[width=3cm]{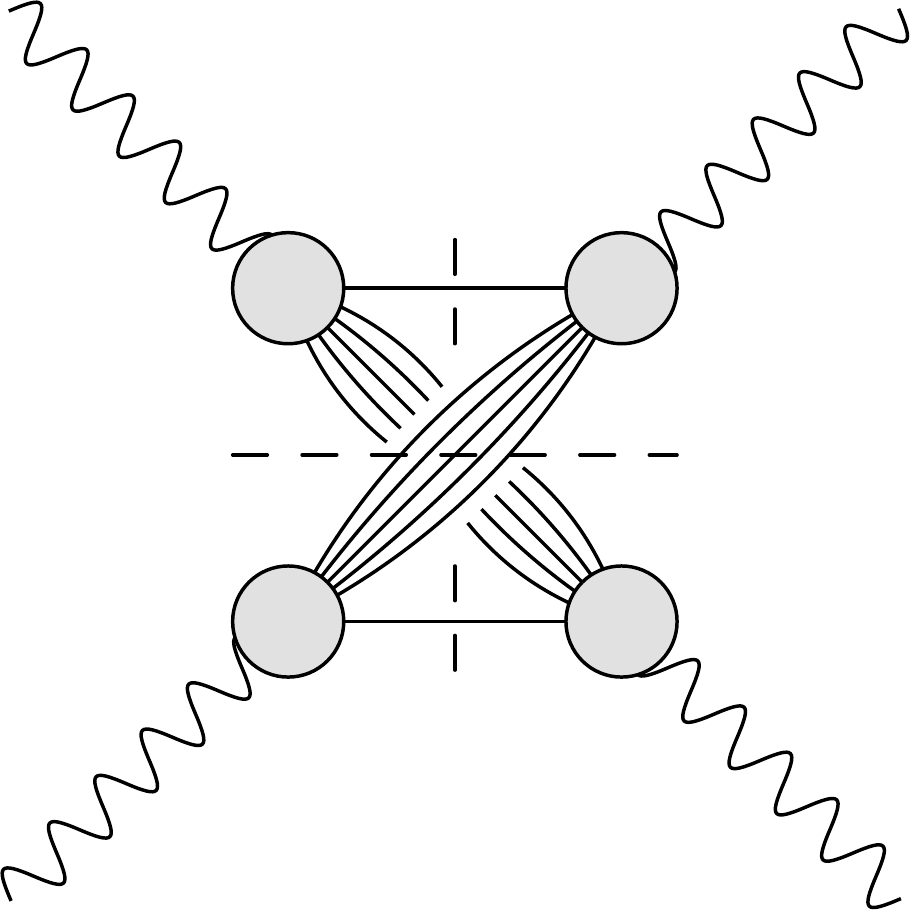}
		\caption{$\rho_{ut}$}
		\label{img:HLbLBox2DiscD}
	\end{subfigure}
	\caption{Unitarity diagrams representing the `2disc'-box contributions that are (partially) accessible through a fixed-$t$ dispersion relation.}
	\label{img:HLbLBox2Disc}
\end{figure}

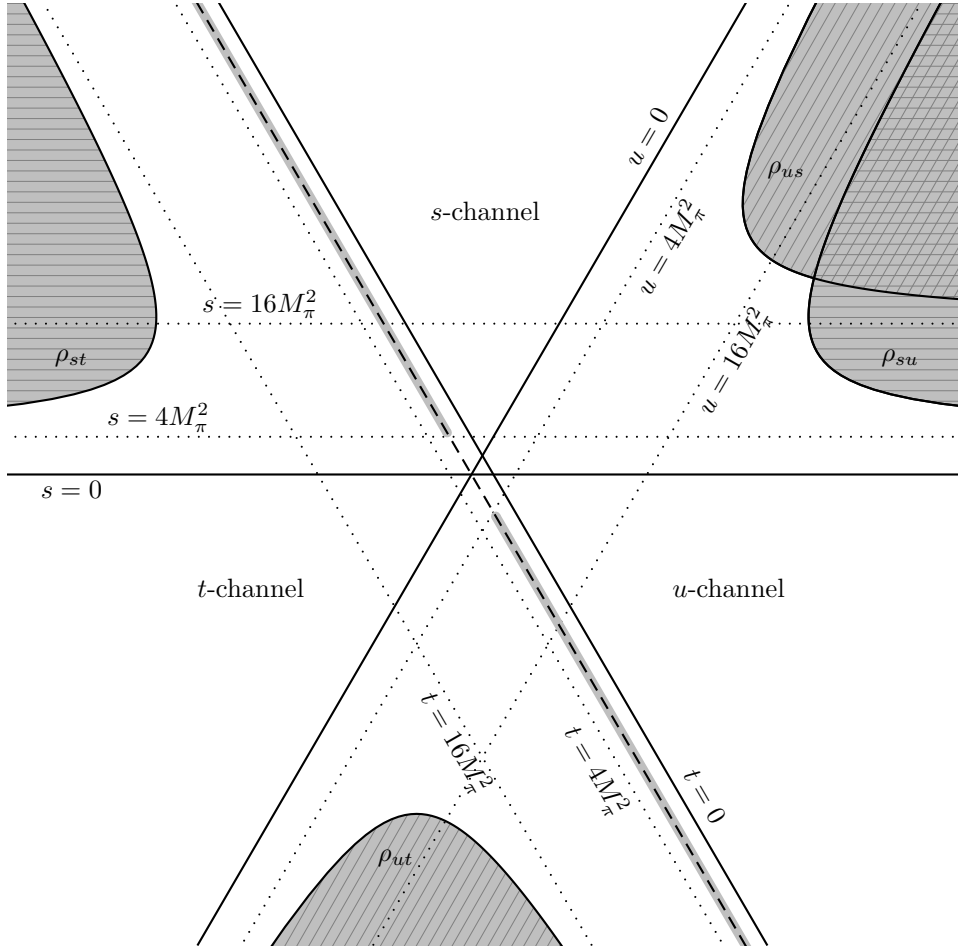
\begin{figure}[t]
	\centering
	\psset{unit=0.125cm}
	\begin{pspicture*}(-50,-50)(50,50)
		\pnode(0,0){orig}
		\pnode(0,2){twomp2}

		\pnode([angle=-30,nodesep=4]twomp2){u4mp2}
		\pnode([angle=-30,nodesep=16]twomp2){u16mp2}

		\pnode(-3.4641,4.){tconst}
		\pnode([angle=210,nodesep=4]twomp2){t4mp2}
		\pnode([angle=210,nodesep=16]twomp2){t16mp2}

		\pnode([angle=120,nodesep=4.618802153517007]t4mp2){stcorner}
		\pnode([angle=60,nodesep=4.618802153517007]u4mp2){sucorner}
		\pnode(0,-6){tucorner}

		
		\psline[fillstyle=hlines*, hatchangle=0, hatchwidth=0.1, hatchsep=1, hatchcolor=gray, fillcolor=lightgray](-67.55,6.)(-58.4856,6.5)(-52.5389,7.)(-48.3737,7.5)(-45.322,8.)(-43.0126,8.5)(-41.2228,9.)(-39.8109,9.5)(-38.6825,10.)(-37.772,10.5)(-37.0329,11.)(-36.4308,11.5)(-35.9401,12.)(-35.541,12.5)(-35.2184,13.)(-34.9601,13.5)(-34.7565,14.)(-34.5998,14.5)(-34.4836,15.)(-34.4025,15.5)(-34.3523,16.)(-34.3292,16.5)(-34.3301,17.)(-34.3523,17.5)(-34.3936,18.)(-34.4519,18.5)(-34.5255,19.)(-34.6131,19.5)(-34.7132,20.)(-34.8247,20.5)(-34.9467,21.)(-35.0782,21.5)(-35.2184,22.)(-35.3666,22.5)(-35.5222,23.)(-35.6847,23.5)(-35.8535,24.)(-36.0281,24.5)(-36.2081,25.)(-36.3932,25.5)(-36.583,26.)(-36.7772,26.5)(-36.9755,27.)(-37.1777,27.5)(-37.3834,28.)(-37.5926,28.5)(-37.8049,29.)(-38.0202,29.5)(-38.2384,30.)(-38.4592,30.5)(-38.6825,31.)(-38.9082,31.5)(-39.1361,32.)(-39.3662,32.5)(-39.5983,33.)(-39.8323,33.5)(-40.0681,34.)(-40.3057,34.5)(-40.5449,35.)(-40.7857,35.5)(-41.028,36.)(-41.2717,36.5)(-41.5167,37.)(-41.7631,37.5)(-42.0107,38.)(-42.2595,38.5)(-42.5095,39.)(-42.7605,39.5)(-43.0126,40.)(-43.2657,40.5)(-43.5197,41.)(-43.7747,41.5)(-44.0306,42.)(-44.2873,42.5)(-44.5448,43.)(-44.8031,43.5)(-45.0622,44.)(-45.322,44.5)(-45.5825,45.)(-45.8437,45.5)(-46.1055,46.)(-46.368,46.5)(-46.6311,47.)(-46.8948,47.5)(-47.159,48.)(-47.4238,48.5)(-47.6891,49.)(-47.955,49.5)(-48.2213,50.)(-48.4881,50.5)(-48.7554,51.)(-49.0231,51.5)(-49.2913,52.)(-49.5599,52.5)(-49.8289,53.)(-50.0983,53.5)(-50.368,54.)(-50.6382,54.5)(-50.9087,55.)(-51.1796,55.5)(-51.4508,56.)(-51.7223,56.5)(-51.9942,57.)(-52.2664,57.5)(-52.5389,58.)(-52.8117,58.5)(-53.0847,59.)(-53.3581,59.5)(-53.6317,60.)
		
		\psline[fillstyle=solid, hatchangle=60, hatchwidth=0.1, hatchsep=1, hatchcolor=gray, fillcolor=lightgray](60.6218,97.)(45.6107,70.)(38.3938,56.5)(34.2946,48.4)(31.7543,43.)(30.1047,39.1429)(29.0119,36.25)(28.2902,34.)(27.8283,32.2)(27.5554,30.7273)(27.4241,29.5)(27.4019,28.4615)(27.4654,27.5714)(27.5973,26.8)(27.785,26.125)(28.0185,25.5294)(28.2902,25.)(28.594,24.5263)(28.9252,24.1)(29.2799,23.7143)(29.6548,23.3636)(30.0473,23.0435)(30.4552,22.75)(30.8767,22.48)(31.3101,22.2308)(31.7543,22.)(32.2079,21.7857)(32.6701,21.5862)(33.1399,21.4)(33.6167,21.2258)(34.0998,21.0625)(34.5885,20.9091)(35.0825,20.7647)(35.5813,20.6286)(36.0844,20.5)(36.5915,20.3784)(37.1024,20.2632)(37.6166,20.1538)(38.134,20.05)(38.6543,19.9512)(39.1773,19.8571)(39.7029,19.7674)(40.2308,19.6818)(40.7609,19.6)(41.2931,19.5217)(41.8272,19.4468)(42.3631,19.375)(42.9007,19.3061)(43.4398,19.24)(43.9805,19.1765)(44.5226,19.1154)(45.066,19.0566)(45.6107,19.)(46.1565,18.9455)(46.7035,18.8929)(47.2516,18.8421)(47.8006,18.7931)(48.3506,18.7458)(48.9016,18.7)(49.4534,18.6557)(50.006,18.6129)(50.5594,18.5714)(51.1135,18.5313)(51.6684,18.4923)(52.224,18.4545)(52.7802,18.4179)(53.337,18.3824)(53.8944,18.3478)(54.4524,18.3143)(55.0109,18.2817)(55.57,18.25)(56.1295,18.2192)(56.6896,18.1892)(57.2501,18.16)(57.811,18.1316)(58.3724,18.1039)(58.9341,18.0769)(59.4963,18.0506)(60.0589,18.025)(60.6218,18.)(61.185,17.9756)(61.7487,17.9518)(62.3126,17.9286)(62.8768,17.9059)(63.4414,17.8837)(64.0062,17.8621)(64.5714,17.8409)(65.1368,17.8202)(65.7025,17.8)(66.2684,17.7802)(66.8346,17.7609)(67.401,17.7419)(67.9676,17.7234)(68.5345,17.7053)(69.1016,17.6875)(69.6689,17.6701)(70.2364,17.6531)(70.8041,17.6364)(71.372,17.62)(71.9401,17.604)(72.5084,17.5882)(73.0768,17.5728)(73.6455,17.5577)(74.2143,17.5429)(74.7832,17.5283)(75.3523,17.514)(75.9216,17.5)(76.491,17.4862)(77.0605,17.4727)(77.6302,17.4595)(78.2,17.4464)

		\psline[fillstyle=hlines*, hatchangle=0, hatchwidth=0.1, hatchsep=1, hatchcolor=gray, fillcolor=lightgray](67.55,6.)(58.4856,6.5)(52.5389,7.)(48.3737,7.5)(45.322,8.)(43.0126,8.5)(41.2228,9.)(39.8109,9.5)(38.6825,10.)(37.772,10.5)(37.0329,11.)(36.4308,11.5)(35.9401,12.)(35.541,12.5)(35.2184,13.)(34.9601,13.5)(34.7565,14.)(34.5998,14.5)(34.4836,15.)(34.4025,15.5)(34.3523,16.)(34.3292,16.5)(34.3301,17.)(34.3523,17.5)(34.3936,18.)(34.4519,18.5)(34.5255,19.)(34.6131,19.5)(34.7132,20.)(34.8247,20.5)(34.9467,21.)(35.0782,21.5)(35.2184,22.)(35.3666,22.5)(35.5222,23.)(35.6847,23.5)(35.8535,24.)(36.0281,24.5)(36.2081,25.)(36.3932,25.5)(36.583,26.)(36.7772,26.5)(36.9755,27.)(37.1777,27.5)(37.3834,28.)(37.5926,28.5)(37.8049,29.)(38.0202,29.5)(38.2384,30.)(38.4592,30.5)(38.6825,31.)(38.9082,31.5)(39.1361,32.)(39.3662,32.5)(39.5983,33.)(39.8323,33.5)(40.0681,34.)(40.3057,34.5)(40.5449,35.)(40.7857,35.5)(41.028,36.)(41.2717,36.5)(41.5167,37.)(41.7631,37.5)(42.0107,38.)(42.2595,38.5)(42.5095,39.)(42.7605,39.5)(43.0126,40.)(43.2657,40.5)(43.5197,41.)(43.7747,41.5)(44.0306,42.)(44.2873,42.5)(44.5448,43.)(44.8031,43.5)(45.0622,44.)(45.322,44.5)(45.5825,45.)(45.8437,45.5)(46.1055,46.)(46.368,46.5)(46.6311,47.)(46.8948,47.5)(47.159,48.)(47.4238,48.5)(47.6891,49.)(47.955,49.5)(48.2213,50.)(48.4881,50.5)(48.7554,51.)(49.0231,51.5)(49.2913,52.)(49.5599,52.5)(49.8289,53.)(50.0983,53.5)(50.368,54.)(50.6382,54.5)(50.9087,55.)(51.1796,55.5)(51.4508,56.)(51.7223,56.5)(51.9942,57.)(52.2664,57.5)(52.5389,58.)(52.8117,58.5)(53.0847,59.)(53.3581,59.5)(53.6317,60.)
		
		\psline[fillstyle=hlines, hatchangle=60, hatchwidth=0.1, hatchsep=1, hatchcolor=gray](60.6218,97.)(45.6107,70.)(38.3938,56.5)(34.2946,48.4)(31.7543,43.)(30.1047,39.1429)(29.0119,36.25)(28.2902,34.)(27.8283,32.2)(27.5554,30.7273)(27.4241,29.5)(27.4019,28.4615)(27.4654,27.5714)(27.5973,26.8)(27.785,26.125)(28.0185,25.5294)(28.2902,25.)(28.594,24.5263)(28.9252,24.1)(29.2799,23.7143)(29.6548,23.3636)(30.0473,23.0435)(30.4552,22.75)(30.8767,22.48)(31.3101,22.2308)(31.7543,22.)(32.2079,21.7857)(32.6701,21.5862)(33.1399,21.4)(33.6167,21.2258)(34.0998,21.0625)(34.5885,20.9091)(35.0825,20.7647)(35.5813,20.6286)(36.0844,20.5)(36.5915,20.3784)(37.1024,20.2632)(37.6166,20.1538)(38.134,20.05)(38.6543,19.9512)(39.1773,19.8571)(39.7029,19.7674)(40.2308,19.6818)(40.7609,19.6)(41.2931,19.5217)(41.8272,19.4468)(42.3631,19.375)(42.9007,19.3061)(43.4398,19.24)(43.9805,19.1765)(44.5226,19.1154)(45.066,19.0566)(45.6107,19.)(46.1565,18.9455)(46.7035,18.8929)(47.2516,18.8421)(47.8006,18.7931)(48.3506,18.7458)(48.9016,18.7)(49.4534,18.6557)(50.006,18.6129)(50.5594,18.5714)(51.1135,18.5313)(51.6684,18.4923)(52.224,18.4545)(52.7802,18.4179)(53.337,18.3824)(53.8944,18.3478)(54.4524,18.3143)(55.0109,18.2817)(55.57,18.25)(56.1295,18.2192)(56.6896,18.1892)(57.2501,18.16)(57.811,18.1316)(58.3724,18.1039)(58.9341,18.0769)(59.4963,18.0506)(60.0589,18.025)(60.6218,18.)(61.185,17.9756)(61.7487,17.9518)(62.3126,17.9286)(62.8768,17.9059)(63.4414,17.8837)(64.0062,17.8621)(64.5714,17.8409)(65.1368,17.8202)(65.7025,17.8)(66.2684,17.7802)(66.8346,17.7609)(67.401,17.7419)(67.9676,17.7234)(68.5345,17.7053)(69.1016,17.6875)(69.6689,17.6701)(70.2364,17.6531)(70.8041,17.6364)(71.372,17.62)(71.9401,17.604)(72.5084,17.5882)(73.0768,17.5728)(73.6455,17.5577)(74.2143,17.5429)(74.7832,17.5283)(75.3523,17.514)(75.9216,17.5)(76.491,17.4862)(77.0605,17.4727)(77.6302,17.4595)(78.2,17.4464)
		
		\psline(67.55,6.)(58.4856,6.5)(52.5389,7.)(48.3737,7.5)(45.322,8.)(43.0126,8.5)(41.2228,9.)(39.8109,9.5)(38.6825,10.)(37.772,10.5)(37.0329,11.)(36.4308,11.5)(35.9401,12.)(35.541,12.5)(35.2184,13.)(34.9601,13.5)(34.7565,14.)(34.5998,14.5)(34.4836,15.)(34.4025,15.5)(34.3523,16.)(34.3292,16.5)(34.3301,17.)(34.3523,17.5)(34.3936,18.)(34.4519,18.5)(34.5255,19.)(34.6131,19.5)(34.7132,20.)(34.8247,20.5)(34.9467,21.)(35.0782,21.5)(35.2184,22.)(35.3666,22.5)(35.5222,23.)(35.6847,23.5)(35.8535,24.)(36.0281,24.5)(36.2081,25.)(36.3932,25.5)(36.583,26.)(36.7772,26.5)(36.9755,27.)(37.1777,27.5)(37.3834,28.)(37.5926,28.5)(37.8049,29.)(38.0202,29.5)(38.2384,30.)(38.4592,30.5)(38.6825,31.)(38.9082,31.5)(39.1361,32.)(39.3662,32.5)(39.5983,33.)(39.8323,33.5)(40.0681,34.)(40.3057,34.5)(40.5449,35.)(40.7857,35.5)(41.028,36.)(41.2717,36.5)(41.5167,37.)(41.7631,37.5)(42.0107,38.)(42.2595,38.5)(42.5095,39.)(42.7605,39.5)(43.0126,40.)(43.2657,40.5)(43.5197,41.)(43.7747,41.5)(44.0306,42.)(44.2873,42.5)(44.5448,43.)(44.8031,43.5)(45.0622,44.)(45.322,44.5)(45.5825,45.)(45.8437,45.5)(46.1055,46.)(46.368,46.5)(46.6311,47.)(46.8948,47.5)(47.159,48.)(47.4238,48.5)(47.6891,49.)(47.955,49.5)(48.2213,50.)(48.4881,50.5)(48.7554,51.)(49.0231,51.5)(49.2913,52.)(49.5599,52.5)(49.8289,53.)(50.0983,53.5)(50.368,54.)(50.6382,54.5)(50.9087,55.)(51.1796,55.5)(51.4508,56.)(51.7223,56.5)(51.9942,57.)(52.2664,57.5)(52.5389,58.)(52.8117,58.5)(53.0847,59.)(53.3581,59.5)(53.6317,60.)
		
		\psline[fillstyle=hlines*, hatchangle=60, hatchwidth=0.1, hatchsep=1, hatchcolor=gray, fillcolor=lightgray](-29.1562,-60.5)(-24.191,-52.9)(-20.7846,-48.)(-18.269,-44.6429)(-16.3101,-42.25)(-14.7224,-40.5)(-13.3945,-39.2)(-12.2556,-38.2273)(-11.2583,-37.5)(-10.3701,-36.9615)(-9.56752,-36.5714)(-8.83346,-36.3)(-8.15507,-36.125)(-7.52253,-36.0294)(-6.9282,-36.)(-6.36605,-36.0263)(-5.83124,-36.1)(-5.31987,-36.2143)(-4.82875,-36.3636)(-4.35523,-36.5435)(-3.89711,-36.75)(-3.45255,-36.98)(-3.01999,-37.2308)(-2.59808,-37.5)(-2.18568,-37.7857)(-1.78182,-38.0862)(-1.38564,-38.4)(-0.996395,-38.7258)(-0.613435,-39.0625)(-0.236189,-39.4091)(0.135847,-39.7647)(0.50312,-40.1286)(0.866025,-40.5)(1.22492,-40.8784)(1.58012,-41.2632)(1.9319,-41.6538)(2.28053,-42.05)(2.62624,-42.4512)(2.96923,-42.8571)(3.30969,-43.2674)(3.6478,-43.6818)(3.98372,-44.1)(4.31758,-44.5217)(4.64951,-44.9468)(4.97965,-45.375)(5.30809,-45.8061)(5.63494,-46.24)(5.96029,-46.6765)(6.28424,-47.1154)(6.60685,-47.5566)(6.9282,-48.)(7.24837,-48.4455)(7.56741,-48.8929)(7.88539,-49.3421)(8.20236,-49.7931)(8.51836,-50.2458)(8.83346,-50.7)(9.14769,-51.1557)(9.46109,-51.6129)(9.77372,-52.0714)(10.0856,-52.5313)(10.3967,-52.9923)(10.7072,-53.4545)(11.017,-53.9179)(11.3263,-54.3824)(11.6349,-54.8478)(11.9429,-55.3143)(12.2504,-55.7817)(12.5574,-56.25)(12.8638,-56.7192)(13.1698,-57.1892)(13.4754,-57.66)(13.7804,-58.1316)(14.0851,-58.6039)(14.3893,-59.0769)(14.6932,-59.5506)(14.9967,-60.025)(15.2998,-60.5)(15.6025,-60.9756)(15.905,-61.4518)(16.207,-61.9286)(16.5088,-62.4059)(16.8103,-62.8837)(17.1115,-63.3621)(17.4124,-63.8409)(17.713,-64.3202)(18.0133,-64.8)(18.3134,-65.2802)(18.6133,-65.7609)(18.9129,-66.2419)(19.2123,-66.7234)(19.5114,-67.2053)(19.8103,-67.6875)(20.1091,-68.1701)(20.4076,-68.6531)(20.7059,-69.1364)(21.004,-69.62)(21.3019,-70.104)(21.5997,-70.5882)(21.8973,-71.0728)(22.1947,-71.5577)(22.4919,-72.0429)(22.789,-72.5283)(23.0859,-73.014)(23.3827,-73.5)(23.6793,-73.9862)(23.9758,-74.4727)(24.2721,-74.9595)(24.5683,-75.4464)
		
		\psline(-100, 0)(100, 0)
		\psline([angle=60,nodesep=-60]twomp2)([angle=60,nodesep=60]twomp2)
		\psline([angle=120,nodesep=-60]twomp2)([angle=120,nodesep=60]twomp2)

		\psline[linestyle=dotted](-60, 4)(60, 4)
		\psline[linestyle=dotted](-60, 16)(60, 16)

		\psline[linestyle=dotted]([angle=60,nodesep=80]u4mp2)([angle=60,nodesep=-60]u4mp2)
		\psline[linestyle=dotted]([angle=60,nodesep=80]u16mp2)([angle=60,nodesep=-60]u16mp2)

		\psline[linestyle=dotted]([angle=120,nodesep=80]t4mp2)([angle=120,nodesep=-60]t4mp2)
		\psline[linestyle=dotted]([angle=120,nodesep=80]t16mp2)([angle=120,nodesep=-60]t16mp2)

		\psline[linestyle=solid,linewidth=1.,linecolor=lightgray,arrows=cc-cc]([angle=120,nodesep=0]tconst)([angle=120,nodesep=60]tconst)
		\psline[linestyle=solid,linewidth=1.,linecolor=lightgray,arrows=cc-cc]([angle=120,nodesep=-9.2376]tconst)([angle=120,nodesep=-80]tconst)
		\psline[linestyle=dashed]([angle=120,nodesep=60]tconst)([angle=120,nodesep=-80]tconst)

		\put(-46.5,-2.5){$s=0$}
		\put(-39.5,5.25){$s=4M_\pi^2$}
		\put(-29.5,17.25){$s=16M_\pi^2$}
		
		\rput{-60}([angle=-57,nodesep=43]twomp2){$t=0$}
		\rput{-60}([angle=-64,nodesep=35]t4mp2){$t=4M_\pi^2$}
		\rput{-60}([angle=-64,nodesep=25]t16mp2){$t=16M_\pi^2$}
		
		\rput{60}([angle=63,nodesep=38]twomp2){$u=0$}
		\rput{60}([angle=56,nodesep=29]u4mp2){$u=4M_\pi^2$}
		\rput{60}([angle=55,nodesep=22]u16mp2){$u=16M_\pi^2$}

		\put(-5.5,27){$s$-channel}
		\put(-30,-13){$t$-channel}
		\put(20,-13){$u$-channel}

		\put(-45,12){$\rho_{st}$}
		\put(42,12){$\rho_{su}$}
		\put(30,32){$\rho_{us}$}
		\put(-11,-41){$\rho_{ut}$}

	\end{pspicture*}
	\caption{Mandelstam diagram for HLbL scattering for the case $q_i^2=0.5M_\pi^2$. Only those double-spectral regions for `2disc'-box topologies are shown that are reconstructed from the fixed-$t$ dispersion relation.}
	\label{img:HLbLMandelstamDiagramBoxDiscFixedT}
\end{figure}

Figure~\ref{img:HLbLMandelstamDiagramBoxDiscFixedT} shows the Mandelstam diagram with the double-spectral regions that are generated if we start in our derivation from the fixed-$t$ dispersion relation. Figure~\ref{img:HLbLBox2Disc} shows the corresponding unitarity diagrams: the diagrams~\ref{img:HLbLBox2DiscA} and \ref{img:HLbLBox2DiscB} generate a cut for $s>4M_\pi^2$, while the diagrams~\ref{img:HLbLBox2DiscC} and \ref{img:HLbLBox2DiscD} are responsible for the left-hand cut for $u>4M_\pi^2$ (note that the first cut is always the one through the pion poles).

In contrast to the case of pure box topologies, where the fixed-$t$ dispersion relation led to a symmetric and complete double-spectral representation, not all the double-spectral contributions are generated in the present case. The reason is of course that we neglect in the primary cut higher intermediate states than two pions. We note that the contributions from $\rho_{st}$ and $\rho_{ut}$ are complete but that the contributions from $\rho_{us}$ and $\rho_{su}$ are not, because only one double-spectral integral for each of these contributions is obtained. However, we learnt in the case of box topologies that in the fixed-$t$ representation, two double-spectral integrals are needed to generate the full contribution of these regions. One of the two integrals starts now at the higher threshold $16M_\pi^2$ and is neglected in the fixed-$t$ representation. Furthermore, two more double-spectral regions $\rho_{ts}$ and $\rho_{tu}$ are completely missing in the fixed-$t$ representation.

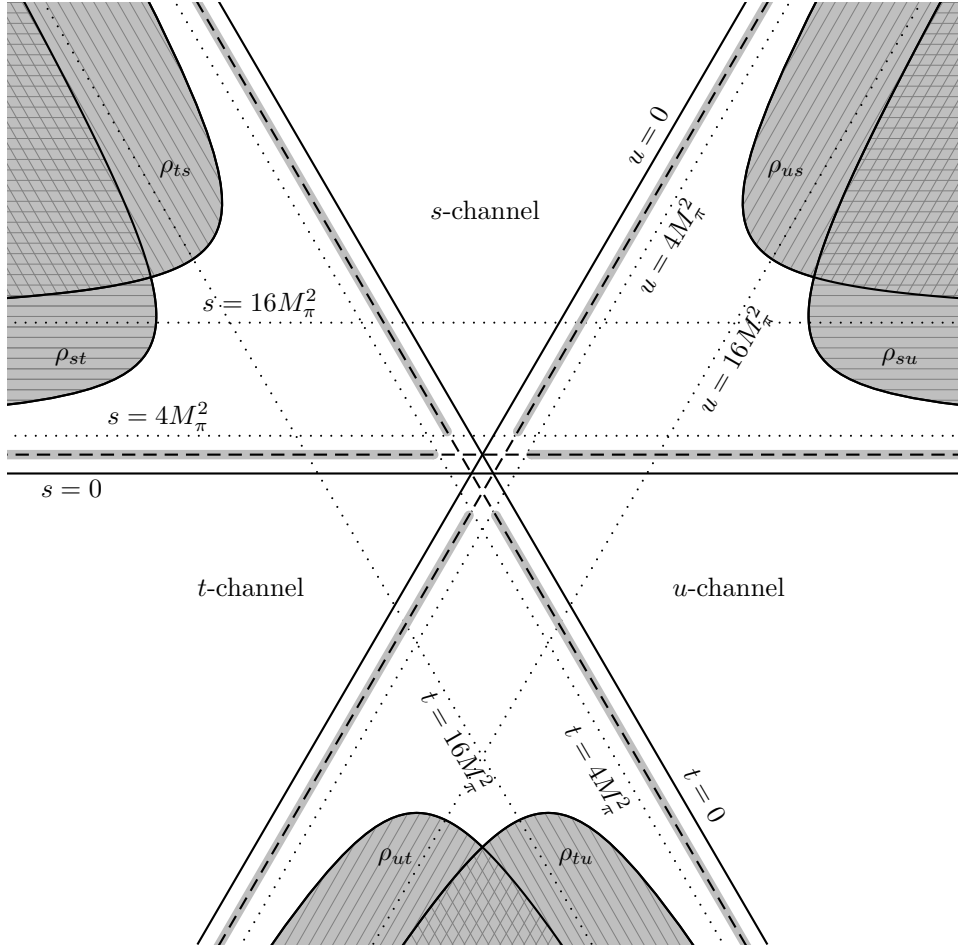
\begin{figure}[H]
	\centering
	\psset{unit=0.125cm}
	\begin{pspicture*}(-50,-50)(50,50)
		\pnode(0,0){orig}
		\pnode(0,2){twomp2}

		\pnode(-4.6188,2.){sconst}
		\pnode(-3.4641,4.){tconst}
		\pnode(3.4641,4.){uconst}

		\pnode([angle=-30,nodesep=4]twomp2){u4mp2}
		\pnode([angle=-30,nodesep=16]twomp2){u16mp2}

		\pnode([angle=210,nodesep=4]twomp2){t4mp2}
		\pnode([angle=210,nodesep=16]twomp2){t16mp2}

		\pnode([angle=120,nodesep=4.618802153517007]t4mp2){stcorner}
		\pnode([angle=60,nodesep=4.618802153517007]u4mp2){sucorner}
		\pnode(0,-6){tucorner}

		
		\psline[fillstyle=hlines*, hatchangle=0, hatchwidth=0.1, hatchsep=1, hatchcolor=gray, fillcolor=lightgray](-67.55,6.)(-58.4856,6.5)(-52.5389,7.)(-48.3737,7.5)(-45.322,8.)(-43.0126,8.5)(-41.2228,9.)(-39.8109,9.5)(-38.6825,10.)(-37.772,10.5)(-37.0329,11.)(-36.4308,11.5)(-35.9401,12.)(-35.541,12.5)(-35.2184,13.)(-34.9601,13.5)(-34.7565,14.)(-34.5998,14.5)(-34.4836,15.)(-34.4025,15.5)(-34.3523,16.)(-34.3292,16.5)(-34.3301,17.)(-34.3523,17.5)(-34.3936,18.)(-34.4519,18.5)(-34.5255,19.)(-34.6131,19.5)(-34.7132,20.)(-34.8247,20.5)(-34.9467,21.)(-35.0782,21.5)(-35.2184,22.)(-35.3666,22.5)(-35.5222,23.)(-35.6847,23.5)(-35.8535,24.)(-36.0281,24.5)(-36.2081,25.)(-36.3932,25.5)(-36.583,26.)(-36.7772,26.5)(-36.9755,27.)(-37.1777,27.5)(-37.3834,28.)(-37.5926,28.5)(-37.8049,29.)(-38.0202,29.5)(-38.2384,30.)(-38.4592,30.5)(-38.6825,31.)(-38.9082,31.5)(-39.1361,32.)(-39.3662,32.5)(-39.5983,33.)(-39.8323,33.5)(-40.0681,34.)(-40.3057,34.5)(-40.5449,35.)(-40.7857,35.5)(-41.028,36.)(-41.2717,36.5)(-41.5167,37.)(-41.7631,37.5)(-42.0107,38.)(-42.2595,38.5)(-42.5095,39.)(-42.7605,39.5)(-43.0126,40.)(-43.2657,40.5)(-43.5197,41.)(-43.7747,41.5)(-44.0306,42.)(-44.2873,42.5)(-44.5448,43.)(-44.8031,43.5)(-45.0622,44.)(-45.322,44.5)(-45.5825,45.)(-45.8437,45.5)(-46.1055,46.)(-46.368,46.5)(-46.6311,47.)(-46.8948,47.5)(-47.159,48.)(-47.4238,48.5)(-47.6891,49.)(-47.955,49.5)(-48.2213,50.)(-48.4881,50.5)(-48.7554,51.)(-49.0231,51.5)(-49.2913,52.)(-49.5599,52.5)(-49.8289,53.)(-50.0983,53.5)(-50.368,54.)(-50.6382,54.5)(-50.9087,55.)(-51.1796,55.5)(-51.4508,56.)(-51.7223,56.5)(-51.9942,57.)(-52.2664,57.5)(-52.5389,58.)(-52.8117,58.5)(-53.0847,59.)(-53.3581,59.5)(-53.6317,60.)
		
		\psline[fillstyle=hlines*, hatchangle=-60, hatchwidth=0.1, hatchsep=1, hatchcolor=gray, fillcolor=lightgray](-60.6218,97.)(-45.6107,70.)(-38.3938,56.5)(-34.2946,48.4)(-31.7543,43.)(-30.1047,39.1429)(-29.0119,36.25)(-28.2902,34.)(-27.8283,32.2)(-27.5554,30.7273)(-27.4241,29.5)(-27.4019,28.4615)(-27.4654,27.5714)(-27.5973,26.8)(-27.785,26.125)(-28.0185,25.5294)(-28.2902,25.)(-28.594,24.5263)(-28.9252,24.1)(-29.2799,23.7143)(-29.6548,23.3636)(-30.0473,23.0435)(-30.4552,22.75)(-30.8767,22.48)(-31.3101,22.2308)(-31.7543,22.)(-32.2079,21.7857)(-32.6701,21.5862)(-33.1399,21.4)(-33.6167,21.2258)(-34.0998,21.0625)(-34.5885,20.9091)(-35.0825,20.7647)(-35.5813,20.6286)(-36.0844,20.5)(-36.5915,20.3784)(-37.1024,20.2632)(-37.6166,20.1538)(-38.134,20.05)(-38.6543,19.9512)(-39.1773,19.8571)(-39.7029,19.7674)(-40.2308,19.6818)(-40.7609,19.6)(-41.2931,19.5217)(-41.8272,19.4468)(-42.3631,19.375)(-42.9007,19.3061)(-43.4398,19.24)(-43.9805,19.1765)(-44.5226,19.1154)(-45.066,19.0566)(-45.6107,19.)(-46.1565,18.9455)(-46.7035,18.8929)(-47.2516,18.8421)(-47.8006,18.7931)(-48.3506,18.7458)(-48.9016,18.7)(-49.4534,18.6557)(-50.006,18.6129)(-50.5594,18.5714)(-51.1135,18.5313)(-51.6684,18.4923)(-52.224,18.4545)(-52.7802,18.4179)(-53.337,18.3824)(-53.8944,18.3478)(-54.4524,18.3143)(-55.0109,18.2817)(-55.57,18.25)(-56.1295,18.2192)(-56.6896,18.1892)(-57.2501,18.16)(-57.811,18.1316)(-58.3724,18.1039)(-58.9341,18.0769)(-59.4963,18.0506)(-60.0589,18.025)(-60.6218,18.)(-61.185,17.9756)(-61.7487,17.9518)(-62.3126,17.9286)(-62.8768,17.9059)(-63.4414,17.8837)(-64.0062,17.8621)(-64.5714,17.8409)(-65.1368,17.8202)(-65.7025,17.8)(-66.2684,17.7802)(-66.8346,17.7609)(-67.401,17.7419)(-67.9676,17.7234)(-68.5345,17.7053)(-69.1016,17.6875)(-69.6689,17.6701)(-70.2364,17.6531)(-70.8041,17.6364)(-71.372,17.62)(-71.9401,17.604)(-72.5084,17.5882)(-73.0768,17.5728)(-73.6455,17.5577)(-74.2143,17.5429)(-74.7832,17.5283)(-75.3523,17.514)(-75.9216,17.5)(-76.491,17.4862)(-77.0605,17.4727)(-77.6302,17.4595)(-78.2,17.4464)

		\psline[fillstyle=hlines, hatchangle=0, hatchwidth=0.1, hatchsep=1, hatchcolor=gray, fillcolor=lightgray](-67.55,6.)(-58.4856,6.5)(-52.5389,7.)(-48.3737,7.5)(-45.322,8.)(-43.0126,8.5)(-41.2228,9.)(-39.8109,9.5)(-38.6825,10.)(-37.772,10.5)(-37.0329,11.)(-36.4308,11.5)(-35.9401,12.)(-35.541,12.5)(-35.2184,13.)(-34.9601,13.5)(-34.7565,14.)(-34.5998,14.5)(-34.4836,15.)(-34.4025,15.5)(-34.3523,16.)(-34.3292,16.5)(-34.3301,17.)(-34.3523,17.5)(-34.3936,18.)(-34.4519,18.5)(-34.5255,19.)(-34.6131,19.5)(-34.7132,20.)(-34.8247,20.5)(-34.9467,21.)(-35.0782,21.5)(-35.2184,22.)(-35.3666,22.5)(-35.5222,23.)(-35.6847,23.5)(-35.8535,24.)(-36.0281,24.5)(-36.2081,25.)(-36.3932,25.5)(-36.583,26.)(-36.7772,26.5)(-36.9755,27.)(-37.1777,27.5)(-37.3834,28.)(-37.5926,28.5)(-37.8049,29.)(-38.0202,29.5)(-38.2384,30.)(-38.4592,30.5)(-38.6825,31.)(-38.9082,31.5)(-39.1361,32.)(-39.3662,32.5)(-39.5983,33.)(-39.8323,33.5)(-40.0681,34.)(-40.3057,34.5)(-40.5449,35.)(-40.7857,35.5)(-41.028,36.)(-41.2717,36.5)(-41.5167,37.)(-41.7631,37.5)(-42.0107,38.)(-42.2595,38.5)(-42.5095,39.)(-42.7605,39.5)(-43.0126,40.)(-43.2657,40.5)(-43.5197,41.)(-43.7747,41.5)(-44.0306,42.)(-44.2873,42.5)(-44.5448,43.)(-44.8031,43.5)(-45.0622,44.)(-45.322,44.5)(-45.5825,45.)(-45.8437,45.5)(-46.1055,46.)(-46.368,46.5)(-46.6311,47.)(-46.8948,47.5)(-47.159,48.)(-47.4238,48.5)(-47.6891,49.)(-47.955,49.5)(-48.2213,50.)(-48.4881,50.5)(-48.7554,51.)(-49.0231,51.5)(-49.2913,52.)(-49.5599,52.5)(-49.8289,53.)(-50.0983,53.5)(-50.368,54.)(-50.6382,54.5)(-50.9087,55.)(-51.1796,55.5)(-51.4508,56.)(-51.7223,56.5)(-51.9942,57.)(-52.2664,57.5)(-52.5389,58.)(-52.8117,58.5)(-53.0847,59.)(-53.3581,59.5)(-53.6317,60.)
		
		\psline(-60.6218,97.)(-45.6107,70.)(-38.3938,56.5)(-34.2946,48.4)(-31.7543,43.)(-30.1047,39.1429)(-29.0119,36.25)(-28.2902,34.)(-27.8283,32.2)(-27.5554,30.7273)(-27.4241,29.5)(-27.4019,28.4615)(-27.4654,27.5714)(-27.5973,26.8)(-27.785,26.125)(-28.0185,25.5294)(-28.2902,25.)(-28.594,24.5263)(-28.9252,24.1)(-29.2799,23.7143)(-29.6548,23.3636)(-30.0473,23.0435)(-30.4552,22.75)(-30.8767,22.48)(-31.3101,22.2308)(-31.7543,22.)(-32.2079,21.7857)(-32.6701,21.5862)(-33.1399,21.4)(-33.6167,21.2258)(-34.0998,21.0625)(-34.5885,20.9091)(-35.0825,20.7647)(-35.5813,20.6286)(-36.0844,20.5)(-36.5915,20.3784)(-37.1024,20.2632)(-37.6166,20.1538)(-38.134,20.05)(-38.6543,19.9512)(-39.1773,19.8571)(-39.7029,19.7674)(-40.2308,19.6818)(-40.7609,19.6)(-41.2931,19.5217)(-41.8272,19.4468)(-42.3631,19.375)(-42.9007,19.3061)(-43.4398,19.24)(-43.9805,19.1765)(-44.5226,19.1154)(-45.066,19.0566)(-45.6107,19.)(-46.1565,18.9455)(-46.7035,18.8929)(-47.2516,18.8421)(-47.8006,18.7931)(-48.3506,18.7458)(-48.9016,18.7)(-49.4534,18.6557)(-50.006,18.6129)(-50.5594,18.5714)(-51.1135,18.5313)(-51.6684,18.4923)(-52.224,18.4545)(-52.7802,18.4179)(-53.337,18.3824)(-53.8944,18.3478)(-54.4524,18.3143)(-55.0109,18.2817)(-55.57,18.25)(-56.1295,18.2192)(-56.6896,18.1892)(-57.2501,18.16)(-57.811,18.1316)(-58.3724,18.1039)(-58.9341,18.0769)(-59.4963,18.0506)(-60.0589,18.025)(-60.6218,18.)(-61.185,17.9756)(-61.7487,17.9518)(-62.3126,17.9286)(-62.8768,17.9059)(-63.4414,17.8837)(-64.0062,17.8621)(-64.5714,17.8409)(-65.1368,17.8202)(-65.7025,17.8)(-66.2684,17.7802)(-66.8346,17.7609)(-67.401,17.7419)(-67.9676,17.7234)(-68.5345,17.7053)(-69.1016,17.6875)(-69.6689,17.6701)(-70.2364,17.6531)(-70.8041,17.6364)(-71.372,17.62)(-71.9401,17.604)(-72.5084,17.5882)(-73.0768,17.5728)(-73.6455,17.5577)(-74.2143,17.5429)(-74.7832,17.5283)(-75.3523,17.514)(-75.9216,17.5)(-76.491,17.4862)(-77.0605,17.4727)(-77.6302,17.4595)(-78.2,17.4464)

		\psline[fillstyle=solid, hatchangle=60, hatchwidth=0.1, hatchsep=1, hatchcolor=gray, fillcolor=lightgray](60.6218,97.)(45.6107,70.)(38.3938,56.5)(34.2946,48.4)(31.7543,43.)(30.1047,39.1429)(29.0119,36.25)(28.2902,34.)(27.8283,32.2)(27.5554,30.7273)(27.4241,29.5)(27.4019,28.4615)(27.4654,27.5714)(27.5973,26.8)(27.785,26.125)(28.0185,25.5294)(28.2902,25.)(28.594,24.5263)(28.9252,24.1)(29.2799,23.7143)(29.6548,23.3636)(30.0473,23.0435)(30.4552,22.75)(30.8767,22.48)(31.3101,22.2308)(31.7543,22.)(32.2079,21.7857)(32.6701,21.5862)(33.1399,21.4)(33.6167,21.2258)(34.0998,21.0625)(34.5885,20.9091)(35.0825,20.7647)(35.5813,20.6286)(36.0844,20.5)(36.5915,20.3784)(37.1024,20.2632)(37.6166,20.1538)(38.134,20.05)(38.6543,19.9512)(39.1773,19.8571)(39.7029,19.7674)(40.2308,19.6818)(40.7609,19.6)(41.2931,19.5217)(41.8272,19.4468)(42.3631,19.375)(42.9007,19.3061)(43.4398,19.24)(43.9805,19.1765)(44.5226,19.1154)(45.066,19.0566)(45.6107,19.)(46.1565,18.9455)(46.7035,18.8929)(47.2516,18.8421)(47.8006,18.7931)(48.3506,18.7458)(48.9016,18.7)(49.4534,18.6557)(50.006,18.6129)(50.5594,18.5714)(51.1135,18.5313)(51.6684,18.4923)(52.224,18.4545)(52.7802,18.4179)(53.337,18.3824)(53.8944,18.3478)(54.4524,18.3143)(55.0109,18.2817)(55.57,18.25)(56.1295,18.2192)(56.6896,18.1892)(57.2501,18.16)(57.811,18.1316)(58.3724,18.1039)(58.9341,18.0769)(59.4963,18.0506)(60.0589,18.025)(60.6218,18.)(61.185,17.9756)(61.7487,17.9518)(62.3126,17.9286)(62.8768,17.9059)(63.4414,17.8837)(64.0062,17.8621)(64.5714,17.8409)(65.1368,17.8202)(65.7025,17.8)(66.2684,17.7802)(66.8346,17.7609)(67.401,17.7419)(67.9676,17.7234)(68.5345,17.7053)(69.1016,17.6875)(69.6689,17.6701)(70.2364,17.6531)(70.8041,17.6364)(71.372,17.62)(71.9401,17.604)(72.5084,17.5882)(73.0768,17.5728)(73.6455,17.5577)(74.2143,17.5429)(74.7832,17.5283)(75.3523,17.514)(75.9216,17.5)(76.491,17.4862)(77.0605,17.4727)(77.6302,17.4595)(78.2,17.4464)
		
		\psline[fillstyle=hlines*, hatchangle=0, hatchwidth=0.1, hatchsep=1, hatchcolor=gray, fillcolor=lightgray](67.55,6.)(58.4856,6.5)(52.5389,7.)(48.3737,7.5)(45.322,8.)(43.0126,8.5)(41.2228,9.)(39.8109,9.5)(38.6825,10.)(37.772,10.5)(37.0329,11.)(36.4308,11.5)(35.9401,12.)(35.541,12.5)(35.2184,13.)(34.9601,13.5)(34.7565,14.)(34.5998,14.5)(34.4836,15.)(34.4025,15.5)(34.3523,16.)(34.3292,16.5)(34.3301,17.)(34.3523,17.5)(34.3936,18.)(34.4519,18.5)(34.5255,19.)(34.6131,19.5)(34.7132,20.)(34.8247,20.5)(34.9467,21.)(35.0782,21.5)(35.2184,22.)(35.3666,22.5)(35.5222,23.)(35.6847,23.5)(35.8535,24.)(36.0281,24.5)(36.2081,25.)(36.3932,25.5)(36.583,26.)(36.7772,26.5)(36.9755,27.)(37.1777,27.5)(37.3834,28.)(37.5926,28.5)(37.8049,29.)(38.0202,29.5)(38.2384,30.)(38.4592,30.5)(38.6825,31.)(38.9082,31.5)(39.1361,32.)(39.3662,32.5)(39.5983,33.)(39.8323,33.5)(40.0681,34.)(40.3057,34.5)(40.5449,35.)(40.7857,35.5)(41.028,36.)(41.2717,36.5)(41.5167,37.)(41.7631,37.5)(42.0107,38.)(42.2595,38.5)(42.5095,39.)(42.7605,39.5)(43.0126,40.)(43.2657,40.5)(43.5197,41.)(43.7747,41.5)(44.0306,42.)(44.2873,42.5)(44.5448,43.)(44.8031,43.5)(45.0622,44.)(45.322,44.5)(45.5825,45.)(45.8437,45.5)(46.1055,46.)(46.368,46.5)(46.6311,47.)(46.8948,47.5)(47.159,48.)(47.4238,48.5)(47.6891,49.)(47.955,49.5)(48.2213,50.)(48.4881,50.5)(48.7554,51.)(49.0231,51.5)(49.2913,52.)(49.5599,52.5)(49.8289,53.)(50.0983,53.5)(50.368,54.)(50.6382,54.5)(50.9087,55.)(51.1796,55.5)(51.4508,56.)(51.7223,56.5)(51.9942,57.)(52.2664,57.5)(52.5389,58.)(52.8117,58.5)(53.0847,59.)(53.3581,59.5)(53.6317,60.)
		
		\psline[fillstyle=hlines, hatchangle=60, hatchwidth=0.1, hatchsep=1, hatchcolor=gray](60.6218,97.)(45.6107,70.)(38.3938,56.5)(34.2946,48.4)(31.7543,43.)(30.1047,39.1429)(29.0119,36.25)(28.2902,34.)(27.8283,32.2)(27.5554,30.7273)(27.4241,29.5)(27.4019,28.4615)(27.4654,27.5714)(27.5973,26.8)(27.785,26.125)(28.0185,25.5294)(28.2902,25.)(28.594,24.5263)(28.9252,24.1)(29.2799,23.7143)(29.6548,23.3636)(30.0473,23.0435)(30.4552,22.75)(30.8767,22.48)(31.3101,22.2308)(31.7543,22.)(32.2079,21.7857)(32.6701,21.5862)(33.1399,21.4)(33.6167,21.2258)(34.0998,21.0625)(34.5885,20.9091)(35.0825,20.7647)(35.5813,20.6286)(36.0844,20.5)(36.5915,20.3784)(37.1024,20.2632)(37.6166,20.1538)(38.134,20.05)(38.6543,19.9512)(39.1773,19.8571)(39.7029,19.7674)(40.2308,19.6818)(40.7609,19.6)(41.2931,19.5217)(41.8272,19.4468)(42.3631,19.375)(42.9007,19.3061)(43.4398,19.24)(43.9805,19.1765)(44.5226,19.1154)(45.066,19.0566)(45.6107,19.)(46.1565,18.9455)(46.7035,18.8929)(47.2516,18.8421)(47.8006,18.7931)(48.3506,18.7458)(48.9016,18.7)(49.4534,18.6557)(50.006,18.6129)(50.5594,18.5714)(51.1135,18.5313)(51.6684,18.4923)(52.224,18.4545)(52.7802,18.4179)(53.337,18.3824)(53.8944,18.3478)(54.4524,18.3143)(55.0109,18.2817)(55.57,18.25)(56.1295,18.2192)(56.6896,18.1892)(57.2501,18.16)(57.811,18.1316)(58.3724,18.1039)(58.9341,18.0769)(59.4963,18.0506)(60.0589,18.025)(60.6218,18.)(61.185,17.9756)(61.7487,17.9518)(62.3126,17.9286)(62.8768,17.9059)(63.4414,17.8837)(64.0062,17.8621)(64.5714,17.8409)(65.1368,17.8202)(65.7025,17.8)(66.2684,17.7802)(66.8346,17.7609)(67.401,17.7419)(67.9676,17.7234)(68.5345,17.7053)(69.1016,17.6875)(69.6689,17.6701)(70.2364,17.6531)(70.8041,17.6364)(71.372,17.62)(71.9401,17.604)(72.5084,17.5882)(73.0768,17.5728)(73.6455,17.5577)(74.2143,17.5429)(74.7832,17.5283)(75.3523,17.514)(75.9216,17.5)(76.491,17.4862)(77.0605,17.4727)(77.6302,17.4595)(78.2,17.4464)
		
		\psline(67.55,6.)(58.4856,6.5)(52.5389,7.)(48.3737,7.5)(45.322,8.)(43.0126,8.5)(41.2228,9.)(39.8109,9.5)(38.6825,10.)(37.772,10.5)(37.0329,11.)(36.4308,11.5)(35.9401,12.)(35.541,12.5)(35.2184,13.)(34.9601,13.5)(34.7565,14.)(34.5998,14.5)(34.4836,15.)(34.4025,15.5)(34.3523,16.)(34.3292,16.5)(34.3301,17.)(34.3523,17.5)(34.3936,18.)(34.4519,18.5)(34.5255,19.)(34.6131,19.5)(34.7132,20.)(34.8247,20.5)(34.9467,21.)(35.0782,21.5)(35.2184,22.)(35.3666,22.5)(35.5222,23.)(35.6847,23.5)(35.8535,24.)(36.0281,24.5)(36.2081,25.)(36.3932,25.5)(36.583,26.)(36.7772,26.5)(36.9755,27.)(37.1777,27.5)(37.3834,28.)(37.5926,28.5)(37.8049,29.)(38.0202,29.5)(38.2384,30.)(38.4592,30.5)(38.6825,31.)(38.9082,31.5)(39.1361,32.)(39.3662,32.5)(39.5983,33.)(39.8323,33.5)(40.0681,34.)(40.3057,34.5)(40.5449,35.)(40.7857,35.5)(41.028,36.)(41.2717,36.5)(41.5167,37.)(41.7631,37.5)(42.0107,38.)(42.2595,38.5)(42.5095,39.)(42.7605,39.5)(43.0126,40.)(43.2657,40.5)(43.5197,41.)(43.7747,41.5)(44.0306,42.)(44.2873,42.5)(44.5448,43.)(44.8031,43.5)(45.0622,44.)(45.322,44.5)(45.5825,45.)(45.8437,45.5)(46.1055,46.)(46.368,46.5)(46.6311,47.)(46.8948,47.5)(47.159,48.)(47.4238,48.5)(47.6891,49.)(47.955,49.5)(48.2213,50.)(48.4881,50.5)(48.7554,51.)(49.0231,51.5)(49.2913,52.)(49.5599,52.5)(49.8289,53.)(50.0983,53.5)(50.368,54.)(50.6382,54.5)(50.9087,55.)(51.1796,55.5)(51.4508,56.)(51.7223,56.5)(51.9942,57.)(52.2664,57.5)(52.5389,58.)(52.8117,58.5)(53.0847,59.)(53.3581,59.5)(53.6317,60.)
		
		\psline[fillstyle=solid, hatchangle=60, hatchwidth=0.1, hatchsep=1, hatchcolor=gray, fillcolor=lightgray](-29.1562,-60.5)(-24.191,-52.9)(-20.7846,-48.)(-18.269,-44.6429)(-16.3101,-42.25)(-14.7224,-40.5)(-13.3945,-39.2)(-12.2556,-38.2273)(-11.2583,-37.5)(-10.3701,-36.9615)(-9.56752,-36.5714)(-8.83346,-36.3)(-8.15507,-36.125)(-7.52253,-36.0294)(-6.9282,-36.)(-6.36605,-36.0263)(-5.83124,-36.1)(-5.31987,-36.2143)(-4.82875,-36.3636)(-4.35523,-36.5435)(-3.89711,-36.75)(-3.45255,-36.98)(-3.01999,-37.2308)(-2.59808,-37.5)(-2.18568,-37.7857)(-1.78182,-38.0862)(-1.38564,-38.4)(-0.996395,-38.7258)(-0.613435,-39.0625)(-0.236189,-39.4091)(0.135847,-39.7647)(0.50312,-40.1286)(0.866025,-40.5)(1.22492,-40.8784)(1.58012,-41.2632)(1.9319,-41.6538)(2.28053,-42.05)(2.62624,-42.4512)(2.96923,-42.8571)(3.30969,-43.2674)(3.6478,-43.6818)(3.98372,-44.1)(4.31758,-44.5217)(4.64951,-44.9468)(4.97965,-45.375)(5.30809,-45.8061)(5.63494,-46.24)(5.96029,-46.6765)(6.28424,-47.1154)(6.60685,-47.5566)(6.9282,-48.)(7.24837,-48.4455)(7.56741,-48.8929)(7.88539,-49.3421)(8.20236,-49.7931)(8.51836,-50.2458)(8.83346,-50.7)(9.14769,-51.1557)(9.46109,-51.6129)(9.77372,-52.0714)(10.0856,-52.5313)(10.3967,-52.9923)(10.7072,-53.4545)(11.017,-53.9179)(11.3263,-54.3824)(11.6349,-54.8478)(11.9429,-55.3143)(12.2504,-55.7817)(12.5574,-56.25)(12.8638,-56.7192)(13.1698,-57.1892)(13.4754,-57.66)(13.7804,-58.1316)(14.0851,-58.6039)(14.3893,-59.0769)(14.6932,-59.5506)(14.9967,-60.025)(15.2998,-60.5)(15.6025,-60.9756)(15.905,-61.4518)(16.207,-61.9286)(16.5088,-62.4059)(16.8103,-62.8837)(17.1115,-63.3621)(17.4124,-63.8409)(17.713,-64.3202)(18.0133,-64.8)(18.3134,-65.2802)(18.6133,-65.7609)(18.9129,-66.2419)(19.2123,-66.7234)(19.5114,-67.2053)(19.8103,-67.6875)(20.1091,-68.1701)(20.4076,-68.6531)(20.7059,-69.1364)(21.004,-69.62)(21.3019,-70.104)(21.5997,-70.5882)(21.8973,-71.0728)(22.1947,-71.5577)(22.4919,-72.0429)(22.789,-72.5283)(23.0859,-73.014)(23.3827,-73.5)(23.6793,-73.9862)(23.9758,-74.4727)(24.2721,-74.9595)(24.5683,-75.4464)
		
		\psline[fillstyle=hlines*, hatchangle=-60, hatchwidth=0.1, hatchsep=1, hatchcolor=gray, fillcolor=lightgray](53.1162,-100.)(37.2391,-73.5)(29.1562,-60.5)(24.191,-52.9)(20.7846,-48.)(18.269,-44.6429)(16.3101,-42.25)(14.7224,-40.5)(13.3945,-39.2)(12.2556,-38.2273)(11.2583,-37.5)(10.3701,-36.9615)(9.56752,-36.5714)(8.83346,-36.3)(8.15507,-36.125)(7.52253,-36.0294)(6.9282,-36.)(6.36605,-36.0263)(5.83124,-36.1)(5.31987,-36.2143)(4.82875,-36.3636)(4.35523,-36.5435)(3.89711,-36.75)(3.45255,-36.98)(3.01999,-37.2308)(2.59808,-37.5)(2.18568,-37.7857)(1.78182,-38.0862)(1.38564,-38.4)(0.996395,-38.7258)(0.613435,-39.0625)(0.236189,-39.4091)(-0.135847,-39.7647)(-0.50312,-40.1286)(-0.866025,-40.5)(-1.22492,-40.8784)(-1.58012,-41.2632)(-1.9319,-41.6538)(-2.28053,-42.05)(-2.62624,-42.4512)(-2.96923,-42.8571)(-3.30969,-43.2674)(-3.6478,-43.6818)(-3.98372,-44.1)(-4.31758,-44.5217)(-4.64951,-44.9468)(-4.97965,-45.375)(-5.30809,-45.8061)(-5.63494,-46.24)(-5.96029,-46.6765)(-6.28424,-47.1154)(-6.60685,-47.5566)(-6.9282,-48.)(-7.24837,-48.4455)(-7.56741,-48.8929)(-7.88539,-49.3421)(-8.20236,-49.7931)(-8.51836,-50.2458)(-8.83346,-50.7)(-9.14769,-51.1557)(-9.46109,-51.6129)(-9.77372,-52.0714)(-10.0856,-52.5313)(-10.3967,-52.9923)(-10.7072,-53.4545)(-11.017,-53.9179)(-11.3263,-54.3824)(-11.6349,-54.8478)(-11.9429,-55.3143)(-12.2504,-55.7817)(-12.5574,-56.25)(-12.8638,-56.7192)(-13.1698,-57.1892)(-13.4754,-57.66)(-13.7804,-58.1316)(-14.0851,-58.6039)(-14.3893,-59.0769)(-14.6932,-59.5506)(-14.9967,-60.025)(-15.2998,-60.5)(-15.6025,-60.9756)(-15.905,-61.4518)(-16.207,-61.9286)(-16.5088,-62.4059)(-16.8103,-62.8837)(-17.1115,-63.3621)(-17.4124,-63.8409)(-17.713,-64.3202)(-18.0133,-64.8)(-18.3134,-65.2802)(-18.6133,-65.7609)(-18.9129,-66.2419)(-19.2123,-66.7234)(-19.5114,-67.2053)(-19.8103,-67.6875)(-20.1091,-68.1701)(-20.4076,-68.6531)(-20.7059,-69.1364)(-21.004,-69.62)(-21.3019,-70.104)(-21.5997,-70.5882)(-21.8973,-71.0728)(-22.1947,-71.5577)(-22.4919,-72.0429)(-22.789,-72.5283)(-23.0859,-73.014)(-23.3827,-73.5)(-23.6793,-73.9862)(-23.9758,-74.4727)(-24.2721,-74.9595)(-24.5683,-75.4464)
		
		\psline[fillstyle=hlines, hatchangle=60, hatchwidth=0.1, hatchsep=1, hatchcolor=gray, fillcolor=lightgray](-29.1562,-60.5)(-24.191,-52.9)(-20.7846,-48.)(-18.269,-44.6429)(-16.3101,-42.25)(-14.7224,-40.5)(-13.3945,-39.2)(-12.2556,-38.2273)(-11.2583,-37.5)(-10.3701,-36.9615)(-9.56752,-36.5714)(-8.83346,-36.3)(-8.15507,-36.125)(-7.52253,-36.0294)(-6.9282,-36.)(-6.36605,-36.0263)(-5.83124,-36.1)(-5.31987,-36.2143)(-4.82875,-36.3636)(-4.35523,-36.5435)(-3.89711,-36.75)(-3.45255,-36.98)(-3.01999,-37.2308)(-2.59808,-37.5)(-2.18568,-37.7857)(-1.78182,-38.0862)(-1.38564,-38.4)(-0.996395,-38.7258)(-0.613435,-39.0625)(-0.236189,-39.4091)(0.135847,-39.7647)(0.50312,-40.1286)(0.866025,-40.5)(1.22492,-40.8784)(1.58012,-41.2632)(1.9319,-41.6538)(2.28053,-42.05)(2.62624,-42.4512)(2.96923,-42.8571)(3.30969,-43.2674)(3.6478,-43.6818)(3.98372,-44.1)(4.31758,-44.5217)(4.64951,-44.9468)(4.97965,-45.375)(5.30809,-45.8061)(5.63494,-46.24)(5.96029,-46.6765)(6.28424,-47.1154)(6.60685,-47.5566)(6.9282,-48.)(7.24837,-48.4455)(7.56741,-48.8929)(7.88539,-49.3421)(8.20236,-49.7931)(8.51836,-50.2458)(8.83346,-50.7)(9.14769,-51.1557)(9.46109,-51.6129)(9.77372,-52.0714)(10.0856,-52.5313)(10.3967,-52.9923)(10.7072,-53.4545)(11.017,-53.9179)(11.3263,-54.3824)(11.6349,-54.8478)(11.9429,-55.3143)(12.2504,-55.7817)(12.5574,-56.25)(12.8638,-56.7192)(13.1698,-57.1892)(13.4754,-57.66)(13.7804,-58.1316)(14.0851,-58.6039)(14.3893,-59.0769)(14.6932,-59.5506)(14.9967,-60.025)(15.2998,-60.5)(15.6025,-60.9756)(15.905,-61.4518)(16.207,-61.9286)(16.5088,-62.4059)(16.8103,-62.8837)(17.1115,-63.3621)(17.4124,-63.8409)(17.713,-64.3202)(18.0133,-64.8)(18.3134,-65.2802)(18.6133,-65.7609)(18.9129,-66.2419)(19.2123,-66.7234)(19.5114,-67.2053)(19.8103,-67.6875)(20.1091,-68.1701)(20.4076,-68.6531)(20.7059,-69.1364)(21.004,-69.62)(21.3019,-70.104)(21.5997,-70.5882)(21.8973,-71.0728)(22.1947,-71.5577)(22.4919,-72.0429)(22.789,-72.5283)(23.0859,-73.014)(23.3827,-73.5)(23.6793,-73.9862)(23.9758,-74.4727)(24.2721,-74.9595)(24.5683,-75.4464)
		
		\psline(53.1162,-100.)(37.2391,-73.5)(29.1562,-60.5)(24.191,-52.9)(20.7846,-48.)(18.269,-44.6429)(16.3101,-42.25)(14.7224,-40.5)(13.3945,-39.2)(12.2556,-38.2273)(11.2583,-37.5)(10.3701,-36.9615)(9.56752,-36.5714)(8.83346,-36.3)(8.15507,-36.125)(7.52253,-36.0294)(6.9282,-36.)(6.36605,-36.0263)(5.83124,-36.1)(5.31987,-36.2143)(4.82875,-36.3636)(4.35523,-36.5435)(3.89711,-36.75)(3.45255,-36.98)(3.01999,-37.2308)(2.59808,-37.5)(2.18568,-37.7857)(1.78182,-38.0862)(1.38564,-38.4)(0.996395,-38.7258)(0.613435,-39.0625)(0.236189,-39.4091)(-0.135847,-39.7647)(-0.50312,-40.1286)(-0.866025,-40.5)(-1.22492,-40.8784)(-1.58012,-41.2632)(-1.9319,-41.6538)(-2.28053,-42.05)(-2.62624,-42.4512)(-2.96923,-42.8571)(-3.30969,-43.2674)(-3.6478,-43.6818)(-3.98372,-44.1)(-4.31758,-44.5217)(-4.64951,-44.9468)(-4.97965,-45.375)(-5.30809,-45.8061)(-5.63494,-46.24)(-5.96029,-46.6765)(-6.28424,-47.1154)(-6.60685,-47.5566)(-6.9282,-48.)(-7.24837,-48.4455)(-7.56741,-48.8929)(-7.88539,-49.3421)(-8.20236,-49.7931)(-8.51836,-50.2458)(-8.83346,-50.7)(-9.14769,-51.1557)(-9.46109,-51.6129)(-9.77372,-52.0714)(-10.0856,-52.5313)(-10.3967,-52.9923)(-10.7072,-53.4545)(-11.017,-53.9179)(-11.3263,-54.3824)(-11.6349,-54.8478)(-11.9429,-55.3143)(-12.2504,-55.7817)(-12.5574,-56.25)(-12.8638,-56.7192)(-13.1698,-57.1892)(-13.4754,-57.66)(-13.7804,-58.1316)(-14.0851,-58.6039)(-14.3893,-59.0769)(-14.6932,-59.5506)(-14.9967,-60.025)(-15.2998,-60.5)(-15.6025,-60.9756)(-15.905,-61.4518)(-16.207,-61.9286)(-16.5088,-62.4059)(-16.8103,-62.8837)(-17.1115,-63.3621)(-17.4124,-63.8409)(-17.713,-64.3202)(-18.0133,-64.8)(-18.3134,-65.2802)(-18.6133,-65.7609)(-18.9129,-66.2419)(-19.2123,-66.7234)(-19.5114,-67.2053)(-19.8103,-67.6875)(-20.1091,-68.1701)(-20.4076,-68.6531)(-20.7059,-69.1364)(-21.004,-69.62)(-21.3019,-70.104)(-21.5997,-70.5882)(-21.8973,-71.0728)(-22.1947,-71.5577)(-22.4919,-72.0429)(-22.789,-72.5283)(-23.0859,-73.014)(-23.3827,-73.5)(-23.6793,-73.9862)(-23.9758,-74.4727)(-24.2721,-74.9595)(-24.5683,-75.4464)
		
		\psline(-100, 0)(100, 0)
		\psline([angle=60,nodesep=-60]twomp2)([angle=60,nodesep=60]twomp2)
		\psline([angle=120,nodesep=-60]twomp2)([angle=120,nodesep=60]twomp2)

		\psline[linestyle=dotted](-60, 4)(60, 4)
		\psline[linestyle=dotted](-60, 16)(60, 16)

		\psline[linestyle=dotted]([angle=60,nodesep=80]u4mp2)([angle=60,nodesep=-60]u4mp2)
		\psline[linestyle=dotted]([angle=60,nodesep=80]u16mp2)([angle=60,nodesep=-60]u16mp2)

		\psline[linestyle=dotted]([angle=120,nodesep=80]t4mp2)([angle=120,nodesep=-60]t4mp2)
		\psline[linestyle=dotted]([angle=120,nodesep=80]t16mp2)([angle=120,nodesep=-60]t16mp2)

		\psline[linestyle=solid,linewidth=1.,linecolor=lightgray,arrows=cc-cc]([angle=0,nodesep=0]sconst)([angle=0,nodesep=-60]sconst)
		\psline[linestyle=solid,linewidth=1.,linecolor=lightgray,arrows=cc-cc]([angle=0,nodesep=9.2376]sconst)([angle=0,nodesep=80]sconst)
		\psline[linestyle=dashed]([angle=0,nodesep=60]sconst)([angle=0,nodesep=-80]sconst)

		\psline[linestyle=solid,linewidth=1.,linecolor=lightgray,arrows=cc-cc]([angle=120,nodesep=0]tconst)([angle=120,nodesep=60]tconst)
		\psline[linestyle=solid,linewidth=1.,linecolor=lightgray,arrows=cc-cc]([angle=120,nodesep=-9.2376]tconst)([angle=120,nodesep=-80]tconst)
		\psline[linestyle=dashed]([angle=120,nodesep=60]tconst)([angle=120,nodesep=-80]tconst)

		\psline[linestyle=solid,linewidth=1.,linecolor=lightgray,arrows=cc-cc]([angle=60,nodesep=0]uconst)([angle=60,nodesep=60]uconst)
		\psline[linestyle=solid,linewidth=1.,linecolor=lightgray,arrows=cc-cc]([angle=60,nodesep=-9.2376]uconst)([angle=60,nodesep=-80]uconst)
		\psline[linestyle=dashed]([angle=60,nodesep=60]uconst)([angle=60,nodesep=-80]uconst)

		\put(-46.5,-2.5){$s=0$}
		\put(-39.5,5.25){$s=4M_\pi^2$}
		\put(-29.5,17.25){$s=16M_\pi^2$}
		
		\rput{-60}([angle=-57,nodesep=43]twomp2){$t=0$}
		\rput{-60}([angle=-64,nodesep=35]t4mp2){$t=4M_\pi^2$}
		\rput{-60}([angle=-64,nodesep=25]t16mp2){$t=16M_\pi^2$}
		
		\rput{60}([angle=63,nodesep=38]twomp2){$u=0$}
		\rput{60}([angle=56,nodesep=29]u4mp2){$u=4M_\pi^2$}
		\rput{60}([angle=55,nodesep=22]u16mp2){$u=16M_\pi^2$}

		\put(-5.5,27){$s$-channel}
		\put(-30,-13){$t$-channel}
		\put(20,-13){$u$-channel}

		\put(-45,12){$\rho_{st}$}
		\put(-34,32){$\rho_{ts}$}

		\put(42,12){$\rho_{su}$}
		\put(30,32){$\rho_{us}$}

		\put(-11,-41){$\rho_{ut}$}
		\put(8,-41){$\rho_{tu}$}

	\end{pspicture*}
	\caption{Mandelstam diagram for HLbL scattering for the case $q_i^2=0.5M_\pi^2$ with all the double-spectral regions for `2disc'-box topologies.}
	\label{img:HLbLMandelstamDiagramBoxDiscSymm}
\end{figure}

The complete set of double-spectral regions, which is obtained after symmetrisation, is shown in figure~\ref{img:HLbLMandelstamDiagramBoxDiscSymm}. In the symmetric version, the double-spectral integrals over $\rho_{st}$ and $\rho_{ut}$ are taken from the fixed-$t$ representation, $\rho_{ts}$ and $\rho_{us}$ come from the fixed-$s$ representation and finally $\rho_{su}$ and $\rho_{tu}$ stem from the fixed-$u$ dispersion relation.

\begin{figure}[t]
	\centering
	\begin{subfigure}[b]{0.15\textwidth}
		\centering
		\includegraphics[width=2cm]{HLbL/images/Box_2Disc}
		\caption{$\rho_{st}$}
	\end{subfigure}
	\begin{subfigure}[b]{0.15\textwidth}
		\centering
		\includegraphics[width=2cm]{HLbL/images/Box_2Disc_Crossed1}
		\caption{$\rho_{su}$}
	\end{subfigure}
	\begin{subfigure}[b]{0.15\textwidth}
		\centering
		\includegraphics[width=2cm,angle=90,origin=c]{HLbL/images/Box_2Disc}
		\caption{$\rho_{ts}$}
	\end{subfigure}
	\begin{subfigure}[b]{0.15\textwidth}
		\centering
		\reflectbox{\includegraphics[width=2cm,angle=90,origin=c]{HLbL/images/Box_2Disc_Crossed1}}
		\caption{$\rho_{tu}$}
	\end{subfigure}
	\begin{subfigure}[b]{0.15\textwidth}
		\centering
		\includegraphics[width=2cm]{HLbL/images/Box_2Disc_Crossed2}
		\caption{$\rho_{us}$}
	\end{subfigure}
	\begin{subfigure}[b]{0.15\textwidth}
		\centering
		\includegraphics[width=2cm]{HLbL/images/Box_2Disc_Crossed3}
		\caption{$\rho_{ut}$}
	\end{subfigure}
	
	\vspace{0.25cm}
	
	$\stackrel{\underbrace{\hspace{4cm}}}{\stackrel{}{\approx}}$ \hspace{0.5cm} $\stackrel{\underbrace{\hspace{4cm}}}{\stackrel{}{\approx}}$  \hspace{0.5cm} $\stackrel{\underbrace{\hspace{4cm}}}{\stackrel{}{\approx}}$
	
	\vspace{0.25cm}
	
	\begin{subfigure}[b]{0.3\textwidth}
		\centering
		\includegraphics[width=2cm]{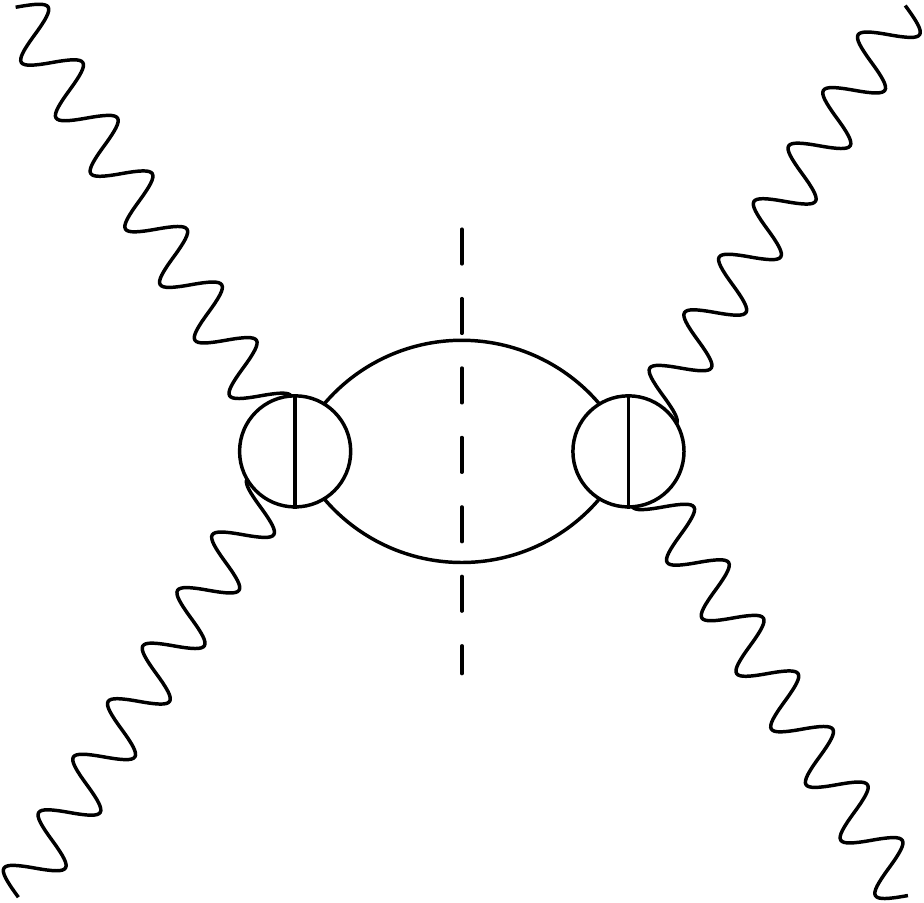}
		\caption{}
	\end{subfigure}
	\begin{subfigure}[b]{0.315\textwidth}
		\centering
		\includegraphics[width=2cm,angle=90,origin=c]{HLbL/images/Bulb}
		\caption{}
	\end{subfigure}
	\begin{subfigure}[b]{0.3\textwidth}
		\centering
		\includegraphics[width=2cm]{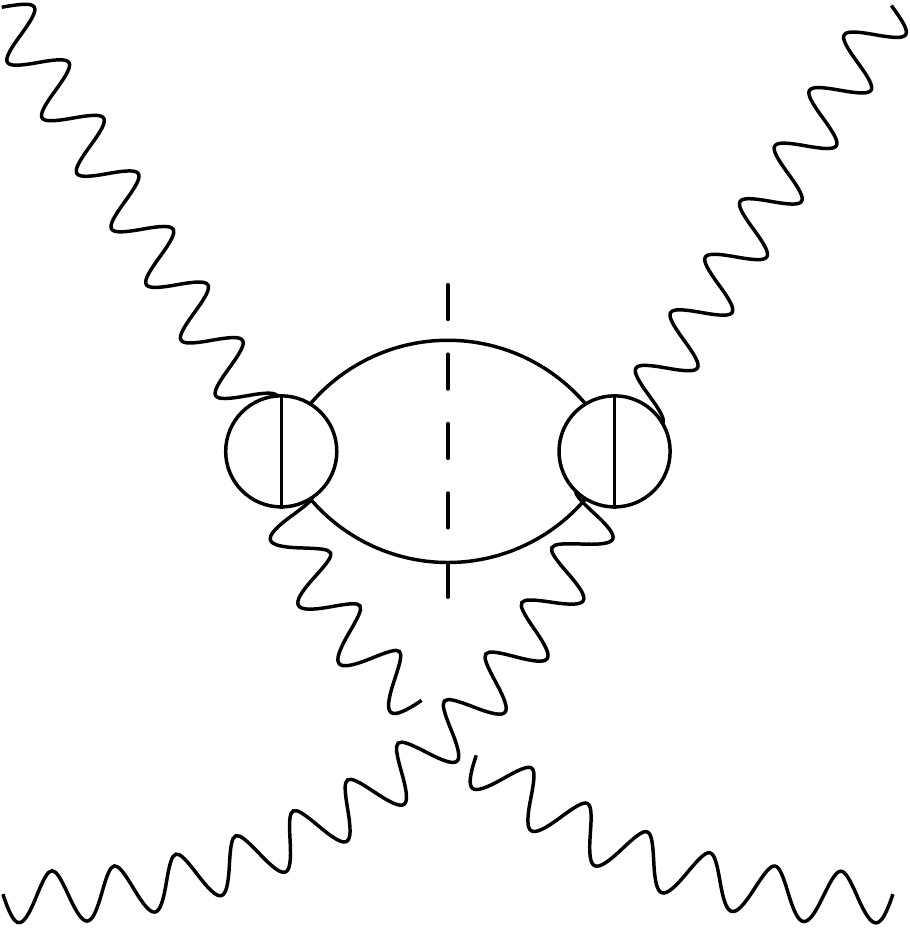}
		\caption{}
	\end{subfigure}
	
	\caption{{\bf (a)--(f)} Unitarity diagrams representing the complete set of `2disc'-box contributions. \newline {\bf (g)--(i)} Partial-wave approximation: the sub-process becomes a polynomial in the crossed variable.}
	\label{img:HLbLBox2DiscPolynomialApproximation}
\end{figure}

In summary, we can write the contribution of higher intermediate states in the secondary channel as a double-spectral representation:
\begin{align}
	\begin{split}
		\label{eq:HLbLDoubleSpectralDisc}
		\Pi_i(s,t,u) \Big|_\mathrm{disc} &= \frac{1}{\pi^2} \int_{4M_\pi^2}^\infty ds^\prime \int_{t^+(s^\prime)}^\infty dt^\prime \frac{\rho_{i;st}(s^\prime,t^\prime)}{(s^\prime - s)(t^\prime - t)} \\
			&\quad + \frac{1}{\pi^2} \int_{4M_\pi^2}^\infty ds^\prime \int_{u^+(s^\prime)}^\infty du^\prime \frac{\rho_{i;su}(s^\prime,u^\prime)}{(s^\prime - s)(u^\prime - u)} \\
			&\quad + \frac{1}{\pi^2} \int_{4M_\pi^2}^\infty dt^\prime \int_{s^+(t^\prime)}^\infty ds^\prime \frac{\rho_{i;ts}(t^\prime,s^\prime)}{(t^\prime - t)(s^\prime - s)} \\
			&\quad + \frac{1}{\pi^2} \int_{4M_\pi^2}^\infty dt^\prime \int_{u^+(t^\prime)}^\infty du^\prime \frac{\rho_{i;tu}(t^\prime,u^\prime)}{(t^\prime - t)(u^\prime - u)} \\
			&\quad + \frac{1}{\pi^2} \int_{4M_\pi^2}^\infty du^\prime \int_{s^+(u^\prime)}^\infty ds^\prime \frac{\rho_{i;us}(u^\prime,s^\prime)}{(u^\prime - u)(s^\prime - s)} \\
			&\quad + \frac{1}{\pi^2} \int_{4M_\pi^2}^\infty du^\prime \int_{t^+(u^\prime)}^\infty dt^\prime \frac{\rho_{i;ut}(u^\prime,t^\prime)}{(u^\prime - u)(t^\prime - t)} .
	\end{split}
\end{align}
The border functions of the double-spectral regions approach asymptotically $t^+(s)\stackrel{s\to\infty}{\longrightarrow}9M_\pi^2$ for the `1disc' contribution or $16M_\pi^2$ for the `2dics' contribution.

In a next step, we apply subtractions to all the secondary dispersion integrals in (\ref{eq:HLbLDoubleSpectralDisc}) by using the relation
\begin{align}
	\begin{split}
		\frac{1}{t^\prime - t} = \frac{1}{t^\prime} + \frac{t}{(t^\prime-t)t^\prime} .
	\end{split}
\end{align}
Using e.g.~two subtractions, this leads to
\begin{align}
	\begin{split}
		\Pi_i(s,t,u) \Big|_\mathrm{disc} &= \frac{1}{\pi} \int_{4M_\pi^2}^\infty ds^\prime \frac{1}{s^\prime-s} \left( \frac{1}{\pi} \int_{t^+(s^\prime)}^\infty dt^\prime \frac{\rho_{i;st}(s^\prime,t^\prime)}{t^\prime} + \frac{t}{\pi} \int_{t^+(s^\prime)}^\infty dt^\prime \frac{\rho_{i;st}(s^\prime,t^\prime)}{{t^\prime}^2} \right) \\
			&\quad + \frac{t^2}{\pi^2} \int_{4M_\pi^2}^\infty ds^\prime \int_{t^+(s^\prime)}^\infty dt^\prime \frac{\rho_{i;st}(s^\prime,t^\prime)}{(s^\prime - s)(t^\prime - t) {t^\prime}^2} \\
			&\quad + \frac{1}{\pi} \int_{4M_\pi^2}^\infty ds^\prime \frac{1}{s^\prime - s} \left( \frac{1}{\pi} \int_{u^+(s^\prime)}^\infty du^\prime \frac{\rho_{i;su}(s^\prime,u^\prime)}{u^\prime} + \frac{u}{\pi} \int_{u^+(s^\prime)}^\infty du^\prime \frac{\rho_{i;su}(s^\prime,u^\prime)}{{u^\prime}^2} \right) \\
			&\quad + \frac{u^2}{\pi^2} \int_{4M_\pi^2}^\infty ds^\prime \int_{u^+(s^\prime)}^\infty du^\prime \frac{\rho_{i;su}(s^\prime,u^\prime)}{(s^\prime - s)(u^\prime - u) {u^\prime}^2} \\
			&\quad + \ldots \\
			&=: \frac{1}{\pi} \int_{4M_\pi^2}^\infty ds^\prime \frac{\rho_{i;st}^0(s^\prime)}{s^\prime-s} + t \frac{1}{\pi} \int_{4M_\pi^2}^\infty ds^\prime \frac{\rho_{i;st}^1(s^\prime)}{s^\prime - s} \\
			&\quad + \frac{t^2}{\pi^2} \int_{4M_\pi^2}^\infty ds^\prime \int_{t^+(s^\prime)}^\infty dt^\prime \frac{\rho_{i;st}(s^\prime,t^\prime)}{(s^\prime - s)(t^\prime - t) {t^\prime}^2} \\
			&\quad + \frac{1}{\pi} \int_{4M_\pi^2}^\infty ds^\prime \frac{\rho_{i;su}^0(s^\prime)}{s^\prime - s} + u \frac{1}{\pi} \int_{4M_\pi^2}^\infty ds^\prime \frac{\rho_{i;su}^1(s^\prime)}{s^\prime - s} \\
			&\quad + \frac{u^2}{\pi^2} \int_{4M_\pi^2}^\infty ds^\prime \int_{u^+(s^\prime)}^\infty du^\prime \frac{\rho_{i;su}(s^\prime,u^\prime)}{(s^\prime - s)(u^\prime - u) {u^\prime}^2} \\
			&\quad + \ldots ,
	\end{split}
\end{align}
where the dots stand for the crossed channels. If we neglect the subtracted double-dispersive integrals, we remove the double-spectral regions and approximate the dispersive structure in the crossed channel just by the subtraction polynomial. In terms of unitarity diagrams, this is illustrated in figure~\ref{img:HLbLBox2DiscPolynomialApproximation}: the blobs with a bar in the middle denote contributions without intermediate states in the direction of the bar. The topologies with a discontinuity in only one of the sub-processes are analogously transformed into triangle topologies.

The approximation based on two subtractions can be written as
\begin{align}
	\begin{split}
		\label{eq:PWApproximateDispersionRelationBTT}
		\Pi_i(s,t,u) \Big|_\mathrm{disc} &= \frac{1}{\pi} \int_{4M_\pi^2}^\infty ds^\prime \frac{1}{s^\prime-s}\left( \rho_{i;st}^0(s^\prime) + \rho_{i;su}^0(s^\prime) + \frac{1}{2}\left( \Sigma - s^\prime\right)\left( \rho_{i;st}^1(s^\prime) + \rho_{i;su}^1(s^\prime) \right) \right) \\
			&\quad + ( t - u ) \frac{1}{\pi} \int_{4M_\pi^2}^\infty ds^\prime \frac{1}{s^\prime - s} \frac{1}{2} \left( \rho_{i;st}^1(s^\prime) - \rho_{i;su}^1(s^\prime) \right) \\
			&\quad + \frac{1}{\pi} \int_{4M_\pi^2}^\infty ds^\prime \frac{1}{2}\left( \rho_{i;st}^1(s^\prime) + \rho_{i;su}^1(s^\prime) \right) \\
			&\quad + \ldots ,
	\end{split}
\end{align}
where the dots stand again for two analogous contributions from the crossed channels.

Because the double-dispersive integrals are neglected, the original asymptotic behaviour is spoilt in this approximation: e.g.~the $s$-channel contribution grows polynomially in $t$ and $u$. We choose to preserve the asymptotic behaviour in $s$ for the $s$-channel contribution along lines of fixed $t-u$ (and analogously for the crossed-channel contributions). This is equivalent to imposing the sum rule
\begin{align}
	\begin{split}
		\int_{4M_\pi^2}^\infty ds^\prime \left( \rho_{i;st}^1(s^\prime) + \rho_{i;su}^1(s^\prime) \right) = 0 .
	\end{split}
\end{align}
Additional sum rules are required if more than two subtractions are used.

We have now obtained a dispersion relation for contributions of two-pion intermediate states beyond pure box topologies, based on a polynomial approximation in the crossed variable. What remains is the determination of the imaginary part along the cuts of the (single-variable) dispersion relation. The $s$-channel imaginary part can be calculated by writing the sub-process in terms of $s$-channel helicity partial waves: the phase-space integration will then result again in a polynomial in the crossed variables, which can be mapped onto the imaginary parts of the derived dispersion relation. Due to the sum rules that we impose, this mapping is unambiguous.

If only $S$-waves are used to describe $\gamma^*\gamma^*\to\pi\pi$, it is sufficient to apply a single subtraction. If $D$-waves are present, most of the scalar functions need three subtractions. Surprisingly, some functions even need four subtractions (the projection of the imaginary part due to $D$-waves produces terms proportional to $z^3$ in some scalar functions). Although we neglect the dispersive structure in the crossed channel of the sub-process, this partial-wave approach takes $\pi\pi$-rescattering effects in the direct channel into account. We expect these to be the most important contributions apart from pion-pole and box topologies. The error that is introduced by the partial-wave approximation can be estimated by applying the same technique to the box topologies, which can be calculated exactly.

Another subtlety is worth being mentioned: the generating set of Lorentz structures (\ref{eq:HLbLBTTStructures}) is of course only unique up to transformations that do not introduce kinematic singularities. Concerning the asymptotic behaviour of the scalar functions $\Pi_i$, this means that not all the functions should show the same behaviour but rather that the functions multiplying Lorentz structures of higher mass dimension should fall down even faster for asymptotic values of the Mandelstam variables. Hence, the asymptotic behaviour of the whole HLbL tensor should be considered; a parameter-free dispersive representation is possible in the case
\begin{align}
	\begin{split}
		\Pi^{\mu\nu\lambda\sigma} \sim s, t, u ,
	\end{split}
\end{align}
which results in the following behaviour of the scalar functions:
\begin{align}
	\begin{split}
		\label{eq:BTTAsymptoticBehaviour}
		\Pi_1, \Pi_4 &\sim \frac{1}{s}, \frac{1}{t}, \frac{1}{u} , \\
		\Pi_7, \Pi_{19}, \Pi_{49} &\sim \frac{1}{s^2}, \frac{1}{t^2}, \frac{1}{u^2} , \\
		\Pi_{31} &\sim \frac{1}{s^3}, \frac{1}{t^3}, \frac{1}{u^3} ,
	\end{split}
\end{align}
and analogous asymptotics for the functions related by crossing symmetry. On the level of the Mandelstam representation, this implies a set of sum rules for the scalar functions $\Pi_7$, $\Pi_{19}$, $\Pi_{31}$ and $\Pi_{49}$, which ensure that the result for the HLbL tensor is independent of the choice of the basis: the difference between the Mandelstam representations for one set of basis functions and a second, equally valid set of functions will vanish as a consequence of the sum rules.

We illustrate this with a simple one-dimensional dispersion relation. Let the function $f(s)$ behave asymptotically like $1/s^2$ and have a right-hand cut. Then, it fulfils a dispersion relation
\begin{align}
	\begin{split}
		f(s) &= \frac{1}{\pi} \int_{s_0}^\infty ds^\prime \frac{\Im f(s^\prime)}{s^\prime - s} .
	\end{split}
\end{align}
Because $s f(s)$ still falls down like $1/s$, the dispersion relation
\begin{align}
	\begin{split}
		s f(s) &= \frac{1}{\pi} \int_{s_0}^\infty ds^\prime \frac{ s^\prime \Im f(s^\prime) }{s^\prime - s}
	\end{split}
\end{align}
is valid as well. Subtracting this dispersion relations leads to
\begin{align}
	\begin{split}
		f(s) &= \frac{1}{s} \frac{1}{\pi} \int_{s_0}^\infty ds^\prime \Im f(s^\prime) + \frac{1}{\pi} \int_{s_0}^\infty ds^\prime \frac{ \Im f(s^\prime) }{s^\prime - s},
	\end{split}
\end{align}
which together with the first dispersion relation implies the sum rule (also known as `superconvergence relation' \cite{Martin1970})
\begin{align}
	\begin{split}
		\int_{s_0}^\infty ds^\prime \Im f(s^\prime) = 0 .
	\end{split}
\end{align}

Since we treat the $\pi\pi$-rescattering contribution in a partial-wave expansion that spoils the asymptotic behaviour, we expect the sum rules to be fulfilled by this contribution only approximately. Hence, a dependence on the choice of basis is introduced, which can be used as another estimate of the uncertainty due to the partial-wave expansion. Alternatively, during the construction of the input, the high-energy behaviour of the partial waves could be tuned such that the sum rules are fulfilled exactly.

\subsubsection{$S$-Wave $\pi\pi$-Rescattering Contribution}

We determine now the imaginary parts with the unitarity relation and the partial-wave representation of the sub-process $\gamma^*\gamma^*\to\pi\pi$. If only $S$-waves are taken into account, the sub-process is given by (\ref{eq:SWavesSubProcessScalarFunctions}).

Because nothing depends in this case on the momenta of the intermediate pions, the phase-space integral in the unitarity relation is trivial:
\begin{align}
	\begin{split}
		\Im_s^{\pi\pi} \Pi^{\mu\nu\lambda\sigma} \Big|_{S\text{-waves}} &= \frac{1}{16 \pi} \sigma_\pi(s) \begin{aligned}[t] 
			& \bigg( W_{+-}^{\mu\nu}(q_1,q_2) {W_{+-}^{\lambda\sigma}}^*(q_3,-q_4) \\
			& + \frac{1}{2} W_{00}^{\mu\nu}(q_1,q_2) {W_{00}^{\lambda\sigma}}^*(q_3,-q_4) \bigg) \end{aligned} \\
			&=  \begin{aligned}[t]
				& \sum_I \frac{1}{S_I} \frac{1}{16 \pi} \sigma_\pi(s) \Bigg( \left( q_1 \cdot q_2 g^{\mu\nu} - q_2^\mu q_1^\nu \right) A_1^{(I)}(q_1,q_2) \\
				& + \left( q_1^2 q_2^2 g^{\mu\nu} + q_1 \cdot q_2 q_1^\mu q_2^\nu - q_1^2 q_2^\mu q_2^\nu - q_2^2 q_1^\mu q_1^\nu \right) A_2^{(I)}(q_1,q_2) \Bigg) \\
				& \cdot \Bigg( \left( -q_3 \cdot q_4 g^{\lambda\sigma} + q_4^\lambda q_3^\sigma \right) {A_1^{(I)}}^*(q_3,-q_4) \\
				& + \left( q_3^2 q_4^2 g^{\lambda\sigma} + q_3 \cdot q_4 q_3^\lambda q_4^\sigma - q_3^2 q_4^\lambda q_4^\sigma - q_4^2 q_3^\lambda q_3^\sigma \right) {A_2^{(I)}}^*(q_3,-q_4) \Bigg) , \end{aligned}
	\end{split}
\end{align}
where $I\in\{+-,00\}$ and $S_{+-} = 1$, $S_{00} = 2$.

We project now this expression for the imaginary part on the basis (\ref{eq:HLbLBasisStructures}):
\begin{align}
	\begin{split}
		\Im_s^{\pi\pi} \tilde\Pi_4 \Big|_{S\text{-waves}} &= - \frac{1}{16\pi} \sigma_\pi(s) \sum_I  \frac{1}{S_I}  A_1^{(I)}(q_1,q_2) {A_1^{(I)}}^*(q_3, -q_4) \\
			&\quad  + \frac{1}{2} ( s + q_1^2 + q_2^2 ) \frac{1}{16\pi} \sigma_\pi(s) \sum_I  \frac{1}{S_I} A_2^{(I)}(q_1,q_2) {A_1^{(I)}}^*(q_3, -q_4) + \ldots, \\
		\Im_s^{\pi\pi} \tilde\Pi_{17} \Big|_{S\text{-waves}} &= \Im_s^{\pi\pi} \tilde\Pi_{18} \Big|_{S\text{-waves}} = \Im_s^{\pi\pi} \tilde\Pi_{25} \Big|_{S\text{-waves}} = \Im_s^{\pi\pi} \tilde\Pi_{26} \Big|_{S\text{-waves}} \\
			&= - \frac{1}{16\pi} \sigma_\pi(s) \sum_I \frac{1}{S_I} A_2^{(I)}(q_1,q_2) {A_1^{(I)}}^*(q_3, -q_4)  + \ldots ,
	\end{split}
\end{align}
where the dots denote terms that stem from the projection of the structure $T_2^{\lambda\sigma}(q_3,-q_4)$. Since the amplitudes $A_i$ are free from kinematics, we can immediately make the step to the redundant set of 54 functions (the imaginary part cannot be reshuffled into $\Pi_{32}$ without introducing kinematic singularities and zeros):
\begin{align}
	\begin{split}
		\label{eq:HLbLSwavesSchannelImaginaryParts}
		\Im_s^{\pi\pi} \Pi_4 \Big|_{S\text{-waves}} &= - \frac{1}{16\pi} \sigma_\pi(s) \sum_I  \frac{1}{S_I}  A_1^{(I)}(q_1,q_2) {A_1^{(I)}}^*(q_3, -q_4) \\
			&\quad  + \frac{1}{2} ( s + q_1^2 + q_2^2 ) \frac{1}{16\pi} \sigma_\pi(s) \sum_I  \frac{1}{S_I} A_2^{(I)}(q_1,q_2) {A_1^{(I)}}^*(q_3, -q_4) + \ldots, \\
		\Im_s^{\pi\pi} \Pi_{17} \Big|_{S\text{-waves}} &= \Im_s^{\pi\pi} \Pi_{18} \Big|_{S\text{-waves}} = \Im_s^{\pi\pi} \Pi_{29} \Big|_{S\text{-waves}} = \Im_s^{\pi\pi} \Pi_{30} \Big|_{S\text{-waves}} \\
			&= - \frac{1}{16\pi} \sigma_\pi(s) \sum_I \frac{1}{S_I} A_2^{(I)}(q_1,q_2) {A_1^{(I)}}^*(q_3, -q_4)  + \ldots .
	\end{split}
\end{align}

We argue now why the structure $T_2^{\lambda\sigma}(q_3,-q_4)$, which consists of terms either proportional to $q_4^2$ or $q_4^\sigma$, does not contribute to the $(g-2)_\mu$.

Consider the projection of the product $T_1^{\mu\nu}(q_1,q_2) T_2^{\lambda\sigma}(q_3, -q_4)$:
\begin{align}
	\begin{split}
		T_1^{\mu\nu}(q_1,q_2) T_2^{\lambda\sigma}(q_3, -q_4) &= \frac{1}{2} ( s + q_3^2 + q_4^2 ) T_4^{\mu\nu\lambda\sigma} - T_7^{\mu\nu\lambda\sigma} - T_8^{\mu\nu\lambda\sigma} - T_{19}^{\mu\nu\lambda\sigma} - T_{20}^{\mu\nu\lambda\sigma} .
	\end{split}
\end{align}
If we write dispersion relations for the functions $\Pi_i$, this results in a contribution to the HLbL tensor of
\begin{align}
	\begin{split}
		\Pi^{\mu\nu\lambda\sigma}_{\pi\pi} \Big|{\scriptsize\begin{aligned} & s\text{-channel} \\[-5pt] & S\text{-waves} \\[-5pt] & T_1 T_2\end{aligned}} &= T_4^{\mu\nu\lambda\sigma} \frac{1}{\pi} \int_{4M_\pi^2}^\infty ds^\prime \frac{1}{s^\prime - s} \frac{1}{2} ( s^\prime + q_3^2 + q_4^2 ) f(s^\prime) \\
			&- \left( T_7^{\mu\nu\lambda\sigma} + T_8^{\mu\nu\lambda\sigma} + T_{19}^{\mu\nu\lambda\sigma} + T_{20}^{\mu\nu\lambda\sigma} \right) \frac{1}{\pi} \int_{4M_\pi^2}^\infty ds^\prime \frac{1}{s^\prime - s} f(s^\prime) ,
	\end{split}
\end{align}
where
\begin{align}
	\begin{split}
		f(s) = \frac{1}{16\pi} \sigma_\pi(s) \sum_I \frac{1}{S_I} A_1^{(I)}(q_1,q_2) {A_2^{(I)}}^*(q_3,-q_4) .
	\end{split}
\end{align}
In terms of the scalar functions $\bar \Pi_i$ that enter the master formula for $a_\mu$ (\ref{eq:MasterFormula3Dim}), this results in
\begin{align}
	\begin{split}
		\lim_{q_4\to0} \bar \Pi_3 \Big|{\scriptsize\begin{aligned} & s\text{-channel} \\[-5pt] & S\text{-waves} \\[-5pt] & T_1 T_2\end{aligned}} &= \frac{1}{\pi} \int_{4M_\pi^2}^\infty ds^\prime \frac{1}{s^\prime - s} \frac{1}{2} ( s^\prime + q_3^2 ) f(s^\prime)  - s \frac{1}{\pi} \int_{4M_\pi^2}^\infty ds^\prime \frac{1}{s^\prime - s} f(s^\prime) \\
			&=  \frac{1}{2 \pi} \int_{4M_\pi^2}^\infty ds^\prime f(s^\prime) = 0 ,
	\end{split}
\end{align}
because $s=q_3^2$ in the considered limit. The last equality is a sum rule following from the fact that $f(s)$ is the contribution of $T_1^{\mu\nu}T_2^{\lambda\sigma}$ to the imaginary part of $\Pi_7$. The result vanishes because we assume according to (\ref{eq:BTTAsymptoticBehaviour}) the asymptotic behaviour $\Pi_7 \sim 1/s^2$.

We could have argued in a different way as well: instead of using the generating set of Lorentz structures (\ref{eq:HLbLBTTStructures}), we could choose a different set $\{\Pi_i^\prime\}$ that is also free of kinematics and defined by ${T_7^\prime}^{\mu\nu\lambda\sigma} = T_7^{\mu\nu\lambda\sigma} - \frac{1}{2}(s+q_3^2+q_4^2) T_4^{\mu\nu\lambda\sigma}$. Writing dispersion relations for the corresponding functions $\Pi_i^\prime$, we find immediately that the $T_1^{\mu\nu}T_2^{\lambda\sigma}$ contribution to $a_\mu$ vanishes. In other words, the sum rules and the independence on the choice of `basis' are equivalent and a consequence of the assumed asymptotic behaviour of the HLbL tensor.

The projection of the structure $T_2^{\mu\nu}(q_1,q_2) T_2^{\lambda\sigma}(q_3,-q_4)$ looks more complicated, but the same arguments apply here, too. Hence, for the $s$-channel $S$-wave contribution, it is sufficient to take the imaginary parts in (\ref{eq:HLbLSwavesSchannelImaginaryParts}) into account. We express them in terms of the helicity partial waves of $\gamma^*\gamma^*\to\pi\pi$ and take the limit $q_4^2\to0$:
{\small
\begin{align}
	\begin{split}
		\lim_{q_4^2\to0} \Im_s^{\pi\pi} \Pi_4 \big|_{S\text{-waves}}^\mathrm{disc} &= \frac{\sigma_\pi(s)}{4\pi} \frac{1}{\lambda(s,q_1^2,q_2^2)(s-q_3^2)} \sum_I \\
			&\quad \cdot \mathcal{S} \begin{aligned}[t]
				& \Bigg[ \left( 2 \sqrt{q_1^2 q_2^2} h_{00}^{0,(I)}(s,q_1^2,q_2^2) - (s - q_1^2 - q_2^2) h_{++}^{0,(I)}(s,q_1^2,q_2^2) \right) h_{++}^{0,(I)}(s,q_3^2,0)^* \\
				& + \frac{1}{2} ( s + q_1^2 + q_2^2 ) \\
				& \cdot \Bigg( \frac{s - q_1^2 - q_2^2}{\sqrt{q_1^2 q_2^2}} h_{00}^{0,(I)}(s,q_1^2,q_2^2) - 2 h_{++}^{0,(I)}(s,q_1^2,q_2^2) \Bigg) h_{++}^{0,(I)}(s,q_3^2,0)^* \Bigg] + \ldots, \end{aligned} \\
		\lim_{q_4^2\to0} \Im_s^{\pi\pi} \Pi_{17} \big|_{S\text{-waves}}^\mathrm{disc} &= \lim_{q_4^2\to0} \Im_s^{\pi\pi} \Pi_{18} \big|_{S\text{-waves}}^\mathrm{disc} = \lim_{q_4^2\to0} \Im_s^{\pi\pi} \Pi_{29} \big|_{S\text{-waves}}^\mathrm{disc} = \lim_{q_4^2\to0} \Im_s^{\pi\pi} \Pi_{30} \big|_{S\text{-waves}}^\mathrm{disc} \\
			&= - \frac{\sigma_\pi(s)}{4\pi} \frac{1}{\lambda(s,q_1^2,q_2^2)(s-q_3^2)} \sum_I \\
			&\quad \cdot \mathcal{S} \Bigg[ \left( \frac{s - q_1^2 - q_2^2}{\sqrt{q_1^2 q_2^2}} h_{00}^{0,(I)}(s,q_1^2,q_2^2) - 2 h_{++}^{0,(I)}(s,q_1^2,q_2^2) \right) h_{++}^{0,(I)}(s,q_3^2,0)^* \Bigg] + \ldots ,
	\end{split}
\end{align} }%
where the operator $\mathcal{S}$ takes care of the symmetry factor and the subtraction of the pole $\times$ pole terms in the products of partial-waves (the box-contributions are treated exactly and not in the partial-wave expansion):
\begin{align}
	\begin{split}
		\mathcal{S}\Big[ h_{\lambda_1\lambda_2}^{l,(+-)}(s,q_1^2,q_2^2) h_{\lambda_3\lambda_4}^{l,(+-)}(s,q_3^2,q_4^2)^* \Big] &:= h_{\lambda_1\lambda_2}^{l,(+-)}(s,q_1^2,q_2^2) h_{\lambda_3\lambda_4}^{l,(+-)}(s,q_3^2,q_4^2)^* \\
			&\quad - N_{\lambda_1\lambda_2}^{l}(s,q_1^2,q_2^2) N_{\lambda_3\lambda_4}^{l}(s,q_3^2,q_3^2)^* , \\
		\mathcal{S}\Big[ h_{\lambda_1\lambda_2}^{l,(00)}(s,q_1^2,q_2^2) h_{\lambda_3\lambda_4}^{l,(00)}(s,q_3^2,q_4^2)^* \Big] &:= \frac{1}{2} h_{\lambda_1\lambda_2}^{l,(00)}(s,q_1^2,q_2^2) h_{\lambda_3\lambda_4}^{l,(00)}(s,q_3^2,q_4^2)^* ,
	\end{split}
\end{align}
where $N_{\lambda_i\lambda_j}^l$ denotes the partial-wave projection of the pure pion-pole term.

The imaginary parts in the $t$- and $u$-channel are easily obtained by crossing symmetry:
{\small
\begin{align}
	\begin{split}
		\lim_{q_4^2\to0} \Im_t^{\pi\pi} \Pi_5 \big|_{S\text{-waves}}^\mathrm{disc} &= \frac{\sigma_\pi(t)}{4\pi} \frac{1}{\lambda(t,q_1^2,q_3^2)(t-q_2^2)} \sum_I \\
			&\quad \cdot \mathcal{S} \begin{aligned}[t]
				& \Bigg[ \left( 2 \sqrt{q_1^2 q_3^2} h_{00}^{0,(I)}(t,q_1^2,q_3^2) - (t - q_1^2 - q_3^2) h_{++}^{0,(I)}(t,q_1^2,q_3^2) \right) h_{++}^{0,(I)}(t,q_2^2,0)^* \\
				& + \frac{1}{2} ( t + q_1^2 + q_3^2 ) \\
				& \cdot \Bigg( \frac{t - q_1^2 - q_3^2}{\sqrt{q_1^2 q_3^2}} h_{00}^{0,(I)}(t,q_1^2,q_3^2) - 2 h_{++}^{0,(I)}(t,q_1^2,q_3^2) \Bigg) h_{++}^{0,(I)}(t,q_2^2,0)^* \Bigg] + \ldots, \end{aligned} \\
		\lim_{q_4^2\to0} \Im_t^{\pi\pi} \Pi_{15} \big|_{S\text{-waves}}^\mathrm{disc} &= \lim_{q_4^2\to0} \Im_t^{\pi\pi} \Pi_{16} \big|_{S\text{-waves}}^\mathrm{disc} = \lim_{q_4^2\to0} \Im_t^{\pi\pi} \Pi_{27} \big|_{S\text{-waves}}^\mathrm{disc} = \lim_{q_4^2\to0} \Im_t^{\pi\pi} \Pi_{28} \big|_{S\text{-waves}}^\mathrm{disc} \\
			&= - \frac{\sigma_\pi(t)}{4\pi} \frac{1}{\lambda(t,q_1^2,q_3^2)(t-q_2^2)} \sum_I \\
			&\quad \cdot \mathcal{S} \Bigg[ \left( \frac{t - q_1^2 - q_3^2}{\sqrt{q_1^2 q_3^2}} h_{00}^{0,(I)}(t,q_1^2,q_3^2) - 2 h_{++}^{0,(I)}(t,q_1^2,q_3^2) \right) h_{++}^{0,(I)}(t,q_2^2,0)^* \Bigg] + \ldots , \\
		\lim_{q_4^2\to0} \Im_u^{\pi\pi} \Pi_6 \big|_{S\text{-waves}}^\mathrm{disc} &= \frac{\sigma_\pi(u)}{4\pi} \frac{1}{\lambda(u,q_2^2,q_3^2)(u-q_1^2)} \sum_I \\
			&\quad \cdot \mathcal{S} \begin{aligned}[t]
				& \Bigg[ \left( 2 \sqrt{q_2^2 q_3^2} h_{00}^{0,(I)}(u,q_2^2,q_3^2) - (u - q_2^2 - q_3^2) h_{++}^{0,(I)}(u,q_2^2,q_3^2) \right) h_{++}^{0,(I)}(u,q_1^2,0)^* \\
				& + \frac{1}{2} ( u + q_2^2 + q_3^2 ) \\
				& \cdot \Bigg( \frac{u - q_2^2 - q_3^2}{\sqrt{q_2^2 q_3^2}} h_{00}^{0,(I)}(u,q_2^2,q_3^2) - 2 h_{++}^{0,(I)}(u,q_2^2,q_3^2) \Bigg) h_{++}^{0,(I)}(u,q_1^2,0)^* \Bigg] + \ldots, \end{aligned} \\
		\lim_{q_4^2\to0} \Im_u^{\pi\pi} \Pi_{11} \big|_{S\text{-waves}}^\mathrm{disc} &= \lim_{q_4^2\to0} \Im_u^{\pi\pi} \Pi_{12} \big|_{S\text{-waves}}^\mathrm{disc} = \lim_{q_4^2\to0} \Im_u^{\pi\pi} \Pi_{23} \big|_{S\text{-waves}}^\mathrm{disc} = \lim_{q_4^2\to0} \Im_u^{\pi\pi} \Pi_{24} \big|_{S\text{-waves}}^\mathrm{disc} \\
			&= - \frac{\sigma_\pi(u)}{4\pi} \frac{1}{\lambda(u,q_2^2,q_3^2)(u-q_1^2)} \sum_I \\
			&\quad \cdot \mathcal{S} \Bigg[ \left( \frac{u - q_2^2 - q_3^2}{\sqrt{q_2^2 q_3^2}} h_{00}^{0,(I)}(u,q_2^2,q_3^2) - 2 h_{++}^{0,(I)}(u,q_2^2,q_3^2) \right) h_{++}^{0,(I)}(u,q_1^2,0)^* \Bigg] + \ldots .
	\end{split}
\end{align} }%

We insert the imaginary parts into the dispersion relation (\ref{eq:PWApproximateDispersionRelationBTT}) and subsequently calculate the contribution to the $(g-2)_\mu$ with the master formula (\ref{eq:MasterFormula3Dim}). The relevant hadronic scalar functions become
{\small
\begin{align}
	\begin{split}
		\bar \Pi_3 \big|_{S\text{-waves}}^\mathrm{disc} &= \frac{1}{\pi} \int_{4M_\pi^2}^\infty ds^\prime \frac{1}{s^\prime - s} \frac{4}{\lambda(s^\prime,q_1^2,q_2^2)(s^\prime-s)} \\
			&\quad \cdot \begin{aligned}[t]
				& \Bigg[ \Big( 2 q_1^2 q_2^2 \Im h_{00,++}^0(s^\prime;q_1^2,q_2^2;s,0) - (s^\prime - q_1^2 - q_2^2) \Im h_{++,++}^0(s^\prime;q_1^2,q_2^2;s,0) \Big) \\
				& + \frac{1}{2} ( s^\prime + q_1^2 + q_2^2 ) \Big( (s^\prime - q_1^2 - q_2^2) \Im h_{00,++}^0(s^\prime;q_1^2,q_2^2;s,0) - 2 \Im h_{++,++}^0(s^\prime;q_1^2,q_2^2;s,0) \Big) \Bigg] , \end{aligned} \\
		\bar \Pi_4 \big|_{S\text{-waves}}^\mathrm{disc} &= \frac{1}{\pi} \int_{4M_\pi^2}^\infty dt^\prime \frac{1}{t^\prime - q_2^2} \frac{4}{\lambda(t^\prime,s,q_1^2)(t^\prime-q_2^2)} \\
			&\quad \cdot \begin{aligned}[t]
				& \Bigg[ \left( 2 s q_1^2 \Im h_{00,++}^0(t^\prime;q_1^2,s;q_2^2,0) - (t^\prime - s - q_1^2) \Im h_{++,++}^0(t^\prime;q_1^2,s;q_2^2,0) \right) \\
				& + \frac{1}{2} ( t^\prime + s + q_1^2 ) \Big( (t^\prime - s - q_1^2) \Im h_{00,++}^0(t^\prime;q_1^2,s;q_2^2,0) - 2 \Im h_{++,++}^0(t^\prime;q_1^2,s;q_2^2,0) \Big) \Bigg] , \end{aligned} \\
		\bar \Pi_8 \big|_{S\text{-waves}}^\mathrm{disc} &= - \frac{1}{\pi} \int_{4M_\pi^2}^\infty dt^\prime \frac{1}{t^\prime - q_2^2} \frac{4}{\lambda(t^\prime,s,q_1^2)(t^\prime-q_2^2)} \\
			&\quad \cdot  \bigg[ (t^\prime - s - q_1^2) \Im h_{00,++}^0(t^\prime;q_1^2,s;q_2^2,0) - 2 \Im h_{++,++}^0(t^\prime;q_1^2,s;q_2^2,0) \bigg] , \\
		\bar \Pi_9 \big|_{S\text{-waves}}^\mathrm{disc} &= - \frac{1}{\pi} \int_{4M_\pi^2}^\infty ds^\prime \frac{1}{s^\prime - s} \frac{4}{\lambda(s^\prime,q_1^2,q_2^2)(s^\prime-s)} \\
			&\quad \cdot \bigg[ (s^\prime - q_1^2 - q_2^2) \Im h_{00,++}^0(s^\prime;q_1^2,q_2^2;s,0) - 2 \Im h_{++,++}^0(s^\prime;q_1^2,q_2^2;s,0) \bigg] , \\
	\end{split}
\end{align} }%
where
\begin{align}
	\begin{split}
		\Im h_{\lambda_1\lambda_2,\lambda_3\lambda_4}^{l}(s;q_1^2,q_2^2;q_3^2,q_4^2) &:= \frac{\sigma_\pi(s)}{16\pi} \frac{1}{\xi_{\lambda_1}\xi_{\lambda_2}\xi_{\lambda_3}\xi_{\lambda_4}} \sum_I \mathcal{S}\Big[ h_{\lambda_1\lambda_2}^{l,(I)}(s,q_1^2,q_2^2)  h_{\lambda_3\lambda_4}^{l,(I)}(s,q_3^2,q_4^2)^* \Big] , \\
		\xi_{\lambda_i} &:= \left\{ \begin{matrix} \sqrt{q_i^2},  & \lambda_i = 0 , \\ 1,  & \lambda_i = \pm . \end{matrix} \right.
	\end{split}
\end{align}
The kinematic variables are evaluated at
\begin{align}
	\begin{split}
		q_1^2 = - Q_1^2,  \quad q_2^2 = - Q_2^2, \quad s = - Q_3^2 = -Q_1^2 - 2 Q_1 Q_2 \tau - Q_2^2 .
	\end{split}
\end{align}

\subsubsection{$D$-Wave $\pi\pi$-Rescattering Contribution}

The partial-wave framework is not limited to $S$-waves, hence we can take into account $D$-waves of the sub-process as well. However, the treatment becomes technically much more involved. First of all, the unitarity relation does not factorise as in the case of $S$-waves. The sub-process is given by (\ref{eq:DWavesSubProcessScalarFunctions}). Now, the Lorentz structures of the sub-process depend explicitly on the momenta of the intermediate pions. Therefore, we need to compute the following phase-space integrals (we abbreviate $Q:=q_1+q_2$):
\begin{align}
	\begin{split}
		I_0 &:= \int \widetilde{dp}_1 \widetilde{dp}_2 (2\pi)^4 \delta^{(4)}\big(Q - p_1 - p_2\big) , \\
		I_1^{\mu} &:= \int \widetilde{dp}_1 \widetilde{dp}_2 (2\pi)^4 \delta^{(4)}\big(Q - p_1 - p_2\big) p_1^\mu, \\
		I_2^{\mu\nu} &:= \int \widetilde{dp}_1 \widetilde{dp}_2 (2\pi)^4 \delta^{(4)}\big(Q - p_1 - p_2\big) p_1^\mu p_1^\nu , \\
		I_3^{\mu\nu\lambda} &:= \int \widetilde{dp}_1 \widetilde{dp}_2 (2\pi)^4 \delta^{(4)}\big(Q - p_1 - p_2\big) p_1^\mu p_1^\nu p_1^\lambda , \\
		I_4^{\mu\nu\lambda\sigma} &:= \int \widetilde{dp}_1 \widetilde{dp}_2 (2\pi)^4 \delta^{(4)}\big(Q - p_1 - p_2\big) p_1^\mu p_1^\nu p_1^\lambda p_1^\sigma .
	\end{split}
\end{align}
We use a tensor decomposition
\begin{align}
	\begin{split}
		 I_1^\mu &= Q^\mu f_1(Q^2) , \\
		 I_2^{\mu\nu} &= g^{\mu\nu} g_1(Q^2) + Q^\mu Q^\nu g_2(Q^2) , \\
		 I_3^{\mu\nu\lambda} &= \left( g^{\mu\nu} Q^\lambda + g^{\mu\lambda} Q^\nu + g^{\nu\lambda} Q^\mu \right) h_1(Q^2) + Q^\mu Q^\nu Q^\lambda h_2(Q^2) , \\
		 I_4^{\mu\nu\lambda\sigma} &= \left( g^{\mu\nu} g^{\lambda\sigma} + g^{\mu\lambda} g^{\nu\sigma} + g^{\mu\sigma} g^{\nu\lambda} \right) i_1(Q^2) \\
		 	& \quad + \begin{aligned}[t]
				& \Big( g^{\mu\nu} Q^\lambda Q^\sigma + g^{\mu\lambda} Q^\nu Q^\sigma + g^{\mu\sigma} Q^\nu Q^\lambda \\
				& + g^{\nu\lambda} Q^\mu Q^\sigma + g^{\nu\sigma} Q^\mu Q^\lambda + g^{\lambda\sigma} Q^\mu Q^\nu \Big) i_2(Q^2) \end{aligned} \\
			& \quad + Q^\mu Q^\nu Q^\lambda Q^\sigma i_3(Q^2)
	\end{split}
\end{align}
and find the results for the scalar integrals:
\begin{align}
	\begin{split}
		I_0 &= \frac{1}{16\pi^2} \int d\Omega \int_0^\infty dp \frac{p^2}{M_\pi^2 + p^2} \delta(Q^0 - 2\sqrt{M_\pi^2 + p^2}) \\
			&= \frac{1}{4\pi} \int_0^\infty dp  \frac{p}{2\sqrt{M_\pi^2 + p^2}} \delta\left(p - \sqrt{\frac{(Q^0)^2}{4} - M_\pi^2}\right) = \frac{1}{8\pi} \sigma_\pi(s) , \\
		f_1(Q^2) &= \frac{Q_\mu}{Q^2} I_1^\mu = \frac{1}{16\pi} \sigma_\pi(s) , \\
		g_1(Q^2) &= \frac{Q^2 g_{\mu\nu} - Q_\mu Q_\nu}{3 Q^2} I_2^{\mu\nu} = \frac{1}{24\pi} \sigma_\pi(s) \left(M_\pi^2 - \frac{s}{4}\right) , \\
		g_2(Q^2) &= \frac{4 Q_\mu Q_\nu - Q^2 g_{\mu\nu}}{3 (Q^2)^2} I_2^{\mu\nu} =  \frac{1}{24\pi} \sigma_\pi(s) \left( 1 - \frac{M_\pi^2}{s} \right) , \\
		h_1(Q^2) &= \frac{1}{48\pi} \sigma_\pi(s) \left( M_\pi^2 - \frac{s}{4} \right) , \\
		h_2(Q^2) &= \frac{1}{32\pi} \sigma_\pi(s) \left( 1 - \frac{2M_\pi^2}{s} \right) , \\
		i_1(Q^2) &= \frac{1}{1920\pi} \sigma_\pi(s) \left( s - 4 M_\pi^2 \right)^2 , \\
		i_2(Q^2) &= - \frac{1}{960\pi} \sigma_\pi(s) \left( \frac{8M_\pi^4}{s} - 14 M_\pi^2 + 3 s \right) , \\
		i_3(Q^2) &= \frac{1}{40\pi} \sigma_\pi(s) \frac{s^2 - 3 M_\pi^2 s + M_\pi^4}{s^2} .
	\end{split}
\end{align}
This allows us to compute the phase-space integral for the unitarity relation. Because the result is very large, we do not present it here. Conceptually, the calculation of the contribution to the $(g-2)_\mu$ is completely analogous to the case of $S$-waves. The imaginary parts can be projected onto the scalar functions. The real parts are then reconstructed by dispersion integrals. The integrand is a linear combination of products of helicity partial waves for $\gamma^*\gamma^*\to\pi\pi$.


\chapter{Discussion and Conclusion}

\label{sec:HLbLDiscussionConclusion}

\section{Summary}

In the previous chapters, we have derived a decomposition of the hadronic light-by-light tensor into kinematic-free scalar functions (\ref{eqn:HLbLTensorKinematicFreeStructures}), which allowed us to derive a master formula (\ref{eq:MasterFormula3Dim}) for the HLbL contribution to the anomalous magnetic moment of the muon $(g-2)_\mu$. The scalar functions fulfil a Mandelstam representation. By considering the lowest-lying intermediate states, we have split the HLbL contribution to the $(g-2)_\mu$ into three parts classified by unitarity topologies of the scalar functions:
\begin{itemize}
	\item pion-pole contribution,
	\item box topologies ($\pi\pi$ intermediate states in two channels),
	\item $\pi\pi$-rescattering contribution, approximated in a partial-wave expansion.
\end{itemize}
This treatment is based on fundamental principles of particle physics: gauge invariance and crossing symmetry is already implemented in the decomposition of the HLbL tensor into scalar functions. The dispersive description uses analyticity and unitarity to establish a relation between the three HLbL contributions and different on-shell quantities. These on-shell quantities are in principle either experimentally accessible or can be reconstructed from data with dispersion methods. They will serve as an input for a numerical evaluation of the HLbL contribution to the $(g-2)_\mu$:
\begin{itemize}
	\item the pion-pole is parametrised by the pion transition form factor $\mathcal{F}_{\pi^0\gamma^*\gamma^*}$ (with an on-shell pion),
	\item the box topologies use pion vector form factors $F_\pi^V$ as an input,
	\item the $\pi\pi$-rescattering contribution is written in terms of $\gamma^*\gamma^*\to\pi\pi$ helicity partial waves.
\end{itemize}
These three input quantities are needed for negative virtualities of the off-shell photons. The presented treatment of HLbL scattering shows a path towards a data-driven and thus less model-dependent evaluation of the $(g-2)_\mu$.

In \cite{Colangelo2014b}, we have discussed which experimentally accessible processes can help to constrain and reconstruct the input of this dispersive treatment in the absence of data on the doubly-virtual photon-pion interactions.

\pagebreak

\section{Possible Extensions of the Dispersive Treatment}

At the beginning of chapter~\ref{sec:MandelstamRepresentation}, we have listed the two approximations that we make in our dispersive framework. Let us recall them:
\begin{enumerate}
	\item limitation to lightest intermediate states ($\pi^0$ and $\pi\pi$) in the direct channel;
	\item partial-wave approximation of the intermediate states higher than $\pi\pi$ in the crossed channels.
\end{enumerate}
An extension of the presented dispersive treatment is possible with respect to both of these approximations, though within certain limits.

Concerning the intermediate states in the direct channel, it is straightforward to include higher pseudoscalar poles, i.e.~the $\eta$ and $\eta^\prime$ mesons by just adding their contribution in complete analogy to the $\pi^0$ pole. The input quantities will be the transition form factors of these heavier pseudoscalars. Although these mesons are unstable in QCD, their decay width is certainly small enough to justify the treatment as a pure pole.

A bit more difficult is the inclusion of higher two-particle intermediate states: an extension to e.g.~$\pi^0\eta$, $K^+K^-$, $K^0\bar K^0$ seems still straightforward, but the situation concerning experimental input will be much worse and the preparation of the partial-wave input will probably require a coupled-channel analysis.

Another possible extension would be the explicit treatment of higher intermediate states in the crossed channels, which are in principle implemented in the Mandelstam representation: with the help of the anomalous process $\gamma^*\pi\to\pi\pi$, one could construct a spectral density serving as an input for the discontinuity in the sub-process. Of course, in the preparation of the partial-wave input for the rescattering contribution, such contributions would have to be subtracted in the way we subtract now the pion-pole.

Contributions of intermediate states of more than two particles in the direct channel cannot be treated by a Mandelstam representation. It is not clear if a model-independent treatment is possible at all. However, here we are talking about contributions with intermediate states of more than two particles in two channels simultaneously, such that we hope that higher thresholds and phase-space suppression lead to such a small numerical relevance that this residual model-dependence will not be an issue.

\section{Outlook}

The presented dispersive formalism certainly opens a new window to a data-driven evaluation of the HLbL contribution to the $(g-2)_\mu$. It should provide a model-independent approach that can be improved systematically, in a similar way as it is possible in the case of hadronic vacuum polarisation already now. Nevertheless, the treatment is not yet complete, and a couple of issues are work in progress or the subject of future investigation. Most importantly, the numerical evaluation of all the contributions has to show which input quantities will be the largest source of uncertainty. Such an analysis will finally show where new experimental data would have the largest impact concerning the improvement of the theoretical prediction of the $(g-2)_\mu$.

\begin{appendices}
	\numberwithin{equation}{chapter}


\chapter{Born Contribution to $\gamma^*\gamma^*\to\pi^+\pi^-$}

\label{sec:AppendixScalarQEDBornggpipi}

Here, we calculate the Born contribution to $\gamma^*\gamma^*\to\pi^+\pi^-$ in scalar QED (tree level in \ChPT{}).

The usual LO \ChPT{} Feynman rules for the vertex $\gamma^*(q,\lambda) \to \pi^+(p_1) \pi^-(p_2)$ and the seagull vertex $\gamma^*(q_1,\lambda_1)\gamma^*(q_2,\lambda_2)\to\pi^+(p_1)\pi^-(p_2)$ read
\begin{align}
	\begin{split}
		\minidiagSize{HLbL}{gpipiVertex}{1.5cm} = i e (p_2^\mu - p_1^\mu) , \qquad \minidiagSize{HLbL}{ggpipiVertex}{1.5cm} = 2 i e^2 g^{\mu\nu} .
	\end{split}
\end{align}
With these rules, we easily calculate the sQED Born contribution to $\gamma^*\gamma^*\to\pi^+\pi^-$:
\begin{align}
	\begin{split}
		i e^2 W^{\mu\nu}_\mathrm{Born} &= \minidiagSize{HLbL}{SubTPoleFeyn}{2cm} +  \minidiagSize{HLbL}{SubUPoleFeyn}{2cm} + \minidiagSize{HLbL}{SubSeagullFeyn}{2cm} \\
			&= i e^2 (2 p_1^\mu - q_1^\mu)(2p_2^\nu - q_2^\nu) \frac{1}{t-M_\pi^2} + i e^2 (2p_2^\mu - q_1^\mu)(2p_1^\nu - q_2^\nu) \frac{1}{u - M_\pi^2} + 2 i e^2 g^{\mu\nu} .
	\end{split}
\end{align}
We can read off the values of the scalar functions:
\begin{align}
	\begin{split}
		A_1^\mathrm{Born} &= - \left( \frac{1}{t-M_\pi^2} + \frac{1}{u-M_\pi^2}\right) , \\
		A_5^\mathrm{Born} &= - \frac{2}{s - q_1^2 - q_2^2} \left( \frac{1}{t-M_\pi^2} + \frac{1}{u-M_\pi^2}\right) , \\
		A_2^\mathrm{Born} &= A_3^\mathrm{Born} = A_4^\mathrm{Born} = A_6^\mathrm{Born} = 0 .
	\end{split}
\end{align}


\chapter{Integral Kernels}

\section{Intermediate Kernels}

\label{sec:AppendixIntermediateKernels}

After calculating the trace and performing the contraction of the Lorentz indices in (\ref{eq:DefinitionIntermediateKernels}), one finds the following integral kernels:
\begin{align}
	\small
	\begin{split}
		\hat T_1(q_1,q_2;p) &= -\frac{8}{3} \left((q_1 \cdot q_2)^2-q_1^2 q_2^2\right) m_\mu^2-\frac{8}{3} q_2^2 (p \cdot q_1)^2-\frac{8}{3} q_1^2 (p \cdot q_2)^2 \\
			&\quad -\frac{4}{3} q_1^2 p \cdot q_2 \left(q_2^2+q_1 \cdot q_2\right)+p \cdot q_1 \left(\frac{4}{3} \left(q_1^2+q_1 \cdot q_2\right) q_2^2+\frac{16}{3} p \cdot q_2 q_1 \cdot q_2\right) , \\
   		\hat T_2(q_1,q_2;p) &= -\frac{8}{3} \left((q_1 \cdot q_2)^2-q_1^2 q_2^2\right) m_\mu^2-\frac{8}{3} q_1^2 (p \cdot q_2)^2+p \cdot q_1\left(\frac{8}{3} p \cdot q_2 q_1 \cdot q_2-\frac{4}{3} q_2^2 q_1 \cdot q_2\right) \\
			&\quad +p \cdot q_2 \left(4 q_1^2 q_2^2-\frac{8}{3} (q_1 \cdot q_2)^2\right) , \\
   		\hat T_3(q_1,q_2;p) &= -\frac{8}{3} \left((q_1 \cdot q_2)^2-q_1^2 q_2^2\right) m_\mu^2-\frac{8}{3} q_2^2 (p \cdot q_1)^2+\frac{4}{3} q_1^2 p \cdot q_2 q_1 \cdot q_2 \\
			&\quad +p \cdot q_1 \left(-4 q_1^2 q_2^2+\frac{8}{3} (q_1 \cdot q_2)^2+\frac{8}{3} p \cdot q_2 q_1 \cdot q_2\right) , \\
		\hat T_4(q_1,q_2;p) &= \frac{8}{3} q_1 \cdot q_2 \left(q_1^2+q_2^2+2 q_1 \cdot q_2\right) m_\mu^2+\frac{8}{3} q_2^2 (p \cdot q_1)^2+\frac{8}{3} q_1^2 (p \cdot q_2)^2 \\
			&\quad -\frac{8}{3} p \cdot q_1 p \cdot q_2 \left(q_1^2+q_2^2+4 q_1 \cdot q_2\right) , \\
		\hat T_5(q_1,q_2;p) &= \frac{8}{3} (p \cdot q_1)^2 q_2^2-\frac{8}{3} m_\mu^2 \left(q_1^2+q_1 \cdot q_2\right) q_2^2+\frac{8}{3} q_1^2 (p \cdot q_2)^2-\frac{4}{3} p \cdot q_2 q_1 \cdot q_2 \left(3 q_1^2+2 q_1 \cdot q_2\right) \\
			&\quad +p \cdot q_1 \left(\frac{4}{3} q_2^2 \left(3 q_1^2+2 q_1 \cdot q_2\right)-\frac{8}{3} p \cdot q_2 \left(q_1 \cdot q_2-q_2^2\right)\right) , \\
		\hat T_6(q_1,q_2;p) &= \frac{8}{3} (p \cdot q_2)^2 q_1^2-\frac{8}{3} m_\mu^2 \left(q_2^2+q_1 \cdot q_2\right) q_1^2-\frac{4}{3} p \cdot q_2 \left(3 q_2^2+2 q_1 \cdot q_2\right) q_1^2 \\
			&\quad +\frac{8}{3} q_2^2 (p \cdot q_1)^2+p \cdot q_1 \left(\frac{8}{3} p \cdot q_2 \left(q_1^2-q_1 \cdot q_2\right)+\frac{4}{3} q_1 \cdot q_2 \left(3 q_2^2+2 q_1 \cdot q_2\right)\right) , \\
		\hat T_7(q_1,q_2;p) &= \frac{4}{3} q_1 \cdot q_2 \left(2 q_1^4+\left(q_2^2+4 q_1 \cdot q_2\right) q_1^2+(q_1 \cdot q_2)^2\right) m_\mu^2-\frac{4}{3} q_2^2 (p \cdot q_1)^2 q_1 \cdot q_2 \\
			&\quad +\frac{4}{3} q_1^2 (p \cdot q_2)^2 \left(2 q_1^2+q_1 \cdot q_2\right)-\frac{8}{3} p \cdot q_1 p \cdot q_2 \left(q_1^4+3 q_1 \cdot q_2 q_1^2+(q_1 \cdot q_2)^2\right) , \\
		\hat T_8(q_1,q_2;p) &= \frac{4}{3} q_1 \cdot q_2 \left(2 q_2^4+\left(q_1^2+4 q_1 \cdot q_2\right) q_2^2+(q_1 \cdot q_2)^2\right) m_\mu^2-\frac{4}{3} q_1^2 (p \cdot q_2)^2 q_1 \cdot q_2 \\
			&\quad +\frac{4}{3} q_2^2 (p \cdot q_1)^2 \left(2 q_2^2+q_1 \cdot q_2\right)-\frac{8}{3} p \cdot q_1 p \cdot q_2 \left(q_2^4+3 q_1 \cdot q_2 q_2^2+(q_1 \cdot q_2)^2\right) , \\
	\end{split}
\end{align}
\begin{align}
	\small
	\begin{split}
		\hat T_9(q_1,q_2;p) &= -\frac{4}{3} \left(q_1^2+q_1 \cdot q_2\right) \left(2 q_2^4+\left(q_1^2+4 q_1 \cdot q_2\right) q_2^2+(q_1 \cdot q_2)^2\right) m_\mu^2 \\
			&\quad +\frac{4}{3} q_1^2 (p \cdot q_2)^2 \left(q_1^2+q_1 \cdot q_2\right) -\frac{4}{3} p \cdot q_2 \left(q_1^2+q_1 \cdot q_2\right) \left(q_2^2+q_1 \cdot q_2\right) \left(3 q_1^2+2 q_1 \cdot q_2\right) \\
			&\quad +\frac{4}{3} (p \cdot q_1)^2 \left(2 q_2^4+\left(q_1^2+5 q_1 \cdot q_2\right) q_2^2+2 (q_1 \cdot q_2)^2\right) \\
			&\quad +p \cdot q_1 \left(\frac{4}{3} \left(3 q_1^2+2 q_1 \cdot q_2\right) \left(q_2^2+q_1 \cdot q_2\right)^2+\frac{8}{3} p \cdot q_2 \left(q_2^2 \left(q_2^2+2 q_1 \cdot q_2\right)-q_1^2 q_1 \cdot q_2\right)\right) , \\
		\hat T_{10}(q_1,q_2;p) &= -\frac{4}{3} \left(q_1^2+q_1 \cdot q_2\right) \left(q_1^2 q_2^2+(q_1 \cdot q_2)^2\right) m_\mu^2+\frac{4}{3} q_1^2 (p \cdot q_2)^2 \left(q_1^2-q_1 \cdot q_2\right) \\
			&\quad -\frac{4}{3} q_1^2 p \cdot q_2 q_1 \cdot q_2 \left(3 q_1^2+2 q_1 \cdot q_2\right)+\frac{4}{3} (p \cdot q_1)^2 \left(\left(q_1^2+q_1 \cdot q_2\right) q_2^2+2 (q_1 \cdot q_2)^2\right) \\
			&\quad +p \cdot q_1 \left(\frac{4}{3} \left(3 q_1^2+2 q_1 \cdot q_2\right) (q_1 \cdot q_2)^2+\frac{8}{3} p \cdot q_2 \left(q_1 \cdot q_2-q_1^2\right) q_1 \cdot q_2\right) , \\
		\hat T_{11}(q_1,q_2;p) &= -\frac{8}{3} m_\mu^2 q_1 \cdot q_2 \left(q_1^2+q_1 \cdot q_2\right) q_1^2-\frac{4}{3} p \cdot q_2 \left(\left(q_2^2+2 q_1 \cdot q_2\right) q_1^2+2 (q_1 \cdot q_2)^2\right) q_1^2 \\
			&\quad +p \cdot q_1 \left(\frac{8}{3} p \cdot q_2 \left(q_1^2+q_1 \cdot q_2\right) q_1^2+\frac{4}{3} q_1 \cdot q_2 \left(\left(q_2^2+2 q_1 \cdot q_2\right) q_1^2+2 (q_1 \cdot q_2)^2\right)\right) , \\
		\hat T_{12}(q_1,q_2;p) &= -\frac{4}{3} \left(q_2^2+q_1 \cdot q_2\right) \left(2 q_1^4+\left(q_2^2+4 q_1 \cdot q_2\right) q_1^2+(q_1 \cdot q_2)^2\right) m_\mu^2 \\
			&\quad +\frac{4}{3} q_2^2 (p \cdot q_1)^2 \left(q_2^2+q_1 \cdot q_2\right)-\frac{4}{3} p \cdot q_2 \left(q_1^2+q_1 \cdot q_2\right)^2 \left(3 q_2^2+2 q_1 \cdot q_2\right) \\
			&\quad +\frac{4}{3} (p \cdot q_2)^2 \left(2 q_1^4+\left(q_2^2+5 q_1 \cdot q_2\right) q_1^2+2 (q_1 \cdot q_2)^2\right) \\
			&\quad +p \cdot q_1 \begin{aligned}[t]
				& \bigg(\frac{4}{3} \left(q_1^2+q_1 \cdot q_2\right) \left(q_2^2+q_1 \cdot q_2\right) \left(3 q_2^2+2 q_1 \cdot q_2\right) \\
				& +\frac{8}{3} p \cdot q_2 \left(q_1^4+2 q_1 \cdot q_2 q_1^2-q_2^2 q_1 \cdot q_2\right)\bigg) , \end{aligned} \\
		\hat T_{13}(q_1,q_2;p) &= -\frac{4}{3} \left(q_2^2+q_1 \cdot q_2\right) \left(q_1^2 q_2^2+(q_1 \cdot q_2)^2\right) m_\mu^2+\frac{4}{3} q_2^2 (p \cdot q_1)^2 \left(q_2^2-q_1 \cdot q_2\right) \\
			&\quad -\frac{4}{3} p \cdot q_2 (q_1 \cdot q_2)^2 \left(3 q_2^2+2 q_1 \cdot q_2\right)+\frac{4}{3} (p \cdot q_2)^2 \left(\left(q_2^2+q_1 \cdot q_2\right) q_1^2+2 (q_1 \cdot q_2)^2\right) \\
			&\quad +p \cdot q_1 \left(\frac{4}{3} q_1 \cdot q_2 \left(3 q_2^2+2 q_1 \cdot q_2\right) q_2^2+\frac{8}{3} p \cdot q_2 q_1 \cdot q_2 \left(q_1 \cdot q_2-q_2^2\right)\right) , \\
		\hat T_{14}(q_1,q_2;p) &= -\frac{8}{3} m_\mu^2 q_1 \cdot q_2 \left(q_2^2+q_1 \cdot q_2\right) q_2^2+p \cdot q_2 \left(-\frac{8}{3} (q_1 \cdot q_2)^3-\frac{4}{3} q_2^2 \left(q_1^2+2 q_1 \cdot q_2\right) q_1 \cdot q_2\right) \\
			&\quad +p \cdot q_1 \left(\frac{8}{3} p \cdot q_2 \left(q_2^2+q_1 \cdot q_2\right) q_2^2+\frac{4}{3} \left(\left(q_1^2+2 q_1 \cdot q_2\right) q_2^2+2 (q_1 \cdot q_2)^2\right) q_2^2\right) , \\
		\hat T_{15}(q_1,q_2;p) &= \frac{8}{3} m_\mu^2 \left(q_1^2+q_1 \cdot q_2\right) \left(q_2^2+q_1 \cdot q_2\right) \left(q_1^2+q_2^2+2 q_1 \cdot q_2\right) -\frac{8}{3} p \cdot q_1 p \cdot q_2 \left(q_1^2+q_2^2+2 q_1 \cdot q_2\right)^2 , \\
		\hat T_{16}(q_1,q_2;p) &= \frac{8}{3} \left(q_1^2+q_2^2+q_1 \cdot q_2\right) \left(q_1^2 q_2^2-(q_1 \cdot q_2)^2\right) m_\mu^2 -\frac{8}{3} q_2^2 (p \cdot q_1)^2 \left(q_1^2+q_2^2+q_1 \cdot q_2\right) \\
			&\quad -\frac{8}{3} q_1^2 (p \cdot q_2)^2 \left(q_1^2+q_2^2+q_1 \cdot q_2\right)+\frac{16}{3} p \cdot q_1 p \cdot q_2 q_1 \cdot q_2 \left(q_1^2+q_2^2+q_1 \cdot q_2\right) , \\
		\hat T_{17}(q_1,q_2;p) &= -\frac{4}{3} q_1^2 p \cdot q_2 \left(q_1^2+q_1 \cdot q_2\right) q_2^2+\frac{4}{3} m_\mu^2 \left(q_1^2 q_2^2-(q_1 \cdot q_2)^2\right) q_2^2 \\
			&\quad -\frac{4}{3} (p \cdot q_1)^2 \left(q_2^2+6 \left(q_1^2+q_1 \cdot q_2\right)\right) q_2^2+\frac{4}{3} q_1^2 (p \cdot q_2)^2 \left(q_2^2+2 q_1 \cdot q_2\right) \\
			&\quad +p \cdot q_1 \left(\frac{4}{3} q_1^2 \left(q_2^2+q_1 \cdot q_2\right) q_2^2+\frac{8}{3} p \cdot q_2 q_1 \cdot q_2 \left(3 q_1^2+2 q_1 \cdot q_2\right)\right) , \\
	\end{split}
\end{align}
\begin{align}
	\small
	\begin{split}
		\hat T_{18}(q_1,q_2;p) &= -\frac{4}{3} q_2^2 p \cdot q_2 \left(q_1^2+q_1 \cdot q_2\right) q_1^2+\frac{4}{3} m_\mu^2 \left(q_1^2 q_2^2-(q_1 \cdot q_2)^2\right) q_1^2 \\
			&\quad -\frac{4}{3} (p \cdot q_2)^2 \left(q_1^2+6 \left(q_2^2+q_1 \cdot q_2\right)\right) q_1^2+\frac{4}{3} q_2^2 (p \cdot q_1)^2 \left(q_1^2+2 q_1 \cdot q_2\right) \\
			&\quad +p \cdot q_1 \left(\frac{4}{3} q_1^2 \left(q_2^2+q_1 \cdot q_2\right) q_2^2+\frac{8}{3} p \cdot q_2 q_1 \cdot q_2 \left(3 q_2^2+2 q_1 \cdot q_2\right)\right) , \\
		\hat T_{19}(q_1,q_2;p) &= -\frac{4}{3} m_\mu^2 \left(q_1^2-q_2^2\right) \left(q_1^2 \left(4 q_1 \cdot q_2+q_2^2\right)+q_1 \cdot q_2 \left(7 q_1 \cdot q_2+4 q_2^2\right)\right) \\
			&\quad -\frac{8}{3} p \cdot q_2 \left(q_1^2-q_2^2\right) q_1 \cdot q_2 \left(q_1^2+q_1 \cdot q_2\right)+\frac{4}{3} q_2^2 (p \cdot q_1)^2 \left(q_1^2+2 q_1 \cdot q_2+q_2^2\right) \\
			&\quad -\frac{4}{3} q_1^2 (p \cdot q_2)^2 \left(q_1^2+2 q_1 \cdot q_2+q_2^2\right) \\
			&\quad +p \cdot q_1 \left(\frac{16}{3} p \cdot q_2 \left(q_1^2-q_2^2\right) \left(q_1^2+2 q_1 \cdot q_2+q_2^2\right)+\frac{8}{3} \left(q_1^2-q_2^2\right) q_1 \cdot q_2 \left(q_1 \cdot q_2+q_2^2\right)\right) .
	\end{split}
\end{align}

\section{Kernels for the Master Formula}

\label{sec:AppendixMasterFormulaKernels}

In the master formula (\ref{eq:MasterFormula3Dim}), the following integral kernels appear:
\begin{align}
	\small
	\begin{split}
		T_1 &= \frac{Q_1^2 \tau  \left(\sigma_1^E-1\right) \left(\sigma_1^E+5\right)+Q_2^2 \tau  \left(\sigma_2^E-1\right) \left(\sigma_2^E+5\right)+4 Q_1 Q_2 \left(\sigma_1^E+\sigma_2^E-2\right)-8 \tau  m_{\mu }^2}{2 Q_1 Q_2 Q_3^2 m_{\mu }^2} \\
			&\quad + X \left(\frac{8 \left(\tau ^2-1\right)}{Q_3^2}-\frac{4}{m_{\mu }^2}\right) , \\
		T_2 &= \frac{Q_1 \left(\sigma _1^E-1\right) \left(Q_1 \tau  \left(\sigma _1^E+1\right)+4 Q_2 \left(\tau ^2-1\right)\right)-4 \tau  m_{\mu}^2}{Q_1 Q_2 Q_3^2 m_{\mu }^2} +  X \frac{8 \left(\tau ^2-1\right) \left(2 m_{\mu }^2-Q_2^2\right)}{Q_3^2 m_{\mu }^2} , \\
		T_3 &= \frac{1}{Q_3^2} \begin{aligned}[t]
			& \bigg( -\frac{2 \left(\sigma _1^E+\sigma _2^E-2\right)}{m_{\mu }^2}-\frac{Q_1 \tau  \left(\sigma _1^E-1\right) \left(\sigma_1^E+7\right)}{2 Q_2 m_{\mu }^2} + \frac{8 \tau }{Q_1Q_2} \\
			& -\frac{Q_2 \tau  \left(\sigma _2^E-1\right) \left(\sigma _2^E+7\right)}{2 Q_1 m_{\mu }^2} + \frac{Q_1^2 \left(1-\sigma _1^E\right)}{Q_2^2 m_{\mu }^2} + \frac{Q_2^2 \left(1-\sigma _2^E\right)}{Q_1^2 m_{\mu}^2}+\frac{2}{Q_1^2}+\frac{2}{Q_2^2}\bigg) \end{aligned} \\
			&\quad + X \left(\frac{4}{m_{\mu }^2}-\frac{8 \tau }{Q_1 Q_2}\right) , \\
		T_4 &= \frac{1}{Q_3^2} \begin{aligned}[t]
			& \Bigg(\frac{4 \left(\tau ^2 \left(\sigma _1^E-1\right)+\sigma _2^E-1\right)}{m_{\mu }^2}-\frac{Q_1 \tau  \left(\sigma _1^E-5\right) \left(\sigma _1^E-1\right)}{Q_2 m_{\mu }^2} + \frac{4 \tau }{Q_1Q_2} \\
			& - \frac{Q_2 \tau  \left(\sigma _2^E-3\right) \left(\sigma_2^E-1\right)}{Q_1m_{\mu }^2} +\frac{2 Q_2^2 \left(\sigma _2^E-1\right)}{Q_1^2m_{\mu }^2}-\frac{4}{Q_1^2} \\
   			& +X \left(-\frac{8 Q_2^2 \tau^2}{m_{\mu }^2}-\frac{16 Q_2 Q_1 \tau }{m_{\mu }^2}-\frac{8 Q_1^2}{m_{\mu }^2}+\frac{16 Q_2 \tau }{Q_1}+16\right)\Bigg) , \end{aligned} \\
		T_5 &= \frac{1}{Q_3^2} \begin{aligned}[t]
			& \Bigg( Q_1^2 \left(\frac{\tau ^2 \left(\sigma _1^E-1\right) \left(\sigma _1^E+3\right)+4 \left(\sigma _1^E+\sigma _2^E-2\right)}{2 m_{\mu }^2} -\frac{4}{Q_2^2}\right) -\frac{Q_2^2 \tau ^2 \left(\sigma _2^E-5\right) \left(\sigma _2^E-1\right)}{2 m_{\mu}^2} \\
			& +\frac{Q_1^3 \tau  \left(\sigma _1^E-1\right) \left(\sigma _1^E+5\right)}{Q_2 m_{\mu }^2} + Q_1 \left(\frac{Q_2 \tau  \left(\sigma_1^E+5 \sigma _2^E-6\right)}{m_{\mu }^2}-\frac{12 \tau }{Q_2}\right)+\frac{2 Q_1^4 \left(\sigma _1^E-1\right)}{Q_2^2 m_{\mu }^2} \\
			&  - 4 \tau ^2 +X \begin{aligned}[t]
				& \Bigg(Q_1 \left(8 Q_2 \left(\tau ^3+\tau \right)-\frac{2 Q_2^3 \tau }{m_{\mu }^2}\right)+Q_1^2 \bigg(32 \tau ^2-\frac{4 Q_2^2 \left(\tau ^2+1\right)}{m_{\mu }^2}\bigg) \\
				& +Q_1^3 \left(\frac{16 \tau }{Q_2}-\frac{10 Q_2 \tau }{m_{\mu }^2}\right)-\frac{4 Q_1^4}{m_{\mu }^2}\Bigg) \Bigg) , \end{aligned} \end{aligned} \\
	\end{split}
\end{align}
\begin{align}
	\small
	\begin{split}
   		T_6 &= \frac{1}{Q_3^2} \begin{aligned}[t]
			& \Bigg(\frac{Q_1^2 \left(\tau ^2 \left(\left(\sigma _1^E-22\right) \sigma _1^E-8 \sigma _2^E+29\right)+2 \left(-5 \sigma _1^E+\sigma_2^E+4\right)\right)}{2 m_{\mu }^2} \\
			& +Q_1 \left(\frac{Q_2 \tau  \left(2 \tau ^2 \left(\left(\sigma _2^E-3\right)^2-4 \sigma_1^E\right)-26 \sigma _1^E+\sigma _2^E \left(\sigma _2^E-12\right)+37\right)}{2 m_{\mu }^2}-\frac{4 \tau }{Q_2}\right) \\
			& +\frac{Q_2^2 \left(\tau ^2 \left(-8 \sigma _1^E+\sigma _2^E \left(5 \sigma _2^E-26\right)+29\right)-4 \left(\sigma _1^E+2 \sigma_2^E-3\right)\right)}{2 m_{\mu }^2}+\frac{Q_1^3 \tau  \left(\sigma _1^E-9\right) \left(\sigma _1^E-1\right)}{2 Q_2 m_{\mu}^2} \\
			& +\frac{Q_2^3 \tau  \left(\sigma _2^E-9\right) \left(\sigma _2^E-1\right)}{Q_1m_{\mu }^2}+\frac{8 Q_2 \tau }{Q_1}+\frac{2 Q_2^4 \left(1- \sigma _2^E\right)}{Q_1^2 m_{\mu }^2}+\frac{4 Q_2^2}{Q_1^2} \\
			& +X \begin{aligned}[t]
				& \Bigg(\frac{Q_2 Q_1^3 \left(8 \tau ^3+22 \tau \right)}{m_{\mu}^2}+\frac{Q_1^4 \left(8 \tau ^2-2\right)}{m_{\mu }^2}+Q_1^2 \left(\frac{Q_2^2 \left(36 \tau ^2+18\right)}{m_{\mu }^2}-8 \left(\tau^2+1\right)\right) \\
				& +\frac{Q_2^4 \left(8 \tau ^2+4\right)}{m_{\mu }^2}+Q_1 \left(\frac{Q_2^3 \left(8 \tau ^3+34 \tau \right)}{m_{\mu}^2}-8 Q_2 \tau  \left(\tau ^2+5\right)\right) \\
				& -16 Q_2^2 \left(2 \tau ^2+1\right)-\frac{16 Q_2^3 \tau }{Q_1}\Bigg)\Bigg) , \end{aligned} \end{aligned} \\
   		T_7 &= \frac{1}{Q_3^2} \begin{aligned}[t]
			& \Bigg( \frac{Q_1^2 \left(2 \left(\sigma _1^E+\sigma _2^E-2\right)-\tau ^2 \left(\left(\sigma _1^E+10\right) \sigma _1^E+8 \sigma_2^E-19\right)\right)}{2 m_{\mu }^2} \\
			& +Q_1 \left(\frac{Q_2 \tau  \left(2 \tau ^2 \left(\sigma _2^E-5\right) \left(\sigma_2^E-1\right)-2 \sigma _1^E+\sigma _2^E \left(\sigma _2^E+4\right)-3\right)}{2 m_{\mu }^2}-\frac{4 \tau }{Q_2}\right) \\
			& +\frac{Q_2^2 \tau ^2 \left(\sigma _2^E-5\right) \left(\sigma _2^E-1\right)}{2 m_{\mu }^2}+\frac{Q_1^3 \tau  \left(\sigma _1^E-9\right) \left(\sigma _1^E-1\right)}{2 Q_2 m_{\mu }^2} + 4 \tau ^2 \\
			& +X \begin{aligned}[t]
				& \Bigg(\frac{Q_2 Q_1^3 \left(8 \tau ^3+6 \tau \right)}{m_{\mu }^2}+Q_1 \left(\frac{2 Q_2^3 \tau }{m_{\mu }^2}-8 Q_2 \left(\tau ^3+\tau \right)\right) \\
				& +\frac{Q_1^4 \left(8 \tau ^2-2\right)}{m_{\mu }^2}+Q_1^2 \left(\frac{2 Q_2^2 \left(6 \tau ^2-1\right)}{m_{\mu }^2}-8 \left(\tau ^2+1\right)\right)\Bigg) \Bigg) , \end{aligned} \end{aligned} \\
   		T_8 &= \frac{1}{Q_3^2} \begin{aligned}[t]
			& \Bigg( Q_1^2 \left(\frac{4}{Q_2^2}-\frac{2 \left(2 \tau ^2+1\right) \left(\sigma _1^E+\sigma _2^E-2\right)}{m_{\mu }^2}\right)+Q_1 \left(\frac{4 \tau }{Q_2}-\frac{4 Q_2 \tau  \left(\tau ^2+1\right) \left(\sigma _2^E-1\right)}{m_{\mu }^2}\right) \\
			& -\frac{6 Q_1^3 \tau \left(\sigma _1^E-1\right)}{Q_2 m_{\mu }^2}+\frac{Q_1^4 \left(2-2 \sigma _1^E\right)}{Q_2^2 m_{\mu }^2} \\
			& +X \left(\frac{Q_1^4 \left(8 \tau ^2+4\right)}{m_{\mu }^2}+Q_1^3 \left(\frac{8 Q_2 \tau  \left(\tau ^2+2\right)}{m_{\mu }^2}-\frac{16 \tau }{Q_2}\right)+Q_1^2 \left(\frac{Q_2^2 \left(8 \tau ^2+4\right)}{m_{\mu }^2}-16 \tau ^2\right)\right)\Bigg) , \end{aligned} \\ 
   		T_9 &=  Q_3^2 \left(\frac{\sigma _1^E-1}{Q_2^2 m_{\mu }^2}+\frac{\sigma _2^E-1}{Q_1^2 m_{\mu }^2}-\frac{2}{Q_1^2Q_2^2} \right)+X
   \left(-\frac{2 Q_3^2}{m_{\mu }^2}+\frac{8 Q_2 \tau }{Q_1}+\frac{8 Q_1 \tau }{Q_2}+8 \left(\tau ^2+1\right)\right) , \\
	\end{split}
\end{align}
\begin{align}
	\small
	\begin{split}
   		T_{10} &= \frac{1}{Q_3^2} \begin{aligned}[t]
			& \Bigg( -\frac{Q_1^2 \left(\tau ^2 \left(\sigma _1^E-1\right) \left(\sigma _1^E+3\right)+2 \left(\sigma _1^E+\sigma_2^E-2\right)\right)}{m_{\mu }^2} -\frac{Q_2^3 \tau  \left(\sigma _2^E-1\right) \left(\sigma_2^E+3\right)}{Q_1m_{\mu }^2} \\
			& -\frac{Q_2^2 \left(\tau ^2 \left(\sigma _2^E-1\right) \left(\sigma _2^E+3\right)+2 \left(\sigma_1^E+\sigma _2^E-2\right)\right)}{m_{\mu }^2}-\frac{Q_1^3 \tau  \left(\sigma _1^E-1\right) \left(\sigma _1^E+3\right)}{Q_2 m_{\mu}^2} \\
			& +Q_1 \left(\frac{8 \tau }{Q_2}-\frac{Q_2 \tau  \left(\left(\sigma _1^E+4\right) \sigma _1^E+\sigma _2^E \left(\sigma_2^E+4\right)-10\right)}{m_{\mu }^2}\right)+\frac{8 Q_2 \tau}{Q_1} \\
			& +8 \tau ^2 +X \left(-16 Q_1^2 \left(\tau ^2-1\right)-16 Q_2 Q_1 \tau  \left(\tau ^2-1\right)-16 Q_2^2 \left(\tau ^2-1\right)\right) \Bigg) \end{aligned} \\
   			&\quad +X \left(\frac{4 Q_2 Q_1 \tau }{m_{\mu }^2}+\frac{4 Q_1^2}{m_{\mu }^2}+\frac{4 Q_2^2}{m_{\mu }^2}\right) , \\
   		T_{11} &= \frac{1}{Q_3^2} \begin{aligned}[t]
			& \Bigg( \frac{Q_1^2 \left(\tau ^2 \left(\sigma _1^E-5\right) \left(\sigma _1^E-1\right)-2 \left(\sigma _1^E+3 \sigma_2^E-4\right)\right)}{m_{\mu }^2} \\
			& +\frac{Q_2^2 \left(\tau ^2 \left(\left(2-3 \sigma _2^E\right) \sigma _2^E+1\right)+\sigma _1^E-3 \sigma _2^E+2\right)}{m_{\mu }^2}-\frac{6 Q_1^3 \tau  \left(\sigma _1^E-1\right)}{Q_2 m_{\mu }^2} \\
			& +Q_1 \left(\frac{Q_2 \tau \left(\left(\sigma _1^E-2\right) \sigma _1^E-2 \sigma _2^E \left(3 \sigma _2^E+8\right)+23\right)}{2 m_{\mu }^2}+\frac{12 \tau}{Q_2}\right)-\frac{Q_2^3 \tau  \left(\sigma _2^E-1\right)^2}{2 Q_1 m_{\mu }^2} \\
			& +X \left(\frac{Q_2^2 Q_1^2 \left(8 \tau^2+10\right)}{m_{\mu }^2}+\frac{28 Q_2 Q_1^3 \tau }{m_{\mu }^2}+\frac{12 Q_1^4}{m_{\mu }^2}-\frac{2 Q_2^4}{m_{\mu }^2}+Q_2^2 \left(8-8 \tau ^2\right)\right)+8 \tau ^2 \Bigg) , \end{aligned} \\
   		T_{12} &= -\frac{Q_1 \tau  \left({\sigma _1^E}^2-7\right)}{4 Q_2 m_{\mu }^2}+\frac{Q_2 \tau  \left({\sigma_2^E}^2 -7\right)}{4 Q_1 m_{\mu}^2}  +\frac{2 Q_1^2}{Q_2^2 m_{\mu }^2} - \frac{2 Q_2^2}{Q_1^2m_{\mu}^2} - \frac{4}{Q_1^2}+\frac{4}{Q_2^2} \\
			&\quad +\frac{1}{Q_3^2} \begin{aligned}[t]
				& \Bigg(\frac{Q_2^2 \left(4 \tau ^2 \sigma _1^E+2 \tau ^2 \sigma _2^E+3 \sigma _1^E+\sigma _2^E\right)}{2 m_{\mu }^2}-\frac{Q_1^2 \left(2 \tau ^2 \sigma _1^E+4 \tau ^2 \sigma _2^E+\sigma _1^E+3 \sigma _2^E\right)}{2 m_{\mu }^2} \\
				& -\frac{11 Q_1^3 \tau  \sigma_1^E}{2 Q_2 m_{\mu }^2}+\frac{11 Q_2^3 \tau  \sigma _2^E}{2 Q_1 m_{\mu }^2}+\frac{11 Q_1 Q_2 \tau  \left(\sigma _1^E-\sigma_2^E\right)}{2 m_{\mu }^2}-\frac{2 Q_1^4 \sigma _1^E}{Q_2^2 m_{\mu }^2}+\frac{2 Q_2^4 \sigma _2^E}{Q_1^2 m_{\mu }^2} \\
				& +X \begin{aligned}[t]
					& \Bigg(\frac{Q_1^4 \left(4 \tau ^2+3\right)}{m_{\mu }^2}-\frac{Q_2^4 \left(4 \tau ^2+3\right)}{m_{\mu }^2}+Q_1^3 \left(\frac{14 Q_2 \tau }{m_{\mu }^2}-\frac{16 \tau }{Q_2}\right)-\frac{14 Q_2^3 Q_1 \tau }{m_{\mu }^2} \\
					& -4 Q_1^2 \left(7 \tau ^2+1\right)+Q_2^2 \left(28 \tau ^2+4\right)+\frac{16 Q_2^3 \tau }{Q_1}\Bigg)\Bigg) , \end{aligned} \end{aligned}
	\end{split}
\end{align}
where
\begin{align}
	\begin{split}
		X &= \frac{1}{Q_1 Q_2 x} \atan\left( \frac{z x}{1 - z \tau} \right) , \quad x = \sqrt{1 - \tau^2} , \\
		z &= \frac{Q_1 Q_2}{4m_\mu^2} (1-\sigma^E_1)(1-\sigma^E_2) , \quad \sigma^E_i = \sqrt{ 1 + \frac{4 m_\mu^2}{Q_i^2} } , \\
		Q_3^2 &= Q_1^2 + 2 Q_1 Q_2 \tau + Q_2^2 .
	\end{split}
\end{align}


\chapter{Projection of the Scalar Functions}

\label{sec:AppendixProjection}

Given any representation of the HLbL tensor $\Pi^{\mu\nu\lambda\sigma}$, the following procedure allows the immediate identification of the basis coefficients $\tilde \Pi_i$ in (\ref{eq:HLbLTensor43Basis}):
\begin{itemize}
	\item write the tensor in terms of the 138 elementary structures (\ref{eq:HLbLTensor138StructuresLSM}) and identify the scalar coefficients $\Xi_i$
	\item take the subset consisting of the following 43 scalar coefficients:
	\begin{align}
		\begin{split}
			\{ \tilde \Xi_i \} = \big\{ \Xi_i | i ={} & 1,2,3,4,5,7,8,16,17,19,20,25,26,28,29,\\
				& 34,35,37,38,43,44,46,47,53,54,56,57,\\
				& 94,95,97,98,103,104,106,107,\\
				& 121,122,124,125,130,131,133,134 \big\}
		\end{split}
	\end{align}
	\item perform a basis change according to
	\begin{align}
		\begin{split}
			\tilde \Pi_j = \sum_{i=1}^{43} \tilde \Xi_i P_{i,j} .
		\end{split}
	\end{align}
	$P$ is an invertible sparse $43\times43$ matrix, defined below.
	\item split the scalar functions $\tilde \Pi_i$ into 54 scalar functions $\Pi_i$ according to (\ref{eq:HLbLBTTProjectedOnBasis}), such that no kinematic singularities are introduced into the functions $\Pi_i$.
\end{itemize}

The matrix describing the change of basis is sparse and has the following non-zero entries:
\begin{align}
	\footnotesize
	\begin{split}
		P_{1,4} &= \frac{1}{q_{12} q_{34}}, \\
		P_{2,3} &= \frac{1}{q_{12} q_{34}}, \quad
		P_{2,4} = \frac{q_{13} q_{24}}{q_{12}^2 q_{34}^2}+\frac{1}{q_{12} q_{34}}, \quad
		P_{2,5} = \frac{1}{q_{12} q_{34}}, \quad
		P_{2,19} = -\frac{1}{q_{12}^2 q_{34}}, \quad
		P_{2,25} = \frac{1}{q_{12} q_{34}^2}, \\
		P_{3,2} &= \frac{1}{q_{12} q_{34}}, \quad
		P_{3,4} = \frac{q_{14} q_{23}}{q_{12}^2 q_{34}^2}+\frac{1}{q_{12} q_{34}}, \quad
		P_{3,6} = \frac{1}{q_{12} q_{34}}, \quad
		P_{3,20} = -\frac{1}{q_{12}^2 q_{34}}, \quad
		P_{3,26} = \frac{1}{q_{12} q_{34}^2}, \\
		P_{4,4} &= -\frac{q_{13} q_{14}}{q_{12} q_{34}^2}, \quad
		P_{4,7} = \frac{1}{q_{12} q_{34}}, \quad
		P_{5,4} = -\frac{q_{13} q_{24}}{q_{12} q_{34}^2}, \quad
		P_{5,19} = \frac{1}{q_{12} q_{34}}, \\
		P_{6,4} &= -\frac{q_{14} q_{23}}{q_{12} q_{34}^2}, \quad
		P_{6,20} = \frac{1}{q_{12} q_{34}}, \quad
		P_{7,4} = -\frac{q_{23} q_{24}}{q_{12} q_{34}^2}, \quad
		P_{7,8} = \frac{1}{q_{12} q_{34}}, \\
		P_{8,2} &= \frac{q_{13}}{2 q_{12} q_{34}}, \quad
		P_{8,4} = \frac{q_{13}}{2 q_{12} q_{34}}+\frac{q_{14} q_{23} q_{13}}{q_{12}^2 q_{34}^2}, \quad
		P_{8,6} = \frac{q_{13}}{2 q_{12} q_{34}}, \quad
		P_{8,7} = -\frac{q_{23}}{q_{12}^2 q_{34}}, \\
		P_{8,17} &= -\frac{q_{14}}{q_{12} q_{34}^2}, \quad
		P_{8,32} = -\frac{1}{2 q_{12} q_{34}}, \quad
		P_{8,35} = -\frac{1}{2 q_{12} q_{34}}, \quad
		P_{8,40} = \frac{1}{2 q_{12} q_{34}}, \\
	\end{split}
\end{align}
\begin{align}
	\footnotesize
	\begin{split}
		P_{9,3} &= \frac{q_{23}}{2 q_{12} q_{34}}, \quad
		P_{9,4} = \frac{q_{23}}{2 q_{12} q_{34}}+\frac{q_{13} q_{24} q_{23}}{q_{12}^2 q_{34}^2}, \quad
		P_{9,5} = \frac{q_{23}}{2 q_{12} q_{34}}, \quad
		P_{9,17} = -\frac{q_{24}}{q_{12} q_{34}^2}, \\
		P_{9,19} &= -\frac{q_{23}}{q_{12}^2 q_{34}}, \quad
		P_{9,33} = \frac{1}{2 q_{12} q_{34}}, \quad
		P_{9,36} = -\frac{1}{2 q_{12} q_{34}}, \quad
		P_{9,41} = \frac{1}{2 q_{12} q_{34}}, \\
		P_{10,3} &= \frac{q_{14}}{2 q_{12} q_{34}}, \quad
		P_{10,4} = \frac{q_{14}}{2 q_{12} q_{34}}+\frac{q_{13} q_{24} q_{14}}{q_{12}^2 q_{34}^2}, \quad
		P_{10,5} = \frac{q_{14}}{2 q_{12} q_{34}}, \quad
		P_{10,7} = -\frac{q_{24}}{q_{12}^2 q_{34}}, \\
		P_{10,25} &= \frac{q_{14}}{q_{12} q_{34}^2}, \quad
		P_{10,31} = \frac{1}{2 q_{12} q_{34}}, \quad
		P_{10,37} = -\frac{1}{2 q_{12} q_{34}}, \quad
		P_{10,42} = -\frac{1}{2 q_{12} q_{34}}, \\
		P_{11,3} &= \frac{q_{24}}{q_{12} q_{34}}, \quad
		P_{11,4} = \frac{q_{13} q_{24}^2}{q_{12}^2 q_{34}^2}+\frac{q_{24}}{q_{12} q_{34}}, \quad
		P_{11,5} = \frac{q_{24}}{q_{12} q_{34}}, \quad
		P_{11,19} = -\frac{q_{24}}{q_{12}^2 q_{34}}, \\
		P_{11,24} &= \frac{1}{q_{12} q_{34}}, \quad
		P_{11,25} = \frac{q_{24}}{q_{12} q_{34}^2}, \\
		P_{12,2} &= \frac{q_{13}}{2 q_{12} q_{34}}, \quad
		P_{12,4} = \frac{q_{13}}{2 q_{12} q_{34}}+\frac{q_{14} q_{23} q_{13}}{q_{12}^2 q_{34}^2}, \quad
		P_{12,6} = \frac{q_{13}}{2 q_{12} q_{34}}, \quad
		P_{12,7} = -\frac{q_{23}}{q_{12}^2 q_{34}}, \\
		P_{12,26} &= \frac{q_{13}}{q_{12} q_{34}^2}, \quad
		P_{12,32} = -\frac{1}{2 q_{12} q_{34}}, \quad
		P_{12,35} = \frac{1}{2 q_{12} q_{34}}, \quad
		P_{12,40} = \frac{1}{2 q_{12} q_{34}}, \\
		P_{13,2} &= \frac{q_{23}}{q_{12} q_{34}}, \quad
		P_{13,4} = \frac{q_{14} q_{23}^2}{q_{12}^2 q_{34}^2}+\frac{q_{23}}{q_{12} q_{34}}, \quad
		P_{13,6} = \frac{q_{23}}{q_{12} q_{34}}, \quad
		P_{13,20} = -\frac{q_{23}}{q_{12}^2 q_{34}}, \\
		P_{13,23} &= -\frac{1}{q_{12} q_{34}}, \quad
		P_{13,26} = \frac{q_{23}}{q_{12} q_{34}^2}, \\
		P_{14,3} &= \frac{q_{14}}{2 q_{12} q_{34}}, \quad
		P_{14,4} = \frac{q_{14}}{2 q_{12} q_{34}}+\frac{q_{13} q_{24} q_{14}}{q_{12}^2 q_{34}^2}, \quad
		P_{14,5} = \frac{q_{14}}{2 q_{12} q_{34}}, \quad
		P_{14,7} = -\frac{q_{24}}{q_{12}^2 q_{34}}, \\
		P_{14,18} &= -\frac{q_{13}}{q_{12} q_{34}^2}, \quad
		P_{14,31} = \frac{1}{2 q_{12} q_{34}}, \quad
		P_{14,37} = \frac{1}{2 q_{12} q_{34}}, \quad
		P_{14,42} = -\frac{1}{2 q_{12} q_{34}}, \\
		P_{15,2} &= \frac{q_{24}}{2 q_{12} q_{34}}, \quad
		P_{15,4} = \frac{q_{24}}{2 q_{12} q_{34}}+\frac{q_{14} q_{23} q_{24}}{q_{12}^2 q_{34}^2}, \quad
		P_{15,6} = \frac{q_{24}}{2 q_{12} q_{34}}, \quad
		P_{15,18} = -\frac{q_{23}}{q_{12} q_{34}^2}, \\
		P_{15,20} &= -\frac{q_{24}}{q_{12}^2 q_{34}}, \quad
		P_{15,34} = -\frac{1}{2 q_{12} q_{34}}, \quad
		P_{15,38} = \frac{1}{2 q_{12} q_{34}}, \quad
		P_{15,43} = -\frac{1}{2 q_{12} q_{34}}, \\
		P_{16,2} &= \frac{q_{13}}{2 q_{12} q_{34}}, \quad
		P_{16,4} = \frac{q_{13}}{2 q_{12} q_{34}}+\frac{q_{14} q_{23} q_{13}}{q_{12}^2 q_{34}^2}, \quad
		P_{16,6} = \frac{q_{13}}{2 q_{12} q_{34}}, \quad
		P_{16,17} = -\frac{q_{14}}{q_{12} q_{34}^2}, \\
		P_{16,20} &= -\frac{q_{13}}{q_{12}^2 q_{34}}, \quad
		P_{16,32} = \frac{1}{2 q_{12} q_{34}}, \quad
		P_{16,35} = -\frac{1}{2 q_{12} q_{34}}, \quad
		P_{16,40} = \frac{1}{2 q_{12} q_{34}}, \\
		P_{17,3} &= \frac{q_{23}}{2 q_{12} q_{34}}, \quad
		P_{17,4} = \frac{q_{23}}{2 q_{12} q_{34}}+\frac{q_{13} q_{24} q_{23}}{q_{12}^2 q_{34}^2}, \quad
		P_{17,5} = \frac{q_{23}}{2 q_{12} q_{34}}, \quad
		P_{17,8} = -\frac{q_{13}}{q_{12}^2 q_{34}}, \\
		P_{17,17} &= -\frac{q_{24}}{q_{12} q_{34}^2}, \quad
		P_{17,33} = -\frac{1}{2 q_{12} q_{34}}, \quad
		P_{17,36} = -\frac{1}{2 q_{12} q_{34}}, \quad
		P_{17,41} = \frac{1}{2 q_{12} q_{34}}, \\
		P_{18,2} &= \frac{q_{14}}{q_{12} q_{34}}, \quad
		P_{18,4} = \frac{q_{23} q_{14}^2}{q_{12}^2 q_{34}^2}+\frac{q_{14}}{q_{12} q_{34}}, \quad
		P_{18,6} = \frac{q_{14}}{q_{12} q_{34}}, \quad
		P_{18,20} = -\frac{q_{14}}{q_{12}^2 q_{34}}, \\
		P_{18,22} &= \frac{1}{q_{12} q_{34}}, \quad
		P_{18,26} = \frac{q_{14}}{q_{12} q_{34}^2}, \\
		P_{19,2} &= \frac{q_{24}}{2 q_{12} q_{34}}, \quad
		P_{19,4} = \frac{q_{24}}{2 q_{12} q_{34}}+\frac{q_{14} q_{23} q_{24}}{q_{12}^2 q_{34}^2}, \quad
		P_{19,6} = \frac{q_{24}}{2 q_{12} q_{34}}, \quad
		P_{19,8} = -\frac{q_{14}}{q_{12}^2 q_{34}}, \\
		P_{19,26} &= \frac{q_{24}}{q_{12} q_{34}^2}, \quad
		P_{19,34} = \frac{1}{2 q_{12} q_{34}}, \quad
		P_{19,38} = -\frac{1}{2 q_{12} q_{34}}, \quad
		P_{19,43} = -\frac{1}{2 q_{12} q_{34}}, \\
		P_{20,3} &= \frac{q_{13}}{q_{12} q_{34}}, \quad
		P_{20,4} = \frac{q_{24} q_{13}^2}{q_{12}^2 q_{34}^2}+\frac{q_{13}}{q_{12} q_{34}}, \quad
		P_{20,5} = \frac{q_{13}}{q_{12} q_{34}}, \quad
		P_{20,19} = -\frac{q_{13}}{q_{12}^2 q_{34}}, \\
		P_{20,21} &= -\frac{1}{q_{12} q_{34}}, \quad
		P_{20,25} = \frac{q_{13}}{q_{12} q_{34}^2}, \\
		P_{21,3} &= \frac{q_{23}}{2 q_{12} q_{34}}, \quad
		P_{21,4} = \frac{q_{23}}{2 q_{12} q_{34}}+\frac{q_{13} q_{24} q_{23}}{q_{12}^2 q_{34}^2}, \quad
		P_{21,5} = \frac{q_{23}}{2 q_{12} q_{34}}, \quad
		P_{21,8} = -\frac{q_{13}}{q_{12}^2 q_{34}}, \\
		P_{21,25} &= \frac{q_{23}}{q_{12} q_{34}^2}, \quad
		P_{21,33} = -\frac{1}{2 q_{12} q_{34}}, \quad
		P_{21,36} = \frac{1}{2 q_{12} q_{34}}, \quad
		P_{21,41} = \frac{1}{2 q_{12} q_{34}}, \\
	\end{split}
\end{align}
\begin{align}
	\footnotesize
	\begin{split}
		P_{22,3} &= \frac{q_{14}}{2 q_{12} q_{34}}, \quad
		P_{22,4} = \frac{q_{14}}{2 q_{12} q_{34}}+\frac{q_{13} q_{24} q_{14}}{q_{12}^2 q_{34}^2}, \quad
		P_{22,5} = \frac{q_{14}}{2 q_{12} q_{34}}, \quad
		P_{22,18} = -\frac{q_{13}}{q_{12} q_{34}^2}, \\
		P_{22,19} &= -\frac{q_{14}}{q_{12}^2 q_{34}}, \quad
		P_{22,31} = -\frac{1}{2 q_{12} q_{34}}, \quad
		P_{22,37} = \frac{1}{2 q_{12} q_{34}}, \quad
		P_{22,42} = -\frac{1}{2 q_{12} q_{34}}, \\
		P_{23,2} &= \frac{q_{24}}{2 q_{12} q_{34}}, \quad
		P_{23,4} = \frac{q_{24}}{2 q_{12} q_{34}}+\frac{q_{14} q_{23} q_{24}}{q_{12}^2 q_{34}^2}, \quad
		P_{23,6} = \frac{q_{24}}{2 q_{12} q_{34}}, \quad
		P_{23,8} = -\frac{q_{14}}{q_{12}^2 q_{34}}, \\
		P_{23,18} &= -\frac{q_{23}}{q_{12} q_{34}^2}, \quad
		P_{23,34} = \frac{1}{2 q_{12} q_{34}}, \quad
		P_{23,38} = \frac{1}{2 q_{12} q_{34}}, \quad
		P_{23,43} = -\frac{1}{2 q_{12} q_{34}}, \\
		P_{24,4} &= -\frac{q_{13} q_{23}}{q_{12}^2 q_{34}}, \quad
		P_{24,17} = \frac{1}{q_{12} q_{34}}, \\
		P_{25,4} &= -\frac{q_{13} q_{24}}{q_{12}^2 q_{34}}, \quad
		P_{25,25} = -\frac{1}{q_{12} q_{34}}, \\
		P_{26,4} &= -\frac{q_{14} q_{23}}{q_{12}^2 q_{34}}, \quad
		P_{26,26} = -\frac{1}{q_{12} q_{34}}, \\
		P_{27,4} &= -\frac{q_{14} q_{24}}{q_{12}^2 q_{34}}, \quad
		P_{27,18} = \frac{1}{q_{12} q_{34}}, \\
		P_{28,2} &= \frac{q_{13}^2}{2 q_{12} q_{34}}, \quad
		P_{28,4} = \frac{q_{13}^2}{2 q_{12} q_{34}}+\frac{q_{14} q_{23} q_{13}^2}{q_{12}^2 q_{34}^2}, \quad
		P_{28,6} = \frac{q_{13}^2}{2 q_{12} q_{34}}, \quad
		P_{28,7} = -\frac{q_{13} q_{23}}{q_{12}^2 q_{34}}, \\
		P_{28,17} &= -\frac{q_{13} q_{14}}{q_{12} q_{34}^2}, \quad
		P_{28,27} = -\frac{1}{q_{12} q_{34}}, \quad
		P_{28,32} = -\frac{q_{13}}{2 q_{12} q_{34}}, \quad
		P_{28,35} = -\frac{q_{13}}{2 q_{12} q_{34}}, \\
		P_{28,40} &= \frac{q_{13}}{2 q_{12} q_{34}}, \\
		P_{29,3} &= \frac{q_{13} q_{23}}{2 q_{12} q_{34}}, \quad
		P_{29,4} = \frac{q_{23} q_{24} q_{13}^2}{q_{12}^2 q_{34}^2}+\frac{q_{23} q_{13}}{2 q_{12} q_{34}}, \quad
		P_{29,5} = \frac{q_{13} q_{23}}{2 q_{12} q_{34}}, \quad
		P_{29,9} = \frac{1}{q_{34}}, \\
		P_{29,17} &= -\frac{q_{13} q_{24}}{q_{12} q_{34}^2}, \quad
		P_{29,19} = -\frac{q_{13} q_{23}}{q_{12}^2 q_{34}}, \quad
		P_{29,21} = -\frac{q_{23}}{q_{12} q_{34}}, \quad
		P_{29,33} = \frac{q_{13}}{2 q_{12} q_{34}}, \\
		P_{29,36} &= -\frac{q_{13}}{2 q_{12} q_{34}}, \quad
		P_{29,41} = \frac{q_{13}}{2 q_{12} q_{34}}, \\
		P_{30,2} &= \frac{q_{13} q_{23}}{2 q_{12} q_{34}}, \quad
		P_{30,4} = \frac{q_{13} q_{14} q_{23}^2}{q_{12}^2 q_{34}^2}+\frac{q_{13} q_{23}}{2 q_{12} q_{34}}, \quad
		P_{30,6} = \frac{q_{13} q_{23}}{2 q_{12} q_{34}}, \quad
		P_{30,13} = \frac{1}{q_{34}}, \\
		P_{30,17} &= -\frac{q_{14} q_{23}}{q_{12} q_{34}^2}, \quad
		P_{30,20} = -\frac{q_{13} q_{23}}{q_{12}^2 q_{34}}, \quad
		P_{30,23} = -\frac{q_{13}}{q_{12} q_{34}}, \quad
		P_{30,32} = \frac{q_{23}}{2 q_{12} q_{34}}, \\
		P_{30,35} &= -\frac{q_{23}}{2 q_{12} q_{34}}, \quad
		P_{30,40} = \frac{q_{23}}{2 q_{12} q_{34}}, \\
		P_{31,3} &= \frac{q_{23}^2}{2 q_{12} q_{34}}, \quad
		P_{31,4} = \frac{q_{23}^2}{2 q_{12} q_{34}}+\frac{q_{13} q_{24} q_{23}^2}{q_{12}^2 q_{34}^2}, \quad
		P_{31,5} = \frac{q_{23}^2}{2 q_{12} q_{34}}, \quad
		P_{31,8} = -\frac{q_{13} q_{23}}{q_{12}^2 q_{34}}, \\
		P_{31,17} &= -\frac{q_{23} q_{24}}{q_{12} q_{34}^2}, \quad
		P_{31,28} = -\frac{1}{q_{12} q_{34}}, \quad
		P_{31,33} = -\frac{q_{23}}{2 q_{12} q_{34}}, \quad
		P_{31,36} = -\frac{q_{23}}{2 q_{12} q_{34}}, \\
		P_{31,41} &= \frac{q_{23}}{2 q_{12} q_{34}}, \\
		P_{32,3} &= \frac{q_{13} q_{14}}{2 q_{12} q_{34}}, \quad
		P_{32,4} = \frac{q_{14} q_{24} q_{13}^2}{q_{12}^2 q_{34}^2}+\frac{q_{14} q_{13}}{2 q_{12} q_{34}}, \quad
		P_{32,5} = \frac{q_{13} q_{14}}{2 q_{12} q_{34}}, \quad
		P_{32,7} = -\frac{q_{13} q_{24}}{q_{12}^2 q_{34}}, \\
		P_{32,10} &= \frac{1}{q_{12}}, \quad
		P_{32,21} = -\frac{q_{14}}{q_{12} q_{34}}, \quad
		P_{32,25} = \frac{q_{13} q_{14}}{q_{12} q_{34}^2}, \quad
		P_{32,31} = \frac{q_{13}}{2 q_{12} q_{34}}, \\
		P_{32,37} &= -\frac{q_{13}}{2 q_{12} q_{34}}, \quad
		P_{32,42} = -\frac{q_{13}}{2 q_{12} q_{34}}, \\
		P_{33,3} &= \frac{q_{13} q_{24}}{q_{12} q_{34}}, \quad
		P_{33,4} = \frac{q_{13}^2 q_{24}^2}{q_{12}^2 q_{34}^2}+\frac{q_{13} q_{24}}{q_{12} q_{34}}, \quad
		P_{33,5} = \frac{q_{13} q_{24}}{q_{12} q_{34}}+1, \quad
		P_{33,19} = -\frac{q_{13} q_{24}}{q_{12}^2 q_{34}}, \\
		P_{33,21} &= -\frac{q_{24}}{q_{12} q_{34}}, \quad
		P_{33,24} = \frac{q_{13}}{q_{12} q_{34}}, \quad
		P_{33,25} = \frac{q_{13} q_{24}}{q_{12} q_{34}^2}, \\
		P_{34,1} &= \frac{1}{2}, \quad
		P_{34,4} = \frac{q_{13} q_{14} q_{23} q_{24}}{q_{12}^2 q_{34}^2}, \quad
		P_{34,5} = \frac{1}{2}, \quad
		P_{34,6} = \frac{1}{2}, \quad
		P_{34,20} = -\frac{q_{13} q_{24}}{q_{12}^2 q_{34}}, \\
		P_{34,25} &= \frac{q_{14} q_{23}}{q_{12} q_{34}^2}, \quad
		P_{34,32} = \frac{q_{24}}{2 q_{12} q_{34}}, \quad
		P_{34,34} = -\frac{q_{13}}{2 q_{12} q_{34}}, \quad
		P_{34,36} = \frac{q_{14}}{2 q_{12} q_{34}}, \\
		P_{34,37} &= -\frac{q_{23}}{2 q_{12} q_{34}}, \quad
		P_{34,39} = -\frac{1}{2 q_{34}}, \quad
		P_{34,40} = \frac{q_{24}}{2 q_{12} q_{34}}, \quad
		P_{34,43} = -\frac{q_{13}}{2 q_{12} q_{34}}, \\
	\end{split}
\end{align}
\begin{align}
	\footnotesize
	\begin{split}
		P_{35,3} &= \frac{q_{23} q_{24}}{2 q_{12} q_{34}}, \quad
		P_{35,4} = \frac{q_{13} q_{23} q_{24}^2}{q_{12}^2 q_{34}^2}+\frac{q_{23} q_{24}}{2 q_{12} q_{34}}, \quad
		P_{35,5} = \frac{q_{23} q_{24}}{2 q_{12} q_{34}}, \quad
		P_{35,8} = -\frac{q_{13} q_{24}}{q_{12}^2 q_{34}}, \\
		P_{35,16} &= \frac{1}{q_{12}}, \quad
		P_{35,24} = \frac{q_{23}}{q_{12} q_{34}}, \quad
		P_{35,25} = \frac{q_{23} q_{24}}{q_{12} q_{34}^2}, \quad
		P_{35,33} = -\frac{q_{24}}{2 q_{12} q_{34}}, \\
		P_{35,36} &= \frac{q_{24}}{2 q_{12} q_{34}}, \quad
		P_{35,41} = \frac{q_{24}}{2 q_{12} q_{34}}, \\
		P_{36,2} &= \frac{q_{13} q_{14}}{2 q_{12} q_{34}}, \quad
		P_{36,4} = \frac{q_{13} q_{23} q_{14}^2}{q_{12}^2 q_{34}^2}+\frac{q_{13} q_{14}}{2 q_{12} q_{34}}, \quad
		P_{36,6} = \frac{q_{13} q_{14}}{2 q_{12} q_{34}}, \quad
		P_{36,7} = -\frac{q_{14} q_{23}}{q_{12}^2 q_{34}}, \\
		P_{36,11} &= \frac{1}{q_{12}}, \quad
		P_{36,22} = \frac{q_{13}}{q_{12} q_{34}}, \quad
		P_{36,26} = \frac{q_{13} q_{14}}{q_{12} q_{34}^2}, \quad
		P_{36,32} = -\frac{q_{14}}{2 q_{12} q_{34}}, \\
		P_{36,35} &= \frac{q_{14}}{2 q_{12} q_{34}}, \quad
		P_{36,40} = \frac{q_{14}}{2 q_{12} q_{34}}, \\
		P_{37,1} &= \frac{1}{2}, \quad
		P_{37,4} = \frac{q_{13} q_{14} q_{23} q_{24}}{q_{12}^2 q_{34}^2}, \quad
		P_{37,5} = \frac{1}{2}, \quad
		P_{37,6} = \frac{1}{2}, \quad
		P_{37,19} = -\frac{q_{14} q_{23}}{q_{12}^2 q_{34}}, \\
		P_{37,26} &= \frac{q_{13} q_{24}}{q_{12} q_{34}^2}, \quad
		P_{37,31} = -\frac{q_{23}}{2 q_{12} q_{34}}, \quad
		P_{37,33} = \frac{q_{14}}{2 q_{12} q_{34}}, \quad
		P_{37,35} = \frac{q_{24}}{2 q_{12} q_{34}}, \\
		P_{37,38} &= -\frac{q_{13}}{2 q_{12} q_{34}}, \quad
		P_{37,39} = \frac{1}{2 q_{34}}, \quad
		P_{37,41} = \frac{q_{14}}{2 q_{12} q_{34}}, \quad
		P_{37,42} = -\frac{q_{23}}{2 q_{12} q_{34}}, \\
		P_{38,2} &= \frac{q_{14} q_{23}}{q_{12} q_{34}}, \quad
		P_{38,4} = \frac{q_{14}^2 q_{23}^2}{q_{12}^2 q_{34}^2}+\frac{q_{14} q_{23}}{q_{12} q_{34}}, \quad
		P_{38,6} = \frac{q_{14} q_{23}}{q_{12} q_{34}}+1, \quad
		P_{38,20} = -\frac{q_{14} q_{23}}{q_{12}^2 q_{34}}, \\
		P_{38,22} &= \frac{q_{23}}{q_{12} q_{34}}, \quad
		P_{38,23} = -\frac{q_{14}}{q_{12} q_{34}}, \quad
		P_{38,26} = \frac{q_{14} q_{23}}{q_{12} q_{34}^2}, \\
		P_{39,2} &= \frac{q_{23} q_{24}}{2 q_{12} q_{34}}, \quad
		P_{39,4} = \frac{q_{14} q_{24} q_{23}^2}{q_{12}^2 q_{34}^2}+\frac{q_{24} q_{23}}{2 q_{12} q_{34}}, \quad
		P_{39,6} = \frac{q_{23} q_{24}}{2 q_{12} q_{34}}, \quad
		P_{39,8} = -\frac{q_{14} q_{23}}{q_{12}^2 q_{34}}, \\
		P_{39,14} &= \frac{1}{q_{12}}, \quad
		P_{39,23} = -\frac{q_{24}}{q_{12} q_{34}}, \quad
		P_{39,26} = \frac{q_{23} q_{24}}{q_{12} q_{34}^2}, \quad
		P_{39,34} = \frac{q_{23}}{2 q_{12} q_{34}}, \\
		P_{39,38} &= -\frac{q_{23}}{2 q_{12} q_{34}}, \quad
		P_{39,43} = -\frac{q_{23}}{2 q_{12} q_{34}}, \\
		P_{40,3} &= \frac{q_{14}^2}{2 q_{12} q_{34}}, \quad
		P_{40,4} = \frac{q_{14}^2}{2 q_{12} q_{34}}+\frac{q_{13} q_{24} q_{14}^2}{q_{12}^2 q_{34}^2}, \quad
		P_{40,5} = \frac{q_{14}^2}{2 q_{12} q_{34}}, \quad
		P_{40,7} = -\frac{q_{14} q_{24}}{q_{12}^2 q_{34}}, \\
		P_{40,18} &= -\frac{q_{13} q_{14}}{q_{12} q_{34}^2}, \quad
		P_{40,29} = -\frac{1}{q_{12} q_{34}}, \quad
		P_{40,31} = \frac{q_{14}}{2 q_{12} q_{34}}, \quad
		P_{40,37} = \frac{q_{14}}{2 q_{12} q_{34}}, \\
		P_{40,42} &= -\frac{q_{14}}{2 q_{12} q_{34}}, \\
		P_{41,3} &= \frac{q_{14} q_{24}}{2 q_{12} q_{34}}, \quad
		P_{41,4} = \frac{q_{13} q_{14} q_{24}^2}{q_{12}^2 q_{34}^2}+\frac{q_{14} q_{24}}{2 q_{12} q_{34}}, \quad
		P_{41,5} = \frac{q_{14} q_{24}}{2 q_{12} q_{34}}, \quad
		P_{41,15} = \frac{1}{q_{34}}, \\
		P_{41,18} &= -\frac{q_{13} q_{24}}{q_{12} q_{34}^2}, \quad
		P_{41,19} = -\frac{q_{14} q_{24}}{q_{12}^2 q_{34}}, \quad
		P_{41,24} = \frac{q_{14}}{q_{12} q_{34}}, \quad
		P_{41,31} = -\frac{q_{24}}{2 q_{12} q_{34}}, \\
		P_{41,37} &= \frac{q_{24}}{2 q_{12} q_{34}}, \quad
		P_{41,42} = -\frac{q_{24}}{2 q_{12} q_{34}}, \\
		P_{42,2} &= \frac{q_{14} q_{24}}{2 q_{12} q_{34}}, \quad
		P_{42,4} = \frac{q_{23} q_{24} q_{14}^2}{q_{12}^2 q_{34}^2}+\frac{q_{24} q_{14}}{2 q_{12} q_{34}}, \quad
		P_{42,6} = \frac{q_{14} q_{24}}{2 q_{12} q_{34}}, \quad
		P_{42,12} = \frac{1}{q_{34}}, \\
		P_{42,18} &= -\frac{q_{14} q_{23}}{q_{12} q_{34}^2}, \quad
		P_{42,20} = -\frac{q_{14} q_{24}}{q_{12}^2 q_{34}}, \quad
		P_{42,22} = \frac{q_{24}}{q_{12} q_{34}}, \quad
		P_{42,34} = -\frac{q_{14}}{2 q_{12} q_{34}}, \\
		P_{42,38} &= \frac{q_{14}}{2 q_{12} q_{34}}, \quad
		P_{42,43} = -\frac{q_{14}}{2 q_{12} q_{34}}, \\
		P_{43,2} &= \frac{q_{24}^2}{2 q_{12} q_{34}}, \quad
		P_{43,4} = \frac{q_{24}^2}{2 q_{12} q_{34}}+\frac{q_{14} q_{23} q_{24}^2}{q_{12}^2 q_{34}^2}, \quad
		P_{43,6} = \frac{q_{24}^2}{2 q_{12} q_{34}}, \quad
		P_{43,8} = -\frac{q_{14} q_{24}}{q_{12}^2 q_{34}}, \\
		P_{43,18} &= -\frac{q_{23} q_{24}}{q_{12} q_{34}^2}, \quad
		P_{43,30} = -\frac{1}{q_{12} q_{34}}, \quad
		P_{43,34} = \frac{q_{24}}{2 q_{12} q_{34}}, \quad
		P_{43,38} = \frac{q_{24}}{2 q_{12} q_{34}}, \quad
		P_{43,43} = -\frac{q_{24}}{2 q_{12} q_{34}} ,
	\end{split}
\end{align}
where $q_{ij} := q_i \cdot q_j$.


\chapter{Phase-Space Integration}

\label{sec:AppendixPhaseSpaceIntegration}

In the derivation of the Mandelstam representation, phase-space integrals with two propagators appear. We consider here the case of the simplest scalar integral:
\begin{align}
	\begin{split}
		\int d\Omega_s^\dprime \frac{1}{a_1 - t^\prime} \frac{1}{a_2 - t^\dprime} ,
	\end{split}
\end{align}
where $a_{1,2}$ are either pion masses in the case of box topologies or integration variables of dispersion integrals in the case of box topologies with discontinuities in the sub-processes.

Let us introduce some explicit kinematics:
\begin{align}
	\begin{split}
		s &= ( q_1 + q_2 )^2 = (p_1 + p_2)^2 = (q_3 - q_4)^2 , \\
		t^\prime &= (q_1 - p_1)^2 = (q_2 - p_2)^2 , \\
		u^\prime &= (q_1 - p_2)^2 = (q_2 - p_1)^2 , \\
		t^\dprime &= (q_3 + p_1)^2 = (q_4 - p_2)^2 , \\
		u^\dprime &= (q_3 + p_2)^2 = (q_4 - p_1)^2 ,
	\end{split}
\end{align}
where the primed variables belong to the sub-process on the left-hand and the double-primed variables to the sub-process on the right-hand side of the cut.

\begin{figure}[ht]
	\centering
	\psset{unit=1cm}
	\begin{pspicture}(-3,-3)(3,3)
		\psline{->}(0,0)(0,3)
		\psline{->}(0,0)(3,0)
		\psline{->}(0,0)(-2,-1)
		\psline[linewidth=1.5pt, arrowsize=8pt]{->}(0,0)(0,2.5)
		\psline[linewidth=1.5pt, arrowsize=8pt]{->}(0,0)(-1,1.75)
		\psline[linewidth=1.5pt, arrowsize=8pt]{->}(0,0)(1.5,1)
		\psline[linestyle=dotted](1.5,1)(1.5,-0.5)
		\psline[linestyle=dotted](0,0)(1.5,-0.5)
		\psline[linestyle=dotted](-1,1.75)(-1,-0.5)
		\psline[linearc=1.75,linestyle=dashed](0,2)(0.75,1.75)(1.2,0.8)
		\psline[linearc=1.5,linestyle=dashed](0,2)(-0.5,1.75)(-0.75,1.31)
		\psline[linearc=3,linestyle=dashed](-0.75,1.31)(0.5,1.25)(1.2,0.8)
		\psline[linearc=4,linestyle=dashed](-0.75,-0.375)(0,-0.5)(0.9,-0.3)
		\rput(-0.5,2){$\theta_s$}
		\rput(0.25,0.85){$\theta_s^\prime$}
		\rput(1.2,1.8){$\theta_s^\dprime$}
		\rput(0,-0.75){$\phi_s^\dprime$}
		\rput(0.5,2.4){$-\vec q_3$}
		\rput(-1.5,2){$\vec q_1$}
		\rput(2,1){$\vec p_1$}
	\end{pspicture}
	\caption{Vectors and angles in the $s$-channel centre-of-mass frame}
	\label{img:HLbLAnglesSChannel}
\end{figure}
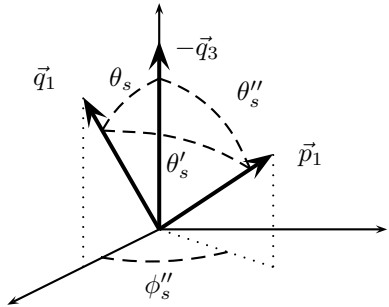

The relation of the vectors and angles in the $s$-channel centre-of-mass frame is displayed in figure~\ref{img:HLbLAnglesSChannel}. The angles are related by
\begin{align}
	\begin{split}
		\cos\theta_s^\prime &= \sin\theta_s \sin\theta_s^\dprime \cos\phi_s^\dprime + \cos\theta_s \cos\theta_s^\dprime , \\
	\end{split}
\end{align}
or equivalently
\begin{align}
	\begin{split}
		z_s^\prime &= \sqrt{1-z_s^2} \sqrt{1-{z_s^\dprime}^2} \cos\phi_s^\dprime + z_s z_s^\dprime , \\
	\end{split}
\end{align}
where $z_s := \cos\theta_s$, $z_s^\prime := \cos\theta_s^\prime$, $z_s^\dprime := \cos\theta_s^\dprime$.

The three-vectors in the $s$-channel centre-of-mass frame are given by
\begin{align}
	\begin{split}
		\vec q_1 &= \frac{\lambda^{1/2}_{12}(s)}{2\sqrt{s}} \left( \sin\theta_s, 0, \cos\theta_s \right) , \\
		\vec p_1 &= \sqrt{\frac{s}{4} - M_\pi^2} \left( \sin\theta_s^\dprime \cos\phi_s^\dprime, \sin\theta_s^\dprime \sin\phi_s^\dprime, \cos\theta_s^\dprime \right) , \\
		-\vec q_3 &= \frac{\lambda^{1/2}_{34}(s)}{2\sqrt{s}} \left( 0, 0, 1 \right) ,
	\end{split}
\end{align}
where $\lambda_{ik}(s) := \lambda(s,q_i^2,q_k^2)$ and $\lambda(a,b,c) = a^2 + b^2 + c^2 - 2(ab+bc+ca)$.

The angles are connected to the different Mandelstam variables by
\begin{align}
	\begin{split}
		t &= \frac{\Sigma - s}{2} - \frac{\Delta_{12}\Delta_{34}}{2s} + \frac{\lambda^{1/2}_{12}(s) \lambda^{1/2}_{34}(s)}{2s} z_s , \\
		u &= \frac{\Sigma - s}{2} + \frac{\Delta_{12}\Delta_{34}}{2s} - \frac{\lambda^{1/2}_{12}(s) \lambda^{1/2}_{34}(s)}{2s} z_s , \\
		t^\prime &= \frac{\Sigma_{12} - s}{2} + \frac{1}{2} \sigma_\pi(s) \lambda^{1/2}_{12}(s) z_s^\prime , \\
		u^\prime &= \frac{\Sigma_{12} - s}{2} - \frac{1}{2} \sigma_\pi(s) \lambda^{1/2}_{12}(s) z_s^\prime , \\
		t^\dprime &= \frac{\Sigma_{34} - s}{2} + \frac{1}{2} \sigma_\pi(s) \lambda^{1/2}_{34}(s) z_s^\dprime , \\
		u^\dprime &= \frac{\Sigma_{34} - s}{2} - \frac{1}{2} \sigma_\pi(s) \lambda^{1/2}_{34}(s) z_s^\dprime , \\
	\end{split}
\end{align}
where $\Delta_{ik} := q_i^2 - q_k^2$, $\Sigma = \sum_i q_i^2$ and $\Sigma_{ik} := 2M_\pi^2 + q_i^2 + q_k^2$.

We express now our integral in terms of the angles:
\begin{align}
	\begin{split}
		\int d\Omega_s^\dprime \frac{1}{a_1 - t^\prime} \frac{1}{a_2 - t^\dprime} &= \frac{4}{\lambda^{1/2}_{12}(s)\lambda^{1/2}_{34}(s) \sigma_\pi^2(s)} \int d\Omega_s^\dprime \frac{1}{\alpha_1 - z_s^\prime} \frac{1}{\alpha_2 - z_s^\dprime} \\
			&=: \frac{4}{\lambda^{1/2}_{12}(s)\lambda^{1/2}_{34}(s) \sigma_\pi^2(s)} I_{\alpha_1,\alpha_2}(s,t) ,
	\end{split}
\end{align}
where
\begin{align}
	\alpha_1 := \frac{2a_1 - \Sigma_{12} + s}{\sigma_\pi(s) \lambda^{1/2}_{12}(s)} \quad , \quad \alpha_2 := \frac{2a_2 - \Sigma_{34} + s}{\sigma_\pi(s)\lambda_{34}^{1/2}(s)} .
\end{align}
Further integrals are
\begin{align}
	\begin{split}
		\int d\Omega_s^\dprime \frac{1}{a_1 - t^\prime} \frac{1}{a_2 - u^\dprime} &= - \frac{4}{\lambda^{1/2}_{12}(s)\lambda^{1/2}_{34}(s) \sigma_\pi^2(s)} I_{\alpha_1,-\alpha_2}(s,t) , \\
		\int d\Omega_s^\dprime \frac{1}{a_1 - u^\prime} \frac{1}{a_2 - t^\dprime} &= - \frac{4}{\lambda^{1/2}_{12}(s)\lambda^{1/2}_{34}(s) \sigma_\pi^2(s)} I_{-\alpha_1,\alpha_2}(s,t) , \\
		\int d\Omega_s^\dprime \frac{1}{a_1 - u^\prime} \frac{1}{a_2 - u^\dprime} &= \frac{4}{\lambda^{1/2}_{12}(s)\lambda^{1/2}_{34}(s) \sigma_\pi^2(s)} I_{-\alpha_1,-\alpha_2}(s,t) .
	\end{split}
\end{align}
The integral $I_{\alpha_1,\alpha_2}$ can be computed with the help of a Feynman parametrisation \cite{Barut1967}:
\begin{align}
	\begin{split}
		I_{\alpha_1,\alpha_2}(s,t) &= \int d\Omega_s^\dprime \frac{1}{\alpha_1 - z_s^\prime} \frac{1}{\alpha_2 - z_s^\dprime} \\
			&= \int d\Omega_s^\dprime \int_0^1 dx \left( x (\alpha_1 - z_s^\prime) + (1-x)(\alpha_2 - z_s^\dprime) \right)^{-2} \\
			&= \int d\Omega_s^\dprime \int_0^1 dx \left( x \alpha_1 + (1-x)\alpha_2 - \left( x z_s^\prime + (1-x) z_s^\dprime \right) \right)^{-2} \\
			&= \int_0^1 dx \int d\Omega_s^\dprime \left( x \alpha_1 + (1-x)\alpha_2 - \left( x \frac{\vec q_1}{|\vec q_1|} - (1-x) \frac{\vec q_3}{|\vec q_3|} \right) \cdot \frac{\vec p_1}{|\vec p_1|} \right)^{-2} \\
			&= 4\pi \int_0^1 dx \left( \left(x \alpha_1 + (1-x)\alpha_2 \right)^2 - \left| x \frac{\vec q_1}{|\vec q_1|} - (1-x) \frac{\vec q_3}{|\vec q_3|} \right|^2 \right)^{-1} \\
			&= 4\pi \int_0^1 dx \left( \left(x \alpha_1 + (1-x)\alpha_2 \right)^2 - \left( x^2 + (1-x)^2 + 2 x(1-x) z_s \right) \right)^{-1} \\
			&= 4\pi \int_0^1 dx \left( \left(x \alpha_1 + (1-x)\alpha_2 \right)^2 - \left( 1 - 2x(1-x)(1 - z_s) \right) \right)^{-1} .
	\end{split}
\end{align}
A dispersive representation of this integral can be found directly by considering its singularity structure:

The denominator of the integrand has two zeros:
\begin{align}
	x_{1,2}(z_s) = \frac{\pm\sqrt{{\alpha_1}^2 - 2 \alpha_1 \alpha_2 z_s + \alpha_2^2 + z_s^2 - 1} + z_s - 1 - \alpha_1 \alpha_2 + \alpha_2^2}{(\alpha_1 - \alpha_2)^2 - 2(1 - z_s)} .
\end{align}
A singularity of the integral can occur if the singularities of the integrand hit an end-point of the integration contour or if they pinch the integration contour. An end-point singularity happens only for $|z_s|\to \infty$. A pinch-singularity can be present if $x_1 = x_2$, i.e. at $z_s = z_s^\pm$, with
\begin{align}
	z_s^\pm = \alpha_1 \alpha_2 \pm \sqrt{\left(\alpha_1^2 - 1\right)\left(\alpha_2^2 - 1\right)} .
\end{align}
There, $x_{1,2}$ take the value
\begin{align}
	x^\pm := x_{1,2}(z_s^\pm) = \frac{1}{1 \pm \sqrt{\frac{\alpha_1^2-1}{\alpha_2^2 - 1}}} .
\end{align}
Obviously, the integration contour will be pinched only at $x^+$. For $z_s^+ < z_s < \infty$, the singularities of the integrand $x_{1,2}$ move from $x^+$ to 0 or 1 respectively. If we give $z_s$ a small imaginary part, they lie slightly beside the integration contour. Therefore, the integral has a branch cut for $z_s^+ < z_s < \infty$. The discontinuity along the branch cut is given by closed integration circles around $x_{1,2}$:
\begin{align}
	I(z_s+i\epsilon) - I(z_s - i\epsilon) = 4\pi \cdot 2\pi i \left( \mathrm{Res}\left( \frac{1}{D}, x_1\right) - \mathrm{Res}\left( \frac{1}{D}, x_2\right) \right),
\end{align}
where $D = \left(x \alpha_1 + (1-x)\alpha_2 \right)^2 - \left( 1 + 2x(1-x)(z_s-1) \right)$, hence
\begin{align}
	\begin{split}
		I(z_s+i\epsilon) - I(z_s - i\epsilon) &=  4\pi \frac{2\pi i}{\sqrt{\alpha_1^2 - 2 \alpha_1 \alpha_2 z_s + \alpha_2^2 + z_s^2 -1}} \\
			&= 4\pi \frac{2\pi i}{\sqrt{(z_s - z_s^+)(z_s - z_s^-)}} .
	\end{split}
\end{align}
We can therefore write a dispersion relation:
\begin{align}
	I_{\alpha_1,\alpha_2}(z_s) = 4\pi \int_{z_s^+}^\infty dz \frac{1}{(z-z_s)\sqrt{(z - z_s^+)(z-z_s^-)}} .
\end{align}
After expressing the angles in terms of the Mandelstam variables
\begin{align}
	\begin{split}
		z_s &= \frac{2s}{\lambda^{1/2}_{12}(s) \lambda^{1/2}_{34}(s)} \left( t -  \frac{\Sigma - s}{2} +  \frac{\Delta_{12}\Delta_{34}}{2s} \right) , \\
		z &=: \frac{2s}{\lambda^{1/2}_{12}(s) \lambda^{1/2}_{34}(s)} \left( \tilde t -  \frac{\Sigma - s}{2} +  \frac{\Delta_{12}\Delta_{34}}{2s} \right) , \\
		z_s^\pm &=: \frac{2s}{\lambda^{1/2}_{12}(s) \lambda^{1/2}_{34}(s)} \left( t^\pm -  \frac{\Sigma - s}{2} +  \frac{\Delta_{12}\Delta_{34}}{2s} \right) ,
	\end{split}
\end{align}
we find
\begin{align}
	I_{\alpha_1,\alpha_2}(s,t) = 2\pi \frac{\lambda_{12}^{1/2}(s) \lambda_{34}^{1/2}(s)}{s} \int_{t^+}^\infty d\tilde t \frac{1}{(\tilde t - t)\sqrt{(\tilde t - t^+)(\tilde t - t^-)}} ,
\end{align}
where the dependence on $\alpha_{1,2}$ is through $t^\pm$ (or $z_s^\pm$).

Let us study the other similar integrals (note that $\alpha_i>1$ for small $q_i^2$):
\begin{align}
	\begin{split}
		I_{-\alpha_1,-\alpha_2}&(s,t) = \int d\Omega_s^\dprime \frac{1}{\alpha_1 + z_s^\prime} \frac{1}{\alpha_2 + z_s^\dprime} \\
			&= \int d\Omega_s^\dprime \int_0^1 dx \left( x \alpha_1 + (1-x)\alpha_2 + \left( x z_s^\prime + (1-x) z_s^\dprime \right) \right)^{-2} \\
			&= \int_0^1 dx \int d\Omega_s^\dprime \left( x \alpha_1 + (1-x)\alpha_2 + \left( x \frac{\vec q_1}{|\vec q_1|} - (1-x) \frac{\vec q_3}{|\vec q_3|} \right) \cdot \frac{\vec p_1}{|\vec p_1|} \right)^{-2} \\
			&= 4\pi \int_0^1 dx \left( \left(x \alpha_1 + (1-x)\alpha_2\right)^2 - \left| x \frac{\vec q_1}{|\vec q_1|} - (1-x) \frac{\vec q_3}{|\vec q_3|} \right|^2 \right)^{-1} \\
			&= 4\pi \int_0^1 dx \left( \left(x \alpha_1 + (1-x)\alpha_2\right)^2 - \left( 1 - 2x(1-x)(1 - z_s) \right) \right)^{-1} \\
			&= I_{\alpha_1,\alpha_2}(s,t) .
	\end{split}
\end{align}
\begin{align}
	\begin{split}
		I_{-\alpha_1,\alpha_2}&(s,t) = -\int d\Omega_s^\dprime \frac{1}{\alpha_1 + z_s^\prime} \frac{1}{\alpha_2 - z_s^\dprime} \\
			&= - \int d\Omega_s^\dprime \int_0^1 dx \left( x \alpha_1 + (1-x)\alpha_2 + \left( x z_s^\prime - (1-x) z_s^\dprime \right) \right)^{-2} \\
			&= - \int_0^1 dx \int d\Omega_s^\dprime \left( x \alpha_1 + (1-x)\alpha_2 + \left( x \frac{\vec q_1}{|\vec q_1|} + (1-x) \frac{\vec q_3}{|\vec q_3|} \right) \cdot \frac{\vec p_1}{|\vec p_1|} \right)^{-2} \\
			&= - 4\pi \int_0^1 dx \left( \left(x \alpha_1 + (1-x)\alpha_2\right)^2 - \left| x \frac{\vec q_1}{|\vec q_1|} + (1-x) \frac{\vec q_3}{|\vec q_3|} \right|^2 \right)^{-1} \\
			&= - 4\pi \int_0^1 dx \left( \left(x \alpha_1 + (1-x)\alpha_2\right)^2 - \left( x^2 + (1-x)^2 - 2 x(1-x) z_s \right) \right)^{-1} \\
			&= - 4\pi \int_0^1 dx \left( \left(x \alpha_1 + (1-x)\alpha_2\right)^2 - \left( 1 - 2x(1-x)(1+z_s) \right) \right)^{-1} \\
			&= - I_{\alpha_1,\alpha_2}(-z_s) \\
			&= - 4\pi \int_{z_s^+}^\infty dz \frac{1}{(z+z_s)\sqrt{(z - z_s^+)(z-z_s^-)}} .
	\end{split}
\end{align}
Using the alternative relations
\begin{align}
	\begin{split}
		z_s &= - \frac{2s}{\lambda^{1/2}_{12}(s) \lambda^{1/2}_{34}(s)} \left( u -  \frac{\Sigma - s}{2} -  \frac{\Delta_{12}\Delta_{34}}{2s} \right) , \\
		z &=: \frac{2s}{\lambda^{1/2}_{12}(s) \lambda^{1/2}_{34}(s)} \left( \tilde u -  \frac{\Sigma - s}{2} -  \frac{\Delta_{12}\Delta_{34}}{2s} \right) , \\
		z_s^\pm &=: \frac{2s}{\lambda^{1/2}_{12}(s) \lambda^{1/2}_{34}(s)} \left( u^\pm -  \frac{\Sigma - s}{2} -  \frac{\Delta_{12}\Delta_{34}}{2s} \right) ,
	\end{split}
\end{align}
we obtain
\begin{align}
	\begin{split}
		-I_{-\alpha_1,\alpha_2}(s,t) &= 2\pi \frac{\lambda_{12}^{1/2}(s) \lambda_{34}^{1/2}(s)}{s} \int_{u^+}^\infty d\tilde u \frac{1}{(\tilde u-u)\sqrt{(\tilde u - u^+)(\tilde u - u^-)}} .
	\end{split}
\end{align}
Finally, the last integral is
\begin{align}
	\begin{split}
		I_{\alpha_1,-\alpha_2}&(s,t) = -\int d\Omega_s^\dprime \frac{1}{\alpha_1 - z_s^\prime} \frac{1}{\alpha_2 + z_s^\dprime} \\
			&= - \int d\Omega_s^\dprime \int_0^1 dx \left( x \alpha_1 + (1-x)\alpha_2 - \left( x z_s^\prime - (1-x) z_s^\dprime \right) \right)^{-2} \\
			&= - \int_0^1 dx \int d\Omega_s^\dprime \left( x \alpha_1 + (1-x)\alpha_2 - \left( x \frac{\vec q_1}{|\vec q_1|} + (1-x) \frac{\vec q_3}{|\vec q_3|} \right) \cdot \frac{\vec p_1}{|\vec p_1|} \right)^{-2} \\
			&= - 4\pi \int_0^1 dx \left( \left(x \alpha_1 + (1-x)\alpha_2\right)^2 - \left| x \frac{\vec q_1}{|\vec q_1|} + (1-x) \frac{\vec q_3}{|\vec q_3|} \right|^2 \right)^{-1} \\
			&= I_{-\alpha_1,\alpha_2}(s,t) .
	\end{split}
\end{align}

In summary, we have transformed the phase-space integrals into dispersive integrals:
\begin{align}
	\begin{split}
		\int d\Omega_s^\dprime \frac{1}{a_1 - t^\prime} \frac{1}{a_2 - t^\dprime} &= \int d\Omega_s^\dprime \frac{1}{a_1 - u^\prime} \frac{1}{a_2 - u^\dprime} \\
			&= \frac{8\pi}{s \sigma_\pi^2(s)} \int_{t^+}^\infty d\tilde t \frac{1}{(\tilde t - t)\sqrt{\vphantom{\big|}(\tilde t - t^+)(\tilde t - t^-)}} , \\
		\int d\Omega_s^\dprime \frac{1}{a_1 - t^\prime} \frac{1}{a_2 - u^\dprime} &= \int d\Omega_s^\dprime \frac{1}{a_1 - u^\prime} \frac{1}{a_2 - t^\dprime} \\
			&= \frac{8\pi}{s \sigma_\pi^2(s)} \int_{u^+}^\infty d\tilde u \frac{1}{(\tilde u-u)\sqrt{(\tilde u - u^+)(\tilde u - u^-)}} .
	\end{split}
\end{align}


\chapter{Dispersive Representation of Loop Functions}

\label{sec:AppendixDispersiveLoops}

\section{Scalar Two-Point Function}

\begin{figure}[ht]
	\centering
	\psset{unit=0.75cm}
	\begin{pspicture*}(0,0)(10,10)
		\psellipticarc[arrows=->,arrowsize=5pt](4.5,4.5)(1.5,1){-90}{98}
		\psellipticarc[arrows=->,arrowsize=5pt](4.5,4.5)(1.5,1){90}{278}
		\psline[ArrowInside=->,arrowsize=5pt](1.0,4.5)(3.0,4.5)
		\psline[ArrowInside=->,arrowsize=5pt](8.0,4.5)(6.0,4.5)
		\rput(4.5,6){$m_1$}
		\rput(4.5,3){$m_2$}
		\rput(2,4){$p$}
		\rput(7.0,4.0){$-p$}
		\rput(4.5,5){$q$}
	\end{pspicture*}
	\caption{Bulb diagram}
	\label{img:BulbDiagram}
\end{figure}
The scalar two-point function \cite{Denner1993} is defined by
\begin{align}
	\begin{split}
		B_0(p^2,m_1^2,m_2^2) = \frac{1}{i} \int \frac{d^4q}{(2\pi)^4} \frac{1}{[q^2-m_1^2] [(q+p)^2-m_2^2] } .
	\end{split}
\end{align}
We define $s:=p^2$. The two-point function has a normal threshold at $s=(m_1+m_2)^2$. According to Cutkosky's rule \cite{Cutkosky1960}, the discontinuity is given by:
\begin{align}
	\begin{split}
		B_0&(s+i\epsilon) - B_0(s-i\epsilon) \\
			&= \frac{1}{i} \int  \frac{d^4q}{(2\pi)^4} \left(-2\pi i \delta(q^2 - m_1^2) \theta(-q^0)\right) \left( - 2\pi i \delta((q+p)^2-m_2^2) \theta(q^0 + p^0) \right) \\
			&= \frac{i}{8\pi^2} \int \frac{d^3q}{|q^0|} \delta((q+p)^2-m_2^2) \theta(q^0 + p^0) \\
			&= \frac{i}{8\pi^2} \int \frac{d^3q}{\sqrt{m_1^2+\vec q^2}} \, \delta\left(m_1^2 + s - m_2^2 - 2\sqrt{m_1^2 + \vec q^2}\sqrt{s}\right) \theta\left(\sqrt{s} - \sqrt{m_1^2 + \vec q^2}\right) \\
			&= \frac{i}{16\pi^2} \int \frac{d^3q}{|\vec q|\sqrt{s}} \delta\left(|\vec q| - \frac{\lambda^{1/2}(s,m_1^2,m_2^2)}{2\sqrt{s}}\right) \theta\left(\sqrt{s} - \sqrt{m_1^2 + \vec q^2}\right) \\
			&= \frac{i}{8\pi} \frac{\lambda^{1/2}(s,m_1^2,m_2^2)}{s} =: 2i \Delta_s B_0(s) .
	\end{split}
\end{align}
$B_0$ is divergent and satisfies a once-subtracted dispersion relation:
\begin{align}
	B_0(s) - B_0(0) = \frac{s}{\pi} \int_{(m_1+m_2)^2}^\infty ds^\prime \frac{\Delta_s B_0(s^\prime)}{(s^\prime - s - i \epsilon)s^\prime} .
\end{align}

\clearpage

\section{Scalar Three-Point Function}

\begin{figure}[ht]
	\centering
	\psset{unit=0.75cm}
	\begin{pspicture*}(0,0)(10,10)
		\pspolygon[ArrowInside=->,arrowsize=5pt,ArrowInsidePos=0.55](3,3)(5.6,4.5)(3,6)
		\psline[ArrowInside=->,arrowsize=5pt](7.0,4.5)(5.6,4.5)
		\psrotate(3.87,4.5){120}{\psline[ArrowInside=->,arrowsize=5pt](7.0,4.5)(5.6,4.5)}
		\psrotate(3.87,4.5){240}{\psline[ArrowInside=->,arrowsize=5pt](7.0,4.5)(5.6,4.5)}
		\rput(4.5,5.75){$m_1$}
		\rput(2.5,4.5){$m_2$}
		\rput(4.5,3.25){$m_3$}
		\rput(2,7.25){$p_1$}
		\rput(2,1.75){$p_2$}
		\rput(7.0,4.0){$p_3$}
		\rput(4,5){$q$}
	\end{pspicture*}
	\caption{Triangle diagram}
	\label{img:TriangleDiagram}
\end{figure}

Consider the scalar three-point function \cite{Denner1993}, defined by
\begin{align}
	\begin{split}
		C_0&(p_1^2,p_2^2,p_3^2,m_1^2,m_2^2,m_3^2) = \\
		& \frac{1}{i} \int \frac{d^4q}{(2\pi)^4} \frac{1}{[q^2-m_1^2] [(q+p_1)^2-m_2^2] [(q+p_1+p_2)^2-m_3^2]} .
	\end{split}
\end{align}
We define $s := (p_1+p_2)^2 = p_3^2$. The scalar three-point function has a normal threshold at $s=(m_1+m_3)^2$. The discontinuity along the corresponding branch cut can be found with Cutkosky's rules \cite{Cutkosky1960}:
\begin{align}
	\begin{split}
		C_0&(s+i\epsilon) - C_0(s-i\epsilon) \\
			&= \frac{1}{i} \int \frac{d^4q}{(2\pi)^4} \frac{1}{(q+p_1)^2-m_2^2} \\
			&\qquad\qquad \cdot \left( - 2\pi i \delta(q^2 - m_1^2) \theta(-q^0) \right) \left( - 2\pi i \delta((q-p_3)^2-m_3^2) \theta(q^0 - p_3^0) \right) \\
			&= \frac{i}{8\pi^2} \int \frac{d^3q}{|q^0|} \frac{1}{(q+p_1)^2-m_2^2} \delta((q-p_3)^2-m_3^2) \theta(q^0 - p_3^0) \\
			&= \frac{i}{8\pi^2} \int \frac{d^3q}{\sqrt{m_1^2+\vec q^2}} \frac{1}{(q+p_1)^2-m_2^2} \\
			&\qquad \qquad \cdot \delta\left(m_1^2 + s - m_3^2 - 2\sqrt{m_1^2 + \vec q^2}\sqrt{s}\right) \theta\left(\sqrt{s} - \sqrt{m_1^2 + \vec q^2}\right) \\
			&= \frac{i}{16\pi^2} \int \frac{d^3q}{|\vec q|\sqrt{s}} \frac{1}{(q+p_1)^2-m_2^2} \delta\left(|\vec q| - \frac{\lambda^{1/2}(s,m_1^2,m_3^2)}{2\sqrt{s}}\right) \theta\left(\sqrt{s} - \sqrt{m_1^2 + \vec q^2}\right) \\
			&= \frac{i}{32\pi^2} \frac{\lambda^{1/2}(s,m_1^2,m_3^2)}{s} \int d\Omega_s^\dprime \frac{1}{(q+p_1)^2-m_2^2} \\
			&=: 2i \Delta_s C_0(s) .
	\end{split}
\end{align}
We perform the phase space integral:
\begin{align}
	\begin{split}
			\Delta_s C_0(s) &= \frac{1}{64\pi^2} \frac{\tilde\lambda_{13}^{1/2}(s)}{s} \int d\Omega_s^\dprime \left( m_1^2 + p_1^2 - m_2^2 + 2q^0 p_1^0 - 2 |\vec q| |\vec p_1| \cos\theta_s^\dprime \right)^{-1} \\
			&= \frac{1}{32\pi} \frac{\tilde\lambda_{13}^{1/2}(s)}{s} \frac{1}{2|\vec q||\vec p_1|} \log\left( \frac{m_2^2 - m_1^2 - p_1^2 -2q^0 p_1^0 - 2 |\vec q||\vec p_1|}{m_2^2 - m_1^2 - p_1^2 -2q^0 p_1^0 + 2 |\vec q||\vec p_1|} \right) \\
			&= \frac{1}{16\pi} \frac{1}{\lambda_{12}^{1/2}(s)} \log\left( \frac{m_2^2 - m_1^2 - p_1^2 -2q^0 p_1^0 - 2 |\vec q||\vec p_1|}{m_2^2 - m_1^2 - p_1^2 -2q^0 p_1^0 + 2 |\vec q||\vec p_1|} \right) \\
			&= \frac{1}{16\pi} \frac{1}{\lambda_{12}^{1/2}(s)} \log\left( \frac{2s(\tilde\Delta_{12}+p_1^2) - (s+\tilde\Delta_{13})(s+\Delta_{12})+\tilde\lambda_{13}^{1/2}(s)\lambda_{12}^{1/2}(s)}{2s(\tilde\Delta_{12}+p_1^2) - (s+\tilde\Delta_{13})(s+\Delta_{12})-\tilde\lambda_{13}^{1/2}(s)\lambda_{12}^{1/2}(s)} \right) ,
	\end{split}
\end{align}
where
\begin{align}
	\begin{split}
		\Delta_{ik} &:= p_i^2 - p_k^2 , \\
		\tilde\Delta_{ik} &:= m_i^2-m_k^2 , \\
		\lambda_{ik}(s) &:= \lambda(s,p_i^2,p_k^2) , \\
		\tilde\lambda_{ik}(s) &:= \lambda(s,m_i^2,m_k^2) .
	\end{split}
\end{align}
$C_0$ satisfies an unsubtracted dispersion relation:
\begin{align}
	C_0(s) = \frac{1}{\pi} \int_{(m_1+m_3)^2}^\infty ds^\prime \frac{\Delta_s C_0(s^\prime)}{s^\prime - s - i \epsilon} .
\end{align}
The validity of this dispersive representation breaks down with the appearance of anomalous thresholds.

\clearpage

\section{Scalar Four-Point Function}

\label{sec:AppendixScalar4PointFunction}

\begin{figure}[ht]
	\centering
	\psset{unit=0.75cm}
	\begin{pspicture*}(0,0)(10,10)
		\pspolygon[ArrowInside=->,arrowsize=5pt,ArrowInsidePos=0.55](3,3)(6,3)(6,6)(3,6)
		\psline[ArrowInside=->,arrowsize=5pt](2,2)(3,3)
		\psrotate(4.5,4.5){90}{\psline[ArrowInside=->,arrowsize=5pt](2,2)(3,3)}
		\psrotate(4.5,4.5){180}{\psline[ArrowInside=->,arrowsize=5pt](2,2)(3,3)}
		\psrotate(4.5,4.5){270}{\psline[ArrowInside=->,arrowsize=5pt](2,2)(3,3)}
		\rput(4.5,6.5){$m_1$}
		\rput(2.5,4.5){$m_2$}
		\rput(4.5,2.5){$m_3$}
		\rput(6.5,4.5){$m_4$}
		\rput(1.75,7.25){$p_1$}
		\rput(1.75,1.75){$p_2$}
		\rput(7.25,1.75){$p_3$}
		\rput(7.25,7.25){$p_4$}
		\rput(4.5,5.5){$q$}
	\end{pspicture*}
	\caption{Box diagram}
	\label{img:BoxDiagram}
\end{figure}

Let us consider the scalar four-point function \cite{Denner1993}, corresponding to the diagram in figure~\ref{img:BoxDiagram}:
\begin{align}
	\begin{split}
		D_0&(p_1^2,p_2^2,p_3^2,p_4^2,(p_1+p_2)^2,(p_2+p_3)^2,m_1^2,m_2^2,m_3^2,m_4^2) = \\
		& \frac{1}{i} \int \frac{d^4q}{(2\pi)^4} \frac{1}{[q^2-m_1^2] [(q+p_1)^2-m_2^2] [(q+p_1+p_2)^2-m_3^2] [(q-p_4)^2-m_4^2]} .
	\end{split}
\end{align}
We define
\begin{align}
	\begin{split}
		s &:= (p_1 + p_2)^2 , \\
		t &:= (p_2 + p_3)^2 , \\
		u &:= (p_1+p_3)^2 , \\
		P &:= p_1 + p_2 = - p_3 - p_4 .
	\end{split}
\end{align}
In the $s$-channel, the diagram has a normal threshold at $s=(m_1+m_3)^2$, in the $t$-channel at $t=(m_2+m_4)^2$. For a fixed value of $t$, we find the discontinuity along the $s$-channel cut using Cutkosky's rules \cite{Cutkosky1960}:

\begin{align}
	\begin{split}
		D_0&(s+i\epsilon,t) - D_0(s-i\epsilon,t) \\
			&= \frac{1}{i} \int \frac{d^4q}{(2\pi)^4} \frac{1}{(q+p_1)^2-m_2^2} \frac{1}{(q-p_4)^2-m_4^2} \\
			&\qquad\qquad \cdot \left( -2\pi i \delta(q^2 - m_1^2) \theta(-q^0) \right) \left( - 2\pi i \delta((q+P)^2-m_3^2) \theta(q^0 + P^0) \right) \\
			&= \frac{i}{8\pi^2} \int \frac{d^3q}{|q^0|} \frac{1}{(q+p_1)^2-m_2^2} \frac{1}{(q-p_4)^2-m_4^2} \delta(m_1^2 + s + 2qP-m_3^2) \theta(q^0 + P^0) \\
			&= \frac{i}{8\pi^2} \int \frac{d^3q}{\sqrt{m_1^2+\vec q^2}} \frac{1}{(q+p_1)^2-m_2^2} \frac{1}{(q-p_4)^2-m_4^2} \\
			&\qquad\qquad \cdot \delta\left(m_1^2 + s - 2\sqrt{s}\sqrt{m_1^2+\vec q^2}-m_3^2\right) \theta\left(\sqrt{s} - \sqrt{m_1^2+\vec q^2}\right) \\
			&= \frac{i}{16\pi^2} \int \frac{d^3q}{|\vec q| \sqrt{s}} \frac{1}{(q+p_1)^2-m_2^2} \frac{1}{(q-p_4)^2-m_4^2} \\
			&\qquad\qquad \cdot \delta\left(|\vec q| - \frac{\lambda^{1/2}(s,m_1^2,m_3^2)}{2\sqrt{s}} \right) \theta\left(\sqrt{s} - \sqrt{m_1^2+\vec q^2}\right) \\
			&= \frac{i}{32\pi^2} \frac{\lambda^{1/2}(s,m_1^2,m_3^2)}{s} \int d\Omega_s^\dprime \frac{1}{(q+p_1)^2-m_2^2} \frac{1}{(q-p_4)^2-m_4^2} \\
			&=: 2i \Delta_s^t D_0(s,t).
	\end{split}
\end{align}

The phase-space integral is exactly of the type discussed in appendix~\ref{sec:AppendixPhaseSpaceIntegration}. Therefore, we can transform it into a dispersive integral:
\begin{align}
	\begin{split}
		\Delta_s^t D_0(s,t) &= \frac{1}{8\pi} \frac{1}{\tilde\lambda_{13}^{1/2}(s)} \int_{t^+(s)}^\infty dt^\prime \frac{1}{(t^\prime-t - i \epsilon)\sqrt{\big(t^\prime - t^+(s)\big)\big(t^\prime-t^-(s)\big)}} ,
	\end{split}
\end{align}
where
\begin{align}
	\begin{split}
		t^\pm(s) &:= \frac{\Sigma - s}{2} + \frac{\Delta_{12}\Delta_{34}}{2s} + \frac{\lambda_{12}^{1/2}(s)\lambda_{34}^{1/2}(s)}{2s} z_s^\pm , \\
		z_s^\pm &:= \alpha_1 \alpha_2 \pm \sqrt{\left(\alpha_1^2 - 1\right)\left(\alpha_2^2 - 1\right)} , \\
		\alpha_1 &:= \frac{s( 2 m_2^2 - \Sigma^\prime) + s^2 + \Delta_{12}\tilde\Delta_{13}}{\lambda_{12}^{1/2}(s)\tilde\lambda_{13}^{1/2}(s)} , \\
		\alpha_2 &:= \frac{s( 2 m_4^2 - \Sigma^\dprime) + s^2 - \Delta_{34}\tilde\Delta_{13}}{\lambda_{34}^{1/2}(s)\tilde\lambda_{13}^{1/2}(s)} .
	\end{split}
\end{align}
and
\begin{align}
	\begin{split}
		\Sigma &:= p_1^2 + p_2^2 + p_3^2 + p_4^2 , \\
		\Sigma^\prime &:= p_1^2 + p_2^2 + m_1^2 + m_3^2 , \\
		\Sigma^\dprime &:= p_3^2+p_4^2+m_1^2+m_3^2 , \\
		\Delta_{ik} &:= p_i^2 - p_k^2 , \\
		\tilde\Delta_{ik} &:= m_i^2-m_k^2 , \\
		\lambda_{ik}(s) &:= \lambda(s,p_i^2,p_k^2) , \\
		 \tilde\lambda_{ik}(s) &:= \lambda(s,m_i^2,m_k^2) .
	\end{split}
\end{align}
Since $D_0$ satisfies an unsubtracted dispersion relation, it can be written as
\begin{align}
	\begin{split}
		D_0(s,t) &= \frac{1}{\pi^2} \int_{(m_1+m_3)^2}^\infty ds^\prime  \int_{t^+(s^\prime)}^\infty dt^\prime \frac{\rho_{st}(s^\prime,t^\prime)}{(s^\prime - s - i \epsilon)(t^\prime-t - i \epsilon)} ,
	\end{split}
\end{align}
where
\begin{align}
	\rho_{st}(s,t) = \frac{1}{8 \tilde\lambda_{13}^{1/2}(s)} \frac{1}{\sqrt{(t - t^+(s))(t-t^-(s))}} .
\end{align}
The validity of the double-spectral representation of the box diagram breaks down with the appearance of anomalous thresholds.

If we do not take a fixed-$t$ dispersion relation as the starting point but a fixed-$u$ dispersion relation, we have to take into account that a line of fixed $u$ in the Mandelstam plane will encounter a right- as well as a left-hand cut:
\begin{align}
	\begin{split}
		D_0(s,t) &= \frac{1}{\pi} \int_{-\infty}^\infty ds^\prime \frac{\Delta_s^u D_0(s^\prime,\Sigma-u-s^\prime)}{s^\prime - s - i\epsilon} ,
	\end{split}
\end{align}
where
\begin{align}
	\begin{split}
		\Delta_s^u &D_0(s^\prime, \Sigma-u-s^\prime) \\
			&= \theta(s^\prime - (m_1+m_3)^2) \frac{D_0(s^\prime + i\epsilon, \Sigma-u-s^\prime)-D_0(s^\prime - i\epsilon, \Sigma-u-s^\prime)}{2i} \\
			&+ \theta(\Sigma-u-s^\prime - (m_2+m_4)^2) \frac{D_0(s^\prime, \Sigma-u-s^\prime - i\epsilon)-D_0(s^\prime, \Sigma-u-s^\prime + i\epsilon)}{2i},
	\end{split}
\end{align}
hence
\begin{align}
	\begin{split}
		D_0(s,t) &= \frac{1}{\pi} \int_{(m_1+m_3)^2}^\infty ds^\prime \frac{\Delta_s^u D_0(s^\prime,\Sigma-u-s^\prime)}{s^\prime - s - i\epsilon} \\
			&\quad +  \frac{1}{\pi} \int_{-\infty}^{\Sigma-u-(m_2+m_4)^2} ds^\prime \frac{\Delta_s^u D_0(s^\prime,\Sigma-u-s^\prime)}{s^\prime - s - i\epsilon} \\
			&= \frac{1}{\pi} \int_{(m_1+m_3)^2}^\infty ds^\prime \frac{\Delta_s^u D_0(s^\prime,\Sigma-u-s^\prime)}{s^\prime - s - i\epsilon} \\
			&\quad +  \frac{1}{\pi} \int_{(m_2+m_4)^2}^\infty dt^\prime \frac{\Delta_t^u D_0(\Sigma - u - t^\prime,t^\prime)}{t^\prime - t - i\epsilon} .
	\end{split}
\end{align}
Cutkosky's rules lead to the following discontinuities:
\begin{align}
	\begin{split}
		\Delta_s^u D_0(s,\Sigma - u - s) &= \frac{1}{8\pi} \frac{1}{\tilde\lambda_{13}^{1/2}(s)} \int_{t^+(s)}^\infty dt^\prime \frac{1}{(t^\prime - \Sigma + u  + s - i \epsilon)\sqrt{(t^\prime - t^+)(t^\prime-t^-)}} , \\
		\Delta_t^u D_0(\Sigma - u - t,t) &= \frac{1}{8\pi} \frac{1}{\tilde\lambda_{24}^{1/2}(t)} \int_{s^+(t)}^\infty ds^\prime \frac{1}{(s^\prime - \Sigma + u  + t - i \epsilon)\sqrt{(s^\prime - s^+)(s^\prime-s^-)}} ,
	\end{split}
\end{align}
where $s^\pm(t)$ are defined in analogy to $t^\pm(s)$ with the proper permutations of the momenta and masses.

Now, we note that
\begin{align}
	\begin{split}
		\int_{(m_2+m_4)^2}^\infty dt^\prime \int_{s^+(t^\prime)}^\infty ds^\prime = \int_{(m_1+m_3)^2}^\infty ds^\prime \int_{t^+(s^\prime)}^\infty dt^\prime
	\end{split}
\end{align}
and find
\begin{align}
	\begin{split}
		D_0(s,t) &= \frac{1}{\pi^2} \int_{(m_1+m_3)^2}^\infty ds^\prime  \int_{t^+(s^\prime)}^\infty dt^\prime X(s,t,s^\prime,t^\prime) ,
	\end{split}
\end{align}
where
\begin{align}
	\begin{split}
		X&(s,t,s^\prime, t^\prime) \\
			&= \frac{1}{8 \tilde \lambda_{13}^{1/2}(s^\prime)} \frac{1}{s^\prime - s - i\epsilon} \frac{1}{t^\prime - \Sigma + u + s^\prime - i \epsilon} \frac{1}{\sqrt{(t^\prime - t^+)(t^\prime - t^-)}} \\
			&\quad + \frac{1}{8 \tilde \lambda_{24}^{1/2}(t^\prime)} \frac{1}{s^\prime - \Sigma + u + t^\prime - i\epsilon} \frac{1}{t^\prime - t - i \epsilon} \frac{1}{\sqrt{(s^\prime - s^+)(s^\prime - s^-)}} \\
			&= \frac{1}{8 \tilde \lambda_{13}^{1/2}(s^\prime)} \frac{1}{s^\prime - s - i\epsilon} \frac{1}{t^\prime - t - i \epsilon} \frac{1}{\sqrt{(t^\prime - t^+)(t^\prime - t^-)}} \frac{1}{s^\prime - s + t^\prime - t - i\epsilon} \\
			&\quad  \cdot \left( t^\prime - t + (s^\prime - s) \frac{\tilde \lambda_{13}^{1/2}(s^\prime)}{\tilde \lambda_{24}^{1/2}(t^\prime)} \sqrt{\frac{(t^\prime - t^+(s^\prime))(t^\prime - t^-(s^\prime))}{(s^\prime - s^+(t^\prime))(s^\prime - s^-(t^\prime))}} \right) .
	\end{split}
\end{align}
With some (computer) algebra, we can check that
\begin{align}
	\begin{split}
		\frac{\tilde \lambda_{13}^{1/2}(s^\prime)}{\tilde \lambda_{24}^{1/2}(t^\prime)} \sqrt{\frac{(t^\prime - t^+(s^\prime))(t^\prime - t^-(s^\prime))}{(s^\prime - s^+(t^\prime))(s^\prime - s^-(t^\prime))}} = 1 ,
	\end{split}
\end{align}
such that we find again
\begin{align}
	X(s,t,s^\prime,t^\prime) =  \frac{\rho_{st}(s^\prime,t^\prime)}{(s^\prime - s - i \epsilon)(t^\prime-t - i \epsilon)} .
\end{align}

\end{appendices}

\clearpage
\renewcommand\bibname{References}
\renewcommand{\bibfont}{\raggedright}
\bibliographystyle{my-physrev}
\phantomsection
\addcontentsline{toc}{chapter}{References}
\bibliography{HLbL/Literature}

\cleardoublepage


\chapter*{Conclusion}
\addcontentsline{toc}{part}{Conclusion}
\addtocontents{toc}{\vskip-6pt\par\noindent\protect\hrulefill\par}
\markboth{CONCLUSION}{CONCLUSION}

In this thesis, I have presented three projects. The first two are closely connected and concern $K_{\ell4}$ decays. The motivation to study this particular decay channel of kaons is that it happens at very low energies and that it is the best source of information on some of the low-energy constants of chiral perturbation theory. With our dispersive framework, we have performed a resummation of final-state $\pi\pi$- and $K\pi$-rescattering effects. This has been achieved by iterating numerically the system of integral equations until convergence. As a result, the dispersive description of the form factors reproduces the measured shapes much better than a pure chiral treatment. By matching the dispersion relation to \ChPT{}, we have extracted values for the low-energy constants $L_1^r$, $L_2^r$ and $L_3^r$. The matching at two-loop level still shows some instability due to the dependence on the very poorly known NNLO LECs. It might be possible that with some additional constraints on the convergence of the chiral expansion the two-loop matching can be stabilised.

The second part of the thesis concerns the isospin-breaking corrections for the $K_{\ell4}$ process due to both electromagnetic effects and mass differences. The results of this project have been used for a correction of the experimental data before the fit within our dispersive treatment.

The third part of the thesis is devoted to a dispersive treatment of hadronic light-by-light scattering. The procedure how to construct scalar functions free of kinematics for arbitrary photon processes has been known for a long time. However, we have applied it probably for the first time to a four-photon process. We have constructed a generating set of Lorentz structures that contain all the kinematics including the constraints from gauge invariance. The corresponding scalar functions are then free of kinematic zeros and singularities and well-suited for a dispersive description. Based on Mandelstam's double-spectral representation, we have derived a dispersive framework that splits the HLbL contribution to the muon anomalous magnetic moment into three parts: pion-pole contributions, the contributions of box topologies and $\pi\pi$-rescattering contributions. Our new dispersive approach to HLbL scattering shows an avenue towards a data-driven and less model-dependent evaluation of the $(g-2)_\mu$.

In both of the presented subjects, some more work can be done and is actually in progress. In $K_{\ell4}$, we will try to stabilise the two-loop matching. Furthermore, we will include the two-dimensional data set that we were provided with only very recently.

Concerning HLbL, a dispersive description of the doubly-virtual sub-process $\gamma^*\gamma^*\to\pi\pi$ is work in progress. Together with a dispersive analysis of the pion transition form factor, this constitutes a crucial input for the numerical evaluation of our dispersion relation. A thorough analysis of the influence of available experimental data and the identification of processes that help to reduce the uncertainties is a major task, which we will tackle in the future.

Dispersion relations have proven very useful in describing hadronic processes at low and intermediate energies, which is crucial not only for a better understanding of the non-perturbative regime of QCD but also in view of hadronic uncertainties that obscure the search for new physics. Some of the techniques presented here could be applied also in other fields where hadronic states play an important role: e.g.~in hadronic $\tau$ decays, where they could provide a more precise description needed for the investigation of $CP$-violating effects.

While in the sixties, dispersion relations were thought to be the key ingredient for constructing a fundamental theory of the strong interaction, the techniques of $S$-matrix theory were certainly too quickly abandoned during the seventies and eighties. Since the nineties, their revival in a more phenomenological context has led to a rich program of applications. It can be expected that they will continue to play an important role, probably with focus on controlling hadronic uncertainties in the search for new physics.

\cleardoublepage

\pagestyle{empty}

\mbox{}

\cleardoublepage

\pagestyle{empty}


\begin{center}
	{ \LARGE \bf Erklärung }
	
	\large
	
	\vspace{0.5cm}
	
	gemäss Art.~28 Abs.~2 RSL 05
	
	\vspace{1.5cm}
	
	{
	\setlength{\extrarowheight}{0.5cm}
	\begin{tabularx}{\textwidth}{p{4cm} p{10cm}}
		Name, Vorname: & Stoffer, Peter \\
		Matrikelnummer: & 05-126-529 \\
		Studiengang: & Physik \\
			& Bachelor $\Box$ \hspace{1cm} Master $\Box$ \hspace{1cm} Dissertation $\boxtimes$ \\
		Titel der Arbeit: & \raggedright \disstitle \tabularnewline
		Leiter der Arbeit: & Prof.~Dr.~Gilberto Colangelo \\
		\\
		\multicolumn{2}{p{\textwidth}}{Ich erkläre hiermit, dass ich diese Arbeit selbständig verfasst und keine anderen als die angegebenen Quellen benutzt habe. Alle Stellen, die wörtlich oder sinngemäss aus Quellen entnommen wurden, habe ich als solche gekennzeichnet. Mir ist bekannt, dass andernfalls der Senat gemäss Artikel 36 Absatz 1 Buchstabe r des Gesetztes vom 5.~September 1996 über die Universität zum Entzug des auf Grund dieser Arbeit verliehenen Titels berechtigt ist.} \\
		\multicolumn{2}{p{\textwidth}}{Ich gewähre hiermit Einsicht in diese Arbeit.} \\
		\\
		Bern, den 06.09.2014 \\
		\\
			& \hspace{4cm} Peter Stoffer
	\end{tabularx}
	}
	
\end{center}

\cleardoublepage


\section*{Curriculum vit\ae}

\subsection*{Personal Data}

{
\renewcommand{\arraystretch}{1.5}
\setlength{\tabcolsep}{0cm}

	\begin{tabularx}{\textwidth}{p{4cm}X}
		Name & \bf Stoffer, Peter \\
		Date of birth & 28th July 1986 \\
		Place of origin & Mägenwil AG \\
		Nationality & Swiss \\
		Marital status & Single
	\end{tabularx}

	\subsection*{Education}
	
	\begin{tabularx}{\textwidth}{p{4cm}X}
		2010 -- 2014 & University of Bern \newline
			Ph.D.~student of Prof.~Dr.~G.~Colangelo \\
		2013 -- 2014 & Los Alamos National Laboratory \newline
			Six month's stay with a mobility grant of the \newline
			Swiss National Science Foundation \newline
			supervised by Dr.~Vincenzo Cirigliano \\
		2009 -- 2012 & Pädagogische Hochschule PHBern \newline
			Teaching certificate for Maturitätsschulen \newline
			in physics and mathematics \\
		Oct.~2010 & University of Bern \newline
			Master of Science in Physics: \newline
			{\it A Dispersive Treatment of $K_{\ell4}$ Decays} \newline
			supervised by Prof.~Dr.~G.~Colangelo \\
		Jul.~2008 & University of Bern \newline
			Bachelor of Science in Physics: \newline
			{\it Feynman-Integrale} \newline
			supervised by Prof.~Dr.~J.~Gasser \\
		2005 -- 2010 & University of Bern \newline
			Studies in physics, mathematics and astronomy \\
		2001 -- 2005 & Gymnasium in Burgdorf \\
		1993 -- 2001 & Primary and secondary school in Biglen
	\end{tabularx}

}

\end{document}